\documentclass[12pt, a4paper, twoside]{report}
\usepackage[pdftex,
        pdfauthor={Chun Tian},
        pdftitle={A Formalization of Unique Solutions of Equations in Process Algebra},
        pdfstartpage=1,
        pdfstartview=FitH,
        bookmarksopen=true]{hyperref}

\usepackage[a4paper,left=3.5cm,right=2.5cm,top=3.5cm,bottom=3.5cm, twoside]{geometry}

\usepackage[english]{babel} %italian

\usepackage{newlfont}
\usepackage[T1]{fontenc}

\usepackage{latexsym}
\usepackage{sansmath}
\usepackage{graphicx}
\usepackage{pdfpages}
\usepackage{bussproofs}
\usepackage{amsmath}
\usepackage{amssymb}
\usepackage{mathrsfs}
\usepackage{multicol}
\usepackage{fancyvrb}

\usepackage{bookmark}
\usepackage{listings}
\usepackage{holindex}
\usepackage{alltt}
\newcommand{\HOLTokenStrongEQ}{$\sim$}
\newcommand{\HOLTokenWeakEQ}{$\approx$}
\newcommand{\HOLTokenObsCongr}{$\approx^c$}
\newcommand{\HOLTokenEPS}{$\overset{\epsilon}{\Rightarrow}$}
\newcommand{\HOLTokenTransBegin}{$-$}
\newcommand{\HOLTokenTransEnd}{$\rightarrow$}
\newcommand{\HOLTokenWeakTransBegin}{$=$}
\newcommand{\HOLTokenWeakTransEnd}{$\Rightarrow$}
\newcommand{\HOLTokenExpands}{$\succeq_e$}
\newcommand{\HOLTokenContracts}{$\succeq_{bis}$}
\newcommand{\HOLTokenObsContracts}{$\succeq^c_{bis}$}
\renewcommand{\HOLTokenImp}{\ensuremath{\Longrightarrow}}

\usepackage{environ}
\NewEnviron{HOLTrans}{\overset{\BODY}{\longrightarrow}}
\NewEnviron{HOLWeakTrans}{\overset{\BODY}{\Longrightarrow}}

\usepackage{amsthm}
\theoremstyle{definition}
\newtheorem{definition}{Definition}[section]
\theoremstyle{proposition}
\newtheorem{proposition}{Proposition}[section]
\theoremstyle{theorem}
\newtheorem{theorem}{Theorem}[section]
\theoremstyle{lemma}
\newtheorem{lemma}{Lemma}[section]
\theoremstyle{remark}
\newtheorem*{remark}{Remark}

\usepackage[all,cmtip]{xy}

\makeindex
\begin{document}

\includepdf{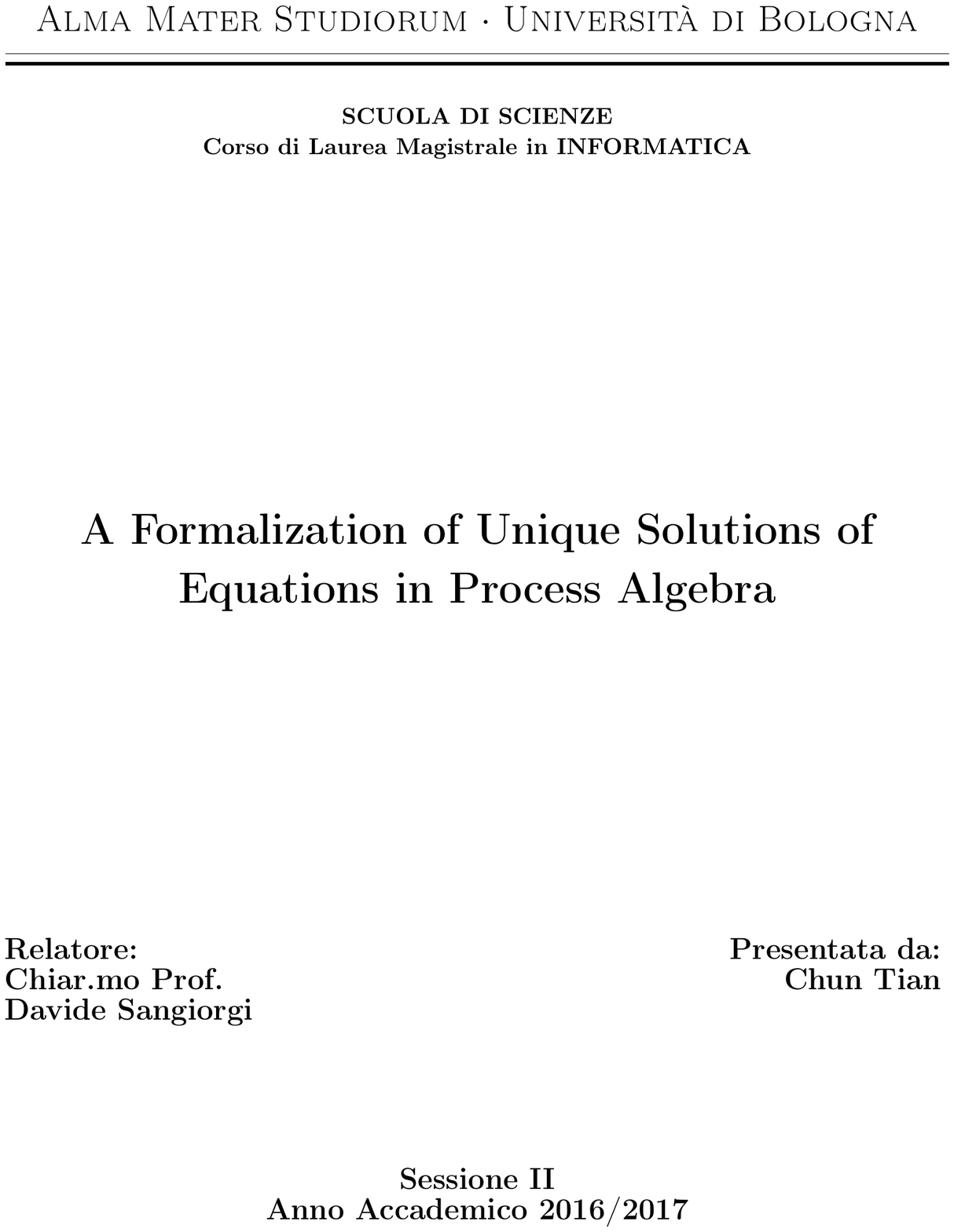}
%\blankpage{}
%\cleardoublepage

\chapter*{Abstract}
In this thesis, a comprehensive formalization of Milner's \emph{Calculus of
Communicating Systems} (also known as CCS) has been done in HOL
theorem prover (HOL4), based on an old work in HOL88.
This includes all classical properties of strong/weak
bisimulation equivalences and observation congruence, a theory of
congruence for CCS, various
versions of ``bisimulation up to'' techniques, and several deep theorems, namely
the ``coarsest congruence contained in $\approx$'', and
three versions of the ``unique solution of equations'' theorem in
Milner's book.

This work is further extended to support recent developments in
Concurrency Theory, namely the ``contraction'' relation and the
related ``unique solutions of contractions'' theorem found by 
Prof.\ Davide Sangiorgi, University of Bologna. As a result, a rather
complete theory of ``contraction'' (and a similar relation called ``expansion'') for CCS is
also formalized in this thesis.  Further more, a new variant of contraction called
``observational contraction'' was found by the author during this work,
based on existing contraction relation. It's formally proved that, this
new relation is preserved by direct sums of CCS processes, and has
a more elegant form of the ``unique solutions of contractions''
theorem without any restriction on the CCS grammar.

The contribution of this thesis project is at least threefold:
First, it can be seen as
a formal verification of the core results in Prof.\ Sangiorgi's paper,
and it provides all details for the informal proof sketches given in the
paper. Second, a large piece of old proof scripts from the time
of Hol88 (1990s) has been ported to HOL4 and made available to all its
users. Third, it's a proof engineering research by itself on the correct
formalization of process algebra, because the work has made extensive uses
of some new features 
(e.g. coinductive relation) provided in recent versions of HOL4 (Kananaskis-11 and
later).
The uses of HOL4's rich theory libraries is also a highlight of this project, and it has
successfully minimized the development efforts in this work.  As a result, 
this project serves as a sample application of HOL4, which seems
a natural choice for the formalization of process algebra.

For the
author himself, after this thesis project he has become a skilled user
of interactive theorem proving techniques, fully prepared at the level
of formal methods in his future research activities in Computer Science.

\chapter*{Acknowledgments}

The author want to give special thanks to Prof.\ \emph{Davide Sangiorgi} for his supervision of this
thesis and the initial proposal of the thesis topic on formalizing his recent
research results beside the classical ones. Without Prof.\ Sangiorgi's
kind approval and guidances the author wouldn't be able to graduate by
doing some formalizations using his favorite software (HOL4).

Thanks to Prof.\ \emph{Monica Nesi} (University of L'Aquila) for finding and sending her old HOL88
proof scripts on the old CCS formalization (during 1992-1995) to the
author in 2016 and 2017. By studying these proof scripts, the author has actually learnt how to
use HOL theorem prover, and the existing work has provided a
good working basis for quickly reaching to the formalization of deep
theorems
in the frontier of Concurrency Theory.

Thanks to Prof.\ \emph{Roberto Gorrieri}, who has taught his \emph{Concurrent
  System and Models} course twice to the author (in Spring 2016 and
2017), It's all from these courses that the author has learnt Concurrency Theory
and CCS. Prof.\ Gorrieri has
also supervised the authors' previous two projects (one for exam, the
other for internship) on the porting of the old CCS formalization of
Monica Nesi from HOL88 to HOL4. These work now
serves as the working basis of this thesis project.

Thanks to Prof.\ \emph{Andrea Asperti}, who has taught the general idea of
Interactive Theorem Proving (ITP)
techniques to the author through several seminars. (Prof.\ Asperti used
to teach formal methods but since 2015 he has changed to teach Machine
Learning under the same course title) Although it's
based on a different theorem
prover (Matita, originally created by himself), the author wouldn't be able to learn HOL4 without
necessary background knowledges learnt from Prof.\ Asperti, who has
also approved the authors' exam project on a formalization of Lambek
Calculus (mostly ported from Coq).

Also thanks to many people from HOL community (Michael Norrish, Thomas Tuerk, 
Ramana Kumar, Konrad Slind, etc.) for their kind help on resolving issues and doubts that the
author has met during this thesis and previous related projects. All
these people are professors and senior researchers in other
universities. Without their help this thesis project won't be finished
in reasonable time. 

At last, thanks to Prof.\ \emph{Mike J. Gordon} (University of
Cambridge, the creator of HOL Theorem
Prover), who has unfortunately passed away in the evening of August
22, 2017. He ever kindly encouraged the author's work in several
private emails, and the author initially learnt the existence of the
CCS formalization on HOL, from an introduction paper
\cite{Gordon:2000vt}
written by Prof.\ Gordon.

The author also wants to thank his parents and all
his friends in China, Italy and other countries, for
all their  supports in the past years. The author won't be 
able to start and focus on his degree study (and finally finish it) without
their supports.

The paper is written in \LaTeX using LNCS template, with theorems
generated automatically by HOL's \TeX
exporting module (\texttt{EmitTex}) from the working proof
scripts.\footnote{The author has also submited many bug reports and small Pull
Requests to help improving HOL on this part. But still there're known
issues which are not reported yet.}

\cleardoublepage

\tableofcontents
\cleardoublepage
\HOLpagestyle

%%%% -*- Mode: LaTeX -*-

\chapter*{Two-page Summary in Italian}

In questo progetto di tesi di laurea, viene proposta una formalizzazione quasi
completa del \emph{Calculus of
Communicating Systems} (\textbf{CCS}) di Robin Milner realizzata in
HOL theorem prover (HOL4).
L'implementazione proposta si basa sul riutilizzo di una formalizzazione gi\`a
esistente realizzata in HOL88 da Monica Nesi. Tale formalizzazione \`e stata quindi
riadattata per essere riutilizzata in HOL4 per l'aggiunta
di nuove funzionalit\`a. 

La formalizzazione include propriet\`a classiche
della bisimulazione, sia forte che debole, e la congruenza fondata
sull'osservazione. Viene inoltre proposta una teoria di congruenza per CCS.
 Sono incluse inoltre varie tecniche di ``bisimulazione fino
a'' (bisimulation up to) e numerosi teoremi profondi, come ad esempio:
\begin{enumerate}
\item Il lemma di Hennessy:
\begin{alltt}
\HOLTokenTurnstile{} \HOLFreeVar{p} \HOLSymConst{\HOLTokenWeakEQ} \HOLFreeVar{q} \HOLSymConst{\HOLTokenEquiv{}} \HOLFreeVar{p} \HOLSymConst{\HOLTokenObsCongr} \HOLFreeVar{q} \HOLSymConst{\HOLTokenDisj{}} \HOLFreeVar{p} \HOLSymConst{\HOLTokenObsCongr} \HOLSymConst{\ensuremath{\tau}}\HOLSymConst{..}\HOLFreeVar{q} \HOLSymConst{\HOLTokenDisj{}} \HOLSymConst{\ensuremath{\tau}}\HOLSymConst{..}\HOLFreeVar{p} \HOLSymConst{\HOLTokenObsCongr} \HOLFreeVar{q}
\end{alltt}
\item Il lemma di Deng:
\begin{alltt}
\HOLTokenTurnstile{} \HOLFreeVar{p} \HOLSymConst{\HOLTokenWeakEQ} \HOLFreeVar{q} \HOLSymConst{\HOLTokenImp{}}
   (\HOLSymConst{\HOLTokenExists{}}\HOLBoundVar{p\sp{\prime}}. \HOLFreeVar{p} \HOLTokenTransBegin\HOLSymConst{\ensuremath{\tau}}\HOLTokenTransEnd \HOLBoundVar{p\sp{\prime}} \HOLSymConst{\HOLTokenConj{}} \HOLBoundVar{p\sp{\prime}} \HOLSymConst{\HOLTokenWeakEQ} \HOLFreeVar{q}) \HOLSymConst{\HOLTokenDisj{}} (\HOLSymConst{\HOLTokenExists{}}\HOLBoundVar{q\sp{\prime}}. \HOLFreeVar{q} \HOLTokenTransBegin\HOLSymConst{\ensuremath{\tau}}\HOLTokenTransEnd \HOLBoundVar{q\sp{\prime}} \HOLSymConst{\HOLTokenConj{}} \HOLFreeVar{p} \HOLSymConst{\HOLTokenWeakEQ} \HOLBoundVar{q\sp{\prime}}) \HOLSymConst{\HOLTokenDisj{}}
   \HOLFreeVar{p} \HOLSymConst{\HOLTokenObsCongr} \HOLFreeVar{q}
\end{alltt}
\item Il teorema ``coarsest congruence contained in $\approx$'':
\begin{alltt}
\HOLTokenTurnstile{} \HOLConst{finite_state} \HOLFreeVar{p} \HOLSymConst{\HOLTokenConj{}} \HOLConst{finite_state} \HOLFreeVar{q} \HOLSymConst{\HOLTokenImp{}}
   (\HOLFreeVar{p} \HOLSymConst{\HOLTokenObsCongr} \HOLFreeVar{q} \HOLSymConst{\HOLTokenEquiv{}} \HOLSymConst{\HOLTokenForall{}}\HOLBoundVar{r}. \HOLFreeVar{p} \HOLSymConst{+} \HOLBoundVar{r} \HOLSymConst{\HOLTokenWeakEQ} \HOLFreeVar{q} \HOLSymConst{+} \HOLBoundVar{r})
\end{alltt}
\item Tre versioni del teorema della ``unica soluzione degli equazioni'' nel libro
di Milner \cite{Milner:1989}:
\begin{alltt}
\HOLTokenTurnstile{} \HOLConst{WG} \HOLFreeVar{E} \HOLSymConst{\HOLTokenImp{}} \HOLSymConst{\HOLTokenForall{}}\HOLBoundVar{P} \HOLBoundVar{Q}. \HOLBoundVar{P} \HOLSymConst{\HOLTokenStrongEQ} \HOLFreeVar{E} \HOLBoundVar{P} \HOLSymConst{\HOLTokenConj{}} \HOLBoundVar{Q} \HOLSymConst{\HOLTokenStrongEQ} \HOLFreeVar{E} \HOLBoundVar{Q} \HOLSymConst{\HOLTokenImp{}} \HOLBoundVar{P} \HOLSymConst{\HOLTokenStrongEQ} \HOLBoundVar{Q}
\HOLTokenTurnstile{} \HOLConst{SG} \HOLFreeVar{E} \HOLSymConst{\HOLTokenConj{}} \HOLConst{GSEQ} \HOLFreeVar{E} \HOLSymConst{\HOLTokenImp{}} \HOLSymConst{\HOLTokenForall{}}\HOLBoundVar{P} \HOLBoundVar{Q}. \HOLBoundVar{P} \HOLSymConst{\HOLTokenWeakEQ} \HOLFreeVar{E} \HOLBoundVar{P} \HOLSymConst{\HOLTokenConj{}} \HOLBoundVar{Q} \HOLSymConst{\HOLTokenWeakEQ} \HOLFreeVar{E} \HOLBoundVar{Q} \HOLSymConst{\HOLTokenImp{}} \HOLBoundVar{P} \HOLSymConst{\HOLTokenWeakEQ} \HOLBoundVar{Q}
\HOLTokenTurnstile{} \HOLConst{SG} \HOLFreeVar{E} \HOLSymConst{\HOLTokenConj{}} \HOLConst{SEQ} \HOLFreeVar{E} \HOLSymConst{\HOLTokenImp{}} \HOLSymConst{\HOLTokenForall{}}\HOLBoundVar{P} \HOLBoundVar{Q}. \HOLBoundVar{P} \HOLSymConst{\HOLTokenObsCongr} \HOLFreeVar{E} \HOLBoundVar{P} \HOLSymConst{\HOLTokenConj{}} \HOLBoundVar{Q} \HOLSymConst{\HOLTokenObsCongr} \HOLFreeVar{E} \HOLBoundVar{Q} \HOLSymConst{\HOLTokenImp{}} \HOLBoundVar{P} \HOLSymConst{\HOLTokenObsCongr} \HOLBoundVar{Q}
\end{alltt}
\end{enumerate}

Successivamente abbiamo esteso questo lavoro a sostegno dello sviluppo della
recente Teoria di Concorrente, ossia della relazione
\emph{contrazione} ($\succeq_{\mathrm{bis}}$) e del suo teorema \emph{``l'unica soluzione degli
equazioni''},  proposta dal Professore Davide Sangiorgi dell'Universit`a
di Bologna:
\begin{alltt}
\HOLTokenTurnstile{} \HOLConst{WGS} \HOLFreeVar{E} \HOLSymConst{\HOLTokenImp{}} \HOLSymConst{\HOLTokenForall{}}\HOLBoundVar{P} \HOLBoundVar{Q}. \HOLBoundVar{P} \HOLSymConst{\HOLTokenContracts{}} \HOLFreeVar{E} \HOLBoundVar{P} \HOLSymConst{\HOLTokenConj{}} \HOLBoundVar{Q} \HOLSymConst{\HOLTokenContracts{}} \HOLFreeVar{E} \HOLBoundVar{Q} \HOLSymConst{\HOLTokenImp{}} \HOLBoundVar{P} \HOLSymConst{\HOLTokenWeakEQ} \HOLBoundVar{Q}
\end{alltt}

Durante la dimostrazione della teorema ``l'unica soluzione degli
equazioni'', una teoria completa della relazione
``contrazione'' (e un'altra relazione ``expansione'') per CCS \`e stata
sviluppata nel progetto. Inoltre, una nuova versione
della relazione ``contrazione''  \`e stata proposta dall'autore.
La nuova
relazione, basata sula relazione ``contrazione'' di Sangiorgi, si chiama
``la contrazione fondato sull'osservazione'' (\emph{observational
  contraction}, $\succeq_{\mathrm{bis}}^c$).  Questa nuova relazione
ha le propriet\`a migliore rispetto alla contrazione normale, infatti il suo teorema ``l'unica soluzione
dei equazioni'' \`e in grado di accettare  processi composti da somme dirette:
\begin{alltt}
\HOLTokenTurnstile{} \HOLConst{WG} \HOLFreeVar{E} \HOLSymConst{\HOLTokenImp{}} \HOLSymConst{\HOLTokenForall{}}\HOLBoundVar{P} \HOLBoundVar{Q}. \HOLBoundVar{P} \HOLSymConst{\HOLTokenObsContracts} \HOLFreeVar{E} \HOLBoundVar{P} \HOLSymConst{\HOLTokenConj{}} \HOLBoundVar{Q} \HOLSymConst{\HOLTokenObsContracts} \HOLFreeVar{E} \HOLBoundVar{Q} \HOLSymConst{\HOLTokenImp{}} \HOLBoundVar{P} \HOLSymConst{\HOLTokenWeakEQ} \HOLBoundVar{Q}
\end{alltt}
Cos\`i non ci sar\`a bisogno di limitare la grammatica CCS con solo le
somme guardinghe, com'\`e trovato nell'articolo di Sangiorgi e gli altri.

La contribuzione di questa tesi di laurea \`e al meno triplo:
\begin{enumerate}
\item Il lavoro ottimo del professor Sangiorgi viene verificato parzialmente
  adesso. Coloro che vogliono sapere tutti i dettagli delle
  dimostrazioni nell'articolo \cite{sangiorgi2015equations} possono
  leggere la nostra formalizione, se si capisca la logica di higher
  order (HOL).
\item Una parte grande di codici vecchi da HOL88 (1990s) (scritto dal
  prof.ssa Monica Nesi) viene
trovato, salvato e
  portato nell'ultima versione di HOL theorem prover. Ora questi
  codici vivano nel deposito codice di HOL officiale (cos\`i \`e
  disponibile ai utenti tutti). Non puo perdersi
  di pi\`u.
\item \`E una ricerca sul modo meglio della formalizzare di
  \emph{process algebra}, e il tipico oggeto (CCS). Abbiamo usato
  tantissimo \emph{new feature} nell'HOL4 per minimizzare il lavoro
  della formalizzazione. Sono inclusi la capacit\`a per definire la
  relazione co-induttiva, e le teorie esistente (\texttt{relation},
  \texttt{list}, etc.) nell'HOL4. Molti codici vecchi viene eliminati,
  mentre tutti i teoremi sono ancora l\`i.
\end{enumerate}

Per l'autore stesso, questo progetto \`e da maggior parte una pratica
per migliorare le sue capacit\`a di usare il theorem prover e il
metodo formale per risolvere i problemi in informatica.

\cleardoublepage

\chapter{Introduction}

In Concurrency Theory \cite{Gorrieri:2015jt}, the initial research
motivation was the modeling of concurrent and reactive systems.
Such systems can usually be abstracted into finite or infinite number
of \emph{states},
together with labelled \emph{transitions} between pairs of states. From a
mathematical view, these states and transitions form a
\emph{edge-labeled directed graph}, in which each edge represents a
possible transition of states in the system. As a model, we gives such
graphs a special name: \emph{Labeled Transition Systems} (LTS), which is
originally introduced by R. Keller \cite{Keller:1976kl}.

The mainline research of Concurrency Theory concerns with a particular class of mathematical
models for concurrent communicating processes, namely the class of
\emph{algebraic calculi} (process algebra). There're different approaches to the algebraic
treatment of processes, the main distinction among them are the
constructions by which processes are assembled.  Milner's CCS is
the simplest treatment with least number of such constructions, while
it's still very powerful, in the sense that it's Turing-complete, although
Turing-completeness is not enough to ensure the solvability of all the
problems in Concurrency Theory. \cite{Gorrieri:2015jt}

On the relationship between CCS and LTS, there seems to be two
approaches: we can either treat CCS as a
compact, algebraic representation of LTS, or treat LTS as the
sequential (or interleaving)
semantic model of CCS\footnote{According to Gorrieri's book \cite{gorrieriprocess}, each
  process algebra has at least three different semantics: sequential semantic (LTS), parallel
  semantics (Step Transition System, STS) and distributed semantics
(Petri nets).}. We think the former approach is not quite fair,
because LTS itself has no computational power at all, while CCS can do
computations due to its ability of synchronizations, even though this
power has been limited between pairs of processes.

In this thesis, we have taken the following approach: we put CCS at the
central position and define all equivalence concepts on CCS only,
while LTS
itself doesn't appear explicitly in the formalization. As the result,
it's impossible to formalize the theorems involving both CCS and LTS,
e.g. the \emph{Representability Theorem}. But it's should be possible
to extend the current work with LTS included as a dedicated type. \footnote{The author has
  recently found that, it's actually possible and meaningful to embed
  LTS into CCS as a special constructor, without hurting almost all
  existing theorems. However he has finaly decided to
  not include this work into the thesis, because such additions may
  decrease the purity of CCS formalization, making people think that
  certain theorems may not be provable without having LTS ranged over
  process variables.}

Although titled as ``A Formalization of Unique Solutions of Equations
in Process Algebra'', this thesis project actually contains a quite
complete
 formalization of Milner's Calculus of
Communicating Systems (CCS). The precise CCS
class is Finitary and Pure CCS (i.e. no Value-Passing) with explicit relabeling
operator. The work has covered almost everything in Milner's classical textbook
\cite{Milner:1989}, and we have also formalized some recent developments in Concurrency Theory, namely the ``contraction'' relation and its ``unique
solution of contractions'' theorem, introduced in the paper \cite{sangiorgi2015equations} of
Prof.\ Davide Sangiorgi.

Nowadays many researchers in Concurrency Theory have turned to other more powerful process
algebras, e.g. $\pi$-calculus and $\psi$-calculus.  Even in scope of
CCS, the research interests were shifted to CCS extensions like
Multi-CCS and Value-Passing CCS. However, the original theory of pure CCS is easy
to understand, thus it's still the main content of many Concurrency Theory
courses. On the other side, many more powerful process algebras were
based on CCS, thus understanding CCS should be a good step for
students who want to study more powerful process algebras in the future.

But even for pure CCS there're still something quite deep when doing
its formalization.
For example, the use of ``Process constants'' in CCS
didn't appears in tother powerful process algebras like $\pi$-calculus
\cite{sangiorgi2003pi}. It's actually a challenging work to correctly formalize the ``process constants'' in CCS.
Many authors has completely ignored process
constants, while they still
claim to successfully formalized CCS. This is the case for the CCS
formalization we found using Coq and Isabelle/HOL
\cite{bengtson2010formalising}.
 What they have formalized is only a sub-language of CCS, in which all theorems for equivalence relations still hold.

The CCS formalization of Monica Nesi \cite{Nesi:1992ve} has included process constants.
In fact, this work has some advanced features beyond theorem proving
in the usual sense: besides having proved many theorems, Monica Nesi
has also implemented two decision procedures as ML functions: one for generating
transitions from any CCS process, the other for automatically checking
the equivalences between two processes. Different with similar
functionalities in dedicated software, the results generated by
theorem provers are trusted, because the outputs are also theorems.
This has almost touched the area of CCS-based model checking with
self-contained trustness.

In the work of Monica Nesi, there's a creative treatment of process constants. In standard literature,
a general CCS process can be defined by the following grammar: (suppose $p$ and $q$ are already CCS processes, $a$ and $b$
are actions)
\begin{equation}
p ::= \mathrm{nil} \quad | \quad p + q \quad | \quad p\,\|\,q \quad | \quad (\nu
a)p \quad | \quad p[b/a] \quad | \quad C
\end{equation}
here $C$ is a process constant denoting another CCS process, in which
the $C$ itself or some other constants may appear. For example:
\begin{equation}
\begin{cases}
A \overset{def}{=} (\nu a)(b.\mathrm{nil}\, | A) \\
B \overset{def}{=} b.B + a.A
\end{cases}
\end{equation}
But this form cannot be formalized directly, because in theorem provers a process must be
represented as single term with all information. The solution of Monica Nesi
is to use process variables and a recursion construction instead:
\begin{equation}
p ::= \mathrm{nil} \quad | \quad p + q \quad | \quad p\,\|\,q \quad | \quad (\nu
a)p \quad | \quad p[b/a] \quad | \quad \mathrm{var}\, X \quad | \quad
\mathrm{rec}\, X\, p
\end{equation}
Suppose $B$ is the root process, with the new grammar
it's possible to represent this process by the following term:
\begin{equation}
\mathrm{rec}\, \mathbb{B}\, (b.(\mathrm{var}\,B) + a.(\mathrm{rec}\, \mathbb{A}\, (\nu
a)(b.\mathrm{nil}\, | \mathrm{var}\,A)))
\end{equation}
This term contains both previous definitions of constants $A$ and
$B$. With a little more imagination, we can see the similarity between
above term and a similar $\lambda$-calculus term (suppose other CCS
operators were available):
\begin{equation}
\lambda \mathbb{B}.\, (b.B + a.(\lambda \mathbb{A}.\, (\nu
a)(b.\mathrm{nil}\, | A)))
\end{equation}
Now it comes to the interesting part: given above CCS grammars, the
following ``process'' is also valid:
\begin{equation}
\mathrm{rec}\, \mathbb{A}.\, (a.A + b.C)
\end{equation}
or in the grammar with constants:
\begin{equation}
A \overset{def}{=} a.A + b.C
\end{equation}
Here the constant $C$ is undefined, but the entire
definition is still valid according to the CCS grammar. In standard
literature, process terms containing undefined constants do not have fully
given semantics, c.f. Definition 3.2 (and the remarks after it) in Gorrieri's book:
\begin{definition}{(Defined constant and fully defined term)}
A process constant $A$ is \emph{defined} if it possesses a defining
equation: $A \overset{def}{=} p$. A process term $q$ is \emph{fully
defined} if all the constants in it are defined.

The requirement that a term must be fully defined is due to the fact that its
semantics cannot be fully given otherwise. For instance, term $a.A$
can execute $a$, but after that we do not know what to do if $A$ is
not equipped with a defining equation.
\end{definition}

The approach is different in our formalization of CCS: we think every
process term has determined semantics, even when it contains
undefined constants! The rules are simple: \emph{undefined constants
  have no transitions.} In another word, an undefined constant $A$
works like a $\mathrm{nil}$, they simply have no
transitions. However, we didn't explicitly define such a rule, instead
it's a natural consequence derived from the transition semantics.

It remains to understand what's exactly an undefined process
variable. Inspired by \cite{milner1990operational}, we have
realized that, an undefined constant is actually a free
variable in the process term. Formally speaking, now we took the
following approach, which is better than those in current standard
literature:
\begin{enumerate}
\item CCS terms may contain countable many process variables:
  $A, B, C, \ldots$;
\item Whenever such a variable $A$ is wrapper by a $\mathrm{rec}$
  construction with the same variable, i.e. $\mathrm{rec}\, A. \ldots
  A \ldots$, it's a \emph{bounded variable}, or \emph{(process)
    constant};
\item Other variables are \emph{free} (process) variables;
\item We call a CCS term with free variables \emph{CCS expressions};
\item A CCS term without free variables is \emph{CCS process}.
\item Free variables have no transitions.
\item Variable substitutions in CCS terms $p$ can only apply to those free
  variables, and a free variable cannot be substituted with a term $q$
  containing free variables having the same name with any surrounding constant.
\end{enumerate}
These rules are very similar with $\lambda$-calculus. For example, in
$(\lambda x. x + y)$, $x$ is bounded, and $y$ is free. We can
substitute $y$ with another $lambda$-term, but this term shouldn't
contain $x$ as free variable (otherwise we may have to rename the
bound variable $x$ into $x'$).

With this approach, there's actually a straightforward way to
formalize CCS equations with multiple variables, at least for finitely
many variables: $\tilde{X} = \tilde{E}[\tilde{X}]$, in which the right
side of each
single equation $E_i[\tilde{X}]$ can be seen as a repeated variable
substitution process: $E_i\{p_1/X_1\}\{p_2/X_2\}...\{p_n/X_n\}$, where
$p_1, p_2, \ldots, p_n$ are solution of the equation as a list of CCS
processes (without free variables). Thus in theory it's actually possible
to formalize the ``unique solution of equations'' theorem with
equations having multiple variables.

Above approach is currently not fully implemented,
although we do have defined functions for retrieving the set of free
and bound variables in CCS terms, and multi-variable substitutions.
In fact, above approach was
realized too late during the thesis, thus we chose to only consider single-variable
equations, which actually coincide with semantics contexts of CCS as unary $\lambda$ functions,
taking one CCS process and returning another.

This choice has guaranteed the quick complete of this thesis project in
reasonable time. But there's one drawback: it's impossible to express
(and prove) the ``Completeness of contractions'' (Theorem 3.13 of
\cite{sangiorgi2015equations}), because it requires an infinite system of contractions.
A formalized theory of "contraction" wouldn't be complete with this completeness theorem, because it's the theorem where the ``expansion''  (a closely related relation) doesn't hold.
But currently we have no choice but to leave it as not formalized.

\section{Limitations}

There's limitation in our formalization: the CCS
summation operator is only binary, while in Milner's original CCS definition, infinite sums of processes are supported.
As the result, only finite
summation can be expressed in our definition of CCS, and the resulting CCS language is finitary.

Infinite sums of processes over a arbitrary infinite
set of processes turns out to be impossible for higher order logic,
i.e. having the following datatype:
\begin{lstlisting}
Datatype `CCS = summ (CCS set) | ...`;
\end{lstlisting}
The reason was explained by Michael Norrish (the HOL maintainer):
\begin{quote}
``You can't define a type that recurses under the set ``constructor''
(your \texttt{summ} constructor has (CCS set) as an argument).  Ignoring the
num set argument, you would then have an injective function (the \texttt{summ}
constructor itself) from sets of CCS values into single CCS
values. This ultimately falls foul of Cantor's proof that the power
set is strictly larger than the set.''
\end{quote}

What's certainly allowed in higher order logic is the summation over functions taking
numerical indexes (numbers or even ordinals) returning CCS processes, i.e. something like:
\begin{lstlisting}
Datatype `CCS = summ (num -> CCS) | ...`;
\end{lstlisting}
However, currently the datatype package in HOL4 cannot support this
kind of ``nesting recursive definitions''. In theory it's possible to define
the desired datatype manually by constructing a bijection from another existing type, but this work is hard and goes beyond the
author's ability.\footnote{In the work of Monica Nesi on formalizing
  Value-Passing CCS \cite{Nesi:2017wo}, there're supports of
  infinite sums in the CCS datatype, however currently we
  don't know how it's actually implemented, as the related proof scripts are
  not available.} The long-term plan is to implement (by the
author) the idea of
Andrei Popescu \cite{traytel2012foundational} based on Category Theory. It has been implemented in Isabelle/HOL.
The author would like to have the same datatype defining ability in HOL4, and once this goal is achieved,
the current CCS formalization will be also updated to support infinite sums.

It's also worth noting that, Milner's CCS has several
different forms, in which the more comprehensive one was introduced in
\cite{milner1990operational} as the last
chapter of the \emph{Handbook of Theoretical Computer Science, Volume
  B}.  In this paper, Robin Milner has defined the set
$\mathscr{E}$ of processes expressions $E, F, \ldots$ (also called
terms) as the smallest set including $\mathscr{E}$ and the following
expressions--when $E$, $E_i$ are already in $\mathscr{E}$:
\begin{quote}
$\alpha.E,\qquad$ a \emph{Prefix} $(\alpha \in Act)$, \\
$\sum_{i\in I} E_i,\qquad$ a \emph{Summation}, \\
$E_0 \| E_1,\qquad$ a \emph{Composition}, \\
$E \setminus L, \qquad$ a \emph{Restriction} ($L \subseteq
\mathscr{L}$), \\
$E [f], \qquad$ a \emph{Relabeling} ($f$ is a relabeling function), \\
$\mathrm{fix}_j \{ X_i = E_i \colon i \in I \},\qquad$ a
\emph{Recursion} ($j \in I$).
\end{quote}
In the final form (Recursion) the variables $X_i$ are \emph{bound}
variables. For any expression $E$, its \emph{free} variables $fv(E)$
are those while occur unbound in $E$. We say $E$ is \emph{closed} if
$fv(E) = \emptyset$; in this case we say $E$ is a \emph{process}, and
we shall use $P, Q, R$ to range over the processes $\mathscr{P}$.

In this way, ``expressions'' (as the core part of an equation) now becomes first-class definitions
in the theory of CCS, and a normal CCS process is nothing but an
expression without any free variable.  Further more, we quote,``for simplicity
purpose we sometimes use an alternative formulation of Recursion:
instead of the \textbf{fix} construction we introduce \emph{process
  constants $A, B, \ldots$ into the language. We then admit a set $\tilde{A}
  \overset{\mathrm{def}}{=} \tilde{P}$ of \emph{defining equations}
  for the constants.}''  So the process constants are actually defined
by processes equations.

Unfortunately, almost all CCS textbooks (including the Milner's own book
\cite{Milner:1989}) have adopted the ``simple'' approach in which
  process constants are part of the CCS grammar, and there's no
  \textbf{fix} construction. 
  
Our work is derived from the work of Monica Nesi, and we didn't touch the definition of her CCS datatype except for the uses of type variables.
In this thesis,
we focused on the formalization of various versions of the ``unique
solution of equations/contractions'' theorem, limited for
the case of single-variable equations, because in this special case an expression
is also a context, i.e. lambda functions taking one CCS process
returning another.  And the concepts of guardedness can also be easily
and recursively defined by lambda functions.  In this way, we actually ignored all free variables in CCS terms and didn't treat them as equation variables
at all. And the resulting ``unique solution of equations'' theorems have very simple and elegant statements.

The proof scripts (about 20,000 lines) of this thesis is currently available in the
`\texttt{example/CCS}' directory of HOL4 source code repository
\footnote{\url{https://github.com/HOL-Theorem-Prover/HOL}}.

The structure of this thesis paper is as follows: (TODO) In Chapter ..

\section{Related work}

Beside the early CCS formalization by Monica Nesi, the author found only two other CCS formalizations. One
is done in Isabelle/HOL (based on Nominal datatypes) in 2010, by Jesper Bengtson as part of his PhD
thesis project \cite{bengtson2010formalising}.
He also formalized $\pi$-calculus and $\psi$-calculus, all available in Isabelle's Archive of Formal
Proofs (AFP) \footnote{\url{https://www.isa-afp.org}}.   Since he
has done 3 big formal systems in one project, it's expected that, for each of
them only very basic results can be done. The part of CCS has indeed formalized
classic properties for strong/weak equivalence and observational
congruence, however his CCS grammar doesn't include process
constants. And his proved theorems also look quite different with their original forms in CCS textbooks.

The other formalization we found is based on Coq, done by Solange
Coupet-Grimal in 1995, which is part of the
``coq-contribs'' archive (Coq's official user contributed code)
\footnote{\url{https://github.com/coq-contribs/ccs}}. This is actually
a formalization of transition systems without any algebraic operator,
and all has been covered is still the very basic properties of strong/weak equivalence and observational
congruence.

Monica Nesi's early work \cite{Nesi:1992ve} on pure CCS covered basically the same
thing, i.e. classic properties for strong/weak equivalence and observational
congruence. However, in her work there's a decision procedure written as ML
program (i.e. a function): given any CCS process, this program can generate a theorem
which precisely captures all possible (direct) transitions leading from
the process. Monica Nesi also described another decision procedure
for checking the bisimulation equivalence between two CCS
processes. She also formalized Hennessy-Milner Logic (HML), although no
decision procedures were implemented.  HOL proof scripts are simply ML
programs (for HOL88 this was Classic ML, now it's Standard ML), so is
the HOL theorem prover itself. So it's very natural for its end users
to write ML programs to extend the automatic proof searching abilities
during the formalization of related theorems.  But this same thing is
not obvious for Isabelle, Coq and many other theorem provers, where
the language for expressing theorems and proofs are usually domain
languages unrelated with the underlying programming language for
implementing the theorem prover itself.

As the audience shall see, the coverage of this project is so far the largest among
all existing CCS formalizations. And our definitions, proved theorems
have  almost the same look as in standard CCS textbooks (thus easily
understandable for students and professors in Concurrency Theory).
Almost all proofs have clear, human readable steps for replay
and learning purposes.

\section{Availability}

It's usually very hard to maintain a stable online URL to guarentee
the availability of the work done in this thesis project for many
years. The author's strategy is to put the entire work into the
official code repository of HOL theorem prover, which is currently
hosted at
GitHub\footnote{\url{https://github.com/HOL-Theorem-Prover/HOL}},
under the folder ``\texttt{examples/CCS}''. This also means that,
every HOL user who has recently installed latest HOL4 from its source
code cloned by Git, or is using a formal HOL4 release since Kananaskis
12 (not released at current moment) should have a copy of our proof
scripts in his computer under the same sub-folder related HOL
top-level folder.

But things may change in the future, and the proof scripts found in
HOL's code base may be further upgraded by the author or others. Thus
it seems also necessary to maintain a copy which is exactly the same
as the one mentioned in this thesis paper. Currently such a copy is also in
Github\footnote{\url{https://github.com/binghe/informatica-public/tree/master/thesis}}
but the author cannot promise its availability for even next 10
years. Whenever this link is not available any more, please find the
``official'' version in HOL theorem prover.

\cleardoublepage

%%%% -*- Mode: LaTeX -*-

\def\HOL{\textsc{Hol}}
\newcommand\fun{{\to}}
\newcommand\prd{{\times}}
\newcommand{\ty}[1]{\textsl{#1}}
\newcommand\conj{\ \wedge\ }
\newcommand\disj{\ \vee\ }
\newcommand\imp{ \Rightarrow }
\newcommand\eqv{\ \equiv\ }
\newcommand\cond{\rightarrow}
\newcommand\vbar{\mid}
\newcommand\turn{\ \vdash\ } % FIXME: "\ " resultgs in extra space
\newcommand\hilbert{\varepsilon}
\newcommand{\uquant}[1]{\forall #1.\ }
\newcommand{\equant}[1]{\exists #1.\ }
\newcommand{\hquant}[1]{\hilbert #1.\ }
\newcommand{\iquant}[1]{\exists ! #1.\ }
\newcommand{\lquant}[1]{\lambda #1.\ }
\newcommand{\ml}[1]{\mbox{{\def\_{\char'137}\texttt{#1}}}}
\newcommand{\con}[1]{\mathrm{#1}}

\newcommand\bool{\ty{bool}}
\newcommand\num{\ty{num}}
\newcommand\ind{\ty{ind}}
\newcommand\lst{\ty{list}}

\providecommand{\T}{\con{T}}
\renewcommand{\T}{\con{T}}
\newcommand\F{\con{F}}
\newcommand\OneOne{\con{One\_One}}
\newcommand\OntoSubset{\con{Onto\_Subset}}
\newcommand\Onto{\con{Onto}}
\newcommand\TyDef{\con{Type\_Definition}}

\chapter{HOL Theorem Prover}

The HOL interactive theorem prover is a proof assistant for
higher-order logic: a programming environment in which logical theorems can be
proved, with proof tools implemented. Built-in decision procedures and
theorem provers can automatically establish many simple theorems
(users may have to prove the hard theorems interactively!) An oracle
mechanism gives access to external programs such as SMT and BDD
engines. HOL is particularly suitable as a platform for implementing
combinations of deduction, execution and property checking. 

The approach to mechanizing formal proof used in HOL is due to Robin
Milner \cite{gordon1979edinburgh}, who also headed the team that
designed and implemented the language ML.

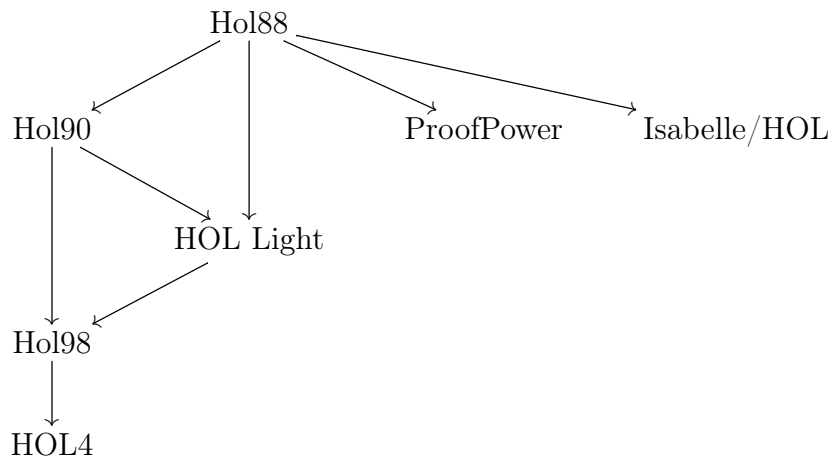
\begin{figure}[htbp]
\begin{center}
\begin{displaymath}
\xymatrix{
{} &  {\text{Hol88}} \ar[ld] \ar[dd] \ar[rd] \ar[rrd] \\
{\text{Hol90}} \ar[dd] \ar[rd] & {} & {\text{ProofPower}} & {\text{Isabelle/HOL}} \\
{} & {\text{HOL Light}} \ar[ld] \\
{\text{Hol98}} \ar[d] \\
{\text{HOL4}}
}
\end{displaymath}
\caption{HOL4 and its relatives}
\label{fig:HOL}
\end{center}
\end{figure}

HOL has several different major versions in history, the latest
version is usually called HOL4.  A relationship between HOL4 and other
HOL versions and derived systems can be seen from Figure
\ref{fig:HOL}.

HOL4 was implemented in Standard ML. This programming language plays three roles here:
\begin{enumerate}
\item It serves as the underlying implementation language for the core HOL engine;
\item It's used to implement tactics (and tacticals) for automatic theorem proving;
\item It's used as the command language of the interactive HOL system.
\end{enumerate}
These roles can be done in separated languages. For example, in the
Coq proof-assistant \footnote{\url{https://coq.inria.fr}}, the
proof-assistant itself is written in OCaml, but the language for
expressing theorems (and their proofs) is another language called
\emph{Gallina}, while the tactic language is again another different
language called \emph{Ltac}. But in HOL-4, all these three languages
are the same (except for the inner language of logic, which is derived from
Classic ML). This fact also allows us to use HOL-4 as a general
programming platform and write an entire model checking software which
uses the theorem prover as a library.
Such advantage is not possible for most other theorem provers, at least
not that straightforward.

\section{Higher Order Logic}

Higher Order Logic (or ``HOL Logic'') means simply typed $\lambda$-calculus plus Hibert
choice operator, axiom of infinity, and rank-1 polymorphim (type
variables). In this section we briefly introduce the HOL Logic, and
please refer to official documents \cite{hollogic} for a complete picture.

The HOL syntax contains syntactic categories of types and terms whose elements are
intended to denote respectively certain sets and elements of sets. This model is given in terms of a fixed set of
sets $\mathcal{U}$, which will be called the universe and which is assumed to have
the following properties:
\begin{enumerate}
\item[\textbf{Inhab}] Each element of $\mathcal{U}$ is a non-empty
  set.
\item[\textbf{Sub}] If $X\in\mathcal{U}$ and $\emptyset \neq
  Y\subseteq X$, then $Y\in\mathcal{U}$.
\item[\textbf{Prod}] If $X\in\mathcal{U}$ and $Y\in\mathcal{U}$, then
  $X\times Y\in \mathcal{U}$. The set $X\times Y$ is the cartesian
  product, consisting of ordered pairs $(x, y)$ with $x\in X$ and
  $y\in Y$.
\item[\textbf{Pow}] If $X\in\mathcal{U}$, then the powerset $\wp(X) =
  \{Y\colon Y \subseteq X\}$ is also an element of $\mathcal{U}$.
\item[\textbf{Infty}] $\mathcal{U}$ contains a distinguished infinite
  set $I$.
\item[\textbf{Choice}] There is a distinguished element $\mathrm{ch}
  \in \prod_{X\in\mathcal{U}} X$. The element of the product
  $\prod_{X\in\mathcal{U}} X$
are (dependently typed) functions: the for all $X\in\mathcal{U}$, $X$
is non-empty by \textbf{Inhab} and $\mathrm{ch}(X) \in X$ witnesses this.
\end{enumerate}

Although HOL also supports non-standard models, above set-theoretic
model is the basis of all builtin theories of HOL theorem prover and
is usually adopted by HOL users when they claim to have formalized
something in HOL.

The above assumptions on $U$ are strictly weaker than those imposed on a
universe of sets by the axioms of ZFC (Zermelo-Frankel set theory with the
Axiom of Choice), principally because $\mathcal{U}$ is not required
to satisfy any form of the Axiom of Replacement. Thus it's possible to
prove the existence of a set $\mathcal{U}$ with the above properties
from the axioms of ZFC\footnote{For example, one could take
  $\mathcal{U}$ to consist of all non-empty sets in the von Neumann
  cumulative hierarchy formed before stage $\omega + \omega$.}, and it's possible in principal to give a
completely formal version within ZFC set theory of the semantics of
the HOL logic.

On the other side, some mathematics theories cannot be formalized in
above standard model of HOL. One notable example is the full theory of
ordinal numbers in the von Neumann hierarchy and the cardinal
numbers. However, HOL can support the formalization of a large
portation of mathematics, including Lebesgue Measures and Probability
Theory. We have also strong witenesses seen in another
theorem prover in HOL family (i.e. Isabelle/HOL).

There are some notable consequences of above assumptions (\textbf{Inhab},
\textbf{Sub}, \textbf{Prod}, \textbf{Pow}, \textbf{Infty} and
\textbf{Choice}). Important one is the concept of functions. In
set theory, functions are identified with their graphs, which are
certain sets of order pairs. Thus the set $X\fun Y$ of all
functions from a set $X$ to a set $Y$ is a subset of $\wp(X \times
Y)$; and it is a non-empty set when $Y$ is non-empty. So \textbf{Sub},
\textbf{Prod} and \textbf{Pow} together imply that $\mathcal{U}$ also satisfies
\begin{enumerate}
\item[\textbf{Fun}] If $X\in\mathcal{U}$ and $Y\in\mathcal{U}$, then
  $X\fun Y \in\mathcal{U}$.
\end{enumerate}
By iterating \textbf{Prod}, one has that the certasian product of any
finite, non-zero number of sets in $\mathcal{U}$ is again in
$\mathcal{U}$. But $\mathcal{U}$ also contains the cartesian product
of no sets, which is to say that it contains a one-element set (by
virtue of \textbf{Sub} applied to any set in
$\mathcal{U}$--\textbf{Infty} guarantees there is one); for
definiteness, a particular one-element set will be singled out:
\begin{enumerate}
\item[\textbf{Unit}] $\mathcal{U}$ contains a distinguished one-element
  set $1 = \{0\}$.
\end{enumerate}
Similarly, because of \textbf{Sub} and \textbf{Infty}, $\mathcal{U}$
contains two-element sets, one of which will be singled out:
\begin{enumerate}
\item[\textbf{Bool}] $\mathcal{U}$ contains a distinguished two-element
  set $2 = \{0, 1\}$.
\end{enumerate}

\subsection{Types and Terms}

The types of the HOL logic are expressions that denote sets (in the
universe $\mathcal{U}$). HOL's type system is much simpler than those
based on dependent types and other type theories. There are four kinds of types in the HOL
logic, as illustrated in Fig. \ref{fig:hol-types} for its BNF
grammar. Noticed that, in HOL the standard atomic types \emph{bool} and \emph{ind}
 denote, respectively, the distinguished two-element set 2 and the
distinguished infinite set $I$.

\newlength{\ttX}
\settowidth{\ttX}{\tt X}
\newcommand{\tyvar}{\setlength{\unitlength}{\ttX}\begin{picture}(1,6)
\put(.5,0){\makebox(0,0)[b]{\footnotesize type variables}}
\put(0,1.5){\vector(0,1){4.5}}
\end{picture}}
\newcommand{\tyatom}{\setlength{\unitlength}{\ttX}\begin{picture}(1,6)
\put(.5,2.3){\makebox(0,0)[b]{\footnotesize atomic types}}
\put(.5,3.3){\vector(0,1){2.6}}
\end{picture}}
\newcommand{\funty}{\setlength{\unitlength}{\ttX}\begin{picture}(1,6)
\put(.5,1.5){\makebox(0,0)[b]{\footnotesize function types}}
\put(.5,0){\makebox(0,0)[b]{\footnotesize (domain $\sigma_1$, codomain $\sigma_2$)}}
\put(1,2.5){\vector(0,1){3.5}}
\end{picture}}
\newcommand{\cmpty}{\setlength{\unitlength}{\ttX}\begin{picture}(1,6)
\put(2,3.3){\makebox(0,0)[b]{\footnotesize compound types}}
\put(1.9,4.5){\vector(0,1){1.5}}
\end{picture}}

\begin{figure}[h]
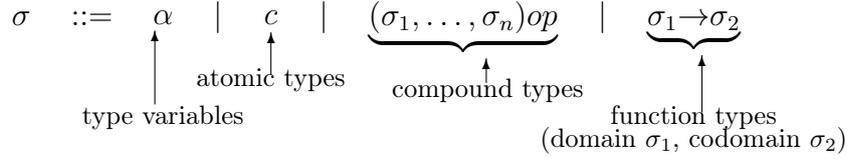

\begin{equation*}
\sigma\quad ::=\quad {\mathord{\mathop{\alpha}\limits_{\tyvar}}}
        \quad\mid\quad{\mathord{\mathop{c}\limits_{\tyatom}}}
        \quad\mid\quad\underbrace{(\sigma_1, \ldots , \sigma_n){op}}_{\cmpty}
        \quad\mid\quad\underbrace{\sigma_1\fun\sigma_2}_{\funty}
\end{equation*}
   \caption{HOL's type grammar}
   \label{fig:hol-types}
\end{figure}

The terms of the HOL logic are expressions that denote elements of the
sets denoted by types. There're four kinds of terms in the HOL
logic. There can be described approximately by the BNF grammar in
Fig. \ref{fig:hol-terms}.

\settowidth{\ttX}{\tt X}
\newcommand{\var}{\setlength{\unitlength}{\ttX}\begin{picture}(1,6)
\put(.5,0){\makebox(0,0)[b]{\footnotesize variables}}
\put(0,1.5){\vector(0,1){4.5}}
\end{picture}}
\newcommand{\const}{\setlength{\unitlength}{\ttX}\begin{picture}(1,6)
\put(.5,2.3){\makebox(0,0)[b]{\footnotesize constants}}
\put(.5,3.5){\vector(0,1){2.4}}
\end{picture}}
\newcommand{\app}{\setlength{\unitlength}{\ttX}\begin{picture}(1,6)
\put(.5,1.5){\makebox(0,0)[b]{\footnotesize function applications}}
\put(.5,0){\makebox(0,0)[b]{\footnotesize (function $t$, argument $t'$)}}
\put(0.5,2.5){\vector(0,1){3.5}}
\end{picture}}
\newcommand{\abs}{\setlength{\unitlength}{\ttX}\begin{picture}(1,6)
\put(1,3.3){\makebox(0,0)[b]{\footnotesize $\lambda$-abstractions}}
\put(0.7,4.5){\vector(0,1){1.5}}
\end{picture}}

\begin{figure}[h]
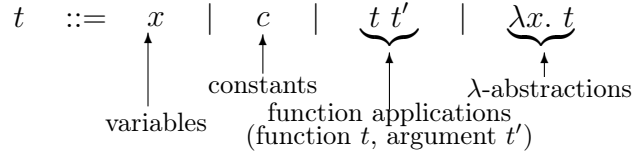

\begin{equation*}
t \quad ::=\quad {\mathord{\mathop{x}\limits_{\var}}}
        \quad\mid\quad{\mathord{\mathop{c}\limits_{\const}}}
        \quad\mid\quad\underbrace{t\ t'}_{\app}
        \quad\mid\quad\underbrace{\lambda x .\ t}_{\abs}
\end{equation*}
   \caption{HOL's term grammar}
   \label{fig:hol-terms}
\end{figure}

So HOL is a deductive system for simply typed
$\lambda$-calculus, as its term syntax has shown, with each term in HOL
associated with a unique simple type.

\subsection{The HOL deductive system}

The deductive system of the HOL logic is specified by the following eight
rules of inference. (The identifiers in square brackets are the names
of the ML functions in the HOL system that implement the corresponding
inference rules.)

\subsubsection*{1. Assumption introduction [{\small\tt
ASSUME}]}
\[
\overline{t \turn t}
\]

\subsubsection*{2. Reflexivity [{\small\tt
REFL}]}
\[
\overline{\turn t = t}
\]

\subsubsection*{3. Beta-conversion [{\small\tt BETA\_CONV}]}
\[
\overline{\turn (\lquant{x}t_1)t_2 = t_1[t_2/x]}
\]
\begin{itemize}
\item Where $t_1[t_2/x]$ is
the result of substituting $t_2$ for $x$
in $t_1$, with suitable renaming of variables to prevent free variables
in $t_2$ becoming bound after substitution.
\end{itemize}

\subsubsection*{4. Substitution [{\small\tt
SUBST}]}
\[
\frac{\Gamma_1\turn t_1 = t_1'\qquad\cdots\qquad\Gamma_n\turn t_n =
t_n'\qquad\qquad \Gamma\turn t[t_1,\ldots,t_n]}
{\Gamma_1\cup\cdots\cup\Gamma_n\cup\Gamma\turn t[t_1',\ldots,t_n']}
\]
\begin{itemize}
\item Where $t[t_1,\ldots,t_n]$ denotes a term $t$ with some free
occurrences of subterms $t_1$, $\ldots$ , $t_n$ singled out and
$t[t_1',\ldots,t_n']$ denotes the result of replacing each selected
occurrence of $t_i$ by $t_i'$ (for $1{\leq}i{\leq}n$), with suitable
renaming of variables to prevent free variables in $t_i'$ becoming
bound after substitution.
\end{itemize}

\subsubsection*{5. Abstraction [{\small\tt ABS}]}
\[
\frac{\Gamma\turn t_1 = t_2}
{\Gamma\turn (\lquant{x}t_1) = (\lquant{x}t_2)}
\]
\begin{itemize}
\item Provided $x$ is not free in $\Gamma$.
\end{itemize}

\subsubsection*{6. Type instantiation [{\small\tt INST\_TYPE}]}
\newcommand{\insttysub}{[\sigma_1,\ldots,\sigma_n/\alpha_1,\ldots,\alpha_n]}
\[
\frac{\Gamma\turn t}
{\Gamma\insttysub\turn t\insttysub}
\]
\begin{itemize}
\item Where $t\insttysub$ is the result of substituting, in parallel,
  the types $\sigma_1$, $\dots$, $\sigma_n$ for type variables
  $\alpha_1$, $\dots$, $\alpha_n$ in $t$, and where $\Gamma\insttysub$
  is the result of performing the same substitution across all of the
  theorem's hypotheses.
\item After the instantiation, variables free in the input can not
  become bound, but distinct free variables in the input may become
  identified.
\end{itemize}

\subsubsection*{7. Discharging an assumption [{\small\tt
DISCH}]}
\[
\frac{\Gamma\turn t_2}
{\Gamma -\{t_1\} \turn t_1 \imp t_2}
\]
\begin{itemize}
\item Where $\Gamma -\{t_1\}$ is the set subtraction of $\{t_1\}$
from $\Gamma$.
\end{itemize}

\subsubsection*{8. Modus Ponens [{\small\tt
MP}]}
\[
\frac{\Gamma_1 \turn t_1 \imp t_2  \qquad\qquad   \Gamma_2\turn t_1}
{\Gamma_1 \cup \Gamma_2 \turn t_2}
\]

In HOL, all the proofs are finally reduced to sequences of
applications of above eight primitive
inference rules. What's also implicitly given by above rules are
the following two foundamental operators:
\begin{enumerate}
\item[\textbf{Equality}] Equality (\texttt{= : 'a -> 'a -> bool}) is an infix
  operator.
\item[\textbf{Implication}] Implication (\texttt{$\Rightarrow$ : bool -> bool -> bool})
  is the \emph{material implication} and is an infix operator that is
  right-associative, i.e. $x \Rightarrow y \Rightarrow z$ parses to the same term as
$x \Rightarrow (y \Rightarrow z)$.
\end{enumerate}

Noticed that, unlike in other formal systems like Propositional Logic,
logical implication ($\Rightarrow$) in
HOL is not defined by other operators (e.g. $p \rightarrow q :=
\neg p \vee q$), instead it's a primitive constant whose meanings are
given completely by above eight primitive deduction rules
(\texttt{DISCH} and \texttt{MP}, to be precise). 

\subsection{Standard Theory of HOL}

It's quite amazing (at least to the author) that, all the rest logical
constants of Propositional Logic and First-order Logic (universal and
existential quantifiers) can be embeded in $\lambda$-calculus, with
equality and logical implication:
\[
\begin{array}{l}
\turn \T       =  ((\lquant{x_{\ty{bool}}}x) =
               (\lquant{x_{\ty{bool}}}x))    \\
\turn \forall  =  \lquant{P_{\alpha\fun\ty{bool}}}\ P =
                    (\lquant{x}\T ) \\
\turn \exists  =  \lquant{P_{\alpha\fun\ty{bool}}}\
                    P({\hilbert}\ P) \\
\turn \F       =  \uquant{b_{\ty{bool}}}\ b  \\
\turn \neg    =  \lquant{b}\ b \imp \F \\
\turn {\wedge}  =  \lquant{b_1\ b_2}\uquant{b}
                     (b_1\imp (b_2 \imp b)) \imp b \\
\turn {\vee}  =  \lquant{b_1\ b_2}\uquant{b}
                   (b_1 \imp b)\imp ((b_2 \imp b) \imp b) \\
\end{array}
\]
with the following special notations:
\begin{center}
\begin{tabular}{|l|l|}\hline
{\rm Notation} & {\rm Meaning}\\ \hline $\uquant{x_{\sigma}}t$ &
$\forall(\lambda x_{\sigma}.\ t)$\\ \hline $\uquant{x_1\ x_2\ \cdots\
x_n}t$ & $\uquant{x_1}(\uquant{x_2} \cdots\ (\uquant{x_n}t)
\ \cdots\ )$\\ \hline
$\equant{x_{\sigma}}t$
  & $\exists(\lambda x_{\sigma}.\ t)$\\ \hline
$\equant{x_1\ x_2\ \cdots\ x_n}t$
  & $\equant{x_1}(\equant{x_2} \cdots\ (\equant{x_n}t)
\ \cdots\ )$\\ \hline
$t_1\ \wedge\ t_2$  & $\wedge\ t_1\ t_2$\\ \hline
$t_1\ \vee\ t_2$  & $\vee\ t_1\ t_2$\\ \hline
\end{tabular}\end{center}

Thus in HOL, an universal quantified term like $\forall x. P(x)$ 
is represented as the functional application of
the constant $\forall$ as a $\lambda$-function and another $\lambda$-function
$\lambda x. P(x)$. At first glance, these definitions seem a
little strange. However, it can be
manually verified that, they're indeed consistent with the usual
semantics in Propositional
Logic and First-order Logic. For the new constant ($\varepsilon$)
used in the definition of existential quantifier ($\exists$), HOL
gives it the following meaning:
\begin{enumerate}
\item[\textbf{Choice}] If $t$ is a term having type
  $\sigma\rightarrow\mathrm{bool}$, then \texttt{@x.t x} (or,
  equivalently, \texttt{\$@t}) denotes \emph{some} member of the set
  whose characteristic function is $t$. If the set is empty, then
  \texttt{@x.t x} denotes an arbitrary member of the set denoted by
  $\sigma$. The constant \texttt{@} is a higher order version of Hilbert's
  $\varepsilon$-operator; it is related to the constant $\iota$ 
in Church's formulation of higher order logic.
\end{enumerate}
However so far above statement is just a wish: its behavior is not
given by any primitive deduction rules, nor have we its definition
upon other more primitive objects. To actually have the choice
operator we need an axiom.  There're totally four such axioms in HOL's
standard model:
\[
\begin{array}{@{}l@{\qquad}l}
\mbox{\small\tt BOOL\_CASES\_AX}&\vdash \uquant{b} (b = \T ) \vee (b = \F )\\
 \\
\mbox{\small\tt ETA\_AX}&
\vdash \uquant{f_{\alpha\fun\beta}}(\lquant{x}f\ x) = f\\
 \\
\mbox{\small\tt SELECT\_AX}&
\vdash \uquant{P_{\alpha\fun\ty{bool}}\ x} P\ x \imp
P({\hilbert}\ P)\\
  \\
\mbox{\small\tt INFINITY\_AX}&
\vdash \equant{f_{\ind\fun \ind}} \OneOne \ f \conj \neg(\Onto \ f)\\
\end{array}
\]

Among these axioms, \texttt{BOOL\_CASES\_AX} represents the Law of
Excluded Middle, \texttt{ETA\_AX} is the $\eta$-conversion rule in
$\lambda$-calculus ($\eta$-conversion cannot be derived from
$\beta$-conversion).
\texttt{SELECT\_AX} gives Hilbert's $\varepsilon$-operator
the actual meaning, it has the position as the Axiom of Choice in ZFC. The last
axiom \texttt{INFINITY\_AX} represents the Axiom of Infinity in ZFC, 
the two other constants appeared in the axiom have the following definitions:
\[
\begin{array}{l}
\turn \OneOne  =  \lquant{f_{\alpha \fun\beta}}\uquant{x_1\ x_2}
                    (f\ x_1 = f\ x_2)  \imp (x_1 = x_2) \\
\turn \Onto  =  \lquant{f_{\alpha\fun\beta}}
                  \uquant{y}\equant{x} y = f\ x \\
\end{array}
\]
This is how the concept of ``infinity'' is formalized: if there's a function
$f\colon I\to I$ which is one-one but not onto, then the set
$I$ has no choice but contains infinite elements. (And starting with
this infinite set, the set of natual numbers, real numbers, ... can be
built in HOL)

The last primitive definition is for defining new primitive types in HOL:
\[
\begin{array}{l}
\turn \TyDef  =   \begin{array}[t]{l}
                  \lambda P_{\alpha\fun\ty{bool}}\
                  rep_{\beta\fun\alpha}.\;
                  \OneOne\ rep\ \ \wedge{}\\
                  \quad(\uquant{x}P\ x \ =\ (\equant{y} x = rep\ y))
                  \end{array}
\end{array}
\]
In HOL, new primitive types are defined by mapping a subset of
existing types into a new type. Such mappings must be
bijections. The constant  $\TyDef$ is used for asserting such
mappings, then new types can be used legally as primitive types like
\texttt{bool} and \texttt{ind}. Notable types defined in this way are
\texttt{num} (natural numbers), \texttt{sum} (sum types),
\texttt{prod} (product types), \texttt{unit} (singleton type), etc.

The next level of types is customized inductive datatypes like
structures in C-like languages. HOL's Datatype package was created by Thomas Melham
\cite{Melham:1989dk}, it's not part of the HOL Logic itself but a
package for automatically creation of new inductive and recursive datatypes with
supporting theorems. In this project, the \texttt{CCS} datatype is
defined by this package.

\section{Relations in HOL}

\subsection{Notations}

From the view of mathematics, a $n$-ary \emph{relation} $R$ is nothing but a
\emph{set} of $n$-tuples, and we say a term (or just a sentence) $R \enspace x_1 \enspace
x_2 \enspace \ldots \enspace x_n$ is true if and only if $(x_1, x_2,
\ldots, x_n) \in R$. In particolar, an unary relation is also called a
\emph{predicate} which usually denote the property of objects. Binary
relations are usually represented in infix ways, e.g., instead of $R\enspace x \enspace y$ we write $x
\enspace R \enspace y$. Notable examples of binary relations include
equality ($x = y$), inequality ($x < y$) and set relations like $A
\subset B$ or even $x \in A$.

In general cases, a (binary) relation needs not to be
transitive at all, nor each parameter has the same type. From the view of higher order logic, or simple-typed lambda
calculus, however, there's a difference between terms like
\begin{equation*}
R \enspace x_1 \enspace x_2 \enspace \ldots \enspace x_n,
\end{equation*}
and
\begin{equation*}
R \enspace (x_1, x_2, \ldots, x_n),
\end{equation*}
in which the latter one equals to $(x_1, x_2, \ldots, x_n) \in R$ in predicate-based set
theory \cite{Anonymous:kDKi-cmu}.

In HOL, the former term has the types like
\begin{equation*}
\alpha \rightarrow \beta \rightarrow \cdots  \rightarrow bool,
\end{equation*}
while the latter term has the types like
\begin{equation*}
(\alpha \text{ \# } \beta \text{ \# } \cdots ) \rightarrow bool
\end{equation*}
where $\alpha, \beta, \ldots $ represent type variables (for each
parameter of the relation) and \# is the
operator for building Cartesian product types. The two types are
equivalent (but not the same) for precisely defining any relation. The process of
converting a relation from the former type to latter type is called
\emph{curry}, and the reverting process is called \emph{uncurry}.

Relations used in logic, formal methods and theorem proving areas are
usually defined in curried forms, because it has extra advantages
that, a relation term with parameters \emph{partially given} could
still has a meaningful type.  On the other side, mathematicians usually use the uncurried forms. In the rest of
this paper, we will ignore the differences between curried and
uncurried forms, and the reader should make implicit type translations in context-switching between
mathematics and theorem proving talks.

\subsection{Ways to define relations}

To precisely define a relation (which is essentially a set), we can
explicitly enumerating all its elements (when the number of them is finite), e.g.
\begin{equation*}
R = \{ (1,2), (2,3), (3,4), \ldots \}
\end{equation*}
or define the set by its characteristic function $f$ (with the proof of the soundness
of such functions given first, as required by ZFC), e.g.
\begin{equation*}
R = \{ (x, y) \colon f(x,y) = \text{true} \}
\end{equation*}
But mostly above two ways are not enough, instead, we have to define a
relation by a set of rules in the following two forms:

\begin{multicols}{2}

\begin{prooftree}
\AxiomC{$\emptyset$}
\UnaryInfC{$R \enspace x_1 \enspace x_2 \enspace \ldots \enspace x_n$}
\end{prooftree}

\begin{prooftree}
\AxiomC{$\text{<hypothesis>}$}
\UnaryInfC{$R \enspace x_1 \enspace x_2 \enspace \ldots \enspace x_n$}
\end{prooftree}

\end{multicols}

Rules in above left form should be understood as a
element $(x_1, x_2, \ldots, x_n)$ belongs to the relation (as set)
\emph{unconditionally} (thus we usually call such rules as \emph{base
  rules} or \emph{axioms}), while rules in above right form should be
understood in either \emph{forward} form
\begin{equation*}
\forall x_1, x_2, \ldots x_n.\; \text{<hypothesis>} \Rightarrow
(x_1, x_2, \ldots, x_n) \in R,
\end{equation*}
and/or \emph{backward} form
\begin{equation*}
\forall x_1, x_2, \ldots x_n.\; (x_1, x_2, \ldots, x_n) \in R
\Rightarrow \text{the same <hypothesis> in existence form},
\end{equation*}
depending on different situations. In general, a hypothesis used in
the above rules could be any term, in which the relation $R$ itself
can also be used, even when it's not fully defined yet.  In such cases, we say the relation is defined
\emph{recursively}, and the previous two rule-defining methods cannot
handle such relation definitions.

But for relations which must be defined recursively, in general we
can't simply define it, say $R$, in the following way:
\begin{equation*}
\forall x_1, x_2, \ldots, x_n. \; (x_1, x_2, \ldots, x_n) \in R
\text{ if and only if $x_1, x_2, \ldots, x_n$ satisfy all the
  rules.}
\end{equation*}
In other words, the relation $R$ is defined as the closure of all
those rules.

The problem is, the resulting relation $R$ of above definition may not be unique (even with
\emph{axiom of extension} (first axiom of ZFC) assumed always), thus
the whole definition cannot be used to precisely describe a mathematic
object in general.

To see the reasons, we can construct a function $F(R)$ from \emph{any}
rule set. The function takes a relation and returns another relation of the
same type. The way to construct such a function is to just
replace all occurences of $R$ in the conclusion (i.e. under the line)
part, into $F(R)$ and keep the (possible) occurences of $R$ in
hypothesis unchanged. So the rules now look like this:
\begin{multicols}{2}

\begin{prooftree}
\AxiomC{$\emptyset$}
\UnaryInfC{$F(R) \enspace x_1 \enspace x_2 \enspace \ldots \enspace x_n$}
\end{prooftree}

\begin{prooftree}
\AxiomC{$\text{<hypothesis>}$}
\UnaryInfC{$F(R) \enspace x_1 \enspace x_2 \enspace \ldots \enspace x_n$}
\end{prooftree}

\end{multicols}
It's important to notice that, the definition of function $F$ (as
above rules) is \emph{not} recursive. Given any $R$ as input (assuming
the type correctness, of course), $F(R)$ is simply the set of tuples
indicated in all base rules, unioned with all tuples which satisfy the
hypothesis of any other rules.   In particolar, if $R = F(R)$, then we
go back to the previous rules, this is to say, $R$ is closed under
those rules if and only if $R$ is a fixed point of the function $F$.

Now we discuss the characteristics of function $F$. The first thing to
notice is: no matter what hypothesis we have in each rule, the function $F$
defined in this way, is always \emph{monotone} (or \emph{order
  preserving}), i.e. $\forall R_1\; R_2.\; R_1 \subset R_2 \Rightarrow F(R_1) \subset F(R_2)$. The
reason is simple, if we consider more than tuples in $R_1$ with the hypothesis, we can only add
more (or at least the same) than tuples in $F(R_1)$ into $F(R_2)$,
there's absolutely no chance to eliminate any existing tuples in
$F(R_1)$ from $F(R_2)$. (The conclusion is still true if all non-axiom
rules were understood in backward)

On the other side, the set $\mathcal{L}$ of all relations $R$ (of the same type)
forms a \emph{complete lattice}, because $\mathcal{L}$ has always a
top element (the relation containing all possible tuples) and a bottom
element (empty relation). Thus $F$ is an order-preserving self-maping function
on a complete lattice. According to Knester-Tarski Fixed Point Theorem
\cite{Davey:2002vo}, $F$ has always a least fixed point and a greatest
fixed point.  To see the two fixed points do not coincide in general,
it's enough to take a sample relation which satisfy only one rule:
\begin{prooftree}
\AxiomC{$n + 1 \in R$}
\UnaryInfC{$n \in R$}
\end{prooftree}
where $n$ has the type of natural numbers (0, 1, \ldots).  Then it's
trivial to see that, the following function $F$
\begin{prooftree}
\AxiomC{$n + 1 \in R$}
\UnaryInfC{$n \in F(R)$}
\end{prooftree}
has a least fixed point $\emptyset$ (no numbers) and a greatest fixed
point $\mathbb{N}$ (all numbers).  Noticed that, it's not true that
all relations between the least and greatest fixed points are also
fixed points (in current case there's no other fixed points), but in
general there may be other fixed points beside the least and greatest ones.

Thus, beside the for-sure existence of fixed points, the number of fixed points
and the usefulness between least and greatest fixed points, are
totally decided by specific relation defining rules. Most of time, one
of least and greatest fixed points is trivial: either the least fixed
point is the bottom element or the greatest fixed point is the top
element, then only the other one is meaningful to be defined.

\subsection{Inductive and co-inductive relation definitions}

Fixing a rule set, depending on the desired relation object,
sometimes we want the least fixed point, sometimes we want the
greatest fixed point. Whenever the least fixed point is used, such a
relation definition is called \emph{inductive} relation definition,
and the other one is called \emph{co-inductive} relation
definition.

Noticed that, in many relation definitions, the fixed points are not explicitly
mentioned, while we still have precise (unique) definition for the
desired relation object. For instance, the following defintion of
an unary relation (or predicate, or set) is a typical inductive
relation definition:
\begin{definition}
The set $R$ is the \emph{smallest} set such that
\begin{enumerate}
\item $0 \in R$;
\item forall $n$, if $n \in R$ then $n+2 \in R$.
\end{enumerate}
\end{definition}
It's not hard to imagine that, the set $R$ is the set of all even numbers,
i.e. $R = \{ 0, 2, 4, 6, \ldots \}$.

Instead, if we required a \emph{greatest} set which is closed under
the same rules, we could have the entire set of natural numbers $\{0,
1, 2, \ldots \}$, even when the axiom rule $0 \in R$ is absent.

Sometimes the greatest and least fixed points coincide, and as a
result there's only a single fixed point for certain rules. For
instance, let $A$ be a set, the following rules can be used to
inductively define all finite lists with elements from $A$:
\begin{multicols}{2}

\begin{prooftree}
\AxiomC{$\emptyset$}
\UnaryInfC{$\text{nil} \in \mathcal{L}$}
\end{prooftree}

\begin{prooftree}
\AxiomC{$s \in \mathcal{L}$}
\AxiomC{$a \in A$}
\BinaryInfC{$\langle a \rangle \bullet s \in \mathcal{L}$}
\end{prooftree}

\end{multicols}

According to \cite{Sangiorgi:2011ut} (p. 36), if above rules were
used co-inductively, the resulting relation will be the set all \emph{finite and
infinite} lists. But if we limit the type of lists to only finite
lists (e.g. the list data structure defined inductively), then the
co-inductive defintion and inductive defintion coincide. This fact can
be proved in HOL4 by the following scripts:
\begin{lstlisting}
val (List_rules, List_ind, List_cases) = Hol_reln
   `(!l. (l = []) ==> List l) /\
    (!l h t. (l = h::t) /\ List t ==> List l)`;

val (coList_rules, coList_coind, coList_cases) = Hol_coreln
   `(!l. (l = []) ==> coList l) /\
    (!l h t. (l = h::t) /\ coList t ==> coList l)`;

val List_imp_coList = store_thm (
   "List_imp_coList", ``!l. List l ==> coList l``,
    HO_MATCH_MP_TAC List_ind
 >> RW_TAC bool_ss [coList_rules]);

val coList_imp_List = store_thm (
   "coList_imp_List", ``!l. coList l ==> List l``,
    Induct_on `l`
 >| [ RW_TAC bool_ss [List_rules, coList_rules],
      STRIP_TAC
   >> ONCE_REWRITE_TAC [coList_cases]
   >> ONCE_REWRITE_TAC [List_cases]
   >> REPEAT STRIP_TAC
   >| [ ASM_REWRITE_TAC [],
	SIMP_TAC list_ss []
     >> `t = l` by PROVE_TAC [CONS_11]
     >> PROVE_TAC [] ] ]);

val List_eq_coList = store_thm (
   "List_eq_coList", ``!l. coList l = List l``,
    PROVE_TAC [List_imp_coList, coList_imp_List]);
\end{lstlisting}
Here we have defined two unary relations (\texttt{List} and
\texttt{coList}), using exactly the same rules (except for the
relation names):
\begin{alltt}
\HOLTokenTurnstile{} (\HOLSymConst{\HOLTokenForall{}}\HOLBoundVar{l}. (\HOLBoundVar{l} \HOLSymConst{=} []) \HOLSymConst{\HOLTokenImp{}} \HOLConst{List} \HOLBoundVar{l}) \HOLSymConst{\HOLTokenConj{}}
   \HOLSymConst{\HOLTokenForall{}}\HOLBoundVar{l} \HOLBoundVar{h} \HOLBoundVar{t}. (\HOLBoundVar{l} \HOLSymConst{=} \HOLBoundVar{h}\HOLSymConst{::}\HOLBoundVar{t}) \HOLSymConst{\HOLTokenConj{}} \HOLConst{List} \HOLBoundVar{t} \HOLSymConst{\HOLTokenImp{}} \HOLConst{List} \HOLBoundVar{l}
\end{alltt}
Besides the rules, there's also a ``case theorem'' generated with the
relation definition. This theorem is also the same for both \texttt{List} and
\texttt{coList}, because it just expressed the same rules from the backward:
\begin{alltt}
\HOLTokenTurnstile{} \HOLConst{List} \HOLFreeVar{a\sb{\mathrm{0}}} \HOLSymConst{\HOLTokenEquiv{}} (\HOLFreeVar{a\sb{\mathrm{0}}} \HOLSymConst{=} []) \HOLSymConst{\HOLTokenDisj{}} \HOLSymConst{\HOLTokenExists{}}\HOLBoundVar{h} \HOLBoundVar{t}. (\HOLFreeVar{a\sb{\mathrm{0}}} \HOLSymConst{=} \HOLBoundVar{h}\HOLSymConst{::}\HOLBoundVar{t}) \HOLSymConst{\HOLTokenConj{}} \HOLConst{List} \HOLBoundVar{t}
\end{alltt}

The only differences between the two relations are the following theorems:
\begin{alltt}
List_ind:
\HOLTokenTurnstile{} (\HOLSymConst{\HOLTokenForall{}}\HOLBoundVar{l}. (\HOLBoundVar{l} \HOLSymConst{=} []) \HOLSymConst{\HOLTokenImp{}} \HOLFreeVar{List\sp{\prime}} \HOLBoundVar{l}) \HOLSymConst{\HOLTokenConj{}}
   (\HOLSymConst{\HOLTokenForall{}}\HOLBoundVar{l} \HOLBoundVar{h} \HOLBoundVar{t}. (\HOLBoundVar{l} \HOLSymConst{=} \HOLBoundVar{h}\HOLSymConst{::}\HOLBoundVar{t}) \HOLSymConst{\HOLTokenConj{}} \HOLFreeVar{List\sp{\prime}} \HOLBoundVar{t} \HOLSymConst{\HOLTokenImp{}} \HOLFreeVar{List\sp{\prime}} \HOLBoundVar{l}) \HOLSymConst{\HOLTokenImp{}}
   \HOLSymConst{\HOLTokenForall{}}\HOLBoundVar{a\sb{\mathrm{0}}}. \HOLConst{List} \HOLBoundVar{a\sb{\mathrm{0}}} \HOLSymConst{\HOLTokenImp{}} \HOLFreeVar{List\sp{\prime}} \HOLBoundVar{a\sb{\mathrm{0}}}
coList_coind:
\HOLTokenTurnstile{} (\HOLSymConst{\HOLTokenForall{}}\HOLBoundVar{a\sb{\mathrm{0}}}.
      \HOLFreeVar{coList\sp{\prime}} \HOLBoundVar{a\sb{\mathrm{0}}} \HOLSymConst{\HOLTokenImp{}} (\HOLBoundVar{a\sb{\mathrm{0}}} \HOLSymConst{=} []) \HOLSymConst{\HOLTokenDisj{}} \HOLSymConst{\HOLTokenExists{}}\HOLBoundVar{h} \HOLBoundVar{t}. (\HOLBoundVar{a\sb{\mathrm{0}}} \HOLSymConst{=} \HOLBoundVar{h}\HOLSymConst{::}\HOLBoundVar{t}) \HOLSymConst{\HOLTokenConj{}} \HOLFreeVar{coList\sp{\prime}} \HOLBoundVar{t}) \HOLSymConst{\HOLTokenImp{}}
   \HOLSymConst{\HOLTokenForall{}}\HOLBoundVar{a\sb{\mathrm{0}}}. \HOLFreeVar{coList\sp{\prime}} \HOLBoundVar{a\sb{\mathrm{0}}} \HOLSymConst{\HOLTokenImp{}} \HOLConst{coList} \HOLBoundVar{a\sb{\mathrm{0}}}
\end{alltt}
From the shapes of above two theorems, it's not hard to see that,
 the purpose of the induction theorem for \texttt{List} is to restrict it
to the least possible closure, while the purpose of the co-induction
theorem for \texttt{coList} is to restrict it to the greatest possible
closure.  Thus, for each defined relation, their characteristic can be
fully decided by three generated theorems:
\begin{enumerate}
\item The original rules;
\item The same rules expressed from backward;
\item The induction or co-induction theorem.
\end{enumerate}

Most of time, it's necessary to define inductive relation, because the induction theorem is necessary to prove many
important results, and those results won't be proved if the same rules
were defined co-inductively. And if for some reasons the inductive and
co-inductive definitions coincide, then in theory both induction co-induction
theorems can be used to prove other results. (although they cannot be
generated together)

What's more intersting is the following game: for a specific relation
defined from a group of rules, if either induction theorem or
co-induction theorem were never used to prove any results in the whole
theory, we want to know, if the relation \emph{becomes larger}, when
switching from the an inductive defintion to co-inductive defintion.
For above List examples, such changes didn't make the resulting
relation any larger, since we can prove their equivalence.

\section{Writing proofs in HOL}

In this section, we use some examples to illustrate the general
techniques for proving theorems in HOL. CCS syntax and SOS (Structural Operational
Semantics) rules appeared in this section will be re-introduced formally in next chapter.

\subsection{Forward and backward proofs}

For simple derivations (as proofs of CCS transition theorems) that we
already know the ``proof'', to formally represent the proof in HOL4,
we can directly \emph{construct} the proof in forward way.
\begin{prooftree}
\AxiomC{}
\LeftLabel{(Pref)}
\UnaryInfC{$a.b.0 \overset{a}\longrightarrow b.0$}
\LeftLabel{($\text{Sum}_1$)}
\UnaryInfC{$a.b.0 + b.a.0 \overset{a}\longrightarrow b.0$}
\end{prooftree}
To proof the derivation theorem:
\begin{alltt}
\HOLTokenTurnstile{} \HOLConst{In} \HOLStringLit{a}\HOLSymConst{..}\HOLConst{In} \HOLStringLit{b}\HOLSymConst{..}\HOLConst{nil} \HOLSymConst{+} \HOLConst{In} \HOLStringLit{b}\HOLSymConst{..}\HOLConst{In} \HOLStringLit{a}\HOLSymConst{..}\HOLConst{nil}
   \HOLTokenTransBegin\HOLConst{In} \HOLStringLit{a}\HOLTokenTransEnd
   \HOLConst{In} \HOLStringLit{b}\HOLSymConst{..}\HOLConst{nil}
\end{alltt}
the most
fundamental way is to use HOL4's standard drivation rule (drule) \texttt{ISPEC}
and primitive inference rule \texttt{MP}.

In SOS rule \texttt{PREFIX}, if we specialize the universally
quantified variable $E$ to $b.0$, we get a new theorem saying 
$\forall u. u.b.0 \overset{u}{\longrightarrow} b.0$:
\begin{lstlisting}
> ISPEC ``prefix (label (name "b")) nil`` PREFIX;
val it =
   |- !u. u..label (name "b")..nil --u-> label (name "b")..nil:
   thm
\end{lstlisting}

Doing this again on the rest universally quantified variable $u$, specialized to Action $a$, then
we get again a new theorem $a.b.0 \overset{a}{\longrightarrow} b.0$
(and saved into a variable \texttt{t1}):
\begin{lstlisting}
> val t1 = ISPEC ``label (name "a")``
                 (ISPEC ``prefix (label (name "b")) nil`` PREFIX);
val t1 =
   |- label (name "a")..label (name "b")..nil
   --label (name "a")->
   label (name "b")..nil:
   thm
\end{lstlisting}

Now we want to use \texttt{SUM1} to reach the final theorem. There two
ways to do this.  The first way is to manually specialize all the four
universally quantified variables in the theorem. To make this work easier, HOL4 has
provided the drule \texttt{ISPECL}, which takes a list of terms and a
theorem, internally it calls \texttt{ISPEC} repeatedly on each
universally quantified variables in the theorem:
\begin{lstlisting}
> val t2 = ISPECL [``prefix (label (name "a"))
                         (prefix (label (name "b")) nil)``,
		    ``label (name "a")``,
		    ``prefix (label (name "b")) nil``,
		    ``prefix (label (name "b"))
                         (prefix (label (name "a")) nil)``]
		   SUM1;
val t2 =
   |- label (name "a")..label (name "b")..nil
   --label (name "a")->
   label (name "b")..nil
  ==>
   label (name "a")..label (name "b")..nil +
   label (name "b")..label (name "a")..nil
   --label (name "a")->
   label (name "b")..nil:
   thm
\end{lstlisting}

Now if we see theorem \texttt{t1} as $A$, then \texttt{t2} looks like
$A \Rightarrow B$. Now we're ready to use HOL4's primitive inference
rule \texttt{MP} (Modus Ponens) to get $B$ from $A$ and $A \Rightarrow
B$:
\begin{lstlisting}
> MP t2 t1;
val it =
   |- label (name "a")..label (name "b")..nil +
   label (name "b")..label (name "a")..nil
   --label (name "a")->
   label (name "b")..nil:
   thm
\end{lstlisting}
This is exactly the target theorem we wanted to prove (or
verify). Putting above code together, we can write down the following
code piece in Standard ML as the procedure to construct out the
theorem:
\begin{lstlisting}
local
    val t1 = ISPEC ``label (name "a")``
                 (ISPEC ``prefix (label (name "b")) nil`` PREFIX)
    and t2 = SPECL [``prefix (label (name "a"))
                                 (prefix (label (name "b")) nil)``,
		    ``label (name "a")``,
		    ``prefix (label (name "b")) nil``,
		    ``prefix (label (name "b"))
                                 (prefix (label (name "a")) nil)``]
		   SUM1;
in
    val ex1 = save_thm ("ex1", MP t2 t1)
end;
\end{lstlisting}

In theory, any formalized theorem in HOL4 can be constructed
in this primitive way manually. However it's quite inconvenient to supply large amount of
specialized parameters to build theorems like above \texttt{t2}. A
slightly smarter way is to use \texttt{MATCH_MP} directly on
\texttt{SUM} and \texttt{t1}, HOL4's drule \texttt{MATCH_MP} will do
Modus Ponens with automatic matching:
\begin{lstlisting}
> val t2 = MATCH_MP SUM1 t1;
val t2 =
   |- !E'.
     label (name "a")..label (name "b")..nil + E'
     --label (name "a")->
     label (name "b")..nil:
   thm
\end{lstlisting}
This theorem looks almost the same as the target theorm, except for
the universally quantified variable $E'$, which takes any CCS
term. Now the only thing left is to specialize it to $b.a.0$:
\begin{lstlisting}
> ISPEC ``prefix (label (name "b"))
                 (prefix (label (name "a")) nil)`` t2;
val it =
   |- label (name "a")..label (name "b")..nil +
   label (name "b")..label (name "a")..nil
   --label (name "a")->
   label (name "b")..nil:
   thm
\end{lstlisting}

The third way to prove this simple theorem is to prove it from
backward. In other words, we first put the final theorem as a
\emph{goal} to prove, then we apply possible theorems to reduce the
goal to smaller (easier) sub-goals, until we arrive to basic logic facts.

Using HOL4's interactive proof management facility, we can just write down the
target theorem and use command ``\texttt{g}'' to put it into the proof
manager as the initial goal:
\begin{lstlisting}
> g `TRANS (sum (prefix (label (name "a"))
                                 (prefix (label (name "b")) nil))
               (prefix (label (name "b"))
                         (prefix (label (name "a")) nil)))
	  (label (name "a"))
	  (prefix (label (name "b")) nil)`;
val it =
   Proof manager status: 1 proof.
1. Incomplete goalstack:
     Initial goal:

     label (name "a")..label (name "b")..nil +
     label (name "b")..label (name "a")..nil
     --label (name "a")->
     label (name "b")..nil
:
   proofs
\end{lstlisting}

To finish the proof, we need to apply the so-called \emph{tacticals},
which translates current goal into new sub-goals. A tactical can be
considered as the reverse object of its correspond derivation rules
(or primitive inference rules). From previous forward proof of the
same theorem, we have known that rules \texttt{SUM1} and \texttt{PREFIX} must
be used. For backward proofs, the order of applying them must be
reverted too.

To benefit from \texttt{SUM1} (or other rules generated from
\texttt{Hol_reln}), the key tactical here is
\texttt{MATCH_MP_TAC}. For any goal $A$, if we know there's a theorem
with forms like $\forall x. B \Rightarrow A$, then
\texttt{MATCH_MP_TAC} could translated this goal into single sub-goal
$B$. Now we apply this tactical using command ``\texttt{e}'':
\begin{lstlisting}
> e (MATCH_MP_TAC SUM1);
OK..
1 subgoal:
val it =
   
label (name "a")..label (name "b")..nil
--label (name "a")->
label (name "b")..nil
:
   proof
\end{lstlisting}
As we expected, now the new goal is simplier: the outside \texttt{sum}
has been removed by rule \texttt{SUM1}. Now we can see this new goal
looks just like our axiom \texttt{PREFIX}, with universally quantified
variables specialized to certain terms. In another word, it's exactly
the same as previous intermediate theorem \texttt{t1}.

To finish the proof, we can either benefit from \texttt{t1}, using
tactical \texttt{ACCEPT_TAC} which simply take a theorem and compare it
with current goal, and when they're the same,  the proof is finished:
\begin{lstlisting}
> e (ACCEPT_TAC t1);
OK..

Goal proved.
|- label (name "a")..label (name "b")..nil
   --label (name "a")->
   label (name "b")..nil
val it =
   Initial goal proved.
|- label (name "a")..label (name "b")..nil +
   label (name "b")..label (name "a")..nil
   --label (name "a")->
   label (name "b")..nil:
   proof
\end{lstlisting}
Of course, if there's no theorem \texttt{t1}, we can also define it
freshly using previous method (call \texttt{ISPEC} or \texttt{ISPECL} on
\texttt{PREFIX}) and then apply it with \texttt{ACCEPT_TAC}. But there's
another better way: we can ask HOL4 to try to \emph{rewrite} current
goal with \texttt{PREFIX} referenced. In this way, HOL4 will rewrite the
goal into \texttt{T}, the logical truth, and the proof is also
finished: (before trying this new way, we can use command ``\texttt{b()}'' to
go back to last step)
\begin{lstlisting}
> b();
val it =
   
label (name "a")..label (name "b")..nil
--label (name "a")->
label (name "b")..nil

:
   proof
> e (REWRITE_TAC [PREFIX]);
OK..

Goal proved.
|- label (name "a")..label (name "b")..nil
   --label (name "a")->
   label (name "b")..nil
val it =
   Initial goal proved.
|- label (name "a")..label (name "b")..nil +
   label (name "b")..label (name "a")..nil
   --label (name "a")->
   label (name "b")..nil:
   proof
\end{lstlisting}

Rewriting is one of the most common used proof techniques, and the
tactical \texttt{REWRITE_TAC} is the most common used rewriting
tactical. It can actually take a list of reference theorems, and it
will repeatedly rewrite current goal until the rewriting process
converge.\footnote{Thus it's possible the process diverges and goes
  into infinite loops. In HOL4, there're many other rewriting tacticals with
  slightly different features.} Usually these theorems are equation theorems like
$A = B$, and whenever there's a $A$ in the goal, it becomes $B$. In
our case, \texttt{PREFIX} can be seen as \HOLinline{\HOLSymConst{\HOLTokenForall{}}\HOLBoundVar{E} \HOLBoundVar{u}. \HOLBoundVar{u}\HOLSymConst{..}\HOLBoundVar{E} \HOLTokenTransBegin\HOLBoundVar{u}\HOLTokenTransEnd \HOLBoundVar{E} \HOLSymConst{\HOLTokenEquiv{}} \HOLConst{T}}, so the rewriting result is \texttt{T}, the logical truth.

In HOL4, a formal proof is just a normal Standard ML code file, the
proof script is just a single function call on
\texttt{store_thm}. Above proof, although is done by interactive
commands, finally we should write it down into a piece of code, and
give the theorem a name. Here it is:
\begin{lstlisting}
(* (a.b.0 + b.a.0) --a-> (b.0) *)
val ex1'' = store_thm ("ex1''",
  ``TRANS (sum (prefix (label (name "a"))
                         (prefix (label (name "b")) nil))
               (prefix (label (name "b"))
                         (prefix (label (name "a")) nil)))
	  (label (name "a"))
	  (prefix (label (name "b")) nil)``,
    MATCH_MP_TAC SUM1
 >> REWRITE_TAC [PREFIX]);
\end{lstlisting}

Here, the symbol ``\texttt{> >}'' is an abbrevation of HOL's tactical
\texttt{THEN}. Sometimes we also use double-backslashs at the end of a
line, it's also an abbrevation of \texttt{THEN}.  The so-called
``tacticals'' have the type
\texttt{tactic -> tactic}, they're for connecting and combining
multiple tactics into a single big one. Thus in above code, the
whole proof step is nothing but a single parameter to ML function
\texttt{store_thm}.

Goal-directed proofs, when written carefully and friendly, even with
some extra code comments, can be human-readable. Actually from above
code, we can see clearly the following information:
\begin{enumerate}
\item The theorem name;
\item The theorem contents;
\item The proof script, including the two key theorems
  (\texttt{SUM1} and \texttt{PREFIX}).
\end{enumerate}

Given such a formal proof, the convinciblity of the correctness of
theorem comes from the following facts:
\begin{enumerate}
\item the theorem proving tools (HOL4 here) is a reliable software
  with minimal, verified logical kernel.
\item our definition of CCS inference rules are clear, simple, and
  same as in the textbook.
\item there's no other axiom defined and involved in the proving
  process.
\item by replaying the proof process (or simply watching it), in
  theory we can construct a corresponding \emph{informal} proof on
  pencil and paper. (Thus the only purpose of using software is to
  finally get this \emph{informal} proof)
\end{enumerate}

The formal proof of \texttt{(a.b.0 + b.a.0) --b-> (a.0)} is similar,
the only difference is to use \texttt{SUM2} instead of \texttt{SUM1}:
\begin{lstlisting}
(* (a.b.0 + b.a.0) --b-> (a.0) *)
val ex2 = store_thm ("ex2",
  ``TRANS (sum (prefix (label (name "a"))
                         (prefix (label (name "b")) nil))
               (prefix (label (name "b"))
                         (prefix (label (name "a")) nil)))
	  (label (name "b"))
	  (prefix (label (name "a")) nil)``,
    MATCH_MP_TAC SUM2
 >> REWRITE_TAC [PREFIX]);
\end{lstlisting}

From now on, we'll only use backward proof techniques, and when
necessary, we still have to construct intermediate theorems in forward
way and use them directly as input of some tacticals.

\subsection{Bigger examples}

In last section we have shown that the tactical \texttt{MATCH_MP_TAC}
is the main tool to benefit from the inference rules we defined for
CCS. However just using \texttt{MATCH_MP_TAC} is not enough to do all
kinds of CCS/SOS derivations. Now we formalize a bigger derivation with more rules involved, and by
showing the formal proof we introduce some other important tacticals.

\begin{prooftree}
\AxiomC{}
\LeftLabel{(Pref)}
\UnaryInfC{$a.c.0 \overset{a}\longrightarrow c.0$}

\AxiomC{}
\LeftLabel{(Pref)}
\UnaryInfC{$\bar{a}.0 \overset{\bar{a}}\longrightarrow 0$}
\LeftLabel{($\text{Sum}_1$)}
\UnaryInfC{$\bar{a}.0 + c.0 \overset{\bar{a}}\longrightarrow 0$}
\LeftLabel{(Com)}
\BinaryInfC{$a.c.0 | \bar{a}.0 + c.0 \overset{\tau}\longrightarrow
  c.0 | 0$}
\LeftLabel{(Res)}
\UnaryInfC{$(\nu c)(a.c.0 | \bar{a}.0 + c.0) \overset{\tau}\longrightarrow
  (\nu c)(c.0 | 0)$}
\end{prooftree}
Here is the formal proof of above derivations:
\begin{lstlisting}
val ex3 = store_thm ("ex3",
  ``TRANS (restr { name "c" }
		 (par (prefix (label (name "a"))
			      (prefix (label (name "c")) nil))
		      (sum (prefix (label (coname "a")) nil)
			   (prefix (label (name "c")) nil))))
	  tau
	  (restr { name "c" }
		 (par (prefix (label (name "c")) nil) nil))``,
    MATCH_MP_TAC RESTR
 >> RW_TAC std_ss []
 >> MATCH_MP_TAC COM
 >> EXISTS_TAC ``name "a"``
 >> CONJ_TAC (* 2 sub-goals here *)
 >- REWRITE_TAC [PREFIX]
 >> MATCH_MP_TAC SUM1
 >> REWRITE_TAC [PREFIX, CCS_COMPL_def]);
\end{lstlisting}

When doing an informal proof for above theorem, it suffices to show
that, inference rules \texttt{RESTR}, \texttt{COM}, and two
\texttt{PREFIX}s must be used sequentially. However in the formal proof
things are not that simple. Let's see what happened after the first
tactical \texttt{(MATCH_MP_TAC RESTR)}:
\begin{lstlisting}
1 subgoal:
val it =
   
?l.
  label (name "a")..label (name "c")..nil ||
  (label (coname "a")..nil + label (name "c")..nil)
  --tau->
  label (name "c")..nil || nil
  /\
  ((tau = tau) \/
   (tau = label l) /\ ~(l IN {name "c"}) /\ ~(COMPL l IN {name "c"}))
:
   proof
\end{lstlisting}
Here, beside the CCS transition with outter \texttt{restr} removed, we
also got an extra part starting with \texttt{(tau = tau)}. This is
 reasonable, because we can't freely remove the outter \texttt{restr},
 unless the current transition action ($\tau$ here) wasn't restricted.
There's also a bounded existential variable $l$, but we can safely
ignore it in this case.

Before applying next rule \texttt{COM}, we must simplify the current goal
 and completely remove these extra parts. Fortunately the first part
 \HOLinline{\HOLSymConst{\ensuremath{\tau}} \HOLSymConst{=} \HOLSymConst{\ensuremath{\tau}}} is always true, so the rest part in the term is not
 important any more. Here we actually need multiple basic logical
 theorems, but writing down the detailed proofs are not rewarding to
 people who is trying to understand the \emph{essential} things in
 this proof. Thus we want HOL4 to simplify the current goal using all
 possible basic logic formulae, which we don't care about the
 details. The tactical \texttt{RW_TAC} with the so-called \emph{simp
   set}, \texttt{std_ss} (standard simplification rule set) does
 exactly this trick. With this tactical we successfully simplified
 current goal into the desired one\footnote{the 3rd empty list
   indicates no other extra special rewriting theorems.}:
\begin{lstlisting}
> e (RW_TAC std_ss []);
OK..
1 subgoal:
val it =
   
label (name "a")..label (name "c")..nil ||
(label (coname "a")..nil + label (name "c")..nil)
--tau->
label (name "c")..nil || nil
:
   proof
\end{lstlisting}

Next tricky thing happens after applying the rule \texttt{COM}:
\begin{lstlisting}
> e (MATCH_MP_TAC COM);
OK..
1 subgoal:
val it =
   
?l.
  label (name "a")..label (name "c")..nil
  --label l->
  label (name "c")..nil /\
  label (coname "a")..nil + label (name "c")..nil
  --label (COMPL l)->
  nil
:
   proof
\end{lstlisting}
Here we got an new goal with existential quatified variable $l$ as the
transition label. This is always the case when appling rule
\texttt{COM}. If we take a closer look at rule \texttt{COM}
\begin{alltt}
  \HOLTokenTurnstile{} \HOLFreeVar{E} \HOLTokenTransBegin\HOLConst{label} \HOLFreeVar{l}\HOLTokenTransEnd \HOLFreeVar{E\sb{\mathrm{1}}} \HOLSymConst{\HOLTokenConj{}} \HOLFreeVar{E\sp{\prime}} \HOLTokenTransBegin\HOLConst{label} (\HOLConst{COMPL} \HOLFreeVar{l})\HOLTokenTransEnd \HOLFreeVar{E\sb{\mathrm{2}}} \HOLSymConst{\HOLTokenImp{}}
   \HOLFreeVar{E} \HOLSymConst{\ensuremath{\parallel}} \HOLFreeVar{E\sp{\prime}} \HOLTokenTransBegin\HOLSymConst{\ensuremath{\tau}}\HOLTokenTransEnd \HOLFreeVar{E\sb{\mathrm{1}}} \HOLSymConst{\ensuremath{\parallel}} \HOLFreeVar{E\sb{\mathrm{2}}}
\end{alltt}
we can see that, the two transition labels, $l$ and $\bar{l}$, only
appear in premiss of the rule. This means, if we went from the
conclusion back to the premiss, we must make a guess about the
transition labels. And for complex CCS terms, such a guess may
required a deep look inside the remain terms. In the terminalogy of
sequent calculus, the rule \texttt{COM} is actually a \emph{cut},
which blocks the automatic inference process.  When doing
semi-automatic theorem proving, what we need is to make a choice and
instantiate the existential variables. In current case, it's easy to
see that, to make further simplification of current goals, we must
choose the label $l$ to be $a$. This is how the tactical
\texttt{EXISTS_TAC} gets used:
\begin{lstlisting}
> e (EXISTS_TAC ``name "a"``);
OK..
1 subgoal:
val it =
   
label (name "a")..label (name "c")..nil
--label (name "a")->
label (name "c")..nil
 /\
label (coname "a")..nil + label (name "c")..nil
--label (COMPL (name "a"))->
nil
:
   proof
\end{lstlisting}
Now we got a new sub-goal with forms like $A \wedge  B$. But the part
$A$ we can easily prove with a simple rewriting using axiom
\texttt{PREFIX}. To actually do this, we need to break the current goal
into two sub-goals. the tactical \texttt{CONJ_TAC} does exactly this
job:
\begin{lstlisting}
> e (CONJ_TAC);
OK..
2 subgoals:
val it =
   
label (coname "a")..nil + label (name "c")..nil
--label (COMPL (name "a"))->
nil



label (name "a")..label (name "c")..nil
--label (name "a")->
label (name "c")..nil


2 subgoals
:
   proof
\end{lstlisting}

Noticed that, when multiple goals appears, the lowest goal on the
screen is the current sub-goal. We already know how to prove this
sub-goal:
\begin{lstlisting}
> e (REWRITE_TAC [PREFIX]);
OK..

Goal proved.
|- label (name "a")..label (name "c")..nil
   --label (name "a")->
   label (name "c")..nil

Remaining subgoals:
val it =
   
label (coname "a")..nil + label (name "c")..nil
--label (COMPL (name "a"))->
nil

:
   proof
\end{lstlisting}

To prove the rest goal, first we need to use \texttt{SUM1} to remove
the right part of the sum:
\begin{lstlisting}
> e (MATCH_MP_TAC SUM1);
OK..
1 subgoal:
val it =
   
label (coname "a")..nil --label (COMPL (name "a"))-> nil

:
   proof
\end{lstlisting}

Then we could do rewriting again using
\texttt{PREFIX}, however there's one problem: the transition label is
not the same as
\texttt{label (coname "a")}, instead, it's \texttt{label (COMPL (name  "a"))}.
\texttt{COMPL} is a function for converting actions and their
corresponding co-actions. It has the following definition:
\begin{lstlisting}
(* Define the complement of a label, COMPL: Label -> Label. *)
val CCS_COMPL_def = Define `(COMPL (name s) = (coname s)) /\
			(COMPL (coname s) = (name s))`;
\end{lstlisting}
Any definition is equtional theorem, therefore is accepted by
HOL's rewriting system. Thus, to finish the whole proof, we need to
rewrite the current goal using both \texttt{PREFIX} and
\texttt{CCS_COMPL_def}:
\begin{lstlisting}
> e (REWRITE_TAC [PREFIX, CCS_COMPL_def]);
OK..

Goal proved.
|- label (coname "a")..nil --label (COMPL (name "a"))-> nil

...

val it =
   Initial goal proved.
|- restr {name "c"}
     (label (name "a")..label (name "c")..nil ||
      (label (coname "a")..nil + label (name "c")..nil))
   --tau->
   restr {name "c"} (label (name "c")..nil || nil):
   proof
\end{lstlisting}

With these new proving tecniques, we're ready to prove CCS transition
theorems using any inference rules.

\cleardoublepage

%%%% -*- Mode: LaTeX -*-

\chapter{Calculus of Communicating Systems}

In this chapter we describe a formalization of Milner's \emph{Calculus
  of Communicating Systems} in HOL theorem prover (HOL4).
It covers basic definitions of CCS and its transition behaviors
based on Structural Operational Semantics (SOS), the concepts and
properties
of strong bisimulation equivalence ($\sim$), weak bisimulation
equivalence ($\approx$) and observational congruence ($\approx^c$,
also called \emph{rooted weak bisimulation equivalence}),
together with their relationships.
We have also formalized the Expansion Law, Hennessy Lemma, Deng Lemma
and several versions of the ``coarsest congruence
contained in $\approx$'' theorem.

The work was initially based on a porting of the old work by Monica
Nesi using Hol88 theorem prover, then the author has made some
modifications and improvements.

\section{Labels and Actions}

In most literature, there's no difference between \emph{Actions} and
\emph{Labels}. A  labeled transition systems (LTS for short) is a
triple $TS = (Q, A, \rightarrow)$ in which $A$ is the 
union of a countable set of input actions $\mathscr{L}$ and output
actions (co-actions) $\overline{\mathscr{L}}$ plus a special invisible
action $\tau \notin \mathscr{L} \cup \overline{\mathscr{L}}$.

In the formalization of CCS, however, it's better to have two distinct
types: the type \HOLinline{\ensuremath{\beta} \HOLTyOp{Label}} ($\beta$ is a type variable.) for visible actions, divided by input
and output actions,
 and the type \HOLinline{\ensuremath{\beta} \HOLTyOp{Action}} is the
union of all visible and invisible actions.  This is a better approach because some
constructions in CCS only accept visible actions as valid parameters,
e.g. the restriction operator. And by having a dedicated type for just
visible actions 
we can directly guarantee the exclusion of invisible actions (because otherwise it's a type mismatch).

In HOL4, the type \HOLinline{\ensuremath{\beta} \HOLTyOp{Label}} can be defined as a simple datatype (using
HOL's datatype package) with a type
variable $\beta$ representing the underlying label types: (the first
type variable $\alpha$ is reserved for other use)
\begin{lstlisting}
Datatype `Label = name 'b | coname 'b`;
\end{lstlisting}
Defined in this way, it's possible to precisely control the
cardinality of labels available for the processes in question. In the
earliest formalization by Monica Nesi, the type variable in above
definition was forcedly instantiated with HOL's \HOLinline{\HOLTyOp{string}} type.\footnote{In her later developments for
Value-Passing CCS (\cite{Nesi:2017wo}) such a limitation was removed,
and one more type variable $\gamma$ was introduced for the type of
indexing set.} For practical 
purposes it's enough to use just strings as labels, but for large
models generated automatically it's easier to use other label types,
e.g. natural number (\HOLinline{\HOLTyOp{num}}).

Noticed that, in HOL every type must
have at least one element. Thus no matter how ``small'' the type
$\beta$ is, there's at least two visible actions (one input and one output) available to the
process. Some deep theorems (e.g. the full version of ``coarsest
congruence contained in $\approx$'' must assume the existence of at
least one visible action for the given processes, such assumptions
can be removed when we're trying to formalize it. Our type \HOLinline{\ensuremath{\beta} \HOLTyOp{Label}} is inductive but not
recursive. What's immediately available from above definitions, is
some supporting theorems, for example: (We should keep in mind that, using a \emph{datatype} is actually using its
supporting theorems.)
\begin{alltt}
Label_induction:
\HOLTokenTurnstile{} (\HOLSymConst{\HOLTokenForall{}}\HOLBoundVar{b}. \HOLFreeVar{P} (\HOLConst{name} \HOLBoundVar{b})) \HOLSymConst{\HOLTokenConj{}} (\HOLSymConst{\HOLTokenForall{}}\HOLBoundVar{b}. \HOLFreeVar{P} (\HOLConst{coname} \HOLBoundVar{b})) \HOLSymConst{\HOLTokenImp{}} \HOLSymConst{\HOLTokenForall{}}\HOLBoundVar{L}. \HOLFreeVar{P} \HOLBoundVar{L}
Label_nchotomy:
\HOLTokenTurnstile{} (\HOLSymConst{\HOLTokenExists{}}\HOLBoundVar{b}. \HOLFreeVar{LL} \HOLSymConst{=} \HOLConst{name} \HOLBoundVar{b}) \HOLSymConst{\HOLTokenDisj{}} \HOLSymConst{\HOLTokenExists{}}\HOLBoundVar{b}. \HOLFreeVar{LL} \HOLSymConst{=} \HOLConst{coname} \HOLBoundVar{b}
Label_distinct:
\HOLTokenTurnstile{} \HOLConst{name} \HOLFreeVar{a} \HOLSymConst{\HOLTokenNotEqual{}} \HOLConst{coname} \HOLFreeVar{a\sp{\prime}}
Label_11:
\HOLTokenTurnstile{} (\HOLSymConst{\HOLTokenForall{}}\HOLBoundVar{a} \HOLBoundVar{a\sp{\prime}}. (\HOLConst{name} \HOLBoundVar{a} \HOLSymConst{=} \HOLConst{name} \HOLBoundVar{a\sp{\prime}}) \HOLSymConst{\HOLTokenEquiv{}} (\HOLBoundVar{a} \HOLSymConst{=} \HOLBoundVar{a\sp{\prime}})) \HOLSymConst{\HOLTokenConj{}}
   \HOLSymConst{\HOLTokenForall{}}\HOLBoundVar{a} \HOLBoundVar{a\sp{\prime}}. (\HOLConst{coname} \HOLBoundVar{a} \HOLSymConst{=} \HOLConst{coname} \HOLBoundVar{a\sp{\prime}}) \HOLSymConst{\HOLTokenEquiv{}} (\HOLBoundVar{a} \HOLSymConst{=} \HOLBoundVar{a\sp{\prime}})
\end{alltt}

The type \HOLinline{\ensuremath{\beta} \HOLTyOp{Action}} must contain all elements from the type
\HOLinline{\ensuremath{\beta} \HOLTyOp{Label}}, plus
the invisible action ($\tau$).  Previously this was defined as another
simple datatype:
\begin{lstlisting}
Datatype `Action = tau | label Label`;
\end{lstlisting}
But recently the author has realized that, it's actually very natural to use HOL's
optionTheory in the definition of Action, because the theory
\texttt{option}
 defines a type operator \texttt{option} that `lifts' its argument type, creating
a type with all of the values of the argument and one other, specially distinguished
value.   For us, that `specially distinguished value' is just
$\tau$. The new definition of Action type doesn't contain any new
logical constants, it's nothing but a type abbreviation plus some
overloading on existing constants:
\begin{lstlisting}
type_abbrev ("Action", ``:'b Label option``);

overload_on ("tau",    ``NONE :'b Action``);
overload_on ("label",  ``SOME :'b Label -> 'b Action``);
\end{lstlisting}

By type instantiation of existing theorems provided by HOL's
\texttt{optionTheory}, now we can easily get the following supporting theorems
for the Action type:
\begin{alltt}
Action_induction:
\HOLTokenTurnstile{} \HOLFreeVar{P} \HOLSymConst{\ensuremath{\tau}} \HOLSymConst{\HOLTokenConj{}} (\HOLSymConst{\HOLTokenForall{}}\HOLBoundVar{a}. \HOLFreeVar{P} (\HOLConst{label} \HOLBoundVar{a})) \HOLSymConst{\HOLTokenImp{}} \HOLSymConst{\HOLTokenForall{}}\HOLBoundVar{x}. \HOLFreeVar{P} \HOLBoundVar{x}
Action_nchotomy:
\HOLTokenTurnstile{} (\HOLFreeVar{opt} \HOLSymConst{=} \HOLSymConst{\ensuremath{\tau}}) \HOLSymConst{\HOLTokenDisj{}} \HOLSymConst{\HOLTokenExists{}}\HOLBoundVar{x}. \HOLFreeVar{opt} \HOLSymConst{=} \HOLConst{label} \HOLBoundVar{x}
Action_distinct:
\HOLTokenTurnstile{} \HOLSymConst{\ensuremath{\tau}} \HOLSymConst{\HOLTokenNotEqual{}} \HOLConst{label} \HOLFreeVar{x}
Action_11:
\HOLTokenTurnstile{} (\HOLConst{label} \HOLFreeVar{x} \HOLSymConst{=} \HOLConst{label} \HOLFreeVar{y}) \HOLSymConst{\HOLTokenEquiv{}} (\HOLFreeVar{x} \HOLSymConst{=} \HOLFreeVar{y})
\end{alltt}

The main operation on the types \HOLinline{\ensuremath{\beta} \HOLTyOp{Label}} and \HOLinline{\ensuremath{\beta} \HOLTyOp{Action}}
is \texttt{COMPL} for getting their complement actions: (for convenience we
also define the complement of $\tau$ as itself)
\begin{alltt}
\HOLTokenTurnstile{} (\HOLSymConst{\HOLTokenForall{}}\HOLBoundVar{s}. \HOLConst{COMPL} (\HOLConst{name} \HOLBoundVar{s}) \HOLSymConst{=} \HOLConst{coname} \HOLBoundVar{s}) \HOLSymConst{\HOLTokenConj{}}
   \HOLSymConst{\HOLTokenForall{}}\HOLBoundVar{s}. \HOLConst{COMPL} (\HOLConst{coname} \HOLBoundVar{s}) \HOLSymConst{=} \HOLConst{name} \HOLBoundVar{s}
\HOLTokenTurnstile{} (\HOLSymConst{\HOLTokenForall{}}\HOLBoundVar{l}. \HOLConst{COMPL} (\HOLConst{label} \HOLBoundVar{l}) \HOLSymConst{=} \HOLConst{label} (\HOLConst{COMPL} \HOLBoundVar{l})) \HOLSymConst{\HOLTokenConj{}} (\HOLConst{COMPL} \HOLSymConst{\ensuremath{\tau}} \HOLSymConst{=} \HOLSymConst{\ensuremath{\tau}})
\end{alltt}
As we know \HOLinline{\ensuremath{\beta} \HOLTyOp{Label}} and \HOLinline{\ensuremath{\beta} \HOLTyOp{Action}} are different types,
the \texttt{COMPL} operator on them are actually overloaded operator
of \texttt{COMPL_LAB} and \texttt{COMPL_ACT}, the complement operator
for \HOLinline{\ensuremath{\beta} \HOLTyOp{Label}} and \HOLinline{\ensuremath{\beta} \HOLTyOp{Action}}.

The key theorem about \HOLinline{\ensuremath{\beta} \HOLTyOp{Label}} says that, doing complements
twice for the same label gets the label itself:
\begin{alltt}
COMPL_COMPL_LAB:
\HOLTokenTurnstile{} \HOLConst{COMPL} (\HOLConst{COMPL} \HOLFreeVar{l}) \HOLSymConst{=} \HOLFreeVar{l}
\end{alltt}
There's also a similar theorem for the double-complements of
\HOLinline{\ensuremath{\beta} \HOLTyOp{Action}}.

Table \ref{tab:ccsactions} listed the notation of various actions:
\begin{table}
\begin{tabular}{|c|c|c|c|}
\hline
\textbf{Action} & \textbf{notation} & \textbf{HOL} &   \textbf{HOL
                                    (alternative)}\\
\hline
invisible action & $\tau$ &  \texttt{tau} & \HOLinline{\HOLSymConst{\ensuremath{\tau}}} \\
visible action & $l$ & \texttt{label l} & \HOLinline{\HOLConst{label} \HOLFreeVar{l}} \\
input action & $a$ & \texttt{label (name "a")} & \HOLinline{\HOLConst{In} \HOLStringLit{a}} \\
output action & $\bar{a}$ & \texttt{label (coname "a")} & \HOLinline{\HOLConst{Out} \HOLStringLit{a}} \\
\hline
\end{tabular}
   \caption{Actions used in CCS}
   \label{tab:ccsactions}
\end{table}

\subsection{Relabeling operator}

In standard literature, Relabeling operator is usually defined as an unary
substitution operator $\_ [b/a]$, which takes a unary substitution $b/a$
(hence, $a\neq b$), and a process $p$ to construct a new process
$p[b/a]$, whose semantics is that of $p$, where action $a (\bar{a})$
is turned into $b(\bar{b})$. And multi-label relabeling can be done by
appending more unary substitution operators to the existing process
after relabeling. The
order of multiple relabelings is relevant, especially when new labels introduced
in previous relabeling operation were further relabeled.

In our formalization, following the work of Monica Nesi, we support
multi-label relabeling in single 
operation, and instead of using a list of substitutions, we have
defined a new primitive type called
``\HOLinline{\ensuremath{\beta} \HOLTyOp{Relabeling}}''.  \HOLinline{\ensuremath{\beta} \HOLTyOp{Relabeling}} is a
is a bijection into a subset of functions
\HOLinline{\ensuremath{\beta} \HOLTyOp{Label} \HOLTokenTransEnd \ensuremath{\beta} \HOLTyOp{Label}}, which is called the \emph{representation} of
the type ``\HOLinline{\ensuremath{\beta} \HOLTyOp{Relabeling}}''. Not all functions of type ``\HOLinline{\ensuremath{\beta} \HOLTyOp{Label} \HOLTokenTransEnd \ensuremath{\beta} \HOLTyOp{Label}}'' are valid representations of ``\HOLinline{\ensuremath{\beta} \HOLTyOp{Relabeling}}'', but only
functions which satisfy the following property:
\begin{alltt}
\HOLTokenTurnstile{} \HOLConst{Is_Relabeling} \HOLFreeVar{f} \HOLSymConst{\HOLTokenEquiv{}} \HOLSymConst{\HOLTokenForall{}}\HOLBoundVar{s}. \HOLFreeVar{f} (\HOLConst{coname} \HOLBoundVar{s}) \HOLSymConst{=} \HOLConst{COMPL} (\HOLFreeVar{f} (\HOLConst{name} \HOLBoundVar{s}))
\end{alltt}

Noticed that, any identify function of type \HOLinline{\ensuremath{\beta} \HOLTyOp{Label} \HOLTokenTransEnd \ensuremath{\beta} \HOLTyOp{Label}}
also satisfies above property. Thus, beside specific substitutions
we wanted, all
relabeling functions are \emph{total}: they must be able to handle all other labels too (just
return the same label as input). (As we'll see later, such
requirements could reduce the two rules for relabelling into just one).

But usually it's more convenient to represent relabeling functions as
a list of substitutions of type \HOLinline{(\ensuremath{\beta} \HOLTyOp{Label} \HOLTokenProd{} \ensuremath{\beta} \HOLTyOp{Label}) \HOLTyOp{list}}. The
operator \HOLinline{\HOLConst{RELAB}} can be used to define such a relabeling
function. For instance, the term
\begin{alltt}
\HOLinline{\HOLConst{RELAB} [(\HOLConst{name} \HOLStringLit{b}\HOLSymConst{,}\HOLConst{name} \HOLStringLit{a}); (\HOLConst{name} \HOLStringLit{d}\HOLSymConst{,}\HOLConst{name} \HOLStringLit{c})]}
\end{alltt}
can be used in place of a relabeling operator
$[b/a, d/c]$, because its type is ``\HOLinline{\ensuremath{\beta} \HOLTyOp{Relabeling}}''. And it must be
understood that, all relabeling functions are total functions: for all other labels except \texttt{a} and
\texttt{c}, the substitution will be themselves (another way to
express ``No relabeling'').

Finally, having the relabeling facility defined as a multi-label
relabeling function and as part of CCS syntax, we can completely
avoid the complexity of the Syntactic Substitution (c.f. p.171 of
\cite{Gorrieri:2015jt}) which has a complicated recursive
definition\footnote{However, syntactic relabeling is still considered
  as an ``economic'' way of doing relabeling, because having one
  native CCS operator will also introduce the corresponding SOS inference rules and
  equivalence laws.} and heavily depends on some other recursive functions
like $fn(\cdot)$ (free names) and $bn(\cdot)$ (bound names) for CCS
processes (in our project, these functions are defined but not used).

\section{The CCS Datatype}

The core datatype ``\HOLinline{(\ensuremath{\alpha}, \ensuremath{\beta}) \HOLTyOp{CCS}}'' in this formalization is defined as an inductive
datatype in HOL, based on its Datatype package, as shown in Fig. \ref{fig:ccsdatatype}.

\begin{figure}[h]
\begin{lstlisting}
val _ = Datatype `CCS = nil
		      | var 'a
		      | prefix ('b Action) CCS
		      | sum CCS CCS
		      | par CCS CCS
		      | restr (('b Label) set) CCS
		      | relab CCS ('b Relabeling)
		      | rec 'a CCS`;
\end{lstlisting}
   \caption{CCS Datatype}
   \label{fig:ccsdatatype}
\end{figure}

Comparing with the original work, now the type of CCS processes has
been extended with two type variables: 
$\alpha$ and $\beta$. $\alpha$ is the type of process constants, and
$\beta$ is the type of actions. In HOL, such a higher order type is represented
as ``\HOLinline{(\ensuremath{\alpha}, \ensuremath{\beta}) \HOLTyOp{CCS}}''.  If both type variables were
instantiated as \HOLinline{\HOLTyOp{string}}, the resulting type ``\HOLinline{(\HOLTyOp{string}, \HOLTyOp{string}) \HOLTyOp{CCS}}'' is equivalent with the \texttt{CCS} datatype in the old CCS formalization.

As we have explained, there's no infinite sums: the sum operator is simply binary. Since the CCS datatype is inductively defined and has no infinite sum operator, it must be Finitary.

We have added some minimal grammar support using HOL's pretty printer, to represent CCS
processes in more readable forms (this was not available in the old work).
Table \ref{tab:ccsoperator} has listed the notation of typical CCS processes and
major operators supported by above definition:

\begin{table}[h]
\begin{tabular}{|c|c|c|c|}
\hline
\textbf{Operator} & \textbf{Notation} & \textbf{HOL} & \textbf{HOL
  (alternative)}\\
\hline
nil & $\textbf{0}$ &  \HOLinline{\HOLConst{nil}} & \HOLinline{\HOLConst{nil}} \\
Rrefix & $a.b.0$ & \texttt{prefix a (prefix b nil)} & \HOLinline{\HOLFreeVar{a}\HOLSymConst{..}\HOLFreeVar{b}\HOLSymConst{..}\HOLConst{nil}} \\
Sum & $p + q$ & \texttt{sum p q} & \HOLinline{\HOLFreeVar{p} \HOLSymConst{+} \HOLFreeVar{q}} \\
Parallel & $p \,\mid\, q$ & \texttt{par p q} & \HOLinline{\HOLFreeVar{p} \HOLSymConst{\ensuremath{\parallel}} \HOLFreeVar{q}} \\
Restriction & $(\nu L)、，p$ & \texttt{restr L p} & \HOLinline{\HOLSymConst{\ensuremath{\nu}} \HOLFreeVar{L} \HOLFreeVar{p}} \\
Constant & $A=a.A$ & \texttt{rec A (prefix a (var A))} & \HOLinline{\HOLConst{rec} \HOLFreeVar{A} (\HOLFreeVar{a}\HOLSymConst{..}\HOLConst{var} \HOLFreeVar{A})} \\
\hline
\end{tabular}
   \caption{Syntax of CCS operators}
   \label{tab:ccsoperator}
\end{table}

For Relabeling, as we described in the last section, to express ``$p[b/a]$'',
it must be written as \HOLinline{\HOLConst{relab} \HOLFreeVar{p} (\HOLConst{RELAB} [(\HOLConst{name} \HOLStringLit{b}\HOLSymConst{,}\HOLConst{name} \HOLStringLit{a})])},
which is a little long literally.

For CCS processes defined by one or more constants, in our formalization
in HOL4, all constants must be written into single term. (This is necessary for
theorem proving, because otherwise there's no way to store all
information into single variable in CCS-related theorems)  The syntax
for defining new constants is \HOLinline{\HOLConst{rec}} and the syntax to actually
use a constant is \HOLinline{\HOLConst{var}}. To see how these operators are actually
used, consider the following CCS process (the famous coffee machine
model from \cite{Gorrieri:2015jt}):
\begin{align*}
VM & \overset{def}{=} coin.(\text{ask-esp}.VM_1 + \text{ask-am}.VM_2) \\
VM_1 & \overset{def}{=} \overline{\text{esp-coffee}}.VM \\
VM_2 & \overset{def}{=} \overline{\text{am-coffee}}.VM
\end{align*}

In our formalization in HOL4, the above CCS process can be represented as
the following single term:
\begin{lstlisting}
``rec "VM"
    (In "coin"
     ..
     (In "ask-esp" .. (rec "VM1" (Out "esp-coffee"..var "VM")) +
      In "ask-am" .. (rec "VM2" (Out "am-coffee"..var "VM"))))``
\end{lstlisting}
That is, for the first time a new constant appears, use \HOLinline{\HOLConst{rec}}
with the name of constants as string to ``declare'' it; when any
constant appears again, use \HOLinline{\HOLConst{var}} to access it.

Finally, although not part of the formal definition, the
\textbf{if-then-else} construct from value-passing CCS is automatically supported by
HOL. This is because, for any boolean value $b$ and two terms $t_1$ and $t_2$ of type
$\alpha$, the term \HOLinline{\HOLKeyword{if} \HOLFreeVar{b} \HOLKeyword{then} \HOLFreeVar{t\sb{\mathrm{1}}} \HOLKeyword{else} \HOLFreeVar{t\sb{\mathrm{2}}}} has also the type
$\alpha$. Thus the conditional term can legally appears inside other
CCS processes as a sub-process. We'll see in next section that it's
necessary for handling transitions of CCS processes containing
constants.

\section{Transition Semantics}

The transition semantics of CCS processes were defined by the following
Structural Operational Semantics (SOS for short) rules:
\begin{multicols}{2}

\begin{prooftree}
\AxiomC{}
\LeftLabel{(Perf)}
\UnaryInfC{$\mu.p \overset{\mu}\longrightarrow p$}
\end{prooftree}

\begin{prooftree}
\AxiomC{$q[\text{\texttt{rec} } x. q\enspace / \enspace x] \overset{\mu}\longrightarrow r$}
\LeftLabel{(Rec)} % Cons
\UnaryInfC{$\text{\texttt{rec} } x. q \overset{\mu}\longrightarrow r$}
\end{prooftree}

\begin{prooftree}
\AxiomC{$p \overset{\mu}\longrightarrow p'$}
\LeftLabel{($\text{Sum}_1$)}
\UnaryInfC{$p + q \overset{\mu}\longrightarrow p'$}
\end{prooftree}

\begin{prooftree}
\AxiomC{$q \overset{\mu}\longrightarrow q'$}
\LeftLabel{($\text{Sum}_2$)}
\UnaryInfC{$p + q \overset{\mu}\longrightarrow q'$}
\end{prooftree}

\begin{prooftree}
\AxiomC{$p \overset{\mu}\longrightarrow p'$}
\LeftLabel{($\text{Par}_1$)}
\UnaryInfC{$p | q \overset{\mu}\longrightarrow p' | q$}
\end{prooftree}

\begin{prooftree}
\AxiomC{$q \overset{\mu}\longrightarrow q'$}
\LeftLabel{($\text{Par}_2$)}
\UnaryInfC{$p | q \overset{\mu}\longrightarrow p | q'$}
\end{prooftree}

\begin{prooftree}
\AxiomC{$p \overset{\alpha}\longrightarrow p'$}
\AxiomC{$q \overset{\bar{\alpha}}\longrightarrow q'$}
\LeftLabel{($\text{Par}_3$)} % Com
\BinaryInfC{$p | q \overset{\tau}\longrightarrow p' | q'$}
\end{prooftree}

\begin{prooftree}
\AxiomC{$p \overset{\mu}\longrightarrow p'$}
\LeftLabel{(Res)}
\RightLabel{$\mu \neq a,\bar{a}$}
\UnaryInfC{$(\nu a)p \overset{\mu}\longrightarrow (\nu a) p'$}
\end{prooftree}
\end{multicols}

Besides, we have a rule for relabeling:
\begin{center}
\begin{prooftree}
\AxiomC{$p \overset{\mu}\longrightarrow p'$}
\LeftLabel{(Rel)}
\UnaryInfC{$p[f] \overset{f(\mu)}\longrightarrow (p'[f]$}
\end{prooftree}
\end{center}

In some literatures \cite{Gorrieri:2015jt}, the rule $\text{Par}_3$ is called ``Com'' (communication), and
the rule ``Rec'' (in a different form based on separated agent
definitions) is also called ``Cons'' (constants). (Here we 
have preserved the rule names in the HOL88 work, because it's easier to
locate for their names in the proof scripts.)

From the view of theorem prover (or just first-order logic), these
inference rules are nothing but an \emph{inductive 
  definition} on 3-ary relation \HOLinline{\HOLConst{TRANS}} (with compact
representation \texttt{--()->})  of type \HOLinline{(\ensuremath{\alpha}, \ensuremath{\beta}) \HOLTyOp{transition}}, generated by HOL4's function
\texttt{Hol_reln} \cite{Anonymous:Bxz1gZYL}. Then we break them into separated theorems as
primitive inference rules\footnote{They're considered as the axioms in
  our logic system, however they're not defined directly as
  axioms. HOL makes sure in such cases the logic system is still consistent.}:
\begin{alltt}
PREFIX: \HOLTokenTurnstile{} \HOLFreeVar{u}\HOLSymConst{..}\HOLFreeVar{E} \HOLTokenTransBegin\HOLFreeVar{u}\HOLTokenTransEnd \HOLFreeVar{E}
REC:    \HOLTokenTurnstile{} \HOLConst{CCS_Subst} \HOLFreeVar{E} (\HOLConst{rec} \HOLFreeVar{X} \HOLFreeVar{E}) \HOLFreeVar{X} \HOLTokenTransBegin\HOLFreeVar{u}\HOLTokenTransEnd \HOLFreeVar{E\sb{\mathrm{1}}} \HOLSymConst{\HOLTokenImp{}} \HOLConst{rec} \HOLFreeVar{X} \HOLFreeVar{E} \HOLTokenTransBegin\HOLFreeVar{u}\HOLTokenTransEnd \HOLFreeVar{E\sb{\mathrm{1}}}
SUM1:   \HOLTokenTurnstile{} \HOLFreeVar{E} \HOLTokenTransBegin\HOLFreeVar{u}\HOLTokenTransEnd \HOLFreeVar{E\sb{\mathrm{1}}} \HOLSymConst{\HOLTokenImp{}} \HOLFreeVar{E} \HOLSymConst{+} \HOLFreeVar{E\sp{\prime}} \HOLTokenTransBegin\HOLFreeVar{u}\HOLTokenTransEnd \HOLFreeVar{E\sb{\mathrm{1}}}
SUM2:   \HOLTokenTurnstile{} \HOLFreeVar{E} \HOLTokenTransBegin\HOLFreeVar{u}\HOLTokenTransEnd \HOLFreeVar{E\sb{\mathrm{1}}} \HOLSymConst{\HOLTokenImp{}} \HOLFreeVar{E\sp{\prime}} \HOLSymConst{+} \HOLFreeVar{E} \HOLTokenTransBegin\HOLFreeVar{u}\HOLTokenTransEnd \HOLFreeVar{E\sb{\mathrm{1}}}
PAR1:   \HOLTokenTurnstile{} \HOLFreeVar{E} \HOLTokenTransBegin\HOLFreeVar{u}\HOLTokenTransEnd \HOLFreeVar{E\sb{\mathrm{1}}} \HOLSymConst{\HOLTokenImp{}} \HOLFreeVar{E} \HOLSymConst{\ensuremath{\parallel}} \HOLFreeVar{E\sp{\prime}} \HOLTokenTransBegin\HOLFreeVar{u}\HOLTokenTransEnd \HOLFreeVar{E\sb{\mathrm{1}}} \HOLSymConst{\ensuremath{\parallel}} \HOLFreeVar{E\sp{\prime}}
PAR2:   \HOLTokenTurnstile{} \HOLFreeVar{E} \HOLTokenTransBegin\HOLFreeVar{u}\HOLTokenTransEnd \HOLFreeVar{E\sb{\mathrm{1}}} \HOLSymConst{\HOLTokenImp{}} \HOLFreeVar{E\sp{\prime}} \HOLSymConst{\ensuremath{\parallel}} \HOLFreeVar{E} \HOLTokenTransBegin\HOLFreeVar{u}\HOLTokenTransEnd \HOLFreeVar{E\sp{\prime}} \HOLSymConst{\ensuremath{\parallel}} \HOLFreeVar{E\sb{\mathrm{1}}}

PAR3:
\HOLTokenTurnstile{} \HOLFreeVar{E} \HOLTokenTransBegin\HOLConst{label} \HOLFreeVar{l}\HOLTokenTransEnd \HOLFreeVar{E\sb{\mathrm{1}}} \HOLSymConst{\HOLTokenConj{}} \HOLFreeVar{E\sp{\prime}} \HOLTokenTransBegin\HOLConst{label} (\HOLConst{COMPL} \HOLFreeVar{l})\HOLTokenTransEnd \HOLFreeVar{E\sb{\mathrm{2}}} \HOLSymConst{\HOLTokenImp{}}
   \HOLFreeVar{E} \HOLSymConst{\ensuremath{\parallel}} \HOLFreeVar{E\sp{\prime}} \HOLTokenTransBegin\HOLSymConst{\ensuremath{\tau}}\HOLTokenTransEnd \HOLFreeVar{E\sb{\mathrm{1}}} \HOLSymConst{\ensuremath{\parallel}} \HOLFreeVar{E\sb{\mathrm{2}}}

RESTR:
\HOLTokenTurnstile{} \HOLFreeVar{E} \HOLTokenTransBegin\HOLFreeVar{u}\HOLTokenTransEnd \HOLFreeVar{E\sp{\prime}} \HOLSymConst{\HOLTokenConj{}}
   ((\HOLFreeVar{u} \HOLSymConst{=} \HOLSymConst{\ensuremath{\tau}}) \HOLSymConst{\HOLTokenDisj{}} (\HOLFreeVar{u} \HOLSymConst{=} \HOLConst{label} \HOLFreeVar{l}) \HOLSymConst{\HOLTokenConj{}} \HOLFreeVar{l} \HOLSymConst{\HOLTokenNotIn{}} \HOLFreeVar{L} \HOLSymConst{\HOLTokenConj{}} \HOLConst{COMPL} \HOLFreeVar{l} \HOLSymConst{\HOLTokenNotIn{}} \HOLFreeVar{L}) \HOLSymConst{\HOLTokenImp{}}
   \HOLSymConst{\ensuremath{\nu}} \HOLFreeVar{L} \HOLFreeVar{E} \HOLTokenTransBegin\HOLFreeVar{u}\HOLTokenTransEnd \HOLSymConst{\ensuremath{\nu}} \HOLFreeVar{L} \HOLFreeVar{E\sp{\prime}}

RELABELING:
\HOLTokenTurnstile{} \HOLFreeVar{E} \HOLTokenTransBegin\HOLFreeVar{u}\HOLTokenTransEnd \HOLFreeVar{E\sp{\prime}} \HOLSymConst{\HOLTokenImp{}} \HOLConst{relab} \HOLFreeVar{E} \HOLFreeVar{rf} \HOLTokenTransBegin\HOLConst{relabel} \HOLFreeVar{rf} \HOLFreeVar{u}\HOLTokenTransEnd \HOLConst{relab} \HOLFreeVar{E\sp{\prime}} \HOLFreeVar{rf}
\end{alltt}

Noticed that, in the rule \texttt{REC}, a recursive function
\HOLinline{\HOLConst{CCS_Subst}} was used. It has the following definition which
depends on the conditional clause (\texttt{if .. then .. else ..}):
\begin{alltt}
\HOLTokenTurnstile{} (\HOLSymConst{\HOLTokenForall{}}\HOLBoundVar{E\sp{\prime}} \HOLBoundVar{X}. \HOLConst{CCS_Subst} \HOLConst{nil} \HOLBoundVar{E\sp{\prime}} \HOLBoundVar{X} \HOLSymConst{=} \HOLConst{nil}) \HOLSymConst{\HOLTokenConj{}}
   (\HOLSymConst{\HOLTokenForall{}}\HOLBoundVar{u} \HOLBoundVar{E} \HOLBoundVar{E\sp{\prime}} \HOLBoundVar{X}. \HOLConst{CCS_Subst} (\HOLBoundVar{u}\HOLSymConst{..}\HOLBoundVar{E}) \HOLBoundVar{E\sp{\prime}} \HOLBoundVar{X} \HOLSymConst{=} \HOLBoundVar{u}\HOLSymConst{..}\HOLConst{CCS_Subst} \HOLBoundVar{E} \HOLBoundVar{E\sp{\prime}} \HOLBoundVar{X}) \HOLSymConst{\HOLTokenConj{}}
   (\HOLSymConst{\HOLTokenForall{}}\HOLBoundVar{E\sb{\mathrm{1}}} \HOLBoundVar{E\sb{\mathrm{2}}} \HOLBoundVar{E\sp{\prime}} \HOLBoundVar{X}.
      \HOLConst{CCS_Subst} (\HOLBoundVar{E\sb{\mathrm{1}}} \HOLSymConst{+} \HOLBoundVar{E\sb{\mathrm{2}}}) \HOLBoundVar{E\sp{\prime}} \HOLBoundVar{X} \HOLSymConst{=}
      \HOLConst{CCS_Subst} \HOLBoundVar{E\sb{\mathrm{1}}} \HOLBoundVar{E\sp{\prime}} \HOLBoundVar{X} \HOLSymConst{+} \HOLConst{CCS_Subst} \HOLBoundVar{E\sb{\mathrm{2}}} \HOLBoundVar{E\sp{\prime}} \HOLBoundVar{X}) \HOLSymConst{\HOLTokenConj{}}
   (\HOLSymConst{\HOLTokenForall{}}\HOLBoundVar{E\sb{\mathrm{1}}} \HOLBoundVar{E\sb{\mathrm{2}}} \HOLBoundVar{E\sp{\prime}} \HOLBoundVar{X}.
      \HOLConst{CCS_Subst} (\HOLBoundVar{E\sb{\mathrm{1}}} \HOLSymConst{\ensuremath{\parallel}} \HOLBoundVar{E\sb{\mathrm{2}}}) \HOLBoundVar{E\sp{\prime}} \HOLBoundVar{X} \HOLSymConst{=}
      \HOLConst{CCS_Subst} \HOLBoundVar{E\sb{\mathrm{1}}} \HOLBoundVar{E\sp{\prime}} \HOLBoundVar{X} \HOLSymConst{\ensuremath{\parallel}} \HOLConst{CCS_Subst} \HOLBoundVar{E\sb{\mathrm{2}}} \HOLBoundVar{E\sp{\prime}} \HOLBoundVar{X}) \HOLSymConst{\HOLTokenConj{}}
   (\HOLSymConst{\HOLTokenForall{}}\HOLBoundVar{L} \HOLBoundVar{E} \HOLBoundVar{E\sp{\prime}} \HOLBoundVar{X}.
      \HOLConst{CCS_Subst} (\HOLSymConst{\ensuremath{\nu}} \HOLBoundVar{L} \HOLBoundVar{E}) \HOLBoundVar{E\sp{\prime}} \HOLBoundVar{X} \HOLSymConst{=} \HOLSymConst{\ensuremath{\nu}} \HOLBoundVar{L} (\HOLConst{CCS_Subst} \HOLBoundVar{E} \HOLBoundVar{E\sp{\prime}} \HOLBoundVar{X})) \HOLSymConst{\HOLTokenConj{}}
   (\HOLSymConst{\HOLTokenForall{}}\HOLBoundVar{E} \HOLBoundVar{f} \HOLBoundVar{E\sp{\prime}} \HOLBoundVar{X}.
      \HOLConst{CCS_Subst} (\HOLConst{relab} \HOLBoundVar{E} \HOLBoundVar{f}) \HOLBoundVar{E\sp{\prime}} \HOLBoundVar{X} \HOLSymConst{=}
      \HOLConst{relab} (\HOLConst{CCS_Subst} \HOLBoundVar{E} \HOLBoundVar{E\sp{\prime}} \HOLBoundVar{X}) \HOLBoundVar{f}) \HOLSymConst{\HOLTokenConj{}}
   (\HOLSymConst{\HOLTokenForall{}}\HOLBoundVar{Y} \HOLBoundVar{E\sp{\prime}} \HOLBoundVar{X}.
      \HOLConst{CCS_Subst} (\HOLConst{var} \HOLBoundVar{Y}) \HOLBoundVar{E\sp{\prime}} \HOLBoundVar{X} \HOLSymConst{=} \HOLKeyword{if} \HOLBoundVar{Y} \HOLSymConst{=} \HOLBoundVar{X} \HOLKeyword{then} \HOLBoundVar{E\sp{\prime}} \HOLKeyword{else} \HOLConst{var} \HOLBoundVar{Y}) \HOLSymConst{\HOLTokenConj{}}
   \HOLSymConst{\HOLTokenForall{}}\HOLBoundVar{Y} \HOLBoundVar{E} \HOLBoundVar{E\sp{\prime}} \HOLBoundVar{X}.
     \HOLConst{CCS_Subst} (\HOLConst{rec} \HOLBoundVar{Y} \HOLBoundVar{E}) \HOLBoundVar{E\sp{\prime}} \HOLBoundVar{X} \HOLSymConst{=}
     \HOLKeyword{if} \HOLBoundVar{Y} \HOLSymConst{=} \HOLBoundVar{X} \HOLKeyword{then} \HOLConst{rec} \HOLBoundVar{Y} \HOLBoundVar{E} \HOLKeyword{else} \HOLConst{rec} \HOLBoundVar{Y} (\HOLConst{CCS_Subst} \HOLBoundVar{E} \HOLBoundVar{E\sp{\prime}} \HOLBoundVar{X})
\end{alltt}
The idea of \HOLinline{\HOLConst{CCS_Subst}} is very close to the variable
substitution for $\lambda$-terms: it only replaced those \emph{free
  variables} with the same name as the input variable.

In HOL4, any inductive relation defined by command \texttt{Hol_reln}
will return with three (well, actually four) theorems: 1) the
rules, 2) the induction (and strong induction) theorem and 3) the ``cases''
theorem. Only with all these theorems, the relation can be precisely
defined. For example, to prove certain CCS transitions are impossible,
the following long ``cases'' theorem (which asserts that the relation is a fixed
point) must be used:
\begin{alltt}
\HOLTokenTurnstile{} \HOLFreeVar{a\sb{\mathrm{0}}} \HOLTokenTransBegin\HOLFreeVar{a\sb{\mathrm{1}}}\HOLTokenTransEnd \HOLFreeVar{a\sb{\mathrm{2}}} \HOLSymConst{\HOLTokenEquiv{}}
   (\HOLFreeVar{a\sb{\mathrm{0}}} \HOLSymConst{=} \HOLFreeVar{a\sb{\mathrm{1}}}\HOLSymConst{..}\HOLFreeVar{a\sb{\mathrm{2}}}) \HOLSymConst{\HOLTokenDisj{}} (\HOLSymConst{\HOLTokenExists{}}\HOLBoundVar{E} \HOLBoundVar{E\sp{\prime}}. (\HOLFreeVar{a\sb{\mathrm{0}}} \HOLSymConst{=} \HOLBoundVar{E} \HOLSymConst{+} \HOLBoundVar{E\sp{\prime}}) \HOLSymConst{\HOLTokenConj{}} \HOLBoundVar{E} \HOLTokenTransBegin\HOLFreeVar{a\sb{\mathrm{1}}}\HOLTokenTransEnd \HOLFreeVar{a\sb{\mathrm{2}}}) \HOLSymConst{\HOLTokenDisj{}}
   (\HOLSymConst{\HOLTokenExists{}}\HOLBoundVar{E} \HOLBoundVar{E\sp{\prime}}. (\HOLFreeVar{a\sb{\mathrm{0}}} \HOLSymConst{=} \HOLBoundVar{E\sp{\prime}} \HOLSymConst{+} \HOLBoundVar{E}) \HOLSymConst{\HOLTokenConj{}} \HOLBoundVar{E} \HOLTokenTransBegin\HOLFreeVar{a\sb{\mathrm{1}}}\HOLTokenTransEnd \HOLFreeVar{a\sb{\mathrm{2}}}) \HOLSymConst{\HOLTokenDisj{}}
   (\HOLSymConst{\HOLTokenExists{}}\HOLBoundVar{E} \HOLBoundVar{E\sb{\mathrm{1}}} \HOLBoundVar{E\sp{\prime}}. (\HOLFreeVar{a\sb{\mathrm{0}}} \HOLSymConst{=} \HOLBoundVar{E} \HOLSymConst{\ensuremath{\parallel}} \HOLBoundVar{E\sp{\prime}}) \HOLSymConst{\HOLTokenConj{}} (\HOLFreeVar{a\sb{\mathrm{2}}} \HOLSymConst{=} \HOLBoundVar{E\sb{\mathrm{1}}} \HOLSymConst{\ensuremath{\parallel}} \HOLBoundVar{E\sp{\prime}}) \HOLSymConst{\HOLTokenConj{}} \HOLBoundVar{E} \HOLTokenTransBegin\HOLFreeVar{a\sb{\mathrm{1}}}\HOLTokenTransEnd \HOLBoundVar{E\sb{\mathrm{1}}}) \HOLSymConst{\HOLTokenDisj{}}
   (\HOLSymConst{\HOLTokenExists{}}\HOLBoundVar{E} \HOLBoundVar{E\sb{\mathrm{1}}} \HOLBoundVar{E\sp{\prime}}. (\HOLFreeVar{a\sb{\mathrm{0}}} \HOLSymConst{=} \HOLBoundVar{E\sp{\prime}} \HOLSymConst{\ensuremath{\parallel}} \HOLBoundVar{E}) \HOLSymConst{\HOLTokenConj{}} (\HOLFreeVar{a\sb{\mathrm{2}}} \HOLSymConst{=} \HOLBoundVar{E\sp{\prime}} \HOLSymConst{\ensuremath{\parallel}} \HOLBoundVar{E\sb{\mathrm{1}}}) \HOLSymConst{\HOLTokenConj{}} \HOLBoundVar{E} \HOLTokenTransBegin\HOLFreeVar{a\sb{\mathrm{1}}}\HOLTokenTransEnd \HOLBoundVar{E\sb{\mathrm{1}}}) \HOLSymConst{\HOLTokenDisj{}}
   (\HOLSymConst{\HOLTokenExists{}}\HOLBoundVar{E} \HOLBoundVar{l} \HOLBoundVar{E\sb{\mathrm{1}}} \HOLBoundVar{E\sp{\prime}} \HOLBoundVar{E\sb{\mathrm{2}}}.
      (\HOLFreeVar{a\sb{\mathrm{0}}} \HOLSymConst{=} \HOLBoundVar{E} \HOLSymConst{\ensuremath{\parallel}} \HOLBoundVar{E\sp{\prime}}) \HOLSymConst{\HOLTokenConj{}} (\HOLFreeVar{a\sb{\mathrm{1}}} \HOLSymConst{=} \HOLSymConst{\ensuremath{\tau}}) \HOLSymConst{\HOLTokenConj{}} (\HOLFreeVar{a\sb{\mathrm{2}}} \HOLSymConst{=} \HOLBoundVar{E\sb{\mathrm{1}}} \HOLSymConst{\ensuremath{\parallel}} \HOLBoundVar{E\sb{\mathrm{2}}}) \HOLSymConst{\HOLTokenConj{}}
      \HOLBoundVar{E} \HOLTokenTransBegin\HOLConst{label} \HOLBoundVar{l}\HOLTokenTransEnd \HOLBoundVar{E\sb{\mathrm{1}}} \HOLSymConst{\HOLTokenConj{}} \HOLBoundVar{E\sp{\prime}} \HOLTokenTransBegin\HOLConst{label} (\HOLConst{COMPL} \HOLBoundVar{l})\HOLTokenTransEnd \HOLBoundVar{E\sb{\mathrm{2}}}) \HOLSymConst{\HOLTokenDisj{}}
   (\HOLSymConst{\HOLTokenExists{}}\HOLBoundVar{E} \HOLBoundVar{E\sp{\prime}} \HOLBoundVar{l} \HOLBoundVar{L}.
      (\HOLFreeVar{a\sb{\mathrm{0}}} \HOLSymConst{=} \HOLSymConst{\ensuremath{\nu}} \HOLBoundVar{L} \HOLBoundVar{E}) \HOLSymConst{\HOLTokenConj{}} (\HOLFreeVar{a\sb{\mathrm{2}}} \HOLSymConst{=} \HOLSymConst{\ensuremath{\nu}} \HOLBoundVar{L} \HOLBoundVar{E\sp{\prime}}) \HOLSymConst{\HOLTokenConj{}} \HOLBoundVar{E} \HOLTokenTransBegin\HOLFreeVar{a\sb{\mathrm{1}}}\HOLTokenTransEnd \HOLBoundVar{E\sp{\prime}} \HOLSymConst{\HOLTokenConj{}}
      ((\HOLFreeVar{a\sb{\mathrm{1}}} \HOLSymConst{=} \HOLSymConst{\ensuremath{\tau}}) \HOLSymConst{\HOLTokenDisj{}} (\HOLFreeVar{a\sb{\mathrm{1}}} \HOLSymConst{=} \HOLConst{label} \HOLBoundVar{l}) \HOLSymConst{\HOLTokenConj{}} \HOLBoundVar{l} \HOLSymConst{\HOLTokenNotIn{}} \HOLBoundVar{L} \HOLSymConst{\HOLTokenConj{}} \HOLConst{COMPL} \HOLBoundVar{l} \HOLSymConst{\HOLTokenNotIn{}} \HOLBoundVar{L})) \HOLSymConst{\HOLTokenDisj{}}
   (\HOLSymConst{\HOLTokenExists{}}\HOLBoundVar{E} \HOLBoundVar{u} \HOLBoundVar{E\sp{\prime}} \HOLBoundVar{rf}.
      (\HOLFreeVar{a\sb{\mathrm{0}}} \HOLSymConst{=} \HOLConst{relab} \HOLBoundVar{E} \HOLBoundVar{rf}) \HOLSymConst{\HOLTokenConj{}} (\HOLFreeVar{a\sb{\mathrm{1}}} \HOLSymConst{=} \HOLConst{relabel} \HOLBoundVar{rf} \HOLBoundVar{u}) \HOLSymConst{\HOLTokenConj{}}
      (\HOLFreeVar{a\sb{\mathrm{2}}} \HOLSymConst{=} \HOLConst{relab} \HOLBoundVar{E\sp{\prime}} \HOLBoundVar{rf}) \HOLSymConst{\HOLTokenConj{}} \HOLBoundVar{E} \HOLTokenTransBegin\HOLBoundVar{u}\HOLTokenTransEnd \HOLBoundVar{E\sp{\prime}}) \HOLSymConst{\HOLTokenDisj{}}
   \HOLSymConst{\HOLTokenExists{}}\HOLBoundVar{E} \HOLBoundVar{X}. (\HOLFreeVar{a\sb{\mathrm{0}}} \HOLSymConst{=} \HOLConst{rec} \HOLBoundVar{X} \HOLBoundVar{E}) \HOLSymConst{\HOLTokenConj{}} \HOLConst{CCS_Subst} \HOLBoundVar{E} (\HOLConst{rec} \HOLBoundVar{X} \HOLBoundVar{E}) \HOLBoundVar{X} \HOLTokenTransBegin\HOLFreeVar{a\sb{\mathrm{1}}}\HOLTokenTransEnd \HOLFreeVar{a\sb{\mathrm{2}}}
\end{alltt}

Here are some results proved using above ``cases'' theorem (i. e. they cannot be
proved with only the SOS inference rules):
\begin{alltt}
NIL_NO_TRANS:   \HOLTokenTurnstile{} \HOLSymConst{\HOLTokenNeg{}}(\HOLConst{nil} \HOLTokenTransBegin\HOLFreeVar{u}\HOLTokenTransEnd \HOLFreeVar{E})
VAR_NO_TRANS:   \HOLTokenTurnstile{} \HOLSymConst{\HOLTokenNeg{}}(\HOLConst{var} \HOLFreeVar{X} \HOLTokenTransBegin\HOLFreeVar{u}\HOLTokenTransEnd \HOLFreeVar{E})

TRANS_IMP_NO_NIL:
\HOLTokenTurnstile{} \HOLFreeVar{E} \HOLTokenTransBegin\HOLFreeVar{u}\HOLTokenTransEnd \HOLFreeVar{E\sp{\prime}} \HOLSymConst{\HOLTokenImp{}} \HOLFreeVar{E} \HOLSymConst{\HOLTokenNotEqual{}} \HOLConst{nil}

TRANS_SUM_EQ:
\HOLTokenTurnstile{} \HOLFreeVar{E} \HOLSymConst{+} \HOLFreeVar{E\sp{\prime}} \HOLTokenTransBegin\HOLFreeVar{u}\HOLTokenTransEnd \HOLFreeVar{E\sp{\prime\prime}} \HOLSymConst{\HOLTokenEquiv{}} \HOLFreeVar{E} \HOLTokenTransBegin\HOLFreeVar{u}\HOLTokenTransEnd \HOLFreeVar{E\sp{\prime\prime}} \HOLSymConst{\HOLTokenDisj{}} \HOLFreeVar{E\sp{\prime}} \HOLTokenTransBegin\HOLFreeVar{u}\HOLTokenTransEnd \HOLFreeVar{E\sp{\prime\prime}}

TRANS_PAR_EQ:
\HOLTokenTurnstile{} \HOLFreeVar{E} \HOLSymConst{\ensuremath{\parallel}} \HOLFreeVar{E\sp{\prime}} \HOLTokenTransBegin\HOLFreeVar{u}\HOLTokenTransEnd \HOLFreeVar{E\sp{\prime\prime}} \HOLSymConst{\HOLTokenEquiv{}}
   (\HOLSymConst{\HOLTokenExists{}}\HOLBoundVar{E\sb{\mathrm{1}}}. (\HOLFreeVar{E\sp{\prime\prime}} \HOLSymConst{=} \HOLBoundVar{E\sb{\mathrm{1}}} \HOLSymConst{\ensuremath{\parallel}} \HOLFreeVar{E\sp{\prime}}) \HOLSymConst{\HOLTokenConj{}} \HOLFreeVar{E} \HOLTokenTransBegin\HOLFreeVar{u}\HOLTokenTransEnd \HOLBoundVar{E\sb{\mathrm{1}}}) \HOLSymConst{\HOLTokenDisj{}}
   (\HOLSymConst{\HOLTokenExists{}}\HOLBoundVar{E\sb{\mathrm{1}}}. (\HOLFreeVar{E\sp{\prime\prime}} \HOLSymConst{=} \HOLFreeVar{E} \HOLSymConst{\ensuremath{\parallel}} \HOLBoundVar{E\sb{\mathrm{1}}}) \HOLSymConst{\HOLTokenConj{}} \HOLFreeVar{E\sp{\prime}} \HOLTokenTransBegin\HOLFreeVar{u}\HOLTokenTransEnd \HOLBoundVar{E\sb{\mathrm{1}}}) \HOLSymConst{\HOLTokenDisj{}}
   \HOLSymConst{\HOLTokenExists{}}\HOLBoundVar{E\sb{\mathrm{1}}} \HOLBoundVar{E\sb{\mathrm{2}}} \HOLBoundVar{l}.
     (\HOLFreeVar{u} \HOLSymConst{=} \HOLSymConst{\ensuremath{\tau}}) \HOLSymConst{\HOLTokenConj{}} (\HOLFreeVar{E\sp{\prime\prime}} \HOLSymConst{=} \HOLBoundVar{E\sb{\mathrm{1}}} \HOLSymConst{\ensuremath{\parallel}} \HOLBoundVar{E\sb{\mathrm{2}}}) \HOLSymConst{\HOLTokenConj{}} \HOLFreeVar{E} \HOLTokenTransBegin\HOLConst{label} \HOLBoundVar{l}\HOLTokenTransEnd \HOLBoundVar{E\sb{\mathrm{1}}} \HOLSymConst{\HOLTokenConj{}}
     \HOLFreeVar{E\sp{\prime}} \HOLTokenTransBegin\HOLConst{label} (\HOLConst{COMPL} \HOLBoundVar{l})\HOLTokenTransEnd \HOLBoundVar{E\sb{\mathrm{2}}}

TRANS_RESTR_EQ:
\HOLTokenTurnstile{} \HOLSymConst{\ensuremath{\nu}} \HOLFreeVar{L} \HOLFreeVar{E} \HOLTokenTransBegin\HOLFreeVar{u}\HOLTokenTransEnd \HOLFreeVar{E\sp{\prime}} \HOLSymConst{\HOLTokenEquiv{}}
   \HOLSymConst{\HOLTokenExists{}}\HOLBoundVar{E\sp{\prime\prime}} \HOLBoundVar{l}.
     (\HOLFreeVar{E\sp{\prime}} \HOLSymConst{=} \HOLSymConst{\ensuremath{\nu}} \HOLFreeVar{L} \HOLBoundVar{E\sp{\prime\prime}}) \HOLSymConst{\HOLTokenConj{}} \HOLFreeVar{E} \HOLTokenTransBegin\HOLFreeVar{u}\HOLTokenTransEnd \HOLBoundVar{E\sp{\prime\prime}} \HOLSymConst{\HOLTokenConj{}}
     ((\HOLFreeVar{u} \HOLSymConst{=} \HOLSymConst{\ensuremath{\tau}}) \HOLSymConst{\HOLTokenDisj{}} (\HOLFreeVar{u} \HOLSymConst{=} \HOLConst{label} \HOLBoundVar{l}) \HOLSymConst{\HOLTokenConj{}} \HOLBoundVar{l} \HOLSymConst{\HOLTokenNotIn{}} \HOLFreeVar{L} \HOLSymConst{\HOLTokenConj{}} \HOLConst{COMPL} \HOLBoundVar{l} \HOLSymConst{\HOLTokenNotIn{}} \HOLFreeVar{L})
\end{alltt}

\subsection{Decision procedure for CCS transitions}

It's possible to use SOS inference rules and theorems derived from
them for proving theorems about the transitions between any two CCS
processes. However, what's more useful is the decision procedure
which automatically decide all possible transitions and formally prove them.

For any CCS process, there is a decision procedure as a recursive
function, which can completely decide all its possible (one-step)
transitions. In HOL, this decision procedure can be implemented as a
normal Standard ML function \texttt{CCS\_TRANS\_CONV} of type
\texttt{term -> theorem}, the returned theorem fully characterize the
possible transitions of the input CCS process.

For instance, we know that the process $(a.0 | \bar{a}.0)$ have three
possible transitions:
\begin{enumerate}
\item $(a.0 | \bar{a}.0) \overset{a}{\longrightarrow} (0 | \bar{a}.0)$;
\item $(a.0 | \bar{a}.0) \overset{\bar{a}}{\longrightarrow} (a.0 | 0)$;
\item $(a.0 | \bar{a}.0) \overset{\tau}{\longrightarrow} (0 | 0)$.
\end{enumerate}
To completely decide all possible transitions, if done manually, the following work should be done:
\begin{enumerate}
\item Prove there exists transitions from $(a.0 | \bar{a}.0)$ (optionally);
\item Prove each of above three transitions using SOS inference rules;
\item Prove there's no other transitions, using the ``cases'' theorems
  generated from the \texttt{TRANS} relation.
\end{enumerate}

Instead, if we use the function \texttt{CCS\_TRANS\_CONV} with the
root process:
\begin{lstlisting}
> CCS_TRANS_CONV
	 ``par (prefix (label (name "a")) nil)
	       (prefix (label (coname "a")) nil)``
\end{lstlisting}
As the result, the following theorem is returned:
From this theorem, we can see there're only three possible transitions
and there's no others. Therefore it contains all information expressed
by previous manually proved 5 theorems (in theory we can also try to
manually prove this single theorem, but it's not easy since the steps
required will be at least the sum of all previous proofs).

As a further example, if we put a restriction on label ``a'' and check
the process $(\nu a)(a.0 |
\bar{a}.0)$ instead, there will be only one possible transition:

It's possible to extract a list of possible transitions together with
the actions, into a list. This work can be done automatically by the
function \texttt{strip_trans}. Finally, if both the theorem and the
list of transitions are needed, the function \texttt{CCS\_TRANS} and
its compact-form variant \texttt{CCS\_TRANS'} can be used. For the
previous example process $(a.0 | \bar{a}.0)$, calling
\texttt{CCS\_TRANS'} on it in HOL's interactive environment has the
following results:
\begin{lstlisting}
> CCS_TRANS ``In "a"..nil || Out "a"..nil``;
val it =
   (|- !u E.
     In "a"..nil || Out "a"..nil --u-> E <=>
     ((u = In "a") /\ (E = nil || Out "a"..nil) \/
      (u = Out "a") /\ (E = In "a"..nil || nil)) \/
     (u = tau) /\ (E = nil || nil),
    [(``In "a"``,
      ``nil || Out "a"..nil``),
     (``Out "a"``,
      ``In "a"..nil || nil``),
     (``'t``,
      ``nil || nil``)]):
   thm * (term * term) list
\end{lstlisting}

The main function \texttt{CCS_TRANS_CONV} is implemented in about 500
lines of Standard ML code, and it depends on many dedicated tactics written for CCS,
and functions to access the internal structure of CCS-related theorem
and terms.  We have tried our best to make sure the correctness of
this function, but certain bugs are still inevitable.\footnote{If the
  internal proof constructed in the function is wrong, then the function won't
  return a theorem. But if the function successfully returns a theorem,
  the proof for this theorem must be correct, because there's no other
  way to return a theorem except for correctly proving it in HOL
  theorem prover.} However, since
it's implemented in theorem prover, and the return value of this
function is a theorem, what we can guarantee is the following things:
\begin{quote}
Whenever the function terminates with a theorem returned, as long as
the theorem has ``correct'' forms, the CCS
transitions indicated in the returned theorem is indeed all possible
transitions from the input process. No matter if there're bugs in our program.
\end{quote}

In another words, any remain bug in the program can only stop the
whole function from returning a result, but as long as the result is
returned, it cannot be wrong! This sounds like a
different kind of trusted computing than the normal senses. In general, for
any algorithm implemented in any normal programming languages, since
the output is just a primitive value or data structure which can be
arbitrary constructed or changed due to potential bugs in the
software, the only way to trust these results, is to have the
entire program carefully modeled and verified. But in our case, the
Standard ML program code is not verified, but the result (once appears) can still be
fully\footnote{Well, HOL theorem prover itself, like any other
  software, contains bugs for sure, but the chances for HOL to
  produce fake theorems are very very low, although such a chance is not
zero in theory. If we went further in this direction, PolyML (which compiles HOL's ML
code into binary executions) may also contain bugs, so even HOL's code is correct it could
still produce fake theorems. The final solution should be using
formally verified ML implementations like CakeML to build HOL and
other ML-based theorem provers to completely eliminate such concerns,
but so far it has't reached such a perfect level. Another way to convince the
audience is that, HOL can also output the primitive
reasoning steps (using those eight primitive rules) behind each
theorem, through its \emph{logging kernel} (and OpenTheory formats), in theory these steps can
be verify by other HOL-family theorem provers  or just by hands. My
professor (Roberto Gorrieri) ever commented that I trusted HOL just
like how he trusted Concurrency Workbench (CWB), I think this is not fair!} trusted,
simply because it's a theorem derived from HOL.

\section{Strong equivalence}

The concept of \emph{bisimulation} and \emph{bisimulation
  equivalence} (bisimilarity) and their variants have the central position in Concurrency Theory.
One major approach in model
checking is to check the bisimulation equivalence between the
specification and implementation of the same system.
Besides, it's well known that,
strong equivalence as a relation, must be defined
\emph{co-inductively}. (And in fact, strong equivalence is one of the
most well-studied co-inductive relation in computer
science. \cite{Sangiorgi:2011ut}) In this section, we study the
definition of strong and weak
bisimulation and (bisimulation) equivalences, and their
possible formalizations in HOL.

Recall the standard definition of strong bisimulation and strong equivalence
(c.f. p.43 of \cite{Gorrieri:2015jt}):
\begin{definition}{((Strong) bisimulation and (strong) bisimulation equivalence)}
Let $TS = (Q, A, \rightarrow)$ be a transition system. A
\emph{bisimulation} is a relation $R \subset Q \times Q$ such that $R$
and its inverse $R^{-1}$ are both simulation relations. More
explicitly, a bisimulation is a relation $R$ such that if
$(q_1,q_2)\in R$ then for all $\mu\in A$
\begin{itemize}
\item $\forall q_1' \text{ such that } q_1
  \overset{\mu}{\longrightarrow} q_1', \exists q_2' \text{ such that }
  q_2 \overset{\mu}{\longrightarrow} q_2' \text{ and } (q_1', q_2')
  \in R$,
\item $\forall q_2' \text{ such that } q_2
  \overset{\mu}{\longrightarrow} q_2', \exists q_1' \text{ such that }
  q_1 \overset{\mu}{\longrightarrow} q_1' \text{ and } (q_1', q_2')
  \in R$.
\end{itemize}
Two states $q$ and $q'$ are \emph{bisimular} (or \emph{bisimulation
  equivalent}), denoted $q \sim q'$, if there exists a bisimulation
$R$ such that $(q, q') \in R$.
\end{definition}
Noticed that, although above definition is expressed in LTS, it's also
applicable to CCS in which each process has the semantic model as a
rooted LTS. Given the fact that, all states involved in above
definition are target states of direct or indirect transition of the
initial pair of states, above definition can be directly used for CCS.

In HOL88, there's no direct way to define co-inductive relation.
However, it's possible to follow above definition literally
and define bisimulation first, then define the bisimulation
equivalence on top of bisimulation. Here are the definitions 
translated from HOL88 to HOL4:
\begin{alltt}
\HOLTokenTurnstile{} \HOLConst{STRONG_BISIM} \HOLFreeVar{Bsm} \HOLSymConst{\HOLTokenEquiv{}}
   \HOLSymConst{\HOLTokenForall{}}\HOLBoundVar{E} \HOLBoundVar{E\sp{\prime}}.
     \HOLFreeVar{Bsm} \HOLBoundVar{E} \HOLBoundVar{E\sp{\prime}} \HOLSymConst{\HOLTokenImp{}}
     \HOLSymConst{\HOLTokenForall{}}\HOLBoundVar{u}.
       (\HOLSymConst{\HOLTokenForall{}}\HOLBoundVar{E\sb{\mathrm{1}}}. \HOLBoundVar{E} \HOLTokenTransBegin\HOLBoundVar{u}\HOLTokenTransEnd \HOLBoundVar{E\sb{\mathrm{1}}} \HOLSymConst{\HOLTokenImp{}} \HOLSymConst{\HOLTokenExists{}}\HOLBoundVar{E\sb{\mathrm{2}}}. \HOLBoundVar{E\sp{\prime}} \HOLTokenTransBegin\HOLBoundVar{u}\HOLTokenTransEnd \HOLBoundVar{E\sb{\mathrm{2}}} \HOLSymConst{\HOLTokenConj{}} \HOLFreeVar{Bsm} \HOLBoundVar{E\sb{\mathrm{1}}} \HOLBoundVar{E\sb{\mathrm{2}}}) \HOLSymConst{\HOLTokenConj{}}
       \HOLSymConst{\HOLTokenForall{}}\HOLBoundVar{E\sb{\mathrm{2}}}. \HOLBoundVar{E\sp{\prime}} \HOLTokenTransBegin\HOLBoundVar{u}\HOLTokenTransEnd \HOLBoundVar{E\sb{\mathrm{2}}} \HOLSymConst{\HOLTokenImp{}} \HOLSymConst{\HOLTokenExists{}}\HOLBoundVar{E\sb{\mathrm{1}}}. \HOLBoundVar{E} \HOLTokenTransBegin\HOLBoundVar{u}\HOLTokenTransEnd \HOLBoundVar{E\sb{\mathrm{1}}} \HOLSymConst{\HOLTokenConj{}} \HOLFreeVar{Bsm} \HOLBoundVar{E\sb{\mathrm{1}}} \HOLBoundVar{E\sb{\mathrm{2}}}
\HOLTokenTurnstile{} \HOLFreeVar{E} \HOLSymConst{\HOLTokenStrongEQ} \HOLFreeVar{E\sp{\prime}} \HOLSymConst{\HOLTokenEquiv{}} \HOLSymConst{\HOLTokenExists{}}\HOLBoundVar{Bsm}. \HOLBoundVar{Bsm} \HOLFreeVar{E} \HOLFreeVar{E\sp{\prime}} \HOLSymConst{\HOLTokenConj{}} \HOLConst{STRONG_BISIM} \HOLBoundVar{Bsm}
\end{alltt}
From the second definition, we can see that, $q \sim q'$ if there
exists a bisimulation containing the pair $(q, q')$. This means that
$\sim$ is the union of all bisimulations, i.e.,
\begin{equation*}
\sim = \bigcup \{ R \subset Q \times Q \colon R \text{ is a
  bisimulation} \}.
\end{equation*}

The other way to define strong equivalence is through
the fixed point of the following function $F$: (c.f. p.72 of \cite{Gorrieri:2015jt})
\begin{definition}
Given an LTS $(Q, A, \rightarrow)$, the function $F\colon \wp(Q\times
Q) \rightarrow \wp(Q\times Q)$ (i.e., a transformer of binary
relations over $Q$) is defined as follows. If $R \subset Q\times Q$,
then $(q_1,q_2) \in F(R)$ if and only if for all $\mu \in A$
\begin{itemize}
\item $\forall q_1' \text{ such that } q_1 \overset{\mu}{\longrightarrow} q_1', \exists
  q_2' \text{ such that } q_2 \overset{\mu}{\longrightarrow} q_2'
  \text{ and } (q_1', q_2') \in R$,
\item $\forall q_2' \text{ such that } q_2 \overset{\mu}{\longrightarrow} q_2', \exists
  q_1' \text{ such that } q_1 \overset{\mu}{\longrightarrow} q_1'
  \text{ and } (q_1', q_2') \in R$.
\end{itemize}
\end{definition}
And we can see by comparing the definition of above function and the
definition of bisimulation that (no formal proofs):
\begin{enumerate}
\item The function $F$ is monotone, i.e. if $R_1 \subset R_2$ then
  $F(R_1) \subset F(R_2)$.
\item A relation $R\subset Q\times Q$ is a bisimulation if and only if
  $R \subset F(R)$.
\end{enumerate}
Then according to Knaster-Tarski Fixed Point theorem, strong
bisimilarity $\sim$ is the greatest fixed point of $F$. And this is
also the definition of co-inductive relation defined by the same rules.

In HOL4, since the release Kananaskis-11, there's a new facility for
defining co-inductive relation. The entry command is
\texttt{Hol_coreln}, which has the same syntax as \texttt{Hol_reln}
for definining inductive relations. Using \texttt{Hol_coreln}, it's
possible to define the bisimulation equivalence \emph{directly} in
this way: (here we has chosen a new relation name
\texttt{STRONG_EQ})\footnote{Whenever ASCII-based HOL proof scripts were directly
  pasted, please understand the letter ``\texttt{!}'' as $\forall$,
  and ``\texttt{?}'' as $\exists$. They're part of HOL's term syntax. \cite{Anonymous:Iu-sOoz1}}

The first theorem is the original rules appearing in the
definition. Roughly speaking, it's kind of rules for building a
bisimulation relation in forward way, however this is impossible
because of the lack of base rules (which exists in most inductive
relation). And it's not original in this case, since it can be derived
from the last theorem \texttt{STRONG_EQ_cases} (RHS $\Rightarrow$ LHS).

The second theorem is the co-induction principle. It says, for what
ever relation which satisfies those rules, it must be contained
in strong equivalence. In another word, it make sure the target
relation is the maximal relation containing all others.

The purpose of the last theorem (also called ``cases'' theorem), is to make sure the target relation
is indeed a fixed point of the function $F$ built by the given
rules. However, it
doesn't give any information about the size of such a fixed point. In
general, if the greatest fixed point and least fixed point doesn't
coincide, without the restriction by co-induction theorem, the rest
two theorems will not give a precise definition for that relation.
For strong equivalence, we already know that, the least fixed point of
$F$ is empty relation $\emptyset$, and the great fixed point is the
strong equivalence $\sim$. And in fact, the ``cases'' theorem has
``defined'' a relation which lies in the middle of the greatest and
least fixed point. To see why this argument is true, we found 
this theorem as an equation could be used as a possible definition of
strong equivalence: (c.f. p. 49 of \cite{Gorrieri:2015jt})
\begin{definition}
Define \emph{recursively} a new behavioral relation $\sim' \in Q
\times Q$ as follows: $q_1 \sim' q_2$ \emph{if and only if} for all
$\mu \in A$
\begin{itemize}
\item $\forall q_1' \text{ such that } q_1
  \overset{\mu}{\longrightarrow} q_1', \exists q_2' \text{ such that }
  q_2\overset{\mu}{\longrightarrow} q_2' \text{ and } q_1' \sim'
  q_2'$,
\item $\forall q_2' \text{ such that } q_2
  \overset{\mu}{\longrightarrow} q_2', \exists q_1' \text{ such that }
  q_1 \overset{\mu}{\longrightarrow} q_1' \text{ and } q_1' \sim'
  q_2'$.
\end{itemize}
\end{definition}
This is exactly the same as above ``cases'' theorem if the theorem
were used as a definition of strong equivalence. Robin Milner calls
this theorem the `` property (*)'' of strong
equivalence. (c.f. p. 88 of \cite{Milner:1989}) But as Prof.\ Gorrieri's book \cite{Gorrieri:2015jt}
already told with examples: ``this does not identify a unique
relation, as many different relations satisfy this recursive
definition.'', and the fact that any mathematical (or logic)
definitions must precisely specify the targeting object (unless the
possible covered range itself is a targeting object).

But why the recursive definition failed to define a largest
bisimulation (i.e. strong equivalence)?
The textbooks didn't give a clear answer, but in the view of theorem
proving, now it's quite clear:
such a recursive definition can only restrict the target relation into the range of all
fixed points, while it's the co-induction theorem who finally restricts the
target relation to the greatest solution. Without any of them, the
solution will not be unique (thus not a valid mathematical definition).

Based on the definition of \HOLinline{\HOLConst{STRONG_EQUIV}} and SOS inference
rules for the \texttt{TRANS} relation, we have proved a large set of
theorems concerning the strong equivalence of CCS processes. Below is
a list of fundamental congruence theorems for strong equivalence:
\begin{alltt}
STRONG_EQUIV_SUBST_PREFIX:
\HOLTokenTurnstile{} \HOLFreeVar{E} \HOLSymConst{\HOLTokenStrongEQ} \HOLFreeVar{E\sp{\prime}} \HOLSymConst{\HOLTokenImp{}} \HOLSymConst{\HOLTokenForall{}}\HOLBoundVar{u}. \HOLBoundVar{u}\HOLSymConst{..}\HOLFreeVar{E} \HOLSymConst{\HOLTokenStrongEQ} \HOLBoundVar{u}\HOLSymConst{..}\HOLFreeVar{E\sp{\prime}}

STRONG_EQUIV_PRESD_BY_SUM:
\HOLTokenTurnstile{} \HOLFreeVar{E\sb{\mathrm{1}}} \HOLSymConst{\HOLTokenStrongEQ} \HOLFreeVar{E\sb{\mathrm{1}}\sp{\prime}} \HOLSymConst{\HOLTokenConj{}} \HOLFreeVar{E\sb{\mathrm{2}}} \HOLSymConst{\HOLTokenStrongEQ} \HOLFreeVar{E\sb{\mathrm{2}}\sp{\prime}} \HOLSymConst{\HOLTokenImp{}} \HOLFreeVar{E\sb{\mathrm{1}}} \HOLSymConst{+} \HOLFreeVar{E\sb{\mathrm{2}}} \HOLSymConst{\HOLTokenStrongEQ} \HOLFreeVar{E\sb{\mathrm{1}}\sp{\prime}} \HOLSymConst{+} \HOLFreeVar{E\sb{\mathrm{2}}\sp{\prime}}

STRONG_EQUIV_PRESD_BY_PAR:
\HOLTokenTurnstile{} \HOLFreeVar{E\sb{\mathrm{1}}} \HOLSymConst{\HOLTokenStrongEQ} \HOLFreeVar{E\sb{\mathrm{1}}\sp{\prime}} \HOLSymConst{\HOLTokenConj{}} \HOLFreeVar{E\sb{\mathrm{2}}} \HOLSymConst{\HOLTokenStrongEQ} \HOLFreeVar{E\sb{\mathrm{2}}\sp{\prime}} \HOLSymConst{\HOLTokenImp{}} \HOLFreeVar{E\sb{\mathrm{1}}} \HOLSymConst{\ensuremath{\parallel}} \HOLFreeVar{E\sb{\mathrm{2}}} \HOLSymConst{\HOLTokenStrongEQ} \HOLFreeVar{E\sb{\mathrm{1}}\sp{\prime}} \HOLSymConst{\ensuremath{\parallel}} \HOLFreeVar{E\sb{\mathrm{2}}\sp{\prime}}

STRONG_EQUIV_SUBST_RESTR:
\HOLTokenTurnstile{} \HOLFreeVar{E} \HOLSymConst{\HOLTokenStrongEQ} \HOLFreeVar{E\sp{\prime}} \HOLSymConst{\HOLTokenImp{}} \HOLSymConst{\HOLTokenForall{}}\HOLBoundVar{L}. \HOLSymConst{\ensuremath{\nu}} \HOLBoundVar{L} \HOLFreeVar{E} \HOLSymConst{\HOLTokenStrongEQ} \HOLSymConst{\ensuremath{\nu}} \HOLBoundVar{L} \HOLFreeVar{E\sp{\prime}}

STRONG_EQUIV_SUBST_RELAB:
\HOLTokenTurnstile{} \HOLFreeVar{E} \HOLSymConst{\HOLTokenStrongEQ} \HOLFreeVar{E\sp{\prime}} \HOLSymConst{\HOLTokenImp{}} \HOLSymConst{\HOLTokenForall{}}\HOLBoundVar{rf}. \HOLConst{relab} \HOLFreeVar{E} \HOLBoundVar{rf} \HOLSymConst{\HOLTokenStrongEQ} \HOLConst{relab} \HOLFreeVar{E\sp{\prime}} \HOLBoundVar{rf}
\end{alltt}

Noticed that, the strong bisimulation equivalence is co-inductively
defined, and two processes are strong equivalent if there's a
bisimulation containing them. Thus, to prove two processes are
strong equivalent, it's enough to find a bisimulation containing
them. To prove the they're not strong equivalent, it's enough to try
to construct a bisimulation starting from them and the proof is
finished whenever the
attempt fails. In any case, there's no need to do induction on the
data type of involved CCS processes.

\subsection{Algebraic Laws for strong equivalence}

Here are the strong laws proved for the sum operator: (noticed that,
the lack of some parentheses is because we have defined the sum and
parallel operators as left-associative)
\begin{alltt}
STRONG_SUM_IDEMP:          \HOLTokenTurnstile{} \HOLFreeVar{E} \HOLSymConst{+} \HOLFreeVar{E} \HOLSymConst{\HOLTokenStrongEQ} \HOLFreeVar{E}
STRONG_SUM_COMM:           \HOLTokenTurnstile{} \HOLFreeVar{E} \HOLSymConst{+} \HOLFreeVar{E\sp{\prime}} \HOLSymConst{\HOLTokenStrongEQ} \HOLFreeVar{E\sp{\prime}} \HOLSymConst{+} \HOLFreeVar{E}
STRONG_SUM_IDENT_L:        \HOLTokenTurnstile{} \HOLConst{nil} \HOLSymConst{+} \HOLFreeVar{E} \HOLSymConst{\HOLTokenStrongEQ} \HOLFreeVar{E}
STRONG_SUM_IDENT_R:        \HOLTokenTurnstile{} \HOLFreeVar{E} \HOLSymConst{+} \HOLConst{nil} \HOLSymConst{\HOLTokenStrongEQ} \HOLFreeVar{E}
STRONG_SUM_ASSOC_R:        \HOLTokenTurnstile{} \HOLFreeVar{E} \HOLSymConst{+} \HOLFreeVar{E\sp{\prime}} \HOLSymConst{+} \HOLFreeVar{E\sp{\prime\prime}} \HOLSymConst{\HOLTokenStrongEQ} \HOLFreeVar{E} \HOLSymConst{+} (\HOLFreeVar{E\sp{\prime}} \HOLSymConst{+} \HOLFreeVar{E\sp{\prime\prime}})
STRONG_SUM_ASSOC_L:        \HOLTokenTurnstile{} \HOLFreeVar{E} \HOLSymConst{+} (\HOLFreeVar{E\sp{\prime}} \HOLSymConst{+} \HOLFreeVar{E\sp{\prime\prime}}) \HOLSymConst{\HOLTokenStrongEQ} \HOLFreeVar{E} \HOLSymConst{+} \HOLFreeVar{E\sp{\prime}} \HOLSymConst{+} \HOLFreeVar{E\sp{\prime\prime}}
STRONG_SUM_MID_IDEMP:      \HOLTokenTurnstile{} \HOLFreeVar{E} \HOLSymConst{+} \HOLFreeVar{E\sp{\prime}} \HOLSymConst{+} \HOLFreeVar{E} \HOLSymConst{\HOLTokenStrongEQ} \HOLFreeVar{E\sp{\prime}} \HOLSymConst{+} \HOLFreeVar{E}
STRONG_LEFT_SUM_MID_IDEMP: \HOLTokenTurnstile{} \HOLFreeVar{E} \HOLSymConst{+} \HOLFreeVar{E\sp{\prime}} \HOLSymConst{+} \HOLFreeVar{E\sp{\prime\prime}} \HOLSymConst{+} \HOLFreeVar{E\sp{\prime}} \HOLSymConst{\HOLTokenStrongEQ} \HOLFreeVar{E} \HOLSymConst{+} \HOLFreeVar{E\sp{\prime\prime}} \HOLSymConst{+} \HOLFreeVar{E\sp{\prime}}
\end{alltt}

Not all above theorems are primitive (in the sense of providing a
minimal axiomatization set for proving all other strong algebraic laws). The
first several theorems must be proved by constructing bisimulation
relations and then verifying the definitions of strong bisimulation
and strong equivalence, and their formal proofs were written in
goal-directed ways. Instead, the
last three ones were all constructed in forward way by applications of
previous proven algebraic laws, without directly using any SOS
inference rules and the definition of strong equivalence. Such
constructions were based on two useful ML functions \texttt{S_SYM} and
\texttt{S_TRANS} which builds new strong laws from the symmetry and
transitivity of strong equivalence:
\begin{lstlisting}
(* Define S_SYM such that, when given a theorem A
   |- STRONG_EQUIV t1 t2,
   returns the theorem A |- STRONG_EQUIV t2 t1. *)
fun S_SYM thm = MATCH_MP STRONG_EQUIV_SYM thm;

(* Define S_TRANS such that, when given the theorems thm1 and
   thm2, applies
   STRONG_EQUIV_TRANS on them, if possible. *)
fun S_TRANS thm1 thm2 =
    if rhs_tm thm1 = lhs_tm thm2 then
       MATCH_MP STRONG_EQUIV_TRANS (CONJ thm1 thm2)
    else
       failwith
        "transitivity of strong equivalence not applicable";
\end{lstlisting}

For instance, to construct the proof of \texttt{STRONG_SUM_MID_IDEMP},
the following code was written:
\begin{lstlisting}
(* STRONG_SUM_MID_IDEMP:
   |- !E E'. STRONG_EQUIV (sum (sum E E') E) (sum E' E)
 *)
val STRONG_SUM_MID_IDEMP = save_thm (
   "STRONG_SUM_MID_IDEMP",
    GEN ``E: CCS``
     (GEN ``E': CCS``
       (S_TRANS
        (SPEC ``E: CCS``
         (MATCH_MP STRONG_EQUIV_SUBST_SUM_R
          (SPECL [``E: CCS``, ``E': CCS``] STRONG_SUM_COMM)))
        (S_TRANS
         (SPECL [``E': CCS``, ``E: CCS``, ``E: CCS``]
                         STRONG_SUM_ASSOC_R)
         (SPEC ``E': CCS``
          (MATCH_MP STRONG_EQUIV_SUBST_SUM_L
           (SPEC ``E: CCS`` STRONG_SUM_IDEMP)))))));
\end{lstlisting}

Here are the strong laws we have proved for the par operator:
\begin{alltt}
STRONG_PAR_IDENT_R:        \HOLTokenTurnstile{} \HOLFreeVar{E} \HOLSymConst{\ensuremath{\parallel}} \HOLConst{nil} \HOLSymConst{\HOLTokenStrongEQ} \HOLFreeVar{E}
STRONG_PAR_COMM:           \HOLTokenTurnstile{} \HOLFreeVar{E} \HOLSymConst{\ensuremath{\parallel}} \HOLFreeVar{E\sp{\prime}} \HOLSymConst{\HOLTokenStrongEQ} \HOLFreeVar{E\sp{\prime}} \HOLSymConst{\ensuremath{\parallel}} \HOLFreeVar{E}
STRONG_PAR_IDENT_L:        \HOLTokenTurnstile{} \HOLConst{nil} \HOLSymConst{\ensuremath{\parallel}} \HOLFreeVar{E} \HOLSymConst{\HOLTokenStrongEQ} \HOLFreeVar{E}
STRONG_PAR_ASSOC:          \HOLTokenTurnstile{} \HOLFreeVar{E} \HOLSymConst{\ensuremath{\parallel}} \HOLFreeVar{E\sp{\prime}} \HOLSymConst{\ensuremath{\parallel}} \HOLFreeVar{E\sp{\prime\prime}} \HOLSymConst{\HOLTokenStrongEQ} \HOLFreeVar{E} \HOLSymConst{\ensuremath{\parallel}} (\HOLFreeVar{E\sp{\prime}} \HOLSymConst{\ensuremath{\parallel}} \HOLFreeVar{E\sp{\prime\prime}})

STRONG_PAR_PREF_TAU:
\HOLTokenTurnstile{} \HOLFreeVar{u}\HOLSymConst{..}\HOLFreeVar{E} \HOLSymConst{\ensuremath{\parallel}} \HOLSymConst{\ensuremath{\tau}}\HOLSymConst{..}\HOLFreeVar{E\sp{\prime}} \HOLSymConst{\HOLTokenStrongEQ} \HOLFreeVar{u}\HOLSymConst{..}(\HOLFreeVar{E} \HOLSymConst{\ensuremath{\parallel}} \HOLSymConst{\ensuremath{\tau}}\HOLSymConst{..}\HOLFreeVar{E\sp{\prime}}) \HOLSymConst{+} \HOLSymConst{\ensuremath{\tau}}\HOLSymConst{..}(\HOLFreeVar{u}\HOLSymConst{..}\HOLFreeVar{E} \HOLSymConst{\ensuremath{\parallel}} \HOLFreeVar{E\sp{\prime}})

STRONG_PAR_TAU_PREF:
\HOLTokenTurnstile{} \HOLSymConst{\ensuremath{\tau}}\HOLSymConst{..}\HOLFreeVar{E} \HOLSymConst{\ensuremath{\parallel}} \HOLFreeVar{u}\HOLSymConst{..}\HOLFreeVar{E\sp{\prime}} \HOLSymConst{\HOLTokenStrongEQ} \HOLSymConst{\ensuremath{\tau}}\HOLSymConst{..}(\HOLFreeVar{E} \HOLSymConst{\ensuremath{\parallel}} \HOLFreeVar{u}\HOLSymConst{..}\HOLFreeVar{E\sp{\prime}}) \HOLSymConst{+} \HOLFreeVar{u}\HOLSymConst{..}(\HOLSymConst{\ensuremath{\tau}}\HOLSymConst{..}\HOLFreeVar{E} \HOLSymConst{\ensuremath{\parallel}} \HOLFreeVar{E\sp{\prime}})

STRONG_PAR_TAU_TAU:
\HOLTokenTurnstile{} \HOLSymConst{\ensuremath{\tau}}\HOLSymConst{..}\HOLFreeVar{E} \HOLSymConst{\ensuremath{\parallel}} \HOLSymConst{\ensuremath{\tau}}\HOLSymConst{..}\HOLFreeVar{E\sp{\prime}} \HOLSymConst{\HOLTokenStrongEQ} \HOLSymConst{\ensuremath{\tau}}\HOLSymConst{..}(\HOLFreeVar{E} \HOLSymConst{\ensuremath{\parallel}} \HOLSymConst{\ensuremath{\tau}}\HOLSymConst{..}\HOLFreeVar{E\sp{\prime}}) \HOLSymConst{+} \HOLSymConst{\ensuremath{\tau}}\HOLSymConst{..}(\HOLSymConst{\ensuremath{\tau}}\HOLSymConst{..}\HOLFreeVar{E} \HOLSymConst{\ensuremath{\parallel}} \HOLFreeVar{E\sp{\prime}})

STRONG_PAR_PREF_NO_SYNCR:
\HOLTokenTurnstile{} \HOLFreeVar{l} \HOLSymConst{\HOLTokenNotEqual{}} \HOLConst{COMPL} \HOLFreeVar{l\sp{\prime}} \HOLSymConst{\HOLTokenImp{}}
   \HOLSymConst{\HOLTokenForall{}}\HOLBoundVar{E} \HOLBoundVar{E\sp{\prime}}.
     \HOLConst{label} \HOLFreeVar{l}\HOLSymConst{..}\HOLBoundVar{E} \HOLSymConst{\ensuremath{\parallel}} \HOLConst{label} \HOLFreeVar{l\sp{\prime}}\HOLSymConst{..}\HOLBoundVar{E\sp{\prime}} \HOLSymConst{\HOLTokenStrongEQ}
     \HOLConst{label} \HOLFreeVar{l}\HOLSymConst{..}(\HOLBoundVar{E} \HOLSymConst{\ensuremath{\parallel}} \HOLConst{label} \HOLFreeVar{l\sp{\prime}}\HOLSymConst{..}\HOLBoundVar{E\sp{\prime}}) \HOLSymConst{+} \HOLConst{label} \HOLFreeVar{l\sp{\prime}}\HOLSymConst{..}(\HOLConst{label} \HOLFreeVar{l}\HOLSymConst{..}\HOLBoundVar{E} \HOLSymConst{\ensuremath{\parallel}} \HOLBoundVar{E\sp{\prime}})

STRONG_PAR_PREF_SYNCR:
\HOLTokenTurnstile{} (\HOLFreeVar{l} \HOLSymConst{=} \HOLConst{COMPL} \HOLFreeVar{l\sp{\prime}}) \HOLSymConst{\HOLTokenImp{}}
   \HOLSymConst{\HOLTokenForall{}}\HOLBoundVar{E} \HOLBoundVar{E\sp{\prime}}.
     \HOLConst{label} \HOLFreeVar{l}\HOLSymConst{..}\HOLBoundVar{E} \HOLSymConst{\ensuremath{\parallel}} \HOLConst{label} \HOLFreeVar{l\sp{\prime}}\HOLSymConst{..}\HOLBoundVar{E\sp{\prime}} \HOLSymConst{\HOLTokenStrongEQ}
     \HOLConst{label} \HOLFreeVar{l}\HOLSymConst{..}(\HOLBoundVar{E} \HOLSymConst{\ensuremath{\parallel}} \HOLConst{label} \HOLFreeVar{l\sp{\prime}}\HOLSymConst{..}\HOLBoundVar{E\sp{\prime}}) \HOLSymConst{+}
     \HOLConst{label} \HOLFreeVar{l\sp{\prime}}\HOLSymConst{..}(\HOLConst{label} \HOLFreeVar{l}\HOLSymConst{..}\HOLBoundVar{E} \HOLSymConst{\ensuremath{\parallel}} \HOLBoundVar{E\sp{\prime}}) \HOLSymConst{+} \HOLSymConst{\ensuremath{\tau}}\HOLSymConst{..}(\HOLBoundVar{E} \HOLSymConst{\ensuremath{\parallel}} \HOLBoundVar{E\sp{\prime}})
\end{alltt}

And the strong laws for the restriction operator:
\begin{alltt}
STRONG_RESTR_NIL:          \HOLTokenTurnstile{} \HOLSymConst{\ensuremath{\nu}} \HOLFreeVar{L} \HOLConst{nil} \HOLSymConst{\HOLTokenStrongEQ} \HOLConst{nil}
STRONG_RESTR_SUM:          \HOLTokenTurnstile{} \HOLSymConst{\ensuremath{\nu}} \HOLFreeVar{L} (\HOLFreeVar{E} \HOLSymConst{+} \HOLFreeVar{E\sp{\prime}}) \HOLSymConst{\HOLTokenStrongEQ} \HOLSymConst{\ensuremath{\nu}} \HOLFreeVar{L} \HOLFreeVar{E} \HOLSymConst{+} \HOLSymConst{\ensuremath{\nu}} \HOLFreeVar{L} \HOLFreeVar{E\sp{\prime}}
STRONG_RESTR_PREFIX_TAU:   \HOLTokenTurnstile{} \HOLSymConst{\ensuremath{\nu}} \HOLFreeVar{L} (\HOLSymConst{\ensuremath{\tau}}\HOLSymConst{..}\HOLFreeVar{E}) \HOLSymConst{\HOLTokenStrongEQ} \HOLSymConst{\ensuremath{\tau}}\HOLSymConst{..}\HOLSymConst{\ensuremath{\nu}} \HOLFreeVar{L} \HOLFreeVar{E}

STRONG_RESTR_PR_LAB_NIL:
\HOLTokenTurnstile{} \HOLFreeVar{l} \HOLSymConst{\HOLTokenIn{}} \HOLFreeVar{L} \HOLSymConst{\HOLTokenDisj{}} \HOLConst{COMPL} \HOLFreeVar{l} \HOLSymConst{\HOLTokenIn{}} \HOLFreeVar{L} \HOLSymConst{\HOLTokenImp{}} \HOLSymConst{\HOLTokenForall{}}\HOLBoundVar{E}. \HOLSymConst{\ensuremath{\nu}} \HOLFreeVar{L} (\HOLConst{label} \HOLFreeVar{l}\HOLSymConst{..}\HOLBoundVar{E}) \HOLSymConst{\HOLTokenStrongEQ} \HOLConst{nil}

STRONG_RESTR_PREFIX_LABEL:
\HOLTokenTurnstile{} \HOLFreeVar{l} \HOLSymConst{\HOLTokenNotIn{}} \HOLFreeVar{L} \HOLSymConst{\HOLTokenConj{}} \HOLConst{COMPL} \HOLFreeVar{l} \HOLSymConst{\HOLTokenNotIn{}} \HOLFreeVar{L} \HOLSymConst{\HOLTokenImp{}}
   \HOLSymConst{\HOLTokenForall{}}\HOLBoundVar{E}. \HOLSymConst{\ensuremath{\nu}} \HOLFreeVar{L} (\HOLConst{label} \HOLFreeVar{l}\HOLSymConst{..}\HOLBoundVar{E}) \HOLSymConst{\HOLTokenStrongEQ} \HOLConst{label} \HOLFreeVar{l}\HOLSymConst{..}\HOLSymConst{\ensuremath{\nu}} \HOLFreeVar{L} \HOLBoundVar{E}
\end{alltt}

The strong laws for the relabeling operator:
\begin{alltt}
STRONG_RELAB_NIL:
\HOLTokenTurnstile{} \HOLConst{relab} \HOLConst{nil} \HOLFreeVar{rf} \HOLSymConst{\HOLTokenStrongEQ} \HOLConst{nil}

STRONG_RELAB_SUM:
\HOLTokenTurnstile{} \HOLConst{relab} (\HOLFreeVar{E} \HOLSymConst{+} \HOLFreeVar{E\sp{\prime}}) \HOLFreeVar{rf} \HOLSymConst{\HOLTokenStrongEQ} \HOLConst{relab} \HOLFreeVar{E} \HOLFreeVar{rf} \HOLSymConst{+} \HOLConst{relab} \HOLFreeVar{E\sp{\prime}} \HOLFreeVar{rf}

STRONG_RELAB_PREFIX:
\HOLTokenTurnstile{} \HOLConst{relab} (\HOLFreeVar{u}\HOLSymConst{..}\HOLFreeVar{E}) (\HOLConst{RELAB} \HOLFreeVar{labl}) \HOLSymConst{\HOLTokenStrongEQ}
   \HOLConst{relabel} (\HOLConst{RELAB} \HOLFreeVar{labl}) \HOLFreeVar{u}\HOLSymConst{..}\HOLConst{relab} \HOLFreeVar{E} (\HOLConst{RELAB} \HOLFreeVar{labl})
\end{alltt}

The strong laws for the recursion operator (for constants):
\begin{alltt}
STRONG_UNFOLDING:
\HOLTokenTurnstile{} \HOLConst{rec} \HOLFreeVar{X} \HOLFreeVar{E} \HOLSymConst{\HOLTokenStrongEQ} \HOLConst{CCS_Subst} \HOLFreeVar{E} (\HOLConst{rec} \HOLFreeVar{X} \HOLFreeVar{E}) \HOLFreeVar{X}

STRONG_PREF_REC_EQUIV:
\HOLTokenTurnstile{} \HOLFreeVar{u}\HOLSymConst{..}\HOLConst{rec} \HOLFreeVar{s} (\HOLFreeVar{v}\HOLSymConst{..}\HOLFreeVar{u}\HOLSymConst{..}\HOLConst{var} \HOLFreeVar{s}) \HOLSymConst{\HOLTokenStrongEQ} \HOLConst{rec} \HOLFreeVar{s} (\HOLFreeVar{u}\HOLSymConst{..}\HOLFreeVar{v}\HOLSymConst{..}\HOLConst{var} \HOLFreeVar{s})

STRONG_REC_ACT2:
\HOLTokenTurnstile{} \HOLConst{rec} \HOLFreeVar{s} (\HOLFreeVar{u}\HOLSymConst{..}\HOLFreeVar{u}\HOLSymConst{..}\HOLConst{var} \HOLFreeVar{s}) \HOLSymConst{\HOLTokenStrongEQ} \HOLConst{rec} \HOLFreeVar{s} (\HOLFreeVar{u}\HOLSymConst{..}\HOLConst{var} \HOLFreeVar{s})
\end{alltt}

All above three theorems for recursion operator were fundamental (in the sense that, they
cannot be proved by just using other strong laws).

Finally, all above strong laws could be used either manually or as part of the
decision procedure for automatically deciding strong equivalences
between two CCS process. However such a decision procedure is not done
in the current project.

\subsection{The Strong Expansion Law}

Another big piece of proof work in this project is the
representation and proof of the following \emph{expansion law} (sometimes
also called the \emph{interleaving law}:
\begin{proposition}{(Expansion Law)}
Let $p = \sum_{i=1}^n \mu_i.p_i$ and $q = \sum_{j=1}^m
\mu'_j.q_j$. Then
\begin{equation}
p | q \sim \sum_{i=1}^n \mu_i.(p_i | q) + \sum_{j=1}^m \mu'_j.(p|q_j)
+ \sum_{i,j:\overline{\mu_i}=\mu'_j} \tau.(p_i | q_j)
\end{equation}
\end{proposition}

Some characteristics made the formal proof very special and different from all other theorems that
we have proved so far. First of all, arithmetic numbers (of type \HOLinline{\HOLTyOp{num}}) were involved for the first
  time, and now our CCS theory depends on elementary mathematical theories provided by HOL,
  namely the \texttt{prim_recTheory} and
  \texttt{arithmeticTheory}. Although arithmetic operations like $+,
  -, \cdot, /$ were not involved (yet), but we do need to compare
  number values and use some related theorems.

Also two CCS accessors were defined and used to access the internal
  structure of CCS processes, namely \HOLinline{\HOLConst{PREF_ACT}} for getting the
  initial action and \HOLinline{\HOLConst{PREF_PROC}} for getting the rest of process
  without the first action. Together there's predicate
  \HOLinline{\HOLConst{Is_Prefix}} for testing if a CCS is a prefixed process:
\begin{alltt}
\HOLTokenTurnstile{} \HOLConst{PREF_ACT} (\HOLFreeVar{u}\HOLSymConst{..}\HOLFreeVar{E}) \HOLSymConst{=} \HOLFreeVar{u}
\HOLTokenTurnstile{} \HOLConst{PREF_PROC} (\HOLFreeVar{u}\HOLSymConst{..}\HOLFreeVar{E}) \HOLSymConst{=} \HOLFreeVar{E}
\HOLTokenTurnstile{} \HOLConst{Is_Prefix} \HOLFreeVar{E} \HOLSymConst{\HOLTokenEquiv{}} \HOLSymConst{\HOLTokenExists{}}\HOLBoundVar{u} \HOLBoundVar{E\sp{\prime}}. \HOLFreeVar{E} \HOLSymConst{=} \HOLBoundVar{u}\HOLSymConst{..}\HOLBoundVar{E\sp{\prime}}
\end{alltt}
They are needed because we're going to represent $\mu_i.p_i$ as the
value of a function: $f(i)$
in which $f$ has the type \HOLinline{\HOLTyOp{num} \HOLTokenTransEnd (\ensuremath{\alpha}, \ensuremath{\beta}) \HOLTyOp{CCS}}. And in
this way, to get $\mu_i$ and $p_i$ we have to use accessors:
``\HOLinline{\HOLConst{PREF_ACT} (\HOLFreeVar{f} \HOLFreeVar{i})}'' and ``\HOLinline{\HOLConst{PREF_PROC} (\HOLFreeVar{f} \HOLFreeVar{i})}''.

The next job is to represent a finite sum of CCS processes. This is
done by the following recursive function \HOLinline{\HOLConst{SIGMA}}:
\begin{alltt}
\HOLConst{SIGMA} \HOLFreeVar{f} \HOLNumLit{0} \HOLSymConst{=} \HOLFreeVar{f} \HOLNumLit{0}
\HOLConst{SIGMA} \HOLFreeVar{f} (\HOLConst{SUC} \HOLFreeVar{n}) \HOLSymConst{=} \HOLConst{SIGMA} \HOLFreeVar{f} \HOLFreeVar{n} \HOLSymConst{+} \HOLFreeVar{f} (\HOLConst{SUC} \HOLFreeVar{n})
\end{alltt}
Thus if there's a function $f$ of type \HOLinline{\HOLTyOp{num} \HOLTokenTransEnd (\ensuremath{\alpha}, \ensuremath{\beta}) \HOLTyOp{CCS}}, we should
be able to represent $\sum_{i=1}^n f(i)$ by HOL term
``\HOLinline{\HOLConst{SIGMA} \HOLFreeVar{f} \HOLFreeVar{n}}''.

Now if we took a deeper look at the last summation of the right side
of the expansion law, i.e. $\sum_{i,j:\overline{\mu_i}=\mu'_j}
\tau.(p_i | q_j)$, we found that such a ``sum'' cannot be represented
directly, because there're two index $i,j$ and their possible value
pairs used in the sum depends on the synchronization of corresponding
actions from each $p_i$ and $q_j$.  What we actually need is a
recursively defined function taking all the $p_i$ and $q_j$ and return
the synchronized process in forms like $\sum \tau.(p_i | q_j)$.

But this is still too complicated, instead we first define functions to synchronize
just one process with another group of processes. This work is
achieved by the function \HOLinline{\HOLConst{SYNC}} of type \HOLinline{\ensuremath{\beta} \HOLTyOp{Action} \HOLTokenTransEnd
           (\ensuremath{\alpha}, \ensuremath{\beta}) \HOLTyOp{CCS} \HOLTokenTransEnd
           (\HOLTyOp{num} \HOLTokenTransEnd (\ensuremath{\alpha}, \ensuremath{\beta}) \HOLTyOp{CCS}) \HOLTokenTransEnd \HOLTyOp{num} \HOLTokenTransEnd (\ensuremath{\alpha}, \ensuremath{\beta}) \HOLTyOp{CCS}}:
\begin{alltt}
\HOLTokenTurnstile{} (\HOLSymConst{\HOLTokenForall{}}\HOLBoundVar{u} \HOLBoundVar{P} \HOLBoundVar{f}.
      \HOLConst{SYNC} \HOLBoundVar{u} \HOLBoundVar{P} \HOLBoundVar{f} \HOLNumLit{0} \HOLSymConst{=}
      \HOLKeyword{if} (\HOLBoundVar{u} \HOLSymConst{=} \HOLSymConst{\ensuremath{\tau}}) \HOLSymConst{\HOLTokenDisj{}} (\HOLConst{PREF_ACT} (\HOLBoundVar{f} \HOLNumLit{0}) \HOLSymConst{=} \HOLSymConst{\ensuremath{\tau}}) \HOLKeyword{then} \HOLConst{nil}
      \HOLKeyword{else} \HOLKeyword{if} \HOLConst{LABEL} \HOLBoundVar{u} \HOLSymConst{=} \HOLConst{COMPL} (\HOLConst{LABEL} (\HOLConst{PREF_ACT} (\HOLBoundVar{f} \HOLNumLit{0}))) \HOLKeyword{then}
        \HOLSymConst{\ensuremath{\tau}}\HOLSymConst{..}(\HOLBoundVar{P} \HOLSymConst{\ensuremath{\parallel}} \HOLConst{PREF_PROC} (\HOLBoundVar{f} \HOLNumLit{0}))
      \HOLKeyword{else} \HOLConst{nil}) \HOLSymConst{\HOLTokenConj{}}
   \HOLSymConst{\HOLTokenForall{}}\HOLBoundVar{u} \HOLBoundVar{P} \HOLBoundVar{f} \HOLBoundVar{n}.
     \HOLConst{SYNC} \HOLBoundVar{u} \HOLBoundVar{P} \HOLBoundVar{f} (\HOLConst{SUC} \HOLBoundVar{n}) \HOLSymConst{=}
     \HOLKeyword{if} (\HOLBoundVar{u} \HOLSymConst{=} \HOLSymConst{\ensuremath{\tau}}) \HOLSymConst{\HOLTokenDisj{}} (\HOLConst{PREF_ACT} (\HOLBoundVar{f} (\HOLConst{SUC} \HOLBoundVar{n})) \HOLSymConst{=} \HOLSymConst{\ensuremath{\tau}}) \HOLKeyword{then}
       \HOLConst{SYNC} \HOLBoundVar{u} \HOLBoundVar{P} \HOLBoundVar{f} \HOLBoundVar{n}
     \HOLKeyword{else} \HOLKeyword{if}
       \HOLConst{LABEL} \HOLBoundVar{u} \HOLSymConst{=} \HOLConst{COMPL} (\HOLConst{LABEL} (\HOLConst{PREF_ACT} (\HOLBoundVar{f} (\HOLConst{SUC} \HOLBoundVar{n}))))
     \HOLKeyword{then}
       \HOLSymConst{\ensuremath{\tau}}\HOLSymConst{..}(\HOLBoundVar{P} \HOLSymConst{\ensuremath{\parallel}} \HOLConst{PREF_PROC} (\HOLBoundVar{f} (\HOLConst{SUC} \HOLBoundVar{n}))) \HOLSymConst{+} \HOLConst{SYNC} \HOLBoundVar{u} \HOLBoundVar{P} \HOLBoundVar{f} \HOLBoundVar{n}
     \HOLKeyword{else} \HOLConst{SYNC} \HOLBoundVar{u} \HOLBoundVar{P} \HOLBoundVar{f} \HOLBoundVar{n}
\end{alltt}

Then the synchronization of two group of processes can be further
defined by another recursive function \HOLinline{\HOLConst{ALL_SYNC}} of type
\HOLinline{(\HOLTyOp{num} \HOLTokenTransEnd (\ensuremath{\alpha}, \ensuremath{\beta}) \HOLTyOp{CCS}) \HOLTokenTransEnd
           \HOLTyOp{num} \HOLTokenTransEnd (\HOLTyOp{num} \HOLTokenTransEnd (\ensuremath{\alpha}, \ensuremath{\beta}) \HOLTyOp{CCS}) \HOLTokenTransEnd \HOLTyOp{num} \HOLTokenTransEnd (\ensuremath{\alpha}, \ensuremath{\beta}) \HOLTyOp{CCS}}:
\begin{alltt}
\HOLTokenTurnstile{} (\HOLSymConst{\HOLTokenForall{}}\HOLBoundVar{f} \HOLBoundVar{f\sp{\prime}} \HOLBoundVar{m}.
      \HOLConst{ALL_SYNC} \HOLBoundVar{f} \HOLNumLit{0} \HOLBoundVar{f\sp{\prime}} \HOLBoundVar{m} \HOLSymConst{=}
      \HOLConst{SYNC} (\HOLConst{PREF_ACT} (\HOLBoundVar{f} \HOLNumLit{0})) (\HOLConst{PREF_PROC} (\HOLBoundVar{f} \HOLNumLit{0})) \HOLBoundVar{f\sp{\prime}} \HOLBoundVar{m}) \HOLSymConst{\HOLTokenConj{}}
   \HOLSymConst{\HOLTokenForall{}}\HOLBoundVar{f} \HOLBoundVar{n} \HOLBoundVar{f\sp{\prime}} \HOLBoundVar{m}.
     \HOLConst{ALL_SYNC} \HOLBoundVar{f} (\HOLConst{SUC} \HOLBoundVar{n}) \HOLBoundVar{f\sp{\prime}} \HOLBoundVar{m} \HOLSymConst{=}
     \HOLConst{ALL_SYNC} \HOLBoundVar{f} \HOLBoundVar{n} \HOLBoundVar{f\sp{\prime}} \HOLBoundVar{m} \HOLSymConst{+}
     \HOLConst{SYNC} (\HOLConst{PREF_ACT} (\HOLBoundVar{f} (\HOLConst{SUC} \HOLBoundVar{n}))) (\HOLConst{PREF_PROC} (\HOLBoundVar{f} (\HOLConst{SUC} \HOLBoundVar{n}))) \HOLBoundVar{f\sp{\prime}} \HOLBoundVar{m}
\end{alltt}

Some lemmas about \HOLinline{\HOLConst{SIGMA}} and the two synchronization
functions were proved first:
\begin{alltt}
SIGMA_TRANS_THM_EQ:
\HOLTokenTurnstile{} \HOLConst{SIGMA} \HOLFreeVar{f} \HOLFreeVar{n} \HOLTokenTransBegin\HOLFreeVar{u}\HOLTokenTransEnd \HOLFreeVar{E} \HOLSymConst{\HOLTokenEquiv{}} \HOLSymConst{\HOLTokenExists{}}\HOLBoundVar{k}. \HOLBoundVar{k} \HOLSymConst{\HOLTokenLeq{}} \HOLFreeVar{n} \HOLSymConst{\HOLTokenConj{}} \HOLFreeVar{f} \HOLBoundVar{k} \HOLTokenTransBegin\HOLFreeVar{u}\HOLTokenTransEnd \HOLFreeVar{E}

SYNC_TRANS_THM_EQ:
\HOLTokenTurnstile{} \HOLConst{SYNC} \HOLFreeVar{u} \HOLFreeVar{P} \HOLFreeVar{f} \HOLFreeVar{m} \HOLTokenTransBegin\HOLFreeVar{v}\HOLTokenTransEnd \HOLFreeVar{Q} \HOLSymConst{\HOLTokenEquiv{}}
   \HOLSymConst{\HOLTokenExists{}}\HOLBoundVar{j} \HOLBoundVar{l}.
     \HOLBoundVar{j} \HOLSymConst{\HOLTokenLeq{}} \HOLFreeVar{m} \HOLSymConst{\HOLTokenConj{}} (\HOLFreeVar{u} \HOLSymConst{=} \HOLConst{label} \HOLBoundVar{l}) \HOLSymConst{\HOLTokenConj{}}
     (\HOLConst{PREF_ACT} (\HOLFreeVar{f} \HOLBoundVar{j}) \HOLSymConst{=} \HOLConst{label} (\HOLConst{COMPL} \HOLBoundVar{l})) \HOLSymConst{\HOLTokenConj{}} (\HOLFreeVar{v} \HOLSymConst{=} \HOLSymConst{\ensuremath{\tau}}) \HOLSymConst{\HOLTokenConj{}}
     (\HOLFreeVar{Q} \HOLSymConst{=} \HOLFreeVar{P} \HOLSymConst{\ensuremath{\parallel}} \HOLConst{PREF_PROC} (\HOLFreeVar{f} \HOLBoundVar{j}))

ALL_SYNC_TRANS_THM_EQ:
\HOLTokenTurnstile{} \HOLConst{ALL_SYNC} \HOLFreeVar{f} \HOLFreeVar{n} \HOLFreeVar{f\sp{\prime}} \HOLFreeVar{m} \HOLTokenTransBegin\HOLFreeVar{u}\HOLTokenTransEnd \HOLFreeVar{E} \HOLSymConst{\HOLTokenEquiv{}}
   \HOLSymConst{\HOLTokenExists{}}\HOLBoundVar{k} \HOLBoundVar{k\sp{\prime}} \HOLBoundVar{l}.
     \HOLBoundVar{k} \HOLSymConst{\HOLTokenLeq{}} \HOLFreeVar{n} \HOLSymConst{\HOLTokenConj{}} \HOLBoundVar{k\sp{\prime}} \HOLSymConst{\HOLTokenLeq{}} \HOLFreeVar{m} \HOLSymConst{\HOLTokenConj{}} (\HOLConst{PREF_ACT} (\HOLFreeVar{f} \HOLBoundVar{k}) \HOLSymConst{=} \HOLConst{label} \HOLBoundVar{l}) \HOLSymConst{\HOLTokenConj{}}
     (\HOLConst{PREF_ACT} (\HOLFreeVar{f\sp{\prime}} \HOLBoundVar{k\sp{\prime}}) \HOLSymConst{=} \HOLConst{label} (\HOLConst{COMPL} \HOLBoundVar{l})) \HOLSymConst{\HOLTokenConj{}} (\HOLFreeVar{u} \HOLSymConst{=} \HOLSymConst{\ensuremath{\tau}}) \HOLSymConst{\HOLTokenConj{}}
     (\HOLFreeVar{E} \HOLSymConst{=} \HOLConst{PREF_PROC} (\HOLFreeVar{f} \HOLBoundVar{k}) \HOLSymConst{\ensuremath{\parallel}} \HOLConst{PREF_PROC} (\HOLFreeVar{f\sp{\prime}} \HOLBoundVar{k\sp{\prime}}))
\end{alltt}

Finally, we have proved the Expansion Law in the following form:
\begin{alltt}
STRONG_EXPANSION_LAW:
\HOLTokenTurnstile{} (\HOLSymConst{\HOLTokenForall{}}\HOLBoundVar{i}. \HOLBoundVar{i} \HOLSymConst{\HOLTokenLeq{}} \HOLFreeVar{n} \HOLSymConst{\HOLTokenImp{}} \HOLConst{Is_Prefix} (\HOLFreeVar{f} \HOLBoundVar{i})) \HOLSymConst{\HOLTokenConj{}}
   (\HOLSymConst{\HOLTokenForall{}}\HOLBoundVar{j}. \HOLBoundVar{j} \HOLSymConst{\HOLTokenLeq{}} \HOLFreeVar{m} \HOLSymConst{\HOLTokenImp{}} \HOLConst{Is_Prefix} (\HOLFreeVar{f\sp{\prime}} \HOLBoundVar{j})) \HOLSymConst{\HOLTokenImp{}}
   \HOLConst{SIGMA} \HOLFreeVar{f} \HOLFreeVar{n} \HOLSymConst{\ensuremath{\parallel}} \HOLConst{SIGMA} \HOLFreeVar{f\sp{\prime}} \HOLFreeVar{m} \HOLSymConst{\HOLTokenStrongEQ}
   \HOLConst{SIGMA} (\HOLTokenLambda{}\HOLBoundVar{i}. \HOLConst{PREF_ACT} (\HOLFreeVar{f} \HOLBoundVar{i})\HOLSymConst{..}(\HOLConst{PREF_PROC} (\HOLFreeVar{f} \HOLBoundVar{i}) \HOLSymConst{\ensuremath{\parallel}} \HOLConst{SIGMA} \HOLFreeVar{f\sp{\prime}} \HOLFreeVar{m}))
     \HOLFreeVar{n} \HOLSymConst{+}
   \HOLConst{SIGMA} (\HOLTokenLambda{}\HOLBoundVar{j}. \HOLConst{PREF_ACT} (\HOLFreeVar{f\sp{\prime}} \HOLBoundVar{j})\HOLSymConst{..}(\HOLConst{SIGMA} \HOLFreeVar{f} \HOLFreeVar{n} \HOLSymConst{\ensuremath{\parallel}} \HOLConst{PREF_PROC} (\HOLFreeVar{f\sp{\prime}} \HOLBoundVar{j})))
     \HOLFreeVar{m} \HOLSymConst{+} \HOLConst{ALL_SYNC} \HOLFreeVar{f} \HOLFreeVar{n} \HOLFreeVar{f\sp{\prime}} \HOLFreeVar{m}
\end{alltt}

\section{Weak transitions and the EPS relation}

In this part, the main purpose
is to define the weak equivalence co-inductively \emph{first} and then
prove the traditional definition (like \texttt{STRONG_EQUIV}) as a
theorem. Using HOL's
coinductive relation module (\texttt{Hol_coreln}), it's much easier to get the same set of theorems
like those for strong equivalence. These works are not part of the old
CCS formalization in Hol88.

There're multiple ways to define the concept of weak transitions used
in the definition of weak bisimulation. In early approach like Milner's
book, the first step is to define a \texttt{EPS} relation, which indicates that between two
processes there's nothing but zero or more $\tau$ transitions. In HOL,
this can be defined through a non-recursive inductive relation and the
RTC (reflexive transitive closure) on top of it:
\begin{definition}{(EPS)}
For any two CCS processes $E, E' \in Q$, define relation $EPS \subseteq
Q\times Q$ as the reflexive transitive closure (RTC) of single-$\tau$
transition between $E$ and $E'$ ($E \overset{\tau}{\longrightarrow} E'$):\footnote{In HOL4's
  \texttt{relationTheory}, the relation types is
  curried: instead of having the same type ``\HOLinline{\ensuremath{\alpha} \HOLTyOp{reln}}'' as the math definition, it has the type ``\HOLinline{\ensuremath{\alpha} \HOLTokenTransEnd \ensuremath{\alpha} \HOLTokenTransEnd \HOLTyOp{bool}}''. And the star(*) notation is for defining RTCs.}
\begin{alltt}
\HOLTokenTurnstile{} \HOLConst{EPS} \HOLSymConst{=} (\HOLTokenLambda{}\HOLBoundVar{E} \HOLBoundVar{E\sp{\prime}}. \HOLBoundVar{E} \HOLTokenTransBegin\HOLSymConst{\ensuremath{\tau}}\HOLTokenTransEnd \HOLBoundVar{E\sp{\prime}})\HOLSymConst{\HOLTokenSupStar{}}
\end{alltt}
Intuitively speaking, \HOLinline{\HOLFreeVar{E} \HOLSymConst{\HOLTokenEPS} \HOLFreeVar{E\sp{\prime}}} (Math notion: $E
\overset{\epsilon}{\Longrightarrow} E'$) means there're zero or more $tau$-transitions from $p$ to $q$.
\end{definition}

Sometimes it's necessary to consider different transition cases when
\HOLinline{\HOLFreeVar{p} \HOLSymConst{\HOLTokenEPS} \HOLFreeVar{q}} holds, or induct on the number of $tau$ transitions
between $p$ and $q$. With such a
definition, beside the obvious reflexive and transitive
properties, a large amount of ``cases'' and induction theorem
already proved in HOL's \texttt{relationTheory} are immediately
available to us:
\begin{proposition}{(The ``cases'' theorem of the \HOLinline{\HOLConst{EPS}} relation)}
\begin{enumerate}
\item Taking one $\tau$-transition at left: \hfill[\texttt{EPS_cases1}]
\begin{alltt}
\HOLTokenTurnstile{} \HOLFreeVar{x} \HOLSymConst{\HOLTokenEPS} \HOLFreeVar{y} \HOLSymConst{\HOLTokenEquiv{}} (\HOLFreeVar{x} \HOLSymConst{=} \HOLFreeVar{y}) \HOLSymConst{\HOLTokenDisj{}} \HOLSymConst{\HOLTokenExists{}}\HOLBoundVar{u}. \HOLFreeVar{x} \HOLTokenTransBegin\HOLSymConst{\ensuremath{\tau}}\HOLTokenTransEnd \HOLBoundVar{u} \HOLSymConst{\HOLTokenConj{}} \HOLBoundVar{u} \HOLSymConst{\HOLTokenEPS} \HOLFreeVar{y}
\end{alltt}
\item Taking one $\tau$-transition at right: \hfill[\texttt{EPS_cases2}]
\begin{alltt}
\HOLTokenTurnstile{} \HOLFreeVar{x} \HOLSymConst{\HOLTokenEPS} \HOLFreeVar{y} \HOLSymConst{\HOLTokenEquiv{}} (\HOLFreeVar{x} \HOLSymConst{=} \HOLFreeVar{y}) \HOLSymConst{\HOLTokenDisj{}} \HOLSymConst{\HOLTokenExists{}}\HOLBoundVar{u}. \HOLFreeVar{x} \HOLSymConst{\HOLTokenEPS} \HOLBoundVar{u} \HOLSymConst{\HOLTokenConj{}} \HOLBoundVar{u} \HOLTokenTransBegin\HOLSymConst{\ensuremath{\tau}}\HOLTokenTransEnd \HOLFreeVar{y}
\end{alltt}
\item Three cases of EPS transition: \hfill[\texttt{EPS_cases}]
\begin{alltt}
\HOLTokenTurnstile{} \HOLFreeVar{E} \HOLSymConst{\HOLTokenEPS} \HOLFreeVar{E\sp{\prime}} \HOLSymConst{\HOLTokenEquiv{}} \HOLFreeVar{E} \HOLTokenTransBegin\HOLSymConst{\ensuremath{\tau}}\HOLTokenTransEnd \HOLFreeVar{E\sp{\prime}} \HOLSymConst{\HOLTokenDisj{}} (\HOLFreeVar{E} \HOLSymConst{=} \HOLFreeVar{E\sp{\prime}}) \HOLSymConst{\HOLTokenDisj{}} \HOLSymConst{\HOLTokenExists{}}\HOLBoundVar{E\sb{\mathrm{1}}}. \HOLFreeVar{E} \HOLSymConst{\HOLTokenEPS} \HOLBoundVar{E\sb{\mathrm{1}}} \HOLSymConst{\HOLTokenConj{}} \HOLBoundVar{E\sb{\mathrm{1}}} \HOLSymConst{\HOLTokenEPS} \HOLFreeVar{E\sp{\prime}}
\end{alltt}
\end{enumerate}
\end{proposition}

\begin{proposition}{(The induction and strong induction principles of
   the \HOLinline{\HOLConst{EPS}} relation)}
\begin{enumerate}
\item Induction from left: \hfill[\texttt{EPS_ind}]
\begin{alltt}
\HOLTokenTurnstile{} (\HOLSymConst{\HOLTokenForall{}}\HOLBoundVar{x}. \HOLFreeVar{P} \HOLBoundVar{x} \HOLBoundVar{x}) \HOLSymConst{\HOLTokenConj{}} (\HOLSymConst{\HOLTokenForall{}}\HOLBoundVar{x} \HOLBoundVar{y} \HOLBoundVar{z}. \HOLBoundVar{x} \HOLTokenTransBegin\HOLSymConst{\ensuremath{\tau}}\HOLTokenTransEnd \HOLBoundVar{y} \HOLSymConst{\HOLTokenConj{}} \HOLFreeVar{P} \HOLBoundVar{y} \HOLBoundVar{z} \HOLSymConst{\HOLTokenImp{}} \HOLFreeVar{P} \HOLBoundVar{x} \HOLBoundVar{z}) \HOLSymConst{\HOLTokenImp{}}
   \HOLSymConst{\HOLTokenForall{}}\HOLBoundVar{x} \HOLBoundVar{y}. \HOLBoundVar{x} \HOLSymConst{\HOLTokenEPS} \HOLBoundVar{y} \HOLSymConst{\HOLTokenImp{}} \HOLFreeVar{P} \HOLBoundVar{x} \HOLBoundVar{y}
\end{alltt}
\item Induction from right: \hfill[\texttt{EPS_ind_right}]
\begin{alltt}
\HOLTokenTurnstile{} (\HOLSymConst{\HOLTokenForall{}}\HOLBoundVar{x}. \HOLFreeVar{P} \HOLBoundVar{x} \HOLBoundVar{x}) \HOLSymConst{\HOLTokenConj{}} (\HOLSymConst{\HOLTokenForall{}}\HOLBoundVar{x} \HOLBoundVar{y} \HOLBoundVar{z}. \HOLFreeVar{P} \HOLBoundVar{x} \HOLBoundVar{y} \HOLSymConst{\HOLTokenConj{}} \HOLBoundVar{y} \HOLTokenTransBegin\HOLSymConst{\ensuremath{\tau}}\HOLTokenTransEnd \HOLBoundVar{z} \HOLSymConst{\HOLTokenImp{}} \HOLFreeVar{P} \HOLBoundVar{x} \HOLBoundVar{z}) \HOLSymConst{\HOLTokenImp{}}
   \HOLSymConst{\HOLTokenForall{}}\HOLBoundVar{x} \HOLBoundVar{y}. \HOLBoundVar{x} \HOLSymConst{\HOLTokenEPS} \HOLBoundVar{y} \HOLSymConst{\HOLTokenImp{}} \HOLFreeVar{P} \HOLBoundVar{x} \HOLBoundVar{y}
\end{alltt}
\item Strong induction from left: \hfill[\texttt{EPS_strongind}]
\begin{alltt}
\HOLTokenTurnstile{} (\HOLSymConst{\HOLTokenForall{}}\HOLBoundVar{x}. \HOLFreeVar{P} \HOLBoundVar{x} \HOLBoundVar{x}) \HOLSymConst{\HOLTokenConj{}} (\HOLSymConst{\HOLTokenForall{}}\HOLBoundVar{x} \HOLBoundVar{y} \HOLBoundVar{z}. \HOLBoundVar{x} \HOLTokenTransBegin\HOLSymConst{\ensuremath{\tau}}\HOLTokenTransEnd \HOLBoundVar{y} \HOLSymConst{\HOLTokenConj{}} \HOLBoundVar{y} \HOLSymConst{\HOLTokenEPS} \HOLBoundVar{z} \HOLSymConst{\HOLTokenConj{}} \HOLFreeVar{P} \HOLBoundVar{y} \HOLBoundVar{z} \HOLSymConst{\HOLTokenImp{}} \HOLFreeVar{P} \HOLBoundVar{x} \HOLBoundVar{z}) \HOLSymConst{\HOLTokenImp{}}
   \HOLSymConst{\HOLTokenForall{}}\HOLBoundVar{x} \HOLBoundVar{y}. \HOLBoundVar{x} \HOLSymConst{\HOLTokenEPS} \HOLBoundVar{y} \HOLSymConst{\HOLTokenImp{}} \HOLFreeVar{P} \HOLBoundVar{x} \HOLBoundVar{y}
\end{alltt}
\item Strong induction from right: \hfill[\texttt{EPS_strongind_right}]
\begin{alltt}
\HOLTokenTurnstile{} (\HOLSymConst{\HOLTokenForall{}}\HOLBoundVar{x}. \HOLFreeVar{P} \HOLBoundVar{x} \HOLBoundVar{x}) \HOLSymConst{\HOLTokenConj{}} (\HOLSymConst{\HOLTokenForall{}}\HOLBoundVar{x} \HOLBoundVar{y} \HOLBoundVar{z}. \HOLFreeVar{P} \HOLBoundVar{x} \HOLBoundVar{y} \HOLSymConst{\HOLTokenConj{}} \HOLBoundVar{x} \HOLSymConst{\HOLTokenEPS} \HOLBoundVar{y} \HOLSymConst{\HOLTokenConj{}} \HOLBoundVar{y} \HOLTokenTransBegin\HOLSymConst{\ensuremath{\tau}}\HOLTokenTransEnd \HOLBoundVar{z} \HOLSymConst{\HOLTokenImp{}} \HOLFreeVar{P} \HOLBoundVar{x} \HOLBoundVar{z}) \HOLSymConst{\HOLTokenImp{}}
   \HOLSymConst{\HOLTokenForall{}}\HOLBoundVar{x} \HOLBoundVar{y}. \HOLBoundVar{x} \HOLSymConst{\HOLTokenEPS} \HOLBoundVar{y} \HOLSymConst{\HOLTokenImp{}} \HOLFreeVar{P} \HOLBoundVar{x} \HOLBoundVar{y}
\end{alltt}
\item Induction from the middle: \hfill[\texttt{EPS_INDUCT}]
\begin{alltt}
\HOLTokenTurnstile{} (\HOLSymConst{\HOLTokenForall{}}\HOLBoundVar{E} \HOLBoundVar{E\sp{\prime}}. \HOLBoundVar{E} \HOLTokenTransBegin\HOLSymConst{\ensuremath{\tau}}\HOLTokenTransEnd \HOLBoundVar{E\sp{\prime}} \HOLSymConst{\HOLTokenImp{}} \HOLFreeVar{P} \HOLBoundVar{E} \HOLBoundVar{E\sp{\prime}}) \HOLSymConst{\HOLTokenConj{}} (\HOLSymConst{\HOLTokenForall{}}\HOLBoundVar{E}. \HOLFreeVar{P} \HOLBoundVar{E} \HOLBoundVar{E}) \HOLSymConst{\HOLTokenConj{}}
   (\HOLSymConst{\HOLTokenForall{}}\HOLBoundVar{E} \HOLBoundVar{E\sb{\mathrm{1}}} \HOLBoundVar{E\sp{\prime}}. \HOLFreeVar{P} \HOLBoundVar{E} \HOLBoundVar{E\sb{\mathrm{1}}} \HOLSymConst{\HOLTokenConj{}} \HOLFreeVar{P} \HOLBoundVar{E\sb{\mathrm{1}}} \HOLBoundVar{E\sp{\prime}} \HOLSymConst{\HOLTokenImp{}} \HOLFreeVar{P} \HOLBoundVar{E} \HOLBoundVar{E\sp{\prime}}) \HOLSymConst{\HOLTokenImp{}}
   \HOLSymConst{\HOLTokenForall{}}\HOLBoundVar{x} \HOLBoundVar{y}. \HOLBoundVar{x} \HOLSymConst{\HOLTokenEPS} \HOLBoundVar{y} \HOLSymConst{\HOLTokenImp{}} \HOLFreeVar{P} \HOLBoundVar{x} \HOLBoundVar{y} 
\end{alltt}
\end{enumerate}
\end{proposition}

Then we define the weak transition between two CCS processes upon the
\HOLinline{\HOLConst{EPS}} relation:
\begin{definition}
For any two CCS processes $E, E' \in Q$, define ``weak transition'' relation $\Longrightarrow \subseteq
Q\times A \times Q$, where A can be $\tau$ or a visible action:
$E \overset{a}{\longrightarrow} E'$ if and only if there exists two processes
$E_1$ and $E_2$ such that $E \overset{\epsilon}{\Longrightarrow} E_1
\overset{a}{\longrightarrow} E_2 \overset{\epsilon}{\Longrightarrow} E'$:
\begin{alltt}
WEAK_TRANS:
\HOLTokenTurnstile{} \HOLFreeVar{E} \HOLTokenWeakTransBegin\HOLFreeVar{u}\HOLTokenWeakTransEnd \HOLFreeVar{E\sp{\prime}} \HOLSymConst{\HOLTokenEquiv{}} \HOLSymConst{\HOLTokenExists{}}\HOLBoundVar{E\sb{\mathrm{1}}} \HOLBoundVar{E\sb{\mathrm{2}}}. \HOLFreeVar{E} \HOLSymConst{\HOLTokenEPS} \HOLBoundVar{E\sb{\mathrm{1}}} \HOLSymConst{\HOLTokenConj{}} \HOLBoundVar{E\sb{\mathrm{1}}} \HOLTokenTransBegin\HOLFreeVar{u}\HOLTokenTransEnd \HOLBoundVar{E\sb{\mathrm{2}}} \HOLSymConst{\HOLTokenConj{}} \HOLBoundVar{E\sb{\mathrm{2}}} \HOLSymConst{\HOLTokenEPS} \HOLFreeVar{E\sp{\prime}}
\end{alltt}
\end{definition}

Using above two definitions and the ``cases'' and induction theorems,
a large amount of properties about \HOLinline{\HOLConst{EPS}} and \HOLinline{\HOLConst{WEAK_TRANS}}
were proved:
\begin{proposition}{(Properties of \HOLinline{\HOLConst{EPS}} and \HOLinline{\HOLConst{WEAK_TRANS}})}
\begin{enumerate}
\item Any transition also implies a weak transition:
\begin{alltt}
\HOLTokenTurnstile{} \HOLFreeVar{E} \HOLTokenTransBegin\HOLFreeVar{u}\HOLTokenTransEnd \HOLFreeVar{E\sp{\prime}} \HOLSymConst{\HOLTokenImp{}} \HOLFreeVar{E} \HOLTokenWeakTransBegin\HOLFreeVar{u}\HOLTokenWeakTransEnd \HOLFreeVar{E\sp{\prime}}\hfill [TRANS_IMP_WEAK_TRANS]
\end{alltt}

\item Weak $\tau$-transition  implies \HOLinline{\HOLConst{EPS}} relation:
\begin{alltt}
\HOLTokenTurnstile{} \HOLFreeVar{E} \HOLTokenWeakTransBegin\HOLSymConst{\ensuremath{\tau}}\HOLTokenWeakTransEnd \HOLFreeVar{E\sb{\mathrm{1}}} \HOLSymConst{\HOLTokenEquiv{}} \HOLSymConst{\HOLTokenExists{}}\HOLBoundVar{E\sp{\prime}}. \HOLFreeVar{E} \HOLTokenTransBegin\HOLSymConst{\ensuremath{\tau}}\HOLTokenTransEnd \HOLBoundVar{E\sp{\prime}} \HOLSymConst{\HOLTokenConj{}} \HOLBoundVar{E\sp{\prime}} \HOLSymConst{\HOLTokenEPS} \HOLFreeVar{E\sb{\mathrm{1}}}\hfill [WEAK_TRANS_TAU]
\end{alltt}

\item $\tau$-transition implies \HOLinline{\HOLConst{EPS}} relation:
\begin{alltt}
\HOLTokenTurnstile{} \HOLFreeVar{E} \HOLTokenTransBegin\HOLSymConst{\ensuremath{\tau}}\HOLTokenTransEnd \HOLFreeVar{E\sp{\prime}} \HOLSymConst{\HOLTokenImp{}} \HOLFreeVar{E} \HOLSymConst{\HOLTokenEPS} \HOLFreeVar{E\sp{\prime}}\hfill [TRANS_TAU_IMP_EPS]
\end{alltt}

\item Weak $\tau$-transition implies an $\tau$ transition followed by
  EPS transition:
\begin{alltt}
WEAK_TRANS_TAU_IMP_TRANS_TAU:
\HOLTokenTurnstile{} \HOLFreeVar{E} \HOLTokenWeakTransBegin\HOLSymConst{\ensuremath{\tau}}\HOLTokenWeakTransEnd \HOLFreeVar{E\sp{\prime}} \HOLSymConst{\HOLTokenImp{}} \HOLSymConst{\HOLTokenExists{}}\HOLBoundVar{E\sb{\mathrm{1}}}. \HOLFreeVar{E} \HOLTokenTransBegin\HOLSymConst{\ensuremath{\tau}}\HOLTokenTransEnd \HOLBoundVar{E\sb{\mathrm{1}}} \HOLSymConst{\HOLTokenConj{}} \HOLBoundVar{E\sb{\mathrm{1}}} \HOLSymConst{\HOLTokenEPS} \HOLFreeVar{E\sp{\prime}}
\end{alltt}

\item \HOLinline{\HOLConst{EPS}} implies $\tau$-prefixed \HOLinline{\HOLConst{EPS}}:
\begin{alltt}
\HOLTokenTurnstile{} \HOLFreeVar{E} \HOLSymConst{\HOLTokenEPS} \HOLFreeVar{E\sp{\prime}} \HOLSymConst{\HOLTokenImp{}} \HOLSymConst{\ensuremath{\tau}}\HOLSymConst{..}\HOLFreeVar{E} \HOLSymConst{\HOLTokenEPS} \HOLFreeVar{E\sp{\prime}}\hfill [TAU_PREFIX_EPS]
\end{alltt}

\item Weak $\tau$-transition implies $\tau$-prefixed weak:
  $\tau$-transition:
\begin{alltt}
\HOLTokenTurnstile{} \HOLFreeVar{E} \HOLTokenWeakTransBegin\HOLFreeVar{u}\HOLTokenWeakTransEnd \HOLFreeVar{E\sp{\prime}} \HOLSymConst{\HOLTokenImp{}} \HOLSymConst{\ensuremath{\tau}}\HOLSymConst{..}\HOLFreeVar{E} \HOLTokenWeakTransBegin\HOLFreeVar{u}\HOLTokenWeakTransEnd \HOLFreeVar{E\sp{\prime}} \hfill [TAU_PREFIX_WEAK_TRANS]
\end{alltt}

\item A weak transition wrapped by EPS transitions is still a weak
  transition:
\begin{alltt}
EPS_WEAK_EPS:
\HOLTokenTurnstile{} \HOLFreeVar{E} \HOLSymConst{\HOLTokenEPS} \HOLFreeVar{E\sb{\mathrm{1}}} \HOLSymConst{\HOLTokenConj{}} \HOLFreeVar{E\sb{\mathrm{1}}} \HOLTokenWeakTransBegin\HOLFreeVar{u}\HOLTokenWeakTransEnd \HOLFreeVar{E\sb{\mathrm{2}}} \HOLSymConst{\HOLTokenConj{}} \HOLFreeVar{E\sb{\mathrm{2}}} \HOLSymConst{\HOLTokenEPS} \HOLFreeVar{E\sp{\prime}} \HOLSymConst{\HOLTokenImp{}} \HOLFreeVar{E} \HOLTokenWeakTransBegin\HOLFreeVar{u}\HOLTokenWeakTransEnd \HOLFreeVar{E\sp{\prime}}
\end{alltt}

\item A weak transition after a $\tau$-transition is still a weak
  transition:
\begin{alltt}
\HOLTokenTurnstile{} \HOLFreeVar{E} \HOLTokenTransBegin\HOLSymConst{\ensuremath{\tau}}\HOLTokenTransEnd \HOLFreeVar{E\sb{\mathrm{1}}} \HOLSymConst{\HOLTokenConj{}} \HOLFreeVar{E\sb{\mathrm{1}}} \HOLTokenWeakTransBegin\HOLFreeVar{u}\HOLTokenWeakTransEnd \HOLFreeVar{E\sp{\prime}} \HOLSymConst{\HOLTokenImp{}} \HOLFreeVar{E} \HOLTokenWeakTransBegin\HOLFreeVar{u}\HOLTokenWeakTransEnd \HOLFreeVar{E\sp{\prime}}\hfill [TRANS_TAU_AND_WEAK]
\end{alltt}

\item Any transition followed by an EPS transition becomes a weak
  transition:
\begin{alltt}
\HOLTokenTurnstile{} \HOLFreeVar{E} \HOLTokenTransBegin\HOLFreeVar{u}\HOLTokenTransEnd \HOLFreeVar{E\sb{\mathrm{1}}} \HOLSymConst{\HOLTokenConj{}} \HOLFreeVar{E\sb{\mathrm{1}}} \HOLSymConst{\HOLTokenEPS} \HOLFreeVar{E\sp{\prime}} \HOLSymConst{\HOLTokenImp{}} \HOLFreeVar{E} \HOLTokenWeakTransBegin\HOLFreeVar{u}\HOLTokenWeakTransEnd \HOLFreeVar{E\sp{\prime}}\hfill [TRANS_AND_EPS]
\end{alltt}

\item An EPS transition implies either no transition or a weak
  $\tau$-transition:
\begin{alltt}
\HOLTokenTurnstile{} \HOLFreeVar{E} \HOLSymConst{\HOLTokenEPS} \HOLFreeVar{E\sp{\prime}} \HOLSymConst{\HOLTokenImp{}} (\HOLFreeVar{E} \HOLSymConst{=} \HOLFreeVar{E\sp{\prime}}) \HOLSymConst{\HOLTokenDisj{}} \HOLFreeVar{E} \HOLTokenWeakTransBegin\HOLSymConst{\ensuremath{\tau}}\HOLTokenWeakTransEnd \HOLFreeVar{E\sp{\prime}}\hfill [EPS_IMP_WEAK_TRANS]
\end{alltt}

\item Two possible cases for the first step of a weak transition:
\begin{alltt}
WEAK_TRANS_cases1:
\HOLTokenTurnstile{} \HOLFreeVar{E} \HOLTokenWeakTransBegin\HOLFreeVar{u}\HOLTokenWeakTransEnd \HOLFreeVar{E\sb{\mathrm{1}}} \HOLSymConst{\HOLTokenImp{}}
   (\HOLSymConst{\HOLTokenExists{}}\HOLBoundVar{E\sp{\prime}}. \HOLFreeVar{E} \HOLTokenTransBegin\HOLSymConst{\ensuremath{\tau}}\HOLTokenTransEnd \HOLBoundVar{E\sp{\prime}} \HOLSymConst{\HOLTokenConj{}} \HOLBoundVar{E\sp{\prime}} \HOLTokenWeakTransBegin\HOLFreeVar{u}\HOLTokenWeakTransEnd \HOLFreeVar{E\sb{\mathrm{1}}}) \HOLSymConst{\HOLTokenDisj{}} \HOLSymConst{\HOLTokenExists{}}\HOLBoundVar{E\sp{\prime}}. \HOLFreeVar{E} \HOLTokenTransBegin\HOLFreeVar{u}\HOLTokenTransEnd \HOLBoundVar{E\sp{\prime}} \HOLSymConst{\HOLTokenConj{}} \HOLBoundVar{E\sp{\prime}} \HOLSymConst{\HOLTokenEPS} \HOLFreeVar{E\sb{\mathrm{1}}}
\end{alltt}

\item The weak transition version of SOS inference rule $(Sum_1)$ and $(Sum_2)$:
\begin{alltt}
\HOLTokenTurnstile{} \HOLFreeVar{E} \HOLTokenWeakTransBegin\HOLFreeVar{u}\HOLTokenWeakTransEnd \HOLFreeVar{E\sb{\mathrm{1}}} \HOLSymConst{\HOLTokenImp{}} \HOLFreeVar{E} \HOLSymConst{+} \HOLFreeVar{E\sp{\prime}} \HOLTokenWeakTransBegin\HOLFreeVar{u}\HOLTokenWeakTransEnd \HOLFreeVar{E\sb{\mathrm{1}}}\hfill [WEAK_SUM1]
\HOLTokenTurnstile{} \HOLFreeVar{E} \HOLTokenWeakTransBegin\HOLFreeVar{u}\HOLTokenWeakTransEnd \HOLFreeVar{E\sb{\mathrm{1}}} \HOLSymConst{\HOLTokenImp{}} \HOLFreeVar{E\sp{\prime}} \HOLSymConst{+} \HOLFreeVar{E} \HOLTokenWeakTransBegin\HOLFreeVar{u}\HOLTokenWeakTransEnd \HOLFreeVar{E\sb{\mathrm{1}}}\hfill [WEAK_SUM2]
\end{alltt}
\end{enumerate}
\end{proposition}

\section{Weak bisimulation equivalence}

The concepts of weak bisimulation and weak bisimulation equivalence
(a.k.a. observational equivalence) and their properties were used all over this project.
Several basic results (Deng lemma,
Hennessy lemma, ``Coarsest congruence contained in weak equivalence'') were
formally proved in this project, they all talk about the
relationship between weak bisimulation equivalence and rooted weak
bisimulation equivalence (a.k.a. observational congruence, we'll use
this shorted names in the rest of the paper).  Also, since  observational congruence is not recursively defined but relies
on the definition of weak equivalence, the
properties of weak equivalence were heavily used in the proof of
properties of observational congruence.

On the other side, it's quite easy to derive almost all the
algebraic laws for weak equivalence (and observational congruence) from strong algebraic laws, because strong equivalence implies weak equivalence (and also
observational congruence). This also reflects the fact that,
although strong equivalence and its algebraic laws were relative less useful in real-world model checking, they do have contributions for
deriving more useful algebraic laws. And from the view of theorem
proving it totally make sense: if we try to prove any algebraic law
for weak equivalence \emph{directly}, the proof will be quite long and
difficult, and the handling of $tau$-transitions will be a common part
in all these proofs. But if we use the strong algebraic laws as
lemmas, the proofs were actually divided into two logical parts: one for
handling the algebraic law itself, the other for handling the $\tau$-transitions.

The definition of weak bisimulation is the same as in
\cite{Gorrieri:2015jt}, except for the use of EPS in case of
$\tau$-transitions:
\begin{definition}{(Weak bisimulation)}
\begin{alltt}
\HOLTokenTurnstile{} \HOLConst{WEAK_BISIM} \HOLFreeVar{Wbsm} \HOLSymConst{\HOLTokenEquiv{}}
   \HOLSymConst{\HOLTokenForall{}}\HOLBoundVar{E} \HOLBoundVar{E\sp{\prime}}.
     \HOLFreeVar{Wbsm} \HOLBoundVar{E} \HOLBoundVar{E\sp{\prime}} \HOLSymConst{\HOLTokenImp{}}
     (\HOLSymConst{\HOLTokenForall{}}\HOLBoundVar{l}.
        (\HOLSymConst{\HOLTokenForall{}}\HOLBoundVar{E\sb{\mathrm{1}}}.
           \HOLBoundVar{E} \HOLTokenTransBegin\HOLConst{label} \HOLBoundVar{l}\HOLTokenTransEnd \HOLBoundVar{E\sb{\mathrm{1}}} \HOLSymConst{\HOLTokenImp{}}
           \HOLSymConst{\HOLTokenExists{}}\HOLBoundVar{E\sb{\mathrm{2}}}. \HOLBoundVar{E\sp{\prime}} \HOLTokenWeakTransBegin\HOLConst{label} \HOLBoundVar{l}\HOLTokenWeakTransEnd \HOLBoundVar{E\sb{\mathrm{2}}} \HOLSymConst{\HOLTokenConj{}} \HOLFreeVar{Wbsm} \HOLBoundVar{E\sb{\mathrm{1}}} \HOLBoundVar{E\sb{\mathrm{2}}}) \HOLSymConst{\HOLTokenConj{}}
        \HOLSymConst{\HOLTokenForall{}}\HOLBoundVar{E\sb{\mathrm{2}}}.
          \HOLBoundVar{E\sp{\prime}} \HOLTokenTransBegin\HOLConst{label} \HOLBoundVar{l}\HOLTokenTransEnd \HOLBoundVar{E\sb{\mathrm{2}}} \HOLSymConst{\HOLTokenImp{}} \HOLSymConst{\HOLTokenExists{}}\HOLBoundVar{E\sb{\mathrm{1}}}. \HOLBoundVar{E} \HOLTokenWeakTransBegin\HOLConst{label} \HOLBoundVar{l}\HOLTokenWeakTransEnd \HOLBoundVar{E\sb{\mathrm{1}}} \HOLSymConst{\HOLTokenConj{}} \HOLFreeVar{Wbsm} \HOLBoundVar{E\sb{\mathrm{1}}} \HOLBoundVar{E\sb{\mathrm{2}}}) \HOLSymConst{\HOLTokenConj{}}
     (\HOLSymConst{\HOLTokenForall{}}\HOLBoundVar{E\sb{\mathrm{1}}}. \HOLBoundVar{E} \HOLTokenTransBegin\HOLSymConst{\ensuremath{\tau}}\HOLTokenTransEnd \HOLBoundVar{E\sb{\mathrm{1}}} \HOLSymConst{\HOLTokenImp{}} \HOLSymConst{\HOLTokenExists{}}\HOLBoundVar{E\sb{\mathrm{2}}}. \HOLBoundVar{E\sp{\prime}} \HOLSymConst{\HOLTokenEPS} \HOLBoundVar{E\sb{\mathrm{2}}} \HOLSymConst{\HOLTokenConj{}} \HOLFreeVar{Wbsm} \HOLBoundVar{E\sb{\mathrm{1}}} \HOLBoundVar{E\sb{\mathrm{2}}}) \HOLSymConst{\HOLTokenConj{}}
     \HOLSymConst{\HOLTokenForall{}}\HOLBoundVar{E\sb{\mathrm{2}}}. \HOLBoundVar{E\sp{\prime}} \HOLTokenTransBegin\HOLSymConst{\ensuremath{\tau}}\HOLTokenTransEnd \HOLBoundVar{E\sb{\mathrm{2}}} \HOLSymConst{\HOLTokenImp{}} \HOLSymConst{\HOLTokenExists{}}\HOLBoundVar{E\sb{\mathrm{1}}}. \HOLBoundVar{E} \HOLSymConst{\HOLTokenEPS} \HOLBoundVar{E\sb{\mathrm{1}}} \HOLSymConst{\HOLTokenConj{}} \HOLFreeVar{Wbsm} \HOLBoundVar{E\sb{\mathrm{1}}} \HOLBoundVar{E\sb{\mathrm{2}}}
\end{alltt}
\end{definition}

Weak bisimulation has some common properties:
\begin{proposition}{Properties of weak bisimulation}
\begin{enumerate}
\item The identity relation is a weak bisimulation:
\begin{alltt}
\HOLTokenTurnstile{} \HOLConst{WEAK_BISIM} (\HOLTokenLambda{}\HOLBoundVar{x} \HOLBoundVar{y}. \HOLBoundVar{x} \HOLSymConst{=} \HOLBoundVar{y})\hfill[IDENTITY_WEAK_BISIM]
\end{alltt}
\item The converse of a weak bisimulation is still a weak
  bisimulation:
\begin{alltt}
\HOLTokenTurnstile{} \HOLConst{WEAK_BISIM} \HOLFreeVar{Wbsm} \HOLSymConst{\HOLTokenImp{}} \HOLConst{WEAK_BISIM} \HOLFreeVar{Wbsm}\HOLSymConst{\HOLTokenRInverse{}}\hfill[IDENTITY_WEAK_BISIM]
\end{alltt}
\item The composition of two weak bisimulations is a weak
  bisimulation:
\begin{alltt}
\HOLTokenTurnstile{} \HOLConst{WEAK_BISIM} \HOLFreeVar{Wbsm\sb{\mathrm{1}}} \HOLSymConst{\HOLTokenConj{}} \HOLConst{WEAK_BISIM} \HOLFreeVar{Wbsm\sb{\mathrm{2}}} \HOLSymConst{\HOLTokenImp{}}
   \HOLConst{WEAK_BISIM} (\HOLFreeVar{Wbsm\sb{\mathrm{2}}} \HOLSymConst{\HOLTokenRCompose{}} \HOLFreeVar{Wbsm\sb{\mathrm{1}}})\hfill[COMP_WEAK_BISIM]
\end{alltt}
\item The union of two weak bisimulations is a weak bisimulation:
\begin{alltt}
\HOLTokenTurnstile{} \HOLConst{WEAK_BISIM} \HOLFreeVar{Wbsm\sb{\mathrm{1}}} \HOLSymConst{\HOLTokenConj{}} \HOLConst{WEAK_BISIM} \HOLFreeVar{Wbsm\sb{\mathrm{2}}} \HOLSymConst{\HOLTokenImp{}}
   \HOLConst{WEAK_BISIM} (\HOLFreeVar{Wbsm\sb{\mathrm{1}}} \HOLSymConst{\HOLTokenRUnion{}} \HOLFreeVar{Wbsm\sb{\mathrm{2}}})\hfill[UNION_WEAK_BISIM]
\end{alltt}
\end{enumerate}
\end{proposition}

There're two ways to define weak bisimulation equivalence in HOL4, one
is to define it as the union of all weak bisimulations:
\begin{definition}{(Alternative definition of weak equivalence)}
For any two CCS processes $E$ and $E'$, they're \emph{weak
  bisimulation equivalent} (or weak bisimilar) if and only if there's
a weak bisimulation relation between $E$ and $E'$:
\begin{alltt}
WEAK_EQUIV:
\HOLTokenTurnstile{} \HOLFreeVar{E} \HOLSymConst{\HOLTokenWeakEQ} \HOLFreeVar{E\sp{\prime}} \HOLSymConst{\HOLTokenEquiv{}} \HOLSymConst{\HOLTokenExists{}}\HOLBoundVar{Wbsm}. \HOLBoundVar{Wbsm} \HOLFreeVar{E} \HOLFreeVar{E\sp{\prime}} \HOLSymConst{\HOLTokenConj{}} \HOLConst{WEAK_BISIM} \HOLBoundVar{Wbsm}
\end{alltt}
\end{definition}
This was the definition used by Monica Nesi in Hol88 in which there was no
direct support for defining co-inductive relations.  The new method we
have used in this project, is to use HOL4's new co-inductive relation
defining facility \texttt{Hol_coreln} to define weak bisimulation
equivalence:
\begin{lstlisting}
val (WEAK_EQUIV_rules, WEAK_EQUIV_coind, WEAK_EQUIV_cases)
 = Hol_coreln `
    (!(E :('a, 'b) CCS) (E' :('a, 'b) CCS).
       (!l.
	 (!E1. TRANS E  (label l) E1 ==>
	       (?E2. WEAK_TRANS E' (label l) E2 /\
                 WEAK_EQUIV E1 E2)) /\
	 (!E2. TRANS E' (label l) E2 ==>
	       (?E1. WEAK_TRANS E  (label l) E1 /\
                 WEAK_EQUIV E1 E2))) /\
       (!E1. TRANS E  tau E1 ==>
                 (?E2. EPS E' E2 /\ WEAK_EQUIV E1 E2)) /\
       (!E2. TRANS E' tau E2 ==>
                 (?E1. EPS E  E1 /\ WEAK_EQUIV E1 E2))
      ==> WEAK_EQUIV E E')`;
\end{lstlisting}

The disadvantage of this new method is that,  the rules used in above
definition actually duplicated the definition of weak bisimulation,
while the advantage is that, HOL4 automatically proved an important
theorem and returned it as the third return value of above
definition. This theorem is also called ``the property (*)'' (in
Milner's book \cite{Milner:1989}:
\begin{proposition}{(The property (*) for weak bisimulation
    equivalence)}
\begin{alltt}
WEAK_PROPERTY_STAR:
\HOLTokenTurnstile{} \HOLFreeVar{a\sb{\mathrm{0}}} \HOLSymConst{\HOLTokenWeakEQ} \HOLFreeVar{a\sb{\mathrm{1}}} \HOLSymConst{\HOLTokenEquiv{}}
   (\HOLSymConst{\HOLTokenForall{}}\HOLBoundVar{l}.
      (\HOLSymConst{\HOLTokenForall{}}\HOLBoundVar{E\sb{\mathrm{1}}}. \HOLFreeVar{a\sb{\mathrm{0}}} \HOLTokenTransBegin\HOLConst{label} \HOLBoundVar{l}\HOLTokenTransEnd \HOLBoundVar{E\sb{\mathrm{1}}} \HOLSymConst{\HOLTokenImp{}} \HOLSymConst{\HOLTokenExists{}}\HOLBoundVar{E\sb{\mathrm{2}}}. \HOLFreeVar{a\sb{\mathrm{1}}} \HOLTokenWeakTransBegin\HOLConst{label} \HOLBoundVar{l}\HOLTokenWeakTransEnd \HOLBoundVar{E\sb{\mathrm{2}}} \HOLSymConst{\HOLTokenConj{}} \HOLBoundVar{E\sb{\mathrm{1}}} \HOLSymConst{\HOLTokenWeakEQ} \HOLBoundVar{E\sb{\mathrm{2}}}) \HOLSymConst{\HOLTokenConj{}}
      \HOLSymConst{\HOLTokenForall{}}\HOLBoundVar{E\sb{\mathrm{2}}}. \HOLFreeVar{a\sb{\mathrm{1}}} \HOLTokenTransBegin\HOLConst{label} \HOLBoundVar{l}\HOLTokenTransEnd \HOLBoundVar{E\sb{\mathrm{2}}} \HOLSymConst{\HOLTokenImp{}} \HOLSymConst{\HOLTokenExists{}}\HOLBoundVar{E\sb{\mathrm{1}}}. \HOLFreeVar{a\sb{\mathrm{0}}} \HOLTokenWeakTransBegin\HOLConst{label} \HOLBoundVar{l}\HOLTokenWeakTransEnd \HOLBoundVar{E\sb{\mathrm{1}}} \HOLSymConst{\HOLTokenConj{}} \HOLBoundVar{E\sb{\mathrm{1}}} \HOLSymConst{\HOLTokenWeakEQ} \HOLBoundVar{E\sb{\mathrm{2}}}) \HOLSymConst{\HOLTokenConj{}}
   (\HOLSymConst{\HOLTokenForall{}}\HOLBoundVar{E\sb{\mathrm{1}}}. \HOLFreeVar{a\sb{\mathrm{0}}} \HOLTokenTransBegin\HOLSymConst{\ensuremath{\tau}}\HOLTokenTransEnd \HOLBoundVar{E\sb{\mathrm{1}}} \HOLSymConst{\HOLTokenImp{}} \HOLSymConst{\HOLTokenExists{}}\HOLBoundVar{E\sb{\mathrm{2}}}. \HOLFreeVar{a\sb{\mathrm{1}}} \HOLSymConst{\HOLTokenEPS} \HOLBoundVar{E\sb{\mathrm{2}}} \HOLSymConst{\HOLTokenConj{}} \HOLBoundVar{E\sb{\mathrm{1}}} \HOLSymConst{\HOLTokenWeakEQ} \HOLBoundVar{E\sb{\mathrm{2}}}) \HOLSymConst{\HOLTokenConj{}}
   \HOLSymConst{\HOLTokenForall{}}\HOLBoundVar{E\sb{\mathrm{2}}}. \HOLFreeVar{a\sb{\mathrm{1}}} \HOLTokenTransBegin\HOLSymConst{\ensuremath{\tau}}\HOLTokenTransEnd \HOLBoundVar{E\sb{\mathrm{2}}} \HOLSymConst{\HOLTokenImp{}} \HOLSymConst{\HOLTokenExists{}}\HOLBoundVar{E\sb{\mathrm{1}}}. \HOLFreeVar{a\sb{\mathrm{0}}} \HOLSymConst{\HOLTokenEPS} \HOLBoundVar{E\sb{\mathrm{1}}} \HOLSymConst{\HOLTokenConj{}} \HOLBoundVar{E\sb{\mathrm{1}}} \HOLSymConst{\HOLTokenWeakEQ} \HOLBoundVar{E\sb{\mathrm{2}}}
\end{alltt}
\end{proposition}

It's known that, above property cannot be used as an alternative
definition of weak equivalence, because it doesn't capture all
possible weak equivalences. But it turns out that, for the proof of
most theorems about weak bisimilarities this property is enough to be
used as a rewrite rule in their proofs. And, if we had used the old
method to define weak equivalence, it's quite difficult to prove above
property (*). In previous project, the property (*) for
  strong equivalence was proved based on the old method, then in this
  project we have completely removed these code and now both strong
  and weak bisimulation equivalences were based on the new method. On
  the other side, the fact that Monica Nesi can define co-inductive relation
  without using \texttt{Hol_coreln} has shown that, the core HOL Logic
doesn't need to be extended to support co-inductive relation, and all
what \texttt{Hol_coreln} does internally is to use the existing HOL
theorems to construct the related proofs.

Using the alternative definition of weak equivalence, it's quite
simple to prove that, the weak equivalence is an equivalence relation:
\begin{proposition}{(Weak equivalence is an equivalence relation)}
\begin{alltt}
\HOLTokenTurnstile{} \HOLConst{equivalence} \HOLConst{WEAK_EQUIV}\hfill[WEAK_EQUIV_equivalence]
\end{alltt}
or
\begin{alltt}
\HOLTokenTurnstile{} \HOLFreeVar{E} \HOLSymConst{\HOLTokenWeakEQ} \HOLFreeVar{E}\hfill[WEAK_EQUIV_REFL]
\HOLTokenTurnstile{} \HOLFreeVar{E} \HOLSymConst{\HOLTokenWeakEQ} \HOLFreeVar{E\sp{\prime}} \HOLSymConst{\HOLTokenImp{}} \HOLFreeVar{E\sp{\prime}} \HOLSymConst{\HOLTokenWeakEQ} \HOLFreeVar{E}\hfill[WEAK_EQUIV_SYM]
\HOLTokenTurnstile{} \HOLFreeVar{E} \HOLSymConst{\HOLTokenWeakEQ} \HOLFreeVar{E\sp{\prime}} \HOLSymConst{\HOLTokenConj{}} \HOLFreeVar{E\sp{\prime}} \HOLSymConst{\HOLTokenWeakEQ} \HOLFreeVar{E\sp{\prime\prime}} \HOLSymConst{\HOLTokenImp{}} \HOLFreeVar{E} \HOLSymConst{\HOLTokenWeakEQ} \HOLFreeVar{E\sp{\prime\prime}}\hfill[WEAK_EQUIV_TRANS]
\end{alltt}
\end{proposition}

The substitutability of weak equivalence under various CCS process
operators were then proved based on above definition and property
(*). However, as we know weak equivalence is not a congruence, in some
of these substitutability theorems we must added extra assumptions on
the processes involved, i.e. the stability of CCS processes:
\begin{definition}{(Stable processes (or agents))}
A process (or agent) is said to be \emph{stable} if there's no $\tau$-transition
coming from it's root:
\begin{alltt}
\HOLTokenTurnstile{} \HOLConst{STABLE} \HOLFreeVar{E} \HOLSymConst{\HOLTokenEquiv{}} \HOLSymConst{\HOLTokenForall{}}\HOLBoundVar{u} \HOLBoundVar{E\sp{\prime}}. \HOLFreeVar{E} \HOLTokenTransBegin\HOLBoundVar{u}\HOLTokenTransEnd \HOLBoundVar{E\sp{\prime}} \HOLSymConst{\HOLTokenImp{}} \HOLBoundVar{u} \HOLSymConst{\HOLTokenNotEqual{}} \HOLSymConst{\ensuremath{\tau}}
\end{alltt}
\end{definition}
Notice that, the stability of a CCS process doesn't imply the
$\tau$-free of all its sub-processes. Instead the definition only
concerns on the first transition leading from the process (root).

Among other small lemmas, we have proved the following properties of weak bisimulation
equivalence:
\begin{proposition}{Properties of weak bisimulation equivalence)}
\begin{enumerate}
\item Weak equivalence is substitutive under prefix operator:
\begin{alltt}
\HOLTokenTurnstile{} \HOLFreeVar{E} \HOLSymConst{\HOLTokenWeakEQ} \HOLFreeVar{E\sp{\prime}} \HOLSymConst{\HOLTokenImp{}} \HOLSymConst{\HOLTokenForall{}}\HOLBoundVar{u}. \HOLBoundVar{u}\HOLSymConst{..}\HOLFreeVar{E} \HOLSymConst{\HOLTokenWeakEQ} \HOLBoundVar{u}\HOLSymConst{..}\HOLFreeVar{E\sp{\prime}}\hfill[WEAK_EQUIV_SUBST_PREFIX]
\end{alltt}
\item Weak equivalence of stable agents is preserved by binary
  summation:
\begin{alltt}
\HOLTokenTurnstile{} \HOLFreeVar{E\sb{\mathrm{1}}} \HOLSymConst{\HOLTokenWeakEQ} \HOLFreeVar{E\sb{\mathrm{1}}\sp{\prime}} \HOLSymConst{\HOLTokenConj{}} \HOLConst{STABLE} \HOLFreeVar{E\sb{\mathrm{1}}} \HOLSymConst{\HOLTokenConj{}} \HOLConst{STABLE} \HOLFreeVar{E\sb{\mathrm{1}}\sp{\prime}} \HOLSymConst{\HOLTokenConj{}} \HOLFreeVar{E\sb{\mathrm{2}}} \HOLSymConst{\HOLTokenWeakEQ} \HOLFreeVar{E\sb{\mathrm{2}}\sp{\prime}} \HOLSymConst{\HOLTokenConj{}} \HOLConst{STABLE} \HOLFreeVar{E\sb{\mathrm{2}}} \HOLSymConst{\HOLTokenConj{}}
   \HOLConst{STABLE} \HOLFreeVar{E\sb{\mathrm{2}}\sp{\prime}} \HOLSymConst{\HOLTokenImp{}}
   \HOLFreeVar{E\sb{\mathrm{1}}} \HOLSymConst{+} \HOLFreeVar{E\sb{\mathrm{2}}} \HOLSymConst{\HOLTokenWeakEQ} \HOLFreeVar{E\sb{\mathrm{1}}\sp{\prime}} \HOLSymConst{+} \HOLFreeVar{E\sb{\mathrm{2}}\sp{\prime}}\hfill[WEAK_EQUIV_PRESD_BY_SUM]
\end{alltt}
\item Weak equivalence is preserved by guarded binary summation:
\begin{alltt}
WEAK_EQUIV_PRESD_BY_GUARDED_SUM:
\HOLTokenTurnstile{} \HOLFreeVar{E\sb{\mathrm{1}}} \HOLSymConst{\HOLTokenWeakEQ} \HOLFreeVar{E\sb{\mathrm{1}}\sp{\prime}} \HOLSymConst{\HOLTokenConj{}} \HOLFreeVar{E\sb{\mathrm{2}}} \HOLSymConst{\HOLTokenWeakEQ} \HOLFreeVar{E\sb{\mathrm{2}}\sp{\prime}} \HOLSymConst{\HOLTokenImp{}} \HOLFreeVar{a\sb{\mathrm{1}}}\HOLSymConst{..}\HOLFreeVar{E\sb{\mathrm{1}}} \HOLSymConst{+} \HOLFreeVar{a\sb{\mathrm{2}}}\HOLSymConst{..}\HOLFreeVar{E\sb{\mathrm{2}}} \HOLSymConst{\HOLTokenWeakEQ} \HOLFreeVar{a\sb{\mathrm{1}}}\HOLSymConst{..}\HOLFreeVar{E\sb{\mathrm{1}}\sp{\prime}} \HOLSymConst{+} \HOLFreeVar{a\sb{\mathrm{2}}}\HOLSymConst{..}\HOLFreeVar{E\sb{\mathrm{2}}\sp{\prime}}
\end{alltt}
\item Weak equivalence of stable agents is substitutive under binary
  summation on the right:
\begin{alltt}
WEAK_EQUIV_SUBST_SUM_R:
\HOLTokenTurnstile{} \HOLFreeVar{E} \HOLSymConst{\HOLTokenWeakEQ} \HOLFreeVar{E\sp{\prime}} \HOLSymConst{\HOLTokenConj{}} \HOLConst{STABLE} \HOLFreeVar{E} \HOLSymConst{\HOLTokenConj{}} \HOLConst{STABLE} \HOLFreeVar{E\sp{\prime}} \HOLSymConst{\HOLTokenImp{}} \HOLSymConst{\HOLTokenForall{}}\HOLBoundVar{E\sp{\prime\prime}}. \HOLFreeVar{E} \HOLSymConst{+} \HOLBoundVar{E\sp{\prime\prime}} \HOLSymConst{\HOLTokenWeakEQ} \HOLFreeVar{E\sp{\prime}} \HOLSymConst{+} \HOLBoundVar{E\sp{\prime\prime}}
\end{alltt}
\item Weak equivalence of stable agents is substitutive under binary
  summation on the left:
\begin{alltt}
WEAK_EQUIV_SUBST_SUM_L:
\HOLTokenTurnstile{} \HOLFreeVar{E} \HOLSymConst{\HOLTokenWeakEQ} \HOLFreeVar{E\sp{\prime}} \HOLSymConst{\HOLTokenConj{}} \HOLConst{STABLE} \HOLFreeVar{E} \HOLSymConst{\HOLTokenConj{}} \HOLConst{STABLE} \HOLFreeVar{E\sp{\prime}} \HOLSymConst{\HOLTokenImp{}} \HOLSymConst{\HOLTokenForall{}}\HOLBoundVar{E\sp{\prime\prime}}. \HOLBoundVar{E\sp{\prime\prime}} \HOLSymConst{+} \HOLFreeVar{E} \HOLSymConst{\HOLTokenWeakEQ} \HOLBoundVar{E\sp{\prime\prime}} \HOLSymConst{+} \HOLFreeVar{E\sp{\prime}}
\end{alltt}
\item Weak equivalence is preserved by parallel operator:
\begin{alltt}
WEAK_EQUIV_PRESD_BY_PAR:
\HOLTokenTurnstile{} \HOLFreeVar{E\sb{\mathrm{1}}} \HOLSymConst{\HOLTokenWeakEQ} \HOLFreeVar{E\sb{\mathrm{1}}\sp{\prime}} \HOLSymConst{\HOLTokenConj{}} \HOLFreeVar{E\sb{\mathrm{2}}} \HOLSymConst{\HOLTokenWeakEQ} \HOLFreeVar{E\sb{\mathrm{2}}\sp{\prime}} \HOLSymConst{\HOLTokenImp{}} \HOLFreeVar{E\sb{\mathrm{1}}} \HOLSymConst{\ensuremath{\parallel}} \HOLFreeVar{E\sb{\mathrm{2}}} \HOLSymConst{\HOLTokenWeakEQ} \HOLFreeVar{E\sb{\mathrm{1}}\sp{\prime}} \HOLSymConst{\ensuremath{\parallel}} \HOLFreeVar{E\sb{\mathrm{2}}\sp{\prime}}
\end{alltt}
\item Weak equivalence is substitutive under restriction operator:
\begin{alltt}
WEAK_EQUIV_SUBST_RESTR:
\HOLTokenTurnstile{} \HOLFreeVar{E} \HOLSymConst{\HOLTokenWeakEQ} \HOLFreeVar{E\sp{\prime}} \HOLSymConst{\HOLTokenImp{}} \HOLSymConst{\HOLTokenForall{}}\HOLBoundVar{L}. \HOLSymConst{\ensuremath{\nu}} \HOLBoundVar{L} \HOLFreeVar{E} \HOLSymConst{\HOLTokenWeakEQ} \HOLSymConst{\ensuremath{\nu}} \HOLBoundVar{L} \HOLFreeVar{E\sp{\prime}}
\end{alltt}
\item Weak equivalence is substitutive under relabelling operator:
\begin{alltt}
WEAK_EQUIV_SUBST_RELAB:
\HOLTokenTurnstile{} \HOLFreeVar{E} \HOLSymConst{\HOLTokenWeakEQ} \HOLFreeVar{E\sp{\prime}} \HOLSymConst{\HOLTokenImp{}} \HOLSymConst{\HOLTokenForall{}}\HOLBoundVar{rf}. \HOLConst{relab} \HOLFreeVar{E} \HOLBoundVar{rf} \HOLSymConst{\HOLTokenWeakEQ} \HOLConst{relab} \HOLFreeVar{E\sp{\prime}} \HOLBoundVar{rf}
\end{alltt}
\end{enumerate}
\end{proposition}

Finally, we have proved that, strong equivalence implies weak
equivalence:
\begin{theorem}{(Strong equivalence implies weak equivalence)}
\begin{alltt}
\HOLTokenTurnstile{} \HOLFreeVar{E} \HOLSymConst{\HOLTokenStrongEQ} \HOLFreeVar{E\sp{\prime}} \HOLSymConst{\HOLTokenImp{}} \HOLFreeVar{E} \HOLSymConst{\HOLTokenWeakEQ} \HOLFreeVar{E\sp{\prime}}\hfill[STRONG_IMP_WEAK_EQUIV]
\end{alltt}
\end{theorem}

Here we omit all the algebraic laws for weak equivalence, because they
were all easily derived from the corresponding algebraic laws for
strong equivalence, except for the following $\tau$-law:
\begin{theorem}{The $\tau$-law for weak equivalence)}
\begin{alltt}
\HOLTokenTurnstile{} \HOLSymConst{\ensuremath{\tau}}\HOLSymConst{..}\HOLFreeVar{E} \HOLSymConst{\HOLTokenWeakEQ} \HOLFreeVar{E}\hfill[TAU_WEAK]
\end{alltt}
\end{theorem}

\section{Observational Congruence}

The concept of rooted weak bisimulation equivalence (also namsed
\emph{observation congruence}) is an ``obvious fix'' to convert weak
bisimulation equivalence into a congruence. Its definition is not
recursive but based on the definition of weak equivalence:
\begin{definition}{(Observation congruence)}
Two CCS processes are observation congruence if and only if for any
transition from one of them, there's a responding weak transition from
the other, and the resulting two sub-processes are weak equivalence:
\begin{alltt}
\HOLTokenTurnstile{} \HOLFreeVar{E} \HOLSymConst{\HOLTokenObsCongr} \HOLFreeVar{E\sp{\prime}} \HOLSymConst{\HOLTokenEquiv{}}
   \HOLSymConst{\HOLTokenForall{}}\HOLBoundVar{u}.
     (\HOLSymConst{\HOLTokenForall{}}\HOLBoundVar{E\sb{\mathrm{1}}}. \HOLFreeVar{E} \HOLTokenTransBegin\HOLBoundVar{u}\HOLTokenTransEnd \HOLBoundVar{E\sb{\mathrm{1}}} \HOLSymConst{\HOLTokenImp{}} \HOLSymConst{\HOLTokenExists{}}\HOLBoundVar{E\sb{\mathrm{2}}}. \HOLFreeVar{E\sp{\prime}} \HOLTokenWeakTransBegin\HOLBoundVar{u}\HOLTokenWeakTransEnd \HOLBoundVar{E\sb{\mathrm{2}}} \HOLSymConst{\HOLTokenConj{}} \HOLBoundVar{E\sb{\mathrm{1}}} \HOLSymConst{\HOLTokenWeakEQ} \HOLBoundVar{E\sb{\mathrm{2}}}) \HOLSymConst{\HOLTokenConj{}}
     \HOLSymConst{\HOLTokenForall{}}\HOLBoundVar{E\sb{\mathrm{2}}}. \HOLFreeVar{E\sp{\prime}} \HOLTokenTransBegin\HOLBoundVar{u}\HOLTokenTransEnd \HOLBoundVar{E\sb{\mathrm{2}}} \HOLSymConst{\HOLTokenImp{}} \HOLSymConst{\HOLTokenExists{}}\HOLBoundVar{E\sb{\mathrm{1}}}. \HOLFreeVar{E} \HOLTokenWeakTransBegin\HOLBoundVar{u}\HOLTokenWeakTransEnd \HOLBoundVar{E\sb{\mathrm{1}}} \HOLSymConst{\HOLTokenConj{}} \HOLBoundVar{E\sb{\mathrm{1}}} \HOLSymConst{\HOLTokenWeakEQ} \HOLBoundVar{E\sb{\mathrm{2}}}\hfill[OBS_CONGR]
\end{alltt}
\end{definition}

By observing the differences between the definition of observation
equivalence (weak equivalence) and congruence, we can see that,
observation equivalence requires a little more: for each $\tau$-transition
from one process, the other process must response with at least one
$\tau$-transition. Thus what's immediately
proven is the following two theorems:
\begin{theorem}{(Observation congruence implies observation
    equivalence)}
\begin{alltt}
\HOLTokenTurnstile{} \HOLFreeVar{E} \HOLSymConst{\HOLTokenObsCongr} \HOLFreeVar{E\sp{\prime}} \HOLSymConst{\HOLTokenImp{}} \HOLFreeVar{E} \HOLSymConst{\HOLTokenWeakEQ} \HOLFreeVar{E\sp{\prime}}\hfill[OBS_CONGR_IMP_WEAK_EQUIV]
\end{alltt}
\end{theorem}

\begin{theorem}{(Observation equivalence on stable agents implies
    observation congruence)}
\begin{alltt}
WEAK_EQUIV_STABLE_IMP_CONGR:
\HOLTokenTurnstile{} \HOLFreeVar{E} \HOLSymConst{\HOLTokenWeakEQ} \HOLFreeVar{E\sp{\prime}} \HOLSymConst{\HOLTokenConj{}} \HOLConst{STABLE} \HOLFreeVar{E} \HOLSymConst{\HOLTokenConj{}} \HOLConst{STABLE} \HOLFreeVar{E\sp{\prime}} \HOLSymConst{\HOLTokenImp{}} \HOLFreeVar{E} \HOLSymConst{\HOLTokenObsCongr} \HOLFreeVar{E\sp{\prime}}
\end{alltt}
\end{theorem}

Surprisingly, it's not trivial to prove that, the observation
equivalence is indeed an equivalence relation. The reflexivity and
symmetry are trivial:
\begin{proposition}{(The reflexivity and symmetry of observation congruence)}
\begin{alltt}
\HOLTokenTurnstile{} \HOLFreeVar{E} \HOLSymConst{\HOLTokenObsCongr} \HOLFreeVar{E}\hfill[OBS_CONGR_REFL]
\HOLTokenTurnstile{} \HOLFreeVar{E} \HOLSymConst{\HOLTokenObsCongr} \HOLFreeVar{E\sp{\prime}} \HOLSymConst{\HOLTokenImp{}} \HOLFreeVar{E\sp{\prime}} \HOLSymConst{\HOLTokenObsCongr} \HOLFreeVar{E}\hfill[OBS_CONGR_SYM]
\end{alltt}
\end{proposition}
But the transitivity is hard to prove.\footnote{Actually it's not
  proven in the old work, the formal proofs that we did in this
  project is completely new.}  Our proof here is based on the following
lemmas:
\begin{lemma}
If two processes $E$ and $E'$ are observation congruence, then for any
EPS transition coming from $E$, there's a corresponding EPS transition
from $E'$, and the resulting two subprocesses are weakly equivalent:
\begin{alltt}
OBS_CONGR_EPS:
\HOLTokenTurnstile{} \HOLFreeVar{E} \HOLSymConst{\HOLTokenObsCongr} \HOLFreeVar{E\sp{\prime}} \HOLSymConst{\HOLTokenImp{}} \HOLSymConst{\HOLTokenForall{}}\HOLBoundVar{E\sb{\mathrm{1}}}. \HOLFreeVar{E} \HOLSymConst{\HOLTokenEPS} \HOLBoundVar{E\sb{\mathrm{1}}} \HOLSymConst{\HOLTokenImp{}} \HOLSymConst{\HOLTokenExists{}}\HOLBoundVar{E\sb{\mathrm{2}}}. \HOLFreeVar{E\sp{\prime}} \HOLSymConst{\HOLTokenEPS} \HOLBoundVar{E\sb{\mathrm{2}}} \HOLSymConst{\HOLTokenConj{}} \HOLBoundVar{E\sb{\mathrm{1}}} \HOLSymConst{\HOLTokenWeakEQ} \HOLBoundVar{E\sb{\mathrm{2}}}
\end{alltt}
\end{lemma}
\begin{proof}
By (right) induction\footnote{The induction theorem used here is \texttt{EPS_ind_right}.} on the number of $\tau$ in the EPS transition of $E$. In
the base case, there's no $\tau$ at all, the $E$ transits to
itself. And in this case $E$' can respond with itself, which is also
an EPS transition:
\begin{displaymath}
\xymatrix{
  {E} \ar@{.}[r]^{\approx^c} \ar@{-}[d]^{=} & {E'} \ar@{-}[d]^{=} \\
  {E} \ar@{.}[r]^{\approx} & {E'}
}
\end{displaymath}

For the induction case, suppose the proposition is true for zero or more $\tau$
transitions except for the last step, that's, $\forall E, \exists E1,
E2$, such that \HOLinline{\HOLFreeVar{E} \HOLSymConst{\HOLTokenEPS} \HOLFreeVar{E\sb{\mathrm{1}}}}, \HOLinline{\HOLFreeVar{E\sp{\prime}} \HOLSymConst{\HOLTokenEPS} \HOLFreeVar{E\sb{\mathrm{2}}}} and
\HOLinline{\HOLFreeVar{E\sb{\mathrm{1}}} \HOLSymConst{\HOLTokenWeakEQ} \HOLFreeVar{E\sb{\mathrm{2}}}}. Now by definition of weak equivalence, if
\HOLinline{\HOLFreeVar{E\sb{\mathrm{1}}} \HOLTokenTransBegin\HOLSymConst{\ensuremath{\tau}}\HOLTokenTransEnd \HOLFreeVar{E\sb{\mathrm{1}}\sp{\prime}}} then there exists $E2'$ such that \HOLinline{\HOLFreeVar{E\sb{\mathrm{2}}} \HOLSymConst{\HOLTokenEPS} \HOLFreeVar{E\sb{\mathrm{2}}\sp{\prime}}} and \HOLinline{\HOLFreeVar{E\sb{\mathrm{1}}\sp{\prime}} \HOLSymConst{\HOLTokenWeakEQ} \HOLFreeVar{E\sb{\mathrm{2}}\sp{\prime}}}. Then by transitivity of EPS,
we have \HOLinline{\HOLFreeVar{E\sp{\prime}} \HOLSymConst{\HOLTokenEPS} \HOLFreeVar{E\sb{\mathrm{2}}} \HOLSymConst{\HOLTokenConj{}} \HOLFreeVar{E\sb{\mathrm{2}}} \HOLSymConst{\HOLTokenEPS} \HOLFreeVar{E\sb{\mathrm{2}}\sp{\prime}} \HOLSymConst{\HOLTokenImp{}} \HOLFreeVar{E\sp{\prime}} \HOLSymConst{\HOLTokenEPS} \HOLFreeVar{E\sb{\mathrm{2}}\sp{\prime}}}, thus $E_2'$ is a valid response required by
observation congruence:
\begin{displaymath}
\xymatrix{
{E} \ar@{.}[r]^{\approx^c} \ar@{=>}[d]^{\epsilon} & {E'}
\ar@{=>}[d]^{\epsilon} \ar@/^4ex/[dd]^{\epsilon} \\
{\forall E_1} \ar@{.}[r]^{\approx} \ar[d]^{\tau} & {\forall E_2} \ar@{=>}[d]^{\epsilon}
\\
{\forall E_1'} \ar@{.}[r]^{\approx} & {\exists E_2'}
}
\end{displaymath}
\qed
\end{proof}

\begin{lemma}
If two processes $E$ and $E'$ are observation congruence, then for any
weak transition coming from $E$, there's a corresponding weak
transition from $E'$, and the resulting two subprocesses are weakly
equivalent:
\begin{alltt}
\HOLTokenTurnstile{} \HOLFreeVar{E} \HOLSymConst{\HOLTokenObsCongr} \HOLFreeVar{E\sp{\prime}} \HOLSymConst{\HOLTokenImp{}} \HOLSymConst{\HOLTokenForall{}}\HOLBoundVar{u} \HOLBoundVar{E\sb{\mathrm{1}}}. \HOLFreeVar{E} \HOLTokenWeakTransBegin\HOLBoundVar{u}\HOLTokenWeakTransEnd \HOLBoundVar{E\sb{\mathrm{1}}} \HOLSymConst{\HOLTokenImp{}} \HOLSymConst{\HOLTokenExists{}}\HOLBoundVar{E\sb{\mathrm{2}}}. \HOLFreeVar{E\sp{\prime}} \HOLTokenWeakTransBegin\HOLBoundVar{u}\HOLTokenWeakTransEnd \HOLBoundVar{E\sb{\mathrm{2}}} \HOLSymConst{\HOLTokenConj{}} \HOLBoundVar{E\sb{\mathrm{1}}} \HOLSymConst{\HOLTokenWeakEQ} \HOLBoundVar{E\sb{\mathrm{2}}}
\end{alltt}
\end{lemma}
\begin{proof}{(sketch}
Consider the two cases when the action is $\tau$ or not
$\tau$. For all weak $\tau$-transitions coming from $E$, the
observation congruence requires that there's at least one $\tau$
following $E'$ and the resulting two sub-processes, say $E_1'$ and
$E_2$ are weak equivalence. Then the desired responses can be found by
using a similar existence lemma for weak equivalence:
\begin{displaymath}
\xymatrix{
{E} \ar@{.}[r]^{\approx^c} \ar[d]^{\tau} \ar@/^-4ex/[dd]^{\tau} & {E'}
\ar@{=>}[d]^{\tau} \ar@/^4ex/[dd]^{\tau} \\
{\exists E_1'} \ar@{.}[r]^{\approx} \ar@{=>}[d]^{\epsilon} & {\exists E_2}
\ar@{=>}[d]^{\epsilon} \\
{\forall E_1} \ar@{.}[r]^{\approx} & {\exists E_2'}
}
\end{displaymath}

For all the non-$\tau$ weak transitions from $E$, the proof follows
from previous lemma and a similar existence lemma for weak
equivalence. The following figure is a sketch for the proof of this
case:
\begin{displaymath}
\xymatrix{
{E} \ar@{.}[r]^{\approx^c} \ar@{=>}[d]^{\epsilon}
\ar@/^-4ex/[ddd]^{\forall L}
& {E'} \ar@{=>}[d]^{\epsilon} \ar@/^4ex/[ddd]^{L} \\
{\exists E_1'} \ar@{.}[r]^{\approx} \ar[d]^{L} & {\exists E_2'} \ar@{=>}[d]^{L} \\
{\exists E_2} \ar@{=>}[d]^{\epsilon} \ar@{.}[r]^{\approx} & {\exists E_2''}
\ar@{=>}[d]^{\epsilon} \\
{\forall E_1} \ar@{.}[r]^{\approx} & {\exists E2'''}
}
\end{displaymath}
In the previous figure, the existence of $E_2'$ follows by previous lemma, the existence of
$E_2''$ follows by the definition of weak equivalence, and the
existence of $E_2'''$ follows by the next existence lemma of weak equivalence.
\end{proof}

The existence lemma for weak equivalences that we mentioned in
previous proof is the following one:
\begin{lemma}
\begin{alltt}
EPS_TRANS_AUX:
\HOLTokenTurnstile{} \HOLFreeVar{E} \HOLSymConst{\HOLTokenEPS} \HOLFreeVar{E\sb{\mathrm{1}}} \HOLSymConst{\HOLTokenImp{}}
   \HOLSymConst{\HOLTokenForall{}}\HOLBoundVar{Wbsm} \HOLBoundVar{E\sp{\prime}}.
     \HOLConst{WEAK_BISIM} \HOLBoundVar{Wbsm} \HOLSymConst{\HOLTokenConj{}} \HOLBoundVar{Wbsm} \HOLFreeVar{E} \HOLBoundVar{E\sp{\prime}} \HOLSymConst{\HOLTokenImp{}} \HOLSymConst{\HOLTokenExists{}}\HOLBoundVar{E\sb{\mathrm{2}}}. \HOLBoundVar{E\sp{\prime}} \HOLSymConst{\HOLTokenEPS} \HOLBoundVar{E\sb{\mathrm{2}}} \HOLSymConst{\HOLTokenConj{}} \HOLBoundVar{Wbsm} \HOLFreeVar{E\sb{\mathrm{1}}} \HOLBoundVar{E\sb{\mathrm{2}}}
\end{alltt}
\end{lemma}

Now we prove the transitivity of Observation Congruence ($\approx^c$):
\begin{theorem}{(Transitivity of Observation Congruence)}
\begin{alltt}
\HOLTokenTurnstile{} \HOLFreeVar{E} \HOLSymConst{\HOLTokenObsCongr} \HOLFreeVar{E\sp{\prime}} \HOLSymConst{\HOLTokenConj{}} \HOLFreeVar{E\sp{\prime}} \HOLSymConst{\HOLTokenObsCongr} \HOLFreeVar{E\sp{\prime\prime}} \HOLSymConst{\HOLTokenImp{}} \HOLFreeVar{E} \HOLSymConst{\HOLTokenObsCongr} \HOLFreeVar{E\sp{\prime\prime}}\hfill[OBS_CONGR_TRANS]
\end{alltt}
\end{theorem}

\begin{proof}
Suppose \HOLinline{\HOLFreeVar{E} \HOLSymConst{\HOLTokenObsCongr} \HOLFreeVar{E\sp{\prime}}} and \HOLinline{\HOLFreeVar{E\sp{\prime}} \HOLSymConst{\HOLTokenObsCongr} \HOLFreeVar{E\sp{\prime\prime}}}, we're
going to prove \HOLinline{\HOLFreeVar{E} \HOLSymConst{\HOLTokenObsCongr} \HOLFreeVar{E\sp{\prime\prime}}} by checking directly the
definition of observation congruence.

For any $u$ and $E_1$ which satisfy \HOLinline{\HOLFreeVar{E} \HOLTokenTransBegin\HOLFreeVar{u}\HOLTokenTransEnd \HOLFreeVar{E\sb{\mathrm{1}}}}, by definition of observation congruence, there exists $E_2$
such that \HOLinline{\HOLFreeVar{E\sp{\prime}} \HOLTokenWeakTransBegin\HOLFreeVar{u}\HOLTokenWeakTransEnd \HOLFreeVar{E\sb{\mathrm{2}}}} with \HOLinline{\HOLFreeVar{E\sb{\mathrm{1}}} \HOLSymConst{\HOLTokenWeakEQ} \HOLFreeVar{E\sb{\mathrm{2}}}}. By above Lemma 2, there exists
another $E_3$ such that \HOLinline{\HOLFreeVar{E\sp{\prime\prime}} \HOLTokenWeakTransBegin\HOLFreeVar{u}\HOLTokenWeakTransEnd \HOLFreeVar{E\sb{\mathrm{3}}}} with
\HOLinline{\HOLFreeVar{E\sb{\mathrm{2}}} \HOLSymConst{\HOLTokenWeakEQ} \HOLFreeVar{E\sb{\mathrm{3}}}}. By the already proven transitivity of weak
equivalence, \HOLinline{\HOLFreeVar{E\sb{\mathrm{1}}} \HOLSymConst{\HOLTokenWeakEQ} \HOLFreeVar{E\sb{\mathrm{3}}}}, thus $E_3$ is the required
process which satisfies the definition of observation
congruence. This proves the first part. The other part is completely symmetric.
\begin{displaymath}
\xymatrix{
{\forall E1} \ar@{.}[r]^{\approx} \ar@/^4ex/[rr]^{\approx (goal)} &
{\exists E_2}
\ar@{.}[r]^{\approx} & {\exists E_3} \\
{\forall E} \ar[u]^{\forall u} \ar@{.}[r]^{\approx^c} & {E'} \ar@{=>}[u]^{u}
\ar@{.}[r]^{\approx^c} & {E''} \ar@{=>}[u]^{u}
}
\end{displaymath}
\end{proof}

Then we have proved the congruence (substitutivity) of observational congruence under
various CCS process operators:
\begin{proposition}{(The substitutivity of Observational Congruence)}
\begin{enumerate}
\item Observation congruence is substitutive under the prefix
  operator:
\begin{alltt}
OBS_CONGR_SUBST_PREFIX:
\HOLTokenTurnstile{} \HOLFreeVar{E} \HOLSymConst{\HOLTokenObsCongr} \HOLFreeVar{E\sp{\prime}} \HOLSymConst{\HOLTokenImp{}} \HOLSymConst{\HOLTokenForall{}}\HOLBoundVar{u}. \HOLBoundVar{u}\HOLSymConst{..}\HOLFreeVar{E} \HOLSymConst{\HOLTokenObsCongr} \HOLBoundVar{u}\HOLSymConst{..}\HOLFreeVar{E\sp{\prime}}
\end{alltt}
\item Observation congruence is substitutive under binary summation:
\begin{alltt}
OBS_CONGR_PRESD_BY_SUM
\HOLTokenTurnstile{} \HOLFreeVar{E\sb{\mathrm{1}}} \HOLSymConst{\HOLTokenObsCongr} \HOLFreeVar{E\sb{\mathrm{1}}\sp{\prime}} \HOLSymConst{\HOLTokenConj{}} \HOLFreeVar{E\sb{\mathrm{2}}} \HOLSymConst{\HOLTokenObsCongr} \HOLFreeVar{E\sb{\mathrm{2}}\sp{\prime}} \HOLSymConst{\HOLTokenImp{}} \HOLFreeVar{E\sb{\mathrm{1}}} \HOLSymConst{+} \HOLFreeVar{E\sb{\mathrm{2}}} \HOLSymConst{\HOLTokenObsCongr} \HOLFreeVar{E\sb{\mathrm{1}}\sp{\prime}} \HOLSymConst{+} \HOLFreeVar{E\sb{\mathrm{2}}\sp{\prime}}
\end{alltt}
\item Observation congruence is preserved by parallel composition:
\begin{alltt}
OBS_CONGR_PRESD_BY_PAR:
\HOLTokenTurnstile{} \HOLFreeVar{E\sb{\mathrm{1}}} \HOLSymConst{\HOLTokenObsCongr} \HOLFreeVar{E\sb{\mathrm{1}}\sp{\prime}} \HOLSymConst{\HOLTokenConj{}} \HOLFreeVar{E\sb{\mathrm{2}}} \HOLSymConst{\HOLTokenObsCongr} \HOLFreeVar{E\sb{\mathrm{2}}\sp{\prime}} \HOLSymConst{\HOLTokenImp{}} \HOLFreeVar{E\sb{\mathrm{1}}} \HOLSymConst{\ensuremath{\parallel}} \HOLFreeVar{E\sb{\mathrm{2}}} \HOLSymConst{\HOLTokenObsCongr} \HOLFreeVar{E\sb{\mathrm{1}}\sp{\prime}} \HOLSymConst{\ensuremath{\parallel}} \HOLFreeVar{E\sb{\mathrm{2}}\sp{\prime}}
\end{alltt}
\item Observation congruence is substitutive under the restriction
  operator:
\begin{alltt}
OBS_CONGR_SUBST_RESTR:
\HOLTokenTurnstile{} \HOLFreeVar{E} \HOLSymConst{\HOLTokenObsCongr} \HOLFreeVar{E\sp{\prime}} \HOLSymConst{\HOLTokenImp{}} \HOLSymConst{\HOLTokenForall{}}\HOLBoundVar{L}. \HOLSymConst{\ensuremath{\nu}} \HOLBoundVar{L} \HOLFreeVar{E} \HOLSymConst{\HOLTokenObsCongr} \HOLSymConst{\ensuremath{\nu}} \HOLBoundVar{L} \HOLFreeVar{E\sp{\prime}}
\end{alltt}
\item Observation congruence is substitutive under the relabeling
  operator:
\begin{alltt}
OBS_CONGR_SUBST_RELAB:
\HOLTokenTurnstile{} \HOLFreeVar{E} \HOLSymConst{\HOLTokenObsCongr} \HOLFreeVar{E\sp{\prime}} \HOLSymConst{\HOLTokenImp{}} \HOLSymConst{\HOLTokenForall{}}\HOLBoundVar{rf}. \HOLConst{relab} \HOLFreeVar{E} \HOLBoundVar{rf} \HOLSymConst{\HOLTokenObsCongr} \HOLConst{relab} \HOLFreeVar{E\sp{\prime}} \HOLBoundVar{rf}
\end{alltt}
\end{enumerate}
\end{proposition}

Finally, like the case for weak equivalence, we can easily prove the
relationship between strong equivalence and observation congruence:
\begin{theorem}{(Strong equivalence implies observation congruence)}
\begin{alltt}
\HOLTokenTurnstile{} \HOLFreeVar{E} \HOLSymConst{\HOLTokenStrongEQ} \HOLFreeVar{E\sp{\prime}} \HOLSymConst{\HOLTokenImp{}} \HOLFreeVar{E} \HOLSymConst{\HOLTokenObsCongr} \HOLFreeVar{E\sp{\prime}}\hfill[STRONG_IMP_OBS_CONGR]
\end{alltt}
\end{theorem}

With this result, all algebraic laws for observation congruence can be
derived from the corresponding algebraic laws of strong
equivalence. Here we omit these theorems, except for the following four
$\tau$-laws:
\begin{theorem}{(The $\tau$-laws for observational congruence)}
\begin{alltt}
\HOLTokenTurnstile{} \HOLFreeVar{u}\HOLSymConst{..}\HOLSymConst{\ensuremath{\tau}}\HOLSymConst{..}\HOLFreeVar{E} \HOLSymConst{\HOLTokenObsCongr} \HOLFreeVar{u}\HOLSymConst{..}\HOLFreeVar{E}\hfill[TAU1]
\HOLTokenTurnstile{} \HOLFreeVar{E} \HOLSymConst{+} \HOLSymConst{\ensuremath{\tau}}\HOLSymConst{..}\HOLFreeVar{E} \HOLSymConst{\HOLTokenObsCongr} \HOLSymConst{\ensuremath{\tau}}\HOLSymConst{..}\HOLFreeVar{E}\hfill[TAU2]
\HOLTokenTurnstile{} \HOLFreeVar{u}\HOLSymConst{..}(\HOLFreeVar{E} \HOLSymConst{+} \HOLSymConst{\ensuremath{\tau}}\HOLSymConst{..}\HOLFreeVar{E\sp{\prime}}) \HOLSymConst{+} \HOLFreeVar{u}\HOLSymConst{..}\HOLFreeVar{E\sp{\prime}} \HOLSymConst{\HOLTokenObsCongr} \HOLFreeVar{u}\HOLSymConst{..}(\HOLFreeVar{E} \HOLSymConst{+} \HOLSymConst{\ensuremath{\tau}}\HOLSymConst{..}\HOLFreeVar{E\sp{\prime}})\hfill[TAU3]
\HOLTokenTurnstile{} \HOLFreeVar{E} \HOLSymConst{+} \HOLSymConst{\ensuremath{\tau}}\HOLSymConst{..}(\HOLFreeVar{E\sp{\prime}} \HOLSymConst{+} \HOLFreeVar{E}) \HOLSymConst{\HOLTokenObsCongr} \HOLSymConst{\ensuremath{\tau}}\HOLSymConst{..}(\HOLFreeVar{E\sp{\prime}} \HOLSymConst{+} \HOLFreeVar{E})\hfill[TAU_STRAT]
\end{alltt}
\end{theorem}

\section{Relationship between Weak Equivalence and Observational
  Congruence}

The relationship between weak equivalence and observation congruence
was an interesting research topic, and there're many deep lemmas
related. In this project, we have proved two such deep lemmas. The
first one is the following Deng Lemma (for weak
bisimularity\footnote{The original Deng lemma is for another kind of
  equivalence relation called \emph{rooted branching bisimularity},
  which is not touched in this project.}):
\begin{theorem}{(Deng lemma for weak bisimilarity)}
If \HOLinline{\HOLFreeVar{p} \HOLSymConst{\HOLTokenWeakEQ} \HOLFreeVar{q}}, then one of the following three cases
holds:
\begin{enumerate}
\item $\exists p'$ such that \HOLinline{\HOLFreeVar{p} \HOLTokenTransBegin\HOLSymConst{\ensuremath{\tau}}\HOLTokenTransEnd \HOLFreeVar{p\sp{\prime}}} and
  \HOLinline{\HOLFreeVar{p\sp{\prime}} \HOLSymConst{\HOLTokenWeakEQ} \HOLFreeVar{q}}, or
\item $\exists q'$ such that \HOLinline{\HOLFreeVar{q} \HOLTokenTransBegin\HOLSymConst{\ensuremath{\tau}}\HOLTokenTransEnd \HOLFreeVar{q\sp{\prime}}} and
  \HOLinline{\HOLFreeVar{p} \HOLSymConst{\HOLTokenWeakEQ} \HOLFreeVar{q\sp{\prime}}}, or
\item \HOLinline{\HOLFreeVar{p} \HOLSymConst{\HOLTokenObsCongr} \HOLFreeVar{q}}.
\end{enumerate}
Formally:
\begin{alltt}
\HOLTokenTurnstile{} \HOLFreeVar{p} \HOLSymConst{\HOLTokenWeakEQ} \HOLFreeVar{q} \HOLSymConst{\HOLTokenImp{}}
   (\HOLSymConst{\HOLTokenExists{}}\HOLBoundVar{p\sp{\prime}}. \HOLFreeVar{p} \HOLTokenTransBegin\HOLSymConst{\ensuremath{\tau}}\HOLTokenTransEnd \HOLBoundVar{p\sp{\prime}} \HOLSymConst{\HOLTokenConj{}} \HOLBoundVar{p\sp{\prime}} \HOLSymConst{\HOLTokenWeakEQ} \HOLFreeVar{q}) \HOLSymConst{\HOLTokenDisj{}} (\HOLSymConst{\HOLTokenExists{}}\HOLBoundVar{q\sp{\prime}}. \HOLFreeVar{q} \HOLTokenTransBegin\HOLSymConst{\ensuremath{\tau}}\HOLTokenTransEnd \HOLBoundVar{q\sp{\prime}} \HOLSymConst{\HOLTokenConj{}} \HOLFreeVar{p} \HOLSymConst{\HOLTokenWeakEQ} \HOLBoundVar{q\sp{\prime}}) \HOLSymConst{\HOLTokenDisj{}}
   \HOLFreeVar{p} \HOLSymConst{\HOLTokenObsCongr} \HOLFreeVar{q}\hfill[DENG_LEMMA]
\end{alltt}
\end{theorem}
\begin{proof}
Actually there's no need to consider thee difference cases. Using the
logical tautology \HOLinline{(\HOLSymConst{\HOLTokenNeg{}}\HOLFreeVar{P} \HOLSymConst{\HOLTokenConj{}} \HOLSymConst{\HOLTokenNeg{}}\HOLFreeVar{Q} \HOLSymConst{\HOLTokenImp{}} \HOLFreeVar{R}) \HOLSymConst{\HOLTokenImp{}} \HOLFreeVar{P} \HOLSymConst{\HOLTokenDisj{}} \HOLFreeVar{Q} \HOLSymConst{\HOLTokenDisj{}} \HOLFreeVar{R}},
the theorem can be reduced to the following goal:
\begin{quote}
Prove \HOLinline{\HOLFreeVar{p} \HOLSymConst{\HOLTokenObsCongr} \HOLFreeVar{q}}, with the following three assumptions:
\begin{enumerate}
\item \HOLinline{\HOLFreeVar{p} \HOLSymConst{\HOLTokenWeakEQ} \HOLFreeVar{q}}
\item \HOLinline{\HOLSymConst{\HOLTokenNeg{}}\HOLSymConst{\HOLTokenExists{}}\HOLBoundVar{p\sp{\prime}}. \HOLFreeVar{p} \HOLTokenTransBegin\HOLSymConst{\ensuremath{\tau}}\HOLTokenTransEnd \HOLBoundVar{p\sp{\prime}} \HOLSymConst{\HOLTokenConj{}} \HOLBoundVar{p\sp{\prime}} \HOLSymConst{\HOLTokenWeakEQ} \HOLFreeVar{q}}
\item \HOLinline{\HOLSymConst{\HOLTokenNeg{}}\HOLSymConst{\HOLTokenExists{}}\HOLBoundVar{q\sp{\prime}}. \HOLFreeVar{q} \HOLTokenTransBegin\HOLSymConst{\ensuremath{\tau}}\HOLTokenTransEnd \HOLBoundVar{q\sp{\prime}} \HOLSymConst{\HOLTokenConj{}} \HOLFreeVar{p} \HOLSymConst{\HOLTokenWeakEQ} \HOLBoundVar{q\sp{\prime}}}
\end{enumerate}
\end{quote}

Now we check the definition of observation congruence: for any
transition from $p$, say \HOLinline{\HOLFreeVar{p} \HOLTokenTransBegin\HOLFreeVar{u}\HOLTokenTransEnd \HOLFreeVar{E\sb{\mathrm{1}}}}, consider the cases when
$u = \tau$ and $u \neq \tau$:
\begin{enumerate}
\item If $u = \tau$, then by \HOLinline{\HOLFreeVar{p} \HOLSymConst{\HOLTokenWeakEQ} \HOLFreeVar{q}} and the definition
  of weak equivalence, there exists $E_2$ such that \HOLinline{\HOLFreeVar{q} \HOLSymConst{\HOLTokenEPS} \HOLFreeVar{E\sb{\mathrm{2}}}}
  and \HOLinline{\HOLFreeVar{E\sb{\mathrm{1}}} \HOLSymConst{\HOLTokenWeakEQ} \HOLFreeVar{E\sb{\mathrm{2}}}}. But by assumption we know $q \neq E2$, thus
  \HOLinline{\HOLFreeVar{q} \HOLSymConst{\HOLTokenEPS} \HOLFreeVar{E\sb{\mathrm{2}}}} contains at least one $\tau$-transition, thus is
  actually \HOLinline{\HOLFreeVar{q} \HOLTokenWeakTransBegin\HOLSymConst{\ensuremath{\tau}}\HOLTokenWeakTransEnd \HOLFreeVar{E\sb{\mathrm{2}}}}, which is required by the definition
  of observation congruence for \HOLinline{\HOLFreeVar{p} \HOLSymConst{\HOLTokenWeakEQ} \HOLFreeVar{q}}.
\begin{displaymath}
\xymatrix{
{p} \ar@{.}[r]^{\approx^c} \ar[d]^{\tau} & {q} \ar@{=>}[d]^{\epsilon
  (\tau)} \\
{\forall E_1} \ar@{.}[ur]^{\not\approx} \ar@{.}[r]^{\approx} &
{\exists E_2}
}
\end{displaymath}
\item If $u = L$, then the requirement of observation congruence is
  directly satisfied.
\end{enumerate}
The other direction is completely symmetric.
\end{proof}

Now we prove the famous Hennessy Lemma:
\begin{theorem}{(Hennessy Lemma)}
For any processes $p$ and $q$, \HOLinline{\HOLFreeVar{p} \HOLSymConst{\HOLTokenWeakEQ} \HOLFreeVar{q}} if and only if
(\HOLinline{\HOLFreeVar{p} \HOLSymConst{\HOLTokenObsCongr} \HOLFreeVar{q}} or \HOLinline{\HOLFreeVar{p} \HOLSymConst{\HOLTokenObsCongr} \HOLSymConst{\ensuremath{\tau}}\HOLSymConst{..}\HOLFreeVar{q}} or
\HOLinline{\HOLSymConst{\ensuremath{\tau}}\HOLSymConst{..}\HOLFreeVar{p} \HOLSymConst{\HOLTokenObsCongr} \HOLFreeVar{q}}):
\begin{alltt}
\HOLTokenTurnstile{} \HOLFreeVar{p} \HOLSymConst{\HOLTokenWeakEQ} \HOLFreeVar{q} \HOLSymConst{\HOLTokenEquiv{}} \HOLFreeVar{p} \HOLSymConst{\HOLTokenObsCongr} \HOLFreeVar{q} \HOLSymConst{\HOLTokenDisj{}} \HOLFreeVar{p} \HOLSymConst{\HOLTokenObsCongr} \HOLSymConst{\ensuremath{\tau}}\HOLSymConst{..}\HOLFreeVar{q} \HOLSymConst{\HOLTokenDisj{}} \HOLSymConst{\ensuremath{\tau}}\HOLSymConst{..}\HOLFreeVar{p} \HOLSymConst{\HOLTokenObsCongr} \HOLFreeVar{q}\hfill[HENNESSY_LEMMA]
\end{alltt}
\end{theorem}

\begin{proof}
The ``if'' part (from right to left) can be easily derived by
applying:
\begin{itemize}
\item \texttt{OBS_CONGR_IMP_WEAK_EQUIV},
\item \texttt{TAU_WEAK},
\item \texttt{WEAK_EQUIV_SYM}, and
\item \texttt{WEAK_EQUIV_TRANS}
\end{itemize}
We'll focus on
the hard ``only if'' part (from left to right). The proof represent
here is slightly simplier than the one in \cite{Gorrieri:2015jt}, but
the idea is the same. The proof is based on creative case analysis:

\begin{enumerate}
\item If there exists an $E$ such that \HOLinline{\HOLFreeVar{p} \HOLTokenTransBegin\HOLSymConst{\ensuremath{\tau}}\HOLTokenTransEnd \HOLFreeVar{E} \HOLSymConst{\HOLTokenConj{}} \HOLFreeVar{E} \HOLSymConst{\HOLTokenWeakEQ} \HOLFreeVar{q}}, we can prove \HOLinline{\HOLFreeVar{p} \HOLSymConst{\HOLTokenObsCongr} \HOLSymConst{\ensuremath{\tau}}\HOLSymConst{..}\HOLFreeVar{q}} by
expanding \HOLinline{\HOLFreeVar{p} \HOLSymConst{\HOLTokenWeakEQ} \HOLFreeVar{q}} by \texttt{WEAK_PROPERTY_STAR}. The
other needed theorems are the definition of weak transition,
\texttt{EPS_REFL}, SOS rule \texttt{PREFIX} and \texttt{TRANS_PREFIX},
\texttt{TAU_PREFIX_WEAK_TRANS} and \texttt{TRANS_IMP_WEAK_TRANS}.

\item If there's no $E$ such that \HOLinline{\HOLFreeVar{p} \HOLTokenTransBegin\HOLSymConst{\ensuremath{\tau}}\HOLTokenTransEnd \HOLFreeVar{E} \HOLSymConst{\HOLTokenConj{}} \HOLFreeVar{E} \HOLSymConst{\HOLTokenWeakEQ} \HOLFreeVar{q}}, we can further check if there exist an $E$ such that
\HOLinline{\HOLFreeVar{q} \HOLTokenTransBegin\HOLSymConst{\ensuremath{\tau}}\HOLTokenTransEnd \HOLFreeVar{E} \HOLSymConst{\HOLTokenConj{}} \HOLFreeVar{p} \HOLSymConst{\HOLTokenWeakEQ} \HOLFreeVar{E}}, and in this case we can prove
\HOLinline{\HOLSymConst{\ensuremath{\tau}}\HOLSymConst{..}\HOLFreeVar{p} \HOLSymConst{\HOLTokenObsCongr} \HOLFreeVar{q}} in the same way as the above case.

\item Otherwise we got exactly the same condition as in Deng Lemma
(after the initial goal reduced in the previous proof), and in this
case we can directly prove that \HOLinline{\HOLFreeVar{p} \HOLSymConst{\HOLTokenObsCongr} \HOLFreeVar{q}}.
\end{enumerate}
\end{proof}

This formal proof has basically shown that, for most
informal proofs in Concurrency Theory which doesn't depend on external
mathematics theories, the author has got the ability to
express it in HOL theorem prover.

\section{The theory of (pre)congruence}

One of the highlight of this project is the formal proofs for various
versions of the ``coarsest congruence contained in weak equivalence'',
\begin{proposition}{(Coarsest congruence contained in $\approx$)}
For any processes $p$ and $q$, we have $p \approx^c q$ if and only if
$\forall r.\; p+r \approx q+r$.
\end{proposition}

At first glance, the name of above theorem doesn't make much
sense. To see the nature of above theorem more clearly, here we
represent a rather complete theory about the congruence of CCS. It's
based on contents from \cite{vanGlabbeek:2005ur}.

To formalize the concept of congruence, we need to define ``semantic context''
first. There're multiple solutions, here we have chosen a simple
solution based on $\lambda$-calculus:
\begin{definition}{(Semantic context of CCS)}
The semantic context (or one-hole context) of CCS is a function
$C[\cdot]$ of type ``\HOLinline{(\ensuremath{\alpha}, \ensuremath{\beta}) \HOLTyOp{context}}'' recursively defined by following rules:
\begin{quote}
\begin{alltt}
\HOLConst{CONTEXT} (\HOLTokenLambda{}\HOLBoundVar{t}. \HOLBoundVar{t})
\HOLConst{CONTEXT} (\HOLTokenLambda{}\HOLBoundVar{t}. \HOLFreeVar{p})
\HOLConst{CONTEXT} \HOLFreeVar{e} \HOLSymConst{\HOLTokenImp{}} \HOLConst{CONTEXT} (\HOLTokenLambda{}\HOLBoundVar{t}. \HOLFreeVar{a}\HOLSymConst{..}\HOLFreeVar{e} \HOLBoundVar{t})
\HOLConst{CONTEXT} \HOLFreeVar{e\sb{\mathrm{1}}} \HOLSymConst{\HOLTokenConj{}} \HOLConst{CONTEXT} \HOLFreeVar{e\sb{\mathrm{2}}} \HOLSymConst{\HOLTokenImp{}} \HOLConst{CONTEXT} (\HOLTokenLambda{}\HOLBoundVar{t}. \HOLFreeVar{e\sb{\mathrm{1}}} \HOLBoundVar{t} \HOLSymConst{+} \HOLFreeVar{e\sb{\mathrm{2}}} \HOLBoundVar{t})
\HOLConst{CONTEXT} \HOLFreeVar{e\sb{\mathrm{1}}} \HOLSymConst{\HOLTokenConj{}} \HOLConst{CONTEXT} \HOLFreeVar{e\sb{\mathrm{2}}} \HOLSymConst{\HOLTokenImp{}} \HOLConst{CONTEXT} (\HOLTokenLambda{}\HOLBoundVar{t}. \HOLFreeVar{e\sb{\mathrm{1}}} \HOLBoundVar{t} \HOLSymConst{\ensuremath{\parallel}} \HOLFreeVar{e\sb{\mathrm{2}}} \HOLBoundVar{t})
\HOLConst{CONTEXT} \HOLFreeVar{e} \HOLSymConst{\HOLTokenImp{}} \HOLConst{CONTEXT} (\HOLTokenLambda{}\HOLBoundVar{t}. \HOLSymConst{\ensuremath{\nu}} \HOLFreeVar{L} (\HOLFreeVar{e} \HOLBoundVar{t}))
\HOLConst{CONTEXT} \HOLFreeVar{e} \HOLSymConst{\HOLTokenImp{}} \HOLConst{CONTEXT} (\HOLTokenLambda{}\HOLBoundVar{t}. \HOLConst{relab} (\HOLFreeVar{e} \HOLBoundVar{t}) \HOLFreeVar{rf})\hfill[CONTEXT_rules]
\end{alltt}
\end{quote}
By repeatedly applying above rules, one can imagine that, any CCS term with zero or more  ``holes''
at any depth, becomes a $\lambda$-function, and by
calling the function with another CCS term, the holes were filled by that term.
\end{definition}

The notable property of semantic context is that, the functional
combination of two contexts is still a context:
\begin{proposition}{(The combination of one-hole contexts)}
If both $c_1$ and $c_2$ are two semantic contexts, then $c_1 \circ
c_2$\footnote{$(c_1 \circ c_2) t := c_1 (c_2 t).$} is
still a one-hole context:
\begin{alltt}
CONTEXT_combin:
\HOLTokenTurnstile{} \HOLConst{CONTEXT} \HOLFreeVar{c\sb{\mathrm{1}}} \HOLSymConst{\HOLTokenConj{}} \HOLConst{CONTEXT} \HOLFreeVar{c\sb{\mathrm{2}}} \HOLSymConst{\HOLTokenImp{}} \HOLConst{CONTEXT} (\HOLFreeVar{c\sb{\mathrm{1}}} \HOLSymConst{\HOLTokenCompose} \HOLFreeVar{c\sb{\mathrm{2}}})
\end{alltt}
\end{proposition}
\begin{proof}
By induction on the first context $c_1$.
\end{proof}

Now we're ready to define the concept of congruence (for CCS):
\begin{definition}{(Congruence of CCS)}
An equivalence relation $\approx$\footnote{The symbol $\approx$ here
  shouldn't be understood as weak equivalence.} on a specific space of
CCS processes is a \emph{congruence} iff for every $n$-ary operator
$f$, one has $g_1 \approx h_1 \wedge \cdots g_n \approx h_n \Rightarrow
f(g_1, \ldots, g_n) \approx f(h_1, \ldots, h_n$. This is the case iff
for every semantic context $C[\cdot]$ on has $g\approx h \Rightarrow
C[g] \approx C[h]$:
\begin{alltt}
congruence_def:
\HOLConst{congruence} \HOLFreeVar{R} \HOLSymConst{\HOLTokenEquiv{}} \HOLConst{equivalence} \HOLFreeVar{R} \HOLSymConst{\HOLTokenConj{}} \HOLConst{precongruence} \HOLFreeVar{R}
\end{alltt}

If we remove the requirement that the relation must be equivalence, we got a similar definition of pre-congruence:
\begin{alltt}
precongruence_def:
\HOLConst{precongruence} \HOLFreeVar{R} \HOLSymConst{\HOLTokenEquiv{}}
\HOLSymConst{\HOLTokenForall{}}\HOLBoundVar{x} \HOLBoundVar{y} \HOLBoundVar{ctx}. \HOLConst{CONTEXT} \HOLBoundVar{ctx} \HOLSymConst{\HOLTokenImp{}} \HOLFreeVar{R} \HOLBoundVar{x} \HOLBoundVar{y} \HOLSymConst{\HOLTokenImp{}} \HOLFreeVar{R} (\HOLBoundVar{ctx} \HOLBoundVar{x}) (\HOLBoundVar{ctx} \HOLBoundVar{y})
\end{alltt}
\end{definition}

We can easily prove that, strong equivalence and observation
congruence is indeed a congruence following above definition, using
the substitutability and preserving properties of these relations:
\begin{theorem}
Strong Equivalence ($\sim$) and Observation Congruence ($\approx^C$)
are both congruence according to the above definition:
\begin{alltt}
STRONG_EQUIV_congruence:
\HOLTokenTurnstile{} \HOLConst{congruence} \HOLConst{STRONG_EQUIV}

OBS_CONGR_congruence:
\HOLTokenTurnstile{} \HOLConst{congruence} \HOLConst{OBS_CONGR}
\end{alltt}
\end{theorem}

For relations which is not congruence, it's possible to ``convert'' them
into congruence:
\begin{definition}{(Constructing congruences from equivalence relation)}
Given an equivalence relation $\sim$\footnote{The Symbol $\sim$ here
  shouldn't be understood as strong equivalence.}, define $\sim^c$ by:
\begin{alltt}
\HOLConst{CC} \HOLFreeVar{R} \HOLSymConst{=} (\HOLTokenLambda{}\HOLBoundVar{g} \HOLBoundVar{h}. \HOLSymConst{\HOLTokenForall{}}\HOLBoundVar{c}. \HOLConst{CONTEXT} \HOLBoundVar{c} \HOLSymConst{\HOLTokenImp{}} \HOLFreeVar{R} (\HOLBoundVar{c} \HOLBoundVar{g}) (\HOLBoundVar{c} \HOLBoundVar{h}))\hfill[CC_def]
\end{alltt}
\end{definition}

This new operator on relations has the following three properties:
\begin{proposition}
For all equivalence relation $R$, $R^c$ is a congruence:
\begin{alltt}
\HOLTokenTurnstile{} \HOLConst{equivalence} \HOLFreeVar{R} \HOLSymConst{\HOLTokenImp{}} \HOLConst{congruence} (\HOLConst{CC} \HOLFreeVar{R})\hfill[CC_congruence]
\end{alltt}
\end{proposition}
\begin{proof}
By construction, $\sim^c$ is a congruence. For if $g \sim^c h$ and
$D[\cdot]$ is a semantic context, then for every semantic context
$C[\cdot]$ also $C[D[\cdot]]$ is a semantic context, so $\forall
C[\cdot].\;(C[D[g]] \sim C[D[h]])$ and hence $D[g] \sim^c D[h]$.
\end{proof}

\begin{proposition}
For all $R$, $R^c$ is finer than $R$:
\begin{alltt}
\HOLTokenTurnstile{} \HOLConst{CC} \HOLFreeVar{R} \HOLSymConst{\HOLTokenRSubset{}} \HOLFreeVar{R}\hfill[CC_is_finer]
\end{alltt}
\end{proposition}
\begin{proof}
The trivial context guarantees that $g \sim^c h \Rightarrow g\sim h$,
so $\sim^c$ is finer than $\sim$.
\end{proof}

\begin{proposition}
For all $R$, $R^c$ is the coarsest congruence finer than $R$, that is,
for any other congruence finer than $R$, it's finer than $R^c$:
\begin{alltt}
\HOLTokenTurnstile{} \HOLConst{congruence} \HOLFreeVar{R\sp{\prime}} \HOLSymConst{\HOLTokenConj{}} \HOLFreeVar{R\sp{\prime}} \HOLSymConst{\HOLTokenRSubset{}} \HOLFreeVar{R} \HOLSymConst{\HOLTokenImp{}} \HOLFreeVar{R\sp{\prime}} \HOLSymConst{\HOLTokenRSubset{}} \HOLConst{CC} \HOLFreeVar{R}\hfill[CC_is_coarsest]
\end{alltt}
\end{proposition}

\begin{proof}
If $\approx$ is any congruence finer than $\sim$, then
\begin{equation}
g \approx h \Rightarrow \forall C[\cdot].\;(C[g] \approx C[h])
\Rightarrow \forall C[\cdot].\;(C[g] \sim C[h]) \Rightarrow g \sim^c h.
\end{equation}
Thus $\approx$ is finer than $\sim^c$. (i.e. $\sim^c$ is coarser than
$\approx$, then the arbitrariness of $\approx$ implies that $\sim^c$ is
coarsest.)
\end{proof}

As we know weak equivalence is not a congruence, and one way to ``fix'' it,
is to use observation congruence which is based on weak equivalence
but have special treatments on the first transitions. The other way is to
build a congruence from existing weak equivalence relation, using
above approach based on one-hole contexts. Such a congruence has a new name:

\begin{definition}{(Weak bisimulation congruence)}
The coarsest congruence that is finer than weak bisimulation equivalence is called \emph{weak bisimulation
  congruence} (notation: $\sim_w^c$):
\begin{alltt}
\HOLTokenTurnstile{} \HOLConst{WEAK_CONGR} \HOLSymConst{=} \HOLConst{CC} \HOLConst{WEAK_EQUIV}\hfill[WEAK_CONGR]
\end{alltt}
or
\begin{alltt}
WEAK_CONGR_THM:
\HOLTokenTurnstile{} \HOLConst{WEAK_CONGR} \HOLSymConst{=} (\HOLTokenLambda{}\HOLBoundVar{g} \HOLBoundVar{h}. \HOLSymConst{\HOLTokenForall{}}\HOLBoundVar{c}. \HOLConst{CONTEXT} \HOLBoundVar{c} \HOLSymConst{\HOLTokenImp{}} \HOLBoundVar{c} \HOLBoundVar{g} \HOLSymConst{\HOLTokenWeakEQ} \HOLBoundVar{c} \HOLBoundVar{h})
\end{alltt}
\end{definition}

So far, the weak bisimulation congruence $\sim_w^c$ defined above is irrelevant with 
rooted weak bisimulation (a.k.a. observation congruence) $\approx^c$,
which has the following standard definition also based on
weak equivalence:
\begin{alltt}
OBS_CONGR:
\HOLTokenTurnstile{} \HOLFreeVar{E} \HOLSymConst{\HOLTokenObsCongr} \HOLFreeVar{E\sp{\prime}} \HOLSymConst{\HOLTokenEquiv{}}
   \HOLSymConst{\HOLTokenForall{}}\HOLBoundVar{u}.
     (\HOLSymConst{\HOLTokenForall{}}\HOLBoundVar{E\sb{\mathrm{1}}}. \HOLFreeVar{E} \HOLTokenTransBegin\HOLBoundVar{u}\HOLTokenTransEnd \HOLBoundVar{E\sb{\mathrm{1}}} \HOLSymConst{\HOLTokenImp{}} \HOLSymConst{\HOLTokenExists{}}\HOLBoundVar{E\sb{\mathrm{2}}}. \HOLFreeVar{E\sp{\prime}} \HOLTokenWeakTransBegin\HOLBoundVar{u}\HOLTokenWeakTransEnd \HOLBoundVar{E\sb{\mathrm{2}}} \HOLSymConst{\HOLTokenConj{}} \HOLBoundVar{E\sb{\mathrm{1}}} \HOLSymConst{\HOLTokenWeakEQ} \HOLBoundVar{E\sb{\mathrm{2}}}) \HOLSymConst{\HOLTokenConj{}}
     \HOLSymConst{\HOLTokenForall{}}\HOLBoundVar{E\sb{\mathrm{2}}}. \HOLFreeVar{E\sp{\prime}} \HOLTokenTransBegin\HOLBoundVar{u}\HOLTokenTransEnd \HOLBoundVar{E\sb{\mathrm{2}}} \HOLSymConst{\HOLTokenImp{}} \HOLSymConst{\HOLTokenExists{}}\HOLBoundVar{E\sb{\mathrm{1}}}. \HOLFreeVar{E} \HOLTokenWeakTransBegin\HOLBoundVar{u}\HOLTokenWeakTransEnd \HOLBoundVar{E\sb{\mathrm{1}}} \HOLSymConst{\HOLTokenConj{}} \HOLBoundVar{E\sb{\mathrm{1}}} \HOLSymConst{\HOLTokenWeakEQ} \HOLBoundVar{E\sb{\mathrm{2}}}
\end{alltt}

But since observational congruence is congruence, it must be finer than
weak bisimulation congruence:
\begin{lemma}
Observational congruence implies weak bisimulation congruence:
\begin{alltt}
OBS_CONGR_IMP_WEAK_CONGR:
\HOLTokenTurnstile{} \HOLFreeVar{p} \HOLSymConst{\HOLTokenObsCongr} \HOLFreeVar{q} \HOLSymConst{\HOLTokenImp{}} \HOLConst{WEAK_CONGR} \HOLFreeVar{p} \HOLFreeVar{q}
\end{alltt}
\end{lemma}
On the other side, by consider the trivial context and sum contexts in the definition
of weak bisimulation congruence, we can easily prove the following
result:
\begin{lemma}
Weak bisimulation congruence implies sum equivalence:
\begin{alltt}
WEAK_CONGR_IMP_SUM_EQUIV:
\HOLTokenTurnstile{} \HOLConst{WEAK_CONGR} \HOLFreeVar{p} \HOLFreeVar{q} \HOLSymConst{\HOLTokenImp{}} \HOLConst{SUM_EQUIV} \HOLFreeVar{p} \HOLFreeVar{q}
\end{alltt}
\end{lemma}
Noticed that, in above theorem, the sum operator can be replaced by
any other operator in CCS, but we know sum is special because it's the
only operator in which the weak equivalence is not preserved after
substitutions.

From above two lemmas, we can easily see that, weak
equivalence is between the observation congruence and an unnamed
relation $\{ (p, q) \colon \forall\,r. p + r \approx q + r \}$ (we can
temporarily call it ``sum equivalence'', because we don't if it's a
congruence, or even if it's contained in weak equivalence). If we
could further prove that ``sum equivalence'' is finer than observation congruence,
then all three congruences (observation congruence, weak equivalence
and the ``sum equivalence'' must all coincide, as illustrated in the
following figure:
\begin{displaymath}
\xymatrix{
{\textrm{Weak equivalence} (\approx)} & {} & {\textrm{Sum
    equivalence}} \ar@/^3ex/[ldd]^{\subseteq ?}\\
{} & {\textrm{Weak bisim. congruence} (\sim_w^c)}
\ar[lu]^{\subseteq} \ar[ru]^{\subseteq} \\
{} & {\textrm{Observational congruence} (\approx^c)} \ar[u]^{\subseteq}
}
\end{displaymath}

This is why the proposition at the beginning of this section is called ``coarsest congruence
contained in weak equivalence'', it's actually trying to prove that the
``sum equivalence'' is finer than ``observation congruence'' therefore
makes ``weak bisimulation congruence'' ($\sim_w^c$) coincide with ``observation
congruence'' ($\approx^c$).

\section{Coarsest congruence contained in $\approx$}
\label{sec:coarsest-congruence}

The easy part (left $\Longrightarrow$ right) is already proven in
previous section by combining
\begin{itemize}
\item \texttt{OBS_CONGR_IMP_WEAK_CONGR}, and
\item \texttt{WEAK_CONGR_IMP_SUM_EQUIV},
\end{itemize}
or it can be proved directly, using:
\begin{itemize}
\item \texttt{OBS_CONGR_IMP_WEAK_EQUIV}, and
\item \texttt{OBS_CONGR_SUBST_SUM_R}
\end{itemize}

\begin{theorem}(The easy part ``Coarsest congruence
  contained in $\approx$'')
\label{thm:easy-part}
\begin{alltt}
\HOLTokenTurnstile{} \HOLFreeVar{p} \HOLSymConst{\HOLTokenObsCongr} \HOLFreeVar{q} \HOLSymConst{\HOLTokenImp{}} \HOLSymConst{\HOLTokenForall{}}\HOLBoundVar{r}. \HOLFreeVar{p} \HOLSymConst{+} \HOLBoundVar{r} \HOLSymConst{\HOLTokenWeakEQ} \HOLFreeVar{q} \HOLSymConst{+} \HOLBoundVar{r} \hfill[COARSEST_CONGR_LR]
\end{alltt}
\end{theorem}

Thus we only focus on the hard part (right $\Longrightarrow$ left) in
the rest of this section.

\subsection{With classical cardinality assumptions}

A classic restriction is to assume cardinality limitations on the two
processes, so that didn't use up all possible labels. Sometimes this
assumption is automatically satisfied, for example: the CCS is
finitrary and the set of all actions is infinite. But in our setting,
the CCS datatype contains twi type variables, and if the set of all possible
labels has only finite cardinalities, this assumtion may not be satisfied.

In \cite{Milner:1989} (Proposition 3 in Chapter 7, p. 153), Robin
Milner simply calls
 this important theorem as ``Proposition 3'':
\begin{proposition}{(Proposition 3 of observation congruence)}
Assume that $\mathcal{L}(P) \cup \mathcal{L}(Q) \neq
\mathcal{L}$. Then $P \approx^c Q$ iff, for all $R$, $P + R \approx Q
+ R$.
\end{proposition}
And in \cite{Gorrieri:2015jt} (Theorem 4.5 in Chapter 4, p. 185),
Prof.\ Roberto Gorrieri
calls it ``Coarsest congruence contained in $\approx$'' (so did
us in this paper):
\begin{theorem}{(Coarsest congruence contained in $\approx$)}
Assume that $\mathrm{fn}(p) \cup \mathrm{fn}(q) \neq \mathscr{L}$. Then $p \approx^c q$
if and only if $p + r \approx q + r$ for all $r \in \mathscr{P}$.
\end{theorem}
Both $\mathcal{L}(\cdot)$ and $\mathrm{fn}(\cdot)$ used in above
theorems mean the set of ``non-$\tau$ actions'' (i.e. labels) used in a given process.

We analyzed the proof of above theorem and have
found that, the assumption that the two processes didn't use up all
available labels. Instead, it can be  weakened to the following
stronger version, which assumes the following properties instead:
\begin{definition}{(Processes having free actions)}
A CCS process is said to have \emph{free actions} if there exists an
non-$\tau$ action such that it doesn't appear in any transition or
weak transition directly leading from the root of the process:
\begin{alltt}
free_action_def:
\HOLTokenTurnstile{} \HOLConst{free_action} \HOLFreeVar{p} \HOLSymConst{\HOLTokenEquiv{}} \HOLSymConst{\HOLTokenExists{}}\HOLBoundVar{a}. \HOLSymConst{\HOLTokenForall{}}\HOLBoundVar{p\sp{\prime}}. \HOLSymConst{\HOLTokenNeg{}}(\HOLFreeVar{p} \HOLTokenWeakTransBegin\HOLConst{label} \HOLBoundVar{a}\HOLTokenWeakTransEnd \HOLBoundVar{p\sp{\prime}})
\end{alltt}
\end{definition}

\begin{theorem}{(Stronger version of ``Coarsest congruence contained
    in $\approx$'', only the hard part)}
Assuming for two processes $p$ and $q$ have free actions, then $p \approx^c q$ 
if $p + r \approx q + r$ for all $r \in \mathscr{P}$:
\begin{alltt}
COARSEST_CONGR_RL:
\HOLTokenTurnstile{} \HOLConst{free_action} \HOLFreeVar{p} \HOLSymConst{\HOLTokenConj{}} \HOLConst{free_action} \HOLFreeVar{q} \HOLSymConst{\HOLTokenImp{}} (\HOLSymConst{\HOLTokenForall{}}\HOLBoundVar{r}. \HOLFreeVar{p} \HOLSymConst{+} \HOLBoundVar{r} \HOLSymConst{\HOLTokenWeakEQ} \HOLFreeVar{q} \HOLSymConst{+} \HOLBoundVar{r}) \HOLSymConst{\HOLTokenImp{}} \HOLFreeVar{p} \HOLSymConst{\HOLTokenObsCongr} \HOLFreeVar{q}
\end{alltt}
\end{theorem}
This new assumption is weaker because, even $p$ and $q$ may have used all possible
actions in their transition graphs, as long as there's one such free
action for their first-step weak transitions, therefore the theorem
still holds.
Also noticed that, the two processes do not have to share the same
free actions, this property focuses on single process.

\begin{proof}{(Proof of the stronger version of ``Coarsest congruence contained
    in $\approx$'')}
The kernel idea in this proof is to use that free action, say $a$, and
have $p + a.0 \approx q + a.0$ as the working basis. Then for any
transition from $p + a.0$, say $p + a.0 \overset{u}{\Longrightarrow}
E_1$, there must be a weak transition of the same action $u$ (or EPS
when $u = \tau$) coming
from $q + a.0$ as the response. We're going to use the free-action
assumptions to conclude that, when $u = \tau$, that EPS must contain
at least one $\tau$ (thus satisfied the definition of observation
congruence):
\begin{displaymath}
\xymatrix{
{p+a.0} \ar@{.}[r]^{\approx} \ar[d]^{u=\tau} & {q + a.0}
\ar@{=>}[d]^{\epsilon} \\
{E_1} \ar@{.}[r]^{\approx} & {E_2}
}
\end{displaymath}
Indeed, if the EPS leading from $q+a.0$ actually contains no
$\tau$-transition, that is, $q+a.0 = E_2$, then $E_1$ and $E_2$ cannot
be weak equivalence: for any $a$-transition from $q+a.0$, $E1$ must
response with a weak $a$-transition as $E_1
\overset{a}{\Longrightarrow} E_1'$, but this means $p
\overset{a}{\Longrightarrow} E_1'$, which is impossible by free-action
assumption on $p$:
\begin{displaymath}
\xymatrix{
{p} \ar[rd]^{\tau} \ar[rdd]^{a} & {p + a.0} \ar@{.}[r]^{\approx} \ar[d]^{\tau}
 & {q+a.0 = E_2} \ar[d]^{a} \\
{} & {E_1} \ar@{.}[ru]^{\approx} \ar@{=>}[d]^{a} & {0} \\
{} & {E_1'} \ar@{.}[ru]^{\approx}
}
\end{displaymath}
Once we have $q + a.0 \overset{\tau}{\Longrightarrow} E2$, the first
$\tau$-transition must comes from $q$, then it's obvious to see that
$E_2$ is a valid response required by observation congruence of $p$
and $q$ in this case.

When $p \overset{L}{\longrightarrow} E_1$, we have $p+a.0
\overset{L}{\longrightarrow} E_1$, then there's an $E_2$ such that
$q+a.0\overset{L}{\Longrightarrow} E_2$. We can further conclude that
$q\overset{L}{\Longrightarrow}E_2$ because by free-action assumption
$L\neq a$. This finishes the first half of the proof, the second half
(for all transition coming from $q$) is completely symmetric. 
\end{proof}

Combining the easy and hard parts, the following theorem is proved:
\begin{theorem}{(Coarsest congruence contained in $\approx$)}
\begin{alltt}
COARSEST_CONGR_THM:
\HOLTokenTurnstile{} \HOLConst{free_action} \HOLFreeVar{p} \HOLSymConst{\HOLTokenConj{}} \HOLConst{free_action} \HOLFreeVar{q} \HOLSymConst{\HOLTokenImp{}} (\HOLFreeVar{p} \HOLSymConst{\HOLTokenObsCongr} \HOLFreeVar{q} \HOLSymConst{\HOLTokenEquiv{}} \HOLSymConst{\HOLTokenForall{}}\HOLBoundVar{r}. \HOLFreeVar{p} \HOLSymConst{+} \HOLBoundVar{r} \HOLSymConst{\HOLTokenWeakEQ} \HOLFreeVar{q} \HOLSymConst{+} \HOLBoundVar{r})
\end{alltt}
\end{theorem}

\subsection{Without cardinality assumptions}

In 2005, Rob J. van Glabbeek published a paper
\cite{vanGlabbeek:2005ur} showing that ``the weak bisimulation
congruence can be characterized as rooted weak bisimulation
equivalence, even without making assumptions on the cardinality of the
sets of states or actions of the process under consideration''. That
is to say, above ``Coarsest congruence contained in $\approx$''
theorem holds even for two arbitrary processes! The idea is actually
from Jan Willem Klop back to the 80s, but it's not published until
that 2005 paper. This proof is not known to Robin Milner in \cite{Milner:1989}.
The author carefully investigated this paper and formalized the proof in it.

The main result is the following version of the hard part of
``Coarsest congruence contained in $\approx$'' theorem under new
assumptions:
\begin{theorem}{(Coarsest congruence contained in $\approx$, new
    assumptions)}
\label{thm:new-assum}
For any two CCS processes $p$ and $q$, if there exists another stable
(i.e. first-step transitions are never $\tau$-transition) process
$k$ which is not weak bisimlar with any sub-process follows from $p$
and $q$ by \emph{one-step} weak transitions, then $p \approx^c q$
if $p + r \approx q + r$ for all $r \in \mathscr{P}$.
\begin{alltt}
\HOLTokenTurnstile{} (\HOLSymConst{\HOLTokenExists{}}\HOLBoundVar{k}.
      \HOLConst{STABLE} \HOLBoundVar{k} \HOLSymConst{\HOLTokenConj{}} (\HOLSymConst{\HOLTokenForall{}}\HOLBoundVar{p\sp{\prime}} \HOLBoundVar{u}. \HOLFreeVar{p} \HOLTokenWeakTransBegin\HOLBoundVar{u}\HOLTokenWeakTransEnd \HOLBoundVar{p\sp{\prime}} \HOLSymConst{\HOLTokenImp{}} \HOLSymConst{\HOLTokenNeg{}}(\HOLBoundVar{p\sp{\prime}} \HOLSymConst{\HOLTokenWeakEQ} \HOLBoundVar{k})) \HOLSymConst{\HOLTokenConj{}}
      \HOLSymConst{\HOLTokenForall{}}\HOLBoundVar{q\sp{\prime}} \HOLBoundVar{u}. \HOLFreeVar{q} \HOLTokenWeakTransBegin\HOLBoundVar{u}\HOLTokenWeakTransEnd \HOLBoundVar{q\sp{\prime}} \HOLSymConst{\HOLTokenImp{}} \HOLSymConst{\HOLTokenNeg{}}(\HOLBoundVar{q\sp{\prime}} \HOLSymConst{\HOLTokenWeakEQ} \HOLBoundVar{k})) \HOLSymConst{\HOLTokenImp{}}
   (\HOLSymConst{\HOLTokenForall{}}\HOLBoundVar{r}. \HOLFreeVar{p} \HOLSymConst{+} \HOLBoundVar{r} \HOLSymConst{\HOLTokenWeakEQ} \HOLFreeVar{q} \HOLSymConst{+} \HOLBoundVar{r}) \HOLSymConst{\HOLTokenImp{}}
   \HOLFreeVar{p} \HOLSymConst{\HOLTokenObsCongr} \HOLFreeVar{q}
\end{alltt}
\end{theorem}

\begin{proof}
Assuming the existence of that special process $k$, and take an
arbitrary non-$\tau$ action, say $a$ (this is always possible in our
setting, because in higher order logic any valid type must contain at
least one value), we'll use the fact that $p + a.k \approx q + a.k$ as
our working basis. For all transitions from $p$, say $p
\overset{u}{\longrightarrow} E_1$, we're going to prove that, there
must be a corresponding weak transition such that $q
\overset{u}{\Longrightarrow} E_2$, and $E_1 \approx E_2$ (thus $p
\approx^c q$. There're three cases to consider:

\begin{enumerate}
\item $\tau$-transitions: $p
\overset{\tau}{\longrightarrow} E_1$. By SOS rule ($\mathrm{Sum}_1$),
we have $p + a.k \overset{\tau}{\longrightarrow} E_1$, now by $p + a.k
\approx q + a.k$ and the property (*) of weak equivalence, there
exists an $E_2$ such that $q + a.k \overset{\epsilon}{\Longrightarrow}
E_2$. We can
use the property of $k$ to assert that, such an EPS transition must
contains at least one $\tau$-transition. Because if it's not, then $q
+a.k = E_2$, and since $E_1 \approx E_2$, for transition $q + a.k
\overset{a}{\longrightarrow} k$, $E_1$ must make a response by $E_1
\overset{a}{\Longrightarrow} E_1'$, and as the result we have $p
\overset{a}{\Longrightarrow} E_1'$ and $E_1' \approx k$, which is
impossible by the special choice of $k$:
\begin{displaymath}
\xymatrix{
{p} \ar[dr]^{\tau} \ar@{=>}[ddr]^{a} & {p+a.k} \ar@{.}[r]^{\approx}
\ar[d]^{\tau} & {q+a.k = E_2} \ar[d]^{a} \\
{} & {E_1} \ar@{.}[ru]^{\approx} \ar@{=>}[d]^{a} & {k} \\
{} & {E_1'} \ar@{.}[ru]^{\not\approx}
}
\end{displaymath}
\item If there's a $a$-transition coming from $p$ (means that the
  arbitrary chosen action $a$ is normally used by processes $p$ and
  $q$), that is, $p \overset{a}{\longrightarrow} E_1$, also
  $p+a.k\overset{a}{\longrightarrow} E_1$,
by property (*) of weak equivalence, there exists $E_2$ such that $q +
a.k \overset{a}{\Longrightarrow} E_2$:
\begin{displaymath}
\xymatrix{
{p} \ar[dr]^{a} & {p+a.k} \ar@{.}[r]^{\approx} \ar[d]^{a} & {q+a.k}
\ar@{=>}[d]^{a} \\
{} & {\forall E_1} \ar@{.}[r]^{\approx} & {\exists E_2}
}
\end{displaymath}
We must further divide this weak transition into two cases based on its first step:
\begin{enumerate}
\item If the first step is a $\tau$-transition, then for sure this
  entire weak transition must come from $q$ (otherwise the first step
  would be an $a$-transition from $a.k$). And in this case we can
  easily conclude $q \overset{a}{\Longrightarrow} E_2$ without using
 the property of $k$:
\begin{displaymath}
\xymatrix{
{p} \ar[dr]^{a} \ar@/^5ex/[rrr]^{\approx^c} & {p+a.k} \ar@{.}[r]^{\approx} \ar[d]^{a} & {q+a.k}
\ar[d]^{\tau} & {q} \ar[ld]^{\tau} \ar@{=>}[ldd]^{a} \\
{} & {\forall E_1} \ar@{.}[rd]^{\approx} & {\exists E'} \ar@{=>}[d]^{a} \\
{} & {} & {\exists E_2}
}
\end{displaymath}
\item If the first step is an $a$-transition, we can prove that, this
  $a$-transition must come from $h$ (then the proof finishes for the
  entire $a$-transition case). Because if it's from the $a.k$, since
  $k$ is stable, then there's no other coice but $E_2 = k$ and $E_1
  \approx E_2$. This is again impossible for the special choice of
  $k$:
\begin{displaymath}
\xymatrix{
{p} \ar[dr]^{a} & {p+a.k} \ar@{.}[r]^{\approx} \ar[d]^{a} & {q+a.k}
\ar[d]^{a} \\
{} & {\forall E_1} \ar@{.}[r]^{\not\approx} & {E2 = k}
}
\end{displaymath}
\end{enumerate}

\item For other $L$-transitions coming from $p$, where $L \neq a$ and
  $L \neq \tau$. As a response to $p + a.k \overset{L}{\longrightarrow}
  E_1$, we have $q + a.k \overset{L}{\Longrightarrow} E_2$ and $E_1
  \approx E_2$. It's obvious that $q \overset{L}{\Longrightarrow} E_2$
  in this case, no matter what the first step is (it can only be $\tau$
  and $L$) and this satisfies the requirement of observation
  congruence natually:
\begin{displaymath}
\xymatrix{
{p} \ar[dr]^{\forall L} \ar@/^5ex/[rrr]^{\approx^c} & {p+a.k} \ar@{.}[r]^{\approx} \ar[d]^{L} & {q+a.k}
\ar@{=>}[d]^{L} & {q} \ar@{=>}[ld]^{L} \\
{} & {\forall E_1} \ar@{.}[r]^{\approx} & {\exists E2}
}
\end{displaymath}
\end{enumerate}

The other direction (for all transitions coming from $q$) is
completely symmetric. Combining all the cases, we have $p \approx^c q$.
\end{proof}
Now it remains to prove the existence of the special process mentioned in the assumption of above theorem.

\subsection{Arbitrary many non-bisimilar processes}

Strong equivalence, weak equivalence, observation congruence, they're
all equivalence relations on CCS process space. General speaking, each
equivalence relation must have \emph{partitioned} all processes into
several disjoint equivalence classes: processes in the same
equivalence class are equivalent, and processes in different
equivalence class are not equivalent.

The assumption in previous Theorem \ref{thm:new-assum} requires the
existence of a special CCS process, which is not weak equivalence to
any sub-process leading from the two root processes by weak
transitions. On worst cases, there may be infinite such
sub-processes\footnote{Even the CCS is finite branching, that's
  because after a weak transition, the end process may have an
  infinite $\tau$-chain, and with each $\tau$-transition added into
  the weak transition, the new end process is still a valid weak
  transition, thus lead to infinite number of weak transitions.} Thus
there's no essential differences to consider all states in the
process group instead.

Then it's natural to ask if there are infinite
equivalence classes of CCS processes. If so, then it should be 
possible to choose one which is not equivalent with all the (finite) states in the
graphs of the two given processes. It turns out that, after Jan Willem
Klop, it's possible to construct such processes, in which each of them forms a new
equivalence class, we call them ``Klop processes'' in this paper:
\begin{definition}{(Klop processes)}
For each ordinal $\lambda$, and an arbitrary chosen non-$\tau$ action $a$,
define a CCS process $k_\lambda$ as follows:
\begin{enumerate}
\item $k_0 = 0$,
\item $k_{\lambda+1} = k_\lambda + a.k_\lambda$ and
\item for $\lambda$ a limit ordinal, $k_\lambda = \sum_{\mu < \lambda}
  k_\mu$, meaning that $k_\lambda$ is constructed from all graphs
  $k_\mu$ for $\mu < \lambda$ by identifying their root.
\end{enumerate}
\end{definition}

Unfortunately, it's impossible to express infinite sums in our CCS
datatype settings\footnote{And such infinite sums seems to go beyond the
ability of the HOL's Datatype package} without introducing new axioms.
Therefore we have followed a two-step approach in this project: first
we consider only the finite-state CCS (no need for axioms), then we turn
to the general case.

\subsection{Finite-state CCS}

If both processes $p$ and $q$ are finite-state CCS processes, that is,
the number of reachable states from $p$ and $q$ are both finite. And
in this case, the following limited version of Klop processes can be
defined as a recursive function (on natural numbers) in HOL4:
\begin{definition}{(Klop processes as recursive function on natural numbers)}
\begin{alltt}
\HOLConst{KLOP} \HOLFreeVar{a} \HOLNumLit{0} \HOLSymConst{=} \HOLConst{nil}
\HOLConst{KLOP} \HOLFreeVar{a} (\HOLConst{SUC} \HOLFreeVar{n}) \HOLSymConst{=} \HOLConst{KLOP} \HOLFreeVar{a} \HOLFreeVar{n} \HOLSymConst{+} \HOLConst{label} \HOLFreeVar{a}\HOLSymConst{..}\HOLConst{KLOP} \HOLFreeVar{a} \HOLFreeVar{n}\hfill[KLOP_def]
\end{alltt}
\end{definition}

By induction on the definition of Klop processes and SOS inference
rules ($\mathrm{Sum}_1$) and ($\mathrm{Sum}_2$), we can easily prove
the following properties of Klop functions:
\begin{proposition}{(Properties of Klop functions and processes)}
\begin{enumerate}
\item All Klop processes are stable:
\begin{alltt}
\HOLTokenTurnstile{} \HOLConst{STABLE} (\HOLConst{KLOP} \HOLFreeVar{a} \HOLFreeVar{n})\hfill[KLOP_PROP0]
\end{alltt}
\item All transitions of a Klop process must lead to another smaller Klop
  process, and any smaller Klop process must be a possible transition
  of a larger Klop process:
\begin{alltt}
\hfill[KLOP_PROP1]
\HOLTokenTurnstile{} \HOLConst{KLOP} \HOLFreeVar{a} \HOLFreeVar{n} \HOLTokenTransBegin\HOLConst{label} \HOLFreeVar{a}\HOLTokenTransEnd \HOLFreeVar{E} \HOLSymConst{\HOLTokenEquiv{}} \HOLSymConst{\HOLTokenExists{}}\HOLBoundVar{m}. \HOLBoundVar{m} \HOLSymConst{\HOLTokenLt{}} \HOLFreeVar{n} \HOLSymConst{\HOLTokenConj{}} (\HOLFreeVar{E} \HOLSymConst{=} \HOLConst{KLOP} \HOLFreeVar{a} \HOLBoundVar{m})
\end{alltt}
\item The weak transition version of above property:
\begin{alltt}
\hfill[KLOP_PROP1']
\HOLTokenTurnstile{} \HOLConst{KLOP} \HOLFreeVar{a} \HOLFreeVar{n} \HOLTokenWeakTransBegin\HOLConst{label} \HOLFreeVar{a}\HOLTokenWeakTransEnd \HOLFreeVar{E} \HOLSymConst{\HOLTokenEquiv{}} \HOLSymConst{\HOLTokenExists{}}\HOLBoundVar{m}. \HOLBoundVar{m} \HOLSymConst{\HOLTokenLt{}} \HOLFreeVar{n} \HOLSymConst{\HOLTokenConj{}} (\HOLFreeVar{E} \HOLSymConst{=} \HOLConst{KLOP} \HOLFreeVar{a} \HOLBoundVar{m})
\end{alltt}
\item All Klop processes are distinct according to strong equivalence:
\begin{alltt}
\HOLTokenTurnstile{} \HOLFreeVar{m} \HOLSymConst{\HOLTokenLt{}} \HOLFreeVar{n} \HOLSymConst{\HOLTokenImp{}} \HOLSymConst{\HOLTokenNeg{}}(\HOLConst{KLOP} \HOLFreeVar{a} \HOLFreeVar{m} \HOLSymConst{\HOLTokenStrongEQ} \HOLConst{KLOP} \HOLFreeVar{a} \HOLFreeVar{n})\hfill[KLOP_PROP2]
\end{alltt}
\item All Klop processes are distinct according to weak equivalence:
\begin{alltt}
\HOLTokenTurnstile{} \HOLFreeVar{m} \HOLSymConst{\HOLTokenLt{}} \HOLFreeVar{n} \HOLSymConst{\HOLTokenImp{}} \HOLSymConst{\HOLTokenNeg{}}(\HOLConst{KLOP} \HOLFreeVar{a} \HOLFreeVar{m} \HOLSymConst{\HOLTokenWeakEQ} \HOLConst{KLOP} \HOLFreeVar{a} \HOLFreeVar{n})\hfill[KLOP_PROP2']
\end{alltt}
\item Klop functions are one-one:
\begin{alltt}
\HOLTokenTurnstile{} \HOLConst{ONE_ONE} (\HOLConst{KLOP} \HOLFreeVar{a})\hfill{KLOP_ONE_ONE}
\end{alltt}
\end{enumerate}
\end{proposition}

Once we have a recursive function defined on all natural numbers $0, 1,
\ldots$, we can map them into a set containing all these Klop processes,
and the set is countable infinite. On the other side, the number of
all states coming from
two finite-state CCS processes $p$ and $q$ is finite. Choosing from an
infinite set for an element distinct with any subprocess leading from
$p$ and $q$, is always possible.  This result is purely mathematical,
completely falling into basic set theory:
\begin{lemma}
Given an equivalence relation $R$ defined on a type, and two sets $A, B$
of elements in this type, $A$ is finite, $B$ is infinite, and all elements
in $B$ are not equivalent, then there exists an element $k$ in $B$
which is not equivalent with any element in $A$:
\begin{alltt}
\HOLTokenTurnstile{} \HOLConst{equivalence} \HOLFreeVar{R} \HOLSymConst{\HOLTokenImp{}}
   \HOLConst{FINITE} \HOLFreeVar{A} \HOLSymConst{\HOLTokenConj{}} \HOLConst{INFINITE} \HOLFreeVar{B} \HOLSymConst{\HOLTokenConj{}}
   (\HOLSymConst{\HOLTokenForall{}}\HOLBoundVar{x} \HOLBoundVar{y}. \HOLBoundVar{x} \HOLSymConst{\HOLTokenIn{}} \HOLFreeVar{B} \HOLSymConst{\HOLTokenConj{}} \HOLBoundVar{y} \HOLSymConst{\HOLTokenIn{}} \HOLFreeVar{B} \HOLSymConst{\HOLTokenConj{}} \HOLBoundVar{x} \HOLSymConst{\HOLTokenNotEqual{}} \HOLBoundVar{y} \HOLSymConst{\HOLTokenImp{}} \HOLSymConst{\HOLTokenNeg{}}\HOLFreeVar{R} \HOLBoundVar{x} \HOLBoundVar{y}) \HOLSymConst{\HOLTokenImp{}}
   \HOLSymConst{\HOLTokenExists{}}\HOLBoundVar{k}. \HOLBoundVar{k} \HOLSymConst{\HOLTokenIn{}} \HOLFreeVar{B} \HOLSymConst{\HOLTokenConj{}} \HOLSymConst{\HOLTokenForall{}}\HOLBoundVar{n}. \HOLBoundVar{n} \HOLSymConst{\HOLTokenIn{}} \HOLFreeVar{A} \HOLSymConst{\HOLTokenImp{}} \HOLSymConst{\HOLTokenNeg{}}\HOLFreeVar{R} \HOLBoundVar{n} \HOLBoundVar{k}\hfill[INFINITE_EXISTS_LEMMA]
\end{alltt}
\end{lemma}
\begin{proof}
  We built an explicit mapping $f$ from $A$ to $B$\footnote{There're
    multiple ways to prove this lemma, a simpler proof is to make a
    reverse mapping from $B$ to the power set of $A$ (or further use
    the Axiom of Choice (AC) to make a mapping from $B$ to $A$), then
    the non-injectivity of this mapping will contradict the fact that
    all elements in the infinite set are distinct. Our proof doesn't
    need AC, and it relies on very simple truths about sets.}, for all
  $x \in A$, $y = f(x)$ if $y \in B$ and $y$ is equivalent with
  $x$. But it's possible that no element in $B$ is equivalent with
  $x$, and in this case we just choose an arbitrary element as
  $f(x)$. Such a mapping is to make sure the range of $f$ always fall
  into $B$.

  Now we can map $A$ to a subset of $B$, say $B_0$, and the
  cardinality of $B_0$ must be equal or smaller than the cardinality
  of $A$, thus finite. Now we choose an element $k$ from the rest part
  of $B$, this element is the desire one, because for any element
  $x \in A$, if it's equivalent with $k$, consider two cases for
  $y = f(x) \in B_0$:
  \begin{enumerate}
  \item $y$ is equivalent with $x$. In this case by transitivity of
    $R$, we have two distinct elements $y$ and $k$, one in $B_0$, the
    other in $B\setminus B_0$, they're equivalent. This violates the
    assumption that all elements in $B$ are distinct.
  \item $y$ is arbitrary chosen because there's no equivalent element
    for $x$ in $B$. But we already know one: $k$.
  \end{enumerate}
  Thus there's no element $x$ (in $A$) which is equivalent with $k$.
\end{proof}

To reason about finite-state CCS, we also need to define the concept
of ``finite-state'':
\begin{definition}{(Definitions related to finite-state CCS)}
\begin{enumerate}
\item Define \emph{reachable} as the RTC of a relation, which
  indicates the existence of a transition between two processes:
\begin{alltt}
\HOLConst{Reach} \HOLSymConst{=} (\HOLTokenLambda{}\HOLBoundVar{E} \HOLBoundVar{E\sp{\prime}}. \HOLSymConst{\HOLTokenExists{}}\HOLBoundVar{u}. \HOLBoundVar{E} \HOLTokenTransBegin\HOLBoundVar{u}\HOLTokenTransEnd \HOLBoundVar{E\sp{\prime}})\HOLSymConst{\HOLTokenSupStar{}}\hfill[Reach_def]
\end{alltt}
\item The ``nodes'' of a process is the set of all processes reachable
  from it:
\begin{alltt}
\HOLTokenTurnstile{} \HOLConst{NODES} \HOLFreeVar{p} \HOLSymConst{=} \HOLTokenLeftbrace{}\HOLBoundVar{q} \HOLTokenBar{} \HOLConst{Reach} \HOLFreeVar{p} \HOLBoundVar{q}\HOLTokenRightbrace{}\hfill[NODES_def]
\end{alltt}
\item A process is finite-state if the set of nodes is finite:
\begin{alltt}
\HOLTokenTurnstile{} \HOLConst{finite_state} \HOLFreeVar{p} \HOLSymConst{\HOLTokenEquiv{}} \HOLConst{FINITE} (\HOLConst{NODES} \HOLFreeVar{p})\hfill[finite_state_def]
\end{alltt}
\end{enumerate}
\end{definition}

Among many properties of above definitions, we mainly rely on the
following ``obvious'' property on weak transitions:
\begin{proposition}
If $p$ weakly transit to $q$, then $q$ must be in the node set of $p$:
\begin{alltt}
\HOLTokenTurnstile{} \HOLFreeVar{p} \HOLTokenWeakTransBegin\HOLFreeVar{u}\HOLTokenWeakTransEnd \HOLFreeVar{q} \HOLSymConst{\HOLTokenImp{}} \HOLFreeVar{q} \HOLSymConst{\HOLTokenIn{}} \HOLConst{NODES} \HOLFreeVar{p}\hfill[WEAK_TRANS_IN_NODES]
\end{alltt}
\end{proposition}

Using all above results, now we can easily prove the following finite
version of ``Klop lemma'':
\begin{lemma}{(Klop lemma, the finite version)}
\label{lem:klop-lemma-finite}
For any two finite-state CCS $p$ and $q$, there exists another process $k$, which
is not weak equivalent with any sub-process weakly transited from $p$
and $q$:
\begin{alltt}
KLOP_LEMMA_FINITE:
\HOLTokenTurnstile{} \HOLSymConst{\HOLTokenForall{}}\HOLBoundVar{p} \HOLBoundVar{q}.
     \HOLConst{finite_state} \HOLBoundVar{p} \HOLSymConst{\HOLTokenConj{}} \HOLConst{finite_state} \HOLBoundVar{q} \HOLSymConst{\HOLTokenImp{}}
     \HOLSymConst{\HOLTokenExists{}}\HOLBoundVar{k}.
       \HOLConst{STABLE} \HOLBoundVar{k} \HOLSymConst{\HOLTokenConj{}} (\HOLSymConst{\HOLTokenForall{}}\HOLBoundVar{p\sp{\prime}} \HOLBoundVar{u}. \HOLBoundVar{p} \HOLTokenWeakTransBegin\HOLBoundVar{u}\HOLTokenWeakTransEnd \HOLBoundVar{p\sp{\prime}} \HOLSymConst{\HOLTokenImp{}} \HOLSymConst{\HOLTokenNeg{}}(\HOLBoundVar{p\sp{\prime}} \HOLSymConst{\HOLTokenWeakEQ} \HOLBoundVar{k})) \HOLSymConst{\HOLTokenConj{}}
       \HOLSymConst{\HOLTokenForall{}}\HOLBoundVar{q\sp{\prime}} \HOLBoundVar{u}. \HOLBoundVar{q} \HOLTokenWeakTransBegin\HOLBoundVar{u}\HOLTokenWeakTransEnd \HOLBoundVar{q\sp{\prime}} \HOLSymConst{\HOLTokenImp{}} \HOLSymConst{\HOLTokenNeg{}}(\HOLBoundVar{q\sp{\prime}} \HOLSymConst{\HOLTokenWeakEQ} \HOLBoundVar{k})
\end{alltt}
\end{lemma}

Combining above lemma with Theorem \ref{thm:new-assum} and Theorem
\ref{thm:easy-part}, we can easily prove the following theorem for finite-state CCS:
\begin{theorem}{(Coarsest congruence contained in $\approx$ for
    finite-state CCS)}
\begin{alltt}
\HOLTokenTurnstile{} \HOLConst{finite_state} \HOLFreeVar{p} \HOLSymConst{\HOLTokenConj{}} \HOLConst{finite_state} \HOLFreeVar{q} \HOLSymConst{\HOLTokenImp{}}
   (\HOLFreeVar{p} \HOLSymConst{\HOLTokenObsCongr} \HOLFreeVar{q} \HOLSymConst{\HOLTokenEquiv{}} \HOLSymConst{\HOLTokenForall{}}\HOLBoundVar{r}. \HOLFreeVar{p} \HOLSymConst{+} \HOLBoundVar{r} \HOLSymConst{\HOLTokenWeakEQ} \HOLFreeVar{q} \HOLSymConst{+} \HOLBoundVar{r})\hfill[COARSEST_CONGR_FINITE]
\end{alltt}
\end{theorem}

\subsection{Finitary CCS and general cases}

For Finitrary CCS with potential infinite states,
 the proof of ``Coarsest congruence contained in $\approx$'' is
 currently not known. Currently we tend to believe the proof doesn't
 exist, i.e. Observational Congruence may not be the coarsest
 congruence contained in weak equivalence if Finitary CCS is considered.

For more general cases in which CCS's sum operator is allowed to take
infinite (not only countable but also arbitrary largee) number of
processes. The proof of ``Coarsest congruence contained in
$\approx$'', according to Rob J. van Glabbeek's paper
\cite{vanGlabbeek:2005ur}, indeed exists. However, this proof involves
arbitrary large ordinals which is not supported in Higher Order Logic
(not the software, but the logic itself).

The limitation also happens in HOL's datatype package: infinite sums
are not directly supported. However, if we're allowed to add one axiom
to enable infinite sums of CCS processes without touching the existing
CCS datatype definition, we can actually precisely formalize the proof
with the same steps as in \cite{vanGlabbeek:2005ur}. For such a work,
please refer to the author's ``internship'' project
report \cite{Tian:2017tk} with proof
scripts available at
\footnote{\url{https://github.com/binghe/informatica-public/tree/master/CCS2}}. This
work is not included into this thesis simply because an axiom is
introduced, which may potentially break the consistency of HOL.

\cleardoublepage

%%%% -*- Mode: LaTeX -*-

\chapter{A Formalization of ``bisimulation up to''}

``Bisimulation up to'' is a powerful proof technique for proving many
difficult results in process algebra. Generally speaking, it's a
technique for reducing the size of the relation needed to define a bisimulation.
By definition, two processes are bisimilar if there exists a
bisimulation relation containing them as a pair. However, in practice
this definition is hardly ever followed plainly; instead, to reduce
the size of the relations exhibited one prefers to define relations
which are bisimulations only when closed up under some specific and
priviledged relation, so to relieve the proof work needed. We call
this an \emph{``up-to'' technique}. It is a pretty general device
which allows a great variety of prssibilities.

According to \cite{Milner:1992vp}, the variety of the up-to techniques
is useful because it allows us each time to make the most convenient
choice, depending upon the equilibrium we want between the size of the
relation to exhibit and the fineness of the closure relation(s).

In this thesis project, we have basically
followed the path of Robin Milner, by using ``Bisimulaition up to
strong equivalence'' to prove the ``Unique solutions of equations'' theorem
(for the strong equivalence case). But all the rest versions of ``Unique
solutions of equations''  theorems didn't use any ``Bisimulation
up to'' techniques. This is mostly because the errors in the original
version of Milner's 1989 book \cite{Milner:1989}, which has lead to the failures when
applying ``Weak bisimulation up to'' techniques in the proof of unique
solutions theorem. Instead, in next chapter we'll present new proofs, which is slightly
longer than the version in Milner's book, but has no dependencies on
any ``Bisimulation up to'' techniques, therefore the overall proof
size is smaller.  In this chapter, we have still formalized all versions of
``Bisimulation up to'' mentioned in Milner's book, its errata, and \cite{Milner:1992vp}, plus
one more variants for observational congruence (but it's too
restrictive for proving the corresponding unique solutions theorem).

\section{Bisimulation up to $\sim$}

Following \cite{Milner:1989}, the concept of ``Bisimulation up to
$\sim$'' starts with a generalization of the notion of strong
bisimulation, which is often more useful in applications.
The following definition and proposition put the idea on a firm
basis. Henceforward we shall oftern write $P\mathcal{R} Q$ to mean
$(P, Q) \in \mathcal{R}$, for any binary relation $\mathcal{R}$. Note
also that $\sim \mathcal{S} \sim$ is a composition of binary
relations, so that $P \sim \mathcal{S} \sim Q$ means that for some
$P'$ and $Q'$ we have $P \sim P'$, $P' \mathcal{S} Q'$ and $Q' \sim Q$.

\begin{definition}{(Bisimulation up to $\sim$)}
$\mathcal{S}$ is a ``\emph{bisimulation up to $\sim$}'' if $P
  \mathcal{S} Q$ implies, for all $\alpha \in Act$,
\begin{enumerate}
\item Whenever $P \overset{\alpha}{\rightarrow} P'$ then, for some
  $Q'$, $Q \overset{\alpha}{\rightarrow} Q'$ and $P' \sim \mathcal{S}
  \sim Q'$,
\item Whenever $Q \overset{\alpha}{\rightarrow} Q'$ then, for some
  $P'$, $P \overset{\alpha}{\rightarrow} P'$ and $P' \sim \mathcal{S}
  \sim Q'$.
\end{enumerate}
Or formally,
\begin{alltt}
STRONG_BISIM_UPTO:
\HOLTokenTurnstile{} \HOLConst{STRONG_BISIM_UPTO} \HOLFreeVar{Bsm} \HOLSymConst{\HOLTokenEquiv{}}
   \HOLSymConst{\HOLTokenForall{}}\HOLBoundVar{E} \HOLBoundVar{E\sp{\prime}}.
     \HOLFreeVar{Bsm} \HOLBoundVar{E} \HOLBoundVar{E\sp{\prime}} \HOLSymConst{\HOLTokenImp{}}
     \HOLSymConst{\HOLTokenForall{}}\HOLBoundVar{u}.
       (\HOLSymConst{\HOLTokenForall{}}\HOLBoundVar{E\sb{\mathrm{1}}}.
          \HOLBoundVar{E} \HOLTokenTransBegin\HOLBoundVar{u}\HOLTokenTransEnd \HOLBoundVar{E\sb{\mathrm{1}}} \HOLSymConst{\HOLTokenImp{}}
          \HOLSymConst{\HOLTokenExists{}}\HOLBoundVar{E\sb{\mathrm{2}}}.
            \HOLBoundVar{E\sp{\prime}} \HOLTokenTransBegin\HOLBoundVar{u}\HOLTokenTransEnd \HOLBoundVar{E\sb{\mathrm{2}}} \HOLSymConst{\HOLTokenConj{}}
            (\HOLConst{STRONG_EQUIV} \HOLSymConst{\HOLTokenRCompose{}} \HOLFreeVar{Bsm} \HOLSymConst{\HOLTokenRCompose{}} \HOLConst{STRONG_EQUIV}) \HOLBoundVar{E\sb{\mathrm{1}}} \HOLBoundVar{E\sb{\mathrm{2}}}) \HOLSymConst{\HOLTokenConj{}}
       \HOLSymConst{\HOLTokenForall{}}\HOLBoundVar{E\sb{\mathrm{2}}}.
         \HOLBoundVar{E\sp{\prime}} \HOLTokenTransBegin\HOLBoundVar{u}\HOLTokenTransEnd \HOLBoundVar{E\sb{\mathrm{2}}} \HOLSymConst{\HOLTokenImp{}}
         \HOLSymConst{\HOLTokenExists{}}\HOLBoundVar{E\sb{\mathrm{1}}}.
           \HOLBoundVar{E} \HOLTokenTransBegin\HOLBoundVar{u}\HOLTokenTransEnd \HOLBoundVar{E\sb{\mathrm{1}}} \HOLSymConst{\HOLTokenConj{}} (\HOLConst{STRONG_EQUIV} \HOLSymConst{\HOLTokenRCompose{}} \HOLFreeVar{Bsm} \HOLSymConst{\HOLTokenRCompose{}} \HOLConst{STRONG_EQUIV}) \HOLBoundVar{E\sb{\mathrm{1}}} \HOLBoundVar{E\sb{\mathrm{2}}}
\end{alltt}
\end{definition}

Pictorially, clause (1) says that if $P \mathcal{S} Q$ and
$P\overset{\alpha} P'$ then we can fill in the following diagram:
\begin{displaymath}
\xymatrix{
{} & {P} \ar[ld]^{\alpha} & {\mathcal{S}} & {Q} \ar[rd]^{\alpha} \\
{P'} \ar@{.}[r]^{\sim} & {P''} & {\mathcal{S}} & {Q''}
\ar@{.}[r]^{\sim} & {Q'}
}
\end{displaymath}

``Bisimulation up to $\sim$'' has the following basic properties:
\begin{proposition}{Properties of ``strong bisimulation up to $\sim$''}
\begin{enumerate}
\item Identity relation is a ``strong bisimulation up to $\sim$'':
\begin{alltt}
\HOLTokenTurnstile{} \HOLConst{STRONG_BISIM_UPTO} (\HOLSymConst{=})\hfill[IDENTITY_STRONG_BISIM_UPTO]
\end{alltt}
\item The converse of  a ``strong bisimulation up to $\sim$'' is still ``strong bisimulation up to $\sim$'':
\begin{alltt}
CONVERSE_STRONG_BISIM_UPTO:
\HOLTokenTurnstile{} \HOLConst{STRONG_BISIM_UPTO} \HOLFreeVar{Wbsm} \HOLSymConst{\HOLTokenImp{}} \HOLConst{STRONG_BISIM_UPTO} \HOLFreeVar{Wbsm}\HOLSymConst{\HOLTokenRInverse{}}
\end{alltt}
\end{enumerate}
\end{proposition}

And we have proved the following lemma, which establishes the
relationship between ``strong bisimulation up to $\sim$'' and ``strong bisimulation'':
\begin{lemma}
If $\mathcal{S}$ is a ``bisimulation up to $\sim$'', then $\sim
\mathcal{S} \sim$ is a strong bisimulation: 
\begin{alltt}
STRONG_BISIM_UPTO_LEMMA:
\HOLTokenTurnstile{} \HOLConst{STRONG_BISIM_UPTO} \HOLFreeVar{Bsm} \HOLSymConst{\HOLTokenImp{}}
   \HOLConst{STRONG_BISIM} (\HOLConst{STRONG_EQUIV} \HOLSymConst{\HOLTokenRCompose{}} \HOLFreeVar{Bsm} \HOLSymConst{\HOLTokenRCompose{}} \HOLConst{STRONG_EQUIV})
\end{alltt}
\end{lemma}
\begin{proof}
The idea is to fix two process $E$ and $E'$, which satisfies $E \sim
\circ Bsm \circ E'$, then check it for the
definition of strong bisimulation: for all $E_1$ such that
\HOLinline{\HOLFreeVar{E} \HOLTokenTransBegin\HOLFreeVar{u}\HOLTokenTransEnd \HOLFreeVar{E\sb{\mathrm{1}}}}, there exists $E_2''$ such that \HOLinline{\HOLFreeVar{E\sp{\prime}} \HOLTokenTransBegin\HOLFreeVar{u}\HOLTokenTransEnd \HOLFreeVar{E\sb{\mathrm{2}}\sp{\prime\prime}}} (the other side is totally symmetric), as shown in the following graph:
\begin{displaymath}
\xymatrix{
{E} \ar@{-}[rr]^{\sim} \ar[d]^{\forall u} & {} & {\exists y'} \ar@{-}[r]^{Bsm}
\ar[ld]^{u} & {\exists y}
\ar@{-}[rr]^{\sim} \ar[rd]^{u} & {} & {E'} \ar[d]^{u} \\
{\forall E_1} \ar@{-}[r]^{\sim} & {\exists E_2} \ar@{-}[r]^{\sim} &
{\exists y'''}
\ar@{-}[r]^{Bsm} & {\exists y''} \ar@{-}[r]^{\sim} & {\exists E_2'}
\ar@{-}[r]^{\sim} & {\exists E_2''}
}
\end{displaymath}
During the proof, needed lemmas are the definition of ``bisimulation
up to $\sim$'' (for expanding ``$y'\text{ Bsm }y$'' into ``$E_2 \sim y'''\text{ Bsm }y''
\sim E_2'$''), plus the property (*) and transitivity of strong equivalence.
\end{proof}

Based on above lemma, we then easily proved the following proposition:
\begin{theorem}
If $\mathcal{S}$ is a ``bisimulation up to $\sim$'', then
$\mathcal{S} \subseteq \sim$:
\begin{alltt}
STRONG_BISIM_UPTO_THM:
\HOLTokenTurnstile{} \HOLConst{STRONG_BISIM_UPTO} \HOLFreeVar{Bsm} \HOLSymConst{\HOLTokenImp{}} \HOLFreeVar{Bsm} \HOLSymConst{\HOLTokenRSubset{}} \HOLConst{STRONG_EQUIV}
\end{alltt}
\end{theorem}
Hence, to prove $P \sim Q$, we only have to find a strong bisimulation
up to $\sim$ which contains $(P, Q)$.

\section{Bisimulation up to $\approx$}

The concept of bisimulation up to $\approx$ is a modified version of
Milner's original definition presented in modern textbooks and
originally in \cite{Milner:1992vp}.
\begin{definition}{(Bisimulation up to $\approx$)}
$\mathcal{S}$ is a ``\emph{bisimulation up to $\approx$}'' if $P
  \mathcal{S} Q$ implies, for all $\alpha \in Act$,
\begin{enumerate}
\item Whenever $P \overset{\alpha}{\rightarrow} P'$ then, for some
  $Q'$, $Q \overset{\hat{\alpha}}{\rightarrow} Q'$ and $P' \sim \mathcal{S}
  \approx Q'$,
\item Whenever $Q \overset{\alpha}{\rightarrow} Q'$ then, for some
  $P'$, $P \overset{\hat{\alpha}}{\rightarrow} P'$ and $P' \approx \mathcal{S}
  \sim Q'$.
\end{enumerate}
Or formally,
\begin{alltt}
WEAK_BISIM_UPTO:
\HOLTokenTurnstile{} \HOLConst{WEAK_BISIM_UPTO} \HOLFreeVar{Wbsm} \HOLSymConst{\HOLTokenEquiv{}}
   \HOLSymConst{\HOLTokenForall{}}\HOLBoundVar{E} \HOLBoundVar{E\sp{\prime}}.
     \HOLFreeVar{Wbsm} \HOLBoundVar{E} \HOLBoundVar{E\sp{\prime}} \HOLSymConst{\HOLTokenImp{}}
     (\HOLSymConst{\HOLTokenForall{}}\HOLBoundVar{l}.
        (\HOLSymConst{\HOLTokenForall{}}\HOLBoundVar{E\sb{\mathrm{1}}}.
           \HOLBoundVar{E} \HOLTokenTransBegin\HOLConst{label} \HOLBoundVar{l}\HOLTokenTransEnd \HOLBoundVar{E\sb{\mathrm{1}}} \HOLSymConst{\HOLTokenImp{}}
           \HOLSymConst{\HOLTokenExists{}}\HOLBoundVar{E\sb{\mathrm{2}}}.
             \HOLBoundVar{E\sp{\prime}} \HOLTokenWeakTransBegin\HOLConst{label} \HOLBoundVar{l}\HOLTokenWeakTransEnd \HOLBoundVar{E\sb{\mathrm{2}}} \HOLSymConst{\HOLTokenConj{}}
             (\HOLConst{WEAK_EQUIV} \HOLSymConst{\HOLTokenRCompose{}} \HOLFreeVar{Wbsm} \HOLSymConst{\HOLTokenRCompose{}} \HOLConst{STRONG_EQUIV}) \HOLBoundVar{E\sb{\mathrm{1}}} \HOLBoundVar{E\sb{\mathrm{2}}}) \HOLSymConst{\HOLTokenConj{}}
        \HOLSymConst{\HOLTokenForall{}}\HOLBoundVar{E\sb{\mathrm{2}}}.
          \HOLBoundVar{E\sp{\prime}} \HOLTokenTransBegin\HOLConst{label} \HOLBoundVar{l}\HOLTokenTransEnd \HOLBoundVar{E\sb{\mathrm{2}}} \HOLSymConst{\HOLTokenImp{}}
          \HOLSymConst{\HOLTokenExists{}}\HOLBoundVar{E\sb{\mathrm{1}}}.
            \HOLBoundVar{E} \HOLTokenWeakTransBegin\HOLConst{label} \HOLBoundVar{l}\HOLTokenWeakTransEnd \HOLBoundVar{E\sb{\mathrm{1}}} \HOLSymConst{\HOLTokenConj{}}
            (\HOLConst{STRONG_EQUIV} \HOLSymConst{\HOLTokenRCompose{}} \HOLFreeVar{Wbsm} \HOLSymConst{\HOLTokenRCompose{}} \HOLConst{WEAK_EQUIV}) \HOLBoundVar{E\sb{\mathrm{1}}} \HOLBoundVar{E\sb{\mathrm{2}}}) \HOLSymConst{\HOLTokenConj{}}
     (\HOLSymConst{\HOLTokenForall{}}\HOLBoundVar{E\sb{\mathrm{1}}}.
        \HOLBoundVar{E} \HOLTokenTransBegin\HOLSymConst{\ensuremath{\tau}}\HOLTokenTransEnd \HOLBoundVar{E\sb{\mathrm{1}}} \HOLSymConst{\HOLTokenImp{}}
        \HOLSymConst{\HOLTokenExists{}}\HOLBoundVar{E\sb{\mathrm{2}}}.
          \HOLBoundVar{E\sp{\prime}} \HOLSymConst{\HOLTokenEPS} \HOLBoundVar{E\sb{\mathrm{2}}} \HOLSymConst{\HOLTokenConj{}} (\HOLConst{WEAK_EQUIV} \HOLSymConst{\HOLTokenRCompose{}} \HOLFreeVar{Wbsm} \HOLSymConst{\HOLTokenRCompose{}} \HOLConst{STRONG_EQUIV}) \HOLBoundVar{E\sb{\mathrm{1}}} \HOLBoundVar{E\sb{\mathrm{2}}}) \HOLSymConst{\HOLTokenConj{}}
     \HOLSymConst{\HOLTokenForall{}}\HOLBoundVar{E\sb{\mathrm{2}}}.
       \HOLBoundVar{E\sp{\prime}} \HOLTokenTransBegin\HOLSymConst{\ensuremath{\tau}}\HOLTokenTransEnd \HOLBoundVar{E\sb{\mathrm{2}}} \HOLSymConst{\HOLTokenImp{}}
       \HOLSymConst{\HOLTokenExists{}}\HOLBoundVar{E\sb{\mathrm{1}}}. \HOLBoundVar{E} \HOLSymConst{\HOLTokenEPS} \HOLBoundVar{E\sb{\mathrm{1}}} \HOLSymConst{\HOLTokenConj{}} (\HOLConst{STRONG_EQUIV} \HOLSymConst{\HOLTokenRCompose{}} \HOLFreeVar{Wbsm} \HOLSymConst{\HOLTokenRCompose{}} \HOLConst{WEAK_EQUIV}) \HOLBoundVar{E\sb{\mathrm{1}}} \HOLBoundVar{E\sb{\mathrm{2}}}
\end{alltt}
\end{definition}
A few things must be noticed:
\begin{enumerate}
\item In HOL4, the big ``O'' notion as relation composition has
  different orders with usual Math notion: \emph{the right-most relation
  takes the input argument first}, which is actually the case for
function composition: $(f \circ g) (x) = f(g(x))$. Thus in all
HOL-generated terms like
\begin{alltt}
\HOLinline{(\HOLConst{WEAK_EQUIV} \HOLSymConst{\HOLTokenRCompose{}} \HOLFreeVar{Wbsm} \HOLSymConst{\HOLTokenRCompose{}} \HOLConst{STRONG_EQUIV}) \HOLFreeVar{E\sb{\mathrm{1}}} \HOLFreeVar{E\sb{\mathrm{2}}}}
\end{alltt}
in this paper, it should be understood
like ``$E_1 \sim y \text{ Wbsm } y' \approx E_2$''.
(There was no such issues for the strong bisimulation cases, because we had $\sim$ on
both side)
\item The original definition in Milner's book \cite{Milner:1989},
  in which he used $\approx \mathcal{S} \approx$ in all
  places in above definition, has been found (by his student, now
  Prof.\ Davide Sangiogi) as problematic. The reason has been explained
  in Gorrieri's book \cite{Gorrieri:2015jt} (page 65), that the
  resulting relation may not be a subset of $\approx$! Thus we have
  used the definition from Gorrieri's book, with the definition in
  Sangiorgi's book \cite{Sangiorgi:2011ut} (page 115) doubly
  confirmed.
\item Some authors (e.g. Prof.\ Davide Sangiorgi) uses the notions like $P
  \overset{\hat{\mu}}{\Rightarrow} P'$ to represent special case that,
  $P \overset{\epsilon}{\Rightarrow} P'$ when $\mu = \tau$ (i.e. it's
  possible that $P = Q$). Such notions are concise, but inconvenient
  to use in formalization work, because the EPS transition, in many
  cases, has very different characteristics. Thus we have above long
  definition but easier to use when proving all needed results.
\end{enumerate}

Two basic properties to help understanding ``bisimulation up to $\approx$'':
\begin{lemma}{(Properties of bisimulation up to $\approx$)}
\begin{enumerate}
\item The identity relation is ``bisimulation up to $\approx$'':
\begin{alltt}
\HOLTokenTurnstile{} \HOLConst{WEAK_BISIM_UPTO} (\HOLSymConst{=})\hfill[IDENTITY_WEAK_BISIM_UPTO]
\end{alltt}
\item The converse of a ``bisimulation up to $\approx$'' is still ``bisimulation up to $\approx$'':
\begin{alltt}
CONVERSE_WEAK_BISIM_UPTO:
\HOLTokenTurnstile{} \HOLConst{WEAK_BISIM_UPTO} \HOLFreeVar{Wbsm} \HOLSymConst{\HOLTokenImp{}} \HOLConst{WEAK_BISIM_UPTO} \HOLFreeVar{Wbsm}\HOLSymConst{\HOLTokenRInverse{}}
\end{alltt}
\end{enumerate}
\end{lemma}

Now we want to prove the following main lemma:
\begin{lemma}
If $\mathcal{S}$ is a ``bisimulation up to $\approx$'', then $\approx
\mathcal{S} \approx$ is a bisimulation.
\begin{alltt}
WEAK_BISIM_UPTO_LEMMA:
\HOLTokenTurnstile{} \HOLConst{WEAK_BISIM_UPTO} \HOLFreeVar{Wbsm} \HOLSymConst{\HOLTokenImp{}}
   \HOLConst{WEAK_BISIM} (\HOLConst{WEAK_EQUIV} \HOLSymConst{\HOLTokenRCompose{}} \HOLFreeVar{Wbsm} \HOLSymConst{\HOLTokenRCompose{}} \HOLConst{WEAK_EQUIV})
\end{alltt}
\end{lemma}
\begin{proof}
Milner's books simply said that the proof is ``analogous'' to the same lemma
for strong bisimulation. This is basically true, from left to right
(for visible transitions):
\begin{displaymath}
\xymatrix{
{E} \ar@{-}[rr]^{\approx} \ar[d]^{\forall l} & {} & {\exists y'} \ar@{-}[r]^{Wbsm}
\ar@{=>}[ld]^{l} & {\exists y}
\ar@{-}[rr]^{\approx} \ar@{=>}[rd]^{l} & {} & {E'} \ar@{=>}[d]^{l} \\
{\forall E_1} \ar@{-}[r]^{\approx} & {\exists E_2} \ar@{-}[r]^{\sim} &
{\exists y'''}
\ar@{-}[r]^{Wbsm} & {\exists y''} \ar@{-}[r]^{\approx} & {\exists E_2'}
\ar@{-}[r]^{\approx} & {\exists E_2''}
}
\end{displaymath}
There's a little difficulty, however. Given $y \approx E'$ and $y
\overset{l}{\Rightarrow} E_2'$, the existence of $E_2''$ doesn't
follow directly from the definition or property (*) of weak
equivalence. Instead, we have to prove a lemma (to be presented below)
to finish this last step.

More difficulties appear from right to left:
\begin{displaymath}
\xymatrix{
{E} \ar@{-}[rr]^{\approx} \ar@{=>}[d]^{l} & {} & {\exists y'} \ar@{-}[r]^{Wbsm}
\ar@{=>}[ld]^{l} & {\exists y}
\ar@{-}[rr]^{\approx} \ar@{=>}[rd]^{l} & {} & {E'} \ar[d]^{\forall l} \\
{\exists E_1''} \ar@{-}[r]^{\approx} & {\exists E_1'} \ar@{-}[r]^{\approx} &
{\exists y'''}
\ar@{-}[r]^{Wbsm} & {\exists y''} \ar@{-}[r]^{\sim} & {\exists E_1}
\ar@{-}[r]^{\approx} & {\forall E_2}
}
\end{displaymath}
The problem is, given $y' \text{ Bsm } y$ and
$y\overset{l}{\Rightarrow} E_1$, the existence of $E_1'$ doesn't
follow directly from the definition of ``bisimulation up to $\approx$'',
instead this result must be proved (to be presented below) and the proof is non-trivial.

The other two cases concerning $\tau$-transitions:
\begin{displaymath}
\xymatrix{
{E} \ar@{-}[rr]^{\approx} \ar[d]^{\tau} & {} & {\exists y'} \ar@{-}[r]^{Wbsm}
\ar@{=>}[ld]^{\epsilon} & {\exists y}
\ar@{-}[rr]^{\approx} \ar@{=>}[rd]^{\epsilon} & {} & {E'} \ar@{=>}[d]^{\epsilon} \\
{\forall E_1} \ar@{-}[r]^{\approx} & {\exists E_2} \ar@{-}[r]^{\sim} &
{\exists y'''}
\ar@{-}[r]^{Wbsm} & {\exists y''} \ar@{-}[r]^{\approx} & {\exists E_2'}
\ar@{-}[r]^{\approx} & {\exists E_2''}
}
\end{displaymath}
in which the EPS transition bypass of weak equivalence (from $E_2'$
to $E_2''$) must be proved, and
\begin{displaymath}
\xymatrix{
{E} \ar@{-}[rr]^{\approx} \ar@{=>}[d]^{\epsilon} & {} & {\exists y'} \ar@{-}[r]^{Wbsm}
\ar@{=>}[ld]^{\epsilon} & {\exists y}
\ar@{-}[rr]^{\approx} \ar@{=>}[rd]^{\epsilon} & {} & {E'} \ar[d]^{\tau} \\
{\exists E_1''} \ar@{-}[r]^{\approx} & {\exists E_1'} \ar@{-}[r]^{\approx} &
{\exists y'''}
\ar@{-}[r]^{Wbsm} & {\exists y''} \ar@{-}[r]^{\sim} & {\exists E_1}
\ar@{-}[r]^{\approx} & {\forall E_2}
}
\end{displaymath}
in which the EPS transition bypass for ``bisimulation up to
$\approx$'' (from $E_1$ to $E_1'$) must be proved as a lemma.
\end{proof}

As a summary of all ``difficulties'', here is a list of lemmas we
have used to prove the previous lemma. Each lemma has also their
``companion lemma'' concerning the other directions (here we omit them):
\begin{lemma}{(Useful lemmas concerning the first weak transitions from
 $\sim$, $\approx$ and ``bisimulation up to $\approx$'')}
\begin{enumerate}
\item \begin{alltt}
STRONG_EQUIV_EPS:
\HOLTokenTurnstile{} \HOLFreeVar{E} \HOLSymConst{\HOLTokenStrongEQ} \HOLFreeVar{E\sp{\prime}} \HOLSymConst{\HOLTokenImp{}} \HOLSymConst{\HOLTokenForall{}}\HOLBoundVar{E\sb{\mathrm{1}}}. \HOLFreeVar{E} \HOLSymConst{\HOLTokenEPS} \HOLBoundVar{E\sb{\mathrm{1}}} \HOLSymConst{\HOLTokenImp{}} \HOLSymConst{\HOLTokenExists{}}\HOLBoundVar{E\sb{\mathrm{2}}}. \HOLFreeVar{E\sp{\prime}} \HOLSymConst{\HOLTokenEPS} \HOLBoundVar{E\sb{\mathrm{2}}} \HOLSymConst{\HOLTokenConj{}} \HOLBoundVar{E\sb{\mathrm{1}}} \HOLSymConst{\HOLTokenStrongEQ} \HOLBoundVar{E\sb{\mathrm{2}}}
\end{alltt}
\item \begin{alltt}
WEAK_EQUIV_EPS:
\HOLTokenTurnstile{} \HOLFreeVar{E} \HOLSymConst{\HOLTokenWeakEQ} \HOLFreeVar{E\sp{\prime}} \HOLSymConst{\HOLTokenImp{}} \HOLSymConst{\HOLTokenForall{}}\HOLBoundVar{E\sb{\mathrm{1}}}. \HOLFreeVar{E} \HOLSymConst{\HOLTokenEPS} \HOLBoundVar{E\sb{\mathrm{1}}} \HOLSymConst{\HOLTokenImp{}} \HOLSymConst{\HOLTokenExists{}}\HOLBoundVar{E\sb{\mathrm{2}}}. \HOLFreeVar{E\sp{\prime}} \HOLSymConst{\HOLTokenEPS} \HOLBoundVar{E\sb{\mathrm{2}}} \HOLSymConst{\HOLTokenConj{}} \HOLBoundVar{E\sb{\mathrm{1}}} \HOLSymConst{\HOLTokenWeakEQ} \HOLBoundVar{E\sb{\mathrm{2}}}
\end{alltt}

\item \begin{alltt}
WEAK_EQUIV_WEAK_TRANS_label:
\HOLTokenTurnstile{} \HOLFreeVar{E} \HOLSymConst{\HOLTokenWeakEQ} \HOLFreeVar{E\sp{\prime}} \HOLSymConst{\HOLTokenImp{}}
   \HOLSymConst{\HOLTokenForall{}}\HOLBoundVar{l} \HOLBoundVar{E\sb{\mathrm{1}}}. \HOLFreeVar{E} \HOLTokenWeakTransBegin\HOLConst{label} \HOLBoundVar{l}\HOLTokenWeakTransEnd \HOLBoundVar{E\sb{\mathrm{1}}} \HOLSymConst{\HOLTokenImp{}} \HOLSymConst{\HOLTokenExists{}}\HOLBoundVar{E\sb{\mathrm{2}}}. \HOLFreeVar{E\sp{\prime}} \HOLTokenWeakTransBegin\HOLConst{label} \HOLBoundVar{l}\HOLTokenWeakTransEnd \HOLBoundVar{E\sb{\mathrm{2}}} \HOLSymConst{\HOLTokenConj{}} \HOLBoundVar{E\sb{\mathrm{1}}} \HOLSymConst{\HOLTokenWeakEQ} \HOLBoundVar{E\sb{\mathrm{2}}}
\end{alltt}

\item \begin{alltt}
WEAK_EQUIV_WEAK_TRANS_tau:
\HOLTokenTurnstile{} \HOLFreeVar{E} \HOLSymConst{\HOLTokenWeakEQ} \HOLFreeVar{E\sp{\prime}} \HOLSymConst{\HOLTokenImp{}} \HOLSymConst{\HOLTokenForall{}}\HOLBoundVar{E\sb{\mathrm{1}}}. \HOLFreeVar{E} \HOLTokenWeakTransBegin\HOLSymConst{\ensuremath{\tau}}\HOLTokenWeakTransEnd \HOLBoundVar{E\sb{\mathrm{1}}} \HOLSymConst{\HOLTokenImp{}} \HOLSymConst{\HOLTokenExists{}}\HOLBoundVar{E\sb{\mathrm{2}}}. \HOLFreeVar{E\sp{\prime}} \HOLSymConst{\HOLTokenEPS} \HOLBoundVar{E\sb{\mathrm{2}}} \HOLSymConst{\HOLTokenConj{}} \HOLBoundVar{E\sb{\mathrm{1}}} \HOLSymConst{\HOLTokenWeakEQ} \HOLBoundVar{E\sb{\mathrm{2}}}
\end{alltt}

\item \begin{alltt}
WEAK_BISIM_UPTO_EPS:
\HOLTokenTurnstile{} \HOLConst{WEAK_BISIM_UPTO} \HOLFreeVar{Wbsm} \HOLSymConst{\HOLTokenImp{}}
   \HOLSymConst{\HOLTokenForall{}}\HOLBoundVar{E} \HOLBoundVar{E\sp{\prime}}.
     \HOLFreeVar{Wbsm} \HOLBoundVar{E} \HOLBoundVar{E\sp{\prime}} \HOLSymConst{\HOLTokenImp{}}
     \HOLSymConst{\HOLTokenForall{}}\HOLBoundVar{E\sb{\mathrm{1}}}.
       \HOLBoundVar{E} \HOLSymConst{\HOLTokenEPS} \HOLBoundVar{E\sb{\mathrm{1}}} \HOLSymConst{\HOLTokenImp{}}
       \HOLSymConst{\HOLTokenExists{}}\HOLBoundVar{E\sb{\mathrm{2}}}. \HOLBoundVar{E\sp{\prime}} \HOLSymConst{\HOLTokenEPS} \HOLBoundVar{E\sb{\mathrm{2}}} \HOLSymConst{\HOLTokenConj{}} (\HOLConst{WEAK_EQUIV} \HOLSymConst{\HOLTokenRCompose{}} \HOLFreeVar{Wbsm} \HOLSymConst{\HOLTokenRCompose{}} \HOLConst{STRONG_EQUIV}) \HOLBoundVar{E\sb{\mathrm{1}}} \HOLBoundVar{E\sb{\mathrm{2}}}
\end{alltt}

\item \begin{alltt}
WEAK_BISIM_UPTO_WEAK_TRANS_label:
\HOLTokenTurnstile{} \HOLConst{WEAK_BISIM_UPTO} \HOLFreeVar{Wbsm} \HOLSymConst{\HOLTokenImp{}}
   \HOLSymConst{\HOLTokenForall{}}\HOLBoundVar{E} \HOLBoundVar{E\sp{\prime}}.
     \HOLFreeVar{Wbsm} \HOLBoundVar{E} \HOLBoundVar{E\sp{\prime}} \HOLSymConst{\HOLTokenImp{}}
     \HOLSymConst{\HOLTokenForall{}}\HOLBoundVar{l} \HOLBoundVar{E\sb{\mathrm{1}}}.
       \HOLBoundVar{E} \HOLTokenWeakTransBegin\HOLConst{label} \HOLBoundVar{l}\HOLTokenWeakTransEnd \HOLBoundVar{E\sb{\mathrm{1}}} \HOLSymConst{\HOLTokenImp{}}
       \HOLSymConst{\HOLTokenExists{}}\HOLBoundVar{E\sb{\mathrm{2}}}.
         \HOLBoundVar{E\sp{\prime}} \HOLTokenWeakTransBegin\HOLConst{label} \HOLBoundVar{l}\HOLTokenWeakTransEnd \HOLBoundVar{E\sb{\mathrm{2}}} \HOLSymConst{\HOLTokenConj{}}
         (\HOLConst{WEAK_EQUIV} \HOLSymConst{\HOLTokenRCompose{}} \HOLFreeVar{Wbsm} \HOLSymConst{\HOLTokenRCompose{}} \HOLConst{STRONG_EQUIV}) \HOLBoundVar{E\sb{\mathrm{1}}} \HOLBoundVar{E\sb{\mathrm{2}}}
\end{alltt}
\end{enumerate}
\end{lemma}
\begin{proof}{(Proof sketch of above lemmas)}
The proof of \texttt{STORNG_EQUIV_EPS} and \texttt{WEAK_EQUIV_EPS}
depends on the following ``right-side induction'' \footnote{Such induction
  theorems are part of HOL's theorem \texttt{relationTheory} for RTCs
  (reflexive transitive closure). The proof of transitivity of
  observation congruence also heavily depends on this induction
  theorem, but in the work of Monica Nesi where the EPS relation is
  manually defined inductively, such an induction theorem is not
  available (and it's not easy to prove it), as a result Monica Nesi
  couldn't finish the proof for transitivity of
  observation congruence, which is incredible hard to prove without
  proving lemmas like \texttt{WEAK_EQUIV_EPS} first.} of the EPS
transition:
\begin{alltt}
\HOLTokenTurnstile{} (\HOLSymConst{\HOLTokenForall{}}\HOLBoundVar{x}. \HOLFreeVar{P} \HOLBoundVar{x} \HOLBoundVar{x}) \HOLSymConst{\HOLTokenConj{}} (\HOLSymConst{\HOLTokenForall{}}\HOLBoundVar{x} \HOLBoundVar{y} \HOLBoundVar{z}. \HOLFreeVar{P} \HOLBoundVar{x} \HOLBoundVar{y} \HOLSymConst{\HOLTokenConj{}} \HOLBoundVar{y} \HOLTokenTransBegin\HOLSymConst{\ensuremath{\tau}}\HOLTokenTransEnd \HOLBoundVar{z} \HOLSymConst{\HOLTokenImp{}} \HOLFreeVar{P} \HOLBoundVar{x} \HOLBoundVar{z}) \HOLSymConst{\HOLTokenImp{}}
   \HOLSymConst{\HOLTokenForall{}}\HOLBoundVar{x} \HOLBoundVar{y}. \HOLBoundVar{x} \HOLSymConst{\HOLTokenEPS} \HOLBoundVar{y} \HOLSymConst{\HOLTokenImp{}} \HOLFreeVar{P} \HOLBoundVar{x} \HOLBoundVar{y}\hfill[EPS_ind_right]
\end{alltt}
Basically, we need to prove that, if the lemma already holds for
$n-1$ $\tau$-transitions, it also holds for $n$ $\tau$-transitions.

The proof of \texttt{WEAK_BISIM_UPTO_EPS} is also based on above
induction theorem. The induction case of this proof can be sketched
using the following graph:
\begin{displaymath}
\xymatrix{
{} & {E} \ar@{-}[rrr]^{Wbsm} \ar@{.}[d] & {} & {} & {E'} \ar@{=>}[d]^{\epsilon} \\
{} & {E_1} \ar@{-}[r]^{\sim} \ar[ld]^{\tau} & {\exists y'} \ar@{-}[r]^{Wbsm}
\ar[ld]^{\tau} & {\exists y}
\ar@{-}[r]^{\approx} \ar@{=>}[rd]^{\epsilon} & {E_2}
\ar@{=>}[rd]^{\epsilon} \\
{\forall E_1'} \ar@{-}[r]^{\approx} & {\exists E_2'} \ar@{-}[r]^{\sim} &
{\exists y'''}
\ar@{-}[r]^{Wbsm} & {\exists y''} \ar@{-}[r]^{\approx} & {\exists E_2''}
\ar@{-}[r]^{\approx} & {\exists E_2'''}
}
\end{displaymath}
The goal is to find $E_2'''$ which satisfy $E'
\overset{\epsilon}{\Rightarrow} E2'''$. As we can see from the graph,
using the induction, now given $y' \text{
  Wbsm } y$ and $y' \overset{\tau}{\rightarrow} E_2'$, we can easily
crossover the ``bisimulation up to $\approx$'' and assert the
existence of $E_2''$ to finish the proof.

The proof of \texttt{WEAK_BISIM_UPTO_WEAK_TRANS_label} is based on
\texttt{WEAK_BISIM_UPTO_EPS}. It is much more
difficult, because an even bigger graph must be step-by-step
constructed:
\begin{displaymath}
\xymatrix{
{} & {E} \ar@{-}[rrr]^{Wbsm} \ar@{=>}[d]^{\epsilon} & {} & {} & {E'} \ar@{=>}[d]^{\epsilon} \\
{} & {\exists E_1'} \ar@{-}[r]^{\sim} \ar[ld]^{l} & {\exists y'} \ar@{-}[r]^{Wbsm}
\ar[ld]^{l} & {\exists y} \ar@{-}[r]^{\approx} \ar@{=>}[rd]^{l} & {E_2'} \ar@{=>}[rd]^{l} \\ 
{\exists E_2} \ar@{-}[r]^{\sim} \ar@{.}[rd] & {\exists E_2''}
\ar@{-}[rrr]^{\sim\text{ Wbsm }\approx} \ar@{.}[rd] & {} & {} &
{\exists E_2'''}
\ar@{-}[r]^{\approx} \ar@{.}[ld] & {\exists E_2^{(4)}} \ar@{.}[ld] \\
{} & {\exists E_2} \ar@{-}[r]^{\sim} \ar@{=>}[ld]^{\epsilon} & {\exists y'''}
\ar@{-}[r]^{Wbsm} \ar@{=>}[ld]^{\epsilon} & {\exists y''}
\ar@{-}[r]^{\approx} \ar@{=>}[rd]^{\epsilon} & {\exists E_2^{(4)}}
\ar@{=>}[rd]^{\epsilon} \\
{\forall E_1} \ar@{-}[r]^{\sim} & {\exists E_2^{(5)}}
\ar@{-}[rrr]^{\sim\text{ Wbsm }\approx}  & {} & {} & {\exists E_2^{(6)}}
\ar@{-}[r]^{\approx} & {\exists E_2^{(7)}}
}
\end{displaymath}
That is, for all $E_1$ such that $E \overset{l}{\Rightarrow} E_1$
(which by definition of weak transitions exists $E_1'$ and $E_2$ such
that $E \overset{\epsilon}{\Rightarrow} E_1'$, $E_1'
\overset{l}{\rightarrow} E_2$ and $E_2 \overset{\epsilon}{\Rightarrow}
E_1$), we would like to finally find an $E_2^{(7)}$ such that $E'
\overset{l}{\Rightarrow} E_2^{(7)}$. This process is long and painful,
and we have to use \texttt{WEAK_BISIM_UPTO_EPS} twice.  The formal
proof tries to build above graph by asserting the existences of each
process step-by-step, until it finally reached to $E_2^(7)$. This
proof is so-far the largest formal proof (in single branch) that the
author ever met, before closing it has 26 assumptions which represents
above graph:
\begin{lstlisting}
?E2. E' ==label l=>> E2 /\ (WEAK_EQUIV O Wbsm O STRONG_EQUIV) E1 E2
------------------------------------
  0.  WEAK_BISIM_UPTO Wbsm
  1.  Wbsm E E'
  2.  E ==label l=>> E1
  3.  EPS E E1'
  4.  E1' --label l-> E2
  5.  EPS E2 E1
  6.  EPS E' E2'
  7.  STRONG_EQUIV E1' y'
  8.  Wbsm y' y
  9.  WEAK_EQUIV y E2'
  10.  y' --label l-> E2''
  11.  STRONG_EQUIV E2 E2''
  12.  y ==label l=>> E2'''
  13.  (WEAK_EQUIV O Wbsm O STRONG_EQUIV) E2'' E2'''
  14.  E2' ==label l=>> E2''''
  15.  WEAK_EQUIV E2''' E2''''
  16.  STRONG_EQUIV E2 y'''
  17.  Wbsm y''' y''
  18.  WEAK_EQUIV y'' E2''''
  19.  EPS y''' E2'''''
  20.  STRONG_EQUIV E1 E2'''''
  21.  EPS y'' E2''''''
  22.  (WEAK_EQUIV O Wbsm O STRONG_EQUIV) E2''''' E2''''''
  23.  EPS E2'''' E2'''''''
  24.  WEAK_EQUIV E2'''''' E2'''''''
  25.  (WEAK_EQUIV O Wbsm O STRONG_EQUIV) E1 E2'''''''
\end{lstlisting}
\end{proof}

All above lemmas concern the cases from left and right (for all $P$,
exists $Q$ such that ...) To prove the
other side (for all $Q$ there exist $P$ such that ...), there's no
need to go over the painful proving process again, instead
we can easily derive the other side by using
\texttt{CONVERSE_WEAK_BISIM_UPTO}. For example, once above lemma
\texttt{WEAK_BISIM_UPTO_WEAK_TRANS_label} is proved, it's trivial to
get the following companion lemma:
\begin{lemma}{(The companion lemma of \texttt{WEAK_BISIM_UPTO_WEAK_TRANS_label})}
\begin{alltt}
WEAK_BISIM_UPTO_WEAK_TRANS_label':
\HOLTokenTurnstile{} \HOLConst{WEAK_BISIM_UPTO} \HOLFreeVar{Wbsm} \HOLSymConst{\HOLTokenImp{}}
   \HOLSymConst{\HOLTokenForall{}}\HOLBoundVar{E} \HOLBoundVar{E\sp{\prime}}.
     \HOLFreeVar{Wbsm} \HOLBoundVar{E} \HOLBoundVar{E\sp{\prime}} \HOLSymConst{\HOLTokenImp{}}
     \HOLSymConst{\HOLTokenForall{}}\HOLBoundVar{l} \HOLBoundVar{E\sb{\mathrm{2}}}.
       \HOLBoundVar{E\sp{\prime}} \HOLTokenWeakTransBegin\HOLConst{label} \HOLBoundVar{l}\HOLTokenWeakTransEnd \HOLBoundVar{E\sb{\mathrm{2}}} \HOLSymConst{\HOLTokenImp{}}
       \HOLSymConst{\HOLTokenExists{}}\HOLBoundVar{E\sb{\mathrm{1}}}.
         \HOLBoundVar{E} \HOLTokenWeakTransBegin\HOLConst{label} \HOLBoundVar{l}\HOLTokenWeakTransEnd \HOLBoundVar{E\sb{\mathrm{1}}} \HOLSymConst{\HOLTokenConj{}}
         (\HOLConst{STRONG_EQUIV} \HOLSymConst{\HOLTokenRCompose{}} \HOLFreeVar{Wbsm} \HOLSymConst{\HOLTokenRCompose{}} \HOLConst{WEAK_EQUIV}) \HOLBoundVar{E\sb{\mathrm{1}}} \HOLBoundVar{E\sb{\mathrm{2}}}
\end{alltt}
\end{lemma}

Finally, once the main lemma \texttt{WEAK_BISIM_UPTO_LEMMA}, the
following final result can be easily proved, following the same idea
in the proof of strong bisimulation cases:
\begin{theorem}
If $\mathcal{S}$ is a bisimulation up to $\approx$, then
$\mathcal{S} \subseteq \approx$:
WEAK_BISIM_UPTO_THM:
\begin{alltt}
\HOLTokenTurnstile{} \HOLConst{WEAK_BISIM_UPTO} \HOLFreeVar{Wbsm} \HOLSymConst{\HOLTokenImp{}} \HOLFreeVar{Wbsm} \HOLSymConst{\HOLTokenRSubset{}} \HOLConst{WEAK_EQUIV}
\end{alltt}
\end{theorem}

As we have told at the beginning of this chapter, above result cannot
be used for proving the ``unique solution of equations'' theorem for
the weak equivalence (and observational congruence) cases. So above
final theorem is not used anywhere in this thesis. It's simply an
interesting result on its own.

\section{Another version of ``Bisimulation upto $\approx$''}

Milner's 1989 book contains an error: the definition of ``bisimulation
upto $\approx$'' cannot be used to prove the ``unique solutions of
equations'' theorem for weak equivalence.  This issue was originally
found by Prof.\ Davide Sangiorgi when he was a student of Robin
Milner. The solution was published as a new paper
\cite{sangiorgi1992problem}, co-authored by them. The solution,
however, is not to fix the original definition but to invent an
\emph{alternative} version of ``bisimulation up to $\approx$'' which
has not relationship with the old one. It's been found that, both
versions were useful.  In our project, this alternative version is
also formalized, to support the proof of ``unique solutions of
equations'' theorem for weak equivalence. Here is the definition:

\begin{definition}{(Another version of bisimulation up to $\approx$)}
$\mathcal{S}$ is a ``\emph{bisimulation up to $\approx$}'' if $P
  \mathcal{S} Q$ implies, for all $\alpha \in Act$,
\begin{enumerate}
\item Whenever $P \overset{\alpha}{\Rightarrow} P'$ then, for some
  $Q'$, $Q \overset{\hat{\alpha}}{\rightarrow} Q'$ and $P' \approx \mathcal{S}
  \approx Q'$,
\item Whenever $Q \overset{\alpha}{\Rightarrow} Q'$ then, for some
  $P'$, $P \overset{\hat{\alpha}}{\rightarrow} P'$ and $P' \approx \mathcal{S}
  \approx Q'$.
\end{enumerate}
Or formally,
\begin{alltt}
WEAK_BISIM_UPTO_ALT:
\HOLTokenTurnstile{} \HOLConst{WEAK_BISIM_UPTO_ALT} \HOLFreeVar{Wbsm} \HOLSymConst{\HOLTokenEquiv{}}
   \HOLSymConst{\HOLTokenForall{}}\HOLBoundVar{E} \HOLBoundVar{E\sp{\prime}}.
     \HOLFreeVar{Wbsm} \HOLBoundVar{E} \HOLBoundVar{E\sp{\prime}} \HOLSymConst{\HOLTokenImp{}}
     (\HOLSymConst{\HOLTokenForall{}}\HOLBoundVar{l}.
        (\HOLSymConst{\HOLTokenForall{}}\HOLBoundVar{E\sb{\mathrm{1}}}.
           \HOLBoundVar{E} \HOLTokenWeakTransBegin\HOLConst{label} \HOLBoundVar{l}\HOLTokenWeakTransEnd \HOLBoundVar{E\sb{\mathrm{1}}} \HOLSymConst{\HOLTokenImp{}}
           \HOLSymConst{\HOLTokenExists{}}\HOLBoundVar{E\sb{\mathrm{2}}}.
             \HOLBoundVar{E\sp{\prime}} \HOLTokenWeakTransBegin\HOLConst{label} \HOLBoundVar{l}\HOLTokenWeakTransEnd \HOLBoundVar{E\sb{\mathrm{2}}} \HOLSymConst{\HOLTokenConj{}}
             (\HOLConst{WEAK_EQUIV} \HOLSymConst{\HOLTokenRCompose{}} \HOLFreeVar{Wbsm} \HOLSymConst{\HOLTokenRCompose{}} \HOLConst{WEAK_EQUIV}) \HOLBoundVar{E\sb{\mathrm{1}}} \HOLBoundVar{E\sb{\mathrm{2}}}) \HOLSymConst{\HOLTokenConj{}}
        \HOLSymConst{\HOLTokenForall{}}\HOLBoundVar{E\sb{\mathrm{2}}}.
          \HOLBoundVar{E\sp{\prime}} \HOLTokenWeakTransBegin\HOLConst{label} \HOLBoundVar{l}\HOLTokenWeakTransEnd \HOLBoundVar{E\sb{\mathrm{2}}} \HOLSymConst{\HOLTokenImp{}}
          \HOLSymConst{\HOLTokenExists{}}\HOLBoundVar{E\sb{\mathrm{1}}}.
            \HOLBoundVar{E} \HOLTokenWeakTransBegin\HOLConst{label} \HOLBoundVar{l}\HOLTokenWeakTransEnd \HOLBoundVar{E\sb{\mathrm{1}}} \HOLSymConst{\HOLTokenConj{}}
            (\HOLConst{WEAK_EQUIV} \HOLSymConst{\HOLTokenRCompose{}} \HOLFreeVar{Wbsm} \HOLSymConst{\HOLTokenRCompose{}} \HOLConst{WEAK_EQUIV}) \HOLBoundVar{E\sb{\mathrm{1}}} \HOLBoundVar{E\sb{\mathrm{2}}}) \HOLSymConst{\HOLTokenConj{}}
     (\HOLSymConst{\HOLTokenForall{}}\HOLBoundVar{E\sb{\mathrm{1}}}.
        \HOLBoundVar{E} \HOLTokenWeakTransBegin\HOLSymConst{\ensuremath{\tau}}\HOLTokenWeakTransEnd \HOLBoundVar{E\sb{\mathrm{1}}} \HOLSymConst{\HOLTokenImp{}}
        \HOLSymConst{\HOLTokenExists{}}\HOLBoundVar{E\sb{\mathrm{2}}}.
          \HOLBoundVar{E\sp{\prime}} \HOLSymConst{\HOLTokenEPS} \HOLBoundVar{E\sb{\mathrm{2}}} \HOLSymConst{\HOLTokenConj{}} (\HOLConst{WEAK_EQUIV} \HOLSymConst{\HOLTokenRCompose{}} \HOLFreeVar{Wbsm} \HOLSymConst{\HOLTokenRCompose{}} \HOLConst{WEAK_EQUIV}) \HOLBoundVar{E\sb{\mathrm{1}}} \HOLBoundVar{E\sb{\mathrm{2}}}) \HOLSymConst{\HOLTokenConj{}}
     \HOLSymConst{\HOLTokenForall{}}\HOLBoundVar{E\sb{\mathrm{2}}}.
       \HOLBoundVar{E\sp{\prime}} \HOLTokenWeakTransBegin\HOLSymConst{\ensuremath{\tau}}\HOLTokenWeakTransEnd \HOLBoundVar{E\sb{\mathrm{2}}} \HOLSymConst{\HOLTokenImp{}}
       \HOLSymConst{\HOLTokenExists{}}\HOLBoundVar{E\sb{\mathrm{1}}}. \HOLBoundVar{E} \HOLSymConst{\HOLTokenEPS} \HOLBoundVar{E\sb{\mathrm{1}}} \HOLSymConst{\HOLTokenConj{}} (\HOLConst{WEAK_EQUIV} \HOLSymConst{\HOLTokenRCompose{}} \HOLFreeVar{Wbsm} \HOLSymConst{\HOLTokenRCompose{}} \HOLConst{WEAK_EQUIV}) \HOLBoundVar{E\sb{\mathrm{1}}} \HOLBoundVar{E\sb{\mathrm{2}}}
\end{alltt}
\end{definition}

Noticed that, now there're two weak equivalences which surround the
central relation. (So instead of calling it ``alternative version'',
maybe ``double-weak version'' is better.)

It turns out that, the proof for properties of this alternative
relation is slightly easier. We have proved exactly the same lemmas
and theorems for it, of which the last two results are the following
ones:

\begin{lemma}
If $\mathcal{S}$ is a ``bisimulation up to $\approx$'' (alternative version), then $\approx
\mathcal{S} \approx$ is a weak bisimulation:
\begin{alltt}
WEAK_BISIM_UPTO_ALT_LEMMA:
\HOLTokenTurnstile{} \HOLConst{WEAK_BISIM_UPTO_ALT} \HOLFreeVar{Wbsm} \HOLSymConst{\HOLTokenImp{}}
   \HOLConst{WEAK_BISIM} (\HOLConst{WEAK_EQUIV} \HOLSymConst{\HOLTokenRCompose{}} \HOLFreeVar{Wbsm} \HOLSymConst{\HOLTokenRCompose{}} \HOLConst{WEAK_EQUIV})
\end{alltt}
\end{lemma}

\begin{theorem}
If $\mathcal{S}$ is a ``bisimulation up to $\approx$'' (alternative version), then
$\mathcal{S} \subseteq \approx$:
\begin{alltt}
WEAK_BISIM_UPTO_ALT_THM:
\HOLTokenTurnstile{} \HOLConst{WEAK_BISIM_UPTO_ALT} \HOLFreeVar{Wbsm} \HOLSymConst{\HOLTokenImp{}} \HOLFreeVar{Wbsm} \HOLSymConst{\HOLTokenRSubset{}} \HOLConst{WEAK_EQUIV}
\end{alltt}
\end{theorem}

In next chapter, we'll use the theorem \texttt{WEAK_BISIM_UPTO_ALT_THM} to prove
Milner's ``unique solutions of equations'' theorems for weak
equivalence, however without it the proof can still be finished (by
constructing a more complex bisimulation), just a little longer.

\section{Observational bisimulation up to $\approx$}

This is a new creation by the author during the early proof attempts for
Milner's ``unique solutions of equations'' theorems for observational
congruence. It turns out that, this new relation is too restrictive,
thus finally it's not used. Here we briefly mention it as an
independent (useless) findings. The definition is based on the
original ``bisimulation up to $\approx$'' with EPS transitions replaced
with normal weak $\tau$ transtions:

\begin{definition}{(Observational bisimulation up to $\approx$)}
$\mathcal{S}$ is a ``\emph{observational bisimulation up to $\approx$}'' if $P
  \mathcal{S} Q$ implies, for all $\alpha \in Act$,
\begin{enumerate}
\item Whenever $P \overset{\alpha}{\rightarrow} P'$ then, for some
  $Q'$, $Q \overset{\alpha}{\rightarrow} Q'$ and $P' \sim \mathcal{S}
  \sim Q'$,
\item Whenever $Q \overset{\alpha}{\rightarrow} Q'$ then, for some
  $P'$, $P \overset{\alpha}{\rightarrow} P'$ and $P' \sim \mathcal{S}
  \sim Q'$.
\end{enumerate}
Or formally,
\begin{alltt}
OBS_BISIM_UPTO:
\HOLTokenTurnstile{} \HOLConst{OBS_BISIM_UPTO} \HOLFreeVar{Obsm} \HOLSymConst{\HOLTokenEquiv{}}
   \HOLSymConst{\HOLTokenForall{}}\HOLBoundVar{E} \HOLBoundVar{E\sp{\prime}}.
     \HOLFreeVar{Obsm} \HOLBoundVar{E} \HOLBoundVar{E\sp{\prime}} \HOLSymConst{\HOLTokenImp{}}
     \HOLSymConst{\HOLTokenForall{}}\HOLBoundVar{u}.
       (\HOLSymConst{\HOLTokenForall{}}\HOLBoundVar{E\sb{\mathrm{1}}}.
          \HOLBoundVar{E} \HOLTokenTransBegin\HOLBoundVar{u}\HOLTokenTransEnd \HOLBoundVar{E\sb{\mathrm{1}}} \HOLSymConst{\HOLTokenImp{}}
          \HOLSymConst{\HOLTokenExists{}}\HOLBoundVar{E\sb{\mathrm{2}}}.
            \HOLBoundVar{E\sp{\prime}} \HOLTokenWeakTransBegin\HOLBoundVar{u}\HOLTokenWeakTransEnd \HOLBoundVar{E\sb{\mathrm{2}}} \HOLSymConst{\HOLTokenConj{}}
            (\HOLConst{WEAK_EQUIV} \HOLSymConst{\HOLTokenRCompose{}} \HOLFreeVar{Obsm} \HOLSymConst{\HOLTokenRCompose{}} \HOLConst{STRONG_EQUIV}) \HOLBoundVar{E\sb{\mathrm{1}}} \HOLBoundVar{E\sb{\mathrm{2}}}) \HOLSymConst{\HOLTokenConj{}}
       \HOLSymConst{\HOLTokenForall{}}\HOLBoundVar{E\sb{\mathrm{2}}}.
         \HOLBoundVar{E\sp{\prime}} \HOLTokenTransBegin\HOLBoundVar{u}\HOLTokenTransEnd \HOLBoundVar{E\sb{\mathrm{2}}} \HOLSymConst{\HOLTokenImp{}}
         \HOLSymConst{\HOLTokenExists{}}\HOLBoundVar{E\sb{\mathrm{1}}}.
           \HOLBoundVar{E} \HOLTokenWeakTransBegin\HOLBoundVar{u}\HOLTokenWeakTransEnd \HOLBoundVar{E\sb{\mathrm{1}}} \HOLSymConst{\HOLTokenConj{}} (\HOLConst{STRONG_EQUIV} \HOLSymConst{\HOLTokenRCompose{}} \HOLFreeVar{Obsm} \HOLSymConst{\HOLTokenRCompose{}} \HOLConst{WEAK_EQUIV}) \HOLBoundVar{E\sb{\mathrm{1}}} \HOLBoundVar{E\sb{\mathrm{2}}}
\end{alltt}
\end{definition}

Notice the removal of ``hats'' in above definition in comparasion with
previous versions of ``bisimulation up to $\approx$'', thus it's the
so far most concise definition among others.

Properties are similar. Then we proved a final theorem (there's no lemma) for proving two processes are
observational congruence: instead of proving it directly from
definition, we only have to find an ``observational bisimulation up
to'' which contains the pair of processes:
\begin{theorem}
If $\mathcal{S}$ is a ``observational bisimulation up to $\approx$'', then
$\mathcal{S} \subseteq \approx^c$:
\begin{alltt}
OBS_BISIM_UPTO_THM:
\HOLTokenTurnstile{} \HOLConst{OBS_BISIM_UPTO} \HOLFreeVar{Obsm} \HOLSymConst{\HOLTokenImp{}} \HOLFreeVar{Obsm} \HOLSymConst{\HOLTokenRSubset{}} \HOLConst{OBS_CONGR}
\end{alltt}
\end{theorem}

It was hard to prove above theorem, but it remains to find potential applications for this beautiful result.

\cleardoublepage

%%%% -*- Mode: LaTeX -*-

\chapter{Unique Solutions of Equations}

To prove two processes are (strong or weak) bisimilar, currently we
have explored two kind of bisimulation proof methods: one is based on
the definition: finding a bisimulation relation containing the given
two processes, which is usually hard, while the problem may become
easier using bisimilation up-to techniques; the other is to use those
algebraic laws to derive the resulting conclusion step by step,
although there's no decision procedure (algorithms) to construct such
proofs automatically.

In Milner's 1989 book \cite{Milner:1989}, he carefully explains that the bisimulation
proof method is not supposed to be the only method for reasoning about
bisimilarity. Indeed, various interesting examples in the book are
handled using other techniques, notably \emph{unique solution of
  equations}, whereby two tuples of processes are component-wise
bisimilar if they are solutions of the same system of equations. The
main conclusion is that, under a certain condition on the expression
$E$, there is an unique $P$ (up to $\sim$) such that
\begin{equation*}
P \sim E\{P/X\}
\end{equation*}
That solution is, naturally, the agent $A$ defined by $A
\overset{\mathrm{def}}{=} E\{A/X\}$.

Clearly this cannot be true for all $E$; in the case where $E$ is just
$X$, for example, \emph{every} agant $P$ satisfies the equation,
because the equation is just $P \sim P$. But we shall see that the
conclusion holds provided that $X$ is weakly guarded in $E$, in which
the concept of \emph{weak guardness} will be explained soon in the following section.

For simplicity purposes we only consider single-variable equation in this
project. There's no much additional work towards to multi-variable
cases from the view of informal proofs in related books (see also
\cite{Gorrieri:2015jt}, page 181--183).  But to formally represent CCS
equations with multiple variables, there may involves modified basic datatypes
and many small theorems for recursively replacing those variables in CCS
expressions. At the end of this thesis, the author has explored some
approaches towards multi-variable equations, however, even the proofs
of single-variable cases reflect the central ideas of this kind of
results, given that even many textbooks and papers only prove these
theorems for single-variable cases.

\section{Guarded Expressions}

If we consider only single-variable equations, it's straightforward to
treat expressions like $E(X)$ as a $\lambda$-function taking any
single CCS process returning another. In this way, no new datatypes
were introduced,
and actually that single variable $X$ never appears alone in any formal
proof, thus no need to represent it as a dedicated object.

In our previous work, to formalize the theory of
congruence for CCS, we have defined the concept of ``context'' based
on $\lambda$-calculus. There're actually two types of contexts:
one-hole and multi-hole (this includes no-hole case):
\begin{definition}{(one-hole context  of CCS)}
The (one-hole) semantic context of CCS is a function
$C[\cdot]$ of type ``\HOLinline{(\ensuremath{\alpha}, \ensuremath{\beta}) \HOLTyOp{context}}'' recursively defined by following rules:
\begin{alltt}
\HOLConst{OH_CONTEXT} (\HOLTokenLambda{}\HOLBoundVar{t}. \HOLBoundVar{t})
\HOLConst{OH_CONTEXT} \HOLFreeVar{c} \HOLSymConst{\HOLTokenImp{}} \HOLConst{OH_CONTEXT} (\HOLTokenLambda{}\HOLBoundVar{t}. \HOLFreeVar{a}\HOLSymConst{..}\HOLFreeVar{c} \HOLBoundVar{t})
\HOLConst{OH_CONTEXT} \HOLFreeVar{c} \HOLSymConst{\HOLTokenImp{}} \HOLConst{OH_CONTEXT} (\HOLTokenLambda{}\HOLBoundVar{t}. \HOLFreeVar{c} \HOLBoundVar{t} \HOLSymConst{+} \HOLFreeVar{x})
\HOLConst{OH_CONTEXT} \HOLFreeVar{c} \HOLSymConst{\HOLTokenImp{}} \HOLConst{OH_CONTEXT} (\HOLTokenLambda{}\HOLBoundVar{t}. \HOLFreeVar{x} \HOLSymConst{+} \HOLFreeVar{c} \HOLBoundVar{t})
\HOLConst{OH_CONTEXT} \HOLFreeVar{c} \HOLSymConst{\HOLTokenImp{}} \HOLConst{OH_CONTEXT} (\HOLTokenLambda{}\HOLBoundVar{t}. \HOLFreeVar{c} \HOLBoundVar{t} \HOLSymConst{\ensuremath{\parallel}} \HOLFreeVar{x})
\HOLConst{OH_CONTEXT} \HOLFreeVar{c} \HOLSymConst{\HOLTokenImp{}} \HOLConst{OH_CONTEXT} (\HOLTokenLambda{}\HOLBoundVar{t}. \HOLFreeVar{x} \HOLSymConst{\ensuremath{\parallel}} \HOLFreeVar{c} \HOLBoundVar{t})
\HOLConst{OH_CONTEXT} \HOLFreeVar{c} \HOLSymConst{\HOLTokenImp{}} \HOLConst{OH_CONTEXT} (\HOLTokenLambda{}\HOLBoundVar{t}. \HOLSymConst{\ensuremath{\nu}} \HOLFreeVar{L} (\HOLFreeVar{c} \HOLBoundVar{t}))
\HOLConst{OH_CONTEXT} \HOLFreeVar{c} \HOLSymConst{\HOLTokenImp{}} \HOLConst{OH_CONTEXT} (\HOLTokenLambda{}\HOLBoundVar{t}. \HOLConst{relab} (\HOLFreeVar{c} \HOLBoundVar{t}) \HOLFreeVar{rf})\hfill[OH_CONTEXT_rules]
\end{alltt}
\end{definition}

\begin{definition}{(multi-hole context  of CCS)}
The semantic context of CCS is a function
$C[\cdot]$ of type ``\HOLinline{(\ensuremath{\alpha}, \ensuremath{\beta}) \HOLTyOp{context}}'' recursively defined by following rules:
\begin{alltt}
\HOLConst{CONTEXT} (\HOLTokenLambda{}\HOLBoundVar{t}. \HOLBoundVar{t})
\HOLConst{CONTEXT} (\HOLTokenLambda{}\HOLBoundVar{t}. \HOLFreeVar{p})
\HOLConst{CONTEXT} \HOLFreeVar{e} \HOLSymConst{\HOLTokenImp{}} \HOLConst{CONTEXT} (\HOLTokenLambda{}\HOLBoundVar{t}. \HOLFreeVar{a}\HOLSymConst{..}\HOLFreeVar{e} \HOLBoundVar{t})
\HOLConst{CONTEXT} \HOLFreeVar{e\sb{\mathrm{1}}} \HOLSymConst{\HOLTokenConj{}} \HOLConst{CONTEXT} \HOLFreeVar{e\sb{\mathrm{2}}} \HOLSymConst{\HOLTokenImp{}} \HOLConst{CONTEXT} (\HOLTokenLambda{}\HOLBoundVar{t}. \HOLFreeVar{e\sb{\mathrm{1}}} \HOLBoundVar{t} \HOLSymConst{+} \HOLFreeVar{e\sb{\mathrm{2}}} \HOLBoundVar{t})
\HOLConst{CONTEXT} \HOLFreeVar{e\sb{\mathrm{1}}} \HOLSymConst{\HOLTokenConj{}} \HOLConst{CONTEXT} \HOLFreeVar{e\sb{\mathrm{2}}} \HOLSymConst{\HOLTokenImp{}} \HOLConst{CONTEXT} (\HOLTokenLambda{}\HOLBoundVar{t}. \HOLFreeVar{e\sb{\mathrm{1}}} \HOLBoundVar{t} \HOLSymConst{\ensuremath{\parallel}} \HOLFreeVar{e\sb{\mathrm{2}}} \HOLBoundVar{t})
\HOLConst{CONTEXT} \HOLFreeVar{e} \HOLSymConst{\HOLTokenImp{}} \HOLConst{CONTEXT} (\HOLTokenLambda{}\HOLBoundVar{t}. \HOLSymConst{\ensuremath{\nu}} \HOLFreeVar{L} (\HOLFreeVar{e} \HOLBoundVar{t}))
\HOLConst{CONTEXT} \HOLFreeVar{e} \HOLSymConst{\HOLTokenImp{}} \HOLConst{CONTEXT} (\HOLTokenLambda{}\HOLBoundVar{t}. \HOLConst{relab} (\HOLFreeVar{e} \HOLBoundVar{t}) \HOLFreeVar{rf})\hfill[CONTEXT_rules]
\end{alltt}
\end{definition}

By repeatedly applying these inductive rules, one can imagine that, the ``holes''
in any CCS expressions at any depth, can be filled by the same
process, when the $\lambda$ function is called with that process.
For CCS equations containing only one variable, we can simply treat a
context as the core part of an equation. Thus, if $C$ is a context,
$P$ is the solution of the equation (in case of strong equivalence),
this actually means \HOLinline{\HOLFreeVar{P} \HOLSymConst{\HOLTokenStrongEQ} \HOLFreeVar{C} \HOLFreeVar{P}}.  Thus the variable itself
never need to be formalized, and it never appears in any related proof.

Notice the difference with one-hole context in the branches of sum
and parallel operators. And also the possibility that, the expression
may finally contains no variable at all (this is necessary, otherwise
we can't finish the proof).

One major drawback of above techniques is, we cannot further define
the weakly guardedness (or sequential property) as a predicate of CCS expressions, simply because
there's no way to recursively check the internal structure of
$\lambda$-functions, as such functions were basically black-boxes once
defined.  A workaround solution is to define it independently and recursively:
\begin{definition}{(Weakly guarded expressions)}
$X$ is \emph{weakly guarded} in $E$ if each occurrence of $X$ is
within some subexpression $\alpha.F$ of $E$:
\begin{alltt}
\HOLConst{WG} (\HOLTokenLambda{}\HOLBoundVar{t}. \HOLFreeVar{p})
\HOLConst{CONTEXT} \HOLFreeVar{e} \HOLSymConst{\HOLTokenImp{}} \HOLConst{WG} (\HOLTokenLambda{}\HOLBoundVar{t}. \HOLFreeVar{a}\HOLSymConst{..}\HOLFreeVar{e} \HOLBoundVar{t})
\HOLConst{WG} \HOLFreeVar{e\sb{\mathrm{1}}} \HOLSymConst{\HOLTokenConj{}} \HOLConst{WG} \HOLFreeVar{e\sb{\mathrm{2}}} \HOLSymConst{\HOLTokenImp{}} \HOLConst{WG} (\HOLTokenLambda{}\HOLBoundVar{t}. \HOLFreeVar{e\sb{\mathrm{1}}} \HOLBoundVar{t} \HOLSymConst{+} \HOLFreeVar{e\sb{\mathrm{2}}} \HOLBoundVar{t})
\HOLConst{WG} \HOLFreeVar{e\sb{\mathrm{1}}} \HOLSymConst{\HOLTokenConj{}} \HOLConst{WG} \HOLFreeVar{e\sb{\mathrm{2}}} \HOLSymConst{\HOLTokenImp{}} \HOLConst{WG} (\HOLTokenLambda{}\HOLBoundVar{t}. \HOLFreeVar{e\sb{\mathrm{1}}} \HOLBoundVar{t} \HOLSymConst{\ensuremath{\parallel}} \HOLFreeVar{e\sb{\mathrm{2}}} \HOLBoundVar{t})
\HOLConst{WG} \HOLFreeVar{e} \HOLSymConst{\HOLTokenImp{}} \HOLConst{WG} (\HOLTokenLambda{}\HOLBoundVar{t}. \HOLSymConst{\ensuremath{\nu}} \HOLFreeVar{L} (\HOLFreeVar{e} \HOLBoundVar{t}))
\HOLConst{WG} \HOLFreeVar{e} \HOLSymConst{\HOLTokenImp{}} \HOLConst{WG} (\HOLTokenLambda{}\HOLBoundVar{t}. \HOLConst{relab} (\HOLFreeVar{e} \HOLBoundVar{t}) \HOLFreeVar{rf})\hfill[WG_rules]
\end{alltt}
\end{definition}

Notice the only difference between weakly guarded expressions and
normal expressions is at their first branch. In this way, a weakly
guarded expression won't expose the variable without a prefixed
action (could be $\tau$).

To make a connection between above two kind of expressions (and the
one-hole context), we have
proved their relationships by induction on their
structures:
\begin{proposition}{(Relationship between one-hole/multi-hold contexts
    and weakly guarded expressions)}
\begin{enumerate}
\item One-hole context is also context:
\begin{alltt}
OH_CONTEXT_IS_CONTEXT:
\HOLTokenTurnstile{} \HOLConst{OH_CONTEXT} \HOLFreeVar{c} \HOLSymConst{\HOLTokenImp{}} \HOLConst{CONTEXT} \HOLFreeVar{c}
\end{alltt}
\item Weakly guarded expressions is also context:
\begin{alltt}
WG_IS_CONTEXT:
\HOLTokenTurnstile{} \HOLConst{WG} \HOLFreeVar{e} \HOLSymConst{\HOLTokenImp{}} \HOLConst{CONTEXT} \HOLFreeVar{e}
\end{alltt}
\end{enumerate}
\end{proposition}
Noticed that, the first result (and one-hole contexts) is never needed in the rest of this
thesis, while the second one will be heavily used.

One limitation in our definitions is the lacking of CCS constants
(i.e. \texttt{var} and
\texttt{rec} operators defined as part of our CCS datatypes) in all
above recursive definitions. This doesn't means the expressions cannot
contains constants, just these constants must be irrelevant with the
variable, that is, \emph{variable substitutions never happen inside
  the body of any CCS constant!}. This restriction can be removed when
we can prove the congruence of equivalences under \HOLinline{\HOLConst{rec}}
operators, however this needs to discuss the ``free variables'' in CCS
process, currently it's beyond the scope of this thesis.

\section{Milner's three ``unique solution of equations'' theorems}

Here we formalized three versions of the ``unique solution of equations'' theorem in Milner's book.

\subsection{For strong equivalence}

Based on results on bisimulation up to $\sim$, we have first proved the following
non-trivial lemma. It states in effect that if $X$ is weakly guarded
in $E$, then the ``first move'' of $E$ is independent of the agent
substituted for $X$:
\begin{lemma}{(Lemma 3.13 of \cite{Milner:1989})}
If the variable $X$ are weakly guarded in $E$, and
$E\{P/X\}\overset{\alpha}{\rightarrow} P'$, then $P'$ takes the form
$E'\{P/X\}$ (for some expression $E'$), and moreover, for any $Q$,
$E\{Q/X\}\overset{\alpha}{\rightarrow} E'\{Q/X\}$:
\begin{alltt}
STRONG_UNIQUE_SOLUTIONS_LEMMA:
\HOLTokenTurnstile{} \HOLConst{WG} \HOLFreeVar{E} \HOLSymConst{\HOLTokenImp{}}
   \HOLSymConst{\HOLTokenForall{}}\HOLBoundVar{P} \HOLBoundVar{a} \HOLBoundVar{P\sp{\prime}}.
     \HOLFreeVar{E} \HOLBoundVar{P} \HOLTokenTransBegin\HOLBoundVar{a}\HOLTokenTransEnd \HOLBoundVar{P\sp{\prime}} \HOLSymConst{\HOLTokenImp{}}
     \HOLSymConst{\HOLTokenExists{}}\HOLBoundVar{E\sp{\prime}}. \HOLConst{CONTEXT} \HOLBoundVar{E\sp{\prime}} \HOLSymConst{\HOLTokenConj{}} (\HOLBoundVar{P\sp{\prime}} \HOLSymConst{=} \HOLBoundVar{E\sp{\prime}} \HOLBoundVar{P}) \HOLSymConst{\HOLTokenConj{}} \HOLSymConst{\HOLTokenForall{}}\HOLBoundVar{Q}. \HOLFreeVar{E} \HOLBoundVar{Q} \HOLTokenTransBegin\HOLBoundVar{a}\HOLTokenTransEnd \HOLBoundVar{E\sp{\prime}} \HOLBoundVar{Q}
\end{alltt}
\end{lemma}

We're now ready to prove the following, the main proposition above the
``unique solution of equations'':
\begin{theorem}{(Proposition 3.14 of \cite{Milner:1989})}
Let the expression $E$ contains at most the variable $X$, and let $X$ be weakly
guarded in $E$, then
\begin{equation}
\text{If } P \sim E\{P/X\} \text{ and } Q \sim E\{Q/X\} \text{ then }
P \sim Q.
\end{equation}
\begin{alltt}
STRONG_UNIQUE_SOLUTIONS:
\HOLTokenTurnstile{} \HOLConst{WG} \HOLFreeVar{E} \HOLSymConst{\HOLTokenImp{}} \HOLSymConst{\HOLTokenForall{}}\HOLBoundVar{P} \HOLBoundVar{Q}. \HOLBoundVar{P} \HOLSymConst{\HOLTokenStrongEQ} \HOLFreeVar{E} \HOLBoundVar{P} \HOLSymConst{\HOLTokenConj{}} \HOLBoundVar{Q} \HOLSymConst{\HOLTokenStrongEQ} \HOLFreeVar{E} \HOLBoundVar{Q} \HOLSymConst{\HOLTokenImp{}} \HOLBoundVar{P} \HOLSymConst{\HOLTokenStrongEQ} \HOLBoundVar{Q}
\end{alltt}
\end{theorem}
In above proof, we have identified 14 major sub-goals, dividing into 7
groups, in which each pairs are symmetric (thus having similar proof
steps). The proof of this last theorem consists of 500 lines (each
line usually have 2 or 3 HOL tactics, to make the proof not too long
in lines).

\subsection{For weak equivalence}

Actually Milner's book contains only two ``unique solutions of
equations'' theorems, one for strong equivalence, the other for
observational congruence (with a wrong proof).   But there's indeed a
version for weak equivalence which shares a large portion of proof
steps with the case for ``observation congruence''.  The problem is, since weak
equivalence is not congruence, to make correct statement of this
theorem, we must slightly modify the concept of sequential expresses
with further restrictions: no direct sums (or only guarded sums):
\begin{definition}{(Sequential expressions restricted with guarded sum)}
\begin{alltt}
\HOLConst{GSEQ} (\HOLTokenLambda{}\HOLBoundVar{t}. \HOLBoundVar{t})
\HOLConst{GSEQ} (\HOLTokenLambda{}\HOLBoundVar{t}. \HOLFreeVar{p})
\HOLConst{GSEQ} \HOLFreeVar{e} \HOLSymConst{\HOLTokenImp{}} \HOLConst{GSEQ} (\HOLTokenLambda{}\HOLBoundVar{t}. \HOLFreeVar{a}\HOLSymConst{..}\HOLFreeVar{e} \HOLBoundVar{t})
\HOLConst{GSEQ} \HOLFreeVar{e\sb{\mathrm{1}}} \HOLSymConst{\HOLTokenConj{}} \HOLConst{GSEQ} \HOLFreeVar{e\sb{\mathrm{2}}} \HOLSymConst{\HOLTokenImp{}} \HOLConst{GSEQ} (\HOLTokenLambda{}\HOLBoundVar{t}. \HOLFreeVar{a\sb{\mathrm{1}}}\HOLSymConst{..}\HOLFreeVar{e\sb{\mathrm{1}}} \HOLBoundVar{t} \HOLSymConst{+} \HOLFreeVar{a\sb{\mathrm{2}}}\HOLSymConst{..}\HOLFreeVar{e\sb{\mathrm{2}}} \HOLBoundVar{t})\hfill[GSEQ_rules]
\end{alltt}
\end{definition}

The ``unique solution of equations'' theorem for weak equivalence
requires more restrictions on the expression, namely, strong guardness:
\begin{definition}{((Strongly) guarded expressions)}
$X$ is (strongly) guarded in $E$ if each occurrence of $X$ is within some subexpression of $E$ of
   the form $l.F$ ($l$ is a visible action):
\begin{alltt}
\HOLConst{SG} (\HOLTokenLambda{}\HOLBoundVar{t}. \HOLFreeVar{p})
\HOLConst{CONTEXT} \HOLFreeVar{e} \HOLSymConst{\HOLTokenImp{}} \HOLConst{SG} (\HOLTokenLambda{}\HOLBoundVar{t}. \HOLConst{label} \HOLFreeVar{l}\HOLSymConst{..}\HOLFreeVar{e} \HOLBoundVar{t})
\HOLConst{SG} \HOLFreeVar{e} \HOLSymConst{\HOLTokenImp{}} \HOLConst{SG} (\HOLTokenLambda{}\HOLBoundVar{t}. \HOLFreeVar{a}\HOLSymConst{..}\HOLFreeVar{e} \HOLBoundVar{t})
\HOLConst{SG} \HOLFreeVar{e\sb{\mathrm{1}}} \HOLSymConst{\HOLTokenConj{}} \HOLConst{SG} \HOLFreeVar{e\sb{\mathrm{2}}} \HOLSymConst{\HOLTokenImp{}} \HOLConst{SG} (\HOLTokenLambda{}\HOLBoundVar{t}. \HOLFreeVar{e\sb{\mathrm{1}}} \HOLBoundVar{t} \HOLSymConst{+} \HOLFreeVar{e\sb{\mathrm{2}}} \HOLBoundVar{t})
\HOLConst{SG} \HOLFreeVar{e\sb{\mathrm{1}}} \HOLSymConst{\HOLTokenConj{}} \HOLConst{SG} \HOLFreeVar{e\sb{\mathrm{2}}} \HOLSymConst{\HOLTokenImp{}} \HOLConst{SG} (\HOLTokenLambda{}\HOLBoundVar{t}. \HOLFreeVar{e\sb{\mathrm{1}}} \HOLBoundVar{t} \HOLSymConst{\ensuremath{\parallel}} \HOLFreeVar{e\sb{\mathrm{2}}} \HOLBoundVar{t})
\HOLConst{SG} \HOLFreeVar{e} \HOLSymConst{\HOLTokenImp{}} \HOLConst{SG} (\HOLTokenLambda{}\HOLBoundVar{t}. \HOLSymConst{\ensuremath{\nu}} \HOLFreeVar{L} (\HOLFreeVar{e} \HOLBoundVar{t}))
\HOLConst{SG} \HOLFreeVar{e} \HOLSymConst{\HOLTokenImp{}} \HOLConst{SG} (\HOLTokenLambda{}\HOLBoundVar{t}. \HOLConst{relab} (\HOLFreeVar{e} \HOLBoundVar{t}) \HOLFreeVar{rf})\hfill[SG_rules]
\end{alltt}
\end{definition}

Now we're ready to state and prove the following lemma:
\begin{lemma}
If the variable $X$ are (strongly) guarded and sequential in $G$, and
$G\{P/X\}\overset{\alpha}{\rightarrow} P'$, then $P'$ takes the form
$H\{P/X\}$ (for some expression $H$), and for any $Q$,
$G\{Q/X\}\overset{\alpha}{\rightarrow} H\{Q/X\}$. Moreover $H$ is
sequential, and if $\alpha = \tau$ then $H$ is also guarded.
\begin{alltt}
WEAK_UNIQUE_SOLUTIONS_LEMMA:
\HOLTokenTurnstile{} \HOLConst{SG} \HOLFreeVar{G} \HOLSymConst{\HOLTokenConj{}} \HOLConst{GSEQ} \HOLFreeVar{G} \HOLSymConst{\HOLTokenImp{}}
   \HOLSymConst{\HOLTokenForall{}}\HOLBoundVar{P} \HOLBoundVar{a} \HOLBoundVar{P\sp{\prime}}.
     \HOLFreeVar{G} \HOLBoundVar{P} \HOLTokenTransBegin\HOLBoundVar{a}\HOLTokenTransEnd \HOLBoundVar{P\sp{\prime}} \HOLSymConst{\HOLTokenImp{}}
     \HOLSymConst{\HOLTokenExists{}}\HOLBoundVar{H}.
       \HOLConst{GSEQ} \HOLBoundVar{H} \HOLSymConst{\HOLTokenConj{}} ((\HOLBoundVar{a} \HOLSymConst{=} \HOLSymConst{\ensuremath{\tau}}) \HOLSymConst{\HOLTokenImp{}} \HOLConst{SG} \HOLBoundVar{H}) \HOLSymConst{\HOLTokenConj{}} (\HOLBoundVar{P\sp{\prime}} \HOLSymConst{=} \HOLBoundVar{H} \HOLBoundVar{P}) \HOLSymConst{\HOLTokenConj{}}
       \HOLSymConst{\HOLTokenForall{}}\HOLBoundVar{Q}. \HOLFreeVar{G} \HOLBoundVar{Q} \HOLTokenTransBegin\HOLBoundVar{a}\HOLTokenTransEnd \HOLBoundVar{H} \HOLBoundVar{Q}
\end{alltt}
\end{lemma}

An important technique in the proof of above lemma is the ability to
do inductions on the structure of $G$ which is both guarded and
sequential. But if we apply the induction theorem generated by
\texttt{SG} and \texttt{GSEQ} separately, the total numer of proof
sub-goals is \emph{huge}. Instead, we have proved the following
combined induction theorems for \texttt{SG+GSEQ} expressions:
\begin{proposition}{(Combined induction principle for guarded and
    sequential expressions)}
\begin{alltt}
SG_GSEQ_strong_induction:
\HOLTokenTurnstile{} (\HOLSymConst{\HOLTokenForall{}}\HOLBoundVar{p}. \HOLFreeVar{R} (\HOLTokenLambda{}\HOLBoundVar{t}. \HOLBoundVar{p})) \HOLSymConst{\HOLTokenConj{}} (\HOLSymConst{\HOLTokenForall{}}\HOLBoundVar{l} \HOLBoundVar{e}. \HOLConst{GSEQ} \HOLBoundVar{e} \HOLSymConst{\HOLTokenImp{}} \HOLFreeVar{R} (\HOLTokenLambda{}\HOLBoundVar{t}. \HOLConst{label} \HOLBoundVar{l}\HOLSymConst{..}\HOLBoundVar{e} \HOLBoundVar{t})) \HOLSymConst{\HOLTokenConj{}}
   (\HOLSymConst{\HOLTokenForall{}}\HOLBoundVar{a} \HOLBoundVar{e}. \HOLConst{SG} \HOLBoundVar{e} \HOLSymConst{\HOLTokenConj{}} \HOLConst{GSEQ} \HOLBoundVar{e} \HOLSymConst{\HOLTokenConj{}} \HOLFreeVar{R} \HOLBoundVar{e} \HOLSymConst{\HOLTokenImp{}} \HOLFreeVar{R} (\HOLTokenLambda{}\HOLBoundVar{t}. \HOLBoundVar{a}\HOLSymConst{..}\HOLBoundVar{e} \HOLBoundVar{t})) \HOLSymConst{\HOLTokenConj{}}
   (\HOLSymConst{\HOLTokenForall{}}\HOLBoundVar{e\sb{\mathrm{1}}} \HOLBoundVar{e\sb{\mathrm{2}}}.
      \HOLConst{SG} \HOLBoundVar{e\sb{\mathrm{1}}} \HOLSymConst{\HOLTokenConj{}} \HOLConst{GSEQ} \HOLBoundVar{e\sb{\mathrm{1}}} \HOLSymConst{\HOLTokenConj{}} \HOLFreeVar{R} \HOLBoundVar{e\sb{\mathrm{1}}} \HOLSymConst{\HOLTokenConj{}} \HOLConst{SG} \HOLBoundVar{e\sb{\mathrm{2}}} \HOLSymConst{\HOLTokenConj{}} \HOLConst{GSEQ} \HOLBoundVar{e\sb{\mathrm{2}}} \HOLSymConst{\HOLTokenConj{}} \HOLFreeVar{R} \HOLBoundVar{e\sb{\mathrm{2}}} \HOLSymConst{\HOLTokenImp{}}
      \HOLFreeVar{R} (\HOLTokenLambda{}\HOLBoundVar{t}. \HOLSymConst{\ensuremath{\tau}}\HOLSymConst{..}\HOLBoundVar{e\sb{\mathrm{1}}} \HOLBoundVar{t} \HOLSymConst{+} \HOLSymConst{\ensuremath{\tau}}\HOLSymConst{..}\HOLBoundVar{e\sb{\mathrm{2}}} \HOLBoundVar{t})) \HOLSymConst{\HOLTokenConj{}}
   (\HOLSymConst{\HOLTokenForall{}}\HOLBoundVar{l\sb{\mathrm{2}}} \HOLBoundVar{e\sb{\mathrm{1}}} \HOLBoundVar{e\sb{\mathrm{2}}}.
      \HOLConst{SG} \HOLBoundVar{e\sb{\mathrm{1}}} \HOLSymConst{\HOLTokenConj{}} \HOLConst{GSEQ} \HOLBoundVar{e\sb{\mathrm{1}}} \HOLSymConst{\HOLTokenConj{}} \HOLFreeVar{R} \HOLBoundVar{e\sb{\mathrm{1}}} \HOLSymConst{\HOLTokenConj{}} \HOLConst{GSEQ} \HOLBoundVar{e\sb{\mathrm{2}}} \HOLSymConst{\HOLTokenImp{}}
      \HOLFreeVar{R} (\HOLTokenLambda{}\HOLBoundVar{t}. \HOLSymConst{\ensuremath{\tau}}\HOLSymConst{..}\HOLBoundVar{e\sb{\mathrm{1}}} \HOLBoundVar{t} \HOLSymConst{+} \HOLConst{label} \HOLBoundVar{l\sb{\mathrm{2}}}\HOLSymConst{..}\HOLBoundVar{e\sb{\mathrm{2}}} \HOLBoundVar{t})) \HOLSymConst{\HOLTokenConj{}}
   (\HOLSymConst{\HOLTokenForall{}}\HOLBoundVar{l\sb{\mathrm{1}}} \HOLBoundVar{e\sb{\mathrm{1}}} \HOLBoundVar{e\sb{\mathrm{2}}}.
      \HOLConst{GSEQ} \HOLBoundVar{e\sb{\mathrm{1}}} \HOLSymConst{\HOLTokenConj{}} \HOLConst{SG} \HOLBoundVar{e\sb{\mathrm{2}}} \HOLSymConst{\HOLTokenConj{}} \HOLConst{GSEQ} \HOLBoundVar{e\sb{\mathrm{2}}} \HOLSymConst{\HOLTokenConj{}} \HOLFreeVar{R} \HOLBoundVar{e\sb{\mathrm{2}}} \HOLSymConst{\HOLTokenImp{}}
      \HOLFreeVar{R} (\HOLTokenLambda{}\HOLBoundVar{t}. \HOLConst{label} \HOLBoundVar{l\sb{\mathrm{1}}}\HOLSymConst{..}\HOLBoundVar{e\sb{\mathrm{1}}} \HOLBoundVar{t} \HOLSymConst{+} \HOLSymConst{\ensuremath{\tau}}\HOLSymConst{..}\HOLBoundVar{e\sb{\mathrm{2}}} \HOLBoundVar{t})) \HOLSymConst{\HOLTokenConj{}}
   (\HOLSymConst{\HOLTokenForall{}}\HOLBoundVar{l\sb{\mathrm{1}}} \HOLBoundVar{l\sb{\mathrm{2}}} \HOLBoundVar{e\sb{\mathrm{1}}} \HOLBoundVar{e\sb{\mathrm{2}}}.
      \HOLConst{GSEQ} \HOLBoundVar{e\sb{\mathrm{1}}} \HOLSymConst{\HOLTokenConj{}} \HOLConst{GSEQ} \HOLBoundVar{e\sb{\mathrm{2}}} \HOLSymConst{\HOLTokenImp{}}
      \HOLFreeVar{R} (\HOLTokenLambda{}\HOLBoundVar{t}. \HOLConst{label} \HOLBoundVar{l\sb{\mathrm{1}}}\HOLSymConst{..}\HOLBoundVar{e\sb{\mathrm{1}}} \HOLBoundVar{t} \HOLSymConst{+} \HOLConst{label} \HOLBoundVar{l\sb{\mathrm{2}}}\HOLSymConst{..}\HOLBoundVar{e\sb{\mathrm{2}}} \HOLBoundVar{t})) \HOLSymConst{\HOLTokenImp{}}
   \HOLSymConst{\HOLTokenForall{}}\HOLBoundVar{e}. \HOLConst{SG} \HOLBoundVar{e} \HOLSymConst{\HOLTokenConj{}} \HOLConst{GSEQ} \HOLBoundVar{e} \HOLSymConst{\HOLTokenImp{}} \HOLFreeVar{R} \HOLBoundVar{e}
\end{alltt}
\end{proposition}

Once above lemma is proved, it's just one small step toward the target theorem:
\begin{theorem}{(Unique solution of equations for weak equivalence)}
Let $E$ be guarded and sequential expressions, and let $P \approx
E\{P/X\}$,
$Q \approx E\{Q/X\}$. Then $P \approx Q$.
\begin{alltt}
WEAK_UNIQUE_SOLUTIONS:
\HOLTokenTurnstile{} \HOLConst{SG} \HOLFreeVar{E} \HOLSymConst{\HOLTokenConj{}} \HOLConst{GSEQ} \HOLFreeVar{E} \HOLSymConst{\HOLTokenImp{}} \HOLSymConst{\HOLTokenForall{}}\HOLBoundVar{P} \HOLBoundVar{Q}. \HOLBoundVar{P} \HOLSymConst{\HOLTokenWeakEQ} \HOLFreeVar{E} \HOLBoundVar{P} \HOLSymConst{\HOLTokenConj{}} \HOLBoundVar{Q} \HOLSymConst{\HOLTokenWeakEQ} \HOLFreeVar{E} \HOLBoundVar{Q} \HOLSymConst{\HOLTokenImp{}} \HOLBoundVar{P} \HOLSymConst{\HOLTokenWeakEQ} \HOLBoundVar{Q}
\end{alltt}
\end{theorem}

\subsection{For observational congruence}

For ``unique solutions of
equations'' theorem of observational congruence, we have used a whole new technique, i.e. the
following lemma:
\begin{lemma}
To prove two processes $E$ and $E'$ are observational congruence, it's
enough to construct a bisimulation with additional observational
transition properties:
\begin{alltt}
OBS_CONGR_BY_WEAK_BISIM:
\HOLTokenTurnstile{} \HOLConst{WEAK_BISIM} \HOLFreeVar{Wbsm} \HOLSymConst{\HOLTokenImp{}}
   \HOLSymConst{\HOLTokenForall{}}\HOLBoundVar{E} \HOLBoundVar{E\sp{\prime}}.
     (\HOLSymConst{\HOLTokenForall{}}\HOLBoundVar{u}.
        (\HOLSymConst{\HOLTokenForall{}}\HOLBoundVar{E\sb{\mathrm{1}}}. \HOLBoundVar{E} \HOLTokenTransBegin\HOLBoundVar{u}\HOLTokenTransEnd \HOLBoundVar{E\sb{\mathrm{1}}} \HOLSymConst{\HOLTokenImp{}} \HOLSymConst{\HOLTokenExists{}}\HOLBoundVar{E\sb{\mathrm{2}}}. \HOLBoundVar{E\sp{\prime}} \HOLTokenWeakTransBegin\HOLBoundVar{u}\HOLTokenWeakTransEnd \HOLBoundVar{E\sb{\mathrm{2}}} \HOLSymConst{\HOLTokenConj{}} \HOLFreeVar{Wbsm} \HOLBoundVar{E\sb{\mathrm{1}}} \HOLBoundVar{E\sb{\mathrm{2}}}) \HOLSymConst{\HOLTokenConj{}}
        \HOLSymConst{\HOLTokenForall{}}\HOLBoundVar{E\sb{\mathrm{2}}}. \HOLBoundVar{E\sp{\prime}} \HOLTokenTransBegin\HOLBoundVar{u}\HOLTokenTransEnd \HOLBoundVar{E\sb{\mathrm{2}}} \HOLSymConst{\HOLTokenImp{}} \HOLSymConst{\HOLTokenExists{}}\HOLBoundVar{E\sb{\mathrm{1}}}. \HOLBoundVar{E} \HOLTokenWeakTransBegin\HOLBoundVar{u}\HOLTokenWeakTransEnd \HOLBoundVar{E\sb{\mathrm{1}}} \HOLSymConst{\HOLTokenConj{}} \HOLFreeVar{Wbsm} \HOLBoundVar{E\sb{\mathrm{1}}} \HOLBoundVar{E\sb{\mathrm{2}}}) \HOLSymConst{\HOLTokenImp{}}
     \HOLBoundVar{E} \HOLSymConst{\HOLTokenObsCongr} \HOLBoundVar{E\sp{\prime}}
\end{alltt}
\end{lemma}

This lemma can be easily proved by definition of observational
congruence and weak equivalence, however it's never mentioned in
Milner's book. Actually we think it's impossible to prove the
corresponding ``unique solutions'' theorem without this result.

What we have proved is the following one. Notice that, we didn't use
any ``bisimulation up-to'' techniques, because they're not applicable
once above theorem \texttt{OBS_CONGR_BY_WEAK_BISIM} is used. Instead
we have directly constructed a bisimulation to finish the proof.

\begin{lemma}
If the variable $X$ are (strongly) guarded and sequential in $G$, and
$G\{P/X\}\overset{\alpha}{\rightarrow} P'$, then $P'$ takes the form
$H\{P/X\}$ (for some expression $H$), and for any $Q$,
$G\{Q/X\}\overset{\alpha}{\rightarrow} H\{Q/X\}$. Moreover $H$ is
sequential, and if $\alpha = \tau$ then $H$ is also guarded.
\begin{alltt}
OBS_UNIQUE_SOLUTIONS_LEMMA:
\HOLTokenTurnstile{} \HOLConst{SG} \HOLFreeVar{G} \HOLSymConst{\HOLTokenConj{}} \HOLConst{SEQ} \HOLFreeVar{G} \HOLSymConst{\HOLTokenImp{}}
   \HOLSymConst{\HOLTokenForall{}}\HOLBoundVar{P} \HOLBoundVar{a} \HOLBoundVar{P\sp{\prime}}.
     \HOLFreeVar{G} \HOLBoundVar{P} \HOLTokenTransBegin\HOLBoundVar{a}\HOLTokenTransEnd \HOLBoundVar{P\sp{\prime}} \HOLSymConst{\HOLTokenImp{}}
     \HOLSymConst{\HOLTokenExists{}}\HOLBoundVar{H}.
       \HOLConst{SEQ} \HOLBoundVar{H} \HOLSymConst{\HOLTokenConj{}} ((\HOLBoundVar{a} \HOLSymConst{=} \HOLSymConst{\ensuremath{\tau}}) \HOLSymConst{\HOLTokenImp{}} \HOLConst{SG} \HOLBoundVar{H}) \HOLSymConst{\HOLTokenConj{}} (\HOLBoundVar{P\sp{\prime}} \HOLSymConst{=} \HOLBoundVar{H} \HOLBoundVar{P}) \HOLSymConst{\HOLTokenConj{}} \HOLSymConst{\HOLTokenForall{}}\HOLBoundVar{Q}. \HOLFreeVar{G} \HOLBoundVar{Q} \HOLTokenTransBegin\HOLBoundVar{a}\HOLTokenTransEnd \HOLBoundVar{H} \HOLBoundVar{Q}
\end{alltt}
\end{lemma}

\begin{theorem}{(Unique solution of equations for observational congruence)}
Let $E$ be guarded and sequential expressions, and let $P \approx^c
E\{P/X\}$,
$Q \approx^c E\{Q/X\}$. Then $P \approx^c Q$.
\begin{alltt}
OBS_UNIQUE_SOLUTIONS:
\HOLTokenTurnstile{} \HOLConst{SG} \HOLFreeVar{E} \HOLSymConst{\HOLTokenConj{}} \HOLConst{SEQ} \HOLFreeVar{E} \HOLSymConst{\HOLTokenImp{}} \HOLSymConst{\HOLTokenForall{}}\HOLBoundVar{P} \HOLBoundVar{Q}. \HOLBoundVar{P} \HOLSymConst{\HOLTokenObsCongr} \HOLFreeVar{E} \HOLBoundVar{P} \HOLSymConst{\HOLTokenConj{}} \HOLBoundVar{Q} \HOLSymConst{\HOLTokenObsCongr} \HOLFreeVar{E} \HOLBoundVar{Q} \HOLSymConst{\HOLTokenImp{}} \HOLBoundVar{P} \HOLSymConst{\HOLTokenObsCongr} \HOLBoundVar{Q}
\end{alltt}
\end{theorem}

\chapter{Equations, Contractions and Unique Solutions}

Here we have used the title of Prof.\ Sangiorgi's paper \cite{sangiorgi2015equations} as the
title of this chapter, as the purpose of all the work mentioned in
this chapter is to prove the key ``unique solutions of contractions''
theorem in that paper. Although the theorem looks similar with
Milner's classical results, the underlying proof idea and techniques
are completely different. And we have to formalize new fundamental
transition concepts (Trace) in order to finish the proof.

In this chapter, we start with the formalization of another relation called
``expansion'', which can be seen as the origin of the ``contraction''
relation. We're going to prove that, the expansion relation between
two processes (called `expands') is
a pre-congruence and pre-order.

Then we introduce the ``contraction'' relation and prove that,
contractions between two processes (called `contracts') have the same
properties as expansion, with something more reasonable.

During the proof attempts of the ``unique solution of
contractions/expansions'' theorem, we found that, sometimes we need to
track the precise number of steps inside certain weak transitions, and
such number will be passed into next proof steps during the
challenging of contractions, expansions and weak equivalences. As a
result, we have to further formalize the (strong) trace transitions
between two CCS processes, and this involves HOL's \texttt{listTheory}
into our formalization project. Many intermediate results using lists
were proved during this work, while the final statement of neither the
unique solutions theorems nor their lemmas need to explicitly mention
lists (or traces). This is very interesting and unexpected, because,
after having formalized so many results, plus Milner's original three
``unique solutions of equations'' theorems, we still need to invent
new devices to finish the proof of Sangiorgi's results. Thus this last
piece (and central piece) of thesis project has really made something
new and goes beyond all previous achievements.

\section{Expansion}

Expansion is a non-symmetric relation for CCS processes. ``$P$ expands
$Q$'' means ``$P$ is at least as fast as $Q$'', or more generally
``$Q$ uses at least as much resources as $P$''. Expansion is studied
by Arun-Kumar and Hennessy \cite{arun1992efficiency} under a different
terminology: they show that expansion is a mathematically tractable
preorder and has a complete proof system for finite terms based on a
modification of the standard $\tau$ laws for CCS. In CCS, strong and
weak bisimilarity are congruence relations (for weak bisimilarity,
guarded sums are required), and expansion is a precongruence.

Its definition and properties are quite like weak equivalence.  To define
expansion, we need to follow a two-step processes, first define a
predicate of relations, called ``expansion'', then define a 2-ary
relation called `contracts' as the maximal relation containing all
expansions.

\begin{definition}{(expansion)}
A processes relation $\mathcal{R}$ is an \emph{expansion} if, whenever
$P\, \mathcal{R}\, Q$,
\begin{enumerate}
\item $P \overset{\mu}{\rightarrow} P'$ implies that there is $Q'$
  with $Q \overset{\hat{\mu}}{\rightarrow} Q'$ and $P'\, \mathcal{R}
  \, Q'$;
\item $Q \overset{\mu}{\rightarrow} Q'$ implies that there is $P'$
  with $P \overset{\mu}{\Rightarrow} P'$ and $P'\,\mathcal{R}\,Q'$.
\end{enumerate}
or formally:
\begin{alltt}
\hfill[EXPANSION]
\HOLTokenTurnstile{} \HOLConst{EXPANSION} \HOLFreeVar{Exp} \HOLSymConst{\HOLTokenEquiv{}}
   \HOLSymConst{\HOLTokenForall{}}\HOLBoundVar{E} \HOLBoundVar{E\sp{\prime}}.
     \HOLFreeVar{Exp} \HOLBoundVar{E} \HOLBoundVar{E\sp{\prime}} \HOLSymConst{\HOLTokenImp{}}
     (\HOLSymConst{\HOLTokenForall{}}\HOLBoundVar{l}.
        (\HOLSymConst{\HOLTokenForall{}}\HOLBoundVar{E\sb{\mathrm{1}}}.
           \HOLBoundVar{E} \HOLTokenTransBegin\HOLConst{label} \HOLBoundVar{l}\HOLTokenTransEnd \HOLBoundVar{E\sb{\mathrm{1}}} \HOLSymConst{\HOLTokenImp{}} \HOLSymConst{\HOLTokenExists{}}\HOLBoundVar{E\sb{\mathrm{2}}}. \HOLBoundVar{E\sp{\prime}} \HOLTokenTransBegin\HOLConst{label} \HOLBoundVar{l}\HOLTokenTransEnd \HOLBoundVar{E\sb{\mathrm{2}}} \HOLSymConst{\HOLTokenConj{}} \HOLFreeVar{Exp} \HOLBoundVar{E\sb{\mathrm{1}}} \HOLBoundVar{E\sb{\mathrm{2}}}) \HOLSymConst{\HOLTokenConj{}}
        \HOLSymConst{\HOLTokenForall{}}\HOLBoundVar{E\sb{\mathrm{2}}}.
          \HOLBoundVar{E\sp{\prime}} \HOLTokenTransBegin\HOLConst{label} \HOLBoundVar{l}\HOLTokenTransEnd \HOLBoundVar{E\sb{\mathrm{2}}} \HOLSymConst{\HOLTokenImp{}} \HOLSymConst{\HOLTokenExists{}}\HOLBoundVar{E\sb{\mathrm{1}}}. \HOLBoundVar{E} \HOLTokenWeakTransBegin\HOLConst{label} \HOLBoundVar{l}\HOLTokenWeakTransEnd \HOLBoundVar{E\sb{\mathrm{1}}} \HOLSymConst{\HOLTokenConj{}} \HOLFreeVar{Exp} \HOLBoundVar{E\sb{\mathrm{1}}} \HOLBoundVar{E\sb{\mathrm{2}}}) \HOLSymConst{\HOLTokenConj{}}
     (\HOLSymConst{\HOLTokenForall{}}\HOLBoundVar{E\sb{\mathrm{1}}}.
        \HOLBoundVar{E} \HOLTokenTransBegin\HOLSymConst{\ensuremath{\tau}}\HOLTokenTransEnd \HOLBoundVar{E\sb{\mathrm{1}}} \HOLSymConst{\HOLTokenImp{}} \HOLFreeVar{Exp} \HOLBoundVar{E\sb{\mathrm{1}}} \HOLBoundVar{E\sp{\prime}} \HOLSymConst{\HOLTokenDisj{}} \HOLSymConst{\HOLTokenExists{}}\HOLBoundVar{E\sb{\mathrm{2}}}. \HOLBoundVar{E\sp{\prime}} \HOLTokenTransBegin\HOLSymConst{\ensuremath{\tau}}\HOLTokenTransEnd \HOLBoundVar{E\sb{\mathrm{2}}} \HOLSymConst{\HOLTokenConj{}} \HOLFreeVar{Exp} \HOLBoundVar{E\sb{\mathrm{1}}} \HOLBoundVar{E\sb{\mathrm{2}}}) \HOLSymConst{\HOLTokenConj{}}
     \HOLSymConst{\HOLTokenForall{}}\HOLBoundVar{E\sb{\mathrm{2}}}. \HOLBoundVar{E\sp{\prime}} \HOLTokenTransBegin\HOLSymConst{\ensuremath{\tau}}\HOLTokenTransEnd \HOLBoundVar{E\sb{\mathrm{2}}} \HOLSymConst{\HOLTokenImp{}} \HOLSymConst{\HOLTokenExists{}}\HOLBoundVar{E\sb{\mathrm{1}}}. \HOLBoundVar{E} \HOLTokenWeakTransBegin\HOLSymConst{\ensuremath{\tau}}\HOLTokenWeakTransEnd \HOLBoundVar{E\sb{\mathrm{1}}} \HOLSymConst{\HOLTokenConj{}} \HOLFreeVar{Exp} \HOLBoundVar{E\sb{\mathrm{1}}} \HOLBoundVar{E\sb{\mathrm{2}}}
\end{alltt}
$P$ \emph{expands} $Q$, written $P \succeq_e Q$, if $P \mathcal{R} Q$
for some expansion $\mathcal{R}$.
\end{definition}

Above definition for `expands' can be simply characterized as the
following statement:
\begin{definition}{(Original definition of `expands')}
\begin{alltt}
\HOLTokenTurnstile{} \HOLFreeVar{P} \HOLSymConst{\HOLTokenExpands{}} \HOLFreeVar{Q} \HOLSymConst{\HOLTokenEquiv{}} \HOLSymConst{\HOLTokenExists{}}\HOLBoundVar{Exp}. \HOLBoundVar{Exp} \HOLFreeVar{P} \HOLFreeVar{Q} \HOLSymConst{\HOLTokenConj{}} \HOLConst{EXPANSION} \HOLBoundVar{Exp}\hfill[expands_thm]
\end{alltt}
\end{definition}
However it's not easy to derive its properties from such a
definition (it's possible for sure). Instead we have used HOL's co-inductive relation package
to define it co-inductively, the same way as in the case of weak
equivalence. As a result, the above ``original'' definition now
becomes a theorem as alternative definition.

We have proved that, expansion is contained in weak bisimulation and
`expands' is between strong and  weak equivalence:
\begin{proposition}{(Relationships between expansion, strong and weak equivalences)}
\begin{enumerate}
\item expansion implies weak bisimulation:
\begin{alltt}
\HOLTokenTurnstile{} \HOLConst{EXPANSION} \HOLFreeVar{Exp} \HOLSymConst{\HOLTokenImp{}} \HOLConst{WEAK_BISIM} \HOLFreeVar{Exp}\hfill[EXPANSION_IMP_WEAK_BISIM]
\end{alltt}
\item `expands' implies weak equivalence:
\begin{alltt}
\HOLTokenTurnstile{} \HOLFreeVar{P} \HOLSymConst{\HOLTokenExpands{}} \HOLFreeVar{Q} \HOLSymConst{\HOLTokenImp{}} \HOLFreeVar{P} \HOLSymConst{\HOLTokenWeakEQ} \HOLFreeVar{Q}\hfill[expands_IMP_WEAK_EQUIV]
\end{alltt}
\item Strong equivalence implies `expands':
\begin{alltt}
\HOLTokenTurnstile{} \HOLFreeVar{P} \HOLSymConst{\HOLTokenStrongEQ} \HOLFreeVar{Q} \HOLSymConst{\HOLTokenImp{}} \HOLFreeVar{P} \HOLSymConst{\HOLTokenExpands{}} \HOLFreeVar{Q}\hfill[STRONG_EQUIV_IMP_expands]
\end{alltt}
\end{enumerate}
\end{proposition}

\subsection{Expansion is pre-order}

It's not hard to prove that, the `expands' relation is pre-order, that
is, transitive and reflexitive. Actually we proved these properties
from the properties of the `expansion' predicate: if the identity
relation is expansion, then `expands' is reflexitive; and if the
composition of any two expansions is still an expansion, then for sure
the `expands' relation is transitive.

\begin{proposition}
The composition of two expansions is still an expansion:
\begin{alltt}
COMP_EXPANSION:
\HOLTokenTurnstile{} \HOLConst{EXPANSION} \HOLFreeVar{Exp\sb{\mathrm{1}}} \HOLSymConst{\HOLTokenConj{}} \HOLConst{EXPANSION} \HOLFreeVar{Exp\sb{\mathrm{2}}} \HOLSymConst{\HOLTokenImp{}} \HOLConst{EXPANSION} (\HOLFreeVar{Exp\sb{\mathrm{2}}} \HOLSymConst{\HOLTokenRCompose{}} \HOLFreeVar{Exp\sb{\mathrm{1}}})
\end{alltt}
\end{proposition}

Using this result, it's trivial to prove the transitivity of `expands' relation:
\begin{proposition}{(Transitivity of `expands' relation)}
\begin{alltt}
\HOLTokenTurnstile{} \HOLFreeVar{x} \HOLSymConst{\HOLTokenExpands{}} \HOLFreeVar{y} \HOLSymConst{\HOLTokenConj{}} \HOLFreeVar{y} \HOLSymConst{\HOLTokenExpands{}} \HOLFreeVar{z} \HOLSymConst{\HOLTokenImp{}} \HOLFreeVar{x} \HOLSymConst{\HOLTokenExpands{}} \HOLFreeVar{z}\hfill[expands_TRANS]
\end{alltt}
\end{proposition}

The `expands' relation is also reflexitive (accordingly the identity
relation is also an expansion):

\begin{proposition}{(Reflexitivity of `expands' relation)}
\begin{alltt}
\HOLTokenTurnstile{} \HOLConst{EXPANSION} (\HOLSymConst{=})\hfill[IDENTITY_EXPANSION]
\HOLTokenTurnstile{} \HOLFreeVar{x} \HOLSymConst{\HOLTokenExpands{}} \HOLFreeVar{x}\hfill[expands_TRANS]
\end{alltt}
\end{proposition}

Combining above two results, we have proved that, `expands' is a
pre-order:
\begin{lemma}
Bisimularity expandsion is a pre-order:
\begin{alltt}
\HOLTokenTurnstile{} \HOLConst{PreOrder} (\HOLSymConst{expands})\hfill[expands_PreOrder]
\end{alltt}
where
\begin{alltt}
\HOLTokenTurnstile{} \HOLConst{PreOrder} \HOLFreeVar{R} \HOLSymConst{\HOLTokenEquiv{}} \HOLConst{reflexive} \HOLFreeVar{R} \HOLSymConst{\HOLTokenConj{}} \HOLConst{transitive} \HOLFreeVar{R}\hfill[PreOrder]
\end{alltt}
\end{lemma}

\subsection{Expansion is precongruence}

Now we prove `contracts' relation is a precongruence (a special
version which requires guarded sum). To get this
result, we have to prove that, the `expands' relation is preserved by
all CCS operators (except for \texttt{REC} which we have ignored in
this project). For the case of sum operator, we can only prove it for
guarded sums:

\begin{proposition}{(`expands' is precongruence)}
\begin{enumerate}
\item `expands' is substitutive by prefix operator:
\begin{alltt}
expands_SUBST_PREFIX:
\HOLTokenTurnstile{} \HOLFreeVar{E} \HOLSymConst{\HOLTokenExpands{}} \HOLFreeVar{E\sp{\prime}} \HOLSymConst{\HOLTokenImp{}} \HOLSymConst{\HOLTokenForall{}}\HOLBoundVar{u}. \HOLBoundVar{u}\HOLSymConst{..}\HOLFreeVar{E} \HOLSymConst{\HOLTokenExpands{}} \HOLBoundVar{u}\HOLSymConst{..}\HOLFreeVar{E\sp{\prime}}
\end{alltt}
\item `expands' is preserved by guarded sums:
\begin{alltt}
expands_PRESD_BY_GUARDED_SUM:
\HOLTokenTurnstile{} \HOLFreeVar{E\sb{\mathrm{1}}} \HOLSymConst{\HOLTokenExpands{}} \HOLFreeVar{E\sb{\mathrm{1}}\sp{\prime}} \HOLSymConst{\HOLTokenConj{}} \HOLFreeVar{E\sb{\mathrm{2}}} \HOLSymConst{\HOLTokenExpands{}} \HOLFreeVar{E\sb{\mathrm{2}}\sp{\prime}} \HOLSymConst{\HOLTokenImp{}} \HOLFreeVar{a\sb{\mathrm{1}}}\HOLSymConst{..}\HOLFreeVar{E\sb{\mathrm{1}}} \HOLSymConst{+} \HOLFreeVar{a\sb{\mathrm{2}}}\HOLSymConst{..}\HOLFreeVar{E\sb{\mathrm{2}}} \HOLSymConst{\HOLTokenExpands{}} (\HOLFreeVar{a\sb{\mathrm{1}}}\HOLSymConst{..}\HOLFreeVar{E\sb{\mathrm{1}}\sp{\prime}} \HOLSymConst{+} \HOLFreeVar{a\sb{\mathrm{2}}}\HOLSymConst{..}\HOLFreeVar{E\sb{\mathrm{2}}\sp{\prime}})
\end{alltt}
\item `expands' is preserved by parallel composition:
\begin{alltt}
expands_PRESD_BY_PAR:
\HOLTokenTurnstile{} \HOLFreeVar{E\sb{\mathrm{1}}} \HOLSymConst{\HOLTokenExpands{}} \HOLFreeVar{E\sb{\mathrm{1}}\sp{\prime}} \HOLSymConst{\HOLTokenConj{}} \HOLFreeVar{E\sb{\mathrm{2}}} \HOLSymConst{\HOLTokenExpands{}} \HOLFreeVar{E\sb{\mathrm{2}}\sp{\prime}} \HOLSymConst{\HOLTokenImp{}} \HOLFreeVar{E\sb{\mathrm{1}}} \HOLSymConst{\ensuremath{\parallel}} \HOLFreeVar{E\sb{\mathrm{2}}} \HOLSymConst{\HOLTokenExpands{}} \HOLFreeVar{E\sb{\mathrm{1}}\sp{\prime}} \HOLSymConst{\ensuremath{\parallel}} \HOLFreeVar{E\sb{\mathrm{2}}\sp{\prime}}
\end{alltt}
\item `expands' is substitutive by restrictions:
\begin{alltt}
expands_SUBST_RESTR:
\HOLTokenTurnstile{} \HOLFreeVar{E} \HOLSymConst{\HOLTokenExpands{}} \HOLFreeVar{E\sp{\prime}} \HOLSymConst{\HOLTokenImp{}} \HOLSymConst{\HOLTokenForall{}}\HOLBoundVar{L}. \HOLSymConst{\ensuremath{\nu}} \HOLBoundVar{L} \HOLFreeVar{E} \HOLSymConst{\HOLTokenExpands{}} \HOLSymConst{\ensuremath{\nu}} \HOLBoundVar{L} \HOLFreeVar{E\sp{\prime}}
\end{alltt}
\item `expands' is substitutive by relabeling operator:
\begin{alltt}
expands_SUBST_RELAB:
\HOLTokenTurnstile{} \HOLFreeVar{E} \HOLSymConst{\HOLTokenExpands{}} \HOLFreeVar{E\sp{\prime}} \HOLSymConst{\HOLTokenImp{}} \HOLSymConst{\HOLTokenForall{}}\HOLBoundVar{rf}. \HOLConst{relab} \HOLFreeVar{E} \HOLBoundVar{rf} \HOLSymConst{\HOLTokenExpands{}} \HOLConst{relab} \HOLFreeVar{E\sp{\prime}} \HOLBoundVar{rf}
\end{alltt}
\end{enumerate}
\end{proposition}

With above results, now we can inductively prove the `expands'
relation is preserved by any context with guarded sums, which by
definition is a precongruce:
\begin{theorem}
Bisimilarity expansion (`expands' relation) is precongruence, i.e. it's substitutive by
semantics contexts (with restrictions of guarded sums):
\begin{alltt}
expands_SUBST_GCONTEXT:
\HOLTokenTurnstile{} \HOLFreeVar{P} \HOLSymConst{\HOLTokenExpands{}} \HOLFreeVar{Q} \HOLSymConst{\HOLTokenImp{}} \HOLSymConst{\HOLTokenForall{}}\HOLBoundVar{E}. \HOLConst{GCONTEXT} \HOLBoundVar{E} \HOLSymConst{\HOLTokenImp{}} \HOLBoundVar{E} \HOLFreeVar{P} \HOLSymConst{\HOLTokenExpands{}} \HOLBoundVar{E} \HOLFreeVar{Q}
\end{alltt}
or
\begin{alltt}
expands_precongruence:
\HOLTokenTurnstile{} \HOLConst{precongruence1} (\HOLSymConst{expands})
\end{alltt}
where
\begin{alltt}
precongruence1_def:
\HOLTokenTurnstile{} \HOLConst{precongruence1} \HOLFreeVar{R} \HOLSymConst{\HOLTokenEquiv{}}
   \HOLSymConst{\HOLTokenForall{}}\HOLBoundVar{x} \HOLBoundVar{y} \HOLBoundVar{ctx}. \HOLConst{GCONTEXT} \HOLBoundVar{ctx} \HOLSymConst{\HOLTokenImp{}} \HOLFreeVar{R} \HOLBoundVar{x} \HOLBoundVar{y} \HOLSymConst{\HOLTokenImp{}} \HOLFreeVar{R} (\HOLBoundVar{ctx} \HOLBoundVar{x}) (\HOLBoundVar{ctx} \HOLBoundVar{y})
\end{alltt}
\end{theorem}

\section{Contraction}

Contraction is the new invention by Prof.\ Davide Sangiorgi. It's another
non-symmetric relation.  Roughly speaking, ``$P$ contracts
$Q$'' holds if ``$P$ is equivalent to $Q$'' and, in addition, ``$Q$
has the possibility of being as efficient as $P$''. That is, $Q$ is
capable of simulating $P$ by performing less internal work. It is
suggicient that $Q$ has one `efficient' path; $Q$ could also have
other paths that are slower than any path in $P$.

Its definition and properties are almost the same as expansions (and
weak equivalence).  To define
contraction, we also need to follow a two-step processes: first we define a
predicate of relations, called ``(bisimulation) contraction'', then we define a binary
relation called `contracts' (bisimilarity contraction) as the union of all contractions.

\begin{definition}{(Bisimulation contraction)}
\label{def:contraction}
A processes relation $\mathcal{R}$ is a \emph{bisimulation contraction} if, whenever
$P\,\mathcal{R}\, Q$,
\begin{enumerate}
\item $P \overset{\mu}{\rightarrow} P'$ implies that there is $Q'$
  with $Q \overset{\hat{\mu}}{\rightarrow} Q'$ and $P'\, \mathcal{R}
  \, Q'$;
\item $Q \overset{\mu}{\rightarrow} Q'$ implies that there is $P'$
  with $P \overset{\mu}{\Rightarrow} P'$ and $P' \approx Q'$.
\end{enumerate}
or formally:
\begin{alltt}
\hfill[CONTRACTION]
\HOLTokenTurnstile{} \HOLConst{CONTRACTION} \HOLFreeVar{Con} \HOLSymConst{\HOLTokenEquiv{}}
   \HOLSymConst{\HOLTokenForall{}}\HOLBoundVar{E} \HOLBoundVar{E\sp{\prime}}.
     \HOLFreeVar{Con} \HOLBoundVar{E} \HOLBoundVar{E\sp{\prime}} \HOLSymConst{\HOLTokenImp{}}
     (\HOLSymConst{\HOLTokenForall{}}\HOLBoundVar{l}.
        (\HOLSymConst{\HOLTokenForall{}}\HOLBoundVar{E\sb{\mathrm{1}}}.
           \HOLBoundVar{E} \HOLTokenTransBegin\HOLConst{label} \HOLBoundVar{l}\HOLTokenTransEnd \HOLBoundVar{E\sb{\mathrm{1}}} \HOLSymConst{\HOLTokenImp{}} \HOLSymConst{\HOLTokenExists{}}\HOLBoundVar{E\sb{\mathrm{2}}}. \HOLBoundVar{E\sp{\prime}} \HOLTokenTransBegin\HOLConst{label} \HOLBoundVar{l}\HOLTokenTransEnd \HOLBoundVar{E\sb{\mathrm{2}}} \HOLSymConst{\HOLTokenConj{}} \HOLFreeVar{Con} \HOLBoundVar{E\sb{\mathrm{1}}} \HOLBoundVar{E\sb{\mathrm{2}}}) \HOLSymConst{\HOLTokenConj{}}
        \HOLSymConst{\HOLTokenForall{}}\HOLBoundVar{E\sb{\mathrm{2}}}. \HOLBoundVar{E\sp{\prime}} \HOLTokenTransBegin\HOLConst{label} \HOLBoundVar{l}\HOLTokenTransEnd \HOLBoundVar{E\sb{\mathrm{2}}} \HOLSymConst{\HOLTokenImp{}} \HOLSymConst{\HOLTokenExists{}}\HOLBoundVar{E\sb{\mathrm{1}}}. \HOLBoundVar{E} \HOLTokenWeakTransBegin\HOLConst{label} \HOLBoundVar{l}\HOLTokenWeakTransEnd \HOLBoundVar{E\sb{\mathrm{1}}} \HOLSymConst{\HOLTokenConj{}} \HOLBoundVar{E\sb{\mathrm{1}}} \HOLSymConst{\HOLTokenWeakEQ} \HOLBoundVar{E\sb{\mathrm{2}}}) \HOLSymConst{\HOLTokenConj{}}
     (\HOLSymConst{\HOLTokenForall{}}\HOLBoundVar{E\sb{\mathrm{1}}}.
        \HOLBoundVar{E} \HOLTokenTransBegin\HOLSymConst{\ensuremath{\tau}}\HOLTokenTransEnd \HOLBoundVar{E\sb{\mathrm{1}}} \HOLSymConst{\HOLTokenImp{}} \HOLFreeVar{Con} \HOLBoundVar{E\sb{\mathrm{1}}} \HOLBoundVar{E\sp{\prime}} \HOLSymConst{\HOLTokenDisj{}} \HOLSymConst{\HOLTokenExists{}}\HOLBoundVar{E\sb{\mathrm{2}}}. \HOLBoundVar{E\sp{\prime}} \HOLTokenTransBegin\HOLSymConst{\ensuremath{\tau}}\HOLTokenTransEnd \HOLBoundVar{E\sb{\mathrm{2}}} \HOLSymConst{\HOLTokenConj{}} \HOLFreeVar{Con} \HOLBoundVar{E\sb{\mathrm{1}}} \HOLBoundVar{E\sb{\mathrm{2}}}) \HOLSymConst{\HOLTokenConj{}}
     \HOLSymConst{\HOLTokenForall{}}\HOLBoundVar{E\sb{\mathrm{2}}}. \HOLBoundVar{E\sp{\prime}} \HOLTokenTransBegin\HOLSymConst{\ensuremath{\tau}}\HOLTokenTransEnd \HOLBoundVar{E\sb{\mathrm{2}}} \HOLSymConst{\HOLTokenImp{}} \HOLSymConst{\HOLTokenExists{}}\HOLBoundVar{E\sb{\mathrm{1}}}. \HOLBoundVar{E} \HOLSymConst{\HOLTokenEPS} \HOLBoundVar{E\sb{\mathrm{1}}} \HOLSymConst{\HOLTokenConj{}} \HOLBoundVar{E\sb{\mathrm{1}}} \HOLSymConst{\HOLTokenWeakEQ} \HOLBoundVar{E\sb{\mathrm{2}}}
\end{alltt}
\emph{Bisimilarity contraction}, written $\succeq_{\mathrm{bis}}$, is
the union of all bisimulation contractions.
\end{definition}

Above definition for `contracts' can be simply characterized as the
following statement:
\begin{definition}{(Original definition of `contracts')}
\begin{alltt}
\HOLTokenTurnstile{} \HOLFreeVar{P} \HOLSymConst{\HOLTokenContracts{}} \HOLFreeVar{Q} \HOLSymConst{\HOLTokenEquiv{}} \HOLSymConst{\HOLTokenExists{}}\HOLBoundVar{Con}. \HOLBoundVar{Con} \HOLFreeVar{P} \HOLFreeVar{Q} \HOLSymConst{\HOLTokenConj{}} \HOLConst{CONTRACTION} \HOLBoundVar{Con}
\end{alltt}
\end{definition}
Like the case of expansions, it's not easy to derive its properties from such a
definition. Instead we have used HOL's co-inductive relation package
to define it co-inductively, the same way as in the case of weak
equivalence. As a result, the above ``original'' definition now
becomes a theorem as alternative definition.

In the first clause $Q$ is required to match $P$'s challenge
transition with at most one transition. This makes sure that $Q$ is
capable of mimicking $P$'s work at least as efficiently as $P$. In
contrast, the second clause of Def. \ref{def:contraction}, on the
challenges from $Q$, entirely ignores efficiency: It is the same
clause of weak bisimulation---the final derivatives are even required
to be related by $\approx$ rather then by $\mathcal{R}$.

We can prove that, `contraction' is contained in weak bisimulation,
and the `contracts' (bisimularity contraction) is between `expands'
and  weak equivalence:
\begin{proposition}{(Relationships between contraction, expansion and weak
    bisimulation)}
\begin{enumerate}
\item `expands' implies `contracts',
\begin{alltt}
\HOLTokenTurnstile{} \HOLFreeVar{P} \HOLSymConst{\HOLTokenExpands{}} \HOLFreeVar{Q} \HOLSymConst{\HOLTokenImp{}} \HOLFreeVar{P} \HOLSymConst{\HOLTokenContracts{}} \HOLFreeVar{Q}\hfill[expands_IMP_contracts]
\end{alltt}
\item `contracts' implies weak equivalence (`contracts' is contained
  in weak equivalence).
\begin{alltt}
\HOLTokenTurnstile{} \HOLFreeVar{P} \HOLSymConst{\HOLTokenContracts{}} \HOLFreeVar{Q} \HOLSymConst{\HOLTokenImp{}} \HOLFreeVar{P} \HOLSymConst{\HOLTokenWeakEQ} \HOLFreeVar{Q}\hfill[contracts_IMP_WEAK_EQUIV]
\end{alltt}
\end{enumerate}
\end{proposition}

The proof of all properties of contraction and `contracts' are slightly
harder than the case of expansion and `expands', and we usually need
to reduce the proof to corresponding properties of weak equivalence.
Also, surprisingly, a contraction doesn't imply weak bisimulation,
i.e. the following property doesn't hold:
\begin{alltt}
\HOLinline{\HOLSymConst{\HOLTokenForall{}}\HOLBoundVar{Con}. \HOLConst{CONTRACTION} \HOLBoundVar{Con} \HOLSymConst{\HOLTokenImp{}} \HOLConst{WEAK_BISIM} \HOLBoundVar{Con}}
\end{alltt}
As a result, to finish the proof of \texttt{contracts_IMP_WEAK_EQUIV}, we do not prove $Con$ itself is a weak
   bisimulation, but rather that $Con$ "union" weak bisimilarity is a
   weak bisimulation. This is quite unexpected.

\subsection{Contraction is pre-order}

Following the same idea in the case of expansion, we prove that
bisimularity contraction is a pre-order.

\begin{proposition}
The composition of two contractions is still a contraction:
\begin{alltt}
\HOLTokenTurnstile{} \HOLConst{CONTRACTION} \HOLFreeVar{Con\sb{\mathrm{1}}} \HOLSymConst{\HOLTokenConj{}} \HOLConst{CONTRACTION} \HOLFreeVar{Con\sb{\mathrm{2}}} \HOLSymConst{\HOLTokenImp{}}
   \HOLConst{CONTRACTION} (\HOLFreeVar{Con\sb{\mathrm{2}}} \HOLSymConst{\HOLTokenRCompose{}} \HOLFreeVar{Con\sb{\mathrm{1}}})\hfill[COMP_CONTRACTION]
\end{alltt}
\end{proposition}

Using this result, it's easy to prove the transitivity of `contracts'
relation:
\begin{proposition}
The bisimularity contraction (`contrats' relation) is transitive:
\begin{alltt}
\HOLTokenTurnstile{} \HOLFreeVar{x} \HOLSymConst{\HOLTokenContracts{}} \HOLFreeVar{y} \HOLSymConst{\HOLTokenConj{}} \HOLFreeVar{y} \HOLSymConst{\HOLTokenContracts{}} \HOLFreeVar{z} \HOLSymConst{\HOLTokenImp{}} \HOLFreeVar{x} \HOLSymConst{\HOLTokenContracts{}} \HOLFreeVar{z}\hfill[contracts_TRANS]
\end{alltt}
\end{proposition}

`contracts' is also reflexitive (accordingly the identity relation is also an
CONTRACTION):
\begin{proposition}
Bisimilarity contraction is reflexitive:
\begin{alltt}
\HOLTokenTurnstile{} \HOLConst{CONTRACTION} (\HOLSymConst{=})\hfill[IDENTITY_CONTRACTION]
\HOLTokenTurnstile{} \HOLFreeVar{x} \HOLSymConst{\HOLTokenContracts{}} \HOLFreeVar{x}\hfill[contracts_TRANS]
\end{alltt}
\end{proposition}

Combining above two results, we got:
\begin{lemma}
Bisimularity contraction is a pre-order:
\begin{alltt}
\HOLTokenTurnstile{} \HOLConst{PreOrder} (\HOLSymConst{contracts})\hfill[contracts_PreOrder]
\end{alltt}
where
\begin{alltt}
\HOLTokenTurnstile{} \HOLConst{PreOrder} \HOLFreeVar{R} \HOLSymConst{\HOLTokenEquiv{}} \HOLConst{reflexive} \HOLFreeVar{R} \HOLSymConst{\HOLTokenConj{}} \HOLConst{transitive} \HOLFreeVar{R}\hfill[PreOrder]
\end{alltt}
\end{lemma}

\subsection{Contraction is precongruence}

Now we prove `contracts' relation is a precongruence (a special
version which requires guarded sum). To get this
result, we have to prove that, the `contracts' relation is preserved by
all CCS operators (except for \texttt{REC} which we have ignored in
this project). For the case of sum operator, we can only prove it for
guarded sums:

\begin{proposition}{(`contracts' is precongruence)}
\begin{enumerate}
\item `contracts' is substitutive by prefix operator:
\begin{alltt}
contracts_SUBST_PREFIX:
\HOLTokenTurnstile{} \HOLFreeVar{E} \HOLSymConst{\HOLTokenContracts{}} \HOLFreeVar{E\sp{\prime}} \HOLSymConst{\HOLTokenImp{}} \HOLSymConst{\HOLTokenForall{}}\HOLBoundVar{u}. \HOLBoundVar{u}\HOLSymConst{..}\HOLFreeVar{E} \HOLSymConst{\HOLTokenContracts{}} \HOLBoundVar{u}\HOLSymConst{..}\HOLFreeVar{E\sp{\prime}}
\end{alltt}
\item `contracts' is preserved by guarded syms:
\begin{alltt}
contracts_PRESD_BY_GUARDED_SUM:
\HOLTokenTurnstile{} \HOLFreeVar{E\sb{\mathrm{1}}} \HOLSymConst{\HOLTokenContracts{}} \HOLFreeVar{E\sb{\mathrm{1}}\sp{\prime}} \HOLSymConst{\HOLTokenConj{}} \HOLFreeVar{E\sb{\mathrm{2}}} \HOLSymConst{\HOLTokenContracts{}} \HOLFreeVar{E\sb{\mathrm{2}}\sp{\prime}} \HOLSymConst{\HOLTokenImp{}} \HOLFreeVar{a\sb{\mathrm{1}}}\HOLSymConst{..}\HOLFreeVar{E\sb{\mathrm{1}}} \HOLSymConst{+} \HOLFreeVar{a\sb{\mathrm{2}}}\HOLSymConst{..}\HOLFreeVar{E\sb{\mathrm{2}}} \HOLSymConst{\HOLTokenContracts{}} \HOLFreeVar{a\sb{\mathrm{1}}}\HOLSymConst{..}\HOLFreeVar{E\sb{\mathrm{1}}\sp{\prime}} \HOLSymConst{+} \HOLFreeVar{a\sb{\mathrm{2}}}\HOLSymConst{..}\HOLFreeVar{E\sb{\mathrm{2}}\sp{\prime}}
\end{alltt}
\item `contracts' is preserved by parallel composition:
\begin{alltt}
contracts_PRESD_BY_PAR:
\HOLTokenTurnstile{} \HOLFreeVar{E\sb{\mathrm{1}}} \HOLSymConst{\HOLTokenContracts{}} \HOLFreeVar{E\sb{\mathrm{1}}\sp{\prime}} \HOLSymConst{\HOLTokenConj{}} \HOLFreeVar{E\sb{\mathrm{2}}} \HOLSymConst{\HOLTokenContracts{}} \HOLFreeVar{E\sb{\mathrm{2}}\sp{\prime}} \HOLSymConst{\HOLTokenImp{}} \HOLFreeVar{E\sb{\mathrm{1}}} \HOLSymConst{\ensuremath{\parallel}} \HOLFreeVar{E\sb{\mathrm{2}}} \HOLSymConst{\HOLTokenContracts{}} \HOLFreeVar{E\sb{\mathrm{1}}\sp{\prime}} \HOLSymConst{\ensuremath{\parallel}} \HOLFreeVar{E\sb{\mathrm{2}}\sp{\prime}}
\end{alltt}
\item `contracts' is substitutive by restrictions:
\begin{alltt}
contracts_SUBST_RESTR:
\HOLTokenTurnstile{} \HOLFreeVar{E} \HOLSymConst{\HOLTokenContracts{}} \HOLFreeVar{E\sp{\prime}} \HOLSymConst{\HOLTokenImp{}} \HOLSymConst{\HOLTokenForall{}}\HOLBoundVar{L}. \HOLSymConst{\ensuremath{\nu}} \HOLBoundVar{L} \HOLFreeVar{E} \HOLSymConst{\HOLTokenContracts{}} \HOLSymConst{\ensuremath{\nu}} \HOLBoundVar{L} \HOLFreeVar{E\sp{\prime}}
\end{alltt}
\item `contracts' is substitutive by relabeling operator:
\begin{alltt}
contracts_SUBST_RELAB:
\HOLTokenTurnstile{} \HOLFreeVar{E} \HOLSymConst{\HOLTokenContracts{}} \HOLFreeVar{E\sp{\prime}} \HOLSymConst{\HOLTokenImp{}} \HOLSymConst{\HOLTokenForall{}}\HOLBoundVar{rf}. \HOLConst{relab} \HOLFreeVar{E} \HOLBoundVar{rf} \HOLSymConst{\HOLTokenContracts{}} \HOLConst{relab} \HOLFreeVar{E\sp{\prime}} \HOLBoundVar{rf}
\end{alltt}
\end{enumerate}
\end{proposition}

With above results, now we can inductively prove the `contracts'
relation is preserved by any context with guarded sums, which by
definition is a precongruence:
\begin{theorem}
Bisimilarity contraction (`contracts' relation) is precongruence, i.e. it's substitutive by
semantics contexts (with restrictions of guarded sums):
\begin{alltt}
contracts_SUBST_GCONTEXT:
\HOLTokenTurnstile{} \HOLFreeVar{P} \HOLSymConst{\HOLTokenContracts{}} \HOLFreeVar{Q} \HOLSymConst{\HOLTokenImp{}} \HOLSymConst{\HOLTokenForall{}}\HOLBoundVar{E}. \HOLConst{GCONTEXT} \HOLBoundVar{E} \HOLSymConst{\HOLTokenImp{}} \HOLBoundVar{E} \HOLFreeVar{P} \HOLSymConst{\HOLTokenContracts{}} \HOLBoundVar{E} \HOLFreeVar{Q}
\end{alltt}
or
\begin{alltt}
contracts_precongruence:
\HOLTokenTurnstile{} \HOLConst{precongruence1} (\HOLSymConst{contracts})
\end{alltt}
where
\begin{alltt}
precongruence1_def:
\HOLTokenTurnstile{} \HOLConst{precongruence1} \HOLFreeVar{R} \HOLSymConst{\HOLTokenEquiv{}}
   \HOLSymConst{\HOLTokenForall{}}\HOLBoundVar{x} \HOLBoundVar{y} \HOLBoundVar{ctx}. \HOLConst{GCONTEXT} \HOLBoundVar{ctx} \HOLSymConst{\HOLTokenImp{}} \HOLFreeVar{R} \HOLBoundVar{x} \HOLBoundVar{y} \HOLSymConst{\HOLTokenImp{}} \HOLFreeVar{R} (\HOLBoundVar{ctx} \HOLBoundVar{x}) (\HOLBoundVar{ctx} \HOLBoundVar{y})
\end{alltt}
\end{theorem}

\section{Step and Trace transitions}

To finish the proof of ``unique solutions of contractions'' theorem,
we need to ability to reason about the ``length'' of weak
transitions. We want to make sure the weak transitions become smaller
(or same length) after passing a weak equivalence or contraction.
 Such a requirement is unusual because no other theorems need
it.  At the beginning we defined a simple relation which only capture
the ``length'' of general transitions between two processes, but it
turns out to be useless, because between any two process there may be
multiple transitions with the same length, and having the facts that
``$P$ weakly transits to $Q$'' and ``there's a $n$-step transition from
$P$ to $Q$'', we actually know nothing about the length of that
specific weak transition. But we kept the definition of the so-called
``step'' transition in case it's needed somehow in the future.

\subsection{$n$-step transitions}

The concept of ``$n$-step transitions'' can be defined as a numbered relation
closure of its single transition:
\begin{definition}{($n$-step transition)}
A $n$-step transition from $P$ to $Q$ is the numbered relation closure
(NRC) of single step transition without respect to transition actions:
\begin{alltt}
STEP_def:
\HOLTokenTurnstile{} \HOLConst{STEP} \HOLFreeVar{P} \HOLFreeVar{n} \HOLFreeVar{Q} \HOLSymConst{\HOLTokenEquiv{}} \HOLConst{NRC} (\HOLTokenLambda{}\HOLBoundVar{E} \HOLBoundVar{E\sp{\prime}}. \HOLSymConst{\HOLTokenExists{}}\HOLBoundVar{u}. \HOLBoundVar{E} \HOLTokenTransBegin\HOLBoundVar{u}\HOLTokenTransEnd \HOLBoundVar{E\sp{\prime}}) \HOLFreeVar{n} \HOLFreeVar{P} \HOLFreeVar{Q}
\end{alltt}
where
\begin{alltt}
NRC:
\HOLConst{NRC} \HOLFreeVar{R} \HOLNumLit{0} \HOLFreeVar{x} \HOLFreeVar{y} \HOLSymConst{\HOLTokenEquiv{}} (\HOLFreeVar{x} \HOLSymConst{=} \HOLFreeVar{y})
\HOLConst{NRC} \HOLFreeVar{R} (\HOLConst{SUC} \HOLFreeVar{n}) \HOLFreeVar{x} \HOLFreeVar{y} \HOLSymConst{\HOLTokenEquiv{}} \HOLSymConst{\HOLTokenExists{}}\HOLBoundVar{z}. \HOLFreeVar{R} \HOLFreeVar{x} \HOLBoundVar{z} \HOLSymConst{\HOLTokenConj{}} \HOLConst{NRC} \HOLFreeVar{R} \HOLFreeVar{n} \HOLBoundVar{z} \HOLFreeVar{y}
\end{alltt}
\end{definition}

For this relation, we have some nice arithmetic-like properties proved
here:
\begin{proposition}{(Properties of $n$-step transitions)}
\begin{enumerate}
\item 0-step transition means equality:
\begin{alltt}
\HOLTokenTurnstile{} \HOLConst{STEP} \HOLFreeVar{x} \HOLNumLit{0} \HOLFreeVar{y} \HOLSymConst{\HOLTokenEquiv{}} (\HOLFreeVar{x} \HOLSymConst{=} \HOLFreeVar{y})\hfill[STEP0]
\end{alltt}
\item 1-step transition means single-step (strong) transition:
\begin{alltt}
\HOLTokenTurnstile{} \HOLConst{STEP} \HOLFreeVar{x} \HOLNumLit{1} \HOLFreeVar{y} \HOLSymConst{\HOLTokenEquiv{}} \HOLSymConst{\HOLTokenExists{}}\HOLBoundVar{u}. \HOLFreeVar{x} \HOLTokenTransBegin\HOLBoundVar{u}\HOLTokenTransEnd \HOLFreeVar{y}\hfill[STEP1]
\end{alltt}
\item Reduce $n+1$-step transition to $n$-step transition:
\begin{alltt}
STEP_SUC:
\HOLTokenTurnstile{} \HOLConst{STEP} \HOLFreeVar{x} (\HOLConst{SUC} \HOLFreeVar{n}) \HOLFreeVar{y} \HOLSymConst{\HOLTokenEquiv{}} \HOLSymConst{\HOLTokenExists{}}\HOLBoundVar{z}. (\HOLSymConst{\HOLTokenExists{}}\HOLBoundVar{u}. \HOLFreeVar{x} \HOLTokenTransBegin\HOLBoundVar{u}\HOLTokenTransEnd \HOLBoundVar{z}) \HOLSymConst{\HOLTokenConj{}} \HOLConst{STEP} \HOLBoundVar{z} \HOLFreeVar{n} \HOLFreeVar{y}
\end{alltt}
\item Reduce $n+1$-step transition to $n$-step transition (another way):
\begin{alltt}
STEP_SUC_LEFT:
\HOLTokenTurnstile{} \HOLConst{STEP} \HOLFreeVar{x} (\HOLConst{SUC} \HOLFreeVar{n}) \HOLFreeVar{y} \HOLSymConst{\HOLTokenEquiv{}} \HOLSymConst{\HOLTokenExists{}}\HOLBoundVar{z}. \HOLConst{STEP} \HOLFreeVar{x} \HOLFreeVar{n} \HOLBoundVar{z} \HOLSymConst{\HOLTokenConj{}} \HOLSymConst{\HOLTokenExists{}}\HOLBoundVar{u}. \HOLBoundVar{z} \HOLTokenTransBegin\HOLBoundVar{u}\HOLTokenTransEnd \HOLFreeVar{y}
\end{alltt}
\item Reduce $m+n$-step transition to $m$-step and $n$-step
  transitions:
\begin{alltt}
STEP_ADD_EQN:
\HOLTokenTurnstile{} \HOLConst{STEP} \HOLFreeVar{x} (\HOLFreeVar{m} \HOLSymConst{+} \HOLFreeVar{n}) \HOLFreeVar{z} \HOLSymConst{\HOLTokenEquiv{}} \HOLSymConst{\HOLTokenExists{}}\HOLBoundVar{y}. \HOLConst{STEP} \HOLFreeVar{x} \HOLFreeVar{m} \HOLBoundVar{y} \HOLSymConst{\HOLTokenConj{}} \HOLConst{STEP} \HOLBoundVar{y} \HOLFreeVar{n} \HOLFreeVar{z}
\end{alltt}
\end{enumerate}
\end{proposition}

What's more important is its relationship with EPS and weak transitions:
\begin{lemma}{(Making $n$-step transitions from EPS and weak transitions)}
\begin{enumerate}
\item EPS implies the existence of $n$-step transition with the same ends:
\begin{alltt}
EPS_AND_STEP:
\HOLTokenTurnstile{} \HOLFreeVar{E} \HOLSymConst{\HOLTokenEPS} \HOLFreeVar{E\sp{\prime}} \HOLSymConst{\HOLTokenImp{}} \HOLSymConst{\HOLTokenExists{}}\HOLBoundVar{n}. \HOLConst{STEP} \HOLFreeVar{E} \HOLBoundVar{n} \HOLFreeVar{E\sp{\prime}}
\end{alltt}
\item Weak transition implies the existence of $n$-step transition with the same ends:
\begin{alltt}
WEAK_TRANS_AND_STEP:
\HOLTokenTurnstile{} \HOLFreeVar{E} \HOLTokenWeakTransBegin\HOLFreeVar{u}\HOLTokenWeakTransEnd \HOLFreeVar{E\sp{\prime}} \HOLSymConst{\HOLTokenImp{}} \HOLSymConst{\HOLTokenExists{}}\HOLBoundVar{n}. \HOLConst{STEP} \HOLFreeVar{E} \HOLBoundVar{n} \HOLFreeVar{E\sp{\prime}}
\end{alltt}
\end{enumerate}
\end{lemma}

However, it must be understood that, there may be multiple paths (with
same or different lengths)
between two processes $P$ and $Q$, thus the $n$-step transitions
derived from above theorems may actually not reflect the path behind
the initial weak transition (or EPS transition). This is why they're
useless for proving the ``unique solution of contractions'' theorem.
The $n$-step transition relation is simple, less powerfull but beautiful. It's never used in any proof
in this thesis.

\subsection{Traces}

The successful formalization of Prof.\ Sangiorgi's ``unique solution of
contractions'' theorem is completely based on successful definition of
``Trace'' in this thesis project.  For an informal proof, it's not
needed because the related arguments are quite obvious.   Noticed
that, trace equivalence is not covered by our projects. Also, the
concept of ``weak trace'' is not defined anywhere.

Mathematically speaking, a trace transition is nothing but a special
reflexive transitive closure (RTC) with a list accumulator, we call
it LRTC:
\begin{definition}{(LRTC)}
A reflexitive transitive closure with list of transitions (LRTC) is a
normal reflexitive transitive closure (RTC) of binary relations
enhanced with a list, which accumulates all the transitions in the path:
\begin{alltt}
\HOLTokenTurnstile{} \HOLConst{LRTC} \HOLFreeVar{R} \HOLFreeVar{a} \HOLFreeVar{l} \HOLFreeVar{b} \HOLSymConst{\HOLTokenEquiv{}}
   \HOLSymConst{\HOLTokenForall{}}\HOLBoundVar{P}.
     (\HOLSymConst{\HOLTokenForall{}}\HOLBoundVar{x}. \HOLBoundVar{P} \HOLBoundVar{x} [] \HOLBoundVar{x}) \HOLSymConst{\HOLTokenConj{}}
     (\HOLSymConst{\HOLTokenForall{}}\HOLBoundVar{x} \HOLBoundVar{h} \HOLBoundVar{y} \HOLBoundVar{t} \HOLBoundVar{z}. \HOLFreeVar{R} \HOLBoundVar{x} \HOLBoundVar{h} \HOLBoundVar{y} \HOLSymConst{\HOLTokenConj{}} \HOLBoundVar{P} \HOLBoundVar{y} \HOLBoundVar{t} \HOLBoundVar{z} \HOLSymConst{\HOLTokenImp{}} \HOLBoundVar{P} \HOLBoundVar{x} (\HOLBoundVar{h}\HOLSymConst{::}\HOLBoundVar{t}) \HOLBoundVar{z}) \HOLSymConst{\HOLTokenImp{}}
     \HOLBoundVar{P} \HOLFreeVar{a} \HOLFreeVar{l} \HOLFreeVar{b}
\end{alltt}
\end{definition}

We have followed the proof idea in HOL's \texttt{relationTheory} and
the proof sketches of normal RTC, and successfully proved many of its
properties, including its induction, strong induction, rules and cases theorems:
\begin{proposition}{(Properties of LRTC)}
\begin{enumerate}
\item Induction principle of LRTC:
\begin{alltt}
\hfill[LRTC_INDUCT]
\HOLTokenTurnstile{} (\HOLSymConst{\HOLTokenForall{}}\HOLBoundVar{x}. \HOLFreeVar{P} \HOLBoundVar{x} [] \HOLBoundVar{x}) \HOLSymConst{\HOLTokenConj{}}
   (\HOLSymConst{\HOLTokenForall{}}\HOLBoundVar{x} \HOLBoundVar{h} \HOLBoundVar{y} \HOLBoundVar{t} \HOLBoundVar{z}. \HOLFreeVar{R} \HOLBoundVar{x} \HOLBoundVar{h} \HOLBoundVar{y} \HOLSymConst{\HOLTokenConj{}} \HOLFreeVar{P} \HOLBoundVar{y} \HOLBoundVar{t} \HOLBoundVar{z} \HOLSymConst{\HOLTokenImp{}} \HOLFreeVar{P} \HOLBoundVar{x} (\HOLBoundVar{h}\HOLSymConst{::}\HOLBoundVar{t}) \HOLBoundVar{z}) \HOLSymConst{\HOLTokenImp{}}
   \HOLSymConst{\HOLTokenForall{}}\HOLBoundVar{x} \HOLBoundVar{l} \HOLBoundVar{y}. \HOLConst{LRTC} \HOLFreeVar{R} \HOLBoundVar{x} \HOLBoundVar{l} \HOLBoundVar{y} \HOLSymConst{\HOLTokenImp{}} \HOLFreeVar{P} \HOLBoundVar{x} \HOLBoundVar{l} \HOLBoundVar{y}
\end{alltt}
\item The `rules' theorem of LRTC:
\begin{alltt}
\hfill[LRTC_RULES]
\HOLConst{LRTC} \HOLFreeVar{R} \HOLFreeVar{x} [] \HOLFreeVar{x}
\HOLFreeVar{R} \HOLFreeVar{x} \HOLFreeVar{h} \HOLFreeVar{y} \HOLSymConst{\HOLTokenConj{}} \HOLConst{LRTC} \HOLFreeVar{R} \HOLFreeVar{y} \HOLFreeVar{t} \HOLFreeVar{z} \HOLSymConst{\HOLTokenImp{}} \HOLConst{LRTC} \HOLFreeVar{R} \HOLFreeVar{x} (\HOLFreeVar{h}\HOLSymConst{::}\HOLFreeVar{t}) \HOLFreeVar{z}
\end{alltt}
\item The strong induction principle of LRTC:
\begin{alltt}
\hfill[LRTC_STRONG_INDUCT]
\HOLTokenTurnstile{} (\HOLSymConst{\HOLTokenForall{}}\HOLBoundVar{x}. \HOLFreeVar{P} \HOLBoundVar{x} [] \HOLBoundVar{x}) \HOLSymConst{\HOLTokenConj{}}
   (\HOLSymConst{\HOLTokenForall{}}\HOLBoundVar{x} \HOLBoundVar{h} \HOLBoundVar{y} \HOLBoundVar{t} \HOLBoundVar{z}.
      \HOLFreeVar{R} \HOLBoundVar{x} \HOLBoundVar{h} \HOLBoundVar{y} \HOLSymConst{\HOLTokenConj{}} \HOLConst{LRTC} \HOLFreeVar{R} \HOLBoundVar{y} \HOLBoundVar{t} \HOLBoundVar{z} \HOLSymConst{\HOLTokenConj{}} \HOLFreeVar{P} \HOLBoundVar{y} \HOLBoundVar{t} \HOLBoundVar{z} \HOLSymConst{\HOLTokenImp{}} \HOLFreeVar{P} \HOLBoundVar{x} (\HOLBoundVar{h}\HOLSymConst{::}\HOLBoundVar{t}) \HOLBoundVar{z}) \HOLSymConst{\HOLTokenImp{}}
   \HOLSymConst{\HOLTokenForall{}}\HOLBoundVar{x} \HOLBoundVar{l} \HOLBoundVar{y}. \HOLConst{LRTC} \HOLFreeVar{R} \HOLBoundVar{x} \HOLBoundVar{l} \HOLBoundVar{y} \HOLSymConst{\HOLTokenImp{}} \HOLFreeVar{P} \HOLBoundVar{x} \HOLBoundVar{l} \HOLBoundVar{y}
\end{alltt}
\item Reduce a $m+n$-step LRTC into $m$-step and $n$-step LRTCs:
\begin{alltt}
\hfill[LRTC_LRTC]
\HOLTokenTurnstile{} \HOLConst{LRTC} \HOLFreeVar{R} \HOLFreeVar{x} \HOLFreeVar{m} \HOLFreeVar{y} \HOLSymConst{\HOLTokenImp{}} \HOLSymConst{\HOLTokenForall{}}\HOLBoundVar{n} \HOLBoundVar{z}. \HOLConst{LRTC} \HOLFreeVar{R} \HOLFreeVar{y} \HOLBoundVar{n} \HOLBoundVar{z} \HOLSymConst{\HOLTokenImp{}} \HOLConst{LRTC} \HOLFreeVar{R} \HOLFreeVar{x} (\HOLFreeVar{m} \HOLSymConst{\HOLTokenDoublePlus} \HOLBoundVar{n}) \HOLBoundVar{z}
\end{alltt}
\item Transitivity of LRTC:
\begin{alltt}
\hfill[LRTC_TRANS]
\HOLTokenTurnstile{} \HOLConst{LRTC} \HOLFreeVar{R} \HOLFreeVar{x} \HOLFreeVar{m} \HOLFreeVar{y} \HOLSymConst{\HOLTokenConj{}} \HOLConst{LRTC} \HOLFreeVar{R} \HOLFreeVar{y} \HOLFreeVar{n} \HOLFreeVar{z} \HOLSymConst{\HOLTokenImp{}} \HOLConst{LRTC} \HOLFreeVar{R} \HOLFreeVar{x} (\HOLFreeVar{m} \HOLSymConst{\HOLTokenDoublePlus} \HOLFreeVar{n}) \HOLFreeVar{z}
\end{alltt}
\item `cases' theorem of LRTC (using \texttt{HD} and \texttt{TL} of
  lists):
\begin{alltt}
\HOLTokenTurnstile{} \HOLConst{LRTC} \HOLFreeVar{R} \HOLFreeVar{x} \HOLFreeVar{l} \HOLFreeVar{y} \HOLSymConst{\HOLTokenEquiv{}}
   \HOLKeyword{if} \HOLConst{NULL} \HOLFreeVar{l} \HOLKeyword{then} \HOLFreeVar{x} \HOLSymConst{=} \HOLFreeVar{y}
   \HOLKeyword{else} \HOLSymConst{\HOLTokenExists{}}\HOLBoundVar{u}. \HOLFreeVar{R} \HOLFreeVar{x} (\HOLConst{HD} \HOLFreeVar{l}) \HOLBoundVar{u} \HOLSymConst{\HOLTokenConj{}} \HOLConst{LRTC} \HOLFreeVar{R} \HOLBoundVar{u} (\HOLConst{TL} \HOLFreeVar{l}) \HOLFreeVar{y}\hfill[LRTC_CASES1]
\end{alltt}
\item `cases' theorem of LRTC (using \texttt{FRONT} and \texttt{LAST} of
  lists):
\begin{alltt}
\HOLTokenTurnstile{} \HOLConst{LRTC} \HOLFreeVar{R} \HOLFreeVar{x} \HOLFreeVar{l} \HOLFreeVar{y} \HOLSymConst{\HOLTokenEquiv{}}
   \HOLKeyword{if} \HOLConst{NULL} \HOLFreeVar{l} \HOLKeyword{then} \HOLFreeVar{x} \HOLSymConst{=} \HOLFreeVar{y}
   \HOLKeyword{else} \HOLSymConst{\HOLTokenExists{}}\HOLBoundVar{u}. \HOLConst{LRTC} \HOLFreeVar{R} \HOLFreeVar{x} (\HOLConst{FRONT} \HOLFreeVar{l}) \HOLBoundVar{u} \HOLSymConst{\HOLTokenConj{}} \HOLFreeVar{R} \HOLBoundVar{u} (\HOLConst{LAST} \HOLFreeVar{l}) \HOLFreeVar{y}\hfill[LRTC_CASES2]
\end{alltt}
\end{enumerate}
\end{proposition}

Now we can simply define a trace as the LRTC of its single step
transition, the ``trace'' between two CCS processes are stored into a
list of actions. This is simpler than most of textbook which defines
it inductively:
\begin{definition}{(Trace)}
A trace transition between two processes is the LRTC of single-step
(strong) transition (with transition action accumulated into a list):
\begin{alltt}
\HOLTokenTurnstile{} \HOLConst{TRACE} \HOLSymConst{=} \HOLConst{LRTC} \HOLConst{TRANS}\hfill[TRACE_def]
\end{alltt}
\end{definition}

Using above results of LRTC, we can easily derives the following
properties of traces (which some of them are hard to get if we define the trace
natively and inductively from the ground):
\begin{proposition}{(Properties of trace)}
\begin{enumerate}
\item The `rules' theorem of trace:
\begin{alltt}
\hfill[TRACE_rules]
\HOLTokenTurnstile{} (\HOLSymConst{\HOLTokenForall{}}\HOLBoundVar{x}. \HOLConst{TRACE} \HOLBoundVar{x} \HOLSymConst{\ensuremath{\epsilon}} \HOLBoundVar{x}) \HOLSymConst{\HOLTokenConj{}}
   \HOLSymConst{\HOLTokenForall{}}\HOLBoundVar{x} \HOLBoundVar{h} \HOLBoundVar{y} \HOLBoundVar{t} \HOLBoundVar{z}. \HOLBoundVar{x} \HOLTokenTransBegin\HOLBoundVar{h}\HOLTokenTransEnd \HOLBoundVar{y} \HOLSymConst{\HOLTokenConj{}} \HOLConst{TRACE} \HOLBoundVar{y} \HOLBoundVar{t} \HOLBoundVar{z} \HOLSymConst{\HOLTokenImp{}} \HOLConst{TRACE} \HOLBoundVar{x} (\HOLBoundVar{h}\HOLSymConst{::}\HOLBoundVar{t}) \HOLBoundVar{z}
\end{alltt}
\item Transitivity of traces:
\begin{alltt}
\hfill[TRACE_trans]
\HOLTokenTurnstile{} \HOLConst{TRACE} \HOLFreeVar{x} \HOLFreeVar{m} \HOLFreeVar{y} \HOLSymConst{\HOLTokenImp{}} \HOLSymConst{\HOLTokenForall{}}\HOLBoundVar{n} \HOLBoundVar{z}. \HOLConst{TRACE} \HOLFreeVar{y} \HOLBoundVar{n} \HOLBoundVar{z} \HOLSymConst{\HOLTokenImp{}} \HOLConst{TRACE} \HOLFreeVar{x} (\HOLFreeVar{m} \HOLSymConst{\HOLTokenDoublePlus} \HOLBoundVar{n}) \HOLBoundVar{z}
\end{alltt}
\item Induction principle of traces:
\begin{alltt}
\hfill[TRACE_ind]
\HOLTokenTurnstile{} (\HOLSymConst{\HOLTokenForall{}}\HOLBoundVar{x}. \HOLFreeVar{P} \HOLBoundVar{x} \HOLSymConst{\ensuremath{\epsilon}} \HOLBoundVar{x}) \HOLSymConst{\HOLTokenConj{}}
   (\HOLSymConst{\HOLTokenForall{}}\HOLBoundVar{x} \HOLBoundVar{h} \HOLBoundVar{y} \HOLBoundVar{t} \HOLBoundVar{z}. \HOLBoundVar{x} \HOLTokenTransBegin\HOLBoundVar{h}\HOLTokenTransEnd \HOLBoundVar{y} \HOLSymConst{\HOLTokenConj{}} \HOLFreeVar{P} \HOLBoundVar{y} \HOLBoundVar{t} \HOLBoundVar{z} \HOLSymConst{\HOLTokenImp{}} \HOLFreeVar{P} \HOLBoundVar{x} (\HOLBoundVar{h}\HOLSymConst{::}\HOLBoundVar{t}) \HOLBoundVar{z}) \HOLSymConst{\HOLTokenImp{}}
   \HOLSymConst{\HOLTokenForall{}}\HOLBoundVar{x} \HOLBoundVar{l} \HOLBoundVar{y}. \HOLConst{TRACE} \HOLBoundVar{x} \HOLBoundVar{l} \HOLBoundVar{y} \HOLSymConst{\HOLTokenImp{}} \HOLFreeVar{P} \HOLBoundVar{x} \HOLBoundVar{l} \HOLBoundVar{y}
\end{alltt}
\item Strong induction principle of traces:
\begin{alltt}
\hfill[TRACE_strongind]
\HOLTokenTurnstile{} (\HOLSymConst{\HOLTokenForall{}}\HOLBoundVar{x}. \HOLFreeVar{P} \HOLBoundVar{x} \HOLSymConst{\ensuremath{\epsilon}} \HOLBoundVar{x}) \HOLSymConst{\HOLTokenConj{}}
   (\HOLSymConst{\HOLTokenForall{}}\HOLBoundVar{x} \HOLBoundVar{h} \HOLBoundVar{y} \HOLBoundVar{t} \HOLBoundVar{z}.
      \HOLBoundVar{x} \HOLTokenTransBegin\HOLBoundVar{h}\HOLTokenTransEnd \HOLBoundVar{y} \HOLSymConst{\HOLTokenConj{}} \HOLConst{TRACE} \HOLBoundVar{y} \HOLBoundVar{t} \HOLBoundVar{z} \HOLSymConst{\HOLTokenConj{}} \HOLFreeVar{P} \HOLBoundVar{y} \HOLBoundVar{t} \HOLBoundVar{z} \HOLSymConst{\HOLTokenImp{}} \HOLFreeVar{P} \HOLBoundVar{x} (\HOLBoundVar{h}\HOLSymConst{::}\HOLBoundVar{t}) \HOLBoundVar{z}) \HOLSymConst{\HOLTokenImp{}}
   \HOLSymConst{\HOLTokenForall{}}\HOLBoundVar{x} \HOLBoundVar{l} \HOLBoundVar{y}. \HOLConst{TRACE} \HOLBoundVar{x} \HOLBoundVar{l} \HOLBoundVar{y} \HOLSymConst{\HOLTokenImp{}} \HOLFreeVar{P} \HOLBoundVar{x} \HOLBoundVar{l} \HOLBoundVar{y}
\end{alltt}
\item `cases' theorem of traces (using \texttt{HD} and \texttt{TL} of
  lists):
\begin{alltt}
\hfill[TRACE_cases1]
\HOLTokenTurnstile{} \HOLConst{TRACE} \HOLFreeVar{x} \HOLFreeVar{l} \HOLFreeVar{y} \HOLSymConst{\HOLTokenEquiv{}}
   \HOLKeyword{if} \HOLConst{NULL} \HOLFreeVar{l} \HOLKeyword{then} \HOLFreeVar{x} \HOLSymConst{=} \HOLFreeVar{y} \HOLKeyword{else} \HOLSymConst{\HOLTokenExists{}}\HOLBoundVar{u}. \HOLFreeVar{x} \HOLTokenTransBegin\HOLConst{HD} \HOLFreeVar{l}\HOLTokenTransEnd \HOLBoundVar{u} \HOLSymConst{\HOLTokenConj{}} \HOLConst{TRACE} \HOLBoundVar{u} (\HOLConst{TL} \HOLFreeVar{l}) \HOLFreeVar{y}
\end{alltt}
\item `cases' theorem of traces (using \texttt{FRONT} and \texttt{LAST} of
  lists):
\begin{alltt}
\hfill[TRACE_cases2]
\HOLTokenTurnstile{} \HOLConst{TRACE} \HOLFreeVar{x} \HOLFreeVar{l} \HOLFreeVar{y} \HOLSymConst{\HOLTokenEquiv{}}
   \HOLKeyword{if} \HOLConst{NULL} \HOLFreeVar{l} \HOLKeyword{then} \HOLFreeVar{x} \HOLSymConst{=} \HOLFreeVar{y}
   \HOLKeyword{else} \HOLSymConst{\HOLTokenExists{}}\HOLBoundVar{u}. \HOLConst{TRACE} \HOLFreeVar{x} (\HOLConst{FRONT} \HOLFreeVar{l}) \HOLBoundVar{u} \HOLSymConst{\HOLTokenConj{}} \HOLBoundVar{u} \HOLTokenTransBegin\HOLConst{LAST} \HOLFreeVar{l}\HOLTokenTransEnd \HOLFreeVar{y}
\end{alltt}
\item Breaking a trace (at any middle position) into two traces:
\begin{alltt}
\hfill[TRACE_cases_twice]
\HOLTokenTurnstile{} \HOLConst{TRACE} \HOLFreeVar{x} \HOLFreeVar{l} \HOLFreeVar{y} \HOLSymConst{\HOLTokenEquiv{}}
   \HOLSymConst{\HOLTokenExists{}}\HOLBoundVar{u} \HOLBoundVar{l\sb{\mathrm{1}}} \HOLBoundVar{l\sb{\mathrm{2}}}. \HOLConst{TRACE} \HOLFreeVar{x} \HOLBoundVar{l\sb{\mathrm{1}}} \HOLBoundVar{u} \HOLSymConst{\HOLTokenConj{}} \HOLConst{TRACE} \HOLBoundVar{u} \HOLBoundVar{l\sb{\mathrm{2}}} \HOLFreeVar{y} \HOLSymConst{\HOLTokenConj{}} (\HOLFreeVar{l} \HOLSymConst{=} \HOLBoundVar{l\sb{\mathrm{1}}} \HOLSymConst{\HOLTokenDoublePlus} \HOLBoundVar{l\sb{\mathrm{2}}})
\end{alltt}
\item Appending two traces:
\begin{alltt}
\hfill[TRACE_APPEND_cases]
\HOLTokenTurnstile{} \HOLConst{TRACE} \HOLFreeVar{x} (\HOLFreeVar{l\sb{\mathrm{1}}} \HOLSymConst{\HOLTokenDoublePlus} \HOLFreeVar{l\sb{\mathrm{2}}}) \HOLFreeVar{y} \HOLSymConst{\HOLTokenEquiv{}} \HOLSymConst{\HOLTokenExists{}}\HOLBoundVar{u}. \HOLConst{TRACE} \HOLFreeVar{x} \HOLFreeVar{l\sb{\mathrm{1}}} \HOLBoundVar{u} \HOLSymConst{\HOLTokenConj{}} \HOLConst{TRACE} \HOLBoundVar{u} \HOLFreeVar{l\sb{\mathrm{2}}} \HOLFreeVar{y}
\end{alltt}
\end{enumerate}
\end{proposition}
Here some list operations: \HOLinline{\HOLFreeVar{h}\HOLSymConst{::}\HOLFreeVar{t}} means the list head $h$
connected with the rest part (tail) $t$ of the list.  \HOLinline{\HOLConst{HD}} and
\HOLinline{\HOLConst{TL}} are operators for getting the head and tail of a list, and
\HOLinline{\HOLFreeVar{L\sb{\mathrm{1}}} \HOLSymConst{\HOLTokenDoublePlus} \HOLFreeVar{L\sb{\mathrm{2}}}} means the connection of two lists.  \HOLinline{\HOLConst{FRONT}}
return the parts of list without the last element, and \HOLinline{\HOLConst{LAST}}
literally returns the last element of the list.

Our goal here is to be able to freely move between an EPS or weak
transition, and the corresponding trace with exactly the same
intermediate actions (thus with the same length). \footnote{Actually,
  even a trace may not be unique if we consider the possibility that,
  there're two different paths in the labeled transition system
  sharing the same initial and terminal nodes, also with the same
  intermediate action list. But to finish the proof we don't need to
  know the uniqueness at all.}  Clearly every \texttt{EPS} or weak transition is also a
trace, but the reverse is not always true.   To become an EPS
transition, all intermediate actions must be $tau$. As for weak
transitions, if there're visible actions in the trace, it must be
unique in the action list.  To capture such properties, we have
defined two helper concepts:

\begin{definition}{(No label and unique lebal)}
\begin{enumerate}
\item
\begin{alltt}
NO_LABEL_def:
\HOLTokenTurnstile{} \HOLConst{NO_LABEL} \HOLFreeVar{L} \HOLSymConst{\HOLTokenEquiv{}} \HOLSymConst{\HOLTokenNeg{}}\HOLSymConst{\HOLTokenExists{}}\HOLBoundVar{l}. \HOLConst{MEM} (\HOLConst{label} \HOLBoundVar{l}) \HOLFreeVar{L}
\end{alltt}
\item
\begin{alltt}
UNIQUE_LABEL_def:
\HOLTokenTurnstile{} \HOLConst{UNIQUE_LABEL} \HOLFreeVar{u} \HOLFreeVar{L} \HOLSymConst{\HOLTokenEquiv{}}
   \HOLSymConst{\HOLTokenExists{}}\HOLBoundVar{L\sb{\mathrm{1}}} \HOLBoundVar{L\sb{\mathrm{2}}}.
     (\HOLBoundVar{L\sb{\mathrm{1}}} \HOLSymConst{\HOLTokenDoublePlus} [\HOLFreeVar{u}] \HOLSymConst{\HOLTokenDoublePlus} \HOLBoundVar{L\sb{\mathrm{2}}} \HOLSymConst{=} \HOLFreeVar{L}) \HOLSymConst{\HOLTokenConj{}}
     \HOLSymConst{\HOLTokenNeg{}}\HOLSymConst{\HOLTokenExists{}}\HOLBoundVar{l}. \HOLConst{MEM} (\HOLConst{label} \HOLBoundVar{l}) \HOLBoundVar{L\sb{\mathrm{1}}} \HOLSymConst{\HOLTokenDisj{}} \HOLConst{MEM} (\HOLConst{label} \HOLBoundVar{l}) \HOLBoundVar{L\sb{\mathrm{2}}}
\end{alltt}
\end{enumerate}
\end{definition}
The definition of ``no label'' seems straightforward, while the definition
of ``unique label'' is quite smart.\footnote{It's actually learnt from Robert
Beers, an experienced HOL user. Now this definition has been added
into HOL's \texttt{listTheory} with three common-used alternative definitions
proved as equivalence theorems. This work was supported by Ramana
Kumar, one of HOL maintainers.} Usually one can simply count the number of
visible actions in the list and assert that number to be one, but the
advantage of above definition is that, once we know that a
visible action is unique in a list (or trace), from the definition we
can immediately conclude that, the same action doesn't appear in the
rest two parts of the list. And the related proofs become very straightforward.

\HOLinline{\HOLConst{NO_LABEL}} has the following lemma concerning its initial
transition:
\begin{lemma}{(Case of ``no (visible) label'')}
If there's no (visible) label in a list of actions, then either the
first action in the list is $\tau$, or the rest of actions has no labels:
\begin{alltt}
NO_LABEL_cases:
\HOLTokenTurnstile{} \HOLConst{NO_LABEL} (\HOLFreeVar{x}\HOLSymConst{::}\HOLFreeVar{xs}) \HOLSymConst{\HOLTokenEquiv{}} (\HOLFreeVar{x} \HOLSymConst{=} \HOLSymConst{\ensuremath{\tau}}) \HOLSymConst{\HOLTokenConj{}} \HOLConst{NO_LABEL} \HOLFreeVar{xs}
\end{alltt}
\end{lemma}

\HOLinline{\HOLConst{UNIQUE_LABEL}} has two useful lemmas concerning its initial
transition: (noticed that how a \HOLinline{\HOLConst{UNIQUE_LABEL}} becomes \HOLinline{\HOLConst{NO_LABEL}} after
the first visible action)
\begin{lemma}{(Two cases of ``unique label'')}
\begin{enumerate}
\item If a (visible) label $l$ is unique in a list of actions starting
  with $\tau$ (invisible action), then $l$ is also unique in the rest
  of the list:
\begin{alltt}
UNIQUE_LABEL_cases1:
\HOLTokenTurnstile{} \HOLConst{UNIQUE_LABEL} (\HOLConst{label} \HOLFreeVar{l}) (\HOLSymConst{\ensuremath{\tau}}\HOLSymConst{::}\HOLFreeVar{xs}) \HOLSymConst{\HOLTokenEquiv{}} \HOLConst{UNIQUE_LABEL} (\HOLConst{label} \HOLFreeVar{l}) \HOLFreeVar{xs}
\end{alltt}
\item If a (visible) label $l$ is unique in a list of actions starting
  with another visible label $l'$, then $l = l'$ and the rest of
  actions has no (visible) labels:
\begin{alltt}
UNIQUE_LABEL_cases2:
\HOLTokenTurnstile{} \HOLConst{UNIQUE_LABEL} (\HOLConst{label} \HOLFreeVar{l}) (\HOLConst{label} \HOLFreeVar{l\sp{\prime}}\HOLSymConst{::}\HOLFreeVar{xs}) \HOLSymConst{\HOLTokenEquiv{}}
   (\HOLFreeVar{l} \HOLSymConst{=} \HOLFreeVar{l\sp{\prime}}) \HOLSymConst{\HOLTokenConj{}} \HOLConst{NO_LABEL} \HOLFreeVar{xs}
\end{alltt}
\end{enumerate}
\end{lemma}

What we have finally established here, is the precise condition for translating an
EPS or weak transition from/to a trace transition:
\begin{theorem}{(Trace, EPS and trace)}
\begin{enumerate}
\item For any EPS transition, there exists a trace in which the action
  list has no labels:
\begin{alltt}
\hfill[EPS_AND_TRACE]
\HOLTokenTurnstile{} \HOLFreeVar{E} \HOLSymConst{\HOLTokenEPS} \HOLFreeVar{E\sp{\prime}} \HOLSymConst{\HOLTokenEquiv{}} \HOLSymConst{\HOLTokenExists{}}\HOLBoundVar{xs}. \HOLConst{TRACE} \HOLFreeVar{E} \HOLBoundVar{xs} \HOLFreeVar{E\sp{\prime}} \HOLSymConst{\HOLTokenConj{}} \HOLConst{NO_LABEL} \HOLBoundVar{xs}
\end{alltt}
\item For any weak transition, there exists a trace in which the
  action list (not null) either has no labels or the label is unique:
\begin{alltt}
\hfill[WEAK_TRANS_AND_TRACE]
\HOLTokenTurnstile{} \HOLFreeVar{E} \HOLTokenWeakTransBegin\HOLFreeVar{u}\HOLTokenWeakTransEnd \HOLFreeVar{E\sp{\prime}} \HOLSymConst{\HOLTokenEquiv{}}
   \HOLSymConst{\HOLTokenExists{}}\HOLBoundVar{us}.
     \HOLConst{TRACE} \HOLFreeVar{E} \HOLBoundVar{us} \HOLFreeVar{E\sp{\prime}} \HOLSymConst{\HOLTokenConj{}} \HOLSymConst{\HOLTokenNeg{}}\HOLConst{NULL} \HOLBoundVar{us} \HOLSymConst{\HOLTokenConj{}}
     \HOLKeyword{if} \HOLFreeVar{u} \HOLSymConst{=} \HOLSymConst{\ensuremath{\tau}} \HOLKeyword{then} \HOLConst{NO_LABEL} \HOLBoundVar{us} \HOLKeyword{else} \HOLConst{UNIQUE_LABEL} \HOLFreeVar{u} \HOLBoundVar{us}
\end{alltt}
\end{enumerate}
\end{theorem}

\subsection{Traces for Expansions and Contractions}

Whenever a trace (corresponding to an EPS or weak transition) passes
an expansion or contraction, its length either remains the same or
becomes shorter.
But for our purposes here, we only care about those traces with no
(visible) labels or unique (visible) labels, and the traces always
pass through the expansion/contraction from left to right (i.e. from
1st to 2nd parameter). For simplicity we call them input and output traces.

\begin{proposition}{(Traces passing expansions and contractions)}
\begin{enumerate}
\item Whenever a no-labeled trace passes through an expansion, its
  length remains the same or becomes shorter, while the output
  trace has still no label:
\begin{alltt}
expands_AND_TRACE_tau:
\HOLTokenTurnstile{} \HOLFreeVar{E} \HOLSymConst{\HOLTokenExpands{}} \HOLFreeVar{E\sp{\prime}} \HOLSymConst{\HOLTokenImp{}}
   \HOLSymConst{\HOLTokenForall{}}\HOLBoundVar{xs} \HOLBoundVar{l} \HOLBoundVar{E\sb{\mathrm{1}}}.
     \HOLConst{TRACE} \HOLFreeVar{E} \HOLBoundVar{xs} \HOLBoundVar{E\sb{\mathrm{1}}} \HOLSymConst{\HOLTokenConj{}} \HOLConst{NO_LABEL} \HOLBoundVar{xs} \HOLSymConst{\HOLTokenImp{}}
     \HOLSymConst{\HOLTokenExists{}}\HOLBoundVar{xs\sp{\prime}} \HOLBoundVar{E\sb{\mathrm{2}}}.
       \HOLConst{TRACE} \HOLFreeVar{E\sp{\prime}} \HOLBoundVar{xs\sp{\prime}} \HOLBoundVar{E\sb{\mathrm{2}}} \HOLSymConst{\HOLTokenConj{}} \HOLBoundVar{E\sb{\mathrm{1}}} \HOLSymConst{\HOLTokenExpands{}} \HOLBoundVar{E\sb{\mathrm{2}}} \HOLSymConst{\HOLTokenConj{}} \HOLConst{LENGTH} \HOLBoundVar{xs\sp{\prime}} \HOLSymConst{\HOLTokenLeq{}} \HOLConst{LENGTH} \HOLBoundVar{xs} \HOLSymConst{\HOLTokenConj{}}
       \HOLConst{NO_LABEL} \HOLBoundVar{xs\sp{\prime}}
\end{alltt}
\item Whenever an unique-labeled trace passes through an expansion, its
  length remains the same or becomes shorter, while the output trace has
still unique label (which is the same one as input trace):
\begin{alltt}
expands_AND_TRACE_label:
\HOLTokenTurnstile{} \HOLFreeVar{E} \HOLSymConst{\HOLTokenExpands{}} \HOLFreeVar{E\sp{\prime}} \HOLSymConst{\HOLTokenImp{}}
   \HOLSymConst{\HOLTokenForall{}}\HOLBoundVar{xs} \HOLBoundVar{l} \HOLBoundVar{E\sb{\mathrm{1}}}.
     \HOLConst{TRACE} \HOLFreeVar{E} \HOLBoundVar{xs} \HOLBoundVar{E\sb{\mathrm{1}}} \HOLSymConst{\HOLTokenConj{}} \HOLConst{UNIQUE_LABEL} (\HOLConst{label} \HOLBoundVar{l}) \HOLBoundVar{xs} \HOLSymConst{\HOLTokenImp{}}
     \HOLSymConst{\HOLTokenExists{}}\HOLBoundVar{xs\sp{\prime}} \HOLBoundVar{E\sb{\mathrm{2}}}.
       \HOLConst{TRACE} \HOLFreeVar{E\sp{\prime}} \HOLBoundVar{xs\sp{\prime}} \HOLBoundVar{E\sb{\mathrm{2}}} \HOLSymConst{\HOLTokenConj{}} \HOLBoundVar{E\sb{\mathrm{1}}} \HOLSymConst{\HOLTokenExpands{}} \HOLBoundVar{E\sb{\mathrm{2}}} \HOLSymConst{\HOLTokenConj{}} \HOLConst{LENGTH} \HOLBoundVar{xs\sp{\prime}} \HOLSymConst{\HOLTokenLeq{}} \HOLConst{LENGTH} \HOLBoundVar{xs} \HOLSymConst{\HOLTokenConj{}}
       \HOLConst{UNIQUE_LABEL} (\HOLConst{label} \HOLBoundVar{l}) \HOLBoundVar{xs\sp{\prime}}
\end{alltt}
\item Whenever an no-labeled trace passes through a contraction, its
  length remains the same or becomes shorter, while the output
  trace has still no label:
\begin{alltt}
contracts_AND_TRACE_tau:
\HOLTokenTurnstile{} \HOLFreeVar{E} \HOLSymConst{\HOLTokenContracts{}} \HOLFreeVar{E\sp{\prime}} \HOLSymConst{\HOLTokenImp{}}
   \HOLSymConst{\HOLTokenForall{}}\HOLBoundVar{xs} \HOLBoundVar{E\sb{\mathrm{1}}}.
     \HOLConst{TRACE} \HOLFreeVar{E} \HOLBoundVar{xs} \HOLBoundVar{E\sb{\mathrm{1}}} \HOLSymConst{\HOLTokenConj{}} \HOLConst{NO_LABEL} \HOLBoundVar{xs} \HOLSymConst{\HOLTokenImp{}}
     \HOLSymConst{\HOLTokenExists{}}\HOLBoundVar{xs\sp{\prime}} \HOLBoundVar{E\sb{\mathrm{2}}}.
       \HOLConst{TRACE} \HOLFreeVar{E\sp{\prime}} \HOLBoundVar{xs\sp{\prime}} \HOLBoundVar{E\sb{\mathrm{2}}} \HOLSymConst{\HOLTokenConj{}} \HOLBoundVar{E\sb{\mathrm{1}}} \HOLSymConst{\HOLTokenContracts{}} \HOLBoundVar{E\sb{\mathrm{2}}} \HOLSymConst{\HOLTokenConj{}} \HOLConst{LENGTH} \HOLBoundVar{xs\sp{\prime}} \HOLSymConst{\HOLTokenLeq{}} \HOLConst{LENGTH} \HOLBoundVar{xs} \HOLSymConst{\HOLTokenConj{}}
       \HOLConst{NO_LABEL} \HOLBoundVar{xs\sp{\prime}}
\end{alltt}
\item Whenever an unique-labeled trace passes through an contraction, its
  length remains the same or becomes shorter, while the output trace has
still unique label (which is the same one as input trace):
\begin{alltt}
contracts_AND_TRACE_label:
\HOLTokenTurnstile{} \HOLFreeVar{E} \HOLSymConst{\HOLTokenContracts{}} \HOLFreeVar{E\sp{\prime}} \HOLSymConst{\HOLTokenImp{}}
   \HOLSymConst{\HOLTokenForall{}}\HOLBoundVar{xs} \HOLBoundVar{l} \HOLBoundVar{E\sb{\mathrm{1}}}.
     \HOLConst{TRACE} \HOLFreeVar{E} \HOLBoundVar{xs} \HOLBoundVar{E\sb{\mathrm{1}}} \HOLSymConst{\HOLTokenConj{}} \HOLConst{UNIQUE_LABEL} (\HOLConst{label} \HOLBoundVar{l}) \HOLBoundVar{xs} \HOLSymConst{\HOLTokenImp{}}
     \HOLSymConst{\HOLTokenExists{}}\HOLBoundVar{xs\sp{\prime}} \HOLBoundVar{E\sb{\mathrm{2}}}.
       \HOLConst{TRACE} \HOLFreeVar{E\sp{\prime}} \HOLBoundVar{xs\sp{\prime}} \HOLBoundVar{E\sb{\mathrm{2}}} \HOLSymConst{\HOLTokenConj{}} \HOLBoundVar{E\sb{\mathrm{1}}} \HOLSymConst{\HOLTokenContracts{}} \HOLBoundVar{E\sb{\mathrm{2}}} \HOLSymConst{\HOLTokenConj{}} \HOLConst{LENGTH} \HOLBoundVar{xs\sp{\prime}} \HOLSymConst{\HOLTokenLeq{}} \HOLConst{LENGTH} \HOLBoundVar{xs} \HOLSymConst{\HOLTokenConj{}}
       \HOLConst{UNIQUE_LABEL} (\HOLConst{label} \HOLBoundVar{l}) \HOLBoundVar{xs\sp{\prime}}
\end{alltt}
\end{enumerate}
\end{proposition}

\begin{proof}
It's only important to be noticed that, depending on the first transition, it's possible that a weak
transition becomes an EPS transition after passing though the
expansion/contraction.
\end{proof}

These results are essentially important for the proof of the ``unique solutions of
expansions (or contractions)'' in next section.

\section{Unique solutions of contractions}

In this section, we describe the formal proof of Theorem 3.10 in
Prof.\ Sangiorgi's paper \cite{sangiorgi2015equations}:
\begin{theorem}{(unique solution of contractions for $\approx$)}
A system of weakly-guarded contractions has an unique solution for
$\approx$.

or formally:
\begin{alltt}
UNIQUE_SOLUTIONS_OF_CONTRACTIONS:
\HOLTokenTurnstile{} \HOLConst{WGS} \HOLFreeVar{E} \HOLSymConst{\HOLTokenImp{}} \HOLSymConst{\HOLTokenForall{}}\HOLBoundVar{P} \HOLBoundVar{Q}. \HOLBoundVar{P} \HOLSymConst{\HOLTokenContracts{}} \HOLFreeVar{E} \HOLBoundVar{P} \HOLSymConst{\HOLTokenConj{}} \HOLBoundVar{Q} \HOLSymConst{\HOLTokenContracts{}} \HOLFreeVar{E} \HOLBoundVar{Q} \HOLSymConst{\HOLTokenImp{}} \HOLBoundVar{P} \HOLSymConst{\HOLTokenWeakEQ} \HOLBoundVar{Q}
\end{alltt}
\end{theorem}

It must be noticed that, in our formal proof there's no ``system of
contractions'', instead there's just one contraction, because
currently we can only formalize equations (or expressions) with single
process variable.  Also, we must understand that ``weakly-guarded''
contractions with only ``guarded sums''.  This is not a problem of the
statement itself, because in Sangiorgi's paper the CCS grammar itself
was defined to have only guarded sums. Our CCS grammar has general
binary sums, instead, that's why we have ``WGS'' instead of ``WG'' in
the formal statement of the theorem.

Here ``WGS'' means ``weakly guarded expression (context) with only
guarded sum'', its recursive definition is similar with ``WG'':
\begin{definition}{(Weakly guarded expression with restriction of
    guarded sums)}
A \emph{weakly guarded expression with restriction of guarded sums} is
defined inductively as a $\lambda$-function:
\begin{alltt}
\HOLConst{WGS} (\HOLTokenLambda{}\HOLBoundVar{t}. \HOLFreeVar{p})
\HOLConst{GCONTEXT} \HOLFreeVar{e} \HOLSymConst{\HOLTokenImp{}} \HOLConst{WGS} (\HOLTokenLambda{}\HOLBoundVar{t}. \HOLFreeVar{a}\HOLSymConst{..}\HOLFreeVar{e} \HOLBoundVar{t})
\HOLConst{GCONTEXT} \HOLFreeVar{e\sb{\mathrm{1}}} \HOLSymConst{\HOLTokenConj{}} \HOLConst{GCONTEXT} \HOLFreeVar{e\sb{\mathrm{2}}} \HOLSymConst{\HOLTokenImp{}} \HOLConst{WGS} (\HOLTokenLambda{}\HOLBoundVar{t}. \HOLFreeVar{a\sb{\mathrm{1}}}\HOLSymConst{..}\HOLFreeVar{e\sb{\mathrm{1}}} \HOLBoundVar{t} \HOLSymConst{+} \HOLFreeVar{a\sb{\mathrm{2}}}\HOLSymConst{..}\HOLFreeVar{e\sb{\mathrm{2}}} \HOLBoundVar{t})
\HOLConst{WGS} \HOLFreeVar{e\sb{\mathrm{1}}} \HOLSymConst{\HOLTokenConj{}} \HOLConst{WGS} \HOLFreeVar{e\sb{\mathrm{2}}} \HOLSymConst{\HOLTokenImp{}} \HOLConst{WGS} (\HOLTokenLambda{}\HOLBoundVar{t}. \HOLFreeVar{e\sb{\mathrm{1}}} \HOLBoundVar{t} \HOLSymConst{\ensuremath{\parallel}} \HOLFreeVar{e\sb{\mathrm{2}}} \HOLBoundVar{t})
\HOLConst{WGS} \HOLFreeVar{e} \HOLSymConst{\HOLTokenImp{}} \HOLConst{WGS} (\HOLTokenLambda{}\HOLBoundVar{t}. \HOLSymConst{\ensuremath{\nu}} \HOLFreeVar{L} (\HOLFreeVar{e} \HOLBoundVar{t}))
\HOLConst{WGS} \HOLFreeVar{e} \HOLSymConst{\HOLTokenImp{}} \HOLConst{WGS} (\HOLTokenLambda{}\HOLBoundVar{t}. \HOLConst{relab} (\HOLFreeVar{e} \HOLBoundVar{t}) \HOLFreeVar{rf})\hfill[WGS_rules]
\end{alltt}
where \texttt{GCONTEXT} is defined by
\begin{alltt}
\HOLConst{GCONTEXT} (\HOLTokenLambda{}\HOLBoundVar{t}. \HOLBoundVar{t})
\HOLConst{GCONTEXT} (\HOLTokenLambda{}\HOLBoundVar{t}. \HOLFreeVar{p})
\HOLConst{GCONTEXT} \HOLFreeVar{e} \HOLSymConst{\HOLTokenImp{}} \HOLConst{GCONTEXT} (\HOLTokenLambda{}\HOLBoundVar{t}. \HOLFreeVar{a}\HOLSymConst{..}\HOLFreeVar{e} \HOLBoundVar{t})
\HOLConst{GCONTEXT} \HOLFreeVar{e\sb{\mathrm{1}}} \HOLSymConst{\HOLTokenConj{}} \HOLConst{GCONTEXT} \HOLFreeVar{e\sb{\mathrm{2}}} \HOLSymConst{\HOLTokenImp{}} \HOLConst{GCONTEXT} (\HOLTokenLambda{}\HOLBoundVar{t}. \HOLFreeVar{a\sb{\mathrm{1}}}\HOLSymConst{..}\HOLFreeVar{e\sb{\mathrm{1}}} \HOLBoundVar{t} \HOLSymConst{+} \HOLFreeVar{a\sb{\mathrm{2}}}\HOLSymConst{..}\HOLFreeVar{e\sb{\mathrm{2}}} \HOLBoundVar{t})
\HOLConst{GCONTEXT} \HOLFreeVar{e\sb{\mathrm{1}}} \HOLSymConst{\HOLTokenConj{}} \HOLConst{GCONTEXT} \HOLFreeVar{e\sb{\mathrm{2}}} \HOLSymConst{\HOLTokenImp{}} \HOLConst{GCONTEXT} (\HOLTokenLambda{}\HOLBoundVar{t}. \HOLFreeVar{e\sb{\mathrm{1}}} \HOLBoundVar{t} \HOLSymConst{\ensuremath{\parallel}} \HOLFreeVar{e\sb{\mathrm{2}}} \HOLBoundVar{t})
\HOLConst{GCONTEXT} \HOLFreeVar{e} \HOLSymConst{\HOLTokenImp{}} \HOLConst{GCONTEXT} (\HOLTokenLambda{}\HOLBoundVar{t}. \HOLSymConst{\ensuremath{\nu}} \HOLFreeVar{L} (\HOLFreeVar{e} \HOLBoundVar{t}))
\HOLConst{GCONTEXT} \HOLFreeVar{e} \HOLSymConst{\HOLTokenImp{}} \HOLConst{GCONTEXT} (\HOLTokenLambda{}\HOLBoundVar{t}. \HOLConst{relab} (\HOLFreeVar{e} \HOLBoundVar{t}) \HOLFreeVar{rf})\hfill[GCONTEXT_rules]
\end{alltt}
\end{definition}

And we have proved the following results:
\begin{proposition}
A ``WGS'' is also an \texttt{GCONTEXT}, and its combination with a \texttt{GCONTEXT} gives another WGS:
\begin{alltt}
\HOLTokenTurnstile{} \HOLConst{WGS} \HOLFreeVar{e} \HOLSymConst{\HOLTokenImp{}} \HOLConst{GCONTEXT} \HOLFreeVar{e}\hfill[WGS_IS_GCONTEXT]
\HOLTokenTurnstile{} \HOLConst{GCONTEXT} \HOLFreeVar{c} \HOLSymConst{\HOLTokenConj{}} \HOLConst{WGS} \HOLFreeVar{e} \HOLSymConst{\HOLTokenImp{}} \HOLConst{WGS} (\HOLFreeVar{c} \HOLSymConst{\HOLTokenCompose} \HOLFreeVar{e})\hfill[GCONTEXT_WGS_combin]
\end{alltt}
\end{proposition}

Actually the proof of above theorem itself is quite simple, what's not
simple is the Lemma 3.9 before it: (also reduced to single-variable case)
\begin{lemma}
\label{lem:uniq-contractions}
Suppose $P$ and $Q$ are solutions for $\approx$ of a weakly-guarded
contraction. For any context $C$, if $C[P] \overset{\mu}{\Rightarrow}
R$, then there is a context $C'$ such that $R \succeq_{bis} C'[P]$ and
$C[Q] \overset{\hat{\mu}} \approx C'[Q]$.

or formally:
\begin{alltt}
UNIQUE_SOLUTIONS_OF_CONTRACTIONS_LEMMA:
\HOLTokenTurnstile{} (\HOLSymConst{\HOLTokenExists{}}\HOLBoundVar{E}. \HOLConst{WGS} \HOLBoundVar{E} \HOLSymConst{\HOLTokenConj{}} \HOLFreeVar{P} \HOLSymConst{\HOLTokenContracts{}} \HOLBoundVar{E} \HOLFreeVar{P} \HOLSymConst{\HOLTokenConj{}} \HOLFreeVar{Q} \HOLSymConst{\HOLTokenContracts{}} \HOLBoundVar{E} \HOLFreeVar{Q}) \HOLSymConst{\HOLTokenImp{}}
   \HOLSymConst{\HOLTokenForall{}}\HOLBoundVar{C}.
     \HOLConst{GCONTEXT} \HOLBoundVar{C} \HOLSymConst{\HOLTokenImp{}}
     (\HOLSymConst{\HOLTokenForall{}}\HOLBoundVar{l} \HOLBoundVar{R}.
        \HOLBoundVar{C} \HOLFreeVar{P} \HOLTokenWeakTransBegin\HOLConst{label} \HOLBoundVar{l}\HOLTokenWeakTransEnd \HOLBoundVar{R} \HOLSymConst{\HOLTokenImp{}}
        \HOLSymConst{\HOLTokenExists{}}\HOLBoundVar{C\sp{\prime}}.
          \HOLConst{GCONTEXT} \HOLBoundVar{C\sp{\prime}} \HOLSymConst{\HOLTokenConj{}} \HOLBoundVar{R} \HOLSymConst{\HOLTokenContracts{}} \HOLBoundVar{C\sp{\prime}} \HOLFreeVar{P} \HOLSymConst{\HOLTokenConj{}}
          (\HOLConst{WEAK_EQUIV} \HOLSymConst{\HOLTokenRCompose{}} (\HOLTokenLambda{}\HOLBoundVar{x} \HOLBoundVar{y}. \HOLBoundVar{x} \HOLTokenWeakTransBegin\HOLConst{label} \HOLBoundVar{l}\HOLTokenWeakTransEnd \HOLBoundVar{y})) (\HOLBoundVar{C} \HOLFreeVar{Q}) (\HOLBoundVar{C\sp{\prime}} \HOLFreeVar{Q})) \HOLSymConst{\HOLTokenConj{}}
     \HOLSymConst{\HOLTokenForall{}}\HOLBoundVar{R}.
       \HOLBoundVar{C} \HOLFreeVar{P} \HOLTokenWeakTransBegin\HOLSymConst{\ensuremath{\tau}}\HOLTokenWeakTransEnd \HOLBoundVar{R} \HOLSymConst{\HOLTokenImp{}}
       \HOLSymConst{\HOLTokenExists{}}\HOLBoundVar{C\sp{\prime}}.
         \HOLConst{GCONTEXT} \HOLBoundVar{C\sp{\prime}} \HOLSymConst{\HOLTokenConj{}} \HOLBoundVar{R} \HOLSymConst{\HOLTokenContracts{}} \HOLBoundVar{C\sp{\prime}} \HOLFreeVar{P} \HOLSymConst{\HOLTokenConj{}}
         (\HOLConst{WEAK_EQUIV} \HOLSymConst{\HOLTokenRCompose{}} \HOLConst{EPS}) (\HOLBoundVar{C} \HOLFreeVar{Q}) (\HOLBoundVar{C\sp{\prime}} \HOLFreeVar{Q})
\end{alltt}
\end{lemma}

It's in this proof that we must use traces to capture the length of
weak transitions.   The formal proof also used the following four
so-called ``unfolding lemmas'' to minimize the proof size for single
theorem (otherwise it's very hard to replay and learn the entire
single proof):

\begin{lemma}{(Four ``unfolding lemmas'' used in proof of Lemma
    \ref{lem:uniq-contractions})}
\begin{alltt}
unfolding_lemma1:
\HOLTokenTurnstile{} \HOLConst{GCONTEXT} \HOLFreeVar{E} \HOLSymConst{\HOLTokenConj{}} \HOLConst{GCONTEXT} \HOLFreeVar{C} \HOLSymConst{\HOLTokenConj{}} \HOLFreeVar{P} \HOLSymConst{\HOLTokenContracts{}} \HOLFreeVar{E} \HOLFreeVar{P} \HOLSymConst{\HOLTokenImp{}}
   \HOLSymConst{\HOLTokenForall{}}\HOLBoundVar{n}. \HOLFreeVar{C} \HOLFreeVar{P} \HOLSymConst{\HOLTokenContracts{}} (\HOLFreeVar{C} \HOLSymConst{\HOLTokenCompose} \HOLConst{FUNPOW} \HOLFreeVar{E} \HOLBoundVar{n}) \HOLFreeVar{P}

unfolding_lemma2:
\HOLTokenTurnstile{} \HOLConst{WGS} \HOLFreeVar{E} \HOLSymConst{\HOLTokenImp{}}
   \HOLSymConst{\HOLTokenForall{}}\HOLBoundVar{P} \HOLBoundVar{u} \HOLBoundVar{P\sp{\prime}}.
     \HOLFreeVar{E} \HOLBoundVar{P} \HOLTokenTransBegin\HOLBoundVar{u}\HOLTokenTransEnd \HOLBoundVar{P\sp{\prime}} \HOLSymConst{\HOLTokenImp{}}
     \HOLSymConst{\HOLTokenExists{}}\HOLBoundVar{C\sp{\prime}}. \HOLConst{GCONTEXT} \HOLBoundVar{C\sp{\prime}} \HOLSymConst{\HOLTokenConj{}} (\HOLBoundVar{P\sp{\prime}} \HOLSymConst{=} \HOLBoundVar{C\sp{\prime}} \HOLBoundVar{P}) \HOLSymConst{\HOLTokenConj{}} \HOLSymConst{\HOLTokenForall{}}\HOLBoundVar{Q}. \HOLFreeVar{E} \HOLBoundVar{Q} \HOLTokenTransBegin\HOLBoundVar{u}\HOLTokenTransEnd \HOLBoundVar{C\sp{\prime}} \HOLBoundVar{Q}

unfolding_lemma3:
\HOLTokenTurnstile{} \HOLConst{GCONTEXT} \HOLFreeVar{C} \HOLSymConst{\HOLTokenConj{}} \HOLConst{WGS} \HOLFreeVar{E} \HOLSymConst{\HOLTokenImp{}}
   \HOLSymConst{\HOLTokenForall{}}\HOLBoundVar{P} \HOLBoundVar{x} \HOLBoundVar{P\sp{\prime}}.
     \HOLFreeVar{C} (\HOLFreeVar{E} \HOLBoundVar{P}) \HOLTokenTransBegin\HOLBoundVar{x}\HOLTokenTransEnd \HOLBoundVar{P\sp{\prime}} \HOLSymConst{\HOLTokenImp{}}
     \HOLSymConst{\HOLTokenExists{}}\HOLBoundVar{C\sp{\prime}}. \HOLConst{GCONTEXT} \HOLBoundVar{C\sp{\prime}} \HOLSymConst{\HOLTokenConj{}} (\HOLBoundVar{P\sp{\prime}} \HOLSymConst{=} \HOLBoundVar{C\sp{\prime}} \HOLBoundVar{P}) \HOLSymConst{\HOLTokenConj{}} \HOLSymConst{\HOLTokenForall{}}\HOLBoundVar{Q}. \HOLFreeVar{C} (\HOLFreeVar{E} \HOLBoundVar{Q}) \HOLTokenTransBegin\HOLBoundVar{x}\HOLTokenTransEnd \HOLBoundVar{C\sp{\prime}} \HOLBoundVar{Q}

unfolding_lemma4:
\HOLTokenTurnstile{} \HOLConst{GCONTEXT} \HOLFreeVar{C} \HOLSymConst{\HOLTokenConj{}} \HOLConst{WGS} \HOLFreeVar{E} \HOLSymConst{\HOLTokenConj{}} \HOLConst{TRACE} ((\HOLFreeVar{C} \HOLSymConst{\HOLTokenCompose} \HOLConst{FUNPOW} \HOLFreeVar{E} \HOLFreeVar{n}) \HOLFreeVar{P}) \HOLFreeVar{xs} \HOLFreeVar{P\sp{\prime}} \HOLSymConst{\HOLTokenConj{}}
   \HOLConst{LENGTH} \HOLFreeVar{xs} \HOLSymConst{\HOLTokenLeq{}} \HOLFreeVar{n} \HOLSymConst{\HOLTokenImp{}}
   \HOLSymConst{\HOLTokenExists{}}\HOLBoundVar{C\sp{\prime}}.
     \HOLConst{GCONTEXT} \HOLBoundVar{C\sp{\prime}} \HOLSymConst{\HOLTokenConj{}} (\HOLFreeVar{P\sp{\prime}} \HOLSymConst{=} \HOLBoundVar{C\sp{\prime}} \HOLFreeVar{P}) \HOLSymConst{\HOLTokenConj{}}
     \HOLSymConst{\HOLTokenForall{}}\HOLBoundVar{Q}. \HOLConst{TRACE} ((\HOLFreeVar{C} \HOLSymConst{\HOLTokenCompose} \HOLConst{FUNPOW} \HOLFreeVar{E} \HOLFreeVar{n}) \HOLBoundVar{Q}) \HOLFreeVar{xs} (\HOLBoundVar{C\sp{\prime}} \HOLBoundVar{Q})
\end{alltt}
\end{lemma}

The purpose of \texttt{unfolding_lemma1} is to make sure the contraction
is preserved by $n$-times wrapping of contexts (\texttt{GCONTEXT}
actually), and the proof only need to precongruence property of the
`contracts' relation. (Thus any precongruence or congruence relation
fits the lemma). This lemma is directly used in the proof of main
lemma, \texttt{UNIQUE_SOLUTIONS_OF_CONTRACTIONS_LEMMA}.

The purpose of \texttt{unfolding_lemma2} is to make sure the first
transition from a weakly guarded expression do not come from the
inside of its variable(s). This is very important, because otherwise
we will not be able to freely change the value of process variable
(and still form a valid transition).  To prove this lemma, we have to
do induction into the structure of a weakly-guarded expressions, and
for the cases of sums, we have to do further inductions to look at
each of its summands.

The purpose of \texttt{unfolding_lemma3} is a simple wrapper for
\texttt{unfolding_lemma2}, to fit the use in
\texttt{unfolding_lemma4}, in which the weakly-guarded expression is
actually a composition of two expressions.

The purpose of \texttt{unfolding_lemma4} is to guarantee that, if the
process variable is hiding deeply enough in the weakly guarded
expressions, then after $n$ times transition, it's still not
touched. This actually completes the first half (most important part)
of the informal proof of Lemma 3.9 of \cite{sangiorgi2015equations}
(\texttt{UNIQUE_SOLUTIONS_OF_CONTRACTIONS_LEMMA}).

With all these lemmas (nicely separated out from the previous big
informal proofs), the whole ``unique solution of contractions''
theorem get proved formally.  Again, although we have only formalized
its single-variable case, extending it to multiple variable cases
informally is quite straightforward, so we can claim that, the formal
verification of this result is a \emph{success}.

In fact, we can also prove the ``unique solutions of expansions''
theorem in the same way, because in above proof, all used properties for
``contracts'' also hold for ``expands''.  Therefore by almost the same
steps, we have also proved the following results for expansion,
together with its main lemma:

\begin{theorem}{(Unique solution of expansions and its lemma)}
\label{thm:uniq-expansion}
\begin{alltt}
UNIQUE_SOLUTIONS_OF_EXPANSIONS_LEMMA:
\HOLTokenTurnstile{} (\HOLSymConst{\HOLTokenExists{}}\HOLBoundVar{E}. \HOLConst{WGS} \HOLBoundVar{E} \HOLSymConst{\HOLTokenConj{}} \HOLFreeVar{P} \HOLSymConst{\HOLTokenExpands{}} \HOLBoundVar{E} \HOLFreeVar{P} \HOLSymConst{\HOLTokenConj{}} \HOLFreeVar{Q} \HOLSymConst{\HOLTokenExpands{}} \HOLBoundVar{E} \HOLFreeVar{Q}) \HOLSymConst{\HOLTokenImp{}}
   \HOLSymConst{\HOLTokenForall{}}\HOLBoundVar{C}.
     \HOLConst{GCONTEXT} \HOLBoundVar{C} \HOLSymConst{\HOLTokenImp{}}
     (\HOLSymConst{\HOLTokenForall{}}\HOLBoundVar{l} \HOLBoundVar{R}.
        \HOLBoundVar{C} \HOLFreeVar{P} \HOLTokenWeakTransBegin\HOLConst{label} \HOLBoundVar{l}\HOLTokenWeakTransEnd \HOLBoundVar{R} \HOLSymConst{\HOLTokenImp{}}
        \HOLSymConst{\HOLTokenExists{}}\HOLBoundVar{C\sp{\prime}}.
          \HOLConst{GCONTEXT} \HOLBoundVar{C\sp{\prime}} \HOLSymConst{\HOLTokenConj{}} \HOLBoundVar{R} \HOLSymConst{\HOLTokenExpands{}} \HOLBoundVar{C\sp{\prime}} \HOLFreeVar{P} \HOLSymConst{\HOLTokenConj{}}
          (\HOLConst{WEAK_EQUIV} \HOLSymConst{\HOLTokenRCompose{}} (\HOLTokenLambda{}\HOLBoundVar{x} \HOLBoundVar{y}. \HOLBoundVar{x} \HOLTokenWeakTransBegin\HOLConst{label} \HOLBoundVar{l}\HOLTokenWeakTransEnd \HOLBoundVar{y})) (\HOLBoundVar{C} \HOLFreeVar{Q}) (\HOLBoundVar{C\sp{\prime}} \HOLFreeVar{Q})) \HOLSymConst{\HOLTokenConj{}}
     \HOLSymConst{\HOLTokenForall{}}\HOLBoundVar{R}.
       \HOLBoundVar{C} \HOLFreeVar{P} \HOLTokenWeakTransBegin\HOLSymConst{\ensuremath{\tau}}\HOLTokenWeakTransEnd \HOLBoundVar{R} \HOLSymConst{\HOLTokenImp{}}
       \HOLSymConst{\HOLTokenExists{}}\HOLBoundVar{C\sp{\prime}}.
         \HOLConst{GCONTEXT} \HOLBoundVar{C\sp{\prime}} \HOLSymConst{\HOLTokenConj{}} \HOLBoundVar{R} \HOLSymConst{\HOLTokenExpands{}} \HOLBoundVar{C\sp{\prime}} \HOLFreeVar{P} \HOLSymConst{\HOLTokenConj{}}
         (\HOLConst{WEAK_EQUIV} \HOLSymConst{\HOLTokenRCompose{}} \HOLConst{EPS}) (\HOLBoundVar{C} \HOLFreeVar{Q}) (\HOLBoundVar{C\sp{\prime}} \HOLFreeVar{Q})

UNIQUE_SOLUTIONS_OF_EXPANSIONS:
\HOLTokenTurnstile{} \HOLConst{WGS} \HOLFreeVar{E} \HOLSymConst{\HOLTokenImp{}} \HOLSymConst{\HOLTokenForall{}}\HOLBoundVar{P} \HOLBoundVar{Q}. \HOLBoundVar{P} \HOLSymConst{\HOLTokenExpands{}} \HOLFreeVar{E} \HOLBoundVar{P} \HOLSymConst{\HOLTokenConj{}} \HOLBoundVar{Q} \HOLSymConst{\HOLTokenExpands{}} \HOLFreeVar{E} \HOLBoundVar{Q} \HOLSymConst{\HOLTokenImp{}} \HOLBoundVar{P} \HOLSymConst{\HOLTokenWeakEQ} \HOLBoundVar{Q}
\end{alltt}
\end{theorem}

We have re-used 3 of the 4 above ``unfolding lemmas'', because in
their statements there's ``contracts'' at all. For the first unfolding
lemma, we have easily duplicated it into the following version for
``expands'':
\begin{alltt}
unfolding_lemma1':
\HOLTokenTurnstile{} \HOLConst{GCONTEXT} \HOLFreeVar{E} \HOLSymConst{\HOLTokenConj{}} \HOLConst{GCONTEXT} \HOLFreeVar{C} \HOLSymConst{\HOLTokenConj{}} \HOLFreeVar{P} \HOLSymConst{\HOLTokenExpands{}} \HOLFreeVar{E} \HOLFreeVar{P} \HOLSymConst{\HOLTokenImp{}}
   \HOLSymConst{\HOLTokenForall{}}\HOLBoundVar{n}. \HOLFreeVar{C} \HOLFreeVar{P} \HOLSymConst{\HOLTokenExpands{}} (\HOLFreeVar{C} \HOLSymConst{\HOLTokenCompose} \HOLConst{FUNPOW} \HOLFreeVar{E} \HOLBoundVar{n}) \HOLFreeVar{P}
\end{alltt}

But actually there's a much easier way to prove ``unique solutions of
expansions'' theorem, or the ``unique solution'' theorem for any
relation contained in ``contracts''. Here is the idea: (also
formalized)
\begin{proof}{(Easy proof for Theorem \ref{thm:uniq-expansion})}
If $P$ and $Q$ are both solutions of an expansion, i.e. $P \succeq_e
E[P]$ and $P \succeq_e E[Q]$, as we know expansion implies
contraction, so we have also $P \succeq_{bis}
E[P]$ and $P \succeq_{bis} E[Q]$.  By ``unique solutions of
contractions'' theorem, immediately we get $P \approx Q$.
\end{proof}

Here a natural question raised up: since we have also the theorem for ``unique
solution of expansions'', then why contractions? Is there any property
which holds for only contraction but expansion?  The answer is the
following completeness theorem (Theorem 3.13 in
\cite{sangiorgi2015equations}) that we cannot formalize right now
(because it holds only for a system of contractions but single
contraction):
\begin{theorem}{(completeness)}
Suppose $\mathcal{R}$ is a bisimulation. Then there is a system of
weakly-guarded pure contractions of which $\mathcal{R_1}$ and
$\mathcal{R_2}$ are solutions for $\succeq_{bis}$. (Here
$\mathcal{R}_i$ indicates the tuple obtained by projecting the pairs
in $\mathcal{R}$ on the $i$-th component ($i=1,2$).
\end{theorem}

We omit the proof here, while only quote the following remarks given
by Prof.\ Sangiorgi in his paper:

\begin{remark}
In the final step of the proof above, relation
$\overset{\hat{\mu_s}}{\Longrightarrow}$ comes from the definition of
weak bisimulation, and could not be replaced by
$\overset{\mu_s}{\Longrightarrow}$.
This explains why \emph{the completeness proof fails with expansion in
  place of contraction}.\footnote{However, Prof.\ Sangiorgi also said
  (on Oct 10, 2017) in a private mail that, ``Is the difference
  important for applications? I doubt. Contractions are a variation of
  expansion. However, completeness to me shows that contractions are
  the `right' relation. 
I wish I had further evidence for that.''}
\end{remark}

\section{Observational contraction}

In this thesis, we have slightly gone beyond the paper of
Prof.\ Sangiorgi \cite{sangiorgi2015equations} by finding a slightly ``better''
contraction relation which is real precongruence beside all
properties of the existing one, such that the
restriction on guarded sums can be removed from all
levels (CCS grammar, weakly guarded contractions, etc.). Fortunately
such a relation does exist and was found in just one day. The quick
finding is based on the following two principles:
\begin{enumerate}
\item Its definition must NOT be recursive (co-inductive), just like the definition of observation congruence ($\approx^c$);
\item It must be based on the existing `contracts' relation, which we
  believe it's the `right' one (become its completeness), just like weak equivalence ($approx$).
\end{enumerate}

We call this new contraction relation ``observational contraction''
with the symbol $\succeq^c_{bis}$ (the small letter $c$ indicates it's
a (pre)congruence):
(\texttt{OBS_contracts} in HOL),
here is its definition:
\begin{definition}{(Observational contraction)}
\label{def:obs-contraction}
Two processes $E$ and $E'$ has observational contraction relation
(called ``$E$ observational contracts $E'$'') if and only if
\begin{enumerate}
\item $E \overset{\mu}{\rightarrow} E_1$ implies that there is $E_2$
  with $E' \overset{\mu}{\rightarrow} E_2$ and $E_1
  \succeq_{\mathrm{bis}} E_2$,
\item $E' \overset{\mu}{\rightarrow} E_2$ implies that there is $E_1$
  with $E \overset{\mu}{\Rightarrow} E_1$ and $E_1 \approx E_2$.
\end{enumerate}
or formally:
\begin{alltt}
\HOLTokenTurnstile{} \HOLFreeVar{E} \HOLSymConst{\HOLTokenObsContracts} \HOLFreeVar{E\sp{\prime}} \HOLSymConst{\HOLTokenEquiv{}}
   \HOLSymConst{\HOLTokenForall{}}\HOLBoundVar{u}.
     (\HOLSymConst{\HOLTokenForall{}}\HOLBoundVar{E\sb{\mathrm{1}}}. \HOLFreeVar{E} \HOLTokenTransBegin\HOLBoundVar{u}\HOLTokenTransEnd \HOLBoundVar{E\sb{\mathrm{1}}} \HOLSymConst{\HOLTokenImp{}} \HOLSymConst{\HOLTokenExists{}}\HOLBoundVar{E\sb{\mathrm{2}}}. \HOLFreeVar{E\sp{\prime}} \HOLTokenTransBegin\HOLBoundVar{u}\HOLTokenTransEnd \HOLBoundVar{E\sb{\mathrm{2}}} \HOLSymConst{\HOLTokenConj{}} \HOLBoundVar{E\sb{\mathrm{1}}} \HOLSymConst{\HOLTokenContracts{}} \HOLBoundVar{E\sb{\mathrm{2}}}) \HOLSymConst{\HOLTokenConj{}}
     \HOLSymConst{\HOLTokenForall{}}\HOLBoundVar{E\sb{\mathrm{2}}}. \HOLFreeVar{E\sp{\prime}} \HOLTokenTransBegin\HOLBoundVar{u}\HOLTokenTransEnd \HOLBoundVar{E\sb{\mathrm{2}}} \HOLSymConst{\HOLTokenImp{}} \HOLSymConst{\HOLTokenExists{}}\HOLBoundVar{E\sb{\mathrm{1}}}. \HOLFreeVar{E} \HOLTokenWeakTransBegin\HOLBoundVar{u}\HOLTokenWeakTransEnd \HOLBoundVar{E\sb{\mathrm{1}}} \HOLSymConst{\HOLTokenConj{}} \HOLBoundVar{E\sb{\mathrm{1}}} \HOLSymConst{\HOLTokenWeakEQ} \HOLBoundVar{E\sb{\mathrm{2}}}
\end{alltt}
\end{definition}

The primary property we wanted from this new relation is to make sure
that it's preserved by direct sums. This is indeed true for this new
relation:
\begin{proposition}
Observational contraction is preserved by (direct) sums:
\begin{alltt}
OBS_contracts_PRESD_BY_SUM:
\HOLTokenTurnstile{} \HOLFreeVar{E\sb{\mathrm{1}}} \HOLSymConst{\HOLTokenObsContracts} \HOLFreeVar{E\sb{\mathrm{1}}\sp{\prime}} \HOLSymConst{\HOLTokenConj{}} \HOLFreeVar{E\sb{\mathrm{2}}} \HOLSymConst{\HOLTokenObsContracts} \HOLFreeVar{E\sb{\mathrm{2}}\sp{\prime}} \HOLSymConst{\HOLTokenImp{}} \HOLFreeVar{E\sb{\mathrm{1}}} \HOLSymConst{+} \HOLFreeVar{E\sb{\mathrm{2}}} \HOLSymConst{\HOLTokenObsContracts} \HOLFreeVar{E\sb{\mathrm{1}}\sp{\prime}} \HOLSymConst{+} \HOLFreeVar{E\sb{\mathrm{2}}\sp{\prime}}
\end{alltt}
\end{proposition}

Following standard techniques (especially the proof ideas for
``observational congruence''), we have also proved its preserving
properties on other CCS constructions:
\begin{proposition}{(Other properties of observational contraction)}
\begin{enumerate}
\item Observational contraction is substitutive by prefix
\begin{alltt}
OBS_contracts_SUBST_PREFIX:
\HOLTokenTurnstile{} \HOLFreeVar{E} \HOLSymConst{\HOLTokenObsContracts} \HOLFreeVar{E\sp{\prime}} \HOLSymConst{\HOLTokenImp{}} \HOLSymConst{\HOLTokenForall{}}\HOLBoundVar{u}. \HOLBoundVar{u}\HOLSymConst{..}\HOLFreeVar{E} \HOLSymConst{\HOLTokenObsContracts} \HOLBoundVar{u}\HOLSymConst{..}\HOLFreeVar{E\sp{\prime}}
\end{alltt}
\item Observational contraction is preserved by parallel composition:
\begin{alltt}
OBS_contracts_PRESD_BY_PAR:
\HOLTokenTurnstile{} \HOLFreeVar{E\sb{\mathrm{1}}} \HOLSymConst{\HOLTokenObsContracts} \HOLFreeVar{E\sb{\mathrm{1}}\sp{\prime}} \HOLSymConst{\HOLTokenConj{}} \HOLFreeVar{E\sb{\mathrm{2}}} \HOLSymConst{\HOLTokenObsContracts} \HOLFreeVar{E\sb{\mathrm{2}}\sp{\prime}} \HOLSymConst{\HOLTokenImp{}} \HOLFreeVar{E\sb{\mathrm{1}}} \HOLSymConst{\ensuremath{\parallel}} \HOLFreeVar{E\sb{\mathrm{2}}} \HOLSymConst{\HOLTokenObsContracts} \HOLFreeVar{E\sb{\mathrm{1}}\sp{\prime}} \HOLSymConst{\ensuremath{\parallel}} \HOLFreeVar{E\sb{\mathrm{2}}\sp{\prime}}
\end{alltt}
\item Observational contraction is substitutive by restrictions:
\begin{alltt}
OBS_contracts_SUBST_RESTR:
\HOLTokenTurnstile{} \HOLFreeVar{E} \HOLSymConst{\HOLTokenObsContracts} \HOLFreeVar{E\sp{\prime}} \HOLSymConst{\HOLTokenImp{}} \HOLSymConst{\HOLTokenForall{}}\HOLBoundVar{L}. \HOLSymConst{\ensuremath{\nu}} \HOLBoundVar{L} \HOLFreeVar{E} \HOLSymConst{\HOLTokenObsContracts} \HOLSymConst{\ensuremath{\nu}} \HOLBoundVar{L} \HOLFreeVar{E\sp{\prime}}
\end{alltt}
\item Observational contraction is substitutive by relabeling
  operators:
\begin{alltt}
OBS_contracts_SUBST_RELAB:
\HOLTokenTurnstile{} \HOLFreeVar{E} \HOLSymConst{\HOLTokenObsContracts} \HOLFreeVar{E\sp{\prime}} \HOLSymConst{\HOLTokenImp{}} \HOLSymConst{\HOLTokenForall{}}\HOLBoundVar{rf}. \HOLConst{relab} \HOLFreeVar{E} \HOLBoundVar{rf} \HOLSymConst{\HOLTokenObsContracts} \HOLConst{relab} \HOLFreeVar{E\sp{\prime}} \HOLBoundVar{rf}
\end{alltt}
\end{enumerate}
\end{proposition}

Putting all together, observational contraction relation is a (real)
precongruence:
\begin{lemma}
Observational contraction is precongruence, i.e. it's substitutive by
semantics contexts:
\begin{alltt}
OBS_contracts_SUBST_CONTEXT:
\HOLTokenTurnstile{} \HOLFreeVar{P} \HOLSymConst{\HOLTokenObsContracts} \HOLFreeVar{Q} \HOLSymConst{\HOLTokenImp{}} \HOLSymConst{\HOLTokenForall{}}\HOLBoundVar{E}. \HOLConst{CONTEXT} \HOLBoundVar{E} \HOLSymConst{\HOLTokenImp{}} \HOLBoundVar{E} \HOLFreeVar{P} \HOLSymConst{\HOLTokenObsContracts} \HOLBoundVar{E} \HOLFreeVar{Q}
\end{alltt}
or
\begin{alltt}
\HOLTokenTurnstile{} \HOLConst{precongruence} \HOLConst{OBS_contracts}\hfill[OBS_contracts_precongruence]
\end{alltt}
where
\begin{alltt}
\HOLTokenTurnstile{} \HOLConst{precongruence} \HOLFreeVar{R} \HOLSymConst{\HOLTokenEquiv{}}
   \HOLSymConst{\HOLTokenForall{}}\HOLBoundVar{x} \HOLBoundVar{y} \HOLBoundVar{ctx}. \HOLConst{CONTEXT} \HOLBoundVar{ctx} \HOLSymConst{\HOLTokenImp{}} \HOLFreeVar{R} \HOLBoundVar{x} \HOLBoundVar{y} \HOLSymConst{\HOLTokenImp{}} \HOLFreeVar{R} (\HOLBoundVar{ctx} \HOLBoundVar{x}) (\HOLBoundVar{ctx} \HOLBoundVar{y})
\end{alltt}
\end{lemma}

The other potential problematic property is the transitivity, but
fortunately this is also proven:
\begin{proposition}{(Transitivity of observational contraction)}
\begin{alltt}
\HOLTokenTurnstile{} \HOLFreeVar{E} \HOLSymConst{\HOLTokenObsContracts} \HOLFreeVar{E\sp{\prime}} \HOLSymConst{\HOLTokenConj{}} \HOLFreeVar{E\sp{\prime}} \HOLSymConst{\HOLTokenObsContracts} \HOLFreeVar{E\sp{\prime\prime}} \HOLSymConst{\HOLTokenImp{}} \HOLFreeVar{E} \HOLSymConst{\HOLTokenObsContracts} \HOLFreeVar{E\sp{\prime\prime}}\hfill[OBS_contracts_TRANS]
\end{alltt}
\end{proposition}
Together with the easily proved reflexivity, we show the
``observational contraction'' is indeed a pre-order:
\begin{proposition}
Observational contraction is a pre-order:
\begin{alltt}
\HOLTokenTurnstile{} \HOLConst{PreOrder} (\HOLSymConst{contracts})\hfill[contracts_PreOrder]
\end{alltt}
\end{proposition}

On the relationship with other relations, we have the following
results:
\begin{proposition}{(Observational contraction and other relations)}
\begin{enumerate}
\item Observational contraction implies contraction:
\begin{alltt}
\HOLTokenTurnstile{} \HOLFreeVar{E} \HOLSymConst{\HOLTokenObsContracts} \HOLFreeVar{E\sp{\prime}} \HOLSymConst{\HOLTokenImp{}} \HOLFreeVar{E} \HOLSymConst{\HOLTokenContracts{}} \HOLFreeVar{E\sp{\prime}}\hfill[OBS_contracts_IMP_contracts]
\end{alltt}
\item Observational contraction implies observational congruence:
\begin{alltt}
\HOLTokenTurnstile{} \HOLFreeVar{E} \HOLSymConst{\HOLTokenObsContracts} \HOLFreeVar{E\sp{\prime}} \HOLSymConst{\HOLTokenImp{}} \HOLFreeVar{E} \HOLSymConst{\HOLTokenObsCongr} \HOLFreeVar{E\sp{\prime}}\hfill[OBS_contracts_IMP_OBS_CONGR]
\end{alltt}
\end{enumerate}
\end{proposition}

Thus observational contraction is also contained in the
existing `contracts' relation, plus it's contained in observational
congruence. The overall relationships we know so far looks like this:
\begin{align*}
\succeq_e &\subset \; \succeq_{bis} \; \subset \; \approx \\
\succeq^c_{bis}  &\subset \; \succeq_{bis} \; \subset \; \approx \\
\succeq^c_{bis}  &\subset \; \approx^c \; \subset \; \approx
\end{align*}
Noticed that, the strict subset relationship is not proved in our
project, but they're indeed true (by showing some
counterexamples). Also, we don't know if there's a subset relationship
between expansion and observational contraction but we believe they
have no such relationship at all.

Also, so far we haven't found any relation which is \emph{coarser}
than $\succeq_{bis}$, this probably means it's the ``coarsest one'' in
some sense (although we don't no such results so far), and since
$\succeq^c_{bis}$ is contained in $\succeq_{bis}$, its ``unique
solution'' theorem is immediately available from the ``unique solution
of contractions'' theorem, just like the case of ``expansion'':
\begin{alltt}
UNIQUE_SOLUTIONS_OF_OBS_CONTRACTIONS':
\HOLTokenTurnstile{} \HOLConst{WGS} \HOLFreeVar{E} \HOLSymConst{\HOLTokenImp{}} \HOLSymConst{\HOLTokenForall{}}\HOLBoundVar{P} \HOLBoundVar{Q}. \HOLBoundVar{P} \HOLSymConst{\HOLTokenObsContracts} \HOLFreeVar{E} \HOLBoundVar{P} \HOLSymConst{\HOLTokenConj{}} \HOLBoundVar{Q} \HOLSymConst{\HOLTokenObsContracts} \HOLFreeVar{E} \HOLBoundVar{Q} \HOLSymConst{\HOLTokenImp{}} \HOLBoundVar{P} \HOLSymConst{\HOLTokenWeakEQ} \HOLBoundVar{Q}
\end{alltt}

However, we can do certainly much better: all restrictions on guarded
sums can be removed from the proof of ``unique solution
of contractions'' theorem, while we still have essentially the same
proof. What we have finally proved is the following one, together with
its main lemma:
\begin{alltt}
UNIQUE_SOLUTIONS_OF_OBS_CONTRACTIONS_LEMMA:
\HOLTokenTurnstile{} (\HOLSymConst{\HOLTokenExists{}}\HOLBoundVar{E}. \HOLConst{WG} \HOLBoundVar{E} \HOLSymConst{\HOLTokenConj{}} \HOLFreeVar{P} \HOLSymConst{\HOLTokenObsContracts} \HOLBoundVar{E} \HOLFreeVar{P} \HOLSymConst{\HOLTokenConj{}} \HOLFreeVar{Q} \HOLSymConst{\HOLTokenObsContracts} \HOLBoundVar{E} \HOLFreeVar{Q}) \HOLSymConst{\HOLTokenImp{}}
   \HOLSymConst{\HOLTokenForall{}}\HOLBoundVar{C}.
     \HOLConst{CONTEXT} \HOLBoundVar{C} \HOLSymConst{\HOLTokenImp{}}
     (\HOLSymConst{\HOLTokenForall{}}\HOLBoundVar{l} \HOLBoundVar{R}.
        \HOLBoundVar{C} \HOLFreeVar{P} \HOLTokenWeakTransBegin\HOLConst{label} \HOLBoundVar{l}\HOLTokenWeakTransEnd \HOLBoundVar{R} \HOLSymConst{\HOLTokenImp{}}
        \HOLSymConst{\HOLTokenExists{}}\HOLBoundVar{C\sp{\prime}}.
          \HOLConst{CONTEXT} \HOLBoundVar{C\sp{\prime}} \HOLSymConst{\HOLTokenConj{}} \HOLBoundVar{R} \HOLSymConst{\HOLTokenContracts{}} \HOLBoundVar{C\sp{\prime}} \HOLFreeVar{P} \HOLSymConst{\HOLTokenConj{}}
          (\HOLConst{WEAK_EQUIV} \HOLSymConst{\HOLTokenRCompose{}} (\HOLTokenLambda{}\HOLBoundVar{x} \HOLBoundVar{y}. \HOLBoundVar{x} \HOLTokenWeakTransBegin\HOLConst{label} \HOLBoundVar{l}\HOLTokenWeakTransEnd \HOLBoundVar{y})) (\HOLBoundVar{C} \HOLFreeVar{Q}) (\HOLBoundVar{C\sp{\prime}} \HOLFreeVar{Q})) \HOLSymConst{\HOLTokenConj{}}
     \HOLSymConst{\HOLTokenForall{}}\HOLBoundVar{R}.
       \HOLBoundVar{C} \HOLFreeVar{P} \HOLTokenWeakTransBegin\HOLSymConst{\ensuremath{\tau}}\HOLTokenWeakTransEnd \HOLBoundVar{R} \HOLSymConst{\HOLTokenImp{}}
       \HOLSymConst{\HOLTokenExists{}}\HOLBoundVar{C\sp{\prime}}.
         \HOLConst{CONTEXT} \HOLBoundVar{C\sp{\prime}} \HOLSymConst{\HOLTokenConj{}} \HOLBoundVar{R} \HOLSymConst{\HOLTokenContracts{}} \HOLBoundVar{C\sp{\prime}} \HOLFreeVar{P} \HOLSymConst{\HOLTokenConj{}}
         (\HOLConst{WEAK_EQUIV} \HOLSymConst{\HOLTokenRCompose{}} \HOLConst{EPS}) (\HOLBoundVar{C} \HOLFreeVar{Q}) (\HOLBoundVar{C\sp{\prime}} \HOLFreeVar{Q})

UNIQUE_SOLUTIONS_OF_OBS_CONTRACTIONS:
\HOLTokenTurnstile{} \HOLConst{WG} \HOLFreeVar{E} \HOLSymConst{\HOLTokenImp{}} \HOLSymConst{\HOLTokenForall{}}\HOLBoundVar{P} \HOLBoundVar{Q}. \HOLBoundVar{P} \HOLSymConst{\HOLTokenObsContracts} \HOLFreeVar{E} \HOLBoundVar{P} \HOLSymConst{\HOLTokenConj{}} \HOLBoundVar{Q} \HOLSymConst{\HOLTokenObsContracts} \HOLFreeVar{E} \HOLBoundVar{Q} \HOLSymConst{\HOLTokenImp{}} \HOLBoundVar{P} \HOLSymConst{\HOLTokenWeakEQ} \HOLBoundVar{Q}
\end{alltt}

The last theorem above is a beautiful and concise result, because it requires only a
normal weakly-guarded expression, which is the same as in Milner's
``unique solution of equations'' theorem for strong equivalence.

To finish this formal proof, we have almost used everything, please
notice that, in the conclusion of above main lemma, it still requires
the normal ``contracts'' relation.  This is because our definition of
``observation contraction'' is not recursive: after passing the first
transition, it ``downgrades'' to the normal contraction, thus we need
to use the properties of normal contractions to finish the rest of the
proof.  What's also needed, is the following properties related to
trace:
\begin{alltt}
OBS_contracts_AND_TRACE_tau:
\HOLTokenTurnstile{} \HOLFreeVar{E} \HOLSymConst{\HOLTokenObsContracts} \HOLFreeVar{E\sp{\prime}} \HOLSymConst{\HOLTokenImp{}}
   \HOLSymConst{\HOLTokenForall{}}\HOLBoundVar{xs} \HOLBoundVar{l} \HOLBoundVar{E\sb{\mathrm{1}}}.
     \HOLConst{TRACE} \HOLFreeVar{E} \HOLBoundVar{xs} \HOLBoundVar{E\sb{\mathrm{1}}} \HOLSymConst{\HOLTokenConj{}} \HOLConst{NO_LABEL} \HOLBoundVar{xs} \HOLSymConst{\HOLTokenImp{}}
     \HOLSymConst{\HOLTokenExists{}}\HOLBoundVar{xs\sp{\prime}} \HOLBoundVar{E\sb{\mathrm{2}}}.
       \HOLConst{TRACE} \HOLFreeVar{E\sp{\prime}} \HOLBoundVar{xs\sp{\prime}} \HOLBoundVar{E\sb{\mathrm{2}}} \HOLSymConst{\HOLTokenConj{}} \HOLBoundVar{E\sb{\mathrm{1}}} \HOLSymConst{\HOLTokenContracts{}} \HOLBoundVar{E\sb{\mathrm{2}}} \HOLSymConst{\HOLTokenConj{}} \HOLConst{LENGTH} \HOLBoundVar{xs\sp{\prime}} \HOLSymConst{\HOLTokenLeq{}} \HOLConst{LENGTH} \HOLBoundVar{xs} \HOLSymConst{\HOLTokenConj{}}
       \HOLConst{NO_LABEL} \HOLBoundVar{xs\sp{\prime}}

OBS_contracts_AND_TRACE_label:
\HOLTokenTurnstile{} \HOLFreeVar{E} \HOLSymConst{\HOLTokenObsContracts} \HOLFreeVar{E\sp{\prime}} \HOLSymConst{\HOLTokenImp{}}
   \HOLSymConst{\HOLTokenForall{}}\HOLBoundVar{xs} \HOLBoundVar{l} \HOLBoundVar{E\sb{\mathrm{1}}}.
     \HOLConst{TRACE} \HOLFreeVar{E} \HOLBoundVar{xs} \HOLBoundVar{E\sb{\mathrm{1}}} \HOLSymConst{\HOLTokenConj{}} \HOLConst{UNIQUE_LABEL} (\HOLConst{label} \HOLBoundVar{l}) \HOLBoundVar{xs} \HOLSymConst{\HOLTokenImp{}}
     \HOLSymConst{\HOLTokenExists{}}\HOLBoundVar{xs\sp{\prime}} \HOLBoundVar{E\sb{\mathrm{2}}}.
       \HOLConst{TRACE} \HOLFreeVar{E\sp{\prime}} \HOLBoundVar{xs\sp{\prime}} \HOLBoundVar{E\sb{\mathrm{2}}} \HOLSymConst{\HOLTokenConj{}} \HOLBoundVar{E\sb{\mathrm{1}}} \HOLSymConst{\HOLTokenContracts{}} \HOLBoundVar{E\sb{\mathrm{2}}} \HOLSymConst{\HOLTokenConj{}} \HOLConst{LENGTH} \HOLBoundVar{xs\sp{\prime}} \HOLSymConst{\HOLTokenLeq{}} \HOLConst{LENGTH} \HOLBoundVar{xs} \HOLSymConst{\HOLTokenConj{}}
       \HOLConst{UNIQUE_LABEL} (\HOLConst{label} \HOLBoundVar{l}) \HOLBoundVar{xs\sp{\prime}}

OBS_contracts_WEAK_TRANS_label':
\HOLTokenTurnstile{} \HOLFreeVar{E} \HOLSymConst{\HOLTokenObsContracts} \HOLFreeVar{E\sp{\prime}} \HOLSymConst{\HOLTokenImp{}}
   \HOLSymConst{\HOLTokenForall{}}\HOLBoundVar{l} \HOLBoundVar{E\sb{\mathrm{2}}}. \HOLFreeVar{E\sp{\prime}} \HOLTokenWeakTransBegin\HOLConst{label} \HOLBoundVar{l}\HOLTokenWeakTransEnd \HOLBoundVar{E\sb{\mathrm{2}}} \HOLSymConst{\HOLTokenImp{}} \HOLSymConst{\HOLTokenExists{}}\HOLBoundVar{E\sb{\mathrm{1}}}. \HOLFreeVar{E} \HOLTokenWeakTransBegin\HOLConst{label} \HOLBoundVar{l}\HOLTokenWeakTransEnd \HOLBoundVar{E\sb{\mathrm{1}}} \HOLSymConst{\HOLTokenConj{}} \HOLBoundVar{E\sb{\mathrm{1}}} \HOLSymConst{\HOLTokenWeakEQ} \HOLBoundVar{E\sb{\mathrm{2}}}

OBS_contracts_EPS':
\HOLTokenTurnstile{} \HOLFreeVar{E} \HOLSymConst{\HOLTokenObsContracts} \HOLFreeVar{E\sp{\prime}} \HOLSymConst{\HOLTokenImp{}} \HOLSymConst{\HOLTokenForall{}}\HOLBoundVar{E\sb{\mathrm{2}}}. \HOLFreeVar{E\sp{\prime}} \HOLSymConst{\HOLTokenEPS} \HOLBoundVar{E\sb{\mathrm{2}}} \HOLSymConst{\HOLTokenImp{}} \HOLSymConst{\HOLTokenExists{}}\HOLBoundVar{E\sb{\mathrm{1}}}. \HOLFreeVar{E} \HOLSymConst{\HOLTokenEPS} \HOLBoundVar{E\sb{\mathrm{1}}} \HOLSymConst{\HOLTokenConj{}} \HOLBoundVar{E\sb{\mathrm{1}}} \HOLSymConst{\HOLTokenWeakEQ} \HOLBoundVar{E\sb{\mathrm{2}}}
\end{alltt}
Noticed how ``observational contraction'' becomes normal contraction
in above statements, after passing the first transition step.

\subsection{coarsest precongruence contained in $\succeq_{bis}$}

So far we haven't examined any practical application in which our new
``observational contraction'' is meaningful and useful. However, it's
natural to ask if there's another relation having all the properties
of ``observational contraction'' while containing more process pairs
(in another word, coarser than $\succeq^c_{bis}$.  We want to prove
our finding is the \emph{coarsest} precongruence contained in $\succeq_{bis}$, just like observation
congruence is the coarsest congruence contained in weak
equivalence. Following the theory of congruence we have built for CCS,
this proposition can be reduce to the following simpler form:
\begin{proposition}
The ``observational contraction'' ($\succeq^c_{bis}$) is the coarsest
precongruence contained in $\succeq_{bis}$ if and only if:
\begin{equation}
\forall p\, q.\, p \succeq^c_{bis} q \Longleftrightarrow \forall r.\, p + r \succeq_{bis} q + r.
\end{equation}
\end{proposition}

In the following formalizations, we only how to reduce the initial
concern to above form, and formally prove above proposition from
left to right (which is quite easy).

First of all, we can define a new relation (\texttt{C_contracts}) as the context closure of
$\succeq_{bis}$ (also recall the definition of context closure),
\begin{alltt}
\HOLTokenTurnstile{} \HOLConst{C_contracts} \HOLSymConst{=} \HOLConst{CC} (\HOLSymConst{contracts})\hfill[C_contracts]

\HOLTokenTurnstile{} \HOLConst{CC} \HOLFreeVar{R} \HOLSymConst{=} (\HOLTokenLambda{}\HOLBoundVar{g} \HOLBoundVar{h}. \HOLSymConst{\HOLTokenForall{}}\HOLBoundVar{c}. \HOLConst{CONTEXT} \HOLBoundVar{c} \HOLSymConst{\HOLTokenImp{}} \HOLFreeVar{R} (\HOLBoundVar{c} \HOLBoundVar{g}) (\HOLBoundVar{c} \HOLBoundVar{h}))\hfill[CC_def]
\end{alltt}
Then we can immediately show that, this new relation is indeed a
(real) precongruence (because any congruence closure is automatically
precongruence):
\begin{alltt}
\HOLTokenTurnstile{} \HOLConst{precongruence} \HOLConst{C_contracts}\hfill[C_contracts_precongruence]
\end{alltt}

We can also easily prove that, above new relation contains
$\succeq^c_{bis}$, because it's the coarsest one (recall a congruence closure is always coarsest):
\begin{alltt}
OBS_contracts_IMP_C_contracts:
\HOLTokenTurnstile{} \HOLFreeVar{p} \HOLSymConst{\HOLTokenObsContracts} \HOLFreeVar{q} \HOLSymConst{\HOLTokenImp{}} \HOLConst{C_contracts} \HOLFreeVar{p} \HOLFreeVar{q}
\end{alltt}

Next, we define another new relation, which seems much larger, as the
closure closed only on direct sums: (here we don't even know if it's a
precongruence)
\begin{alltt}
SUM_contracts:
\HOLTokenTurnstile{} \HOLConst{SUM_contracts} \HOLSymConst{=} (\HOLTokenLambda{}\HOLBoundVar{p} \HOLBoundVar{q}. \HOLSymConst{\HOLTokenForall{}}\HOLBoundVar{r}. \HOLBoundVar{p} \HOLSymConst{+} \HOLBoundVar{r} \HOLSymConst{\HOLTokenContracts{}} \HOLBoundVar{q} \HOLSymConst{+} \HOLBoundVar{r})
\end{alltt}
However, what we do know is, this new relation contains the previous
one: (this actually proves the ``left-to-right part'' of our earlier proposition!)
\begin{alltt}
C_contracts_IMP_SUM_contracts:
\HOLTokenTurnstile{} \HOLConst{C_contracts} \HOLFreeVar{p} \HOLFreeVar{q} \HOLSymConst{\HOLTokenImp{}} \HOLConst{SUM_contracts} \HOLFreeVar{p} \HOLFreeVar{q}
\end{alltt}

Now we have a relation chain: $\text{OBS_contracts}
\subseteq \text{C_contracts} \subseteq \text{SUM_contracts}$. To prove
our ``observational contraction''  is the ``coarsest one'', i.e. it
coincides with ``C_contracts'', it's sufficient to prove
``SUM_contracts'' implies ``OBS_contracts'':
\begin{alltt}
\HOLinline{\HOLSymConst{\HOLTokenForall{}}\HOLBoundVar{p} \HOLBoundVar{q}. \HOLConst{SUM_contracts} \HOLBoundVar{p} \HOLBoundVar{q} \HOLSymConst{\HOLTokenImp{}} \HOLBoundVar{p} \HOLSymConst{\HOLTokenObsContracts} \HOLBoundVar{q}}
\end{alltt}
which equals to the ``right-to-left'' part of our earlier
proposition.   This proof is actually quite like the case of
``coarsest congruence contained in $\approx$'' (Section
\ref{sec:coarsest-congruence}). Whenever the two process didn't use up
all available actions, we can prove the following easy version:
\begin{alltt}
COARSEST_PRECONGR_THM:
\HOLTokenTurnstile{} \HOLConst{free_action} \HOLFreeVar{p} \HOLSymConst{\HOLTokenConj{}} \HOLConst{free_action} \HOLFreeVar{q} \HOLSymConst{\HOLTokenImp{}} (\HOLFreeVar{p} \HOLSymConst{\HOLTokenObsContracts} \HOLFreeVar{q} \HOLSymConst{\HOLTokenEquiv{}} \HOLConst{SUM_contracts} \HOLFreeVar{p} \HOLFreeVar{q})
\end{alltt}
\begin{alltt}
COARSEST_PRECONGR_THM':
\HOLTokenTurnstile{} \HOLConst{free_action} \HOLFreeVar{p} \HOLSymConst{\HOLTokenConj{}} \HOLConst{free_action} \HOLFreeVar{q} \HOLSymConst{\HOLTokenImp{}} (\HOLFreeVar{p} \HOLSymConst{\HOLTokenObsContracts} \HOLFreeVar{q} \HOLSymConst{\HOLTokenEquiv{}} \HOLSymConst{\HOLTokenForall{}}\HOLBoundVar{r}. \HOLFreeVar{p} \HOLSymConst{+} \HOLBoundVar{r} \HOLSymConst{\HOLTokenContracts{}} \HOLFreeVar{q} \HOLSymConst{+} \HOLBoundVar{r})
\end{alltt}

A further extension of this result for finite state CCS (based on Klop
functions) should be also possible, but in this project we didn't try
to prove it due to time limits.

\cleardoublepage

\chapter{Towards multi-variable equations}

As we have explained in the Introduction chapter, the existing CCS
datatype in our formalization project can be seen as
\emph{expressions} with some free variables. Suppose $E$ is such an
expression with free variables $X_1, X_2, \ldots, X_n$ (there must be
finite number of them because any CCS expression can only be finitary
given there's no infinite sums, which is the current case),
abbreviated as $E[\tilde{X}]$, then $X_1 \sim E[\tilde{X}]$ is CCS
equations for strong equivalence.  And suppose we have a system of
expressions $\tilde{E}$, then $\tilde{X} \sim \tilde{E}[\tilde{X}]$ is
a system of CCS equations.

To actually prove things about these equations, we need to at least
two devices: 1) the ability to find all free variables in any CCS
term, and 2) the ability to replace any free variable with specific
CCS process (which shouldn't further contain any free variable).

The following recursive function can be used to retrieve a set of free
variables in any CCS term:
\begin{alltt}
\HOLConst{FV} \HOLConst{nil} \HOLSymConst{=} \HOLSymConst{\HOLTokenEmpty{}}
\HOLConst{FV} (\HOLFreeVar{u}\HOLSymConst{..}\HOLFreeVar{p}) \HOLSymConst{=} \HOLConst{FV} \HOLFreeVar{p}
\HOLConst{FV} (\HOLFreeVar{p} \HOLSymConst{+} \HOLFreeVar{q}) \HOLSymConst{=} \HOLConst{FV} \HOLFreeVar{p} \HOLSymConst{\HOLTokenUnion{}} \HOLConst{FV} \HOLFreeVar{q}
\HOLConst{FV} (\HOLFreeVar{p} \HOLSymConst{\ensuremath{\parallel}} \HOLFreeVar{q}) \HOLSymConst{=} \HOLConst{FV} \HOLFreeVar{p} \HOLSymConst{\HOLTokenUnion{}} \HOLConst{FV} \HOLFreeVar{q}
\HOLConst{FV} (\HOLSymConst{\ensuremath{\nu}} \HOLFreeVar{L} \HOLFreeVar{p}) \HOLSymConst{=} \HOLConst{FV} \HOLFreeVar{p}
\HOLConst{FV} (\HOLConst{relab} \HOLFreeVar{p} \HOLFreeVar{rf}) \HOLSymConst{=} \HOLConst{FV} \HOLFreeVar{p}
\HOLConst{FV} (\HOLConst{var} \HOLFreeVar{X}) \HOLSymConst{=} \HOLTokenLeftbrace{}\HOLFreeVar{X}\HOLTokenRightbrace{}
\HOLConst{FV} (\HOLConst{rec} \HOLFreeVar{X} \HOLFreeVar{p}) \HOLSymConst{=} \HOLConst{FV} \HOLFreeVar{p} \HOLConst{DIFF} \HOLTokenLeftbrace{}\HOLFreeVar{X}\HOLTokenRightbrace{}
\end{alltt}
Similarity, we can also define a function for retrieving the set of
all bound variables:
\begin{alltt}
\HOLConst{BV} \HOLConst{nil} \HOLSymConst{=} \HOLSymConst{\HOLTokenEmpty{}}
\HOLConst{BV} (\HOLFreeVar{u}\HOLSymConst{..}\HOLFreeVar{p}) \HOLSymConst{=} \HOLConst{BV} \HOLFreeVar{p}
\HOLConst{BV} (\HOLFreeVar{p} \HOLSymConst{+} \HOLFreeVar{q}) \HOLSymConst{=} \HOLConst{BV} \HOLFreeVar{p} \HOLSymConst{\HOLTokenUnion{}} \HOLConst{BV} \HOLFreeVar{q}
\HOLConst{BV} (\HOLFreeVar{p} \HOLSymConst{\ensuremath{\parallel}} \HOLFreeVar{q}) \HOLSymConst{=} \HOLConst{BV} \HOLFreeVar{p} \HOLSymConst{\HOLTokenUnion{}} \HOLConst{BV} \HOLFreeVar{q}
\HOLConst{BV} (\HOLSymConst{\ensuremath{\nu}} \HOLFreeVar{L} \HOLFreeVar{p}) \HOLSymConst{=} \HOLConst{BV} \HOLFreeVar{p}
\HOLConst{BV} (\HOLConst{relab} \HOLFreeVar{p} \HOLFreeVar{rf}) \HOLSymConst{=} \HOLConst{BV} \HOLFreeVar{p}
\HOLConst{BV} (\HOLConst{var} \HOLFreeVar{X}) \HOLSymConst{=} \HOLSymConst{\HOLTokenEmpty{}}
\HOLConst{BV} (\HOLConst{rec} \HOLFreeVar{X} \HOLFreeVar{p}) \HOLSymConst{=} \HOLFreeVar{X} \HOLConst{INSERT} \HOLConst{BV} \HOLFreeVar{p}
\end{alltt}
However, the set of bound variables are not quite interesting, because a
variable bounded somewhere in the term may be also free at another
position in the term, thus the set of free and bound variables are not disjoint.

Sometimes we need to if a CCS term is an expression or process, which
can be checked by the following function based on \HOLinline{\HOLConst{FV}}:
\begin{alltt}
IS_PROC_def:
\HOLTokenTurnstile{} \HOLConst{IS_PROC} \HOLFreeVar{E} \HOLSymConst{\HOLTokenEquiv{}} (\HOLConst{FV} \HOLFreeVar{E} \HOLSymConst{=} \HOLSymConst{\HOLTokenEmpty{}})
\end{alltt}

To replace the free variables in CCS terms with processes, we must
clearly specify the correspondence between each variable and its
replacement, thus the resulting (single) CCS process after all its free variables
have been replaced, should be in forms like
$E\{\tilde{p}/\tilde{X}\}$. In case some variables in $\tilde{X}$
didn't appear in $E$, it's natural to simple ignore the replacement
operation, or the operation has no effects.  Using constants defined
in HOL's \texttt{listTheory}, such a substitution operation could be
defined simply upon the existing \HOLinline{\HOLConst{CCS_Subst}}. However there will
be all kinds of issues regarding the order of substitutions and it's
hard to prove that the resulting process is unique for whatever
permutations of the variable list.  Another solution is to use finite
maps to represent variable substitutions, to get rid of the ordering
issues (suggested by Konrad Slind from HOL community):
\begin{alltt}
\HOLConst{CCS_Subst1} \HOLConst{nil} \HOLFreeVar{fm} \HOLSymConst{=} \HOLConst{nil}
\HOLConst{CCS_Subst1} (\HOLFreeVar{u}\HOLSymConst{..}\HOLFreeVar{E}) \HOLFreeVar{fm} \HOLSymConst{=} \HOLFreeVar{u}\HOLSymConst{..}\HOLConst{CCS_Subst1} \HOLFreeVar{E} \HOLFreeVar{fm}
\HOLConst{CCS_Subst1} (\HOLFreeVar{E\sb{\mathrm{1}}} \HOLSymConst{+} \HOLFreeVar{E\sb{\mathrm{2}}}) \HOLFreeVar{fm} \HOLSymConst{=} \HOLConst{CCS_Subst1} \HOLFreeVar{E\sb{\mathrm{1}}} \HOLFreeVar{fm} \HOLSymConst{+} \HOLConst{CCS_Subst1} \HOLFreeVar{E\sb{\mathrm{2}}} \HOLFreeVar{fm}
\HOLConst{CCS_Subst1} (\HOLFreeVar{E\sb{\mathrm{1}}} \HOLSymConst{\ensuremath{\parallel}} \HOLFreeVar{E\sb{\mathrm{2}}}) \HOLFreeVar{fm} \HOLSymConst{=} \HOLConst{CCS_Subst1} \HOLFreeVar{E\sb{\mathrm{1}}} \HOLFreeVar{fm} \HOLSymConst{\ensuremath{\parallel}} \HOLConst{CCS_Subst1} \HOLFreeVar{E\sb{\mathrm{2}}} \HOLFreeVar{fm}
\HOLConst{CCS_Subst1} (\HOLSymConst{\ensuremath{\nu}} \HOLFreeVar{L} \HOLFreeVar{E}) \HOLFreeVar{fm} \HOLSymConst{=} \HOLSymConst{\ensuremath{\nu}} \HOLFreeVar{L} (\HOLConst{CCS_Subst1} \HOLFreeVar{E} \HOLFreeVar{fm})
\HOLConst{CCS_Subst1} (\HOLConst{relab} \HOLFreeVar{E} \HOLFreeVar{rf}) \HOLFreeVar{fm} \HOLSymConst{=} \HOLConst{relab} (\HOLConst{CCS_Subst1} \HOLFreeVar{E} \HOLFreeVar{fm}) \HOLFreeVar{rf}
\HOLConst{CCS_Subst1} (\HOLConst{var} \HOLFreeVar{Y}) \HOLFreeVar{fm} \HOLSymConst{=} \HOLKeyword{if} \HOLFreeVar{Y} \HOLSymConst{\HOLTokenIn{}} \HOLConst{FDOM} \HOLFreeVar{fm} \HOLKeyword{then} \HOLFreeVar{fm} \HOLConst{'} \HOLFreeVar{Y} \HOLKeyword{else} \HOLConst{var} \HOLFreeVar{Y}
\HOLConst{CCS_Subst1} (\HOLConst{rec} \HOLFreeVar{Y} \HOLFreeVar{E}) \HOLFreeVar{fm} \HOLSymConst{=}
\HOLKeyword{if} \HOLFreeVar{Y} \HOLSymConst{\HOLTokenIn{}} \HOLConst{FDOM} \HOLFreeVar{fm} \HOLKeyword{then} \HOLConst{rec} \HOLFreeVar{Y} \HOLFreeVar{E} \HOLKeyword{else} \HOLConst{rec} \HOLFreeVar{Y} (\HOLConst{CCS_Subst1} \HOLFreeVar{E} \HOLFreeVar{fm})
\end{alltt}

To further do substitutions on a list of CCS terms, we can define
\HOLinline{\HOLConst{CCS_Subst2}} which depends on \HOLinline{\HOLConst{CCS_Subst1}}:
\begin{alltt}
CCS_Subst2_def:
\HOLTokenTurnstile{} \HOLConst{CCS_Subst2} \HOLFreeVar{Es} \HOLFreeVar{fm} \HOLSymConst{=} \HOLConst{MAP} (\HOLTokenLambda{}\HOLBoundVar{e}. \HOLConst{CCS_Subst1} \HOLBoundVar{e} \HOLFreeVar{fm}) \HOLFreeVar{Es}
\end{alltt}
Here we have used the constant \HOLinline{\HOLConst{MAP}}, which has the following
recursive definition:
\begin{alltt}
\HOLConst{MAP} \HOLFreeVar{f} [] \HOLSymConst{=} []
\HOLConst{MAP} \HOLFreeVar{f} (\HOLFreeVar{h}\HOLSymConst{::}\HOLFreeVar{t}) \HOLSymConst{=} \HOLFreeVar{f} \HOLFreeVar{h}\HOLSymConst{::}\HOLConst{MAP} \HOLFreeVar{f} \HOLFreeVar{t}
\end{alltt}
Thus it will call \HOLinline{\HOLConst{CCS_Subst1}} with the same list of variables
and their substitutions on each given CCS term, and then collect the
resulting processes into a returning list. The type of
\HOLinline{\HOLConst{CCS_Subst2}} is
\begin{alltt}
\texttt{:('a, 'b) CCS list -> ('a, 'b) CCS list -> 'a list -> ('a, 'b) CCS list}
\end{alltt}

To define weakly guarded expressions, say $E$, we must make sure that all free
variables in $E$ are weakly guarded. If we ignore all other free
variables and focus on just one of them, say, $X$, then it's possible
to define weak guardedness of any CCS term in the following way:
\begin{alltt}
weakly_guarded1_def:
\HOLTokenTurnstile{} \HOLConst{weakly_guarded1} \HOLFreeVar{E} \HOLSymConst{\HOLTokenEquiv{}}
   \HOLSymConst{\HOLTokenForall{}}\HOLBoundVar{X}. \HOLBoundVar{X} \HOLSymConst{\HOLTokenIn{}} \HOLConst{FV} \HOLFreeVar{E} \HOLSymConst{\HOLTokenImp{}} \HOLSymConst{\HOLTokenForall{}}\HOLBoundVar{e}. \HOLConst{CONTEXT} \HOLBoundVar{e} \HOLSymConst{\HOLTokenConj{}} (\HOLBoundVar{e} (\HOLConst{var} \HOLBoundVar{X}) \HOLSymConst{=} \HOLFreeVar{E}) \HOLSymConst{\HOLTokenImp{}} \HOLConst{WG} \HOLBoundVar{e}
\end{alltt}
Similarily we can also define other guardedness concepts in the same
way. With these new devices, it's possible to formalize all ``unique solution of
equations'' theorems with multi-variable equations, without touching
existing formalization framework.

\chapter{Conclusions}

In this thesis project, we have further formalized Milner's Calculus of
Communicating Systems (CCS) using HOL theorem prover (HOL4).
Beside classical results like the properties and algebraic laws of three
equivalence relations (strong/weak bisimilarities and observational
congruence), this work also includes a comprehensive formalization of
 ``bisimulation up to'', expansions and contractions, a theory of
 congruence for CCS, up to several versions of the ``unique solution
 of equations (or contractions)''
theorem.
Some of the concepts and theorems were introduced by
Prof.\ Davide Sangiorgi in his recent paper
\cite{sangiorgi2015equations}, and during this thesis we have further
introduced new concept (observational contraction) and proved an more
elegant form of its ``unique solution of contraction'' theorem.
Therefore, to some extent, this thesis work has touched
the current frontier in the research of concurrency theory and process algebra.

Although we chose to focus on single-variable equations in all
``unique solutions'' theorems, to minimize the development efforts,
the resulting
formal proofs actually have the same steps with the informal proofs,
therefore we have confidence that the related informal proofs were all
correct (so is our formalization).  An extension
to multi-variable cases is possible, without touching the CCS grammar (datatype)
definition, as the free variables in CCS terms can be treated as
equation variables, although this is not the current standard viewpoint.

Our formalization is based on the work done by Prof.\ Monica Nesi
during 1992-1995 on Hol88 theorem prover, and by porting them into
latest HOL4, we have successfully preserved this important work and
make sure it's availability in the future (by submitting them into
HOL's official code base). The use of HOL4's co-inductive relation and
many builtin theories has demonstrated that HOL4's advantages over
other theorem provers on the formalization of CCS or other process algebras.

\section{Further directions}

The \emph{proof engineering} research in Concurrency Theory is far
from fully resolved. Although we have quickly achieve a more-than-ever
depth on the formalization of a simple process algebra (CCS), by
covering several deep theorems, some important foundations in the work are still
not perfectly built. And the author is planning to make further
perfections in the future. Below is a list of plans:
\begin{enumerate}
\item \textbf{Infinite sum of CCS processes}. We want to precisely
  implement Milner's original CCS theory in which infinite sum of processes is
  included. Infinite sum is necessary for the
  proof of some theoretical results, and it's also needed for the
  encoding of Value-Passing CCS into Pure CCS. Currently this is
  limited by the datatype package of HOL4.  Instead of manually define
  an CCS datatype with infinite sums in higher order logic, the author
  would like to improve HOL4's datatype defining abilities by
  implementing \cite{traytel2012foundational} (or porting its existing
  implementation in Isabelle/HOL), a compositional (co) datatype
  support based on category theory.
\item \textbf{Extending to multi-variable equations}. Currently we
  have only proved all ``unique solutions of equations/contractions''
  theorems for single-variable equations. It is possible to prove the
  same theorems for systems of multi-variable equations in
  current framework.  The congruence of equivalence with respect to
  recursive constructions, and supporting lemma for multi-variable
  substitutions must be formally proved first.
\item \textbf{Decision procedures for equivalence checking}. 
We want to implement the symbolic equivalence checking for CCS
processes, and turn theorem prover into a model checking utility like the
Concurrency Workbench (CWB). Instead of implementing the underlying
algorithms directly in Standard ML, we would like to reduce the
problem into BDD (Binary Decision Diagram) problems and call HOL's BDD
library for the actual model checking.
\end{enumerate}

\chapter{References}

This chapter contains all definitions and theorems we have proved in
this thesis project. They're separated by theories with reasonable
names, ordered by their logical dependencies. 
 
Some theorems are generated from existing
theorems in forward way, they're also exported because from the view
of HOL there's no difference between them and other manually proved
theorems in backward ways. Thus there're some duplications in the
sense that some theorems are just other theorems in separated (or
combined) forms.

On the other side, each theorem has only its statement in HOL presented
here. Their formal proofs are NOT presented here. (This is different
with \TeX exporting facility found in Isabelle, which exports both the
statements and (usually human-readable) proofs into papers) This
project has roughly 20,000 lines of proof scripts, which are exported
into about 100 pages. Please keep in mind that, the number of these
pages has no strict relationship with the complexity of the theorems
being proved: it's possible for a very complicated theorem to have
several thousands of line of proof code (which is not a good style
though) but when exporting it into \TeX pages, it's just a few lines!
In this thesis project, the author tends to break large theorems into
small pieces and prove them increasingly, and most of the meaningful
intermediate theorems were also saved and exported.

Also noticed that, due to issues in HOL's \TeX exporting facility,
some theorems have slightly different forms with the same theroems
appeared in previous chapters, with Unicode symbols replaced by their
original ASCII-based names.

\newcommand{\HOLCCSDate}{01 Dicembre 2017}
\newcommand{\HOLCCSTime}{16:53}
\begin{SaveVerbatim}{HOLCCSDatatypesCCS}
\HOLFreeVar{CCS} =
    \HOLConst{nil}
  \HOLTokenBar{} \HOLConst{var} 'a
  \HOLTokenBar{} \HOLConst{prefix} ('b \HOLTyOp{Action}) (('a, 'b) \HOLTyOp{CCS})
  \HOLTokenBar{} (\HOLSymConst{+}) (('a, 'b) \HOLTyOp{CCS}) (('a, 'b) \HOLTyOp{CCS})
  \HOLTokenBar{} \HOLConst{par} (('a, 'b) \HOLTyOp{CCS}) (('a, 'b) \HOLTyOp{CCS})
  \HOLTokenBar{} \HOLConst{\ensuremath{\nu}} ('b \HOLTyOp{Label} \HOLTokenTransEnd \HOLTyOp{bool}) (('a, 'b) \HOLTyOp{CCS})
  \HOLTokenBar{} \HOLConst{relab} (('a, 'b) \HOLTyOp{CCS}) ('b \HOLTyOp{Relabeling})
  \HOLTokenBar{} \HOLConst{rec} 'a (('a, 'b) \HOLTyOp{CCS})
\end{SaveVerbatim}
\newcommand{\HOLCCSDatatypesCCS}{\UseVerbatim{HOLCCSDatatypesCCS}}
\begin{SaveVerbatim}{HOLCCSDatatypesLabel}
\HOLFreeVar{Label} = \HOLConst{name} 'b \HOLTokenBar{} \HOLConst{coname} 'b
\end{SaveVerbatim}
\newcommand{\HOLCCSDatatypesLabel}{\UseVerbatim{HOLCCSDatatypesLabel}}
\newcommand{\HOLCCSDatatypes}{
\HOLCCSDatatypesCCS\HOLCCSDatatypesLabel}
\begin{SaveVerbatim}{HOLCCSDefinitionsALLXXIDENTICALXXdef}
\HOLTokenTurnstile{} \HOLSymConst{\HOLTokenForall{}}\HOLBoundVar{t}. \HOLConst{ALL_IDENTICAL} \HOLBoundVar{t} \HOLSymConst{\HOLTokenEquiv{}} \HOLSymConst{\HOLTokenExists{}}\HOLBoundVar{x}. \HOLSymConst{\HOLTokenForall{}}\HOLBoundVar{y}. \HOLConst{MEM} \HOLBoundVar{y} \HOLBoundVar{t} \HOLSymConst{\HOLTokenImp{}} (\HOLBoundVar{y} \HOLSymConst{=} \HOLBoundVar{x})
\end{SaveVerbatim}
\newcommand{\HOLCCSDefinitionsALLXXIDENTICALXXdef}{\UseVerbatim{HOLCCSDefinitionsALLXXIDENTICALXXdef}}
\begin{SaveVerbatim}{HOLCCSDefinitionsALLXXPROCXXdef}
\HOLTokenTurnstile{} \HOLSymConst{\HOLTokenForall{}}\HOLBoundVar{Es}. \HOLConst{ALL_PROC} \HOLBoundVar{Es} \HOLSymConst{\HOLTokenEquiv{}} \HOLConst{EVERY} \HOLConst{IS_PROC} \HOLBoundVar{Es}
\end{SaveVerbatim}
\newcommand{\HOLCCSDefinitionsALLXXPROCXXdef}{\UseVerbatim{HOLCCSDefinitionsALLXXPROCXXdef}}
\begin{SaveVerbatim}{HOLCCSDefinitionsApplyXXRelabXXdef}
\HOLTokenTurnstile{} (\HOLSymConst{\HOLTokenForall{}}\HOLBoundVar{l}. \HOLConst{Apply_Relab} [] \HOLBoundVar{l} \HOLSymConst{=} \HOLBoundVar{l}) \HOLSymConst{\HOLTokenConj{}}
   \HOLSymConst{\HOLTokenForall{}}\HOLBoundVar{newold} \HOLBoundVar{ls} \HOLBoundVar{l}.
     \HOLConst{Apply_Relab} (\HOLBoundVar{newold}\HOLSymConst{::}\HOLBoundVar{ls}) \HOLBoundVar{l} \HOLSymConst{=}
     \HOLKeyword{if} \HOLConst{SND} \HOLBoundVar{newold} \HOLSymConst{=} \HOLBoundVar{l} \HOLKeyword{then} \HOLConst{FST} \HOLBoundVar{newold}
     \HOLKeyword{else} \HOLKeyword{if} \HOLConst{COMPL} (\HOLConst{SND} \HOLBoundVar{newold}) \HOLSymConst{=} \HOLBoundVar{l} \HOLKeyword{then} \HOLConst{COMPL} (\HOLConst{FST} \HOLBoundVar{newold})
     \HOLKeyword{else} \HOLConst{Apply_Relab} \HOLBoundVar{ls} \HOLBoundVar{l}
\end{SaveVerbatim}
\newcommand{\HOLCCSDefinitionsApplyXXRelabXXdef}{\UseVerbatim{HOLCCSDefinitionsApplyXXRelabXXdef}}
\begin{SaveVerbatim}{HOLCCSDefinitionsBNXXdef}
\HOLTokenTurnstile{} (\HOLSymConst{\HOLTokenForall{}}\HOLBoundVar{J}. \HOLConst{BN} \HOLConst{nil} \HOLBoundVar{J} \HOLSymConst{=} \HOLTokenLeftbrace{}\HOLTokenRightbrace{}) \HOLSymConst{\HOLTokenConj{}} (\HOLSymConst{\HOLTokenForall{}}\HOLBoundVar{u} \HOLBoundVar{p} \HOLBoundVar{J}. \HOLConst{BN} (\HOLBoundVar{u}\HOLSymConst{..}\HOLBoundVar{p}) \HOLBoundVar{J} \HOLSymConst{=} \HOLConst{BN} \HOLBoundVar{p} \HOLBoundVar{J}) \HOLSymConst{\HOLTokenConj{}}
   (\HOLSymConst{\HOLTokenForall{}}\HOLBoundVar{p} \HOLBoundVar{q} \HOLBoundVar{J}. \HOLConst{BN} (\HOLBoundVar{p} \HOLSymConst{+} \HOLBoundVar{q}) \HOLBoundVar{J} \HOLSymConst{=} \HOLConst{BN} \HOLBoundVar{p} \HOLBoundVar{J} \HOLConst{\HOLTokenUnion{}} \HOLConst{BN} \HOLBoundVar{q} \HOLBoundVar{J}) \HOLSymConst{\HOLTokenConj{}}
   (\HOLSymConst{\HOLTokenForall{}}\HOLBoundVar{p} \HOLBoundVar{q} \HOLBoundVar{J}. \HOLConst{BN} (\HOLBoundVar{p} \HOLSymConst{\ensuremath{\parallel}} \HOLBoundVar{q}) \HOLBoundVar{J} \HOLSymConst{=} \HOLConst{BN} \HOLBoundVar{p} \HOLBoundVar{J} \HOLConst{\HOLTokenUnion{}} \HOLConst{BN} \HOLBoundVar{q} \HOLBoundVar{J}) \HOLSymConst{\HOLTokenConj{}}
   (\HOLSymConst{\HOLTokenForall{}}\HOLBoundVar{L} \HOLBoundVar{p} \HOLBoundVar{J}. \HOLConst{BN} (\HOLConst{\ensuremath{\nu}} \HOLBoundVar{L} \HOLBoundVar{p}) \HOLBoundVar{J} \HOLSymConst{=} \HOLConst{BN} \HOLBoundVar{p} \HOLBoundVar{J} \HOLConst{\HOLTokenUnion{}} \HOLBoundVar{L}) \HOLSymConst{\HOLTokenConj{}}
   (\HOLSymConst{\HOLTokenForall{}}\HOLBoundVar{p} \HOLBoundVar{rf} \HOLBoundVar{J}. \HOLConst{BN} (\HOLConst{relab} \HOLBoundVar{p} \HOLBoundVar{rf}) \HOLBoundVar{J} \HOLSymConst{=} \HOLConst{BN} \HOLBoundVar{p} \HOLBoundVar{J}) \HOLSymConst{\HOLTokenConj{}}
   (\HOLSymConst{\HOLTokenForall{}}\HOLBoundVar{X} \HOLBoundVar{J}. \HOLConst{BN} (\HOLConst{var} \HOLBoundVar{X}) \HOLBoundVar{J} \HOLSymConst{=} \HOLTokenLeftbrace{}\HOLTokenRightbrace{}) \HOLSymConst{\HOLTokenConj{}}
   \HOLSymConst{\HOLTokenForall{}}\HOLBoundVar{X} \HOLBoundVar{p} \HOLBoundVar{J}.
     \HOLConst{BN} (\HOLConst{rec} \HOLBoundVar{X} \HOLBoundVar{p}) \HOLBoundVar{J} \HOLSymConst{=}
     \HOLKeyword{if} \HOLConst{MEM} \HOLBoundVar{X} \HOLBoundVar{J} \HOLKeyword{then}
       \HOLConst{FN} (\HOLConst{CCS_Subst} \HOLBoundVar{p} (\HOLConst{rec} \HOLBoundVar{X} \HOLBoundVar{p}) \HOLBoundVar{X}) (\HOLConst{DELETE_ELEMENT} \HOLBoundVar{X} \HOLBoundVar{J})
     \HOLKeyword{else} \HOLTokenLeftbrace{}\HOLTokenRightbrace{}
\end{SaveVerbatim}
\newcommand{\HOLCCSDefinitionsBNXXdef}{\UseVerbatim{HOLCCSDefinitionsBNXXdef}}
\begin{SaveVerbatim}{HOLCCSDefinitionsboundXXnamesXXdef}
\HOLTokenTurnstile{} \HOLSymConst{\HOLTokenForall{}}\HOLBoundVar{p}. \HOLConst{bound_names} \HOLBoundVar{p} \HOLSymConst{=} \HOLConst{BN} \HOLBoundVar{p} (\HOLConst{SET_TO_LIST} (\HOLConst{BV} \HOLBoundVar{p}))
\end{SaveVerbatim}
\newcommand{\HOLCCSDefinitionsboundXXnamesXXdef}{\UseVerbatim{HOLCCSDefinitionsboundXXnamesXXdef}}
\begin{SaveVerbatim}{HOLCCSDefinitionsBVXXdef}
\HOLTokenTurnstile{} (\HOLConst{BV} \HOLConst{nil} \HOLSymConst{=} \HOLTokenLeftbrace{}\HOLTokenRightbrace{}) \HOLSymConst{\HOLTokenConj{}} (\HOLSymConst{\HOLTokenForall{}}\HOLBoundVar{u} \HOLBoundVar{p}. \HOLConst{BV} (\HOLBoundVar{u}\HOLSymConst{..}\HOLBoundVar{p}) \HOLSymConst{=} \HOLConst{BV} \HOLBoundVar{p}) \HOLSymConst{\HOLTokenConj{}}
   (\HOLSymConst{\HOLTokenForall{}}\HOLBoundVar{p} \HOLBoundVar{q}. \HOLConst{BV} (\HOLBoundVar{p} \HOLSymConst{+} \HOLBoundVar{q}) \HOLSymConst{=} \HOLConst{BV} \HOLBoundVar{p} \HOLConst{\HOLTokenUnion{}} \HOLConst{BV} \HOLBoundVar{q}) \HOLSymConst{\HOLTokenConj{}}
   (\HOLSymConst{\HOLTokenForall{}}\HOLBoundVar{p} \HOLBoundVar{q}. \HOLConst{BV} (\HOLBoundVar{p} \HOLSymConst{\ensuremath{\parallel}} \HOLBoundVar{q}) \HOLSymConst{=} \HOLConst{BV} \HOLBoundVar{p} \HOLConst{\HOLTokenUnion{}} \HOLConst{BV} \HOLBoundVar{q}) \HOLSymConst{\HOLTokenConj{}}
   (\HOLSymConst{\HOLTokenForall{}}\HOLBoundVar{L} \HOLBoundVar{p}. \HOLConst{BV} (\HOLConst{\ensuremath{\nu}} \HOLBoundVar{L} \HOLBoundVar{p}) \HOLSymConst{=} \HOLConst{BV} \HOLBoundVar{p}) \HOLSymConst{\HOLTokenConj{}} (\HOLSymConst{\HOLTokenForall{}}\HOLBoundVar{p} \HOLBoundVar{rf}. \HOLConst{BV} (\HOLConst{relab} \HOLBoundVar{p} \HOLBoundVar{rf}) \HOLSymConst{=} \HOLConst{BV} \HOLBoundVar{p}) \HOLSymConst{\HOLTokenConj{}}
   (\HOLSymConst{\HOLTokenForall{}}\HOLBoundVar{X}. \HOLConst{BV} (\HOLConst{var} \HOLBoundVar{X}) \HOLSymConst{=} \HOLTokenLeftbrace{}\HOLTokenRightbrace{}) \HOLSymConst{\HOLTokenConj{}} \HOLSymConst{\HOLTokenForall{}}\HOLBoundVar{X} \HOLBoundVar{p}. \HOLConst{BV} (\HOLConst{rec} \HOLBoundVar{X} \HOLBoundVar{p}) \HOLSymConst{=} \HOLBoundVar{X} \HOLConst{INSERT} \HOLConst{BV} \HOLBoundVar{p}
\end{SaveVerbatim}
\newcommand{\HOLCCSDefinitionsBVXXdef}{\UseVerbatim{HOLCCSDefinitionsBVXXdef}}
\begin{SaveVerbatim}{HOLCCSDefinitionsCCSXXSubstOneXXdef}
\HOLTokenTurnstile{} (\HOLSymConst{\HOLTokenForall{}}\HOLBoundVar{fm}. \HOLConst{CCS_Subst1} \HOLConst{nil} \HOLBoundVar{fm} \HOLSymConst{=} \HOLConst{nil}) \HOLSymConst{\HOLTokenConj{}}
   (\HOLSymConst{\HOLTokenForall{}}\HOLBoundVar{u} \HOLBoundVar{E} \HOLBoundVar{fm}. \HOLConst{CCS_Subst1} (\HOLBoundVar{u}\HOLSymConst{..}\HOLBoundVar{E}) \HOLBoundVar{fm} \HOLSymConst{=} \HOLBoundVar{u}\HOLSymConst{..}\HOLConst{CCS_Subst1} \HOLBoundVar{E} \HOLBoundVar{fm}) \HOLSymConst{\HOLTokenConj{}}
   (\HOLSymConst{\HOLTokenForall{}}\HOLBoundVar{E\sb{\mathrm{1}}} \HOLBoundVar{E\sb{\mathrm{2}}} \HOLBoundVar{fm}.
      \HOLConst{CCS_Subst1} (\HOLBoundVar{E\sb{\mathrm{1}}} \HOLSymConst{+} \HOLBoundVar{E\sb{\mathrm{2}}}) \HOLBoundVar{fm} \HOLSymConst{=}
      \HOLConst{CCS_Subst1} \HOLBoundVar{E\sb{\mathrm{1}}} \HOLBoundVar{fm} \HOLSymConst{+} \HOLConst{CCS_Subst1} \HOLBoundVar{E\sb{\mathrm{2}}} \HOLBoundVar{fm}) \HOLSymConst{\HOLTokenConj{}}
   (\HOLSymConst{\HOLTokenForall{}}\HOLBoundVar{E\sb{\mathrm{1}}} \HOLBoundVar{E\sb{\mathrm{2}}} \HOLBoundVar{fm}.
      \HOLConst{CCS_Subst1} (\HOLBoundVar{E\sb{\mathrm{1}}} \HOLSymConst{\ensuremath{\parallel}} \HOLBoundVar{E\sb{\mathrm{2}}}) \HOLBoundVar{fm} \HOLSymConst{=}
      \HOLConst{CCS_Subst1} \HOLBoundVar{E\sb{\mathrm{1}}} \HOLBoundVar{fm} \HOLSymConst{\ensuremath{\parallel}} \HOLConst{CCS_Subst1} \HOLBoundVar{E\sb{\mathrm{2}}} \HOLBoundVar{fm}) \HOLSymConst{\HOLTokenConj{}}
   (\HOLSymConst{\HOLTokenForall{}}\HOLBoundVar{L} \HOLBoundVar{E} \HOLBoundVar{fm}. \HOLConst{CCS_Subst1} (\HOLConst{\ensuremath{\nu}} \HOLBoundVar{L} \HOLBoundVar{E}) \HOLBoundVar{fm} \HOLSymConst{=} \HOLConst{\ensuremath{\nu}} \HOLBoundVar{L} (\HOLConst{CCS_Subst1} \HOLBoundVar{E} \HOLBoundVar{fm})) \HOLSymConst{\HOLTokenConj{}}
   (\HOLSymConst{\HOLTokenForall{}}\HOLBoundVar{E} \HOLBoundVar{rf} \HOLBoundVar{fm}.
      \HOLConst{CCS_Subst1} (\HOLConst{relab} \HOLBoundVar{E} \HOLBoundVar{rf}) \HOLBoundVar{fm} \HOLSymConst{=} \HOLConst{relab} (\HOLConst{CCS_Subst1} \HOLBoundVar{E} \HOLBoundVar{fm}) \HOLBoundVar{rf}) \HOLSymConst{\HOLTokenConj{}}
   (\HOLSymConst{\HOLTokenForall{}}\HOLBoundVar{Y} \HOLBoundVar{fm}.
      \HOLConst{CCS_Subst1} (\HOLConst{var} \HOLBoundVar{Y}) \HOLBoundVar{fm} \HOLSymConst{=}
      \HOLKeyword{if} \HOLBoundVar{Y} \HOLConst{\HOLTokenIn{}} \HOLConst{FDOM} \HOLBoundVar{fm} \HOLKeyword{then} \HOLBoundVar{fm} \HOLConst{'} \HOLBoundVar{Y} \HOLKeyword{else} \HOLConst{var} \HOLBoundVar{Y}) \HOLSymConst{\HOLTokenConj{}}
   \HOLSymConst{\HOLTokenForall{}}\HOLBoundVar{Y} \HOLBoundVar{E} \HOLBoundVar{fm}.
     \HOLConst{CCS_Subst1} (\HOLConst{rec} \HOLBoundVar{Y} \HOLBoundVar{E}) \HOLBoundVar{fm} \HOLSymConst{=}
     \HOLKeyword{if} \HOLBoundVar{Y} \HOLConst{\HOLTokenIn{}} \HOLConst{FDOM} \HOLBoundVar{fm} \HOLKeyword{then} \HOLConst{rec} \HOLBoundVar{Y} \HOLBoundVar{E} \HOLKeyword{else} \HOLConst{rec} \HOLBoundVar{Y} (\HOLConst{CCS_Subst1} \HOLBoundVar{E} \HOLBoundVar{fm})
\end{SaveVerbatim}
\newcommand{\HOLCCSDefinitionsCCSXXSubstOneXXdef}{\UseVerbatim{HOLCCSDefinitionsCCSXXSubstOneXXdef}}
\begin{SaveVerbatim}{HOLCCSDefinitionsCCSXXSubstTwoXXdef}
\HOLTokenTurnstile{} \HOLSymConst{\HOLTokenForall{}}\HOLBoundVar{Es} \HOLBoundVar{fm}. \HOLConst{CCS_Subst2} \HOLBoundVar{Es} \HOLBoundVar{fm} \HOLSymConst{=} \HOLConst{MAP} (\HOLTokenLambda{}\HOLBoundVar{e}. \HOLConst{CCS_Subst1} \HOLBoundVar{e} \HOLBoundVar{fm}) \HOLBoundVar{Es}
\end{SaveVerbatim}
\newcommand{\HOLCCSDefinitionsCCSXXSubstTwoXXdef}{\UseVerbatim{HOLCCSDefinitionsCCSXXSubstTwoXXdef}}
\begin{SaveVerbatim}{HOLCCSDefinitionsCCSXXSubstXXdef}
\HOLTokenTurnstile{} (\HOLSymConst{\HOLTokenForall{}}\HOLBoundVar{E\sp{\prime}} \HOLBoundVar{X}. \HOLConst{CCS_Subst} \HOLConst{nil} \HOLBoundVar{E\sp{\prime}} \HOLBoundVar{X} \HOLSymConst{=} \HOLConst{nil}) \HOLSymConst{\HOLTokenConj{}}
   (\HOLSymConst{\HOLTokenForall{}}\HOLBoundVar{u} \HOLBoundVar{E} \HOLBoundVar{E\sp{\prime}} \HOLBoundVar{X}. \HOLConst{CCS_Subst} (\HOLBoundVar{u}\HOLSymConst{..}\HOLBoundVar{E}) \HOLBoundVar{E\sp{\prime}} \HOLBoundVar{X} \HOLSymConst{=} \HOLBoundVar{u}\HOLSymConst{..}\HOLConst{CCS_Subst} \HOLBoundVar{E} \HOLBoundVar{E\sp{\prime}} \HOLBoundVar{X}) \HOLSymConst{\HOLTokenConj{}}
   (\HOLSymConst{\HOLTokenForall{}}\HOLBoundVar{E\sb{\mathrm{1}}} \HOLBoundVar{E\sb{\mathrm{2}}} \HOLBoundVar{E\sp{\prime}} \HOLBoundVar{X}.
      \HOLConst{CCS_Subst} (\HOLBoundVar{E\sb{\mathrm{1}}} \HOLSymConst{+} \HOLBoundVar{E\sb{\mathrm{2}}}) \HOLBoundVar{E\sp{\prime}} \HOLBoundVar{X} \HOLSymConst{=}
      \HOLConst{CCS_Subst} \HOLBoundVar{E\sb{\mathrm{1}}} \HOLBoundVar{E\sp{\prime}} \HOLBoundVar{X} \HOLSymConst{+} \HOLConst{CCS_Subst} \HOLBoundVar{E\sb{\mathrm{2}}} \HOLBoundVar{E\sp{\prime}} \HOLBoundVar{X}) \HOLSymConst{\HOLTokenConj{}}
   (\HOLSymConst{\HOLTokenForall{}}\HOLBoundVar{E\sb{\mathrm{1}}} \HOLBoundVar{E\sb{\mathrm{2}}} \HOLBoundVar{E\sp{\prime}} \HOLBoundVar{X}.
      \HOLConst{CCS_Subst} (\HOLBoundVar{E\sb{\mathrm{1}}} \HOLSymConst{\ensuremath{\parallel}} \HOLBoundVar{E\sb{\mathrm{2}}}) \HOLBoundVar{E\sp{\prime}} \HOLBoundVar{X} \HOLSymConst{=}
      \HOLConst{CCS_Subst} \HOLBoundVar{E\sb{\mathrm{1}}} \HOLBoundVar{E\sp{\prime}} \HOLBoundVar{X} \HOLSymConst{\ensuremath{\parallel}} \HOLConst{CCS_Subst} \HOLBoundVar{E\sb{\mathrm{2}}} \HOLBoundVar{E\sp{\prime}} \HOLBoundVar{X}) \HOLSymConst{\HOLTokenConj{}}
   (\HOLSymConst{\HOLTokenForall{}}\HOLBoundVar{L} \HOLBoundVar{E} \HOLBoundVar{E\sp{\prime}} \HOLBoundVar{X}.
      \HOLConst{CCS_Subst} (\HOLConst{\ensuremath{\nu}} \HOLBoundVar{L} \HOLBoundVar{E}) \HOLBoundVar{E\sp{\prime}} \HOLBoundVar{X} \HOLSymConst{=} \HOLConst{\ensuremath{\nu}} \HOLBoundVar{L} (\HOLConst{CCS_Subst} \HOLBoundVar{E} \HOLBoundVar{E\sp{\prime}} \HOLBoundVar{X})) \HOLSymConst{\HOLTokenConj{}}
   (\HOLSymConst{\HOLTokenForall{}}\HOLBoundVar{E} \HOLBoundVar{f} \HOLBoundVar{E\sp{\prime}} \HOLBoundVar{X}.
      \HOLConst{CCS_Subst} (\HOLConst{relab} \HOLBoundVar{E} \HOLBoundVar{f}) \HOLBoundVar{E\sp{\prime}} \HOLBoundVar{X} \HOLSymConst{=} \HOLConst{relab} (\HOLConst{CCS_Subst} \HOLBoundVar{E} \HOLBoundVar{E\sp{\prime}} \HOLBoundVar{X}) \HOLBoundVar{f}) \HOLSymConst{\HOLTokenConj{}}
   (\HOLSymConst{\HOLTokenForall{}}\HOLBoundVar{Y} \HOLBoundVar{E\sp{\prime}} \HOLBoundVar{X}.
      \HOLConst{CCS_Subst} (\HOLConst{var} \HOLBoundVar{Y}) \HOLBoundVar{E\sp{\prime}} \HOLBoundVar{X} \HOLSymConst{=} \HOLKeyword{if} \HOLBoundVar{Y} \HOLSymConst{=} \HOLBoundVar{X} \HOLKeyword{then} \HOLBoundVar{E\sp{\prime}} \HOLKeyword{else} \HOLConst{var} \HOLBoundVar{Y}) \HOLSymConst{\HOLTokenConj{}}
   \HOLSymConst{\HOLTokenForall{}}\HOLBoundVar{Y} \HOLBoundVar{E} \HOLBoundVar{E\sp{\prime}} \HOLBoundVar{X}.
     \HOLConst{CCS_Subst} (\HOLConst{rec} \HOLBoundVar{Y} \HOLBoundVar{E}) \HOLBoundVar{E\sp{\prime}} \HOLBoundVar{X} \HOLSymConst{=}
     \HOLKeyword{if} \HOLBoundVar{Y} \HOLSymConst{=} \HOLBoundVar{X} \HOLKeyword{then} \HOLConst{rec} \HOLBoundVar{Y} \HOLBoundVar{E} \HOLKeyword{else} \HOLConst{rec} \HOLBoundVar{Y} (\HOLConst{CCS_Subst} \HOLBoundVar{E} \HOLBoundVar{E\sp{\prime}} \HOLBoundVar{X})
\end{SaveVerbatim}
\newcommand{\HOLCCSDefinitionsCCSXXSubstXXdef}{\UseVerbatim{HOLCCSDefinitionsCCSXXSubstXXdef}}
\begin{SaveVerbatim}{HOLCCSDefinitionsCOMPLXXACTXXdef}
\HOLTokenTurnstile{} (\HOLSymConst{\HOLTokenForall{}}\HOLBoundVar{l}. \HOLConst{COMPL} (\HOLConst{label} \HOLBoundVar{l}) \HOLSymConst{=} \HOLConst{label} (\HOLConst{COMPL} \HOLBoundVar{l})) \HOLSymConst{\HOLTokenConj{}} (\HOLConst{COMPL} \HOLConst{\ensuremath{\tau}} \HOLSymConst{=} \HOLConst{\ensuremath{\tau}})
\end{SaveVerbatim}
\newcommand{\HOLCCSDefinitionsCOMPLXXACTXXdef}{\UseVerbatim{HOLCCSDefinitionsCOMPLXXACTXXdef}}
\begin{SaveVerbatim}{HOLCCSDefinitionsCOMPLXXLABXXdef}
\HOLTokenTurnstile{} (\HOLSymConst{\HOLTokenForall{}}\HOLBoundVar{s}. \HOLConst{COMPL} (\HOLConst{name} \HOLBoundVar{s}) \HOLSymConst{=} \HOLConst{coname} \HOLBoundVar{s}) \HOLSymConst{\HOLTokenConj{}}
   \HOLSymConst{\HOLTokenForall{}}\HOLBoundVar{s}. \HOLConst{COMPL} (\HOLConst{coname} \HOLBoundVar{s}) \HOLSymConst{=} \HOLConst{name} \HOLBoundVar{s}
\end{SaveVerbatim}
\newcommand{\HOLCCSDefinitionsCOMPLXXLABXXdef}{\UseVerbatim{HOLCCSDefinitionsCOMPLXXLABXXdef}}
\begin{SaveVerbatim}{HOLCCSDefinitionsDELETEXXELEMENTXXdef}
\HOLTokenTurnstile{} (\HOLSymConst{\HOLTokenForall{}}\HOLBoundVar{e}. \HOLConst{DELETE_ELEMENT} \HOLBoundVar{e} [] \HOLSymConst{=} []) \HOLSymConst{\HOLTokenConj{}}
   \HOLSymConst{\HOLTokenForall{}}\HOLBoundVar{e} \HOLBoundVar{x} \HOLBoundVar{l}.
     \HOLConst{DELETE_ELEMENT} \HOLBoundVar{e} (\HOLBoundVar{x}\HOLSymConst{::}\HOLBoundVar{l}) \HOLSymConst{=}
     \HOLKeyword{if} \HOLBoundVar{e} \HOLSymConst{=} \HOLBoundVar{x} \HOLKeyword{then} \HOLConst{DELETE_ELEMENT} \HOLBoundVar{e} \HOLBoundVar{l} \HOLKeyword{else} \HOLBoundVar{x}\HOLSymConst{::}\HOLConst{DELETE_ELEMENT} \HOLBoundVar{e} \HOLBoundVar{l}
\end{SaveVerbatim}
\newcommand{\HOLCCSDefinitionsDELETEXXELEMENTXXdef}{\UseVerbatim{HOLCCSDefinitionsDELETEXXELEMENTXXdef}}
\begin{SaveVerbatim}{HOLCCSDefinitionsfreeXXnamesXXdef}
\HOLTokenTurnstile{} \HOLSymConst{\HOLTokenForall{}}\HOLBoundVar{p}. \HOLConst{free_names} \HOLBoundVar{p} \HOLSymConst{=} \HOLConst{FN} \HOLBoundVar{p} (\HOLConst{SET_TO_LIST} (\HOLConst{BV} \HOLBoundVar{p}))
\end{SaveVerbatim}
\newcommand{\HOLCCSDefinitionsfreeXXnamesXXdef}{\UseVerbatim{HOLCCSDefinitionsfreeXXnamesXXdef}}
\begin{SaveVerbatim}{HOLCCSDefinitionsFVXXdef}
\HOLTokenTurnstile{} (\HOLConst{FV} \HOLConst{nil} \HOLSymConst{=} \HOLTokenLeftbrace{}\HOLTokenRightbrace{}) \HOLSymConst{\HOLTokenConj{}} (\HOLSymConst{\HOLTokenForall{}}\HOLBoundVar{u} \HOLBoundVar{p}. \HOLConst{FV} (\HOLBoundVar{u}\HOLSymConst{..}\HOLBoundVar{p}) \HOLSymConst{=} \HOLConst{FV} \HOLBoundVar{p}) \HOLSymConst{\HOLTokenConj{}}
   (\HOLSymConst{\HOLTokenForall{}}\HOLBoundVar{p} \HOLBoundVar{q}. \HOLConst{FV} (\HOLBoundVar{p} \HOLSymConst{+} \HOLBoundVar{q}) \HOLSymConst{=} \HOLConst{FV} \HOLBoundVar{p} \HOLConst{\HOLTokenUnion{}} \HOLConst{FV} \HOLBoundVar{q}) \HOLSymConst{\HOLTokenConj{}}
   (\HOLSymConst{\HOLTokenForall{}}\HOLBoundVar{p} \HOLBoundVar{q}. \HOLConst{FV} (\HOLBoundVar{p} \HOLSymConst{\ensuremath{\parallel}} \HOLBoundVar{q}) \HOLSymConst{=} \HOLConst{FV} \HOLBoundVar{p} \HOLConst{\HOLTokenUnion{}} \HOLConst{FV} \HOLBoundVar{q}) \HOLSymConst{\HOLTokenConj{}}
   (\HOLSymConst{\HOLTokenForall{}}\HOLBoundVar{L} \HOLBoundVar{p}. \HOLConst{FV} (\HOLConst{\ensuremath{\nu}} \HOLBoundVar{L} \HOLBoundVar{p}) \HOLSymConst{=} \HOLConst{FV} \HOLBoundVar{p}) \HOLSymConst{\HOLTokenConj{}} (\HOLSymConst{\HOLTokenForall{}}\HOLBoundVar{p} \HOLBoundVar{rf}. \HOLConst{FV} (\HOLConst{relab} \HOLBoundVar{p} \HOLBoundVar{rf}) \HOLSymConst{=} \HOLConst{FV} \HOLBoundVar{p}) \HOLSymConst{\HOLTokenConj{}}
   (\HOLSymConst{\HOLTokenForall{}}\HOLBoundVar{X}. \HOLConst{FV} (\HOLConst{var} \HOLBoundVar{X}) \HOLSymConst{=} \HOLTokenLeftbrace{}\HOLBoundVar{X}\HOLTokenRightbrace{}) \HOLSymConst{\HOLTokenConj{}} \HOLSymConst{\HOLTokenForall{}}\HOLBoundVar{X} \HOLBoundVar{p}. \HOLConst{FV} (\HOLConst{rec} \HOLBoundVar{X} \HOLBoundVar{p}) \HOLSymConst{=} \HOLConst{FV} \HOLBoundVar{p} \HOLConst{DIFF} \HOLTokenLeftbrace{}\HOLBoundVar{X}\HOLTokenRightbrace{}
\end{SaveVerbatim}
\newcommand{\HOLCCSDefinitionsFVXXdef}{\UseVerbatim{HOLCCSDefinitionsFVXXdef}}
\begin{SaveVerbatim}{HOLCCSDefinitionsISXXPROCXXdef}
\HOLTokenTurnstile{} \HOLSymConst{\HOLTokenForall{}}\HOLBoundVar{E}. \HOLConst{IS_PROC} \HOLBoundVar{E} \HOLSymConst{\HOLTokenEquiv{}} (\HOLConst{FV} \HOLBoundVar{E} \HOLSymConst{=} \HOLTokenLeftbrace{}\HOLTokenRightbrace{})
\end{SaveVerbatim}
\newcommand{\HOLCCSDefinitionsISXXPROCXXdef}{\UseVerbatim{HOLCCSDefinitionsISXXPROCXXdef}}
\begin{SaveVerbatim}{HOLCCSDefinitionsIsXXRelabelingXXdef}
\HOLTokenTurnstile{} \HOLSymConst{\HOLTokenForall{}}\HOLBoundVar{f}. \HOLConst{Is_Relabeling} \HOLBoundVar{f} \HOLSymConst{\HOLTokenEquiv{}} \HOLSymConst{\HOLTokenForall{}}\HOLBoundVar{s}. \HOLBoundVar{f} (\HOLConst{coname} \HOLBoundVar{s}) \HOLSymConst{=} \HOLConst{COMPL} (\HOLBoundVar{f} (\HOLConst{name} \HOLBoundVar{s}))
\end{SaveVerbatim}
\newcommand{\HOLCCSDefinitionsIsXXRelabelingXXdef}{\UseVerbatim{HOLCCSDefinitionsIsXXRelabelingXXdef}}
\begin{SaveVerbatim}{HOLCCSDefinitionsRELABXXdef}
\HOLTokenTurnstile{} \HOLSymConst{\HOLTokenForall{}}\HOLBoundVar{labl}. \HOLConst{RELAB} \HOLBoundVar{labl} \HOLSymConst{=} \HOLConst{ABS_Relabeling} (\HOLConst{Apply_Relab} \HOLBoundVar{labl})
\end{SaveVerbatim}
\newcommand{\HOLCCSDefinitionsRELABXXdef}{\UseVerbatim{HOLCCSDefinitionsRELABXXdef}}
\begin{SaveVerbatim}{HOLCCSDefinitionsrelabelXXdef}
\HOLTokenTurnstile{} (\HOLSymConst{\HOLTokenForall{}}\HOLBoundVar{rf}. \HOLConst{relabel} \HOLBoundVar{rf} \HOLConst{\ensuremath{\tau}} \HOLSymConst{=} \HOLConst{\ensuremath{\tau}}) \HOLSymConst{\HOLTokenConj{}}
   \HOLSymConst{\HOLTokenForall{}}\HOLBoundVar{rf} \HOLBoundVar{l}. \HOLConst{relabel} \HOLBoundVar{rf} (\HOLConst{label} \HOLBoundVar{l}) \HOLSymConst{=} \HOLConst{label} (\HOLConst{REP_Relabeling} \HOLBoundVar{rf} \HOLBoundVar{l})
\end{SaveVerbatim}
\newcommand{\HOLCCSDefinitionsrelabelXXdef}{\UseVerbatim{HOLCCSDefinitionsrelabelXXdef}}
\begin{SaveVerbatim}{HOLCCSDefinitionsRelabelingXXISOXXDEF}
\HOLTokenTurnstile{} (\HOLSymConst{\HOLTokenForall{}}\HOLBoundVar{a}. \HOLConst{ABS_Relabeling} (\HOLConst{REP_Relabeling} \HOLBoundVar{a}) \HOLSymConst{=} \HOLBoundVar{a}) \HOLSymConst{\HOLTokenConj{}}
   \HOLSymConst{\HOLTokenForall{}}\HOLBoundVar{r}.
     \HOLConst{Is_Relabeling} \HOLBoundVar{r} \HOLSymConst{\HOLTokenEquiv{}} (\HOLConst{REP_Relabeling} (\HOLConst{ABS_Relabeling} \HOLBoundVar{r}) \HOLSymConst{=} \HOLBoundVar{r})
\end{SaveVerbatim}
\newcommand{\HOLCCSDefinitionsRelabelingXXISOXXDEF}{\UseVerbatim{HOLCCSDefinitionsRelabelingXXISOXXDEF}}
\begin{SaveVerbatim}{HOLCCSDefinitionsRelabelingXXTYXXDEF}
\HOLTokenTurnstile{} \HOLSymConst{\HOLTokenExists{}}\HOLBoundVar{rep}. \HOLConst{TYPE_DEFINITION} \HOLConst{Is_Relabeling} \HOLBoundVar{rep}
\end{SaveVerbatim}
\newcommand{\HOLCCSDefinitionsRelabelingXXTYXXDEF}{\UseVerbatim{HOLCCSDefinitionsRelabelingXXTYXXDEF}}
\begin{SaveVerbatim}{HOLCCSDefinitionssizeXXdef}
\HOLTokenTurnstile{} \HOLSymConst{\HOLTokenForall{}}\HOLBoundVar{p}. \HOLConst{size} \HOLBoundVar{p} \HOLSymConst{=} \HOLConst{CCS_size} (\HOLTokenLambda{}\HOLBoundVar{x}. \HOLNumLit{0}) (\HOLTokenLambda{}\HOLBoundVar{x}. \HOLNumLit{0}) \HOLBoundVar{p}
\end{SaveVerbatim}
\newcommand{\HOLCCSDefinitionssizeXXdef}{\UseVerbatim{HOLCCSDefinitionssizeXXdef}}
\begin{SaveVerbatim}{HOLCCSDefinitionsTRANSXXdef}
\HOLTokenTurnstile{} \HOLConst{TRANS} \HOLSymConst{=}
   (\HOLTokenLambda{}\HOLBoundVar{a\sb{\mathrm{0}}} \HOLBoundVar{a\sb{\mathrm{1}}} \HOLBoundVar{a\sb{\mathrm{2}}}.
      \HOLSymConst{\HOLTokenForall{}}\HOLBoundVar{TRANS\sp{\prime}}.
        (\HOLSymConst{\HOLTokenForall{}}\HOLBoundVar{a\sb{\mathrm{0}}} \HOLBoundVar{a\sb{\mathrm{1}}} \HOLBoundVar{a\sb{\mathrm{2}}}.
           (\HOLBoundVar{a\sb{\mathrm{0}}} \HOLSymConst{=} \HOLBoundVar{a\sb{\mathrm{1}}}\HOLSymConst{..}\HOLBoundVar{a\sb{\mathrm{2}}}) \HOLSymConst{\HOLTokenDisj{}}
           (\HOLSymConst{\HOLTokenExists{}}\HOLBoundVar{E} \HOLBoundVar{E\sp{\prime}}. (\HOLBoundVar{a\sb{\mathrm{0}}} \HOLSymConst{=} \HOLBoundVar{E} \HOLSymConst{+} \HOLBoundVar{E\sp{\prime}}) \HOLSymConst{\HOLTokenConj{}} \HOLBoundVar{TRANS\sp{\prime}} \HOLBoundVar{E} \HOLBoundVar{a\sb{\mathrm{1}}} \HOLBoundVar{a\sb{\mathrm{2}}}) \HOLSymConst{\HOLTokenDisj{}}
           (\HOLSymConst{\HOLTokenExists{}}\HOLBoundVar{E} \HOLBoundVar{E\sp{\prime}}. (\HOLBoundVar{a\sb{\mathrm{0}}} \HOLSymConst{=} \HOLBoundVar{E\sp{\prime}} \HOLSymConst{+} \HOLBoundVar{E}) \HOLSymConst{\HOLTokenConj{}} \HOLBoundVar{TRANS\sp{\prime}} \HOLBoundVar{E} \HOLBoundVar{a\sb{\mathrm{1}}} \HOLBoundVar{a\sb{\mathrm{2}}}) \HOLSymConst{\HOLTokenDisj{}}
           (\HOLSymConst{\HOLTokenExists{}}\HOLBoundVar{E} \HOLBoundVar{E\sb{\mathrm{1}}} \HOLBoundVar{E\sp{\prime}}.
              (\HOLBoundVar{a\sb{\mathrm{0}}} \HOLSymConst{=} \HOLBoundVar{E} \HOLSymConst{\ensuremath{\parallel}} \HOLBoundVar{E\sp{\prime}}) \HOLSymConst{\HOLTokenConj{}} (\HOLBoundVar{a\sb{\mathrm{2}}} \HOLSymConst{=} \HOLBoundVar{E\sb{\mathrm{1}}} \HOLSymConst{\ensuremath{\parallel}} \HOLBoundVar{E\sp{\prime}}) \HOLSymConst{\HOLTokenConj{}} \HOLBoundVar{TRANS\sp{\prime}} \HOLBoundVar{E} \HOLBoundVar{a\sb{\mathrm{1}}} \HOLBoundVar{E\sb{\mathrm{1}}}) \HOLSymConst{\HOLTokenDisj{}}
           (\HOLSymConst{\HOLTokenExists{}}\HOLBoundVar{E} \HOLBoundVar{E\sb{\mathrm{1}}} \HOLBoundVar{E\sp{\prime}}.
              (\HOLBoundVar{a\sb{\mathrm{0}}} \HOLSymConst{=} \HOLBoundVar{E\sp{\prime}} \HOLSymConst{\ensuremath{\parallel}} \HOLBoundVar{E}) \HOLSymConst{\HOLTokenConj{}} (\HOLBoundVar{a\sb{\mathrm{2}}} \HOLSymConst{=} \HOLBoundVar{E\sp{\prime}} \HOLSymConst{\ensuremath{\parallel}} \HOLBoundVar{E\sb{\mathrm{1}}}) \HOLSymConst{\HOLTokenConj{}} \HOLBoundVar{TRANS\sp{\prime}} \HOLBoundVar{E} \HOLBoundVar{a\sb{\mathrm{1}}} \HOLBoundVar{E\sb{\mathrm{1}}}) \HOLSymConst{\HOLTokenDisj{}}
           (\HOLSymConst{\HOLTokenExists{}}\HOLBoundVar{E} \HOLBoundVar{l} \HOLBoundVar{E\sb{\mathrm{1}}} \HOLBoundVar{E\sp{\prime}} \HOLBoundVar{E\sb{\mathrm{2}}}.
              (\HOLBoundVar{a\sb{\mathrm{0}}} \HOLSymConst{=} \HOLBoundVar{E} \HOLSymConst{\ensuremath{\parallel}} \HOLBoundVar{E\sp{\prime}}) \HOLSymConst{\HOLTokenConj{}} (\HOLBoundVar{a\sb{\mathrm{1}}} \HOLSymConst{=} \HOLConst{\ensuremath{\tau}}) \HOLSymConst{\HOLTokenConj{}} (\HOLBoundVar{a\sb{\mathrm{2}}} \HOLSymConst{=} \HOLBoundVar{E\sb{\mathrm{1}}} \HOLSymConst{\ensuremath{\parallel}} \HOLBoundVar{E\sb{\mathrm{2}}}) \HOLSymConst{\HOLTokenConj{}}
              \HOLBoundVar{TRANS\sp{\prime}} \HOLBoundVar{E} (\HOLConst{label} \HOLBoundVar{l}) \HOLBoundVar{E\sb{\mathrm{1}}} \HOLSymConst{\HOLTokenConj{}}
              \HOLBoundVar{TRANS\sp{\prime}} \HOLBoundVar{E\sp{\prime}} (\HOLConst{label} (\HOLConst{COMPL} \HOLBoundVar{l})) \HOLBoundVar{E\sb{\mathrm{2}}}) \HOLSymConst{\HOLTokenDisj{}}
           (\HOLSymConst{\HOLTokenExists{}}\HOLBoundVar{E} \HOLBoundVar{E\sp{\prime}} \HOLBoundVar{l} \HOLBoundVar{L}.
              (\HOLBoundVar{a\sb{\mathrm{0}}} \HOLSymConst{=} \HOLConst{\ensuremath{\nu}} \HOLBoundVar{L} \HOLBoundVar{E}) \HOLSymConst{\HOLTokenConj{}} (\HOLBoundVar{a\sb{\mathrm{2}}} \HOLSymConst{=} \HOLConst{\ensuremath{\nu}} \HOLBoundVar{L} \HOLBoundVar{E\sp{\prime}}) \HOLSymConst{\HOLTokenConj{}} \HOLBoundVar{TRANS\sp{\prime}} \HOLBoundVar{E} \HOLBoundVar{a\sb{\mathrm{1}}} \HOLBoundVar{E\sp{\prime}} \HOLSymConst{\HOLTokenConj{}}
              ((\HOLBoundVar{a\sb{\mathrm{1}}} \HOLSymConst{=} \HOLConst{\ensuremath{\tau}}) \HOLSymConst{\HOLTokenDisj{}}
               (\HOLBoundVar{a\sb{\mathrm{1}}} \HOLSymConst{=} \HOLConst{label} \HOLBoundVar{l}) \HOLSymConst{\HOLTokenConj{}} \HOLBoundVar{l} \HOLConst{\HOLTokenNotIn{}} \HOLBoundVar{L} \HOLSymConst{\HOLTokenConj{}} \HOLConst{COMPL} \HOLBoundVar{l} \HOLConst{\HOLTokenNotIn{}} \HOLBoundVar{L})) \HOLSymConst{\HOLTokenDisj{}}
           (\HOLSymConst{\HOLTokenExists{}}\HOLBoundVar{E} \HOLBoundVar{u} \HOLBoundVar{E\sp{\prime}} \HOLBoundVar{rf}.
              (\HOLBoundVar{a\sb{\mathrm{0}}} \HOLSymConst{=} \HOLConst{relab} \HOLBoundVar{E} \HOLBoundVar{rf}) \HOLSymConst{\HOLTokenConj{}} (\HOLBoundVar{a\sb{\mathrm{1}}} \HOLSymConst{=} \HOLConst{relabel} \HOLBoundVar{rf} \HOLBoundVar{u}) \HOLSymConst{\HOLTokenConj{}}
              (\HOLBoundVar{a\sb{\mathrm{2}}} \HOLSymConst{=} \HOLConst{relab} \HOLBoundVar{E\sp{\prime}} \HOLBoundVar{rf}) \HOLSymConst{\HOLTokenConj{}} \HOLBoundVar{TRANS\sp{\prime}} \HOLBoundVar{E} \HOLBoundVar{u} \HOLBoundVar{E\sp{\prime}}) \HOLSymConst{\HOLTokenDisj{}}
           (\HOLSymConst{\HOLTokenExists{}}\HOLBoundVar{E} \HOLBoundVar{X}.
              (\HOLBoundVar{a\sb{\mathrm{0}}} \HOLSymConst{=} \HOLConst{rec} \HOLBoundVar{X} \HOLBoundVar{E}) \HOLSymConst{\HOLTokenConj{}}
              \HOLBoundVar{TRANS\sp{\prime}} (\HOLConst{CCS_Subst} \HOLBoundVar{E} (\HOLConst{rec} \HOLBoundVar{X} \HOLBoundVar{E}) \HOLBoundVar{X}) \HOLBoundVar{a\sb{\mathrm{1}}} \HOLBoundVar{a\sb{\mathrm{2}}}) \HOLSymConst{\HOLTokenImp{}}
           \HOLBoundVar{TRANS\sp{\prime}} \HOLBoundVar{a\sb{\mathrm{0}}} \HOLBoundVar{a\sb{\mathrm{1}}} \HOLBoundVar{a\sb{\mathrm{2}}}) \HOLSymConst{\HOLTokenImp{}}
        \HOLBoundVar{TRANS\sp{\prime}} \HOLBoundVar{a\sb{\mathrm{0}}} \HOLBoundVar{a\sb{\mathrm{1}}} \HOLBoundVar{a\sb{\mathrm{2}}})
\end{SaveVerbatim}
\newcommand{\HOLCCSDefinitionsTRANSXXdef}{\UseVerbatim{HOLCCSDefinitionsTRANSXXdef}}
\newcommand{\HOLCCSDefinitions}{
\HOLDfnTag{CCS}{ALL_IDENTICAL_def}\HOLCCSDefinitionsALLXXIDENTICALXXdef
\HOLDfnTag{CCS}{ALL_PROC_def}\HOLCCSDefinitionsALLXXPROCXXdef
\HOLDfnTag{CCS}{Apply_Relab_def}\HOLCCSDefinitionsApplyXXRelabXXdef
\HOLDfnTag{CCS}{BN_def}\HOLCCSDefinitionsBNXXdef
\HOLDfnTag{CCS}{bound_names_def}\HOLCCSDefinitionsboundXXnamesXXdef
\HOLDfnTag{CCS}{BV_def}\HOLCCSDefinitionsBVXXdef
\HOLDfnTag{CCS}{CCS_Subst1_def}\HOLCCSDefinitionsCCSXXSubstOneXXdef
\HOLDfnTag{CCS}{CCS_Subst2_def}\HOLCCSDefinitionsCCSXXSubstTwoXXdef
\HOLDfnTag{CCS}{CCS_Subst_def}\HOLCCSDefinitionsCCSXXSubstXXdef
\HOLDfnTag{CCS}{COMPL_ACT_def}\HOLCCSDefinitionsCOMPLXXACTXXdef
\HOLDfnTag{CCS}{COMPL_LAB_def}\HOLCCSDefinitionsCOMPLXXLABXXdef
\HOLDfnTag{CCS}{DELETE_ELEMENT_def}\HOLCCSDefinitionsDELETEXXELEMENTXXdef
\HOLDfnTag{CCS}{free_names_def}\HOLCCSDefinitionsfreeXXnamesXXdef
\HOLDfnTag{CCS}{FV_def}\HOLCCSDefinitionsFVXXdef
\HOLDfnTag{CCS}{IS_PROC_def}\HOLCCSDefinitionsISXXPROCXXdef
\HOLDfnTag{CCS}{Is_Relabeling_def}\HOLCCSDefinitionsIsXXRelabelingXXdef
\HOLDfnTag{CCS}{RELAB_def}\HOLCCSDefinitionsRELABXXdef
\HOLDfnTag{CCS}{relabel_def}\HOLCCSDefinitionsrelabelXXdef
\HOLDfnTag{CCS}{Relabeling_ISO_DEF}\HOLCCSDefinitionsRelabelingXXISOXXDEF
\HOLDfnTag{CCS}{Relabeling_TY_DEF}\HOLCCSDefinitionsRelabelingXXTYXXDEF
\HOLDfnTag{CCS}{size_def}\HOLCCSDefinitionssizeXXdef
\HOLDfnTag{CCS}{TRANS_def}\HOLCCSDefinitionsTRANSXXdef
}
\begin{SaveVerbatim}{HOLCCSTheoremsActionXXOneOne}
\HOLTokenTurnstile{} \HOLSymConst{\HOLTokenForall{}}\HOLBoundVar{x} \HOLBoundVar{y}. (\HOLConst{label} \HOLBoundVar{x} \HOLSymConst{=} \HOLConst{label} \HOLBoundVar{y}) \HOLSymConst{\HOLTokenEquiv{}} (\HOLBoundVar{x} \HOLSymConst{=} \HOLBoundVar{y})
\end{SaveVerbatim}
\newcommand{\HOLCCSTheoremsActionXXOneOne}{\UseVerbatim{HOLCCSTheoremsActionXXOneOne}}
\begin{SaveVerbatim}{HOLCCSTheoremsActionXXdistinct}
\HOLTokenTurnstile{} \HOLSymConst{\HOLTokenForall{}}\HOLBoundVar{x}. \HOLConst{\ensuremath{\tau}} \HOLSymConst{\HOLTokenNotEqual{}} \HOLConst{label} \HOLBoundVar{x}
\end{SaveVerbatim}
\newcommand{\HOLCCSTheoremsActionXXdistinct}{\UseVerbatim{HOLCCSTheoremsActionXXdistinct}}
\begin{SaveVerbatim}{HOLCCSTheoremsActionXXdistinctXXlabel}
\HOLTokenTurnstile{} \HOLSymConst{\HOLTokenForall{}}\HOLBoundVar{x}. \HOLConst{label} \HOLBoundVar{x} \HOLSymConst{\HOLTokenNotEqual{}} \HOLConst{\ensuremath{\tau}}
\end{SaveVerbatim}
\newcommand{\HOLCCSTheoremsActionXXdistinctXXlabel}{\UseVerbatim{HOLCCSTheoremsActionXXdistinctXXlabel}}
\begin{SaveVerbatim}{HOLCCSTheoremsActionXXinduction}
\HOLTokenTurnstile{} \HOLSymConst{\HOLTokenForall{}}\HOLBoundVar{P}. \HOLBoundVar{P} \HOLConst{\ensuremath{\tau}} \HOLSymConst{\HOLTokenConj{}} (\HOLSymConst{\HOLTokenForall{}}\HOLBoundVar{a}. \HOLBoundVar{P} (\HOLConst{label} \HOLBoundVar{a})) \HOLSymConst{\HOLTokenImp{}} \HOLSymConst{\HOLTokenForall{}}\HOLBoundVar{x}. \HOLBoundVar{P} \HOLBoundVar{x}
\end{SaveVerbatim}
\newcommand{\HOLCCSTheoremsActionXXinduction}{\UseVerbatim{HOLCCSTheoremsActionXXinduction}}
\begin{SaveVerbatim}{HOLCCSTheoremsActionXXnchotomy}
\HOLTokenTurnstile{} \HOLSymConst{\HOLTokenForall{}}\HOLBoundVar{opt}. (\HOLBoundVar{opt} \HOLSymConst{=} \HOLConst{\ensuremath{\tau}}) \HOLSymConst{\HOLTokenDisj{}} \HOLSymConst{\HOLTokenExists{}}\HOLBoundVar{x}. \HOLBoundVar{opt} \HOLSymConst{=} \HOLConst{label} \HOLBoundVar{x}
\end{SaveVerbatim}
\newcommand{\HOLCCSTheoremsActionXXnchotomy}{\UseVerbatim{HOLCCSTheoremsActionXXnchotomy}}
\begin{SaveVerbatim}{HOLCCSTheoremsActionXXnoXXtauXXisXXLabel}
\HOLTokenTurnstile{} \HOLSymConst{\HOLTokenForall{}}\HOLBoundVar{A}. \HOLBoundVar{A} \HOLSymConst{\HOLTokenNotEqual{}} \HOLConst{\ensuremath{\tau}} \HOLSymConst{\HOLTokenImp{}} \HOLSymConst{\HOLTokenExists{}}\HOLBoundVar{x}. \HOLBoundVar{A} \HOLSymConst{=} \HOLConst{label} \HOLBoundVar{x}
\end{SaveVerbatim}
\newcommand{\HOLCCSTheoremsActionXXnoXXtauXXisXXLabel}{\UseVerbatim{HOLCCSTheoremsActionXXnoXXtauXXisXXLabel}}
\begin{SaveVerbatim}{HOLCCSTheoremsApplyXXRelabXXCOMPLXXTHM}
\HOLTokenTurnstile{} \HOLSymConst{\HOLTokenForall{}}\HOLBoundVar{labl} \HOLBoundVar{s}.
     \HOLConst{Apply_Relab} \HOLBoundVar{labl} (\HOLConst{coname} \HOLBoundVar{s}) \HOLSymConst{=}
     \HOLConst{COMPL} (\HOLConst{Apply_Relab} \HOLBoundVar{labl} (\HOLConst{name} \HOLBoundVar{s}))
\end{SaveVerbatim}
\newcommand{\HOLCCSTheoremsApplyXXRelabXXCOMPLXXTHM}{\UseVerbatim{HOLCCSTheoremsApplyXXRelabXXCOMPLXXTHM}}
\begin{SaveVerbatim}{HOLCCSTheoremsAPPLYXXRELABXXTHM}
\HOLTokenTurnstile{} \HOLSymConst{\HOLTokenForall{}}\HOLBoundVar{labl\sp{\prime}} \HOLBoundVar{labl}.
     (\HOLConst{RELAB} \HOLBoundVar{labl\sp{\prime}} \HOLSymConst{=} \HOLConst{RELAB} \HOLBoundVar{labl}) \HOLSymConst{\HOLTokenEquiv{}}
     (\HOLConst{Apply_Relab} \HOLBoundVar{labl\sp{\prime}} \HOLSymConst{=} \HOLConst{Apply_Relab} \HOLBoundVar{labl})
\end{SaveVerbatim}
\newcommand{\HOLCCSTheoremsAPPLYXXRELABXXTHM}{\UseVerbatim{HOLCCSTheoremsAPPLYXXRELABXXTHM}}
\begin{SaveVerbatim}{HOLCCSTheoremsCCSXXcaseeq}
\HOLTokenTurnstile{} (\HOLConst{CCS_CASE} \HOLFreeVar{x} \HOLFreeVar{v} \HOLFreeVar{f} \HOLFreeVar{f\sb{\mathrm{1}}} \HOLFreeVar{f\sb{\mathrm{2}}} \HOLFreeVar{f\sb{\mathrm{3}}} \HOLFreeVar{f\sb{\mathrm{4}}} \HOLFreeVar{f\sb{\mathrm{5}}} \HOLFreeVar{f\sb{\mathrm{6}}} \HOLSymConst{=} \HOLFreeVar{v\sp{\prime}}) \HOLSymConst{\HOLTokenEquiv{}}
   (\HOLFreeVar{x} \HOLSymConst{=} \HOLConst{nil}) \HOLSymConst{\HOLTokenConj{}} (\HOLFreeVar{v} \HOLSymConst{=} \HOLFreeVar{v\sp{\prime}}) \HOLSymConst{\HOLTokenDisj{}} (\HOLSymConst{\HOLTokenExists{}}\HOLBoundVar{a}. (\HOLFreeVar{x} \HOLSymConst{=} \HOLConst{var} \HOLBoundVar{a}) \HOLSymConst{\HOLTokenConj{}} (\HOLFreeVar{f} \HOLBoundVar{a} \HOLSymConst{=} \HOLFreeVar{v\sp{\prime}})) \HOLSymConst{\HOLTokenDisj{}}
   (\HOLSymConst{\HOLTokenExists{}}\HOLBoundVar{o\sp{\prime}} \HOLBoundVar{C\sp{\prime}}. (\HOLFreeVar{x} \HOLSymConst{=} \HOLBoundVar{o\sp{\prime}}\HOLSymConst{..}\HOLBoundVar{C\sp{\prime}}) \HOLSymConst{\HOLTokenConj{}} (\HOLFreeVar{f\sb{\mathrm{1}}} \HOLBoundVar{o\sp{\prime}} \HOLBoundVar{C\sp{\prime}} \HOLSymConst{=} \HOLFreeVar{v\sp{\prime}})) \HOLSymConst{\HOLTokenDisj{}}
   (\HOLSymConst{\HOLTokenExists{}}\HOLBoundVar{C\sp{\prime}} \HOLBoundVar{C\sb{\mathrm{0}}}. (\HOLFreeVar{x} \HOLSymConst{=} \HOLBoundVar{C\sp{\prime}} \HOLSymConst{+} \HOLBoundVar{C\sb{\mathrm{0}}}) \HOLSymConst{\HOLTokenConj{}} (\HOLFreeVar{f\sb{\mathrm{2}}} \HOLBoundVar{C\sp{\prime}} \HOLBoundVar{C\sb{\mathrm{0}}} \HOLSymConst{=} \HOLFreeVar{v\sp{\prime}})) \HOLSymConst{\HOLTokenDisj{}}
   (\HOLSymConst{\HOLTokenExists{}}\HOLBoundVar{C\sp{\prime}} \HOLBoundVar{C\sb{\mathrm{0}}}. (\HOLFreeVar{x} \HOLSymConst{=} \HOLBoundVar{C\sp{\prime}} \HOLSymConst{\ensuremath{\parallel}} \HOLBoundVar{C\sb{\mathrm{0}}}) \HOLSymConst{\HOLTokenConj{}} (\HOLFreeVar{f\sb{\mathrm{3}}} \HOLBoundVar{C\sp{\prime}} \HOLBoundVar{C\sb{\mathrm{0}}} \HOLSymConst{=} \HOLFreeVar{v\sp{\prime}})) \HOLSymConst{\HOLTokenDisj{}}
   (\HOLSymConst{\HOLTokenExists{}}\HOLBoundVar{f\sp{\prime}} \HOLBoundVar{C\sp{\prime}}. (\HOLFreeVar{x} \HOLSymConst{=} \HOLConst{\ensuremath{\nu}} \HOLBoundVar{f\sp{\prime}} \HOLBoundVar{C\sp{\prime}}) \HOLSymConst{\HOLTokenConj{}} (\HOLFreeVar{f\sb{\mathrm{4}}} \HOLBoundVar{f\sp{\prime}} \HOLBoundVar{C\sp{\prime}} \HOLSymConst{=} \HOLFreeVar{v\sp{\prime}})) \HOLSymConst{\HOLTokenDisj{}}
   (\HOLSymConst{\HOLTokenExists{}}\HOLBoundVar{C\sp{\prime}} \HOLBoundVar{R}. (\HOLFreeVar{x} \HOLSymConst{=} \HOLConst{relab} \HOLBoundVar{C\sp{\prime}} \HOLBoundVar{R}) \HOLSymConst{\HOLTokenConj{}} (\HOLFreeVar{f\sb{\mathrm{5}}} \HOLBoundVar{C\sp{\prime}} \HOLBoundVar{R} \HOLSymConst{=} \HOLFreeVar{v\sp{\prime}})) \HOLSymConst{\HOLTokenDisj{}}
   \HOLSymConst{\HOLTokenExists{}}\HOLBoundVar{a} \HOLBoundVar{C\sp{\prime}}. (\HOLFreeVar{x} \HOLSymConst{=} \HOLConst{rec} \HOLBoundVar{a} \HOLBoundVar{C\sp{\prime}}) \HOLSymConst{\HOLTokenConj{}} (\HOLFreeVar{f\sb{\mathrm{6}}} \HOLBoundVar{a} \HOLBoundVar{C\sp{\prime}} \HOLSymConst{=} \HOLFreeVar{v\sp{\prime}})
\end{SaveVerbatim}
\newcommand{\HOLCCSTheoremsCCSXXcaseeq}{\UseVerbatim{HOLCCSTheoremsCCSXXcaseeq}}
\begin{SaveVerbatim}{HOLCCSTheoremsCCSXXCONDXXCLAUSES}
\HOLTokenTurnstile{} \HOLSymConst{\HOLTokenForall{}}\HOLBoundVar{t\sb{\mathrm{1}}} \HOLBoundVar{t\sb{\mathrm{2}}}.
     ((\HOLKeyword{if} \HOLConst{T} \HOLKeyword{then} \HOLBoundVar{t\sb{\mathrm{1}}} \HOLKeyword{else} \HOLBoundVar{t\sb{\mathrm{2}}}) \HOLSymConst{=} \HOLBoundVar{t\sb{\mathrm{1}}}) \HOLSymConst{\HOLTokenConj{}}
     ((\HOLKeyword{if} \HOLConst{F} \HOLKeyword{then} \HOLBoundVar{t\sb{\mathrm{1}}} \HOLKeyword{else} \HOLBoundVar{t\sb{\mathrm{2}}}) \HOLSymConst{=} \HOLBoundVar{t\sb{\mathrm{2}}})
\end{SaveVerbatim}
\newcommand{\HOLCCSTheoremsCCSXXCONDXXCLAUSES}{\UseVerbatim{HOLCCSTheoremsCCSXXCONDXXCLAUSES}}
\begin{SaveVerbatim}{HOLCCSTheoremsCCSXXdistinctYY}
\HOLTokenTurnstile{} (\HOLSymConst{\HOLTokenForall{}}\HOLBoundVar{a}. \HOLConst{nil} \HOLSymConst{\HOLTokenNotEqual{}} \HOLConst{var} \HOLBoundVar{a}) \HOLSymConst{\HOLTokenConj{}} (\HOLSymConst{\HOLTokenForall{}}\HOLBoundVar{a\sb{\mathrm{1}}} \HOLBoundVar{a\sb{\mathrm{0}}}. \HOLConst{nil} \HOLSymConst{\HOLTokenNotEqual{}} \HOLBoundVar{a\sb{\mathrm{0}}}\HOLSymConst{..}\HOLBoundVar{a\sb{\mathrm{1}}}) \HOLSymConst{\HOLTokenConj{}}
   (\HOLSymConst{\HOLTokenForall{}}\HOLBoundVar{a\sb{\mathrm{1}}} \HOLBoundVar{a\sb{\mathrm{0}}}. \HOLConst{nil} \HOLSymConst{\HOLTokenNotEqual{}} \HOLBoundVar{a\sb{\mathrm{0}}} \HOLSymConst{+} \HOLBoundVar{a\sb{\mathrm{1}}}) \HOLSymConst{\HOLTokenConj{}} (\HOLSymConst{\HOLTokenForall{}}\HOLBoundVar{a\sb{\mathrm{1}}} \HOLBoundVar{a\sb{\mathrm{0}}}. \HOLConst{nil} \HOLSymConst{\HOLTokenNotEqual{}} \HOLBoundVar{a\sb{\mathrm{0}}} \HOLSymConst{\ensuremath{\parallel}} \HOLBoundVar{a\sb{\mathrm{1}}}) \HOLSymConst{\HOLTokenConj{}}
   (\HOLSymConst{\HOLTokenForall{}}\HOLBoundVar{a\sb{\mathrm{1}}} \HOLBoundVar{a\sb{\mathrm{0}}}. \HOLConst{nil} \HOLSymConst{\HOLTokenNotEqual{}} \HOLConst{\ensuremath{\nu}} \HOLBoundVar{a\sb{\mathrm{0}}} \HOLBoundVar{a\sb{\mathrm{1}}}) \HOLSymConst{\HOLTokenConj{}} (\HOLSymConst{\HOLTokenForall{}}\HOLBoundVar{a\sb{\mathrm{1}}} \HOLBoundVar{a\sb{\mathrm{0}}}. \HOLConst{nil} \HOLSymConst{\HOLTokenNotEqual{}} \HOLConst{relab} \HOLBoundVar{a\sb{\mathrm{0}}} \HOLBoundVar{a\sb{\mathrm{1}}}) \HOLSymConst{\HOLTokenConj{}}
   (\HOLSymConst{\HOLTokenForall{}}\HOLBoundVar{a\sb{\mathrm{1}}} \HOLBoundVar{a\sb{\mathrm{0}}}. \HOLConst{nil} \HOLSymConst{\HOLTokenNotEqual{}} \HOLConst{rec} \HOLBoundVar{a\sb{\mathrm{0}}} \HOLBoundVar{a\sb{\mathrm{1}}}) \HOLSymConst{\HOLTokenConj{}} (\HOLSymConst{\HOLTokenForall{}}\HOLBoundVar{a\sb{\mathrm{1}}} \HOLBoundVar{a\sb{\mathrm{0}}} \HOLBoundVar{a}. \HOLConst{var} \HOLBoundVar{a} \HOLSymConst{\HOLTokenNotEqual{}} \HOLBoundVar{a\sb{\mathrm{0}}}\HOLSymConst{..}\HOLBoundVar{a\sb{\mathrm{1}}}) \HOLSymConst{\HOLTokenConj{}}
   (\HOLSymConst{\HOLTokenForall{}}\HOLBoundVar{a\sb{\mathrm{1}}} \HOLBoundVar{a\sb{\mathrm{0}}} \HOLBoundVar{a}. \HOLConst{var} \HOLBoundVar{a} \HOLSymConst{\HOLTokenNotEqual{}} \HOLBoundVar{a\sb{\mathrm{0}}} \HOLSymConst{+} \HOLBoundVar{a\sb{\mathrm{1}}}) \HOLSymConst{\HOLTokenConj{}} (\HOLSymConst{\HOLTokenForall{}}\HOLBoundVar{a\sb{\mathrm{1}}} \HOLBoundVar{a\sb{\mathrm{0}}} \HOLBoundVar{a}. \HOLConst{var} \HOLBoundVar{a} \HOLSymConst{\HOLTokenNotEqual{}} \HOLBoundVar{a\sb{\mathrm{0}}} \HOLSymConst{\ensuremath{\parallel}} \HOLBoundVar{a\sb{\mathrm{1}}}) \HOLSymConst{\HOLTokenConj{}}
   (\HOLSymConst{\HOLTokenForall{}}\HOLBoundVar{a\sb{\mathrm{1}}} \HOLBoundVar{a\sb{\mathrm{0}}} \HOLBoundVar{a}. \HOLConst{var} \HOLBoundVar{a} \HOLSymConst{\HOLTokenNotEqual{}} \HOLConst{\ensuremath{\nu}} \HOLBoundVar{a\sb{\mathrm{0}}} \HOLBoundVar{a\sb{\mathrm{1}}}) \HOLSymConst{\HOLTokenConj{}}
   (\HOLSymConst{\HOLTokenForall{}}\HOLBoundVar{a\sb{\mathrm{1}}} \HOLBoundVar{a\sb{\mathrm{0}}} \HOLBoundVar{a}. \HOLConst{var} \HOLBoundVar{a} \HOLSymConst{\HOLTokenNotEqual{}} \HOLConst{relab} \HOLBoundVar{a\sb{\mathrm{0}}} \HOLBoundVar{a\sb{\mathrm{1}}}) \HOLSymConst{\HOLTokenConj{}}
   (\HOLSymConst{\HOLTokenForall{}}\HOLBoundVar{a\sb{\mathrm{1}}} \HOLBoundVar{a\sb{\mathrm{0}}} \HOLBoundVar{a}. \HOLConst{var} \HOLBoundVar{a} \HOLSymConst{\HOLTokenNotEqual{}} \HOLConst{rec} \HOLBoundVar{a\sb{\mathrm{0}}} \HOLBoundVar{a\sb{\mathrm{1}}}) \HOLSymConst{\HOLTokenConj{}}
   (\HOLSymConst{\HOLTokenForall{}}\HOLBoundVar{a\sb{\mathrm{1}}\sp{\prime}} \HOLBoundVar{a\sb{\mathrm{1}}} \HOLBoundVar{a\sb{\mathrm{0}}\sp{\prime}} \HOLBoundVar{a\sb{\mathrm{0}}}. \HOLBoundVar{a\sb{\mathrm{0}}}\HOLSymConst{..}\HOLBoundVar{a\sb{\mathrm{1}}} \HOLSymConst{\HOLTokenNotEqual{}} \HOLBoundVar{a\sb{\mathrm{0}}\sp{\prime}} \HOLSymConst{+} \HOLBoundVar{a\sb{\mathrm{1}}\sp{\prime}}) \HOLSymConst{\HOLTokenConj{}}
   (\HOLSymConst{\HOLTokenForall{}}\HOLBoundVar{a\sb{\mathrm{1}}\sp{\prime}} \HOLBoundVar{a\sb{\mathrm{1}}} \HOLBoundVar{a\sb{\mathrm{0}}\sp{\prime}} \HOLBoundVar{a\sb{\mathrm{0}}}. \HOLBoundVar{a\sb{\mathrm{0}}}\HOLSymConst{..}\HOLBoundVar{a\sb{\mathrm{1}}} \HOLSymConst{\HOLTokenNotEqual{}} \HOLBoundVar{a\sb{\mathrm{0}}\sp{\prime}} \HOLSymConst{\ensuremath{\parallel}} \HOLBoundVar{a\sb{\mathrm{1}}\sp{\prime}}) \HOLSymConst{\HOLTokenConj{}}
   (\HOLSymConst{\HOLTokenForall{}}\HOLBoundVar{a\sb{\mathrm{1}}\sp{\prime}} \HOLBoundVar{a\sb{\mathrm{1}}} \HOLBoundVar{a\sb{\mathrm{0}}\sp{\prime}} \HOLBoundVar{a\sb{\mathrm{0}}}. \HOLBoundVar{a\sb{\mathrm{0}}}\HOLSymConst{..}\HOLBoundVar{a\sb{\mathrm{1}}} \HOLSymConst{\HOLTokenNotEqual{}} \HOLConst{\ensuremath{\nu}} \HOLBoundVar{a\sb{\mathrm{0}}\sp{\prime}} \HOLBoundVar{a\sb{\mathrm{1}}\sp{\prime}}) \HOLSymConst{\HOLTokenConj{}}
   (\HOLSymConst{\HOLTokenForall{}}\HOLBoundVar{a\sb{\mathrm{1}}\sp{\prime}} \HOLBoundVar{a\sb{\mathrm{1}}} \HOLBoundVar{a\sb{\mathrm{0}}\sp{\prime}} \HOLBoundVar{a\sb{\mathrm{0}}}. \HOLBoundVar{a\sb{\mathrm{0}}}\HOLSymConst{..}\HOLBoundVar{a\sb{\mathrm{1}}} \HOLSymConst{\HOLTokenNotEqual{}} \HOLConst{relab} \HOLBoundVar{a\sb{\mathrm{0}}\sp{\prime}} \HOLBoundVar{a\sb{\mathrm{1}}\sp{\prime}}) \HOLSymConst{\HOLTokenConj{}}
   (\HOLSymConst{\HOLTokenForall{}}\HOLBoundVar{a\sb{\mathrm{1}}\sp{\prime}} \HOLBoundVar{a\sb{\mathrm{1}}} \HOLBoundVar{a\sb{\mathrm{0}}\sp{\prime}} \HOLBoundVar{a\sb{\mathrm{0}}}. \HOLBoundVar{a\sb{\mathrm{0}}}\HOLSymConst{..}\HOLBoundVar{a\sb{\mathrm{1}}} \HOLSymConst{\HOLTokenNotEqual{}} \HOLConst{rec} \HOLBoundVar{a\sb{\mathrm{0}}\sp{\prime}} \HOLBoundVar{a\sb{\mathrm{1}}\sp{\prime}}) \HOLSymConst{\HOLTokenConj{}}
   (\HOLSymConst{\HOLTokenForall{}}\HOLBoundVar{a\sb{\mathrm{1}}\sp{\prime}} \HOLBoundVar{a\sb{\mathrm{1}}} \HOLBoundVar{a\sb{\mathrm{0}}\sp{\prime}} \HOLBoundVar{a\sb{\mathrm{0}}}. \HOLBoundVar{a\sb{\mathrm{0}}} \HOLSymConst{+} \HOLBoundVar{a\sb{\mathrm{1}}} \HOLSymConst{\HOLTokenNotEqual{}} \HOLBoundVar{a\sb{\mathrm{0}}\sp{\prime}} \HOLSymConst{\ensuremath{\parallel}} \HOLBoundVar{a\sb{\mathrm{1}}\sp{\prime}}) \HOLSymConst{\HOLTokenConj{}}
   (\HOLSymConst{\HOLTokenForall{}}\HOLBoundVar{a\sb{\mathrm{1}}\sp{\prime}} \HOLBoundVar{a\sb{\mathrm{1}}} \HOLBoundVar{a\sb{\mathrm{0}}\sp{\prime}} \HOLBoundVar{a\sb{\mathrm{0}}}. \HOLBoundVar{a\sb{\mathrm{0}}} \HOLSymConst{+} \HOLBoundVar{a\sb{\mathrm{1}}} \HOLSymConst{\HOLTokenNotEqual{}} \HOLConst{\ensuremath{\nu}} \HOLBoundVar{a\sb{\mathrm{0}}\sp{\prime}} \HOLBoundVar{a\sb{\mathrm{1}}\sp{\prime}}) \HOLSymConst{\HOLTokenConj{}}
   (\HOLSymConst{\HOLTokenForall{}}\HOLBoundVar{a\sb{\mathrm{1}}\sp{\prime}} \HOLBoundVar{a\sb{\mathrm{1}}} \HOLBoundVar{a\sb{\mathrm{0}}\sp{\prime}} \HOLBoundVar{a\sb{\mathrm{0}}}. \HOLBoundVar{a\sb{\mathrm{0}}} \HOLSymConst{+} \HOLBoundVar{a\sb{\mathrm{1}}} \HOLSymConst{\HOLTokenNotEqual{}} \HOLConst{relab} \HOLBoundVar{a\sb{\mathrm{0}}\sp{\prime}} \HOLBoundVar{a\sb{\mathrm{1}}\sp{\prime}}) \HOLSymConst{\HOLTokenConj{}}
   (\HOLSymConst{\HOLTokenForall{}}\HOLBoundVar{a\sb{\mathrm{1}}\sp{\prime}} \HOLBoundVar{a\sb{\mathrm{1}}} \HOLBoundVar{a\sb{\mathrm{0}}\sp{\prime}} \HOLBoundVar{a\sb{\mathrm{0}}}. \HOLBoundVar{a\sb{\mathrm{0}}} \HOLSymConst{+} \HOLBoundVar{a\sb{\mathrm{1}}} \HOLSymConst{\HOLTokenNotEqual{}} \HOLConst{rec} \HOLBoundVar{a\sb{\mathrm{0}}\sp{\prime}} \HOLBoundVar{a\sb{\mathrm{1}}\sp{\prime}}) \HOLSymConst{\HOLTokenConj{}}
   (\HOLSymConst{\HOLTokenForall{}}\HOLBoundVar{a\sb{\mathrm{1}}\sp{\prime}} \HOLBoundVar{a\sb{\mathrm{1}}} \HOLBoundVar{a\sb{\mathrm{0}}\sp{\prime}} \HOLBoundVar{a\sb{\mathrm{0}}}. \HOLBoundVar{a\sb{\mathrm{0}}} \HOLSymConst{\ensuremath{\parallel}} \HOLBoundVar{a\sb{\mathrm{1}}} \HOLSymConst{\HOLTokenNotEqual{}} \HOLConst{\ensuremath{\nu}} \HOLBoundVar{a\sb{\mathrm{0}}\sp{\prime}} \HOLBoundVar{a\sb{\mathrm{1}}\sp{\prime}}) \HOLSymConst{\HOLTokenConj{}}
   (\HOLSymConst{\HOLTokenForall{}}\HOLBoundVar{a\sb{\mathrm{1}}\sp{\prime}} \HOLBoundVar{a\sb{\mathrm{1}}} \HOLBoundVar{a\sb{\mathrm{0}}\sp{\prime}} \HOLBoundVar{a\sb{\mathrm{0}}}. \HOLBoundVar{a\sb{\mathrm{0}}} \HOLSymConst{\ensuremath{\parallel}} \HOLBoundVar{a\sb{\mathrm{1}}} \HOLSymConst{\HOLTokenNotEqual{}} \HOLConst{relab} \HOLBoundVar{a\sb{\mathrm{0}}\sp{\prime}} \HOLBoundVar{a\sb{\mathrm{1}}\sp{\prime}}) \HOLSymConst{\HOLTokenConj{}}
   (\HOLSymConst{\HOLTokenForall{}}\HOLBoundVar{a\sb{\mathrm{1}}\sp{\prime}} \HOLBoundVar{a\sb{\mathrm{1}}} \HOLBoundVar{a\sb{\mathrm{0}}\sp{\prime}} \HOLBoundVar{a\sb{\mathrm{0}}}. \HOLBoundVar{a\sb{\mathrm{0}}} \HOLSymConst{\ensuremath{\parallel}} \HOLBoundVar{a\sb{\mathrm{1}}} \HOLSymConst{\HOLTokenNotEqual{}} \HOLConst{rec} \HOLBoundVar{a\sb{\mathrm{0}}\sp{\prime}} \HOLBoundVar{a\sb{\mathrm{1}}\sp{\prime}}) \HOLSymConst{\HOLTokenConj{}}
   (\HOLSymConst{\HOLTokenForall{}}\HOLBoundVar{a\sb{\mathrm{1}}\sp{\prime}} \HOLBoundVar{a\sb{\mathrm{1}}} \HOLBoundVar{a\sb{\mathrm{0}}\sp{\prime}} \HOLBoundVar{a\sb{\mathrm{0}}}. \HOLConst{\ensuremath{\nu}} \HOLBoundVar{a\sb{\mathrm{0}}} \HOLBoundVar{a\sb{\mathrm{1}}} \HOLSymConst{\HOLTokenNotEqual{}} \HOLConst{relab} \HOLBoundVar{a\sb{\mathrm{0}}\sp{\prime}} \HOLBoundVar{a\sb{\mathrm{1}}\sp{\prime}}) \HOLSymConst{\HOLTokenConj{}}
   (\HOLSymConst{\HOLTokenForall{}}\HOLBoundVar{a\sb{\mathrm{1}}\sp{\prime}} \HOLBoundVar{a\sb{\mathrm{1}}} \HOLBoundVar{a\sb{\mathrm{0}}\sp{\prime}} \HOLBoundVar{a\sb{\mathrm{0}}}. \HOLConst{\ensuremath{\nu}} \HOLBoundVar{a\sb{\mathrm{0}}} \HOLBoundVar{a\sb{\mathrm{1}}} \HOLSymConst{\HOLTokenNotEqual{}} \HOLConst{rec} \HOLBoundVar{a\sb{\mathrm{0}}\sp{\prime}} \HOLBoundVar{a\sb{\mathrm{1}}\sp{\prime}}) \HOLSymConst{\HOLTokenConj{}}
   (\HOLSymConst{\HOLTokenForall{}}\HOLBoundVar{a\sb{\mathrm{1}}\sp{\prime}} \HOLBoundVar{a\sb{\mathrm{1}}} \HOLBoundVar{a\sb{\mathrm{0}}\sp{\prime}} \HOLBoundVar{a\sb{\mathrm{0}}}. \HOLConst{relab} \HOLBoundVar{a\sb{\mathrm{0}}} \HOLBoundVar{a\sb{\mathrm{1}}} \HOLSymConst{\HOLTokenNotEqual{}} \HOLConst{rec} \HOLBoundVar{a\sb{\mathrm{0}}\sp{\prime}} \HOLBoundVar{a\sb{\mathrm{1}}\sp{\prime}}) \HOLSymConst{\HOLTokenConj{}}
   (\HOLSymConst{\HOLTokenForall{}}\HOLBoundVar{a}. \HOLConst{var} \HOLBoundVar{a} \HOLSymConst{\HOLTokenNotEqual{}} \HOLConst{nil}) \HOLSymConst{\HOLTokenConj{}} (\HOLSymConst{\HOLTokenForall{}}\HOLBoundVar{a\sb{\mathrm{1}}} \HOLBoundVar{a\sb{\mathrm{0}}}. \HOLBoundVar{a\sb{\mathrm{0}}}\HOLSymConst{..}\HOLBoundVar{a\sb{\mathrm{1}}} \HOLSymConst{\HOLTokenNotEqual{}} \HOLConst{nil}) \HOLSymConst{\HOLTokenConj{}}
   (\HOLSymConst{\HOLTokenForall{}}\HOLBoundVar{a\sb{\mathrm{1}}} \HOLBoundVar{a\sb{\mathrm{0}}}. \HOLBoundVar{a\sb{\mathrm{0}}} \HOLSymConst{+} \HOLBoundVar{a\sb{\mathrm{1}}} \HOLSymConst{\HOLTokenNotEqual{}} \HOLConst{nil}) \HOLSymConst{\HOLTokenConj{}} (\HOLSymConst{\HOLTokenForall{}}\HOLBoundVar{a\sb{\mathrm{1}}} \HOLBoundVar{a\sb{\mathrm{0}}}. \HOLBoundVar{a\sb{\mathrm{0}}} \HOLSymConst{\ensuremath{\parallel}} \HOLBoundVar{a\sb{\mathrm{1}}} \HOLSymConst{\HOLTokenNotEqual{}} \HOLConst{nil}) \HOLSymConst{\HOLTokenConj{}}
   (\HOLSymConst{\HOLTokenForall{}}\HOLBoundVar{a\sb{\mathrm{1}}} \HOLBoundVar{a\sb{\mathrm{0}}}. \HOLConst{\ensuremath{\nu}} \HOLBoundVar{a\sb{\mathrm{0}}} \HOLBoundVar{a\sb{\mathrm{1}}} \HOLSymConst{\HOLTokenNotEqual{}} \HOLConst{nil}) \HOLSymConst{\HOLTokenConj{}} (\HOLSymConst{\HOLTokenForall{}}\HOLBoundVar{a\sb{\mathrm{1}}} \HOLBoundVar{a\sb{\mathrm{0}}}. \HOLConst{relab} \HOLBoundVar{a\sb{\mathrm{0}}} \HOLBoundVar{a\sb{\mathrm{1}}} \HOLSymConst{\HOLTokenNotEqual{}} \HOLConst{nil}) \HOLSymConst{\HOLTokenConj{}}
   (\HOLSymConst{\HOLTokenForall{}}\HOLBoundVar{a\sb{\mathrm{1}}} \HOLBoundVar{a\sb{\mathrm{0}}}. \HOLConst{rec} \HOLBoundVar{a\sb{\mathrm{0}}} \HOLBoundVar{a\sb{\mathrm{1}}} \HOLSymConst{\HOLTokenNotEqual{}} \HOLConst{nil}) \HOLSymConst{\HOLTokenConj{}} (\HOLSymConst{\HOLTokenForall{}}\HOLBoundVar{a\sb{\mathrm{1}}} \HOLBoundVar{a\sb{\mathrm{0}}} \HOLBoundVar{a}. \HOLBoundVar{a\sb{\mathrm{0}}}\HOLSymConst{..}\HOLBoundVar{a\sb{\mathrm{1}}} \HOLSymConst{\HOLTokenNotEqual{}} \HOLConst{var} \HOLBoundVar{a}) \HOLSymConst{\HOLTokenConj{}}
   (\HOLSymConst{\HOLTokenForall{}}\HOLBoundVar{a\sb{\mathrm{1}}} \HOLBoundVar{a\sb{\mathrm{0}}} \HOLBoundVar{a}. \HOLBoundVar{a\sb{\mathrm{0}}} \HOLSymConst{+} \HOLBoundVar{a\sb{\mathrm{1}}} \HOLSymConst{\HOLTokenNotEqual{}} \HOLConst{var} \HOLBoundVar{a}) \HOLSymConst{\HOLTokenConj{}} (\HOLSymConst{\HOLTokenForall{}}\HOLBoundVar{a\sb{\mathrm{1}}} \HOLBoundVar{a\sb{\mathrm{0}}} \HOLBoundVar{a}. \HOLBoundVar{a\sb{\mathrm{0}}} \HOLSymConst{\ensuremath{\parallel}} \HOLBoundVar{a\sb{\mathrm{1}}} \HOLSymConst{\HOLTokenNotEqual{}} \HOLConst{var} \HOLBoundVar{a}) \HOLSymConst{\HOLTokenConj{}}
   (\HOLSymConst{\HOLTokenForall{}}\HOLBoundVar{a\sb{\mathrm{1}}} \HOLBoundVar{a\sb{\mathrm{0}}} \HOLBoundVar{a}. \HOLConst{\ensuremath{\nu}} \HOLBoundVar{a\sb{\mathrm{0}}} \HOLBoundVar{a\sb{\mathrm{1}}} \HOLSymConst{\HOLTokenNotEqual{}} \HOLConst{var} \HOLBoundVar{a}) \HOLSymConst{\HOLTokenConj{}}
   (\HOLSymConst{\HOLTokenForall{}}\HOLBoundVar{a\sb{\mathrm{1}}} \HOLBoundVar{a\sb{\mathrm{0}}} \HOLBoundVar{a}. \HOLConst{relab} \HOLBoundVar{a\sb{\mathrm{0}}} \HOLBoundVar{a\sb{\mathrm{1}}} \HOLSymConst{\HOLTokenNotEqual{}} \HOLConst{var} \HOLBoundVar{a}) \HOLSymConst{\HOLTokenConj{}}
   (\HOLSymConst{\HOLTokenForall{}}\HOLBoundVar{a\sb{\mathrm{1}}} \HOLBoundVar{a\sb{\mathrm{0}}} \HOLBoundVar{a}. \HOLConst{rec} \HOLBoundVar{a\sb{\mathrm{0}}} \HOLBoundVar{a\sb{\mathrm{1}}} \HOLSymConst{\HOLTokenNotEqual{}} \HOLConst{var} \HOLBoundVar{a}) \HOLSymConst{\HOLTokenConj{}}
   (\HOLSymConst{\HOLTokenForall{}}\HOLBoundVar{a\sb{\mathrm{1}}\sp{\prime}} \HOLBoundVar{a\sb{\mathrm{1}}} \HOLBoundVar{a\sb{\mathrm{0}}\sp{\prime}} \HOLBoundVar{a\sb{\mathrm{0}}}. \HOLBoundVar{a\sb{\mathrm{0}}\sp{\prime}} \HOLSymConst{+} \HOLBoundVar{a\sb{\mathrm{1}}\sp{\prime}} \HOLSymConst{\HOLTokenNotEqual{}} \HOLBoundVar{a\sb{\mathrm{0}}}\HOLSymConst{..}\HOLBoundVar{a\sb{\mathrm{1}}}) \HOLSymConst{\HOLTokenConj{}}
   (\HOLSymConst{\HOLTokenForall{}}\HOLBoundVar{a\sb{\mathrm{1}}\sp{\prime}} \HOLBoundVar{a\sb{\mathrm{1}}} \HOLBoundVar{a\sb{\mathrm{0}}\sp{\prime}} \HOLBoundVar{a\sb{\mathrm{0}}}. \HOLBoundVar{a\sb{\mathrm{0}}\sp{\prime}} \HOLSymConst{\ensuremath{\parallel}} \HOLBoundVar{a\sb{\mathrm{1}}\sp{\prime}} \HOLSymConst{\HOLTokenNotEqual{}} \HOLBoundVar{a\sb{\mathrm{0}}}\HOLSymConst{..}\HOLBoundVar{a\sb{\mathrm{1}}}) \HOLSymConst{\HOLTokenConj{}}
   (\HOLSymConst{\HOLTokenForall{}}\HOLBoundVar{a\sb{\mathrm{1}}\sp{\prime}} \HOLBoundVar{a\sb{\mathrm{1}}} \HOLBoundVar{a\sb{\mathrm{0}}\sp{\prime}} \HOLBoundVar{a\sb{\mathrm{0}}}. \HOLConst{\ensuremath{\nu}} \HOLBoundVar{a\sb{\mathrm{0}}\sp{\prime}} \HOLBoundVar{a\sb{\mathrm{1}}\sp{\prime}} \HOLSymConst{\HOLTokenNotEqual{}} \HOLBoundVar{a\sb{\mathrm{0}}}\HOLSymConst{..}\HOLBoundVar{a\sb{\mathrm{1}}}) \HOLSymConst{\HOLTokenConj{}}
   (\HOLSymConst{\HOLTokenForall{}}\HOLBoundVar{a\sb{\mathrm{1}}\sp{\prime}} \HOLBoundVar{a\sb{\mathrm{1}}} \HOLBoundVar{a\sb{\mathrm{0}}\sp{\prime}} \HOLBoundVar{a\sb{\mathrm{0}}}. \HOLConst{relab} \HOLBoundVar{a\sb{\mathrm{0}}\sp{\prime}} \HOLBoundVar{a\sb{\mathrm{1}}\sp{\prime}} \HOLSymConst{\HOLTokenNotEqual{}} \HOLBoundVar{a\sb{\mathrm{0}}}\HOLSymConst{..}\HOLBoundVar{a\sb{\mathrm{1}}}) \HOLSymConst{\HOLTokenConj{}}
   (\HOLSymConst{\HOLTokenForall{}}\HOLBoundVar{a\sb{\mathrm{1}}\sp{\prime}} \HOLBoundVar{a\sb{\mathrm{1}}} \HOLBoundVar{a\sb{\mathrm{0}}\sp{\prime}} \HOLBoundVar{a\sb{\mathrm{0}}}. \HOLConst{rec} \HOLBoundVar{a\sb{\mathrm{0}}\sp{\prime}} \HOLBoundVar{a\sb{\mathrm{1}}\sp{\prime}} \HOLSymConst{\HOLTokenNotEqual{}} \HOLBoundVar{a\sb{\mathrm{0}}}\HOLSymConst{..}\HOLBoundVar{a\sb{\mathrm{1}}}) \HOLSymConst{\HOLTokenConj{}}
   (\HOLSymConst{\HOLTokenForall{}}\HOLBoundVar{a\sb{\mathrm{1}}\sp{\prime}} \HOLBoundVar{a\sb{\mathrm{1}}} \HOLBoundVar{a\sb{\mathrm{0}}\sp{\prime}} \HOLBoundVar{a\sb{\mathrm{0}}}. \HOLBoundVar{a\sb{\mathrm{0}}\sp{\prime}} \HOLSymConst{\ensuremath{\parallel}} \HOLBoundVar{a\sb{\mathrm{1}}\sp{\prime}} \HOLSymConst{\HOLTokenNotEqual{}} \HOLBoundVar{a\sb{\mathrm{0}}} \HOLSymConst{+} \HOLBoundVar{a\sb{\mathrm{1}}}) \HOLSymConst{\HOLTokenConj{}}
   (\HOLSymConst{\HOLTokenForall{}}\HOLBoundVar{a\sb{\mathrm{1}}\sp{\prime}} \HOLBoundVar{a\sb{\mathrm{1}}} \HOLBoundVar{a\sb{\mathrm{0}}\sp{\prime}} \HOLBoundVar{a\sb{\mathrm{0}}}. \HOLConst{\ensuremath{\nu}} \HOLBoundVar{a\sb{\mathrm{0}}\sp{\prime}} \HOLBoundVar{a\sb{\mathrm{1}}\sp{\prime}} \HOLSymConst{\HOLTokenNotEqual{}} \HOLBoundVar{a\sb{\mathrm{0}}} \HOLSymConst{+} \HOLBoundVar{a\sb{\mathrm{1}}}) \HOLSymConst{\HOLTokenConj{}}
   (\HOLSymConst{\HOLTokenForall{}}\HOLBoundVar{a\sb{\mathrm{1}}\sp{\prime}} \HOLBoundVar{a\sb{\mathrm{1}}} \HOLBoundVar{a\sb{\mathrm{0}}\sp{\prime}} \HOLBoundVar{a\sb{\mathrm{0}}}. \HOLConst{relab} \HOLBoundVar{a\sb{\mathrm{0}}\sp{\prime}} \HOLBoundVar{a\sb{\mathrm{1}}\sp{\prime}} \HOLSymConst{\HOLTokenNotEqual{}} \HOLBoundVar{a\sb{\mathrm{0}}} \HOLSymConst{+} \HOLBoundVar{a\sb{\mathrm{1}}}) \HOLSymConst{\HOLTokenConj{}}
   (\HOLSymConst{\HOLTokenForall{}}\HOLBoundVar{a\sb{\mathrm{1}}\sp{\prime}} \HOLBoundVar{a\sb{\mathrm{1}}} \HOLBoundVar{a\sb{\mathrm{0}}\sp{\prime}} \HOLBoundVar{a\sb{\mathrm{0}}}. \HOLConst{rec} \HOLBoundVar{a\sb{\mathrm{0}}\sp{\prime}} \HOLBoundVar{a\sb{\mathrm{1}}\sp{\prime}} \HOLSymConst{\HOLTokenNotEqual{}} \HOLBoundVar{a\sb{\mathrm{0}}} \HOLSymConst{+} \HOLBoundVar{a\sb{\mathrm{1}}}) \HOLSymConst{\HOLTokenConj{}}
   (\HOLSymConst{\HOLTokenForall{}}\HOLBoundVar{a\sb{\mathrm{1}}\sp{\prime}} \HOLBoundVar{a\sb{\mathrm{1}}} \HOLBoundVar{a\sb{\mathrm{0}}\sp{\prime}} \HOLBoundVar{a\sb{\mathrm{0}}}. \HOLConst{\ensuremath{\nu}} \HOLBoundVar{a\sb{\mathrm{0}}\sp{\prime}} \HOLBoundVar{a\sb{\mathrm{1}}\sp{\prime}} \HOLSymConst{\HOLTokenNotEqual{}} \HOLBoundVar{a\sb{\mathrm{0}}} \HOLSymConst{\ensuremath{\parallel}} \HOLBoundVar{a\sb{\mathrm{1}}}) \HOLSymConst{\HOLTokenConj{}}
   (\HOLSymConst{\HOLTokenForall{}}\HOLBoundVar{a\sb{\mathrm{1}}\sp{\prime}} \HOLBoundVar{a\sb{\mathrm{1}}} \HOLBoundVar{a\sb{\mathrm{0}}\sp{\prime}} \HOLBoundVar{a\sb{\mathrm{0}}}. \HOLConst{relab} \HOLBoundVar{a\sb{\mathrm{0}}\sp{\prime}} \HOLBoundVar{a\sb{\mathrm{1}}\sp{\prime}} \HOLSymConst{\HOLTokenNotEqual{}} \HOLBoundVar{a\sb{\mathrm{0}}} \HOLSymConst{\ensuremath{\parallel}} \HOLBoundVar{a\sb{\mathrm{1}}}) \HOLSymConst{\HOLTokenConj{}}
   (\HOLSymConst{\HOLTokenForall{}}\HOLBoundVar{a\sb{\mathrm{1}}\sp{\prime}} \HOLBoundVar{a\sb{\mathrm{1}}} \HOLBoundVar{a\sb{\mathrm{0}}\sp{\prime}} \HOLBoundVar{a\sb{\mathrm{0}}}. \HOLConst{rec} \HOLBoundVar{a\sb{\mathrm{0}}\sp{\prime}} \HOLBoundVar{a\sb{\mathrm{1}}\sp{\prime}} \HOLSymConst{\HOLTokenNotEqual{}} \HOLBoundVar{a\sb{\mathrm{0}}} \HOLSymConst{\ensuremath{\parallel}} \HOLBoundVar{a\sb{\mathrm{1}}}) \HOLSymConst{\HOLTokenConj{}}
   (\HOLSymConst{\HOLTokenForall{}}\HOLBoundVar{a\sb{\mathrm{1}}\sp{\prime}} \HOLBoundVar{a\sb{\mathrm{1}}} \HOLBoundVar{a\sb{\mathrm{0}}\sp{\prime}} \HOLBoundVar{a\sb{\mathrm{0}}}. \HOLConst{relab} \HOLBoundVar{a\sb{\mathrm{0}}\sp{\prime}} \HOLBoundVar{a\sb{\mathrm{1}}\sp{\prime}} \HOLSymConst{\HOLTokenNotEqual{}} \HOLConst{\ensuremath{\nu}} \HOLBoundVar{a\sb{\mathrm{0}}} \HOLBoundVar{a\sb{\mathrm{1}}}) \HOLSymConst{\HOLTokenConj{}}
   (\HOLSymConst{\HOLTokenForall{}}\HOLBoundVar{a\sb{\mathrm{1}}\sp{\prime}} \HOLBoundVar{a\sb{\mathrm{1}}} \HOLBoundVar{a\sb{\mathrm{0}}\sp{\prime}} \HOLBoundVar{a\sb{\mathrm{0}}}. \HOLConst{rec} \HOLBoundVar{a\sb{\mathrm{0}}\sp{\prime}} \HOLBoundVar{a\sb{\mathrm{1}}\sp{\prime}} \HOLSymConst{\HOLTokenNotEqual{}} \HOLConst{\ensuremath{\nu}} \HOLBoundVar{a\sb{\mathrm{0}}} \HOLBoundVar{a\sb{\mathrm{1}}}) \HOLSymConst{\HOLTokenConj{}}
   \HOLSymConst{\HOLTokenForall{}}\HOLBoundVar{a\sb{\mathrm{1}}\sp{\prime}} \HOLBoundVar{a\sb{\mathrm{1}}} \HOLBoundVar{a\sb{\mathrm{0}}\sp{\prime}} \HOLBoundVar{a\sb{\mathrm{0}}}. \HOLConst{rec} \HOLBoundVar{a\sb{\mathrm{0}}\sp{\prime}} \HOLBoundVar{a\sb{\mathrm{1}}\sp{\prime}} \HOLSymConst{\HOLTokenNotEqual{}} \HOLConst{relab} \HOLBoundVar{a\sb{\mathrm{0}}} \HOLBoundVar{a\sb{\mathrm{1}}}
\end{SaveVerbatim}
\newcommand{\HOLCCSTheoremsCCSXXdistinctYY}{\UseVerbatim{HOLCCSTheoremsCCSXXdistinctYY}}
\begin{SaveVerbatim}{HOLCCSTheoremsCCSXXSubstXXrec}
\HOLTokenTurnstile{} \HOLSymConst{\HOLTokenForall{}}\HOLBoundVar{X} \HOLBoundVar{E} \HOLBoundVar{E\sp{\prime}}. \HOLConst{CCS_Subst} (\HOLConst{rec} \HOLBoundVar{X} \HOLBoundVar{E}) \HOLBoundVar{E\sp{\prime}} \HOLBoundVar{X} \HOLSymConst{=} \HOLConst{rec} \HOLBoundVar{X} \HOLBoundVar{E}
\end{SaveVerbatim}
\newcommand{\HOLCCSTheoremsCCSXXSubstXXrec}{\UseVerbatim{HOLCCSTheoremsCCSXXSubstXXrec}}
\begin{SaveVerbatim}{HOLCCSTheoremsCCSXXSubstXXvar}
\HOLTokenTurnstile{} \HOLSymConst{\HOLTokenForall{}}\HOLBoundVar{X} \HOLBoundVar{E}. \HOLConst{CCS_Subst} (\HOLConst{var} \HOLBoundVar{X}) \HOLBoundVar{E} \HOLBoundVar{X} \HOLSymConst{=} \HOLBoundVar{E}
\end{SaveVerbatim}
\newcommand{\HOLCCSTheoremsCCSXXSubstXXvar}{\UseVerbatim{HOLCCSTheoremsCCSXXSubstXXvar}}
\begin{SaveVerbatim}{HOLCCSTheoremsCOMPLXXCOMPLXXACT}
\HOLTokenTurnstile{} \HOLSymConst{\HOLTokenForall{}}\HOLBoundVar{a}. \HOLConst{COMPL} (\HOLConst{COMPL} \HOLBoundVar{a}) \HOLSymConst{=} \HOLBoundVar{a}
\end{SaveVerbatim}
\newcommand{\HOLCCSTheoremsCOMPLXXCOMPLXXACT}{\UseVerbatim{HOLCCSTheoremsCOMPLXXCOMPLXXACT}}
\begin{SaveVerbatim}{HOLCCSTheoremsCOMPLXXCOMPLXXLAB}
\HOLTokenTurnstile{} \HOLSymConst{\HOLTokenForall{}}\HOLBoundVar{l}. \HOLConst{COMPL} (\HOLConst{COMPL} \HOLBoundVar{l}) \HOLSymConst{=} \HOLBoundVar{l}
\end{SaveVerbatim}
\newcommand{\HOLCCSTheoremsCOMPLXXCOMPLXXLAB}{\UseVerbatim{HOLCCSTheoremsCOMPLXXCOMPLXXLAB}}
\begin{SaveVerbatim}{HOLCCSTheoremsCOMPLXXTHM}
\HOLTokenTurnstile{} \HOLSymConst{\HOLTokenForall{}}\HOLBoundVar{l} \HOLBoundVar{s}.
     (\HOLBoundVar{l} \HOLSymConst{\HOLTokenNotEqual{}} \HOLConst{name} \HOLBoundVar{s} \HOLSymConst{\HOLTokenImp{}} \HOLConst{COMPL} \HOLBoundVar{l} \HOLSymConst{\HOLTokenNotEqual{}} \HOLConst{coname} \HOLBoundVar{s}) \HOLSymConst{\HOLTokenConj{}}
     (\HOLBoundVar{l} \HOLSymConst{\HOLTokenNotEqual{}} \HOLConst{coname} \HOLBoundVar{s} \HOLSymConst{\HOLTokenImp{}} \HOLConst{COMPL} \HOLBoundVar{l} \HOLSymConst{\HOLTokenNotEqual{}} \HOLConst{name} \HOLBoundVar{s})
\end{SaveVerbatim}
\newcommand{\HOLCCSTheoremsCOMPLXXTHM}{\UseVerbatim{HOLCCSTheoremsCOMPLXXTHM}}
\begin{SaveVerbatim}{HOLCCSTheoremsconameXXCOMPL}
\HOLTokenTurnstile{} \HOLSymConst{\HOLTokenForall{}}\HOLBoundVar{s}. \HOLConst{coname} \HOLBoundVar{s} \HOLSymConst{=} \HOLConst{COMPL} (\HOLConst{name} \HOLBoundVar{s})
\end{SaveVerbatim}
\newcommand{\HOLCCSTheoremsconameXXCOMPL}{\UseVerbatim{HOLCCSTheoremsconameXXCOMPL}}
\begin{SaveVerbatim}{HOLCCSTheoremsDELETEXXELEMENTXXAPPEND}
\HOLTokenTurnstile{} \HOLSymConst{\HOLTokenForall{}}\HOLBoundVar{a} \HOLBoundVar{L} \HOLBoundVar{L\sp{\prime}}.
     \HOLConst{DELETE_ELEMENT} \HOLBoundVar{a} (\HOLBoundVar{L} \HOLSymConst{++} \HOLBoundVar{L\sp{\prime}}) \HOLSymConst{=}
     \HOLConst{DELETE_ELEMENT} \HOLBoundVar{a} \HOLBoundVar{L} \HOLSymConst{++} \HOLConst{DELETE_ELEMENT} \HOLBoundVar{a} \HOLBoundVar{L\sp{\prime}}
\end{SaveVerbatim}
\newcommand{\HOLCCSTheoremsDELETEXXELEMENTXXAPPEND}{\UseVerbatim{HOLCCSTheoremsDELETEXXELEMENTXXAPPEND}}
\begin{SaveVerbatim}{HOLCCSTheoremsDELETEXXELEMENTXXFILTER}
\HOLTokenTurnstile{} \HOLSymConst{\HOLTokenForall{}}\HOLBoundVar{e} \HOLBoundVar{L}. \HOLConst{DELETE_ELEMENT} \HOLBoundVar{e} \HOLBoundVar{L} \HOLSymConst{=} \HOLConst{FILTER} (\HOLTokenLambda{}\HOLBoundVar{y}. \HOLBoundVar{e} \HOLSymConst{\HOLTokenNotEqual{}} \HOLBoundVar{y}) \HOLBoundVar{L}
\end{SaveVerbatim}
\newcommand{\HOLCCSTheoremsDELETEXXELEMENTXXFILTER}{\UseVerbatim{HOLCCSTheoremsDELETEXXELEMENTXXFILTER}}
\begin{SaveVerbatim}{HOLCCSTheoremsEVERYXXDELETEXXELEMENT}
\HOLTokenTurnstile{} \HOLSymConst{\HOLTokenForall{}}\HOLBoundVar{e} \HOLBoundVar{L} \HOLBoundVar{P}. \HOLBoundVar{P} \HOLBoundVar{e} \HOLSymConst{\HOLTokenConj{}} \HOLConst{EVERY} \HOLBoundVar{P} (\HOLConst{DELETE_ELEMENT} \HOLBoundVar{e} \HOLBoundVar{L}) \HOLSymConst{\HOLTokenImp{}} \HOLConst{EVERY} \HOLBoundVar{P} \HOLBoundVar{L}
\end{SaveVerbatim}
\newcommand{\HOLCCSTheoremsEVERYXXDELETEXXELEMENT}{\UseVerbatim{HOLCCSTheoremsEVERYXXDELETEXXELEMENT}}
\begin{SaveVerbatim}{HOLCCSTheoremsEXISTSXXRelabeling}
\HOLTokenTurnstile{} \HOLSymConst{\HOLTokenExists{}}\HOLBoundVar{f}. \HOLConst{Is_Relabeling} \HOLBoundVar{f}
\end{SaveVerbatim}
\newcommand{\HOLCCSTheoremsEXISTSXXRelabeling}{\UseVerbatim{HOLCCSTheoremsEXISTSXXRelabeling}}
\begin{SaveVerbatim}{HOLCCSTheoremsFNXXdef}
\HOLTokenTurnstile{} (\HOLSymConst{\HOLTokenForall{}}\HOLBoundVar{J}. \HOLConst{FN} \HOLConst{nil} \HOLBoundVar{J} \HOLSymConst{=} \HOLTokenLeftbrace{}\HOLTokenRightbrace{}) \HOLSymConst{\HOLTokenConj{}}
   (\HOLSymConst{\HOLTokenForall{}}\HOLBoundVar{p} \HOLBoundVar{l} \HOLBoundVar{J}. \HOLConst{FN} (\HOLConst{label} \HOLBoundVar{l}\HOLSymConst{..}\HOLBoundVar{p}) \HOLBoundVar{J} \HOLSymConst{=} \HOLBoundVar{l} \HOLConst{INSERT} \HOLConst{FN} \HOLBoundVar{p} \HOLBoundVar{J}) \HOLSymConst{\HOLTokenConj{}}
   (\HOLSymConst{\HOLTokenForall{}}\HOLBoundVar{p} \HOLBoundVar{J}. \HOLConst{FN} (\HOLConst{\ensuremath{\tau}}\HOLSymConst{..}\HOLBoundVar{p}) \HOLBoundVar{J} \HOLSymConst{=} \HOLConst{FN} \HOLBoundVar{p} \HOLBoundVar{J}) \HOLSymConst{\HOLTokenConj{}}
   (\HOLSymConst{\HOLTokenForall{}}\HOLBoundVar{q} \HOLBoundVar{p} \HOLBoundVar{J}. \HOLConst{FN} (\HOLBoundVar{p} \HOLSymConst{+} \HOLBoundVar{q}) \HOLBoundVar{J} \HOLSymConst{=} \HOLConst{FN} \HOLBoundVar{p} \HOLBoundVar{J} \HOLConst{\HOLTokenUnion{}} \HOLConst{FN} \HOLBoundVar{q} \HOLBoundVar{J}) \HOLSymConst{\HOLTokenConj{}}
   (\HOLSymConst{\HOLTokenForall{}}\HOLBoundVar{q} \HOLBoundVar{p} \HOLBoundVar{J}. \HOLConst{FN} (\HOLBoundVar{p} \HOLSymConst{\ensuremath{\parallel}} \HOLBoundVar{q}) \HOLBoundVar{J} \HOLSymConst{=} \HOLConst{FN} \HOLBoundVar{p} \HOLBoundVar{J} \HOLConst{\HOLTokenUnion{}} \HOLConst{FN} \HOLBoundVar{q} \HOLBoundVar{J}) \HOLSymConst{\HOLTokenConj{}}
   (\HOLSymConst{\HOLTokenForall{}}\HOLBoundVar{p} \HOLBoundVar{L} \HOLBoundVar{J}. \HOLConst{FN} (\HOLConst{\ensuremath{\nu}} \HOLBoundVar{L} \HOLBoundVar{p}) \HOLBoundVar{J} \HOLSymConst{=} \HOLConst{FN} \HOLBoundVar{p} \HOLBoundVar{J} \HOLConst{DIFF} (\HOLBoundVar{L} \HOLConst{\HOLTokenUnion{}} \HOLConst{IMAGE} \HOLConst{COMPL} \HOLBoundVar{L})) \HOLSymConst{\HOLTokenConj{}}
   (\HOLSymConst{\HOLTokenForall{}}\HOLBoundVar{rf} \HOLBoundVar{p} \HOLBoundVar{J}.
      \HOLConst{FN} (\HOLConst{relab} \HOLBoundVar{p} \HOLBoundVar{rf}) \HOLBoundVar{J} \HOLSymConst{=} \HOLConst{IMAGE} (\HOLConst{REP_Relabeling} \HOLBoundVar{rf}) (\HOLConst{FN} \HOLBoundVar{p} \HOLBoundVar{J})) \HOLSymConst{\HOLTokenConj{}}
   (\HOLSymConst{\HOLTokenForall{}}\HOLBoundVar{X} \HOLBoundVar{J}. \HOLConst{FN} (\HOLConst{var} \HOLBoundVar{X}) \HOLBoundVar{J} \HOLSymConst{=} \HOLTokenLeftbrace{}\HOLTokenRightbrace{}) \HOLSymConst{\HOLTokenConj{}}
   \HOLSymConst{\HOLTokenForall{}}\HOLBoundVar{p} \HOLBoundVar{X} \HOLBoundVar{J}.
     \HOLConst{FN} (\HOLConst{rec} \HOLBoundVar{X} \HOLBoundVar{p}) \HOLBoundVar{J} \HOLSymConst{=}
     \HOLKeyword{if} \HOLConst{MEM} \HOLBoundVar{X} \HOLBoundVar{J} \HOLKeyword{then}
       \HOLConst{FN} (\HOLConst{CCS_Subst} \HOLBoundVar{p} (\HOLConst{rec} \HOLBoundVar{X} \HOLBoundVar{p}) \HOLBoundVar{X}) (\HOLConst{DELETE_ELEMENT} \HOLBoundVar{X} \HOLBoundVar{J})
     \HOLKeyword{else} \HOLTokenLeftbrace{}\HOLTokenRightbrace{}
\end{SaveVerbatim}
\newcommand{\HOLCCSTheoremsFNXXdef}{\UseVerbatim{HOLCCSTheoremsFNXXdef}}
\begin{SaveVerbatim}{HOLCCSTheoremsFNXXind}
\HOLTokenTurnstile{} \HOLSymConst{\HOLTokenForall{}}\HOLBoundVar{P}.
     (\HOLSymConst{\HOLTokenForall{}}\HOLBoundVar{J}. \HOLBoundVar{P} \HOLConst{nil} \HOLBoundVar{J}) \HOLSymConst{\HOLTokenConj{}} (\HOLSymConst{\HOLTokenForall{}}\HOLBoundVar{l} \HOLBoundVar{p} \HOLBoundVar{J}. \HOLBoundVar{P} \HOLBoundVar{p} \HOLBoundVar{J} \HOLSymConst{\HOLTokenImp{}} \HOLBoundVar{P} (\HOLConst{label} \HOLBoundVar{l}\HOLSymConst{..}\HOLBoundVar{p}) \HOLBoundVar{J}) \HOLSymConst{\HOLTokenConj{}}
     (\HOLSymConst{\HOLTokenForall{}}\HOLBoundVar{p} \HOLBoundVar{J}. \HOLBoundVar{P} \HOLBoundVar{p} \HOLBoundVar{J} \HOLSymConst{\HOLTokenImp{}} \HOLBoundVar{P} (\HOLConst{\ensuremath{\tau}}\HOLSymConst{..}\HOLBoundVar{p}) \HOLBoundVar{J}) \HOLSymConst{\HOLTokenConj{}}
     (\HOLSymConst{\HOLTokenForall{}}\HOLBoundVar{p} \HOLBoundVar{q} \HOLBoundVar{J}. \HOLBoundVar{P} \HOLBoundVar{p} \HOLBoundVar{J} \HOLSymConst{\HOLTokenConj{}} \HOLBoundVar{P} \HOLBoundVar{q} \HOLBoundVar{J} \HOLSymConst{\HOLTokenImp{}} \HOLBoundVar{P} (\HOLBoundVar{p} \HOLSymConst{+} \HOLBoundVar{q}) \HOLBoundVar{J}) \HOLSymConst{\HOLTokenConj{}}
     (\HOLSymConst{\HOLTokenForall{}}\HOLBoundVar{p} \HOLBoundVar{q} \HOLBoundVar{J}. \HOLBoundVar{P} \HOLBoundVar{p} \HOLBoundVar{J} \HOLSymConst{\HOLTokenConj{}} \HOLBoundVar{P} \HOLBoundVar{q} \HOLBoundVar{J} \HOLSymConst{\HOLTokenImp{}} \HOLBoundVar{P} (\HOLBoundVar{p} \HOLSymConst{\ensuremath{\parallel}} \HOLBoundVar{q}) \HOLBoundVar{J}) \HOLSymConst{\HOLTokenConj{}}
     (\HOLSymConst{\HOLTokenForall{}}\HOLBoundVar{L} \HOLBoundVar{p} \HOLBoundVar{J}. \HOLBoundVar{P} \HOLBoundVar{p} \HOLBoundVar{J} \HOLSymConst{\HOLTokenImp{}} \HOLBoundVar{P} (\HOLConst{\ensuremath{\nu}} \HOLBoundVar{L} \HOLBoundVar{p}) \HOLBoundVar{J}) \HOLSymConst{\HOLTokenConj{}}
     (\HOLSymConst{\HOLTokenForall{}}\HOLBoundVar{p} \HOLBoundVar{rf} \HOLBoundVar{J}. \HOLBoundVar{P} \HOLBoundVar{p} \HOLBoundVar{J} \HOLSymConst{\HOLTokenImp{}} \HOLBoundVar{P} (\HOLConst{relab} \HOLBoundVar{p} \HOLBoundVar{rf}) \HOLBoundVar{J}) \HOLSymConst{\HOLTokenConj{}} (\HOLSymConst{\HOLTokenForall{}}\HOLBoundVar{X} \HOLBoundVar{J}. \HOLBoundVar{P} (\HOLConst{var} \HOLBoundVar{X}) \HOLBoundVar{J}) \HOLSymConst{\HOLTokenConj{}}
     (\HOLSymConst{\HOLTokenForall{}}\HOLBoundVar{X} \HOLBoundVar{p} \HOLBoundVar{J}.
        (\HOLConst{MEM} \HOLBoundVar{X} \HOLBoundVar{J} \HOLSymConst{\HOLTokenImp{}}
         \HOLBoundVar{P} (\HOLConst{CCS_Subst} \HOLBoundVar{p} (\HOLConst{rec} \HOLBoundVar{X} \HOLBoundVar{p}) \HOLBoundVar{X}) (\HOLConst{DELETE_ELEMENT} \HOLBoundVar{X} \HOLBoundVar{J})) \HOLSymConst{\HOLTokenImp{}}
        \HOLBoundVar{P} (\HOLConst{rec} \HOLBoundVar{X} \HOLBoundVar{p}) \HOLBoundVar{J}) \HOLSymConst{\HOLTokenImp{}}
     \HOLSymConst{\HOLTokenForall{}}\HOLBoundVar{v} \HOLBoundVar{v\sb{\mathrm{1}}}. \HOLBoundVar{P} \HOLBoundVar{v} \HOLBoundVar{v\sb{\mathrm{1}}}
\end{SaveVerbatim}
\newcommand{\HOLCCSTheoremsFNXXind}{\UseVerbatim{HOLCCSTheoremsFNXXind}}
\begin{SaveVerbatim}{HOLCCSTheoremsFNXXUNIVOne}
\HOLTokenTurnstile{} \HOLSymConst{\HOLTokenForall{}}\HOLBoundVar{p}. \HOLConst{free_names} \HOLBoundVar{p} \HOLSymConst{\HOLTokenNotEqual{}} \ensuremath{\cal{U}}(:'b \HOLTyOp{Label}) \HOLSymConst{\HOLTokenImp{}} \HOLSymConst{\HOLTokenExists{}}\HOLBoundVar{a}. \HOLBoundVar{a} \HOLConst{\HOLTokenNotIn{}} \HOLConst{free_names} \HOLBoundVar{p}
\end{SaveVerbatim}
\newcommand{\HOLCCSTheoremsFNXXUNIVOne}{\UseVerbatim{HOLCCSTheoremsFNXXUNIVOne}}
\begin{SaveVerbatim}{HOLCCSTheoremsFNXXUNIVTwo}
\HOLTokenTurnstile{} \HOLSymConst{\HOLTokenForall{}}\HOLBoundVar{p} \HOLBoundVar{q}.
     \HOLConst{free_names} \HOLBoundVar{p} \HOLConst{\HOLTokenUnion{}} \HOLConst{free_names} \HOLBoundVar{q} \HOLSymConst{\HOLTokenNotEqual{}} \ensuremath{\cal{U}}(:'b \HOLTyOp{Label}) \HOLSymConst{\HOLTokenImp{}}
     \HOLSymConst{\HOLTokenExists{}}\HOLBoundVar{a}. \HOLBoundVar{a} \HOLConst{\HOLTokenNotIn{}} \HOLConst{free_names} \HOLBoundVar{p} \HOLSymConst{\HOLTokenConj{}} \HOLBoundVar{a} \HOLConst{\HOLTokenNotIn{}} \HOLConst{free_names} \HOLBoundVar{q}
\end{SaveVerbatim}
\newcommand{\HOLCCSTheoremsFNXXUNIVTwo}{\UseVerbatim{HOLCCSTheoremsFNXXUNIVTwo}}
\begin{SaveVerbatim}{HOLCCSTheoremsISXXLABELXXdef}
\HOLTokenTurnstile{} (\HOLSymConst{\HOLTokenForall{}}\HOLBoundVar{x}. \HOLConst{IS_SOME} (\HOLConst{label} \HOLBoundVar{x}) \HOLSymConst{\HOLTokenEquiv{}} \HOLConst{T}) \HOLSymConst{\HOLTokenConj{}} (\HOLConst{IS_SOME} \HOLConst{\ensuremath{\tau}} \HOLSymConst{\HOLTokenEquiv{}} \HOLConst{F})
\end{SaveVerbatim}
\newcommand{\HOLCCSTheoremsISXXLABELXXdef}{\UseVerbatim{HOLCCSTheoremsISXXLABELXXdef}}
\begin{SaveVerbatim}{HOLCCSTheoremsISXXRELABELING}
\HOLTokenTurnstile{} \HOLSymConst{\HOLTokenForall{}}\HOLBoundVar{labl}. \HOLConst{Is_Relabeling} (\HOLConst{Apply_Relab} \HOLBoundVar{labl})
\end{SaveVerbatim}
\newcommand{\HOLCCSTheoremsISXXRELABELING}{\UseVerbatim{HOLCCSTheoremsISXXRELABELING}}
\begin{SaveVerbatim}{HOLCCSTheoremsLabelXXcaseeq}
\HOLTokenTurnstile{} (\HOLConst{Label_CASE} \HOLFreeVar{x} \HOLFreeVar{f} \HOLFreeVar{f\sb{\mathrm{1}}} \HOLSymConst{=} \HOLFreeVar{v}) \HOLSymConst{\HOLTokenEquiv{}}
   (\HOLSymConst{\HOLTokenExists{}}\HOLBoundVar{b}. (\HOLFreeVar{x} \HOLSymConst{=} \HOLConst{name} \HOLBoundVar{b}) \HOLSymConst{\HOLTokenConj{}} (\HOLFreeVar{f} \HOLBoundVar{b} \HOLSymConst{=} \HOLFreeVar{v})) \HOLSymConst{\HOLTokenDisj{}}
   \HOLSymConst{\HOLTokenExists{}}\HOLBoundVar{b}. (\HOLFreeVar{x} \HOLSymConst{=} \HOLConst{coname} \HOLBoundVar{b}) \HOLSymConst{\HOLTokenConj{}} (\HOLFreeVar{f\sb{\mathrm{1}}} \HOLBoundVar{b} \HOLSymConst{=} \HOLFreeVar{v})
\end{SaveVerbatim}
\newcommand{\HOLCCSTheoremsLabelXXcaseeq}{\UseVerbatim{HOLCCSTheoremsLabelXXcaseeq}}
\begin{SaveVerbatim}{HOLCCSTheoremsLABELXXdef}
\HOLTokenTurnstile{} \HOLSymConst{\HOLTokenForall{}}\HOLBoundVar{x}. \HOLConst{LABEL} (\HOLConst{label} \HOLBoundVar{x}) \HOLSymConst{=} \HOLBoundVar{x}
\end{SaveVerbatim}
\newcommand{\HOLCCSTheoremsLABELXXdef}{\UseVerbatim{HOLCCSTheoremsLABELXXdef}}
\begin{SaveVerbatim}{HOLCCSTheoremsLabelXXdistinctYY}
\HOLTokenTurnstile{} \HOLSymConst{\HOLTokenForall{}}\HOLBoundVar{a\sp{\prime}} \HOLBoundVar{a}. \HOLConst{coname} \HOLBoundVar{a\sp{\prime}} \HOLSymConst{\HOLTokenNotEqual{}} \HOLConst{name} \HOLBoundVar{a}
\end{SaveVerbatim}
\newcommand{\HOLCCSTheoremsLabelXXdistinctYY}{\UseVerbatim{HOLCCSTheoremsLabelXXdistinctYY}}
\begin{SaveVerbatim}{HOLCCSTheoremsLabelXXnotXXeq}
\HOLTokenTurnstile{} \HOLSymConst{\HOLTokenForall{}}\HOLBoundVar{a\sp{\prime}} \HOLBoundVar{a}. (\HOLConst{name} \HOLBoundVar{a} \HOLSymConst{=} \HOLConst{coname} \HOLBoundVar{a\sp{\prime}}) \HOLSymConst{\HOLTokenEquiv{}} \HOLConst{F}
\end{SaveVerbatim}
\newcommand{\HOLCCSTheoremsLabelXXnotXXeq}{\UseVerbatim{HOLCCSTheoremsLabelXXnotXXeq}}
\begin{SaveVerbatim}{HOLCCSTheoremsLabelXXnotXXeqYY}
\HOLTokenTurnstile{} \HOLSymConst{\HOLTokenForall{}}\HOLBoundVar{a\sp{\prime}} \HOLBoundVar{a}. (\HOLConst{coname} \HOLBoundVar{a\sp{\prime}} \HOLSymConst{=} \HOLConst{name} \HOLBoundVar{a}) \HOLSymConst{\HOLTokenEquiv{}} \HOLConst{F}
\end{SaveVerbatim}
\newcommand{\HOLCCSTheoremsLabelXXnotXXeqYY}{\UseVerbatim{HOLCCSTheoremsLabelXXnotXXeqYY}}
\begin{SaveVerbatim}{HOLCCSTheoremsLENGTHXXDELETEXXELEMENTXXLE}
\HOLTokenTurnstile{} \HOLSymConst{\HOLTokenForall{}}\HOLBoundVar{e} \HOLBoundVar{L}. \HOLConst{MEM} \HOLBoundVar{e} \HOLBoundVar{L} \HOLSymConst{\HOLTokenImp{}} \HOLConst{LENGTH} (\HOLConst{DELETE_ELEMENT} \HOLBoundVar{e} \HOLBoundVar{L}) \HOLSymConst{\HOLTokenLt{}} \HOLConst{LENGTH} \HOLBoundVar{L}
\end{SaveVerbatim}
\newcommand{\HOLCCSTheoremsLENGTHXXDELETEXXELEMENTXXLE}{\UseVerbatim{HOLCCSTheoremsLENGTHXXDELETEXXELEMENTXXLE}}
\begin{SaveVerbatim}{HOLCCSTheoremsLENGTHXXDELETEXXELEMENTXXLEQ}
\HOLTokenTurnstile{} \HOLSymConst{\HOLTokenForall{}}\HOLBoundVar{e} \HOLBoundVar{L}. \HOLConst{LENGTH} (\HOLConst{DELETE_ELEMENT} \HOLBoundVar{e} \HOLBoundVar{L}) \HOLSymConst{\HOLTokenLeq{}} \HOLConst{LENGTH} \HOLBoundVar{L}
\end{SaveVerbatim}
\newcommand{\HOLCCSTheoremsLENGTHXXDELETEXXELEMENTXXLEQ}{\UseVerbatim{HOLCCSTheoremsLENGTHXXDELETEXXELEMENTXXLEQ}}
\begin{SaveVerbatim}{HOLCCSTheoremsNILXXNOXXTRANS}
\HOLTokenTurnstile{} \HOLSymConst{\HOLTokenForall{}}\HOLBoundVar{u} \HOLBoundVar{E}. \HOLSymConst{\HOLTokenNeg{}}(\HOLConst{nil} \HOLTokenTransBegin\HOLBoundVar{u}\HOLTokenTransEnd \HOLBoundVar{E})
\end{SaveVerbatim}
\newcommand{\HOLCCSTheoremsNILXXNOXXTRANS}{\UseVerbatim{HOLCCSTheoremsNILXXNOXXTRANS}}
\begin{SaveVerbatim}{HOLCCSTheoremsNILXXNOXXTRANSXXEQF}
\HOLTokenTurnstile{} \HOLSymConst{\HOLTokenForall{}}\HOLBoundVar{u} \HOLBoundVar{E}. \HOLConst{nil} \HOLTokenTransBegin\HOLBoundVar{u}\HOLTokenTransEnd \HOLBoundVar{E} \HOLSymConst{\HOLTokenEquiv{}} \HOLConst{F}
\end{SaveVerbatim}
\newcommand{\HOLCCSTheoremsNILXXNOXXTRANSXXEQF}{\UseVerbatim{HOLCCSTheoremsNILXXNOXXTRANSXXEQF}}
\begin{SaveVerbatim}{HOLCCSTheoremsNOTXXINXXDELETEXXELEMENT}
\HOLTokenTurnstile{} \HOLSymConst{\HOLTokenForall{}}\HOLBoundVar{e} \HOLBoundVar{L}. \HOLSymConst{\HOLTokenNeg{}}\HOLConst{MEM} \HOLBoundVar{e} (\HOLConst{DELETE_ELEMENT} \HOLBoundVar{e} \HOLBoundVar{L})
\end{SaveVerbatim}
\newcommand{\HOLCCSTheoremsNOTXXINXXDELETEXXELEMENT}{\UseVerbatim{HOLCCSTheoremsNOTXXINXXDELETEXXELEMENT}}
\begin{SaveVerbatim}{HOLCCSTheoremsPAROne}
\HOLTokenTurnstile{} \HOLSymConst{\HOLTokenForall{}}\HOLBoundVar{E} \HOLBoundVar{u} \HOLBoundVar{E\sb{\mathrm{1}}} \HOLBoundVar{E\sp{\prime}}. \HOLBoundVar{E} \HOLTokenTransBegin\HOLBoundVar{u}\HOLTokenTransEnd \HOLBoundVar{E\sb{\mathrm{1}}} \HOLSymConst{\HOLTokenImp{}} \HOLBoundVar{E} \HOLSymConst{\ensuremath{\parallel}} \HOLBoundVar{E\sp{\prime}} \HOLTokenTransBegin\HOLBoundVar{u}\HOLTokenTransEnd \HOLBoundVar{E\sb{\mathrm{1}}} \HOLSymConst{\ensuremath{\parallel}} \HOLBoundVar{E\sp{\prime}}
\end{SaveVerbatim}
\newcommand{\HOLCCSTheoremsPAROne}{\UseVerbatim{HOLCCSTheoremsPAROne}}
\begin{SaveVerbatim}{HOLCCSTheoremsPARTwo}
\HOLTokenTurnstile{} \HOLSymConst{\HOLTokenForall{}}\HOLBoundVar{E} \HOLBoundVar{u} \HOLBoundVar{E\sb{\mathrm{1}}} \HOLBoundVar{E\sp{\prime}}. \HOLBoundVar{E} \HOLTokenTransBegin\HOLBoundVar{u}\HOLTokenTransEnd \HOLBoundVar{E\sb{\mathrm{1}}} \HOLSymConst{\HOLTokenImp{}} \HOLBoundVar{E\sp{\prime}} \HOLSymConst{\ensuremath{\parallel}} \HOLBoundVar{E} \HOLTokenTransBegin\HOLBoundVar{u}\HOLTokenTransEnd \HOLBoundVar{E\sp{\prime}} \HOLSymConst{\ensuremath{\parallel}} \HOLBoundVar{E\sb{\mathrm{1}}}
\end{SaveVerbatim}
\newcommand{\HOLCCSTheoremsPARTwo}{\UseVerbatim{HOLCCSTheoremsPARTwo}}
\begin{SaveVerbatim}{HOLCCSTheoremsPARThree}
\HOLTokenTurnstile{} \HOLSymConst{\HOLTokenForall{}}\HOLBoundVar{E} \HOLBoundVar{l} \HOLBoundVar{E\sb{\mathrm{1}}} \HOLBoundVar{E\sp{\prime}} \HOLBoundVar{E\sb{\mathrm{2}}}.
     \HOLBoundVar{E} \HOLTokenTransBegin\HOLConst{label} \HOLBoundVar{l}\HOLTokenTransEnd \HOLBoundVar{E\sb{\mathrm{1}}} \HOLSymConst{\HOLTokenConj{}} \HOLBoundVar{E\sp{\prime}} \HOLTokenTransBegin\HOLConst{label} (\HOLConst{COMPL} \HOLBoundVar{l})\HOLTokenTransEnd \HOLBoundVar{E\sb{\mathrm{2}}} \HOLSymConst{\HOLTokenImp{}}
     \HOLBoundVar{E} \HOLSymConst{\ensuremath{\parallel}} \HOLBoundVar{E\sp{\prime}} \HOLTokenTransBegin\HOLConst{\ensuremath{\tau}}\HOLTokenTransEnd \HOLBoundVar{E\sb{\mathrm{1}}} \HOLSymConst{\ensuremath{\parallel}} \HOLBoundVar{E\sb{\mathrm{2}}}
\end{SaveVerbatim}
\newcommand{\HOLCCSTheoremsPARThree}{\UseVerbatim{HOLCCSTheoremsPARThree}}
\begin{SaveVerbatim}{HOLCCSTheoremsPARXXcases}
\HOLTokenTurnstile{} \HOLSymConst{\HOLTokenForall{}}\HOLBoundVar{D} \HOLBoundVar{D\sp{\prime}} \HOLBoundVar{u} \HOLBoundVar{D\sp{\prime\prime}}.
     \HOLBoundVar{D} \HOLSymConst{\ensuremath{\parallel}} \HOLBoundVar{D\sp{\prime}} \HOLTokenTransBegin\HOLBoundVar{u}\HOLTokenTransEnd \HOLBoundVar{D\sp{\prime\prime}} \HOLSymConst{\HOLTokenImp{}}
     (\HOLSymConst{\HOLTokenExists{}}\HOLBoundVar{E} \HOLBoundVar{E\sb{\mathrm{1}}} \HOLBoundVar{E\sp{\prime}}.
        ((\HOLBoundVar{D} \HOLSymConst{=} \HOLBoundVar{E}) \HOLSymConst{\HOLTokenConj{}} (\HOLBoundVar{D\sp{\prime}} \HOLSymConst{=} \HOLBoundVar{E\sp{\prime}})) \HOLSymConst{\HOLTokenConj{}} (\HOLBoundVar{D\sp{\prime\prime}} \HOLSymConst{=} \HOLBoundVar{E\sb{\mathrm{1}}} \HOLSymConst{\ensuremath{\parallel}} \HOLBoundVar{E\sp{\prime}}) \HOLSymConst{\HOLTokenConj{}} \HOLBoundVar{E} \HOLTokenTransBegin\HOLBoundVar{u}\HOLTokenTransEnd \HOLBoundVar{E\sb{\mathrm{1}}}) \HOLSymConst{\HOLTokenDisj{}}
     (\HOLSymConst{\HOLTokenExists{}}\HOLBoundVar{E} \HOLBoundVar{E\sb{\mathrm{1}}} \HOLBoundVar{E\sp{\prime}}.
        ((\HOLBoundVar{D} \HOLSymConst{=} \HOLBoundVar{E\sp{\prime}}) \HOLSymConst{\HOLTokenConj{}} (\HOLBoundVar{D\sp{\prime}} \HOLSymConst{=} \HOLBoundVar{E})) \HOLSymConst{\HOLTokenConj{}} (\HOLBoundVar{D\sp{\prime\prime}} \HOLSymConst{=} \HOLBoundVar{E\sp{\prime}} \HOLSymConst{\ensuremath{\parallel}} \HOLBoundVar{E\sb{\mathrm{1}}}) \HOLSymConst{\HOLTokenConj{}} \HOLBoundVar{E} \HOLTokenTransBegin\HOLBoundVar{u}\HOLTokenTransEnd \HOLBoundVar{E\sb{\mathrm{1}}}) \HOLSymConst{\HOLTokenDisj{}}
     \HOLSymConst{\HOLTokenExists{}}\HOLBoundVar{E} \HOLBoundVar{l} \HOLBoundVar{E\sb{\mathrm{1}}} \HOLBoundVar{E\sp{\prime}} \HOLBoundVar{E\sb{\mathrm{2}}}.
       ((\HOLBoundVar{D} \HOLSymConst{=} \HOLBoundVar{E}) \HOLSymConst{\HOLTokenConj{}} (\HOLBoundVar{D\sp{\prime}} \HOLSymConst{=} \HOLBoundVar{E\sp{\prime}})) \HOLSymConst{\HOLTokenConj{}} (\HOLBoundVar{u} \HOLSymConst{=} \HOLConst{\ensuremath{\tau}}) \HOLSymConst{\HOLTokenConj{}} (\HOLBoundVar{D\sp{\prime\prime}} \HOLSymConst{=} \HOLBoundVar{E\sb{\mathrm{1}}} \HOLSymConst{\ensuremath{\parallel}} \HOLBoundVar{E\sb{\mathrm{2}}}) \HOLSymConst{\HOLTokenConj{}}
       \HOLBoundVar{E} \HOLTokenTransBegin\HOLConst{label} \HOLBoundVar{l}\HOLTokenTransEnd \HOLBoundVar{E\sb{\mathrm{1}}} \HOLSymConst{\HOLTokenConj{}} \HOLBoundVar{E\sp{\prime}} \HOLTokenTransBegin\HOLConst{label} (\HOLConst{COMPL} \HOLBoundVar{l})\HOLTokenTransEnd \HOLBoundVar{E\sb{\mathrm{2}}}
\end{SaveVerbatim}
\newcommand{\HOLCCSTheoremsPARXXcases}{\UseVerbatim{HOLCCSTheoremsPARXXcases}}
\begin{SaveVerbatim}{HOLCCSTheoremsPARXXcasesXXEQ}
\HOLTokenTurnstile{} \HOLSymConst{\HOLTokenForall{}}\HOLBoundVar{D} \HOLBoundVar{D\sp{\prime}} \HOLBoundVar{u} \HOLBoundVar{D\sp{\prime\prime}}.
     \HOLBoundVar{D} \HOLSymConst{\ensuremath{\parallel}} \HOLBoundVar{D\sp{\prime}} \HOLTokenTransBegin\HOLBoundVar{u}\HOLTokenTransEnd \HOLBoundVar{D\sp{\prime\prime}} \HOLSymConst{\HOLTokenEquiv{}}
     (\HOLSymConst{\HOLTokenExists{}}\HOLBoundVar{E} \HOLBoundVar{E\sb{\mathrm{1}}} \HOLBoundVar{E\sp{\prime}}.
        ((\HOLBoundVar{D} \HOLSymConst{=} \HOLBoundVar{E}) \HOLSymConst{\HOLTokenConj{}} (\HOLBoundVar{D\sp{\prime}} \HOLSymConst{=} \HOLBoundVar{E\sp{\prime}})) \HOLSymConst{\HOLTokenConj{}} (\HOLBoundVar{D\sp{\prime\prime}} \HOLSymConst{=} \HOLBoundVar{E\sb{\mathrm{1}}} \HOLSymConst{\ensuremath{\parallel}} \HOLBoundVar{E\sp{\prime}}) \HOLSymConst{\HOLTokenConj{}} \HOLBoundVar{E} \HOLTokenTransBegin\HOLBoundVar{u}\HOLTokenTransEnd \HOLBoundVar{E\sb{\mathrm{1}}}) \HOLSymConst{\HOLTokenDisj{}}
     (\HOLSymConst{\HOLTokenExists{}}\HOLBoundVar{E} \HOLBoundVar{E\sb{\mathrm{1}}} \HOLBoundVar{E\sp{\prime}}.
        ((\HOLBoundVar{D} \HOLSymConst{=} \HOLBoundVar{E\sp{\prime}}) \HOLSymConst{\HOLTokenConj{}} (\HOLBoundVar{D\sp{\prime}} \HOLSymConst{=} \HOLBoundVar{E})) \HOLSymConst{\HOLTokenConj{}} (\HOLBoundVar{D\sp{\prime\prime}} \HOLSymConst{=} \HOLBoundVar{E\sp{\prime}} \HOLSymConst{\ensuremath{\parallel}} \HOLBoundVar{E\sb{\mathrm{1}}}) \HOLSymConst{\HOLTokenConj{}} \HOLBoundVar{E} \HOLTokenTransBegin\HOLBoundVar{u}\HOLTokenTransEnd \HOLBoundVar{E\sb{\mathrm{1}}}) \HOLSymConst{\HOLTokenDisj{}}
     \HOLSymConst{\HOLTokenExists{}}\HOLBoundVar{E} \HOLBoundVar{l} \HOLBoundVar{E\sb{\mathrm{1}}} \HOLBoundVar{E\sp{\prime}} \HOLBoundVar{E\sb{\mathrm{2}}}.
       ((\HOLBoundVar{D} \HOLSymConst{=} \HOLBoundVar{E}) \HOLSymConst{\HOLTokenConj{}} (\HOLBoundVar{D\sp{\prime}} \HOLSymConst{=} \HOLBoundVar{E\sp{\prime}})) \HOLSymConst{\HOLTokenConj{}} (\HOLBoundVar{u} \HOLSymConst{=} \HOLConst{\ensuremath{\tau}}) \HOLSymConst{\HOLTokenConj{}} (\HOLBoundVar{D\sp{\prime\prime}} \HOLSymConst{=} \HOLBoundVar{E\sb{\mathrm{1}}} \HOLSymConst{\ensuremath{\parallel}} \HOLBoundVar{E\sb{\mathrm{2}}}) \HOLSymConst{\HOLTokenConj{}}
       \HOLBoundVar{E} \HOLTokenTransBegin\HOLConst{label} \HOLBoundVar{l}\HOLTokenTransEnd \HOLBoundVar{E\sb{\mathrm{1}}} \HOLSymConst{\HOLTokenConj{}} \HOLBoundVar{E\sp{\prime}} \HOLTokenTransBegin\HOLConst{label} (\HOLConst{COMPL} \HOLBoundVar{l})\HOLTokenTransEnd \HOLBoundVar{E\sb{\mathrm{2}}}
\end{SaveVerbatim}
\newcommand{\HOLCCSTheoremsPARXXcasesXXEQ}{\UseVerbatim{HOLCCSTheoremsPARXXcasesXXEQ}}
\begin{SaveVerbatim}{HOLCCSTheoremsPREFIX}
\HOLTokenTurnstile{} \HOLSymConst{\HOLTokenForall{}}\HOLBoundVar{E} \HOLBoundVar{u}. \HOLBoundVar{u}\HOLSymConst{..}\HOLBoundVar{E} \HOLTokenTransBegin\HOLBoundVar{u}\HOLTokenTransEnd \HOLBoundVar{E}
\end{SaveVerbatim}
\newcommand{\HOLCCSTheoremsPREFIX}{\UseVerbatim{HOLCCSTheoremsPREFIX}}
\begin{SaveVerbatim}{HOLCCSTheoremsREC}
\HOLTokenTurnstile{} \HOLSymConst{\HOLTokenForall{}}\HOLBoundVar{E} \HOLBoundVar{u} \HOLBoundVar{X} \HOLBoundVar{E\sb{\mathrm{1}}}. \HOLConst{CCS_Subst} \HOLBoundVar{E} (\HOLConst{rec} \HOLBoundVar{X} \HOLBoundVar{E}) \HOLBoundVar{X} \HOLTokenTransBegin\HOLBoundVar{u}\HOLTokenTransEnd \HOLBoundVar{E\sb{\mathrm{1}}} \HOLSymConst{\HOLTokenImp{}} \HOLConst{rec} \HOLBoundVar{X} \HOLBoundVar{E} \HOLTokenTransBegin\HOLBoundVar{u}\HOLTokenTransEnd \HOLBoundVar{E\sb{\mathrm{1}}}
\end{SaveVerbatim}
\newcommand{\HOLCCSTheoremsREC}{\UseVerbatim{HOLCCSTheoremsREC}}
\begin{SaveVerbatim}{HOLCCSTheoremsRECXXcases}
\HOLTokenTurnstile{} \HOLSymConst{\HOLTokenForall{}}\HOLBoundVar{X} \HOLBoundVar{E} \HOLBoundVar{u} \HOLBoundVar{E\sp{\prime\prime}}.
     \HOLConst{rec} \HOLBoundVar{X} \HOLBoundVar{E} \HOLTokenTransBegin\HOLBoundVar{u}\HOLTokenTransEnd \HOLBoundVar{E\sp{\prime\prime}} \HOLSymConst{\HOLTokenImp{}}
     \HOLSymConst{\HOLTokenExists{}}\HOLBoundVar{E\sp{\prime}} \HOLBoundVar{X\sp{\prime}}.
       ((\HOLBoundVar{X} \HOLSymConst{=} \HOLBoundVar{X\sp{\prime}}) \HOLSymConst{\HOLTokenConj{}} (\HOLBoundVar{E} \HOLSymConst{=} \HOLBoundVar{E\sp{\prime}})) \HOLSymConst{\HOLTokenConj{}}
       \HOLConst{CCS_Subst} \HOLBoundVar{E\sp{\prime}} (\HOLConst{rec} \HOLBoundVar{X\sp{\prime}} \HOLBoundVar{E\sp{\prime}}) \HOLBoundVar{X\sp{\prime}} \HOLTokenTransBegin\HOLBoundVar{u}\HOLTokenTransEnd \HOLBoundVar{E\sp{\prime\prime}}
\end{SaveVerbatim}
\newcommand{\HOLCCSTheoremsRECXXcases}{\UseVerbatim{HOLCCSTheoremsRECXXcases}}
\begin{SaveVerbatim}{HOLCCSTheoremsRECXXcasesXXEQ}
\HOLTokenTurnstile{} \HOLSymConst{\HOLTokenForall{}}\HOLBoundVar{X} \HOLBoundVar{E} \HOLBoundVar{u} \HOLBoundVar{E\sp{\prime\prime}}.
     \HOLConst{rec} \HOLBoundVar{X} \HOLBoundVar{E} \HOLTokenTransBegin\HOLBoundVar{u}\HOLTokenTransEnd \HOLBoundVar{E\sp{\prime\prime}} \HOLSymConst{\HOLTokenEquiv{}}
     \HOLSymConst{\HOLTokenExists{}}\HOLBoundVar{E\sp{\prime}} \HOLBoundVar{X\sp{\prime}}.
       ((\HOLBoundVar{X} \HOLSymConst{=} \HOLBoundVar{X\sp{\prime}}) \HOLSymConst{\HOLTokenConj{}} (\HOLBoundVar{E} \HOLSymConst{=} \HOLBoundVar{E\sp{\prime}})) \HOLSymConst{\HOLTokenConj{}}
       \HOLConst{CCS_Subst} \HOLBoundVar{E\sp{\prime}} (\HOLConst{rec} \HOLBoundVar{X\sp{\prime}} \HOLBoundVar{E\sp{\prime}}) \HOLBoundVar{X\sp{\prime}} \HOLTokenTransBegin\HOLBoundVar{u}\HOLTokenTransEnd \HOLBoundVar{E\sp{\prime\prime}}
\end{SaveVerbatim}
\newcommand{\HOLCCSTheoremsRECXXcasesXXEQ}{\UseVerbatim{HOLCCSTheoremsRECXXcasesXXEQ}}
\begin{SaveVerbatim}{HOLCCSTheoremsRELABXXcases}
\HOLTokenTurnstile{} \HOLSymConst{\HOLTokenForall{}}\HOLBoundVar{E} \HOLBoundVar{rf} \HOLBoundVar{a\sb{\mathrm{1}}} \HOLBoundVar{a\sb{\mathrm{2}}}.
     \HOLConst{relab} \HOLBoundVar{E} \HOLBoundVar{rf} \HOLTokenTransBegin\HOLBoundVar{a\sb{\mathrm{1}}}\HOLTokenTransEnd \HOLBoundVar{a\sb{\mathrm{2}}} \HOLSymConst{\HOLTokenImp{}}
     \HOLSymConst{\HOLTokenExists{}}\HOLBoundVar{E\sp{\prime}} \HOLBoundVar{u} \HOLBoundVar{E\sp{\prime\prime}} \HOLBoundVar{rf\sp{\prime}}.
       ((\HOLBoundVar{E} \HOLSymConst{=} \HOLBoundVar{E\sp{\prime}}) \HOLSymConst{\HOLTokenConj{}} (\HOLBoundVar{rf} \HOLSymConst{=} \HOLBoundVar{rf\sp{\prime}})) \HOLSymConst{\HOLTokenConj{}} (\HOLBoundVar{a\sb{\mathrm{1}}} \HOLSymConst{=} \HOLConst{relabel} \HOLBoundVar{rf\sp{\prime}} \HOLBoundVar{u}) \HOLSymConst{\HOLTokenConj{}}
       (\HOLBoundVar{a\sb{\mathrm{2}}} \HOLSymConst{=} \HOLConst{relab} \HOLBoundVar{E\sp{\prime\prime}} \HOLBoundVar{rf\sp{\prime}}) \HOLSymConst{\HOLTokenConj{}} \HOLBoundVar{E\sp{\prime}} \HOLTokenTransBegin\HOLBoundVar{u}\HOLTokenTransEnd \HOLBoundVar{E\sp{\prime\prime}}
\end{SaveVerbatim}
\newcommand{\HOLCCSTheoremsRELABXXcases}{\UseVerbatim{HOLCCSTheoremsRELABXXcases}}
\begin{SaveVerbatim}{HOLCCSTheoremsRELABXXcasesXXEQ}
\HOLTokenTurnstile{} \HOLSymConst{\HOLTokenForall{}}\HOLBoundVar{E} \HOLBoundVar{rf} \HOLBoundVar{a\sb{\mathrm{1}}} \HOLBoundVar{a\sb{\mathrm{2}}}.
     \HOLConst{relab} \HOLBoundVar{E} \HOLBoundVar{rf} \HOLTokenTransBegin\HOLBoundVar{a\sb{\mathrm{1}}}\HOLTokenTransEnd \HOLBoundVar{a\sb{\mathrm{2}}} \HOLSymConst{\HOLTokenEquiv{}}
     \HOLSymConst{\HOLTokenExists{}}\HOLBoundVar{E\sp{\prime}} \HOLBoundVar{u} \HOLBoundVar{E\sp{\prime\prime}} \HOLBoundVar{rf\sp{\prime}}.
       ((\HOLBoundVar{E} \HOLSymConst{=} \HOLBoundVar{E\sp{\prime}}) \HOLSymConst{\HOLTokenConj{}} (\HOLBoundVar{rf} \HOLSymConst{=} \HOLBoundVar{rf\sp{\prime}})) \HOLSymConst{\HOLTokenConj{}} (\HOLBoundVar{a\sb{\mathrm{1}}} \HOLSymConst{=} \HOLConst{relabel} \HOLBoundVar{rf\sp{\prime}} \HOLBoundVar{u}) \HOLSymConst{\HOLTokenConj{}}
       (\HOLBoundVar{a\sb{\mathrm{2}}} \HOLSymConst{=} \HOLConst{relab} \HOLBoundVar{E\sp{\prime\prime}} \HOLBoundVar{rf\sp{\prime}}) \HOLSymConst{\HOLTokenConj{}} \HOLBoundVar{E\sp{\prime}} \HOLTokenTransBegin\HOLBoundVar{u}\HOLTokenTransEnd \HOLBoundVar{E\sp{\prime\prime}}
\end{SaveVerbatim}
\newcommand{\HOLCCSTheoremsRELABXXcasesXXEQ}{\UseVerbatim{HOLCCSTheoremsRELABXXcasesXXEQ}}
\begin{SaveVerbatim}{HOLCCSTheoremsRelabXXlabel}
\HOLTokenTurnstile{} \HOLSymConst{\HOLTokenForall{}}\HOLBoundVar{rf} \HOLBoundVar{u} \HOLBoundVar{l}. (\HOLConst{relabel} \HOLBoundVar{rf} \HOLBoundVar{u} \HOLSymConst{=} \HOLConst{label} \HOLBoundVar{l}) \HOLSymConst{\HOLTokenImp{}} \HOLSymConst{\HOLTokenExists{}}\HOLBoundVar{l\sp{\prime}}. \HOLBoundVar{u} \HOLSymConst{=} \HOLConst{label} \HOLBoundVar{l\sp{\prime}}
\end{SaveVerbatim}
\newcommand{\HOLCCSTheoremsRelabXXlabel}{\UseVerbatim{HOLCCSTheoremsRelabXXlabel}}
\begin{SaveVerbatim}{HOLCCSTheoremsRELABXXNILXXNOXXTRANS}
\HOLTokenTurnstile{} \HOLSymConst{\HOLTokenForall{}}\HOLBoundVar{rf} \HOLBoundVar{u} \HOLBoundVar{E}. \HOLSymConst{\HOLTokenNeg{}}(\HOLConst{relab} \HOLConst{nil} \HOLBoundVar{rf} \HOLTokenTransBegin\HOLBoundVar{u}\HOLTokenTransEnd \HOLBoundVar{E})
\end{SaveVerbatim}
\newcommand{\HOLCCSTheoremsRELABXXNILXXNOXXTRANS}{\UseVerbatim{HOLCCSTheoremsRELABXXNILXXNOXXTRANS}}
\begin{SaveVerbatim}{HOLCCSTheoremsRelabXXtau}
\HOLTokenTurnstile{} \HOLSymConst{\HOLTokenForall{}}\HOLBoundVar{rf} \HOLBoundVar{u}. (\HOLConst{relabel} \HOLBoundVar{rf} \HOLBoundVar{u} \HOLSymConst{=} \HOLConst{\ensuremath{\tau}}) \HOLSymConst{\HOLTokenImp{}} (\HOLBoundVar{u} \HOLSymConst{=} \HOLConst{\ensuremath{\tau}})
\end{SaveVerbatim}
\newcommand{\HOLCCSTheoremsRelabXXtau}{\UseVerbatim{HOLCCSTheoremsRelabXXtau}}
\begin{SaveVerbatim}{HOLCCSTheoremsRELABELING}
\HOLTokenTurnstile{} \HOLSymConst{\HOLTokenForall{}}\HOLBoundVar{E} \HOLBoundVar{u} \HOLBoundVar{E\sp{\prime}} \HOLBoundVar{rf}. \HOLBoundVar{E} \HOLTokenTransBegin\HOLBoundVar{u}\HOLTokenTransEnd \HOLBoundVar{E\sp{\prime}} \HOLSymConst{\HOLTokenImp{}} \HOLConst{relab} \HOLBoundVar{E} \HOLBoundVar{rf} \HOLTokenTransBegin\HOLConst{relabel} \HOLBoundVar{rf} \HOLBoundVar{u}\HOLTokenTransEnd \HOLConst{relab} \HOLBoundVar{E\sp{\prime}} \HOLBoundVar{rf}
\end{SaveVerbatim}
\newcommand{\HOLCCSTheoremsRELABELING}{\UseVerbatim{HOLCCSTheoremsRELABELING}}
\begin{SaveVerbatim}{HOLCCSTheoremsREPXXRelabelingXXTHM}
\HOLTokenTurnstile{} \HOLSymConst{\HOLTokenForall{}}\HOLBoundVar{rf}. \HOLConst{Is_Relabeling} (\HOLConst{REP_Relabeling} \HOLBoundVar{rf})
\end{SaveVerbatim}
\newcommand{\HOLCCSTheoremsREPXXRelabelingXXTHM}{\UseVerbatim{HOLCCSTheoremsREPXXRelabelingXXTHM}}
\begin{SaveVerbatim}{HOLCCSTheoremsRESTR}
\HOLTokenTurnstile{} \HOLSymConst{\HOLTokenForall{}}\HOLBoundVar{E} \HOLBoundVar{u} \HOLBoundVar{E\sp{\prime}} \HOLBoundVar{l} \HOLBoundVar{L}.
     \HOLBoundVar{E} \HOLTokenTransBegin\HOLBoundVar{u}\HOLTokenTransEnd \HOLBoundVar{E\sp{\prime}} \HOLSymConst{\HOLTokenConj{}}
     ((\HOLBoundVar{u} \HOLSymConst{=} \HOLConst{\ensuremath{\tau}}) \HOLSymConst{\HOLTokenDisj{}} (\HOLBoundVar{u} \HOLSymConst{=} \HOLConst{label} \HOLBoundVar{l}) \HOLSymConst{\HOLTokenConj{}} \HOLBoundVar{l} \HOLConst{\HOLTokenNotIn{}} \HOLBoundVar{L} \HOLSymConst{\HOLTokenConj{}} \HOLConst{COMPL} \HOLBoundVar{l} \HOLConst{\HOLTokenNotIn{}} \HOLBoundVar{L}) \HOLSymConst{\HOLTokenImp{}}
     \HOLConst{\ensuremath{\nu}} \HOLBoundVar{L} \HOLBoundVar{E} \HOLTokenTransBegin\HOLBoundVar{u}\HOLTokenTransEnd \HOLConst{\ensuremath{\nu}} \HOLBoundVar{L} \HOLBoundVar{E\sp{\prime}}
\end{SaveVerbatim}
\newcommand{\HOLCCSTheoremsRESTR}{\UseVerbatim{HOLCCSTheoremsRESTR}}
\begin{SaveVerbatim}{HOLCCSTheoremsRESTRXXcases}
\HOLTokenTurnstile{} \HOLSymConst{\HOLTokenForall{}}\HOLBoundVar{D\sp{\prime}} \HOLBoundVar{u} \HOLBoundVar{L} \HOLBoundVar{D}.
     \HOLConst{\ensuremath{\nu}} \HOLBoundVar{L} \HOLBoundVar{D} \HOLTokenTransBegin\HOLBoundVar{u}\HOLTokenTransEnd \HOLBoundVar{D\sp{\prime}} \HOLSymConst{\HOLTokenImp{}}
     \HOLSymConst{\HOLTokenExists{}}\HOLBoundVar{E} \HOLBoundVar{E\sp{\prime}} \HOLBoundVar{l} \HOLBoundVar{L\sp{\prime}}.
       ((\HOLBoundVar{L} \HOLSymConst{=} \HOLBoundVar{L\sp{\prime}}) \HOLSymConst{\HOLTokenConj{}} (\HOLBoundVar{D} \HOLSymConst{=} \HOLBoundVar{E})) \HOLSymConst{\HOLTokenConj{}} (\HOLBoundVar{D\sp{\prime}} \HOLSymConst{=} \HOLConst{\ensuremath{\nu}} \HOLBoundVar{L\sp{\prime}} \HOLBoundVar{E\sp{\prime}}) \HOLSymConst{\HOLTokenConj{}} \HOLBoundVar{E} \HOLTokenTransBegin\HOLBoundVar{u}\HOLTokenTransEnd \HOLBoundVar{E\sp{\prime}} \HOLSymConst{\HOLTokenConj{}}
       ((\HOLBoundVar{u} \HOLSymConst{=} \HOLConst{\ensuremath{\tau}}) \HOLSymConst{\HOLTokenDisj{}} (\HOLBoundVar{u} \HOLSymConst{=} \HOLConst{label} \HOLBoundVar{l}) \HOLSymConst{\HOLTokenConj{}} \HOLBoundVar{l} \HOLConst{\HOLTokenNotIn{}} \HOLBoundVar{L\sp{\prime}} \HOLSymConst{\HOLTokenConj{}} \HOLConst{COMPL} \HOLBoundVar{l} \HOLConst{\HOLTokenNotIn{}} \HOLBoundVar{L\sp{\prime}})
\end{SaveVerbatim}
\newcommand{\HOLCCSTheoremsRESTRXXcases}{\UseVerbatim{HOLCCSTheoremsRESTRXXcases}}
\begin{SaveVerbatim}{HOLCCSTheoremsRESTRXXcasesXXEQ}
\HOLTokenTurnstile{} \HOLSymConst{\HOLTokenForall{}}\HOLBoundVar{D\sp{\prime}} \HOLBoundVar{u} \HOLBoundVar{L} \HOLBoundVar{D}.
     \HOLConst{\ensuremath{\nu}} \HOLBoundVar{L} \HOLBoundVar{D} \HOLTokenTransBegin\HOLBoundVar{u}\HOLTokenTransEnd \HOLBoundVar{D\sp{\prime}} \HOLSymConst{\HOLTokenEquiv{}}
     \HOLSymConst{\HOLTokenExists{}}\HOLBoundVar{E} \HOLBoundVar{E\sp{\prime}} \HOLBoundVar{l} \HOLBoundVar{L\sp{\prime}}.
       ((\HOLBoundVar{L} \HOLSymConst{=} \HOLBoundVar{L\sp{\prime}}) \HOLSymConst{\HOLTokenConj{}} (\HOLBoundVar{D} \HOLSymConst{=} \HOLBoundVar{E})) \HOLSymConst{\HOLTokenConj{}} (\HOLBoundVar{D\sp{\prime}} \HOLSymConst{=} \HOLConst{\ensuremath{\nu}} \HOLBoundVar{L\sp{\prime}} \HOLBoundVar{E\sp{\prime}}) \HOLSymConst{\HOLTokenConj{}} \HOLBoundVar{E} \HOLTokenTransBegin\HOLBoundVar{u}\HOLTokenTransEnd \HOLBoundVar{E\sp{\prime}} \HOLSymConst{\HOLTokenConj{}}
       ((\HOLBoundVar{u} \HOLSymConst{=} \HOLConst{\ensuremath{\tau}}) \HOLSymConst{\HOLTokenDisj{}} (\HOLBoundVar{u} \HOLSymConst{=} \HOLConst{label} \HOLBoundVar{l}) \HOLSymConst{\HOLTokenConj{}} \HOLBoundVar{l} \HOLConst{\HOLTokenNotIn{}} \HOLBoundVar{L\sp{\prime}} \HOLSymConst{\HOLTokenConj{}} \HOLConst{COMPL} \HOLBoundVar{l} \HOLConst{\HOLTokenNotIn{}} \HOLBoundVar{L\sp{\prime}})
\end{SaveVerbatim}
\newcommand{\HOLCCSTheoremsRESTRXXcasesXXEQ}{\UseVerbatim{HOLCCSTheoremsRESTRXXcasesXXEQ}}
\begin{SaveVerbatim}{HOLCCSTheoremsRESTRXXLABELXXNOXXTRANS}
\HOLTokenTurnstile{} \HOLSymConst{\HOLTokenForall{}}\HOLBoundVar{l} \HOLBoundVar{L}.
     \HOLBoundVar{l} \HOLConst{\HOLTokenIn{}} \HOLBoundVar{L} \HOLSymConst{\HOLTokenDisj{}} \HOLConst{COMPL} \HOLBoundVar{l} \HOLConst{\HOLTokenIn{}} \HOLBoundVar{L} \HOLSymConst{\HOLTokenImp{}} \HOLSymConst{\HOLTokenForall{}}\HOLBoundVar{E} \HOLBoundVar{u} \HOLBoundVar{E\sp{\prime}}. \HOLSymConst{\HOLTokenNeg{}}(\HOLConst{\ensuremath{\nu}} \HOLBoundVar{L} (\HOLConst{label} \HOLBoundVar{l}\HOLSymConst{..}\HOLBoundVar{E}) \HOLTokenTransBegin\HOLBoundVar{u}\HOLTokenTransEnd \HOLBoundVar{E\sp{\prime}})
\end{SaveVerbatim}
\newcommand{\HOLCCSTheoremsRESTRXXLABELXXNOXXTRANS}{\UseVerbatim{HOLCCSTheoremsRESTRXXLABELXXNOXXTRANS}}
\begin{SaveVerbatim}{HOLCCSTheoremsRESTRXXNILXXNOXXTRANS}
\HOLTokenTurnstile{} \HOLSymConst{\HOLTokenForall{}}\HOLBoundVar{L} \HOLBoundVar{u} \HOLBoundVar{E}. \HOLSymConst{\HOLTokenNeg{}}(\HOLConst{\ensuremath{\nu}} \HOLBoundVar{L} \HOLConst{nil} \HOLTokenTransBegin\HOLBoundVar{u}\HOLTokenTransEnd \HOLBoundVar{E})
\end{SaveVerbatim}
\newcommand{\HOLCCSTheoremsRESTRXXNILXXNOXXTRANS}{\UseVerbatim{HOLCCSTheoremsRESTRXXNILXXNOXXTRANS}}
\begin{SaveVerbatim}{HOLCCSTheoremsSUMOne}
\HOLTokenTurnstile{} \HOLSymConst{\HOLTokenForall{}}\HOLBoundVar{E} \HOLBoundVar{u} \HOLBoundVar{E\sb{\mathrm{1}}} \HOLBoundVar{E\sp{\prime}}. \HOLBoundVar{E} \HOLTokenTransBegin\HOLBoundVar{u}\HOLTokenTransEnd \HOLBoundVar{E\sb{\mathrm{1}}} \HOLSymConst{\HOLTokenImp{}} \HOLBoundVar{E} \HOLSymConst{+} \HOLBoundVar{E\sp{\prime}} \HOLTokenTransBegin\HOLBoundVar{u}\HOLTokenTransEnd \HOLBoundVar{E\sb{\mathrm{1}}}
\end{SaveVerbatim}
\newcommand{\HOLCCSTheoremsSUMOne}{\UseVerbatim{HOLCCSTheoremsSUMOne}}
\begin{SaveVerbatim}{HOLCCSTheoremsSUMTwo}
\HOLTokenTurnstile{} \HOLSymConst{\HOLTokenForall{}}\HOLBoundVar{E} \HOLBoundVar{u} \HOLBoundVar{E\sb{\mathrm{1}}} \HOLBoundVar{E\sp{\prime}}. \HOLBoundVar{E} \HOLTokenTransBegin\HOLBoundVar{u}\HOLTokenTransEnd \HOLBoundVar{E\sb{\mathrm{1}}} \HOLSymConst{\HOLTokenImp{}} \HOLBoundVar{E\sp{\prime}} \HOLSymConst{+} \HOLBoundVar{E} \HOLTokenTransBegin\HOLBoundVar{u}\HOLTokenTransEnd \HOLBoundVar{E\sb{\mathrm{1}}}
\end{SaveVerbatim}
\newcommand{\HOLCCSTheoremsSUMTwo}{\UseVerbatim{HOLCCSTheoremsSUMTwo}}
\begin{SaveVerbatim}{HOLCCSTheoremsSUMXXcases}
\HOLTokenTurnstile{} \HOLSymConst{\HOLTokenForall{}}\HOLBoundVar{D} \HOLBoundVar{D\sp{\prime}} \HOLBoundVar{u} \HOLBoundVar{D\sp{\prime\prime}}.
     \HOLBoundVar{D} \HOLSymConst{+} \HOLBoundVar{D\sp{\prime}} \HOLTokenTransBegin\HOLBoundVar{u}\HOLTokenTransEnd \HOLBoundVar{D\sp{\prime\prime}} \HOLSymConst{\HOLTokenImp{}}
     (\HOLSymConst{\HOLTokenExists{}}\HOLBoundVar{E} \HOLBoundVar{E\sp{\prime}}. ((\HOLBoundVar{D} \HOLSymConst{=} \HOLBoundVar{E}) \HOLSymConst{\HOLTokenConj{}} (\HOLBoundVar{D\sp{\prime}} \HOLSymConst{=} \HOLBoundVar{E\sp{\prime}})) \HOLSymConst{\HOLTokenConj{}} \HOLBoundVar{E} \HOLTokenTransBegin\HOLBoundVar{u}\HOLTokenTransEnd \HOLBoundVar{D\sp{\prime\prime}}) \HOLSymConst{\HOLTokenDisj{}}
     \HOLSymConst{\HOLTokenExists{}}\HOLBoundVar{E} \HOLBoundVar{E\sp{\prime}}. ((\HOLBoundVar{D} \HOLSymConst{=} \HOLBoundVar{E\sp{\prime}}) \HOLSymConst{\HOLTokenConj{}} (\HOLBoundVar{D\sp{\prime}} \HOLSymConst{=} \HOLBoundVar{E})) \HOLSymConst{\HOLTokenConj{}} \HOLBoundVar{E} \HOLTokenTransBegin\HOLBoundVar{u}\HOLTokenTransEnd \HOLBoundVar{D\sp{\prime\prime}}
\end{SaveVerbatim}
\newcommand{\HOLCCSTheoremsSUMXXcases}{\UseVerbatim{HOLCCSTheoremsSUMXXcases}}
\begin{SaveVerbatim}{HOLCCSTheoremsSUMXXcasesXXEQ}
\HOLTokenTurnstile{} \HOLSymConst{\HOLTokenForall{}}\HOLBoundVar{D} \HOLBoundVar{D\sp{\prime}} \HOLBoundVar{u} \HOLBoundVar{D\sp{\prime\prime}}.
     \HOLBoundVar{D} \HOLSymConst{+} \HOLBoundVar{D\sp{\prime}} \HOLTokenTransBegin\HOLBoundVar{u}\HOLTokenTransEnd \HOLBoundVar{D\sp{\prime\prime}} \HOLSymConst{\HOLTokenEquiv{}}
     (\HOLSymConst{\HOLTokenExists{}}\HOLBoundVar{E} \HOLBoundVar{E\sp{\prime}}. ((\HOLBoundVar{D} \HOLSymConst{=} \HOLBoundVar{E}) \HOLSymConst{\HOLTokenConj{}} (\HOLBoundVar{D\sp{\prime}} \HOLSymConst{=} \HOLBoundVar{E\sp{\prime}})) \HOLSymConst{\HOLTokenConj{}} \HOLBoundVar{E} \HOLTokenTransBegin\HOLBoundVar{u}\HOLTokenTransEnd \HOLBoundVar{D\sp{\prime\prime}}) \HOLSymConst{\HOLTokenDisj{}}
     \HOLSymConst{\HOLTokenExists{}}\HOLBoundVar{E} \HOLBoundVar{E\sp{\prime}}. ((\HOLBoundVar{D} \HOLSymConst{=} \HOLBoundVar{E\sp{\prime}}) \HOLSymConst{\HOLTokenConj{}} (\HOLBoundVar{D\sp{\prime}} \HOLSymConst{=} \HOLBoundVar{E})) \HOLSymConst{\HOLTokenConj{}} \HOLBoundVar{E} \HOLTokenTransBegin\HOLBoundVar{u}\HOLTokenTransEnd \HOLBoundVar{D\sp{\prime\prime}}
\end{SaveVerbatim}
\newcommand{\HOLCCSTheoremsSUMXXcasesXXEQ}{\UseVerbatim{HOLCCSTheoremsSUMXXcasesXXEQ}}
\begin{SaveVerbatim}{HOLCCSTheoremsTRANSXXASSOCXXEQ}
\HOLTokenTurnstile{} \HOLSymConst{\HOLTokenForall{}}\HOLBoundVar{E} \HOLBoundVar{E\sp{\prime}} \HOLBoundVar{E\sp{\prime\prime}} \HOLBoundVar{E\sb{\mathrm{1}}} \HOLBoundVar{u}. \HOLBoundVar{E} \HOLSymConst{+} \HOLBoundVar{E\sp{\prime}} \HOLSymConst{+} \HOLBoundVar{E\sp{\prime\prime}} \HOLTokenTransBegin\HOLBoundVar{u}\HOLTokenTransEnd \HOLBoundVar{E\sb{\mathrm{1}}} \HOLSymConst{\HOLTokenEquiv{}} \HOLBoundVar{E} \HOLSymConst{+} (\HOLBoundVar{E\sp{\prime}} \HOLSymConst{+} \HOLBoundVar{E\sp{\prime\prime}}) \HOLTokenTransBegin\HOLBoundVar{u}\HOLTokenTransEnd \HOLBoundVar{E\sb{\mathrm{1}}}
\end{SaveVerbatim}
\newcommand{\HOLCCSTheoremsTRANSXXASSOCXXEQ}{\UseVerbatim{HOLCCSTheoremsTRANSXXASSOCXXEQ}}
\begin{SaveVerbatim}{HOLCCSTheoremsTRANSXXASSOCXXRL}
\HOLTokenTurnstile{} \HOLSymConst{\HOLTokenForall{}}\HOLBoundVar{E} \HOLBoundVar{E\sp{\prime}} \HOLBoundVar{E\sp{\prime\prime}} \HOLBoundVar{E\sb{\mathrm{1}}} \HOLBoundVar{u}. \HOLBoundVar{E} \HOLSymConst{+} (\HOLBoundVar{E\sp{\prime}} \HOLSymConst{+} \HOLBoundVar{E\sp{\prime\prime}}) \HOLTokenTransBegin\HOLBoundVar{u}\HOLTokenTransEnd \HOLBoundVar{E\sb{\mathrm{1}}} \HOLSymConst{\HOLTokenImp{}} \HOLBoundVar{E} \HOLSymConst{+} \HOLBoundVar{E\sp{\prime}} \HOLSymConst{+} \HOLBoundVar{E\sp{\prime\prime}} \HOLTokenTransBegin\HOLBoundVar{u}\HOLTokenTransEnd \HOLBoundVar{E\sb{\mathrm{1}}}
\end{SaveVerbatim}
\newcommand{\HOLCCSTheoremsTRANSXXASSOCXXRL}{\UseVerbatim{HOLCCSTheoremsTRANSXXASSOCXXRL}}
\begin{SaveVerbatim}{HOLCCSTheoremsTRANSXXcases}
\HOLTokenTurnstile{} \HOLSymConst{\HOLTokenForall{}}\HOLBoundVar{a\sb{\mathrm{0}}} \HOLBoundVar{a\sb{\mathrm{1}}} \HOLBoundVar{a\sb{\mathrm{2}}}.
     \HOLBoundVar{a\sb{\mathrm{0}}} \HOLTokenTransBegin\HOLBoundVar{a\sb{\mathrm{1}}}\HOLTokenTransEnd \HOLBoundVar{a\sb{\mathrm{2}}} \HOLSymConst{\HOLTokenEquiv{}}
     (\HOLBoundVar{a\sb{\mathrm{0}}} \HOLSymConst{=} \HOLBoundVar{a\sb{\mathrm{1}}}\HOLSymConst{..}\HOLBoundVar{a\sb{\mathrm{2}}}) \HOLSymConst{\HOLTokenDisj{}} (\HOLSymConst{\HOLTokenExists{}}\HOLBoundVar{E} \HOLBoundVar{E\sp{\prime}}. (\HOLBoundVar{a\sb{\mathrm{0}}} \HOLSymConst{=} \HOLBoundVar{E} \HOLSymConst{+} \HOLBoundVar{E\sp{\prime}}) \HOLSymConst{\HOLTokenConj{}} \HOLBoundVar{E} \HOLTokenTransBegin\HOLBoundVar{a\sb{\mathrm{1}}}\HOLTokenTransEnd \HOLBoundVar{a\sb{\mathrm{2}}}) \HOLSymConst{\HOLTokenDisj{}}
     (\HOLSymConst{\HOLTokenExists{}}\HOLBoundVar{E} \HOLBoundVar{E\sp{\prime}}. (\HOLBoundVar{a\sb{\mathrm{0}}} \HOLSymConst{=} \HOLBoundVar{E\sp{\prime}} \HOLSymConst{+} \HOLBoundVar{E}) \HOLSymConst{\HOLTokenConj{}} \HOLBoundVar{E} \HOLTokenTransBegin\HOLBoundVar{a\sb{\mathrm{1}}}\HOLTokenTransEnd \HOLBoundVar{a\sb{\mathrm{2}}}) \HOLSymConst{\HOLTokenDisj{}}
     (\HOLSymConst{\HOLTokenExists{}}\HOLBoundVar{E} \HOLBoundVar{E\sb{\mathrm{1}}} \HOLBoundVar{E\sp{\prime}}. (\HOLBoundVar{a\sb{\mathrm{0}}} \HOLSymConst{=} \HOLBoundVar{E} \HOLSymConst{\ensuremath{\parallel}} \HOLBoundVar{E\sp{\prime}}) \HOLSymConst{\HOLTokenConj{}} (\HOLBoundVar{a\sb{\mathrm{2}}} \HOLSymConst{=} \HOLBoundVar{E\sb{\mathrm{1}}} \HOLSymConst{\ensuremath{\parallel}} \HOLBoundVar{E\sp{\prime}}) \HOLSymConst{\HOLTokenConj{}} \HOLBoundVar{E} \HOLTokenTransBegin\HOLBoundVar{a\sb{\mathrm{1}}}\HOLTokenTransEnd \HOLBoundVar{E\sb{\mathrm{1}}}) \HOLSymConst{\HOLTokenDisj{}}
     (\HOLSymConst{\HOLTokenExists{}}\HOLBoundVar{E} \HOLBoundVar{E\sb{\mathrm{1}}} \HOLBoundVar{E\sp{\prime}}. (\HOLBoundVar{a\sb{\mathrm{0}}} \HOLSymConst{=} \HOLBoundVar{E\sp{\prime}} \HOLSymConst{\ensuremath{\parallel}} \HOLBoundVar{E}) \HOLSymConst{\HOLTokenConj{}} (\HOLBoundVar{a\sb{\mathrm{2}}} \HOLSymConst{=} \HOLBoundVar{E\sp{\prime}} \HOLSymConst{\ensuremath{\parallel}} \HOLBoundVar{E\sb{\mathrm{1}}}) \HOLSymConst{\HOLTokenConj{}} \HOLBoundVar{E} \HOLTokenTransBegin\HOLBoundVar{a\sb{\mathrm{1}}}\HOLTokenTransEnd \HOLBoundVar{E\sb{\mathrm{1}}}) \HOLSymConst{\HOLTokenDisj{}}
     (\HOLSymConst{\HOLTokenExists{}}\HOLBoundVar{E} \HOLBoundVar{l} \HOLBoundVar{E\sb{\mathrm{1}}} \HOLBoundVar{E\sp{\prime}} \HOLBoundVar{E\sb{\mathrm{2}}}.
        (\HOLBoundVar{a\sb{\mathrm{0}}} \HOLSymConst{=} \HOLBoundVar{E} \HOLSymConst{\ensuremath{\parallel}} \HOLBoundVar{E\sp{\prime}}) \HOLSymConst{\HOLTokenConj{}} (\HOLBoundVar{a\sb{\mathrm{1}}} \HOLSymConst{=} \HOLConst{\ensuremath{\tau}}) \HOLSymConst{\HOLTokenConj{}} (\HOLBoundVar{a\sb{\mathrm{2}}} \HOLSymConst{=} \HOLBoundVar{E\sb{\mathrm{1}}} \HOLSymConst{\ensuremath{\parallel}} \HOLBoundVar{E\sb{\mathrm{2}}}) \HOLSymConst{\HOLTokenConj{}}
        \HOLBoundVar{E} \HOLTokenTransBegin\HOLConst{label} \HOLBoundVar{l}\HOLTokenTransEnd \HOLBoundVar{E\sb{\mathrm{1}}} \HOLSymConst{\HOLTokenConj{}} \HOLBoundVar{E\sp{\prime}} \HOLTokenTransBegin\HOLConst{label} (\HOLConst{COMPL} \HOLBoundVar{l})\HOLTokenTransEnd \HOLBoundVar{E\sb{\mathrm{2}}}) \HOLSymConst{\HOLTokenDisj{}}
     (\HOLSymConst{\HOLTokenExists{}}\HOLBoundVar{E} \HOLBoundVar{E\sp{\prime}} \HOLBoundVar{l} \HOLBoundVar{L}.
        (\HOLBoundVar{a\sb{\mathrm{0}}} \HOLSymConst{=} \HOLConst{\ensuremath{\nu}} \HOLBoundVar{L} \HOLBoundVar{E}) \HOLSymConst{\HOLTokenConj{}} (\HOLBoundVar{a\sb{\mathrm{2}}} \HOLSymConst{=} \HOLConst{\ensuremath{\nu}} \HOLBoundVar{L} \HOLBoundVar{E\sp{\prime}}) \HOLSymConst{\HOLTokenConj{}} \HOLBoundVar{E} \HOLTokenTransBegin\HOLBoundVar{a\sb{\mathrm{1}}}\HOLTokenTransEnd \HOLBoundVar{E\sp{\prime}} \HOLSymConst{\HOLTokenConj{}}
        ((\HOLBoundVar{a\sb{\mathrm{1}}} \HOLSymConst{=} \HOLConst{\ensuremath{\tau}}) \HOLSymConst{\HOLTokenDisj{}} (\HOLBoundVar{a\sb{\mathrm{1}}} \HOLSymConst{=} \HOLConst{label} \HOLBoundVar{l}) \HOLSymConst{\HOLTokenConj{}} \HOLBoundVar{l} \HOLConst{\HOLTokenNotIn{}} \HOLBoundVar{L} \HOLSymConst{\HOLTokenConj{}} \HOLConst{COMPL} \HOLBoundVar{l} \HOLConst{\HOLTokenNotIn{}} \HOLBoundVar{L})) \HOLSymConst{\HOLTokenDisj{}}
     (\HOLSymConst{\HOLTokenExists{}}\HOLBoundVar{E} \HOLBoundVar{u} \HOLBoundVar{E\sp{\prime}} \HOLBoundVar{rf}.
        (\HOLBoundVar{a\sb{\mathrm{0}}} \HOLSymConst{=} \HOLConst{relab} \HOLBoundVar{E} \HOLBoundVar{rf}) \HOLSymConst{\HOLTokenConj{}} (\HOLBoundVar{a\sb{\mathrm{1}}} \HOLSymConst{=} \HOLConst{relabel} \HOLBoundVar{rf} \HOLBoundVar{u}) \HOLSymConst{\HOLTokenConj{}}
        (\HOLBoundVar{a\sb{\mathrm{2}}} \HOLSymConst{=} \HOLConst{relab} \HOLBoundVar{E\sp{\prime}} \HOLBoundVar{rf}) \HOLSymConst{\HOLTokenConj{}} \HOLBoundVar{E} \HOLTokenTransBegin\HOLBoundVar{u}\HOLTokenTransEnd \HOLBoundVar{E\sp{\prime}}) \HOLSymConst{\HOLTokenDisj{}}
     \HOLSymConst{\HOLTokenExists{}}\HOLBoundVar{E} \HOLBoundVar{X}. (\HOLBoundVar{a\sb{\mathrm{0}}} \HOLSymConst{=} \HOLConst{rec} \HOLBoundVar{X} \HOLBoundVar{E}) \HOLSymConst{\HOLTokenConj{}} \HOLConst{CCS_Subst} \HOLBoundVar{E} (\HOLConst{rec} \HOLBoundVar{X} \HOLBoundVar{E}) \HOLBoundVar{X} \HOLTokenTransBegin\HOLBoundVar{a\sb{\mathrm{1}}}\HOLTokenTransEnd \HOLBoundVar{a\sb{\mathrm{2}}}
\end{SaveVerbatim}
\newcommand{\HOLCCSTheoremsTRANSXXcases}{\UseVerbatim{HOLCCSTheoremsTRANSXXcases}}
\begin{SaveVerbatim}{HOLCCSTheoremsTRANSXXCOMMXXEQ}
\HOLTokenTurnstile{} \HOLSymConst{\HOLTokenForall{}}\HOLBoundVar{E} \HOLBoundVar{E\sp{\prime}} \HOLBoundVar{E\sp{\prime\prime}} \HOLBoundVar{u}. \HOLBoundVar{E} \HOLSymConst{+} \HOLBoundVar{E\sp{\prime}} \HOLTokenTransBegin\HOLBoundVar{u}\HOLTokenTransEnd \HOLBoundVar{E\sp{\prime\prime}} \HOLSymConst{\HOLTokenEquiv{}} \HOLBoundVar{E\sp{\prime}} \HOLSymConst{+} \HOLBoundVar{E} \HOLTokenTransBegin\HOLBoundVar{u}\HOLTokenTransEnd \HOLBoundVar{E\sp{\prime\prime}}
\end{SaveVerbatim}
\newcommand{\HOLCCSTheoremsTRANSXXCOMMXXEQ}{\UseVerbatim{HOLCCSTheoremsTRANSXXCOMMXXEQ}}
\begin{SaveVerbatim}{HOLCCSTheoremsTRANSXXIMPXXNOXXNIL}
\HOLTokenTurnstile{} \HOLSymConst{\HOLTokenForall{}}\HOLBoundVar{E} \HOLBoundVar{u} \HOLBoundVar{E\sp{\prime}}. \HOLBoundVar{E} \HOLTokenTransBegin\HOLBoundVar{u}\HOLTokenTransEnd \HOLBoundVar{E\sp{\prime}} \HOLSymConst{\HOLTokenImp{}} \HOLBoundVar{E} \HOLSymConst{\HOLTokenNotEqual{}} \HOLConst{nil}
\end{SaveVerbatim}
\newcommand{\HOLCCSTheoremsTRANSXXIMPXXNOXXNIL}{\UseVerbatim{HOLCCSTheoremsTRANSXXIMPXXNOXXNIL}}
\begin{SaveVerbatim}{HOLCCSTheoremsTRANSXXIMPXXNOXXNILYY}
\HOLTokenTurnstile{} \HOLSymConst{\HOLTokenForall{}}\HOLBoundVar{E} \HOLBoundVar{u} \HOLBoundVar{E\sp{\prime}}. \HOLBoundVar{E} \HOLTokenTransBegin\HOLBoundVar{u}\HOLTokenTransEnd \HOLBoundVar{E\sp{\prime}} \HOLSymConst{\HOLTokenImp{}} \HOLBoundVar{E} \HOLSymConst{\HOLTokenNotEqual{}} \HOLConst{nil}
\end{SaveVerbatim}
\newcommand{\HOLCCSTheoremsTRANSXXIMPXXNOXXNILYY}{\UseVerbatim{HOLCCSTheoremsTRANSXXIMPXXNOXXNILYY}}
\begin{SaveVerbatim}{HOLCCSTheoremsTRANSXXIMPXXNOXXRESTRXXNIL}
\HOLTokenTurnstile{} \HOLSymConst{\HOLTokenForall{}}\HOLBoundVar{E} \HOLBoundVar{u} \HOLBoundVar{E\sp{\prime}}. \HOLBoundVar{E} \HOLTokenTransBegin\HOLBoundVar{u}\HOLTokenTransEnd \HOLBoundVar{E\sp{\prime}} \HOLSymConst{\HOLTokenImp{}} \HOLSymConst{\HOLTokenForall{}}\HOLBoundVar{L}. \HOLBoundVar{E} \HOLSymConst{\HOLTokenNotEqual{}} \HOLConst{\ensuremath{\nu}} \HOLBoundVar{L} \HOLConst{nil}
\end{SaveVerbatim}
\newcommand{\HOLCCSTheoremsTRANSXXIMPXXNOXXRESTRXXNIL}{\UseVerbatim{HOLCCSTheoremsTRANSXXIMPXXNOXXRESTRXXNIL}}
\begin{SaveVerbatim}{HOLCCSTheoremsTRANSXXind}
\HOLTokenTurnstile{} \HOLSymConst{\HOLTokenForall{}}\HOLBoundVar{TRANS\sp{\prime}}.
     (\HOLSymConst{\HOLTokenForall{}}\HOLBoundVar{E} \HOLBoundVar{u}. \HOLBoundVar{TRANS\sp{\prime}} (\HOLBoundVar{u}\HOLSymConst{..}\HOLBoundVar{E}) \HOLBoundVar{u} \HOLBoundVar{E}) \HOLSymConst{\HOLTokenConj{}}
     (\HOLSymConst{\HOLTokenForall{}}\HOLBoundVar{E} \HOLBoundVar{u} \HOLBoundVar{E\sb{\mathrm{1}}} \HOLBoundVar{E\sp{\prime}}. \HOLBoundVar{TRANS\sp{\prime}} \HOLBoundVar{E} \HOLBoundVar{u} \HOLBoundVar{E\sb{\mathrm{1}}} \HOLSymConst{\HOLTokenImp{}} \HOLBoundVar{TRANS\sp{\prime}} (\HOLBoundVar{E} \HOLSymConst{+} \HOLBoundVar{E\sp{\prime}}) \HOLBoundVar{u} \HOLBoundVar{E\sb{\mathrm{1}}}) \HOLSymConst{\HOLTokenConj{}}
     (\HOLSymConst{\HOLTokenForall{}}\HOLBoundVar{E} \HOLBoundVar{u} \HOLBoundVar{E\sb{\mathrm{1}}} \HOLBoundVar{E\sp{\prime}}. \HOLBoundVar{TRANS\sp{\prime}} \HOLBoundVar{E} \HOLBoundVar{u} \HOLBoundVar{E\sb{\mathrm{1}}} \HOLSymConst{\HOLTokenImp{}} \HOLBoundVar{TRANS\sp{\prime}} (\HOLBoundVar{E\sp{\prime}} \HOLSymConst{+} \HOLBoundVar{E}) \HOLBoundVar{u} \HOLBoundVar{E\sb{\mathrm{1}}}) \HOLSymConst{\HOLTokenConj{}}
     (\HOLSymConst{\HOLTokenForall{}}\HOLBoundVar{E} \HOLBoundVar{u} \HOLBoundVar{E\sb{\mathrm{1}}} \HOLBoundVar{E\sp{\prime}}. \HOLBoundVar{TRANS\sp{\prime}} \HOLBoundVar{E} \HOLBoundVar{u} \HOLBoundVar{E\sb{\mathrm{1}}} \HOLSymConst{\HOLTokenImp{}} \HOLBoundVar{TRANS\sp{\prime}} (\HOLBoundVar{E} \HOLSymConst{\ensuremath{\parallel}} \HOLBoundVar{E\sp{\prime}}) \HOLBoundVar{u} (\HOLBoundVar{E\sb{\mathrm{1}}} \HOLSymConst{\ensuremath{\parallel}} \HOLBoundVar{E\sp{\prime}})) \HOLSymConst{\HOLTokenConj{}}
     (\HOLSymConst{\HOLTokenForall{}}\HOLBoundVar{E} \HOLBoundVar{u} \HOLBoundVar{E\sb{\mathrm{1}}} \HOLBoundVar{E\sp{\prime}}. \HOLBoundVar{TRANS\sp{\prime}} \HOLBoundVar{E} \HOLBoundVar{u} \HOLBoundVar{E\sb{\mathrm{1}}} \HOLSymConst{\HOLTokenImp{}} \HOLBoundVar{TRANS\sp{\prime}} (\HOLBoundVar{E\sp{\prime}} \HOLSymConst{\ensuremath{\parallel}} \HOLBoundVar{E}) \HOLBoundVar{u} (\HOLBoundVar{E\sp{\prime}} \HOLSymConst{\ensuremath{\parallel}} \HOLBoundVar{E\sb{\mathrm{1}}})) \HOLSymConst{\HOLTokenConj{}}
     (\HOLSymConst{\HOLTokenForall{}}\HOLBoundVar{E} \HOLBoundVar{l} \HOLBoundVar{E\sb{\mathrm{1}}} \HOLBoundVar{E\sp{\prime}} \HOLBoundVar{E\sb{\mathrm{2}}}.
        \HOLBoundVar{TRANS\sp{\prime}} \HOLBoundVar{E} (\HOLConst{label} \HOLBoundVar{l}) \HOLBoundVar{E\sb{\mathrm{1}}} \HOLSymConst{\HOLTokenConj{}} \HOLBoundVar{TRANS\sp{\prime}} \HOLBoundVar{E\sp{\prime}} (\HOLConst{label} (\HOLConst{COMPL} \HOLBoundVar{l})) \HOLBoundVar{E\sb{\mathrm{2}}} \HOLSymConst{\HOLTokenImp{}}
        \HOLBoundVar{TRANS\sp{\prime}} (\HOLBoundVar{E} \HOLSymConst{\ensuremath{\parallel}} \HOLBoundVar{E\sp{\prime}}) \HOLConst{\ensuremath{\tau}} (\HOLBoundVar{E\sb{\mathrm{1}}} \HOLSymConst{\ensuremath{\parallel}} \HOLBoundVar{E\sb{\mathrm{2}}})) \HOLSymConst{\HOLTokenConj{}}
     (\HOLSymConst{\HOLTokenForall{}}\HOLBoundVar{E} \HOLBoundVar{u} \HOLBoundVar{E\sp{\prime}} \HOLBoundVar{l} \HOLBoundVar{L}.
        \HOLBoundVar{TRANS\sp{\prime}} \HOLBoundVar{E} \HOLBoundVar{u} \HOLBoundVar{E\sp{\prime}} \HOLSymConst{\HOLTokenConj{}}
        ((\HOLBoundVar{u} \HOLSymConst{=} \HOLConst{\ensuremath{\tau}}) \HOLSymConst{\HOLTokenDisj{}} (\HOLBoundVar{u} \HOLSymConst{=} \HOLConst{label} \HOLBoundVar{l}) \HOLSymConst{\HOLTokenConj{}} \HOLBoundVar{l} \HOLConst{\HOLTokenNotIn{}} \HOLBoundVar{L} \HOLSymConst{\HOLTokenConj{}} \HOLConst{COMPL} \HOLBoundVar{l} \HOLConst{\HOLTokenNotIn{}} \HOLBoundVar{L}) \HOLSymConst{\HOLTokenImp{}}
        \HOLBoundVar{TRANS\sp{\prime}} (\HOLConst{\ensuremath{\nu}} \HOLBoundVar{L} \HOLBoundVar{E}) \HOLBoundVar{u} (\HOLConst{\ensuremath{\nu}} \HOLBoundVar{L} \HOLBoundVar{E\sp{\prime}})) \HOLSymConst{\HOLTokenConj{}}
     (\HOLSymConst{\HOLTokenForall{}}\HOLBoundVar{E} \HOLBoundVar{u} \HOLBoundVar{E\sp{\prime}} \HOLBoundVar{rf}.
        \HOLBoundVar{TRANS\sp{\prime}} \HOLBoundVar{E} \HOLBoundVar{u} \HOLBoundVar{E\sp{\prime}} \HOLSymConst{\HOLTokenImp{}}
        \HOLBoundVar{TRANS\sp{\prime}} (\HOLConst{relab} \HOLBoundVar{E} \HOLBoundVar{rf}) (\HOLConst{relabel} \HOLBoundVar{rf} \HOLBoundVar{u}) (\HOLConst{relab} \HOLBoundVar{E\sp{\prime}} \HOLBoundVar{rf})) \HOLSymConst{\HOLTokenConj{}}
     (\HOLSymConst{\HOLTokenForall{}}\HOLBoundVar{E} \HOLBoundVar{u} \HOLBoundVar{X} \HOLBoundVar{E\sb{\mathrm{1}}}.
        \HOLBoundVar{TRANS\sp{\prime}} (\HOLConst{CCS_Subst} \HOLBoundVar{E} (\HOLConst{rec} \HOLBoundVar{X} \HOLBoundVar{E}) \HOLBoundVar{X}) \HOLBoundVar{u} \HOLBoundVar{E\sb{\mathrm{1}}} \HOLSymConst{\HOLTokenImp{}}
        \HOLBoundVar{TRANS\sp{\prime}} (\HOLConst{rec} \HOLBoundVar{X} \HOLBoundVar{E}) \HOLBoundVar{u} \HOLBoundVar{E\sb{\mathrm{1}}}) \HOLSymConst{\HOLTokenImp{}}
     \HOLSymConst{\HOLTokenForall{}}\HOLBoundVar{a\sb{\mathrm{0}}} \HOLBoundVar{a\sb{\mathrm{1}}} \HOLBoundVar{a\sb{\mathrm{2}}}. \HOLBoundVar{a\sb{\mathrm{0}}} \HOLTokenTransBegin\HOLBoundVar{a\sb{\mathrm{1}}}\HOLTokenTransEnd \HOLBoundVar{a\sb{\mathrm{2}}} \HOLSymConst{\HOLTokenImp{}} \HOLBoundVar{TRANS\sp{\prime}} \HOLBoundVar{a\sb{\mathrm{0}}} \HOLBoundVar{a\sb{\mathrm{1}}} \HOLBoundVar{a\sb{\mathrm{2}}}
\end{SaveVerbatim}
\newcommand{\HOLCCSTheoremsTRANSXXind}{\UseVerbatim{HOLCCSTheoremsTRANSXXind}}
\begin{SaveVerbatim}{HOLCCSTheoremsTRANSXXPXXRESTR}
\HOLTokenTurnstile{} \HOLSymConst{\HOLTokenForall{}}\HOLBoundVar{E} \HOLBoundVar{u} \HOLBoundVar{E\sp{\prime}} \HOLBoundVar{L}. \HOLConst{\ensuremath{\nu}} \HOLBoundVar{L} \HOLBoundVar{E} \HOLTokenTransBegin\HOLBoundVar{u}\HOLTokenTransEnd \HOLConst{\ensuremath{\nu}} \HOLBoundVar{L} \HOLBoundVar{E\sp{\prime}} \HOLSymConst{\HOLTokenImp{}} \HOLBoundVar{E} \HOLTokenTransBegin\HOLBoundVar{u}\HOLTokenTransEnd \HOLBoundVar{E\sp{\prime}}
\end{SaveVerbatim}
\newcommand{\HOLCCSTheoremsTRANSXXPXXRESTR}{\UseVerbatim{HOLCCSTheoremsTRANSXXPXXRESTR}}
\begin{SaveVerbatim}{HOLCCSTheoremsTRANSXXPXXSUMXXP}
\HOLTokenTurnstile{} \HOLSymConst{\HOLTokenForall{}}\HOLBoundVar{E} \HOLBoundVar{u} \HOLBoundVar{E\sp{\prime}}. \HOLBoundVar{E} \HOLSymConst{+} \HOLBoundVar{E} \HOLTokenTransBegin\HOLBoundVar{u}\HOLTokenTransEnd \HOLBoundVar{E\sp{\prime}} \HOLSymConst{\HOLTokenImp{}} \HOLBoundVar{E} \HOLTokenTransBegin\HOLBoundVar{u}\HOLTokenTransEnd \HOLBoundVar{E\sp{\prime}}
\end{SaveVerbatim}
\newcommand{\HOLCCSTheoremsTRANSXXPXXSUMXXP}{\UseVerbatim{HOLCCSTheoremsTRANSXXPXXSUMXXP}}
\begin{SaveVerbatim}{HOLCCSTheoremsTRANSXXPXXSUMXXPXXEQ}
\HOLTokenTurnstile{} \HOLSymConst{\HOLTokenForall{}}\HOLBoundVar{E} \HOLBoundVar{u} \HOLBoundVar{E\sp{\prime}}. \HOLBoundVar{E} \HOLSymConst{+} \HOLBoundVar{E} \HOLTokenTransBegin\HOLBoundVar{u}\HOLTokenTransEnd \HOLBoundVar{E\sp{\prime}} \HOLSymConst{\HOLTokenEquiv{}} \HOLBoundVar{E} \HOLTokenTransBegin\HOLBoundVar{u}\HOLTokenTransEnd \HOLBoundVar{E\sp{\prime}}
\end{SaveVerbatim}
\newcommand{\HOLCCSTheoremsTRANSXXPXXSUMXXPXXEQ}{\UseVerbatim{HOLCCSTheoremsTRANSXXPXXSUMXXPXXEQ}}
\begin{SaveVerbatim}{HOLCCSTheoremsTRANSXXPAR}
\HOLTokenTurnstile{} \HOLSymConst{\HOLTokenForall{}}\HOLBoundVar{E} \HOLBoundVar{E\sp{\prime}} \HOLBoundVar{u} \HOLBoundVar{E\sp{\prime\prime}}.
     \HOLBoundVar{E} \HOLSymConst{\ensuremath{\parallel}} \HOLBoundVar{E\sp{\prime}} \HOLTokenTransBegin\HOLBoundVar{u}\HOLTokenTransEnd \HOLBoundVar{E\sp{\prime\prime}} \HOLSymConst{\HOLTokenImp{}}
     (\HOLSymConst{\HOLTokenExists{}}\HOLBoundVar{E\sb{\mathrm{1}}}. (\HOLBoundVar{E\sp{\prime\prime}} \HOLSymConst{=} \HOLBoundVar{E\sb{\mathrm{1}}} \HOLSymConst{\ensuremath{\parallel}} \HOLBoundVar{E\sp{\prime}}) \HOLSymConst{\HOLTokenConj{}} \HOLBoundVar{E} \HOLTokenTransBegin\HOLBoundVar{u}\HOLTokenTransEnd \HOLBoundVar{E\sb{\mathrm{1}}}) \HOLSymConst{\HOLTokenDisj{}}
     (\HOLSymConst{\HOLTokenExists{}}\HOLBoundVar{E\sb{\mathrm{1}}}. (\HOLBoundVar{E\sp{\prime\prime}} \HOLSymConst{=} \HOLBoundVar{E} \HOLSymConst{\ensuremath{\parallel}} \HOLBoundVar{E\sb{\mathrm{1}}}) \HOLSymConst{\HOLTokenConj{}} \HOLBoundVar{E\sp{\prime}} \HOLTokenTransBegin\HOLBoundVar{u}\HOLTokenTransEnd \HOLBoundVar{E\sb{\mathrm{1}}}) \HOLSymConst{\HOLTokenDisj{}}
     \HOLSymConst{\HOLTokenExists{}}\HOLBoundVar{E\sb{\mathrm{1}}} \HOLBoundVar{E\sb{\mathrm{2}}} \HOLBoundVar{l}.
       (\HOLBoundVar{u} \HOLSymConst{=} \HOLConst{\ensuremath{\tau}}) \HOLSymConst{\HOLTokenConj{}} (\HOLBoundVar{E\sp{\prime\prime}} \HOLSymConst{=} \HOLBoundVar{E\sb{\mathrm{1}}} \HOLSymConst{\ensuremath{\parallel}} \HOLBoundVar{E\sb{\mathrm{2}}}) \HOLSymConst{\HOLTokenConj{}} \HOLBoundVar{E} \HOLTokenTransBegin\HOLConst{label} \HOLBoundVar{l}\HOLTokenTransEnd \HOLBoundVar{E\sb{\mathrm{1}}} \HOLSymConst{\HOLTokenConj{}}
       \HOLBoundVar{E\sp{\prime}} \HOLTokenTransBegin\HOLConst{label} (\HOLConst{COMPL} \HOLBoundVar{l})\HOLTokenTransEnd \HOLBoundVar{E\sb{\mathrm{2}}}
\end{SaveVerbatim}
\newcommand{\HOLCCSTheoremsTRANSXXPAR}{\UseVerbatim{HOLCCSTheoremsTRANSXXPAR}}
\begin{SaveVerbatim}{HOLCCSTheoremsTRANSXXPARXXEQ}
\HOLTokenTurnstile{} \HOLSymConst{\HOLTokenForall{}}\HOLBoundVar{E} \HOLBoundVar{E\sp{\prime}} \HOLBoundVar{u} \HOLBoundVar{E\sp{\prime\prime}}.
     \HOLBoundVar{E} \HOLSymConst{\ensuremath{\parallel}} \HOLBoundVar{E\sp{\prime}} \HOLTokenTransBegin\HOLBoundVar{u}\HOLTokenTransEnd \HOLBoundVar{E\sp{\prime\prime}} \HOLSymConst{\HOLTokenEquiv{}}
     (\HOLSymConst{\HOLTokenExists{}}\HOLBoundVar{E\sb{\mathrm{1}}}. (\HOLBoundVar{E\sp{\prime\prime}} \HOLSymConst{=} \HOLBoundVar{E\sb{\mathrm{1}}} \HOLSymConst{\ensuremath{\parallel}} \HOLBoundVar{E\sp{\prime}}) \HOLSymConst{\HOLTokenConj{}} \HOLBoundVar{E} \HOLTokenTransBegin\HOLBoundVar{u}\HOLTokenTransEnd \HOLBoundVar{E\sb{\mathrm{1}}}) \HOLSymConst{\HOLTokenDisj{}}
     (\HOLSymConst{\HOLTokenExists{}}\HOLBoundVar{E\sb{\mathrm{1}}}. (\HOLBoundVar{E\sp{\prime\prime}} \HOLSymConst{=} \HOLBoundVar{E} \HOLSymConst{\ensuremath{\parallel}} \HOLBoundVar{E\sb{\mathrm{1}}}) \HOLSymConst{\HOLTokenConj{}} \HOLBoundVar{E\sp{\prime}} \HOLTokenTransBegin\HOLBoundVar{u}\HOLTokenTransEnd \HOLBoundVar{E\sb{\mathrm{1}}}) \HOLSymConst{\HOLTokenDisj{}}
     \HOLSymConst{\HOLTokenExists{}}\HOLBoundVar{E\sb{\mathrm{1}}} \HOLBoundVar{E\sb{\mathrm{2}}} \HOLBoundVar{l}.
       (\HOLBoundVar{u} \HOLSymConst{=} \HOLConst{\ensuremath{\tau}}) \HOLSymConst{\HOLTokenConj{}} (\HOLBoundVar{E\sp{\prime\prime}} \HOLSymConst{=} \HOLBoundVar{E\sb{\mathrm{1}}} \HOLSymConst{\ensuremath{\parallel}} \HOLBoundVar{E\sb{\mathrm{2}}}) \HOLSymConst{\HOLTokenConj{}} \HOLBoundVar{E} \HOLTokenTransBegin\HOLConst{label} \HOLBoundVar{l}\HOLTokenTransEnd \HOLBoundVar{E\sb{\mathrm{1}}} \HOLSymConst{\HOLTokenConj{}}
       \HOLBoundVar{E\sp{\prime}} \HOLTokenTransBegin\HOLConst{label} (\HOLConst{COMPL} \HOLBoundVar{l})\HOLTokenTransEnd \HOLBoundVar{E\sb{\mathrm{2}}}
\end{SaveVerbatim}
\newcommand{\HOLCCSTheoremsTRANSXXPARXXEQ}{\UseVerbatim{HOLCCSTheoremsTRANSXXPARXXEQ}}
\begin{SaveVerbatim}{HOLCCSTheoremsTRANSXXPARXXNOXXSYNCR}
\HOLTokenTurnstile{} \HOLSymConst{\HOLTokenForall{}}\HOLBoundVar{l} \HOLBoundVar{l\sp{\prime}}.
     \HOLBoundVar{l} \HOLSymConst{\HOLTokenNotEqual{}} \HOLConst{COMPL} \HOLBoundVar{l\sp{\prime}} \HOLSymConst{\HOLTokenImp{}}
     \HOLSymConst{\HOLTokenForall{}}\HOLBoundVar{E} \HOLBoundVar{E\sp{\prime}} \HOLBoundVar{E\sp{\prime\prime}}. \HOLSymConst{\HOLTokenNeg{}}(\HOLConst{label} \HOLBoundVar{l}\HOLSymConst{..}\HOLBoundVar{E} \HOLSymConst{\ensuremath{\parallel}} \HOLConst{label} \HOLBoundVar{l\sp{\prime}}\HOLSymConst{..}\HOLBoundVar{E\sp{\prime}} \HOLTokenTransBegin\HOLConst{\ensuremath{\tau}}\HOLTokenTransEnd \HOLBoundVar{E\sp{\prime\prime}})
\end{SaveVerbatim}
\newcommand{\HOLCCSTheoremsTRANSXXPARXXNOXXSYNCR}{\UseVerbatim{HOLCCSTheoremsTRANSXXPARXXNOXXSYNCR}}
\begin{SaveVerbatim}{HOLCCSTheoremsTRANSXXPARXXPXXNIL}
\HOLTokenTurnstile{} \HOLSymConst{\HOLTokenForall{}}\HOLBoundVar{E} \HOLBoundVar{u} \HOLBoundVar{E\sp{\prime}}. \HOLBoundVar{E} \HOLSymConst{\ensuremath{\parallel}} \HOLConst{nil} \HOLTokenTransBegin\HOLBoundVar{u}\HOLTokenTransEnd \HOLBoundVar{E\sp{\prime}} \HOLSymConst{\HOLTokenImp{}} \HOLSymConst{\HOLTokenExists{}}\HOLBoundVar{E\sp{\prime\prime}}. \HOLBoundVar{E} \HOLTokenTransBegin\HOLBoundVar{u}\HOLTokenTransEnd \HOLBoundVar{E\sp{\prime\prime}} \HOLSymConst{\HOLTokenConj{}} (\HOLBoundVar{E\sp{\prime}} \HOLSymConst{=} \HOLBoundVar{E\sp{\prime\prime}} \HOLSymConst{\ensuremath{\parallel}} \HOLConst{nil})
\end{SaveVerbatim}
\newcommand{\HOLCCSTheoremsTRANSXXPARXXPXXNIL}{\UseVerbatim{HOLCCSTheoremsTRANSXXPARXXPXXNIL}}
\begin{SaveVerbatim}{HOLCCSTheoremsTRANSXXPREFIX}
\HOLTokenTurnstile{} \HOLSymConst{\HOLTokenForall{}}\HOLBoundVar{u} \HOLBoundVar{E} \HOLBoundVar{u\sp{\prime}} \HOLBoundVar{E\sp{\prime}}. \HOLBoundVar{u}\HOLSymConst{..}\HOLBoundVar{E} \HOLTokenTransBegin\HOLBoundVar{u\sp{\prime}}\HOLTokenTransEnd \HOLBoundVar{E\sp{\prime}} \HOLSymConst{\HOLTokenImp{}} (\HOLBoundVar{u\sp{\prime}} \HOLSymConst{=} \HOLBoundVar{u}) \HOLSymConst{\HOLTokenConj{}} (\HOLBoundVar{E\sp{\prime}} \HOLSymConst{=} \HOLBoundVar{E})
\end{SaveVerbatim}
\newcommand{\HOLCCSTheoremsTRANSXXPREFIX}{\UseVerbatim{HOLCCSTheoremsTRANSXXPREFIX}}
\begin{SaveVerbatim}{HOLCCSTheoremsTRANSXXPREFIXXXEQ}
\HOLTokenTurnstile{} \HOLSymConst{\HOLTokenForall{}}\HOLBoundVar{u} \HOLBoundVar{E} \HOLBoundVar{u\sp{\prime}} \HOLBoundVar{E\sp{\prime}}. \HOLBoundVar{u}\HOLSymConst{..}\HOLBoundVar{E} \HOLTokenTransBegin\HOLBoundVar{u\sp{\prime}}\HOLTokenTransEnd \HOLBoundVar{E\sp{\prime}} \HOLSymConst{\HOLTokenEquiv{}} (\HOLBoundVar{u\sp{\prime}} \HOLSymConst{=} \HOLBoundVar{u}) \HOLSymConst{\HOLTokenConj{}} (\HOLBoundVar{E\sp{\prime}} \HOLSymConst{=} \HOLBoundVar{E})
\end{SaveVerbatim}
\newcommand{\HOLCCSTheoremsTRANSXXPREFIXXXEQ}{\UseVerbatim{HOLCCSTheoremsTRANSXXPREFIXXXEQ}}
\begin{SaveVerbatim}{HOLCCSTheoremsTRANSXXREC}
\HOLTokenTurnstile{} \HOLSymConst{\HOLTokenForall{}}\HOLBoundVar{X} \HOLBoundVar{E} \HOLBoundVar{u} \HOLBoundVar{E\sp{\prime}}. \HOLConst{rec} \HOLBoundVar{X} \HOLBoundVar{E} \HOLTokenTransBegin\HOLBoundVar{u}\HOLTokenTransEnd \HOLBoundVar{E\sp{\prime}} \HOLSymConst{\HOLTokenImp{}} \HOLConst{CCS_Subst} \HOLBoundVar{E} (\HOLConst{rec} \HOLBoundVar{X} \HOLBoundVar{E}) \HOLBoundVar{X} \HOLTokenTransBegin\HOLBoundVar{u}\HOLTokenTransEnd \HOLBoundVar{E\sp{\prime}}
\end{SaveVerbatim}
\newcommand{\HOLCCSTheoremsTRANSXXREC}{\UseVerbatim{HOLCCSTheoremsTRANSXXREC}}
\begin{SaveVerbatim}{HOLCCSTheoremsTRANSXXRECXXEQ}
\HOLTokenTurnstile{} \HOLSymConst{\HOLTokenForall{}}\HOLBoundVar{X} \HOLBoundVar{E} \HOLBoundVar{u} \HOLBoundVar{E\sp{\prime}}. \HOLConst{rec} \HOLBoundVar{X} \HOLBoundVar{E} \HOLTokenTransBegin\HOLBoundVar{u}\HOLTokenTransEnd \HOLBoundVar{E\sp{\prime}} \HOLSymConst{\HOLTokenEquiv{}} \HOLConst{CCS_Subst} \HOLBoundVar{E} (\HOLConst{rec} \HOLBoundVar{X} \HOLBoundVar{E}) \HOLBoundVar{X} \HOLTokenTransBegin\HOLBoundVar{u}\HOLTokenTransEnd \HOLBoundVar{E\sp{\prime}}
\end{SaveVerbatim}
\newcommand{\HOLCCSTheoremsTRANSXXRECXXEQ}{\UseVerbatim{HOLCCSTheoremsTRANSXXRECXXEQ}}
\begin{SaveVerbatim}{HOLCCSTheoremsTRANSXXRELAB}
\HOLTokenTurnstile{} \HOLSymConst{\HOLTokenForall{}}\HOLBoundVar{E} \HOLBoundVar{rf} \HOLBoundVar{u} \HOLBoundVar{E\sp{\prime}}.
     \HOLConst{relab} \HOLBoundVar{E} \HOLBoundVar{rf} \HOLTokenTransBegin\HOLBoundVar{u}\HOLTokenTransEnd \HOLBoundVar{E\sp{\prime}} \HOLSymConst{\HOLTokenImp{}}
     \HOLSymConst{\HOLTokenExists{}}\HOLBoundVar{u\sp{\prime}} \HOLBoundVar{E\sp{\prime\prime}}.
       (\HOLBoundVar{u} \HOLSymConst{=} \HOLConst{relabel} \HOLBoundVar{rf} \HOLBoundVar{u\sp{\prime}}) \HOLSymConst{\HOLTokenConj{}} (\HOLBoundVar{E\sp{\prime}} \HOLSymConst{=} \HOLConst{relab} \HOLBoundVar{E\sp{\prime\prime}} \HOLBoundVar{rf}) \HOLSymConst{\HOLTokenConj{}} \HOLBoundVar{E} \HOLTokenTransBegin\HOLBoundVar{u\sp{\prime}}\HOLTokenTransEnd \HOLBoundVar{E\sp{\prime\prime}}
\end{SaveVerbatim}
\newcommand{\HOLCCSTheoremsTRANSXXRELAB}{\UseVerbatim{HOLCCSTheoremsTRANSXXRELAB}}
\begin{SaveVerbatim}{HOLCCSTheoremsTRANSXXRELABXXEQ}
\HOLTokenTurnstile{} \HOLSymConst{\HOLTokenForall{}}\HOLBoundVar{E} \HOLBoundVar{rf} \HOLBoundVar{u} \HOLBoundVar{E\sp{\prime}}.
     \HOLConst{relab} \HOLBoundVar{E} \HOLBoundVar{rf} \HOLTokenTransBegin\HOLBoundVar{u}\HOLTokenTransEnd \HOLBoundVar{E\sp{\prime}} \HOLSymConst{\HOLTokenEquiv{}}
     \HOLSymConst{\HOLTokenExists{}}\HOLBoundVar{u\sp{\prime}} \HOLBoundVar{E\sp{\prime\prime}}.
       (\HOLBoundVar{u} \HOLSymConst{=} \HOLConst{relabel} \HOLBoundVar{rf} \HOLBoundVar{u\sp{\prime}}) \HOLSymConst{\HOLTokenConj{}} (\HOLBoundVar{E\sp{\prime}} \HOLSymConst{=} \HOLConst{relab} \HOLBoundVar{E\sp{\prime\prime}} \HOLBoundVar{rf}) \HOLSymConst{\HOLTokenConj{}} \HOLBoundVar{E} \HOLTokenTransBegin\HOLBoundVar{u\sp{\prime}}\HOLTokenTransEnd \HOLBoundVar{E\sp{\prime\prime}}
\end{SaveVerbatim}
\newcommand{\HOLCCSTheoremsTRANSXXRELABXXEQ}{\UseVerbatim{HOLCCSTheoremsTRANSXXRELABXXEQ}}
\begin{SaveVerbatim}{HOLCCSTheoremsTRANSXXRELABXXlabl}
\HOLTokenTurnstile{} \HOLSymConst{\HOLTokenForall{}}\HOLBoundVar{E} \HOLBoundVar{labl} \HOLBoundVar{u} \HOLBoundVar{E\sp{\prime}}.
     \HOLConst{relab} \HOLBoundVar{E} (\HOLConst{RELAB} \HOLBoundVar{labl}) \HOLTokenTransBegin\HOLBoundVar{u}\HOLTokenTransEnd \HOLBoundVar{E\sp{\prime}} \HOLSymConst{\HOLTokenImp{}}
     \HOLSymConst{\HOLTokenExists{}}\HOLBoundVar{u\sp{\prime}} \HOLBoundVar{E\sp{\prime\prime}}.
       (\HOLBoundVar{u} \HOLSymConst{=} \HOLConst{relabel} (\HOLConst{RELAB} \HOLBoundVar{labl}) \HOLBoundVar{u\sp{\prime}}) \HOLSymConst{\HOLTokenConj{}}
       (\HOLBoundVar{E\sp{\prime}} \HOLSymConst{=} \HOLConst{relab} \HOLBoundVar{E\sp{\prime\prime}} (\HOLConst{RELAB} \HOLBoundVar{labl})) \HOLSymConst{\HOLTokenConj{}} \HOLBoundVar{E} \HOLTokenTransBegin\HOLBoundVar{u\sp{\prime}}\HOLTokenTransEnd \HOLBoundVar{E\sp{\prime\prime}}
\end{SaveVerbatim}
\newcommand{\HOLCCSTheoremsTRANSXXRELABXXlabl}{\UseVerbatim{HOLCCSTheoremsTRANSXXRELABXXlabl}}
\begin{SaveVerbatim}{HOLCCSTheoremsTRANSXXRESTR}
\HOLTokenTurnstile{} \HOLSymConst{\HOLTokenForall{}}\HOLBoundVar{E} \HOLBoundVar{L} \HOLBoundVar{u} \HOLBoundVar{E\sp{\prime}}.
     \HOLConst{\ensuremath{\nu}} \HOLBoundVar{L} \HOLBoundVar{E} \HOLTokenTransBegin\HOLBoundVar{u}\HOLTokenTransEnd \HOLBoundVar{E\sp{\prime}} \HOLSymConst{\HOLTokenImp{}}
     \HOLSymConst{\HOLTokenExists{}}\HOLBoundVar{E\sp{\prime\prime}} \HOLBoundVar{l}.
       (\HOLBoundVar{E\sp{\prime}} \HOLSymConst{=} \HOLConst{\ensuremath{\nu}} \HOLBoundVar{L} \HOLBoundVar{E\sp{\prime\prime}}) \HOLSymConst{\HOLTokenConj{}} \HOLBoundVar{E} \HOLTokenTransBegin\HOLBoundVar{u}\HOLTokenTransEnd \HOLBoundVar{E\sp{\prime\prime}} \HOLSymConst{\HOLTokenConj{}}
       ((\HOLBoundVar{u} \HOLSymConst{=} \HOLConst{\ensuremath{\tau}}) \HOLSymConst{\HOLTokenDisj{}} (\HOLBoundVar{u} \HOLSymConst{=} \HOLConst{label} \HOLBoundVar{l}) \HOLSymConst{\HOLTokenConj{}} \HOLBoundVar{l} \HOLConst{\HOLTokenNotIn{}} \HOLBoundVar{L} \HOLSymConst{\HOLTokenConj{}} \HOLConst{COMPL} \HOLBoundVar{l} \HOLConst{\HOLTokenNotIn{}} \HOLBoundVar{L})
\end{SaveVerbatim}
\newcommand{\HOLCCSTheoremsTRANSXXRESTR}{\UseVerbatim{HOLCCSTheoremsTRANSXXRESTR}}
\begin{SaveVerbatim}{HOLCCSTheoremsTRANSXXRESTRXXEQ}
\HOLTokenTurnstile{} \HOLSymConst{\HOLTokenForall{}}\HOLBoundVar{E} \HOLBoundVar{L} \HOLBoundVar{u} \HOLBoundVar{E\sp{\prime}}.
     \HOLConst{\ensuremath{\nu}} \HOLBoundVar{L} \HOLBoundVar{E} \HOLTokenTransBegin\HOLBoundVar{u}\HOLTokenTransEnd \HOLBoundVar{E\sp{\prime}} \HOLSymConst{\HOLTokenEquiv{}}
     \HOLSymConst{\HOLTokenExists{}}\HOLBoundVar{E\sp{\prime\prime}} \HOLBoundVar{l}.
       (\HOLBoundVar{E\sp{\prime}} \HOLSymConst{=} \HOLConst{\ensuremath{\nu}} \HOLBoundVar{L} \HOLBoundVar{E\sp{\prime\prime}}) \HOLSymConst{\HOLTokenConj{}} \HOLBoundVar{E} \HOLTokenTransBegin\HOLBoundVar{u}\HOLTokenTransEnd \HOLBoundVar{E\sp{\prime\prime}} \HOLSymConst{\HOLTokenConj{}}
       ((\HOLBoundVar{u} \HOLSymConst{=} \HOLConst{\ensuremath{\tau}}) \HOLSymConst{\HOLTokenDisj{}} (\HOLBoundVar{u} \HOLSymConst{=} \HOLConst{label} \HOLBoundVar{l}) \HOLSymConst{\HOLTokenConj{}} \HOLBoundVar{l} \HOLConst{\HOLTokenNotIn{}} \HOLBoundVar{L} \HOLSymConst{\HOLTokenConj{}} \HOLConst{COMPL} \HOLBoundVar{l} \HOLConst{\HOLTokenNotIn{}} \HOLBoundVar{L})
\end{SaveVerbatim}
\newcommand{\HOLCCSTheoremsTRANSXXRESTRXXEQ}{\UseVerbatim{HOLCCSTheoremsTRANSXXRESTRXXEQ}}
\begin{SaveVerbatim}{HOLCCSTheoremsTRANSXXRESTRXXNOXXNIL}
\HOLTokenTurnstile{} \HOLSymConst{\HOLTokenForall{}}\HOLBoundVar{E} \HOLBoundVar{L} \HOLBoundVar{u} \HOLBoundVar{E\sp{\prime}}. \HOLConst{\ensuremath{\nu}} \HOLBoundVar{L} \HOLBoundVar{E} \HOLTokenTransBegin\HOLBoundVar{u}\HOLTokenTransEnd \HOLConst{\ensuremath{\nu}} \HOLBoundVar{L} \HOLBoundVar{E\sp{\prime}} \HOLSymConst{\HOLTokenImp{}} \HOLBoundVar{E} \HOLSymConst{\HOLTokenNotEqual{}} \HOLConst{nil}
\end{SaveVerbatim}
\newcommand{\HOLCCSTheoremsTRANSXXRESTRXXNOXXNIL}{\UseVerbatim{HOLCCSTheoremsTRANSXXRESTRXXNOXXNIL}}
\begin{SaveVerbatim}{HOLCCSTheoremsTRANSXXrules}
\HOLTokenTurnstile{} (\HOLSymConst{\HOLTokenForall{}}\HOLBoundVar{E} \HOLBoundVar{u}. \HOLBoundVar{u}\HOLSymConst{..}\HOLBoundVar{E} \HOLTokenTransBegin\HOLBoundVar{u}\HOLTokenTransEnd \HOLBoundVar{E}) \HOLSymConst{\HOLTokenConj{}} (\HOLSymConst{\HOLTokenForall{}}\HOLBoundVar{E} \HOLBoundVar{u} \HOLBoundVar{E\sb{\mathrm{1}}} \HOLBoundVar{E\sp{\prime}}. \HOLBoundVar{E} \HOLTokenTransBegin\HOLBoundVar{u}\HOLTokenTransEnd \HOLBoundVar{E\sb{\mathrm{1}}} \HOLSymConst{\HOLTokenImp{}} \HOLBoundVar{E} \HOLSymConst{+} \HOLBoundVar{E\sp{\prime}} \HOLTokenTransBegin\HOLBoundVar{u}\HOLTokenTransEnd \HOLBoundVar{E\sb{\mathrm{1}}}) \HOLSymConst{\HOLTokenConj{}}
   (\HOLSymConst{\HOLTokenForall{}}\HOLBoundVar{E} \HOLBoundVar{u} \HOLBoundVar{E\sb{\mathrm{1}}} \HOLBoundVar{E\sp{\prime}}. \HOLBoundVar{E} \HOLTokenTransBegin\HOLBoundVar{u}\HOLTokenTransEnd \HOLBoundVar{E\sb{\mathrm{1}}} \HOLSymConst{\HOLTokenImp{}} \HOLBoundVar{E\sp{\prime}} \HOLSymConst{+} \HOLBoundVar{E} \HOLTokenTransBegin\HOLBoundVar{u}\HOLTokenTransEnd \HOLBoundVar{E\sb{\mathrm{1}}}) \HOLSymConst{\HOLTokenConj{}}
   (\HOLSymConst{\HOLTokenForall{}}\HOLBoundVar{E} \HOLBoundVar{u} \HOLBoundVar{E\sb{\mathrm{1}}} \HOLBoundVar{E\sp{\prime}}. \HOLBoundVar{E} \HOLTokenTransBegin\HOLBoundVar{u}\HOLTokenTransEnd \HOLBoundVar{E\sb{\mathrm{1}}} \HOLSymConst{\HOLTokenImp{}} \HOLBoundVar{E} \HOLSymConst{\ensuremath{\parallel}} \HOLBoundVar{E\sp{\prime}} \HOLTokenTransBegin\HOLBoundVar{u}\HOLTokenTransEnd \HOLBoundVar{E\sb{\mathrm{1}}} \HOLSymConst{\ensuremath{\parallel}} \HOLBoundVar{E\sp{\prime}}) \HOLSymConst{\HOLTokenConj{}}
   (\HOLSymConst{\HOLTokenForall{}}\HOLBoundVar{E} \HOLBoundVar{u} \HOLBoundVar{E\sb{\mathrm{1}}} \HOLBoundVar{E\sp{\prime}}. \HOLBoundVar{E} \HOLTokenTransBegin\HOLBoundVar{u}\HOLTokenTransEnd \HOLBoundVar{E\sb{\mathrm{1}}} \HOLSymConst{\HOLTokenImp{}} \HOLBoundVar{E\sp{\prime}} \HOLSymConst{\ensuremath{\parallel}} \HOLBoundVar{E} \HOLTokenTransBegin\HOLBoundVar{u}\HOLTokenTransEnd \HOLBoundVar{E\sp{\prime}} \HOLSymConst{\ensuremath{\parallel}} \HOLBoundVar{E\sb{\mathrm{1}}}) \HOLSymConst{\HOLTokenConj{}}
   (\HOLSymConst{\HOLTokenForall{}}\HOLBoundVar{E} \HOLBoundVar{l} \HOLBoundVar{E\sb{\mathrm{1}}} \HOLBoundVar{E\sp{\prime}} \HOLBoundVar{E\sb{\mathrm{2}}}.
      \HOLBoundVar{E} \HOLTokenTransBegin\HOLConst{label} \HOLBoundVar{l}\HOLTokenTransEnd \HOLBoundVar{E\sb{\mathrm{1}}} \HOLSymConst{\HOLTokenConj{}} \HOLBoundVar{E\sp{\prime}} \HOLTokenTransBegin\HOLConst{label} (\HOLConst{COMPL} \HOLBoundVar{l})\HOLTokenTransEnd \HOLBoundVar{E\sb{\mathrm{2}}} \HOLSymConst{\HOLTokenImp{}}
      \HOLBoundVar{E} \HOLSymConst{\ensuremath{\parallel}} \HOLBoundVar{E\sp{\prime}} \HOLTokenTransBegin\HOLConst{\ensuremath{\tau}}\HOLTokenTransEnd \HOLBoundVar{E\sb{\mathrm{1}}} \HOLSymConst{\ensuremath{\parallel}} \HOLBoundVar{E\sb{\mathrm{2}}}) \HOLSymConst{\HOLTokenConj{}}
   (\HOLSymConst{\HOLTokenForall{}}\HOLBoundVar{E} \HOLBoundVar{u} \HOLBoundVar{E\sp{\prime}} \HOLBoundVar{l} \HOLBoundVar{L}.
      \HOLBoundVar{E} \HOLTokenTransBegin\HOLBoundVar{u}\HOLTokenTransEnd \HOLBoundVar{E\sp{\prime}} \HOLSymConst{\HOLTokenConj{}}
      ((\HOLBoundVar{u} \HOLSymConst{=} \HOLConst{\ensuremath{\tau}}) \HOLSymConst{\HOLTokenDisj{}} (\HOLBoundVar{u} \HOLSymConst{=} \HOLConst{label} \HOLBoundVar{l}) \HOLSymConst{\HOLTokenConj{}} \HOLBoundVar{l} \HOLConst{\HOLTokenNotIn{}} \HOLBoundVar{L} \HOLSymConst{\HOLTokenConj{}} \HOLConst{COMPL} \HOLBoundVar{l} \HOLConst{\HOLTokenNotIn{}} \HOLBoundVar{L}) \HOLSymConst{\HOLTokenImp{}}
      \HOLConst{\ensuremath{\nu}} \HOLBoundVar{L} \HOLBoundVar{E} \HOLTokenTransBegin\HOLBoundVar{u}\HOLTokenTransEnd \HOLConst{\ensuremath{\nu}} \HOLBoundVar{L} \HOLBoundVar{E\sp{\prime}}) \HOLSymConst{\HOLTokenConj{}}
   (\HOLSymConst{\HOLTokenForall{}}\HOLBoundVar{E} \HOLBoundVar{u} \HOLBoundVar{E\sp{\prime}} \HOLBoundVar{rf}.
      \HOLBoundVar{E} \HOLTokenTransBegin\HOLBoundVar{u}\HOLTokenTransEnd \HOLBoundVar{E\sp{\prime}} \HOLSymConst{\HOLTokenImp{}} \HOLConst{relab} \HOLBoundVar{E} \HOLBoundVar{rf} \HOLTokenTransBegin\HOLConst{relabel} \HOLBoundVar{rf} \HOLBoundVar{u}\HOLTokenTransEnd \HOLConst{relab} \HOLBoundVar{E\sp{\prime}} \HOLBoundVar{rf}) \HOLSymConst{\HOLTokenConj{}}
   \HOLSymConst{\HOLTokenForall{}}\HOLBoundVar{E} \HOLBoundVar{u} \HOLBoundVar{X} \HOLBoundVar{E\sb{\mathrm{1}}}. \HOLConst{CCS_Subst} \HOLBoundVar{E} (\HOLConst{rec} \HOLBoundVar{X} \HOLBoundVar{E}) \HOLBoundVar{X} \HOLTokenTransBegin\HOLBoundVar{u}\HOLTokenTransEnd \HOLBoundVar{E\sb{\mathrm{1}}} \HOLSymConst{\HOLTokenImp{}} \HOLConst{rec} \HOLBoundVar{X} \HOLBoundVar{E} \HOLTokenTransBegin\HOLBoundVar{u}\HOLTokenTransEnd \HOLBoundVar{E\sb{\mathrm{1}}}
\end{SaveVerbatim}
\newcommand{\HOLCCSTheoremsTRANSXXrules}{\UseVerbatim{HOLCCSTheoremsTRANSXXrules}}
\begin{SaveVerbatim}{HOLCCSTheoremsTRANSXXstrongind}
\HOLTokenTurnstile{} \HOLSymConst{\HOLTokenForall{}}\HOLBoundVar{TRANS\sp{\prime}}.
     (\HOLSymConst{\HOLTokenForall{}}\HOLBoundVar{E} \HOLBoundVar{u}. \HOLBoundVar{TRANS\sp{\prime}} (\HOLBoundVar{u}\HOLSymConst{..}\HOLBoundVar{E}) \HOLBoundVar{u} \HOLBoundVar{E}) \HOLSymConst{\HOLTokenConj{}}
     (\HOLSymConst{\HOLTokenForall{}}\HOLBoundVar{E} \HOLBoundVar{u} \HOLBoundVar{E\sb{\mathrm{1}}} \HOLBoundVar{E\sp{\prime}}.
        \HOLBoundVar{E} \HOLTokenTransBegin\HOLBoundVar{u}\HOLTokenTransEnd \HOLBoundVar{E\sb{\mathrm{1}}} \HOLSymConst{\HOLTokenConj{}} \HOLBoundVar{TRANS\sp{\prime}} \HOLBoundVar{E} \HOLBoundVar{u} \HOLBoundVar{E\sb{\mathrm{1}}} \HOLSymConst{\HOLTokenImp{}} \HOLBoundVar{TRANS\sp{\prime}} (\HOLBoundVar{E} \HOLSymConst{+} \HOLBoundVar{E\sp{\prime}}) \HOLBoundVar{u} \HOLBoundVar{E\sb{\mathrm{1}}}) \HOLSymConst{\HOLTokenConj{}}
     (\HOLSymConst{\HOLTokenForall{}}\HOLBoundVar{E} \HOLBoundVar{u} \HOLBoundVar{E\sb{\mathrm{1}}} \HOLBoundVar{E\sp{\prime}}.
        \HOLBoundVar{E} \HOLTokenTransBegin\HOLBoundVar{u}\HOLTokenTransEnd \HOLBoundVar{E\sb{\mathrm{1}}} \HOLSymConst{\HOLTokenConj{}} \HOLBoundVar{TRANS\sp{\prime}} \HOLBoundVar{E} \HOLBoundVar{u} \HOLBoundVar{E\sb{\mathrm{1}}} \HOLSymConst{\HOLTokenImp{}} \HOLBoundVar{TRANS\sp{\prime}} (\HOLBoundVar{E\sp{\prime}} \HOLSymConst{+} \HOLBoundVar{E}) \HOLBoundVar{u} \HOLBoundVar{E\sb{\mathrm{1}}}) \HOLSymConst{\HOLTokenConj{}}
     (\HOLSymConst{\HOLTokenForall{}}\HOLBoundVar{E} \HOLBoundVar{u} \HOLBoundVar{E\sb{\mathrm{1}}} \HOLBoundVar{E\sp{\prime}}.
        \HOLBoundVar{E} \HOLTokenTransBegin\HOLBoundVar{u}\HOLTokenTransEnd \HOLBoundVar{E\sb{\mathrm{1}}} \HOLSymConst{\HOLTokenConj{}} \HOLBoundVar{TRANS\sp{\prime}} \HOLBoundVar{E} \HOLBoundVar{u} \HOLBoundVar{E\sb{\mathrm{1}}} \HOLSymConst{\HOLTokenImp{}}
        \HOLBoundVar{TRANS\sp{\prime}} (\HOLBoundVar{E} \HOLSymConst{\ensuremath{\parallel}} \HOLBoundVar{E\sp{\prime}}) \HOLBoundVar{u} (\HOLBoundVar{E\sb{\mathrm{1}}} \HOLSymConst{\ensuremath{\parallel}} \HOLBoundVar{E\sp{\prime}})) \HOLSymConst{\HOLTokenConj{}}
     (\HOLSymConst{\HOLTokenForall{}}\HOLBoundVar{E} \HOLBoundVar{u} \HOLBoundVar{E\sb{\mathrm{1}}} \HOLBoundVar{E\sp{\prime}}.
        \HOLBoundVar{E} \HOLTokenTransBegin\HOLBoundVar{u}\HOLTokenTransEnd \HOLBoundVar{E\sb{\mathrm{1}}} \HOLSymConst{\HOLTokenConj{}} \HOLBoundVar{TRANS\sp{\prime}} \HOLBoundVar{E} \HOLBoundVar{u} \HOLBoundVar{E\sb{\mathrm{1}}} \HOLSymConst{\HOLTokenImp{}}
        \HOLBoundVar{TRANS\sp{\prime}} (\HOLBoundVar{E\sp{\prime}} \HOLSymConst{\ensuremath{\parallel}} \HOLBoundVar{E}) \HOLBoundVar{u} (\HOLBoundVar{E\sp{\prime}} \HOLSymConst{\ensuremath{\parallel}} \HOLBoundVar{E\sb{\mathrm{1}}})) \HOLSymConst{\HOLTokenConj{}}
     (\HOLSymConst{\HOLTokenForall{}}\HOLBoundVar{E} \HOLBoundVar{l} \HOLBoundVar{E\sb{\mathrm{1}}} \HOLBoundVar{E\sp{\prime}} \HOLBoundVar{E\sb{\mathrm{2}}}.
        \HOLBoundVar{E} \HOLTokenTransBegin\HOLConst{label} \HOLBoundVar{l}\HOLTokenTransEnd \HOLBoundVar{E\sb{\mathrm{1}}} \HOLSymConst{\HOLTokenConj{}} \HOLBoundVar{TRANS\sp{\prime}} \HOLBoundVar{E} (\HOLConst{label} \HOLBoundVar{l}) \HOLBoundVar{E\sb{\mathrm{1}}} \HOLSymConst{\HOLTokenConj{}}
        \HOLBoundVar{E\sp{\prime}} \HOLTokenTransBegin\HOLConst{label} (\HOLConst{COMPL} \HOLBoundVar{l})\HOLTokenTransEnd \HOLBoundVar{E\sb{\mathrm{2}}} \HOLSymConst{\HOLTokenConj{}}
        \HOLBoundVar{TRANS\sp{\prime}} \HOLBoundVar{E\sp{\prime}} (\HOLConst{label} (\HOLConst{COMPL} \HOLBoundVar{l})) \HOLBoundVar{E\sb{\mathrm{2}}} \HOLSymConst{\HOLTokenImp{}}
        \HOLBoundVar{TRANS\sp{\prime}} (\HOLBoundVar{E} \HOLSymConst{\ensuremath{\parallel}} \HOLBoundVar{E\sp{\prime}}) \HOLConst{\ensuremath{\tau}} (\HOLBoundVar{E\sb{\mathrm{1}}} \HOLSymConst{\ensuremath{\parallel}} \HOLBoundVar{E\sb{\mathrm{2}}})) \HOLSymConst{\HOLTokenConj{}}
     (\HOLSymConst{\HOLTokenForall{}}\HOLBoundVar{E} \HOLBoundVar{u} \HOLBoundVar{E\sp{\prime}} \HOLBoundVar{l} \HOLBoundVar{L}.
        \HOLBoundVar{E} \HOLTokenTransBegin\HOLBoundVar{u}\HOLTokenTransEnd \HOLBoundVar{E\sp{\prime}} \HOLSymConst{\HOLTokenConj{}} \HOLBoundVar{TRANS\sp{\prime}} \HOLBoundVar{E} \HOLBoundVar{u} \HOLBoundVar{E\sp{\prime}} \HOLSymConst{\HOLTokenConj{}}
        ((\HOLBoundVar{u} \HOLSymConst{=} \HOLConst{\ensuremath{\tau}}) \HOLSymConst{\HOLTokenDisj{}} (\HOLBoundVar{u} \HOLSymConst{=} \HOLConst{label} \HOLBoundVar{l}) \HOLSymConst{\HOLTokenConj{}} \HOLBoundVar{l} \HOLConst{\HOLTokenNotIn{}} \HOLBoundVar{L} \HOLSymConst{\HOLTokenConj{}} \HOLConst{COMPL} \HOLBoundVar{l} \HOLConst{\HOLTokenNotIn{}} \HOLBoundVar{L}) \HOLSymConst{\HOLTokenImp{}}
        \HOLBoundVar{TRANS\sp{\prime}} (\HOLConst{\ensuremath{\nu}} \HOLBoundVar{L} \HOLBoundVar{E}) \HOLBoundVar{u} (\HOLConst{\ensuremath{\nu}} \HOLBoundVar{L} \HOLBoundVar{E\sp{\prime}})) \HOLSymConst{\HOLTokenConj{}}
     (\HOLSymConst{\HOLTokenForall{}}\HOLBoundVar{E} \HOLBoundVar{u} \HOLBoundVar{E\sp{\prime}} \HOLBoundVar{rf}.
        \HOLBoundVar{E} \HOLTokenTransBegin\HOLBoundVar{u}\HOLTokenTransEnd \HOLBoundVar{E\sp{\prime}} \HOLSymConst{\HOLTokenConj{}} \HOLBoundVar{TRANS\sp{\prime}} \HOLBoundVar{E} \HOLBoundVar{u} \HOLBoundVar{E\sp{\prime}} \HOLSymConst{\HOLTokenImp{}}
        \HOLBoundVar{TRANS\sp{\prime}} (\HOLConst{relab} \HOLBoundVar{E} \HOLBoundVar{rf}) (\HOLConst{relabel} \HOLBoundVar{rf} \HOLBoundVar{u}) (\HOLConst{relab} \HOLBoundVar{E\sp{\prime}} \HOLBoundVar{rf})) \HOLSymConst{\HOLTokenConj{}}
     (\HOLSymConst{\HOLTokenForall{}}\HOLBoundVar{E} \HOLBoundVar{u} \HOLBoundVar{X} \HOLBoundVar{E\sb{\mathrm{1}}}.
        \HOLConst{CCS_Subst} \HOLBoundVar{E} (\HOLConst{rec} \HOLBoundVar{X} \HOLBoundVar{E}) \HOLBoundVar{X} \HOLTokenTransBegin\HOLBoundVar{u}\HOLTokenTransEnd \HOLBoundVar{E\sb{\mathrm{1}}} \HOLSymConst{\HOLTokenConj{}}
        \HOLBoundVar{TRANS\sp{\prime}} (\HOLConst{CCS_Subst} \HOLBoundVar{E} (\HOLConst{rec} \HOLBoundVar{X} \HOLBoundVar{E}) \HOLBoundVar{X}) \HOLBoundVar{u} \HOLBoundVar{E\sb{\mathrm{1}}} \HOLSymConst{\HOLTokenImp{}}
        \HOLBoundVar{TRANS\sp{\prime}} (\HOLConst{rec} \HOLBoundVar{X} \HOLBoundVar{E}) \HOLBoundVar{u} \HOLBoundVar{E\sb{\mathrm{1}}}) \HOLSymConst{\HOLTokenImp{}}
     \HOLSymConst{\HOLTokenForall{}}\HOLBoundVar{a\sb{\mathrm{0}}} \HOLBoundVar{a\sb{\mathrm{1}}} \HOLBoundVar{a\sb{\mathrm{2}}}. \HOLBoundVar{a\sb{\mathrm{0}}} \HOLTokenTransBegin\HOLBoundVar{a\sb{\mathrm{1}}}\HOLTokenTransEnd \HOLBoundVar{a\sb{\mathrm{2}}} \HOLSymConst{\HOLTokenImp{}} \HOLBoundVar{TRANS\sp{\prime}} \HOLBoundVar{a\sb{\mathrm{0}}} \HOLBoundVar{a\sb{\mathrm{1}}} \HOLBoundVar{a\sb{\mathrm{2}}}
\end{SaveVerbatim}
\newcommand{\HOLCCSTheoremsTRANSXXstrongind}{\UseVerbatim{HOLCCSTheoremsTRANSXXstrongind}}
\begin{SaveVerbatim}{HOLCCSTheoremsTRANSXXSUM}
\HOLTokenTurnstile{} \HOLSymConst{\HOLTokenForall{}}\HOLBoundVar{E} \HOLBoundVar{E\sp{\prime}} \HOLBoundVar{u} \HOLBoundVar{E\sp{\prime\prime}}. \HOLBoundVar{E} \HOLSymConst{+} \HOLBoundVar{E\sp{\prime}} \HOLTokenTransBegin\HOLBoundVar{u}\HOLTokenTransEnd \HOLBoundVar{E\sp{\prime\prime}} \HOLSymConst{\HOLTokenImp{}} \HOLBoundVar{E} \HOLTokenTransBegin\HOLBoundVar{u}\HOLTokenTransEnd \HOLBoundVar{E\sp{\prime\prime}} \HOLSymConst{\HOLTokenDisj{}} \HOLBoundVar{E\sp{\prime}} \HOLTokenTransBegin\HOLBoundVar{u}\HOLTokenTransEnd \HOLBoundVar{E\sp{\prime\prime}}
\end{SaveVerbatim}
\newcommand{\HOLCCSTheoremsTRANSXXSUM}{\UseVerbatim{HOLCCSTheoremsTRANSXXSUM}}
\begin{SaveVerbatim}{HOLCCSTheoremsTRANSXXSUMXXEQ}
\HOLTokenTurnstile{} \HOLSymConst{\HOLTokenForall{}}\HOLBoundVar{E} \HOLBoundVar{E\sp{\prime}} \HOLBoundVar{u} \HOLBoundVar{E\sp{\prime\prime}}. \HOLBoundVar{E} \HOLSymConst{+} \HOLBoundVar{E\sp{\prime}} \HOLTokenTransBegin\HOLBoundVar{u}\HOLTokenTransEnd \HOLBoundVar{E\sp{\prime\prime}} \HOLSymConst{\HOLTokenEquiv{}} \HOLBoundVar{E} \HOLTokenTransBegin\HOLBoundVar{u}\HOLTokenTransEnd \HOLBoundVar{E\sp{\prime\prime}} \HOLSymConst{\HOLTokenDisj{}} \HOLBoundVar{E\sp{\prime}} \HOLTokenTransBegin\HOLBoundVar{u}\HOLTokenTransEnd \HOLBoundVar{E\sp{\prime\prime}}
\end{SaveVerbatim}
\newcommand{\HOLCCSTheoremsTRANSXXSUMXXEQ}{\UseVerbatim{HOLCCSTheoremsTRANSXXSUMXXEQ}}
\begin{SaveVerbatim}{HOLCCSTheoremsTRANSXXSUMXXEQYY}
\HOLTokenTurnstile{} \HOLSymConst{\HOLTokenForall{}}\HOLBoundVar{E\sb{\mathrm{1}}} \HOLBoundVar{E\sb{\mathrm{2}}} \HOLBoundVar{u} \HOLBoundVar{E}. \HOLBoundVar{E\sb{\mathrm{1}}} \HOLSymConst{+} \HOLBoundVar{E\sb{\mathrm{2}}} \HOLTokenTransBegin\HOLBoundVar{u}\HOLTokenTransEnd \HOLBoundVar{E} \HOLSymConst{\HOLTokenEquiv{}} \HOLBoundVar{E\sb{\mathrm{1}}} \HOLTokenTransBegin\HOLBoundVar{u}\HOLTokenTransEnd \HOLBoundVar{E} \HOLSymConst{\HOLTokenDisj{}} \HOLBoundVar{E\sb{\mathrm{2}}} \HOLTokenTransBegin\HOLBoundVar{u}\HOLTokenTransEnd \HOLBoundVar{E}
\end{SaveVerbatim}
\newcommand{\HOLCCSTheoremsTRANSXXSUMXXEQYY}{\UseVerbatim{HOLCCSTheoremsTRANSXXSUMXXEQYY}}
\begin{SaveVerbatim}{HOLCCSTheoremsTRANSXXSUMXXNIL}
\HOLTokenTurnstile{} \HOLSymConst{\HOLTokenForall{}}\HOLBoundVar{E} \HOLBoundVar{u} \HOLBoundVar{E\sp{\prime}}. \HOLBoundVar{E} \HOLSymConst{+} \HOLConst{nil} \HOLTokenTransBegin\HOLBoundVar{u}\HOLTokenTransEnd \HOLBoundVar{E\sp{\prime}} \HOLSymConst{\HOLTokenImp{}} \HOLBoundVar{E} \HOLTokenTransBegin\HOLBoundVar{u}\HOLTokenTransEnd \HOLBoundVar{E\sp{\prime}}
\end{SaveVerbatim}
\newcommand{\HOLCCSTheoremsTRANSXXSUMXXNIL}{\UseVerbatim{HOLCCSTheoremsTRANSXXSUMXXNIL}}
\begin{SaveVerbatim}{HOLCCSTheoremsTRANSXXSUMXXNILXXEQ}
\HOLTokenTurnstile{} \HOLSymConst{\HOLTokenForall{}}\HOLBoundVar{E} \HOLBoundVar{u} \HOLBoundVar{E\sp{\prime}}. \HOLBoundVar{E} \HOLSymConst{+} \HOLConst{nil} \HOLTokenTransBegin\HOLBoundVar{u}\HOLTokenTransEnd \HOLBoundVar{E\sp{\prime}} \HOLSymConst{\HOLTokenEquiv{}} \HOLBoundVar{E} \HOLTokenTransBegin\HOLBoundVar{u}\HOLTokenTransEnd \HOLBoundVar{E\sp{\prime}}
\end{SaveVerbatim}
\newcommand{\HOLCCSTheoremsTRANSXXSUMXXNILXXEQ}{\UseVerbatim{HOLCCSTheoremsTRANSXXSUMXXNILXXEQ}}
\begin{SaveVerbatim}{HOLCCSTheoremsVARXXNOXXTRANS}
\HOLTokenTurnstile{} \HOLSymConst{\HOLTokenForall{}}\HOLBoundVar{X} \HOLBoundVar{u} \HOLBoundVar{E}. \HOLSymConst{\HOLTokenNeg{}}(\HOLConst{var} \HOLBoundVar{X} \HOLTokenTransBegin\HOLBoundVar{u}\HOLTokenTransEnd \HOLBoundVar{E})
\end{SaveVerbatim}
\newcommand{\HOLCCSTheoremsVARXXNOXXTRANS}{\UseVerbatim{HOLCCSTheoremsVARXXNOXXTRANS}}
\newcommand{\HOLCCSTheorems}{
\HOLThmTag{CCS}{Action_11}\HOLCCSTheoremsActionXXOneOne
\HOLThmTag{CCS}{Action_distinct}\HOLCCSTheoremsActionXXdistinct
\HOLThmTag{CCS}{Action_distinct_label}\HOLCCSTheoremsActionXXdistinctXXlabel
\HOLThmTag{CCS}{Action_induction}\HOLCCSTheoremsActionXXinduction
\HOLThmTag{CCS}{Action_nchotomy}\HOLCCSTheoremsActionXXnchotomy
\HOLThmTag{CCS}{Action_no_tau_is_Label}\HOLCCSTheoremsActionXXnoXXtauXXisXXLabel
\HOLThmTag{CCS}{Apply_Relab_COMPL_THM}\HOLCCSTheoremsApplyXXRelabXXCOMPLXXTHM
\HOLThmTag{CCS}{APPLY_RELAB_THM}\HOLCCSTheoremsAPPLYXXRELABXXTHM
\HOLThmTag{CCS}{CCS_caseeq}\HOLCCSTheoremsCCSXXcaseeq
\HOLThmTag{CCS}{CCS_COND_CLAUSES}\HOLCCSTheoremsCCSXXCONDXXCLAUSES
\HOLThmTag{CCS}{CCS_distinct'}\HOLCCSTheoremsCCSXXdistinctYY
\HOLThmTag{CCS}{CCS_Subst_rec}\HOLCCSTheoremsCCSXXSubstXXrec
\HOLThmTag{CCS}{CCS_Subst_var}\HOLCCSTheoremsCCSXXSubstXXvar
\HOLThmTag{CCS}{COMPL_COMPL_ACT}\HOLCCSTheoremsCOMPLXXCOMPLXXACT
\HOLThmTag{CCS}{COMPL_COMPL_LAB}\HOLCCSTheoremsCOMPLXXCOMPLXXLAB
\HOLThmTag{CCS}{COMPL_THM}\HOLCCSTheoremsCOMPLXXTHM
\HOLThmTag{CCS}{coname_COMPL}\HOLCCSTheoremsconameXXCOMPL
\HOLThmTag{CCS}{DELETE_ELEMENT_APPEND}\HOLCCSTheoremsDELETEXXELEMENTXXAPPEND
\HOLThmTag{CCS}{DELETE_ELEMENT_FILTER}\HOLCCSTheoremsDELETEXXELEMENTXXFILTER
\HOLThmTag{CCS}{EVERY_DELETE_ELEMENT}\HOLCCSTheoremsEVERYXXDELETEXXELEMENT
\HOLThmTag{CCS}{EXISTS_Relabeling}\HOLCCSTheoremsEXISTSXXRelabeling
\HOLThmTag{CCS}{FN_def}\HOLCCSTheoremsFNXXdef
\HOLThmTag{CCS}{FN_ind}\HOLCCSTheoremsFNXXind
\HOLThmTag{CCS}{FN_UNIV1}\HOLCCSTheoremsFNXXUNIVOne
\HOLThmTag{CCS}{FN_UNIV2}\HOLCCSTheoremsFNXXUNIVTwo
\HOLThmTag{CCS}{IS_LABEL_def}\HOLCCSTheoremsISXXLABELXXdef
\HOLThmTag{CCS}{IS_RELABELING}\HOLCCSTheoremsISXXRELABELING
\HOLThmTag{CCS}{Label_caseeq}\HOLCCSTheoremsLabelXXcaseeq
\HOLThmTag{CCS}{LABEL_def}\HOLCCSTheoremsLABELXXdef
\HOLThmTag{CCS}{Label_distinct'}\HOLCCSTheoremsLabelXXdistinctYY
\HOLThmTag{CCS}{Label_not_eq}\HOLCCSTheoremsLabelXXnotXXeq
\HOLThmTag{CCS}{Label_not_eq'}\HOLCCSTheoremsLabelXXnotXXeqYY
\HOLThmTag{CCS}{LENGTH_DELETE_ELEMENT_LE}\HOLCCSTheoremsLENGTHXXDELETEXXELEMENTXXLE
\HOLThmTag{CCS}{LENGTH_DELETE_ELEMENT_LEQ}\HOLCCSTheoremsLENGTHXXDELETEXXELEMENTXXLEQ
\HOLThmTag{CCS}{NIL_NO_TRANS}\HOLCCSTheoremsNILXXNOXXTRANS
\HOLThmTag{CCS}{NIL_NO_TRANS_EQF}\HOLCCSTheoremsNILXXNOXXTRANSXXEQF
\HOLThmTag{CCS}{NOT_IN_DELETE_ELEMENT}\HOLCCSTheoremsNOTXXINXXDELETEXXELEMENT
\HOLThmTag{CCS}{PAR1}\HOLCCSTheoremsPAROne
\HOLThmTag{CCS}{PAR2}\HOLCCSTheoremsPARTwo
\HOLThmTag{CCS}{PAR3}\HOLCCSTheoremsPARThree
\HOLThmTag{CCS}{PAR_cases}\HOLCCSTheoremsPARXXcases
\HOLThmTag{CCS}{PAR_cases_EQ}\HOLCCSTheoremsPARXXcasesXXEQ
\HOLThmTag{CCS}{PREFIX}\HOLCCSTheoremsPREFIX
\HOLThmTag{CCS}{REC}\HOLCCSTheoremsREC
\HOLThmTag{CCS}{REC_cases}\HOLCCSTheoremsRECXXcases
\HOLThmTag{CCS}{REC_cases_EQ}\HOLCCSTheoremsRECXXcasesXXEQ
\HOLThmTag{CCS}{RELAB_cases}\HOLCCSTheoremsRELABXXcases
\HOLThmTag{CCS}{RELAB_cases_EQ}\HOLCCSTheoremsRELABXXcasesXXEQ
\HOLThmTag{CCS}{Relab_label}\HOLCCSTheoremsRelabXXlabel
\HOLThmTag{CCS}{RELAB_NIL_NO_TRANS}\HOLCCSTheoremsRELABXXNILXXNOXXTRANS
\HOLThmTag{CCS}{Relab_tau}\HOLCCSTheoremsRelabXXtau
\HOLThmTag{CCS}{RELABELING}\HOLCCSTheoremsRELABELING
\HOLThmTag{CCS}{REP_Relabeling_THM}\HOLCCSTheoremsREPXXRelabelingXXTHM
\HOLThmTag{CCS}{RESTR}\HOLCCSTheoremsRESTR
\HOLThmTag{CCS}{RESTR_cases}\HOLCCSTheoremsRESTRXXcases
\HOLThmTag{CCS}{RESTR_cases_EQ}\HOLCCSTheoremsRESTRXXcasesXXEQ
\HOLThmTag{CCS}{RESTR_LABEL_NO_TRANS}\HOLCCSTheoremsRESTRXXLABELXXNOXXTRANS
\HOLThmTag{CCS}{RESTR_NIL_NO_TRANS}\HOLCCSTheoremsRESTRXXNILXXNOXXTRANS
\HOLThmTag{CCS}{SUM1}\HOLCCSTheoremsSUMOne
\HOLThmTag{CCS}{SUM2}\HOLCCSTheoremsSUMTwo
\HOLThmTag{CCS}{SUM_cases}\HOLCCSTheoremsSUMXXcases
\HOLThmTag{CCS}{SUM_cases_EQ}\HOLCCSTheoremsSUMXXcasesXXEQ
\HOLThmTag{CCS}{TRANS_ASSOC_EQ}\HOLCCSTheoremsTRANSXXASSOCXXEQ
\HOLThmTag{CCS}{TRANS_ASSOC_RL}\HOLCCSTheoremsTRANSXXASSOCXXRL
\HOLThmTag{CCS}{TRANS_cases}\HOLCCSTheoremsTRANSXXcases
\HOLThmTag{CCS}{TRANS_COMM_EQ}\HOLCCSTheoremsTRANSXXCOMMXXEQ
\HOLThmTag{CCS}{TRANS_IMP_NO_NIL}\HOLCCSTheoremsTRANSXXIMPXXNOXXNIL
\HOLThmTag{CCS}{TRANS_IMP_NO_NIL'}\HOLCCSTheoremsTRANSXXIMPXXNOXXNILYY
\HOLThmTag{CCS}{TRANS_IMP_NO_RESTR_NIL}\HOLCCSTheoremsTRANSXXIMPXXNOXXRESTRXXNIL
\HOLThmTag{CCS}{TRANS_ind}\HOLCCSTheoremsTRANSXXind
\HOLThmTag{CCS}{TRANS_P_RESTR}\HOLCCSTheoremsTRANSXXPXXRESTR
\HOLThmTag{CCS}{TRANS_P_SUM_P}\HOLCCSTheoremsTRANSXXPXXSUMXXP
\HOLThmTag{CCS}{TRANS_P_SUM_P_EQ}\HOLCCSTheoremsTRANSXXPXXSUMXXPXXEQ
\HOLThmTag{CCS}{TRANS_PAR}\HOLCCSTheoremsTRANSXXPAR
\HOLThmTag{CCS}{TRANS_PAR_EQ}\HOLCCSTheoremsTRANSXXPARXXEQ
\HOLThmTag{CCS}{TRANS_PAR_NO_SYNCR}\HOLCCSTheoremsTRANSXXPARXXNOXXSYNCR
\HOLThmTag{CCS}{TRANS_PAR_P_NIL}\HOLCCSTheoremsTRANSXXPARXXPXXNIL
\HOLThmTag{CCS}{TRANS_PREFIX}\HOLCCSTheoremsTRANSXXPREFIX
\HOLThmTag{CCS}{TRANS_PREFIX_EQ}\HOLCCSTheoremsTRANSXXPREFIXXXEQ
\HOLThmTag{CCS}{TRANS_REC}\HOLCCSTheoremsTRANSXXREC
\HOLThmTag{CCS}{TRANS_REC_EQ}\HOLCCSTheoremsTRANSXXRECXXEQ
\HOLThmTag{CCS}{TRANS_RELAB}\HOLCCSTheoremsTRANSXXRELAB
\HOLThmTag{CCS}{TRANS_RELAB_EQ}\HOLCCSTheoremsTRANSXXRELABXXEQ
\HOLThmTag{CCS}{TRANS_RELAB_labl}\HOLCCSTheoremsTRANSXXRELABXXlabl
\HOLThmTag{CCS}{TRANS_RESTR}\HOLCCSTheoremsTRANSXXRESTR
\HOLThmTag{CCS}{TRANS_RESTR_EQ}\HOLCCSTheoremsTRANSXXRESTRXXEQ
\HOLThmTag{CCS}{TRANS_RESTR_NO_NIL}\HOLCCSTheoremsTRANSXXRESTRXXNOXXNIL
\HOLThmTag{CCS}{TRANS_rules}\HOLCCSTheoremsTRANSXXrules
\HOLThmTag{CCS}{TRANS_strongind}\HOLCCSTheoremsTRANSXXstrongind
\HOLThmTag{CCS}{TRANS_SUM}\HOLCCSTheoremsTRANSXXSUM
\HOLThmTag{CCS}{TRANS_SUM_EQ}\HOLCCSTheoremsTRANSXXSUMXXEQ
\HOLThmTag{CCS}{TRANS_SUM_EQ'}\HOLCCSTheoremsTRANSXXSUMXXEQYY
\HOLThmTag{CCS}{TRANS_SUM_NIL}\HOLCCSTheoremsTRANSXXSUMXXNIL
\HOLThmTag{CCS}{TRANS_SUM_NIL_EQ}\HOLCCSTheoremsTRANSXXSUMXXNILXXEQ
\HOLThmTag{CCS}{VAR_NO_TRANS}\HOLCCSTheoremsVARXXNOXXTRANS
}

\newcommand{\HOLStrongEQDate}{02 Dicembre 2017}
\newcommand{\HOLStrongEQTime}{13:31}
\begin{SaveVerbatim}{HOLStrongEQDefinitionsSTRONGXXBISIMXXdef}
\HOLTokenTurnstile{} \HOLSymConst{\HOLTokenForall{}}\HOLBoundVar{R}. \HOLConst{STRONG_BISIM} \HOLBoundVar{R} \HOLSymConst{\HOLTokenEquiv{}} \HOLConst{STRONG_SIM} \HOLBoundVar{R} \HOLSymConst{\HOLTokenConj{}} \HOLConst{STRONG_SIM} (\HOLConst{relinv} \HOLBoundVar{R})
\end{SaveVerbatim}
\newcommand{\HOLStrongEQDefinitionsSTRONGXXBISIMXXdef}{\UseVerbatim{HOLStrongEQDefinitionsSTRONGXXBISIMXXdef}}
\begin{SaveVerbatim}{HOLStrongEQDefinitionsSTRONGXXEQUIVXXdef}
\HOLTokenTurnstile{} \HOLConst{STRONG_EQUIV} \HOLSymConst{=}
   (\HOLTokenLambda{}\HOLBoundVar{a\sb{\mathrm{0}}} \HOLBoundVar{a\sb{\mathrm{1}}}.
      \HOLSymConst{\HOLTokenExists{}}\HOLBoundVar{STRONG\HOLTokenUnderscore{}EQUIV\sp{\prime}}.
        \HOLBoundVar{STRONG\HOLTokenUnderscore{}EQUIV\sp{\prime}} \HOLBoundVar{a\sb{\mathrm{0}}} \HOLBoundVar{a\sb{\mathrm{1}}} \HOLSymConst{\HOLTokenConj{}}
        \HOLSymConst{\HOLTokenForall{}}\HOLBoundVar{a\sb{\mathrm{0}}} \HOLBoundVar{a\sb{\mathrm{1}}}.
          \HOLBoundVar{STRONG\HOLTokenUnderscore{}EQUIV\sp{\prime}} \HOLBoundVar{a\sb{\mathrm{0}}} \HOLBoundVar{a\sb{\mathrm{1}}} \HOLSymConst{\HOLTokenImp{}}
          \HOLSymConst{\HOLTokenForall{}}\HOLBoundVar{u}.
            (\HOLSymConst{\HOLTokenForall{}}\HOLBoundVar{E\sb{\mathrm{1}}}.
               \HOLBoundVar{a\sb{\mathrm{0}}} \HOLTokenTransBegin\HOLBoundVar{u}\HOLTokenTransEnd \HOLBoundVar{E\sb{\mathrm{1}}} \HOLSymConst{\HOLTokenImp{}}
               \HOLSymConst{\HOLTokenExists{}}\HOLBoundVar{E\sb{\mathrm{2}}}. \HOLBoundVar{a\sb{\mathrm{1}}} \HOLTokenTransBegin\HOLBoundVar{u}\HOLTokenTransEnd \HOLBoundVar{E\sb{\mathrm{2}}} \HOLSymConst{\HOLTokenConj{}} \HOLBoundVar{STRONG\HOLTokenUnderscore{}EQUIV\sp{\prime}} \HOLBoundVar{E\sb{\mathrm{1}}} \HOLBoundVar{E\sb{\mathrm{2}}}) \HOLSymConst{\HOLTokenConj{}}
            \HOLSymConst{\HOLTokenForall{}}\HOLBoundVar{E\sb{\mathrm{2}}}.
              \HOLBoundVar{a\sb{\mathrm{1}}} \HOLTokenTransBegin\HOLBoundVar{u}\HOLTokenTransEnd \HOLBoundVar{E\sb{\mathrm{2}}} \HOLSymConst{\HOLTokenImp{}} \HOLSymConst{\HOLTokenExists{}}\HOLBoundVar{E\sb{\mathrm{1}}}. \HOLBoundVar{a\sb{\mathrm{0}}} \HOLTokenTransBegin\HOLBoundVar{u}\HOLTokenTransEnd \HOLBoundVar{E\sb{\mathrm{1}}} \HOLSymConst{\HOLTokenConj{}} \HOLBoundVar{STRONG\HOLTokenUnderscore{}EQUIV\sp{\prime}} \HOLBoundVar{E\sb{\mathrm{1}}} \HOLBoundVar{E\sb{\mathrm{2}}})
\end{SaveVerbatim}
\newcommand{\HOLStrongEQDefinitionsSTRONGXXEQUIVXXdef}{\UseVerbatim{HOLStrongEQDefinitionsSTRONGXXEQUIVXXdef}}
\begin{SaveVerbatim}{HOLStrongEQDefinitionsSTRONGXXSIMXXdef}
\HOLTokenTurnstile{} \HOLSymConst{\HOLTokenForall{}}\HOLBoundVar{R}.
     \HOLConst{STRONG_SIM} \HOLBoundVar{R} \HOLSymConst{\HOLTokenEquiv{}}
     \HOLSymConst{\HOLTokenForall{}}\HOLBoundVar{E} \HOLBoundVar{E\sp{\prime}}. \HOLBoundVar{R} \HOLBoundVar{E} \HOLBoundVar{E\sp{\prime}} \HOLSymConst{\HOLTokenImp{}} \HOLSymConst{\HOLTokenForall{}}\HOLBoundVar{u} \HOLBoundVar{E\sb{\mathrm{1}}}. \HOLBoundVar{E} \HOLTokenTransBegin\HOLBoundVar{u}\HOLTokenTransEnd \HOLBoundVar{E\sb{\mathrm{1}}} \HOLSymConst{\HOLTokenImp{}} \HOLSymConst{\HOLTokenExists{}}\HOLBoundVar{E\sb{\mathrm{2}}}. \HOLBoundVar{E\sp{\prime}} \HOLTokenTransBegin\HOLBoundVar{u}\HOLTokenTransEnd \HOLBoundVar{E\sb{\mathrm{2}}} \HOLSymConst{\HOLTokenConj{}} \HOLBoundVar{R} \HOLBoundVar{E\sb{\mathrm{1}}} \HOLBoundVar{E\sb{\mathrm{2}}}
\end{SaveVerbatim}
\newcommand{\HOLStrongEQDefinitionsSTRONGXXSIMXXdef}{\UseVerbatim{HOLStrongEQDefinitionsSTRONGXXSIMXXdef}}
\newcommand{\HOLStrongEQDefinitions}{
\HOLDfnTag{StrongEQ}{STRONG_BISIM_def}\HOLStrongEQDefinitionsSTRONGXXBISIMXXdef
\HOLDfnTag{StrongEQ}{STRONG_EQUIV_def}\HOLStrongEQDefinitionsSTRONGXXEQUIVXXdef
\HOLDfnTag{StrongEQ}{STRONG_SIM_def}\HOLStrongEQDefinitionsSTRONGXXSIMXXdef
}
\begin{SaveVerbatim}{HOLStrongEQTheoremsCOMPXXSTRONGXXBISIM}
\HOLTokenTurnstile{} \HOLSymConst{\HOLTokenForall{}}\HOLBoundVar{Bsm\sb{\mathrm{1}}} \HOLBoundVar{Bsm\sb{\mathrm{2}}}.
     \HOLConst{STRONG_BISIM} \HOLBoundVar{Bsm\sb{\mathrm{1}}} \HOLSymConst{\HOLTokenConj{}} \HOLConst{STRONG_BISIM} \HOLBoundVar{Bsm\sb{\mathrm{2}}} \HOLSymConst{\HOLTokenImp{}}
     \HOLConst{STRONG_BISIM} (\HOLBoundVar{Bsm\sb{\mathrm{2}}} \HOLConst{O} \HOLBoundVar{Bsm\sb{\mathrm{1}}})
\end{SaveVerbatim}
\newcommand{\HOLStrongEQTheoremsCOMPXXSTRONGXXBISIM}{\UseVerbatim{HOLStrongEQTheoremsCOMPXXSTRONGXXBISIM}}
\begin{SaveVerbatim}{HOLStrongEQTheoremsCONVERSEXXSTRONGXXBISIM}
\HOLTokenTurnstile{} \HOLSymConst{\HOLTokenForall{}}\HOLBoundVar{Bsm}. \HOLConst{STRONG_BISIM} \HOLBoundVar{Bsm} \HOLSymConst{\HOLTokenImp{}} \HOLConst{STRONG_BISIM} (\HOLConst{relinv} \HOLBoundVar{Bsm})
\end{SaveVerbatim}
\newcommand{\HOLStrongEQTheoremsCONVERSEXXSTRONGXXBISIM}{\UseVerbatim{HOLStrongEQTheoremsCONVERSEXXSTRONGXXBISIM}}
\begin{SaveVerbatim}{HOLStrongEQTheoremsEQUALXXIMPXXSTRONGXXEQUIV}
\HOLTokenTurnstile{} \HOLSymConst{\HOLTokenForall{}}\HOLBoundVar{E} \HOLBoundVar{E\sp{\prime}}. (\HOLBoundVar{E} \HOLSymConst{=} \HOLBoundVar{E\sp{\prime}}) \HOLSymConst{\HOLTokenImp{}} \HOLConst{STRONG_EQUIV} \HOLBoundVar{E} \HOLBoundVar{E\sp{\prime}}
\end{SaveVerbatim}
\newcommand{\HOLStrongEQTheoremsEQUALXXIMPXXSTRONGXXEQUIV}{\UseVerbatim{HOLStrongEQTheoremsEQUALXXIMPXXSTRONGXXEQUIV}}
\begin{SaveVerbatim}{HOLStrongEQTheoremsIDENTITYXXSTRONGXXBISIM}
\HOLTokenTurnstile{} \HOLConst{STRONG_BISIM} (\HOLSymConst{=})
\end{SaveVerbatim}
\newcommand{\HOLStrongEQTheoremsIDENTITYXXSTRONGXXBISIM}{\UseVerbatim{HOLStrongEQTheoremsIDENTITYXXSTRONGXXBISIM}}
\begin{SaveVerbatim}{HOLStrongEQTheoremsPROPERTYXXSTAR}
\HOLTokenTurnstile{} \HOLSymConst{\HOLTokenForall{}}\HOLBoundVar{a\sb{\mathrm{0}}} \HOLBoundVar{a\sb{\mathrm{1}}}.
     \HOLConst{STRONG_EQUIV} \HOLBoundVar{a\sb{\mathrm{0}}} \HOLBoundVar{a\sb{\mathrm{1}}} \HOLSymConst{\HOLTokenEquiv{}}
     \HOLSymConst{\HOLTokenForall{}}\HOLBoundVar{u}.
       (\HOLSymConst{\HOLTokenForall{}}\HOLBoundVar{E\sb{\mathrm{1}}}. \HOLBoundVar{a\sb{\mathrm{0}}} \HOLTokenTransBegin\HOLBoundVar{u}\HOLTokenTransEnd \HOLBoundVar{E\sb{\mathrm{1}}} \HOLSymConst{\HOLTokenImp{}} \HOLSymConst{\HOLTokenExists{}}\HOLBoundVar{E\sb{\mathrm{2}}}. \HOLBoundVar{a\sb{\mathrm{1}}} \HOLTokenTransBegin\HOLBoundVar{u}\HOLTokenTransEnd \HOLBoundVar{E\sb{\mathrm{2}}} \HOLSymConst{\HOLTokenConj{}} \HOLConst{STRONG_EQUIV} \HOLBoundVar{E\sb{\mathrm{1}}} \HOLBoundVar{E\sb{\mathrm{2}}}) \HOLSymConst{\HOLTokenConj{}}
       \HOLSymConst{\HOLTokenForall{}}\HOLBoundVar{E\sb{\mathrm{2}}}. \HOLBoundVar{a\sb{\mathrm{1}}} \HOLTokenTransBegin\HOLBoundVar{u}\HOLTokenTransEnd \HOLBoundVar{E\sb{\mathrm{2}}} \HOLSymConst{\HOLTokenImp{}} \HOLSymConst{\HOLTokenExists{}}\HOLBoundVar{E\sb{\mathrm{1}}}. \HOLBoundVar{a\sb{\mathrm{0}}} \HOLTokenTransBegin\HOLBoundVar{u}\HOLTokenTransEnd \HOLBoundVar{E\sb{\mathrm{1}}} \HOLSymConst{\HOLTokenConj{}} \HOLConst{STRONG_EQUIV} \HOLBoundVar{E\sb{\mathrm{1}}} \HOLBoundVar{E\sb{\mathrm{2}}}
\end{SaveVerbatim}
\newcommand{\HOLStrongEQTheoremsPROPERTYXXSTAR}{\UseVerbatim{HOLStrongEQTheoremsPROPERTYXXSTAR}}
\begin{SaveVerbatim}{HOLStrongEQTheoremsPROPERTYXXSTARXXLEFT}
\HOLTokenTurnstile{} \HOLSymConst{\HOLTokenForall{}}\HOLBoundVar{E} \HOLBoundVar{E\sp{\prime}}.
     \HOLConst{STRONG_EQUIV} \HOLBoundVar{E} \HOLBoundVar{E\sp{\prime}} \HOLSymConst{\HOLTokenImp{}}
     \HOLSymConst{\HOLTokenForall{}}\HOLBoundVar{u} \HOLBoundVar{E\sb{\mathrm{1}}}. \HOLBoundVar{E} \HOLTokenTransBegin\HOLBoundVar{u}\HOLTokenTransEnd \HOLBoundVar{E\sb{\mathrm{1}}} \HOLSymConst{\HOLTokenImp{}} \HOLSymConst{\HOLTokenExists{}}\HOLBoundVar{E\sb{\mathrm{2}}}. \HOLBoundVar{E\sp{\prime}} \HOLTokenTransBegin\HOLBoundVar{u}\HOLTokenTransEnd \HOLBoundVar{E\sb{\mathrm{2}}} \HOLSymConst{\HOLTokenConj{}} \HOLConst{STRONG_EQUIV} \HOLBoundVar{E\sb{\mathrm{1}}} \HOLBoundVar{E\sb{\mathrm{2}}}
\end{SaveVerbatim}
\newcommand{\HOLStrongEQTheoremsPROPERTYXXSTARXXLEFT}{\UseVerbatim{HOLStrongEQTheoremsPROPERTYXXSTARXXLEFT}}
\begin{SaveVerbatim}{HOLStrongEQTheoremsPROPERTYXXSTARXXRIGHT}
\HOLTokenTurnstile{} \HOLSymConst{\HOLTokenForall{}}\HOLBoundVar{E} \HOLBoundVar{E\sp{\prime}}.
     \HOLConst{STRONG_EQUIV} \HOLBoundVar{E} \HOLBoundVar{E\sp{\prime}} \HOLSymConst{\HOLTokenImp{}}
     \HOLSymConst{\HOLTokenForall{}}\HOLBoundVar{u} \HOLBoundVar{E\sb{\mathrm{2}}}. \HOLBoundVar{E\sp{\prime}} \HOLTokenTransBegin\HOLBoundVar{u}\HOLTokenTransEnd \HOLBoundVar{E\sb{\mathrm{2}}} \HOLSymConst{\HOLTokenImp{}} \HOLSymConst{\HOLTokenExists{}}\HOLBoundVar{E\sb{\mathrm{1}}}. \HOLBoundVar{E} \HOLTokenTransBegin\HOLBoundVar{u}\HOLTokenTransEnd \HOLBoundVar{E\sb{\mathrm{1}}} \HOLSymConst{\HOLTokenConj{}} \HOLConst{STRONG_EQUIV} \HOLBoundVar{E\sb{\mathrm{1}}} \HOLBoundVar{E\sb{\mathrm{2}}}
\end{SaveVerbatim}
\newcommand{\HOLStrongEQTheoremsPROPERTYXXSTARXXRIGHT}{\UseVerbatim{HOLStrongEQTheoremsPROPERTYXXSTARXXRIGHT}}
\begin{SaveVerbatim}{HOLStrongEQTheoremsSTRONGXXBISIM}
\HOLTokenTurnstile{} \HOLConst{STRONG_BISIM} \HOLFreeVar{Bsm} \HOLSymConst{\HOLTokenEquiv{}}
   \HOLSymConst{\HOLTokenForall{}}\HOLBoundVar{E} \HOLBoundVar{E\sp{\prime}}.
     \HOLFreeVar{Bsm} \HOLBoundVar{E} \HOLBoundVar{E\sp{\prime}} \HOLSymConst{\HOLTokenImp{}}
     \HOLSymConst{\HOLTokenForall{}}\HOLBoundVar{u}.
       (\HOLSymConst{\HOLTokenForall{}}\HOLBoundVar{E\sb{\mathrm{1}}}. \HOLBoundVar{E} \HOLTokenTransBegin\HOLBoundVar{u}\HOLTokenTransEnd \HOLBoundVar{E\sb{\mathrm{1}}} \HOLSymConst{\HOLTokenImp{}} \HOLSymConst{\HOLTokenExists{}}\HOLBoundVar{E\sb{\mathrm{2}}}. \HOLBoundVar{E\sp{\prime}} \HOLTokenTransBegin\HOLBoundVar{u}\HOLTokenTransEnd \HOLBoundVar{E\sb{\mathrm{2}}} \HOLSymConst{\HOLTokenConj{}} \HOLFreeVar{Bsm} \HOLBoundVar{E\sb{\mathrm{1}}} \HOLBoundVar{E\sb{\mathrm{2}}}) \HOLSymConst{\HOLTokenConj{}}
       \HOLSymConst{\HOLTokenForall{}}\HOLBoundVar{E\sb{\mathrm{2}}}. \HOLBoundVar{E\sp{\prime}} \HOLTokenTransBegin\HOLBoundVar{u}\HOLTokenTransEnd \HOLBoundVar{E\sb{\mathrm{2}}} \HOLSymConst{\HOLTokenImp{}} \HOLSymConst{\HOLTokenExists{}}\HOLBoundVar{E\sb{\mathrm{1}}}. \HOLBoundVar{E} \HOLTokenTransBegin\HOLBoundVar{u}\HOLTokenTransEnd \HOLBoundVar{E\sb{\mathrm{1}}} \HOLSymConst{\HOLTokenConj{}} \HOLFreeVar{Bsm} \HOLBoundVar{E\sb{\mathrm{1}}} \HOLBoundVar{E\sb{\mathrm{2}}}
\end{SaveVerbatim}
\newcommand{\HOLStrongEQTheoremsSTRONGXXBISIM}{\UseVerbatim{HOLStrongEQTheoremsSTRONGXXBISIM}}
\begin{SaveVerbatim}{HOLStrongEQTheoremsSTRONGXXBISIMXXSUBSETXXSTRONGXXEQUIV}
\HOLTokenTurnstile{} \HOLSymConst{\HOLTokenForall{}}\HOLBoundVar{Bsm}. \HOLConst{STRONG_BISIM} \HOLBoundVar{Bsm} \HOLSymConst{\HOLTokenImp{}} \HOLBoundVar{Bsm} \HOLConst{RSUBSET} \HOLConst{STRONG_EQUIV}
\end{SaveVerbatim}
\newcommand{\HOLStrongEQTheoremsSTRONGXXBISIMXXSUBSETXXSTRONGXXEQUIV}{\UseVerbatim{HOLStrongEQTheoremsSTRONGXXBISIMXXSUBSETXXSTRONGXXEQUIV}}
\begin{SaveVerbatim}{HOLStrongEQTheoremsSTRONGXXEQUIV}
\HOLTokenTurnstile{} \HOLSymConst{\HOLTokenForall{}}\HOLBoundVar{E} \HOLBoundVar{E\sp{\prime}}. \HOLConst{STRONG_EQUIV} \HOLBoundVar{E} \HOLBoundVar{E\sp{\prime}} \HOLSymConst{\HOLTokenEquiv{}} \HOLSymConst{\HOLTokenExists{}}\HOLBoundVar{Bsm}. \HOLBoundVar{Bsm} \HOLBoundVar{E} \HOLBoundVar{E\sp{\prime}} \HOLSymConst{\HOLTokenConj{}} \HOLConst{STRONG_BISIM} \HOLBoundVar{Bsm}
\end{SaveVerbatim}
\newcommand{\HOLStrongEQTheoremsSTRONGXXEQUIV}{\UseVerbatim{HOLStrongEQTheoremsSTRONGXXEQUIV}}
\begin{SaveVerbatim}{HOLStrongEQTheoremsSTRONGXXEQUIVXXcases}
\HOLTokenTurnstile{} \HOLSymConst{\HOLTokenForall{}}\HOLBoundVar{a\sb{\mathrm{0}}} \HOLBoundVar{a\sb{\mathrm{1}}}.
     \HOLConst{STRONG_EQUIV} \HOLBoundVar{a\sb{\mathrm{0}}} \HOLBoundVar{a\sb{\mathrm{1}}} \HOLSymConst{\HOLTokenEquiv{}}
     \HOLSymConst{\HOLTokenForall{}}\HOLBoundVar{u}.
       (\HOLSymConst{\HOLTokenForall{}}\HOLBoundVar{E\sb{\mathrm{1}}}. \HOLBoundVar{a\sb{\mathrm{0}}} \HOLTokenTransBegin\HOLBoundVar{u}\HOLTokenTransEnd \HOLBoundVar{E\sb{\mathrm{1}}} \HOLSymConst{\HOLTokenImp{}} \HOLSymConst{\HOLTokenExists{}}\HOLBoundVar{E\sb{\mathrm{2}}}. \HOLBoundVar{a\sb{\mathrm{1}}} \HOLTokenTransBegin\HOLBoundVar{u}\HOLTokenTransEnd \HOLBoundVar{E\sb{\mathrm{2}}} \HOLSymConst{\HOLTokenConj{}} \HOLConst{STRONG_EQUIV} \HOLBoundVar{E\sb{\mathrm{1}}} \HOLBoundVar{E\sb{\mathrm{2}}}) \HOLSymConst{\HOLTokenConj{}}
       \HOLSymConst{\HOLTokenForall{}}\HOLBoundVar{E\sb{\mathrm{2}}}. \HOLBoundVar{a\sb{\mathrm{1}}} \HOLTokenTransBegin\HOLBoundVar{u}\HOLTokenTransEnd \HOLBoundVar{E\sb{\mathrm{2}}} \HOLSymConst{\HOLTokenImp{}} \HOLSymConst{\HOLTokenExists{}}\HOLBoundVar{E\sb{\mathrm{1}}}. \HOLBoundVar{a\sb{\mathrm{0}}} \HOLTokenTransBegin\HOLBoundVar{u}\HOLTokenTransEnd \HOLBoundVar{E\sb{\mathrm{1}}} \HOLSymConst{\HOLTokenConj{}} \HOLConst{STRONG_EQUIV} \HOLBoundVar{E\sb{\mathrm{1}}} \HOLBoundVar{E\sb{\mathrm{2}}}
\end{SaveVerbatim}
\newcommand{\HOLStrongEQTheoremsSTRONGXXEQUIVXXcases}{\UseVerbatim{HOLStrongEQTheoremsSTRONGXXEQUIVXXcases}}
\begin{SaveVerbatim}{HOLStrongEQTheoremsSTRONGXXEQUIVXXcoind}
\HOLTokenTurnstile{} \HOLSymConst{\HOLTokenForall{}}\HOLBoundVar{STRONG\HOLTokenUnderscore{}EQUIV\sp{\prime}}.
     (\HOLSymConst{\HOLTokenForall{}}\HOLBoundVar{a\sb{\mathrm{0}}} \HOLBoundVar{a\sb{\mathrm{1}}}.
        \HOLBoundVar{STRONG\HOLTokenUnderscore{}EQUIV\sp{\prime}} \HOLBoundVar{a\sb{\mathrm{0}}} \HOLBoundVar{a\sb{\mathrm{1}}} \HOLSymConst{\HOLTokenImp{}}
        \HOLSymConst{\HOLTokenForall{}}\HOLBoundVar{u}.
          (\HOLSymConst{\HOLTokenForall{}}\HOLBoundVar{E\sb{\mathrm{1}}}.
             \HOLBoundVar{a\sb{\mathrm{0}}} \HOLTokenTransBegin\HOLBoundVar{u}\HOLTokenTransEnd \HOLBoundVar{E\sb{\mathrm{1}}} \HOLSymConst{\HOLTokenImp{}} \HOLSymConst{\HOLTokenExists{}}\HOLBoundVar{E\sb{\mathrm{2}}}. \HOLBoundVar{a\sb{\mathrm{1}}} \HOLTokenTransBegin\HOLBoundVar{u}\HOLTokenTransEnd \HOLBoundVar{E\sb{\mathrm{2}}} \HOLSymConst{\HOLTokenConj{}} \HOLBoundVar{STRONG\HOLTokenUnderscore{}EQUIV\sp{\prime}} \HOLBoundVar{E\sb{\mathrm{1}}} \HOLBoundVar{E\sb{\mathrm{2}}}) \HOLSymConst{\HOLTokenConj{}}
          \HOLSymConst{\HOLTokenForall{}}\HOLBoundVar{E\sb{\mathrm{2}}}.
            \HOLBoundVar{a\sb{\mathrm{1}}} \HOLTokenTransBegin\HOLBoundVar{u}\HOLTokenTransEnd \HOLBoundVar{E\sb{\mathrm{2}}} \HOLSymConst{\HOLTokenImp{}} \HOLSymConst{\HOLTokenExists{}}\HOLBoundVar{E\sb{\mathrm{1}}}. \HOLBoundVar{a\sb{\mathrm{0}}} \HOLTokenTransBegin\HOLBoundVar{u}\HOLTokenTransEnd \HOLBoundVar{E\sb{\mathrm{1}}} \HOLSymConst{\HOLTokenConj{}} \HOLBoundVar{STRONG\HOLTokenUnderscore{}EQUIV\sp{\prime}} \HOLBoundVar{E\sb{\mathrm{1}}} \HOLBoundVar{E\sb{\mathrm{2}}}) \HOLSymConst{\HOLTokenImp{}}
     \HOLSymConst{\HOLTokenForall{}}\HOLBoundVar{a\sb{\mathrm{0}}} \HOLBoundVar{a\sb{\mathrm{1}}}. \HOLBoundVar{STRONG\HOLTokenUnderscore{}EQUIV\sp{\prime}} \HOLBoundVar{a\sb{\mathrm{0}}} \HOLBoundVar{a\sb{\mathrm{1}}} \HOLSymConst{\HOLTokenImp{}} \HOLConst{STRONG_EQUIV} \HOLBoundVar{a\sb{\mathrm{0}}} \HOLBoundVar{a\sb{\mathrm{1}}}
\end{SaveVerbatim}
\newcommand{\HOLStrongEQTheoremsSTRONGXXEQUIVXXcoind}{\UseVerbatim{HOLStrongEQTheoremsSTRONGXXEQUIVXXcoind}}
\begin{SaveVerbatim}{HOLStrongEQTheoremsSTRONGXXEQUIVXXequivalence}
\HOLTokenTurnstile{} \HOLConst{equivalence} \HOLConst{STRONG_EQUIV}
\end{SaveVerbatim}
\newcommand{\HOLStrongEQTheoremsSTRONGXXEQUIVXXequivalence}{\UseVerbatim{HOLStrongEQTheoremsSTRONGXXEQUIVXXequivalence}}
\begin{SaveVerbatim}{HOLStrongEQTheoremsSTRONGXXEQUIVXXISXXSTRONGXXBISIM}
\HOLTokenTurnstile{} \HOLConst{STRONG_BISIM} \HOLConst{STRONG_EQUIV}
\end{SaveVerbatim}
\newcommand{\HOLStrongEQTheoremsSTRONGXXEQUIVXXISXXSTRONGXXBISIM}{\UseVerbatim{HOLStrongEQTheoremsSTRONGXXEQUIVXXISXXSTRONGXXBISIM}}
\begin{SaveVerbatim}{HOLStrongEQTheoremsSTRONGXXEQUIVXXPRESDXXBYXXPAR}
\HOLTokenTurnstile{} \HOLSymConst{\HOLTokenForall{}}\HOLBoundVar{E\sb{\mathrm{1}}} \HOLBoundVar{E\sb{\mathrm{1}}\sp{\prime}} \HOLBoundVar{E\sb{\mathrm{2}}} \HOLBoundVar{E\sb{\mathrm{2}}\sp{\prime}}.
     \HOLConst{STRONG_EQUIV} \HOLBoundVar{E\sb{\mathrm{1}}} \HOLBoundVar{E\sb{\mathrm{1}}\sp{\prime}} \HOLSymConst{\HOLTokenConj{}} \HOLConst{STRONG_EQUIV} \HOLBoundVar{E\sb{\mathrm{2}}} \HOLBoundVar{E\sb{\mathrm{2}}\sp{\prime}} \HOLSymConst{\HOLTokenImp{}}
     \HOLConst{STRONG_EQUIV} (\HOLBoundVar{E\sb{\mathrm{1}}} \HOLSymConst{\ensuremath{\parallel}} \HOLBoundVar{E\sb{\mathrm{2}}}) (\HOLBoundVar{E\sb{\mathrm{1}}\sp{\prime}} \HOLSymConst{\ensuremath{\parallel}} \HOLBoundVar{E\sb{\mathrm{2}}\sp{\prime}})
\end{SaveVerbatim}
\newcommand{\HOLStrongEQTheoremsSTRONGXXEQUIVXXPRESDXXBYXXPAR}{\UseVerbatim{HOLStrongEQTheoremsSTRONGXXEQUIVXXPRESDXXBYXXPAR}}
\begin{SaveVerbatim}{HOLStrongEQTheoremsSTRONGXXEQUIVXXPRESDXXBYXXSUM}
\HOLTokenTurnstile{} \HOLSymConst{\HOLTokenForall{}}\HOLBoundVar{E\sb{\mathrm{1}}} \HOLBoundVar{E\sb{\mathrm{1}}\sp{\prime}} \HOLBoundVar{E\sb{\mathrm{2}}} \HOLBoundVar{E\sb{\mathrm{2}}\sp{\prime}}.
     \HOLConst{STRONG_EQUIV} \HOLBoundVar{E\sb{\mathrm{1}}} \HOLBoundVar{E\sb{\mathrm{1}}\sp{\prime}} \HOLSymConst{\HOLTokenConj{}} \HOLConst{STRONG_EQUIV} \HOLBoundVar{E\sb{\mathrm{2}}} \HOLBoundVar{E\sb{\mathrm{2}}\sp{\prime}} \HOLSymConst{\HOLTokenImp{}}
     \HOLConst{STRONG_EQUIV} (\HOLBoundVar{E\sb{\mathrm{1}}} \HOLSymConst{+} \HOLBoundVar{E\sb{\mathrm{2}}}) (\HOLBoundVar{E\sb{\mathrm{1}}\sp{\prime}} \HOLSymConst{+} \HOLBoundVar{E\sb{\mathrm{2}}\sp{\prime}})
\end{SaveVerbatim}
\newcommand{\HOLStrongEQTheoremsSTRONGXXEQUIVXXPRESDXXBYXXSUM}{\UseVerbatim{HOLStrongEQTheoremsSTRONGXXEQUIVXXPRESDXXBYXXSUM}}
\begin{SaveVerbatim}{HOLStrongEQTheoremsSTRONGXXEQUIVXXREFL}
\HOLTokenTurnstile{} \HOLSymConst{\HOLTokenForall{}}\HOLBoundVar{E}. \HOLConst{STRONG_EQUIV} \HOLBoundVar{E} \HOLBoundVar{E}
\end{SaveVerbatim}
\newcommand{\HOLStrongEQTheoremsSTRONGXXEQUIVXXREFL}{\UseVerbatim{HOLStrongEQTheoremsSTRONGXXEQUIVXXREFL}}
\begin{SaveVerbatim}{HOLStrongEQTheoremsSTRONGXXEQUIVXXrules}
\HOLTokenTurnstile{} \HOLSymConst{\HOLTokenForall{}}\HOLBoundVar{E} \HOLBoundVar{E\sp{\prime}}.
     (\HOLSymConst{\HOLTokenForall{}}\HOLBoundVar{u}.
        (\HOLSymConst{\HOLTokenForall{}}\HOLBoundVar{E\sb{\mathrm{1}}}. \HOLBoundVar{E} \HOLTokenTransBegin\HOLBoundVar{u}\HOLTokenTransEnd \HOLBoundVar{E\sb{\mathrm{1}}} \HOLSymConst{\HOLTokenImp{}} \HOLSymConst{\HOLTokenExists{}}\HOLBoundVar{E\sb{\mathrm{2}}}. \HOLBoundVar{E\sp{\prime}} \HOLTokenTransBegin\HOLBoundVar{u}\HOLTokenTransEnd \HOLBoundVar{E\sb{\mathrm{2}}} \HOLSymConst{\HOLTokenConj{}} \HOLConst{STRONG_EQUIV} \HOLBoundVar{E\sb{\mathrm{1}}} \HOLBoundVar{E\sb{\mathrm{2}}}) \HOLSymConst{\HOLTokenConj{}}
        \HOLSymConst{\HOLTokenForall{}}\HOLBoundVar{E\sb{\mathrm{2}}}. \HOLBoundVar{E\sp{\prime}} \HOLTokenTransBegin\HOLBoundVar{u}\HOLTokenTransEnd \HOLBoundVar{E\sb{\mathrm{2}}} \HOLSymConst{\HOLTokenImp{}} \HOLSymConst{\HOLTokenExists{}}\HOLBoundVar{E\sb{\mathrm{1}}}. \HOLBoundVar{E} \HOLTokenTransBegin\HOLBoundVar{u}\HOLTokenTransEnd \HOLBoundVar{E\sb{\mathrm{1}}} \HOLSymConst{\HOLTokenConj{}} \HOLConst{STRONG_EQUIV} \HOLBoundVar{E\sb{\mathrm{1}}} \HOLBoundVar{E\sb{\mathrm{2}}}) \HOLSymConst{\HOLTokenImp{}}
     \HOLConst{STRONG_EQUIV} \HOLBoundVar{E} \HOLBoundVar{E\sp{\prime}}
\end{SaveVerbatim}
\newcommand{\HOLStrongEQTheoremsSTRONGXXEQUIVXXrules}{\UseVerbatim{HOLStrongEQTheoremsSTRONGXXEQUIVXXrules}}
\begin{SaveVerbatim}{HOLStrongEQTheoremsSTRONGXXEQUIVXXSUBSTXXPARXXL}
\HOLTokenTurnstile{} \HOLSymConst{\HOLTokenForall{}}\HOLBoundVar{E} \HOLBoundVar{E\sp{\prime}}.
     \HOLConst{STRONG_EQUIV} \HOLBoundVar{E} \HOLBoundVar{E\sp{\prime}} \HOLSymConst{\HOLTokenImp{}} \HOLSymConst{\HOLTokenForall{}}\HOLBoundVar{E\sp{\prime\prime}}. \HOLConst{STRONG_EQUIV} (\HOLBoundVar{E\sp{\prime\prime}} \HOLSymConst{\ensuremath{\parallel}} \HOLBoundVar{E}) (\HOLBoundVar{E\sp{\prime\prime}} \HOLSymConst{\ensuremath{\parallel}} \HOLBoundVar{E\sp{\prime}})
\end{SaveVerbatim}
\newcommand{\HOLStrongEQTheoremsSTRONGXXEQUIVXXSUBSTXXPARXXL}{\UseVerbatim{HOLStrongEQTheoremsSTRONGXXEQUIVXXSUBSTXXPARXXL}}
\begin{SaveVerbatim}{HOLStrongEQTheoremsSTRONGXXEQUIVXXSUBSTXXPARXXR}
\HOLTokenTurnstile{} \HOLSymConst{\HOLTokenForall{}}\HOLBoundVar{E} \HOLBoundVar{E\sp{\prime}}.
     \HOLConst{STRONG_EQUIV} \HOLBoundVar{E} \HOLBoundVar{E\sp{\prime}} \HOLSymConst{\HOLTokenImp{}} \HOLSymConst{\HOLTokenForall{}}\HOLBoundVar{E\sp{\prime\prime}}. \HOLConst{STRONG_EQUIV} (\HOLBoundVar{E} \HOLSymConst{\ensuremath{\parallel}} \HOLBoundVar{E\sp{\prime\prime}}) (\HOLBoundVar{E\sp{\prime}} \HOLSymConst{\ensuremath{\parallel}} \HOLBoundVar{E\sp{\prime\prime}})
\end{SaveVerbatim}
\newcommand{\HOLStrongEQTheoremsSTRONGXXEQUIVXXSUBSTXXPARXXR}{\UseVerbatim{HOLStrongEQTheoremsSTRONGXXEQUIVXXSUBSTXXPARXXR}}
\begin{SaveVerbatim}{HOLStrongEQTheoremsSTRONGXXEQUIVXXSUBSTXXPREFIX}
\HOLTokenTurnstile{} \HOLSymConst{\HOLTokenForall{}}\HOLBoundVar{E} \HOLBoundVar{E\sp{\prime}}. \HOLConst{STRONG_EQUIV} \HOLBoundVar{E} \HOLBoundVar{E\sp{\prime}} \HOLSymConst{\HOLTokenImp{}} \HOLSymConst{\HOLTokenForall{}}\HOLBoundVar{u}. \HOLConst{STRONG_EQUIV} (\HOLBoundVar{u}\HOLSymConst{..}\HOLBoundVar{E}) (\HOLBoundVar{u}\HOLSymConst{..}\HOLBoundVar{E\sp{\prime}})
\end{SaveVerbatim}
\newcommand{\HOLStrongEQTheoremsSTRONGXXEQUIVXXSUBSTXXPREFIX}{\UseVerbatim{HOLStrongEQTheoremsSTRONGXXEQUIVXXSUBSTXXPREFIX}}
\begin{SaveVerbatim}{HOLStrongEQTheoremsSTRONGXXEQUIVXXSUBSTXXRELAB}
\HOLTokenTurnstile{} \HOLSymConst{\HOLTokenForall{}}\HOLBoundVar{E} \HOLBoundVar{E\sp{\prime}}.
     \HOLConst{STRONG_EQUIV} \HOLBoundVar{E} \HOLBoundVar{E\sp{\prime}} \HOLSymConst{\HOLTokenImp{}}
     \HOLSymConst{\HOLTokenForall{}}\HOLBoundVar{rf}. \HOLConst{STRONG_EQUIV} (\HOLConst{relab} \HOLBoundVar{E} \HOLBoundVar{rf}) (\HOLConst{relab} \HOLBoundVar{E\sp{\prime}} \HOLBoundVar{rf})
\end{SaveVerbatim}
\newcommand{\HOLStrongEQTheoremsSTRONGXXEQUIVXXSUBSTXXRELAB}{\UseVerbatim{HOLStrongEQTheoremsSTRONGXXEQUIVXXSUBSTXXRELAB}}
\begin{SaveVerbatim}{HOLStrongEQTheoremsSTRONGXXEQUIVXXSUBSTXXRESTR}
\HOLTokenTurnstile{} \HOLSymConst{\HOLTokenForall{}}\HOLBoundVar{E} \HOLBoundVar{E\sp{\prime}}. \HOLConst{STRONG_EQUIV} \HOLBoundVar{E} \HOLBoundVar{E\sp{\prime}} \HOLSymConst{\HOLTokenImp{}} \HOLSymConst{\HOLTokenForall{}}\HOLBoundVar{L}. \HOLConst{STRONG_EQUIV} (\HOLConst{\ensuremath{\nu}} \HOLBoundVar{L} \HOLBoundVar{E}) (\HOLConst{\ensuremath{\nu}} \HOLBoundVar{L} \HOLBoundVar{E\sp{\prime}})
\end{SaveVerbatim}
\newcommand{\HOLStrongEQTheoremsSTRONGXXEQUIVXXSUBSTXXRESTR}{\UseVerbatim{HOLStrongEQTheoremsSTRONGXXEQUIVXXSUBSTXXRESTR}}
\begin{SaveVerbatim}{HOLStrongEQTheoremsSTRONGXXEQUIVXXSUBSTXXSUMXXL}
\HOLTokenTurnstile{} \HOLSymConst{\HOLTokenForall{}}\HOLBoundVar{E\sp{\prime}} \HOLBoundVar{E}.
     \HOLConst{STRONG_EQUIV} \HOLBoundVar{E} \HOLBoundVar{E\sp{\prime}} \HOLSymConst{\HOLTokenImp{}} \HOLSymConst{\HOLTokenForall{}}\HOLBoundVar{E\sp{\prime\prime}}. \HOLConst{STRONG_EQUIV} (\HOLBoundVar{E\sp{\prime\prime}} \HOLSymConst{+} \HOLBoundVar{E}) (\HOLBoundVar{E\sp{\prime\prime}} \HOLSymConst{+} \HOLBoundVar{E\sp{\prime}})
\end{SaveVerbatim}
\newcommand{\HOLStrongEQTheoremsSTRONGXXEQUIVXXSUBSTXXSUMXXL}{\UseVerbatim{HOLStrongEQTheoremsSTRONGXXEQUIVXXSUBSTXXSUMXXL}}
\begin{SaveVerbatim}{HOLStrongEQTheoremsSTRONGXXEQUIVXXSUBSTXXSUMXXR}
\HOLTokenTurnstile{} \HOLSymConst{\HOLTokenForall{}}\HOLBoundVar{E\sp{\prime}} \HOLBoundVar{E}.
     \HOLConst{STRONG_EQUIV} \HOLBoundVar{E} \HOLBoundVar{E\sp{\prime}} \HOLSymConst{\HOLTokenImp{}} \HOLSymConst{\HOLTokenForall{}}\HOLBoundVar{E\sp{\prime\prime}}. \HOLConst{STRONG_EQUIV} (\HOLBoundVar{E} \HOLSymConst{+} \HOLBoundVar{E\sp{\prime\prime}}) (\HOLBoundVar{E\sp{\prime}} \HOLSymConst{+} \HOLBoundVar{E\sp{\prime\prime}})
\end{SaveVerbatim}
\newcommand{\HOLStrongEQTheoremsSTRONGXXEQUIVXXSUBSTXXSUMXXR}{\UseVerbatim{HOLStrongEQTheoremsSTRONGXXEQUIVXXSUBSTXXSUMXXR}}
\begin{SaveVerbatim}{HOLStrongEQTheoremsSTRONGXXEQUIVXXSYM}
\HOLTokenTurnstile{} \HOLSymConst{\HOLTokenForall{}}\HOLBoundVar{E} \HOLBoundVar{E\sp{\prime}}. \HOLConst{STRONG_EQUIV} \HOLBoundVar{E} \HOLBoundVar{E\sp{\prime}} \HOLSymConst{\HOLTokenImp{}} \HOLConst{STRONG_EQUIV} \HOLBoundVar{E\sp{\prime}} \HOLBoundVar{E}
\end{SaveVerbatim}
\newcommand{\HOLStrongEQTheoremsSTRONGXXEQUIVXXSYM}{\UseVerbatim{HOLStrongEQTheoremsSTRONGXXEQUIVXXSYM}}
\begin{SaveVerbatim}{HOLStrongEQTheoremsSTRONGXXEQUIVXXTRANS}
\HOLTokenTurnstile{} \HOLSymConst{\HOLTokenForall{}}\HOLBoundVar{E} \HOLBoundVar{E\sp{\prime}} \HOLBoundVar{E\sp{\prime\prime}}.
     \HOLConst{STRONG_EQUIV} \HOLBoundVar{E} \HOLBoundVar{E\sp{\prime}} \HOLSymConst{\HOLTokenConj{}} \HOLConst{STRONG_EQUIV} \HOLBoundVar{E\sp{\prime}} \HOLBoundVar{E\sp{\prime\prime}} \HOLSymConst{\HOLTokenImp{}}
     \HOLConst{STRONG_EQUIV} \HOLBoundVar{E} \HOLBoundVar{E\sp{\prime\prime}}
\end{SaveVerbatim}
\newcommand{\HOLStrongEQTheoremsSTRONGXXEQUIVXXTRANS}{\UseVerbatim{HOLStrongEQTheoremsSTRONGXXEQUIVXXTRANS}}
\begin{SaveVerbatim}{HOLStrongEQTheoremsUNIONXXSTRONGXXBISIM}
\HOLTokenTurnstile{} \HOLSymConst{\HOLTokenForall{}}\HOLBoundVar{Bsm\sb{\mathrm{1}}} \HOLBoundVar{Bsm\sb{\mathrm{2}}}.
     \HOLConst{STRONG_BISIM} \HOLBoundVar{Bsm\sb{\mathrm{1}}} \HOLSymConst{\HOLTokenConj{}} \HOLConst{STRONG_BISIM} \HOLBoundVar{Bsm\sb{\mathrm{2}}} \HOLSymConst{\HOLTokenImp{}}
     \HOLConst{STRONG_BISIM} (\HOLBoundVar{Bsm\sb{\mathrm{1}}} \HOLConst{RUNION} \HOLBoundVar{Bsm\sb{\mathrm{2}}})
\end{SaveVerbatim}
\newcommand{\HOLStrongEQTheoremsUNIONXXSTRONGXXBISIM}{\UseVerbatim{HOLStrongEQTheoremsUNIONXXSTRONGXXBISIM}}
\newcommand{\HOLStrongEQTheorems}{
\HOLThmTag{StrongEQ}{COMP_STRONG_BISIM}\HOLStrongEQTheoremsCOMPXXSTRONGXXBISIM
\HOLThmTag{StrongEQ}{CONVERSE_STRONG_BISIM}\HOLStrongEQTheoremsCONVERSEXXSTRONGXXBISIM
\HOLThmTag{StrongEQ}{EQUAL_IMP_STRONG_EQUIV}\HOLStrongEQTheoremsEQUALXXIMPXXSTRONGXXEQUIV
\HOLThmTag{StrongEQ}{IDENTITY_STRONG_BISIM}\HOLStrongEQTheoremsIDENTITYXXSTRONGXXBISIM
\HOLThmTag{StrongEQ}{PROPERTY_STAR}\HOLStrongEQTheoremsPROPERTYXXSTAR
\HOLThmTag{StrongEQ}{PROPERTY_STAR_LEFT}\HOLStrongEQTheoremsPROPERTYXXSTARXXLEFT
\HOLThmTag{StrongEQ}{PROPERTY_STAR_RIGHT}\HOLStrongEQTheoremsPROPERTYXXSTARXXRIGHT
\HOLThmTag{StrongEQ}{STRONG_BISIM}\HOLStrongEQTheoremsSTRONGXXBISIM
\HOLThmTag{StrongEQ}{STRONG_BISIM_SUBSET_STRONG_EQUIV}\HOLStrongEQTheoremsSTRONGXXBISIMXXSUBSETXXSTRONGXXEQUIV
\HOLThmTag{StrongEQ}{STRONG_EQUIV}\HOLStrongEQTheoremsSTRONGXXEQUIV
\HOLThmTag{StrongEQ}{STRONG_EQUIV_cases}\HOLStrongEQTheoremsSTRONGXXEQUIVXXcases
\HOLThmTag{StrongEQ}{STRONG_EQUIV_coind}\HOLStrongEQTheoremsSTRONGXXEQUIVXXcoind
\HOLThmTag{StrongEQ}{STRONG_EQUIV_equivalence}\HOLStrongEQTheoremsSTRONGXXEQUIVXXequivalence
\HOLThmTag{StrongEQ}{STRONG_EQUIV_IS_STRONG_BISIM}\HOLStrongEQTheoremsSTRONGXXEQUIVXXISXXSTRONGXXBISIM
\HOLThmTag{StrongEQ}{STRONG_EQUIV_PRESD_BY_PAR}\HOLStrongEQTheoremsSTRONGXXEQUIVXXPRESDXXBYXXPAR
\HOLThmTag{StrongEQ}{STRONG_EQUIV_PRESD_BY_SUM}\HOLStrongEQTheoremsSTRONGXXEQUIVXXPRESDXXBYXXSUM
\HOLThmTag{StrongEQ}{STRONG_EQUIV_REFL}\HOLStrongEQTheoremsSTRONGXXEQUIVXXREFL
\HOLThmTag{StrongEQ}{STRONG_EQUIV_rules}\HOLStrongEQTheoremsSTRONGXXEQUIVXXrules
\HOLThmTag{StrongEQ}{STRONG_EQUIV_SUBST_PAR_L}\HOLStrongEQTheoremsSTRONGXXEQUIVXXSUBSTXXPARXXL
\HOLThmTag{StrongEQ}{STRONG_EQUIV_SUBST_PAR_R}\HOLStrongEQTheoremsSTRONGXXEQUIVXXSUBSTXXPARXXR
\HOLThmTag{StrongEQ}{STRONG_EQUIV_SUBST_PREFIX}\HOLStrongEQTheoremsSTRONGXXEQUIVXXSUBSTXXPREFIX
\HOLThmTag{StrongEQ}{STRONG_EQUIV_SUBST_RELAB}\HOLStrongEQTheoremsSTRONGXXEQUIVXXSUBSTXXRELAB
\HOLThmTag{StrongEQ}{STRONG_EQUIV_SUBST_RESTR}\HOLStrongEQTheoremsSTRONGXXEQUIVXXSUBSTXXRESTR
\HOLThmTag{StrongEQ}{STRONG_EQUIV_SUBST_SUM_L}\HOLStrongEQTheoremsSTRONGXXEQUIVXXSUBSTXXSUMXXL
\HOLThmTag{StrongEQ}{STRONG_EQUIV_SUBST_SUM_R}\HOLStrongEQTheoremsSTRONGXXEQUIVXXSUBSTXXSUMXXR
\HOLThmTag{StrongEQ}{STRONG_EQUIV_SYM}\HOLStrongEQTheoremsSTRONGXXEQUIVXXSYM
\HOLThmTag{StrongEQ}{STRONG_EQUIV_TRANS}\HOLStrongEQTheoremsSTRONGXXEQUIVXXTRANS
\HOLThmTag{StrongEQ}{UNION_STRONG_BISIM}\HOLStrongEQTheoremsUNIONXXSTRONGXXBISIM
}

\newcommand{\HOLStrongLawsDate}{02 Dicembre 2017}
\newcommand{\HOLStrongLawsTime}{13:31}
\begin{SaveVerbatim}{HOLStrongLawsDefinitionsALLXXSYNCXXdef}
\HOLTokenTurnstile{} (\HOLSymConst{\HOLTokenForall{}}\HOLBoundVar{f} \HOLBoundVar{f\sp{\prime}} \HOLBoundVar{m}.
      \HOLConst{ALL_SYNC} \HOLBoundVar{f} \HOLNumLit{0} \HOLBoundVar{f\sp{\prime}} \HOLBoundVar{m} \HOLSymConst{=}
      \HOLConst{SYNC} (\HOLConst{PREF_ACT} (\HOLBoundVar{f} \HOLNumLit{0})) (\HOLConst{PREF_PROC} (\HOLBoundVar{f} \HOLNumLit{0})) \HOLBoundVar{f\sp{\prime}} \HOLBoundVar{m}) \HOLSymConst{\HOLTokenConj{}}
   \HOLSymConst{\HOLTokenForall{}}\HOLBoundVar{f} \HOLBoundVar{n} \HOLBoundVar{f\sp{\prime}} \HOLBoundVar{m}.
     \HOLConst{ALL_SYNC} \HOLBoundVar{f} (\HOLConst{SUC} \HOLBoundVar{n}) \HOLBoundVar{f\sp{\prime}} \HOLBoundVar{m} \HOLSymConst{=}
     \HOLConst{ALL_SYNC} \HOLBoundVar{f} \HOLBoundVar{n} \HOLBoundVar{f\sp{\prime}} \HOLBoundVar{m} \HOLSymConst{+}
     \HOLConst{SYNC} (\HOLConst{PREF_ACT} (\HOLBoundVar{f} (\HOLConst{SUC} \HOLBoundVar{n}))) (\HOLConst{PREF_PROC} (\HOLBoundVar{f} (\HOLConst{SUC} \HOLBoundVar{n}))) \HOLBoundVar{f\sp{\prime}} \HOLBoundVar{m}
\end{SaveVerbatim}
\newcommand{\HOLStrongLawsDefinitionsALLXXSYNCXXdef}{\UseVerbatim{HOLStrongLawsDefinitionsALLXXSYNCXXdef}}
\begin{SaveVerbatim}{HOLStrongLawsDefinitionsCCSXXCOMPXXdef}
\HOLTokenTurnstile{} (\HOLSymConst{\HOLTokenForall{}}\HOLBoundVar{f}. \HOLConst{PI} \HOLBoundVar{f} \HOLNumLit{0} \HOLSymConst{=} \HOLBoundVar{f} \HOLNumLit{0}) \HOLSymConst{\HOLTokenConj{}} \HOLSymConst{\HOLTokenForall{}}\HOLBoundVar{f} \HOLBoundVar{n}. \HOLConst{PI} \HOLBoundVar{f} (\HOLConst{SUC} \HOLBoundVar{n}) \HOLSymConst{=} \HOLConst{PI} \HOLBoundVar{f} \HOLBoundVar{n} \HOLSymConst{\ensuremath{\parallel}} \HOLBoundVar{f} (\HOLConst{SUC} \HOLBoundVar{n})
\end{SaveVerbatim}
\newcommand{\HOLStrongLawsDefinitionsCCSXXCOMPXXdef}{\UseVerbatim{HOLStrongLawsDefinitionsCCSXXCOMPXXdef}}
\begin{SaveVerbatim}{HOLStrongLawsDefinitionsCCSXXSIGMAXXdef}
\HOLTokenTurnstile{} (\HOLSymConst{\HOLTokenForall{}}\HOLBoundVar{f}. \HOLConst{SIGMA} \HOLBoundVar{f} \HOLNumLit{0} \HOLSymConst{=} \HOLBoundVar{f} \HOLNumLit{0}) \HOLSymConst{\HOLTokenConj{}}
   \HOLSymConst{\HOLTokenForall{}}\HOLBoundVar{f} \HOLBoundVar{n}. \HOLConst{SIGMA} \HOLBoundVar{f} (\HOLConst{SUC} \HOLBoundVar{n}) \HOLSymConst{=} \HOLConst{SIGMA} \HOLBoundVar{f} \HOLBoundVar{n} \HOLSymConst{+} \HOLBoundVar{f} (\HOLConst{SUC} \HOLBoundVar{n})
\end{SaveVerbatim}
\newcommand{\HOLStrongLawsDefinitionsCCSXXSIGMAXXdef}{\UseVerbatim{HOLStrongLawsDefinitionsCCSXXSIGMAXXdef}}
\begin{SaveVerbatim}{HOLStrongLawsDefinitionsIsXXPrefixXXdef}
\HOLTokenTurnstile{} \HOLSymConst{\HOLTokenForall{}}\HOLBoundVar{E}. \HOLConst{Is_Prefix} \HOLBoundVar{E} \HOLSymConst{\HOLTokenEquiv{}} \HOLSymConst{\HOLTokenExists{}}\HOLBoundVar{u} \HOLBoundVar{E\sp{\prime}}. \HOLBoundVar{E} \HOLSymConst{=} \HOLBoundVar{u}\HOLSymConst{..}\HOLBoundVar{E\sp{\prime}}
\end{SaveVerbatim}
\newcommand{\HOLStrongLawsDefinitionsIsXXPrefixXXdef}{\UseVerbatim{HOLStrongLawsDefinitionsIsXXPrefixXXdef}}
\begin{SaveVerbatim}{HOLStrongLawsDefinitionsPREFXXACTXXdef}
\HOLTokenTurnstile{} \HOLSymConst{\HOLTokenForall{}}\HOLBoundVar{u} \HOLBoundVar{E}. \HOLConst{PREF_ACT} (\HOLBoundVar{u}\HOLSymConst{..}\HOLBoundVar{E}) \HOLSymConst{=} \HOLBoundVar{u}
\end{SaveVerbatim}
\newcommand{\HOLStrongLawsDefinitionsPREFXXACTXXdef}{\UseVerbatim{HOLStrongLawsDefinitionsPREFXXACTXXdef}}
\begin{SaveVerbatim}{HOLStrongLawsDefinitionsPREFXXPROCXXdef}
\HOLTokenTurnstile{} \HOLSymConst{\HOLTokenForall{}}\HOLBoundVar{u} \HOLBoundVar{E}. \HOLConst{PREF_PROC} (\HOLBoundVar{u}\HOLSymConst{..}\HOLBoundVar{E}) \HOLSymConst{=} \HOLBoundVar{E}
\end{SaveVerbatim}
\newcommand{\HOLStrongLawsDefinitionsPREFXXPROCXXdef}{\UseVerbatim{HOLStrongLawsDefinitionsPREFXXPROCXXdef}}
\begin{SaveVerbatim}{HOLStrongLawsDefinitionsSYNCXXdef}
\HOLTokenTurnstile{} (\HOLSymConst{\HOLTokenForall{}}\HOLBoundVar{u} \HOLBoundVar{P} \HOLBoundVar{f}.
      \HOLConst{SYNC} \HOLBoundVar{u} \HOLBoundVar{P} \HOLBoundVar{f} \HOLNumLit{0} \HOLSymConst{=}
      \HOLKeyword{if} (\HOLBoundVar{u} \HOLSymConst{=} \HOLConst{\ensuremath{\tau}}) \HOLSymConst{\HOLTokenDisj{}} (\HOLConst{PREF_ACT} (\HOLBoundVar{f} \HOLNumLit{0}) \HOLSymConst{=} \HOLConst{\ensuremath{\tau}}) \HOLKeyword{then} \HOLConst{nil}
      \HOLKeyword{else} \HOLKeyword{if} \HOLConst{LABEL} \HOLBoundVar{u} \HOLSymConst{=} \HOLConst{COMPL} (\HOLConst{LABEL} (\HOLConst{PREF_ACT} (\HOLBoundVar{f} \HOLNumLit{0}))) \HOLKeyword{then}
        \HOLConst{\ensuremath{\tau}}\HOLSymConst{..}(\HOLBoundVar{P} \HOLSymConst{\ensuremath{\parallel}} \HOLConst{PREF_PROC} (\HOLBoundVar{f} \HOLNumLit{0}))
      \HOLKeyword{else} \HOLConst{nil}) \HOLSymConst{\HOLTokenConj{}}
   \HOLSymConst{\HOLTokenForall{}}\HOLBoundVar{u} \HOLBoundVar{P} \HOLBoundVar{f} \HOLBoundVar{n}.
     \HOLConst{SYNC} \HOLBoundVar{u} \HOLBoundVar{P} \HOLBoundVar{f} (\HOLConst{SUC} \HOLBoundVar{n}) \HOLSymConst{=}
     \HOLKeyword{if} (\HOLBoundVar{u} \HOLSymConst{=} \HOLConst{\ensuremath{\tau}}) \HOLSymConst{\HOLTokenDisj{}} (\HOLConst{PREF_ACT} (\HOLBoundVar{f} (\HOLConst{SUC} \HOLBoundVar{n})) \HOLSymConst{=} \HOLConst{\ensuremath{\tau}}) \HOLKeyword{then} \HOLConst{SYNC} \HOLBoundVar{u} \HOLBoundVar{P} \HOLBoundVar{f} \HOLBoundVar{n}
     \HOLKeyword{else} \HOLKeyword{if} \HOLConst{LABEL} \HOLBoundVar{u} \HOLSymConst{=} \HOLConst{COMPL} (\HOLConst{LABEL} (\HOLConst{PREF_ACT} (\HOLBoundVar{f} (\HOLConst{SUC} \HOLBoundVar{n})))) \HOLKeyword{then}
       \HOLConst{\ensuremath{\tau}}\HOLSymConst{..}(\HOLBoundVar{P} \HOLSymConst{\ensuremath{\parallel}} \HOLConst{PREF_PROC} (\HOLBoundVar{f} (\HOLConst{SUC} \HOLBoundVar{n}))) \HOLSymConst{+} \HOLConst{SYNC} \HOLBoundVar{u} \HOLBoundVar{P} \HOLBoundVar{f} \HOLBoundVar{n}
     \HOLKeyword{else} \HOLConst{SYNC} \HOLBoundVar{u} \HOLBoundVar{P} \HOLBoundVar{f} \HOLBoundVar{n}
\end{SaveVerbatim}
\newcommand{\HOLStrongLawsDefinitionsSYNCXXdef}{\UseVerbatim{HOLStrongLawsDefinitionsSYNCXXdef}}
\newcommand{\HOLStrongLawsDefinitions}{
\HOLDfnTag{StrongLaws}{ALL_SYNC_def}\HOLStrongLawsDefinitionsALLXXSYNCXXdef
\HOLDfnTag{StrongLaws}{CCS_COMP_def}\HOLStrongLawsDefinitionsCCSXXCOMPXXdef
\HOLDfnTag{StrongLaws}{CCS_SIGMA_def}\HOLStrongLawsDefinitionsCCSXXSIGMAXXdef
\HOLDfnTag{StrongLaws}{Is_Prefix_def}\HOLStrongLawsDefinitionsIsXXPrefixXXdef
\HOLDfnTag{StrongLaws}{PREF_ACT_def}\HOLStrongLawsDefinitionsPREFXXACTXXdef
\HOLDfnTag{StrongLaws}{PREF_PROC_def}\HOLStrongLawsDefinitionsPREFXXPROCXXdef
\HOLDfnTag{StrongLaws}{SYNC_def}\HOLStrongLawsDefinitionsSYNCXXdef
}
\begin{SaveVerbatim}{HOLStrongLawsTheoremsALLXXSYNCXXBASE}
\HOLTokenTurnstile{} \HOLSymConst{\HOLTokenForall{}}\HOLBoundVar{f} \HOLBoundVar{f\sp{\prime}} \HOLBoundVar{m}.
     \HOLConst{ALL_SYNC} \HOLBoundVar{f} \HOLNumLit{0} \HOLBoundVar{f\sp{\prime}} \HOLBoundVar{m} \HOLSymConst{=}
     \HOLConst{SYNC} (\HOLConst{PREF_ACT} (\HOLBoundVar{f} \HOLNumLit{0})) (\HOLConst{PREF_PROC} (\HOLBoundVar{f} \HOLNumLit{0})) \HOLBoundVar{f\sp{\prime}} \HOLBoundVar{m}
\end{SaveVerbatim}
\newcommand{\HOLStrongLawsTheoremsALLXXSYNCXXBASE}{\UseVerbatim{HOLStrongLawsTheoremsALLXXSYNCXXBASE}}
\begin{SaveVerbatim}{HOLStrongLawsTheoremsALLXXSYNCXXdefXXcompute}
\HOLTokenTurnstile{} (\HOLSymConst{\HOLTokenForall{}}\HOLBoundVar{f} \HOLBoundVar{f\sp{\prime}} \HOLBoundVar{m}.
      \HOLConst{ALL_SYNC} \HOLBoundVar{f} \HOLNumLit{0} \HOLBoundVar{f\sp{\prime}} \HOLBoundVar{m} \HOLSymConst{=}
      \HOLConst{SYNC} (\HOLConst{PREF_ACT} (\HOLBoundVar{f} \HOLNumLit{0})) (\HOLConst{PREF_PROC} (\HOLBoundVar{f} \HOLNumLit{0})) \HOLBoundVar{f\sp{\prime}} \HOLBoundVar{m}) \HOLSymConst{\HOLTokenConj{}}
   (\HOLSymConst{\HOLTokenForall{}}\HOLBoundVar{f} \HOLBoundVar{n} \HOLBoundVar{f\sp{\prime}} \HOLBoundVar{m}.
      \HOLConst{ALL_SYNC} \HOLBoundVar{f} (\HOLConst{NUMERAL} (\HOLConst{BIT1} \HOLBoundVar{n})) \HOLBoundVar{f\sp{\prime}} \HOLBoundVar{m} \HOLSymConst{=}
      \HOLConst{ALL_SYNC} \HOLBoundVar{f} (\HOLConst{NUMERAL} (\HOLConst{BIT1} \HOLBoundVar{n}) \HOLSymConst{-} \HOLNumLit{1}) \HOLBoundVar{f\sp{\prime}} \HOLBoundVar{m} \HOLSymConst{+}
      \HOLConst{SYNC} (\HOLConst{PREF_ACT} (\HOLBoundVar{f} (\HOLConst{NUMERAL} (\HOLConst{BIT1} \HOLBoundVar{n}))))
        (\HOLConst{PREF_PROC} (\HOLBoundVar{f} (\HOLConst{NUMERAL} (\HOLConst{BIT1} \HOLBoundVar{n})))) \HOLBoundVar{f\sp{\prime}} \HOLBoundVar{m}) \HOLSymConst{\HOLTokenConj{}}
   \HOLSymConst{\HOLTokenForall{}}\HOLBoundVar{f} \HOLBoundVar{n} \HOLBoundVar{f\sp{\prime}} \HOLBoundVar{m}.
     \HOLConst{ALL_SYNC} \HOLBoundVar{f} (\HOLConst{NUMERAL} (\HOLConst{BIT2} \HOLBoundVar{n})) \HOLBoundVar{f\sp{\prime}} \HOLBoundVar{m} \HOLSymConst{=}
     \HOLConst{ALL_SYNC} \HOLBoundVar{f} (\HOLConst{NUMERAL} (\HOLConst{BIT1} \HOLBoundVar{n})) \HOLBoundVar{f\sp{\prime}} \HOLBoundVar{m} \HOLSymConst{+}
     \HOLConst{SYNC} (\HOLConst{PREF_ACT} (\HOLBoundVar{f} (\HOLConst{NUMERAL} (\HOLConst{BIT2} \HOLBoundVar{n}))))
       (\HOLConst{PREF_PROC} (\HOLBoundVar{f} (\HOLConst{NUMERAL} (\HOLConst{BIT2} \HOLBoundVar{n})))) \HOLBoundVar{f\sp{\prime}} \HOLBoundVar{m}
\end{SaveVerbatim}
\newcommand{\HOLStrongLawsTheoremsALLXXSYNCXXdefXXcompute}{\UseVerbatim{HOLStrongLawsTheoremsALLXXSYNCXXdefXXcompute}}
\begin{SaveVerbatim}{HOLStrongLawsTheoremsALLXXSYNCXXINDUCT}
\HOLTokenTurnstile{} \HOLSymConst{\HOLTokenForall{}}\HOLBoundVar{f} \HOLBoundVar{n} \HOLBoundVar{f\sp{\prime}} \HOLBoundVar{m}.
     \HOLConst{ALL_SYNC} \HOLBoundVar{f} (\HOLConst{SUC} \HOLBoundVar{n}) \HOLBoundVar{f\sp{\prime}} \HOLBoundVar{m} \HOLSymConst{=}
     \HOLConst{ALL_SYNC} \HOLBoundVar{f} \HOLBoundVar{n} \HOLBoundVar{f\sp{\prime}} \HOLBoundVar{m} \HOLSymConst{+}
     \HOLConst{SYNC} (\HOLConst{PREF_ACT} (\HOLBoundVar{f} (\HOLConst{SUC} \HOLBoundVar{n}))) (\HOLConst{PREF_PROC} (\HOLBoundVar{f} (\HOLConst{SUC} \HOLBoundVar{n}))) \HOLBoundVar{f\sp{\prime}} \HOLBoundVar{m}
\end{SaveVerbatim}
\newcommand{\HOLStrongLawsTheoremsALLXXSYNCXXINDUCT}{\UseVerbatim{HOLStrongLawsTheoremsALLXXSYNCXXINDUCT}}
\begin{SaveVerbatim}{HOLStrongLawsTheoremsALLXXSYNCXXTRANSXXTHM}
\HOLTokenTurnstile{} \HOLSymConst{\HOLTokenForall{}}\HOLBoundVar{n} \HOLBoundVar{m} \HOLBoundVar{f} \HOLBoundVar{f\sp{\prime}} \HOLBoundVar{u} \HOLBoundVar{E}.
     \HOLConst{ALL_SYNC} \HOLBoundVar{f} \HOLBoundVar{n} \HOLBoundVar{f\sp{\prime}} \HOLBoundVar{m} \HOLTokenTransBegin\HOLBoundVar{u}\HOLTokenTransEnd \HOLBoundVar{E} \HOLSymConst{\HOLTokenImp{}}
     \HOLSymConst{\HOLTokenExists{}}\HOLBoundVar{k} \HOLBoundVar{k\sp{\prime}} \HOLBoundVar{l}.
       \HOLBoundVar{k} \HOLSymConst{\HOLTokenLeq{}} \HOLBoundVar{n} \HOLSymConst{\HOLTokenConj{}} \HOLBoundVar{k\sp{\prime}} \HOLSymConst{\HOLTokenLeq{}} \HOLBoundVar{m} \HOLSymConst{\HOLTokenConj{}} (\HOLConst{PREF_ACT} (\HOLBoundVar{f} \HOLBoundVar{k}) \HOLSymConst{=} \HOLConst{label} \HOLBoundVar{l}) \HOLSymConst{\HOLTokenConj{}}
       (\HOLConst{PREF_ACT} (\HOLBoundVar{f\sp{\prime}} \HOLBoundVar{k\sp{\prime}}) \HOLSymConst{=} \HOLConst{label} (\HOLConst{COMPL} \HOLBoundVar{l})) \HOLSymConst{\HOLTokenConj{}} (\HOLBoundVar{u} \HOLSymConst{=} \HOLConst{\ensuremath{\tau}}) \HOLSymConst{\HOLTokenConj{}}
       (\HOLBoundVar{E} \HOLSymConst{=} \HOLConst{PREF_PROC} (\HOLBoundVar{f} \HOLBoundVar{k}) \HOLSymConst{\ensuremath{\parallel}} \HOLConst{PREF_PROC} (\HOLBoundVar{f\sp{\prime}} \HOLBoundVar{k\sp{\prime}}))
\end{SaveVerbatim}
\newcommand{\HOLStrongLawsTheoremsALLXXSYNCXXTRANSXXTHM}{\UseVerbatim{HOLStrongLawsTheoremsALLXXSYNCXXTRANSXXTHM}}
\begin{SaveVerbatim}{HOLStrongLawsTheoremsALLXXSYNCXXTRANSXXTHMXXEQ}
\HOLTokenTurnstile{} \HOLSymConst{\HOLTokenForall{}}\HOLBoundVar{n} \HOLBoundVar{m} \HOLBoundVar{f} \HOLBoundVar{f\sp{\prime}} \HOLBoundVar{u} \HOLBoundVar{E}.
     \HOLConst{ALL_SYNC} \HOLBoundVar{f} \HOLBoundVar{n} \HOLBoundVar{f\sp{\prime}} \HOLBoundVar{m} \HOLTokenTransBegin\HOLBoundVar{u}\HOLTokenTransEnd \HOLBoundVar{E} \HOLSymConst{\HOLTokenEquiv{}}
     \HOLSymConst{\HOLTokenExists{}}\HOLBoundVar{k} \HOLBoundVar{k\sp{\prime}} \HOLBoundVar{l}.
       \HOLBoundVar{k} \HOLSymConst{\HOLTokenLeq{}} \HOLBoundVar{n} \HOLSymConst{\HOLTokenConj{}} \HOLBoundVar{k\sp{\prime}} \HOLSymConst{\HOLTokenLeq{}} \HOLBoundVar{m} \HOLSymConst{\HOLTokenConj{}} (\HOLConst{PREF_ACT} (\HOLBoundVar{f} \HOLBoundVar{k}) \HOLSymConst{=} \HOLConst{label} \HOLBoundVar{l}) \HOLSymConst{\HOLTokenConj{}}
       (\HOLConst{PREF_ACT} (\HOLBoundVar{f\sp{\prime}} \HOLBoundVar{k\sp{\prime}}) \HOLSymConst{=} \HOLConst{label} (\HOLConst{COMPL} \HOLBoundVar{l})) \HOLSymConst{\HOLTokenConj{}} (\HOLBoundVar{u} \HOLSymConst{=} \HOLConst{\ensuremath{\tau}}) \HOLSymConst{\HOLTokenConj{}}
       (\HOLBoundVar{E} \HOLSymConst{=} \HOLConst{PREF_PROC} (\HOLBoundVar{f} \HOLBoundVar{k}) \HOLSymConst{\ensuremath{\parallel}} \HOLConst{PREF_PROC} (\HOLBoundVar{f\sp{\prime}} \HOLBoundVar{k\sp{\prime}}))
\end{SaveVerbatim}
\newcommand{\HOLStrongLawsTheoremsALLXXSYNCXXTRANSXXTHMXXEQ}{\UseVerbatim{HOLStrongLawsTheoremsALLXXSYNCXXTRANSXXTHMXXEQ}}
\begin{SaveVerbatim}{HOLStrongLawsTheoremsCCSXXCOMPXXdefXXcompute}
\HOLTokenTurnstile{} (\HOLSymConst{\HOLTokenForall{}}\HOLBoundVar{f}. \HOLConst{PI} \HOLBoundVar{f} \HOLNumLit{0} \HOLSymConst{=} \HOLBoundVar{f} \HOLNumLit{0}) \HOLSymConst{\HOLTokenConj{}}
   (\HOLSymConst{\HOLTokenForall{}}\HOLBoundVar{f} \HOLBoundVar{n}.
      \HOLConst{PI} \HOLBoundVar{f} (\HOLConst{NUMERAL} (\HOLConst{BIT1} \HOLBoundVar{n})) \HOLSymConst{=}
      \HOLConst{PI} \HOLBoundVar{f} (\HOLConst{NUMERAL} (\HOLConst{BIT1} \HOLBoundVar{n}) \HOLSymConst{-} \HOLNumLit{1}) \HOLSymConst{\ensuremath{\parallel}} \HOLBoundVar{f} (\HOLConst{NUMERAL} (\HOLConst{BIT1} \HOLBoundVar{n}))) \HOLSymConst{\HOLTokenConj{}}
   \HOLSymConst{\HOLTokenForall{}}\HOLBoundVar{f} \HOLBoundVar{n}.
     \HOLConst{PI} \HOLBoundVar{f} (\HOLConst{NUMERAL} (\HOLConst{BIT2} \HOLBoundVar{n})) \HOLSymConst{=}
     \HOLConst{PI} \HOLBoundVar{f} (\HOLConst{NUMERAL} (\HOLConst{BIT1} \HOLBoundVar{n})) \HOLSymConst{\ensuremath{\parallel}} \HOLBoundVar{f} (\HOLConst{NUMERAL} (\HOLConst{BIT2} \HOLBoundVar{n}))
\end{SaveVerbatim}
\newcommand{\HOLStrongLawsTheoremsCCSXXCOMPXXdefXXcompute}{\UseVerbatim{HOLStrongLawsTheoremsCCSXXCOMPXXdefXXcompute}}
\begin{SaveVerbatim}{HOLStrongLawsTheoremsCCSXXSIGMAXXdefXXcompute}
\HOLTokenTurnstile{} (\HOLSymConst{\HOLTokenForall{}}\HOLBoundVar{f}. \HOLConst{SIGMA} \HOLBoundVar{f} \HOLNumLit{0} \HOLSymConst{=} \HOLBoundVar{f} \HOLNumLit{0}) \HOLSymConst{\HOLTokenConj{}}
   (\HOLSymConst{\HOLTokenForall{}}\HOLBoundVar{f} \HOLBoundVar{n}.
      \HOLConst{SIGMA} \HOLBoundVar{f} (\HOLConst{NUMERAL} (\HOLConst{BIT1} \HOLBoundVar{n})) \HOLSymConst{=}
      \HOLConst{SIGMA} \HOLBoundVar{f} (\HOLConst{NUMERAL} (\HOLConst{BIT1} \HOLBoundVar{n}) \HOLSymConst{-} \HOLNumLit{1}) \HOLSymConst{+} \HOLBoundVar{f} (\HOLConst{NUMERAL} (\HOLConst{BIT1} \HOLBoundVar{n}))) \HOLSymConst{\HOLTokenConj{}}
   \HOLSymConst{\HOLTokenForall{}}\HOLBoundVar{f} \HOLBoundVar{n}.
     \HOLConst{SIGMA} \HOLBoundVar{f} (\HOLConst{NUMERAL} (\HOLConst{BIT2} \HOLBoundVar{n})) \HOLSymConst{=}
     \HOLConst{SIGMA} \HOLBoundVar{f} (\HOLConst{NUMERAL} (\HOLConst{BIT1} \HOLBoundVar{n})) \HOLSymConst{+} \HOLBoundVar{f} (\HOLConst{NUMERAL} (\HOLConst{BIT2} \HOLBoundVar{n}))
\end{SaveVerbatim}
\newcommand{\HOLStrongLawsTheoremsCCSXXSIGMAXXdefXXcompute}{\UseVerbatim{HOLStrongLawsTheoremsCCSXXSIGMAXXdefXXcompute}}
\begin{SaveVerbatim}{HOLStrongLawsTheoremsCOMPXXBASE}
\HOLTokenTurnstile{} \HOLSymConst{\HOLTokenForall{}}\HOLBoundVar{f}. \HOLConst{PI} \HOLBoundVar{f} \HOLNumLit{0} \HOLSymConst{=} \HOLBoundVar{f} \HOLNumLit{0}
\end{SaveVerbatim}
\newcommand{\HOLStrongLawsTheoremsCOMPXXBASE}{\UseVerbatim{HOLStrongLawsTheoremsCOMPXXBASE}}
\begin{SaveVerbatim}{HOLStrongLawsTheoremsCOMPXXINDUCT}
\HOLTokenTurnstile{} \HOLSymConst{\HOLTokenForall{}}\HOLBoundVar{f} \HOLBoundVar{n}. \HOLConst{PI} \HOLBoundVar{f} (\HOLConst{SUC} \HOLBoundVar{n}) \HOLSymConst{=} \HOLConst{PI} \HOLBoundVar{f} \HOLBoundVar{n} \HOLSymConst{\ensuremath{\parallel}} \HOLBoundVar{f} (\HOLConst{SUC} \HOLBoundVar{n})
\end{SaveVerbatim}
\newcommand{\HOLStrongLawsTheoremsCOMPXXINDUCT}{\UseVerbatim{HOLStrongLawsTheoremsCOMPXXINDUCT}}
\begin{SaveVerbatim}{HOLStrongLawsTheoremsPREFXXISXXPREFIX}
\HOLTokenTurnstile{} \HOLSymConst{\HOLTokenForall{}}\HOLBoundVar{u} \HOLBoundVar{E}. \HOLConst{Is_Prefix} (\HOLBoundVar{u}\HOLSymConst{..}\HOLBoundVar{E})
\end{SaveVerbatim}
\newcommand{\HOLStrongLawsTheoremsPREFXXISXXPREFIX}{\UseVerbatim{HOLStrongLawsTheoremsPREFXXISXXPREFIX}}
\begin{SaveVerbatim}{HOLStrongLawsTheoremsSIGMAXXBASE}
\HOLTokenTurnstile{} \HOLSymConst{\HOLTokenForall{}}\HOLBoundVar{f}. \HOLConst{SIGMA} \HOLBoundVar{f} \HOLNumLit{0} \HOLSymConst{=} \HOLBoundVar{f} \HOLNumLit{0}
\end{SaveVerbatim}
\newcommand{\HOLStrongLawsTheoremsSIGMAXXBASE}{\UseVerbatim{HOLStrongLawsTheoremsSIGMAXXBASE}}
\begin{SaveVerbatim}{HOLStrongLawsTheoremsSIGMAXXINDUCT}
\HOLTokenTurnstile{} \HOLSymConst{\HOLTokenForall{}}\HOLBoundVar{f} \HOLBoundVar{n}. \HOLConst{SIGMA} \HOLBoundVar{f} (\HOLConst{SUC} \HOLBoundVar{n}) \HOLSymConst{=} \HOLConst{SIGMA} \HOLBoundVar{f} \HOLBoundVar{n} \HOLSymConst{+} \HOLBoundVar{f} (\HOLConst{SUC} \HOLBoundVar{n})
\end{SaveVerbatim}
\newcommand{\HOLStrongLawsTheoremsSIGMAXXINDUCT}{\UseVerbatim{HOLStrongLawsTheoremsSIGMAXXINDUCT}}
\begin{SaveVerbatim}{HOLStrongLawsTheoremsSIGMAXXTRANSXXTHM}
\HOLTokenTurnstile{} \HOLSymConst{\HOLTokenForall{}}\HOLBoundVar{n} \HOLBoundVar{f} \HOLBoundVar{u} \HOLBoundVar{E}. \HOLConst{SIGMA} \HOLBoundVar{f} \HOLBoundVar{n} \HOLTokenTransBegin\HOLBoundVar{u}\HOLTokenTransEnd \HOLBoundVar{E} \HOLSymConst{\HOLTokenImp{}} \HOLSymConst{\HOLTokenExists{}}\HOLBoundVar{k}. \HOLBoundVar{k} \HOLSymConst{\HOLTokenLeq{}} \HOLBoundVar{n} \HOLSymConst{\HOLTokenConj{}} \HOLBoundVar{f} \HOLBoundVar{k} \HOLTokenTransBegin\HOLBoundVar{u}\HOLTokenTransEnd \HOLBoundVar{E}
\end{SaveVerbatim}
\newcommand{\HOLStrongLawsTheoremsSIGMAXXTRANSXXTHM}{\UseVerbatim{HOLStrongLawsTheoremsSIGMAXXTRANSXXTHM}}
\begin{SaveVerbatim}{HOLStrongLawsTheoremsSIGMAXXTRANSXXTHMXXEQ}
\HOLTokenTurnstile{} \HOLSymConst{\HOLTokenForall{}}\HOLBoundVar{n} \HOLBoundVar{f} \HOLBoundVar{u} \HOLBoundVar{E}. \HOLConst{SIGMA} \HOLBoundVar{f} \HOLBoundVar{n} \HOLTokenTransBegin\HOLBoundVar{u}\HOLTokenTransEnd \HOLBoundVar{E} \HOLSymConst{\HOLTokenEquiv{}} \HOLSymConst{\HOLTokenExists{}}\HOLBoundVar{k}. \HOLBoundVar{k} \HOLSymConst{\HOLTokenLeq{}} \HOLBoundVar{n} \HOLSymConst{\HOLTokenConj{}} \HOLBoundVar{f} \HOLBoundVar{k} \HOLTokenTransBegin\HOLBoundVar{u}\HOLTokenTransEnd \HOLBoundVar{E}
\end{SaveVerbatim}
\newcommand{\HOLStrongLawsTheoremsSIGMAXXTRANSXXTHMXXEQ}{\UseVerbatim{HOLStrongLawsTheoremsSIGMAXXTRANSXXTHMXXEQ}}
\begin{SaveVerbatim}{HOLStrongLawsTheoremsSTRONGXXEXPANSIONXXLAW}
\HOLTokenTurnstile{} \HOLSymConst{\HOLTokenForall{}}\HOLBoundVar{f} \HOLBoundVar{n} \HOLBoundVar{f\sp{\prime}} \HOLBoundVar{m}.
     (\HOLSymConst{\HOLTokenForall{}}\HOLBoundVar{i}. \HOLBoundVar{i} \HOLSymConst{\HOLTokenLeq{}} \HOLBoundVar{n} \HOLSymConst{\HOLTokenImp{}} \HOLConst{Is_Prefix} (\HOLBoundVar{f} \HOLBoundVar{i})) \HOLSymConst{\HOLTokenConj{}}
     (\HOLSymConst{\HOLTokenForall{}}\HOLBoundVar{j}. \HOLBoundVar{j} \HOLSymConst{\HOLTokenLeq{}} \HOLBoundVar{m} \HOLSymConst{\HOLTokenImp{}} \HOLConst{Is_Prefix} (\HOLBoundVar{f\sp{\prime}} \HOLBoundVar{j})) \HOLSymConst{\HOLTokenImp{}}
     \HOLConst{STRONG_EQUIV} (\HOLConst{SIGMA} \HOLBoundVar{f} \HOLBoundVar{n} \HOLSymConst{\ensuremath{\parallel}} \HOLConst{SIGMA} \HOLBoundVar{f\sp{\prime}} \HOLBoundVar{m})
       (\HOLConst{SIGMA}
          (\HOLTokenLambda{}\HOLBoundVar{i}. \HOLConst{PREF_ACT} (\HOLBoundVar{f} \HOLBoundVar{i})\HOLSymConst{..}(\HOLConst{PREF_PROC} (\HOLBoundVar{f} \HOLBoundVar{i}) \HOLSymConst{\ensuremath{\parallel}} \HOLConst{SIGMA} \HOLBoundVar{f\sp{\prime}} \HOLBoundVar{m}))
          \HOLBoundVar{n} \HOLSymConst{+}
        \HOLConst{SIGMA}
          (\HOLTokenLambda{}\HOLBoundVar{j}. \HOLConst{PREF_ACT} (\HOLBoundVar{f\sp{\prime}} \HOLBoundVar{j})\HOLSymConst{..}(\HOLConst{SIGMA} \HOLBoundVar{f} \HOLBoundVar{n} \HOLSymConst{\ensuremath{\parallel}} \HOLConst{PREF_PROC} (\HOLBoundVar{f\sp{\prime}} \HOLBoundVar{j})))
          \HOLBoundVar{m} \HOLSymConst{+} \HOLConst{ALL_SYNC} \HOLBoundVar{f} \HOLBoundVar{n} \HOLBoundVar{f\sp{\prime}} \HOLBoundVar{m})
\end{SaveVerbatim}
\newcommand{\HOLStrongLawsTheoremsSTRONGXXEXPANSIONXXLAW}{\UseVerbatim{HOLStrongLawsTheoremsSTRONGXXEXPANSIONXXLAW}}
\begin{SaveVerbatim}{HOLStrongLawsTheoremsSTRONGXXLEFTXXSUMXXMIDXXIDEMP}
\HOLTokenTurnstile{} \HOLSymConst{\HOLTokenForall{}}\HOLBoundVar{E} \HOLBoundVar{E\sp{\prime}} \HOLBoundVar{E\sp{\prime\prime}}. \HOLConst{STRONG_EQUIV} (\HOLBoundVar{E} \HOLSymConst{+} \HOLBoundVar{E\sp{\prime}} \HOLSymConst{+} \HOLBoundVar{E\sp{\prime\prime}} \HOLSymConst{+} \HOLBoundVar{E\sp{\prime}}) (\HOLBoundVar{E} \HOLSymConst{+} \HOLBoundVar{E\sp{\prime\prime}} \HOLSymConst{+} \HOLBoundVar{E\sp{\prime}})
\end{SaveVerbatim}
\newcommand{\HOLStrongLawsTheoremsSTRONGXXLEFTXXSUMXXMIDXXIDEMP}{\UseVerbatim{HOLStrongLawsTheoremsSTRONGXXLEFTXXSUMXXMIDXXIDEMP}}
\begin{SaveVerbatim}{HOLStrongLawsTheoremsSTRONGXXPARXXASSOC}
\HOLTokenTurnstile{} \HOLSymConst{\HOLTokenForall{}}\HOLBoundVar{E} \HOLBoundVar{E\sp{\prime}} \HOLBoundVar{E\sp{\prime\prime}}. \HOLConst{STRONG_EQUIV} (\HOLBoundVar{E} \HOLSymConst{\ensuremath{\parallel}} \HOLBoundVar{E\sp{\prime}} \HOLSymConst{\ensuremath{\parallel}} \HOLBoundVar{E\sp{\prime\prime}}) (\HOLBoundVar{E} \HOLSymConst{\ensuremath{\parallel}} (\HOLBoundVar{E\sp{\prime}} \HOLSymConst{\ensuremath{\parallel}} \HOLBoundVar{E\sp{\prime\prime}}))
\end{SaveVerbatim}
\newcommand{\HOLStrongLawsTheoremsSTRONGXXPARXXASSOC}{\UseVerbatim{HOLStrongLawsTheoremsSTRONGXXPARXXASSOC}}
\begin{SaveVerbatim}{HOLStrongLawsTheoremsSTRONGXXPARXXCOMM}
\HOLTokenTurnstile{} \HOLSymConst{\HOLTokenForall{}}\HOLBoundVar{E} \HOLBoundVar{E\sp{\prime}}. \HOLConst{STRONG_EQUIV} (\HOLBoundVar{E} \HOLSymConst{\ensuremath{\parallel}} \HOLBoundVar{E\sp{\prime}}) (\HOLBoundVar{E\sp{\prime}} \HOLSymConst{\ensuremath{\parallel}} \HOLBoundVar{E})
\end{SaveVerbatim}
\newcommand{\HOLStrongLawsTheoremsSTRONGXXPARXXCOMM}{\UseVerbatim{HOLStrongLawsTheoremsSTRONGXXPARXXCOMM}}
\begin{SaveVerbatim}{HOLStrongLawsTheoremsSTRONGXXPARXXIDENTXXL}
\HOLTokenTurnstile{} \HOLSymConst{\HOLTokenForall{}}\HOLBoundVar{E}. \HOLConst{STRONG_EQUIV} (\HOLConst{nil} \HOLSymConst{\ensuremath{\parallel}} \HOLBoundVar{E}) \HOLBoundVar{E}
\end{SaveVerbatim}
\newcommand{\HOLStrongLawsTheoremsSTRONGXXPARXXIDENTXXL}{\UseVerbatim{HOLStrongLawsTheoremsSTRONGXXPARXXIDENTXXL}}
\begin{SaveVerbatim}{HOLStrongLawsTheoremsSTRONGXXPARXXIDENTXXR}
\HOLTokenTurnstile{} \HOLSymConst{\HOLTokenForall{}}\HOLBoundVar{E}. \HOLConst{STRONG_EQUIV} (\HOLBoundVar{E} \HOLSymConst{\ensuremath{\parallel}} \HOLConst{nil}) \HOLBoundVar{E}
\end{SaveVerbatim}
\newcommand{\HOLStrongLawsTheoremsSTRONGXXPARXXIDENTXXR}{\UseVerbatim{HOLStrongLawsTheoremsSTRONGXXPARXXIDENTXXR}}
\begin{SaveVerbatim}{HOLStrongLawsTheoremsSTRONGXXPARXXPREFXXNOXXSYNCR}
\HOLTokenTurnstile{} \HOLSymConst{\HOLTokenForall{}}\HOLBoundVar{l} \HOLBoundVar{l\sp{\prime}}.
     \HOLBoundVar{l} \HOLSymConst{\HOLTokenNotEqual{}} \HOLConst{COMPL} \HOLBoundVar{l\sp{\prime}} \HOLSymConst{\HOLTokenImp{}}
     \HOLSymConst{\HOLTokenForall{}}\HOLBoundVar{E} \HOLBoundVar{E\sp{\prime}}.
       \HOLConst{STRONG_EQUIV} (\HOLConst{label} \HOLBoundVar{l}\HOLSymConst{..}\HOLBoundVar{E} \HOLSymConst{\ensuremath{\parallel}} \HOLConst{label} \HOLBoundVar{l\sp{\prime}}\HOLSymConst{..}\HOLBoundVar{E\sp{\prime}})
         (\HOLConst{label} \HOLBoundVar{l}\HOLSymConst{..}(\HOLBoundVar{E} \HOLSymConst{\ensuremath{\parallel}} \HOLConst{label} \HOLBoundVar{l\sp{\prime}}\HOLSymConst{..}\HOLBoundVar{E\sp{\prime}}) \HOLSymConst{+}
          \HOLConst{label} \HOLBoundVar{l\sp{\prime}}\HOLSymConst{..}(\HOLConst{label} \HOLBoundVar{l}\HOLSymConst{..}\HOLBoundVar{E} \HOLSymConst{\ensuremath{\parallel}} \HOLBoundVar{E\sp{\prime}}))
\end{SaveVerbatim}
\newcommand{\HOLStrongLawsTheoremsSTRONGXXPARXXPREFXXNOXXSYNCR}{\UseVerbatim{HOLStrongLawsTheoremsSTRONGXXPARXXPREFXXNOXXSYNCR}}
\begin{SaveVerbatim}{HOLStrongLawsTheoremsSTRONGXXPARXXPREFXXSYNCR}
\HOLTokenTurnstile{} \HOLSymConst{\HOLTokenForall{}}\HOLBoundVar{l} \HOLBoundVar{l\sp{\prime}}.
     (\HOLBoundVar{l} \HOLSymConst{=} \HOLConst{COMPL} \HOLBoundVar{l\sp{\prime}}) \HOLSymConst{\HOLTokenImp{}}
     \HOLSymConst{\HOLTokenForall{}}\HOLBoundVar{E} \HOLBoundVar{E\sp{\prime}}.
       \HOLConst{STRONG_EQUIV} (\HOLConst{label} \HOLBoundVar{l}\HOLSymConst{..}\HOLBoundVar{E} \HOLSymConst{\ensuremath{\parallel}} \HOLConst{label} \HOLBoundVar{l\sp{\prime}}\HOLSymConst{..}\HOLBoundVar{E\sp{\prime}})
         (\HOLConst{label} \HOLBoundVar{l}\HOLSymConst{..}(\HOLBoundVar{E} \HOLSymConst{\ensuremath{\parallel}} \HOLConst{label} \HOLBoundVar{l\sp{\prime}}\HOLSymConst{..}\HOLBoundVar{E\sp{\prime}}) \HOLSymConst{+}
          \HOLConst{label} \HOLBoundVar{l\sp{\prime}}\HOLSymConst{..}(\HOLConst{label} \HOLBoundVar{l}\HOLSymConst{..}\HOLBoundVar{E} \HOLSymConst{\ensuremath{\parallel}} \HOLBoundVar{E\sp{\prime}}) \HOLSymConst{+} \HOLConst{\ensuremath{\tau}}\HOLSymConst{..}(\HOLBoundVar{E} \HOLSymConst{\ensuremath{\parallel}} \HOLBoundVar{E\sp{\prime}}))
\end{SaveVerbatim}
\newcommand{\HOLStrongLawsTheoremsSTRONGXXPARXXPREFXXSYNCR}{\UseVerbatim{HOLStrongLawsTheoremsSTRONGXXPARXXPREFXXSYNCR}}
\begin{SaveVerbatim}{HOLStrongLawsTheoremsSTRONGXXPARXXPREFXXTAU}
\HOLTokenTurnstile{} \HOLSymConst{\HOLTokenForall{}}\HOLBoundVar{u} \HOLBoundVar{E} \HOLBoundVar{E\sp{\prime}}.
     \HOLConst{STRONG_EQUIV} (\HOLBoundVar{u}\HOLSymConst{..}\HOLBoundVar{E} \HOLSymConst{\ensuremath{\parallel}} \HOLConst{\ensuremath{\tau}}\HOLSymConst{..}\HOLBoundVar{E\sp{\prime}})
       (\HOLBoundVar{u}\HOLSymConst{..}(\HOLBoundVar{E} \HOLSymConst{\ensuremath{\parallel}} \HOLConst{\ensuremath{\tau}}\HOLSymConst{..}\HOLBoundVar{E\sp{\prime}}) \HOLSymConst{+} \HOLConst{\ensuremath{\tau}}\HOLSymConst{..}(\HOLBoundVar{u}\HOLSymConst{..}\HOLBoundVar{E} \HOLSymConst{\ensuremath{\parallel}} \HOLBoundVar{E\sp{\prime}}))
\end{SaveVerbatim}
\newcommand{\HOLStrongLawsTheoremsSTRONGXXPARXXPREFXXTAU}{\UseVerbatim{HOLStrongLawsTheoremsSTRONGXXPARXXPREFXXTAU}}
\begin{SaveVerbatim}{HOLStrongLawsTheoremsSTRONGXXPARXXTAUXXPREF}
\HOLTokenTurnstile{} \HOLSymConst{\HOLTokenForall{}}\HOLBoundVar{E} \HOLBoundVar{u} \HOLBoundVar{E\sp{\prime}}.
     \HOLConst{STRONG_EQUIV} (\HOLConst{\ensuremath{\tau}}\HOLSymConst{..}\HOLBoundVar{E} \HOLSymConst{\ensuremath{\parallel}} \HOLBoundVar{u}\HOLSymConst{..}\HOLBoundVar{E\sp{\prime}})
       (\HOLConst{\ensuremath{\tau}}\HOLSymConst{..}(\HOLBoundVar{E} \HOLSymConst{\ensuremath{\parallel}} \HOLBoundVar{u}\HOLSymConst{..}\HOLBoundVar{E\sp{\prime}}) \HOLSymConst{+} \HOLBoundVar{u}\HOLSymConst{..}(\HOLConst{\ensuremath{\tau}}\HOLSymConst{..}\HOLBoundVar{E} \HOLSymConst{\ensuremath{\parallel}} \HOLBoundVar{E\sp{\prime}}))
\end{SaveVerbatim}
\newcommand{\HOLStrongLawsTheoremsSTRONGXXPARXXTAUXXPREF}{\UseVerbatim{HOLStrongLawsTheoremsSTRONGXXPARXXTAUXXPREF}}
\begin{SaveVerbatim}{HOLStrongLawsTheoremsSTRONGXXPARXXTAUXXTAU}
\HOLTokenTurnstile{} \HOLSymConst{\HOLTokenForall{}}\HOLBoundVar{E} \HOLBoundVar{E\sp{\prime}}.
     \HOLConst{STRONG_EQUIV} (\HOLConst{\ensuremath{\tau}}\HOLSymConst{..}\HOLBoundVar{E} \HOLSymConst{\ensuremath{\parallel}} \HOLConst{\ensuremath{\tau}}\HOLSymConst{..}\HOLBoundVar{E\sp{\prime}})
       (\HOLConst{\ensuremath{\tau}}\HOLSymConst{..}(\HOLBoundVar{E} \HOLSymConst{\ensuremath{\parallel}} \HOLConst{\ensuremath{\tau}}\HOLSymConst{..}\HOLBoundVar{E\sp{\prime}}) \HOLSymConst{+} \HOLConst{\ensuremath{\tau}}\HOLSymConst{..}(\HOLConst{\ensuremath{\tau}}\HOLSymConst{..}\HOLBoundVar{E} \HOLSymConst{\ensuremath{\parallel}} \HOLBoundVar{E\sp{\prime}}))
\end{SaveVerbatim}
\newcommand{\HOLStrongLawsTheoremsSTRONGXXPARXXTAUXXTAU}{\UseVerbatim{HOLStrongLawsTheoremsSTRONGXXPARXXTAUXXTAU}}
\begin{SaveVerbatim}{HOLStrongLawsTheoremsSTRONGXXPREFXXRECXXEQUIV}
\HOLTokenTurnstile{} \HOLSymConst{\HOLTokenForall{}}\HOLBoundVar{u} \HOLBoundVar{s} \HOLBoundVar{v}.
     \HOLConst{STRONG_EQUIV} (\HOLBoundVar{u}\HOLSymConst{..}\HOLConst{rec} \HOLBoundVar{s} (\HOLBoundVar{v}\HOLSymConst{..}\HOLBoundVar{u}\HOLSymConst{..}\HOLConst{var} \HOLBoundVar{s})) (\HOLConst{rec} \HOLBoundVar{s} (\HOLBoundVar{u}\HOLSymConst{..}\HOLBoundVar{v}\HOLSymConst{..}\HOLConst{var} \HOLBoundVar{s}))
\end{SaveVerbatim}
\newcommand{\HOLStrongLawsTheoremsSTRONGXXPREFXXRECXXEQUIV}{\UseVerbatim{HOLStrongLawsTheoremsSTRONGXXPREFXXRECXXEQUIV}}
\begin{SaveVerbatim}{HOLStrongLawsTheoremsSTRONGXXRECXXACTTwo}
\HOLTokenTurnstile{} \HOLSymConst{\HOLTokenForall{}}\HOLBoundVar{s} \HOLBoundVar{u}. \HOLConst{STRONG_EQUIV} (\HOLConst{rec} \HOLBoundVar{s} (\HOLBoundVar{u}\HOLSymConst{..}\HOLBoundVar{u}\HOLSymConst{..}\HOLConst{var} \HOLBoundVar{s})) (\HOLConst{rec} \HOLBoundVar{s} (\HOLBoundVar{u}\HOLSymConst{..}\HOLConst{var} \HOLBoundVar{s}))
\end{SaveVerbatim}
\newcommand{\HOLStrongLawsTheoremsSTRONGXXRECXXACTTwo}{\UseVerbatim{HOLStrongLawsTheoremsSTRONGXXRECXXACTTwo}}
\begin{SaveVerbatim}{HOLStrongLawsTheoremsSTRONGXXRELABXXNIL}
\HOLTokenTurnstile{} \HOLSymConst{\HOLTokenForall{}}\HOLBoundVar{rf}. \HOLConst{STRONG_EQUIV} (\HOLConst{relab} \HOLConst{nil} \HOLBoundVar{rf}) \HOLConst{nil}
\end{SaveVerbatim}
\newcommand{\HOLStrongLawsTheoremsSTRONGXXRELABXXNIL}{\UseVerbatim{HOLStrongLawsTheoremsSTRONGXXRELABXXNIL}}
\begin{SaveVerbatim}{HOLStrongLawsTheoremsSTRONGXXRELABXXPREFIX}
\HOLTokenTurnstile{} \HOLSymConst{\HOLTokenForall{}}\HOLBoundVar{u} \HOLBoundVar{E} \HOLBoundVar{labl}.
     \HOLConst{STRONG_EQUIV} (\HOLConst{relab} (\HOLBoundVar{u}\HOLSymConst{..}\HOLBoundVar{E}) (\HOLConst{RELAB} \HOLBoundVar{labl}))
       (\HOLConst{relabel} (\HOLConst{RELAB} \HOLBoundVar{labl}) \HOLBoundVar{u}\HOLSymConst{..}\HOLConst{relab} \HOLBoundVar{E} (\HOLConst{RELAB} \HOLBoundVar{labl}))
\end{SaveVerbatim}
\newcommand{\HOLStrongLawsTheoremsSTRONGXXRELABXXPREFIX}{\UseVerbatim{HOLStrongLawsTheoremsSTRONGXXRELABXXPREFIX}}
\begin{SaveVerbatim}{HOLStrongLawsTheoremsSTRONGXXRELABXXSUM}
\HOLTokenTurnstile{} \HOLSymConst{\HOLTokenForall{}}\HOLBoundVar{E} \HOLBoundVar{E\sp{\prime}} \HOLBoundVar{rf}.
     \HOLConst{STRONG_EQUIV} (\HOLConst{relab} (\HOLBoundVar{E} \HOLSymConst{+} \HOLBoundVar{E\sp{\prime}}) \HOLBoundVar{rf}) (\HOLConst{relab} \HOLBoundVar{E} \HOLBoundVar{rf} \HOLSymConst{+} \HOLConst{relab} \HOLBoundVar{E\sp{\prime}} \HOLBoundVar{rf})
\end{SaveVerbatim}
\newcommand{\HOLStrongLawsTheoremsSTRONGXXRELABXXSUM}{\UseVerbatim{HOLStrongLawsTheoremsSTRONGXXRELABXXSUM}}
\begin{SaveVerbatim}{HOLStrongLawsTheoremsSTRONGXXRESTRXXNIL}
\HOLTokenTurnstile{} \HOLSymConst{\HOLTokenForall{}}\HOLBoundVar{L}. \HOLConst{STRONG_EQUIV} (\HOLConst{\ensuremath{\nu}} \HOLBoundVar{L} \HOLConst{nil}) \HOLConst{nil}
\end{SaveVerbatim}
\newcommand{\HOLStrongLawsTheoremsSTRONGXXRESTRXXNIL}{\UseVerbatim{HOLStrongLawsTheoremsSTRONGXXRESTRXXNIL}}
\begin{SaveVerbatim}{HOLStrongLawsTheoremsSTRONGXXRESTRXXPRXXLABXXNIL}
\HOLTokenTurnstile{} \HOLSymConst{\HOLTokenForall{}}\HOLBoundVar{l} \HOLBoundVar{L}.
     \HOLBoundVar{l} \HOLConst{\HOLTokenIn{}} \HOLBoundVar{L} \HOLSymConst{\HOLTokenDisj{}} \HOLConst{COMPL} \HOLBoundVar{l} \HOLConst{\HOLTokenIn{}} \HOLBoundVar{L} \HOLSymConst{\HOLTokenImp{}}
     \HOLSymConst{\HOLTokenForall{}}\HOLBoundVar{E}. \HOLConst{STRONG_EQUIV} (\HOLConst{\ensuremath{\nu}} \HOLBoundVar{L} (\HOLConst{label} \HOLBoundVar{l}\HOLSymConst{..}\HOLBoundVar{E})) \HOLConst{nil}
\end{SaveVerbatim}
\newcommand{\HOLStrongLawsTheoremsSTRONGXXRESTRXXPRXXLABXXNIL}{\UseVerbatim{HOLStrongLawsTheoremsSTRONGXXRESTRXXPRXXLABXXNIL}}
\begin{SaveVerbatim}{HOLStrongLawsTheoremsSTRONGXXRESTRXXPREFIXXXLABEL}
\HOLTokenTurnstile{} \HOLSymConst{\HOLTokenForall{}}\HOLBoundVar{l} \HOLBoundVar{L}.
     \HOLBoundVar{l} \HOLConst{\HOLTokenNotIn{}} \HOLBoundVar{L} \HOLSymConst{\HOLTokenConj{}} \HOLConst{COMPL} \HOLBoundVar{l} \HOLConst{\HOLTokenNotIn{}} \HOLBoundVar{L} \HOLSymConst{\HOLTokenImp{}}
     \HOLSymConst{\HOLTokenForall{}}\HOLBoundVar{E}. \HOLConst{STRONG_EQUIV} (\HOLConst{\ensuremath{\nu}} \HOLBoundVar{L} (\HOLConst{label} \HOLBoundVar{l}\HOLSymConst{..}\HOLBoundVar{E})) (\HOLConst{label} \HOLBoundVar{l}\HOLSymConst{..}\HOLConst{\ensuremath{\nu}} \HOLBoundVar{L} \HOLBoundVar{E})
\end{SaveVerbatim}
\newcommand{\HOLStrongLawsTheoremsSTRONGXXRESTRXXPREFIXXXLABEL}{\UseVerbatim{HOLStrongLawsTheoremsSTRONGXXRESTRXXPREFIXXXLABEL}}
\begin{SaveVerbatim}{HOLStrongLawsTheoremsSTRONGXXRESTRXXPREFIXXXTAU}
\HOLTokenTurnstile{} \HOLSymConst{\HOLTokenForall{}}\HOLBoundVar{E} \HOLBoundVar{L}. \HOLConst{STRONG_EQUIV} (\HOLConst{\ensuremath{\nu}} \HOLBoundVar{L} (\HOLConst{\ensuremath{\tau}}\HOLSymConst{..}\HOLBoundVar{E})) (\HOLConst{\ensuremath{\tau}}\HOLSymConst{..}\HOLConst{\ensuremath{\nu}} \HOLBoundVar{L} \HOLBoundVar{E})
\end{SaveVerbatim}
\newcommand{\HOLStrongLawsTheoremsSTRONGXXRESTRXXPREFIXXXTAU}{\UseVerbatim{HOLStrongLawsTheoremsSTRONGXXRESTRXXPREFIXXXTAU}}
\begin{SaveVerbatim}{HOLStrongLawsTheoremsSTRONGXXRESTRXXSUM}
\HOLTokenTurnstile{} \HOLSymConst{\HOLTokenForall{}}\HOLBoundVar{E} \HOLBoundVar{E\sp{\prime}} \HOLBoundVar{L}. \HOLConst{STRONG_EQUIV} (\HOLConst{\ensuremath{\nu}} \HOLBoundVar{L} (\HOLBoundVar{E} \HOLSymConst{+} \HOLBoundVar{E\sp{\prime}})) (\HOLConst{\ensuremath{\nu}} \HOLBoundVar{L} \HOLBoundVar{E} \HOLSymConst{+} \HOLConst{\ensuremath{\nu}} \HOLBoundVar{L} \HOLBoundVar{E\sp{\prime}})
\end{SaveVerbatim}
\newcommand{\HOLStrongLawsTheoremsSTRONGXXRESTRXXSUM}{\UseVerbatim{HOLStrongLawsTheoremsSTRONGXXRESTRXXSUM}}
\begin{SaveVerbatim}{HOLStrongLawsTheoremsSTRONGXXSUMXXASSOCXXL}
\HOLTokenTurnstile{} \HOLSymConst{\HOLTokenForall{}}\HOLBoundVar{E} \HOLBoundVar{E\sp{\prime}} \HOLBoundVar{E\sp{\prime\prime}}. \HOLConst{STRONG_EQUIV} (\HOLBoundVar{E} \HOLSymConst{+} (\HOLBoundVar{E\sp{\prime}} \HOLSymConst{+} \HOLBoundVar{E\sp{\prime\prime}})) (\HOLBoundVar{E} \HOLSymConst{+} \HOLBoundVar{E\sp{\prime}} \HOLSymConst{+} \HOLBoundVar{E\sp{\prime\prime}})
\end{SaveVerbatim}
\newcommand{\HOLStrongLawsTheoremsSTRONGXXSUMXXASSOCXXL}{\UseVerbatim{HOLStrongLawsTheoremsSTRONGXXSUMXXASSOCXXL}}
\begin{SaveVerbatim}{HOLStrongLawsTheoremsSTRONGXXSUMXXASSOCXXR}
\HOLTokenTurnstile{} \HOLSymConst{\HOLTokenForall{}}\HOLBoundVar{E} \HOLBoundVar{E\sp{\prime}} \HOLBoundVar{E\sp{\prime\prime}}. \HOLConst{STRONG_EQUIV} (\HOLBoundVar{E} \HOLSymConst{+} \HOLBoundVar{E\sp{\prime}} \HOLSymConst{+} \HOLBoundVar{E\sp{\prime\prime}}) (\HOLBoundVar{E} \HOLSymConst{+} (\HOLBoundVar{E\sp{\prime}} \HOLSymConst{+} \HOLBoundVar{E\sp{\prime\prime}}))
\end{SaveVerbatim}
\newcommand{\HOLStrongLawsTheoremsSTRONGXXSUMXXASSOCXXR}{\UseVerbatim{HOLStrongLawsTheoremsSTRONGXXSUMXXASSOCXXR}}
\begin{SaveVerbatim}{HOLStrongLawsTheoremsSTRONGXXSUMXXCOMM}
\HOLTokenTurnstile{} \HOLSymConst{\HOLTokenForall{}}\HOLBoundVar{E} \HOLBoundVar{E\sp{\prime}}. \HOLConst{STRONG_EQUIV} (\HOLBoundVar{E} \HOLSymConst{+} \HOLBoundVar{E\sp{\prime}}) (\HOLBoundVar{E\sp{\prime}} \HOLSymConst{+} \HOLBoundVar{E})
\end{SaveVerbatim}
\newcommand{\HOLStrongLawsTheoremsSTRONGXXSUMXXCOMM}{\UseVerbatim{HOLStrongLawsTheoremsSTRONGXXSUMXXCOMM}}
\begin{SaveVerbatim}{HOLStrongLawsTheoremsSTRONGXXSUMXXIDEMP}
\HOLTokenTurnstile{} \HOLSymConst{\HOLTokenForall{}}\HOLBoundVar{E}. \HOLConst{STRONG_EQUIV} (\HOLBoundVar{E} \HOLSymConst{+} \HOLBoundVar{E}) \HOLBoundVar{E}
\end{SaveVerbatim}
\newcommand{\HOLStrongLawsTheoremsSTRONGXXSUMXXIDEMP}{\UseVerbatim{HOLStrongLawsTheoremsSTRONGXXSUMXXIDEMP}}
\begin{SaveVerbatim}{HOLStrongLawsTheoremsSTRONGXXSUMXXIDENTXXL}
\HOLTokenTurnstile{} \HOLSymConst{\HOLTokenForall{}}\HOLBoundVar{E}. \HOLConst{STRONG_EQUIV} (\HOLConst{nil} \HOLSymConst{+} \HOLBoundVar{E}) \HOLBoundVar{E}
\end{SaveVerbatim}
\newcommand{\HOLStrongLawsTheoremsSTRONGXXSUMXXIDENTXXL}{\UseVerbatim{HOLStrongLawsTheoremsSTRONGXXSUMXXIDENTXXL}}
\begin{SaveVerbatim}{HOLStrongLawsTheoremsSTRONGXXSUMXXIDENTXXR}
\HOLTokenTurnstile{} \HOLSymConst{\HOLTokenForall{}}\HOLBoundVar{E}. \HOLConst{STRONG_EQUIV} (\HOLBoundVar{E} \HOLSymConst{+} \HOLConst{nil}) \HOLBoundVar{E}
\end{SaveVerbatim}
\newcommand{\HOLStrongLawsTheoremsSTRONGXXSUMXXIDENTXXR}{\UseVerbatim{HOLStrongLawsTheoremsSTRONGXXSUMXXIDENTXXR}}
\begin{SaveVerbatim}{HOLStrongLawsTheoremsSTRONGXXSUMXXMIDXXIDEMP}
\HOLTokenTurnstile{} \HOLSymConst{\HOLTokenForall{}}\HOLBoundVar{E} \HOLBoundVar{E\sp{\prime}}. \HOLConst{STRONG_EQUIV} (\HOLBoundVar{E} \HOLSymConst{+} \HOLBoundVar{E\sp{\prime}} \HOLSymConst{+} \HOLBoundVar{E}) (\HOLBoundVar{E\sp{\prime}} \HOLSymConst{+} \HOLBoundVar{E})
\end{SaveVerbatim}
\newcommand{\HOLStrongLawsTheoremsSTRONGXXSUMXXMIDXXIDEMP}{\UseVerbatim{HOLStrongLawsTheoremsSTRONGXXSUMXXMIDXXIDEMP}}
\begin{SaveVerbatim}{HOLStrongLawsTheoremsSTRONGXXUNFOLDING}
\HOLTokenTurnstile{} \HOLSymConst{\HOLTokenForall{}}\HOLBoundVar{X} \HOLBoundVar{E}. \HOLConst{STRONG_EQUIV} (\HOLConst{rec} \HOLBoundVar{X} \HOLBoundVar{E}) (\HOLConst{CCS_Subst} \HOLBoundVar{E} (\HOLConst{rec} \HOLBoundVar{X} \HOLBoundVar{E}) \HOLBoundVar{X})
\end{SaveVerbatim}
\newcommand{\HOLStrongLawsTheoremsSTRONGXXUNFOLDING}{\UseVerbatim{HOLStrongLawsTheoremsSTRONGXXUNFOLDING}}
\begin{SaveVerbatim}{HOLStrongLawsTheoremsSYNCXXBASE}
\HOLTokenTurnstile{} \HOLSymConst{\HOLTokenForall{}}\HOLBoundVar{u} \HOLBoundVar{P} \HOLBoundVar{f}.
     \HOLConst{SYNC} \HOLBoundVar{u} \HOLBoundVar{P} \HOLBoundVar{f} \HOLNumLit{0} \HOLSymConst{=}
     \HOLKeyword{if} (\HOLBoundVar{u} \HOLSymConst{=} \HOLConst{\ensuremath{\tau}}) \HOLSymConst{\HOLTokenDisj{}} (\HOLConst{PREF_ACT} (\HOLBoundVar{f} \HOLNumLit{0}) \HOLSymConst{=} \HOLConst{\ensuremath{\tau}}) \HOLKeyword{then} \HOLConst{nil}
     \HOLKeyword{else} \HOLKeyword{if} \HOLConst{LABEL} \HOLBoundVar{u} \HOLSymConst{=} \HOLConst{COMPL} (\HOLConst{LABEL} (\HOLConst{PREF_ACT} (\HOLBoundVar{f} \HOLNumLit{0}))) \HOLKeyword{then}
       \HOLConst{\ensuremath{\tau}}\HOLSymConst{..}(\HOLBoundVar{P} \HOLSymConst{\ensuremath{\parallel}} \HOLConst{PREF_PROC} (\HOLBoundVar{f} \HOLNumLit{0}))
     \HOLKeyword{else} \HOLConst{nil}
\end{SaveVerbatim}
\newcommand{\HOLStrongLawsTheoremsSYNCXXBASE}{\UseVerbatim{HOLStrongLawsTheoremsSYNCXXBASE}}
\begin{SaveVerbatim}{HOLStrongLawsTheoremsSYNCXXdefXXcompute}
\HOLTokenTurnstile{} (\HOLSymConst{\HOLTokenForall{}}\HOLBoundVar{u} \HOLBoundVar{P} \HOLBoundVar{f}.
      \HOLConst{SYNC} \HOLBoundVar{u} \HOLBoundVar{P} \HOLBoundVar{f} \HOLNumLit{0} \HOLSymConst{=}
      \HOLKeyword{if} (\HOLBoundVar{u} \HOLSymConst{=} \HOLConst{\ensuremath{\tau}}) \HOLSymConst{\HOLTokenDisj{}} (\HOLConst{PREF_ACT} (\HOLBoundVar{f} \HOLNumLit{0}) \HOLSymConst{=} \HOLConst{\ensuremath{\tau}}) \HOLKeyword{then} \HOLConst{nil}
      \HOLKeyword{else} \HOLKeyword{if} \HOLConst{LABEL} \HOLBoundVar{u} \HOLSymConst{=} \HOLConst{COMPL} (\HOLConst{LABEL} (\HOLConst{PREF_ACT} (\HOLBoundVar{f} \HOLNumLit{0}))) \HOLKeyword{then}
        \HOLConst{\ensuremath{\tau}}\HOLSymConst{..}(\HOLBoundVar{P} \HOLSymConst{\ensuremath{\parallel}} \HOLConst{PREF_PROC} (\HOLBoundVar{f} \HOLNumLit{0}))
      \HOLKeyword{else} \HOLConst{nil}) \HOLSymConst{\HOLTokenConj{}}
   (\HOLSymConst{\HOLTokenForall{}}\HOLBoundVar{u} \HOLBoundVar{P} \HOLBoundVar{f} \HOLBoundVar{n}.
      \HOLConst{SYNC} \HOLBoundVar{u} \HOLBoundVar{P} \HOLBoundVar{f} (\HOLConst{NUMERAL} (\HOLConst{BIT1} \HOLBoundVar{n})) \HOLSymConst{=}
      \HOLKeyword{if} (\HOLBoundVar{u} \HOLSymConst{=} \HOLConst{\ensuremath{\tau}}) \HOLSymConst{\HOLTokenDisj{}} (\HOLConst{PREF_ACT} (\HOLBoundVar{f} (\HOLConst{NUMERAL} (\HOLConst{BIT1} \HOLBoundVar{n}))) \HOLSymConst{=} \HOLConst{\ensuremath{\tau}}) \HOLKeyword{then}
        \HOLConst{SYNC} \HOLBoundVar{u} \HOLBoundVar{P} \HOLBoundVar{f} (\HOLConst{NUMERAL} (\HOLConst{BIT1} \HOLBoundVar{n}) \HOLSymConst{-} \HOLNumLit{1})
      \HOLKeyword{else} \HOLKeyword{if}
        \HOLConst{LABEL} \HOLBoundVar{u} \HOLSymConst{=}
        \HOLConst{COMPL} (\HOLConst{LABEL} (\HOLConst{PREF_ACT} (\HOLBoundVar{f} (\HOLConst{NUMERAL} (\HOLConst{BIT1} \HOLBoundVar{n})))))
      \HOLKeyword{then}
        \HOLConst{\ensuremath{\tau}}\HOLSymConst{..}(\HOLBoundVar{P} \HOLSymConst{\ensuremath{\parallel}} \HOLConst{PREF_PROC} (\HOLBoundVar{f} (\HOLConst{NUMERAL} (\HOLConst{BIT1} \HOLBoundVar{n})))) \HOLSymConst{+}
        \HOLConst{SYNC} \HOLBoundVar{u} \HOLBoundVar{P} \HOLBoundVar{f} (\HOLConst{NUMERAL} (\HOLConst{BIT1} \HOLBoundVar{n}) \HOLSymConst{-} \HOLNumLit{1})
      \HOLKeyword{else} \HOLConst{SYNC} \HOLBoundVar{u} \HOLBoundVar{P} \HOLBoundVar{f} (\HOLConst{NUMERAL} (\HOLConst{BIT1} \HOLBoundVar{n}) \HOLSymConst{-} \HOLNumLit{1})) \HOLSymConst{\HOLTokenConj{}}
   \HOLSymConst{\HOLTokenForall{}}\HOLBoundVar{u} \HOLBoundVar{P} \HOLBoundVar{f} \HOLBoundVar{n}.
     \HOLConst{SYNC} \HOLBoundVar{u} \HOLBoundVar{P} \HOLBoundVar{f} (\HOLConst{NUMERAL} (\HOLConst{BIT2} \HOLBoundVar{n})) \HOLSymConst{=}
     \HOLKeyword{if} (\HOLBoundVar{u} \HOLSymConst{=} \HOLConst{\ensuremath{\tau}}) \HOLSymConst{\HOLTokenDisj{}} (\HOLConst{PREF_ACT} (\HOLBoundVar{f} (\HOLConst{NUMERAL} (\HOLConst{BIT2} \HOLBoundVar{n}))) \HOLSymConst{=} \HOLConst{\ensuremath{\tau}}) \HOLKeyword{then}
       \HOLConst{SYNC} \HOLBoundVar{u} \HOLBoundVar{P} \HOLBoundVar{f} (\HOLConst{NUMERAL} (\HOLConst{BIT1} \HOLBoundVar{n}))
     \HOLKeyword{else} \HOLKeyword{if}
       \HOLConst{LABEL} \HOLBoundVar{u} \HOLSymConst{=} \HOLConst{COMPL} (\HOLConst{LABEL} (\HOLConst{PREF_ACT} (\HOLBoundVar{f} (\HOLConst{NUMERAL} (\HOLConst{BIT2} \HOLBoundVar{n})))))
     \HOLKeyword{then}
       \HOLConst{\ensuremath{\tau}}\HOLSymConst{..}(\HOLBoundVar{P} \HOLSymConst{\ensuremath{\parallel}} \HOLConst{PREF_PROC} (\HOLBoundVar{f} (\HOLConst{NUMERAL} (\HOLConst{BIT2} \HOLBoundVar{n})))) \HOLSymConst{+}
       \HOLConst{SYNC} \HOLBoundVar{u} \HOLBoundVar{P} \HOLBoundVar{f} (\HOLConst{NUMERAL} (\HOLConst{BIT1} \HOLBoundVar{n}))
     \HOLKeyword{else} \HOLConst{SYNC} \HOLBoundVar{u} \HOLBoundVar{P} \HOLBoundVar{f} (\HOLConst{NUMERAL} (\HOLConst{BIT1} \HOLBoundVar{n}))
\end{SaveVerbatim}
\newcommand{\HOLStrongLawsTheoremsSYNCXXdefXXcompute}{\UseVerbatim{HOLStrongLawsTheoremsSYNCXXdefXXcompute}}
\begin{SaveVerbatim}{HOLStrongLawsTheoremsSYNCXXINDUCT}
\HOLTokenTurnstile{} \HOLSymConst{\HOLTokenForall{}}\HOLBoundVar{u} \HOLBoundVar{P} \HOLBoundVar{f} \HOLBoundVar{n}.
     \HOLConst{SYNC} \HOLBoundVar{u} \HOLBoundVar{P} \HOLBoundVar{f} (\HOLConst{SUC} \HOLBoundVar{n}) \HOLSymConst{=}
     \HOLKeyword{if} (\HOLBoundVar{u} \HOLSymConst{=} \HOLConst{\ensuremath{\tau}}) \HOLSymConst{\HOLTokenDisj{}} (\HOLConst{PREF_ACT} (\HOLBoundVar{f} (\HOLConst{SUC} \HOLBoundVar{n})) \HOLSymConst{=} \HOLConst{\ensuremath{\tau}}) \HOLKeyword{then} \HOLConst{SYNC} \HOLBoundVar{u} \HOLBoundVar{P} \HOLBoundVar{f} \HOLBoundVar{n}
     \HOLKeyword{else} \HOLKeyword{if} \HOLConst{LABEL} \HOLBoundVar{u} \HOLSymConst{=} \HOLConst{COMPL} (\HOLConst{LABEL} (\HOLConst{PREF_ACT} (\HOLBoundVar{f} (\HOLConst{SUC} \HOLBoundVar{n})))) \HOLKeyword{then}
       \HOLConst{\ensuremath{\tau}}\HOLSymConst{..}(\HOLBoundVar{P} \HOLSymConst{\ensuremath{\parallel}} \HOLConst{PREF_PROC} (\HOLBoundVar{f} (\HOLConst{SUC} \HOLBoundVar{n}))) \HOLSymConst{+} \HOLConst{SYNC} \HOLBoundVar{u} \HOLBoundVar{P} \HOLBoundVar{f} \HOLBoundVar{n}
     \HOLKeyword{else} \HOLConst{SYNC} \HOLBoundVar{u} \HOLBoundVar{P} \HOLBoundVar{f} \HOLBoundVar{n}
\end{SaveVerbatim}
\newcommand{\HOLStrongLawsTheoremsSYNCXXINDUCT}{\UseVerbatim{HOLStrongLawsTheoremsSYNCXXINDUCT}}
\begin{SaveVerbatim}{HOLStrongLawsTheoremsSYNCXXTRANSXXTHM}
\HOLTokenTurnstile{} \HOLSymConst{\HOLTokenForall{}}\HOLBoundVar{m} \HOLBoundVar{u} \HOLBoundVar{P} \HOLBoundVar{f} \HOLBoundVar{v} \HOLBoundVar{Q}.
     \HOLConst{SYNC} \HOLBoundVar{u} \HOLBoundVar{P} \HOLBoundVar{f} \HOLBoundVar{m} \HOLTokenTransBegin\HOLBoundVar{v}\HOLTokenTransEnd \HOLBoundVar{Q} \HOLSymConst{\HOLTokenImp{}}
     \HOLSymConst{\HOLTokenExists{}}\HOLBoundVar{j} \HOLBoundVar{l}.
       \HOLBoundVar{j} \HOLSymConst{\HOLTokenLeq{}} \HOLBoundVar{m} \HOLSymConst{\HOLTokenConj{}} (\HOLBoundVar{u} \HOLSymConst{=} \HOLConst{label} \HOLBoundVar{l}) \HOLSymConst{\HOLTokenConj{}}
       (\HOLConst{PREF_ACT} (\HOLBoundVar{f} \HOLBoundVar{j}) \HOLSymConst{=} \HOLConst{label} (\HOLConst{COMPL} \HOLBoundVar{l})) \HOLSymConst{\HOLTokenConj{}} (\HOLBoundVar{v} \HOLSymConst{=} \HOLConst{\ensuremath{\tau}}) \HOLSymConst{\HOLTokenConj{}}
       (\HOLBoundVar{Q} \HOLSymConst{=} \HOLBoundVar{P} \HOLSymConst{\ensuremath{\parallel}} \HOLConst{PREF_PROC} (\HOLBoundVar{f} \HOLBoundVar{j}))
\end{SaveVerbatim}
\newcommand{\HOLStrongLawsTheoremsSYNCXXTRANSXXTHM}{\UseVerbatim{HOLStrongLawsTheoremsSYNCXXTRANSXXTHM}}
\begin{SaveVerbatim}{HOLStrongLawsTheoremsSYNCXXTRANSXXTHMXXEQ}
\HOLTokenTurnstile{} \HOLSymConst{\HOLTokenForall{}}\HOLBoundVar{m} \HOLBoundVar{u} \HOLBoundVar{P} \HOLBoundVar{f} \HOLBoundVar{v} \HOLBoundVar{Q}.
     \HOLConst{SYNC} \HOLBoundVar{u} \HOLBoundVar{P} \HOLBoundVar{f} \HOLBoundVar{m} \HOLTokenTransBegin\HOLBoundVar{v}\HOLTokenTransEnd \HOLBoundVar{Q} \HOLSymConst{\HOLTokenEquiv{}}
     \HOLSymConst{\HOLTokenExists{}}\HOLBoundVar{j} \HOLBoundVar{l}.
       \HOLBoundVar{j} \HOLSymConst{\HOLTokenLeq{}} \HOLBoundVar{m} \HOLSymConst{\HOLTokenConj{}} (\HOLBoundVar{u} \HOLSymConst{=} \HOLConst{label} \HOLBoundVar{l}) \HOLSymConst{\HOLTokenConj{}}
       (\HOLConst{PREF_ACT} (\HOLBoundVar{f} \HOLBoundVar{j}) \HOLSymConst{=} \HOLConst{label} (\HOLConst{COMPL} \HOLBoundVar{l})) \HOLSymConst{\HOLTokenConj{}} (\HOLBoundVar{v} \HOLSymConst{=} \HOLConst{\ensuremath{\tau}}) \HOLSymConst{\HOLTokenConj{}}
       (\HOLBoundVar{Q} \HOLSymConst{=} \HOLBoundVar{P} \HOLSymConst{\ensuremath{\parallel}} \HOLConst{PREF_PROC} (\HOLBoundVar{f} \HOLBoundVar{j}))
\end{SaveVerbatim}
\newcommand{\HOLStrongLawsTheoremsSYNCXXTRANSXXTHMXXEQ}{\UseVerbatim{HOLStrongLawsTheoremsSYNCXXTRANSXXTHMXXEQ}}
\newcommand{\HOLStrongLawsTheorems}{
\HOLThmTag{StrongLaws}{ALL_SYNC_BASE}\HOLStrongLawsTheoremsALLXXSYNCXXBASE
\HOLThmTag{StrongLaws}{ALL_SYNC_def_compute}\HOLStrongLawsTheoremsALLXXSYNCXXdefXXcompute
\HOLThmTag{StrongLaws}{ALL_SYNC_INDUCT}\HOLStrongLawsTheoremsALLXXSYNCXXINDUCT
\HOLThmTag{StrongLaws}{ALL_SYNC_TRANS_THM}\HOLStrongLawsTheoremsALLXXSYNCXXTRANSXXTHM
\HOLThmTag{StrongLaws}{ALL_SYNC_TRANS_THM_EQ}\HOLStrongLawsTheoremsALLXXSYNCXXTRANSXXTHMXXEQ
\HOLThmTag{StrongLaws}{CCS_COMP_def_compute}\HOLStrongLawsTheoremsCCSXXCOMPXXdefXXcompute
\HOLThmTag{StrongLaws}{CCS_SIGMA_def_compute}\HOLStrongLawsTheoremsCCSXXSIGMAXXdefXXcompute
\HOLThmTag{StrongLaws}{COMP_BASE}\HOLStrongLawsTheoremsCOMPXXBASE
\HOLThmTag{StrongLaws}{COMP_INDUCT}\HOLStrongLawsTheoremsCOMPXXINDUCT
\HOLThmTag{StrongLaws}{PREF_IS_PREFIX}\HOLStrongLawsTheoremsPREFXXISXXPREFIX
\HOLThmTag{StrongLaws}{SIGMA_BASE}\HOLStrongLawsTheoremsSIGMAXXBASE
\HOLThmTag{StrongLaws}{SIGMA_INDUCT}\HOLStrongLawsTheoremsSIGMAXXINDUCT
\HOLThmTag{StrongLaws}{SIGMA_TRANS_THM}\HOLStrongLawsTheoremsSIGMAXXTRANSXXTHM
\HOLThmTag{StrongLaws}{SIGMA_TRANS_THM_EQ}\HOLStrongLawsTheoremsSIGMAXXTRANSXXTHMXXEQ
\HOLThmTag{StrongLaws}{STRONG_EXPANSION_LAW}\HOLStrongLawsTheoremsSTRONGXXEXPANSIONXXLAW
\HOLThmTag{StrongLaws}{STRONG_LEFT_SUM_MID_IDEMP}\HOLStrongLawsTheoremsSTRONGXXLEFTXXSUMXXMIDXXIDEMP
\HOLThmTag{StrongLaws}{STRONG_PAR_ASSOC}\HOLStrongLawsTheoremsSTRONGXXPARXXASSOC
\HOLThmTag{StrongLaws}{STRONG_PAR_COMM}\HOLStrongLawsTheoremsSTRONGXXPARXXCOMM
\HOLThmTag{StrongLaws}{STRONG_PAR_IDENT_L}\HOLStrongLawsTheoremsSTRONGXXPARXXIDENTXXL
\HOLThmTag{StrongLaws}{STRONG_PAR_IDENT_R}\HOLStrongLawsTheoremsSTRONGXXPARXXIDENTXXR
\HOLThmTag{StrongLaws}{STRONG_PAR_PREF_NO_SYNCR}\HOLStrongLawsTheoremsSTRONGXXPARXXPREFXXNOXXSYNCR
\HOLThmTag{StrongLaws}{STRONG_PAR_PREF_SYNCR}\HOLStrongLawsTheoremsSTRONGXXPARXXPREFXXSYNCR
\HOLThmTag{StrongLaws}{STRONG_PAR_PREF_TAU}\HOLStrongLawsTheoremsSTRONGXXPARXXPREFXXTAU
\HOLThmTag{StrongLaws}{STRONG_PAR_TAU_PREF}\HOLStrongLawsTheoremsSTRONGXXPARXXTAUXXPREF
\HOLThmTag{StrongLaws}{STRONG_PAR_TAU_TAU}\HOLStrongLawsTheoremsSTRONGXXPARXXTAUXXTAU
\HOLThmTag{StrongLaws}{STRONG_PREF_REC_EQUIV}\HOLStrongLawsTheoremsSTRONGXXPREFXXRECXXEQUIV
\HOLThmTag{StrongLaws}{STRONG_REC_ACT2}\HOLStrongLawsTheoremsSTRONGXXRECXXACTTwo
\HOLThmTag{StrongLaws}{STRONG_RELAB_NIL}\HOLStrongLawsTheoremsSTRONGXXRELABXXNIL
\HOLThmTag{StrongLaws}{STRONG_RELAB_PREFIX}\HOLStrongLawsTheoremsSTRONGXXRELABXXPREFIX
\HOLThmTag{StrongLaws}{STRONG_RELAB_SUM}\HOLStrongLawsTheoremsSTRONGXXRELABXXSUM
\HOLThmTag{StrongLaws}{STRONG_RESTR_NIL}\HOLStrongLawsTheoremsSTRONGXXRESTRXXNIL
\HOLThmTag{StrongLaws}{STRONG_RESTR_PR_LAB_NIL}\HOLStrongLawsTheoremsSTRONGXXRESTRXXPRXXLABXXNIL
\HOLThmTag{StrongLaws}{STRONG_RESTR_PREFIX_LABEL}\HOLStrongLawsTheoremsSTRONGXXRESTRXXPREFIXXXLABEL
\HOLThmTag{StrongLaws}{STRONG_RESTR_PREFIX_TAU}\HOLStrongLawsTheoremsSTRONGXXRESTRXXPREFIXXXTAU
\HOLThmTag{StrongLaws}{STRONG_RESTR_SUM}\HOLStrongLawsTheoremsSTRONGXXRESTRXXSUM
\HOLThmTag{StrongLaws}{STRONG_SUM_ASSOC_L}\HOLStrongLawsTheoremsSTRONGXXSUMXXASSOCXXL
\HOLThmTag{StrongLaws}{STRONG_SUM_ASSOC_R}\HOLStrongLawsTheoremsSTRONGXXSUMXXASSOCXXR
\HOLThmTag{StrongLaws}{STRONG_SUM_COMM}\HOLStrongLawsTheoremsSTRONGXXSUMXXCOMM
\HOLThmTag{StrongLaws}{STRONG_SUM_IDEMP}\HOLStrongLawsTheoremsSTRONGXXSUMXXIDEMP
\HOLThmTag{StrongLaws}{STRONG_SUM_IDENT_L}\HOLStrongLawsTheoremsSTRONGXXSUMXXIDENTXXL
\HOLThmTag{StrongLaws}{STRONG_SUM_IDENT_R}\HOLStrongLawsTheoremsSTRONGXXSUMXXIDENTXXR
\HOLThmTag{StrongLaws}{STRONG_SUM_MID_IDEMP}\HOLStrongLawsTheoremsSTRONGXXSUMXXMIDXXIDEMP
\HOLThmTag{StrongLaws}{STRONG_UNFOLDING}\HOLStrongLawsTheoremsSTRONGXXUNFOLDING
\HOLThmTag{StrongLaws}{SYNC_BASE}\HOLStrongLawsTheoremsSYNCXXBASE
\HOLThmTag{StrongLaws}{SYNC_def_compute}\HOLStrongLawsTheoremsSYNCXXdefXXcompute
\HOLThmTag{StrongLaws}{SYNC_INDUCT}\HOLStrongLawsTheoremsSYNCXXINDUCT
\HOLThmTag{StrongLaws}{SYNC_TRANS_THM}\HOLStrongLawsTheoremsSYNCXXTRANSXXTHM
\HOLThmTag{StrongLaws}{SYNC_TRANS_THM_EQ}\HOLStrongLawsTheoremsSYNCXXTRANSXXTHMXXEQ
}

\newcommand{\HOLWeakEQDate}{02 Dicembre 2017}
\newcommand{\HOLWeakEQTime}{13:31}
\begin{SaveVerbatim}{HOLWeakEQDefinitionsEPSXXdef}
\HOLTokenTurnstile{} \HOLConst{EPS} \HOLSymConst{=} (\HOLTokenLambda{}\HOLBoundVar{E} \HOLBoundVar{E\sp{\prime}}. \HOLBoundVar{E} \HOLTokenTransBegin\HOLConst{\ensuremath{\tau}}\HOLTokenTransEnd \HOLBoundVar{E\sp{\prime}})\HOLSymConst{\HOLTokenSupStar{}}
\end{SaveVerbatim}
\newcommand{\HOLWeakEQDefinitionsEPSXXdef}{\UseVerbatim{HOLWeakEQDefinitionsEPSXXdef}}
\begin{SaveVerbatim}{HOLWeakEQDefinitionsSTABLE}
\HOLTokenTurnstile{} \HOLSymConst{\HOLTokenForall{}}\HOLBoundVar{E}. \HOLConst{STABLE} \HOLBoundVar{E} \HOLSymConst{\HOLTokenEquiv{}} \HOLSymConst{\HOLTokenForall{}}\HOLBoundVar{u} \HOLBoundVar{E\sp{\prime}}. \HOLBoundVar{E} \HOLTokenTransBegin\HOLBoundVar{u}\HOLTokenTransEnd \HOLBoundVar{E\sp{\prime}} \HOLSymConst{\HOLTokenImp{}} \HOLBoundVar{u} \HOLSymConst{\HOLTokenNotEqual{}} \HOLConst{\ensuremath{\tau}}
\end{SaveVerbatim}
\newcommand{\HOLWeakEQDefinitionsSTABLE}{\UseVerbatim{HOLWeakEQDefinitionsSTABLE}}
\begin{SaveVerbatim}{HOLWeakEQDefinitionsWEAKXXBISIMXXdef}
\HOLTokenTurnstile{} \HOLSymConst{\HOLTokenForall{}}\HOLBoundVar{R}. \HOLConst{WEAK_BISIM} \HOLBoundVar{R} \HOLSymConst{\HOLTokenEquiv{}} \HOLConst{WEAK_SIM} \HOLBoundVar{R} \HOLSymConst{\HOLTokenConj{}} \HOLConst{WEAK_SIM} (\HOLConst{relinv} \HOLBoundVar{R})
\end{SaveVerbatim}
\newcommand{\HOLWeakEQDefinitionsWEAKXXBISIMXXdef}{\UseVerbatim{HOLWeakEQDefinitionsWEAKXXBISIMXXdef}}
\begin{SaveVerbatim}{HOLWeakEQDefinitionsWEAKXXEQUIVXXdef}
\HOLTokenTurnstile{} \HOLConst{WEAK_EQUIV} \HOLSymConst{=}
   (\HOLTokenLambda{}\HOLBoundVar{a\sb{\mathrm{0}}} \HOLBoundVar{a\sb{\mathrm{1}}}.
      \HOLSymConst{\HOLTokenExists{}}\HOLBoundVar{WEAK\HOLTokenUnderscore{}EQUIV\sp{\prime}}.
        \HOLBoundVar{WEAK\HOLTokenUnderscore{}EQUIV\sp{\prime}} \HOLBoundVar{a\sb{\mathrm{0}}} \HOLBoundVar{a\sb{\mathrm{1}}} \HOLSymConst{\HOLTokenConj{}}
        \HOLSymConst{\HOLTokenForall{}}\HOLBoundVar{a\sb{\mathrm{0}}} \HOLBoundVar{a\sb{\mathrm{1}}}.
          \HOLBoundVar{WEAK\HOLTokenUnderscore{}EQUIV\sp{\prime}} \HOLBoundVar{a\sb{\mathrm{0}}} \HOLBoundVar{a\sb{\mathrm{1}}} \HOLSymConst{\HOLTokenImp{}}
          (\HOLSymConst{\HOLTokenForall{}}\HOLBoundVar{l}.
             (\HOLSymConst{\HOLTokenForall{}}\HOLBoundVar{E\sb{\mathrm{1}}}.
                \HOLBoundVar{a\sb{\mathrm{0}}} \HOLTokenTransBegin\HOLConst{label} \HOLBoundVar{l}\HOLTokenTransEnd \HOLBoundVar{E\sb{\mathrm{1}}} \HOLSymConst{\HOLTokenImp{}}
                \HOLSymConst{\HOLTokenExists{}}\HOLBoundVar{E\sb{\mathrm{2}}}. \HOLBoundVar{a\sb{\mathrm{1}}} \HOLTokenWeakTransBegin\HOLConst{label} \HOLBoundVar{l}\HOLTokenWeakTransEnd \HOLBoundVar{E\sb{\mathrm{2}}} \HOLSymConst{\HOLTokenConj{}} \HOLBoundVar{WEAK\HOLTokenUnderscore{}EQUIV\sp{\prime}} \HOLBoundVar{E\sb{\mathrm{1}}} \HOLBoundVar{E\sb{\mathrm{2}}}) \HOLSymConst{\HOLTokenConj{}}
             \HOLSymConst{\HOLTokenForall{}}\HOLBoundVar{E\sb{\mathrm{2}}}.
               \HOLBoundVar{a\sb{\mathrm{1}}} \HOLTokenTransBegin\HOLConst{label} \HOLBoundVar{l}\HOLTokenTransEnd \HOLBoundVar{E\sb{\mathrm{2}}} \HOLSymConst{\HOLTokenImp{}}
               \HOLSymConst{\HOLTokenExists{}}\HOLBoundVar{E\sb{\mathrm{1}}}. \HOLBoundVar{a\sb{\mathrm{0}}} \HOLTokenWeakTransBegin\HOLConst{label} \HOLBoundVar{l}\HOLTokenWeakTransEnd \HOLBoundVar{E\sb{\mathrm{1}}} \HOLSymConst{\HOLTokenConj{}} \HOLBoundVar{WEAK\HOLTokenUnderscore{}EQUIV\sp{\prime}} \HOLBoundVar{E\sb{\mathrm{1}}} \HOLBoundVar{E\sb{\mathrm{2}}}) \HOLSymConst{\HOLTokenConj{}}
          (\HOLSymConst{\HOLTokenForall{}}\HOLBoundVar{E\sb{\mathrm{1}}}.
             \HOLBoundVar{a\sb{\mathrm{0}}} \HOLTokenTransBegin\HOLConst{\ensuremath{\tau}}\HOLTokenTransEnd \HOLBoundVar{E\sb{\mathrm{1}}} \HOLSymConst{\HOLTokenImp{}} \HOLSymConst{\HOLTokenExists{}}\HOLBoundVar{E\sb{\mathrm{2}}}. \HOLConst{EPS} \HOLBoundVar{a\sb{\mathrm{1}}} \HOLBoundVar{E\sb{\mathrm{2}}} \HOLSymConst{\HOLTokenConj{}} \HOLBoundVar{WEAK\HOLTokenUnderscore{}EQUIV\sp{\prime}} \HOLBoundVar{E\sb{\mathrm{1}}} \HOLBoundVar{E\sb{\mathrm{2}}}) \HOLSymConst{\HOLTokenConj{}}
          \HOLSymConst{\HOLTokenForall{}}\HOLBoundVar{E\sb{\mathrm{2}}}. \HOLBoundVar{a\sb{\mathrm{1}}} \HOLTokenTransBegin\HOLConst{\ensuremath{\tau}}\HOLTokenTransEnd \HOLBoundVar{E\sb{\mathrm{2}}} \HOLSymConst{\HOLTokenImp{}} \HOLSymConst{\HOLTokenExists{}}\HOLBoundVar{E\sb{\mathrm{1}}}. \HOLConst{EPS} \HOLBoundVar{a\sb{\mathrm{0}}} \HOLBoundVar{E\sb{\mathrm{1}}} \HOLSymConst{\HOLTokenConj{}} \HOLBoundVar{WEAK\HOLTokenUnderscore{}EQUIV\sp{\prime}} \HOLBoundVar{E\sb{\mathrm{1}}} \HOLBoundVar{E\sb{\mathrm{2}}})
\end{SaveVerbatim}
\newcommand{\HOLWeakEQDefinitionsWEAKXXEQUIVXXdef}{\UseVerbatim{HOLWeakEQDefinitionsWEAKXXEQUIVXXdef}}
\begin{SaveVerbatim}{HOLWeakEQDefinitionsWEAKXXSIMXXdef}
\HOLTokenTurnstile{} \HOLSymConst{\HOLTokenForall{}}\HOLBoundVar{R}.
     \HOLConst{WEAK_SIM} \HOLBoundVar{R} \HOLSymConst{\HOLTokenEquiv{}}
     \HOLSymConst{\HOLTokenForall{}}\HOLBoundVar{E} \HOLBoundVar{E\sp{\prime}}.
       \HOLBoundVar{R} \HOLBoundVar{E} \HOLBoundVar{E\sp{\prime}} \HOLSymConst{\HOLTokenImp{}}
       (\HOLSymConst{\HOLTokenForall{}}\HOLBoundVar{l} \HOLBoundVar{E\sb{\mathrm{1}}}.
          \HOLBoundVar{E} \HOLTokenTransBegin\HOLConst{label} \HOLBoundVar{l}\HOLTokenTransEnd \HOLBoundVar{E\sb{\mathrm{1}}} \HOLSymConst{\HOLTokenImp{}} \HOLSymConst{\HOLTokenExists{}}\HOLBoundVar{E\sb{\mathrm{2}}}. \HOLBoundVar{E\sp{\prime}} \HOLTokenWeakTransBegin\HOLConst{label} \HOLBoundVar{l}\HOLTokenWeakTransEnd \HOLBoundVar{E\sb{\mathrm{2}}} \HOLSymConst{\HOLTokenConj{}} \HOLBoundVar{R} \HOLBoundVar{E\sb{\mathrm{1}}} \HOLBoundVar{E\sb{\mathrm{2}}}) \HOLSymConst{\HOLTokenConj{}}
       \HOLSymConst{\HOLTokenForall{}}\HOLBoundVar{E\sb{\mathrm{1}}}. \HOLBoundVar{E} \HOLTokenTransBegin\HOLConst{\ensuremath{\tau}}\HOLTokenTransEnd \HOLBoundVar{E\sb{\mathrm{1}}} \HOLSymConst{\HOLTokenImp{}} \HOLSymConst{\HOLTokenExists{}}\HOLBoundVar{E\sb{\mathrm{2}}}. \HOLConst{EPS} \HOLBoundVar{E\sp{\prime}} \HOLBoundVar{E\sb{\mathrm{2}}} \HOLSymConst{\HOLTokenConj{}} \HOLBoundVar{R} \HOLBoundVar{E\sb{\mathrm{1}}} \HOLBoundVar{E\sb{\mathrm{2}}}
\end{SaveVerbatim}
\newcommand{\HOLWeakEQDefinitionsWEAKXXSIMXXdef}{\UseVerbatim{HOLWeakEQDefinitionsWEAKXXSIMXXdef}}
\begin{SaveVerbatim}{HOLWeakEQDefinitionsWEAKXXTRANS}
\HOLTokenTurnstile{} \HOLSymConst{\HOLTokenForall{}}\HOLBoundVar{E} \HOLBoundVar{u} \HOLBoundVar{E\sp{\prime}}. \HOLBoundVar{E} \HOLTokenWeakTransBegin\HOLBoundVar{u}\HOLTokenWeakTransEnd \HOLBoundVar{E\sp{\prime}} \HOLSymConst{\HOLTokenEquiv{}} \HOLSymConst{\HOLTokenExists{}}\HOLBoundVar{E\sb{\mathrm{1}}} \HOLBoundVar{E\sb{\mathrm{2}}}. \HOLConst{EPS} \HOLBoundVar{E} \HOLBoundVar{E\sb{\mathrm{1}}} \HOLSymConst{\HOLTokenConj{}} \HOLBoundVar{E\sb{\mathrm{1}}} \HOLTokenTransBegin\HOLBoundVar{u}\HOLTokenTransEnd \HOLBoundVar{E\sb{\mathrm{2}}} \HOLSymConst{\HOLTokenConj{}} \HOLConst{EPS} \HOLBoundVar{E\sb{\mathrm{2}}} \HOLBoundVar{E\sp{\prime}}
\end{SaveVerbatim}
\newcommand{\HOLWeakEQDefinitionsWEAKXXTRANS}{\UseVerbatim{HOLWeakEQDefinitionsWEAKXXTRANS}}
\newcommand{\HOLWeakEQDefinitions}{
\HOLDfnTag{WeakEQ}{EPS_def}\HOLWeakEQDefinitionsEPSXXdef
\HOLDfnTag{WeakEQ}{STABLE}\HOLWeakEQDefinitionsSTABLE
\HOLDfnTag{WeakEQ}{WEAK_BISIM_def}\HOLWeakEQDefinitionsWEAKXXBISIMXXdef
\HOLDfnTag{WeakEQ}{WEAK_EQUIV_def}\HOLWeakEQDefinitionsWEAKXXEQUIVXXdef
\HOLDfnTag{WeakEQ}{WEAK_SIM_def}\HOLWeakEQDefinitionsWEAKXXSIMXXdef
\HOLDfnTag{WeakEQ}{WEAK_TRANS}\HOLWeakEQDefinitionsWEAKXXTRANS
}
\begin{SaveVerbatim}{HOLWeakEQTheoremsCOMPXXWEAKXXBISIM}
\HOLTokenTurnstile{} \HOLSymConst{\HOLTokenForall{}}\HOLBoundVar{Wbsm\sb{\mathrm{1}}} \HOLBoundVar{Wbsm\sb{\mathrm{2}}}.
     \HOLConst{WEAK_BISIM} \HOLBoundVar{Wbsm\sb{\mathrm{1}}} \HOLSymConst{\HOLTokenConj{}} \HOLConst{WEAK_BISIM} \HOLBoundVar{Wbsm\sb{\mathrm{2}}} \HOLSymConst{\HOLTokenImp{}}
     \HOLConst{WEAK_BISIM} (\HOLBoundVar{Wbsm\sb{\mathrm{2}}} \HOLConst{O} \HOLBoundVar{Wbsm\sb{\mathrm{1}}})
\end{SaveVerbatim}
\newcommand{\HOLWeakEQTheoremsCOMPXXWEAKXXBISIM}{\UseVerbatim{HOLWeakEQTheoremsCOMPXXWEAKXXBISIM}}
\begin{SaveVerbatim}{HOLWeakEQTheoremsCONVERSEXXWEAKXXBISIM}
\HOLTokenTurnstile{} \HOLSymConst{\HOLTokenForall{}}\HOLBoundVar{Wbsm}. \HOLConst{WEAK_BISIM} \HOLBoundVar{Wbsm} \HOLSymConst{\HOLTokenImp{}} \HOLConst{WEAK_BISIM} (\HOLConst{relinv} \HOLBoundVar{Wbsm})
\end{SaveVerbatim}
\newcommand{\HOLWeakEQTheoremsCONVERSEXXWEAKXXBISIM}{\UseVerbatim{HOLWeakEQTheoremsCONVERSEXXWEAKXXBISIM}}
\begin{SaveVerbatim}{HOLWeakEQTheoremsEPSXXANDXXWEAKXXTRANS}
\HOLTokenTurnstile{} \HOLSymConst{\HOLTokenForall{}}\HOLBoundVar{E} \HOLBoundVar{E\sb{\mathrm{1}}} \HOLBoundVar{u} \HOLBoundVar{E\sb{\mathrm{2}}}. \HOLConst{EPS} \HOLBoundVar{E} \HOLBoundVar{E\sb{\mathrm{1}}} \HOLSymConst{\HOLTokenConj{}} \HOLBoundVar{E\sb{\mathrm{1}}} \HOLTokenWeakTransBegin\HOLBoundVar{u}\HOLTokenWeakTransEnd \HOLBoundVar{E\sb{\mathrm{2}}} \HOLSymConst{\HOLTokenImp{}} \HOLBoundVar{E} \HOLTokenWeakTransBegin\HOLBoundVar{u}\HOLTokenWeakTransEnd \HOLBoundVar{E\sb{\mathrm{2}}}
\end{SaveVerbatim}
\newcommand{\HOLWeakEQTheoremsEPSXXANDXXWEAKXXTRANS}{\UseVerbatim{HOLWeakEQTheoremsEPSXXANDXXWEAKXXTRANS}}
\begin{SaveVerbatim}{HOLWeakEQTheoremsEPSXXcases}
\HOLTokenTurnstile{} \HOLSymConst{\HOLTokenForall{}}\HOLBoundVar{E} \HOLBoundVar{E\sp{\prime}}.
     \HOLConst{EPS} \HOLBoundVar{E} \HOLBoundVar{E\sp{\prime}} \HOLSymConst{\HOLTokenEquiv{}} \HOLBoundVar{E} \HOLTokenTransBegin\HOLConst{\ensuremath{\tau}}\HOLTokenTransEnd \HOLBoundVar{E\sp{\prime}} \HOLSymConst{\HOLTokenDisj{}} (\HOLBoundVar{E} \HOLSymConst{=} \HOLBoundVar{E\sp{\prime}}) \HOLSymConst{\HOLTokenDisj{}} \HOLSymConst{\HOLTokenExists{}}\HOLBoundVar{E\sb{\mathrm{1}}}. \HOLConst{EPS} \HOLBoundVar{E} \HOLBoundVar{E\sb{\mathrm{1}}} \HOLSymConst{\HOLTokenConj{}} \HOLConst{EPS} \HOLBoundVar{E\sb{\mathrm{1}}} \HOLBoundVar{E\sp{\prime}}
\end{SaveVerbatim}
\newcommand{\HOLWeakEQTheoremsEPSXXcases}{\UseVerbatim{HOLWeakEQTheoremsEPSXXcases}}
\begin{SaveVerbatim}{HOLWeakEQTheoremsEPSXXcasesOne}
\HOLTokenTurnstile{} \HOLSymConst{\HOLTokenForall{}}\HOLBoundVar{x} \HOLBoundVar{y}. \HOLConst{EPS} \HOLBoundVar{x} \HOLBoundVar{y} \HOLSymConst{\HOLTokenEquiv{}} (\HOLBoundVar{x} \HOLSymConst{=} \HOLBoundVar{y}) \HOLSymConst{\HOLTokenDisj{}} \HOLSymConst{\HOLTokenExists{}}\HOLBoundVar{u}. \HOLBoundVar{x} \HOLTokenTransBegin\HOLConst{\ensuremath{\tau}}\HOLTokenTransEnd \HOLBoundVar{u} \HOLSymConst{\HOLTokenConj{}} \HOLConst{EPS} \HOLBoundVar{u} \HOLBoundVar{y}
\end{SaveVerbatim}
\newcommand{\HOLWeakEQTheoremsEPSXXcasesOne}{\UseVerbatim{HOLWeakEQTheoremsEPSXXcasesOne}}
\begin{SaveVerbatim}{HOLWeakEQTheoremsEPSXXcasesTwo}
\HOLTokenTurnstile{} \HOLSymConst{\HOLTokenForall{}}\HOLBoundVar{x} \HOLBoundVar{y}. \HOLConst{EPS} \HOLBoundVar{x} \HOLBoundVar{y} \HOLSymConst{\HOLTokenEquiv{}} (\HOLBoundVar{x} \HOLSymConst{=} \HOLBoundVar{y}) \HOLSymConst{\HOLTokenDisj{}} \HOLSymConst{\HOLTokenExists{}}\HOLBoundVar{u}. \HOLConst{EPS} \HOLBoundVar{x} \HOLBoundVar{u} \HOLSymConst{\HOLTokenConj{}} \HOLBoundVar{u} \HOLTokenTransBegin\HOLConst{\ensuremath{\tau}}\HOLTokenTransEnd \HOLBoundVar{y}
\end{SaveVerbatim}
\newcommand{\HOLWeakEQTheoremsEPSXXcasesTwo}{\UseVerbatim{HOLWeakEQTheoremsEPSXXcasesTwo}}
\begin{SaveVerbatim}{HOLWeakEQTheoremsEPSXXIMPXXWEAKXXTRANS}
\HOLTokenTurnstile{} \HOLSymConst{\HOLTokenForall{}}\HOLBoundVar{E} \HOLBoundVar{E\sp{\prime}}. \HOLConst{EPS} \HOLBoundVar{E} \HOLBoundVar{E\sp{\prime}} \HOLSymConst{\HOLTokenImp{}} (\HOLBoundVar{E} \HOLSymConst{=} \HOLBoundVar{E\sp{\prime}}) \HOLSymConst{\HOLTokenDisj{}} \HOLBoundVar{E} \HOLTokenWeakTransBegin\HOLConst{\ensuremath{\tau}}\HOLTokenWeakTransEnd \HOLBoundVar{E\sp{\prime}}
\end{SaveVerbatim}
\newcommand{\HOLWeakEQTheoremsEPSXXIMPXXWEAKXXTRANS}{\UseVerbatim{HOLWeakEQTheoremsEPSXXIMPXXWEAKXXTRANS}}
\begin{SaveVerbatim}{HOLWeakEQTheoremsEPSXXind}
\HOLTokenTurnstile{} \HOLSymConst{\HOLTokenForall{}}\HOLBoundVar{P}.
     (\HOLSymConst{\HOLTokenForall{}}\HOLBoundVar{x}. \HOLBoundVar{P} \HOLBoundVar{x} \HOLBoundVar{x}) \HOLSymConst{\HOLTokenConj{}} (\HOLSymConst{\HOLTokenForall{}}\HOLBoundVar{x} \HOLBoundVar{y} \HOLBoundVar{z}. \HOLBoundVar{x} \HOLTokenTransBegin\HOLConst{\ensuremath{\tau}}\HOLTokenTransEnd \HOLBoundVar{y} \HOLSymConst{\HOLTokenConj{}} \HOLBoundVar{P} \HOLBoundVar{y} \HOLBoundVar{z} \HOLSymConst{\HOLTokenImp{}} \HOLBoundVar{P} \HOLBoundVar{x} \HOLBoundVar{z}) \HOLSymConst{\HOLTokenImp{}}
     \HOLSymConst{\HOLTokenForall{}}\HOLBoundVar{x} \HOLBoundVar{y}. \HOLConst{EPS} \HOLBoundVar{x} \HOLBoundVar{y} \HOLSymConst{\HOLTokenImp{}} \HOLBoundVar{P} \HOLBoundVar{x} \HOLBoundVar{y}
\end{SaveVerbatim}
\newcommand{\HOLWeakEQTheoremsEPSXXind}{\UseVerbatim{HOLWeakEQTheoremsEPSXXind}}
\begin{SaveVerbatim}{HOLWeakEQTheoremsEPSXXindXXright}
\HOLTokenTurnstile{} \HOLSymConst{\HOLTokenForall{}}\HOLBoundVar{P}.
     (\HOLSymConst{\HOLTokenForall{}}\HOLBoundVar{x}. \HOLBoundVar{P} \HOLBoundVar{x} \HOLBoundVar{x}) \HOLSymConst{\HOLTokenConj{}} (\HOLSymConst{\HOLTokenForall{}}\HOLBoundVar{x} \HOLBoundVar{y} \HOLBoundVar{z}. \HOLBoundVar{P} \HOLBoundVar{x} \HOLBoundVar{y} \HOLSymConst{\HOLTokenConj{}} \HOLBoundVar{y} \HOLTokenTransBegin\HOLConst{\ensuremath{\tau}}\HOLTokenTransEnd \HOLBoundVar{z} \HOLSymConst{\HOLTokenImp{}} \HOLBoundVar{P} \HOLBoundVar{x} \HOLBoundVar{z}) \HOLSymConst{\HOLTokenImp{}}
     \HOLSymConst{\HOLTokenForall{}}\HOLBoundVar{x} \HOLBoundVar{y}. \HOLConst{EPS} \HOLBoundVar{x} \HOLBoundVar{y} \HOLSymConst{\HOLTokenImp{}} \HOLBoundVar{P} \HOLBoundVar{x} \HOLBoundVar{y}
\end{SaveVerbatim}
\newcommand{\HOLWeakEQTheoremsEPSXXindXXright}{\UseVerbatim{HOLWeakEQTheoremsEPSXXindXXright}}
\begin{SaveVerbatim}{HOLWeakEQTheoremsEPSXXINDUCT}
\HOLTokenTurnstile{} \HOLSymConst{\HOLTokenForall{}}\HOLBoundVar{P}.
     (\HOLSymConst{\HOLTokenForall{}}\HOLBoundVar{E} \HOLBoundVar{E\sp{\prime}}. \HOLBoundVar{E} \HOLTokenTransBegin\HOLConst{\ensuremath{\tau}}\HOLTokenTransEnd \HOLBoundVar{E\sp{\prime}} \HOLSymConst{\HOLTokenImp{}} \HOLBoundVar{P} \HOLBoundVar{E} \HOLBoundVar{E\sp{\prime}}) \HOLSymConst{\HOLTokenConj{}} (\HOLSymConst{\HOLTokenForall{}}\HOLBoundVar{E}. \HOLBoundVar{P} \HOLBoundVar{E} \HOLBoundVar{E}) \HOLSymConst{\HOLTokenConj{}}
     (\HOLSymConst{\HOLTokenForall{}}\HOLBoundVar{E} \HOLBoundVar{E\sb{\mathrm{1}}} \HOLBoundVar{E\sp{\prime}}. \HOLBoundVar{P} \HOLBoundVar{E} \HOLBoundVar{E\sb{\mathrm{1}}} \HOLSymConst{\HOLTokenConj{}} \HOLBoundVar{P} \HOLBoundVar{E\sb{\mathrm{1}}} \HOLBoundVar{E\sp{\prime}} \HOLSymConst{\HOLTokenImp{}} \HOLBoundVar{P} \HOLBoundVar{E} \HOLBoundVar{E\sp{\prime}}) \HOLSymConst{\HOLTokenImp{}}
     \HOLSymConst{\HOLTokenForall{}}\HOLBoundVar{x} \HOLBoundVar{y}. \HOLConst{EPS} \HOLBoundVar{x} \HOLBoundVar{y} \HOLSymConst{\HOLTokenImp{}} \HOLBoundVar{P} \HOLBoundVar{x} \HOLBoundVar{y}
\end{SaveVerbatim}
\newcommand{\HOLWeakEQTheoremsEPSXXINDUCT}{\UseVerbatim{HOLWeakEQTheoremsEPSXXINDUCT}}
\begin{SaveVerbatim}{HOLWeakEQTheoremsEPSXXPAR}
\HOLTokenTurnstile{} \HOLSymConst{\HOLTokenForall{}}\HOLBoundVar{E} \HOLBoundVar{E\sp{\prime}}.
     \HOLConst{EPS} \HOLBoundVar{E} \HOLBoundVar{E\sp{\prime}} \HOLSymConst{\HOLTokenImp{}}
     \HOLSymConst{\HOLTokenForall{}}\HOLBoundVar{E\sp{\prime\prime}}. \HOLConst{EPS} (\HOLBoundVar{E} \HOLSymConst{\ensuremath{\parallel}} \HOLBoundVar{E\sp{\prime\prime}}) (\HOLBoundVar{E\sp{\prime}} \HOLSymConst{\ensuremath{\parallel}} \HOLBoundVar{E\sp{\prime\prime}}) \HOLSymConst{\HOLTokenConj{}} \HOLConst{EPS} (\HOLBoundVar{E\sp{\prime\prime}} \HOLSymConst{\ensuremath{\parallel}} \HOLBoundVar{E}) (\HOLBoundVar{E\sp{\prime\prime}} \HOLSymConst{\ensuremath{\parallel}} \HOLBoundVar{E\sp{\prime}})
\end{SaveVerbatim}
\newcommand{\HOLWeakEQTheoremsEPSXXPAR}{\UseVerbatim{HOLWeakEQTheoremsEPSXXPAR}}
\begin{SaveVerbatim}{HOLWeakEQTheoremsEPSXXPARXXPAR}
\HOLTokenTurnstile{} \HOLSymConst{\HOLTokenForall{}}\HOLBoundVar{E\sb{\mathrm{1}}} \HOLBoundVar{E\sb{\mathrm{2}}} \HOLBoundVar{F\sb{\mathrm{1}}} \HOLBoundVar{F\sb{\mathrm{2}}}. \HOLConst{EPS} \HOLBoundVar{E\sb{\mathrm{1}}} \HOLBoundVar{E\sb{\mathrm{2}}} \HOLSymConst{\HOLTokenConj{}} \HOLConst{EPS} \HOLBoundVar{F\sb{\mathrm{1}}} \HOLBoundVar{F\sb{\mathrm{2}}} \HOLSymConst{\HOLTokenImp{}} \HOLConst{EPS} (\HOLBoundVar{E\sb{\mathrm{1}}} \HOLSymConst{\ensuremath{\parallel}} \HOLBoundVar{F\sb{\mathrm{1}}}) (\HOLBoundVar{E\sb{\mathrm{2}}} \HOLSymConst{\ensuremath{\parallel}} \HOLBoundVar{F\sb{\mathrm{2}}})
\end{SaveVerbatim}
\newcommand{\HOLWeakEQTheoremsEPSXXPARXXPAR}{\UseVerbatim{HOLWeakEQTheoremsEPSXXPARXXPAR}}
\begin{SaveVerbatim}{HOLWeakEQTheoremsEPSXXREFL}
\HOLTokenTurnstile{} \HOLSymConst{\HOLTokenForall{}}\HOLBoundVar{E}. \HOLConst{EPS} \HOLBoundVar{E} \HOLBoundVar{E}
\end{SaveVerbatim}
\newcommand{\HOLWeakEQTheoremsEPSXXREFL}{\UseVerbatim{HOLWeakEQTheoremsEPSXXREFL}}
\begin{SaveVerbatim}{HOLWeakEQTheoremsEPSXXRELAB}
\HOLTokenTurnstile{} \HOLSymConst{\HOLTokenForall{}}\HOLBoundVar{E} \HOLBoundVar{E\sp{\prime}}.
     \HOLConst{EPS} \HOLBoundVar{E} \HOLBoundVar{E\sp{\prime}} \HOLSymConst{\HOLTokenImp{}}
     \HOLSymConst{\HOLTokenForall{}}\HOLBoundVar{labl}. \HOLConst{EPS} (\HOLConst{relab} \HOLBoundVar{E} (\HOLConst{RELAB} \HOLBoundVar{labl})) (\HOLConst{relab} \HOLBoundVar{E\sp{\prime}} (\HOLConst{RELAB} \HOLBoundVar{labl}))
\end{SaveVerbatim}
\newcommand{\HOLWeakEQTheoremsEPSXXRELAB}{\UseVerbatim{HOLWeakEQTheoremsEPSXXRELAB}}
\begin{SaveVerbatim}{HOLWeakEQTheoremsEPSXXRELABXXrf}
\HOLTokenTurnstile{} \HOLSymConst{\HOLTokenForall{}}\HOLBoundVar{E} \HOLBoundVar{E\sp{\prime}}. \HOLConst{EPS} \HOLBoundVar{E} \HOLBoundVar{E\sp{\prime}} \HOLSymConst{\HOLTokenImp{}} \HOLSymConst{\HOLTokenForall{}}\HOLBoundVar{rf}. \HOLConst{EPS} (\HOLConst{relab} \HOLBoundVar{E} \HOLBoundVar{rf}) (\HOLConst{relab} \HOLBoundVar{E\sp{\prime}} \HOLBoundVar{rf})
\end{SaveVerbatim}
\newcommand{\HOLWeakEQTheoremsEPSXXRELABXXrf}{\UseVerbatim{HOLWeakEQTheoremsEPSXXRELABXXrf}}
\begin{SaveVerbatim}{HOLWeakEQTheoremsEPSXXRESTR}
\HOLTokenTurnstile{} \HOLSymConst{\HOLTokenForall{}}\HOLBoundVar{E} \HOLBoundVar{E\sp{\prime}}. \HOLConst{EPS} \HOLBoundVar{E} \HOLBoundVar{E\sp{\prime}} \HOLSymConst{\HOLTokenImp{}} \HOLSymConst{\HOLTokenForall{}}\HOLBoundVar{L}. \HOLConst{EPS} (\HOLConst{\ensuremath{\nu}} \HOLBoundVar{L} \HOLBoundVar{E}) (\HOLConst{\ensuremath{\nu}} \HOLBoundVar{L} \HOLBoundVar{E\sp{\prime}})
\end{SaveVerbatim}
\newcommand{\HOLWeakEQTheoremsEPSXXRESTR}{\UseVerbatim{HOLWeakEQTheoremsEPSXXRESTR}}
\begin{SaveVerbatim}{HOLWeakEQTheoremsEPSXXSTABLE}
\HOLTokenTurnstile{} \HOLSymConst{\HOLTokenForall{}}\HOLBoundVar{E} \HOLBoundVar{E\sp{\prime}}. \HOLConst{EPS} \HOLBoundVar{E} \HOLBoundVar{E\sp{\prime}} \HOLSymConst{\HOLTokenImp{}} \HOLConst{STABLE} \HOLBoundVar{E} \HOLSymConst{\HOLTokenImp{}} (\HOLBoundVar{E\sp{\prime}} \HOLSymConst{=} \HOLBoundVar{E})
\end{SaveVerbatim}
\newcommand{\HOLWeakEQTheoremsEPSXXSTABLE}{\UseVerbatim{HOLWeakEQTheoremsEPSXXSTABLE}}
\begin{SaveVerbatim}{HOLWeakEQTheoremsEPSXXSTABLEYY}
\HOLTokenTurnstile{} \HOLSymConst{\HOLTokenForall{}}\HOLBoundVar{E} \HOLBoundVar{E\sp{\prime}}. \HOLConst{EPS} \HOLBoundVar{E} \HOLBoundVar{E\sp{\prime}} \HOLSymConst{\HOLTokenConj{}} \HOLConst{STABLE} \HOLBoundVar{E} \HOLSymConst{\HOLTokenImp{}} (\HOLBoundVar{E\sp{\prime}} \HOLSymConst{=} \HOLBoundVar{E})
\end{SaveVerbatim}
\newcommand{\HOLWeakEQTheoremsEPSXXSTABLEYY}{\UseVerbatim{HOLWeakEQTheoremsEPSXXSTABLEYY}}
\begin{SaveVerbatim}{HOLWeakEQTheoremsEPSXXstrongind}
\HOLTokenTurnstile{} \HOLSymConst{\HOLTokenForall{}}\HOLBoundVar{P}.
     (\HOLSymConst{\HOLTokenForall{}}\HOLBoundVar{x}. \HOLBoundVar{P} \HOLBoundVar{x} \HOLBoundVar{x}) \HOLSymConst{\HOLTokenConj{}} (\HOLSymConst{\HOLTokenForall{}}\HOLBoundVar{x} \HOLBoundVar{y} \HOLBoundVar{z}. \HOLBoundVar{x} \HOLTokenTransBegin\HOLConst{\ensuremath{\tau}}\HOLTokenTransEnd \HOLBoundVar{y} \HOLSymConst{\HOLTokenConj{}} \HOLConst{EPS} \HOLBoundVar{y} \HOLBoundVar{z} \HOLSymConst{\HOLTokenConj{}} \HOLBoundVar{P} \HOLBoundVar{y} \HOLBoundVar{z} \HOLSymConst{\HOLTokenImp{}} \HOLBoundVar{P} \HOLBoundVar{x} \HOLBoundVar{z}) \HOLSymConst{\HOLTokenImp{}}
     \HOLSymConst{\HOLTokenForall{}}\HOLBoundVar{x} \HOLBoundVar{y}. \HOLConst{EPS} \HOLBoundVar{x} \HOLBoundVar{y} \HOLSymConst{\HOLTokenImp{}} \HOLBoundVar{P} \HOLBoundVar{x} \HOLBoundVar{y}
\end{SaveVerbatim}
\newcommand{\HOLWeakEQTheoremsEPSXXstrongind}{\UseVerbatim{HOLWeakEQTheoremsEPSXXstrongind}}
\begin{SaveVerbatim}{HOLWeakEQTheoremsEPSXXstrongindXXright}
\HOLTokenTurnstile{} \HOLSymConst{\HOLTokenForall{}}\HOLBoundVar{P}.
     (\HOLSymConst{\HOLTokenForall{}}\HOLBoundVar{x}. \HOLBoundVar{P} \HOLBoundVar{x} \HOLBoundVar{x}) \HOLSymConst{\HOLTokenConj{}} (\HOLSymConst{\HOLTokenForall{}}\HOLBoundVar{x} \HOLBoundVar{y} \HOLBoundVar{z}. \HOLBoundVar{P} \HOLBoundVar{x} \HOLBoundVar{y} \HOLSymConst{\HOLTokenConj{}} \HOLConst{EPS} \HOLBoundVar{x} \HOLBoundVar{y} \HOLSymConst{\HOLTokenConj{}} \HOLBoundVar{y} \HOLTokenTransBegin\HOLConst{\ensuremath{\tau}}\HOLTokenTransEnd \HOLBoundVar{z} \HOLSymConst{\HOLTokenImp{}} \HOLBoundVar{P} \HOLBoundVar{x} \HOLBoundVar{z}) \HOLSymConst{\HOLTokenImp{}}
     \HOLSymConst{\HOLTokenForall{}}\HOLBoundVar{x} \HOLBoundVar{y}. \HOLConst{EPS} \HOLBoundVar{x} \HOLBoundVar{y} \HOLSymConst{\HOLTokenImp{}} \HOLBoundVar{P} \HOLBoundVar{x} \HOLBoundVar{y}
\end{SaveVerbatim}
\newcommand{\HOLWeakEQTheoremsEPSXXstrongindXXright}{\UseVerbatim{HOLWeakEQTheoremsEPSXXstrongindXXright}}
\begin{SaveVerbatim}{HOLWeakEQTheoremsEPSXXTRANS}
\HOLTokenTurnstile{} \HOLSymConst{\HOLTokenForall{}}\HOLBoundVar{x} \HOLBoundVar{y} \HOLBoundVar{z}. \HOLConst{EPS} \HOLBoundVar{x} \HOLBoundVar{y} \HOLSymConst{\HOLTokenConj{}} \HOLConst{EPS} \HOLBoundVar{y} \HOLBoundVar{z} \HOLSymConst{\HOLTokenImp{}} \HOLConst{EPS} \HOLBoundVar{x} \HOLBoundVar{z}
\end{SaveVerbatim}
\newcommand{\HOLWeakEQTheoremsEPSXXTRANS}{\UseVerbatim{HOLWeakEQTheoremsEPSXXTRANS}}
\begin{SaveVerbatim}{HOLWeakEQTheoremsEPSXXTRANSXXAUX}
\HOLTokenTurnstile{} \HOLSymConst{\HOLTokenForall{}}\HOLBoundVar{E} \HOLBoundVar{E\sb{\mathrm{1}}}.
     \HOLConst{EPS} \HOLBoundVar{E} \HOLBoundVar{E\sb{\mathrm{1}}} \HOLSymConst{\HOLTokenImp{}}
     \HOLSymConst{\HOLTokenForall{}}\HOLBoundVar{Wbsm} \HOLBoundVar{E\sp{\prime}}.
       \HOLConst{WEAK_BISIM} \HOLBoundVar{Wbsm} \HOLSymConst{\HOLTokenConj{}} \HOLBoundVar{Wbsm} \HOLBoundVar{E} \HOLBoundVar{E\sp{\prime}} \HOLSymConst{\HOLTokenImp{}} \HOLSymConst{\HOLTokenExists{}}\HOLBoundVar{E\sb{\mathrm{2}}}. \HOLConst{EPS} \HOLBoundVar{E\sp{\prime}} \HOLBoundVar{E\sb{\mathrm{2}}} \HOLSymConst{\HOLTokenConj{}} \HOLBoundVar{Wbsm} \HOLBoundVar{E\sb{\mathrm{1}}} \HOLBoundVar{E\sb{\mathrm{2}}}
\end{SaveVerbatim}
\newcommand{\HOLWeakEQTheoremsEPSXXTRANSXXAUX}{\UseVerbatim{HOLWeakEQTheoremsEPSXXTRANSXXAUX}}
\begin{SaveVerbatim}{HOLWeakEQTheoremsEPSXXTRANSXXAUXXXSYM}
\HOLTokenTurnstile{} \HOLSymConst{\HOLTokenForall{}}\HOLBoundVar{E\sp{\prime}} \HOLBoundVar{E\sb{\mathrm{1}}}.
     \HOLConst{EPS} \HOLBoundVar{E\sp{\prime}} \HOLBoundVar{E\sb{\mathrm{1}}} \HOLSymConst{\HOLTokenImp{}}
     \HOLSymConst{\HOLTokenForall{}}\HOLBoundVar{Wbsm} \HOLBoundVar{E}.
       \HOLConst{WEAK_BISIM} \HOLBoundVar{Wbsm} \HOLSymConst{\HOLTokenConj{}} \HOLBoundVar{Wbsm} \HOLBoundVar{E} \HOLBoundVar{E\sp{\prime}} \HOLSymConst{\HOLTokenImp{}} \HOLSymConst{\HOLTokenExists{}}\HOLBoundVar{E\sb{\mathrm{2}}}. \HOLConst{EPS} \HOLBoundVar{E} \HOLBoundVar{E\sb{\mathrm{2}}} \HOLSymConst{\HOLTokenConj{}} \HOLBoundVar{Wbsm} \HOLBoundVar{E\sb{\mathrm{2}}} \HOLBoundVar{E\sb{\mathrm{1}}}
\end{SaveVerbatim}
\newcommand{\HOLWeakEQTheoremsEPSXXTRANSXXAUXXXSYM}{\UseVerbatim{HOLWeakEQTheoremsEPSXXTRANSXXAUXXXSYM}}
\begin{SaveVerbatim}{HOLWeakEQTheoremsEPSXXWEAKXXEPS}
\HOLTokenTurnstile{} \HOLSymConst{\HOLTokenForall{}}\HOLBoundVar{E} \HOLBoundVar{E\sb{\mathrm{1}}} \HOLBoundVar{u} \HOLBoundVar{E\sb{\mathrm{2}}} \HOLBoundVar{E\sp{\prime}}. \HOLConst{EPS} \HOLBoundVar{E} \HOLBoundVar{E\sb{\mathrm{1}}} \HOLSymConst{\HOLTokenConj{}} \HOLBoundVar{E\sb{\mathrm{1}}} \HOLTokenWeakTransBegin\HOLBoundVar{u}\HOLTokenWeakTransEnd \HOLBoundVar{E\sb{\mathrm{2}}} \HOLSymConst{\HOLTokenConj{}} \HOLConst{EPS} \HOLBoundVar{E\sb{\mathrm{2}}} \HOLBoundVar{E\sp{\prime}} \HOLSymConst{\HOLTokenImp{}} \HOLBoundVar{E} \HOLTokenWeakTransBegin\HOLBoundVar{u}\HOLTokenWeakTransEnd \HOLBoundVar{E\sp{\prime}}
\end{SaveVerbatim}
\newcommand{\HOLWeakEQTheoremsEPSXXWEAKXXEPS}{\UseVerbatim{HOLWeakEQTheoremsEPSXXWEAKXXEPS}}
\begin{SaveVerbatim}{HOLWeakEQTheoremsEQUALXXIMPXXWEAKXXEQUIV}
\HOLTokenTurnstile{} \HOLSymConst{\HOLTokenForall{}}\HOLBoundVar{E} \HOLBoundVar{E\sp{\prime}}. (\HOLBoundVar{E} \HOLSymConst{=} \HOLBoundVar{E\sp{\prime}}) \HOLSymConst{\HOLTokenImp{}} \HOLConst{WEAK_EQUIV} \HOLBoundVar{E} \HOLBoundVar{E\sp{\prime}}
\end{SaveVerbatim}
\newcommand{\HOLWeakEQTheoremsEQUALXXIMPXXWEAKXXEQUIV}{\UseVerbatim{HOLWeakEQTheoremsEQUALXXIMPXXWEAKXXEQUIV}}
\begin{SaveVerbatim}{HOLWeakEQTheoremsIDENTITYXXWEAKXXBISIM}
\HOLTokenTurnstile{} \HOLConst{WEAK_BISIM} (\HOLTokenLambda{}\HOLBoundVar{x} \HOLBoundVar{y}. \HOLBoundVar{x} \HOLSymConst{=} \HOLBoundVar{y})
\end{SaveVerbatim}
\newcommand{\HOLWeakEQTheoremsIDENTITYXXWEAKXXBISIM}{\UseVerbatim{HOLWeakEQTheoremsIDENTITYXXWEAKXXBISIM}}
\begin{SaveVerbatim}{HOLWeakEQTheoremsINVERSEXXREL}
\HOLTokenTurnstile{} \HOLSymConst{\HOLTokenForall{}}\HOLBoundVar{R} \HOLBoundVar{x} \HOLBoundVar{y}. \HOLConst{relinv} \HOLBoundVar{R} \HOLBoundVar{x} \HOLBoundVar{y} \HOLSymConst{\HOLTokenEquiv{}} \HOLBoundVar{R} \HOLBoundVar{y} \HOLBoundVar{x}
\end{SaveVerbatim}
\newcommand{\HOLWeakEQTheoremsINVERSEXXREL}{\UseVerbatim{HOLWeakEQTheoremsINVERSEXXREL}}
\begin{SaveVerbatim}{HOLWeakEQTheoremsONEXXTAU}
\HOLTokenTurnstile{} \HOLSymConst{\HOLTokenForall{}}\HOLBoundVar{E} \HOLBoundVar{E\sp{\prime}}. \HOLBoundVar{E} \HOLTokenTransBegin\HOLConst{\ensuremath{\tau}}\HOLTokenTransEnd \HOLBoundVar{E\sp{\prime}} \HOLSymConst{\HOLTokenImp{}} \HOLConst{EPS} \HOLBoundVar{E} \HOLBoundVar{E\sp{\prime}}
\end{SaveVerbatim}
\newcommand{\HOLWeakEQTheoremsONEXXTAU}{\UseVerbatim{HOLWeakEQTheoremsONEXXTAU}}
\begin{SaveVerbatim}{HOLWeakEQTheoremsSTABLEXXcases}
\HOLTokenTurnstile{} \HOLSymConst{\HOLTokenForall{}}\HOLBoundVar{E}. \HOLConst{STABLE} \HOLBoundVar{E} \HOLSymConst{\HOLTokenDisj{}} \HOLSymConst{\HOLTokenNeg{}}\HOLConst{STABLE} \HOLBoundVar{E}
\end{SaveVerbatim}
\newcommand{\HOLWeakEQTheoremsSTABLEXXcases}{\UseVerbatim{HOLWeakEQTheoremsSTABLEXXcases}}
\begin{SaveVerbatim}{HOLWeakEQTheoremsSTABLEXXNOXXTRANSXXTAU}
\HOLTokenTurnstile{} \HOLSymConst{\HOLTokenForall{}}\HOLBoundVar{E}. \HOLConst{STABLE} \HOLBoundVar{E} \HOLSymConst{\HOLTokenImp{}} \HOLSymConst{\HOLTokenForall{}}\HOLBoundVar{E\sp{\prime}}. \HOLSymConst{\HOLTokenNeg{}}(\HOLBoundVar{E} \HOLTokenTransBegin\HOLConst{\ensuremath{\tau}}\HOLTokenTransEnd \HOLBoundVar{E\sp{\prime}})
\end{SaveVerbatim}
\newcommand{\HOLWeakEQTheoremsSTABLEXXNOXXTRANSXXTAU}{\UseVerbatim{HOLWeakEQTheoremsSTABLEXXNOXXTRANSXXTAU}}
\begin{SaveVerbatim}{HOLWeakEQTheoremsSTABLEXXNOXXWEAKXXTRANSXXTAU}
\HOLTokenTurnstile{} \HOLSymConst{\HOLTokenForall{}}\HOLBoundVar{E}. \HOLConst{STABLE} \HOLBoundVar{E} \HOLSymConst{\HOLTokenImp{}} \HOLSymConst{\HOLTokenForall{}}\HOLBoundVar{E\sp{\prime}}. \HOLSymConst{\HOLTokenNeg{}}(\HOLBoundVar{E} \HOLTokenWeakTransBegin\HOLConst{\ensuremath{\tau}}\HOLTokenWeakTransEnd \HOLBoundVar{E\sp{\prime}})
\end{SaveVerbatim}
\newcommand{\HOLWeakEQTheoremsSTABLEXXNOXXWEAKXXTRANSXXTAU}{\UseVerbatim{HOLWeakEQTheoremsSTABLEXXNOXXWEAKXXTRANSXXTAU}}
\begin{SaveVerbatim}{HOLWeakEQTheoremsSTRONGXXEQUIVXXEPS}
\HOLTokenTurnstile{} \HOLSymConst{\HOLTokenForall{}}\HOLBoundVar{E} \HOLBoundVar{E\sp{\prime}}.
     \HOLConst{STRONG_EQUIV} \HOLBoundVar{E} \HOLBoundVar{E\sp{\prime}} \HOLSymConst{\HOLTokenImp{}}
     \HOLSymConst{\HOLTokenForall{}}\HOLBoundVar{E\sb{\mathrm{1}}}. \HOLConst{EPS} \HOLBoundVar{E} \HOLBoundVar{E\sb{\mathrm{1}}} \HOLSymConst{\HOLTokenImp{}} \HOLSymConst{\HOLTokenExists{}}\HOLBoundVar{E\sb{\mathrm{2}}}. \HOLConst{EPS} \HOLBoundVar{E\sp{\prime}} \HOLBoundVar{E\sb{\mathrm{2}}} \HOLSymConst{\HOLTokenConj{}} \HOLConst{STRONG_EQUIV} \HOLBoundVar{E\sb{\mathrm{1}}} \HOLBoundVar{E\sb{\mathrm{2}}}
\end{SaveVerbatim}
\newcommand{\HOLWeakEQTheoremsSTRONGXXEQUIVXXEPS}{\UseVerbatim{HOLWeakEQTheoremsSTRONGXXEQUIVXXEPS}}
\begin{SaveVerbatim}{HOLWeakEQTheoremsSTRONGXXEQUIVXXEPSYY}
\HOLTokenTurnstile{} \HOLSymConst{\HOLTokenForall{}}\HOLBoundVar{E} \HOLBoundVar{E\sp{\prime}}.
     \HOLConst{STRONG_EQUIV} \HOLBoundVar{E} \HOLBoundVar{E\sp{\prime}} \HOLSymConst{\HOLTokenImp{}}
     \HOLSymConst{\HOLTokenForall{}}\HOLBoundVar{E\sb{\mathrm{2}}}. \HOLConst{EPS} \HOLBoundVar{E\sp{\prime}} \HOLBoundVar{E\sb{\mathrm{2}}} \HOLSymConst{\HOLTokenImp{}} \HOLSymConst{\HOLTokenExists{}}\HOLBoundVar{E\sb{\mathrm{1}}}. \HOLConst{EPS} \HOLBoundVar{E} \HOLBoundVar{E\sb{\mathrm{1}}} \HOLSymConst{\HOLTokenConj{}} \HOLConst{STRONG_EQUIV} \HOLBoundVar{E\sb{\mathrm{1}}} \HOLBoundVar{E\sb{\mathrm{2}}}
\end{SaveVerbatim}
\newcommand{\HOLWeakEQTheoremsSTRONGXXEQUIVXXEPSYY}{\UseVerbatim{HOLWeakEQTheoremsSTRONGXXEQUIVXXEPSYY}}
\begin{SaveVerbatim}{HOLWeakEQTheoremsSTRONGXXEQUIVXXWEAKXXTRANS}
\HOLTokenTurnstile{} \HOLSymConst{\HOLTokenForall{}}\HOLBoundVar{E} \HOLBoundVar{E\sp{\prime}}.
     \HOLConst{STRONG_EQUIV} \HOLBoundVar{E} \HOLBoundVar{E\sp{\prime}} \HOLSymConst{\HOLTokenImp{}}
     \HOLSymConst{\HOLTokenForall{}}\HOLBoundVar{u} \HOLBoundVar{E\sb{\mathrm{1}}}. \HOLBoundVar{E} \HOLTokenWeakTransBegin\HOLBoundVar{u}\HOLTokenWeakTransEnd \HOLBoundVar{E\sb{\mathrm{1}}} \HOLSymConst{\HOLTokenImp{}} \HOLSymConst{\HOLTokenExists{}}\HOLBoundVar{E\sb{\mathrm{2}}}. \HOLBoundVar{E\sp{\prime}} \HOLTokenWeakTransBegin\HOLBoundVar{u}\HOLTokenWeakTransEnd \HOLBoundVar{E\sb{\mathrm{2}}} \HOLSymConst{\HOLTokenConj{}} \HOLConst{STRONG_EQUIV} \HOLBoundVar{E\sb{\mathrm{1}}} \HOLBoundVar{E\sb{\mathrm{2}}}
\end{SaveVerbatim}
\newcommand{\HOLWeakEQTheoremsSTRONGXXEQUIVXXWEAKXXTRANS}{\UseVerbatim{HOLWeakEQTheoremsSTRONGXXEQUIVXXWEAKXXTRANS}}
\begin{SaveVerbatim}{HOLWeakEQTheoremsSTRONGXXEQUIVXXWEAKXXTRANSYY}
\HOLTokenTurnstile{} \HOLSymConst{\HOLTokenForall{}}\HOLBoundVar{E} \HOLBoundVar{E\sp{\prime}}.
     \HOLConst{STRONG_EQUIV} \HOLBoundVar{E} \HOLBoundVar{E\sp{\prime}} \HOLSymConst{\HOLTokenImp{}}
     \HOLSymConst{\HOLTokenForall{}}\HOLBoundVar{u} \HOLBoundVar{E\sb{\mathrm{2}}}. \HOLBoundVar{E\sp{\prime}} \HOLTokenWeakTransBegin\HOLBoundVar{u}\HOLTokenWeakTransEnd \HOLBoundVar{E\sb{\mathrm{2}}} \HOLSymConst{\HOLTokenImp{}} \HOLSymConst{\HOLTokenExists{}}\HOLBoundVar{E\sb{\mathrm{1}}}. \HOLBoundVar{E} \HOLTokenWeakTransBegin\HOLBoundVar{u}\HOLTokenWeakTransEnd \HOLBoundVar{E\sb{\mathrm{1}}} \HOLSymConst{\HOLTokenConj{}} \HOLConst{STRONG_EQUIV} \HOLBoundVar{E\sb{\mathrm{1}}} \HOLBoundVar{E\sb{\mathrm{2}}}
\end{SaveVerbatim}
\newcommand{\HOLWeakEQTheoremsSTRONGXXEQUIVXXWEAKXXTRANSYY}{\UseVerbatim{HOLWeakEQTheoremsSTRONGXXEQUIVXXWEAKXXTRANSYY}}
\begin{SaveVerbatim}{HOLWeakEQTheoremsSTRONGXXIMPXXWEAKXXBISIM}
\HOLTokenTurnstile{} \HOLSymConst{\HOLTokenForall{}}\HOLBoundVar{Bsm}. \HOLConst{STRONG_BISIM} \HOLBoundVar{Bsm} \HOLSymConst{\HOLTokenImp{}} \HOLConst{WEAK_BISIM} \HOLBoundVar{Bsm}
\end{SaveVerbatim}
\newcommand{\HOLWeakEQTheoremsSTRONGXXIMPXXWEAKXXBISIM}{\UseVerbatim{HOLWeakEQTheoremsSTRONGXXIMPXXWEAKXXBISIM}}
\begin{SaveVerbatim}{HOLWeakEQTheoremsSTRONGXXIMPXXWEAKXXEQUIV}
\HOLTokenTurnstile{} \HOLSymConst{\HOLTokenForall{}}\HOLBoundVar{E} \HOLBoundVar{E\sp{\prime}}. \HOLConst{STRONG_EQUIV} \HOLBoundVar{E} \HOLBoundVar{E\sp{\prime}} \HOLSymConst{\HOLTokenImp{}} \HOLConst{WEAK_EQUIV} \HOLBoundVar{E} \HOLBoundVar{E\sp{\prime}}
\end{SaveVerbatim}
\newcommand{\HOLWeakEQTheoremsSTRONGXXIMPXXWEAKXXEQUIV}{\UseVerbatim{HOLWeakEQTheoremsSTRONGXXIMPXXWEAKXXEQUIV}}
\begin{SaveVerbatim}{HOLWeakEQTheoremsTAUXXPREFIXXXEPS}
\HOLTokenTurnstile{} \HOLSymConst{\HOLTokenForall{}}\HOLBoundVar{E} \HOLBoundVar{E\sp{\prime}}. \HOLConst{EPS} \HOLBoundVar{E} \HOLBoundVar{E\sp{\prime}} \HOLSymConst{\HOLTokenImp{}} \HOLConst{EPS} (\HOLConst{\ensuremath{\tau}}\HOLSymConst{..}\HOLBoundVar{E}) \HOLBoundVar{E\sp{\prime}}
\end{SaveVerbatim}
\newcommand{\HOLWeakEQTheoremsTAUXXPREFIXXXEPS}{\UseVerbatim{HOLWeakEQTheoremsTAUXXPREFIXXXEPS}}
\begin{SaveVerbatim}{HOLWeakEQTheoremsTAUXXPREFIXXXWEAKXXTRANS}
\HOLTokenTurnstile{} \HOLSymConst{\HOLTokenForall{}}\HOLBoundVar{E} \HOLBoundVar{u} \HOLBoundVar{E\sp{\prime}}. \HOLBoundVar{E} \HOLTokenWeakTransBegin\HOLBoundVar{u}\HOLTokenWeakTransEnd \HOLBoundVar{E\sp{\prime}} \HOLSymConst{\HOLTokenImp{}} \HOLConst{\ensuremath{\tau}}\HOLSymConst{..}\HOLBoundVar{E} \HOLTokenWeakTransBegin\HOLBoundVar{u}\HOLTokenWeakTransEnd \HOLBoundVar{E\sp{\prime}}
\end{SaveVerbatim}
\newcommand{\HOLWeakEQTheoremsTAUXXPREFIXXXWEAKXXTRANS}{\UseVerbatim{HOLWeakEQTheoremsTAUXXPREFIXXXWEAKXXTRANS}}
\begin{SaveVerbatim}{HOLWeakEQTheoremsTRANSXXANDXXEPS}
\HOLTokenTurnstile{} \HOLSymConst{\HOLTokenForall{}}\HOLBoundVar{E} \HOLBoundVar{u} \HOLBoundVar{E\sb{\mathrm{1}}} \HOLBoundVar{E\sp{\prime}}. \HOLBoundVar{E} \HOLTokenTransBegin\HOLBoundVar{u}\HOLTokenTransEnd \HOLBoundVar{E\sb{\mathrm{1}}} \HOLSymConst{\HOLTokenConj{}} \HOLConst{EPS} \HOLBoundVar{E\sb{\mathrm{1}}} \HOLBoundVar{E\sp{\prime}} \HOLSymConst{\HOLTokenImp{}} \HOLBoundVar{E} \HOLTokenWeakTransBegin\HOLBoundVar{u}\HOLTokenWeakTransEnd \HOLBoundVar{E\sp{\prime}}
\end{SaveVerbatim}
\newcommand{\HOLWeakEQTheoremsTRANSXXANDXXEPS}{\UseVerbatim{HOLWeakEQTheoremsTRANSXXANDXXEPS}}
\begin{SaveVerbatim}{HOLWeakEQTheoremsTRANSXXIMPXXWEAKXXTRANS}
\HOLTokenTurnstile{} \HOLSymConst{\HOLTokenForall{}}\HOLBoundVar{E} \HOLBoundVar{u} \HOLBoundVar{E\sp{\prime}}. \HOLBoundVar{E} \HOLTokenTransBegin\HOLBoundVar{u}\HOLTokenTransEnd \HOLBoundVar{E\sp{\prime}} \HOLSymConst{\HOLTokenImp{}} \HOLBoundVar{E} \HOLTokenWeakTransBegin\HOLBoundVar{u}\HOLTokenWeakTransEnd \HOLBoundVar{E\sp{\prime}}
\end{SaveVerbatim}
\newcommand{\HOLWeakEQTheoremsTRANSXXIMPXXWEAKXXTRANS}{\UseVerbatim{HOLWeakEQTheoremsTRANSXXIMPXXWEAKXXTRANS}}
\begin{SaveVerbatim}{HOLWeakEQTheoremsTRANSXXTAUXXANDXXWEAK}
\HOLTokenTurnstile{} \HOLSymConst{\HOLTokenForall{}}\HOLBoundVar{E} \HOLBoundVar{E\sb{\mathrm{1}}} \HOLBoundVar{u} \HOLBoundVar{E\sp{\prime}}. \HOLBoundVar{E} \HOLTokenTransBegin\HOLConst{\ensuremath{\tau}}\HOLTokenTransEnd \HOLBoundVar{E\sb{\mathrm{1}}} \HOLSymConst{\HOLTokenConj{}} \HOLBoundVar{E\sb{\mathrm{1}}} \HOLTokenWeakTransBegin\HOLBoundVar{u}\HOLTokenWeakTransEnd \HOLBoundVar{E\sp{\prime}} \HOLSymConst{\HOLTokenImp{}} \HOLBoundVar{E} \HOLTokenWeakTransBegin\HOLBoundVar{u}\HOLTokenWeakTransEnd \HOLBoundVar{E\sp{\prime}}
\end{SaveVerbatim}
\newcommand{\HOLWeakEQTheoremsTRANSXXTAUXXANDXXWEAK}{\UseVerbatim{HOLWeakEQTheoremsTRANSXXTAUXXANDXXWEAK}}
\begin{SaveVerbatim}{HOLWeakEQTheoremsTRANSXXTAUXXIMPXXEPS}
\HOLTokenTurnstile{} \HOLSymConst{\HOLTokenForall{}}\HOLBoundVar{E} \HOLBoundVar{E\sp{\prime}}. \HOLBoundVar{E} \HOLTokenTransBegin\HOLConst{\ensuremath{\tau}}\HOLTokenTransEnd \HOLBoundVar{E\sp{\prime}} \HOLSymConst{\HOLTokenImp{}} \HOLConst{EPS} \HOLBoundVar{E} \HOLBoundVar{E\sp{\prime}}
\end{SaveVerbatim}
\newcommand{\HOLWeakEQTheoremsTRANSXXTAUXXIMPXXEPS}{\UseVerbatim{HOLWeakEQTheoremsTRANSXXTAUXXIMPXXEPS}}
\begin{SaveVerbatim}{HOLWeakEQTheoremsUNIONXXWEAKXXBISIM}
\HOLTokenTurnstile{} \HOLSymConst{\HOLTokenForall{}}\HOLBoundVar{Wbsm\sb{\mathrm{1}}} \HOLBoundVar{Wbsm\sb{\mathrm{2}}}.
     \HOLConst{WEAK_BISIM} \HOLBoundVar{Wbsm\sb{\mathrm{1}}} \HOLSymConst{\HOLTokenConj{}} \HOLConst{WEAK_BISIM} \HOLBoundVar{Wbsm\sb{\mathrm{2}}} \HOLSymConst{\HOLTokenImp{}}
     \HOLConst{WEAK_BISIM} (\HOLBoundVar{Wbsm\sb{\mathrm{1}}} \HOLConst{RUNION} \HOLBoundVar{Wbsm\sb{\mathrm{2}}})
\end{SaveVerbatim}
\newcommand{\HOLWeakEQTheoremsUNIONXXWEAKXXBISIM}{\UseVerbatim{HOLWeakEQTheoremsUNIONXXWEAKXXBISIM}}
\begin{SaveVerbatim}{HOLWeakEQTheoremsWEAKXXBISIM}
\HOLTokenTurnstile{} \HOLConst{WEAK_BISIM} \HOLFreeVar{Wbsm} \HOLSymConst{\HOLTokenEquiv{}}
   \HOLSymConst{\HOLTokenForall{}}\HOLBoundVar{E} \HOLBoundVar{E\sp{\prime}}.
     \HOLFreeVar{Wbsm} \HOLBoundVar{E} \HOLBoundVar{E\sp{\prime}} \HOLSymConst{\HOLTokenImp{}}
     (\HOLSymConst{\HOLTokenForall{}}\HOLBoundVar{l}.
        (\HOLSymConst{\HOLTokenForall{}}\HOLBoundVar{E\sb{\mathrm{1}}}.
           \HOLBoundVar{E} \HOLTokenTransBegin\HOLConst{label} \HOLBoundVar{l}\HOLTokenTransEnd \HOLBoundVar{E\sb{\mathrm{1}}} \HOLSymConst{\HOLTokenImp{}} \HOLSymConst{\HOLTokenExists{}}\HOLBoundVar{E\sb{\mathrm{2}}}. \HOLBoundVar{E\sp{\prime}} \HOLTokenWeakTransBegin\HOLConst{label} \HOLBoundVar{l}\HOLTokenWeakTransEnd \HOLBoundVar{E\sb{\mathrm{2}}} \HOLSymConst{\HOLTokenConj{}} \HOLFreeVar{Wbsm} \HOLBoundVar{E\sb{\mathrm{1}}} \HOLBoundVar{E\sb{\mathrm{2}}}) \HOLSymConst{\HOLTokenConj{}}
        \HOLSymConst{\HOLTokenForall{}}\HOLBoundVar{E\sb{\mathrm{2}}}.
          \HOLBoundVar{E\sp{\prime}} \HOLTokenTransBegin\HOLConst{label} \HOLBoundVar{l}\HOLTokenTransEnd \HOLBoundVar{E\sb{\mathrm{2}}} \HOLSymConst{\HOLTokenImp{}} \HOLSymConst{\HOLTokenExists{}}\HOLBoundVar{E\sb{\mathrm{1}}}. \HOLBoundVar{E} \HOLTokenWeakTransBegin\HOLConst{label} \HOLBoundVar{l}\HOLTokenWeakTransEnd \HOLBoundVar{E\sb{\mathrm{1}}} \HOLSymConst{\HOLTokenConj{}} \HOLFreeVar{Wbsm} \HOLBoundVar{E\sb{\mathrm{1}}} \HOLBoundVar{E\sb{\mathrm{2}}}) \HOLSymConst{\HOLTokenConj{}}
     (\HOLSymConst{\HOLTokenForall{}}\HOLBoundVar{E\sb{\mathrm{1}}}. \HOLBoundVar{E} \HOLTokenTransBegin\HOLConst{\ensuremath{\tau}}\HOLTokenTransEnd \HOLBoundVar{E\sb{\mathrm{1}}} \HOLSymConst{\HOLTokenImp{}} \HOLSymConst{\HOLTokenExists{}}\HOLBoundVar{E\sb{\mathrm{2}}}. \HOLConst{EPS} \HOLBoundVar{E\sp{\prime}} \HOLBoundVar{E\sb{\mathrm{2}}} \HOLSymConst{\HOLTokenConj{}} \HOLFreeVar{Wbsm} \HOLBoundVar{E\sb{\mathrm{1}}} \HOLBoundVar{E\sb{\mathrm{2}}}) \HOLSymConst{\HOLTokenConj{}}
     \HOLSymConst{\HOLTokenForall{}}\HOLBoundVar{E\sb{\mathrm{2}}}. \HOLBoundVar{E\sp{\prime}} \HOLTokenTransBegin\HOLConst{\ensuremath{\tau}}\HOLTokenTransEnd \HOLBoundVar{E\sb{\mathrm{2}}} \HOLSymConst{\HOLTokenImp{}} \HOLSymConst{\HOLTokenExists{}}\HOLBoundVar{E\sb{\mathrm{1}}}. \HOLConst{EPS} \HOLBoundVar{E} \HOLBoundVar{E\sb{\mathrm{1}}} \HOLSymConst{\HOLTokenConj{}} \HOLFreeVar{Wbsm} \HOLBoundVar{E\sb{\mathrm{1}}} \HOLBoundVar{E\sb{\mathrm{2}}}
\end{SaveVerbatim}
\newcommand{\HOLWeakEQTheoremsWEAKXXBISIM}{\UseVerbatim{HOLWeakEQTheoremsWEAKXXBISIM}}
\begin{SaveVerbatim}{HOLWeakEQTheoremsWEAKXXBISIMXXSUBSETXXWEAKXXEQUIV}
\HOLTokenTurnstile{} \HOLSymConst{\HOLTokenForall{}}\HOLBoundVar{Wbsm}. \HOLConst{WEAK_BISIM} \HOLBoundVar{Wbsm} \HOLSymConst{\HOLTokenImp{}} \HOLBoundVar{Wbsm} \HOLConst{RSUBSET} \HOLConst{WEAK_EQUIV}
\end{SaveVerbatim}
\newcommand{\HOLWeakEQTheoremsWEAKXXBISIMXXSUBSETXXWEAKXXEQUIV}{\UseVerbatim{HOLWeakEQTheoremsWEAKXXBISIMXXSUBSETXXWEAKXXEQUIV}}
\begin{SaveVerbatim}{HOLWeakEQTheoremsWEAKXXEQUIV}
\HOLTokenTurnstile{} \HOLSymConst{\HOLTokenForall{}}\HOLBoundVar{E} \HOLBoundVar{E\sp{\prime}}. \HOLConst{WEAK_EQUIV} \HOLBoundVar{E} \HOLBoundVar{E\sp{\prime}} \HOLSymConst{\HOLTokenEquiv{}} \HOLSymConst{\HOLTokenExists{}}\HOLBoundVar{Wbsm}. \HOLBoundVar{Wbsm} \HOLBoundVar{E} \HOLBoundVar{E\sp{\prime}} \HOLSymConst{\HOLTokenConj{}} \HOLConst{WEAK_BISIM} \HOLBoundVar{Wbsm}
\end{SaveVerbatim}
\newcommand{\HOLWeakEQTheoremsWEAKXXEQUIV}{\UseVerbatim{HOLWeakEQTheoremsWEAKXXEQUIV}}
\begin{SaveVerbatim}{HOLWeakEQTheoremsWEAKXXEQUIVXXcases}
\HOLTokenTurnstile{} \HOLSymConst{\HOLTokenForall{}}\HOLBoundVar{a\sb{\mathrm{0}}} \HOLBoundVar{a\sb{\mathrm{1}}}.
     \HOLConst{WEAK_EQUIV} \HOLBoundVar{a\sb{\mathrm{0}}} \HOLBoundVar{a\sb{\mathrm{1}}} \HOLSymConst{\HOLTokenEquiv{}}
     (\HOLSymConst{\HOLTokenForall{}}\HOLBoundVar{l}.
        (\HOLSymConst{\HOLTokenForall{}}\HOLBoundVar{E\sb{\mathrm{1}}}.
           \HOLBoundVar{a\sb{\mathrm{0}}} \HOLTokenTransBegin\HOLConst{label} \HOLBoundVar{l}\HOLTokenTransEnd \HOLBoundVar{E\sb{\mathrm{1}}} \HOLSymConst{\HOLTokenImp{}}
           \HOLSymConst{\HOLTokenExists{}}\HOLBoundVar{E\sb{\mathrm{2}}}. \HOLBoundVar{a\sb{\mathrm{1}}} \HOLTokenWeakTransBegin\HOLConst{label} \HOLBoundVar{l}\HOLTokenWeakTransEnd \HOLBoundVar{E\sb{\mathrm{2}}} \HOLSymConst{\HOLTokenConj{}} \HOLConst{WEAK_EQUIV} \HOLBoundVar{E\sb{\mathrm{1}}} \HOLBoundVar{E\sb{\mathrm{2}}}) \HOLSymConst{\HOLTokenConj{}}
        \HOLSymConst{\HOLTokenForall{}}\HOLBoundVar{E\sb{\mathrm{2}}}.
          \HOLBoundVar{a\sb{\mathrm{1}}} \HOLTokenTransBegin\HOLConst{label} \HOLBoundVar{l}\HOLTokenTransEnd \HOLBoundVar{E\sb{\mathrm{2}}} \HOLSymConst{\HOLTokenImp{}}
          \HOLSymConst{\HOLTokenExists{}}\HOLBoundVar{E\sb{\mathrm{1}}}. \HOLBoundVar{a\sb{\mathrm{0}}} \HOLTokenWeakTransBegin\HOLConst{label} \HOLBoundVar{l}\HOLTokenWeakTransEnd \HOLBoundVar{E\sb{\mathrm{1}}} \HOLSymConst{\HOLTokenConj{}} \HOLConst{WEAK_EQUIV} \HOLBoundVar{E\sb{\mathrm{1}}} \HOLBoundVar{E\sb{\mathrm{2}}}) \HOLSymConst{\HOLTokenConj{}}
     (\HOLSymConst{\HOLTokenForall{}}\HOLBoundVar{E\sb{\mathrm{1}}}. \HOLBoundVar{a\sb{\mathrm{0}}} \HOLTokenTransBegin\HOLConst{\ensuremath{\tau}}\HOLTokenTransEnd \HOLBoundVar{E\sb{\mathrm{1}}} \HOLSymConst{\HOLTokenImp{}} \HOLSymConst{\HOLTokenExists{}}\HOLBoundVar{E\sb{\mathrm{2}}}. \HOLConst{EPS} \HOLBoundVar{a\sb{\mathrm{1}}} \HOLBoundVar{E\sb{\mathrm{2}}} \HOLSymConst{\HOLTokenConj{}} \HOLConst{WEAK_EQUIV} \HOLBoundVar{E\sb{\mathrm{1}}} \HOLBoundVar{E\sb{\mathrm{2}}}) \HOLSymConst{\HOLTokenConj{}}
     \HOLSymConst{\HOLTokenForall{}}\HOLBoundVar{E\sb{\mathrm{2}}}. \HOLBoundVar{a\sb{\mathrm{1}}} \HOLTokenTransBegin\HOLConst{\ensuremath{\tau}}\HOLTokenTransEnd \HOLBoundVar{E\sb{\mathrm{2}}} \HOLSymConst{\HOLTokenImp{}} \HOLSymConst{\HOLTokenExists{}}\HOLBoundVar{E\sb{\mathrm{1}}}. \HOLConst{EPS} \HOLBoundVar{a\sb{\mathrm{0}}} \HOLBoundVar{E\sb{\mathrm{1}}} \HOLSymConst{\HOLTokenConj{}} \HOLConst{WEAK_EQUIV} \HOLBoundVar{E\sb{\mathrm{1}}} \HOLBoundVar{E\sb{\mathrm{2}}}
\end{SaveVerbatim}
\newcommand{\HOLWeakEQTheoremsWEAKXXEQUIVXXcases}{\UseVerbatim{HOLWeakEQTheoremsWEAKXXEQUIVXXcases}}
\begin{SaveVerbatim}{HOLWeakEQTheoremsWEAKXXEQUIVXXcoind}
\HOLTokenTurnstile{} \HOLSymConst{\HOLTokenForall{}}\HOLBoundVar{WEAK\HOLTokenUnderscore{}EQUIV\sp{\prime}}.
     (\HOLSymConst{\HOLTokenForall{}}\HOLBoundVar{a\sb{\mathrm{0}}} \HOLBoundVar{a\sb{\mathrm{1}}}.
        \HOLBoundVar{WEAK\HOLTokenUnderscore{}EQUIV\sp{\prime}} \HOLBoundVar{a\sb{\mathrm{0}}} \HOLBoundVar{a\sb{\mathrm{1}}} \HOLSymConst{\HOLTokenImp{}}
        (\HOLSymConst{\HOLTokenForall{}}\HOLBoundVar{l}.
           (\HOLSymConst{\HOLTokenForall{}}\HOLBoundVar{E\sb{\mathrm{1}}}.
              \HOLBoundVar{a\sb{\mathrm{0}}} \HOLTokenTransBegin\HOLConst{label} \HOLBoundVar{l}\HOLTokenTransEnd \HOLBoundVar{E\sb{\mathrm{1}}} \HOLSymConst{\HOLTokenImp{}}
              \HOLSymConst{\HOLTokenExists{}}\HOLBoundVar{E\sb{\mathrm{2}}}. \HOLBoundVar{a\sb{\mathrm{1}}} \HOLTokenWeakTransBegin\HOLConst{label} \HOLBoundVar{l}\HOLTokenWeakTransEnd \HOLBoundVar{E\sb{\mathrm{2}}} \HOLSymConst{\HOLTokenConj{}} \HOLBoundVar{WEAK\HOLTokenUnderscore{}EQUIV\sp{\prime}} \HOLBoundVar{E\sb{\mathrm{1}}} \HOLBoundVar{E\sb{\mathrm{2}}}) \HOLSymConst{\HOLTokenConj{}}
           \HOLSymConst{\HOLTokenForall{}}\HOLBoundVar{E\sb{\mathrm{2}}}.
             \HOLBoundVar{a\sb{\mathrm{1}}} \HOLTokenTransBegin\HOLConst{label} \HOLBoundVar{l}\HOLTokenTransEnd \HOLBoundVar{E\sb{\mathrm{2}}} \HOLSymConst{\HOLTokenImp{}}
             \HOLSymConst{\HOLTokenExists{}}\HOLBoundVar{E\sb{\mathrm{1}}}. \HOLBoundVar{a\sb{\mathrm{0}}} \HOLTokenWeakTransBegin\HOLConst{label} \HOLBoundVar{l}\HOLTokenWeakTransEnd \HOLBoundVar{E\sb{\mathrm{1}}} \HOLSymConst{\HOLTokenConj{}} \HOLBoundVar{WEAK\HOLTokenUnderscore{}EQUIV\sp{\prime}} \HOLBoundVar{E\sb{\mathrm{1}}} \HOLBoundVar{E\sb{\mathrm{2}}}) \HOLSymConst{\HOLTokenConj{}}
        (\HOLSymConst{\HOLTokenForall{}}\HOLBoundVar{E\sb{\mathrm{1}}}. \HOLBoundVar{a\sb{\mathrm{0}}} \HOLTokenTransBegin\HOLConst{\ensuremath{\tau}}\HOLTokenTransEnd \HOLBoundVar{E\sb{\mathrm{1}}} \HOLSymConst{\HOLTokenImp{}} \HOLSymConst{\HOLTokenExists{}}\HOLBoundVar{E\sb{\mathrm{2}}}. \HOLConst{EPS} \HOLBoundVar{a\sb{\mathrm{1}}} \HOLBoundVar{E\sb{\mathrm{2}}} \HOLSymConst{\HOLTokenConj{}} \HOLBoundVar{WEAK\HOLTokenUnderscore{}EQUIV\sp{\prime}} \HOLBoundVar{E\sb{\mathrm{1}}} \HOLBoundVar{E\sb{\mathrm{2}}}) \HOLSymConst{\HOLTokenConj{}}
        \HOLSymConst{\HOLTokenForall{}}\HOLBoundVar{E\sb{\mathrm{2}}}. \HOLBoundVar{a\sb{\mathrm{1}}} \HOLTokenTransBegin\HOLConst{\ensuremath{\tau}}\HOLTokenTransEnd \HOLBoundVar{E\sb{\mathrm{2}}} \HOLSymConst{\HOLTokenImp{}} \HOLSymConst{\HOLTokenExists{}}\HOLBoundVar{E\sb{\mathrm{1}}}. \HOLConst{EPS} \HOLBoundVar{a\sb{\mathrm{0}}} \HOLBoundVar{E\sb{\mathrm{1}}} \HOLSymConst{\HOLTokenConj{}} \HOLBoundVar{WEAK\HOLTokenUnderscore{}EQUIV\sp{\prime}} \HOLBoundVar{E\sb{\mathrm{1}}} \HOLBoundVar{E\sb{\mathrm{2}}}) \HOLSymConst{\HOLTokenImp{}}
     \HOLSymConst{\HOLTokenForall{}}\HOLBoundVar{a\sb{\mathrm{0}}} \HOLBoundVar{a\sb{\mathrm{1}}}. \HOLBoundVar{WEAK\HOLTokenUnderscore{}EQUIV\sp{\prime}} \HOLBoundVar{a\sb{\mathrm{0}}} \HOLBoundVar{a\sb{\mathrm{1}}} \HOLSymConst{\HOLTokenImp{}} \HOLConst{WEAK_EQUIV} \HOLBoundVar{a\sb{\mathrm{0}}} \HOLBoundVar{a\sb{\mathrm{1}}}
\end{SaveVerbatim}
\newcommand{\HOLWeakEQTheoremsWEAKXXEQUIVXXcoind}{\UseVerbatim{HOLWeakEQTheoremsWEAKXXEQUIVXXcoind}}
\begin{SaveVerbatim}{HOLWeakEQTheoremsWEAKXXEQUIVXXEPS}
\HOLTokenTurnstile{} \HOLSymConst{\HOLTokenForall{}}\HOLBoundVar{E} \HOLBoundVar{E\sp{\prime}}.
     \HOLConst{WEAK_EQUIV} \HOLBoundVar{E} \HOLBoundVar{E\sp{\prime}} \HOLSymConst{\HOLTokenImp{}}
     \HOLSymConst{\HOLTokenForall{}}\HOLBoundVar{E\sb{\mathrm{1}}}. \HOLConst{EPS} \HOLBoundVar{E} \HOLBoundVar{E\sb{\mathrm{1}}} \HOLSymConst{\HOLTokenImp{}} \HOLSymConst{\HOLTokenExists{}}\HOLBoundVar{E\sb{\mathrm{2}}}. \HOLConst{EPS} \HOLBoundVar{E\sp{\prime}} \HOLBoundVar{E\sb{\mathrm{2}}} \HOLSymConst{\HOLTokenConj{}} \HOLConst{WEAK_EQUIV} \HOLBoundVar{E\sb{\mathrm{1}}} \HOLBoundVar{E\sb{\mathrm{2}}}
\end{SaveVerbatim}
\newcommand{\HOLWeakEQTheoremsWEAKXXEQUIVXXEPS}{\UseVerbatim{HOLWeakEQTheoremsWEAKXXEQUIVXXEPS}}
\begin{SaveVerbatim}{HOLWeakEQTheoremsWEAKXXEQUIVXXEPSYY}
\HOLTokenTurnstile{} \HOLSymConst{\HOLTokenForall{}}\HOLBoundVar{E} \HOLBoundVar{E\sp{\prime}}.
     \HOLConst{WEAK_EQUIV} \HOLBoundVar{E} \HOLBoundVar{E\sp{\prime}} \HOLSymConst{\HOLTokenImp{}}
     \HOLSymConst{\HOLTokenForall{}}\HOLBoundVar{E\sb{\mathrm{2}}}. \HOLConst{EPS} \HOLBoundVar{E\sp{\prime}} \HOLBoundVar{E\sb{\mathrm{2}}} \HOLSymConst{\HOLTokenImp{}} \HOLSymConst{\HOLTokenExists{}}\HOLBoundVar{E\sb{\mathrm{1}}}. \HOLConst{EPS} \HOLBoundVar{E} \HOLBoundVar{E\sb{\mathrm{1}}} \HOLSymConst{\HOLTokenConj{}} \HOLConst{WEAK_EQUIV} \HOLBoundVar{E\sb{\mathrm{1}}} \HOLBoundVar{E\sb{\mathrm{2}}}
\end{SaveVerbatim}
\newcommand{\HOLWeakEQTheoremsWEAKXXEQUIVXXEPSYY}{\UseVerbatim{HOLWeakEQTheoremsWEAKXXEQUIVXXEPSYY}}
\begin{SaveVerbatim}{HOLWeakEQTheoremsWEAKXXEQUIVXXequivalence}
\HOLTokenTurnstile{} \HOLConst{equivalence} \HOLConst{WEAK_EQUIV}
\end{SaveVerbatim}
\newcommand{\HOLWeakEQTheoremsWEAKXXEQUIVXXequivalence}{\UseVerbatim{HOLWeakEQTheoremsWEAKXXEQUIVXXequivalence}}
\begin{SaveVerbatim}{HOLWeakEQTheoremsWEAKXXEQUIVXXISXXWEAKXXBISIM}
\HOLTokenTurnstile{} \HOLConst{WEAK_BISIM} \HOLConst{WEAK_EQUIV}
\end{SaveVerbatim}
\newcommand{\HOLWeakEQTheoremsWEAKXXEQUIVXXISXXWEAKXXBISIM}{\UseVerbatim{HOLWeakEQTheoremsWEAKXXEQUIVXXISXXWEAKXXBISIM}}
\begin{SaveVerbatim}{HOLWeakEQTheoremsWEAKXXEQUIVXXPRESDXXBYXXGUARDEDXXSUM}
\HOLTokenTurnstile{} \HOLSymConst{\HOLTokenForall{}}\HOLBoundVar{E\sb{\mathrm{1}}} \HOLBoundVar{E\sb{\mathrm{1}}\sp{\prime}} \HOLBoundVar{E\sb{\mathrm{2}}} \HOLBoundVar{E\sb{\mathrm{2}}\sp{\prime}} \HOLBoundVar{a\sb{\mathrm{1}}} \HOLBoundVar{a\sb{\mathrm{2}}}.
     \HOLConst{WEAK_EQUIV} \HOLBoundVar{E\sb{\mathrm{1}}} \HOLBoundVar{E\sb{\mathrm{1}}\sp{\prime}} \HOLSymConst{\HOLTokenConj{}} \HOLConst{WEAK_EQUIV} \HOLBoundVar{E\sb{\mathrm{2}}} \HOLBoundVar{E\sb{\mathrm{2}}\sp{\prime}} \HOLSymConst{\HOLTokenImp{}}
     \HOLConst{WEAK_EQUIV} (\HOLBoundVar{a\sb{\mathrm{1}}}\HOLSymConst{..}\HOLBoundVar{E\sb{\mathrm{1}}} \HOLSymConst{+} \HOLBoundVar{a\sb{\mathrm{2}}}\HOLSymConst{..}\HOLBoundVar{E\sb{\mathrm{2}}}) (\HOLBoundVar{a\sb{\mathrm{1}}}\HOLSymConst{..}\HOLBoundVar{E\sb{\mathrm{1}}\sp{\prime}} \HOLSymConst{+} \HOLBoundVar{a\sb{\mathrm{2}}}\HOLSymConst{..}\HOLBoundVar{E\sb{\mathrm{2}}\sp{\prime}})
\end{SaveVerbatim}
\newcommand{\HOLWeakEQTheoremsWEAKXXEQUIVXXPRESDXXBYXXGUARDEDXXSUM}{\UseVerbatim{HOLWeakEQTheoremsWEAKXXEQUIVXXPRESDXXBYXXGUARDEDXXSUM}}
\begin{SaveVerbatim}{HOLWeakEQTheoremsWEAKXXEQUIVXXPRESDXXBYXXPAR}
\HOLTokenTurnstile{} \HOLSymConst{\HOLTokenForall{}}\HOLBoundVar{E\sb{\mathrm{1}}} \HOLBoundVar{E\sb{\mathrm{1}}\sp{\prime}} \HOLBoundVar{E\sb{\mathrm{2}}} \HOLBoundVar{E\sb{\mathrm{2}}\sp{\prime}}.
     \HOLConst{WEAK_EQUIV} \HOLBoundVar{E\sb{\mathrm{1}}} \HOLBoundVar{E\sb{\mathrm{1}}\sp{\prime}} \HOLSymConst{\HOLTokenConj{}} \HOLConst{WEAK_EQUIV} \HOLBoundVar{E\sb{\mathrm{2}}} \HOLBoundVar{E\sb{\mathrm{2}}\sp{\prime}} \HOLSymConst{\HOLTokenImp{}}
     \HOLConst{WEAK_EQUIV} (\HOLBoundVar{E\sb{\mathrm{1}}} \HOLSymConst{\ensuremath{\parallel}} \HOLBoundVar{E\sb{\mathrm{2}}}) (\HOLBoundVar{E\sb{\mathrm{1}}\sp{\prime}} \HOLSymConst{\ensuremath{\parallel}} \HOLBoundVar{E\sb{\mathrm{2}}\sp{\prime}})
\end{SaveVerbatim}
\newcommand{\HOLWeakEQTheoremsWEAKXXEQUIVXXPRESDXXBYXXPAR}{\UseVerbatim{HOLWeakEQTheoremsWEAKXXEQUIVXXPRESDXXBYXXPAR}}
\begin{SaveVerbatim}{HOLWeakEQTheoremsWEAKXXEQUIVXXPRESDXXBYXXSUM}
\HOLTokenTurnstile{} \HOLSymConst{\HOLTokenForall{}}\HOLBoundVar{E\sb{\mathrm{1}}} \HOLBoundVar{E\sb{\mathrm{1}}\sp{\prime}} \HOLBoundVar{E\sb{\mathrm{2}}} \HOLBoundVar{E\sb{\mathrm{2}}\sp{\prime}}.
     \HOLConst{WEAK_EQUIV} \HOLBoundVar{E\sb{\mathrm{1}}} \HOLBoundVar{E\sb{\mathrm{1}}\sp{\prime}} \HOLSymConst{\HOLTokenConj{}} \HOLConst{STABLE} \HOLBoundVar{E\sb{\mathrm{1}}} \HOLSymConst{\HOLTokenConj{}} \HOLConst{STABLE} \HOLBoundVar{E\sb{\mathrm{1}}\sp{\prime}} \HOLSymConst{\HOLTokenConj{}}
     \HOLConst{WEAK_EQUIV} \HOLBoundVar{E\sb{\mathrm{2}}} \HOLBoundVar{E\sb{\mathrm{2}}\sp{\prime}} \HOLSymConst{\HOLTokenConj{}} \HOLConst{STABLE} \HOLBoundVar{E\sb{\mathrm{2}}} \HOLSymConst{\HOLTokenConj{}} \HOLConst{STABLE} \HOLBoundVar{E\sb{\mathrm{2}}\sp{\prime}} \HOLSymConst{\HOLTokenImp{}}
     \HOLConst{WEAK_EQUIV} (\HOLBoundVar{E\sb{\mathrm{1}}} \HOLSymConst{+} \HOLBoundVar{E\sb{\mathrm{2}}}) (\HOLBoundVar{E\sb{\mathrm{1}}\sp{\prime}} \HOLSymConst{+} \HOLBoundVar{E\sb{\mathrm{2}}\sp{\prime}})
\end{SaveVerbatim}
\newcommand{\HOLWeakEQTheoremsWEAKXXEQUIVXXPRESDXXBYXXSUM}{\UseVerbatim{HOLWeakEQTheoremsWEAKXXEQUIVXXPRESDXXBYXXSUM}}
\begin{SaveVerbatim}{HOLWeakEQTheoremsWEAKXXEQUIVXXREFL}
\HOLTokenTurnstile{} \HOLSymConst{\HOLTokenForall{}}\HOLBoundVar{E}. \HOLConst{WEAK_EQUIV} \HOLBoundVar{E} \HOLBoundVar{E}
\end{SaveVerbatim}
\newcommand{\HOLWeakEQTheoremsWEAKXXEQUIVXXREFL}{\UseVerbatim{HOLWeakEQTheoremsWEAKXXEQUIVXXREFL}}
\begin{SaveVerbatim}{HOLWeakEQTheoremsWEAKXXEQUIVXXrules}
\HOLTokenTurnstile{} \HOLSymConst{\HOLTokenForall{}}\HOLBoundVar{E} \HOLBoundVar{E\sp{\prime}}.
     (\HOLSymConst{\HOLTokenForall{}}\HOLBoundVar{l}.
        (\HOLSymConst{\HOLTokenForall{}}\HOLBoundVar{E\sb{\mathrm{1}}}.
           \HOLBoundVar{E} \HOLTokenTransBegin\HOLConst{label} \HOLBoundVar{l}\HOLTokenTransEnd \HOLBoundVar{E\sb{\mathrm{1}}} \HOLSymConst{\HOLTokenImp{}}
           \HOLSymConst{\HOLTokenExists{}}\HOLBoundVar{E\sb{\mathrm{2}}}. \HOLBoundVar{E\sp{\prime}} \HOLTokenWeakTransBegin\HOLConst{label} \HOLBoundVar{l}\HOLTokenWeakTransEnd \HOLBoundVar{E\sb{\mathrm{2}}} \HOLSymConst{\HOLTokenConj{}} \HOLConst{WEAK_EQUIV} \HOLBoundVar{E\sb{\mathrm{1}}} \HOLBoundVar{E\sb{\mathrm{2}}}) \HOLSymConst{\HOLTokenConj{}}
        \HOLSymConst{\HOLTokenForall{}}\HOLBoundVar{E\sb{\mathrm{2}}}.
          \HOLBoundVar{E\sp{\prime}} \HOLTokenTransBegin\HOLConst{label} \HOLBoundVar{l}\HOLTokenTransEnd \HOLBoundVar{E\sb{\mathrm{2}}} \HOLSymConst{\HOLTokenImp{}}
          \HOLSymConst{\HOLTokenExists{}}\HOLBoundVar{E\sb{\mathrm{1}}}. \HOLBoundVar{E} \HOLTokenWeakTransBegin\HOLConst{label} \HOLBoundVar{l}\HOLTokenWeakTransEnd \HOLBoundVar{E\sb{\mathrm{1}}} \HOLSymConst{\HOLTokenConj{}} \HOLConst{WEAK_EQUIV} \HOLBoundVar{E\sb{\mathrm{1}}} \HOLBoundVar{E\sb{\mathrm{2}}}) \HOLSymConst{\HOLTokenConj{}}
     (\HOLSymConst{\HOLTokenForall{}}\HOLBoundVar{E\sb{\mathrm{1}}}. \HOLBoundVar{E} \HOLTokenTransBegin\HOLConst{\ensuremath{\tau}}\HOLTokenTransEnd \HOLBoundVar{E\sb{\mathrm{1}}} \HOLSymConst{\HOLTokenImp{}} \HOLSymConst{\HOLTokenExists{}}\HOLBoundVar{E\sb{\mathrm{2}}}. \HOLConst{EPS} \HOLBoundVar{E\sp{\prime}} \HOLBoundVar{E\sb{\mathrm{2}}} \HOLSymConst{\HOLTokenConj{}} \HOLConst{WEAK_EQUIV} \HOLBoundVar{E\sb{\mathrm{1}}} \HOLBoundVar{E\sb{\mathrm{2}}}) \HOLSymConst{\HOLTokenConj{}}
     (\HOLSymConst{\HOLTokenForall{}}\HOLBoundVar{E\sb{\mathrm{2}}}. \HOLBoundVar{E\sp{\prime}} \HOLTokenTransBegin\HOLConst{\ensuremath{\tau}}\HOLTokenTransEnd \HOLBoundVar{E\sb{\mathrm{2}}} \HOLSymConst{\HOLTokenImp{}} \HOLSymConst{\HOLTokenExists{}}\HOLBoundVar{E\sb{\mathrm{1}}}. \HOLConst{EPS} \HOLBoundVar{E} \HOLBoundVar{E\sb{\mathrm{1}}} \HOLSymConst{\HOLTokenConj{}} \HOLConst{WEAK_EQUIV} \HOLBoundVar{E\sb{\mathrm{1}}} \HOLBoundVar{E\sb{\mathrm{2}}}) \HOLSymConst{\HOLTokenImp{}}
     \HOLConst{WEAK_EQUIV} \HOLBoundVar{E} \HOLBoundVar{E\sp{\prime}}
\end{SaveVerbatim}
\newcommand{\HOLWeakEQTheoremsWEAKXXEQUIVXXrules}{\UseVerbatim{HOLWeakEQTheoremsWEAKXXEQUIVXXrules}}
\begin{SaveVerbatim}{HOLWeakEQTheoremsWEAKXXEQUIVXXSUBSTXXPARXXL}
\HOLTokenTurnstile{} \HOLSymConst{\HOLTokenForall{}}\HOLBoundVar{E} \HOLBoundVar{E\sp{\prime}}.
     \HOLConst{WEAK_EQUIV} \HOLBoundVar{E} \HOLBoundVar{E\sp{\prime}} \HOLSymConst{\HOLTokenImp{}} \HOLSymConst{\HOLTokenForall{}}\HOLBoundVar{E\sp{\prime\prime}}. \HOLConst{WEAK_EQUIV} (\HOLBoundVar{E\sp{\prime\prime}} \HOLSymConst{\ensuremath{\parallel}} \HOLBoundVar{E}) (\HOLBoundVar{E\sp{\prime\prime}} \HOLSymConst{\ensuremath{\parallel}} \HOLBoundVar{E\sp{\prime}})
\end{SaveVerbatim}
\newcommand{\HOLWeakEQTheoremsWEAKXXEQUIVXXSUBSTXXPARXXL}{\UseVerbatim{HOLWeakEQTheoremsWEAKXXEQUIVXXSUBSTXXPARXXL}}
\begin{SaveVerbatim}{HOLWeakEQTheoremsWEAKXXEQUIVXXSUBSTXXPARXXR}
\HOLTokenTurnstile{} \HOLSymConst{\HOLTokenForall{}}\HOLBoundVar{E} \HOLBoundVar{E\sp{\prime}}.
     \HOLConst{WEAK_EQUIV} \HOLBoundVar{E} \HOLBoundVar{E\sp{\prime}} \HOLSymConst{\HOLTokenImp{}} \HOLSymConst{\HOLTokenForall{}}\HOLBoundVar{E\sp{\prime\prime}}. \HOLConst{WEAK_EQUIV} (\HOLBoundVar{E} \HOLSymConst{\ensuremath{\parallel}} \HOLBoundVar{E\sp{\prime\prime}}) (\HOLBoundVar{E\sp{\prime}} \HOLSymConst{\ensuremath{\parallel}} \HOLBoundVar{E\sp{\prime\prime}})
\end{SaveVerbatim}
\newcommand{\HOLWeakEQTheoremsWEAKXXEQUIVXXSUBSTXXPARXXR}{\UseVerbatim{HOLWeakEQTheoremsWEAKXXEQUIVXXSUBSTXXPARXXR}}
\begin{SaveVerbatim}{HOLWeakEQTheoremsWEAKXXEQUIVXXSUBSTXXPREFIX}
\HOLTokenTurnstile{} \HOLSymConst{\HOLTokenForall{}}\HOLBoundVar{E} \HOLBoundVar{E\sp{\prime}}. \HOLConst{WEAK_EQUIV} \HOLBoundVar{E} \HOLBoundVar{E\sp{\prime}} \HOLSymConst{\HOLTokenImp{}} \HOLSymConst{\HOLTokenForall{}}\HOLBoundVar{u}. \HOLConst{WEAK_EQUIV} (\HOLBoundVar{u}\HOLSymConst{..}\HOLBoundVar{E}) (\HOLBoundVar{u}\HOLSymConst{..}\HOLBoundVar{E\sp{\prime}})
\end{SaveVerbatim}
\newcommand{\HOLWeakEQTheoremsWEAKXXEQUIVXXSUBSTXXPREFIX}{\UseVerbatim{HOLWeakEQTheoremsWEAKXXEQUIVXXSUBSTXXPREFIX}}
\begin{SaveVerbatim}{HOLWeakEQTheoremsWEAKXXEQUIVXXSUBSTXXRELAB}
\HOLTokenTurnstile{} \HOLSymConst{\HOLTokenForall{}}\HOLBoundVar{E} \HOLBoundVar{E\sp{\prime}}.
     \HOLConst{WEAK_EQUIV} \HOLBoundVar{E} \HOLBoundVar{E\sp{\prime}} \HOLSymConst{\HOLTokenImp{}}
     \HOLSymConst{\HOLTokenForall{}}\HOLBoundVar{rf}. \HOLConst{WEAK_EQUIV} (\HOLConst{relab} \HOLBoundVar{E} \HOLBoundVar{rf}) (\HOLConst{relab} \HOLBoundVar{E\sp{\prime}} \HOLBoundVar{rf})
\end{SaveVerbatim}
\newcommand{\HOLWeakEQTheoremsWEAKXXEQUIVXXSUBSTXXRELAB}{\UseVerbatim{HOLWeakEQTheoremsWEAKXXEQUIVXXSUBSTXXRELAB}}
\begin{SaveVerbatim}{HOLWeakEQTheoremsWEAKXXEQUIVXXSUBSTXXRESTR}
\HOLTokenTurnstile{} \HOLSymConst{\HOLTokenForall{}}\HOLBoundVar{E} \HOLBoundVar{E\sp{\prime}}. \HOLConst{WEAK_EQUIV} \HOLBoundVar{E} \HOLBoundVar{E\sp{\prime}} \HOLSymConst{\HOLTokenImp{}} \HOLSymConst{\HOLTokenForall{}}\HOLBoundVar{L}. \HOLConst{WEAK_EQUIV} (\HOLConst{\ensuremath{\nu}} \HOLBoundVar{L} \HOLBoundVar{E}) (\HOLConst{\ensuremath{\nu}} \HOLBoundVar{L} \HOLBoundVar{E\sp{\prime}})
\end{SaveVerbatim}
\newcommand{\HOLWeakEQTheoremsWEAKXXEQUIVXXSUBSTXXRESTR}{\UseVerbatim{HOLWeakEQTheoremsWEAKXXEQUIVXXSUBSTXXRESTR}}
\begin{SaveVerbatim}{HOLWeakEQTheoremsWEAKXXEQUIVXXSUBSTXXSUMXXR}
\HOLTokenTurnstile{} \HOLSymConst{\HOLTokenForall{}}\HOLBoundVar{E} \HOLBoundVar{E\sp{\prime}}.
     \HOLConst{WEAK_EQUIV} \HOLBoundVar{E} \HOLBoundVar{E\sp{\prime}} \HOLSymConst{\HOLTokenConj{}} \HOLConst{STABLE} \HOLBoundVar{E} \HOLSymConst{\HOLTokenConj{}} \HOLConst{STABLE} \HOLBoundVar{E\sp{\prime}} \HOLSymConst{\HOLTokenImp{}}
     \HOLSymConst{\HOLTokenForall{}}\HOLBoundVar{E\sp{\prime\prime}}. \HOLConst{WEAK_EQUIV} (\HOLBoundVar{E} \HOLSymConst{+} \HOLBoundVar{E\sp{\prime\prime}}) (\HOLBoundVar{E\sp{\prime}} \HOLSymConst{+} \HOLBoundVar{E\sp{\prime\prime}})
\end{SaveVerbatim}
\newcommand{\HOLWeakEQTheoremsWEAKXXEQUIVXXSUBSTXXSUMXXR}{\UseVerbatim{HOLWeakEQTheoremsWEAKXXEQUIVXXSUBSTXXSUMXXR}}
\begin{SaveVerbatim}{HOLWeakEQTheoremsWEAKXXEQUIVXXSYM}
\HOLTokenTurnstile{} \HOLSymConst{\HOLTokenForall{}}\HOLBoundVar{E} \HOLBoundVar{E\sp{\prime}}. \HOLConst{WEAK_EQUIV} \HOLBoundVar{E} \HOLBoundVar{E\sp{\prime}} \HOLSymConst{\HOLTokenImp{}} \HOLConst{WEAK_EQUIV} \HOLBoundVar{E\sp{\prime}} \HOLBoundVar{E}
\end{SaveVerbatim}
\newcommand{\HOLWeakEQTheoremsWEAKXXEQUIVXXSYM}{\UseVerbatim{HOLWeakEQTheoremsWEAKXXEQUIVXXSYM}}
\begin{SaveVerbatim}{HOLWeakEQTheoremsWEAKXXEQUIVXXSYMYY}
\HOLTokenTurnstile{} \HOLSymConst{\HOLTokenForall{}}\HOLBoundVar{E} \HOLBoundVar{E\sp{\prime}}. \HOLConst{WEAK_EQUIV} \HOLBoundVar{E} \HOLBoundVar{E\sp{\prime}} \HOLSymConst{\HOLTokenEquiv{}} \HOLConst{WEAK_EQUIV} \HOLBoundVar{E\sp{\prime}} \HOLBoundVar{E}
\end{SaveVerbatim}
\newcommand{\HOLWeakEQTheoremsWEAKXXEQUIVXXSYMYY}{\UseVerbatim{HOLWeakEQTheoremsWEAKXXEQUIVXXSYMYY}}
\begin{SaveVerbatim}{HOLWeakEQTheoremsWEAKXXEQUIVXXTRANS}
\HOLTokenTurnstile{} \HOLSymConst{\HOLTokenForall{}}\HOLBoundVar{E} \HOLBoundVar{E\sp{\prime}} \HOLBoundVar{E\sp{\prime\prime}}.
     \HOLConst{WEAK_EQUIV} \HOLBoundVar{E} \HOLBoundVar{E\sp{\prime}} \HOLSymConst{\HOLTokenConj{}} \HOLConst{WEAK_EQUIV} \HOLBoundVar{E\sp{\prime}} \HOLBoundVar{E\sp{\prime\prime}} \HOLSymConst{\HOLTokenImp{}} \HOLConst{WEAK_EQUIV} \HOLBoundVar{E} \HOLBoundVar{E\sp{\prime\prime}}
\end{SaveVerbatim}
\newcommand{\HOLWeakEQTheoremsWEAKXXEQUIVXXTRANS}{\UseVerbatim{HOLWeakEQTheoremsWEAKXXEQUIVXXTRANS}}
\begin{SaveVerbatim}{HOLWeakEQTheoremsWEAKXXEQUIVXXTRANSXXlabel}
\HOLTokenTurnstile{} \HOLSymConst{\HOLTokenForall{}}\HOLBoundVar{E} \HOLBoundVar{E\sp{\prime}}.
     \HOLConst{WEAK_EQUIV} \HOLBoundVar{E} \HOLBoundVar{E\sp{\prime}} \HOLSymConst{\HOLTokenImp{}}
     \HOLSymConst{\HOLTokenForall{}}\HOLBoundVar{l} \HOLBoundVar{E\sb{\mathrm{1}}}.
       \HOLBoundVar{E} \HOLTokenTransBegin\HOLConst{label} \HOLBoundVar{l}\HOLTokenTransEnd \HOLBoundVar{E\sb{\mathrm{1}}} \HOLSymConst{\HOLTokenImp{}} \HOLSymConst{\HOLTokenExists{}}\HOLBoundVar{E\sb{\mathrm{2}}}. \HOLBoundVar{E\sp{\prime}} \HOLTokenWeakTransBegin\HOLConst{label} \HOLBoundVar{l}\HOLTokenWeakTransEnd \HOLBoundVar{E\sb{\mathrm{2}}} \HOLSymConst{\HOLTokenConj{}} \HOLConst{WEAK_EQUIV} \HOLBoundVar{E\sb{\mathrm{1}}} \HOLBoundVar{E\sb{\mathrm{2}}}
\end{SaveVerbatim}
\newcommand{\HOLWeakEQTheoremsWEAKXXEQUIVXXTRANSXXlabel}{\UseVerbatim{HOLWeakEQTheoremsWEAKXXEQUIVXXTRANSXXlabel}}
\begin{SaveVerbatim}{HOLWeakEQTheoremsWEAKXXEQUIVXXTRANSXXlabelYY}
\HOLTokenTurnstile{} \HOLSymConst{\HOLTokenForall{}}\HOLBoundVar{E} \HOLBoundVar{E\sp{\prime}}.
     \HOLConst{WEAK_EQUIV} \HOLBoundVar{E} \HOLBoundVar{E\sp{\prime}} \HOLSymConst{\HOLTokenImp{}}
     \HOLSymConst{\HOLTokenForall{}}\HOLBoundVar{l} \HOLBoundVar{E\sb{\mathrm{2}}}.
       \HOLBoundVar{E\sp{\prime}} \HOLTokenTransBegin\HOLConst{label} \HOLBoundVar{l}\HOLTokenTransEnd \HOLBoundVar{E\sb{\mathrm{2}}} \HOLSymConst{\HOLTokenImp{}} \HOLSymConst{\HOLTokenExists{}}\HOLBoundVar{E\sb{\mathrm{1}}}. \HOLBoundVar{E} \HOLTokenWeakTransBegin\HOLConst{label} \HOLBoundVar{l}\HOLTokenWeakTransEnd \HOLBoundVar{E\sb{\mathrm{1}}} \HOLSymConst{\HOLTokenConj{}} \HOLConst{WEAK_EQUIV} \HOLBoundVar{E\sb{\mathrm{1}}} \HOLBoundVar{E\sb{\mathrm{2}}}
\end{SaveVerbatim}
\newcommand{\HOLWeakEQTheoremsWEAKXXEQUIVXXTRANSXXlabelYY}{\UseVerbatim{HOLWeakEQTheoremsWEAKXXEQUIVXXTRANSXXlabelYY}}
\begin{SaveVerbatim}{HOLWeakEQTheoremsWEAKXXEQUIVXXTRANSXXtau}
\HOLTokenTurnstile{} \HOLSymConst{\HOLTokenForall{}}\HOLBoundVar{E} \HOLBoundVar{E\sp{\prime}}.
     \HOLConst{WEAK_EQUIV} \HOLBoundVar{E} \HOLBoundVar{E\sp{\prime}} \HOLSymConst{\HOLTokenImp{}}
     \HOLSymConst{\HOLTokenForall{}}\HOLBoundVar{E\sb{\mathrm{1}}}. \HOLBoundVar{E} \HOLTokenTransBegin\HOLConst{\ensuremath{\tau}}\HOLTokenTransEnd \HOLBoundVar{E\sb{\mathrm{1}}} \HOLSymConst{\HOLTokenImp{}} \HOLSymConst{\HOLTokenExists{}}\HOLBoundVar{E\sb{\mathrm{2}}}. \HOLConst{EPS} \HOLBoundVar{E\sp{\prime}} \HOLBoundVar{E\sb{\mathrm{2}}} \HOLSymConst{\HOLTokenConj{}} \HOLConst{WEAK_EQUIV} \HOLBoundVar{E\sb{\mathrm{1}}} \HOLBoundVar{E\sb{\mathrm{2}}}
\end{SaveVerbatim}
\newcommand{\HOLWeakEQTheoremsWEAKXXEQUIVXXTRANSXXtau}{\UseVerbatim{HOLWeakEQTheoremsWEAKXXEQUIVXXTRANSXXtau}}
\begin{SaveVerbatim}{HOLWeakEQTheoremsWEAKXXEQUIVXXTRANSXXtauYY}
\HOLTokenTurnstile{} \HOLSymConst{\HOLTokenForall{}}\HOLBoundVar{E} \HOLBoundVar{E\sp{\prime}}.
     \HOLConst{WEAK_EQUIV} \HOLBoundVar{E} \HOLBoundVar{E\sp{\prime}} \HOLSymConst{\HOLTokenImp{}}
     \HOLSymConst{\HOLTokenForall{}}\HOLBoundVar{E\sb{\mathrm{2}}}. \HOLBoundVar{E\sp{\prime}} \HOLTokenTransBegin\HOLConst{\ensuremath{\tau}}\HOLTokenTransEnd \HOLBoundVar{E\sb{\mathrm{2}}} \HOLSymConst{\HOLTokenImp{}} \HOLSymConst{\HOLTokenExists{}}\HOLBoundVar{E\sb{\mathrm{1}}}. \HOLConst{EPS} \HOLBoundVar{E} \HOLBoundVar{E\sb{\mathrm{1}}} \HOLSymConst{\HOLTokenConj{}} \HOLConst{WEAK_EQUIV} \HOLBoundVar{E\sb{\mathrm{1}}} \HOLBoundVar{E\sb{\mathrm{2}}}
\end{SaveVerbatim}
\newcommand{\HOLWeakEQTheoremsWEAKXXEQUIVXXTRANSXXtauYY}{\UseVerbatim{HOLWeakEQTheoremsWEAKXXEQUIVXXTRANSXXtauYY}}
\begin{SaveVerbatim}{HOLWeakEQTheoremsWEAKXXEQUIVXXWEAKXXTRANSXXlabel}
\HOLTokenTurnstile{} \HOLSymConst{\HOLTokenForall{}}\HOLBoundVar{E} \HOLBoundVar{E\sp{\prime}}.
     \HOLConst{WEAK_EQUIV} \HOLBoundVar{E} \HOLBoundVar{E\sp{\prime}} \HOLSymConst{\HOLTokenImp{}}
     \HOLSymConst{\HOLTokenForall{}}\HOLBoundVar{l} \HOLBoundVar{E\sb{\mathrm{1}}}.
       \HOLBoundVar{E} \HOLTokenWeakTransBegin\HOLConst{label} \HOLBoundVar{l}\HOLTokenWeakTransEnd \HOLBoundVar{E\sb{\mathrm{1}}} \HOLSymConst{\HOLTokenImp{}} \HOLSymConst{\HOLTokenExists{}}\HOLBoundVar{E\sb{\mathrm{2}}}. \HOLBoundVar{E\sp{\prime}} \HOLTokenWeakTransBegin\HOLConst{label} \HOLBoundVar{l}\HOLTokenWeakTransEnd \HOLBoundVar{E\sb{\mathrm{2}}} \HOLSymConst{\HOLTokenConj{}} \HOLConst{WEAK_EQUIV} \HOLBoundVar{E\sb{\mathrm{1}}} \HOLBoundVar{E\sb{\mathrm{2}}}
\end{SaveVerbatim}
\newcommand{\HOLWeakEQTheoremsWEAKXXEQUIVXXWEAKXXTRANSXXlabel}{\UseVerbatim{HOLWeakEQTheoremsWEAKXXEQUIVXXWEAKXXTRANSXXlabel}}
\begin{SaveVerbatim}{HOLWeakEQTheoremsWEAKXXEQUIVXXWEAKXXTRANSXXlabelYY}
\HOLTokenTurnstile{} \HOLSymConst{\HOLTokenForall{}}\HOLBoundVar{E} \HOLBoundVar{E\sp{\prime}}.
     \HOLConst{WEAK_EQUIV} \HOLBoundVar{E} \HOLBoundVar{E\sp{\prime}} \HOLSymConst{\HOLTokenImp{}}
     \HOLSymConst{\HOLTokenForall{}}\HOLBoundVar{l} \HOLBoundVar{E\sb{\mathrm{2}}}.
       \HOLBoundVar{E\sp{\prime}} \HOLTokenWeakTransBegin\HOLConst{label} \HOLBoundVar{l}\HOLTokenWeakTransEnd \HOLBoundVar{E\sb{\mathrm{2}}} \HOLSymConst{\HOLTokenImp{}} \HOLSymConst{\HOLTokenExists{}}\HOLBoundVar{E\sb{\mathrm{1}}}. \HOLBoundVar{E} \HOLTokenWeakTransBegin\HOLConst{label} \HOLBoundVar{l}\HOLTokenWeakTransEnd \HOLBoundVar{E\sb{\mathrm{1}}} \HOLSymConst{\HOLTokenConj{}} \HOLConst{WEAK_EQUIV} \HOLBoundVar{E\sb{\mathrm{1}}} \HOLBoundVar{E\sb{\mathrm{2}}}
\end{SaveVerbatim}
\newcommand{\HOLWeakEQTheoremsWEAKXXEQUIVXXWEAKXXTRANSXXlabelYY}{\UseVerbatim{HOLWeakEQTheoremsWEAKXXEQUIVXXWEAKXXTRANSXXlabelYY}}
\begin{SaveVerbatim}{HOLWeakEQTheoremsWEAKXXEQUIVXXWEAKXXTRANSXXtau}
\HOLTokenTurnstile{} \HOLSymConst{\HOLTokenForall{}}\HOLBoundVar{E} \HOLBoundVar{E\sp{\prime}}.
     \HOLConst{WEAK_EQUIV} \HOLBoundVar{E} \HOLBoundVar{E\sp{\prime}} \HOLSymConst{\HOLTokenImp{}}
     \HOLSymConst{\HOLTokenForall{}}\HOLBoundVar{E\sb{\mathrm{1}}}. \HOLBoundVar{E} \HOLTokenWeakTransBegin\HOLConst{\ensuremath{\tau}}\HOLTokenWeakTransEnd \HOLBoundVar{E\sb{\mathrm{1}}} \HOLSymConst{\HOLTokenImp{}} \HOLSymConst{\HOLTokenExists{}}\HOLBoundVar{E\sb{\mathrm{2}}}. \HOLConst{EPS} \HOLBoundVar{E\sp{\prime}} \HOLBoundVar{E\sb{\mathrm{2}}} \HOLSymConst{\HOLTokenConj{}} \HOLConst{WEAK_EQUIV} \HOLBoundVar{E\sb{\mathrm{1}}} \HOLBoundVar{E\sb{\mathrm{2}}}
\end{SaveVerbatim}
\newcommand{\HOLWeakEQTheoremsWEAKXXEQUIVXXWEAKXXTRANSXXtau}{\UseVerbatim{HOLWeakEQTheoremsWEAKXXEQUIVXXWEAKXXTRANSXXtau}}
\begin{SaveVerbatim}{HOLWeakEQTheoremsWEAKXXEQUIVXXWEAKXXTRANSXXtauYY}
\HOLTokenTurnstile{} \HOLSymConst{\HOLTokenForall{}}\HOLBoundVar{E} \HOLBoundVar{E\sp{\prime}}.
     \HOLConst{WEAK_EQUIV} \HOLBoundVar{E} \HOLBoundVar{E\sp{\prime}} \HOLSymConst{\HOLTokenImp{}}
     \HOLSymConst{\HOLTokenForall{}}\HOLBoundVar{E\sb{\mathrm{2}}}. \HOLBoundVar{E\sp{\prime}} \HOLTokenWeakTransBegin\HOLConst{\ensuremath{\tau}}\HOLTokenWeakTransEnd \HOLBoundVar{E\sb{\mathrm{2}}} \HOLSymConst{\HOLTokenImp{}} \HOLSymConst{\HOLTokenExists{}}\HOLBoundVar{E\sb{\mathrm{1}}}. \HOLConst{EPS} \HOLBoundVar{E} \HOLBoundVar{E\sb{\mathrm{1}}} \HOLSymConst{\HOLTokenConj{}} \HOLConst{WEAK_EQUIV} \HOLBoundVar{E\sb{\mathrm{1}}} \HOLBoundVar{E\sb{\mathrm{2}}}
\end{SaveVerbatim}
\newcommand{\HOLWeakEQTheoremsWEAKXXEQUIVXXWEAKXXTRANSXXtauYY}{\UseVerbatim{HOLWeakEQTheoremsWEAKXXEQUIVXXWEAKXXTRANSXXtauYY}}
\begin{SaveVerbatim}{HOLWeakEQTheoremsWEAKXXPAR}
\HOLTokenTurnstile{} \HOLSymConst{\HOLTokenForall{}}\HOLBoundVar{E} \HOLBoundVar{u} \HOLBoundVar{E\sp{\prime}}.
     \HOLBoundVar{E} \HOLTokenWeakTransBegin\HOLBoundVar{u}\HOLTokenWeakTransEnd \HOLBoundVar{E\sp{\prime}} \HOLSymConst{\HOLTokenImp{}}
     \HOLSymConst{\HOLTokenForall{}}\HOLBoundVar{E\sp{\prime\prime}}. \HOLBoundVar{E} \HOLSymConst{\ensuremath{\parallel}} \HOLBoundVar{E\sp{\prime\prime}} \HOLTokenWeakTransBegin\HOLBoundVar{u}\HOLTokenWeakTransEnd \HOLBoundVar{E\sp{\prime}} \HOLSymConst{\ensuremath{\parallel}} \HOLBoundVar{E\sp{\prime\prime}} \HOLSymConst{\HOLTokenConj{}} \HOLBoundVar{E\sp{\prime\prime}} \HOLSymConst{\ensuremath{\parallel}} \HOLBoundVar{E} \HOLTokenWeakTransBegin\HOLBoundVar{u}\HOLTokenWeakTransEnd \HOLBoundVar{E\sp{\prime\prime}} \HOLSymConst{\ensuremath{\parallel}} \HOLBoundVar{E\sp{\prime}}
\end{SaveVerbatim}
\newcommand{\HOLWeakEQTheoremsWEAKXXPAR}{\UseVerbatim{HOLWeakEQTheoremsWEAKXXPAR}}
\begin{SaveVerbatim}{HOLWeakEQTheoremsWEAKXXPREFIX}
\HOLTokenTurnstile{} \HOLSymConst{\HOLTokenForall{}}\HOLBoundVar{E} \HOLBoundVar{u}. \HOLBoundVar{u}\HOLSymConst{..}\HOLBoundVar{E} \HOLTokenWeakTransBegin\HOLBoundVar{u}\HOLTokenWeakTransEnd \HOLBoundVar{E}
\end{SaveVerbatim}
\newcommand{\HOLWeakEQTheoremsWEAKXXPREFIX}{\UseVerbatim{HOLWeakEQTheoremsWEAKXXPREFIX}}
\begin{SaveVerbatim}{HOLWeakEQTheoremsWEAKXXPROPERTYXXSTAR}
\HOLTokenTurnstile{} \HOLSymConst{\HOLTokenForall{}}\HOLBoundVar{a\sb{\mathrm{0}}} \HOLBoundVar{a\sb{\mathrm{1}}}.
     \HOLConst{WEAK_EQUIV} \HOLBoundVar{a\sb{\mathrm{0}}} \HOLBoundVar{a\sb{\mathrm{1}}} \HOLSymConst{\HOLTokenEquiv{}}
     (\HOLSymConst{\HOLTokenForall{}}\HOLBoundVar{l}.
        (\HOLSymConst{\HOLTokenForall{}}\HOLBoundVar{E\sb{\mathrm{1}}}.
           \HOLBoundVar{a\sb{\mathrm{0}}} \HOLTokenTransBegin\HOLConst{label} \HOLBoundVar{l}\HOLTokenTransEnd \HOLBoundVar{E\sb{\mathrm{1}}} \HOLSymConst{\HOLTokenImp{}}
           \HOLSymConst{\HOLTokenExists{}}\HOLBoundVar{E\sb{\mathrm{2}}}. \HOLBoundVar{a\sb{\mathrm{1}}} \HOLTokenWeakTransBegin\HOLConst{label} \HOLBoundVar{l}\HOLTokenWeakTransEnd \HOLBoundVar{E\sb{\mathrm{2}}} \HOLSymConst{\HOLTokenConj{}} \HOLConst{WEAK_EQUIV} \HOLBoundVar{E\sb{\mathrm{1}}} \HOLBoundVar{E\sb{\mathrm{2}}}) \HOLSymConst{\HOLTokenConj{}}
        \HOLSymConst{\HOLTokenForall{}}\HOLBoundVar{E\sb{\mathrm{2}}}.
          \HOLBoundVar{a\sb{\mathrm{1}}} \HOLTokenTransBegin\HOLConst{label} \HOLBoundVar{l}\HOLTokenTransEnd \HOLBoundVar{E\sb{\mathrm{2}}} \HOLSymConst{\HOLTokenImp{}}
          \HOLSymConst{\HOLTokenExists{}}\HOLBoundVar{E\sb{\mathrm{1}}}. \HOLBoundVar{a\sb{\mathrm{0}}} \HOLTokenWeakTransBegin\HOLConst{label} \HOLBoundVar{l}\HOLTokenWeakTransEnd \HOLBoundVar{E\sb{\mathrm{1}}} \HOLSymConst{\HOLTokenConj{}} \HOLConst{WEAK_EQUIV} \HOLBoundVar{E\sb{\mathrm{1}}} \HOLBoundVar{E\sb{\mathrm{2}}}) \HOLSymConst{\HOLTokenConj{}}
     (\HOLSymConst{\HOLTokenForall{}}\HOLBoundVar{E\sb{\mathrm{1}}}. \HOLBoundVar{a\sb{\mathrm{0}}} \HOLTokenTransBegin\HOLConst{\ensuremath{\tau}}\HOLTokenTransEnd \HOLBoundVar{E\sb{\mathrm{1}}} \HOLSymConst{\HOLTokenImp{}} \HOLSymConst{\HOLTokenExists{}}\HOLBoundVar{E\sb{\mathrm{2}}}. \HOLConst{EPS} \HOLBoundVar{a\sb{\mathrm{1}}} \HOLBoundVar{E\sb{\mathrm{2}}} \HOLSymConst{\HOLTokenConj{}} \HOLConst{WEAK_EQUIV} \HOLBoundVar{E\sb{\mathrm{1}}} \HOLBoundVar{E\sb{\mathrm{2}}}) \HOLSymConst{\HOLTokenConj{}}
     \HOLSymConst{\HOLTokenForall{}}\HOLBoundVar{E\sb{\mathrm{2}}}. \HOLBoundVar{a\sb{\mathrm{1}}} \HOLTokenTransBegin\HOLConst{\ensuremath{\tau}}\HOLTokenTransEnd \HOLBoundVar{E\sb{\mathrm{2}}} \HOLSymConst{\HOLTokenImp{}} \HOLSymConst{\HOLTokenExists{}}\HOLBoundVar{E\sb{\mathrm{1}}}. \HOLConst{EPS} \HOLBoundVar{a\sb{\mathrm{0}}} \HOLBoundVar{E\sb{\mathrm{1}}} \HOLSymConst{\HOLTokenConj{}} \HOLConst{WEAK_EQUIV} \HOLBoundVar{E\sb{\mathrm{1}}} \HOLBoundVar{E\sb{\mathrm{2}}}
\end{SaveVerbatim}
\newcommand{\HOLWeakEQTheoremsWEAKXXPROPERTYXXSTAR}{\UseVerbatim{HOLWeakEQTheoremsWEAKXXPROPERTYXXSTAR}}
\begin{SaveVerbatim}{HOLWeakEQTheoremsWEAKXXRELAB}
\HOLTokenTurnstile{} \HOLSymConst{\HOLTokenForall{}}\HOLBoundVar{E} \HOLBoundVar{u} \HOLBoundVar{E\sp{\prime}}.
     \HOLBoundVar{E} \HOLTokenWeakTransBegin\HOLBoundVar{u}\HOLTokenWeakTransEnd \HOLBoundVar{E\sp{\prime}} \HOLSymConst{\HOLTokenImp{}}
     \HOLSymConst{\HOLTokenForall{}}\HOLBoundVar{labl}.
       \HOLConst{relab} \HOLBoundVar{E} (\HOLConst{RELAB} \HOLBoundVar{labl})
       \HOLTokenWeakTransBegin\HOLConst{relabel} (\HOLConst{RELAB} \HOLBoundVar{labl}) \HOLBoundVar{u}\HOLTokenWeakTransEnd
       \HOLConst{relab} \HOLBoundVar{E\sp{\prime}} (\HOLConst{RELAB} \HOLBoundVar{labl})
\end{SaveVerbatim}
\newcommand{\HOLWeakEQTheoremsWEAKXXRELAB}{\UseVerbatim{HOLWeakEQTheoremsWEAKXXRELAB}}
\begin{SaveVerbatim}{HOLWeakEQTheoremsWEAKXXRELABXXrf}
\HOLTokenTurnstile{} \HOLSymConst{\HOLTokenForall{}}\HOLBoundVar{E} \HOLBoundVar{u} \HOLBoundVar{E\sp{\prime}}.
     \HOLBoundVar{E} \HOLTokenWeakTransBegin\HOLBoundVar{u}\HOLTokenWeakTransEnd \HOLBoundVar{E\sp{\prime}} \HOLSymConst{\HOLTokenImp{}} \HOLSymConst{\HOLTokenForall{}}\HOLBoundVar{rf}. \HOLConst{relab} \HOLBoundVar{E} \HOLBoundVar{rf} \HOLTokenWeakTransBegin\HOLConst{relabel} \HOLBoundVar{rf} \HOLBoundVar{u}\HOLTokenWeakTransEnd \HOLConst{relab} \HOLBoundVar{E\sp{\prime}} \HOLBoundVar{rf}
\end{SaveVerbatim}
\newcommand{\HOLWeakEQTheoremsWEAKXXRELABXXrf}{\UseVerbatim{HOLWeakEQTheoremsWEAKXXRELABXXrf}}
\begin{SaveVerbatim}{HOLWeakEQTheoremsWEAKXXRESTRXXlabel}
\HOLTokenTurnstile{} \HOLSymConst{\HOLTokenForall{}}\HOLBoundVar{l} \HOLBoundVar{L} \HOLBoundVar{E} \HOLBoundVar{E\sp{\prime}}.
     \HOLBoundVar{l} \HOLConst{\HOLTokenNotIn{}} \HOLBoundVar{L} \HOLSymConst{\HOLTokenConj{}} \HOLConst{COMPL} \HOLBoundVar{l} \HOLConst{\HOLTokenNotIn{}} \HOLBoundVar{L} \HOLSymConst{\HOLTokenConj{}} \HOLBoundVar{E} \HOLTokenWeakTransBegin\HOLConst{label} \HOLBoundVar{l}\HOLTokenWeakTransEnd \HOLBoundVar{E\sp{\prime}} \HOLSymConst{\HOLTokenImp{}}
     \HOLConst{\ensuremath{\nu}} \HOLBoundVar{L} \HOLBoundVar{E} \HOLTokenWeakTransBegin\HOLConst{label} \HOLBoundVar{l}\HOLTokenWeakTransEnd \HOLConst{\ensuremath{\nu}} \HOLBoundVar{L} \HOLBoundVar{E\sp{\prime}}
\end{SaveVerbatim}
\newcommand{\HOLWeakEQTheoremsWEAKXXRESTRXXlabel}{\UseVerbatim{HOLWeakEQTheoremsWEAKXXRESTRXXlabel}}
\begin{SaveVerbatim}{HOLWeakEQTheoremsWEAKXXRESTRXXtau}
\HOLTokenTurnstile{} \HOLSymConst{\HOLTokenForall{}}\HOLBoundVar{E} \HOLBoundVar{E\sp{\prime}}. \HOLBoundVar{E} \HOLTokenWeakTransBegin\HOLConst{\ensuremath{\tau}}\HOLTokenWeakTransEnd \HOLBoundVar{E\sp{\prime}} \HOLSymConst{\HOLTokenImp{}} \HOLSymConst{\HOLTokenForall{}}\HOLBoundVar{L}. \HOLConst{\ensuremath{\nu}} \HOLBoundVar{L} \HOLBoundVar{E} \HOLTokenWeakTransBegin\HOLConst{\ensuremath{\tau}}\HOLTokenWeakTransEnd \HOLConst{\ensuremath{\nu}} \HOLBoundVar{L} \HOLBoundVar{E\sp{\prime}}
\end{SaveVerbatim}
\newcommand{\HOLWeakEQTheoremsWEAKXXRESTRXXtau}{\UseVerbatim{HOLWeakEQTheoremsWEAKXXRESTRXXtau}}
\begin{SaveVerbatim}{HOLWeakEQTheoremsWEAKXXSUMOne}
\HOLTokenTurnstile{} \HOLSymConst{\HOLTokenForall{}}\HOLBoundVar{E} \HOLBoundVar{u} \HOLBoundVar{E\sb{\mathrm{1}}} \HOLBoundVar{E\sp{\prime}}. \HOLBoundVar{E} \HOLTokenWeakTransBegin\HOLBoundVar{u}\HOLTokenWeakTransEnd \HOLBoundVar{E\sb{\mathrm{1}}} \HOLSymConst{\HOLTokenImp{}} \HOLBoundVar{E} \HOLSymConst{+} \HOLBoundVar{E\sp{\prime}} \HOLTokenWeakTransBegin\HOLBoundVar{u}\HOLTokenWeakTransEnd \HOLBoundVar{E\sb{\mathrm{1}}}
\end{SaveVerbatim}
\newcommand{\HOLWeakEQTheoremsWEAKXXSUMOne}{\UseVerbatim{HOLWeakEQTheoremsWEAKXXSUMOne}}
\begin{SaveVerbatim}{HOLWeakEQTheoremsWEAKXXSUMTwo}
\HOLTokenTurnstile{} \HOLSymConst{\HOLTokenForall{}}\HOLBoundVar{E} \HOLBoundVar{u} \HOLBoundVar{E\sb{\mathrm{1}}} \HOLBoundVar{E\sp{\prime}}. \HOLBoundVar{E} \HOLTokenWeakTransBegin\HOLBoundVar{u}\HOLTokenWeakTransEnd \HOLBoundVar{E\sb{\mathrm{1}}} \HOLSymConst{\HOLTokenImp{}} \HOLBoundVar{E\sp{\prime}} \HOLSymConst{+} \HOLBoundVar{E} \HOLTokenWeakTransBegin\HOLBoundVar{u}\HOLTokenWeakTransEnd \HOLBoundVar{E\sb{\mathrm{1}}}
\end{SaveVerbatim}
\newcommand{\HOLWeakEQTheoremsWEAKXXSUMTwo}{\UseVerbatim{HOLWeakEQTheoremsWEAKXXSUMTwo}}
\begin{SaveVerbatim}{HOLWeakEQTheoremsWEAKXXTRANSXXANDXXEPS}
\HOLTokenTurnstile{} \HOLSymConst{\HOLTokenForall{}}\HOLBoundVar{E\sb{\mathrm{1}}} \HOLBoundVar{u} \HOLBoundVar{E\sb{\mathrm{2}}} \HOLBoundVar{E\sp{\prime}}. \HOLBoundVar{E\sb{\mathrm{1}}} \HOLTokenWeakTransBegin\HOLBoundVar{u}\HOLTokenWeakTransEnd \HOLBoundVar{E\sb{\mathrm{2}}} \HOLSymConst{\HOLTokenConj{}} \HOLConst{EPS} \HOLBoundVar{E\sb{\mathrm{2}}} \HOLBoundVar{E\sp{\prime}} \HOLSymConst{\HOLTokenImp{}} \HOLBoundVar{E\sb{\mathrm{1}}} \HOLTokenWeakTransBegin\HOLBoundVar{u}\HOLTokenWeakTransEnd \HOLBoundVar{E\sp{\prime}}
\end{SaveVerbatim}
\newcommand{\HOLWeakEQTheoremsWEAKXXTRANSXXANDXXEPS}{\UseVerbatim{HOLWeakEQTheoremsWEAKXXTRANSXXANDXXEPS}}
\begin{SaveVerbatim}{HOLWeakEQTheoremsWEAKXXTRANSXXAUX}
\HOLTokenTurnstile{} \HOLSymConst{\HOLTokenForall{}}\HOLBoundVar{E} \HOLBoundVar{l} \HOLBoundVar{E\sb{\mathrm{1}}}.
     \HOLBoundVar{E} \HOLTokenWeakTransBegin\HOLConst{label} \HOLBoundVar{l}\HOLTokenWeakTransEnd \HOLBoundVar{E\sb{\mathrm{1}}} \HOLSymConst{\HOLTokenImp{}}
     \HOLSymConst{\HOLTokenForall{}}\HOLBoundVar{Wbsm} \HOLBoundVar{E\sp{\prime}}.
       \HOLConst{WEAK_BISIM} \HOLBoundVar{Wbsm} \HOLSymConst{\HOLTokenConj{}} \HOLBoundVar{Wbsm} \HOLBoundVar{E} \HOLBoundVar{E\sp{\prime}} \HOLSymConst{\HOLTokenImp{}}
       \HOLSymConst{\HOLTokenExists{}}\HOLBoundVar{E\sb{\mathrm{2}}}. \HOLBoundVar{E\sp{\prime}} \HOLTokenWeakTransBegin\HOLConst{label} \HOLBoundVar{l}\HOLTokenWeakTransEnd \HOLBoundVar{E\sb{\mathrm{2}}} \HOLSymConst{\HOLTokenConj{}} \HOLBoundVar{Wbsm} \HOLBoundVar{E\sb{\mathrm{1}}} \HOLBoundVar{E\sb{\mathrm{2}}}
\end{SaveVerbatim}
\newcommand{\HOLWeakEQTheoremsWEAKXXTRANSXXAUX}{\UseVerbatim{HOLWeakEQTheoremsWEAKXXTRANSXXAUX}}
\begin{SaveVerbatim}{HOLWeakEQTheoremsWEAKXXTRANSXXAUXXXSYM}
\HOLTokenTurnstile{} \HOLSymConst{\HOLTokenForall{}}\HOLBoundVar{E\sp{\prime}} \HOLBoundVar{l} \HOLBoundVar{E\sb{\mathrm{1}}}.
     \HOLBoundVar{E\sp{\prime}} \HOLTokenWeakTransBegin\HOLConst{label} \HOLBoundVar{l}\HOLTokenWeakTransEnd \HOLBoundVar{E\sb{\mathrm{1}}} \HOLSymConst{\HOLTokenImp{}}
     \HOLSymConst{\HOLTokenForall{}}\HOLBoundVar{Wbsm} \HOLBoundVar{E}.
       \HOLConst{WEAK_BISIM} \HOLBoundVar{Wbsm} \HOLSymConst{\HOLTokenConj{}} \HOLBoundVar{Wbsm} \HOLBoundVar{E} \HOLBoundVar{E\sp{\prime}} \HOLSymConst{\HOLTokenImp{}}
       \HOLSymConst{\HOLTokenExists{}}\HOLBoundVar{E\sb{\mathrm{2}}}. \HOLBoundVar{E} \HOLTokenWeakTransBegin\HOLConst{label} \HOLBoundVar{l}\HOLTokenWeakTransEnd \HOLBoundVar{E\sb{\mathrm{2}}} \HOLSymConst{\HOLTokenConj{}} \HOLBoundVar{Wbsm} \HOLBoundVar{E\sb{\mathrm{2}}} \HOLBoundVar{E\sb{\mathrm{1}}}
\end{SaveVerbatim}
\newcommand{\HOLWeakEQTheoremsWEAKXXTRANSXXAUXXXSYM}{\UseVerbatim{HOLWeakEQTheoremsWEAKXXTRANSXXAUXXXSYM}}
\begin{SaveVerbatim}{HOLWeakEQTheoremsWEAKXXTRANSXXcasesOne}
\HOLTokenTurnstile{} \HOLSymConst{\HOLTokenForall{}}\HOLBoundVar{E} \HOLBoundVar{u} \HOLBoundVar{E\sb{\mathrm{1}}}.
     \HOLBoundVar{E} \HOLTokenWeakTransBegin\HOLBoundVar{u}\HOLTokenWeakTransEnd \HOLBoundVar{E\sb{\mathrm{1}}} \HOLSymConst{\HOLTokenImp{}}
     (\HOLSymConst{\HOLTokenExists{}}\HOLBoundVar{E\sp{\prime}}. \HOLBoundVar{E} \HOLTokenTransBegin\HOLConst{\ensuremath{\tau}}\HOLTokenTransEnd \HOLBoundVar{E\sp{\prime}} \HOLSymConst{\HOLTokenConj{}} \HOLBoundVar{E\sp{\prime}} \HOLTokenWeakTransBegin\HOLBoundVar{u}\HOLTokenWeakTransEnd \HOLBoundVar{E\sb{\mathrm{1}}}) \HOLSymConst{\HOLTokenDisj{}} \HOLSymConst{\HOLTokenExists{}}\HOLBoundVar{E\sp{\prime}}. \HOLBoundVar{E} \HOLTokenTransBegin\HOLBoundVar{u}\HOLTokenTransEnd \HOLBoundVar{E\sp{\prime}} \HOLSymConst{\HOLTokenConj{}} \HOLConst{EPS} \HOLBoundVar{E\sp{\prime}} \HOLBoundVar{E\sb{\mathrm{1}}}
\end{SaveVerbatim}
\newcommand{\HOLWeakEQTheoremsWEAKXXTRANSXXcasesOne}{\UseVerbatim{HOLWeakEQTheoremsWEAKXXTRANSXXcasesOne}}
\begin{SaveVerbatim}{HOLWeakEQTheoremsWEAKXXTRANSXXcasesTwo}
\HOLTokenTurnstile{} \HOLSymConst{\HOLTokenForall{}}\HOLBoundVar{E} \HOLBoundVar{l} \HOLBoundVar{E\sb{\mathrm{1}}}.
     \HOLBoundVar{E} \HOLTokenWeakTransBegin\HOLConst{label} \HOLBoundVar{l}\HOLTokenWeakTransEnd \HOLBoundVar{E\sb{\mathrm{1}}} \HOLSymConst{\HOLTokenImp{}}
     (\HOLSymConst{\HOLTokenExists{}}\HOLBoundVar{E\sp{\prime}}. \HOLBoundVar{E} \HOLTokenTransBegin\HOLConst{\ensuremath{\tau}}\HOLTokenTransEnd \HOLBoundVar{E\sp{\prime}} \HOLSymConst{\HOLTokenConj{}} \HOLBoundVar{E\sp{\prime}} \HOLTokenWeakTransBegin\HOLConst{label} \HOLBoundVar{l}\HOLTokenWeakTransEnd \HOLBoundVar{E\sb{\mathrm{1}}}) \HOLSymConst{\HOLTokenDisj{}}
     \HOLSymConst{\HOLTokenExists{}}\HOLBoundVar{E\sp{\prime}}. \HOLBoundVar{E} \HOLTokenTransBegin\HOLConst{label} \HOLBoundVar{l}\HOLTokenTransEnd \HOLBoundVar{E\sp{\prime}} \HOLSymConst{\HOLTokenConj{}} \HOLConst{EPS} \HOLBoundVar{E\sp{\prime}} \HOLBoundVar{E\sb{\mathrm{1}}}
\end{SaveVerbatim}
\newcommand{\HOLWeakEQTheoremsWEAKXXTRANSXXcasesTwo}{\UseVerbatim{HOLWeakEQTheoremsWEAKXXTRANSXXcasesTwo}}
\begin{SaveVerbatim}{HOLWeakEQTheoremsWEAKXXTRANSXXIMPXXEPS}
\HOLTokenTurnstile{} \HOLSymConst{\HOLTokenForall{}}\HOLBoundVar{E} \HOLBoundVar{E\sp{\prime}}. \HOLBoundVar{E} \HOLTokenWeakTransBegin\HOLConst{\ensuremath{\tau}}\HOLTokenWeakTransEnd \HOLBoundVar{E\sp{\prime}} \HOLSymConst{\HOLTokenImp{}} \HOLConst{EPS} \HOLBoundVar{E} \HOLBoundVar{E\sp{\prime}}
\end{SaveVerbatim}
\newcommand{\HOLWeakEQTheoremsWEAKXXTRANSXXIMPXXEPS}{\UseVerbatim{HOLWeakEQTheoremsWEAKXXTRANSXXIMPXXEPS}}
\begin{SaveVerbatim}{HOLWeakEQTheoremsWEAKXXTRANSXXSTABLE}
\HOLTokenTurnstile{} \HOLSymConst{\HOLTokenForall{}}\HOLBoundVar{E} \HOLBoundVar{l} \HOLBoundVar{E\sp{\prime}}.
     \HOLBoundVar{E} \HOLTokenWeakTransBegin\HOLConst{label} \HOLBoundVar{l}\HOLTokenWeakTransEnd \HOLBoundVar{E\sp{\prime}} \HOLSymConst{\HOLTokenConj{}} \HOLConst{STABLE} \HOLBoundVar{E} \HOLSymConst{\HOLTokenImp{}}
     \HOLSymConst{\HOLTokenExists{}}\HOLBoundVar{E\sp{\prime\prime}}. \HOLBoundVar{E} \HOLTokenTransBegin\HOLConst{label} \HOLBoundVar{l}\HOLTokenTransEnd \HOLBoundVar{E\sp{\prime\prime}} \HOLSymConst{\HOLTokenConj{}} \HOLConst{EPS} \HOLBoundVar{E\sp{\prime\prime}} \HOLBoundVar{E\sp{\prime}}
\end{SaveVerbatim}
\newcommand{\HOLWeakEQTheoremsWEAKXXTRANSXXSTABLE}{\UseVerbatim{HOLWeakEQTheoremsWEAKXXTRANSXXSTABLE}}
\begin{SaveVerbatim}{HOLWeakEQTheoremsWEAKXXTRANSXXTAU}
\HOLTokenTurnstile{} \HOLSymConst{\HOLTokenForall{}}\HOLBoundVar{E} \HOLBoundVar{E\sb{\mathrm{1}}}. \HOLBoundVar{E} \HOLTokenWeakTransBegin\HOLConst{\ensuremath{\tau}}\HOLTokenWeakTransEnd \HOLBoundVar{E\sb{\mathrm{1}}} \HOLSymConst{\HOLTokenEquiv{}} \HOLSymConst{\HOLTokenExists{}}\HOLBoundVar{E\sp{\prime}}. \HOLBoundVar{E} \HOLTokenTransBegin\HOLConst{\ensuremath{\tau}}\HOLTokenTransEnd \HOLBoundVar{E\sp{\prime}} \HOLSymConst{\HOLTokenConj{}} \HOLConst{EPS} \HOLBoundVar{E\sp{\prime}} \HOLBoundVar{E\sb{\mathrm{1}}}
\end{SaveVerbatim}
\newcommand{\HOLWeakEQTheoremsWEAKXXTRANSXXTAU}{\UseVerbatim{HOLWeakEQTheoremsWEAKXXTRANSXXTAU}}
\begin{SaveVerbatim}{HOLWeakEQTheoremsWEAKXXTRANSXXTAUXXIMPXXTRANSXXTAU}
\HOLTokenTurnstile{} \HOLSymConst{\HOLTokenForall{}}\HOLBoundVar{E} \HOLBoundVar{E\sp{\prime}}. \HOLBoundVar{E} \HOLTokenWeakTransBegin\HOLConst{\ensuremath{\tau}}\HOLTokenWeakTransEnd \HOLBoundVar{E\sp{\prime}} \HOLSymConst{\HOLTokenImp{}} \HOLSymConst{\HOLTokenExists{}}\HOLBoundVar{E\sb{\mathrm{1}}}. \HOLBoundVar{E} \HOLTokenTransBegin\HOLConst{\ensuremath{\tau}}\HOLTokenTransEnd \HOLBoundVar{E\sb{\mathrm{1}}} \HOLSymConst{\HOLTokenConj{}} \HOLConst{EPS} \HOLBoundVar{E\sb{\mathrm{1}}} \HOLBoundVar{E\sp{\prime}}
\end{SaveVerbatim}
\newcommand{\HOLWeakEQTheoremsWEAKXXTRANSXXTAUXXIMPXXTRANSXXTAU}{\UseVerbatim{HOLWeakEQTheoremsWEAKXXTRANSXXTAUXXIMPXXTRANSXXTAU}}
\newcommand{\HOLWeakEQTheorems}{
\HOLThmTag{WeakEQ}{COMP_WEAK_BISIM}\HOLWeakEQTheoremsCOMPXXWEAKXXBISIM
\HOLThmTag{WeakEQ}{CONVERSE_WEAK_BISIM}\HOLWeakEQTheoremsCONVERSEXXWEAKXXBISIM
\HOLThmTag{WeakEQ}{EPS_AND_WEAK_TRANS}\HOLWeakEQTheoremsEPSXXANDXXWEAKXXTRANS
\HOLThmTag{WeakEQ}{EPS_cases}\HOLWeakEQTheoremsEPSXXcases
\HOLThmTag{WeakEQ}{EPS_cases1}\HOLWeakEQTheoremsEPSXXcasesOne
\HOLThmTag{WeakEQ}{EPS_cases2}\HOLWeakEQTheoremsEPSXXcasesTwo
\HOLThmTag{WeakEQ}{EPS_IMP_WEAK_TRANS}\HOLWeakEQTheoremsEPSXXIMPXXWEAKXXTRANS
\HOLThmTag{WeakEQ}{EPS_ind}\HOLWeakEQTheoremsEPSXXind
\HOLThmTag{WeakEQ}{EPS_ind_right}\HOLWeakEQTheoremsEPSXXindXXright
\HOLThmTag{WeakEQ}{EPS_INDUCT}\HOLWeakEQTheoremsEPSXXINDUCT
\HOLThmTag{WeakEQ}{EPS_PAR}\HOLWeakEQTheoremsEPSXXPAR
\HOLThmTag{WeakEQ}{EPS_PAR_PAR}\HOLWeakEQTheoremsEPSXXPARXXPAR
\HOLThmTag{WeakEQ}{EPS_REFL}\HOLWeakEQTheoremsEPSXXREFL
\HOLThmTag{WeakEQ}{EPS_RELAB}\HOLWeakEQTheoremsEPSXXRELAB
\HOLThmTag{WeakEQ}{EPS_RELAB_rf}\HOLWeakEQTheoremsEPSXXRELABXXrf
\HOLThmTag{WeakEQ}{EPS_RESTR}\HOLWeakEQTheoremsEPSXXRESTR
\HOLThmTag{WeakEQ}{EPS_STABLE}\HOLWeakEQTheoremsEPSXXSTABLE
\HOLThmTag{WeakEQ}{EPS_STABLE'}\HOLWeakEQTheoremsEPSXXSTABLEYY
\HOLThmTag{WeakEQ}{EPS_strongind}\HOLWeakEQTheoremsEPSXXstrongind
\HOLThmTag{WeakEQ}{EPS_strongind_right}\HOLWeakEQTheoremsEPSXXstrongindXXright
\HOLThmTag{WeakEQ}{EPS_TRANS}\HOLWeakEQTheoremsEPSXXTRANS
\HOLThmTag{WeakEQ}{EPS_TRANS_AUX}\HOLWeakEQTheoremsEPSXXTRANSXXAUX
\HOLThmTag{WeakEQ}{EPS_TRANS_AUX_SYM}\HOLWeakEQTheoremsEPSXXTRANSXXAUXXXSYM
\HOLThmTag{WeakEQ}{EPS_WEAK_EPS}\HOLWeakEQTheoremsEPSXXWEAKXXEPS
\HOLThmTag{WeakEQ}{EQUAL_IMP_WEAK_EQUIV}\HOLWeakEQTheoremsEQUALXXIMPXXWEAKXXEQUIV
\HOLThmTag{WeakEQ}{IDENTITY_WEAK_BISIM}\HOLWeakEQTheoremsIDENTITYXXWEAKXXBISIM
\HOLThmTag{WeakEQ}{INVERSE_REL}\HOLWeakEQTheoremsINVERSEXXREL
\HOLThmTag{WeakEQ}{ONE_TAU}\HOLWeakEQTheoremsONEXXTAU
\HOLThmTag{WeakEQ}{STABLE_cases}\HOLWeakEQTheoremsSTABLEXXcases
\HOLThmTag{WeakEQ}{STABLE_NO_TRANS_TAU}\HOLWeakEQTheoremsSTABLEXXNOXXTRANSXXTAU
\HOLThmTag{WeakEQ}{STABLE_NO_WEAK_TRANS_TAU}\HOLWeakEQTheoremsSTABLEXXNOXXWEAKXXTRANSXXTAU
\HOLThmTag{WeakEQ}{STRONG_EQUIV_EPS}\HOLWeakEQTheoremsSTRONGXXEQUIVXXEPS
\HOLThmTag{WeakEQ}{STRONG_EQUIV_EPS'}\HOLWeakEQTheoremsSTRONGXXEQUIVXXEPSYY
\HOLThmTag{WeakEQ}{STRONG_EQUIV_WEAK_TRANS}\HOLWeakEQTheoremsSTRONGXXEQUIVXXWEAKXXTRANS
\HOLThmTag{WeakEQ}{STRONG_EQUIV_WEAK_TRANS'}\HOLWeakEQTheoremsSTRONGXXEQUIVXXWEAKXXTRANSYY
\HOLThmTag{WeakEQ}{STRONG_IMP_WEAK_BISIM}\HOLWeakEQTheoremsSTRONGXXIMPXXWEAKXXBISIM
\HOLThmTag{WeakEQ}{STRONG_IMP_WEAK_EQUIV}\HOLWeakEQTheoremsSTRONGXXIMPXXWEAKXXEQUIV
\HOLThmTag{WeakEQ}{TAU_PREFIX_EPS}\HOLWeakEQTheoremsTAUXXPREFIXXXEPS
\HOLThmTag{WeakEQ}{TAU_PREFIX_WEAK_TRANS}\HOLWeakEQTheoremsTAUXXPREFIXXXWEAKXXTRANS
\HOLThmTag{WeakEQ}{TRANS_AND_EPS}\HOLWeakEQTheoremsTRANSXXANDXXEPS
\HOLThmTag{WeakEQ}{TRANS_IMP_WEAK_TRANS}\HOLWeakEQTheoremsTRANSXXIMPXXWEAKXXTRANS
\HOLThmTag{WeakEQ}{TRANS_TAU_AND_WEAK}\HOLWeakEQTheoremsTRANSXXTAUXXANDXXWEAK
\HOLThmTag{WeakEQ}{TRANS_TAU_IMP_EPS}\HOLWeakEQTheoremsTRANSXXTAUXXIMPXXEPS
\HOLThmTag{WeakEQ}{UNION_WEAK_BISIM}\HOLWeakEQTheoremsUNIONXXWEAKXXBISIM
\HOLThmTag{WeakEQ}{WEAK_BISIM}\HOLWeakEQTheoremsWEAKXXBISIM
\HOLThmTag{WeakEQ}{WEAK_BISIM_SUBSET_WEAK_EQUIV}\HOLWeakEQTheoremsWEAKXXBISIMXXSUBSETXXWEAKXXEQUIV
\HOLThmTag{WeakEQ}{WEAK_EQUIV}\HOLWeakEQTheoremsWEAKXXEQUIV
\HOLThmTag{WeakEQ}{WEAK_EQUIV_cases}\HOLWeakEQTheoremsWEAKXXEQUIVXXcases
\HOLThmTag{WeakEQ}{WEAK_EQUIV_coind}\HOLWeakEQTheoremsWEAKXXEQUIVXXcoind
\HOLThmTag{WeakEQ}{WEAK_EQUIV_EPS}\HOLWeakEQTheoremsWEAKXXEQUIVXXEPS
\HOLThmTag{WeakEQ}{WEAK_EQUIV_EPS'}\HOLWeakEQTheoremsWEAKXXEQUIVXXEPSYY
\HOLThmTag{WeakEQ}{WEAK_EQUIV_equivalence}\HOLWeakEQTheoremsWEAKXXEQUIVXXequivalence
\HOLThmTag{WeakEQ}{WEAK_EQUIV_IS_WEAK_BISIM}\HOLWeakEQTheoremsWEAKXXEQUIVXXISXXWEAKXXBISIM
\HOLThmTag{WeakEQ}{WEAK_EQUIV_PRESD_BY_GUARDED_SUM}\HOLWeakEQTheoremsWEAKXXEQUIVXXPRESDXXBYXXGUARDEDXXSUM
\HOLThmTag{WeakEQ}{WEAK_EQUIV_PRESD_BY_PAR}\HOLWeakEQTheoremsWEAKXXEQUIVXXPRESDXXBYXXPAR
\HOLThmTag{WeakEQ}{WEAK_EQUIV_PRESD_BY_SUM}\HOLWeakEQTheoremsWEAKXXEQUIVXXPRESDXXBYXXSUM
\HOLThmTag{WeakEQ}{WEAK_EQUIV_REFL}\HOLWeakEQTheoremsWEAKXXEQUIVXXREFL
\HOLThmTag{WeakEQ}{WEAK_EQUIV_rules}\HOLWeakEQTheoremsWEAKXXEQUIVXXrules
\HOLThmTag{WeakEQ}{WEAK_EQUIV_SUBST_PAR_L}\HOLWeakEQTheoremsWEAKXXEQUIVXXSUBSTXXPARXXL
\HOLThmTag{WeakEQ}{WEAK_EQUIV_SUBST_PAR_R}\HOLWeakEQTheoremsWEAKXXEQUIVXXSUBSTXXPARXXR
\HOLThmTag{WeakEQ}{WEAK_EQUIV_SUBST_PREFIX}\HOLWeakEQTheoremsWEAKXXEQUIVXXSUBSTXXPREFIX
\HOLThmTag{WeakEQ}{WEAK_EQUIV_SUBST_RELAB}\HOLWeakEQTheoremsWEAKXXEQUIVXXSUBSTXXRELAB
\HOLThmTag{WeakEQ}{WEAK_EQUIV_SUBST_RESTR}\HOLWeakEQTheoremsWEAKXXEQUIVXXSUBSTXXRESTR
\HOLThmTag{WeakEQ}{WEAK_EQUIV_SUBST_SUM_R}\HOLWeakEQTheoremsWEAKXXEQUIVXXSUBSTXXSUMXXR
\HOLThmTag{WeakEQ}{WEAK_EQUIV_SYM}\HOLWeakEQTheoremsWEAKXXEQUIVXXSYM
\HOLThmTag{WeakEQ}{WEAK_EQUIV_SYM'}\HOLWeakEQTheoremsWEAKXXEQUIVXXSYMYY
\HOLThmTag{WeakEQ}{WEAK_EQUIV_TRANS}\HOLWeakEQTheoremsWEAKXXEQUIVXXTRANS
\HOLThmTag{WeakEQ}{WEAK_EQUIV_TRANS_label}\HOLWeakEQTheoremsWEAKXXEQUIVXXTRANSXXlabel
\HOLThmTag{WeakEQ}{WEAK_EQUIV_TRANS_label'}\HOLWeakEQTheoremsWEAKXXEQUIVXXTRANSXXlabelYY
\HOLThmTag{WeakEQ}{WEAK_EQUIV_TRANS_tau}\HOLWeakEQTheoremsWEAKXXEQUIVXXTRANSXXtau
\HOLThmTag{WeakEQ}{WEAK_EQUIV_TRANS_tau'}\HOLWeakEQTheoremsWEAKXXEQUIVXXTRANSXXtauYY
\HOLThmTag{WeakEQ}{WEAK_EQUIV_WEAK_TRANS_label}\HOLWeakEQTheoremsWEAKXXEQUIVXXWEAKXXTRANSXXlabel
\HOLThmTag{WeakEQ}{WEAK_EQUIV_WEAK_TRANS_label'}\HOLWeakEQTheoremsWEAKXXEQUIVXXWEAKXXTRANSXXlabelYY
\HOLThmTag{WeakEQ}{WEAK_EQUIV_WEAK_TRANS_tau}\HOLWeakEQTheoremsWEAKXXEQUIVXXWEAKXXTRANSXXtau
\HOLThmTag{WeakEQ}{WEAK_EQUIV_WEAK_TRANS_tau'}\HOLWeakEQTheoremsWEAKXXEQUIVXXWEAKXXTRANSXXtauYY
\HOLThmTag{WeakEQ}{WEAK_PAR}\HOLWeakEQTheoremsWEAKXXPAR
\HOLThmTag{WeakEQ}{WEAK_PREFIX}\HOLWeakEQTheoremsWEAKXXPREFIX
\HOLThmTag{WeakEQ}{WEAK_PROPERTY_STAR}\HOLWeakEQTheoremsWEAKXXPROPERTYXXSTAR
\HOLThmTag{WeakEQ}{WEAK_RELAB}\HOLWeakEQTheoremsWEAKXXRELAB
\HOLThmTag{WeakEQ}{WEAK_RELAB_rf}\HOLWeakEQTheoremsWEAKXXRELABXXrf
\HOLThmTag{WeakEQ}{WEAK_RESTR_label}\HOLWeakEQTheoremsWEAKXXRESTRXXlabel
\HOLThmTag{WeakEQ}{WEAK_RESTR_tau}\HOLWeakEQTheoremsWEAKXXRESTRXXtau
\HOLThmTag{WeakEQ}{WEAK_SUM1}\HOLWeakEQTheoremsWEAKXXSUMOne
\HOLThmTag{WeakEQ}{WEAK_SUM2}\HOLWeakEQTheoremsWEAKXXSUMTwo
\HOLThmTag{WeakEQ}{WEAK_TRANS_AND_EPS}\HOLWeakEQTheoremsWEAKXXTRANSXXANDXXEPS
\HOLThmTag{WeakEQ}{WEAK_TRANS_AUX}\HOLWeakEQTheoremsWEAKXXTRANSXXAUX
\HOLThmTag{WeakEQ}{WEAK_TRANS_AUX_SYM}\HOLWeakEQTheoremsWEAKXXTRANSXXAUXXXSYM
\HOLThmTag{WeakEQ}{WEAK_TRANS_cases1}\HOLWeakEQTheoremsWEAKXXTRANSXXcasesOne
\HOLThmTag{WeakEQ}{WEAK_TRANS_cases2}\HOLWeakEQTheoremsWEAKXXTRANSXXcasesTwo
\HOLThmTag{WeakEQ}{WEAK_TRANS_IMP_EPS}\HOLWeakEQTheoremsWEAKXXTRANSXXIMPXXEPS
\HOLThmTag{WeakEQ}{WEAK_TRANS_STABLE}\HOLWeakEQTheoremsWEAKXXTRANSXXSTABLE
\HOLThmTag{WeakEQ}{WEAK_TRANS_TAU}\HOLWeakEQTheoremsWEAKXXTRANSXXTAU
\HOLThmTag{WeakEQ}{WEAK_TRANS_TAU_IMP_TRANS_TAU}\HOLWeakEQTheoremsWEAKXXTRANSXXTAUXXIMPXXTRANSXXTAU
}

\newcommand{\HOLWeakLawsDate}{02 Dicembre 2017}
\newcommand{\HOLWeakLawsTime}{13:31}
\begin{SaveVerbatim}{HOLWeakLawsTheoremsTAUXXWEAK}
\HOLTokenTurnstile{} \HOLSymConst{\HOLTokenForall{}}\HOLBoundVar{E}. \HOLConst{WEAK_EQUIV} (\HOLConst{\ensuremath{\tau}}\HOLSymConst{..}\HOLBoundVar{E}) \HOLBoundVar{E}
\end{SaveVerbatim}
\newcommand{\HOLWeakLawsTheoremsTAUXXWEAK}{\UseVerbatim{HOLWeakLawsTheoremsTAUXXWEAK}}
\begin{SaveVerbatim}{HOLWeakLawsTheoremsWEAKXXEQUIVXXSUBSTXXSUMXXL}
\HOLTokenTurnstile{} \HOLSymConst{\HOLTokenForall{}}\HOLBoundVar{E} \HOLBoundVar{E\sp{\prime}}.
     \HOLConst{WEAK_EQUIV} \HOLBoundVar{E} \HOLBoundVar{E\sp{\prime}} \HOLSymConst{\HOLTokenConj{}} \HOLConst{STABLE} \HOLBoundVar{E} \HOLSymConst{\HOLTokenConj{}} \HOLConst{STABLE} \HOLBoundVar{E\sp{\prime}} \HOLSymConst{\HOLTokenImp{}}
     \HOLSymConst{\HOLTokenForall{}}\HOLBoundVar{E\sp{\prime\prime}}. \HOLConst{WEAK_EQUIV} (\HOLBoundVar{E\sp{\prime\prime}} \HOLSymConst{+} \HOLBoundVar{E}) (\HOLBoundVar{E\sp{\prime\prime}} \HOLSymConst{+} \HOLBoundVar{E\sp{\prime}})
\end{SaveVerbatim}
\newcommand{\HOLWeakLawsTheoremsWEAKXXEQUIVXXSUBSTXXSUMXXL}{\UseVerbatim{HOLWeakLawsTheoremsWEAKXXEQUIVXXSUBSTXXSUMXXL}}
\begin{SaveVerbatim}{HOLWeakLawsTheoremsWEAKXXEXPANSIONXXLAW}
\HOLTokenTurnstile{} \HOLSymConst{\HOLTokenForall{}}\HOLBoundVar{f} \HOLBoundVar{n} \HOLBoundVar{f\sp{\prime}} \HOLBoundVar{m}.
     (\HOLSymConst{\HOLTokenForall{}}\HOLBoundVar{i}. \HOLBoundVar{i} \HOLSymConst{\HOLTokenLeq{}} \HOLBoundVar{n} \HOLSymConst{\HOLTokenImp{}} \HOLConst{Is_Prefix} (\HOLBoundVar{f} \HOLBoundVar{i})) \HOLSymConst{\HOLTokenConj{}}
     (\HOLSymConst{\HOLTokenForall{}}\HOLBoundVar{j}. \HOLBoundVar{j} \HOLSymConst{\HOLTokenLeq{}} \HOLBoundVar{m} \HOLSymConst{\HOLTokenImp{}} \HOLConst{Is_Prefix} (\HOLBoundVar{f\sp{\prime}} \HOLBoundVar{j})) \HOLSymConst{\HOLTokenImp{}}
     \HOLConst{WEAK_EQUIV} (\HOLConst{SIGMA} \HOLBoundVar{f} \HOLBoundVar{n} \HOLSymConst{\ensuremath{\parallel}} \HOLConst{SIGMA} \HOLBoundVar{f\sp{\prime}} \HOLBoundVar{m})
       (\HOLConst{SIGMA}
          (\HOLTokenLambda{}\HOLBoundVar{i}. \HOLConst{PREF_ACT} (\HOLBoundVar{f} \HOLBoundVar{i})\HOLSymConst{..}(\HOLConst{PREF_PROC} (\HOLBoundVar{f} \HOLBoundVar{i}) \HOLSymConst{\ensuremath{\parallel}} \HOLConst{SIGMA} \HOLBoundVar{f\sp{\prime}} \HOLBoundVar{m}))
          \HOLBoundVar{n} \HOLSymConst{+}
        \HOLConst{SIGMA}
          (\HOLTokenLambda{}\HOLBoundVar{j}. \HOLConst{PREF_ACT} (\HOLBoundVar{f\sp{\prime}} \HOLBoundVar{j})\HOLSymConst{..}(\HOLConst{SIGMA} \HOLBoundVar{f} \HOLBoundVar{n} \HOLSymConst{\ensuremath{\parallel}} \HOLConst{PREF_PROC} (\HOLBoundVar{f\sp{\prime}} \HOLBoundVar{j})))
          \HOLBoundVar{m} \HOLSymConst{+} \HOLConst{ALL_SYNC} \HOLBoundVar{f} \HOLBoundVar{n} \HOLBoundVar{f\sp{\prime}} \HOLBoundVar{m})
\end{SaveVerbatim}
\newcommand{\HOLWeakLawsTheoremsWEAKXXEXPANSIONXXLAW}{\UseVerbatim{HOLWeakLawsTheoremsWEAKXXEXPANSIONXXLAW}}
\begin{SaveVerbatim}{HOLWeakLawsTheoremsWEAKXXPARXXASSOC}
\HOLTokenTurnstile{} \HOLSymConst{\HOLTokenForall{}}\HOLBoundVar{E} \HOLBoundVar{E\sp{\prime}} \HOLBoundVar{E\sp{\prime\prime}}. \HOLConst{WEAK_EQUIV} (\HOLBoundVar{E} \HOLSymConst{\ensuremath{\parallel}} \HOLBoundVar{E\sp{\prime}} \HOLSymConst{\ensuremath{\parallel}} \HOLBoundVar{E\sp{\prime\prime}}) (\HOLBoundVar{E} \HOLSymConst{\ensuremath{\parallel}} (\HOLBoundVar{E\sp{\prime}} \HOLSymConst{\ensuremath{\parallel}} \HOLBoundVar{E\sp{\prime\prime}}))
\end{SaveVerbatim}
\newcommand{\HOLWeakLawsTheoremsWEAKXXPARXXASSOC}{\UseVerbatim{HOLWeakLawsTheoremsWEAKXXPARXXASSOC}}
\begin{SaveVerbatim}{HOLWeakLawsTheoremsWEAKXXPARXXCOMM}
\HOLTokenTurnstile{} \HOLSymConst{\HOLTokenForall{}}\HOLBoundVar{E} \HOLBoundVar{E\sp{\prime}}. \HOLConst{WEAK_EQUIV} (\HOLBoundVar{E} \HOLSymConst{\ensuremath{\parallel}} \HOLBoundVar{E\sp{\prime}}) (\HOLBoundVar{E\sp{\prime}} \HOLSymConst{\ensuremath{\parallel}} \HOLBoundVar{E})
\end{SaveVerbatim}
\newcommand{\HOLWeakLawsTheoremsWEAKXXPARXXCOMM}{\UseVerbatim{HOLWeakLawsTheoremsWEAKXXPARXXCOMM}}
\begin{SaveVerbatim}{HOLWeakLawsTheoremsWEAKXXPARXXIDENTXXL}
\HOLTokenTurnstile{} \HOLSymConst{\HOLTokenForall{}}\HOLBoundVar{E}. \HOLConst{WEAK_EQUIV} (\HOLConst{nil} \HOLSymConst{\ensuremath{\parallel}} \HOLBoundVar{E}) \HOLBoundVar{E}
\end{SaveVerbatim}
\newcommand{\HOLWeakLawsTheoremsWEAKXXPARXXIDENTXXL}{\UseVerbatim{HOLWeakLawsTheoremsWEAKXXPARXXIDENTXXL}}
\begin{SaveVerbatim}{HOLWeakLawsTheoremsWEAKXXPARXXIDENTXXR}
\HOLTokenTurnstile{} \HOLSymConst{\HOLTokenForall{}}\HOLBoundVar{E}. \HOLConst{WEAK_EQUIV} (\HOLBoundVar{E} \HOLSymConst{\ensuremath{\parallel}} \HOLConst{nil}) \HOLBoundVar{E}
\end{SaveVerbatim}
\newcommand{\HOLWeakLawsTheoremsWEAKXXPARXXIDENTXXR}{\UseVerbatim{HOLWeakLawsTheoremsWEAKXXPARXXIDENTXXR}}
\begin{SaveVerbatim}{HOLWeakLawsTheoremsWEAKXXPARXXPREFXXNOXXSYNCR}
\HOLTokenTurnstile{} \HOLSymConst{\HOLTokenForall{}}\HOLBoundVar{l} \HOLBoundVar{l\sp{\prime}}.
     \HOLBoundVar{l} \HOLSymConst{\HOLTokenNotEqual{}} \HOLConst{COMPL} \HOLBoundVar{l\sp{\prime}} \HOLSymConst{\HOLTokenImp{}}
     \HOLSymConst{\HOLTokenForall{}}\HOLBoundVar{E} \HOLBoundVar{E\sp{\prime}}.
       \HOLConst{WEAK_EQUIV} (\HOLConst{label} \HOLBoundVar{l}\HOLSymConst{..}\HOLBoundVar{E} \HOLSymConst{\ensuremath{\parallel}} \HOLConst{label} \HOLBoundVar{l\sp{\prime}}\HOLSymConst{..}\HOLBoundVar{E\sp{\prime}})
         (\HOLConst{label} \HOLBoundVar{l}\HOLSymConst{..}(\HOLBoundVar{E} \HOLSymConst{\ensuremath{\parallel}} \HOLConst{label} \HOLBoundVar{l\sp{\prime}}\HOLSymConst{..}\HOLBoundVar{E\sp{\prime}}) \HOLSymConst{+}
          \HOLConst{label} \HOLBoundVar{l\sp{\prime}}\HOLSymConst{..}(\HOLConst{label} \HOLBoundVar{l}\HOLSymConst{..}\HOLBoundVar{E} \HOLSymConst{\ensuremath{\parallel}} \HOLBoundVar{E\sp{\prime}}))
\end{SaveVerbatim}
\newcommand{\HOLWeakLawsTheoremsWEAKXXPARXXPREFXXNOXXSYNCR}{\UseVerbatim{HOLWeakLawsTheoremsWEAKXXPARXXPREFXXNOXXSYNCR}}
\begin{SaveVerbatim}{HOLWeakLawsTheoremsWEAKXXPARXXPREFXXSYNCR}
\HOLTokenTurnstile{} \HOLSymConst{\HOLTokenForall{}}\HOLBoundVar{l} \HOLBoundVar{l\sp{\prime}}.
     (\HOLBoundVar{l} \HOLSymConst{=} \HOLConst{COMPL} \HOLBoundVar{l\sp{\prime}}) \HOLSymConst{\HOLTokenImp{}}
     \HOLSymConst{\HOLTokenForall{}}\HOLBoundVar{E} \HOLBoundVar{E\sp{\prime}}.
       \HOLConst{WEAK_EQUIV} (\HOLConst{label} \HOLBoundVar{l}\HOLSymConst{..}\HOLBoundVar{E} \HOLSymConst{\ensuremath{\parallel}} \HOLConst{label} \HOLBoundVar{l\sp{\prime}}\HOLSymConst{..}\HOLBoundVar{E\sp{\prime}})
         (\HOLConst{label} \HOLBoundVar{l}\HOLSymConst{..}(\HOLBoundVar{E} \HOLSymConst{\ensuremath{\parallel}} \HOLConst{label} \HOLBoundVar{l\sp{\prime}}\HOLSymConst{..}\HOLBoundVar{E\sp{\prime}}) \HOLSymConst{+}
          \HOLConst{label} \HOLBoundVar{l\sp{\prime}}\HOLSymConst{..}(\HOLConst{label} \HOLBoundVar{l}\HOLSymConst{..}\HOLBoundVar{E} \HOLSymConst{\ensuremath{\parallel}} \HOLBoundVar{E\sp{\prime}}) \HOLSymConst{+} \HOLConst{\ensuremath{\tau}}\HOLSymConst{..}(\HOLBoundVar{E} \HOLSymConst{\ensuremath{\parallel}} \HOLBoundVar{E\sp{\prime}}))
\end{SaveVerbatim}
\newcommand{\HOLWeakLawsTheoremsWEAKXXPARXXPREFXXSYNCR}{\UseVerbatim{HOLWeakLawsTheoremsWEAKXXPARXXPREFXXSYNCR}}
\begin{SaveVerbatim}{HOLWeakLawsTheoremsWEAKXXPARXXPREFXXTAU}
\HOLTokenTurnstile{} \HOLSymConst{\HOLTokenForall{}}\HOLBoundVar{u} \HOLBoundVar{E} \HOLBoundVar{E\sp{\prime}}.
     \HOLConst{WEAK_EQUIV} (\HOLBoundVar{u}\HOLSymConst{..}\HOLBoundVar{E} \HOLSymConst{\ensuremath{\parallel}} \HOLConst{\ensuremath{\tau}}\HOLSymConst{..}\HOLBoundVar{E\sp{\prime}}) (\HOLBoundVar{u}\HOLSymConst{..}(\HOLBoundVar{E} \HOLSymConst{\ensuremath{\parallel}} \HOLConst{\ensuremath{\tau}}\HOLSymConst{..}\HOLBoundVar{E\sp{\prime}}) \HOLSymConst{+} \HOLConst{\ensuremath{\tau}}\HOLSymConst{..}(\HOLBoundVar{u}\HOLSymConst{..}\HOLBoundVar{E} \HOLSymConst{\ensuremath{\parallel}} \HOLBoundVar{E\sp{\prime}}))
\end{SaveVerbatim}
\newcommand{\HOLWeakLawsTheoremsWEAKXXPARXXPREFXXTAU}{\UseVerbatim{HOLWeakLawsTheoremsWEAKXXPARXXPREFXXTAU}}
\begin{SaveVerbatim}{HOLWeakLawsTheoremsWEAKXXPARXXTAUXXPREF}
\HOLTokenTurnstile{} \HOLSymConst{\HOLTokenForall{}}\HOLBoundVar{E} \HOLBoundVar{u} \HOLBoundVar{E\sp{\prime}}.
     \HOLConst{WEAK_EQUIV} (\HOLConst{\ensuremath{\tau}}\HOLSymConst{..}\HOLBoundVar{E} \HOLSymConst{\ensuremath{\parallel}} \HOLBoundVar{u}\HOLSymConst{..}\HOLBoundVar{E\sp{\prime}}) (\HOLConst{\ensuremath{\tau}}\HOLSymConst{..}(\HOLBoundVar{E} \HOLSymConst{\ensuremath{\parallel}} \HOLBoundVar{u}\HOLSymConst{..}\HOLBoundVar{E\sp{\prime}}) \HOLSymConst{+} \HOLBoundVar{u}\HOLSymConst{..}(\HOLConst{\ensuremath{\tau}}\HOLSymConst{..}\HOLBoundVar{E} \HOLSymConst{\ensuremath{\parallel}} \HOLBoundVar{E\sp{\prime}}))
\end{SaveVerbatim}
\newcommand{\HOLWeakLawsTheoremsWEAKXXPARXXTAUXXPREF}{\UseVerbatim{HOLWeakLawsTheoremsWEAKXXPARXXTAUXXPREF}}
\begin{SaveVerbatim}{HOLWeakLawsTheoremsWEAKXXPARXXTAUXXTAU}
\HOLTokenTurnstile{} \HOLSymConst{\HOLTokenForall{}}\HOLBoundVar{E} \HOLBoundVar{E\sp{\prime}}.
     \HOLConst{WEAK_EQUIV} (\HOLConst{\ensuremath{\tau}}\HOLSymConst{..}\HOLBoundVar{E} \HOLSymConst{\ensuremath{\parallel}} \HOLConst{\ensuremath{\tau}}\HOLSymConst{..}\HOLBoundVar{E\sp{\prime}}) (\HOLConst{\ensuremath{\tau}}\HOLSymConst{..}(\HOLBoundVar{E} \HOLSymConst{\ensuremath{\parallel}} \HOLConst{\ensuremath{\tau}}\HOLSymConst{..}\HOLBoundVar{E\sp{\prime}}) \HOLSymConst{+} \HOLConst{\ensuremath{\tau}}\HOLSymConst{..}(\HOLConst{\ensuremath{\tau}}\HOLSymConst{..}\HOLBoundVar{E} \HOLSymConst{\ensuremath{\parallel}} \HOLBoundVar{E\sp{\prime}}))
\end{SaveVerbatim}
\newcommand{\HOLWeakLawsTheoremsWEAKXXPARXXTAUXXTAU}{\UseVerbatim{HOLWeakLawsTheoremsWEAKXXPARXXTAUXXTAU}}
\begin{SaveVerbatim}{HOLWeakLawsTheoremsWEAKXXPREFXXRECXXEQUIV}
\HOLTokenTurnstile{} \HOLSymConst{\HOLTokenForall{}}\HOLBoundVar{u} \HOLBoundVar{s} \HOLBoundVar{v}.
     \HOLConst{WEAK_EQUIV} (\HOLBoundVar{u}\HOLSymConst{..}\HOLConst{rec} \HOLBoundVar{s} (\HOLBoundVar{v}\HOLSymConst{..}\HOLBoundVar{u}\HOLSymConst{..}\HOLConst{var} \HOLBoundVar{s})) (\HOLConst{rec} \HOLBoundVar{s} (\HOLBoundVar{u}\HOLSymConst{..}\HOLBoundVar{v}\HOLSymConst{..}\HOLConst{var} \HOLBoundVar{s}))
\end{SaveVerbatim}
\newcommand{\HOLWeakLawsTheoremsWEAKXXPREFXXRECXXEQUIV}{\UseVerbatim{HOLWeakLawsTheoremsWEAKXXPREFXXRECXXEQUIV}}
\begin{SaveVerbatim}{HOLWeakLawsTheoremsWEAKXXRELABXXNIL}
\HOLTokenTurnstile{} \HOLSymConst{\HOLTokenForall{}}\HOLBoundVar{rf}. \HOLConst{WEAK_EQUIV} (\HOLConst{relab} \HOLConst{nil} \HOLBoundVar{rf}) \HOLConst{nil}
\end{SaveVerbatim}
\newcommand{\HOLWeakLawsTheoremsWEAKXXRELABXXNIL}{\UseVerbatim{HOLWeakLawsTheoremsWEAKXXRELABXXNIL}}
\begin{SaveVerbatim}{HOLWeakLawsTheoremsWEAKXXRELABXXPREFIX}
\HOLTokenTurnstile{} \HOLSymConst{\HOLTokenForall{}}\HOLBoundVar{u} \HOLBoundVar{E} \HOLBoundVar{labl}.
     \HOLConst{WEAK_EQUIV} (\HOLConst{relab} (\HOLBoundVar{u}\HOLSymConst{..}\HOLBoundVar{E}) (\HOLConst{RELAB} \HOLBoundVar{labl}))
       (\HOLConst{relabel} (\HOLConst{RELAB} \HOLBoundVar{labl}) \HOLBoundVar{u}\HOLSymConst{..}\HOLConst{relab} \HOLBoundVar{E} (\HOLConst{RELAB} \HOLBoundVar{labl}))
\end{SaveVerbatim}
\newcommand{\HOLWeakLawsTheoremsWEAKXXRELABXXPREFIX}{\UseVerbatim{HOLWeakLawsTheoremsWEAKXXRELABXXPREFIX}}
\begin{SaveVerbatim}{HOLWeakLawsTheoremsWEAKXXRELABXXSUM}
\HOLTokenTurnstile{} \HOLSymConst{\HOLTokenForall{}}\HOLBoundVar{E} \HOLBoundVar{E\sp{\prime}} \HOLBoundVar{rf}.
     \HOLConst{WEAK_EQUIV} (\HOLConst{relab} (\HOLBoundVar{E} \HOLSymConst{+} \HOLBoundVar{E\sp{\prime}}) \HOLBoundVar{rf}) (\HOLConst{relab} \HOLBoundVar{E} \HOLBoundVar{rf} \HOLSymConst{+} \HOLConst{relab} \HOLBoundVar{E\sp{\prime}} \HOLBoundVar{rf})
\end{SaveVerbatim}
\newcommand{\HOLWeakLawsTheoremsWEAKXXRELABXXSUM}{\UseVerbatim{HOLWeakLawsTheoremsWEAKXXRELABXXSUM}}
\begin{SaveVerbatim}{HOLWeakLawsTheoremsWEAKXXRESTRXXNIL}
\HOLTokenTurnstile{} \HOLSymConst{\HOLTokenForall{}}\HOLBoundVar{L}. \HOLConst{WEAK_EQUIV} (\HOLConst{\ensuremath{\nu}} \HOLBoundVar{L} \HOLConst{nil}) \HOLConst{nil}
\end{SaveVerbatim}
\newcommand{\HOLWeakLawsTheoremsWEAKXXRESTRXXNIL}{\UseVerbatim{HOLWeakLawsTheoremsWEAKXXRESTRXXNIL}}
\begin{SaveVerbatim}{HOLWeakLawsTheoremsWEAKXXRESTRXXPRXXLABXXNIL}
\HOLTokenTurnstile{} \HOLSymConst{\HOLTokenForall{}}\HOLBoundVar{l} \HOLBoundVar{L}.
     \HOLBoundVar{l} \HOLConst{\HOLTokenIn{}} \HOLBoundVar{L} \HOLSymConst{\HOLTokenDisj{}} \HOLConst{COMPL} \HOLBoundVar{l} \HOLConst{\HOLTokenIn{}} \HOLBoundVar{L} \HOLSymConst{\HOLTokenImp{}} \HOLSymConst{\HOLTokenForall{}}\HOLBoundVar{E}. \HOLConst{WEAK_EQUIV} (\HOLConst{\ensuremath{\nu}} \HOLBoundVar{L} (\HOLConst{label} \HOLBoundVar{l}\HOLSymConst{..}\HOLBoundVar{E})) \HOLConst{nil}
\end{SaveVerbatim}
\newcommand{\HOLWeakLawsTheoremsWEAKXXRESTRXXPRXXLABXXNIL}{\UseVerbatim{HOLWeakLawsTheoremsWEAKXXRESTRXXPRXXLABXXNIL}}
\begin{SaveVerbatim}{HOLWeakLawsTheoremsWEAKXXRESTRXXPREFIXXXLABEL}
\HOLTokenTurnstile{} \HOLSymConst{\HOLTokenForall{}}\HOLBoundVar{l} \HOLBoundVar{L}.
     \HOLBoundVar{l} \HOLConst{\HOLTokenNotIn{}} \HOLBoundVar{L} \HOLSymConst{\HOLTokenConj{}} \HOLConst{COMPL} \HOLBoundVar{l} \HOLConst{\HOLTokenNotIn{}} \HOLBoundVar{L} \HOLSymConst{\HOLTokenImp{}}
     \HOLSymConst{\HOLTokenForall{}}\HOLBoundVar{E}. \HOLConst{WEAK_EQUIV} (\HOLConst{\ensuremath{\nu}} \HOLBoundVar{L} (\HOLConst{label} \HOLBoundVar{l}\HOLSymConst{..}\HOLBoundVar{E})) (\HOLConst{label} \HOLBoundVar{l}\HOLSymConst{..}\HOLConst{\ensuremath{\nu}} \HOLBoundVar{L} \HOLBoundVar{E})
\end{SaveVerbatim}
\newcommand{\HOLWeakLawsTheoremsWEAKXXRESTRXXPREFIXXXLABEL}{\UseVerbatim{HOLWeakLawsTheoremsWEAKXXRESTRXXPREFIXXXLABEL}}
\begin{SaveVerbatim}{HOLWeakLawsTheoremsWEAKXXRESTRXXPREFIXXXTAU}
\HOLTokenTurnstile{} \HOLSymConst{\HOLTokenForall{}}\HOLBoundVar{E} \HOLBoundVar{L}. \HOLConst{WEAK_EQUIV} (\HOLConst{\ensuremath{\nu}} \HOLBoundVar{L} (\HOLConst{\ensuremath{\tau}}\HOLSymConst{..}\HOLBoundVar{E})) (\HOLConst{\ensuremath{\tau}}\HOLSymConst{..}\HOLConst{\ensuremath{\nu}} \HOLBoundVar{L} \HOLBoundVar{E})
\end{SaveVerbatim}
\newcommand{\HOLWeakLawsTheoremsWEAKXXRESTRXXPREFIXXXTAU}{\UseVerbatim{HOLWeakLawsTheoremsWEAKXXRESTRXXPREFIXXXTAU}}
\begin{SaveVerbatim}{HOLWeakLawsTheoremsWEAKXXRESTRXXSUM}
\HOLTokenTurnstile{} \HOLSymConst{\HOLTokenForall{}}\HOLBoundVar{E} \HOLBoundVar{E\sp{\prime}} \HOLBoundVar{L}. \HOLConst{WEAK_EQUIV} (\HOLConst{\ensuremath{\nu}} \HOLBoundVar{L} (\HOLBoundVar{E} \HOLSymConst{+} \HOLBoundVar{E\sp{\prime}})) (\HOLConst{\ensuremath{\nu}} \HOLBoundVar{L} \HOLBoundVar{E} \HOLSymConst{+} \HOLConst{\ensuremath{\nu}} \HOLBoundVar{L} \HOLBoundVar{E\sp{\prime}})
\end{SaveVerbatim}
\newcommand{\HOLWeakLawsTheoremsWEAKXXRESTRXXSUM}{\UseVerbatim{HOLWeakLawsTheoremsWEAKXXRESTRXXSUM}}
\begin{SaveVerbatim}{HOLWeakLawsTheoremsWEAKXXSUMXXASSOCXXL}
\HOLTokenTurnstile{} \HOLSymConst{\HOLTokenForall{}}\HOLBoundVar{E} \HOLBoundVar{E\sp{\prime}} \HOLBoundVar{E\sp{\prime\prime}}. \HOLConst{WEAK_EQUIV} (\HOLBoundVar{E} \HOLSymConst{+} (\HOLBoundVar{E\sp{\prime}} \HOLSymConst{+} \HOLBoundVar{E\sp{\prime\prime}})) (\HOLBoundVar{E} \HOLSymConst{+} \HOLBoundVar{E\sp{\prime}} \HOLSymConst{+} \HOLBoundVar{E\sp{\prime\prime}})
\end{SaveVerbatim}
\newcommand{\HOLWeakLawsTheoremsWEAKXXSUMXXASSOCXXL}{\UseVerbatim{HOLWeakLawsTheoremsWEAKXXSUMXXASSOCXXL}}
\begin{SaveVerbatim}{HOLWeakLawsTheoremsWEAKXXSUMXXASSOCXXR}
\HOLTokenTurnstile{} \HOLSymConst{\HOLTokenForall{}}\HOLBoundVar{E} \HOLBoundVar{E\sp{\prime}} \HOLBoundVar{E\sp{\prime\prime}}. \HOLConst{WEAK_EQUIV} (\HOLBoundVar{E} \HOLSymConst{+} \HOLBoundVar{E\sp{\prime}} \HOLSymConst{+} \HOLBoundVar{E\sp{\prime\prime}}) (\HOLBoundVar{E} \HOLSymConst{+} (\HOLBoundVar{E\sp{\prime}} \HOLSymConst{+} \HOLBoundVar{E\sp{\prime\prime}}))
\end{SaveVerbatim}
\newcommand{\HOLWeakLawsTheoremsWEAKXXSUMXXASSOCXXR}{\UseVerbatim{HOLWeakLawsTheoremsWEAKXXSUMXXASSOCXXR}}
\begin{SaveVerbatim}{HOLWeakLawsTheoremsWEAKXXSUMXXCOMM}
\HOLTokenTurnstile{} \HOLSymConst{\HOLTokenForall{}}\HOLBoundVar{E} \HOLBoundVar{E\sp{\prime}}. \HOLConst{WEAK_EQUIV} (\HOLBoundVar{E} \HOLSymConst{+} \HOLBoundVar{E\sp{\prime}}) (\HOLBoundVar{E\sp{\prime}} \HOLSymConst{+} \HOLBoundVar{E})
\end{SaveVerbatim}
\newcommand{\HOLWeakLawsTheoremsWEAKXXSUMXXCOMM}{\UseVerbatim{HOLWeakLawsTheoremsWEAKXXSUMXXCOMM}}
\begin{SaveVerbatim}{HOLWeakLawsTheoremsWEAKXXSUMXXIDEMP}
\HOLTokenTurnstile{} \HOLSymConst{\HOLTokenForall{}}\HOLBoundVar{E}. \HOLConst{WEAK_EQUIV} (\HOLBoundVar{E} \HOLSymConst{+} \HOLBoundVar{E}) \HOLBoundVar{E}
\end{SaveVerbatim}
\newcommand{\HOLWeakLawsTheoremsWEAKXXSUMXXIDEMP}{\UseVerbatim{HOLWeakLawsTheoremsWEAKXXSUMXXIDEMP}}
\begin{SaveVerbatim}{HOLWeakLawsTheoremsWEAKXXSUMXXIDENTXXL}
\HOLTokenTurnstile{} \HOLSymConst{\HOLTokenForall{}}\HOLBoundVar{E}. \HOLConst{WEAK_EQUIV} (\HOLConst{nil} \HOLSymConst{+} \HOLBoundVar{E}) \HOLBoundVar{E}
\end{SaveVerbatim}
\newcommand{\HOLWeakLawsTheoremsWEAKXXSUMXXIDENTXXL}{\UseVerbatim{HOLWeakLawsTheoremsWEAKXXSUMXXIDENTXXL}}
\begin{SaveVerbatim}{HOLWeakLawsTheoremsWEAKXXSUMXXIDENTXXR}
\HOLTokenTurnstile{} \HOLSymConst{\HOLTokenForall{}}\HOLBoundVar{E}. \HOLConst{WEAK_EQUIV} (\HOLBoundVar{E} \HOLSymConst{+} \HOLConst{nil}) \HOLBoundVar{E}
\end{SaveVerbatim}
\newcommand{\HOLWeakLawsTheoremsWEAKXXSUMXXIDENTXXR}{\UseVerbatim{HOLWeakLawsTheoremsWEAKXXSUMXXIDENTXXR}}
\begin{SaveVerbatim}{HOLWeakLawsTheoremsWEAKXXUNFOLDING}
\HOLTokenTurnstile{} \HOLSymConst{\HOLTokenForall{}}\HOLBoundVar{X} \HOLBoundVar{E}. \HOLConst{WEAK_EQUIV} (\HOLConst{rec} \HOLBoundVar{X} \HOLBoundVar{E}) (\HOLConst{CCS_Subst} \HOLBoundVar{E} (\HOLConst{rec} \HOLBoundVar{X} \HOLBoundVar{E}) \HOLBoundVar{X})
\end{SaveVerbatim}
\newcommand{\HOLWeakLawsTheoremsWEAKXXUNFOLDING}{\UseVerbatim{HOLWeakLawsTheoremsWEAKXXUNFOLDING}}
\newcommand{\HOLWeakLawsTheorems}{
\HOLThmTag{WeakLaws}{TAU_WEAK}\HOLWeakLawsTheoremsTAUXXWEAK
\HOLThmTag{WeakLaws}{WEAK_EQUIV_SUBST_SUM_L}\HOLWeakLawsTheoremsWEAKXXEQUIVXXSUBSTXXSUMXXL
\HOLThmTag{WeakLaws}{WEAK_EXPANSION_LAW}\HOLWeakLawsTheoremsWEAKXXEXPANSIONXXLAW
\HOLThmTag{WeakLaws}{WEAK_PAR_ASSOC}\HOLWeakLawsTheoremsWEAKXXPARXXASSOC
\HOLThmTag{WeakLaws}{WEAK_PAR_COMM}\HOLWeakLawsTheoremsWEAKXXPARXXCOMM
\HOLThmTag{WeakLaws}{WEAK_PAR_IDENT_L}\HOLWeakLawsTheoremsWEAKXXPARXXIDENTXXL
\HOLThmTag{WeakLaws}{WEAK_PAR_IDENT_R}\HOLWeakLawsTheoremsWEAKXXPARXXIDENTXXR
\HOLThmTag{WeakLaws}{WEAK_PAR_PREF_NO_SYNCR}\HOLWeakLawsTheoremsWEAKXXPARXXPREFXXNOXXSYNCR
\HOLThmTag{WeakLaws}{WEAK_PAR_PREF_SYNCR}\HOLWeakLawsTheoremsWEAKXXPARXXPREFXXSYNCR
\HOLThmTag{WeakLaws}{WEAK_PAR_PREF_TAU}\HOLWeakLawsTheoremsWEAKXXPARXXPREFXXTAU
\HOLThmTag{WeakLaws}{WEAK_PAR_TAU_PREF}\HOLWeakLawsTheoremsWEAKXXPARXXTAUXXPREF
\HOLThmTag{WeakLaws}{WEAK_PAR_TAU_TAU}\HOLWeakLawsTheoremsWEAKXXPARXXTAUXXTAU
\HOLThmTag{WeakLaws}{WEAK_PREF_REC_EQUIV}\HOLWeakLawsTheoremsWEAKXXPREFXXRECXXEQUIV
\HOLThmTag{WeakLaws}{WEAK_RELAB_NIL}\HOLWeakLawsTheoremsWEAKXXRELABXXNIL
\HOLThmTag{WeakLaws}{WEAK_RELAB_PREFIX}\HOLWeakLawsTheoremsWEAKXXRELABXXPREFIX
\HOLThmTag{WeakLaws}{WEAK_RELAB_SUM}\HOLWeakLawsTheoremsWEAKXXRELABXXSUM
\HOLThmTag{WeakLaws}{WEAK_RESTR_NIL}\HOLWeakLawsTheoremsWEAKXXRESTRXXNIL
\HOLThmTag{WeakLaws}{WEAK_RESTR_PR_LAB_NIL}\HOLWeakLawsTheoremsWEAKXXRESTRXXPRXXLABXXNIL
\HOLThmTag{WeakLaws}{WEAK_RESTR_PREFIX_LABEL}\HOLWeakLawsTheoremsWEAKXXRESTRXXPREFIXXXLABEL
\HOLThmTag{WeakLaws}{WEAK_RESTR_PREFIX_TAU}\HOLWeakLawsTheoremsWEAKXXRESTRXXPREFIXXXTAU
\HOLThmTag{WeakLaws}{WEAK_RESTR_SUM}\HOLWeakLawsTheoremsWEAKXXRESTRXXSUM
\HOLThmTag{WeakLaws}{WEAK_SUM_ASSOC_L}\HOLWeakLawsTheoremsWEAKXXSUMXXASSOCXXL
\HOLThmTag{WeakLaws}{WEAK_SUM_ASSOC_R}\HOLWeakLawsTheoremsWEAKXXSUMXXASSOCXXR
\HOLThmTag{WeakLaws}{WEAK_SUM_COMM}\HOLWeakLawsTheoremsWEAKXXSUMXXCOMM
\HOLThmTag{WeakLaws}{WEAK_SUM_IDEMP}\HOLWeakLawsTheoremsWEAKXXSUMXXIDEMP
\HOLThmTag{WeakLaws}{WEAK_SUM_IDENT_L}\HOLWeakLawsTheoremsWEAKXXSUMXXIDENTXXL
\HOLThmTag{WeakLaws}{WEAK_SUM_IDENT_R}\HOLWeakLawsTheoremsWEAKXXSUMXXIDENTXXR
\HOLThmTag{WeakLaws}{WEAK_UNFOLDING}\HOLWeakLawsTheoremsWEAKXXUNFOLDING
}

\newcommand{\HOLObsCongrDate}{02 Dicembre 2017}
\newcommand{\HOLObsCongrTime}{13:31}
\begin{SaveVerbatim}{HOLObsCongrDefinitionsOBSXXCONGR}
\HOLTokenTurnstile{} \HOLSymConst{\HOLTokenForall{}}\HOLBoundVar{E} \HOLBoundVar{E\sp{\prime}}.
     \HOLConst{OBS_CONGR} \HOLBoundVar{E} \HOLBoundVar{E\sp{\prime}} \HOLSymConst{\HOLTokenEquiv{}}
     \HOLSymConst{\HOLTokenForall{}}\HOLBoundVar{u}.
       (\HOLSymConst{\HOLTokenForall{}}\HOLBoundVar{E\sb{\mathrm{1}}}. \HOLBoundVar{E} \HOLTokenTransBegin\HOLBoundVar{u}\HOLTokenTransEnd \HOLBoundVar{E\sb{\mathrm{1}}} \HOLSymConst{\HOLTokenImp{}} \HOLSymConst{\HOLTokenExists{}}\HOLBoundVar{E\sb{\mathrm{2}}}. \HOLBoundVar{E\sp{\prime}} \HOLTokenWeakTransBegin\HOLBoundVar{u}\HOLTokenWeakTransEnd \HOLBoundVar{E\sb{\mathrm{2}}} \HOLSymConst{\HOLTokenConj{}} \HOLConst{WEAK_EQUIV} \HOLBoundVar{E\sb{\mathrm{1}}} \HOLBoundVar{E\sb{\mathrm{2}}}) \HOLSymConst{\HOLTokenConj{}}
       \HOLSymConst{\HOLTokenForall{}}\HOLBoundVar{E\sb{\mathrm{2}}}. \HOLBoundVar{E\sp{\prime}} \HOLTokenTransBegin\HOLBoundVar{u}\HOLTokenTransEnd \HOLBoundVar{E\sb{\mathrm{2}}} \HOLSymConst{\HOLTokenImp{}} \HOLSymConst{\HOLTokenExists{}}\HOLBoundVar{E\sb{\mathrm{1}}}. \HOLBoundVar{E} \HOLTokenWeakTransBegin\HOLBoundVar{u}\HOLTokenWeakTransEnd \HOLBoundVar{E\sb{\mathrm{1}}} \HOLSymConst{\HOLTokenConj{}} \HOLConst{WEAK_EQUIV} \HOLBoundVar{E\sb{\mathrm{1}}} \HOLBoundVar{E\sb{\mathrm{2}}}
\end{SaveVerbatim}
\newcommand{\HOLObsCongrDefinitionsOBSXXCONGR}{\UseVerbatim{HOLObsCongrDefinitionsOBSXXCONGR}}
\newcommand{\HOLObsCongrDefinitions}{
\HOLDfnTag{ObsCongr}{OBS_CONGR}\HOLObsCongrDefinitionsOBSXXCONGR
}
\begin{SaveVerbatim}{HOLObsCongrTheoremsEQUALXXIMPXXOBSXXCONGR}
\HOLTokenTurnstile{} \HOLSymConst{\HOLTokenForall{}}\HOLBoundVar{E} \HOLBoundVar{E\sp{\prime}}. (\HOLBoundVar{E} \HOLSymConst{=} \HOLBoundVar{E\sp{\prime}}) \HOLSymConst{\HOLTokenImp{}} \HOLConst{OBS_CONGR} \HOLBoundVar{E} \HOLBoundVar{E\sp{\prime}}
\end{SaveVerbatim}
\newcommand{\HOLObsCongrTheoremsEQUALXXIMPXXOBSXXCONGR}{\UseVerbatim{HOLObsCongrTheoremsEQUALXXIMPXXOBSXXCONGR}}
\begin{SaveVerbatim}{HOLObsCongrTheoremsOBSXXCONGRXXBYXXWEAKXXBISIM}
\HOLTokenTurnstile{} \HOLSymConst{\HOLTokenForall{}}\HOLBoundVar{Wbsm}.
     \HOLConst{WEAK_BISIM} \HOLBoundVar{Wbsm} \HOLSymConst{\HOLTokenImp{}}
     \HOLSymConst{\HOLTokenForall{}}\HOLBoundVar{E} \HOLBoundVar{E\sp{\prime}}.
       (\HOLSymConst{\HOLTokenForall{}}\HOLBoundVar{u}.
          (\HOLSymConst{\HOLTokenForall{}}\HOLBoundVar{E\sb{\mathrm{1}}}. \HOLBoundVar{E} \HOLTokenTransBegin\HOLBoundVar{u}\HOLTokenTransEnd \HOLBoundVar{E\sb{\mathrm{1}}} \HOLSymConst{\HOLTokenImp{}} \HOLSymConst{\HOLTokenExists{}}\HOLBoundVar{E\sb{\mathrm{2}}}. \HOLBoundVar{E\sp{\prime}} \HOLTokenWeakTransBegin\HOLBoundVar{u}\HOLTokenWeakTransEnd \HOLBoundVar{E\sb{\mathrm{2}}} \HOLSymConst{\HOLTokenConj{}} \HOLBoundVar{Wbsm} \HOLBoundVar{E\sb{\mathrm{1}}} \HOLBoundVar{E\sb{\mathrm{2}}}) \HOLSymConst{\HOLTokenConj{}}
          \HOLSymConst{\HOLTokenForall{}}\HOLBoundVar{E\sb{\mathrm{2}}}. \HOLBoundVar{E\sp{\prime}} \HOLTokenTransBegin\HOLBoundVar{u}\HOLTokenTransEnd \HOLBoundVar{E\sb{\mathrm{2}}} \HOLSymConst{\HOLTokenImp{}} \HOLSymConst{\HOLTokenExists{}}\HOLBoundVar{E\sb{\mathrm{1}}}. \HOLBoundVar{E} \HOLTokenWeakTransBegin\HOLBoundVar{u}\HOLTokenWeakTransEnd \HOLBoundVar{E\sb{\mathrm{1}}} \HOLSymConst{\HOLTokenConj{}} \HOLBoundVar{Wbsm} \HOLBoundVar{E\sb{\mathrm{1}}} \HOLBoundVar{E\sb{\mathrm{2}}}) \HOLSymConst{\HOLTokenImp{}}
       \HOLConst{OBS_CONGR} \HOLBoundVar{E} \HOLBoundVar{E\sp{\prime}}
\end{SaveVerbatim}
\newcommand{\HOLObsCongrTheoremsOBSXXCONGRXXBYXXWEAKXXBISIM}{\UseVerbatim{HOLObsCongrTheoremsOBSXXCONGRXXBYXXWEAKXXBISIM}}
\begin{SaveVerbatim}{HOLObsCongrTheoremsOBSXXCONGRXXEPS}
\HOLTokenTurnstile{} \HOLSymConst{\HOLTokenForall{}}\HOLBoundVar{E} \HOLBoundVar{E\sp{\prime}}.
     \HOLConst{OBS_CONGR} \HOLBoundVar{E} \HOLBoundVar{E\sp{\prime}} \HOLSymConst{\HOLTokenImp{}}
     \HOLSymConst{\HOLTokenForall{}}\HOLBoundVar{E\sb{\mathrm{1}}}. \HOLConst{EPS} \HOLBoundVar{E} \HOLBoundVar{E\sb{\mathrm{1}}} \HOLSymConst{\HOLTokenImp{}} \HOLSymConst{\HOLTokenExists{}}\HOLBoundVar{E\sb{\mathrm{2}}}. \HOLConst{EPS} \HOLBoundVar{E\sp{\prime}} \HOLBoundVar{E\sb{\mathrm{2}}} \HOLSymConst{\HOLTokenConj{}} \HOLConst{WEAK_EQUIV} \HOLBoundVar{E\sb{\mathrm{1}}} \HOLBoundVar{E\sb{\mathrm{2}}}
\end{SaveVerbatim}
\newcommand{\HOLObsCongrTheoremsOBSXXCONGRXXEPS}{\UseVerbatim{HOLObsCongrTheoremsOBSXXCONGRXXEPS}}
\begin{SaveVerbatim}{HOLObsCongrTheoremsOBSXXCONGRXXEPSYY}
\HOLTokenTurnstile{} \HOLSymConst{\HOLTokenForall{}}\HOLBoundVar{E} \HOLBoundVar{E\sp{\prime}}.
     \HOLConst{OBS_CONGR} \HOLBoundVar{E} \HOLBoundVar{E\sp{\prime}} \HOLSymConst{\HOLTokenImp{}}
     \HOLSymConst{\HOLTokenForall{}}\HOLBoundVar{E\sb{\mathrm{2}}}. \HOLConst{EPS} \HOLBoundVar{E\sp{\prime}} \HOLBoundVar{E\sb{\mathrm{2}}} \HOLSymConst{\HOLTokenImp{}} \HOLSymConst{\HOLTokenExists{}}\HOLBoundVar{E\sb{\mathrm{1}}}. \HOLConst{EPS} \HOLBoundVar{E} \HOLBoundVar{E\sb{\mathrm{1}}} \HOLSymConst{\HOLTokenConj{}} \HOLConst{WEAK_EQUIV} \HOLBoundVar{E\sb{\mathrm{1}}} \HOLBoundVar{E\sb{\mathrm{2}}}
\end{SaveVerbatim}
\newcommand{\HOLObsCongrTheoremsOBSXXCONGRXXEPSYY}{\UseVerbatim{HOLObsCongrTheoremsOBSXXCONGRXXEPSYY}}
\begin{SaveVerbatim}{HOLObsCongrTheoremsOBSXXCONGRXXequivalence}
\HOLTokenTurnstile{} \HOLConst{equivalence} \HOLConst{OBS_CONGR}
\end{SaveVerbatim}
\newcommand{\HOLObsCongrTheoremsOBSXXCONGRXXequivalence}{\UseVerbatim{HOLObsCongrTheoremsOBSXXCONGRXXequivalence}}
\begin{SaveVerbatim}{HOLObsCongrTheoremsOBSXXCONGRXXIMPXXWEAKXXEQUIV}
\HOLTokenTurnstile{} \HOLSymConst{\HOLTokenForall{}}\HOLBoundVar{E} \HOLBoundVar{E\sp{\prime}}. \HOLConst{OBS_CONGR} \HOLBoundVar{E} \HOLBoundVar{E\sp{\prime}} \HOLSymConst{\HOLTokenImp{}} \HOLConst{WEAK_EQUIV} \HOLBoundVar{E} \HOLBoundVar{E\sp{\prime}}
\end{SaveVerbatim}
\newcommand{\HOLObsCongrTheoremsOBSXXCONGRXXIMPXXWEAKXXEQUIV}{\UseVerbatim{HOLObsCongrTheoremsOBSXXCONGRXXIMPXXWEAKXXEQUIV}}
\begin{SaveVerbatim}{HOLObsCongrTheoremsOBSXXCONGRXXPRESDXXBYXXPAR}
\HOLTokenTurnstile{} \HOLSymConst{\HOLTokenForall{}}\HOLBoundVar{E\sb{\mathrm{1}}} \HOLBoundVar{E\sb{\mathrm{1}}\sp{\prime}} \HOLBoundVar{E\sb{\mathrm{2}}} \HOLBoundVar{E\sb{\mathrm{2}}\sp{\prime}}.
     \HOLConst{OBS_CONGR} \HOLBoundVar{E\sb{\mathrm{1}}} \HOLBoundVar{E\sb{\mathrm{1}}\sp{\prime}} \HOLSymConst{\HOLTokenConj{}} \HOLConst{OBS_CONGR} \HOLBoundVar{E\sb{\mathrm{2}}} \HOLBoundVar{E\sb{\mathrm{2}}\sp{\prime}} \HOLSymConst{\HOLTokenImp{}}
     \HOLConst{OBS_CONGR} (\HOLBoundVar{E\sb{\mathrm{1}}} \HOLSymConst{\ensuremath{\parallel}} \HOLBoundVar{E\sb{\mathrm{2}}}) (\HOLBoundVar{E\sb{\mathrm{1}}\sp{\prime}} \HOLSymConst{\ensuremath{\parallel}} \HOLBoundVar{E\sb{\mathrm{2}}\sp{\prime}})
\end{SaveVerbatim}
\newcommand{\HOLObsCongrTheoremsOBSXXCONGRXXPRESDXXBYXXPAR}{\UseVerbatim{HOLObsCongrTheoremsOBSXXCONGRXXPRESDXXBYXXPAR}}
\begin{SaveVerbatim}{HOLObsCongrTheoremsOBSXXCONGRXXPRESDXXBYXXSUM}
\HOLTokenTurnstile{} \HOLSymConst{\HOLTokenForall{}}\HOLBoundVar{E\sb{\mathrm{1}}} \HOLBoundVar{E\sb{\mathrm{1}}\sp{\prime}} \HOLBoundVar{E\sb{\mathrm{2}}} \HOLBoundVar{E\sb{\mathrm{2}}\sp{\prime}}.
     \HOLConst{OBS_CONGR} \HOLBoundVar{E\sb{\mathrm{1}}} \HOLBoundVar{E\sb{\mathrm{1}}\sp{\prime}} \HOLSymConst{\HOLTokenConj{}} \HOLConst{OBS_CONGR} \HOLBoundVar{E\sb{\mathrm{2}}} \HOLBoundVar{E\sb{\mathrm{2}}\sp{\prime}} \HOLSymConst{\HOLTokenImp{}}
     \HOLConst{OBS_CONGR} (\HOLBoundVar{E\sb{\mathrm{1}}} \HOLSymConst{+} \HOLBoundVar{E\sb{\mathrm{2}}}) (\HOLBoundVar{E\sb{\mathrm{1}}\sp{\prime}} \HOLSymConst{+} \HOLBoundVar{E\sb{\mathrm{2}}\sp{\prime}})
\end{SaveVerbatim}
\newcommand{\HOLObsCongrTheoremsOBSXXCONGRXXPRESDXXBYXXSUM}{\UseVerbatim{HOLObsCongrTheoremsOBSXXCONGRXXPRESDXXBYXXSUM}}
\begin{SaveVerbatim}{HOLObsCongrTheoremsOBSXXCONGRXXREFL}
\HOLTokenTurnstile{} \HOLSymConst{\HOLTokenForall{}}\HOLBoundVar{E}. \HOLConst{OBS_CONGR} \HOLBoundVar{E} \HOLBoundVar{E}
\end{SaveVerbatim}
\newcommand{\HOLObsCongrTheoremsOBSXXCONGRXXREFL}{\UseVerbatim{HOLObsCongrTheoremsOBSXXCONGRXXREFL}}
\begin{SaveVerbatim}{HOLObsCongrTheoremsOBSXXCONGRXXSUBSTXXPARXXL}
\HOLTokenTurnstile{} \HOLSymConst{\HOLTokenForall{}}\HOLBoundVar{E} \HOLBoundVar{E\sp{\prime}}. \HOLConst{OBS_CONGR} \HOLBoundVar{E} \HOLBoundVar{E\sp{\prime}} \HOLSymConst{\HOLTokenImp{}} \HOLSymConst{\HOLTokenForall{}}\HOLBoundVar{E\sp{\prime\prime}}. \HOLConst{OBS_CONGR} (\HOLBoundVar{E\sp{\prime\prime}} \HOLSymConst{\ensuremath{\parallel}} \HOLBoundVar{E}) (\HOLBoundVar{E\sp{\prime\prime}} \HOLSymConst{\ensuremath{\parallel}} \HOLBoundVar{E\sp{\prime}})
\end{SaveVerbatim}
\newcommand{\HOLObsCongrTheoremsOBSXXCONGRXXSUBSTXXPARXXL}{\UseVerbatim{HOLObsCongrTheoremsOBSXXCONGRXXSUBSTXXPARXXL}}
\begin{SaveVerbatim}{HOLObsCongrTheoremsOBSXXCONGRXXSUBSTXXPARXXR}
\HOLTokenTurnstile{} \HOLSymConst{\HOLTokenForall{}}\HOLBoundVar{E} \HOLBoundVar{E\sp{\prime}}. \HOLConst{OBS_CONGR} \HOLBoundVar{E} \HOLBoundVar{E\sp{\prime}} \HOLSymConst{\HOLTokenImp{}} \HOLSymConst{\HOLTokenForall{}}\HOLBoundVar{E\sp{\prime\prime}}. \HOLConst{OBS_CONGR} (\HOLBoundVar{E} \HOLSymConst{\ensuremath{\parallel}} \HOLBoundVar{E\sp{\prime\prime}}) (\HOLBoundVar{E\sp{\prime}} \HOLSymConst{\ensuremath{\parallel}} \HOLBoundVar{E\sp{\prime\prime}})
\end{SaveVerbatim}
\newcommand{\HOLObsCongrTheoremsOBSXXCONGRXXSUBSTXXPARXXR}{\UseVerbatim{HOLObsCongrTheoremsOBSXXCONGRXXSUBSTXXPARXXR}}
\begin{SaveVerbatim}{HOLObsCongrTheoremsOBSXXCONGRXXSUBSTXXPREFIX}
\HOLTokenTurnstile{} \HOLSymConst{\HOLTokenForall{}}\HOLBoundVar{E} \HOLBoundVar{E\sp{\prime}}. \HOLConst{OBS_CONGR} \HOLBoundVar{E} \HOLBoundVar{E\sp{\prime}} \HOLSymConst{\HOLTokenImp{}} \HOLSymConst{\HOLTokenForall{}}\HOLBoundVar{u}. \HOLConst{OBS_CONGR} (\HOLBoundVar{u}\HOLSymConst{..}\HOLBoundVar{E}) (\HOLBoundVar{u}\HOLSymConst{..}\HOLBoundVar{E\sp{\prime}})
\end{SaveVerbatim}
\newcommand{\HOLObsCongrTheoremsOBSXXCONGRXXSUBSTXXPREFIX}{\UseVerbatim{HOLObsCongrTheoremsOBSXXCONGRXXSUBSTXXPREFIX}}
\begin{SaveVerbatim}{HOLObsCongrTheoremsOBSXXCONGRXXSUBSTXXRELAB}
\HOLTokenTurnstile{} \HOLSymConst{\HOLTokenForall{}}\HOLBoundVar{E} \HOLBoundVar{E\sp{\prime}}.
     \HOLConst{OBS_CONGR} \HOLBoundVar{E} \HOLBoundVar{E\sp{\prime}} \HOLSymConst{\HOLTokenImp{}} \HOLSymConst{\HOLTokenForall{}}\HOLBoundVar{rf}. \HOLConst{OBS_CONGR} (\HOLConst{relab} \HOLBoundVar{E} \HOLBoundVar{rf}) (\HOLConst{relab} \HOLBoundVar{E\sp{\prime}} \HOLBoundVar{rf})
\end{SaveVerbatim}
\newcommand{\HOLObsCongrTheoremsOBSXXCONGRXXSUBSTXXRELAB}{\UseVerbatim{HOLObsCongrTheoremsOBSXXCONGRXXSUBSTXXRELAB}}
\begin{SaveVerbatim}{HOLObsCongrTheoremsOBSXXCONGRXXSUBSTXXRESTR}
\HOLTokenTurnstile{} \HOLSymConst{\HOLTokenForall{}}\HOLBoundVar{E} \HOLBoundVar{E\sp{\prime}}. \HOLConst{OBS_CONGR} \HOLBoundVar{E} \HOLBoundVar{E\sp{\prime}} \HOLSymConst{\HOLTokenImp{}} \HOLSymConst{\HOLTokenForall{}}\HOLBoundVar{L}. \HOLConst{OBS_CONGR} (\HOLConst{\ensuremath{\nu}} \HOLBoundVar{L} \HOLBoundVar{E}) (\HOLConst{\ensuremath{\nu}} \HOLBoundVar{L} \HOLBoundVar{E\sp{\prime}})
\end{SaveVerbatim}
\newcommand{\HOLObsCongrTheoremsOBSXXCONGRXXSUBSTXXRESTR}{\UseVerbatim{HOLObsCongrTheoremsOBSXXCONGRXXSUBSTXXRESTR}}
\begin{SaveVerbatim}{HOLObsCongrTheoremsOBSXXCONGRXXSUBSTXXSUMXXL}
\HOLTokenTurnstile{} \HOLSymConst{\HOLTokenForall{}}\HOLBoundVar{E} \HOLBoundVar{E\sp{\prime}}. \HOLConst{OBS_CONGR} \HOLBoundVar{E} \HOLBoundVar{E\sp{\prime}} \HOLSymConst{\HOLTokenImp{}} \HOLSymConst{\HOLTokenForall{}}\HOLBoundVar{E\sp{\prime\prime}}. \HOLConst{OBS_CONGR} (\HOLBoundVar{E\sp{\prime\prime}} \HOLSymConst{+} \HOLBoundVar{E}) (\HOLBoundVar{E\sp{\prime\prime}} \HOLSymConst{+} \HOLBoundVar{E\sp{\prime}})
\end{SaveVerbatim}
\newcommand{\HOLObsCongrTheoremsOBSXXCONGRXXSUBSTXXSUMXXL}{\UseVerbatim{HOLObsCongrTheoremsOBSXXCONGRXXSUBSTXXSUMXXL}}
\begin{SaveVerbatim}{HOLObsCongrTheoremsOBSXXCONGRXXSUBSTXXSUMXXR}
\HOLTokenTurnstile{} \HOLSymConst{\HOLTokenForall{}}\HOLBoundVar{E} \HOLBoundVar{E\sp{\prime}}. \HOLConst{OBS_CONGR} \HOLBoundVar{E} \HOLBoundVar{E\sp{\prime}} \HOLSymConst{\HOLTokenImp{}} \HOLSymConst{\HOLTokenForall{}}\HOLBoundVar{E\sp{\prime\prime}}. \HOLConst{OBS_CONGR} (\HOLBoundVar{E} \HOLSymConst{+} \HOLBoundVar{E\sp{\prime\prime}}) (\HOLBoundVar{E\sp{\prime}} \HOLSymConst{+} \HOLBoundVar{E\sp{\prime\prime}})
\end{SaveVerbatim}
\newcommand{\HOLObsCongrTheoremsOBSXXCONGRXXSUBSTXXSUMXXR}{\UseVerbatim{HOLObsCongrTheoremsOBSXXCONGRXXSUBSTXXSUMXXR}}
\begin{SaveVerbatim}{HOLObsCongrTheoremsOBSXXCONGRXXSYM}
\HOLTokenTurnstile{} \HOLSymConst{\HOLTokenForall{}}\HOLBoundVar{E} \HOLBoundVar{E\sp{\prime}}. \HOLConst{OBS_CONGR} \HOLBoundVar{E} \HOLBoundVar{E\sp{\prime}} \HOLSymConst{\HOLTokenImp{}} \HOLConst{OBS_CONGR} \HOLBoundVar{E\sp{\prime}} \HOLBoundVar{E}
\end{SaveVerbatim}
\newcommand{\HOLObsCongrTheoremsOBSXXCONGRXXSYM}{\UseVerbatim{HOLObsCongrTheoremsOBSXXCONGRXXSYM}}
\begin{SaveVerbatim}{HOLObsCongrTheoremsOBSXXCONGRXXTRANS}
\HOLTokenTurnstile{} \HOLSymConst{\HOLTokenForall{}}\HOLBoundVar{E} \HOLBoundVar{E\sp{\prime}} \HOLBoundVar{E\sp{\prime\prime}}.
     \HOLConst{OBS_CONGR} \HOLBoundVar{E} \HOLBoundVar{E\sp{\prime}} \HOLSymConst{\HOLTokenConj{}} \HOLConst{OBS_CONGR} \HOLBoundVar{E\sp{\prime}} \HOLBoundVar{E\sp{\prime\prime}} \HOLSymConst{\HOLTokenImp{}} \HOLConst{OBS_CONGR} \HOLBoundVar{E} \HOLBoundVar{E\sp{\prime\prime}}
\end{SaveVerbatim}
\newcommand{\HOLObsCongrTheoremsOBSXXCONGRXXTRANS}{\UseVerbatim{HOLObsCongrTheoremsOBSXXCONGRXXTRANS}}
\begin{SaveVerbatim}{HOLObsCongrTheoremsOBSXXCONGRXXTRANSXXLEFT}
\HOLTokenTurnstile{} \HOLSymConst{\HOLTokenForall{}}\HOLBoundVar{E} \HOLBoundVar{E\sp{\prime}}.
     \HOLConst{OBS_CONGR} \HOLBoundVar{E} \HOLBoundVar{E\sp{\prime}} \HOLSymConst{\HOLTokenImp{}}
     \HOLSymConst{\HOLTokenForall{}}\HOLBoundVar{u} \HOLBoundVar{E\sb{\mathrm{1}}}. \HOLBoundVar{E} \HOLTokenTransBegin\HOLBoundVar{u}\HOLTokenTransEnd \HOLBoundVar{E\sb{\mathrm{1}}} \HOLSymConst{\HOLTokenImp{}} \HOLSymConst{\HOLTokenExists{}}\HOLBoundVar{E\sb{\mathrm{2}}}. \HOLBoundVar{E\sp{\prime}} \HOLTokenWeakTransBegin\HOLBoundVar{u}\HOLTokenWeakTransEnd \HOLBoundVar{E\sb{\mathrm{2}}} \HOLSymConst{\HOLTokenConj{}} \HOLConst{WEAK_EQUIV} \HOLBoundVar{E\sb{\mathrm{1}}} \HOLBoundVar{E\sb{\mathrm{2}}}
\end{SaveVerbatim}
\newcommand{\HOLObsCongrTheoremsOBSXXCONGRXXTRANSXXLEFT}{\UseVerbatim{HOLObsCongrTheoremsOBSXXCONGRXXTRANSXXLEFT}}
\begin{SaveVerbatim}{HOLObsCongrTheoremsOBSXXCONGRXXTRANSXXRIGHT}
\HOLTokenTurnstile{} \HOLSymConst{\HOLTokenForall{}}\HOLBoundVar{E} \HOLBoundVar{E\sp{\prime}}.
     \HOLConst{OBS_CONGR} \HOLBoundVar{E} \HOLBoundVar{E\sp{\prime}} \HOLSymConst{\HOLTokenImp{}}
     \HOLSymConst{\HOLTokenForall{}}\HOLBoundVar{u} \HOLBoundVar{E\sb{\mathrm{2}}}. \HOLBoundVar{E\sp{\prime}} \HOLTokenTransBegin\HOLBoundVar{u}\HOLTokenTransEnd \HOLBoundVar{E\sb{\mathrm{2}}} \HOLSymConst{\HOLTokenImp{}} \HOLSymConst{\HOLTokenExists{}}\HOLBoundVar{E\sb{\mathrm{1}}}. \HOLBoundVar{E} \HOLTokenWeakTransBegin\HOLBoundVar{u}\HOLTokenWeakTransEnd \HOLBoundVar{E\sb{\mathrm{1}}} \HOLSymConst{\HOLTokenConj{}} \HOLConst{WEAK_EQUIV} \HOLBoundVar{E\sb{\mathrm{1}}} \HOLBoundVar{E\sb{\mathrm{2}}}
\end{SaveVerbatim}
\newcommand{\HOLObsCongrTheoremsOBSXXCONGRXXTRANSXXRIGHT}{\UseVerbatim{HOLObsCongrTheoremsOBSXXCONGRXXTRANSXXRIGHT}}
\begin{SaveVerbatim}{HOLObsCongrTheoremsOBSXXCONGRXXWEAKXXTRANS}
\HOLTokenTurnstile{} \HOLSymConst{\HOLTokenForall{}}\HOLBoundVar{E} \HOLBoundVar{E\sp{\prime}}.
     \HOLConst{OBS_CONGR} \HOLBoundVar{E} \HOLBoundVar{E\sp{\prime}} \HOLSymConst{\HOLTokenImp{}}
     \HOLSymConst{\HOLTokenForall{}}\HOLBoundVar{u} \HOLBoundVar{E\sb{\mathrm{1}}}. \HOLBoundVar{E} \HOLTokenWeakTransBegin\HOLBoundVar{u}\HOLTokenWeakTransEnd \HOLBoundVar{E\sb{\mathrm{1}}} \HOLSymConst{\HOLTokenImp{}} \HOLSymConst{\HOLTokenExists{}}\HOLBoundVar{E\sb{\mathrm{2}}}. \HOLBoundVar{E\sp{\prime}} \HOLTokenWeakTransBegin\HOLBoundVar{u}\HOLTokenWeakTransEnd \HOLBoundVar{E\sb{\mathrm{2}}} \HOLSymConst{\HOLTokenConj{}} \HOLConst{WEAK_EQUIV} \HOLBoundVar{E\sb{\mathrm{1}}} \HOLBoundVar{E\sb{\mathrm{2}}}
\end{SaveVerbatim}
\newcommand{\HOLObsCongrTheoremsOBSXXCONGRXXWEAKXXTRANS}{\UseVerbatim{HOLObsCongrTheoremsOBSXXCONGRXXWEAKXXTRANS}}
\begin{SaveVerbatim}{HOLObsCongrTheoremsOBSXXCONGRXXWEAKXXTRANSYY}
\HOLTokenTurnstile{} \HOLSymConst{\HOLTokenForall{}}\HOLBoundVar{E} \HOLBoundVar{E\sp{\prime}}.
     \HOLConst{OBS_CONGR} \HOLBoundVar{E} \HOLBoundVar{E\sp{\prime}} \HOLSymConst{\HOLTokenImp{}}
     \HOLSymConst{\HOLTokenForall{}}\HOLBoundVar{u} \HOLBoundVar{E\sb{\mathrm{2}}}. \HOLBoundVar{E\sp{\prime}} \HOLTokenWeakTransBegin\HOLBoundVar{u}\HOLTokenWeakTransEnd \HOLBoundVar{E\sb{\mathrm{2}}} \HOLSymConst{\HOLTokenImp{}} \HOLSymConst{\HOLTokenExists{}}\HOLBoundVar{E\sb{\mathrm{1}}}. \HOLBoundVar{E} \HOLTokenWeakTransBegin\HOLBoundVar{u}\HOLTokenWeakTransEnd \HOLBoundVar{E\sb{\mathrm{1}}} \HOLSymConst{\HOLTokenConj{}} \HOLConst{WEAK_EQUIV} \HOLBoundVar{E\sb{\mathrm{1}}} \HOLBoundVar{E\sb{\mathrm{2}}}
\end{SaveVerbatim}
\newcommand{\HOLObsCongrTheoremsOBSXXCONGRXXWEAKXXTRANSYY}{\UseVerbatim{HOLObsCongrTheoremsOBSXXCONGRXXWEAKXXTRANSYY}}
\begin{SaveVerbatim}{HOLObsCongrTheoremsPROPSix}
\HOLTokenTurnstile{} \HOLSymConst{\HOLTokenForall{}}\HOLBoundVar{E} \HOLBoundVar{E\sp{\prime}}. \HOLConst{WEAK_EQUIV} \HOLBoundVar{E} \HOLBoundVar{E\sp{\prime}} \HOLSymConst{\HOLTokenImp{}} \HOLSymConst{\HOLTokenForall{}}\HOLBoundVar{u}. \HOLConst{OBS_CONGR} (\HOLBoundVar{u}\HOLSymConst{..}\HOLBoundVar{E}) (\HOLBoundVar{u}\HOLSymConst{..}\HOLBoundVar{E\sp{\prime}})
\end{SaveVerbatim}
\newcommand{\HOLObsCongrTheoremsPROPSix}{\UseVerbatim{HOLObsCongrTheoremsPROPSix}}
\begin{SaveVerbatim}{HOLObsCongrTheoremsSTRONGXXIMPXXOBSXXCONGR}
\HOLTokenTurnstile{} \HOLSymConst{\HOLTokenForall{}}\HOLBoundVar{E} \HOLBoundVar{E\sp{\prime}}. \HOLConst{STRONG_EQUIV} \HOLBoundVar{E} \HOLBoundVar{E\sp{\prime}} \HOLSymConst{\HOLTokenImp{}} \HOLConst{OBS_CONGR} \HOLBoundVar{E} \HOLBoundVar{E\sp{\prime}}
\end{SaveVerbatim}
\newcommand{\HOLObsCongrTheoremsSTRONGXXIMPXXOBSXXCONGR}{\UseVerbatim{HOLObsCongrTheoremsSTRONGXXIMPXXOBSXXCONGR}}
\begin{SaveVerbatim}{HOLObsCongrTheoremsWEAKXXEQUIVXXSTABLEXXIMPXXCONGR}
\HOLTokenTurnstile{} \HOLSymConst{\HOLTokenForall{}}\HOLBoundVar{E} \HOLBoundVar{E\sp{\prime}}.
     \HOLConst{WEAK_EQUIV} \HOLBoundVar{E} \HOLBoundVar{E\sp{\prime}} \HOLSymConst{\HOLTokenConj{}} \HOLConst{STABLE} \HOLBoundVar{E} \HOLSymConst{\HOLTokenConj{}} \HOLConst{STABLE} \HOLBoundVar{E\sp{\prime}} \HOLSymConst{\HOLTokenImp{}} \HOLConst{OBS_CONGR} \HOLBoundVar{E} \HOLBoundVar{E\sp{\prime}}
\end{SaveVerbatim}
\newcommand{\HOLObsCongrTheoremsWEAKXXEQUIVXXSTABLEXXIMPXXCONGR}{\UseVerbatim{HOLObsCongrTheoremsWEAKXXEQUIVXXSTABLEXXIMPXXCONGR}}
\newcommand{\HOLObsCongrTheorems}{
\HOLThmTag{ObsCongr}{EQUAL_IMP_OBS_CONGR}\HOLObsCongrTheoremsEQUALXXIMPXXOBSXXCONGR
\HOLThmTag{ObsCongr}{OBS_CONGR_BY_WEAK_BISIM}\HOLObsCongrTheoremsOBSXXCONGRXXBYXXWEAKXXBISIM
\HOLThmTag{ObsCongr}{OBS_CONGR_EPS}\HOLObsCongrTheoremsOBSXXCONGRXXEPS
\HOLThmTag{ObsCongr}{OBS_CONGR_EPS'}\HOLObsCongrTheoremsOBSXXCONGRXXEPSYY
\HOLThmTag{ObsCongr}{OBS_CONGR_equivalence}\HOLObsCongrTheoremsOBSXXCONGRXXequivalence
\HOLThmTag{ObsCongr}{OBS_CONGR_IMP_WEAK_EQUIV}\HOLObsCongrTheoremsOBSXXCONGRXXIMPXXWEAKXXEQUIV
\HOLThmTag{ObsCongr}{OBS_CONGR_PRESD_BY_PAR}\HOLObsCongrTheoremsOBSXXCONGRXXPRESDXXBYXXPAR
\HOLThmTag{ObsCongr}{OBS_CONGR_PRESD_BY_SUM}\HOLObsCongrTheoremsOBSXXCONGRXXPRESDXXBYXXSUM
\HOLThmTag{ObsCongr}{OBS_CONGR_REFL}\HOLObsCongrTheoremsOBSXXCONGRXXREFL
\HOLThmTag{ObsCongr}{OBS_CONGR_SUBST_PAR_L}\HOLObsCongrTheoremsOBSXXCONGRXXSUBSTXXPARXXL
\HOLThmTag{ObsCongr}{OBS_CONGR_SUBST_PAR_R}\HOLObsCongrTheoremsOBSXXCONGRXXSUBSTXXPARXXR
\HOLThmTag{ObsCongr}{OBS_CONGR_SUBST_PREFIX}\HOLObsCongrTheoremsOBSXXCONGRXXSUBSTXXPREFIX
\HOLThmTag{ObsCongr}{OBS_CONGR_SUBST_RELAB}\HOLObsCongrTheoremsOBSXXCONGRXXSUBSTXXRELAB
\HOLThmTag{ObsCongr}{OBS_CONGR_SUBST_RESTR}\HOLObsCongrTheoremsOBSXXCONGRXXSUBSTXXRESTR
\HOLThmTag{ObsCongr}{OBS_CONGR_SUBST_SUM_L}\HOLObsCongrTheoremsOBSXXCONGRXXSUBSTXXSUMXXL
\HOLThmTag{ObsCongr}{OBS_CONGR_SUBST_SUM_R}\HOLObsCongrTheoremsOBSXXCONGRXXSUBSTXXSUMXXR
\HOLThmTag{ObsCongr}{OBS_CONGR_SYM}\HOLObsCongrTheoremsOBSXXCONGRXXSYM
\HOLThmTag{ObsCongr}{OBS_CONGR_TRANS}\HOLObsCongrTheoremsOBSXXCONGRXXTRANS
\HOLThmTag{ObsCongr}{OBS_CONGR_TRANS_LEFT}\HOLObsCongrTheoremsOBSXXCONGRXXTRANSXXLEFT
\HOLThmTag{ObsCongr}{OBS_CONGR_TRANS_RIGHT}\HOLObsCongrTheoremsOBSXXCONGRXXTRANSXXRIGHT
\HOLThmTag{ObsCongr}{OBS_CONGR_WEAK_TRANS}\HOLObsCongrTheoremsOBSXXCONGRXXWEAKXXTRANS
\HOLThmTag{ObsCongr}{OBS_CONGR_WEAK_TRANS'}\HOLObsCongrTheoremsOBSXXCONGRXXWEAKXXTRANSYY
\HOLThmTag{ObsCongr}{PROP6}\HOLObsCongrTheoremsPROPSix
\HOLThmTag{ObsCongr}{STRONG_IMP_OBS_CONGR}\HOLObsCongrTheoremsSTRONGXXIMPXXOBSXXCONGR
\HOLThmTag{ObsCongr}{WEAK_EQUIV_STABLE_IMP_CONGR}\HOLObsCongrTheoremsWEAKXXEQUIVXXSTABLEXXIMPXXCONGR
}

\newcommand{\HOLObsCongrLawsDate}{02 Dicembre 2017}
\newcommand{\HOLObsCongrLawsTime}{13:31}
\begin{SaveVerbatim}{HOLObsCongrLawsTheoremsOBSXXEXPANSIONXXLAW}
\HOLTokenTurnstile{} \HOLSymConst{\HOLTokenForall{}}\HOLBoundVar{f} \HOLBoundVar{n} \HOLBoundVar{f\sp{\prime}} \HOLBoundVar{m}.
     (\HOLSymConst{\HOLTokenForall{}}\HOLBoundVar{i}. \HOLBoundVar{i} \HOLSymConst{\HOLTokenLeq{}} \HOLBoundVar{n} \HOLSymConst{\HOLTokenImp{}} \HOLConst{Is_Prefix} (\HOLBoundVar{f} \HOLBoundVar{i})) \HOLSymConst{\HOLTokenConj{}}
     (\HOLSymConst{\HOLTokenForall{}}\HOLBoundVar{j}. \HOLBoundVar{j} \HOLSymConst{\HOLTokenLeq{}} \HOLBoundVar{m} \HOLSymConst{\HOLTokenImp{}} \HOLConst{Is_Prefix} (\HOLBoundVar{f\sp{\prime}} \HOLBoundVar{j})) \HOLSymConst{\HOLTokenImp{}}
     \HOLConst{OBS_CONGR} (\HOLConst{SIGMA} \HOLBoundVar{f} \HOLBoundVar{n} \HOLSymConst{\ensuremath{\parallel}} \HOLConst{SIGMA} \HOLBoundVar{f\sp{\prime}} \HOLBoundVar{m})
       (\HOLConst{SIGMA}
          (\HOLTokenLambda{}\HOLBoundVar{i}. \HOLConst{PREF_ACT} (\HOLBoundVar{f} \HOLBoundVar{i})\HOLSymConst{..}(\HOLConst{PREF_PROC} (\HOLBoundVar{f} \HOLBoundVar{i}) \HOLSymConst{\ensuremath{\parallel}} \HOLConst{SIGMA} \HOLBoundVar{f\sp{\prime}} \HOLBoundVar{m}))
          \HOLBoundVar{n} \HOLSymConst{+}
        \HOLConst{SIGMA}
          (\HOLTokenLambda{}\HOLBoundVar{j}. \HOLConst{PREF_ACT} (\HOLBoundVar{f\sp{\prime}} \HOLBoundVar{j})\HOLSymConst{..}(\HOLConst{SIGMA} \HOLBoundVar{f} \HOLBoundVar{n} \HOLSymConst{\ensuremath{\parallel}} \HOLConst{PREF_PROC} (\HOLBoundVar{f\sp{\prime}} \HOLBoundVar{j})))
          \HOLBoundVar{m} \HOLSymConst{+} \HOLConst{ALL_SYNC} \HOLBoundVar{f} \HOLBoundVar{n} \HOLBoundVar{f\sp{\prime}} \HOLBoundVar{m})
\end{SaveVerbatim}
\newcommand{\HOLObsCongrLawsTheoremsOBSXXEXPANSIONXXLAW}{\UseVerbatim{HOLObsCongrLawsTheoremsOBSXXEXPANSIONXXLAW}}
\begin{SaveVerbatim}{HOLObsCongrLawsTheoremsOBSXXPARXXASSOC}
\HOLTokenTurnstile{} \HOLSymConst{\HOLTokenForall{}}\HOLBoundVar{E} \HOLBoundVar{E\sp{\prime}} \HOLBoundVar{E\sp{\prime\prime}}. \HOLConst{OBS_CONGR} (\HOLBoundVar{E} \HOLSymConst{\ensuremath{\parallel}} \HOLBoundVar{E\sp{\prime}} \HOLSymConst{\ensuremath{\parallel}} \HOLBoundVar{E\sp{\prime\prime}}) (\HOLBoundVar{E} \HOLSymConst{\ensuremath{\parallel}} (\HOLBoundVar{E\sp{\prime}} \HOLSymConst{\ensuremath{\parallel}} \HOLBoundVar{E\sp{\prime\prime}}))
\end{SaveVerbatim}
\newcommand{\HOLObsCongrLawsTheoremsOBSXXPARXXASSOC}{\UseVerbatim{HOLObsCongrLawsTheoremsOBSXXPARXXASSOC}}
\begin{SaveVerbatim}{HOLObsCongrLawsTheoremsOBSXXPARXXCOMM}
\HOLTokenTurnstile{} \HOLSymConst{\HOLTokenForall{}}\HOLBoundVar{E} \HOLBoundVar{E\sp{\prime}}. \HOLConst{OBS_CONGR} (\HOLBoundVar{E} \HOLSymConst{\ensuremath{\parallel}} \HOLBoundVar{E\sp{\prime}}) (\HOLBoundVar{E\sp{\prime}} \HOLSymConst{\ensuremath{\parallel}} \HOLBoundVar{E})
\end{SaveVerbatim}
\newcommand{\HOLObsCongrLawsTheoremsOBSXXPARXXCOMM}{\UseVerbatim{HOLObsCongrLawsTheoremsOBSXXPARXXCOMM}}
\begin{SaveVerbatim}{HOLObsCongrLawsTheoremsOBSXXPARXXIDENTXXL}
\HOLTokenTurnstile{} \HOLSymConst{\HOLTokenForall{}}\HOLBoundVar{E}. \HOLConst{OBS_CONGR} (\HOLConst{nil} \HOLSymConst{\ensuremath{\parallel}} \HOLBoundVar{E}) \HOLBoundVar{E}
\end{SaveVerbatim}
\newcommand{\HOLObsCongrLawsTheoremsOBSXXPARXXIDENTXXL}{\UseVerbatim{HOLObsCongrLawsTheoremsOBSXXPARXXIDENTXXL}}
\begin{SaveVerbatim}{HOLObsCongrLawsTheoremsOBSXXPARXXIDENTXXR}
\HOLTokenTurnstile{} \HOLSymConst{\HOLTokenForall{}}\HOLBoundVar{E}. \HOLConst{OBS_CONGR} (\HOLBoundVar{E} \HOLSymConst{\ensuremath{\parallel}} \HOLConst{nil}) \HOLBoundVar{E}
\end{SaveVerbatim}
\newcommand{\HOLObsCongrLawsTheoremsOBSXXPARXXIDENTXXR}{\UseVerbatim{HOLObsCongrLawsTheoremsOBSXXPARXXIDENTXXR}}
\begin{SaveVerbatim}{HOLObsCongrLawsTheoremsOBSXXPARXXPREFXXNOXXSYNCR}
\HOLTokenTurnstile{} \HOLSymConst{\HOLTokenForall{}}\HOLBoundVar{l} \HOLBoundVar{l\sp{\prime}}.
     \HOLBoundVar{l} \HOLSymConst{\HOLTokenNotEqual{}} \HOLConst{COMPL} \HOLBoundVar{l\sp{\prime}} \HOLSymConst{\HOLTokenImp{}}
     \HOLSymConst{\HOLTokenForall{}}\HOLBoundVar{E} \HOLBoundVar{E\sp{\prime}}.
       \HOLConst{OBS_CONGR} (\HOLConst{label} \HOLBoundVar{l}\HOLSymConst{..}\HOLBoundVar{E} \HOLSymConst{\ensuremath{\parallel}} \HOLConst{label} \HOLBoundVar{l\sp{\prime}}\HOLSymConst{..}\HOLBoundVar{E\sp{\prime}})
         (\HOLConst{label} \HOLBoundVar{l}\HOLSymConst{..}(\HOLBoundVar{E} \HOLSymConst{\ensuremath{\parallel}} \HOLConst{label} \HOLBoundVar{l\sp{\prime}}\HOLSymConst{..}\HOLBoundVar{E\sp{\prime}}) \HOLSymConst{+}
          \HOLConst{label} \HOLBoundVar{l\sp{\prime}}\HOLSymConst{..}(\HOLConst{label} \HOLBoundVar{l}\HOLSymConst{..}\HOLBoundVar{E} \HOLSymConst{\ensuremath{\parallel}} \HOLBoundVar{E\sp{\prime}}))
\end{SaveVerbatim}
\newcommand{\HOLObsCongrLawsTheoremsOBSXXPARXXPREFXXNOXXSYNCR}{\UseVerbatim{HOLObsCongrLawsTheoremsOBSXXPARXXPREFXXNOXXSYNCR}}
\begin{SaveVerbatim}{HOLObsCongrLawsTheoremsOBSXXPARXXPREFXXSYNCR}
\HOLTokenTurnstile{} \HOLSymConst{\HOLTokenForall{}}\HOLBoundVar{l} \HOLBoundVar{l\sp{\prime}}.
     (\HOLBoundVar{l} \HOLSymConst{=} \HOLConst{COMPL} \HOLBoundVar{l\sp{\prime}}) \HOLSymConst{\HOLTokenImp{}}
     \HOLSymConst{\HOLTokenForall{}}\HOLBoundVar{E} \HOLBoundVar{E\sp{\prime}}.
       \HOLConst{OBS_CONGR} (\HOLConst{label} \HOLBoundVar{l}\HOLSymConst{..}\HOLBoundVar{E} \HOLSymConst{\ensuremath{\parallel}} \HOLConst{label} \HOLBoundVar{l\sp{\prime}}\HOLSymConst{..}\HOLBoundVar{E\sp{\prime}})
         (\HOLConst{label} \HOLBoundVar{l}\HOLSymConst{..}(\HOLBoundVar{E} \HOLSymConst{\ensuremath{\parallel}} \HOLConst{label} \HOLBoundVar{l\sp{\prime}}\HOLSymConst{..}\HOLBoundVar{E\sp{\prime}}) \HOLSymConst{+}
          \HOLConst{label} \HOLBoundVar{l\sp{\prime}}\HOLSymConst{..}(\HOLConst{label} \HOLBoundVar{l}\HOLSymConst{..}\HOLBoundVar{E} \HOLSymConst{\ensuremath{\parallel}} \HOLBoundVar{E\sp{\prime}}) \HOLSymConst{+} \HOLConst{\ensuremath{\tau}}\HOLSymConst{..}(\HOLBoundVar{E} \HOLSymConst{\ensuremath{\parallel}} \HOLBoundVar{E\sp{\prime}}))
\end{SaveVerbatim}
\newcommand{\HOLObsCongrLawsTheoremsOBSXXPARXXPREFXXSYNCR}{\UseVerbatim{HOLObsCongrLawsTheoremsOBSXXPARXXPREFXXSYNCR}}
\begin{SaveVerbatim}{HOLObsCongrLawsTheoremsOBSXXPARXXPREFXXTAU}
\HOLTokenTurnstile{} \HOLSymConst{\HOLTokenForall{}}\HOLBoundVar{u} \HOLBoundVar{E} \HOLBoundVar{E\sp{\prime}}.
     \HOLConst{OBS_CONGR} (\HOLBoundVar{u}\HOLSymConst{..}\HOLBoundVar{E} \HOLSymConst{\ensuremath{\parallel}} \HOLConst{\ensuremath{\tau}}\HOLSymConst{..}\HOLBoundVar{E\sp{\prime}}) (\HOLBoundVar{u}\HOLSymConst{..}(\HOLBoundVar{E} \HOLSymConst{\ensuremath{\parallel}} \HOLConst{\ensuremath{\tau}}\HOLSymConst{..}\HOLBoundVar{E\sp{\prime}}) \HOLSymConst{+} \HOLConst{\ensuremath{\tau}}\HOLSymConst{..}(\HOLBoundVar{u}\HOLSymConst{..}\HOLBoundVar{E} \HOLSymConst{\ensuremath{\parallel}} \HOLBoundVar{E\sp{\prime}}))
\end{SaveVerbatim}
\newcommand{\HOLObsCongrLawsTheoremsOBSXXPARXXPREFXXTAU}{\UseVerbatim{HOLObsCongrLawsTheoremsOBSXXPARXXPREFXXTAU}}
\begin{SaveVerbatim}{HOLObsCongrLawsTheoremsOBSXXPARXXTAUXXPREF}
\HOLTokenTurnstile{} \HOLSymConst{\HOLTokenForall{}}\HOLBoundVar{E} \HOLBoundVar{u} \HOLBoundVar{E\sp{\prime}}.
     \HOLConst{OBS_CONGR} (\HOLConst{\ensuremath{\tau}}\HOLSymConst{..}\HOLBoundVar{E} \HOLSymConst{\ensuremath{\parallel}} \HOLBoundVar{u}\HOLSymConst{..}\HOLBoundVar{E\sp{\prime}}) (\HOLConst{\ensuremath{\tau}}\HOLSymConst{..}(\HOLBoundVar{E} \HOLSymConst{\ensuremath{\parallel}} \HOLBoundVar{u}\HOLSymConst{..}\HOLBoundVar{E\sp{\prime}}) \HOLSymConst{+} \HOLBoundVar{u}\HOLSymConst{..}(\HOLConst{\ensuremath{\tau}}\HOLSymConst{..}\HOLBoundVar{E} \HOLSymConst{\ensuremath{\parallel}} \HOLBoundVar{E\sp{\prime}}))
\end{SaveVerbatim}
\newcommand{\HOLObsCongrLawsTheoremsOBSXXPARXXTAUXXPREF}{\UseVerbatim{HOLObsCongrLawsTheoremsOBSXXPARXXTAUXXPREF}}
\begin{SaveVerbatim}{HOLObsCongrLawsTheoremsOBSXXPARXXTAUXXTAU}
\HOLTokenTurnstile{} \HOLSymConst{\HOLTokenForall{}}\HOLBoundVar{E} \HOLBoundVar{E\sp{\prime}}.
     \HOLConst{OBS_CONGR} (\HOLConst{\ensuremath{\tau}}\HOLSymConst{..}\HOLBoundVar{E} \HOLSymConst{\ensuremath{\parallel}} \HOLConst{\ensuremath{\tau}}\HOLSymConst{..}\HOLBoundVar{E\sp{\prime}}) (\HOLConst{\ensuremath{\tau}}\HOLSymConst{..}(\HOLBoundVar{E} \HOLSymConst{\ensuremath{\parallel}} \HOLConst{\ensuremath{\tau}}\HOLSymConst{..}\HOLBoundVar{E\sp{\prime}}) \HOLSymConst{+} \HOLConst{\ensuremath{\tau}}\HOLSymConst{..}(\HOLConst{\ensuremath{\tau}}\HOLSymConst{..}\HOLBoundVar{E} \HOLSymConst{\ensuremath{\parallel}} \HOLBoundVar{E\sp{\prime}}))
\end{SaveVerbatim}
\newcommand{\HOLObsCongrLawsTheoremsOBSXXPARXXTAUXXTAU}{\UseVerbatim{HOLObsCongrLawsTheoremsOBSXXPARXXTAUXXTAU}}
\begin{SaveVerbatim}{HOLObsCongrLawsTheoremsOBSXXPREFXXRECXXEQUIV}
\HOLTokenTurnstile{} \HOLSymConst{\HOLTokenForall{}}\HOLBoundVar{u} \HOLBoundVar{s} \HOLBoundVar{v}.
     \HOLConst{OBS_CONGR} (\HOLBoundVar{u}\HOLSymConst{..}\HOLConst{rec} \HOLBoundVar{s} (\HOLBoundVar{v}\HOLSymConst{..}\HOLBoundVar{u}\HOLSymConst{..}\HOLConst{var} \HOLBoundVar{s})) (\HOLConst{rec} \HOLBoundVar{s} (\HOLBoundVar{u}\HOLSymConst{..}\HOLBoundVar{v}\HOLSymConst{..}\HOLConst{var} \HOLBoundVar{s}))
\end{SaveVerbatim}
\newcommand{\HOLObsCongrLawsTheoremsOBSXXPREFXXRECXXEQUIV}{\UseVerbatim{HOLObsCongrLawsTheoremsOBSXXPREFXXRECXXEQUIV}}
\begin{SaveVerbatim}{HOLObsCongrLawsTheoremsOBSXXRELABXXNIL}
\HOLTokenTurnstile{} \HOLSymConst{\HOLTokenForall{}}\HOLBoundVar{rf}. \HOLConst{OBS_CONGR} (\HOLConst{relab} \HOLConst{nil} \HOLBoundVar{rf}) \HOLConst{nil}
\end{SaveVerbatim}
\newcommand{\HOLObsCongrLawsTheoremsOBSXXRELABXXNIL}{\UseVerbatim{HOLObsCongrLawsTheoremsOBSXXRELABXXNIL}}
\begin{SaveVerbatim}{HOLObsCongrLawsTheoremsOBSXXRELABXXPREFIX}
\HOLTokenTurnstile{} \HOLSymConst{\HOLTokenForall{}}\HOLBoundVar{u} \HOLBoundVar{E} \HOLBoundVar{labl}.
     \HOLConst{OBS_CONGR} (\HOLConst{relab} (\HOLBoundVar{u}\HOLSymConst{..}\HOLBoundVar{E}) (\HOLConst{RELAB} \HOLBoundVar{labl}))
       (\HOLConst{relabel} (\HOLConst{RELAB} \HOLBoundVar{labl}) \HOLBoundVar{u}\HOLSymConst{..}\HOLConst{relab} \HOLBoundVar{E} (\HOLConst{RELAB} \HOLBoundVar{labl}))
\end{SaveVerbatim}
\newcommand{\HOLObsCongrLawsTheoremsOBSXXRELABXXPREFIX}{\UseVerbatim{HOLObsCongrLawsTheoremsOBSXXRELABXXPREFIX}}
\begin{SaveVerbatim}{HOLObsCongrLawsTheoremsOBSXXRELABXXSUM}
\HOLTokenTurnstile{} \HOLSymConst{\HOLTokenForall{}}\HOLBoundVar{E} \HOLBoundVar{E\sp{\prime}} \HOLBoundVar{rf}.
     \HOLConst{OBS_CONGR} (\HOLConst{relab} (\HOLBoundVar{E} \HOLSymConst{+} \HOLBoundVar{E\sp{\prime}}) \HOLBoundVar{rf}) (\HOLConst{relab} \HOLBoundVar{E} \HOLBoundVar{rf} \HOLSymConst{+} \HOLConst{relab} \HOLBoundVar{E\sp{\prime}} \HOLBoundVar{rf})
\end{SaveVerbatim}
\newcommand{\HOLObsCongrLawsTheoremsOBSXXRELABXXSUM}{\UseVerbatim{HOLObsCongrLawsTheoremsOBSXXRELABXXSUM}}
\begin{SaveVerbatim}{HOLObsCongrLawsTheoremsOBSXXRESTRXXNIL}
\HOLTokenTurnstile{} \HOLSymConst{\HOLTokenForall{}}\HOLBoundVar{L}. \HOLConst{OBS_CONGR} (\HOLConst{\ensuremath{\nu}} \HOLBoundVar{L} \HOLConst{nil}) \HOLConst{nil}
\end{SaveVerbatim}
\newcommand{\HOLObsCongrLawsTheoremsOBSXXRESTRXXNIL}{\UseVerbatim{HOLObsCongrLawsTheoremsOBSXXRESTRXXNIL}}
\begin{SaveVerbatim}{HOLObsCongrLawsTheoremsOBSXXRESTRXXPRXXLABXXNIL}
\HOLTokenTurnstile{} \HOLSymConst{\HOLTokenForall{}}\HOLBoundVar{l} \HOLBoundVar{L}.
     \HOLBoundVar{l} \HOLConst{\HOLTokenIn{}} \HOLBoundVar{L} \HOLSymConst{\HOLTokenDisj{}} \HOLConst{COMPL} \HOLBoundVar{l} \HOLConst{\HOLTokenIn{}} \HOLBoundVar{L} \HOLSymConst{\HOLTokenImp{}} \HOLSymConst{\HOLTokenForall{}}\HOLBoundVar{E}. \HOLConst{OBS_CONGR} (\HOLConst{\ensuremath{\nu}} \HOLBoundVar{L} (\HOLConst{label} \HOLBoundVar{l}\HOLSymConst{..}\HOLBoundVar{E})) \HOLConst{nil}
\end{SaveVerbatim}
\newcommand{\HOLObsCongrLawsTheoremsOBSXXRESTRXXPRXXLABXXNIL}{\UseVerbatim{HOLObsCongrLawsTheoremsOBSXXRESTRXXPRXXLABXXNIL}}
\begin{SaveVerbatim}{HOLObsCongrLawsTheoremsOBSXXRESTRXXPREFIXXXLABEL}
\HOLTokenTurnstile{} \HOLSymConst{\HOLTokenForall{}}\HOLBoundVar{l} \HOLBoundVar{L}.
     \HOLBoundVar{l} \HOLConst{\HOLTokenNotIn{}} \HOLBoundVar{L} \HOLSymConst{\HOLTokenConj{}} \HOLConst{COMPL} \HOLBoundVar{l} \HOLConst{\HOLTokenNotIn{}} \HOLBoundVar{L} \HOLSymConst{\HOLTokenImp{}}
     \HOLSymConst{\HOLTokenForall{}}\HOLBoundVar{E}. \HOLConst{OBS_CONGR} (\HOLConst{\ensuremath{\nu}} \HOLBoundVar{L} (\HOLConst{label} \HOLBoundVar{l}\HOLSymConst{..}\HOLBoundVar{E})) (\HOLConst{label} \HOLBoundVar{l}\HOLSymConst{..}\HOLConst{\ensuremath{\nu}} \HOLBoundVar{L} \HOLBoundVar{E})
\end{SaveVerbatim}
\newcommand{\HOLObsCongrLawsTheoremsOBSXXRESTRXXPREFIXXXLABEL}{\UseVerbatim{HOLObsCongrLawsTheoremsOBSXXRESTRXXPREFIXXXLABEL}}
\begin{SaveVerbatim}{HOLObsCongrLawsTheoremsOBSXXRESTRXXPREFIXXXTAU}
\HOLTokenTurnstile{} \HOLSymConst{\HOLTokenForall{}}\HOLBoundVar{E} \HOLBoundVar{L}. \HOLConst{OBS_CONGR} (\HOLConst{\ensuremath{\nu}} \HOLBoundVar{L} (\HOLConst{\ensuremath{\tau}}\HOLSymConst{..}\HOLBoundVar{E})) (\HOLConst{\ensuremath{\tau}}\HOLSymConst{..}\HOLConst{\ensuremath{\nu}} \HOLBoundVar{L} \HOLBoundVar{E})
\end{SaveVerbatim}
\newcommand{\HOLObsCongrLawsTheoremsOBSXXRESTRXXPREFIXXXTAU}{\UseVerbatim{HOLObsCongrLawsTheoremsOBSXXRESTRXXPREFIXXXTAU}}
\begin{SaveVerbatim}{HOLObsCongrLawsTheoremsOBSXXRESTRXXSUM}
\HOLTokenTurnstile{} \HOLSymConst{\HOLTokenForall{}}\HOLBoundVar{E} \HOLBoundVar{E\sp{\prime}} \HOLBoundVar{L}. \HOLConst{OBS_CONGR} (\HOLConst{\ensuremath{\nu}} \HOLBoundVar{L} (\HOLBoundVar{E} \HOLSymConst{+} \HOLBoundVar{E\sp{\prime}})) (\HOLConst{\ensuremath{\nu}} \HOLBoundVar{L} \HOLBoundVar{E} \HOLSymConst{+} \HOLConst{\ensuremath{\nu}} \HOLBoundVar{L} \HOLBoundVar{E\sp{\prime}})
\end{SaveVerbatim}
\newcommand{\HOLObsCongrLawsTheoremsOBSXXRESTRXXSUM}{\UseVerbatim{HOLObsCongrLawsTheoremsOBSXXRESTRXXSUM}}
\begin{SaveVerbatim}{HOLObsCongrLawsTheoremsOBSXXSUMXXASSOCXXL}
\HOLTokenTurnstile{} \HOLSymConst{\HOLTokenForall{}}\HOLBoundVar{E} \HOLBoundVar{E\sp{\prime}} \HOLBoundVar{E\sp{\prime\prime}}. \HOLConst{OBS_CONGR} (\HOLBoundVar{E} \HOLSymConst{+} (\HOLBoundVar{E\sp{\prime}} \HOLSymConst{+} \HOLBoundVar{E\sp{\prime\prime}})) (\HOLBoundVar{E} \HOLSymConst{+} \HOLBoundVar{E\sp{\prime}} \HOLSymConst{+} \HOLBoundVar{E\sp{\prime\prime}})
\end{SaveVerbatim}
\newcommand{\HOLObsCongrLawsTheoremsOBSXXSUMXXASSOCXXL}{\UseVerbatim{HOLObsCongrLawsTheoremsOBSXXSUMXXASSOCXXL}}
\begin{SaveVerbatim}{HOLObsCongrLawsTheoremsOBSXXSUMXXASSOCXXR}
\HOLTokenTurnstile{} \HOLSymConst{\HOLTokenForall{}}\HOLBoundVar{E} \HOLBoundVar{E\sp{\prime}} \HOLBoundVar{E\sp{\prime\prime}}. \HOLConst{OBS_CONGR} (\HOLBoundVar{E} \HOLSymConst{+} \HOLBoundVar{E\sp{\prime}} \HOLSymConst{+} \HOLBoundVar{E\sp{\prime\prime}}) (\HOLBoundVar{E} \HOLSymConst{+} (\HOLBoundVar{E\sp{\prime}} \HOLSymConst{+} \HOLBoundVar{E\sp{\prime\prime}}))
\end{SaveVerbatim}
\newcommand{\HOLObsCongrLawsTheoremsOBSXXSUMXXASSOCXXR}{\UseVerbatim{HOLObsCongrLawsTheoremsOBSXXSUMXXASSOCXXR}}
\begin{SaveVerbatim}{HOLObsCongrLawsTheoremsOBSXXSUMXXCOMM}
\HOLTokenTurnstile{} \HOLSymConst{\HOLTokenForall{}}\HOLBoundVar{E} \HOLBoundVar{E\sp{\prime}}. \HOLConst{OBS_CONGR} (\HOLBoundVar{E} \HOLSymConst{+} \HOLBoundVar{E\sp{\prime}}) (\HOLBoundVar{E\sp{\prime}} \HOLSymConst{+} \HOLBoundVar{E})
\end{SaveVerbatim}
\newcommand{\HOLObsCongrLawsTheoremsOBSXXSUMXXCOMM}{\UseVerbatim{HOLObsCongrLawsTheoremsOBSXXSUMXXCOMM}}
\begin{SaveVerbatim}{HOLObsCongrLawsTheoremsOBSXXSUMXXIDEMP}
\HOLTokenTurnstile{} \HOLSymConst{\HOLTokenForall{}}\HOLBoundVar{E}. \HOLConst{OBS_CONGR} (\HOLBoundVar{E} \HOLSymConst{+} \HOLBoundVar{E}) \HOLBoundVar{E}
\end{SaveVerbatim}
\newcommand{\HOLObsCongrLawsTheoremsOBSXXSUMXXIDEMP}{\UseVerbatim{HOLObsCongrLawsTheoremsOBSXXSUMXXIDEMP}}
\begin{SaveVerbatim}{HOLObsCongrLawsTheoremsOBSXXSUMXXIDENTXXL}
\HOLTokenTurnstile{} \HOLSymConst{\HOLTokenForall{}}\HOLBoundVar{E}. \HOLConst{OBS_CONGR} (\HOLConst{nil} \HOLSymConst{+} \HOLBoundVar{E}) \HOLBoundVar{E}
\end{SaveVerbatim}
\newcommand{\HOLObsCongrLawsTheoremsOBSXXSUMXXIDENTXXL}{\UseVerbatim{HOLObsCongrLawsTheoremsOBSXXSUMXXIDENTXXL}}
\begin{SaveVerbatim}{HOLObsCongrLawsTheoremsOBSXXSUMXXIDENTXXR}
\HOLTokenTurnstile{} \HOLSymConst{\HOLTokenForall{}}\HOLBoundVar{E}. \HOLConst{OBS_CONGR} (\HOLBoundVar{E} \HOLSymConst{+} \HOLConst{nil}) \HOLBoundVar{E}
\end{SaveVerbatim}
\newcommand{\HOLObsCongrLawsTheoremsOBSXXSUMXXIDENTXXR}{\UseVerbatim{HOLObsCongrLawsTheoremsOBSXXSUMXXIDENTXXR}}
\begin{SaveVerbatim}{HOLObsCongrLawsTheoremsOBSXXUNFOLDING}
\HOLTokenTurnstile{} \HOLSymConst{\HOLTokenForall{}}\HOLBoundVar{X} \HOLBoundVar{E}. \HOLConst{OBS_CONGR} (\HOLConst{rec} \HOLBoundVar{X} \HOLBoundVar{E}) (\HOLConst{CCS_Subst} \HOLBoundVar{E} (\HOLConst{rec} \HOLBoundVar{X} \HOLBoundVar{E}) \HOLBoundVar{X})
\end{SaveVerbatim}
\newcommand{\HOLObsCongrLawsTheoremsOBSXXUNFOLDING}{\UseVerbatim{HOLObsCongrLawsTheoremsOBSXXUNFOLDING}}
\begin{SaveVerbatim}{HOLObsCongrLawsTheoremsTAUOne}
\HOLTokenTurnstile{} \HOLSymConst{\HOLTokenForall{}}\HOLBoundVar{u} \HOLBoundVar{E}. \HOLConst{OBS_CONGR} (\HOLBoundVar{u}\HOLSymConst{..}\HOLConst{\ensuremath{\tau}}\HOLSymConst{..}\HOLBoundVar{E}) (\HOLBoundVar{u}\HOLSymConst{..}\HOLBoundVar{E})
\end{SaveVerbatim}
\newcommand{\HOLObsCongrLawsTheoremsTAUOne}{\UseVerbatim{HOLObsCongrLawsTheoremsTAUOne}}
\begin{SaveVerbatim}{HOLObsCongrLawsTheoremsTAUTwo}
\HOLTokenTurnstile{} \HOLSymConst{\HOLTokenForall{}}\HOLBoundVar{E}. \HOLConst{OBS_CONGR} (\HOLBoundVar{E} \HOLSymConst{+} \HOLConst{\ensuremath{\tau}}\HOLSymConst{..}\HOLBoundVar{E}) (\HOLConst{\ensuremath{\tau}}\HOLSymConst{..}\HOLBoundVar{E})
\end{SaveVerbatim}
\newcommand{\HOLObsCongrLawsTheoremsTAUTwo}{\UseVerbatim{HOLObsCongrLawsTheoremsTAUTwo}}
\begin{SaveVerbatim}{HOLObsCongrLawsTheoremsTAUThree}
\HOLTokenTurnstile{} \HOLSymConst{\HOLTokenForall{}}\HOLBoundVar{u} \HOLBoundVar{E} \HOLBoundVar{E\sp{\prime}}. \HOLConst{OBS_CONGR} (\HOLBoundVar{u}\HOLSymConst{..}(\HOLBoundVar{E} \HOLSymConst{+} \HOLConst{\ensuremath{\tau}}\HOLSymConst{..}\HOLBoundVar{E\sp{\prime}}) \HOLSymConst{+} \HOLBoundVar{u}\HOLSymConst{..}\HOLBoundVar{E\sp{\prime}}) (\HOLBoundVar{u}\HOLSymConst{..}(\HOLBoundVar{E} \HOLSymConst{+} \HOLConst{\ensuremath{\tau}}\HOLSymConst{..}\HOLBoundVar{E\sp{\prime}}))
\end{SaveVerbatim}
\newcommand{\HOLObsCongrLawsTheoremsTAUThree}{\UseVerbatim{HOLObsCongrLawsTheoremsTAUThree}}
\begin{SaveVerbatim}{HOLObsCongrLawsTheoremsWEAKXXTAUOne}
\HOLTokenTurnstile{} \HOLSymConst{\HOLTokenForall{}}\HOLBoundVar{u} \HOLBoundVar{E}. \HOLConst{WEAK_EQUIV} (\HOLBoundVar{u}\HOLSymConst{..}\HOLConst{\ensuremath{\tau}}\HOLSymConst{..}\HOLBoundVar{E}) (\HOLBoundVar{u}\HOLSymConst{..}\HOLBoundVar{E})
\end{SaveVerbatim}
\newcommand{\HOLObsCongrLawsTheoremsWEAKXXTAUOne}{\UseVerbatim{HOLObsCongrLawsTheoremsWEAKXXTAUOne}}
\begin{SaveVerbatim}{HOLObsCongrLawsTheoremsWEAKXXTAUTwo}
\HOLTokenTurnstile{} \HOLSymConst{\HOLTokenForall{}}\HOLBoundVar{E}. \HOLConst{WEAK_EQUIV} (\HOLBoundVar{E} \HOLSymConst{+} \HOLConst{\ensuremath{\tau}}\HOLSymConst{..}\HOLBoundVar{E}) (\HOLConst{\ensuremath{\tau}}\HOLSymConst{..}\HOLBoundVar{E})
\end{SaveVerbatim}
\newcommand{\HOLObsCongrLawsTheoremsWEAKXXTAUTwo}{\UseVerbatim{HOLObsCongrLawsTheoremsWEAKXXTAUTwo}}
\begin{SaveVerbatim}{HOLObsCongrLawsTheoremsWEAKXXTAUThree}
\HOLTokenTurnstile{} \HOLSymConst{\HOLTokenForall{}}\HOLBoundVar{u} \HOLBoundVar{E} \HOLBoundVar{E\sp{\prime}}. \HOLConst{WEAK_EQUIV} (\HOLBoundVar{u}\HOLSymConst{..}(\HOLBoundVar{E} \HOLSymConst{+} \HOLConst{\ensuremath{\tau}}\HOLSymConst{..}\HOLBoundVar{E\sp{\prime}}) \HOLSymConst{+} \HOLBoundVar{u}\HOLSymConst{..}\HOLBoundVar{E\sp{\prime}}) (\HOLBoundVar{u}\HOLSymConst{..}(\HOLBoundVar{E} \HOLSymConst{+} \HOLConst{\ensuremath{\tau}}\HOLSymConst{..}\HOLBoundVar{E\sp{\prime}}))
\end{SaveVerbatim}
\newcommand{\HOLObsCongrLawsTheoremsWEAKXXTAUThree}{\UseVerbatim{HOLObsCongrLawsTheoremsWEAKXXTAUThree}}
\newcommand{\HOLObsCongrLawsTheorems}{
\HOLThmTag{ObsCongrLaws}{OBS_EXPANSION_LAW}\HOLObsCongrLawsTheoremsOBSXXEXPANSIONXXLAW
\HOLThmTag{ObsCongrLaws}{OBS_PAR_ASSOC}\HOLObsCongrLawsTheoremsOBSXXPARXXASSOC
\HOLThmTag{ObsCongrLaws}{OBS_PAR_COMM}\HOLObsCongrLawsTheoremsOBSXXPARXXCOMM
\HOLThmTag{ObsCongrLaws}{OBS_PAR_IDENT_L}\HOLObsCongrLawsTheoremsOBSXXPARXXIDENTXXL
\HOLThmTag{ObsCongrLaws}{OBS_PAR_IDENT_R}\HOLObsCongrLawsTheoremsOBSXXPARXXIDENTXXR
\HOLThmTag{ObsCongrLaws}{OBS_PAR_PREF_NO_SYNCR}\HOLObsCongrLawsTheoremsOBSXXPARXXPREFXXNOXXSYNCR
\HOLThmTag{ObsCongrLaws}{OBS_PAR_PREF_SYNCR}\HOLObsCongrLawsTheoremsOBSXXPARXXPREFXXSYNCR
\HOLThmTag{ObsCongrLaws}{OBS_PAR_PREF_TAU}\HOLObsCongrLawsTheoremsOBSXXPARXXPREFXXTAU
\HOLThmTag{ObsCongrLaws}{OBS_PAR_TAU_PREF}\HOLObsCongrLawsTheoremsOBSXXPARXXTAUXXPREF
\HOLThmTag{ObsCongrLaws}{OBS_PAR_TAU_TAU}\HOLObsCongrLawsTheoremsOBSXXPARXXTAUXXTAU
\HOLThmTag{ObsCongrLaws}{OBS_PREF_REC_EQUIV}\HOLObsCongrLawsTheoremsOBSXXPREFXXRECXXEQUIV
\HOLThmTag{ObsCongrLaws}{OBS_RELAB_NIL}\HOLObsCongrLawsTheoremsOBSXXRELABXXNIL
\HOLThmTag{ObsCongrLaws}{OBS_RELAB_PREFIX}\HOLObsCongrLawsTheoremsOBSXXRELABXXPREFIX
\HOLThmTag{ObsCongrLaws}{OBS_RELAB_SUM}\HOLObsCongrLawsTheoremsOBSXXRELABXXSUM
\HOLThmTag{ObsCongrLaws}{OBS_RESTR_NIL}\HOLObsCongrLawsTheoremsOBSXXRESTRXXNIL
\HOLThmTag{ObsCongrLaws}{OBS_RESTR_PR_LAB_NIL}\HOLObsCongrLawsTheoremsOBSXXRESTRXXPRXXLABXXNIL
\HOLThmTag{ObsCongrLaws}{OBS_RESTR_PREFIX_LABEL}\HOLObsCongrLawsTheoremsOBSXXRESTRXXPREFIXXXLABEL
\HOLThmTag{ObsCongrLaws}{OBS_RESTR_PREFIX_TAU}\HOLObsCongrLawsTheoremsOBSXXRESTRXXPREFIXXXTAU
\HOLThmTag{ObsCongrLaws}{OBS_RESTR_SUM}\HOLObsCongrLawsTheoremsOBSXXRESTRXXSUM
\HOLThmTag{ObsCongrLaws}{OBS_SUM_ASSOC_L}\HOLObsCongrLawsTheoremsOBSXXSUMXXASSOCXXL
\HOLThmTag{ObsCongrLaws}{OBS_SUM_ASSOC_R}\HOLObsCongrLawsTheoremsOBSXXSUMXXASSOCXXR
\HOLThmTag{ObsCongrLaws}{OBS_SUM_COMM}\HOLObsCongrLawsTheoremsOBSXXSUMXXCOMM
\HOLThmTag{ObsCongrLaws}{OBS_SUM_IDEMP}\HOLObsCongrLawsTheoremsOBSXXSUMXXIDEMP
\HOLThmTag{ObsCongrLaws}{OBS_SUM_IDENT_L}\HOLObsCongrLawsTheoremsOBSXXSUMXXIDENTXXL
\HOLThmTag{ObsCongrLaws}{OBS_SUM_IDENT_R}\HOLObsCongrLawsTheoremsOBSXXSUMXXIDENTXXR
\HOLThmTag{ObsCongrLaws}{OBS_UNFOLDING}\HOLObsCongrLawsTheoremsOBSXXUNFOLDING
\HOLThmTag{ObsCongrLaws}{TAU1}\HOLObsCongrLawsTheoremsTAUOne
\HOLThmTag{ObsCongrLaws}{TAU2}\HOLObsCongrLawsTheoremsTAUTwo
\HOLThmTag{ObsCongrLaws}{TAU3}\HOLObsCongrLawsTheoremsTAUThree
\HOLThmTag{ObsCongrLaws}{WEAK_TAU1}\HOLObsCongrLawsTheoremsWEAKXXTAUOne
\HOLThmTag{ObsCongrLaws}{WEAK_TAU2}\HOLObsCongrLawsTheoremsWEAKXXTAUTwo
\HOLThmTag{ObsCongrLaws}{WEAK_TAU3}\HOLObsCongrLawsTheoremsWEAKXXTAUThree
}

\newcommand{\HOLCongruenceDate}{02 Dicembre 2017}
\newcommand{\HOLCongruenceTime}{13:31}
\begin{SaveVerbatim}{HOLCongruenceDefinitionsCCXXdef}
\HOLTokenTurnstile{} \HOLSymConst{\HOLTokenForall{}}\HOLBoundVar{R}. \HOLConst{CC} \HOLBoundVar{R} \HOLSymConst{=} (\HOLTokenLambda{}\HOLBoundVar{g} \HOLBoundVar{h}. \HOLSymConst{\HOLTokenForall{}}\HOLBoundVar{c}. \HOLConst{CONTEXT} \HOLBoundVar{c} \HOLSymConst{\HOLTokenImp{}} \HOLBoundVar{R} (\HOLBoundVar{c} \HOLBoundVar{g}) (\HOLBoundVar{c} \HOLBoundVar{h}))
\end{SaveVerbatim}
\newcommand{\HOLCongruenceDefinitionsCCXXdef}{\UseVerbatim{HOLCongruenceDefinitionsCCXXdef}}
\begin{SaveVerbatim}{HOLCongruenceDefinitionscongruenceOneXXdef}
\HOLTokenTurnstile{} \HOLSymConst{\HOLTokenForall{}}\HOLBoundVar{R}. \HOLConst{congruence1} \HOLBoundVar{R} \HOLSymConst{\HOLTokenEquiv{}} \HOLConst{equivalence} \HOLBoundVar{R} \HOLSymConst{\HOLTokenConj{}} \HOLConst{precongruence1} \HOLBoundVar{R}
\end{SaveVerbatim}
\newcommand{\HOLCongruenceDefinitionscongruenceOneXXdef}{\UseVerbatim{HOLCongruenceDefinitionscongruenceOneXXdef}}
\begin{SaveVerbatim}{HOLCongruenceDefinitionscongruenceXXdef}
\HOLTokenTurnstile{} \HOLSymConst{\HOLTokenForall{}}\HOLBoundVar{R}. \HOLConst{congruence} \HOLBoundVar{R} \HOLSymConst{\HOLTokenEquiv{}} \HOLConst{equivalence} \HOLBoundVar{R} \HOLSymConst{\HOLTokenConj{}} \HOLConst{precongruence} \HOLBoundVar{R}
\end{SaveVerbatim}
\newcommand{\HOLCongruenceDefinitionscongruenceXXdef}{\UseVerbatim{HOLCongruenceDefinitionscongruenceXXdef}}
\begin{SaveVerbatim}{HOLCongruenceDefinitionsCONTEXTXXdef}
\HOLTokenTurnstile{} \HOLConst{CONTEXT} \HOLSymConst{=}
   (\HOLTokenLambda{}\HOLBoundVar{a\sb{\mathrm{0}}}.
      \HOLSymConst{\HOLTokenForall{}}\HOLBoundVar{CONTEXT\sp{\prime}}.
        (\HOLSymConst{\HOLTokenForall{}}\HOLBoundVar{a\sb{\mathrm{0}}}.
           (\HOLBoundVar{a\sb{\mathrm{0}}} \HOLSymConst{=} (\HOLTokenLambda{}\HOLBoundVar{t}. \HOLBoundVar{t})) \HOLSymConst{\HOLTokenDisj{}} (\HOLSymConst{\HOLTokenExists{}}\HOLBoundVar{p}. \HOLBoundVar{a\sb{\mathrm{0}}} \HOLSymConst{=} (\HOLTokenLambda{}\HOLBoundVar{t}. \HOLBoundVar{p})) \HOLSymConst{\HOLTokenDisj{}}
           (\HOLSymConst{\HOLTokenExists{}}\HOLBoundVar{a} \HOLBoundVar{e}. (\HOLBoundVar{a\sb{\mathrm{0}}} \HOLSymConst{=} (\HOLTokenLambda{}\HOLBoundVar{t}. \HOLBoundVar{a}\HOLSymConst{..}\HOLBoundVar{e} \HOLBoundVar{t})) \HOLSymConst{\HOLTokenConj{}} \HOLBoundVar{CONTEXT\sp{\prime}} \HOLBoundVar{e}) \HOLSymConst{\HOLTokenDisj{}}
           (\HOLSymConst{\HOLTokenExists{}}\HOLBoundVar{e\sb{\mathrm{1}}} \HOLBoundVar{e\sb{\mathrm{2}}}.
              (\HOLBoundVar{a\sb{\mathrm{0}}} \HOLSymConst{=} (\HOLTokenLambda{}\HOLBoundVar{t}. \HOLBoundVar{e\sb{\mathrm{1}}} \HOLBoundVar{t} \HOLSymConst{+} \HOLBoundVar{e\sb{\mathrm{2}}} \HOLBoundVar{t})) \HOLSymConst{\HOLTokenConj{}} \HOLBoundVar{CONTEXT\sp{\prime}} \HOLBoundVar{e\sb{\mathrm{1}}} \HOLSymConst{\HOLTokenConj{}}
              \HOLBoundVar{CONTEXT\sp{\prime}} \HOLBoundVar{e\sb{\mathrm{2}}}) \HOLSymConst{\HOLTokenDisj{}}
           (\HOLSymConst{\HOLTokenExists{}}\HOLBoundVar{e\sb{\mathrm{1}}} \HOLBoundVar{e\sb{\mathrm{2}}}.
              (\HOLBoundVar{a\sb{\mathrm{0}}} \HOLSymConst{=} (\HOLTokenLambda{}\HOLBoundVar{t}. \HOLBoundVar{e\sb{\mathrm{1}}} \HOLBoundVar{t} \HOLSymConst{\ensuremath{\parallel}} \HOLBoundVar{e\sb{\mathrm{2}}} \HOLBoundVar{t})) \HOLSymConst{\HOLTokenConj{}} \HOLBoundVar{CONTEXT\sp{\prime}} \HOLBoundVar{e\sb{\mathrm{1}}} \HOLSymConst{\HOLTokenConj{}}
              \HOLBoundVar{CONTEXT\sp{\prime}} \HOLBoundVar{e\sb{\mathrm{2}}}) \HOLSymConst{\HOLTokenDisj{}}
           (\HOLSymConst{\HOLTokenExists{}}\HOLBoundVar{L} \HOLBoundVar{e}. (\HOLBoundVar{a\sb{\mathrm{0}}} \HOLSymConst{=} (\HOLTokenLambda{}\HOLBoundVar{t}. \HOLConst{\ensuremath{\nu}} \HOLBoundVar{L} (\HOLBoundVar{e} \HOLBoundVar{t}))) \HOLSymConst{\HOLTokenConj{}} \HOLBoundVar{CONTEXT\sp{\prime}} \HOLBoundVar{e}) \HOLSymConst{\HOLTokenDisj{}}
           (\HOLSymConst{\HOLTokenExists{}}\HOLBoundVar{rf} \HOLBoundVar{e}. (\HOLBoundVar{a\sb{\mathrm{0}}} \HOLSymConst{=} (\HOLTokenLambda{}\HOLBoundVar{t}. \HOLConst{relab} (\HOLBoundVar{e} \HOLBoundVar{t}) \HOLBoundVar{rf})) \HOLSymConst{\HOLTokenConj{}} \HOLBoundVar{CONTEXT\sp{\prime}} \HOLBoundVar{e}) \HOLSymConst{\HOLTokenImp{}}
           \HOLBoundVar{CONTEXT\sp{\prime}} \HOLBoundVar{a\sb{\mathrm{0}}}) \HOLSymConst{\HOLTokenImp{}}
        \HOLBoundVar{CONTEXT\sp{\prime}} \HOLBoundVar{a\sb{\mathrm{0}}})
\end{SaveVerbatim}
\newcommand{\HOLCongruenceDefinitionsCONTEXTXXdef}{\UseVerbatim{HOLCongruenceDefinitionsCONTEXTXXdef}}
\begin{SaveVerbatim}{HOLCongruenceDefinitionsGCCXXdef}
\HOLTokenTurnstile{} \HOLSymConst{\HOLTokenForall{}}\HOLBoundVar{R}. \HOLConst{GCC} \HOLBoundVar{R} \HOLSymConst{=} (\HOLTokenLambda{}\HOLBoundVar{g} \HOLBoundVar{h}. \HOLSymConst{\HOLTokenForall{}}\HOLBoundVar{c}. \HOLConst{GCONTEXT} \HOLBoundVar{c} \HOLSymConst{\HOLTokenImp{}} \HOLBoundVar{R} (\HOLBoundVar{c} \HOLBoundVar{g}) (\HOLBoundVar{c} \HOLBoundVar{h}))
\end{SaveVerbatim}
\newcommand{\HOLCongruenceDefinitionsGCCXXdef}{\UseVerbatim{HOLCongruenceDefinitionsGCCXXdef}}
\begin{SaveVerbatim}{HOLCongruenceDefinitionsGCONTEXTXXdef}
\HOLTokenTurnstile{} \HOLConst{GCONTEXT} \HOLSymConst{=}
   (\HOLTokenLambda{}\HOLBoundVar{a\sb{\mathrm{0}}}.
      \HOLSymConst{\HOLTokenForall{}}\HOLBoundVar{GCONTEXT\sp{\prime}}.
        (\HOLSymConst{\HOLTokenForall{}}\HOLBoundVar{a\sb{\mathrm{0}}}.
           (\HOLBoundVar{a\sb{\mathrm{0}}} \HOLSymConst{=} (\HOLTokenLambda{}\HOLBoundVar{t}. \HOLBoundVar{t})) \HOLSymConst{\HOLTokenDisj{}} (\HOLSymConst{\HOLTokenExists{}}\HOLBoundVar{p}. \HOLBoundVar{a\sb{\mathrm{0}}} \HOLSymConst{=} (\HOLTokenLambda{}\HOLBoundVar{t}. \HOLBoundVar{p})) \HOLSymConst{\HOLTokenDisj{}}
           (\HOLSymConst{\HOLTokenExists{}}\HOLBoundVar{a} \HOLBoundVar{e}. (\HOLBoundVar{a\sb{\mathrm{0}}} \HOLSymConst{=} (\HOLTokenLambda{}\HOLBoundVar{t}. \HOLBoundVar{a}\HOLSymConst{..}\HOLBoundVar{e} \HOLBoundVar{t})) \HOLSymConst{\HOLTokenConj{}} \HOLBoundVar{GCONTEXT\sp{\prime}} \HOLBoundVar{e}) \HOLSymConst{\HOLTokenDisj{}}
           (\HOLSymConst{\HOLTokenExists{}}\HOLBoundVar{a\sb{\mathrm{1}}} \HOLBoundVar{a\sb{\mathrm{2}}} \HOLBoundVar{e\sb{\mathrm{1}}} \HOLBoundVar{e\sb{\mathrm{2}}}.
              (\HOLBoundVar{a\sb{\mathrm{0}}} \HOLSymConst{=} (\HOLTokenLambda{}\HOLBoundVar{t}. \HOLBoundVar{a\sb{\mathrm{1}}}\HOLSymConst{..}\HOLBoundVar{e\sb{\mathrm{1}}} \HOLBoundVar{t} \HOLSymConst{+} \HOLBoundVar{a\sb{\mathrm{2}}}\HOLSymConst{..}\HOLBoundVar{e\sb{\mathrm{2}}} \HOLBoundVar{t})) \HOLSymConst{\HOLTokenConj{}} \HOLBoundVar{GCONTEXT\sp{\prime}} \HOLBoundVar{e\sb{\mathrm{1}}} \HOLSymConst{\HOLTokenConj{}}
              \HOLBoundVar{GCONTEXT\sp{\prime}} \HOLBoundVar{e\sb{\mathrm{2}}}) \HOLSymConst{\HOLTokenDisj{}}
           (\HOLSymConst{\HOLTokenExists{}}\HOLBoundVar{e\sb{\mathrm{1}}} \HOLBoundVar{e\sb{\mathrm{2}}}.
              (\HOLBoundVar{a\sb{\mathrm{0}}} \HOLSymConst{=} (\HOLTokenLambda{}\HOLBoundVar{t}. \HOLBoundVar{e\sb{\mathrm{1}}} \HOLBoundVar{t} \HOLSymConst{\ensuremath{\parallel}} \HOLBoundVar{e\sb{\mathrm{2}}} \HOLBoundVar{t})) \HOLSymConst{\HOLTokenConj{}} \HOLBoundVar{GCONTEXT\sp{\prime}} \HOLBoundVar{e\sb{\mathrm{1}}} \HOLSymConst{\HOLTokenConj{}}
              \HOLBoundVar{GCONTEXT\sp{\prime}} \HOLBoundVar{e\sb{\mathrm{2}}}) \HOLSymConst{\HOLTokenDisj{}}
           (\HOLSymConst{\HOLTokenExists{}}\HOLBoundVar{L} \HOLBoundVar{e}. (\HOLBoundVar{a\sb{\mathrm{0}}} \HOLSymConst{=} (\HOLTokenLambda{}\HOLBoundVar{t}. \HOLConst{\ensuremath{\nu}} \HOLBoundVar{L} (\HOLBoundVar{e} \HOLBoundVar{t}))) \HOLSymConst{\HOLTokenConj{}} \HOLBoundVar{GCONTEXT\sp{\prime}} \HOLBoundVar{e}) \HOLSymConst{\HOLTokenDisj{}}
           (\HOLSymConst{\HOLTokenExists{}}\HOLBoundVar{rf} \HOLBoundVar{e}. (\HOLBoundVar{a\sb{\mathrm{0}}} \HOLSymConst{=} (\HOLTokenLambda{}\HOLBoundVar{t}. \HOLConst{relab} (\HOLBoundVar{e} \HOLBoundVar{t}) \HOLBoundVar{rf})) \HOLSymConst{\HOLTokenConj{}} \HOLBoundVar{GCONTEXT\sp{\prime}} \HOLBoundVar{e}) \HOLSymConst{\HOLTokenImp{}}
           \HOLBoundVar{GCONTEXT\sp{\prime}} \HOLBoundVar{a\sb{\mathrm{0}}}) \HOLSymConst{\HOLTokenImp{}}
        \HOLBoundVar{GCONTEXT\sp{\prime}} \HOLBoundVar{a\sb{\mathrm{0}}})
\end{SaveVerbatim}
\newcommand{\HOLCongruenceDefinitionsGCONTEXTXXdef}{\UseVerbatim{HOLCongruenceDefinitionsGCONTEXTXXdef}}
\begin{SaveVerbatim}{HOLCongruenceDefinitionsGSEQXXdef}
\HOLTokenTurnstile{} \HOLConst{GSEQ} \HOLSymConst{=}
   (\HOLTokenLambda{}\HOLBoundVar{a\sb{\mathrm{0}}}.
      \HOLSymConst{\HOLTokenForall{}}\HOLBoundVar{GSEQ\sp{\prime}}.
        (\HOLSymConst{\HOLTokenForall{}}\HOLBoundVar{a\sb{\mathrm{0}}}.
           (\HOLBoundVar{a\sb{\mathrm{0}}} \HOLSymConst{=} (\HOLTokenLambda{}\HOLBoundVar{t}. \HOLBoundVar{t})) \HOLSymConst{\HOLTokenDisj{}} (\HOLSymConst{\HOLTokenExists{}}\HOLBoundVar{p}. \HOLBoundVar{a\sb{\mathrm{0}}} \HOLSymConst{=} (\HOLTokenLambda{}\HOLBoundVar{t}. \HOLBoundVar{p})) \HOLSymConst{\HOLTokenDisj{}}
           (\HOLSymConst{\HOLTokenExists{}}\HOLBoundVar{a} \HOLBoundVar{e}. (\HOLBoundVar{a\sb{\mathrm{0}}} \HOLSymConst{=} (\HOLTokenLambda{}\HOLBoundVar{t}. \HOLBoundVar{a}\HOLSymConst{..}\HOLBoundVar{e} \HOLBoundVar{t})) \HOLSymConst{\HOLTokenConj{}} \HOLBoundVar{GSEQ\sp{\prime}} \HOLBoundVar{e}) \HOLSymConst{\HOLTokenDisj{}}
           (\HOLSymConst{\HOLTokenExists{}}\HOLBoundVar{a\sb{\mathrm{1}}} \HOLBoundVar{a\sb{\mathrm{2}}} \HOLBoundVar{e\sb{\mathrm{1}}} \HOLBoundVar{e\sb{\mathrm{2}}}.
              (\HOLBoundVar{a\sb{\mathrm{0}}} \HOLSymConst{=} (\HOLTokenLambda{}\HOLBoundVar{t}. \HOLBoundVar{a\sb{\mathrm{1}}}\HOLSymConst{..}\HOLBoundVar{e\sb{\mathrm{1}}} \HOLBoundVar{t} \HOLSymConst{+} \HOLBoundVar{a\sb{\mathrm{2}}}\HOLSymConst{..}\HOLBoundVar{e\sb{\mathrm{2}}} \HOLBoundVar{t})) \HOLSymConst{\HOLTokenConj{}} \HOLBoundVar{GSEQ\sp{\prime}} \HOLBoundVar{e\sb{\mathrm{1}}} \HOLSymConst{\HOLTokenConj{}}
              \HOLBoundVar{GSEQ\sp{\prime}} \HOLBoundVar{e\sb{\mathrm{2}}}) \HOLSymConst{\HOLTokenImp{}}
           \HOLBoundVar{GSEQ\sp{\prime}} \HOLBoundVar{a\sb{\mathrm{0}}}) \HOLSymConst{\HOLTokenImp{}}
        \HOLBoundVar{GSEQ\sp{\prime}} \HOLBoundVar{a\sb{\mathrm{0}}})
\end{SaveVerbatim}
\newcommand{\HOLCongruenceDefinitionsGSEQXXdef}{\UseVerbatim{HOLCongruenceDefinitionsGSEQXXdef}}
\begin{SaveVerbatim}{HOLCongruenceDefinitionsOHXXCONTEXTXXdef}
\HOLTokenTurnstile{} \HOLConst{OH_CONTEXT} \HOLSymConst{=}
   (\HOLTokenLambda{}\HOLBoundVar{a\sb{\mathrm{0}}}.
      \HOLSymConst{\HOLTokenForall{}}\HOLBoundVar{OH\HOLTokenUnderscore{}CONTEXT\sp{\prime}}.
        (\HOLSymConst{\HOLTokenForall{}}\HOLBoundVar{a\sb{\mathrm{0}}}.
           (\HOLBoundVar{a\sb{\mathrm{0}}} \HOLSymConst{=} (\HOLTokenLambda{}\HOLBoundVar{t}. \HOLBoundVar{t})) \HOLSymConst{\HOLTokenDisj{}}
           (\HOLSymConst{\HOLTokenExists{}}\HOLBoundVar{a} \HOLBoundVar{c}. (\HOLBoundVar{a\sb{\mathrm{0}}} \HOLSymConst{=} (\HOLTokenLambda{}\HOLBoundVar{t}. \HOLBoundVar{a}\HOLSymConst{..}\HOLBoundVar{c} \HOLBoundVar{t})) \HOLSymConst{\HOLTokenConj{}} \HOLBoundVar{OH\HOLTokenUnderscore{}CONTEXT\sp{\prime}} \HOLBoundVar{c}) \HOLSymConst{\HOLTokenDisj{}}
           (\HOLSymConst{\HOLTokenExists{}}\HOLBoundVar{x} \HOLBoundVar{c}. (\HOLBoundVar{a\sb{\mathrm{0}}} \HOLSymConst{=} (\HOLTokenLambda{}\HOLBoundVar{t}. \HOLBoundVar{c} \HOLBoundVar{t} \HOLSymConst{+} \HOLBoundVar{x})) \HOLSymConst{\HOLTokenConj{}} \HOLBoundVar{OH\HOLTokenUnderscore{}CONTEXT\sp{\prime}} \HOLBoundVar{c}) \HOLSymConst{\HOLTokenDisj{}}
           (\HOLSymConst{\HOLTokenExists{}}\HOLBoundVar{x} \HOLBoundVar{c}. (\HOLBoundVar{a\sb{\mathrm{0}}} \HOLSymConst{=} (\HOLTokenLambda{}\HOLBoundVar{t}. \HOLBoundVar{x} \HOLSymConst{+} \HOLBoundVar{c} \HOLBoundVar{t})) \HOLSymConst{\HOLTokenConj{}} \HOLBoundVar{OH\HOLTokenUnderscore{}CONTEXT\sp{\prime}} \HOLBoundVar{c}) \HOLSymConst{\HOLTokenDisj{}}
           (\HOLSymConst{\HOLTokenExists{}}\HOLBoundVar{x} \HOLBoundVar{c}. (\HOLBoundVar{a\sb{\mathrm{0}}} \HOLSymConst{=} (\HOLTokenLambda{}\HOLBoundVar{t}. \HOLBoundVar{c} \HOLBoundVar{t} \HOLSymConst{\ensuremath{\parallel}} \HOLBoundVar{x})) \HOLSymConst{\HOLTokenConj{}} \HOLBoundVar{OH\HOLTokenUnderscore{}CONTEXT\sp{\prime}} \HOLBoundVar{c}) \HOLSymConst{\HOLTokenDisj{}}
           (\HOLSymConst{\HOLTokenExists{}}\HOLBoundVar{x} \HOLBoundVar{c}. (\HOLBoundVar{a\sb{\mathrm{0}}} \HOLSymConst{=} (\HOLTokenLambda{}\HOLBoundVar{t}. \HOLBoundVar{x} \HOLSymConst{\ensuremath{\parallel}} \HOLBoundVar{c} \HOLBoundVar{t})) \HOLSymConst{\HOLTokenConj{}} \HOLBoundVar{OH\HOLTokenUnderscore{}CONTEXT\sp{\prime}} \HOLBoundVar{c}) \HOLSymConst{\HOLTokenDisj{}}
           (\HOLSymConst{\HOLTokenExists{}}\HOLBoundVar{L} \HOLBoundVar{c}. (\HOLBoundVar{a\sb{\mathrm{0}}} \HOLSymConst{=} (\HOLTokenLambda{}\HOLBoundVar{t}. \HOLConst{\ensuremath{\nu}} \HOLBoundVar{L} (\HOLBoundVar{c} \HOLBoundVar{t}))) \HOLSymConst{\HOLTokenConj{}} \HOLBoundVar{OH\HOLTokenUnderscore{}CONTEXT\sp{\prime}} \HOLBoundVar{c}) \HOLSymConst{\HOLTokenDisj{}}
           (\HOLSymConst{\HOLTokenExists{}}\HOLBoundVar{rf} \HOLBoundVar{c}.
              (\HOLBoundVar{a\sb{\mathrm{0}}} \HOLSymConst{=} (\HOLTokenLambda{}\HOLBoundVar{t}. \HOLConst{relab} (\HOLBoundVar{c} \HOLBoundVar{t}) \HOLBoundVar{rf})) \HOLSymConst{\HOLTokenConj{}} \HOLBoundVar{OH\HOLTokenUnderscore{}CONTEXT\sp{\prime}} \HOLBoundVar{c}) \HOLSymConst{\HOLTokenImp{}}
           \HOLBoundVar{OH\HOLTokenUnderscore{}CONTEXT\sp{\prime}} \HOLBoundVar{a\sb{\mathrm{0}}}) \HOLSymConst{\HOLTokenImp{}}
        \HOLBoundVar{OH\HOLTokenUnderscore{}CONTEXT\sp{\prime}} \HOLBoundVar{a\sb{\mathrm{0}}})
\end{SaveVerbatim}
\newcommand{\HOLCongruenceDefinitionsOHXXCONTEXTXXdef}{\UseVerbatim{HOLCongruenceDefinitionsOHXXCONTEXTXXdef}}
\begin{SaveVerbatim}{HOLCongruenceDefinitionsprecongruenceOneXXdef}
\HOLTokenTurnstile{} \HOLSymConst{\HOLTokenForall{}}\HOLBoundVar{R}.
     \HOLConst{precongruence1} \HOLBoundVar{R} \HOLSymConst{\HOLTokenEquiv{}}
     \HOLSymConst{\HOLTokenForall{}}\HOLBoundVar{x} \HOLBoundVar{y} \HOLBoundVar{ctx}. \HOLConst{GCONTEXT} \HOLBoundVar{ctx} \HOLSymConst{\HOLTokenImp{}} \HOLBoundVar{R} \HOLBoundVar{x} \HOLBoundVar{y} \HOLSymConst{\HOLTokenImp{}} \HOLBoundVar{R} (\HOLBoundVar{ctx} \HOLBoundVar{x}) (\HOLBoundVar{ctx} \HOLBoundVar{y})
\end{SaveVerbatim}
\newcommand{\HOLCongruenceDefinitionsprecongruenceOneXXdef}{\UseVerbatim{HOLCongruenceDefinitionsprecongruenceOneXXdef}}
\begin{SaveVerbatim}{HOLCongruenceDefinitionsprecongruenceXXdef}
\HOLTokenTurnstile{} \HOLSymConst{\HOLTokenForall{}}\HOLBoundVar{R}.
     \HOLConst{precongruence} \HOLBoundVar{R} \HOLSymConst{\HOLTokenEquiv{}}
     \HOLSymConst{\HOLTokenForall{}}\HOLBoundVar{x} \HOLBoundVar{y} \HOLBoundVar{ctx}. \HOLConst{CONTEXT} \HOLBoundVar{ctx} \HOLSymConst{\HOLTokenImp{}} \HOLBoundVar{R} \HOLBoundVar{x} \HOLBoundVar{y} \HOLSymConst{\HOLTokenImp{}} \HOLBoundVar{R} (\HOLBoundVar{ctx} \HOLBoundVar{x}) (\HOLBoundVar{ctx} \HOLBoundVar{y})
\end{SaveVerbatim}
\newcommand{\HOLCongruenceDefinitionsprecongruenceXXdef}{\UseVerbatim{HOLCongruenceDefinitionsprecongruenceXXdef}}
\begin{SaveVerbatim}{HOLCongruenceDefinitionsSEQXXdef}
\HOLTokenTurnstile{} \HOLConst{SEQ} \HOLSymConst{=}
   (\HOLTokenLambda{}\HOLBoundVar{a\sb{\mathrm{0}}}.
      \HOLSymConst{\HOLTokenForall{}}\HOLBoundVar{SEQ\sp{\prime}}.
        (\HOLSymConst{\HOLTokenForall{}}\HOLBoundVar{a\sb{\mathrm{0}}}.
           (\HOLBoundVar{a\sb{\mathrm{0}}} \HOLSymConst{=} (\HOLTokenLambda{}\HOLBoundVar{t}. \HOLBoundVar{t})) \HOLSymConst{\HOLTokenDisj{}} (\HOLSymConst{\HOLTokenExists{}}\HOLBoundVar{p}. \HOLBoundVar{a\sb{\mathrm{0}}} \HOLSymConst{=} (\HOLTokenLambda{}\HOLBoundVar{t}. \HOLBoundVar{p})) \HOLSymConst{\HOLTokenDisj{}}
           (\HOLSymConst{\HOLTokenExists{}}\HOLBoundVar{a} \HOLBoundVar{e}. (\HOLBoundVar{a\sb{\mathrm{0}}} \HOLSymConst{=} (\HOLTokenLambda{}\HOLBoundVar{t}. \HOLBoundVar{a}\HOLSymConst{..}\HOLBoundVar{e} \HOLBoundVar{t})) \HOLSymConst{\HOLTokenConj{}} \HOLBoundVar{SEQ\sp{\prime}} \HOLBoundVar{e}) \HOLSymConst{\HOLTokenDisj{}}
           (\HOLSymConst{\HOLTokenExists{}}\HOLBoundVar{e\sb{\mathrm{1}}} \HOLBoundVar{e\sb{\mathrm{2}}}.
              (\HOLBoundVar{a\sb{\mathrm{0}}} \HOLSymConst{=} (\HOLTokenLambda{}\HOLBoundVar{t}. \HOLBoundVar{e\sb{\mathrm{1}}} \HOLBoundVar{t} \HOLSymConst{+} \HOLBoundVar{e\sb{\mathrm{2}}} \HOLBoundVar{t})) \HOLSymConst{\HOLTokenConj{}} \HOLBoundVar{SEQ\sp{\prime}} \HOLBoundVar{e\sb{\mathrm{1}}} \HOLSymConst{\HOLTokenConj{}} \HOLBoundVar{SEQ\sp{\prime}} \HOLBoundVar{e\sb{\mathrm{2}}}) \HOLSymConst{\HOLTokenImp{}}
           \HOLBoundVar{SEQ\sp{\prime}} \HOLBoundVar{a\sb{\mathrm{0}}}) \HOLSymConst{\HOLTokenImp{}}
        \HOLBoundVar{SEQ\sp{\prime}} \HOLBoundVar{a\sb{\mathrm{0}}})
\end{SaveVerbatim}
\newcommand{\HOLCongruenceDefinitionsSEQXXdef}{\UseVerbatim{HOLCongruenceDefinitionsSEQXXdef}}
\begin{SaveVerbatim}{HOLCongruenceDefinitionsSGXXdef}
\HOLTokenTurnstile{} \HOLConst{SG} \HOLSymConst{=}
   (\HOLTokenLambda{}\HOLBoundVar{a\sb{\mathrm{0}}}.
      \HOLSymConst{\HOLTokenForall{}}\HOLBoundVar{SG\sp{\prime}}.
        (\HOLSymConst{\HOLTokenForall{}}\HOLBoundVar{a\sb{\mathrm{0}}}.
           (\HOLSymConst{\HOLTokenExists{}}\HOLBoundVar{p}. \HOLBoundVar{a\sb{\mathrm{0}}} \HOLSymConst{=} (\HOLTokenLambda{}\HOLBoundVar{t}. \HOLBoundVar{p})) \HOLSymConst{\HOLTokenDisj{}}
           (\HOLSymConst{\HOLTokenExists{}}\HOLBoundVar{l} \HOLBoundVar{e}. (\HOLBoundVar{a\sb{\mathrm{0}}} \HOLSymConst{=} (\HOLTokenLambda{}\HOLBoundVar{t}. \HOLConst{label} \HOLBoundVar{l}\HOLSymConst{..}\HOLBoundVar{e} \HOLBoundVar{t})) \HOLSymConst{\HOLTokenConj{}} \HOLConst{CONTEXT} \HOLBoundVar{e}) \HOLSymConst{\HOLTokenDisj{}}
           (\HOLSymConst{\HOLTokenExists{}}\HOLBoundVar{a} \HOLBoundVar{e}. (\HOLBoundVar{a\sb{\mathrm{0}}} \HOLSymConst{=} (\HOLTokenLambda{}\HOLBoundVar{t}. \HOLBoundVar{a}\HOLSymConst{..}\HOLBoundVar{e} \HOLBoundVar{t})) \HOLSymConst{\HOLTokenConj{}} \HOLBoundVar{SG\sp{\prime}} \HOLBoundVar{e}) \HOLSymConst{\HOLTokenDisj{}}
           (\HOLSymConst{\HOLTokenExists{}}\HOLBoundVar{e\sb{\mathrm{1}}} \HOLBoundVar{e\sb{\mathrm{2}}}.
              (\HOLBoundVar{a\sb{\mathrm{0}}} \HOLSymConst{=} (\HOLTokenLambda{}\HOLBoundVar{t}. \HOLBoundVar{e\sb{\mathrm{1}}} \HOLBoundVar{t} \HOLSymConst{+} \HOLBoundVar{e\sb{\mathrm{2}}} \HOLBoundVar{t})) \HOLSymConst{\HOLTokenConj{}} \HOLBoundVar{SG\sp{\prime}} \HOLBoundVar{e\sb{\mathrm{1}}} \HOLSymConst{\HOLTokenConj{}} \HOLBoundVar{SG\sp{\prime}} \HOLBoundVar{e\sb{\mathrm{2}}}) \HOLSymConst{\HOLTokenDisj{}}
           (\HOLSymConst{\HOLTokenExists{}}\HOLBoundVar{e\sb{\mathrm{1}}} \HOLBoundVar{e\sb{\mathrm{2}}}.
              (\HOLBoundVar{a\sb{\mathrm{0}}} \HOLSymConst{=} (\HOLTokenLambda{}\HOLBoundVar{t}. \HOLBoundVar{e\sb{\mathrm{1}}} \HOLBoundVar{t} \HOLSymConst{\ensuremath{\parallel}} \HOLBoundVar{e\sb{\mathrm{2}}} \HOLBoundVar{t})) \HOLSymConst{\HOLTokenConj{}} \HOLBoundVar{SG\sp{\prime}} \HOLBoundVar{e\sb{\mathrm{1}}} \HOLSymConst{\HOLTokenConj{}} \HOLBoundVar{SG\sp{\prime}} \HOLBoundVar{e\sb{\mathrm{2}}}) \HOLSymConst{\HOLTokenDisj{}}
           (\HOLSymConst{\HOLTokenExists{}}\HOLBoundVar{L} \HOLBoundVar{e}. (\HOLBoundVar{a\sb{\mathrm{0}}} \HOLSymConst{=} (\HOLTokenLambda{}\HOLBoundVar{t}. \HOLConst{\ensuremath{\nu}} \HOLBoundVar{L} (\HOLBoundVar{e} \HOLBoundVar{t}))) \HOLSymConst{\HOLTokenConj{}} \HOLBoundVar{SG\sp{\prime}} \HOLBoundVar{e}) \HOLSymConst{\HOLTokenDisj{}}
           (\HOLSymConst{\HOLTokenExists{}}\HOLBoundVar{rf} \HOLBoundVar{e}. (\HOLBoundVar{a\sb{\mathrm{0}}} \HOLSymConst{=} (\HOLTokenLambda{}\HOLBoundVar{t}. \HOLConst{relab} (\HOLBoundVar{e} \HOLBoundVar{t}) \HOLBoundVar{rf})) \HOLSymConst{\HOLTokenConj{}} \HOLBoundVar{SG\sp{\prime}} \HOLBoundVar{e}) \HOLSymConst{\HOLTokenImp{}}
           \HOLBoundVar{SG\sp{\prime}} \HOLBoundVar{a\sb{\mathrm{0}}}) \HOLSymConst{\HOLTokenImp{}}
        \HOLBoundVar{SG\sp{\prime}} \HOLBoundVar{a\sb{\mathrm{0}}})
\end{SaveVerbatim}
\newcommand{\HOLCongruenceDefinitionsSGXXdef}{\UseVerbatim{HOLCongruenceDefinitionsSGXXdef}}
\begin{SaveVerbatim}{HOLCongruenceDefinitionsweaklyXXguardedOneXXdef}
\HOLTokenTurnstile{} \HOLSymConst{\HOLTokenForall{}}\HOLBoundVar{E}.
     \HOLConst{weakly_guarded1} \HOLBoundVar{E} \HOLSymConst{\HOLTokenEquiv{}}
     \HOLSymConst{\HOLTokenForall{}}\HOLBoundVar{X}. \HOLBoundVar{X} \HOLConst{\HOLTokenIn{}} \HOLConst{FV} \HOLBoundVar{E} \HOLSymConst{\HOLTokenImp{}} \HOLSymConst{\HOLTokenForall{}}\HOLBoundVar{e}. \HOLConst{CONTEXT} \HOLBoundVar{e} \HOLSymConst{\HOLTokenConj{}} (\HOLBoundVar{e} (\HOLConst{var} \HOLBoundVar{X}) \HOLSymConst{=} \HOLBoundVar{E}) \HOLSymConst{\HOLTokenImp{}} \HOLConst{WG} \HOLBoundVar{e}
\end{SaveVerbatim}
\newcommand{\HOLCongruenceDefinitionsweaklyXXguardedOneXXdef}{\UseVerbatim{HOLCongruenceDefinitionsweaklyXXguardedOneXXdef}}
\begin{SaveVerbatim}{HOLCongruenceDefinitionsweaklyXXguardedXXdef}
\HOLTokenTurnstile{} \HOLSymConst{\HOLTokenForall{}}\HOLBoundVar{Es}. \HOLConst{weakly_guarded} \HOLBoundVar{Es} \HOLSymConst{\HOLTokenEquiv{}} \HOLConst{EVERY} \HOLConst{weakly_guarded1} \HOLBoundVar{Es}
\end{SaveVerbatim}
\newcommand{\HOLCongruenceDefinitionsweaklyXXguardedXXdef}{\UseVerbatim{HOLCongruenceDefinitionsweaklyXXguardedXXdef}}
\begin{SaveVerbatim}{HOLCongruenceDefinitionsWGXXdef}
\HOLTokenTurnstile{} \HOLConst{WG} \HOLSymConst{=}
   (\HOLTokenLambda{}\HOLBoundVar{a\sb{\mathrm{0}}}.
      \HOLSymConst{\HOLTokenForall{}}\HOLBoundVar{WG\sp{\prime}}.
        (\HOLSymConst{\HOLTokenForall{}}\HOLBoundVar{a\sb{\mathrm{0}}}.
           (\HOLSymConst{\HOLTokenExists{}}\HOLBoundVar{p}. \HOLBoundVar{a\sb{\mathrm{0}}} \HOLSymConst{=} (\HOLTokenLambda{}\HOLBoundVar{t}. \HOLBoundVar{p})) \HOLSymConst{\HOLTokenDisj{}}
           (\HOLSymConst{\HOLTokenExists{}}\HOLBoundVar{a} \HOLBoundVar{e}. (\HOLBoundVar{a\sb{\mathrm{0}}} \HOLSymConst{=} (\HOLTokenLambda{}\HOLBoundVar{t}. \HOLBoundVar{a}\HOLSymConst{..}\HOLBoundVar{e} \HOLBoundVar{t})) \HOLSymConst{\HOLTokenConj{}} \HOLConst{CONTEXT} \HOLBoundVar{e}) \HOLSymConst{\HOLTokenDisj{}}
           (\HOLSymConst{\HOLTokenExists{}}\HOLBoundVar{e\sb{\mathrm{1}}} \HOLBoundVar{e\sb{\mathrm{2}}}.
              (\HOLBoundVar{a\sb{\mathrm{0}}} \HOLSymConst{=} (\HOLTokenLambda{}\HOLBoundVar{t}. \HOLBoundVar{e\sb{\mathrm{1}}} \HOLBoundVar{t} \HOLSymConst{+} \HOLBoundVar{e\sb{\mathrm{2}}} \HOLBoundVar{t})) \HOLSymConst{\HOLTokenConj{}} \HOLBoundVar{WG\sp{\prime}} \HOLBoundVar{e\sb{\mathrm{1}}} \HOLSymConst{\HOLTokenConj{}} \HOLBoundVar{WG\sp{\prime}} \HOLBoundVar{e\sb{\mathrm{2}}}) \HOLSymConst{\HOLTokenDisj{}}
           (\HOLSymConst{\HOLTokenExists{}}\HOLBoundVar{e\sb{\mathrm{1}}} \HOLBoundVar{e\sb{\mathrm{2}}}.
              (\HOLBoundVar{a\sb{\mathrm{0}}} \HOLSymConst{=} (\HOLTokenLambda{}\HOLBoundVar{t}. \HOLBoundVar{e\sb{\mathrm{1}}} \HOLBoundVar{t} \HOLSymConst{\ensuremath{\parallel}} \HOLBoundVar{e\sb{\mathrm{2}}} \HOLBoundVar{t})) \HOLSymConst{\HOLTokenConj{}} \HOLBoundVar{WG\sp{\prime}} \HOLBoundVar{e\sb{\mathrm{1}}} \HOLSymConst{\HOLTokenConj{}} \HOLBoundVar{WG\sp{\prime}} \HOLBoundVar{e\sb{\mathrm{2}}}) \HOLSymConst{\HOLTokenDisj{}}
           (\HOLSymConst{\HOLTokenExists{}}\HOLBoundVar{L} \HOLBoundVar{e}. (\HOLBoundVar{a\sb{\mathrm{0}}} \HOLSymConst{=} (\HOLTokenLambda{}\HOLBoundVar{t}. \HOLConst{\ensuremath{\nu}} \HOLBoundVar{L} (\HOLBoundVar{e} \HOLBoundVar{t}))) \HOLSymConst{\HOLTokenConj{}} \HOLBoundVar{WG\sp{\prime}} \HOLBoundVar{e}) \HOLSymConst{\HOLTokenDisj{}}
           (\HOLSymConst{\HOLTokenExists{}}\HOLBoundVar{rf} \HOLBoundVar{e}. (\HOLBoundVar{a\sb{\mathrm{0}}} \HOLSymConst{=} (\HOLTokenLambda{}\HOLBoundVar{t}. \HOLConst{relab} (\HOLBoundVar{e} \HOLBoundVar{t}) \HOLBoundVar{rf})) \HOLSymConst{\HOLTokenConj{}} \HOLBoundVar{WG\sp{\prime}} \HOLBoundVar{e}) \HOLSymConst{\HOLTokenImp{}}
           \HOLBoundVar{WG\sp{\prime}} \HOLBoundVar{a\sb{\mathrm{0}}}) \HOLSymConst{\HOLTokenImp{}}
        \HOLBoundVar{WG\sp{\prime}} \HOLBoundVar{a\sb{\mathrm{0}}})
\end{SaveVerbatim}
\newcommand{\HOLCongruenceDefinitionsWGXXdef}{\UseVerbatim{HOLCongruenceDefinitionsWGXXdef}}
\begin{SaveVerbatim}{HOLCongruenceDefinitionsWGSXXdef}
\HOLTokenTurnstile{} \HOLConst{WGS} \HOLSymConst{=}
   (\HOLTokenLambda{}\HOLBoundVar{a\sb{\mathrm{0}}}.
      \HOLSymConst{\HOLTokenForall{}}\HOLBoundVar{WGS\sp{\prime}}.
        (\HOLSymConst{\HOLTokenForall{}}\HOLBoundVar{a\sb{\mathrm{0}}}.
           (\HOLSymConst{\HOLTokenExists{}}\HOLBoundVar{p}. \HOLBoundVar{a\sb{\mathrm{0}}} \HOLSymConst{=} (\HOLTokenLambda{}\HOLBoundVar{t}. \HOLBoundVar{p})) \HOLSymConst{\HOLTokenDisj{}}
           (\HOLSymConst{\HOLTokenExists{}}\HOLBoundVar{a} \HOLBoundVar{e}. (\HOLBoundVar{a\sb{\mathrm{0}}} \HOLSymConst{=} (\HOLTokenLambda{}\HOLBoundVar{t}. \HOLBoundVar{a}\HOLSymConst{..}\HOLBoundVar{e} \HOLBoundVar{t})) \HOLSymConst{\HOLTokenConj{}} \HOLConst{GCONTEXT} \HOLBoundVar{e}) \HOLSymConst{\HOLTokenDisj{}}
           (\HOLSymConst{\HOLTokenExists{}}\HOLBoundVar{a\sb{\mathrm{1}}} \HOLBoundVar{a\sb{\mathrm{2}}} \HOLBoundVar{e\sb{\mathrm{1}}} \HOLBoundVar{e\sb{\mathrm{2}}}.
              (\HOLBoundVar{a\sb{\mathrm{0}}} \HOLSymConst{=} (\HOLTokenLambda{}\HOLBoundVar{t}. \HOLBoundVar{a\sb{\mathrm{1}}}\HOLSymConst{..}\HOLBoundVar{e\sb{\mathrm{1}}} \HOLBoundVar{t} \HOLSymConst{+} \HOLBoundVar{a\sb{\mathrm{2}}}\HOLSymConst{..}\HOLBoundVar{e\sb{\mathrm{2}}} \HOLBoundVar{t})) \HOLSymConst{\HOLTokenConj{}} \HOLConst{GCONTEXT} \HOLBoundVar{e\sb{\mathrm{1}}} \HOLSymConst{\HOLTokenConj{}}
              \HOLConst{GCONTEXT} \HOLBoundVar{e\sb{\mathrm{2}}}) \HOLSymConst{\HOLTokenDisj{}}
           (\HOLSymConst{\HOLTokenExists{}}\HOLBoundVar{e\sb{\mathrm{1}}} \HOLBoundVar{e\sb{\mathrm{2}}}.
              (\HOLBoundVar{a\sb{\mathrm{0}}} \HOLSymConst{=} (\HOLTokenLambda{}\HOLBoundVar{t}. \HOLBoundVar{e\sb{\mathrm{1}}} \HOLBoundVar{t} \HOLSymConst{\ensuremath{\parallel}} \HOLBoundVar{e\sb{\mathrm{2}}} \HOLBoundVar{t})) \HOLSymConst{\HOLTokenConj{}} \HOLBoundVar{WGS\sp{\prime}} \HOLBoundVar{e\sb{\mathrm{1}}} \HOLSymConst{\HOLTokenConj{}} \HOLBoundVar{WGS\sp{\prime}} \HOLBoundVar{e\sb{\mathrm{2}}}) \HOLSymConst{\HOLTokenDisj{}}
           (\HOLSymConst{\HOLTokenExists{}}\HOLBoundVar{L} \HOLBoundVar{e}. (\HOLBoundVar{a\sb{\mathrm{0}}} \HOLSymConst{=} (\HOLTokenLambda{}\HOLBoundVar{t}. \HOLConst{\ensuremath{\nu}} \HOLBoundVar{L} (\HOLBoundVar{e} \HOLBoundVar{t}))) \HOLSymConst{\HOLTokenConj{}} \HOLBoundVar{WGS\sp{\prime}} \HOLBoundVar{e}) \HOLSymConst{\HOLTokenDisj{}}
           (\HOLSymConst{\HOLTokenExists{}}\HOLBoundVar{rf} \HOLBoundVar{e}. (\HOLBoundVar{a\sb{\mathrm{0}}} \HOLSymConst{=} (\HOLTokenLambda{}\HOLBoundVar{t}. \HOLConst{relab} (\HOLBoundVar{e} \HOLBoundVar{t}) \HOLBoundVar{rf})) \HOLSymConst{\HOLTokenConj{}} \HOLBoundVar{WGS\sp{\prime}} \HOLBoundVar{e}) \HOLSymConst{\HOLTokenImp{}}
           \HOLBoundVar{WGS\sp{\prime}} \HOLBoundVar{a\sb{\mathrm{0}}}) \HOLSymConst{\HOLTokenImp{}}
        \HOLBoundVar{WGS\sp{\prime}} \HOLBoundVar{a\sb{\mathrm{0}}})
\end{SaveVerbatim}
\newcommand{\HOLCongruenceDefinitionsWGSXXdef}{\UseVerbatim{HOLCongruenceDefinitionsWGSXXdef}}
\newcommand{\HOLCongruenceDefinitions}{
\HOLDfnTag{Congruence}{CC_def}\HOLCongruenceDefinitionsCCXXdef
\HOLDfnTag{Congruence}{congruence1_def}\HOLCongruenceDefinitionscongruenceOneXXdef
\HOLDfnTag{Congruence}{congruence_def}\HOLCongruenceDefinitionscongruenceXXdef
\HOLDfnTag{Congruence}{CONTEXT_def}\HOLCongruenceDefinitionsCONTEXTXXdef
\HOLDfnTag{Congruence}{GCC_def}\HOLCongruenceDefinitionsGCCXXdef
\HOLDfnTag{Congruence}{GCONTEXT_def}\HOLCongruenceDefinitionsGCONTEXTXXdef
\HOLDfnTag{Congruence}{GSEQ_def}\HOLCongruenceDefinitionsGSEQXXdef
\HOLDfnTag{Congruence}{OH_CONTEXT_def}\HOLCongruenceDefinitionsOHXXCONTEXTXXdef
\HOLDfnTag{Congruence}{precongruence1_def}\HOLCongruenceDefinitionsprecongruenceOneXXdef
\HOLDfnTag{Congruence}{precongruence_def}\HOLCongruenceDefinitionsprecongruenceXXdef
\HOLDfnTag{Congruence}{SEQ_def}\HOLCongruenceDefinitionsSEQXXdef
\HOLDfnTag{Congruence}{SG_def}\HOLCongruenceDefinitionsSGXXdef
\HOLDfnTag{Congruence}{weakly_guarded1_def}\HOLCongruenceDefinitionsweaklyXXguardedOneXXdef
\HOLDfnTag{Congruence}{weakly_guarded_def}\HOLCongruenceDefinitionsweaklyXXguardedXXdef
\HOLDfnTag{Congruence}{WG_def}\HOLCongruenceDefinitionsWGXXdef
\HOLDfnTag{Congruence}{WGS_def}\HOLCongruenceDefinitionsWGSXXdef
}
\begin{SaveVerbatim}{HOLCongruenceTheoremsCCXXcongruence}
\HOLTokenTurnstile{} \HOLSymConst{\HOLTokenForall{}}\HOLBoundVar{R}. \HOLConst{equivalence} \HOLBoundVar{R} \HOLSymConst{\HOLTokenImp{}} \HOLConst{congruence} (\HOLConst{CC} \HOLBoundVar{R})
\end{SaveVerbatim}
\newcommand{\HOLCongruenceTheoremsCCXXcongruence}{\UseVerbatim{HOLCongruenceTheoremsCCXXcongruence}}
\begin{SaveVerbatim}{HOLCongruenceTheoremsCCXXisXXcoarsest}
\HOLTokenTurnstile{} \HOLSymConst{\HOLTokenForall{}}\HOLBoundVar{R} \HOLBoundVar{R\sp{\prime}}. \HOLConst{congruence} \HOLBoundVar{R\sp{\prime}} \HOLSymConst{\HOLTokenConj{}} \HOLBoundVar{R\sp{\prime}} \HOLConst{RSUBSET} \HOLBoundVar{R} \HOLSymConst{\HOLTokenImp{}} \HOLBoundVar{R\sp{\prime}} \HOLConst{RSUBSET} \HOLConst{CC} \HOLBoundVar{R}
\end{SaveVerbatim}
\newcommand{\HOLCongruenceTheoremsCCXXisXXcoarsest}{\UseVerbatim{HOLCongruenceTheoremsCCXXisXXcoarsest}}
\begin{SaveVerbatim}{HOLCongruenceTheoremsCCXXisXXcoarsestYY}
\HOLTokenTurnstile{} \HOLSymConst{\HOLTokenForall{}}\HOLBoundVar{R} \HOLBoundVar{R\sp{\prime}}. \HOLConst{precongruence} \HOLBoundVar{R\sp{\prime}} \HOLSymConst{\HOLTokenConj{}} \HOLBoundVar{R\sp{\prime}} \HOLConst{RSUBSET} \HOLBoundVar{R} \HOLSymConst{\HOLTokenImp{}} \HOLBoundVar{R\sp{\prime}} \HOLConst{RSUBSET} \HOLConst{CC} \HOLBoundVar{R}
\end{SaveVerbatim}
\newcommand{\HOLCongruenceTheoremsCCXXisXXcoarsestYY}{\UseVerbatim{HOLCongruenceTheoremsCCXXisXXcoarsestYY}}
\begin{SaveVerbatim}{HOLCongruenceTheoremsCCXXisXXfiner}
\HOLTokenTurnstile{} \HOLSymConst{\HOLTokenForall{}}\HOLBoundVar{R}. \HOLConst{CC} \HOLBoundVar{R} \HOLConst{RSUBSET} \HOLBoundVar{R}
\end{SaveVerbatim}
\newcommand{\HOLCongruenceTheoremsCCXXisXXfiner}{\UseVerbatim{HOLCongruenceTheoremsCCXXisXXfiner}}
\begin{SaveVerbatim}{HOLCongruenceTheoremsCCXXprecongruence}
\HOLTokenTurnstile{} \HOLSymConst{\HOLTokenForall{}}\HOLBoundVar{R}. \HOLConst{precongruence} (\HOLConst{CC} \HOLBoundVar{R})
\end{SaveVerbatim}
\newcommand{\HOLCongruenceTheoremsCCXXprecongruence}{\UseVerbatim{HOLCongruenceTheoremsCCXXprecongruence}}
\begin{SaveVerbatim}{HOLCongruenceTheoremsCONTEXTOne}
\HOLTokenTurnstile{} \HOLConst{CONTEXT} (\HOLTokenLambda{}\HOLBoundVar{t}. \HOLBoundVar{t})
\end{SaveVerbatim}
\newcommand{\HOLCongruenceTheoremsCONTEXTOne}{\UseVerbatim{HOLCongruenceTheoremsCONTEXTOne}}
\begin{SaveVerbatim}{HOLCongruenceTheoremsCONTEXTTwo}
\HOLTokenTurnstile{} \HOLSymConst{\HOLTokenForall{}}\HOLBoundVar{p}. \HOLConst{CONTEXT} (\HOLTokenLambda{}\HOLBoundVar{t}. \HOLBoundVar{p})
\end{SaveVerbatim}
\newcommand{\HOLCongruenceTheoremsCONTEXTTwo}{\UseVerbatim{HOLCongruenceTheoremsCONTEXTTwo}}
\begin{SaveVerbatim}{HOLCongruenceTheoremsCONTEXTThree}
\HOLTokenTurnstile{} \HOLSymConst{\HOLTokenForall{}}\HOLBoundVar{a} \HOLBoundVar{e}. \HOLConst{CONTEXT} \HOLBoundVar{e} \HOLSymConst{\HOLTokenImp{}} \HOLConst{CONTEXT} (\HOLTokenLambda{}\HOLBoundVar{t}. \HOLBoundVar{a}\HOLSymConst{..}\HOLBoundVar{e} \HOLBoundVar{t})
\end{SaveVerbatim}
\newcommand{\HOLCongruenceTheoremsCONTEXTThree}{\UseVerbatim{HOLCongruenceTheoremsCONTEXTThree}}
\begin{SaveVerbatim}{HOLCongruenceTheoremsCONTEXTThreea}
\HOLTokenTurnstile{} \HOLSymConst{\HOLTokenForall{}}\HOLBoundVar{a}. \HOLConst{CONTEXT} (\HOLTokenLambda{}\HOLBoundVar{t}. \HOLBoundVar{a}\HOLSymConst{..}\HOLBoundVar{t})
\end{SaveVerbatim}
\newcommand{\HOLCongruenceTheoremsCONTEXTThreea}{\UseVerbatim{HOLCongruenceTheoremsCONTEXTThreea}}
\begin{SaveVerbatim}{HOLCongruenceTheoremsCONTEXTFour}
\HOLTokenTurnstile{} \HOLSymConst{\HOLTokenForall{}}\HOLBoundVar{e\sb{\mathrm{1}}} \HOLBoundVar{e\sb{\mathrm{2}}}. \HOLConst{CONTEXT} \HOLBoundVar{e\sb{\mathrm{1}}} \HOLSymConst{\HOLTokenConj{}} \HOLConst{CONTEXT} \HOLBoundVar{e\sb{\mathrm{2}}} \HOLSymConst{\HOLTokenImp{}} \HOLConst{CONTEXT} (\HOLTokenLambda{}\HOLBoundVar{t}. \HOLBoundVar{e\sb{\mathrm{1}}} \HOLBoundVar{t} \HOLSymConst{+} \HOLBoundVar{e\sb{\mathrm{2}}} \HOLBoundVar{t})
\end{SaveVerbatim}
\newcommand{\HOLCongruenceTheoremsCONTEXTFour}{\UseVerbatim{HOLCongruenceTheoremsCONTEXTFour}}
\begin{SaveVerbatim}{HOLCongruenceTheoremsCONTEXTFive}
\HOLTokenTurnstile{} \HOLSymConst{\HOLTokenForall{}}\HOLBoundVar{e\sb{\mathrm{1}}} \HOLBoundVar{e\sb{\mathrm{2}}}. \HOLConst{CONTEXT} \HOLBoundVar{e\sb{\mathrm{1}}} \HOLSymConst{\HOLTokenConj{}} \HOLConst{CONTEXT} \HOLBoundVar{e\sb{\mathrm{2}}} \HOLSymConst{\HOLTokenImp{}} \HOLConst{CONTEXT} (\HOLTokenLambda{}\HOLBoundVar{t}. \HOLBoundVar{e\sb{\mathrm{1}}} \HOLBoundVar{t} \HOLSymConst{\ensuremath{\parallel}} \HOLBoundVar{e\sb{\mathrm{2}}} \HOLBoundVar{t})
\end{SaveVerbatim}
\newcommand{\HOLCongruenceTheoremsCONTEXTFive}{\UseVerbatim{HOLCongruenceTheoremsCONTEXTFive}}
\begin{SaveVerbatim}{HOLCongruenceTheoremsCONTEXTSix}
\HOLTokenTurnstile{} \HOLSymConst{\HOLTokenForall{}}\HOLBoundVar{L} \HOLBoundVar{e}. \HOLConst{CONTEXT} \HOLBoundVar{e} \HOLSymConst{\HOLTokenImp{}} \HOLConst{CONTEXT} (\HOLTokenLambda{}\HOLBoundVar{t}. \HOLConst{\ensuremath{\nu}} \HOLBoundVar{L} (\HOLBoundVar{e} \HOLBoundVar{t}))
\end{SaveVerbatim}
\newcommand{\HOLCongruenceTheoremsCONTEXTSix}{\UseVerbatim{HOLCongruenceTheoremsCONTEXTSix}}
\begin{SaveVerbatim}{HOLCongruenceTheoremsCONTEXTSeven}
\HOLTokenTurnstile{} \HOLSymConst{\HOLTokenForall{}}\HOLBoundVar{rf} \HOLBoundVar{e}. \HOLConst{CONTEXT} \HOLBoundVar{e} \HOLSymConst{\HOLTokenImp{}} \HOLConst{CONTEXT} (\HOLTokenLambda{}\HOLBoundVar{t}. \HOLConst{relab} (\HOLBoundVar{e} \HOLBoundVar{t}) \HOLBoundVar{rf})
\end{SaveVerbatim}
\newcommand{\HOLCongruenceTheoremsCONTEXTSeven}{\UseVerbatim{HOLCongruenceTheoremsCONTEXTSeven}}
\begin{SaveVerbatim}{HOLCongruenceTheoremsCONTEXTXXcases}
\HOLTokenTurnstile{} \HOLSymConst{\HOLTokenForall{}}\HOLBoundVar{a\sb{\mathrm{0}}}.
     \HOLConst{CONTEXT} \HOLBoundVar{a\sb{\mathrm{0}}} \HOLSymConst{\HOLTokenEquiv{}}
     (\HOLBoundVar{a\sb{\mathrm{0}}} \HOLSymConst{=} (\HOLTokenLambda{}\HOLBoundVar{t}. \HOLBoundVar{t})) \HOLSymConst{\HOLTokenDisj{}} (\HOLSymConst{\HOLTokenExists{}}\HOLBoundVar{p}. \HOLBoundVar{a\sb{\mathrm{0}}} \HOLSymConst{=} (\HOLTokenLambda{}\HOLBoundVar{t}. \HOLBoundVar{p})) \HOLSymConst{\HOLTokenDisj{}}
     (\HOLSymConst{\HOLTokenExists{}}\HOLBoundVar{a} \HOLBoundVar{e}. (\HOLBoundVar{a\sb{\mathrm{0}}} \HOLSymConst{=} (\HOLTokenLambda{}\HOLBoundVar{t}. \HOLBoundVar{a}\HOLSymConst{..}\HOLBoundVar{e} \HOLBoundVar{t})) \HOLSymConst{\HOLTokenConj{}} \HOLConst{CONTEXT} \HOLBoundVar{e}) \HOLSymConst{\HOLTokenDisj{}}
     (\HOLSymConst{\HOLTokenExists{}}\HOLBoundVar{e\sb{\mathrm{1}}} \HOLBoundVar{e\sb{\mathrm{2}}}.
        (\HOLBoundVar{a\sb{\mathrm{0}}} \HOLSymConst{=} (\HOLTokenLambda{}\HOLBoundVar{t}. \HOLBoundVar{e\sb{\mathrm{1}}} \HOLBoundVar{t} \HOLSymConst{+} \HOLBoundVar{e\sb{\mathrm{2}}} \HOLBoundVar{t})) \HOLSymConst{\HOLTokenConj{}} \HOLConst{CONTEXT} \HOLBoundVar{e\sb{\mathrm{1}}} \HOLSymConst{\HOLTokenConj{}} \HOLConst{CONTEXT} \HOLBoundVar{e\sb{\mathrm{2}}}) \HOLSymConst{\HOLTokenDisj{}}
     (\HOLSymConst{\HOLTokenExists{}}\HOLBoundVar{e\sb{\mathrm{1}}} \HOLBoundVar{e\sb{\mathrm{2}}}.
        (\HOLBoundVar{a\sb{\mathrm{0}}} \HOLSymConst{=} (\HOLTokenLambda{}\HOLBoundVar{t}. \HOLBoundVar{e\sb{\mathrm{1}}} \HOLBoundVar{t} \HOLSymConst{\ensuremath{\parallel}} \HOLBoundVar{e\sb{\mathrm{2}}} \HOLBoundVar{t})) \HOLSymConst{\HOLTokenConj{}} \HOLConst{CONTEXT} \HOLBoundVar{e\sb{\mathrm{1}}} \HOLSymConst{\HOLTokenConj{}} \HOLConst{CONTEXT} \HOLBoundVar{e\sb{\mathrm{2}}}) \HOLSymConst{\HOLTokenDisj{}}
     (\HOLSymConst{\HOLTokenExists{}}\HOLBoundVar{L} \HOLBoundVar{e}. (\HOLBoundVar{a\sb{\mathrm{0}}} \HOLSymConst{=} (\HOLTokenLambda{}\HOLBoundVar{t}. \HOLConst{\ensuremath{\nu}} \HOLBoundVar{L} (\HOLBoundVar{e} \HOLBoundVar{t}))) \HOLSymConst{\HOLTokenConj{}} \HOLConst{CONTEXT} \HOLBoundVar{e}) \HOLSymConst{\HOLTokenDisj{}}
     \HOLSymConst{\HOLTokenExists{}}\HOLBoundVar{rf} \HOLBoundVar{e}. (\HOLBoundVar{a\sb{\mathrm{0}}} \HOLSymConst{=} (\HOLTokenLambda{}\HOLBoundVar{t}. \HOLConst{relab} (\HOLBoundVar{e} \HOLBoundVar{t}) \HOLBoundVar{rf})) \HOLSymConst{\HOLTokenConj{}} \HOLConst{CONTEXT} \HOLBoundVar{e}
\end{SaveVerbatim}
\newcommand{\HOLCongruenceTheoremsCONTEXTXXcases}{\UseVerbatim{HOLCongruenceTheoremsCONTEXTXXcases}}
\begin{SaveVerbatim}{HOLCongruenceTheoremsCONTEXTXXcombin}
\HOLTokenTurnstile{} \HOLSymConst{\HOLTokenForall{}}\HOLBoundVar{c\sb{\mathrm{1}}} \HOLBoundVar{c\sb{\mathrm{2}}}. \HOLConst{CONTEXT} \HOLBoundVar{c\sb{\mathrm{1}}} \HOLSymConst{\HOLTokenConj{}} \HOLConst{CONTEXT} \HOLBoundVar{c\sb{\mathrm{2}}} \HOLSymConst{\HOLTokenImp{}} \HOLConst{CONTEXT} (\HOLBoundVar{c\sb{\mathrm{1}}} \HOLConst{\HOLTokenCompose} \HOLBoundVar{c\sb{\mathrm{2}}})
\end{SaveVerbatim}
\newcommand{\HOLCongruenceTheoremsCONTEXTXXcombin}{\UseVerbatim{HOLCongruenceTheoremsCONTEXTXXcombin}}
\begin{SaveVerbatim}{HOLCongruenceTheoremsCONTEXTXXind}
\HOLTokenTurnstile{} \HOLSymConst{\HOLTokenForall{}}\HOLBoundVar{CONTEXT\sp{\prime}}.
     \HOLBoundVar{CONTEXT\sp{\prime}} (\HOLTokenLambda{}\HOLBoundVar{t}. \HOLBoundVar{t}) \HOLSymConst{\HOLTokenConj{}} (\HOLSymConst{\HOLTokenForall{}}\HOLBoundVar{p}. \HOLBoundVar{CONTEXT\sp{\prime}} (\HOLTokenLambda{}\HOLBoundVar{t}. \HOLBoundVar{p})) \HOLSymConst{\HOLTokenConj{}}
     (\HOLSymConst{\HOLTokenForall{}}\HOLBoundVar{a} \HOLBoundVar{e}. \HOLBoundVar{CONTEXT\sp{\prime}} \HOLBoundVar{e} \HOLSymConst{\HOLTokenImp{}} \HOLBoundVar{CONTEXT\sp{\prime}} (\HOLTokenLambda{}\HOLBoundVar{t}. \HOLBoundVar{a}\HOLSymConst{..}\HOLBoundVar{e} \HOLBoundVar{t})) \HOLSymConst{\HOLTokenConj{}}
     (\HOLSymConst{\HOLTokenForall{}}\HOLBoundVar{e\sb{\mathrm{1}}} \HOLBoundVar{e\sb{\mathrm{2}}}.
        \HOLBoundVar{CONTEXT\sp{\prime}} \HOLBoundVar{e\sb{\mathrm{1}}} \HOLSymConst{\HOLTokenConj{}} \HOLBoundVar{CONTEXT\sp{\prime}} \HOLBoundVar{e\sb{\mathrm{2}}} \HOLSymConst{\HOLTokenImp{}}
        \HOLBoundVar{CONTEXT\sp{\prime}} (\HOLTokenLambda{}\HOLBoundVar{t}. \HOLBoundVar{e\sb{\mathrm{1}}} \HOLBoundVar{t} \HOLSymConst{+} \HOLBoundVar{e\sb{\mathrm{2}}} \HOLBoundVar{t})) \HOLSymConst{\HOLTokenConj{}}
     (\HOLSymConst{\HOLTokenForall{}}\HOLBoundVar{e\sb{\mathrm{1}}} \HOLBoundVar{e\sb{\mathrm{2}}}.
        \HOLBoundVar{CONTEXT\sp{\prime}} \HOLBoundVar{e\sb{\mathrm{1}}} \HOLSymConst{\HOLTokenConj{}} \HOLBoundVar{CONTEXT\sp{\prime}} \HOLBoundVar{e\sb{\mathrm{2}}} \HOLSymConst{\HOLTokenImp{}}
        \HOLBoundVar{CONTEXT\sp{\prime}} (\HOLTokenLambda{}\HOLBoundVar{t}. \HOLBoundVar{e\sb{\mathrm{1}}} \HOLBoundVar{t} \HOLSymConst{\ensuremath{\parallel}} \HOLBoundVar{e\sb{\mathrm{2}}} \HOLBoundVar{t})) \HOLSymConst{\HOLTokenConj{}}
     (\HOLSymConst{\HOLTokenForall{}}\HOLBoundVar{L} \HOLBoundVar{e}. \HOLBoundVar{CONTEXT\sp{\prime}} \HOLBoundVar{e} \HOLSymConst{\HOLTokenImp{}} \HOLBoundVar{CONTEXT\sp{\prime}} (\HOLTokenLambda{}\HOLBoundVar{t}. \HOLConst{\ensuremath{\nu}} \HOLBoundVar{L} (\HOLBoundVar{e} \HOLBoundVar{t}))) \HOLSymConst{\HOLTokenConj{}}
     (\HOLSymConst{\HOLTokenForall{}}\HOLBoundVar{rf} \HOLBoundVar{e}. \HOLBoundVar{CONTEXT\sp{\prime}} \HOLBoundVar{e} \HOLSymConst{\HOLTokenImp{}} \HOLBoundVar{CONTEXT\sp{\prime}} (\HOLTokenLambda{}\HOLBoundVar{t}. \HOLConst{relab} (\HOLBoundVar{e} \HOLBoundVar{t}) \HOLBoundVar{rf})) \HOLSymConst{\HOLTokenImp{}}
     \HOLSymConst{\HOLTokenForall{}}\HOLBoundVar{a\sb{\mathrm{0}}}. \HOLConst{CONTEXT} \HOLBoundVar{a\sb{\mathrm{0}}} \HOLSymConst{\HOLTokenImp{}} \HOLBoundVar{CONTEXT\sp{\prime}} \HOLBoundVar{a\sb{\mathrm{0}}}
\end{SaveVerbatim}
\newcommand{\HOLCongruenceTheoremsCONTEXTXXind}{\UseVerbatim{HOLCongruenceTheoremsCONTEXTXXind}}
\begin{SaveVerbatim}{HOLCongruenceTheoremsCONTEXTXXrules}
\HOLTokenTurnstile{} \HOLConst{CONTEXT} (\HOLTokenLambda{}\HOLBoundVar{t}. \HOLBoundVar{t}) \HOLSymConst{\HOLTokenConj{}} (\HOLSymConst{\HOLTokenForall{}}\HOLBoundVar{p}. \HOLConst{CONTEXT} (\HOLTokenLambda{}\HOLBoundVar{t}. \HOLBoundVar{p})) \HOLSymConst{\HOLTokenConj{}}
   (\HOLSymConst{\HOLTokenForall{}}\HOLBoundVar{a} \HOLBoundVar{e}. \HOLConst{CONTEXT} \HOLBoundVar{e} \HOLSymConst{\HOLTokenImp{}} \HOLConst{CONTEXT} (\HOLTokenLambda{}\HOLBoundVar{t}. \HOLBoundVar{a}\HOLSymConst{..}\HOLBoundVar{e} \HOLBoundVar{t})) \HOLSymConst{\HOLTokenConj{}}
   (\HOLSymConst{\HOLTokenForall{}}\HOLBoundVar{e\sb{\mathrm{1}}} \HOLBoundVar{e\sb{\mathrm{2}}}.
      \HOLConst{CONTEXT} \HOLBoundVar{e\sb{\mathrm{1}}} \HOLSymConst{\HOLTokenConj{}} \HOLConst{CONTEXT} \HOLBoundVar{e\sb{\mathrm{2}}} \HOLSymConst{\HOLTokenImp{}} \HOLConst{CONTEXT} (\HOLTokenLambda{}\HOLBoundVar{t}. \HOLBoundVar{e\sb{\mathrm{1}}} \HOLBoundVar{t} \HOLSymConst{+} \HOLBoundVar{e\sb{\mathrm{2}}} \HOLBoundVar{t})) \HOLSymConst{\HOLTokenConj{}}
   (\HOLSymConst{\HOLTokenForall{}}\HOLBoundVar{e\sb{\mathrm{1}}} \HOLBoundVar{e\sb{\mathrm{2}}}.
      \HOLConst{CONTEXT} \HOLBoundVar{e\sb{\mathrm{1}}} \HOLSymConst{\HOLTokenConj{}} \HOLConst{CONTEXT} \HOLBoundVar{e\sb{\mathrm{2}}} \HOLSymConst{\HOLTokenImp{}} \HOLConst{CONTEXT} (\HOLTokenLambda{}\HOLBoundVar{t}. \HOLBoundVar{e\sb{\mathrm{1}}} \HOLBoundVar{t} \HOLSymConst{\ensuremath{\parallel}} \HOLBoundVar{e\sb{\mathrm{2}}} \HOLBoundVar{t})) \HOLSymConst{\HOLTokenConj{}}
   (\HOLSymConst{\HOLTokenForall{}}\HOLBoundVar{L} \HOLBoundVar{e}. \HOLConst{CONTEXT} \HOLBoundVar{e} \HOLSymConst{\HOLTokenImp{}} \HOLConst{CONTEXT} (\HOLTokenLambda{}\HOLBoundVar{t}. \HOLConst{\ensuremath{\nu}} \HOLBoundVar{L} (\HOLBoundVar{e} \HOLBoundVar{t}))) \HOLSymConst{\HOLTokenConj{}}
   \HOLSymConst{\HOLTokenForall{}}\HOLBoundVar{rf} \HOLBoundVar{e}. \HOLConst{CONTEXT} \HOLBoundVar{e} \HOLSymConst{\HOLTokenImp{}} \HOLConst{CONTEXT} (\HOLTokenLambda{}\HOLBoundVar{t}. \HOLConst{relab} (\HOLBoundVar{e} \HOLBoundVar{t}) \HOLBoundVar{rf})
\end{SaveVerbatim}
\newcommand{\HOLCongruenceTheoremsCONTEXTXXrules}{\UseVerbatim{HOLCongruenceTheoremsCONTEXTXXrules}}
\begin{SaveVerbatim}{HOLCongruenceTheoremsCONTEXTXXstrongind}
\HOLTokenTurnstile{} \HOLSymConst{\HOLTokenForall{}}\HOLBoundVar{CONTEXT\sp{\prime}}.
     \HOLBoundVar{CONTEXT\sp{\prime}} (\HOLTokenLambda{}\HOLBoundVar{t}. \HOLBoundVar{t}) \HOLSymConst{\HOLTokenConj{}} (\HOLSymConst{\HOLTokenForall{}}\HOLBoundVar{p}. \HOLBoundVar{CONTEXT\sp{\prime}} (\HOLTokenLambda{}\HOLBoundVar{t}. \HOLBoundVar{p})) \HOLSymConst{\HOLTokenConj{}}
     (\HOLSymConst{\HOLTokenForall{}}\HOLBoundVar{a} \HOLBoundVar{e}. \HOLConst{CONTEXT} \HOLBoundVar{e} \HOLSymConst{\HOLTokenConj{}} \HOLBoundVar{CONTEXT\sp{\prime}} \HOLBoundVar{e} \HOLSymConst{\HOLTokenImp{}} \HOLBoundVar{CONTEXT\sp{\prime}} (\HOLTokenLambda{}\HOLBoundVar{t}. \HOLBoundVar{a}\HOLSymConst{..}\HOLBoundVar{e} \HOLBoundVar{t})) \HOLSymConst{\HOLTokenConj{}}
     (\HOLSymConst{\HOLTokenForall{}}\HOLBoundVar{e\sb{\mathrm{1}}} \HOLBoundVar{e\sb{\mathrm{2}}}.
        \HOLConst{CONTEXT} \HOLBoundVar{e\sb{\mathrm{1}}} \HOLSymConst{\HOLTokenConj{}} \HOLBoundVar{CONTEXT\sp{\prime}} \HOLBoundVar{e\sb{\mathrm{1}}} \HOLSymConst{\HOLTokenConj{}} \HOLConst{CONTEXT} \HOLBoundVar{e\sb{\mathrm{2}}} \HOLSymConst{\HOLTokenConj{}} \HOLBoundVar{CONTEXT\sp{\prime}} \HOLBoundVar{e\sb{\mathrm{2}}} \HOLSymConst{\HOLTokenImp{}}
        \HOLBoundVar{CONTEXT\sp{\prime}} (\HOLTokenLambda{}\HOLBoundVar{t}. \HOLBoundVar{e\sb{\mathrm{1}}} \HOLBoundVar{t} \HOLSymConst{+} \HOLBoundVar{e\sb{\mathrm{2}}} \HOLBoundVar{t})) \HOLSymConst{\HOLTokenConj{}}
     (\HOLSymConst{\HOLTokenForall{}}\HOLBoundVar{e\sb{\mathrm{1}}} \HOLBoundVar{e\sb{\mathrm{2}}}.
        \HOLConst{CONTEXT} \HOLBoundVar{e\sb{\mathrm{1}}} \HOLSymConst{\HOLTokenConj{}} \HOLBoundVar{CONTEXT\sp{\prime}} \HOLBoundVar{e\sb{\mathrm{1}}} \HOLSymConst{\HOLTokenConj{}} \HOLConst{CONTEXT} \HOLBoundVar{e\sb{\mathrm{2}}} \HOLSymConst{\HOLTokenConj{}} \HOLBoundVar{CONTEXT\sp{\prime}} \HOLBoundVar{e\sb{\mathrm{2}}} \HOLSymConst{\HOLTokenImp{}}
        \HOLBoundVar{CONTEXT\sp{\prime}} (\HOLTokenLambda{}\HOLBoundVar{t}. \HOLBoundVar{e\sb{\mathrm{1}}} \HOLBoundVar{t} \HOLSymConst{\ensuremath{\parallel}} \HOLBoundVar{e\sb{\mathrm{2}}} \HOLBoundVar{t})) \HOLSymConst{\HOLTokenConj{}}
     (\HOLSymConst{\HOLTokenForall{}}\HOLBoundVar{L} \HOLBoundVar{e}. \HOLConst{CONTEXT} \HOLBoundVar{e} \HOLSymConst{\HOLTokenConj{}} \HOLBoundVar{CONTEXT\sp{\prime}} \HOLBoundVar{e} \HOLSymConst{\HOLTokenImp{}} \HOLBoundVar{CONTEXT\sp{\prime}} (\HOLTokenLambda{}\HOLBoundVar{t}. \HOLConst{\ensuremath{\nu}} \HOLBoundVar{L} (\HOLBoundVar{e} \HOLBoundVar{t}))) \HOLSymConst{\HOLTokenConj{}}
     (\HOLSymConst{\HOLTokenForall{}}\HOLBoundVar{rf} \HOLBoundVar{e}.
        \HOLConst{CONTEXT} \HOLBoundVar{e} \HOLSymConst{\HOLTokenConj{}} \HOLBoundVar{CONTEXT\sp{\prime}} \HOLBoundVar{e} \HOLSymConst{\HOLTokenImp{}}
        \HOLBoundVar{CONTEXT\sp{\prime}} (\HOLTokenLambda{}\HOLBoundVar{t}. \HOLConst{relab} (\HOLBoundVar{e} \HOLBoundVar{t}) \HOLBoundVar{rf})) \HOLSymConst{\HOLTokenImp{}}
     \HOLSymConst{\HOLTokenForall{}}\HOLBoundVar{a\sb{\mathrm{0}}}. \HOLConst{CONTEXT} \HOLBoundVar{a\sb{\mathrm{0}}} \HOLSymConst{\HOLTokenImp{}} \HOLBoundVar{CONTEXT\sp{\prime}} \HOLBoundVar{a\sb{\mathrm{0}}}
\end{SaveVerbatim}
\newcommand{\HOLCongruenceTheoremsCONTEXTXXstrongind}{\UseVerbatim{HOLCongruenceTheoremsCONTEXTXXstrongind}}
\begin{SaveVerbatim}{HOLCongruenceTheoremsCONTEXTXXWGXXcombin}
\HOLTokenTurnstile{} \HOLSymConst{\HOLTokenForall{}}\HOLBoundVar{c} \HOLBoundVar{e}. \HOLConst{CONTEXT} \HOLBoundVar{c} \HOLSymConst{\HOLTokenConj{}} \HOLConst{WG} \HOLBoundVar{e} \HOLSymConst{\HOLTokenImp{}} \HOLConst{WG} (\HOLBoundVar{c} \HOLConst{\HOLTokenCompose} \HOLBoundVar{e})
\end{SaveVerbatim}
\newcommand{\HOLCongruenceTheoremsCONTEXTXXWGXXcombin}{\UseVerbatim{HOLCongruenceTheoremsCONTEXTXXWGXXcombin}}
\begin{SaveVerbatim}{HOLCongruenceTheoremsGCONTEXTOne}
\HOLTokenTurnstile{} \HOLConst{GCONTEXT} (\HOLTokenLambda{}\HOLBoundVar{t}. \HOLBoundVar{t})
\end{SaveVerbatim}
\newcommand{\HOLCongruenceTheoremsGCONTEXTOne}{\UseVerbatim{HOLCongruenceTheoremsGCONTEXTOne}}
\begin{SaveVerbatim}{HOLCongruenceTheoremsGCONTEXTTwo}
\HOLTokenTurnstile{} \HOLSymConst{\HOLTokenForall{}}\HOLBoundVar{p}. \HOLConst{GCONTEXT} (\HOLTokenLambda{}\HOLBoundVar{t}. \HOLBoundVar{p})
\end{SaveVerbatim}
\newcommand{\HOLCongruenceTheoremsGCONTEXTTwo}{\UseVerbatim{HOLCongruenceTheoremsGCONTEXTTwo}}
\begin{SaveVerbatim}{HOLCongruenceTheoremsGCONTEXTThree}
\HOLTokenTurnstile{} \HOLSymConst{\HOLTokenForall{}}\HOLBoundVar{a} \HOLBoundVar{e}. \HOLConst{GCONTEXT} \HOLBoundVar{e} \HOLSymConst{\HOLTokenImp{}} \HOLConst{GCONTEXT} (\HOLTokenLambda{}\HOLBoundVar{t}. \HOLBoundVar{a}\HOLSymConst{..}\HOLBoundVar{e} \HOLBoundVar{t})
\end{SaveVerbatim}
\newcommand{\HOLCongruenceTheoremsGCONTEXTThree}{\UseVerbatim{HOLCongruenceTheoremsGCONTEXTThree}}
\begin{SaveVerbatim}{HOLCongruenceTheoremsGCONTEXTThreea}
\HOLTokenTurnstile{} \HOLSymConst{\HOLTokenForall{}}\HOLBoundVar{a}. \HOLConst{GCONTEXT} (\HOLTokenLambda{}\HOLBoundVar{t}. \HOLBoundVar{a}\HOLSymConst{..}\HOLBoundVar{t})
\end{SaveVerbatim}
\newcommand{\HOLCongruenceTheoremsGCONTEXTThreea}{\UseVerbatim{HOLCongruenceTheoremsGCONTEXTThreea}}
\begin{SaveVerbatim}{HOLCongruenceTheoremsGCONTEXTFour}
\HOLTokenTurnstile{} \HOLSymConst{\HOLTokenForall{}}\HOLBoundVar{a\sb{\mathrm{1}}} \HOLBoundVar{a\sb{\mathrm{2}}} \HOLBoundVar{e\sb{\mathrm{1}}} \HOLBoundVar{e\sb{\mathrm{2}}}.
     \HOLConst{GCONTEXT} \HOLBoundVar{e\sb{\mathrm{1}}} \HOLSymConst{\HOLTokenConj{}} \HOLConst{GCONTEXT} \HOLBoundVar{e\sb{\mathrm{2}}} \HOLSymConst{\HOLTokenImp{}}
     \HOLConst{GCONTEXT} (\HOLTokenLambda{}\HOLBoundVar{t}. \HOLBoundVar{a\sb{\mathrm{1}}}\HOLSymConst{..}\HOLBoundVar{e\sb{\mathrm{1}}} \HOLBoundVar{t} \HOLSymConst{+} \HOLBoundVar{a\sb{\mathrm{2}}}\HOLSymConst{..}\HOLBoundVar{e\sb{\mathrm{2}}} \HOLBoundVar{t})
\end{SaveVerbatim}
\newcommand{\HOLCongruenceTheoremsGCONTEXTFour}{\UseVerbatim{HOLCongruenceTheoremsGCONTEXTFour}}
\begin{SaveVerbatim}{HOLCongruenceTheoremsGCONTEXTFive}
\HOLTokenTurnstile{} \HOLSymConst{\HOLTokenForall{}}\HOLBoundVar{e\sb{\mathrm{1}}} \HOLBoundVar{e\sb{\mathrm{2}}}.
     \HOLConst{GCONTEXT} \HOLBoundVar{e\sb{\mathrm{1}}} \HOLSymConst{\HOLTokenConj{}} \HOLConst{GCONTEXT} \HOLBoundVar{e\sb{\mathrm{2}}} \HOLSymConst{\HOLTokenImp{}} \HOLConst{GCONTEXT} (\HOLTokenLambda{}\HOLBoundVar{t}. \HOLBoundVar{e\sb{\mathrm{1}}} \HOLBoundVar{t} \HOLSymConst{\ensuremath{\parallel}} \HOLBoundVar{e\sb{\mathrm{2}}} \HOLBoundVar{t})
\end{SaveVerbatim}
\newcommand{\HOLCongruenceTheoremsGCONTEXTFive}{\UseVerbatim{HOLCongruenceTheoremsGCONTEXTFive}}
\begin{SaveVerbatim}{HOLCongruenceTheoremsGCONTEXTSix}
\HOLTokenTurnstile{} \HOLSymConst{\HOLTokenForall{}}\HOLBoundVar{L} \HOLBoundVar{e}. \HOLConst{GCONTEXT} \HOLBoundVar{e} \HOLSymConst{\HOLTokenImp{}} \HOLConst{GCONTEXT} (\HOLTokenLambda{}\HOLBoundVar{t}. \HOLConst{\ensuremath{\nu}} \HOLBoundVar{L} (\HOLBoundVar{e} \HOLBoundVar{t}))
\end{SaveVerbatim}
\newcommand{\HOLCongruenceTheoremsGCONTEXTSix}{\UseVerbatim{HOLCongruenceTheoremsGCONTEXTSix}}
\begin{SaveVerbatim}{HOLCongruenceTheoremsGCONTEXTSeven}
\HOLTokenTurnstile{} \HOLSymConst{\HOLTokenForall{}}\HOLBoundVar{rf} \HOLBoundVar{e}. \HOLConst{GCONTEXT} \HOLBoundVar{e} \HOLSymConst{\HOLTokenImp{}} \HOLConst{GCONTEXT} (\HOLTokenLambda{}\HOLBoundVar{t}. \HOLConst{relab} (\HOLBoundVar{e} \HOLBoundVar{t}) \HOLBoundVar{rf})
\end{SaveVerbatim}
\newcommand{\HOLCongruenceTheoremsGCONTEXTSeven}{\UseVerbatim{HOLCongruenceTheoremsGCONTEXTSeven}}
\begin{SaveVerbatim}{HOLCongruenceTheoremsGCONTEXTXXcases}
\HOLTokenTurnstile{} \HOLSymConst{\HOLTokenForall{}}\HOLBoundVar{a\sb{\mathrm{0}}}.
     \HOLConst{GCONTEXT} \HOLBoundVar{a\sb{\mathrm{0}}} \HOLSymConst{\HOLTokenEquiv{}}
     (\HOLBoundVar{a\sb{\mathrm{0}}} \HOLSymConst{=} (\HOLTokenLambda{}\HOLBoundVar{t}. \HOLBoundVar{t})) \HOLSymConst{\HOLTokenDisj{}} (\HOLSymConst{\HOLTokenExists{}}\HOLBoundVar{p}. \HOLBoundVar{a\sb{\mathrm{0}}} \HOLSymConst{=} (\HOLTokenLambda{}\HOLBoundVar{t}. \HOLBoundVar{p})) \HOLSymConst{\HOLTokenDisj{}}
     (\HOLSymConst{\HOLTokenExists{}}\HOLBoundVar{a} \HOLBoundVar{e}. (\HOLBoundVar{a\sb{\mathrm{0}}} \HOLSymConst{=} (\HOLTokenLambda{}\HOLBoundVar{t}. \HOLBoundVar{a}\HOLSymConst{..}\HOLBoundVar{e} \HOLBoundVar{t})) \HOLSymConst{\HOLTokenConj{}} \HOLConst{GCONTEXT} \HOLBoundVar{e}) \HOLSymConst{\HOLTokenDisj{}}
     (\HOLSymConst{\HOLTokenExists{}}\HOLBoundVar{a\sb{\mathrm{1}}} \HOLBoundVar{a\sb{\mathrm{2}}} \HOLBoundVar{e\sb{\mathrm{1}}} \HOLBoundVar{e\sb{\mathrm{2}}}.
        (\HOLBoundVar{a\sb{\mathrm{0}}} \HOLSymConst{=} (\HOLTokenLambda{}\HOLBoundVar{t}. \HOLBoundVar{a\sb{\mathrm{1}}}\HOLSymConst{..}\HOLBoundVar{e\sb{\mathrm{1}}} \HOLBoundVar{t} \HOLSymConst{+} \HOLBoundVar{a\sb{\mathrm{2}}}\HOLSymConst{..}\HOLBoundVar{e\sb{\mathrm{2}}} \HOLBoundVar{t})) \HOLSymConst{\HOLTokenConj{}} \HOLConst{GCONTEXT} \HOLBoundVar{e\sb{\mathrm{1}}} \HOLSymConst{\HOLTokenConj{}}
        \HOLConst{GCONTEXT} \HOLBoundVar{e\sb{\mathrm{2}}}) \HOLSymConst{\HOLTokenDisj{}}
     (\HOLSymConst{\HOLTokenExists{}}\HOLBoundVar{e\sb{\mathrm{1}}} \HOLBoundVar{e\sb{\mathrm{2}}}.
        (\HOLBoundVar{a\sb{\mathrm{0}}} \HOLSymConst{=} (\HOLTokenLambda{}\HOLBoundVar{t}. \HOLBoundVar{e\sb{\mathrm{1}}} \HOLBoundVar{t} \HOLSymConst{\ensuremath{\parallel}} \HOLBoundVar{e\sb{\mathrm{2}}} \HOLBoundVar{t})) \HOLSymConst{\HOLTokenConj{}} \HOLConst{GCONTEXT} \HOLBoundVar{e\sb{\mathrm{1}}} \HOLSymConst{\HOLTokenConj{}} \HOLConst{GCONTEXT} \HOLBoundVar{e\sb{\mathrm{2}}}) \HOLSymConst{\HOLTokenDisj{}}
     (\HOLSymConst{\HOLTokenExists{}}\HOLBoundVar{L} \HOLBoundVar{e}. (\HOLBoundVar{a\sb{\mathrm{0}}} \HOLSymConst{=} (\HOLTokenLambda{}\HOLBoundVar{t}. \HOLConst{\ensuremath{\nu}} \HOLBoundVar{L} (\HOLBoundVar{e} \HOLBoundVar{t}))) \HOLSymConst{\HOLTokenConj{}} \HOLConst{GCONTEXT} \HOLBoundVar{e}) \HOLSymConst{\HOLTokenDisj{}}
     \HOLSymConst{\HOLTokenExists{}}\HOLBoundVar{rf} \HOLBoundVar{e}. (\HOLBoundVar{a\sb{\mathrm{0}}} \HOLSymConst{=} (\HOLTokenLambda{}\HOLBoundVar{t}. \HOLConst{relab} (\HOLBoundVar{e} \HOLBoundVar{t}) \HOLBoundVar{rf})) \HOLSymConst{\HOLTokenConj{}} \HOLConst{GCONTEXT} \HOLBoundVar{e}
\end{SaveVerbatim}
\newcommand{\HOLCongruenceTheoremsGCONTEXTXXcases}{\UseVerbatim{HOLCongruenceTheoremsGCONTEXTXXcases}}
\begin{SaveVerbatim}{HOLCongruenceTheoremsGCONTEXTXXcombin}
\HOLTokenTurnstile{} \HOLSymConst{\HOLTokenForall{}}\HOLBoundVar{c\sb{\mathrm{1}}} \HOLBoundVar{c\sb{\mathrm{2}}}. \HOLConst{GCONTEXT} \HOLBoundVar{c\sb{\mathrm{1}}} \HOLSymConst{\HOLTokenConj{}} \HOLConst{GCONTEXT} \HOLBoundVar{c\sb{\mathrm{2}}} \HOLSymConst{\HOLTokenImp{}} \HOLConst{GCONTEXT} (\HOLBoundVar{c\sb{\mathrm{1}}} \HOLConst{\HOLTokenCompose} \HOLBoundVar{c\sb{\mathrm{2}}})
\end{SaveVerbatim}
\newcommand{\HOLCongruenceTheoremsGCONTEXTXXcombin}{\UseVerbatim{HOLCongruenceTheoremsGCONTEXTXXcombin}}
\begin{SaveVerbatim}{HOLCongruenceTheoremsGCONTEXTXXind}
\HOLTokenTurnstile{} \HOLSymConst{\HOLTokenForall{}}\HOLBoundVar{GCONTEXT\sp{\prime}}.
     \HOLBoundVar{GCONTEXT\sp{\prime}} (\HOLTokenLambda{}\HOLBoundVar{t}. \HOLBoundVar{t}) \HOLSymConst{\HOLTokenConj{}} (\HOLSymConst{\HOLTokenForall{}}\HOLBoundVar{p}. \HOLBoundVar{GCONTEXT\sp{\prime}} (\HOLTokenLambda{}\HOLBoundVar{t}. \HOLBoundVar{p})) \HOLSymConst{\HOLTokenConj{}}
     (\HOLSymConst{\HOLTokenForall{}}\HOLBoundVar{a} \HOLBoundVar{e}. \HOLBoundVar{GCONTEXT\sp{\prime}} \HOLBoundVar{e} \HOLSymConst{\HOLTokenImp{}} \HOLBoundVar{GCONTEXT\sp{\prime}} (\HOLTokenLambda{}\HOLBoundVar{t}. \HOLBoundVar{a}\HOLSymConst{..}\HOLBoundVar{e} \HOLBoundVar{t})) \HOLSymConst{\HOLTokenConj{}}
     (\HOLSymConst{\HOLTokenForall{}}\HOLBoundVar{a\sb{\mathrm{1}}} \HOLBoundVar{a\sb{\mathrm{2}}} \HOLBoundVar{e\sb{\mathrm{1}}} \HOLBoundVar{e\sb{\mathrm{2}}}.
        \HOLBoundVar{GCONTEXT\sp{\prime}} \HOLBoundVar{e\sb{\mathrm{1}}} \HOLSymConst{\HOLTokenConj{}} \HOLBoundVar{GCONTEXT\sp{\prime}} \HOLBoundVar{e\sb{\mathrm{2}}} \HOLSymConst{\HOLTokenImp{}}
        \HOLBoundVar{GCONTEXT\sp{\prime}} (\HOLTokenLambda{}\HOLBoundVar{t}. \HOLBoundVar{a\sb{\mathrm{1}}}\HOLSymConst{..}\HOLBoundVar{e\sb{\mathrm{1}}} \HOLBoundVar{t} \HOLSymConst{+} \HOLBoundVar{a\sb{\mathrm{2}}}\HOLSymConst{..}\HOLBoundVar{e\sb{\mathrm{2}}} \HOLBoundVar{t})) \HOLSymConst{\HOLTokenConj{}}
     (\HOLSymConst{\HOLTokenForall{}}\HOLBoundVar{e\sb{\mathrm{1}}} \HOLBoundVar{e\sb{\mathrm{2}}}.
        \HOLBoundVar{GCONTEXT\sp{\prime}} \HOLBoundVar{e\sb{\mathrm{1}}} \HOLSymConst{\HOLTokenConj{}} \HOLBoundVar{GCONTEXT\sp{\prime}} \HOLBoundVar{e\sb{\mathrm{2}}} \HOLSymConst{\HOLTokenImp{}}
        \HOLBoundVar{GCONTEXT\sp{\prime}} (\HOLTokenLambda{}\HOLBoundVar{t}. \HOLBoundVar{e\sb{\mathrm{1}}} \HOLBoundVar{t} \HOLSymConst{\ensuremath{\parallel}} \HOLBoundVar{e\sb{\mathrm{2}}} \HOLBoundVar{t})) \HOLSymConst{\HOLTokenConj{}}
     (\HOLSymConst{\HOLTokenForall{}}\HOLBoundVar{L} \HOLBoundVar{e}. \HOLBoundVar{GCONTEXT\sp{\prime}} \HOLBoundVar{e} \HOLSymConst{\HOLTokenImp{}} \HOLBoundVar{GCONTEXT\sp{\prime}} (\HOLTokenLambda{}\HOLBoundVar{t}. \HOLConst{\ensuremath{\nu}} \HOLBoundVar{L} (\HOLBoundVar{e} \HOLBoundVar{t}))) \HOLSymConst{\HOLTokenConj{}}
     (\HOLSymConst{\HOLTokenForall{}}\HOLBoundVar{rf} \HOLBoundVar{e}. \HOLBoundVar{GCONTEXT\sp{\prime}} \HOLBoundVar{e} \HOLSymConst{\HOLTokenImp{}} \HOLBoundVar{GCONTEXT\sp{\prime}} (\HOLTokenLambda{}\HOLBoundVar{t}. \HOLConst{relab} (\HOLBoundVar{e} \HOLBoundVar{t}) \HOLBoundVar{rf})) \HOLSymConst{\HOLTokenImp{}}
     \HOLSymConst{\HOLTokenForall{}}\HOLBoundVar{a\sb{\mathrm{0}}}. \HOLConst{GCONTEXT} \HOLBoundVar{a\sb{\mathrm{0}}} \HOLSymConst{\HOLTokenImp{}} \HOLBoundVar{GCONTEXT\sp{\prime}} \HOLBoundVar{a\sb{\mathrm{0}}}
\end{SaveVerbatim}
\newcommand{\HOLCongruenceTheoremsGCONTEXTXXind}{\UseVerbatim{HOLCongruenceTheoremsGCONTEXTXXind}}
\begin{SaveVerbatim}{HOLCongruenceTheoremsGCONTEXTXXISXXCONTEXT}
\HOLTokenTurnstile{} \HOLSymConst{\HOLTokenForall{}}\HOLBoundVar{c}. \HOLConst{GCONTEXT} \HOLBoundVar{c} \HOLSymConst{\HOLTokenImp{}} \HOLConst{CONTEXT} \HOLBoundVar{c}
\end{SaveVerbatim}
\newcommand{\HOLCongruenceTheoremsGCONTEXTXXISXXCONTEXT}{\UseVerbatim{HOLCongruenceTheoremsGCONTEXTXXISXXCONTEXT}}
\begin{SaveVerbatim}{HOLCongruenceTheoremsGCONTEXTXXrules}
\HOLTokenTurnstile{} \HOLConst{GCONTEXT} (\HOLTokenLambda{}\HOLBoundVar{t}. \HOLBoundVar{t}) \HOLSymConst{\HOLTokenConj{}} (\HOLSymConst{\HOLTokenForall{}}\HOLBoundVar{p}. \HOLConst{GCONTEXT} (\HOLTokenLambda{}\HOLBoundVar{t}. \HOLBoundVar{p})) \HOLSymConst{\HOLTokenConj{}}
   (\HOLSymConst{\HOLTokenForall{}}\HOLBoundVar{a} \HOLBoundVar{e}. \HOLConst{GCONTEXT} \HOLBoundVar{e} \HOLSymConst{\HOLTokenImp{}} \HOLConst{GCONTEXT} (\HOLTokenLambda{}\HOLBoundVar{t}. \HOLBoundVar{a}\HOLSymConst{..}\HOLBoundVar{e} \HOLBoundVar{t})) \HOLSymConst{\HOLTokenConj{}}
   (\HOLSymConst{\HOLTokenForall{}}\HOLBoundVar{a\sb{\mathrm{1}}} \HOLBoundVar{a\sb{\mathrm{2}}} \HOLBoundVar{e\sb{\mathrm{1}}} \HOLBoundVar{e\sb{\mathrm{2}}}.
      \HOLConst{GCONTEXT} \HOLBoundVar{e\sb{\mathrm{1}}} \HOLSymConst{\HOLTokenConj{}} \HOLConst{GCONTEXT} \HOLBoundVar{e\sb{\mathrm{2}}} \HOLSymConst{\HOLTokenImp{}}
      \HOLConst{GCONTEXT} (\HOLTokenLambda{}\HOLBoundVar{t}. \HOLBoundVar{a\sb{\mathrm{1}}}\HOLSymConst{..}\HOLBoundVar{e\sb{\mathrm{1}}} \HOLBoundVar{t} \HOLSymConst{+} \HOLBoundVar{a\sb{\mathrm{2}}}\HOLSymConst{..}\HOLBoundVar{e\sb{\mathrm{2}}} \HOLBoundVar{t})) \HOLSymConst{\HOLTokenConj{}}
   (\HOLSymConst{\HOLTokenForall{}}\HOLBoundVar{e\sb{\mathrm{1}}} \HOLBoundVar{e\sb{\mathrm{2}}}.
      \HOLConst{GCONTEXT} \HOLBoundVar{e\sb{\mathrm{1}}} \HOLSymConst{\HOLTokenConj{}} \HOLConst{GCONTEXT} \HOLBoundVar{e\sb{\mathrm{2}}} \HOLSymConst{\HOLTokenImp{}} \HOLConst{GCONTEXT} (\HOLTokenLambda{}\HOLBoundVar{t}. \HOLBoundVar{e\sb{\mathrm{1}}} \HOLBoundVar{t} \HOLSymConst{\ensuremath{\parallel}} \HOLBoundVar{e\sb{\mathrm{2}}} \HOLBoundVar{t})) \HOLSymConst{\HOLTokenConj{}}
   (\HOLSymConst{\HOLTokenForall{}}\HOLBoundVar{L} \HOLBoundVar{e}. \HOLConst{GCONTEXT} \HOLBoundVar{e} \HOLSymConst{\HOLTokenImp{}} \HOLConst{GCONTEXT} (\HOLTokenLambda{}\HOLBoundVar{t}. \HOLConst{\ensuremath{\nu}} \HOLBoundVar{L} (\HOLBoundVar{e} \HOLBoundVar{t}))) \HOLSymConst{\HOLTokenConj{}}
   \HOLSymConst{\HOLTokenForall{}}\HOLBoundVar{rf} \HOLBoundVar{e}. \HOLConst{GCONTEXT} \HOLBoundVar{e} \HOLSymConst{\HOLTokenImp{}} \HOLConst{GCONTEXT} (\HOLTokenLambda{}\HOLBoundVar{t}. \HOLConst{relab} (\HOLBoundVar{e} \HOLBoundVar{t}) \HOLBoundVar{rf})
\end{SaveVerbatim}
\newcommand{\HOLCongruenceTheoremsGCONTEXTXXrules}{\UseVerbatim{HOLCongruenceTheoremsGCONTEXTXXrules}}
\begin{SaveVerbatim}{HOLCongruenceTheoremsGCONTEXTXXstrongind}
\HOLTokenTurnstile{} \HOLSymConst{\HOLTokenForall{}}\HOLBoundVar{GCONTEXT\sp{\prime}}.
     \HOLBoundVar{GCONTEXT\sp{\prime}} (\HOLTokenLambda{}\HOLBoundVar{t}. \HOLBoundVar{t}) \HOLSymConst{\HOLTokenConj{}} (\HOLSymConst{\HOLTokenForall{}}\HOLBoundVar{p}. \HOLBoundVar{GCONTEXT\sp{\prime}} (\HOLTokenLambda{}\HOLBoundVar{t}. \HOLBoundVar{p})) \HOLSymConst{\HOLTokenConj{}}
     (\HOLSymConst{\HOLTokenForall{}}\HOLBoundVar{a} \HOLBoundVar{e}. \HOLConst{GCONTEXT} \HOLBoundVar{e} \HOLSymConst{\HOLTokenConj{}} \HOLBoundVar{GCONTEXT\sp{\prime}} \HOLBoundVar{e} \HOLSymConst{\HOLTokenImp{}} \HOLBoundVar{GCONTEXT\sp{\prime}} (\HOLTokenLambda{}\HOLBoundVar{t}. \HOLBoundVar{a}\HOLSymConst{..}\HOLBoundVar{e} \HOLBoundVar{t})) \HOLSymConst{\HOLTokenConj{}}
     (\HOLSymConst{\HOLTokenForall{}}\HOLBoundVar{a\sb{\mathrm{1}}} \HOLBoundVar{a\sb{\mathrm{2}}} \HOLBoundVar{e\sb{\mathrm{1}}} \HOLBoundVar{e\sb{\mathrm{2}}}.
        \HOLConst{GCONTEXT} \HOLBoundVar{e\sb{\mathrm{1}}} \HOLSymConst{\HOLTokenConj{}} \HOLBoundVar{GCONTEXT\sp{\prime}} \HOLBoundVar{e\sb{\mathrm{1}}} \HOLSymConst{\HOLTokenConj{}} \HOLConst{GCONTEXT} \HOLBoundVar{e\sb{\mathrm{2}}} \HOLSymConst{\HOLTokenConj{}}
        \HOLBoundVar{GCONTEXT\sp{\prime}} \HOLBoundVar{e\sb{\mathrm{2}}} \HOLSymConst{\HOLTokenImp{}}
        \HOLBoundVar{GCONTEXT\sp{\prime}} (\HOLTokenLambda{}\HOLBoundVar{t}. \HOLBoundVar{a\sb{\mathrm{1}}}\HOLSymConst{..}\HOLBoundVar{e\sb{\mathrm{1}}} \HOLBoundVar{t} \HOLSymConst{+} \HOLBoundVar{a\sb{\mathrm{2}}}\HOLSymConst{..}\HOLBoundVar{e\sb{\mathrm{2}}} \HOLBoundVar{t})) \HOLSymConst{\HOLTokenConj{}}
     (\HOLSymConst{\HOLTokenForall{}}\HOLBoundVar{e\sb{\mathrm{1}}} \HOLBoundVar{e\sb{\mathrm{2}}}.
        \HOLConst{GCONTEXT} \HOLBoundVar{e\sb{\mathrm{1}}} \HOLSymConst{\HOLTokenConj{}} \HOLBoundVar{GCONTEXT\sp{\prime}} \HOLBoundVar{e\sb{\mathrm{1}}} \HOLSymConst{\HOLTokenConj{}} \HOLConst{GCONTEXT} \HOLBoundVar{e\sb{\mathrm{2}}} \HOLSymConst{\HOLTokenConj{}}
        \HOLBoundVar{GCONTEXT\sp{\prime}} \HOLBoundVar{e\sb{\mathrm{2}}} \HOLSymConst{\HOLTokenImp{}}
        \HOLBoundVar{GCONTEXT\sp{\prime}} (\HOLTokenLambda{}\HOLBoundVar{t}. \HOLBoundVar{e\sb{\mathrm{1}}} \HOLBoundVar{t} \HOLSymConst{\ensuremath{\parallel}} \HOLBoundVar{e\sb{\mathrm{2}}} \HOLBoundVar{t})) \HOLSymConst{\HOLTokenConj{}}
     (\HOLSymConst{\HOLTokenForall{}}\HOLBoundVar{L} \HOLBoundVar{e}.
        \HOLConst{GCONTEXT} \HOLBoundVar{e} \HOLSymConst{\HOLTokenConj{}} \HOLBoundVar{GCONTEXT\sp{\prime}} \HOLBoundVar{e} \HOLSymConst{\HOLTokenImp{}} \HOLBoundVar{GCONTEXT\sp{\prime}} (\HOLTokenLambda{}\HOLBoundVar{t}. \HOLConst{\ensuremath{\nu}} \HOLBoundVar{L} (\HOLBoundVar{e} \HOLBoundVar{t}))) \HOLSymConst{\HOLTokenConj{}}
     (\HOLSymConst{\HOLTokenForall{}}\HOLBoundVar{rf} \HOLBoundVar{e}.
        \HOLConst{GCONTEXT} \HOLBoundVar{e} \HOLSymConst{\HOLTokenConj{}} \HOLBoundVar{GCONTEXT\sp{\prime}} \HOLBoundVar{e} \HOLSymConst{\HOLTokenImp{}}
        \HOLBoundVar{GCONTEXT\sp{\prime}} (\HOLTokenLambda{}\HOLBoundVar{t}. \HOLConst{relab} (\HOLBoundVar{e} \HOLBoundVar{t}) \HOLBoundVar{rf})) \HOLSymConst{\HOLTokenImp{}}
     \HOLSymConst{\HOLTokenForall{}}\HOLBoundVar{a\sb{\mathrm{0}}}. \HOLConst{GCONTEXT} \HOLBoundVar{a\sb{\mathrm{0}}} \HOLSymConst{\HOLTokenImp{}} \HOLBoundVar{GCONTEXT\sp{\prime}} \HOLBoundVar{a\sb{\mathrm{0}}}
\end{SaveVerbatim}
\newcommand{\HOLCongruenceTheoremsGCONTEXTXXstrongind}{\UseVerbatim{HOLCongruenceTheoremsGCONTEXTXXstrongind}}
\begin{SaveVerbatim}{HOLCongruenceTheoremsGCONTEXTXXWGSXXcombin}
\HOLTokenTurnstile{} \HOLSymConst{\HOLTokenForall{}}\HOLBoundVar{c} \HOLBoundVar{e}. \HOLConst{GCONTEXT} \HOLBoundVar{c} \HOLSymConst{\HOLTokenConj{}} \HOLConst{WGS} \HOLBoundVar{e} \HOLSymConst{\HOLTokenImp{}} \HOLConst{WGS} (\HOLBoundVar{c} \HOLConst{\HOLTokenCompose} \HOLBoundVar{e})
\end{SaveVerbatim}
\newcommand{\HOLCongruenceTheoremsGCONTEXTXXWGSXXcombin}{\UseVerbatim{HOLCongruenceTheoremsGCONTEXTXXWGSXXcombin}}
\begin{SaveVerbatim}{HOLCongruenceTheoremsGSEQOne}
\HOLTokenTurnstile{} \HOLConst{GSEQ} (\HOLTokenLambda{}\HOLBoundVar{t}. \HOLBoundVar{t})
\end{SaveVerbatim}
\newcommand{\HOLCongruenceTheoremsGSEQOne}{\UseVerbatim{HOLCongruenceTheoremsGSEQOne}}
\begin{SaveVerbatim}{HOLCongruenceTheoremsGSEQTwo}
\HOLTokenTurnstile{} \HOLSymConst{\HOLTokenForall{}}\HOLBoundVar{p}. \HOLConst{GSEQ} (\HOLTokenLambda{}\HOLBoundVar{t}. \HOLBoundVar{p})
\end{SaveVerbatim}
\newcommand{\HOLCongruenceTheoremsGSEQTwo}{\UseVerbatim{HOLCongruenceTheoremsGSEQTwo}}
\begin{SaveVerbatim}{HOLCongruenceTheoremsGSEQThree}
\HOLTokenTurnstile{} \HOLSymConst{\HOLTokenForall{}}\HOLBoundVar{a} \HOLBoundVar{e}. \HOLConst{GSEQ} \HOLBoundVar{e} \HOLSymConst{\HOLTokenImp{}} \HOLConst{GSEQ} (\HOLTokenLambda{}\HOLBoundVar{t}. \HOLBoundVar{a}\HOLSymConst{..}\HOLBoundVar{e} \HOLBoundVar{t})
\end{SaveVerbatim}
\newcommand{\HOLCongruenceTheoremsGSEQThree}{\UseVerbatim{HOLCongruenceTheoremsGSEQThree}}
\begin{SaveVerbatim}{HOLCongruenceTheoremsGSEQThreea}
\HOLTokenTurnstile{} \HOLSymConst{\HOLTokenForall{}}\HOLBoundVar{a}. \HOLConst{GSEQ} (\HOLTokenLambda{}\HOLBoundVar{t}. \HOLBoundVar{a}\HOLSymConst{..}\HOLBoundVar{t})
\end{SaveVerbatim}
\newcommand{\HOLCongruenceTheoremsGSEQThreea}{\UseVerbatim{HOLCongruenceTheoremsGSEQThreea}}
\begin{SaveVerbatim}{HOLCongruenceTheoremsGSEQFour}
\HOLTokenTurnstile{} \HOLSymConst{\HOLTokenForall{}}\HOLBoundVar{a\sb{\mathrm{1}}} \HOLBoundVar{a\sb{\mathrm{2}}} \HOLBoundVar{e\sb{\mathrm{1}}} \HOLBoundVar{e\sb{\mathrm{2}}}.
     \HOLConst{GSEQ} \HOLBoundVar{e\sb{\mathrm{1}}} \HOLSymConst{\HOLTokenConj{}} \HOLConst{GSEQ} \HOLBoundVar{e\sb{\mathrm{2}}} \HOLSymConst{\HOLTokenImp{}} \HOLConst{GSEQ} (\HOLTokenLambda{}\HOLBoundVar{t}. \HOLBoundVar{a\sb{\mathrm{1}}}\HOLSymConst{..}\HOLBoundVar{e\sb{\mathrm{1}}} \HOLBoundVar{t} \HOLSymConst{+} \HOLBoundVar{a\sb{\mathrm{2}}}\HOLSymConst{..}\HOLBoundVar{e\sb{\mathrm{2}}} \HOLBoundVar{t})
\end{SaveVerbatim}
\newcommand{\HOLCongruenceTheoremsGSEQFour}{\UseVerbatim{HOLCongruenceTheoremsGSEQFour}}
\begin{SaveVerbatim}{HOLCongruenceTheoremsGSEQXXcases}
\HOLTokenTurnstile{} \HOLSymConst{\HOLTokenForall{}}\HOLBoundVar{a\sb{\mathrm{0}}}.
     \HOLConst{GSEQ} \HOLBoundVar{a\sb{\mathrm{0}}} \HOLSymConst{\HOLTokenEquiv{}}
     (\HOLBoundVar{a\sb{\mathrm{0}}} \HOLSymConst{=} (\HOLTokenLambda{}\HOLBoundVar{t}. \HOLBoundVar{t})) \HOLSymConst{\HOLTokenDisj{}} (\HOLSymConst{\HOLTokenExists{}}\HOLBoundVar{p}. \HOLBoundVar{a\sb{\mathrm{0}}} \HOLSymConst{=} (\HOLTokenLambda{}\HOLBoundVar{t}. \HOLBoundVar{p})) \HOLSymConst{\HOLTokenDisj{}}
     (\HOLSymConst{\HOLTokenExists{}}\HOLBoundVar{a} \HOLBoundVar{e}. (\HOLBoundVar{a\sb{\mathrm{0}}} \HOLSymConst{=} (\HOLTokenLambda{}\HOLBoundVar{t}. \HOLBoundVar{a}\HOLSymConst{..}\HOLBoundVar{e} \HOLBoundVar{t})) \HOLSymConst{\HOLTokenConj{}} \HOLConst{GSEQ} \HOLBoundVar{e}) \HOLSymConst{\HOLTokenDisj{}}
     \HOLSymConst{\HOLTokenExists{}}\HOLBoundVar{a\sb{\mathrm{1}}} \HOLBoundVar{a\sb{\mathrm{2}}} \HOLBoundVar{e\sb{\mathrm{1}}} \HOLBoundVar{e\sb{\mathrm{2}}}.
       (\HOLBoundVar{a\sb{\mathrm{0}}} \HOLSymConst{=} (\HOLTokenLambda{}\HOLBoundVar{t}. \HOLBoundVar{a\sb{\mathrm{1}}}\HOLSymConst{..}\HOLBoundVar{e\sb{\mathrm{1}}} \HOLBoundVar{t} \HOLSymConst{+} \HOLBoundVar{a\sb{\mathrm{2}}}\HOLSymConst{..}\HOLBoundVar{e\sb{\mathrm{2}}} \HOLBoundVar{t})) \HOLSymConst{\HOLTokenConj{}} \HOLConst{GSEQ} \HOLBoundVar{e\sb{\mathrm{1}}} \HOLSymConst{\HOLTokenConj{}} \HOLConst{GSEQ} \HOLBoundVar{e\sb{\mathrm{2}}}
\end{SaveVerbatim}
\newcommand{\HOLCongruenceTheoremsGSEQXXcases}{\UseVerbatim{HOLCongruenceTheoremsGSEQXXcases}}
\begin{SaveVerbatim}{HOLCongruenceTheoremsGSEQXXcombin}
\HOLTokenTurnstile{} \HOLSymConst{\HOLTokenForall{}}\HOLBoundVar{E}. \HOLConst{GSEQ} \HOLBoundVar{E} \HOLSymConst{\HOLTokenImp{}} \HOLSymConst{\HOLTokenForall{}}\HOLBoundVar{E\sp{\prime}}. \HOLConst{GSEQ} \HOLBoundVar{E\sp{\prime}} \HOLSymConst{\HOLTokenImp{}} \HOLConst{GSEQ} (\HOLBoundVar{E} \HOLConst{\HOLTokenCompose} \HOLBoundVar{E\sp{\prime}})
\end{SaveVerbatim}
\newcommand{\HOLCongruenceTheoremsGSEQXXcombin}{\UseVerbatim{HOLCongruenceTheoremsGSEQXXcombin}}
\begin{SaveVerbatim}{HOLCongruenceTheoremsGSEQXXind}
\HOLTokenTurnstile{} \HOLSymConst{\HOLTokenForall{}}\HOLBoundVar{GSEQ\sp{\prime}}.
     \HOLBoundVar{GSEQ\sp{\prime}} (\HOLTokenLambda{}\HOLBoundVar{t}. \HOLBoundVar{t}) \HOLSymConst{\HOLTokenConj{}} (\HOLSymConst{\HOLTokenForall{}}\HOLBoundVar{p}. \HOLBoundVar{GSEQ\sp{\prime}} (\HOLTokenLambda{}\HOLBoundVar{t}. \HOLBoundVar{p})) \HOLSymConst{\HOLTokenConj{}}
     (\HOLSymConst{\HOLTokenForall{}}\HOLBoundVar{a} \HOLBoundVar{e}. \HOLBoundVar{GSEQ\sp{\prime}} \HOLBoundVar{e} \HOLSymConst{\HOLTokenImp{}} \HOLBoundVar{GSEQ\sp{\prime}} (\HOLTokenLambda{}\HOLBoundVar{t}. \HOLBoundVar{a}\HOLSymConst{..}\HOLBoundVar{e} \HOLBoundVar{t})) \HOLSymConst{\HOLTokenConj{}}
     (\HOLSymConst{\HOLTokenForall{}}\HOLBoundVar{a\sb{\mathrm{1}}} \HOLBoundVar{a\sb{\mathrm{2}}} \HOLBoundVar{e\sb{\mathrm{1}}} \HOLBoundVar{e\sb{\mathrm{2}}}.
        \HOLBoundVar{GSEQ\sp{\prime}} \HOLBoundVar{e\sb{\mathrm{1}}} \HOLSymConst{\HOLTokenConj{}} \HOLBoundVar{GSEQ\sp{\prime}} \HOLBoundVar{e\sb{\mathrm{2}}} \HOLSymConst{\HOLTokenImp{}} \HOLBoundVar{GSEQ\sp{\prime}} (\HOLTokenLambda{}\HOLBoundVar{t}. \HOLBoundVar{a\sb{\mathrm{1}}}\HOLSymConst{..}\HOLBoundVar{e\sb{\mathrm{1}}} \HOLBoundVar{t} \HOLSymConst{+} \HOLBoundVar{a\sb{\mathrm{2}}}\HOLSymConst{..}\HOLBoundVar{e\sb{\mathrm{2}}} \HOLBoundVar{t})) \HOLSymConst{\HOLTokenImp{}}
     \HOLSymConst{\HOLTokenForall{}}\HOLBoundVar{a\sb{\mathrm{0}}}. \HOLConst{GSEQ} \HOLBoundVar{a\sb{\mathrm{0}}} \HOLSymConst{\HOLTokenImp{}} \HOLBoundVar{GSEQ\sp{\prime}} \HOLBoundVar{a\sb{\mathrm{0}}}
\end{SaveVerbatim}
\newcommand{\HOLCongruenceTheoremsGSEQXXind}{\UseVerbatim{HOLCongruenceTheoremsGSEQXXind}}
\begin{SaveVerbatim}{HOLCongruenceTheoremsGSEQXXISXXCONTEXT}
\HOLTokenTurnstile{} \HOLSymConst{\HOLTokenForall{}}\HOLBoundVar{e}. \HOLConst{GSEQ} \HOLBoundVar{e} \HOLSymConst{\HOLTokenImp{}} \HOLConst{CONTEXT} \HOLBoundVar{e}
\end{SaveVerbatim}
\newcommand{\HOLCongruenceTheoremsGSEQXXISXXCONTEXT}{\UseVerbatim{HOLCongruenceTheoremsGSEQXXISXXCONTEXT}}
\begin{SaveVerbatim}{HOLCongruenceTheoremsGSEQXXrules}
\HOLTokenTurnstile{} \HOLConst{GSEQ} (\HOLTokenLambda{}\HOLBoundVar{t}. \HOLBoundVar{t}) \HOLSymConst{\HOLTokenConj{}} (\HOLSymConst{\HOLTokenForall{}}\HOLBoundVar{p}. \HOLConst{GSEQ} (\HOLTokenLambda{}\HOLBoundVar{t}. \HOLBoundVar{p})) \HOLSymConst{\HOLTokenConj{}}
   (\HOLSymConst{\HOLTokenForall{}}\HOLBoundVar{a} \HOLBoundVar{e}. \HOLConst{GSEQ} \HOLBoundVar{e} \HOLSymConst{\HOLTokenImp{}} \HOLConst{GSEQ} (\HOLTokenLambda{}\HOLBoundVar{t}. \HOLBoundVar{a}\HOLSymConst{..}\HOLBoundVar{e} \HOLBoundVar{t})) \HOLSymConst{\HOLTokenConj{}}
   \HOLSymConst{\HOLTokenForall{}}\HOLBoundVar{a\sb{\mathrm{1}}} \HOLBoundVar{a\sb{\mathrm{2}}} \HOLBoundVar{e\sb{\mathrm{1}}} \HOLBoundVar{e\sb{\mathrm{2}}}.
     \HOLConst{GSEQ} \HOLBoundVar{e\sb{\mathrm{1}}} \HOLSymConst{\HOLTokenConj{}} \HOLConst{GSEQ} \HOLBoundVar{e\sb{\mathrm{2}}} \HOLSymConst{\HOLTokenImp{}} \HOLConst{GSEQ} (\HOLTokenLambda{}\HOLBoundVar{t}. \HOLBoundVar{a\sb{\mathrm{1}}}\HOLSymConst{..}\HOLBoundVar{e\sb{\mathrm{1}}} \HOLBoundVar{t} \HOLSymConst{+} \HOLBoundVar{a\sb{\mathrm{2}}}\HOLSymConst{..}\HOLBoundVar{e\sb{\mathrm{2}}} \HOLBoundVar{t})
\end{SaveVerbatim}
\newcommand{\HOLCongruenceTheoremsGSEQXXrules}{\UseVerbatim{HOLCongruenceTheoremsGSEQXXrules}}
\begin{SaveVerbatim}{HOLCongruenceTheoremsGSEQXXstrongind}
\HOLTokenTurnstile{} \HOLSymConst{\HOLTokenForall{}}\HOLBoundVar{GSEQ\sp{\prime}}.
     \HOLBoundVar{GSEQ\sp{\prime}} (\HOLTokenLambda{}\HOLBoundVar{t}. \HOLBoundVar{t}) \HOLSymConst{\HOLTokenConj{}} (\HOLSymConst{\HOLTokenForall{}}\HOLBoundVar{p}. \HOLBoundVar{GSEQ\sp{\prime}} (\HOLTokenLambda{}\HOLBoundVar{t}. \HOLBoundVar{p})) \HOLSymConst{\HOLTokenConj{}}
     (\HOLSymConst{\HOLTokenForall{}}\HOLBoundVar{a} \HOLBoundVar{e}. \HOLConst{GSEQ} \HOLBoundVar{e} \HOLSymConst{\HOLTokenConj{}} \HOLBoundVar{GSEQ\sp{\prime}} \HOLBoundVar{e} \HOLSymConst{\HOLTokenImp{}} \HOLBoundVar{GSEQ\sp{\prime}} (\HOLTokenLambda{}\HOLBoundVar{t}. \HOLBoundVar{a}\HOLSymConst{..}\HOLBoundVar{e} \HOLBoundVar{t})) \HOLSymConst{\HOLTokenConj{}}
     (\HOLSymConst{\HOLTokenForall{}}\HOLBoundVar{a\sb{\mathrm{1}}} \HOLBoundVar{a\sb{\mathrm{2}}} \HOLBoundVar{e\sb{\mathrm{1}}} \HOLBoundVar{e\sb{\mathrm{2}}}.
        \HOLConst{GSEQ} \HOLBoundVar{e\sb{\mathrm{1}}} \HOLSymConst{\HOLTokenConj{}} \HOLBoundVar{GSEQ\sp{\prime}} \HOLBoundVar{e\sb{\mathrm{1}}} \HOLSymConst{\HOLTokenConj{}} \HOLConst{GSEQ} \HOLBoundVar{e\sb{\mathrm{2}}} \HOLSymConst{\HOLTokenConj{}} \HOLBoundVar{GSEQ\sp{\prime}} \HOLBoundVar{e\sb{\mathrm{2}}} \HOLSymConst{\HOLTokenImp{}}
        \HOLBoundVar{GSEQ\sp{\prime}} (\HOLTokenLambda{}\HOLBoundVar{t}. \HOLBoundVar{a\sb{\mathrm{1}}}\HOLSymConst{..}\HOLBoundVar{e\sb{\mathrm{1}}} \HOLBoundVar{t} \HOLSymConst{+} \HOLBoundVar{a\sb{\mathrm{2}}}\HOLSymConst{..}\HOLBoundVar{e\sb{\mathrm{2}}} \HOLBoundVar{t})) \HOLSymConst{\HOLTokenImp{}}
     \HOLSymConst{\HOLTokenForall{}}\HOLBoundVar{a\sb{\mathrm{0}}}. \HOLConst{GSEQ} \HOLBoundVar{a\sb{\mathrm{0}}} \HOLSymConst{\HOLTokenImp{}} \HOLBoundVar{GSEQ\sp{\prime}} \HOLBoundVar{a\sb{\mathrm{0}}}
\end{SaveVerbatim}
\newcommand{\HOLCongruenceTheoremsGSEQXXstrongind}{\UseVerbatim{HOLCongruenceTheoremsGSEQXXstrongind}}
\begin{SaveVerbatim}{HOLCongruenceTheoremsOBSXXCONGRXXcongruence}
\HOLTokenTurnstile{} \HOLConst{congruence} \HOLConst{OBS_CONGR}
\end{SaveVerbatim}
\newcommand{\HOLCongruenceTheoremsOBSXXCONGRXXcongruence}{\UseVerbatim{HOLCongruenceTheoremsOBSXXCONGRXXcongruence}}
\begin{SaveVerbatim}{HOLCongruenceTheoremsOBSXXCONGRXXSUBSTXXCONTEXT}
\HOLTokenTurnstile{} \HOLSymConst{\HOLTokenForall{}}\HOLBoundVar{P} \HOLBoundVar{Q}. \HOLConst{OBS_CONGR} \HOLBoundVar{P} \HOLBoundVar{Q} \HOLSymConst{\HOLTokenImp{}} \HOLSymConst{\HOLTokenForall{}}\HOLBoundVar{E}. \HOLConst{CONTEXT} \HOLBoundVar{E} \HOLSymConst{\HOLTokenImp{}} \HOLConst{OBS_CONGR} (\HOLBoundVar{E} \HOLBoundVar{P}) (\HOLBoundVar{E} \HOLBoundVar{Q})
\end{SaveVerbatim}
\newcommand{\HOLCongruenceTheoremsOBSXXCONGRXXSUBSTXXCONTEXT}{\UseVerbatim{HOLCongruenceTheoremsOBSXXCONGRXXSUBSTXXCONTEXT}}
\begin{SaveVerbatim}{HOLCongruenceTheoremsOBSXXCONGRXXSUBSTXXSEQ}
\HOLTokenTurnstile{} \HOLSymConst{\HOLTokenForall{}}\HOLBoundVar{P} \HOLBoundVar{Q}. \HOLConst{OBS_CONGR} \HOLBoundVar{P} \HOLBoundVar{Q} \HOLSymConst{\HOLTokenImp{}} \HOLSymConst{\HOLTokenForall{}}\HOLBoundVar{E}. \HOLConst{SEQ} \HOLBoundVar{E} \HOLSymConst{\HOLTokenImp{}} \HOLConst{OBS_CONGR} (\HOLBoundVar{E} \HOLBoundVar{P}) (\HOLBoundVar{E} \HOLBoundVar{Q})
\end{SaveVerbatim}
\newcommand{\HOLCongruenceTheoremsOBSXXCONGRXXSUBSTXXSEQ}{\UseVerbatim{HOLCongruenceTheoremsOBSXXCONGRXXSUBSTXXSEQ}}
\begin{SaveVerbatim}{HOLCongruenceTheoremsOHXXCONTEXTOne}
\HOLTokenTurnstile{} \HOLConst{OH_CONTEXT} (\HOLTokenLambda{}\HOLBoundVar{t}. \HOLBoundVar{t})
\end{SaveVerbatim}
\newcommand{\HOLCongruenceTheoremsOHXXCONTEXTOne}{\UseVerbatim{HOLCongruenceTheoremsOHXXCONTEXTOne}}
\begin{SaveVerbatim}{HOLCongruenceTheoremsOHXXCONTEXTTwo}
\HOLTokenTurnstile{} \HOLSymConst{\HOLTokenForall{}}\HOLBoundVar{a} \HOLBoundVar{c}. \HOLConst{OH_CONTEXT} \HOLBoundVar{c} \HOLSymConst{\HOLTokenImp{}} \HOLConst{OH_CONTEXT} (\HOLTokenLambda{}\HOLBoundVar{t}. \HOLBoundVar{a}\HOLSymConst{..}\HOLBoundVar{c} \HOLBoundVar{t})
\end{SaveVerbatim}
\newcommand{\HOLCongruenceTheoremsOHXXCONTEXTTwo}{\UseVerbatim{HOLCongruenceTheoremsOHXXCONTEXTTwo}}
\begin{SaveVerbatim}{HOLCongruenceTheoremsOHXXCONTEXTThree}
\HOLTokenTurnstile{} \HOLSymConst{\HOLTokenForall{}}\HOLBoundVar{x} \HOLBoundVar{c}. \HOLConst{OH_CONTEXT} \HOLBoundVar{c} \HOLSymConst{\HOLTokenImp{}} \HOLConst{OH_CONTEXT} (\HOLTokenLambda{}\HOLBoundVar{t}. \HOLBoundVar{c} \HOLBoundVar{t} \HOLSymConst{+} \HOLBoundVar{x})
\end{SaveVerbatim}
\newcommand{\HOLCongruenceTheoremsOHXXCONTEXTThree}{\UseVerbatim{HOLCongruenceTheoremsOHXXCONTEXTThree}}
\begin{SaveVerbatim}{HOLCongruenceTheoremsOHXXCONTEXTFour}
\HOLTokenTurnstile{} \HOLSymConst{\HOLTokenForall{}}\HOLBoundVar{x} \HOLBoundVar{c}. \HOLConst{OH_CONTEXT} \HOLBoundVar{c} \HOLSymConst{\HOLTokenImp{}} \HOLConst{OH_CONTEXT} (\HOLTokenLambda{}\HOLBoundVar{t}. \HOLBoundVar{x} \HOLSymConst{+} \HOLBoundVar{c} \HOLBoundVar{t})
\end{SaveVerbatim}
\newcommand{\HOLCongruenceTheoremsOHXXCONTEXTFour}{\UseVerbatim{HOLCongruenceTheoremsOHXXCONTEXTFour}}
\begin{SaveVerbatim}{HOLCongruenceTheoremsOHXXCONTEXTFive}
\HOLTokenTurnstile{} \HOLSymConst{\HOLTokenForall{}}\HOLBoundVar{x} \HOLBoundVar{c}. \HOLConst{OH_CONTEXT} \HOLBoundVar{c} \HOLSymConst{\HOLTokenImp{}} \HOLConst{OH_CONTEXT} (\HOLTokenLambda{}\HOLBoundVar{t}. \HOLBoundVar{c} \HOLBoundVar{t} \HOLSymConst{\ensuremath{\parallel}} \HOLBoundVar{x})
\end{SaveVerbatim}
\newcommand{\HOLCongruenceTheoremsOHXXCONTEXTFive}{\UseVerbatim{HOLCongruenceTheoremsOHXXCONTEXTFive}}
\begin{SaveVerbatim}{HOLCongruenceTheoremsOHXXCONTEXTSix}
\HOLTokenTurnstile{} \HOLSymConst{\HOLTokenForall{}}\HOLBoundVar{x} \HOLBoundVar{c}. \HOLConst{OH_CONTEXT} \HOLBoundVar{c} \HOLSymConst{\HOLTokenImp{}} \HOLConst{OH_CONTEXT} (\HOLTokenLambda{}\HOLBoundVar{t}. \HOLBoundVar{x} \HOLSymConst{\ensuremath{\parallel}} \HOLBoundVar{c} \HOLBoundVar{t})
\end{SaveVerbatim}
\newcommand{\HOLCongruenceTheoremsOHXXCONTEXTSix}{\UseVerbatim{HOLCongruenceTheoremsOHXXCONTEXTSix}}
\begin{SaveVerbatim}{HOLCongruenceTheoremsOHXXCONTEXTSeven}
\HOLTokenTurnstile{} \HOLSymConst{\HOLTokenForall{}}\HOLBoundVar{L} \HOLBoundVar{c}. \HOLConst{OH_CONTEXT} \HOLBoundVar{c} \HOLSymConst{\HOLTokenImp{}} \HOLConst{OH_CONTEXT} (\HOLTokenLambda{}\HOLBoundVar{t}. \HOLConst{\ensuremath{\nu}} \HOLBoundVar{L} (\HOLBoundVar{c} \HOLBoundVar{t}))
\end{SaveVerbatim}
\newcommand{\HOLCongruenceTheoremsOHXXCONTEXTSeven}{\UseVerbatim{HOLCongruenceTheoremsOHXXCONTEXTSeven}}
\begin{SaveVerbatim}{HOLCongruenceTheoremsOHXXCONTEXTEight}
\HOLTokenTurnstile{} \HOLSymConst{\HOLTokenForall{}}\HOLBoundVar{rf} \HOLBoundVar{c}. \HOLConst{OH_CONTEXT} \HOLBoundVar{c} \HOLSymConst{\HOLTokenImp{}} \HOLConst{OH_CONTEXT} (\HOLTokenLambda{}\HOLBoundVar{t}. \HOLConst{relab} (\HOLBoundVar{c} \HOLBoundVar{t}) \HOLBoundVar{rf})
\end{SaveVerbatim}
\newcommand{\HOLCongruenceTheoremsOHXXCONTEXTEight}{\UseVerbatim{HOLCongruenceTheoremsOHXXCONTEXTEight}}
\begin{SaveVerbatim}{HOLCongruenceTheoremsOHXXCONTEXTXXcases}
\HOLTokenTurnstile{} \HOLSymConst{\HOLTokenForall{}}\HOLBoundVar{a\sb{\mathrm{0}}}.
     \HOLConst{OH_CONTEXT} \HOLBoundVar{a\sb{\mathrm{0}}} \HOLSymConst{\HOLTokenEquiv{}}
     (\HOLBoundVar{a\sb{\mathrm{0}}} \HOLSymConst{=} (\HOLTokenLambda{}\HOLBoundVar{t}. \HOLBoundVar{t})) \HOLSymConst{\HOLTokenDisj{}}
     (\HOLSymConst{\HOLTokenExists{}}\HOLBoundVar{a} \HOLBoundVar{c}. (\HOLBoundVar{a\sb{\mathrm{0}}} \HOLSymConst{=} (\HOLTokenLambda{}\HOLBoundVar{t}. \HOLBoundVar{a}\HOLSymConst{..}\HOLBoundVar{c} \HOLBoundVar{t})) \HOLSymConst{\HOLTokenConj{}} \HOLConst{OH_CONTEXT} \HOLBoundVar{c}) \HOLSymConst{\HOLTokenDisj{}}
     (\HOLSymConst{\HOLTokenExists{}}\HOLBoundVar{x} \HOLBoundVar{c}. (\HOLBoundVar{a\sb{\mathrm{0}}} \HOLSymConst{=} (\HOLTokenLambda{}\HOLBoundVar{t}. \HOLBoundVar{c} \HOLBoundVar{t} \HOLSymConst{+} \HOLBoundVar{x})) \HOLSymConst{\HOLTokenConj{}} \HOLConst{OH_CONTEXT} \HOLBoundVar{c}) \HOLSymConst{\HOLTokenDisj{}}
     (\HOLSymConst{\HOLTokenExists{}}\HOLBoundVar{x} \HOLBoundVar{c}. (\HOLBoundVar{a\sb{\mathrm{0}}} \HOLSymConst{=} (\HOLTokenLambda{}\HOLBoundVar{t}. \HOLBoundVar{x} \HOLSymConst{+} \HOLBoundVar{c} \HOLBoundVar{t})) \HOLSymConst{\HOLTokenConj{}} \HOLConst{OH_CONTEXT} \HOLBoundVar{c}) \HOLSymConst{\HOLTokenDisj{}}
     (\HOLSymConst{\HOLTokenExists{}}\HOLBoundVar{x} \HOLBoundVar{c}. (\HOLBoundVar{a\sb{\mathrm{0}}} \HOLSymConst{=} (\HOLTokenLambda{}\HOLBoundVar{t}. \HOLBoundVar{c} \HOLBoundVar{t} \HOLSymConst{\ensuremath{\parallel}} \HOLBoundVar{x})) \HOLSymConst{\HOLTokenConj{}} \HOLConst{OH_CONTEXT} \HOLBoundVar{c}) \HOLSymConst{\HOLTokenDisj{}}
     (\HOLSymConst{\HOLTokenExists{}}\HOLBoundVar{x} \HOLBoundVar{c}. (\HOLBoundVar{a\sb{\mathrm{0}}} \HOLSymConst{=} (\HOLTokenLambda{}\HOLBoundVar{t}. \HOLBoundVar{x} \HOLSymConst{\ensuremath{\parallel}} \HOLBoundVar{c} \HOLBoundVar{t})) \HOLSymConst{\HOLTokenConj{}} \HOLConst{OH_CONTEXT} \HOLBoundVar{c}) \HOLSymConst{\HOLTokenDisj{}}
     (\HOLSymConst{\HOLTokenExists{}}\HOLBoundVar{L} \HOLBoundVar{c}. (\HOLBoundVar{a\sb{\mathrm{0}}} \HOLSymConst{=} (\HOLTokenLambda{}\HOLBoundVar{t}. \HOLConst{\ensuremath{\nu}} \HOLBoundVar{L} (\HOLBoundVar{c} \HOLBoundVar{t}))) \HOLSymConst{\HOLTokenConj{}} \HOLConst{OH_CONTEXT} \HOLBoundVar{c}) \HOLSymConst{\HOLTokenDisj{}}
     \HOLSymConst{\HOLTokenExists{}}\HOLBoundVar{rf} \HOLBoundVar{c}. (\HOLBoundVar{a\sb{\mathrm{0}}} \HOLSymConst{=} (\HOLTokenLambda{}\HOLBoundVar{t}. \HOLConst{relab} (\HOLBoundVar{c} \HOLBoundVar{t}) \HOLBoundVar{rf})) \HOLSymConst{\HOLTokenConj{}} \HOLConst{OH_CONTEXT} \HOLBoundVar{c}
\end{SaveVerbatim}
\newcommand{\HOLCongruenceTheoremsOHXXCONTEXTXXcases}{\UseVerbatim{HOLCongruenceTheoremsOHXXCONTEXTXXcases}}
\begin{SaveVerbatim}{HOLCongruenceTheoremsOHXXCONTEXTXXcombin}
\HOLTokenTurnstile{} \HOLSymConst{\HOLTokenForall{}}\HOLBoundVar{c\sb{\mathrm{1}}} \HOLBoundVar{c\sb{\mathrm{2}}}. \HOLConst{OH_CONTEXT} \HOLBoundVar{c\sb{\mathrm{1}}} \HOLSymConst{\HOLTokenConj{}} \HOLConst{OH_CONTEXT} \HOLBoundVar{c\sb{\mathrm{2}}} \HOLSymConst{\HOLTokenImp{}} \HOLConst{OH_CONTEXT} (\HOLBoundVar{c\sb{\mathrm{1}}} \HOLConst{\HOLTokenCompose} \HOLBoundVar{c\sb{\mathrm{2}}})
\end{SaveVerbatim}
\newcommand{\HOLCongruenceTheoremsOHXXCONTEXTXXcombin}{\UseVerbatim{HOLCongruenceTheoremsOHXXCONTEXTXXcombin}}
\begin{SaveVerbatim}{HOLCongruenceTheoremsOHXXCONTEXTXXind}
\HOLTokenTurnstile{} \HOLSymConst{\HOLTokenForall{}}\HOLBoundVar{OH\HOLTokenUnderscore{}CONTEXT\sp{\prime}}.
     \HOLBoundVar{OH\HOLTokenUnderscore{}CONTEXT\sp{\prime}} (\HOLTokenLambda{}\HOLBoundVar{t}. \HOLBoundVar{t}) \HOLSymConst{\HOLTokenConj{}}
     (\HOLSymConst{\HOLTokenForall{}}\HOLBoundVar{a} \HOLBoundVar{c}. \HOLBoundVar{OH\HOLTokenUnderscore{}CONTEXT\sp{\prime}} \HOLBoundVar{c} \HOLSymConst{\HOLTokenImp{}} \HOLBoundVar{OH\HOLTokenUnderscore{}CONTEXT\sp{\prime}} (\HOLTokenLambda{}\HOLBoundVar{t}. \HOLBoundVar{a}\HOLSymConst{..}\HOLBoundVar{c} \HOLBoundVar{t})) \HOLSymConst{\HOLTokenConj{}}
     (\HOLSymConst{\HOLTokenForall{}}\HOLBoundVar{x} \HOLBoundVar{c}. \HOLBoundVar{OH\HOLTokenUnderscore{}CONTEXT\sp{\prime}} \HOLBoundVar{c} \HOLSymConst{\HOLTokenImp{}} \HOLBoundVar{OH\HOLTokenUnderscore{}CONTEXT\sp{\prime}} (\HOLTokenLambda{}\HOLBoundVar{t}. \HOLBoundVar{c} \HOLBoundVar{t} \HOLSymConst{+} \HOLBoundVar{x})) \HOLSymConst{\HOLTokenConj{}}
     (\HOLSymConst{\HOLTokenForall{}}\HOLBoundVar{x} \HOLBoundVar{c}. \HOLBoundVar{OH\HOLTokenUnderscore{}CONTEXT\sp{\prime}} \HOLBoundVar{c} \HOLSymConst{\HOLTokenImp{}} \HOLBoundVar{OH\HOLTokenUnderscore{}CONTEXT\sp{\prime}} (\HOLTokenLambda{}\HOLBoundVar{t}. \HOLBoundVar{x} \HOLSymConst{+} \HOLBoundVar{c} \HOLBoundVar{t})) \HOLSymConst{\HOLTokenConj{}}
     (\HOLSymConst{\HOLTokenForall{}}\HOLBoundVar{x} \HOLBoundVar{c}. \HOLBoundVar{OH\HOLTokenUnderscore{}CONTEXT\sp{\prime}} \HOLBoundVar{c} \HOLSymConst{\HOLTokenImp{}} \HOLBoundVar{OH\HOLTokenUnderscore{}CONTEXT\sp{\prime}} (\HOLTokenLambda{}\HOLBoundVar{t}. \HOLBoundVar{c} \HOLBoundVar{t} \HOLSymConst{\ensuremath{\parallel}} \HOLBoundVar{x})) \HOLSymConst{\HOLTokenConj{}}
     (\HOLSymConst{\HOLTokenForall{}}\HOLBoundVar{x} \HOLBoundVar{c}. \HOLBoundVar{OH\HOLTokenUnderscore{}CONTEXT\sp{\prime}} \HOLBoundVar{c} \HOLSymConst{\HOLTokenImp{}} \HOLBoundVar{OH\HOLTokenUnderscore{}CONTEXT\sp{\prime}} (\HOLTokenLambda{}\HOLBoundVar{t}. \HOLBoundVar{x} \HOLSymConst{\ensuremath{\parallel}} \HOLBoundVar{c} \HOLBoundVar{t})) \HOLSymConst{\HOLTokenConj{}}
     (\HOLSymConst{\HOLTokenForall{}}\HOLBoundVar{L} \HOLBoundVar{c}. \HOLBoundVar{OH\HOLTokenUnderscore{}CONTEXT\sp{\prime}} \HOLBoundVar{c} \HOLSymConst{\HOLTokenImp{}} \HOLBoundVar{OH\HOLTokenUnderscore{}CONTEXT\sp{\prime}} (\HOLTokenLambda{}\HOLBoundVar{t}. \HOLConst{\ensuremath{\nu}} \HOLBoundVar{L} (\HOLBoundVar{c} \HOLBoundVar{t}))) \HOLSymConst{\HOLTokenConj{}}
     (\HOLSymConst{\HOLTokenForall{}}\HOLBoundVar{rf} \HOLBoundVar{c}. \HOLBoundVar{OH\HOLTokenUnderscore{}CONTEXT\sp{\prime}} \HOLBoundVar{c} \HOLSymConst{\HOLTokenImp{}} \HOLBoundVar{OH\HOLTokenUnderscore{}CONTEXT\sp{\prime}} (\HOLTokenLambda{}\HOLBoundVar{t}. \HOLConst{relab} (\HOLBoundVar{c} \HOLBoundVar{t}) \HOLBoundVar{rf})) \HOLSymConst{\HOLTokenImp{}}
     \HOLSymConst{\HOLTokenForall{}}\HOLBoundVar{a\sb{\mathrm{0}}}. \HOLConst{OH_CONTEXT} \HOLBoundVar{a\sb{\mathrm{0}}} \HOLSymConst{\HOLTokenImp{}} \HOLBoundVar{OH\HOLTokenUnderscore{}CONTEXT\sp{\prime}} \HOLBoundVar{a\sb{\mathrm{0}}}
\end{SaveVerbatim}
\newcommand{\HOLCongruenceTheoremsOHXXCONTEXTXXind}{\UseVerbatim{HOLCongruenceTheoremsOHXXCONTEXTXXind}}
\begin{SaveVerbatim}{HOLCongruenceTheoremsOHXXCONTEXTXXISXXCONTEXT}
\HOLTokenTurnstile{} \HOLSymConst{\HOLTokenForall{}}\HOLBoundVar{c}. \HOLConst{OH_CONTEXT} \HOLBoundVar{c} \HOLSymConst{\HOLTokenImp{}} \HOLConst{CONTEXT} \HOLBoundVar{c}
\end{SaveVerbatim}
\newcommand{\HOLCongruenceTheoremsOHXXCONTEXTXXISXXCONTEXT}{\UseVerbatim{HOLCongruenceTheoremsOHXXCONTEXTXXISXXCONTEXT}}
\begin{SaveVerbatim}{HOLCongruenceTheoremsOHXXCONTEXTXXrules}
\HOLTokenTurnstile{} \HOLConst{OH_CONTEXT} (\HOLTokenLambda{}\HOLBoundVar{t}. \HOLBoundVar{t}) \HOLSymConst{\HOLTokenConj{}}
   (\HOLSymConst{\HOLTokenForall{}}\HOLBoundVar{a} \HOLBoundVar{c}. \HOLConst{OH_CONTEXT} \HOLBoundVar{c} \HOLSymConst{\HOLTokenImp{}} \HOLConst{OH_CONTEXT} (\HOLTokenLambda{}\HOLBoundVar{t}. \HOLBoundVar{a}\HOLSymConst{..}\HOLBoundVar{c} \HOLBoundVar{t})) \HOLSymConst{\HOLTokenConj{}}
   (\HOLSymConst{\HOLTokenForall{}}\HOLBoundVar{x} \HOLBoundVar{c}. \HOLConst{OH_CONTEXT} \HOLBoundVar{c} \HOLSymConst{\HOLTokenImp{}} \HOLConst{OH_CONTEXT} (\HOLTokenLambda{}\HOLBoundVar{t}. \HOLBoundVar{c} \HOLBoundVar{t} \HOLSymConst{+} \HOLBoundVar{x})) \HOLSymConst{\HOLTokenConj{}}
   (\HOLSymConst{\HOLTokenForall{}}\HOLBoundVar{x} \HOLBoundVar{c}. \HOLConst{OH_CONTEXT} \HOLBoundVar{c} \HOLSymConst{\HOLTokenImp{}} \HOLConst{OH_CONTEXT} (\HOLTokenLambda{}\HOLBoundVar{t}. \HOLBoundVar{x} \HOLSymConst{+} \HOLBoundVar{c} \HOLBoundVar{t})) \HOLSymConst{\HOLTokenConj{}}
   (\HOLSymConst{\HOLTokenForall{}}\HOLBoundVar{x} \HOLBoundVar{c}. \HOLConst{OH_CONTEXT} \HOLBoundVar{c} \HOLSymConst{\HOLTokenImp{}} \HOLConst{OH_CONTEXT} (\HOLTokenLambda{}\HOLBoundVar{t}. \HOLBoundVar{c} \HOLBoundVar{t} \HOLSymConst{\ensuremath{\parallel}} \HOLBoundVar{x})) \HOLSymConst{\HOLTokenConj{}}
   (\HOLSymConst{\HOLTokenForall{}}\HOLBoundVar{x} \HOLBoundVar{c}. \HOLConst{OH_CONTEXT} \HOLBoundVar{c} \HOLSymConst{\HOLTokenImp{}} \HOLConst{OH_CONTEXT} (\HOLTokenLambda{}\HOLBoundVar{t}. \HOLBoundVar{x} \HOLSymConst{\ensuremath{\parallel}} \HOLBoundVar{c} \HOLBoundVar{t})) \HOLSymConst{\HOLTokenConj{}}
   (\HOLSymConst{\HOLTokenForall{}}\HOLBoundVar{L} \HOLBoundVar{c}. \HOLConst{OH_CONTEXT} \HOLBoundVar{c} \HOLSymConst{\HOLTokenImp{}} \HOLConst{OH_CONTEXT} (\HOLTokenLambda{}\HOLBoundVar{t}. \HOLConst{\ensuremath{\nu}} \HOLBoundVar{L} (\HOLBoundVar{c} \HOLBoundVar{t}))) \HOLSymConst{\HOLTokenConj{}}
   \HOLSymConst{\HOLTokenForall{}}\HOLBoundVar{rf} \HOLBoundVar{c}. \HOLConst{OH_CONTEXT} \HOLBoundVar{c} \HOLSymConst{\HOLTokenImp{}} \HOLConst{OH_CONTEXT} (\HOLTokenLambda{}\HOLBoundVar{t}. \HOLConst{relab} (\HOLBoundVar{c} \HOLBoundVar{t}) \HOLBoundVar{rf})
\end{SaveVerbatim}
\newcommand{\HOLCongruenceTheoremsOHXXCONTEXTXXrules}{\UseVerbatim{HOLCongruenceTheoremsOHXXCONTEXTXXrules}}
\begin{SaveVerbatim}{HOLCongruenceTheoremsOHXXCONTEXTXXstrongind}
\HOLTokenTurnstile{} \HOLSymConst{\HOLTokenForall{}}\HOLBoundVar{OH\HOLTokenUnderscore{}CONTEXT\sp{\prime}}.
     \HOLBoundVar{OH\HOLTokenUnderscore{}CONTEXT\sp{\prime}} (\HOLTokenLambda{}\HOLBoundVar{t}. \HOLBoundVar{t}) \HOLSymConst{\HOLTokenConj{}}
     (\HOLSymConst{\HOLTokenForall{}}\HOLBoundVar{a} \HOLBoundVar{c}.
        \HOLConst{OH_CONTEXT} \HOLBoundVar{c} \HOLSymConst{\HOLTokenConj{}} \HOLBoundVar{OH\HOLTokenUnderscore{}CONTEXT\sp{\prime}} \HOLBoundVar{c} \HOLSymConst{\HOLTokenImp{}}
        \HOLBoundVar{OH\HOLTokenUnderscore{}CONTEXT\sp{\prime}} (\HOLTokenLambda{}\HOLBoundVar{t}. \HOLBoundVar{a}\HOLSymConst{..}\HOLBoundVar{c} \HOLBoundVar{t})) \HOLSymConst{\HOLTokenConj{}}
     (\HOLSymConst{\HOLTokenForall{}}\HOLBoundVar{x} \HOLBoundVar{c}.
        \HOLConst{OH_CONTEXT} \HOLBoundVar{c} \HOLSymConst{\HOLTokenConj{}} \HOLBoundVar{OH\HOLTokenUnderscore{}CONTEXT\sp{\prime}} \HOLBoundVar{c} \HOLSymConst{\HOLTokenImp{}}
        \HOLBoundVar{OH\HOLTokenUnderscore{}CONTEXT\sp{\prime}} (\HOLTokenLambda{}\HOLBoundVar{t}. \HOLBoundVar{c} \HOLBoundVar{t} \HOLSymConst{+} \HOLBoundVar{x})) \HOLSymConst{\HOLTokenConj{}}
     (\HOLSymConst{\HOLTokenForall{}}\HOLBoundVar{x} \HOLBoundVar{c}.
        \HOLConst{OH_CONTEXT} \HOLBoundVar{c} \HOLSymConst{\HOLTokenConj{}} \HOLBoundVar{OH\HOLTokenUnderscore{}CONTEXT\sp{\prime}} \HOLBoundVar{c} \HOLSymConst{\HOLTokenImp{}}
        \HOLBoundVar{OH\HOLTokenUnderscore{}CONTEXT\sp{\prime}} (\HOLTokenLambda{}\HOLBoundVar{t}. \HOLBoundVar{x} \HOLSymConst{+} \HOLBoundVar{c} \HOLBoundVar{t})) \HOLSymConst{\HOLTokenConj{}}
     (\HOLSymConst{\HOLTokenForall{}}\HOLBoundVar{x} \HOLBoundVar{c}.
        \HOLConst{OH_CONTEXT} \HOLBoundVar{c} \HOLSymConst{\HOLTokenConj{}} \HOLBoundVar{OH\HOLTokenUnderscore{}CONTEXT\sp{\prime}} \HOLBoundVar{c} \HOLSymConst{\HOLTokenImp{}}
        \HOLBoundVar{OH\HOLTokenUnderscore{}CONTEXT\sp{\prime}} (\HOLTokenLambda{}\HOLBoundVar{t}. \HOLBoundVar{c} \HOLBoundVar{t} \HOLSymConst{\ensuremath{\parallel}} \HOLBoundVar{x})) \HOLSymConst{\HOLTokenConj{}}
     (\HOLSymConst{\HOLTokenForall{}}\HOLBoundVar{x} \HOLBoundVar{c}.
        \HOLConst{OH_CONTEXT} \HOLBoundVar{c} \HOLSymConst{\HOLTokenConj{}} \HOLBoundVar{OH\HOLTokenUnderscore{}CONTEXT\sp{\prime}} \HOLBoundVar{c} \HOLSymConst{\HOLTokenImp{}}
        \HOLBoundVar{OH\HOLTokenUnderscore{}CONTEXT\sp{\prime}} (\HOLTokenLambda{}\HOLBoundVar{t}. \HOLBoundVar{x} \HOLSymConst{\ensuremath{\parallel}} \HOLBoundVar{c} \HOLBoundVar{t})) \HOLSymConst{\HOLTokenConj{}}
     (\HOLSymConst{\HOLTokenForall{}}\HOLBoundVar{L} \HOLBoundVar{c}.
        \HOLConst{OH_CONTEXT} \HOLBoundVar{c} \HOLSymConst{\HOLTokenConj{}} \HOLBoundVar{OH\HOLTokenUnderscore{}CONTEXT\sp{\prime}} \HOLBoundVar{c} \HOLSymConst{\HOLTokenImp{}}
        \HOLBoundVar{OH\HOLTokenUnderscore{}CONTEXT\sp{\prime}} (\HOLTokenLambda{}\HOLBoundVar{t}. \HOLConst{\ensuremath{\nu}} \HOLBoundVar{L} (\HOLBoundVar{c} \HOLBoundVar{t}))) \HOLSymConst{\HOLTokenConj{}}
     (\HOLSymConst{\HOLTokenForall{}}\HOLBoundVar{rf} \HOLBoundVar{c}.
        \HOLConst{OH_CONTEXT} \HOLBoundVar{c} \HOLSymConst{\HOLTokenConj{}} \HOLBoundVar{OH\HOLTokenUnderscore{}CONTEXT\sp{\prime}} \HOLBoundVar{c} \HOLSymConst{\HOLTokenImp{}}
        \HOLBoundVar{OH\HOLTokenUnderscore{}CONTEXT\sp{\prime}} (\HOLTokenLambda{}\HOLBoundVar{t}. \HOLConst{relab} (\HOLBoundVar{c} \HOLBoundVar{t}) \HOLBoundVar{rf})) \HOLSymConst{\HOLTokenImp{}}
     \HOLSymConst{\HOLTokenForall{}}\HOLBoundVar{a\sb{\mathrm{0}}}. \HOLConst{OH_CONTEXT} \HOLBoundVar{a\sb{\mathrm{0}}} \HOLSymConst{\HOLTokenImp{}} \HOLBoundVar{OH\HOLTokenUnderscore{}CONTEXT\sp{\prime}} \HOLBoundVar{a\sb{\mathrm{0}}}
\end{SaveVerbatim}
\newcommand{\HOLCongruenceTheoremsOHXXCONTEXTXXstrongind}{\UseVerbatim{HOLCongruenceTheoremsOHXXCONTEXTXXstrongind}}
\begin{SaveVerbatim}{HOLCongruenceTheoremsSEQOne}
\HOLTokenTurnstile{} \HOLConst{SEQ} (\HOLTokenLambda{}\HOLBoundVar{t}. \HOLBoundVar{t})
\end{SaveVerbatim}
\newcommand{\HOLCongruenceTheoremsSEQOne}{\UseVerbatim{HOLCongruenceTheoremsSEQOne}}
\begin{SaveVerbatim}{HOLCongruenceTheoremsSEQTwo}
\HOLTokenTurnstile{} \HOLSymConst{\HOLTokenForall{}}\HOLBoundVar{p}. \HOLConst{SEQ} (\HOLTokenLambda{}\HOLBoundVar{t}. \HOLBoundVar{p})
\end{SaveVerbatim}
\newcommand{\HOLCongruenceTheoremsSEQTwo}{\UseVerbatim{HOLCongruenceTheoremsSEQTwo}}
\begin{SaveVerbatim}{HOLCongruenceTheoremsSEQThree}
\HOLTokenTurnstile{} \HOLSymConst{\HOLTokenForall{}}\HOLBoundVar{a} \HOLBoundVar{e}. \HOLConst{SEQ} \HOLBoundVar{e} \HOLSymConst{\HOLTokenImp{}} \HOLConst{SEQ} (\HOLTokenLambda{}\HOLBoundVar{t}. \HOLBoundVar{a}\HOLSymConst{..}\HOLBoundVar{e} \HOLBoundVar{t})
\end{SaveVerbatim}
\newcommand{\HOLCongruenceTheoremsSEQThree}{\UseVerbatim{HOLCongruenceTheoremsSEQThree}}
\begin{SaveVerbatim}{HOLCongruenceTheoremsSEQThreea}
\HOLTokenTurnstile{} \HOLSymConst{\HOLTokenForall{}}\HOLBoundVar{a}. \HOLConst{SEQ} (\HOLTokenLambda{}\HOLBoundVar{t}. \HOLBoundVar{a}\HOLSymConst{..}\HOLBoundVar{t})
\end{SaveVerbatim}
\newcommand{\HOLCongruenceTheoremsSEQThreea}{\UseVerbatim{HOLCongruenceTheoremsSEQThreea}}
\begin{SaveVerbatim}{HOLCongruenceTheoremsSEQFour}
\HOLTokenTurnstile{} \HOLSymConst{\HOLTokenForall{}}\HOLBoundVar{e\sb{\mathrm{1}}} \HOLBoundVar{e\sb{\mathrm{2}}}. \HOLConst{SEQ} \HOLBoundVar{e\sb{\mathrm{1}}} \HOLSymConst{\HOLTokenConj{}} \HOLConst{SEQ} \HOLBoundVar{e\sb{\mathrm{2}}} \HOLSymConst{\HOLTokenImp{}} \HOLConst{SEQ} (\HOLTokenLambda{}\HOLBoundVar{t}. \HOLBoundVar{e\sb{\mathrm{1}}} \HOLBoundVar{t} \HOLSymConst{+} \HOLBoundVar{e\sb{\mathrm{2}}} \HOLBoundVar{t})
\end{SaveVerbatim}
\newcommand{\HOLCongruenceTheoremsSEQFour}{\UseVerbatim{HOLCongruenceTheoremsSEQFour}}
\begin{SaveVerbatim}{HOLCongruenceTheoremsSEQXXcases}
\HOLTokenTurnstile{} \HOLSymConst{\HOLTokenForall{}}\HOLBoundVar{a\sb{\mathrm{0}}}.
     \HOLConst{SEQ} \HOLBoundVar{a\sb{\mathrm{0}}} \HOLSymConst{\HOLTokenEquiv{}}
     (\HOLBoundVar{a\sb{\mathrm{0}}} \HOLSymConst{=} (\HOLTokenLambda{}\HOLBoundVar{t}. \HOLBoundVar{t})) \HOLSymConst{\HOLTokenDisj{}} (\HOLSymConst{\HOLTokenExists{}}\HOLBoundVar{p}. \HOLBoundVar{a\sb{\mathrm{0}}} \HOLSymConst{=} (\HOLTokenLambda{}\HOLBoundVar{t}. \HOLBoundVar{p})) \HOLSymConst{\HOLTokenDisj{}}
     (\HOLSymConst{\HOLTokenExists{}}\HOLBoundVar{a} \HOLBoundVar{e}. (\HOLBoundVar{a\sb{\mathrm{0}}} \HOLSymConst{=} (\HOLTokenLambda{}\HOLBoundVar{t}. \HOLBoundVar{a}\HOLSymConst{..}\HOLBoundVar{e} \HOLBoundVar{t})) \HOLSymConst{\HOLTokenConj{}} \HOLConst{SEQ} \HOLBoundVar{e}) \HOLSymConst{\HOLTokenDisj{}}
     \HOLSymConst{\HOLTokenExists{}}\HOLBoundVar{e\sb{\mathrm{1}}} \HOLBoundVar{e\sb{\mathrm{2}}}. (\HOLBoundVar{a\sb{\mathrm{0}}} \HOLSymConst{=} (\HOLTokenLambda{}\HOLBoundVar{t}. \HOLBoundVar{e\sb{\mathrm{1}}} \HOLBoundVar{t} \HOLSymConst{+} \HOLBoundVar{e\sb{\mathrm{2}}} \HOLBoundVar{t})) \HOLSymConst{\HOLTokenConj{}} \HOLConst{SEQ} \HOLBoundVar{e\sb{\mathrm{1}}} \HOLSymConst{\HOLTokenConj{}} \HOLConst{SEQ} \HOLBoundVar{e\sb{\mathrm{2}}}
\end{SaveVerbatim}
\newcommand{\HOLCongruenceTheoremsSEQXXcases}{\UseVerbatim{HOLCongruenceTheoremsSEQXXcases}}
\begin{SaveVerbatim}{HOLCongruenceTheoremsSEQXXcombin}
\HOLTokenTurnstile{} \HOLSymConst{\HOLTokenForall{}}\HOLBoundVar{E}. \HOLConst{SEQ} \HOLBoundVar{E} \HOLSymConst{\HOLTokenImp{}} \HOLSymConst{\HOLTokenForall{}}\HOLBoundVar{E\sp{\prime}}. \HOLConst{SEQ} \HOLBoundVar{E\sp{\prime}} \HOLSymConst{\HOLTokenImp{}} \HOLConst{SEQ} (\HOLBoundVar{E} \HOLConst{\HOLTokenCompose} \HOLBoundVar{E\sp{\prime}})
\end{SaveVerbatim}
\newcommand{\HOLCongruenceTheoremsSEQXXcombin}{\UseVerbatim{HOLCongruenceTheoremsSEQXXcombin}}
\begin{SaveVerbatim}{HOLCongruenceTheoremsSEQXXind}
\HOLTokenTurnstile{} \HOLSymConst{\HOLTokenForall{}}\HOLBoundVar{SEQ\sp{\prime}}.
     \HOLBoundVar{SEQ\sp{\prime}} (\HOLTokenLambda{}\HOLBoundVar{t}. \HOLBoundVar{t}) \HOLSymConst{\HOLTokenConj{}} (\HOLSymConst{\HOLTokenForall{}}\HOLBoundVar{p}. \HOLBoundVar{SEQ\sp{\prime}} (\HOLTokenLambda{}\HOLBoundVar{t}. \HOLBoundVar{p})) \HOLSymConst{\HOLTokenConj{}}
     (\HOLSymConst{\HOLTokenForall{}}\HOLBoundVar{a} \HOLBoundVar{e}. \HOLBoundVar{SEQ\sp{\prime}} \HOLBoundVar{e} \HOLSymConst{\HOLTokenImp{}} \HOLBoundVar{SEQ\sp{\prime}} (\HOLTokenLambda{}\HOLBoundVar{t}. \HOLBoundVar{a}\HOLSymConst{..}\HOLBoundVar{e} \HOLBoundVar{t})) \HOLSymConst{\HOLTokenConj{}}
     (\HOLSymConst{\HOLTokenForall{}}\HOLBoundVar{e\sb{\mathrm{1}}} \HOLBoundVar{e\sb{\mathrm{2}}}. \HOLBoundVar{SEQ\sp{\prime}} \HOLBoundVar{e\sb{\mathrm{1}}} \HOLSymConst{\HOLTokenConj{}} \HOLBoundVar{SEQ\sp{\prime}} \HOLBoundVar{e\sb{\mathrm{2}}} \HOLSymConst{\HOLTokenImp{}} \HOLBoundVar{SEQ\sp{\prime}} (\HOLTokenLambda{}\HOLBoundVar{t}. \HOLBoundVar{e\sb{\mathrm{1}}} \HOLBoundVar{t} \HOLSymConst{+} \HOLBoundVar{e\sb{\mathrm{2}}} \HOLBoundVar{t})) \HOLSymConst{\HOLTokenImp{}}
     \HOLSymConst{\HOLTokenForall{}}\HOLBoundVar{a\sb{\mathrm{0}}}. \HOLConst{SEQ} \HOLBoundVar{a\sb{\mathrm{0}}} \HOLSymConst{\HOLTokenImp{}} \HOLBoundVar{SEQ\sp{\prime}} \HOLBoundVar{a\sb{\mathrm{0}}}
\end{SaveVerbatim}
\newcommand{\HOLCongruenceTheoremsSEQXXind}{\UseVerbatim{HOLCongruenceTheoremsSEQXXind}}
\begin{SaveVerbatim}{HOLCongruenceTheoremsSEQXXISXXCONTEXT}
\HOLTokenTurnstile{} \HOLSymConst{\HOLTokenForall{}}\HOLBoundVar{e}. \HOLConst{SEQ} \HOLBoundVar{e} \HOLSymConst{\HOLTokenImp{}} \HOLConst{CONTEXT} \HOLBoundVar{e}
\end{SaveVerbatim}
\newcommand{\HOLCongruenceTheoremsSEQXXISXXCONTEXT}{\UseVerbatim{HOLCongruenceTheoremsSEQXXISXXCONTEXT}}
\begin{SaveVerbatim}{HOLCongruenceTheoremsSEQXXrules}
\HOLTokenTurnstile{} \HOLConst{SEQ} (\HOLTokenLambda{}\HOLBoundVar{t}. \HOLBoundVar{t}) \HOLSymConst{\HOLTokenConj{}} (\HOLSymConst{\HOLTokenForall{}}\HOLBoundVar{p}. \HOLConst{SEQ} (\HOLTokenLambda{}\HOLBoundVar{t}. \HOLBoundVar{p})) \HOLSymConst{\HOLTokenConj{}}
   (\HOLSymConst{\HOLTokenForall{}}\HOLBoundVar{a} \HOLBoundVar{e}. \HOLConst{SEQ} \HOLBoundVar{e} \HOLSymConst{\HOLTokenImp{}} \HOLConst{SEQ} (\HOLTokenLambda{}\HOLBoundVar{t}. \HOLBoundVar{a}\HOLSymConst{..}\HOLBoundVar{e} \HOLBoundVar{t})) \HOLSymConst{\HOLTokenConj{}}
   \HOLSymConst{\HOLTokenForall{}}\HOLBoundVar{e\sb{\mathrm{1}}} \HOLBoundVar{e\sb{\mathrm{2}}}. \HOLConst{SEQ} \HOLBoundVar{e\sb{\mathrm{1}}} \HOLSymConst{\HOLTokenConj{}} \HOLConst{SEQ} \HOLBoundVar{e\sb{\mathrm{2}}} \HOLSymConst{\HOLTokenImp{}} \HOLConst{SEQ} (\HOLTokenLambda{}\HOLBoundVar{t}. \HOLBoundVar{e\sb{\mathrm{1}}} \HOLBoundVar{t} \HOLSymConst{+} \HOLBoundVar{e\sb{\mathrm{2}}} \HOLBoundVar{t})
\end{SaveVerbatim}
\newcommand{\HOLCongruenceTheoremsSEQXXrules}{\UseVerbatim{HOLCongruenceTheoremsSEQXXrules}}
\begin{SaveVerbatim}{HOLCongruenceTheoremsSEQXXstrongind}
\HOLTokenTurnstile{} \HOLSymConst{\HOLTokenForall{}}\HOLBoundVar{SEQ\sp{\prime}}.
     \HOLBoundVar{SEQ\sp{\prime}} (\HOLTokenLambda{}\HOLBoundVar{t}. \HOLBoundVar{t}) \HOLSymConst{\HOLTokenConj{}} (\HOLSymConst{\HOLTokenForall{}}\HOLBoundVar{p}. \HOLBoundVar{SEQ\sp{\prime}} (\HOLTokenLambda{}\HOLBoundVar{t}. \HOLBoundVar{p})) \HOLSymConst{\HOLTokenConj{}}
     (\HOLSymConst{\HOLTokenForall{}}\HOLBoundVar{a} \HOLBoundVar{e}. \HOLConst{SEQ} \HOLBoundVar{e} \HOLSymConst{\HOLTokenConj{}} \HOLBoundVar{SEQ\sp{\prime}} \HOLBoundVar{e} \HOLSymConst{\HOLTokenImp{}} \HOLBoundVar{SEQ\sp{\prime}} (\HOLTokenLambda{}\HOLBoundVar{t}. \HOLBoundVar{a}\HOLSymConst{..}\HOLBoundVar{e} \HOLBoundVar{t})) \HOLSymConst{\HOLTokenConj{}}
     (\HOLSymConst{\HOLTokenForall{}}\HOLBoundVar{e\sb{\mathrm{1}}} \HOLBoundVar{e\sb{\mathrm{2}}}.
        \HOLConst{SEQ} \HOLBoundVar{e\sb{\mathrm{1}}} \HOLSymConst{\HOLTokenConj{}} \HOLBoundVar{SEQ\sp{\prime}} \HOLBoundVar{e\sb{\mathrm{1}}} \HOLSymConst{\HOLTokenConj{}} \HOLConst{SEQ} \HOLBoundVar{e\sb{\mathrm{2}}} \HOLSymConst{\HOLTokenConj{}} \HOLBoundVar{SEQ\sp{\prime}} \HOLBoundVar{e\sb{\mathrm{2}}} \HOLSymConst{\HOLTokenImp{}}
        \HOLBoundVar{SEQ\sp{\prime}} (\HOLTokenLambda{}\HOLBoundVar{t}. \HOLBoundVar{e\sb{\mathrm{1}}} \HOLBoundVar{t} \HOLSymConst{+} \HOLBoundVar{e\sb{\mathrm{2}}} \HOLBoundVar{t})) \HOLSymConst{\HOLTokenImp{}}
     \HOLSymConst{\HOLTokenForall{}}\HOLBoundVar{a\sb{\mathrm{0}}}. \HOLConst{SEQ} \HOLBoundVar{a\sb{\mathrm{0}}} \HOLSymConst{\HOLTokenImp{}} \HOLBoundVar{SEQ\sp{\prime}} \HOLBoundVar{a\sb{\mathrm{0}}}
\end{SaveVerbatim}
\newcommand{\HOLCongruenceTheoremsSEQXXstrongind}{\UseVerbatim{HOLCongruenceTheoremsSEQXXstrongind}}
\begin{SaveVerbatim}{HOLCongruenceTheoremsSGOne}
\HOLTokenTurnstile{} \HOLSymConst{\HOLTokenForall{}}\HOLBoundVar{p}. \HOLConst{SG} (\HOLTokenLambda{}\HOLBoundVar{t}. \HOLBoundVar{p})
\end{SaveVerbatim}
\newcommand{\HOLCongruenceTheoremsSGOne}{\UseVerbatim{HOLCongruenceTheoremsSGOne}}
\begin{SaveVerbatim}{HOLCongruenceTheoremsSGOneZero}
\HOLTokenTurnstile{} \HOLSymConst{\HOLTokenForall{}}\HOLBoundVar{e} \HOLBoundVar{e\sp{\prime}}. \HOLConst{SG} (\HOLTokenLambda{}\HOLBoundVar{t}. \HOLConst{\ensuremath{\tau}}\HOLSymConst{..}\HOLBoundVar{e} \HOLBoundVar{t} \HOLSymConst{+} \HOLConst{\ensuremath{\tau}}\HOLSymConst{..}\HOLBoundVar{e\sp{\prime}} \HOLBoundVar{t}) \HOLSymConst{\HOLTokenImp{}} \HOLConst{SG} \HOLBoundVar{e} \HOLSymConst{\HOLTokenConj{}} \HOLConst{SG} \HOLBoundVar{e\sp{\prime}}
\end{SaveVerbatim}
\newcommand{\HOLCongruenceTheoremsSGOneZero}{\UseVerbatim{HOLCongruenceTheoremsSGOneZero}}
\begin{SaveVerbatim}{HOLCongruenceTheoremsSGOneOne}
\HOLTokenTurnstile{} \HOLSymConst{\HOLTokenForall{}}\HOLBoundVar{e} \HOLBoundVar{e\sp{\prime}} \HOLBoundVar{L}. \HOLConst{SG} (\HOLTokenLambda{}\HOLBoundVar{t}. \HOLConst{\ensuremath{\tau}}\HOLSymConst{..}\HOLBoundVar{e} \HOLBoundVar{t} \HOLSymConst{+} \HOLConst{label} \HOLBoundVar{L}\HOLSymConst{..}\HOLBoundVar{e\sp{\prime}} \HOLBoundVar{t}) \HOLSymConst{\HOLTokenImp{}} \HOLConst{SG} \HOLBoundVar{e}
\end{SaveVerbatim}
\newcommand{\HOLCongruenceTheoremsSGOneOne}{\UseVerbatim{HOLCongruenceTheoremsSGOneOne}}
\begin{SaveVerbatim}{HOLCongruenceTheoremsSGOneOneYY}
\HOLTokenTurnstile{} \HOLSymConst{\HOLTokenForall{}}\HOLBoundVar{e} \HOLBoundVar{e\sp{\prime}} \HOLBoundVar{L}. \HOLConst{SG} (\HOLTokenLambda{}\HOLBoundVar{t}. \HOLConst{label} \HOLBoundVar{L}\HOLSymConst{..}\HOLBoundVar{e} \HOLBoundVar{t} \HOLSymConst{+} \HOLConst{\ensuremath{\tau}}\HOLSymConst{..}\HOLBoundVar{e\sp{\prime}} \HOLBoundVar{t}) \HOLSymConst{\HOLTokenImp{}} \HOLConst{SG} \HOLBoundVar{e\sp{\prime}}
\end{SaveVerbatim}
\newcommand{\HOLCongruenceTheoremsSGOneOneYY}{\UseVerbatim{HOLCongruenceTheoremsSGOneOneYY}}
\begin{SaveVerbatim}{HOLCongruenceTheoremsSGTwo}
\HOLTokenTurnstile{} \HOLSymConst{\HOLTokenForall{}}\HOLBoundVar{l} \HOLBoundVar{e}. \HOLConst{CONTEXT} \HOLBoundVar{e} \HOLSymConst{\HOLTokenImp{}} \HOLConst{SG} (\HOLTokenLambda{}\HOLBoundVar{t}. \HOLConst{label} \HOLBoundVar{l}\HOLSymConst{..}\HOLBoundVar{e} \HOLBoundVar{t})
\end{SaveVerbatim}
\newcommand{\HOLCongruenceTheoremsSGTwo}{\UseVerbatim{HOLCongruenceTheoremsSGTwo}}
\begin{SaveVerbatim}{HOLCongruenceTheoremsSGThree}
\HOLTokenTurnstile{} \HOLSymConst{\HOLTokenForall{}}\HOLBoundVar{a} \HOLBoundVar{e}. \HOLConst{SG} \HOLBoundVar{e} \HOLSymConst{\HOLTokenImp{}} \HOLConst{SG} (\HOLTokenLambda{}\HOLBoundVar{t}. \HOLBoundVar{a}\HOLSymConst{..}\HOLBoundVar{e} \HOLBoundVar{t})
\end{SaveVerbatim}
\newcommand{\HOLCongruenceTheoremsSGThree}{\UseVerbatim{HOLCongruenceTheoremsSGThree}}
\begin{SaveVerbatim}{HOLCongruenceTheoremsSGFour}
\HOLTokenTurnstile{} \HOLSymConst{\HOLTokenForall{}}\HOLBoundVar{e\sb{\mathrm{1}}} \HOLBoundVar{e\sb{\mathrm{2}}}. \HOLConst{SG} \HOLBoundVar{e\sb{\mathrm{1}}} \HOLSymConst{\HOLTokenConj{}} \HOLConst{SG} \HOLBoundVar{e\sb{\mathrm{2}}} \HOLSymConst{\HOLTokenImp{}} \HOLConst{SG} (\HOLTokenLambda{}\HOLBoundVar{t}. \HOLBoundVar{e\sb{\mathrm{1}}} \HOLBoundVar{t} \HOLSymConst{+} \HOLBoundVar{e\sb{\mathrm{2}}} \HOLBoundVar{t})
\end{SaveVerbatim}
\newcommand{\HOLCongruenceTheoremsSGFour}{\UseVerbatim{HOLCongruenceTheoremsSGFour}}
\begin{SaveVerbatim}{HOLCongruenceTheoremsSGFive}
\HOLTokenTurnstile{} \HOLSymConst{\HOLTokenForall{}}\HOLBoundVar{e\sb{\mathrm{1}}} \HOLBoundVar{e\sb{\mathrm{2}}}. \HOLConst{SG} \HOLBoundVar{e\sb{\mathrm{1}}} \HOLSymConst{\HOLTokenConj{}} \HOLConst{SG} \HOLBoundVar{e\sb{\mathrm{2}}} \HOLSymConst{\HOLTokenImp{}} \HOLConst{SG} (\HOLTokenLambda{}\HOLBoundVar{t}. \HOLBoundVar{e\sb{\mathrm{1}}} \HOLBoundVar{t} \HOLSymConst{\ensuremath{\parallel}} \HOLBoundVar{e\sb{\mathrm{2}}} \HOLBoundVar{t})
\end{SaveVerbatim}
\newcommand{\HOLCongruenceTheoremsSGFive}{\UseVerbatim{HOLCongruenceTheoremsSGFive}}
\begin{SaveVerbatim}{HOLCongruenceTheoremsSGSix}
\HOLTokenTurnstile{} \HOLSymConst{\HOLTokenForall{}}\HOLBoundVar{L} \HOLBoundVar{e}. \HOLConst{SG} \HOLBoundVar{e} \HOLSymConst{\HOLTokenImp{}} \HOLConst{SG} (\HOLTokenLambda{}\HOLBoundVar{t}. \HOLConst{\ensuremath{\nu}} \HOLBoundVar{L} (\HOLBoundVar{e} \HOLBoundVar{t}))
\end{SaveVerbatim}
\newcommand{\HOLCongruenceTheoremsSGSix}{\UseVerbatim{HOLCongruenceTheoremsSGSix}}
\begin{SaveVerbatim}{HOLCongruenceTheoremsSGSeven}
\HOLTokenTurnstile{} \HOLSymConst{\HOLTokenForall{}}\HOLBoundVar{rf} \HOLBoundVar{e}. \HOLConst{SG} \HOLBoundVar{e} \HOLSymConst{\HOLTokenImp{}} \HOLConst{SG} (\HOLTokenLambda{}\HOLBoundVar{t}. \HOLConst{relab} (\HOLBoundVar{e} \HOLBoundVar{t}) \HOLBoundVar{rf})
\end{SaveVerbatim}
\newcommand{\HOLCongruenceTheoremsSGSeven}{\UseVerbatim{HOLCongruenceTheoremsSGSeven}}
\begin{SaveVerbatim}{HOLCongruenceTheoremsSGEight}
\HOLTokenTurnstile{} \HOLSymConst{\HOLTokenForall{}}\HOLBoundVar{e}. \HOLConst{SG} (\HOLTokenLambda{}\HOLBoundVar{t}. \HOLConst{\ensuremath{\tau}}\HOLSymConst{..}\HOLBoundVar{e} \HOLBoundVar{t}) \HOLSymConst{\HOLTokenImp{}} \HOLConst{SG} \HOLBoundVar{e}
\end{SaveVerbatim}
\newcommand{\HOLCongruenceTheoremsSGEight}{\UseVerbatim{HOLCongruenceTheoremsSGEight}}
\begin{SaveVerbatim}{HOLCongruenceTheoremsSGNine}
\HOLTokenTurnstile{} \HOLSymConst{\HOLTokenForall{}}\HOLBoundVar{e} \HOLBoundVar{e\sp{\prime}}. \HOLConst{SG} (\HOLTokenLambda{}\HOLBoundVar{t}. \HOLBoundVar{e} \HOLBoundVar{t} \HOLSymConst{+} \HOLBoundVar{e\sp{\prime}} \HOLBoundVar{t}) \HOLSymConst{\HOLTokenImp{}} \HOLConst{SG} \HOLBoundVar{e} \HOLSymConst{\HOLTokenConj{}} \HOLConst{SG} \HOLBoundVar{e\sp{\prime}}
\end{SaveVerbatim}
\newcommand{\HOLCongruenceTheoremsSGNine}{\UseVerbatim{HOLCongruenceTheoremsSGNine}}
\begin{SaveVerbatim}{HOLCongruenceTheoremsSGXXcases}
\HOLTokenTurnstile{} \HOLSymConst{\HOLTokenForall{}}\HOLBoundVar{a\sb{\mathrm{0}}}.
     \HOLConst{SG} \HOLBoundVar{a\sb{\mathrm{0}}} \HOLSymConst{\HOLTokenEquiv{}}
     (\HOLSymConst{\HOLTokenExists{}}\HOLBoundVar{p}. \HOLBoundVar{a\sb{\mathrm{0}}} \HOLSymConst{=} (\HOLTokenLambda{}\HOLBoundVar{t}. \HOLBoundVar{p})) \HOLSymConst{\HOLTokenDisj{}}
     (\HOLSymConst{\HOLTokenExists{}}\HOLBoundVar{l} \HOLBoundVar{e}. (\HOLBoundVar{a\sb{\mathrm{0}}} \HOLSymConst{=} (\HOLTokenLambda{}\HOLBoundVar{t}. \HOLConst{label} \HOLBoundVar{l}\HOLSymConst{..}\HOLBoundVar{e} \HOLBoundVar{t})) \HOLSymConst{\HOLTokenConj{}} \HOLConst{CONTEXT} \HOLBoundVar{e}) \HOLSymConst{\HOLTokenDisj{}}
     (\HOLSymConst{\HOLTokenExists{}}\HOLBoundVar{a} \HOLBoundVar{e}. (\HOLBoundVar{a\sb{\mathrm{0}}} \HOLSymConst{=} (\HOLTokenLambda{}\HOLBoundVar{t}. \HOLBoundVar{a}\HOLSymConst{..}\HOLBoundVar{e} \HOLBoundVar{t})) \HOLSymConst{\HOLTokenConj{}} \HOLConst{SG} \HOLBoundVar{e}) \HOLSymConst{\HOLTokenDisj{}}
     (\HOLSymConst{\HOLTokenExists{}}\HOLBoundVar{e\sb{\mathrm{1}}} \HOLBoundVar{e\sb{\mathrm{2}}}. (\HOLBoundVar{a\sb{\mathrm{0}}} \HOLSymConst{=} (\HOLTokenLambda{}\HOLBoundVar{t}. \HOLBoundVar{e\sb{\mathrm{1}}} \HOLBoundVar{t} \HOLSymConst{+} \HOLBoundVar{e\sb{\mathrm{2}}} \HOLBoundVar{t})) \HOLSymConst{\HOLTokenConj{}} \HOLConst{SG} \HOLBoundVar{e\sb{\mathrm{1}}} \HOLSymConst{\HOLTokenConj{}} \HOLConst{SG} \HOLBoundVar{e\sb{\mathrm{2}}}) \HOLSymConst{\HOLTokenDisj{}}
     (\HOLSymConst{\HOLTokenExists{}}\HOLBoundVar{e\sb{\mathrm{1}}} \HOLBoundVar{e\sb{\mathrm{2}}}. (\HOLBoundVar{a\sb{\mathrm{0}}} \HOLSymConst{=} (\HOLTokenLambda{}\HOLBoundVar{t}. \HOLBoundVar{e\sb{\mathrm{1}}} \HOLBoundVar{t} \HOLSymConst{\ensuremath{\parallel}} \HOLBoundVar{e\sb{\mathrm{2}}} \HOLBoundVar{t})) \HOLSymConst{\HOLTokenConj{}} \HOLConst{SG} \HOLBoundVar{e\sb{\mathrm{1}}} \HOLSymConst{\HOLTokenConj{}} \HOLConst{SG} \HOLBoundVar{e\sb{\mathrm{2}}}) \HOLSymConst{\HOLTokenDisj{}}
     (\HOLSymConst{\HOLTokenExists{}}\HOLBoundVar{L} \HOLBoundVar{e}. (\HOLBoundVar{a\sb{\mathrm{0}}} \HOLSymConst{=} (\HOLTokenLambda{}\HOLBoundVar{t}. \HOLConst{\ensuremath{\nu}} \HOLBoundVar{L} (\HOLBoundVar{e} \HOLBoundVar{t}))) \HOLSymConst{\HOLTokenConj{}} \HOLConst{SG} \HOLBoundVar{e}) \HOLSymConst{\HOLTokenDisj{}}
     \HOLSymConst{\HOLTokenExists{}}\HOLBoundVar{rf} \HOLBoundVar{e}. (\HOLBoundVar{a\sb{\mathrm{0}}} \HOLSymConst{=} (\HOLTokenLambda{}\HOLBoundVar{t}. \HOLConst{relab} (\HOLBoundVar{e} \HOLBoundVar{t}) \HOLBoundVar{rf})) \HOLSymConst{\HOLTokenConj{}} \HOLConst{SG} \HOLBoundVar{e}
\end{SaveVerbatim}
\newcommand{\HOLCongruenceTheoremsSGXXcases}{\UseVerbatim{HOLCongruenceTheoremsSGXXcases}}
\begin{SaveVerbatim}{HOLCongruenceTheoremsSGXXGSEQXXcombin}
\HOLTokenTurnstile{} \HOLSymConst{\HOLTokenForall{}}\HOLBoundVar{E}. \HOLConst{SG} \HOLBoundVar{E} \HOLSymConst{\HOLTokenConj{}} \HOLConst{GSEQ} \HOLBoundVar{E} \HOLSymConst{\HOLTokenImp{}} \HOLSymConst{\HOLTokenForall{}}\HOLBoundVar{H}. \HOLConst{GSEQ} \HOLBoundVar{H} \HOLSymConst{\HOLTokenImp{}} \HOLConst{SG} (\HOLBoundVar{H} \HOLConst{\HOLTokenCompose} \HOLBoundVar{E}) \HOLSymConst{\HOLTokenConj{}} \HOLConst{GSEQ} (\HOLBoundVar{H} \HOLConst{\HOLTokenCompose} \HOLBoundVar{E})
\end{SaveVerbatim}
\newcommand{\HOLCongruenceTheoremsSGXXGSEQXXcombin}{\UseVerbatim{HOLCongruenceTheoremsSGXXGSEQXXcombin}}
\begin{SaveVerbatim}{HOLCongruenceTheoremsSGXXGSEQXXstrongXXinduction}
\HOLTokenTurnstile{} \HOLSymConst{\HOLTokenForall{}}\HOLBoundVar{R}.
     (\HOLSymConst{\HOLTokenForall{}}\HOLBoundVar{p}. \HOLBoundVar{R} (\HOLTokenLambda{}\HOLBoundVar{t}. \HOLBoundVar{p})) \HOLSymConst{\HOLTokenConj{}} (\HOLSymConst{\HOLTokenForall{}}\HOLBoundVar{l} \HOLBoundVar{e}. \HOLConst{GSEQ} \HOLBoundVar{e} \HOLSymConst{\HOLTokenImp{}} \HOLBoundVar{R} (\HOLTokenLambda{}\HOLBoundVar{t}. \HOLConst{label} \HOLBoundVar{l}\HOLSymConst{..}\HOLBoundVar{e} \HOLBoundVar{t})) \HOLSymConst{\HOLTokenConj{}}
     (\HOLSymConst{\HOLTokenForall{}}\HOLBoundVar{a} \HOLBoundVar{e}. \HOLConst{SG} \HOLBoundVar{e} \HOLSymConst{\HOLTokenConj{}} \HOLConst{GSEQ} \HOLBoundVar{e} \HOLSymConst{\HOLTokenConj{}} \HOLBoundVar{R} \HOLBoundVar{e} \HOLSymConst{\HOLTokenImp{}} \HOLBoundVar{R} (\HOLTokenLambda{}\HOLBoundVar{t}. \HOLBoundVar{a}\HOLSymConst{..}\HOLBoundVar{e} \HOLBoundVar{t})) \HOLSymConst{\HOLTokenConj{}}
     (\HOLSymConst{\HOLTokenForall{}}\HOLBoundVar{e\sb{\mathrm{1}}} \HOLBoundVar{e\sb{\mathrm{2}}}.
        \HOLConst{SG} \HOLBoundVar{e\sb{\mathrm{1}}} \HOLSymConst{\HOLTokenConj{}} \HOLConst{GSEQ} \HOLBoundVar{e\sb{\mathrm{1}}} \HOLSymConst{\HOLTokenConj{}} \HOLBoundVar{R} \HOLBoundVar{e\sb{\mathrm{1}}} \HOLSymConst{\HOLTokenConj{}} \HOLConst{SG} \HOLBoundVar{e\sb{\mathrm{2}}} \HOLSymConst{\HOLTokenConj{}} \HOLConst{GSEQ} \HOLBoundVar{e\sb{\mathrm{2}}} \HOLSymConst{\HOLTokenConj{}} \HOLBoundVar{R} \HOLBoundVar{e\sb{\mathrm{2}}} \HOLSymConst{\HOLTokenImp{}}
        \HOLBoundVar{R} (\HOLTokenLambda{}\HOLBoundVar{t}. \HOLConst{\ensuremath{\tau}}\HOLSymConst{..}\HOLBoundVar{e\sb{\mathrm{1}}} \HOLBoundVar{t} \HOLSymConst{+} \HOLConst{\ensuremath{\tau}}\HOLSymConst{..}\HOLBoundVar{e\sb{\mathrm{2}}} \HOLBoundVar{t})) \HOLSymConst{\HOLTokenConj{}}
     (\HOLSymConst{\HOLTokenForall{}}\HOLBoundVar{l\sb{\mathrm{2}}} \HOLBoundVar{e\sb{\mathrm{1}}} \HOLBoundVar{e\sb{\mathrm{2}}}.
        \HOLConst{SG} \HOLBoundVar{e\sb{\mathrm{1}}} \HOLSymConst{\HOLTokenConj{}} \HOLConst{GSEQ} \HOLBoundVar{e\sb{\mathrm{1}}} \HOLSymConst{\HOLTokenConj{}} \HOLBoundVar{R} \HOLBoundVar{e\sb{\mathrm{1}}} \HOLSymConst{\HOLTokenConj{}} \HOLConst{GSEQ} \HOLBoundVar{e\sb{\mathrm{2}}} \HOLSymConst{\HOLTokenImp{}}
        \HOLBoundVar{R} (\HOLTokenLambda{}\HOLBoundVar{t}. \HOLConst{\ensuremath{\tau}}\HOLSymConst{..}\HOLBoundVar{e\sb{\mathrm{1}}} \HOLBoundVar{t} \HOLSymConst{+} \HOLConst{label} \HOLBoundVar{l\sb{\mathrm{2}}}\HOLSymConst{..}\HOLBoundVar{e\sb{\mathrm{2}}} \HOLBoundVar{t})) \HOLSymConst{\HOLTokenConj{}}
     (\HOLSymConst{\HOLTokenForall{}}\HOLBoundVar{l\sb{\mathrm{1}}} \HOLBoundVar{e\sb{\mathrm{1}}} \HOLBoundVar{e\sb{\mathrm{2}}}.
        \HOLConst{GSEQ} \HOLBoundVar{e\sb{\mathrm{1}}} \HOLSymConst{\HOLTokenConj{}} \HOLConst{SG} \HOLBoundVar{e\sb{\mathrm{2}}} \HOLSymConst{\HOLTokenConj{}} \HOLConst{GSEQ} \HOLBoundVar{e\sb{\mathrm{2}}} \HOLSymConst{\HOLTokenConj{}} \HOLBoundVar{R} \HOLBoundVar{e\sb{\mathrm{2}}} \HOLSymConst{\HOLTokenImp{}}
        \HOLBoundVar{R} (\HOLTokenLambda{}\HOLBoundVar{t}. \HOLConst{label} \HOLBoundVar{l\sb{\mathrm{1}}}\HOLSymConst{..}\HOLBoundVar{e\sb{\mathrm{1}}} \HOLBoundVar{t} \HOLSymConst{+} \HOLConst{\ensuremath{\tau}}\HOLSymConst{..}\HOLBoundVar{e\sb{\mathrm{2}}} \HOLBoundVar{t})) \HOLSymConst{\HOLTokenConj{}}
     (\HOLSymConst{\HOLTokenForall{}}\HOLBoundVar{l\sb{\mathrm{1}}} \HOLBoundVar{l\sb{\mathrm{2}}} \HOLBoundVar{e\sb{\mathrm{1}}} \HOLBoundVar{e\sb{\mathrm{2}}}.
        \HOLConst{GSEQ} \HOLBoundVar{e\sb{\mathrm{1}}} \HOLSymConst{\HOLTokenConj{}} \HOLConst{GSEQ} \HOLBoundVar{e\sb{\mathrm{2}}} \HOLSymConst{\HOLTokenImp{}}
        \HOLBoundVar{R} (\HOLTokenLambda{}\HOLBoundVar{t}. \HOLConst{label} \HOLBoundVar{l\sb{\mathrm{1}}}\HOLSymConst{..}\HOLBoundVar{e\sb{\mathrm{1}}} \HOLBoundVar{t} \HOLSymConst{+} \HOLConst{label} \HOLBoundVar{l\sb{\mathrm{2}}}\HOLSymConst{..}\HOLBoundVar{e\sb{\mathrm{2}}} \HOLBoundVar{t})) \HOLSymConst{\HOLTokenImp{}}
     \HOLSymConst{\HOLTokenForall{}}\HOLBoundVar{e}. \HOLConst{SG} \HOLBoundVar{e} \HOLSymConst{\HOLTokenConj{}} \HOLConst{GSEQ} \HOLBoundVar{e} \HOLSymConst{\HOLTokenImp{}} \HOLBoundVar{R} \HOLBoundVar{e}
\end{SaveVerbatim}
\newcommand{\HOLCongruenceTheoremsSGXXGSEQXXstrongXXinduction}{\UseVerbatim{HOLCongruenceTheoremsSGXXGSEQXXstrongXXinduction}}
\begin{SaveVerbatim}{HOLCongruenceTheoremsSGXXIMPXXWG}
\HOLTokenTurnstile{} \HOLSymConst{\HOLTokenForall{}}\HOLBoundVar{e}. \HOLConst{SG} \HOLBoundVar{e} \HOLSymConst{\HOLTokenImp{}} \HOLConst{WG} \HOLBoundVar{e}
\end{SaveVerbatim}
\newcommand{\HOLCongruenceTheoremsSGXXIMPXXWG}{\UseVerbatim{HOLCongruenceTheoremsSGXXIMPXXWG}}
\begin{SaveVerbatim}{HOLCongruenceTheoremsSGXXind}
\HOLTokenTurnstile{} \HOLSymConst{\HOLTokenForall{}}\HOLBoundVar{SG\sp{\prime}}.
     (\HOLSymConst{\HOLTokenForall{}}\HOLBoundVar{p}. \HOLBoundVar{SG\sp{\prime}} (\HOLTokenLambda{}\HOLBoundVar{t}. \HOLBoundVar{p})) \HOLSymConst{\HOLTokenConj{}}
     (\HOLSymConst{\HOLTokenForall{}}\HOLBoundVar{l} \HOLBoundVar{e}. \HOLConst{CONTEXT} \HOLBoundVar{e} \HOLSymConst{\HOLTokenImp{}} \HOLBoundVar{SG\sp{\prime}} (\HOLTokenLambda{}\HOLBoundVar{t}. \HOLConst{label} \HOLBoundVar{l}\HOLSymConst{..}\HOLBoundVar{e} \HOLBoundVar{t})) \HOLSymConst{\HOLTokenConj{}}
     (\HOLSymConst{\HOLTokenForall{}}\HOLBoundVar{a} \HOLBoundVar{e}. \HOLBoundVar{SG\sp{\prime}} \HOLBoundVar{e} \HOLSymConst{\HOLTokenImp{}} \HOLBoundVar{SG\sp{\prime}} (\HOLTokenLambda{}\HOLBoundVar{t}. \HOLBoundVar{a}\HOLSymConst{..}\HOLBoundVar{e} \HOLBoundVar{t})) \HOLSymConst{\HOLTokenConj{}}
     (\HOLSymConst{\HOLTokenForall{}}\HOLBoundVar{e\sb{\mathrm{1}}} \HOLBoundVar{e\sb{\mathrm{2}}}. \HOLBoundVar{SG\sp{\prime}} \HOLBoundVar{e\sb{\mathrm{1}}} \HOLSymConst{\HOLTokenConj{}} \HOLBoundVar{SG\sp{\prime}} \HOLBoundVar{e\sb{\mathrm{2}}} \HOLSymConst{\HOLTokenImp{}} \HOLBoundVar{SG\sp{\prime}} (\HOLTokenLambda{}\HOLBoundVar{t}. \HOLBoundVar{e\sb{\mathrm{1}}} \HOLBoundVar{t} \HOLSymConst{+} \HOLBoundVar{e\sb{\mathrm{2}}} \HOLBoundVar{t})) \HOLSymConst{\HOLTokenConj{}}
     (\HOLSymConst{\HOLTokenForall{}}\HOLBoundVar{e\sb{\mathrm{1}}} \HOLBoundVar{e\sb{\mathrm{2}}}. \HOLBoundVar{SG\sp{\prime}} \HOLBoundVar{e\sb{\mathrm{1}}} \HOLSymConst{\HOLTokenConj{}} \HOLBoundVar{SG\sp{\prime}} \HOLBoundVar{e\sb{\mathrm{2}}} \HOLSymConst{\HOLTokenImp{}} \HOLBoundVar{SG\sp{\prime}} (\HOLTokenLambda{}\HOLBoundVar{t}. \HOLBoundVar{e\sb{\mathrm{1}}} \HOLBoundVar{t} \HOLSymConst{\ensuremath{\parallel}} \HOLBoundVar{e\sb{\mathrm{2}}} \HOLBoundVar{t})) \HOLSymConst{\HOLTokenConj{}}
     (\HOLSymConst{\HOLTokenForall{}}\HOLBoundVar{L} \HOLBoundVar{e}. \HOLBoundVar{SG\sp{\prime}} \HOLBoundVar{e} \HOLSymConst{\HOLTokenImp{}} \HOLBoundVar{SG\sp{\prime}} (\HOLTokenLambda{}\HOLBoundVar{t}. \HOLConst{\ensuremath{\nu}} \HOLBoundVar{L} (\HOLBoundVar{e} \HOLBoundVar{t}))) \HOLSymConst{\HOLTokenConj{}}
     (\HOLSymConst{\HOLTokenForall{}}\HOLBoundVar{rf} \HOLBoundVar{e}. \HOLBoundVar{SG\sp{\prime}} \HOLBoundVar{e} \HOLSymConst{\HOLTokenImp{}} \HOLBoundVar{SG\sp{\prime}} (\HOLTokenLambda{}\HOLBoundVar{t}. \HOLConst{relab} (\HOLBoundVar{e} \HOLBoundVar{t}) \HOLBoundVar{rf})) \HOLSymConst{\HOLTokenImp{}}
     \HOLSymConst{\HOLTokenForall{}}\HOLBoundVar{a\sb{\mathrm{0}}}. \HOLConst{SG} \HOLBoundVar{a\sb{\mathrm{0}}} \HOLSymConst{\HOLTokenImp{}} \HOLBoundVar{SG\sp{\prime}} \HOLBoundVar{a\sb{\mathrm{0}}}
\end{SaveVerbatim}
\newcommand{\HOLCongruenceTheoremsSGXXind}{\UseVerbatim{HOLCongruenceTheoremsSGXXind}}
\begin{SaveVerbatim}{HOLCongruenceTheoremsSGXXISXXCONTEXT}
\HOLTokenTurnstile{} \HOLSymConst{\HOLTokenForall{}}\HOLBoundVar{e}. \HOLConst{SG} \HOLBoundVar{e} \HOLSymConst{\HOLTokenImp{}} \HOLConst{CONTEXT} \HOLBoundVar{e}
\end{SaveVerbatim}
\newcommand{\HOLCongruenceTheoremsSGXXISXXCONTEXT}{\UseVerbatim{HOLCongruenceTheoremsSGXXISXXCONTEXT}}
\begin{SaveVerbatim}{HOLCongruenceTheoremsSGXXrules}
\HOLTokenTurnstile{} (\HOLSymConst{\HOLTokenForall{}}\HOLBoundVar{p}. \HOLConst{SG} (\HOLTokenLambda{}\HOLBoundVar{t}. \HOLBoundVar{p})) \HOLSymConst{\HOLTokenConj{}}
   (\HOLSymConst{\HOLTokenForall{}}\HOLBoundVar{l} \HOLBoundVar{e}. \HOLConst{CONTEXT} \HOLBoundVar{e} \HOLSymConst{\HOLTokenImp{}} \HOLConst{SG} (\HOLTokenLambda{}\HOLBoundVar{t}. \HOLConst{label} \HOLBoundVar{l}\HOLSymConst{..}\HOLBoundVar{e} \HOLBoundVar{t})) \HOLSymConst{\HOLTokenConj{}}
   (\HOLSymConst{\HOLTokenForall{}}\HOLBoundVar{a} \HOLBoundVar{e}. \HOLConst{SG} \HOLBoundVar{e} \HOLSymConst{\HOLTokenImp{}} \HOLConst{SG} (\HOLTokenLambda{}\HOLBoundVar{t}. \HOLBoundVar{a}\HOLSymConst{..}\HOLBoundVar{e} \HOLBoundVar{t})) \HOLSymConst{\HOLTokenConj{}}
   (\HOLSymConst{\HOLTokenForall{}}\HOLBoundVar{e\sb{\mathrm{1}}} \HOLBoundVar{e\sb{\mathrm{2}}}. \HOLConst{SG} \HOLBoundVar{e\sb{\mathrm{1}}} \HOLSymConst{\HOLTokenConj{}} \HOLConst{SG} \HOLBoundVar{e\sb{\mathrm{2}}} \HOLSymConst{\HOLTokenImp{}} \HOLConst{SG} (\HOLTokenLambda{}\HOLBoundVar{t}. \HOLBoundVar{e\sb{\mathrm{1}}} \HOLBoundVar{t} \HOLSymConst{+} \HOLBoundVar{e\sb{\mathrm{2}}} \HOLBoundVar{t})) \HOLSymConst{\HOLTokenConj{}}
   (\HOLSymConst{\HOLTokenForall{}}\HOLBoundVar{e\sb{\mathrm{1}}} \HOLBoundVar{e\sb{\mathrm{2}}}. \HOLConst{SG} \HOLBoundVar{e\sb{\mathrm{1}}} \HOLSymConst{\HOLTokenConj{}} \HOLConst{SG} \HOLBoundVar{e\sb{\mathrm{2}}} \HOLSymConst{\HOLTokenImp{}} \HOLConst{SG} (\HOLTokenLambda{}\HOLBoundVar{t}. \HOLBoundVar{e\sb{\mathrm{1}}} \HOLBoundVar{t} \HOLSymConst{\ensuremath{\parallel}} \HOLBoundVar{e\sb{\mathrm{2}}} \HOLBoundVar{t})) \HOLSymConst{\HOLTokenConj{}}
   (\HOLSymConst{\HOLTokenForall{}}\HOLBoundVar{L} \HOLBoundVar{e}. \HOLConst{SG} \HOLBoundVar{e} \HOLSymConst{\HOLTokenImp{}} \HOLConst{SG} (\HOLTokenLambda{}\HOLBoundVar{t}. \HOLConst{\ensuremath{\nu}} \HOLBoundVar{L} (\HOLBoundVar{e} \HOLBoundVar{t}))) \HOLSymConst{\HOLTokenConj{}}
   \HOLSymConst{\HOLTokenForall{}}\HOLBoundVar{rf} \HOLBoundVar{e}. \HOLConst{SG} \HOLBoundVar{e} \HOLSymConst{\HOLTokenImp{}} \HOLConst{SG} (\HOLTokenLambda{}\HOLBoundVar{t}. \HOLConst{relab} (\HOLBoundVar{e} \HOLBoundVar{t}) \HOLBoundVar{rf})
\end{SaveVerbatim}
\newcommand{\HOLCongruenceTheoremsSGXXrules}{\UseVerbatim{HOLCongruenceTheoremsSGXXrules}}
\begin{SaveVerbatim}{HOLCongruenceTheoremsSGXXSEQXXcombin}
\HOLTokenTurnstile{} \HOLSymConst{\HOLTokenForall{}}\HOLBoundVar{E}. \HOLConst{SG} \HOLBoundVar{E} \HOLSymConst{\HOLTokenConj{}} \HOLConst{SEQ} \HOLBoundVar{E} \HOLSymConst{\HOLTokenImp{}} \HOLSymConst{\HOLTokenForall{}}\HOLBoundVar{H}. \HOLConst{SEQ} \HOLBoundVar{H} \HOLSymConst{\HOLTokenImp{}} \HOLConst{SG} (\HOLBoundVar{H} \HOLConst{\HOLTokenCompose} \HOLBoundVar{E}) \HOLSymConst{\HOLTokenConj{}} \HOLConst{SEQ} (\HOLBoundVar{H} \HOLConst{\HOLTokenCompose} \HOLBoundVar{E})
\end{SaveVerbatim}
\newcommand{\HOLCongruenceTheoremsSGXXSEQXXcombin}{\UseVerbatim{HOLCongruenceTheoremsSGXXSEQXXcombin}}
\begin{SaveVerbatim}{HOLCongruenceTheoremsSGXXSEQXXstrongXXinduction}
\HOLTokenTurnstile{} \HOLSymConst{\HOLTokenForall{}}\HOLBoundVar{R}.
     (\HOLSymConst{\HOLTokenForall{}}\HOLBoundVar{p}. \HOLBoundVar{R} (\HOLTokenLambda{}\HOLBoundVar{t}. \HOLBoundVar{p})) \HOLSymConst{\HOLTokenConj{}} (\HOLSymConst{\HOLTokenForall{}}\HOLBoundVar{l} \HOLBoundVar{e}. \HOLConst{SEQ} \HOLBoundVar{e} \HOLSymConst{\HOLTokenImp{}} \HOLBoundVar{R} (\HOLTokenLambda{}\HOLBoundVar{t}. \HOLConst{label} \HOLBoundVar{l}\HOLSymConst{..}\HOLBoundVar{e} \HOLBoundVar{t})) \HOLSymConst{\HOLTokenConj{}}
     (\HOLSymConst{\HOLTokenForall{}}\HOLBoundVar{a} \HOLBoundVar{e}. \HOLConst{SG} \HOLBoundVar{e} \HOLSymConst{\HOLTokenConj{}} \HOLConst{SEQ} \HOLBoundVar{e} \HOLSymConst{\HOLTokenConj{}} \HOLBoundVar{R} \HOLBoundVar{e} \HOLSymConst{\HOLTokenImp{}} \HOLBoundVar{R} (\HOLTokenLambda{}\HOLBoundVar{t}. \HOLBoundVar{a}\HOLSymConst{..}\HOLBoundVar{e} \HOLBoundVar{t})) \HOLSymConst{\HOLTokenConj{}}
     (\HOLSymConst{\HOLTokenForall{}}\HOLBoundVar{e\sb{\mathrm{1}}} \HOLBoundVar{e\sb{\mathrm{2}}}.
        \HOLConst{SG} \HOLBoundVar{e\sb{\mathrm{1}}} \HOLSymConst{\HOLTokenConj{}} \HOLConst{SEQ} \HOLBoundVar{e\sb{\mathrm{1}}} \HOLSymConst{\HOLTokenConj{}} \HOLBoundVar{R} \HOLBoundVar{e\sb{\mathrm{1}}} \HOLSymConst{\HOLTokenConj{}} \HOLConst{SG} \HOLBoundVar{e\sb{\mathrm{2}}} \HOLSymConst{\HOLTokenConj{}} \HOLConst{SEQ} \HOLBoundVar{e\sb{\mathrm{2}}} \HOLSymConst{\HOLTokenConj{}} \HOLBoundVar{R} \HOLBoundVar{e\sb{\mathrm{2}}} \HOLSymConst{\HOLTokenImp{}}
        \HOLBoundVar{R} (\HOLTokenLambda{}\HOLBoundVar{t}. \HOLBoundVar{e\sb{\mathrm{1}}} \HOLBoundVar{t} \HOLSymConst{+} \HOLBoundVar{e\sb{\mathrm{2}}} \HOLBoundVar{t})) \HOLSymConst{\HOLTokenImp{}}
     \HOLSymConst{\HOLTokenForall{}}\HOLBoundVar{e}. \HOLConst{SG} \HOLBoundVar{e} \HOLSymConst{\HOLTokenConj{}} \HOLConst{SEQ} \HOLBoundVar{e} \HOLSymConst{\HOLTokenImp{}} \HOLBoundVar{R} \HOLBoundVar{e}
\end{SaveVerbatim}
\newcommand{\HOLCongruenceTheoremsSGXXSEQXXstrongXXinduction}{\UseVerbatim{HOLCongruenceTheoremsSGXXSEQXXstrongXXinduction}}
\begin{SaveVerbatim}{HOLCongruenceTheoremsSGXXstrongind}
\HOLTokenTurnstile{} \HOLSymConst{\HOLTokenForall{}}\HOLBoundVar{SG\sp{\prime}}.
     (\HOLSymConst{\HOLTokenForall{}}\HOLBoundVar{p}. \HOLBoundVar{SG\sp{\prime}} (\HOLTokenLambda{}\HOLBoundVar{t}. \HOLBoundVar{p})) \HOLSymConst{\HOLTokenConj{}}
     (\HOLSymConst{\HOLTokenForall{}}\HOLBoundVar{l} \HOLBoundVar{e}. \HOLConst{CONTEXT} \HOLBoundVar{e} \HOLSymConst{\HOLTokenImp{}} \HOLBoundVar{SG\sp{\prime}} (\HOLTokenLambda{}\HOLBoundVar{t}. \HOLConst{label} \HOLBoundVar{l}\HOLSymConst{..}\HOLBoundVar{e} \HOLBoundVar{t})) \HOLSymConst{\HOLTokenConj{}}
     (\HOLSymConst{\HOLTokenForall{}}\HOLBoundVar{a} \HOLBoundVar{e}. \HOLConst{SG} \HOLBoundVar{e} \HOLSymConst{\HOLTokenConj{}} \HOLBoundVar{SG\sp{\prime}} \HOLBoundVar{e} \HOLSymConst{\HOLTokenImp{}} \HOLBoundVar{SG\sp{\prime}} (\HOLTokenLambda{}\HOLBoundVar{t}. \HOLBoundVar{a}\HOLSymConst{..}\HOLBoundVar{e} \HOLBoundVar{t})) \HOLSymConst{\HOLTokenConj{}}
     (\HOLSymConst{\HOLTokenForall{}}\HOLBoundVar{e\sb{\mathrm{1}}} \HOLBoundVar{e\sb{\mathrm{2}}}.
        \HOLConst{SG} \HOLBoundVar{e\sb{\mathrm{1}}} \HOLSymConst{\HOLTokenConj{}} \HOLBoundVar{SG\sp{\prime}} \HOLBoundVar{e\sb{\mathrm{1}}} \HOLSymConst{\HOLTokenConj{}} \HOLConst{SG} \HOLBoundVar{e\sb{\mathrm{2}}} \HOLSymConst{\HOLTokenConj{}} \HOLBoundVar{SG\sp{\prime}} \HOLBoundVar{e\sb{\mathrm{2}}} \HOLSymConst{\HOLTokenImp{}}
        \HOLBoundVar{SG\sp{\prime}} (\HOLTokenLambda{}\HOLBoundVar{t}. \HOLBoundVar{e\sb{\mathrm{1}}} \HOLBoundVar{t} \HOLSymConst{+} \HOLBoundVar{e\sb{\mathrm{2}}} \HOLBoundVar{t})) \HOLSymConst{\HOLTokenConj{}}
     (\HOLSymConst{\HOLTokenForall{}}\HOLBoundVar{e\sb{\mathrm{1}}} \HOLBoundVar{e\sb{\mathrm{2}}}.
        \HOLConst{SG} \HOLBoundVar{e\sb{\mathrm{1}}} \HOLSymConst{\HOLTokenConj{}} \HOLBoundVar{SG\sp{\prime}} \HOLBoundVar{e\sb{\mathrm{1}}} \HOLSymConst{\HOLTokenConj{}} \HOLConst{SG} \HOLBoundVar{e\sb{\mathrm{2}}} \HOLSymConst{\HOLTokenConj{}} \HOLBoundVar{SG\sp{\prime}} \HOLBoundVar{e\sb{\mathrm{2}}} \HOLSymConst{\HOLTokenImp{}}
        \HOLBoundVar{SG\sp{\prime}} (\HOLTokenLambda{}\HOLBoundVar{t}. \HOLBoundVar{e\sb{\mathrm{1}}} \HOLBoundVar{t} \HOLSymConst{\ensuremath{\parallel}} \HOLBoundVar{e\sb{\mathrm{2}}} \HOLBoundVar{t})) \HOLSymConst{\HOLTokenConj{}}
     (\HOLSymConst{\HOLTokenForall{}}\HOLBoundVar{L} \HOLBoundVar{e}. \HOLConst{SG} \HOLBoundVar{e} \HOLSymConst{\HOLTokenConj{}} \HOLBoundVar{SG\sp{\prime}} \HOLBoundVar{e} \HOLSymConst{\HOLTokenImp{}} \HOLBoundVar{SG\sp{\prime}} (\HOLTokenLambda{}\HOLBoundVar{t}. \HOLConst{\ensuremath{\nu}} \HOLBoundVar{L} (\HOLBoundVar{e} \HOLBoundVar{t}))) \HOLSymConst{\HOLTokenConj{}}
     (\HOLSymConst{\HOLTokenForall{}}\HOLBoundVar{rf} \HOLBoundVar{e}. \HOLConst{SG} \HOLBoundVar{e} \HOLSymConst{\HOLTokenConj{}} \HOLBoundVar{SG\sp{\prime}} \HOLBoundVar{e} \HOLSymConst{\HOLTokenImp{}} \HOLBoundVar{SG\sp{\prime}} (\HOLTokenLambda{}\HOLBoundVar{t}. \HOLConst{relab} (\HOLBoundVar{e} \HOLBoundVar{t}) \HOLBoundVar{rf})) \HOLSymConst{\HOLTokenImp{}}
     \HOLSymConst{\HOLTokenForall{}}\HOLBoundVar{a\sb{\mathrm{0}}}. \HOLConst{SG} \HOLBoundVar{a\sb{\mathrm{0}}} \HOLSymConst{\HOLTokenImp{}} \HOLBoundVar{SG\sp{\prime}} \HOLBoundVar{a\sb{\mathrm{0}}}
\end{SaveVerbatim}
\newcommand{\HOLCongruenceTheoremsSGXXstrongind}{\UseVerbatim{HOLCongruenceTheoremsSGXXstrongind}}
\begin{SaveVerbatim}{HOLCongruenceTheoremsSTRONGXXEQUIVXXcongruence}
\HOLTokenTurnstile{} \HOLConst{congruence} \HOLConst{STRONG_EQUIV}
\end{SaveVerbatim}
\newcommand{\HOLCongruenceTheoremsSTRONGXXEQUIVXXcongruence}{\UseVerbatim{HOLCongruenceTheoremsSTRONGXXEQUIVXXcongruence}}
\begin{SaveVerbatim}{HOLCongruenceTheoremsSTRONGXXEQUIVXXSUBSTXXCONTEXT}
\HOLTokenTurnstile{} \HOLSymConst{\HOLTokenForall{}}\HOLBoundVar{P} \HOLBoundVar{Q}.
     \HOLConst{STRONG_EQUIV} \HOLBoundVar{P} \HOLBoundVar{Q} \HOLSymConst{\HOLTokenImp{}} \HOLSymConst{\HOLTokenForall{}}\HOLBoundVar{E}. \HOLConst{CONTEXT} \HOLBoundVar{E} \HOLSymConst{\HOLTokenImp{}} \HOLConst{STRONG_EQUIV} (\HOLBoundVar{E} \HOLBoundVar{P}) (\HOLBoundVar{E} \HOLBoundVar{Q})
\end{SaveVerbatim}
\newcommand{\HOLCongruenceTheoremsSTRONGXXEQUIVXXSUBSTXXCONTEXT}{\UseVerbatim{HOLCongruenceTheoremsSTRONGXXEQUIVXXSUBSTXXCONTEXT}}
\begin{SaveVerbatim}{HOLCongruenceTheoremsWEAKXXEQUIVXXcongruence}
\HOLTokenTurnstile{} \HOLConst{congruence1} \HOLConst{WEAK_EQUIV}
\end{SaveVerbatim}
\newcommand{\HOLCongruenceTheoremsWEAKXXEQUIVXXcongruence}{\UseVerbatim{HOLCongruenceTheoremsWEAKXXEQUIVXXcongruence}}
\begin{SaveVerbatim}{HOLCongruenceTheoremsWEAKXXEQUIVXXSUBSTXXGCONTEXT}
\HOLTokenTurnstile{} \HOLSymConst{\HOLTokenForall{}}\HOLBoundVar{P} \HOLBoundVar{Q}.
     \HOLConst{WEAK_EQUIV} \HOLBoundVar{P} \HOLBoundVar{Q} \HOLSymConst{\HOLTokenImp{}} \HOLSymConst{\HOLTokenForall{}}\HOLBoundVar{E}. \HOLConst{GCONTEXT} \HOLBoundVar{E} \HOLSymConst{\HOLTokenImp{}} \HOLConst{WEAK_EQUIV} (\HOLBoundVar{E} \HOLBoundVar{P}) (\HOLBoundVar{E} \HOLBoundVar{Q})
\end{SaveVerbatim}
\newcommand{\HOLCongruenceTheoremsWEAKXXEQUIVXXSUBSTXXGCONTEXT}{\UseVerbatim{HOLCongruenceTheoremsWEAKXXEQUIVXXSUBSTXXGCONTEXT}}
\begin{SaveVerbatim}{HOLCongruenceTheoremsWEAKXXEQUIVXXSUBSTXXGSEQ}
\HOLTokenTurnstile{} \HOLSymConst{\HOLTokenForall{}}\HOLBoundVar{P} \HOLBoundVar{Q}. \HOLConst{WEAK_EQUIV} \HOLBoundVar{P} \HOLBoundVar{Q} \HOLSymConst{\HOLTokenImp{}} \HOLSymConst{\HOLTokenForall{}}\HOLBoundVar{E}. \HOLConst{GSEQ} \HOLBoundVar{E} \HOLSymConst{\HOLTokenImp{}} \HOLConst{WEAK_EQUIV} (\HOLBoundVar{E} \HOLBoundVar{P}) (\HOLBoundVar{E} \HOLBoundVar{Q})
\end{SaveVerbatim}
\newcommand{\HOLCongruenceTheoremsWEAKXXEQUIVXXSUBSTXXGSEQ}{\UseVerbatim{HOLCongruenceTheoremsWEAKXXEQUIVXXSUBSTXXGSEQ}}
\begin{SaveVerbatim}{HOLCongruenceTheoremsWGOne}
\HOLTokenTurnstile{} \HOLSymConst{\HOLTokenForall{}}\HOLBoundVar{a}. \HOLConst{WG} (\HOLTokenLambda{}\HOLBoundVar{t}. \HOLBoundVar{a}\HOLSymConst{..}\HOLBoundVar{t})
\end{SaveVerbatim}
\newcommand{\HOLCongruenceTheoremsWGOne}{\UseVerbatim{HOLCongruenceTheoremsWGOne}}
\begin{SaveVerbatim}{HOLCongruenceTheoremsWGTwo}
\HOLTokenTurnstile{} \HOLSymConst{\HOLTokenForall{}}\HOLBoundVar{p}. \HOLConst{WG} (\HOLTokenLambda{}\HOLBoundVar{t}. \HOLBoundVar{p})
\end{SaveVerbatim}
\newcommand{\HOLCongruenceTheoremsWGTwo}{\UseVerbatim{HOLCongruenceTheoremsWGTwo}}
\begin{SaveVerbatim}{HOLCongruenceTheoremsWGThree}
\HOLTokenTurnstile{} \HOLSymConst{\HOLTokenForall{}}\HOLBoundVar{a} \HOLBoundVar{e}. \HOLConst{CONTEXT} \HOLBoundVar{e} \HOLSymConst{\HOLTokenImp{}} \HOLConst{WG} (\HOLTokenLambda{}\HOLBoundVar{t}. \HOLBoundVar{a}\HOLSymConst{..}\HOLBoundVar{e} \HOLBoundVar{t})
\end{SaveVerbatim}
\newcommand{\HOLCongruenceTheoremsWGThree}{\UseVerbatim{HOLCongruenceTheoremsWGThree}}
\begin{SaveVerbatim}{HOLCongruenceTheoremsWGFour}
\HOLTokenTurnstile{} \HOLSymConst{\HOLTokenForall{}}\HOLBoundVar{e\sb{\mathrm{1}}} \HOLBoundVar{e\sb{\mathrm{2}}}. \HOLConst{WG} \HOLBoundVar{e\sb{\mathrm{1}}} \HOLSymConst{\HOLTokenConj{}} \HOLConst{WG} \HOLBoundVar{e\sb{\mathrm{2}}} \HOLSymConst{\HOLTokenImp{}} \HOLConst{WG} (\HOLTokenLambda{}\HOLBoundVar{t}. \HOLBoundVar{e\sb{\mathrm{1}}} \HOLBoundVar{t} \HOLSymConst{+} \HOLBoundVar{e\sb{\mathrm{2}}} \HOLBoundVar{t})
\end{SaveVerbatim}
\newcommand{\HOLCongruenceTheoremsWGFour}{\UseVerbatim{HOLCongruenceTheoremsWGFour}}
\begin{SaveVerbatim}{HOLCongruenceTheoremsWGFive}
\HOLTokenTurnstile{} \HOLSymConst{\HOLTokenForall{}}\HOLBoundVar{e\sb{\mathrm{1}}} \HOLBoundVar{e\sb{\mathrm{2}}}. \HOLConst{WG} \HOLBoundVar{e\sb{\mathrm{1}}} \HOLSymConst{\HOLTokenConj{}} \HOLConst{WG} \HOLBoundVar{e\sb{\mathrm{2}}} \HOLSymConst{\HOLTokenImp{}} \HOLConst{WG} (\HOLTokenLambda{}\HOLBoundVar{t}. \HOLBoundVar{e\sb{\mathrm{1}}} \HOLBoundVar{t} \HOLSymConst{\ensuremath{\parallel}} \HOLBoundVar{e\sb{\mathrm{2}}} \HOLBoundVar{t})
\end{SaveVerbatim}
\newcommand{\HOLCongruenceTheoremsWGFive}{\UseVerbatim{HOLCongruenceTheoremsWGFive}}
\begin{SaveVerbatim}{HOLCongruenceTheoremsWGSix}
\HOLTokenTurnstile{} \HOLSymConst{\HOLTokenForall{}}\HOLBoundVar{L} \HOLBoundVar{e}. \HOLConst{WG} \HOLBoundVar{e} \HOLSymConst{\HOLTokenImp{}} \HOLConst{WG} (\HOLTokenLambda{}\HOLBoundVar{t}. \HOLConst{\ensuremath{\nu}} \HOLBoundVar{L} (\HOLBoundVar{e} \HOLBoundVar{t}))
\end{SaveVerbatim}
\newcommand{\HOLCongruenceTheoremsWGSix}{\UseVerbatim{HOLCongruenceTheoremsWGSix}}
\begin{SaveVerbatim}{HOLCongruenceTheoremsWGSeven}
\HOLTokenTurnstile{} \HOLSymConst{\HOLTokenForall{}}\HOLBoundVar{rf} \HOLBoundVar{e}. \HOLConst{WG} \HOLBoundVar{e} \HOLSymConst{\HOLTokenImp{}} \HOLConst{WG} (\HOLTokenLambda{}\HOLBoundVar{t}. \HOLConst{relab} (\HOLBoundVar{e} \HOLBoundVar{t}) \HOLBoundVar{rf})
\end{SaveVerbatim}
\newcommand{\HOLCongruenceTheoremsWGSeven}{\UseVerbatim{HOLCongruenceTheoremsWGSeven}}
\begin{SaveVerbatim}{HOLCongruenceTheoremsWGXXcases}
\HOLTokenTurnstile{} \HOLSymConst{\HOLTokenForall{}}\HOLBoundVar{a\sb{\mathrm{0}}}.
     \HOLConst{WG} \HOLBoundVar{a\sb{\mathrm{0}}} \HOLSymConst{\HOLTokenEquiv{}}
     (\HOLSymConst{\HOLTokenExists{}}\HOLBoundVar{p}. \HOLBoundVar{a\sb{\mathrm{0}}} \HOLSymConst{=} (\HOLTokenLambda{}\HOLBoundVar{t}. \HOLBoundVar{p})) \HOLSymConst{\HOLTokenDisj{}}
     (\HOLSymConst{\HOLTokenExists{}}\HOLBoundVar{a} \HOLBoundVar{e}. (\HOLBoundVar{a\sb{\mathrm{0}}} \HOLSymConst{=} (\HOLTokenLambda{}\HOLBoundVar{t}. \HOLBoundVar{a}\HOLSymConst{..}\HOLBoundVar{e} \HOLBoundVar{t})) \HOLSymConst{\HOLTokenConj{}} \HOLConst{CONTEXT} \HOLBoundVar{e}) \HOLSymConst{\HOLTokenDisj{}}
     (\HOLSymConst{\HOLTokenExists{}}\HOLBoundVar{e\sb{\mathrm{1}}} \HOLBoundVar{e\sb{\mathrm{2}}}. (\HOLBoundVar{a\sb{\mathrm{0}}} \HOLSymConst{=} (\HOLTokenLambda{}\HOLBoundVar{t}. \HOLBoundVar{e\sb{\mathrm{1}}} \HOLBoundVar{t} \HOLSymConst{+} \HOLBoundVar{e\sb{\mathrm{2}}} \HOLBoundVar{t})) \HOLSymConst{\HOLTokenConj{}} \HOLConst{WG} \HOLBoundVar{e\sb{\mathrm{1}}} \HOLSymConst{\HOLTokenConj{}} \HOLConst{WG} \HOLBoundVar{e\sb{\mathrm{2}}}) \HOLSymConst{\HOLTokenDisj{}}
     (\HOLSymConst{\HOLTokenExists{}}\HOLBoundVar{e\sb{\mathrm{1}}} \HOLBoundVar{e\sb{\mathrm{2}}}. (\HOLBoundVar{a\sb{\mathrm{0}}} \HOLSymConst{=} (\HOLTokenLambda{}\HOLBoundVar{t}. \HOLBoundVar{e\sb{\mathrm{1}}} \HOLBoundVar{t} \HOLSymConst{\ensuremath{\parallel}} \HOLBoundVar{e\sb{\mathrm{2}}} \HOLBoundVar{t})) \HOLSymConst{\HOLTokenConj{}} \HOLConst{WG} \HOLBoundVar{e\sb{\mathrm{1}}} \HOLSymConst{\HOLTokenConj{}} \HOLConst{WG} \HOLBoundVar{e\sb{\mathrm{2}}}) \HOLSymConst{\HOLTokenDisj{}}
     (\HOLSymConst{\HOLTokenExists{}}\HOLBoundVar{L} \HOLBoundVar{e}. (\HOLBoundVar{a\sb{\mathrm{0}}} \HOLSymConst{=} (\HOLTokenLambda{}\HOLBoundVar{t}. \HOLConst{\ensuremath{\nu}} \HOLBoundVar{L} (\HOLBoundVar{e} \HOLBoundVar{t}))) \HOLSymConst{\HOLTokenConj{}} \HOLConst{WG} \HOLBoundVar{e}) \HOLSymConst{\HOLTokenDisj{}}
     \HOLSymConst{\HOLTokenExists{}}\HOLBoundVar{rf} \HOLBoundVar{e}. (\HOLBoundVar{a\sb{\mathrm{0}}} \HOLSymConst{=} (\HOLTokenLambda{}\HOLBoundVar{t}. \HOLConst{relab} (\HOLBoundVar{e} \HOLBoundVar{t}) \HOLBoundVar{rf})) \HOLSymConst{\HOLTokenConj{}} \HOLConst{WG} \HOLBoundVar{e}
\end{SaveVerbatim}
\newcommand{\HOLCongruenceTheoremsWGXXcases}{\UseVerbatim{HOLCongruenceTheoremsWGXXcases}}
\begin{SaveVerbatim}{HOLCongruenceTheoremsWGXXind}
\HOLTokenTurnstile{} \HOLSymConst{\HOLTokenForall{}}\HOLBoundVar{WG\sp{\prime}}.
     (\HOLSymConst{\HOLTokenForall{}}\HOLBoundVar{p}. \HOLBoundVar{WG\sp{\prime}} (\HOLTokenLambda{}\HOLBoundVar{t}. \HOLBoundVar{p})) \HOLSymConst{\HOLTokenConj{}} (\HOLSymConst{\HOLTokenForall{}}\HOLBoundVar{a} \HOLBoundVar{e}. \HOLConst{CONTEXT} \HOLBoundVar{e} \HOLSymConst{\HOLTokenImp{}} \HOLBoundVar{WG\sp{\prime}} (\HOLTokenLambda{}\HOLBoundVar{t}. \HOLBoundVar{a}\HOLSymConst{..}\HOLBoundVar{e} \HOLBoundVar{t})) \HOLSymConst{\HOLTokenConj{}}
     (\HOLSymConst{\HOLTokenForall{}}\HOLBoundVar{e\sb{\mathrm{1}}} \HOLBoundVar{e\sb{\mathrm{2}}}. \HOLBoundVar{WG\sp{\prime}} \HOLBoundVar{e\sb{\mathrm{1}}} \HOLSymConst{\HOLTokenConj{}} \HOLBoundVar{WG\sp{\prime}} \HOLBoundVar{e\sb{\mathrm{2}}} \HOLSymConst{\HOLTokenImp{}} \HOLBoundVar{WG\sp{\prime}} (\HOLTokenLambda{}\HOLBoundVar{t}. \HOLBoundVar{e\sb{\mathrm{1}}} \HOLBoundVar{t} \HOLSymConst{+} \HOLBoundVar{e\sb{\mathrm{2}}} \HOLBoundVar{t})) \HOLSymConst{\HOLTokenConj{}}
     (\HOLSymConst{\HOLTokenForall{}}\HOLBoundVar{e\sb{\mathrm{1}}} \HOLBoundVar{e\sb{\mathrm{2}}}. \HOLBoundVar{WG\sp{\prime}} \HOLBoundVar{e\sb{\mathrm{1}}} \HOLSymConst{\HOLTokenConj{}} \HOLBoundVar{WG\sp{\prime}} \HOLBoundVar{e\sb{\mathrm{2}}} \HOLSymConst{\HOLTokenImp{}} \HOLBoundVar{WG\sp{\prime}} (\HOLTokenLambda{}\HOLBoundVar{t}. \HOLBoundVar{e\sb{\mathrm{1}}} \HOLBoundVar{t} \HOLSymConst{\ensuremath{\parallel}} \HOLBoundVar{e\sb{\mathrm{2}}} \HOLBoundVar{t})) \HOLSymConst{\HOLTokenConj{}}
     (\HOLSymConst{\HOLTokenForall{}}\HOLBoundVar{L} \HOLBoundVar{e}. \HOLBoundVar{WG\sp{\prime}} \HOLBoundVar{e} \HOLSymConst{\HOLTokenImp{}} \HOLBoundVar{WG\sp{\prime}} (\HOLTokenLambda{}\HOLBoundVar{t}. \HOLConst{\ensuremath{\nu}} \HOLBoundVar{L} (\HOLBoundVar{e} \HOLBoundVar{t}))) \HOLSymConst{\HOLTokenConj{}}
     (\HOLSymConst{\HOLTokenForall{}}\HOLBoundVar{rf} \HOLBoundVar{e}. \HOLBoundVar{WG\sp{\prime}} \HOLBoundVar{e} \HOLSymConst{\HOLTokenImp{}} \HOLBoundVar{WG\sp{\prime}} (\HOLTokenLambda{}\HOLBoundVar{t}. \HOLConst{relab} (\HOLBoundVar{e} \HOLBoundVar{t}) \HOLBoundVar{rf})) \HOLSymConst{\HOLTokenImp{}}
     \HOLSymConst{\HOLTokenForall{}}\HOLBoundVar{a\sb{\mathrm{0}}}. \HOLConst{WG} \HOLBoundVar{a\sb{\mathrm{0}}} \HOLSymConst{\HOLTokenImp{}} \HOLBoundVar{WG\sp{\prime}} \HOLBoundVar{a\sb{\mathrm{0}}}
\end{SaveVerbatim}
\newcommand{\HOLCongruenceTheoremsWGXXind}{\UseVerbatim{HOLCongruenceTheoremsWGXXind}}
\begin{SaveVerbatim}{HOLCongruenceTheoremsWGXXISXXCONTEXT}
\HOLTokenTurnstile{} \HOLSymConst{\HOLTokenForall{}}\HOLBoundVar{e}. \HOLConst{WG} \HOLBoundVar{e} \HOLSymConst{\HOLTokenImp{}} \HOLConst{CONTEXT} \HOLBoundVar{e}
\end{SaveVerbatim}
\newcommand{\HOLCongruenceTheoremsWGXXISXXCONTEXT}{\UseVerbatim{HOLCongruenceTheoremsWGXXISXXCONTEXT}}
\begin{SaveVerbatim}{HOLCongruenceTheoremsWGXXrules}
\HOLTokenTurnstile{} (\HOLSymConst{\HOLTokenForall{}}\HOLBoundVar{p}. \HOLConst{WG} (\HOLTokenLambda{}\HOLBoundVar{t}. \HOLBoundVar{p})) \HOLSymConst{\HOLTokenConj{}} (\HOLSymConst{\HOLTokenForall{}}\HOLBoundVar{a} \HOLBoundVar{e}. \HOLConst{CONTEXT} \HOLBoundVar{e} \HOLSymConst{\HOLTokenImp{}} \HOLConst{WG} (\HOLTokenLambda{}\HOLBoundVar{t}. \HOLBoundVar{a}\HOLSymConst{..}\HOLBoundVar{e} \HOLBoundVar{t})) \HOLSymConst{\HOLTokenConj{}}
   (\HOLSymConst{\HOLTokenForall{}}\HOLBoundVar{e\sb{\mathrm{1}}} \HOLBoundVar{e\sb{\mathrm{2}}}. \HOLConst{WG} \HOLBoundVar{e\sb{\mathrm{1}}} \HOLSymConst{\HOLTokenConj{}} \HOLConst{WG} \HOLBoundVar{e\sb{\mathrm{2}}} \HOLSymConst{\HOLTokenImp{}} \HOLConst{WG} (\HOLTokenLambda{}\HOLBoundVar{t}. \HOLBoundVar{e\sb{\mathrm{1}}} \HOLBoundVar{t} \HOLSymConst{+} \HOLBoundVar{e\sb{\mathrm{2}}} \HOLBoundVar{t})) \HOLSymConst{\HOLTokenConj{}}
   (\HOLSymConst{\HOLTokenForall{}}\HOLBoundVar{e\sb{\mathrm{1}}} \HOLBoundVar{e\sb{\mathrm{2}}}. \HOLConst{WG} \HOLBoundVar{e\sb{\mathrm{1}}} \HOLSymConst{\HOLTokenConj{}} \HOLConst{WG} \HOLBoundVar{e\sb{\mathrm{2}}} \HOLSymConst{\HOLTokenImp{}} \HOLConst{WG} (\HOLTokenLambda{}\HOLBoundVar{t}. \HOLBoundVar{e\sb{\mathrm{1}}} \HOLBoundVar{t} \HOLSymConst{\ensuremath{\parallel}} \HOLBoundVar{e\sb{\mathrm{2}}} \HOLBoundVar{t})) \HOLSymConst{\HOLTokenConj{}}
   (\HOLSymConst{\HOLTokenForall{}}\HOLBoundVar{L} \HOLBoundVar{e}. \HOLConst{WG} \HOLBoundVar{e} \HOLSymConst{\HOLTokenImp{}} \HOLConst{WG} (\HOLTokenLambda{}\HOLBoundVar{t}. \HOLConst{\ensuremath{\nu}} \HOLBoundVar{L} (\HOLBoundVar{e} \HOLBoundVar{t}))) \HOLSymConst{\HOLTokenConj{}}
   \HOLSymConst{\HOLTokenForall{}}\HOLBoundVar{rf} \HOLBoundVar{e}. \HOLConst{WG} \HOLBoundVar{e} \HOLSymConst{\HOLTokenImp{}} \HOLConst{WG} (\HOLTokenLambda{}\HOLBoundVar{t}. \HOLConst{relab} (\HOLBoundVar{e} \HOLBoundVar{t}) \HOLBoundVar{rf})
\end{SaveVerbatim}
\newcommand{\HOLCongruenceTheoremsWGXXrules}{\UseVerbatim{HOLCongruenceTheoremsWGXXrules}}
\begin{SaveVerbatim}{HOLCongruenceTheoremsWGXXstrongind}
\HOLTokenTurnstile{} \HOLSymConst{\HOLTokenForall{}}\HOLBoundVar{WG\sp{\prime}}.
     (\HOLSymConst{\HOLTokenForall{}}\HOLBoundVar{p}. \HOLBoundVar{WG\sp{\prime}} (\HOLTokenLambda{}\HOLBoundVar{t}. \HOLBoundVar{p})) \HOLSymConst{\HOLTokenConj{}} (\HOLSymConst{\HOLTokenForall{}}\HOLBoundVar{a} \HOLBoundVar{e}. \HOLConst{CONTEXT} \HOLBoundVar{e} \HOLSymConst{\HOLTokenImp{}} \HOLBoundVar{WG\sp{\prime}} (\HOLTokenLambda{}\HOLBoundVar{t}. \HOLBoundVar{a}\HOLSymConst{..}\HOLBoundVar{e} \HOLBoundVar{t})) \HOLSymConst{\HOLTokenConj{}}
     (\HOLSymConst{\HOLTokenForall{}}\HOLBoundVar{e\sb{\mathrm{1}}} \HOLBoundVar{e\sb{\mathrm{2}}}.
        \HOLConst{WG} \HOLBoundVar{e\sb{\mathrm{1}}} \HOLSymConst{\HOLTokenConj{}} \HOLBoundVar{WG\sp{\prime}} \HOLBoundVar{e\sb{\mathrm{1}}} \HOLSymConst{\HOLTokenConj{}} \HOLConst{WG} \HOLBoundVar{e\sb{\mathrm{2}}} \HOLSymConst{\HOLTokenConj{}} \HOLBoundVar{WG\sp{\prime}} \HOLBoundVar{e\sb{\mathrm{2}}} \HOLSymConst{\HOLTokenImp{}}
        \HOLBoundVar{WG\sp{\prime}} (\HOLTokenLambda{}\HOLBoundVar{t}. \HOLBoundVar{e\sb{\mathrm{1}}} \HOLBoundVar{t} \HOLSymConst{+} \HOLBoundVar{e\sb{\mathrm{2}}} \HOLBoundVar{t})) \HOLSymConst{\HOLTokenConj{}}
     (\HOLSymConst{\HOLTokenForall{}}\HOLBoundVar{e\sb{\mathrm{1}}} \HOLBoundVar{e\sb{\mathrm{2}}}.
        \HOLConst{WG} \HOLBoundVar{e\sb{\mathrm{1}}} \HOLSymConst{\HOLTokenConj{}} \HOLBoundVar{WG\sp{\prime}} \HOLBoundVar{e\sb{\mathrm{1}}} \HOLSymConst{\HOLTokenConj{}} \HOLConst{WG} \HOLBoundVar{e\sb{\mathrm{2}}} \HOLSymConst{\HOLTokenConj{}} \HOLBoundVar{WG\sp{\prime}} \HOLBoundVar{e\sb{\mathrm{2}}} \HOLSymConst{\HOLTokenImp{}}
        \HOLBoundVar{WG\sp{\prime}} (\HOLTokenLambda{}\HOLBoundVar{t}. \HOLBoundVar{e\sb{\mathrm{1}}} \HOLBoundVar{t} \HOLSymConst{\ensuremath{\parallel}} \HOLBoundVar{e\sb{\mathrm{2}}} \HOLBoundVar{t})) \HOLSymConst{\HOLTokenConj{}}
     (\HOLSymConst{\HOLTokenForall{}}\HOLBoundVar{L} \HOLBoundVar{e}. \HOLConst{WG} \HOLBoundVar{e} \HOLSymConst{\HOLTokenConj{}} \HOLBoundVar{WG\sp{\prime}} \HOLBoundVar{e} \HOLSymConst{\HOLTokenImp{}} \HOLBoundVar{WG\sp{\prime}} (\HOLTokenLambda{}\HOLBoundVar{t}. \HOLConst{\ensuremath{\nu}} \HOLBoundVar{L} (\HOLBoundVar{e} \HOLBoundVar{t}))) \HOLSymConst{\HOLTokenConj{}}
     (\HOLSymConst{\HOLTokenForall{}}\HOLBoundVar{rf} \HOLBoundVar{e}. \HOLConst{WG} \HOLBoundVar{e} \HOLSymConst{\HOLTokenConj{}} \HOLBoundVar{WG\sp{\prime}} \HOLBoundVar{e} \HOLSymConst{\HOLTokenImp{}} \HOLBoundVar{WG\sp{\prime}} (\HOLTokenLambda{}\HOLBoundVar{t}. \HOLConst{relab} (\HOLBoundVar{e} \HOLBoundVar{t}) \HOLBoundVar{rf})) \HOLSymConst{\HOLTokenImp{}}
     \HOLSymConst{\HOLTokenForall{}}\HOLBoundVar{a\sb{\mathrm{0}}}. \HOLConst{WG} \HOLBoundVar{a\sb{\mathrm{0}}} \HOLSymConst{\HOLTokenImp{}} \HOLBoundVar{WG\sp{\prime}} \HOLBoundVar{a\sb{\mathrm{0}}}
\end{SaveVerbatim}
\newcommand{\HOLCongruenceTheoremsWGXXstrongind}{\UseVerbatim{HOLCongruenceTheoremsWGXXstrongind}}
\begin{SaveVerbatim}{HOLCongruenceTheoremsWGSOne}
\HOLTokenTurnstile{} \HOLSymConst{\HOLTokenForall{}}\HOLBoundVar{a}. \HOLConst{WGS} (\HOLTokenLambda{}\HOLBoundVar{t}. \HOLBoundVar{a}\HOLSymConst{..}\HOLBoundVar{t})
\end{SaveVerbatim}
\newcommand{\HOLCongruenceTheoremsWGSOne}{\UseVerbatim{HOLCongruenceTheoremsWGSOne}}
\begin{SaveVerbatim}{HOLCongruenceTheoremsWGSTwo}
\HOLTokenTurnstile{} \HOLSymConst{\HOLTokenForall{}}\HOLBoundVar{p}. \HOLConst{WGS} (\HOLTokenLambda{}\HOLBoundVar{t}. \HOLBoundVar{p})
\end{SaveVerbatim}
\newcommand{\HOLCongruenceTheoremsWGSTwo}{\UseVerbatim{HOLCongruenceTheoremsWGSTwo}}
\begin{SaveVerbatim}{HOLCongruenceTheoremsWGSThree}
\HOLTokenTurnstile{} \HOLSymConst{\HOLTokenForall{}}\HOLBoundVar{a} \HOLBoundVar{e}. \HOLConst{GCONTEXT} \HOLBoundVar{e} \HOLSymConst{\HOLTokenImp{}} \HOLConst{WGS} (\HOLTokenLambda{}\HOLBoundVar{t}. \HOLBoundVar{a}\HOLSymConst{..}\HOLBoundVar{e} \HOLBoundVar{t})
\end{SaveVerbatim}
\newcommand{\HOLCongruenceTheoremsWGSThree}{\UseVerbatim{HOLCongruenceTheoremsWGSThree}}
\begin{SaveVerbatim}{HOLCongruenceTheoremsWGSFour}
\HOLTokenTurnstile{} \HOLSymConst{\HOLTokenForall{}}\HOLBoundVar{a\sb{\mathrm{1}}} \HOLBoundVar{a\sb{\mathrm{2}}} \HOLBoundVar{e\sb{\mathrm{1}}} \HOLBoundVar{e\sb{\mathrm{2}}}.
     \HOLConst{GCONTEXT} \HOLBoundVar{e\sb{\mathrm{1}}} \HOLSymConst{\HOLTokenConj{}} \HOLConst{GCONTEXT} \HOLBoundVar{e\sb{\mathrm{2}}} \HOLSymConst{\HOLTokenImp{}} \HOLConst{WGS} (\HOLTokenLambda{}\HOLBoundVar{t}. \HOLBoundVar{a\sb{\mathrm{1}}}\HOLSymConst{..}\HOLBoundVar{e\sb{\mathrm{1}}} \HOLBoundVar{t} \HOLSymConst{+} \HOLBoundVar{a\sb{\mathrm{2}}}\HOLSymConst{..}\HOLBoundVar{e\sb{\mathrm{2}}} \HOLBoundVar{t})
\end{SaveVerbatim}
\newcommand{\HOLCongruenceTheoremsWGSFour}{\UseVerbatim{HOLCongruenceTheoremsWGSFour}}
\begin{SaveVerbatim}{HOLCongruenceTheoremsWGSFive}
\HOLTokenTurnstile{} \HOLSymConst{\HOLTokenForall{}}\HOLBoundVar{e\sb{\mathrm{1}}} \HOLBoundVar{e\sb{\mathrm{2}}}. \HOLConst{WGS} \HOLBoundVar{e\sb{\mathrm{1}}} \HOLSymConst{\HOLTokenConj{}} \HOLConst{WGS} \HOLBoundVar{e\sb{\mathrm{2}}} \HOLSymConst{\HOLTokenImp{}} \HOLConst{WGS} (\HOLTokenLambda{}\HOLBoundVar{t}. \HOLBoundVar{e\sb{\mathrm{1}}} \HOLBoundVar{t} \HOLSymConst{\ensuremath{\parallel}} \HOLBoundVar{e\sb{\mathrm{2}}} \HOLBoundVar{t})
\end{SaveVerbatim}
\newcommand{\HOLCongruenceTheoremsWGSFive}{\UseVerbatim{HOLCongruenceTheoremsWGSFive}}
\begin{SaveVerbatim}{HOLCongruenceTheoremsWGSSix}
\HOLTokenTurnstile{} \HOLSymConst{\HOLTokenForall{}}\HOLBoundVar{L} \HOLBoundVar{e}. \HOLConst{WGS} \HOLBoundVar{e} \HOLSymConst{\HOLTokenImp{}} \HOLConst{WGS} (\HOLTokenLambda{}\HOLBoundVar{t}. \HOLConst{\ensuremath{\nu}} \HOLBoundVar{L} (\HOLBoundVar{e} \HOLBoundVar{t}))
\end{SaveVerbatim}
\newcommand{\HOLCongruenceTheoremsWGSSix}{\UseVerbatim{HOLCongruenceTheoremsWGSSix}}
\begin{SaveVerbatim}{HOLCongruenceTheoremsWGSSeven}
\HOLTokenTurnstile{} \HOLSymConst{\HOLTokenForall{}}\HOLBoundVar{rf} \HOLBoundVar{e}. \HOLConst{WGS} \HOLBoundVar{e} \HOLSymConst{\HOLTokenImp{}} \HOLConst{WGS} (\HOLTokenLambda{}\HOLBoundVar{t}. \HOLConst{relab} (\HOLBoundVar{e} \HOLBoundVar{t}) \HOLBoundVar{rf})
\end{SaveVerbatim}
\newcommand{\HOLCongruenceTheoremsWGSSeven}{\UseVerbatim{HOLCongruenceTheoremsWGSSeven}}
\begin{SaveVerbatim}{HOLCongruenceTheoremsWGSXXcases}
\HOLTokenTurnstile{} \HOLSymConst{\HOLTokenForall{}}\HOLBoundVar{a\sb{\mathrm{0}}}.
     \HOLConst{WGS} \HOLBoundVar{a\sb{\mathrm{0}}} \HOLSymConst{\HOLTokenEquiv{}}
     (\HOLSymConst{\HOLTokenExists{}}\HOLBoundVar{p}. \HOLBoundVar{a\sb{\mathrm{0}}} \HOLSymConst{=} (\HOLTokenLambda{}\HOLBoundVar{t}. \HOLBoundVar{p})) \HOLSymConst{\HOLTokenDisj{}}
     (\HOLSymConst{\HOLTokenExists{}}\HOLBoundVar{a} \HOLBoundVar{e}. (\HOLBoundVar{a\sb{\mathrm{0}}} \HOLSymConst{=} (\HOLTokenLambda{}\HOLBoundVar{t}. \HOLBoundVar{a}\HOLSymConst{..}\HOLBoundVar{e} \HOLBoundVar{t})) \HOLSymConst{\HOLTokenConj{}} \HOLConst{GCONTEXT} \HOLBoundVar{e}) \HOLSymConst{\HOLTokenDisj{}}
     (\HOLSymConst{\HOLTokenExists{}}\HOLBoundVar{a\sb{\mathrm{1}}} \HOLBoundVar{a\sb{\mathrm{2}}} \HOLBoundVar{e\sb{\mathrm{1}}} \HOLBoundVar{e\sb{\mathrm{2}}}.
        (\HOLBoundVar{a\sb{\mathrm{0}}} \HOLSymConst{=} (\HOLTokenLambda{}\HOLBoundVar{t}. \HOLBoundVar{a\sb{\mathrm{1}}}\HOLSymConst{..}\HOLBoundVar{e\sb{\mathrm{1}}} \HOLBoundVar{t} \HOLSymConst{+} \HOLBoundVar{a\sb{\mathrm{2}}}\HOLSymConst{..}\HOLBoundVar{e\sb{\mathrm{2}}} \HOLBoundVar{t})) \HOLSymConst{\HOLTokenConj{}} \HOLConst{GCONTEXT} \HOLBoundVar{e\sb{\mathrm{1}}} \HOLSymConst{\HOLTokenConj{}}
        \HOLConst{GCONTEXT} \HOLBoundVar{e\sb{\mathrm{2}}}) \HOLSymConst{\HOLTokenDisj{}}
     (\HOLSymConst{\HOLTokenExists{}}\HOLBoundVar{e\sb{\mathrm{1}}} \HOLBoundVar{e\sb{\mathrm{2}}}. (\HOLBoundVar{a\sb{\mathrm{0}}} \HOLSymConst{=} (\HOLTokenLambda{}\HOLBoundVar{t}. \HOLBoundVar{e\sb{\mathrm{1}}} \HOLBoundVar{t} \HOLSymConst{\ensuremath{\parallel}} \HOLBoundVar{e\sb{\mathrm{2}}} \HOLBoundVar{t})) \HOLSymConst{\HOLTokenConj{}} \HOLConst{WGS} \HOLBoundVar{e\sb{\mathrm{1}}} \HOLSymConst{\HOLTokenConj{}} \HOLConst{WGS} \HOLBoundVar{e\sb{\mathrm{2}}}) \HOLSymConst{\HOLTokenDisj{}}
     (\HOLSymConst{\HOLTokenExists{}}\HOLBoundVar{L} \HOLBoundVar{e}. (\HOLBoundVar{a\sb{\mathrm{0}}} \HOLSymConst{=} (\HOLTokenLambda{}\HOLBoundVar{t}. \HOLConst{\ensuremath{\nu}} \HOLBoundVar{L} (\HOLBoundVar{e} \HOLBoundVar{t}))) \HOLSymConst{\HOLTokenConj{}} \HOLConst{WGS} \HOLBoundVar{e}) \HOLSymConst{\HOLTokenDisj{}}
     \HOLSymConst{\HOLTokenExists{}}\HOLBoundVar{rf} \HOLBoundVar{e}. (\HOLBoundVar{a\sb{\mathrm{0}}} \HOLSymConst{=} (\HOLTokenLambda{}\HOLBoundVar{t}. \HOLConst{relab} (\HOLBoundVar{e} \HOLBoundVar{t}) \HOLBoundVar{rf})) \HOLSymConst{\HOLTokenConj{}} \HOLConst{WGS} \HOLBoundVar{e}
\end{SaveVerbatim}
\newcommand{\HOLCongruenceTheoremsWGSXXcases}{\UseVerbatim{HOLCongruenceTheoremsWGSXXcases}}
\begin{SaveVerbatim}{HOLCongruenceTheoremsWGSXXind}
\HOLTokenTurnstile{} \HOLSymConst{\HOLTokenForall{}}\HOLBoundVar{WGS\sp{\prime}}.
     (\HOLSymConst{\HOLTokenForall{}}\HOLBoundVar{p}. \HOLBoundVar{WGS\sp{\prime}} (\HOLTokenLambda{}\HOLBoundVar{t}. \HOLBoundVar{p})) \HOLSymConst{\HOLTokenConj{}}
     (\HOLSymConst{\HOLTokenForall{}}\HOLBoundVar{a} \HOLBoundVar{e}. \HOLConst{GCONTEXT} \HOLBoundVar{e} \HOLSymConst{\HOLTokenImp{}} \HOLBoundVar{WGS\sp{\prime}} (\HOLTokenLambda{}\HOLBoundVar{t}. \HOLBoundVar{a}\HOLSymConst{..}\HOLBoundVar{e} \HOLBoundVar{t})) \HOLSymConst{\HOLTokenConj{}}
     (\HOLSymConst{\HOLTokenForall{}}\HOLBoundVar{a\sb{\mathrm{1}}} \HOLBoundVar{a\sb{\mathrm{2}}} \HOLBoundVar{e\sb{\mathrm{1}}} \HOLBoundVar{e\sb{\mathrm{2}}}.
        \HOLConst{GCONTEXT} \HOLBoundVar{e\sb{\mathrm{1}}} \HOLSymConst{\HOLTokenConj{}} \HOLConst{GCONTEXT} \HOLBoundVar{e\sb{\mathrm{2}}} \HOLSymConst{\HOLTokenImp{}}
        \HOLBoundVar{WGS\sp{\prime}} (\HOLTokenLambda{}\HOLBoundVar{t}. \HOLBoundVar{a\sb{\mathrm{1}}}\HOLSymConst{..}\HOLBoundVar{e\sb{\mathrm{1}}} \HOLBoundVar{t} \HOLSymConst{+} \HOLBoundVar{a\sb{\mathrm{2}}}\HOLSymConst{..}\HOLBoundVar{e\sb{\mathrm{2}}} \HOLBoundVar{t})) \HOLSymConst{\HOLTokenConj{}}
     (\HOLSymConst{\HOLTokenForall{}}\HOLBoundVar{e\sb{\mathrm{1}}} \HOLBoundVar{e\sb{\mathrm{2}}}. \HOLBoundVar{WGS\sp{\prime}} \HOLBoundVar{e\sb{\mathrm{1}}} \HOLSymConst{\HOLTokenConj{}} \HOLBoundVar{WGS\sp{\prime}} \HOLBoundVar{e\sb{\mathrm{2}}} \HOLSymConst{\HOLTokenImp{}} \HOLBoundVar{WGS\sp{\prime}} (\HOLTokenLambda{}\HOLBoundVar{t}. \HOLBoundVar{e\sb{\mathrm{1}}} \HOLBoundVar{t} \HOLSymConst{\ensuremath{\parallel}} \HOLBoundVar{e\sb{\mathrm{2}}} \HOLBoundVar{t})) \HOLSymConst{\HOLTokenConj{}}
     (\HOLSymConst{\HOLTokenForall{}}\HOLBoundVar{L} \HOLBoundVar{e}. \HOLBoundVar{WGS\sp{\prime}} \HOLBoundVar{e} \HOLSymConst{\HOLTokenImp{}} \HOLBoundVar{WGS\sp{\prime}} (\HOLTokenLambda{}\HOLBoundVar{t}. \HOLConst{\ensuremath{\nu}} \HOLBoundVar{L} (\HOLBoundVar{e} \HOLBoundVar{t}))) \HOLSymConst{\HOLTokenConj{}}
     (\HOLSymConst{\HOLTokenForall{}}\HOLBoundVar{rf} \HOLBoundVar{e}. \HOLBoundVar{WGS\sp{\prime}} \HOLBoundVar{e} \HOLSymConst{\HOLTokenImp{}} \HOLBoundVar{WGS\sp{\prime}} (\HOLTokenLambda{}\HOLBoundVar{t}. \HOLConst{relab} (\HOLBoundVar{e} \HOLBoundVar{t}) \HOLBoundVar{rf})) \HOLSymConst{\HOLTokenImp{}}
     \HOLSymConst{\HOLTokenForall{}}\HOLBoundVar{a\sb{\mathrm{0}}}. \HOLConst{WGS} \HOLBoundVar{a\sb{\mathrm{0}}} \HOLSymConst{\HOLTokenImp{}} \HOLBoundVar{WGS\sp{\prime}} \HOLBoundVar{a\sb{\mathrm{0}}}
\end{SaveVerbatim}
\newcommand{\HOLCongruenceTheoremsWGSXXind}{\UseVerbatim{HOLCongruenceTheoremsWGSXXind}}
\begin{SaveVerbatim}{HOLCongruenceTheoremsWGSXXISXXCONTEXT}
\HOLTokenTurnstile{} \HOLSymConst{\HOLTokenForall{}}\HOLBoundVar{e}. \HOLConst{WGS} \HOLBoundVar{e} \HOLSymConst{\HOLTokenImp{}} \HOLConst{CONTEXT} \HOLBoundVar{e}
\end{SaveVerbatim}
\newcommand{\HOLCongruenceTheoremsWGSXXISXXCONTEXT}{\UseVerbatim{HOLCongruenceTheoremsWGSXXISXXCONTEXT}}
\begin{SaveVerbatim}{HOLCongruenceTheoremsWGSXXISXXGCONTEXT}
\HOLTokenTurnstile{} \HOLSymConst{\HOLTokenForall{}}\HOLBoundVar{e}. \HOLConst{WGS} \HOLBoundVar{e} \HOLSymConst{\HOLTokenImp{}} \HOLConst{GCONTEXT} \HOLBoundVar{e}
\end{SaveVerbatim}
\newcommand{\HOLCongruenceTheoremsWGSXXISXXGCONTEXT}{\UseVerbatim{HOLCongruenceTheoremsWGSXXISXXGCONTEXT}}
\begin{SaveVerbatim}{HOLCongruenceTheoremsWGSXXrules}
\HOLTokenTurnstile{} (\HOLSymConst{\HOLTokenForall{}}\HOLBoundVar{p}. \HOLConst{WGS} (\HOLTokenLambda{}\HOLBoundVar{t}. \HOLBoundVar{p})) \HOLSymConst{\HOLTokenConj{}} (\HOLSymConst{\HOLTokenForall{}}\HOLBoundVar{a} \HOLBoundVar{e}. \HOLConst{GCONTEXT} \HOLBoundVar{e} \HOLSymConst{\HOLTokenImp{}} \HOLConst{WGS} (\HOLTokenLambda{}\HOLBoundVar{t}. \HOLBoundVar{a}\HOLSymConst{..}\HOLBoundVar{e} \HOLBoundVar{t})) \HOLSymConst{\HOLTokenConj{}}
   (\HOLSymConst{\HOLTokenForall{}}\HOLBoundVar{a\sb{\mathrm{1}}} \HOLBoundVar{a\sb{\mathrm{2}}} \HOLBoundVar{e\sb{\mathrm{1}}} \HOLBoundVar{e\sb{\mathrm{2}}}.
      \HOLConst{GCONTEXT} \HOLBoundVar{e\sb{\mathrm{1}}} \HOLSymConst{\HOLTokenConj{}} \HOLConst{GCONTEXT} \HOLBoundVar{e\sb{\mathrm{2}}} \HOLSymConst{\HOLTokenImp{}}
      \HOLConst{WGS} (\HOLTokenLambda{}\HOLBoundVar{t}. \HOLBoundVar{a\sb{\mathrm{1}}}\HOLSymConst{..}\HOLBoundVar{e\sb{\mathrm{1}}} \HOLBoundVar{t} \HOLSymConst{+} \HOLBoundVar{a\sb{\mathrm{2}}}\HOLSymConst{..}\HOLBoundVar{e\sb{\mathrm{2}}} \HOLBoundVar{t})) \HOLSymConst{\HOLTokenConj{}}
   (\HOLSymConst{\HOLTokenForall{}}\HOLBoundVar{e\sb{\mathrm{1}}} \HOLBoundVar{e\sb{\mathrm{2}}}. \HOLConst{WGS} \HOLBoundVar{e\sb{\mathrm{1}}} \HOLSymConst{\HOLTokenConj{}} \HOLConst{WGS} \HOLBoundVar{e\sb{\mathrm{2}}} \HOLSymConst{\HOLTokenImp{}} \HOLConst{WGS} (\HOLTokenLambda{}\HOLBoundVar{t}. \HOLBoundVar{e\sb{\mathrm{1}}} \HOLBoundVar{t} \HOLSymConst{\ensuremath{\parallel}} \HOLBoundVar{e\sb{\mathrm{2}}} \HOLBoundVar{t})) \HOLSymConst{\HOLTokenConj{}}
   (\HOLSymConst{\HOLTokenForall{}}\HOLBoundVar{L} \HOLBoundVar{e}. \HOLConst{WGS} \HOLBoundVar{e} \HOLSymConst{\HOLTokenImp{}} \HOLConst{WGS} (\HOLTokenLambda{}\HOLBoundVar{t}. \HOLConst{\ensuremath{\nu}} \HOLBoundVar{L} (\HOLBoundVar{e} \HOLBoundVar{t}))) \HOLSymConst{\HOLTokenConj{}}
   \HOLSymConst{\HOLTokenForall{}}\HOLBoundVar{rf} \HOLBoundVar{e}. \HOLConst{WGS} \HOLBoundVar{e} \HOLSymConst{\HOLTokenImp{}} \HOLConst{WGS} (\HOLTokenLambda{}\HOLBoundVar{t}. \HOLConst{relab} (\HOLBoundVar{e} \HOLBoundVar{t}) \HOLBoundVar{rf})
\end{SaveVerbatim}
\newcommand{\HOLCongruenceTheoremsWGSXXrules}{\UseVerbatim{HOLCongruenceTheoremsWGSXXrules}}
\begin{SaveVerbatim}{HOLCongruenceTheoremsWGSXXstrongind}
\HOLTokenTurnstile{} \HOLSymConst{\HOLTokenForall{}}\HOLBoundVar{WGS\sp{\prime}}.
     (\HOLSymConst{\HOLTokenForall{}}\HOLBoundVar{p}. \HOLBoundVar{WGS\sp{\prime}} (\HOLTokenLambda{}\HOLBoundVar{t}. \HOLBoundVar{p})) \HOLSymConst{\HOLTokenConj{}}
     (\HOLSymConst{\HOLTokenForall{}}\HOLBoundVar{a} \HOLBoundVar{e}. \HOLConst{GCONTEXT} \HOLBoundVar{e} \HOLSymConst{\HOLTokenImp{}} \HOLBoundVar{WGS\sp{\prime}} (\HOLTokenLambda{}\HOLBoundVar{t}. \HOLBoundVar{a}\HOLSymConst{..}\HOLBoundVar{e} \HOLBoundVar{t})) \HOLSymConst{\HOLTokenConj{}}
     (\HOLSymConst{\HOLTokenForall{}}\HOLBoundVar{a\sb{\mathrm{1}}} \HOLBoundVar{a\sb{\mathrm{2}}} \HOLBoundVar{e\sb{\mathrm{1}}} \HOLBoundVar{e\sb{\mathrm{2}}}.
        \HOLConst{GCONTEXT} \HOLBoundVar{e\sb{\mathrm{1}}} \HOLSymConst{\HOLTokenConj{}} \HOLConst{GCONTEXT} \HOLBoundVar{e\sb{\mathrm{2}}} \HOLSymConst{\HOLTokenImp{}}
        \HOLBoundVar{WGS\sp{\prime}} (\HOLTokenLambda{}\HOLBoundVar{t}. \HOLBoundVar{a\sb{\mathrm{1}}}\HOLSymConst{..}\HOLBoundVar{e\sb{\mathrm{1}}} \HOLBoundVar{t} \HOLSymConst{+} \HOLBoundVar{a\sb{\mathrm{2}}}\HOLSymConst{..}\HOLBoundVar{e\sb{\mathrm{2}}} \HOLBoundVar{t})) \HOLSymConst{\HOLTokenConj{}}
     (\HOLSymConst{\HOLTokenForall{}}\HOLBoundVar{e\sb{\mathrm{1}}} \HOLBoundVar{e\sb{\mathrm{2}}}.
        \HOLConst{WGS} \HOLBoundVar{e\sb{\mathrm{1}}} \HOLSymConst{\HOLTokenConj{}} \HOLBoundVar{WGS\sp{\prime}} \HOLBoundVar{e\sb{\mathrm{1}}} \HOLSymConst{\HOLTokenConj{}} \HOLConst{WGS} \HOLBoundVar{e\sb{\mathrm{2}}} \HOLSymConst{\HOLTokenConj{}} \HOLBoundVar{WGS\sp{\prime}} \HOLBoundVar{e\sb{\mathrm{2}}} \HOLSymConst{\HOLTokenImp{}}
        \HOLBoundVar{WGS\sp{\prime}} (\HOLTokenLambda{}\HOLBoundVar{t}. \HOLBoundVar{e\sb{\mathrm{1}}} \HOLBoundVar{t} \HOLSymConst{\ensuremath{\parallel}} \HOLBoundVar{e\sb{\mathrm{2}}} \HOLBoundVar{t})) \HOLSymConst{\HOLTokenConj{}}
     (\HOLSymConst{\HOLTokenForall{}}\HOLBoundVar{L} \HOLBoundVar{e}. \HOLConst{WGS} \HOLBoundVar{e} \HOLSymConst{\HOLTokenConj{}} \HOLBoundVar{WGS\sp{\prime}} \HOLBoundVar{e} \HOLSymConst{\HOLTokenImp{}} \HOLBoundVar{WGS\sp{\prime}} (\HOLTokenLambda{}\HOLBoundVar{t}. \HOLConst{\ensuremath{\nu}} \HOLBoundVar{L} (\HOLBoundVar{e} \HOLBoundVar{t}))) \HOLSymConst{\HOLTokenConj{}}
     (\HOLSymConst{\HOLTokenForall{}}\HOLBoundVar{rf} \HOLBoundVar{e}. \HOLConst{WGS} \HOLBoundVar{e} \HOLSymConst{\HOLTokenConj{}} \HOLBoundVar{WGS\sp{\prime}} \HOLBoundVar{e} \HOLSymConst{\HOLTokenImp{}} \HOLBoundVar{WGS\sp{\prime}} (\HOLTokenLambda{}\HOLBoundVar{t}. \HOLConst{relab} (\HOLBoundVar{e} \HOLBoundVar{t}) \HOLBoundVar{rf})) \HOLSymConst{\HOLTokenImp{}}
     \HOLSymConst{\HOLTokenForall{}}\HOLBoundVar{a\sb{\mathrm{0}}}. \HOLConst{WGS} \HOLBoundVar{a\sb{\mathrm{0}}} \HOLSymConst{\HOLTokenImp{}} \HOLBoundVar{WGS\sp{\prime}} \HOLBoundVar{a\sb{\mathrm{0}}}
\end{SaveVerbatim}
\newcommand{\HOLCongruenceTheoremsWGSXXstrongind}{\UseVerbatim{HOLCongruenceTheoremsWGSXXstrongind}}
\newcommand{\HOLCongruenceTheorems}{
\HOLThmTag{Congruence}{CC_congruence}\HOLCongruenceTheoremsCCXXcongruence
\HOLThmTag{Congruence}{CC_is_coarsest}\HOLCongruenceTheoremsCCXXisXXcoarsest
\HOLThmTag{Congruence}{CC_is_coarsest'}\HOLCongruenceTheoremsCCXXisXXcoarsestYY
\HOLThmTag{Congruence}{CC_is_finer}\HOLCongruenceTheoremsCCXXisXXfiner
\HOLThmTag{Congruence}{CC_precongruence}\HOLCongruenceTheoremsCCXXprecongruence
\HOLThmTag{Congruence}{CONTEXT1}\HOLCongruenceTheoremsCONTEXTOne
\HOLThmTag{Congruence}{CONTEXT2}\HOLCongruenceTheoremsCONTEXTTwo
\HOLThmTag{Congruence}{CONTEXT3}\HOLCongruenceTheoremsCONTEXTThree
\HOLThmTag{Congruence}{CONTEXT3a}\HOLCongruenceTheoremsCONTEXTThreea
\HOLThmTag{Congruence}{CONTEXT4}\HOLCongruenceTheoremsCONTEXTFour
\HOLThmTag{Congruence}{CONTEXT5}\HOLCongruenceTheoremsCONTEXTFive
\HOLThmTag{Congruence}{CONTEXT6}\HOLCongruenceTheoremsCONTEXTSix
\HOLThmTag{Congruence}{CONTEXT7}\HOLCongruenceTheoremsCONTEXTSeven
\HOLThmTag{Congruence}{CONTEXT_cases}\HOLCongruenceTheoremsCONTEXTXXcases
\HOLThmTag{Congruence}{CONTEXT_combin}\HOLCongruenceTheoremsCONTEXTXXcombin
\HOLThmTag{Congruence}{CONTEXT_ind}\HOLCongruenceTheoremsCONTEXTXXind
\HOLThmTag{Congruence}{CONTEXT_rules}\HOLCongruenceTheoremsCONTEXTXXrules
\HOLThmTag{Congruence}{CONTEXT_strongind}\HOLCongruenceTheoremsCONTEXTXXstrongind
\HOLThmTag{Congruence}{CONTEXT_WG_combin}\HOLCongruenceTheoremsCONTEXTXXWGXXcombin
\HOLThmTag{Congruence}{GCONTEXT1}\HOLCongruenceTheoremsGCONTEXTOne
\HOLThmTag{Congruence}{GCONTEXT2}\HOLCongruenceTheoremsGCONTEXTTwo
\HOLThmTag{Congruence}{GCONTEXT3}\HOLCongruenceTheoremsGCONTEXTThree
\HOLThmTag{Congruence}{GCONTEXT3a}\HOLCongruenceTheoremsGCONTEXTThreea
\HOLThmTag{Congruence}{GCONTEXT4}\HOLCongruenceTheoremsGCONTEXTFour
\HOLThmTag{Congruence}{GCONTEXT5}\HOLCongruenceTheoremsGCONTEXTFive
\HOLThmTag{Congruence}{GCONTEXT6}\HOLCongruenceTheoremsGCONTEXTSix
\HOLThmTag{Congruence}{GCONTEXT7}\HOLCongruenceTheoremsGCONTEXTSeven
\HOLThmTag{Congruence}{GCONTEXT_cases}\HOLCongruenceTheoremsGCONTEXTXXcases
\HOLThmTag{Congruence}{GCONTEXT_combin}\HOLCongruenceTheoremsGCONTEXTXXcombin
\HOLThmTag{Congruence}{GCONTEXT_ind}\HOLCongruenceTheoremsGCONTEXTXXind
\HOLThmTag{Congruence}{GCONTEXT_IS_CONTEXT}\HOLCongruenceTheoremsGCONTEXTXXISXXCONTEXT
\HOLThmTag{Congruence}{GCONTEXT_rules}\HOLCongruenceTheoremsGCONTEXTXXrules
\HOLThmTag{Congruence}{GCONTEXT_strongind}\HOLCongruenceTheoremsGCONTEXTXXstrongind
\HOLThmTag{Congruence}{GCONTEXT_WGS_combin}\HOLCongruenceTheoremsGCONTEXTXXWGSXXcombin
\HOLThmTag{Congruence}{GSEQ1}\HOLCongruenceTheoremsGSEQOne
\HOLThmTag{Congruence}{GSEQ2}\HOLCongruenceTheoremsGSEQTwo
\HOLThmTag{Congruence}{GSEQ3}\HOLCongruenceTheoremsGSEQThree
\HOLThmTag{Congruence}{GSEQ3a}\HOLCongruenceTheoremsGSEQThreea
\HOLThmTag{Congruence}{GSEQ4}\HOLCongruenceTheoremsGSEQFour
\HOLThmTag{Congruence}{GSEQ_cases}\HOLCongruenceTheoremsGSEQXXcases
\HOLThmTag{Congruence}{GSEQ_combin}\HOLCongruenceTheoremsGSEQXXcombin
\HOLThmTag{Congruence}{GSEQ_ind}\HOLCongruenceTheoremsGSEQXXind
\HOLThmTag{Congruence}{GSEQ_IS_CONTEXT}\HOLCongruenceTheoremsGSEQXXISXXCONTEXT
\HOLThmTag{Congruence}{GSEQ_rules}\HOLCongruenceTheoremsGSEQXXrules
\HOLThmTag{Congruence}{GSEQ_strongind}\HOLCongruenceTheoremsGSEQXXstrongind
\HOLThmTag{Congruence}{OBS_CONGR_congruence}\HOLCongruenceTheoremsOBSXXCONGRXXcongruence
\HOLThmTag{Congruence}{OBS_CONGR_SUBST_CONTEXT}\HOLCongruenceTheoremsOBSXXCONGRXXSUBSTXXCONTEXT
\HOLThmTag{Congruence}{OBS_CONGR_SUBST_SEQ}\HOLCongruenceTheoremsOBSXXCONGRXXSUBSTXXSEQ
\HOLThmTag{Congruence}{OH_CONTEXT1}\HOLCongruenceTheoremsOHXXCONTEXTOne
\HOLThmTag{Congruence}{OH_CONTEXT2}\HOLCongruenceTheoremsOHXXCONTEXTTwo
\HOLThmTag{Congruence}{OH_CONTEXT3}\HOLCongruenceTheoremsOHXXCONTEXTThree
\HOLThmTag{Congruence}{OH_CONTEXT4}\HOLCongruenceTheoremsOHXXCONTEXTFour
\HOLThmTag{Congruence}{OH_CONTEXT5}\HOLCongruenceTheoremsOHXXCONTEXTFive
\HOLThmTag{Congruence}{OH_CONTEXT6}\HOLCongruenceTheoremsOHXXCONTEXTSix
\HOLThmTag{Congruence}{OH_CONTEXT7}\HOLCongruenceTheoremsOHXXCONTEXTSeven
\HOLThmTag{Congruence}{OH_CONTEXT8}\HOLCongruenceTheoremsOHXXCONTEXTEight
\HOLThmTag{Congruence}{OH_CONTEXT_cases}\HOLCongruenceTheoremsOHXXCONTEXTXXcases
\HOLThmTag{Congruence}{OH_CONTEXT_combin}\HOLCongruenceTheoremsOHXXCONTEXTXXcombin
\HOLThmTag{Congruence}{OH_CONTEXT_ind}\HOLCongruenceTheoremsOHXXCONTEXTXXind
\HOLThmTag{Congruence}{OH_CONTEXT_IS_CONTEXT}\HOLCongruenceTheoremsOHXXCONTEXTXXISXXCONTEXT
\HOLThmTag{Congruence}{OH_CONTEXT_rules}\HOLCongruenceTheoremsOHXXCONTEXTXXrules
\HOLThmTag{Congruence}{OH_CONTEXT_strongind}\HOLCongruenceTheoremsOHXXCONTEXTXXstrongind
\HOLThmTag{Congruence}{SEQ1}\HOLCongruenceTheoremsSEQOne
\HOLThmTag{Congruence}{SEQ2}\HOLCongruenceTheoremsSEQTwo
\HOLThmTag{Congruence}{SEQ3}\HOLCongruenceTheoremsSEQThree
\HOLThmTag{Congruence}{SEQ3a}\HOLCongruenceTheoremsSEQThreea
\HOLThmTag{Congruence}{SEQ4}\HOLCongruenceTheoremsSEQFour
\HOLThmTag{Congruence}{SEQ_cases}\HOLCongruenceTheoremsSEQXXcases
\HOLThmTag{Congruence}{SEQ_combin}\HOLCongruenceTheoremsSEQXXcombin
\HOLThmTag{Congruence}{SEQ_ind}\HOLCongruenceTheoremsSEQXXind
\HOLThmTag{Congruence}{SEQ_IS_CONTEXT}\HOLCongruenceTheoremsSEQXXISXXCONTEXT
\HOLThmTag{Congruence}{SEQ_rules}\HOLCongruenceTheoremsSEQXXrules
\HOLThmTag{Congruence}{SEQ_strongind}\HOLCongruenceTheoremsSEQXXstrongind
\HOLThmTag{Congruence}{SG1}\HOLCongruenceTheoremsSGOne
\HOLThmTag{Congruence}{SG10}\HOLCongruenceTheoremsSGOneZero
\HOLThmTag{Congruence}{SG11}\HOLCongruenceTheoremsSGOneOne
\HOLThmTag{Congruence}{SG11'}\HOLCongruenceTheoremsSGOneOneYY
\HOLThmTag{Congruence}{SG2}\HOLCongruenceTheoremsSGTwo
\HOLThmTag{Congruence}{SG3}\HOLCongruenceTheoremsSGThree
\HOLThmTag{Congruence}{SG4}\HOLCongruenceTheoremsSGFour
\HOLThmTag{Congruence}{SG5}\HOLCongruenceTheoremsSGFive
\HOLThmTag{Congruence}{SG6}\HOLCongruenceTheoremsSGSix
\HOLThmTag{Congruence}{SG7}\HOLCongruenceTheoremsSGSeven
\HOLThmTag{Congruence}{SG8}\HOLCongruenceTheoremsSGEight
\HOLThmTag{Congruence}{SG9}\HOLCongruenceTheoremsSGNine
\HOLThmTag{Congruence}{SG_cases}\HOLCongruenceTheoremsSGXXcases
\HOLThmTag{Congruence}{SG_GSEQ_combin}\HOLCongruenceTheoremsSGXXGSEQXXcombin
\HOLThmTag{Congruence}{SG_GSEQ_strong_induction}\HOLCongruenceTheoremsSGXXGSEQXXstrongXXinduction
\HOLThmTag{Congruence}{SG_IMP_WG}\HOLCongruenceTheoremsSGXXIMPXXWG
\HOLThmTag{Congruence}{SG_ind}\HOLCongruenceTheoremsSGXXind
\HOLThmTag{Congruence}{SG_IS_CONTEXT}\HOLCongruenceTheoremsSGXXISXXCONTEXT
\HOLThmTag{Congruence}{SG_rules}\HOLCongruenceTheoremsSGXXrules
\HOLThmTag{Congruence}{SG_SEQ_combin}\HOLCongruenceTheoremsSGXXSEQXXcombin
\HOLThmTag{Congruence}{SG_SEQ_strong_induction}\HOLCongruenceTheoremsSGXXSEQXXstrongXXinduction
\HOLThmTag{Congruence}{SG_strongind}\HOLCongruenceTheoremsSGXXstrongind
\HOLThmTag{Congruence}{STRONG_EQUIV_congruence}\HOLCongruenceTheoremsSTRONGXXEQUIVXXcongruence
\HOLThmTag{Congruence}{STRONG_EQUIV_SUBST_CONTEXT}\HOLCongruenceTheoremsSTRONGXXEQUIVXXSUBSTXXCONTEXT
\HOLThmTag{Congruence}{WEAK_EQUIV_congruence}\HOLCongruenceTheoremsWEAKXXEQUIVXXcongruence
\HOLThmTag{Congruence}{WEAK_EQUIV_SUBST_GCONTEXT}\HOLCongruenceTheoremsWEAKXXEQUIVXXSUBSTXXGCONTEXT
\HOLThmTag{Congruence}{WEAK_EQUIV_SUBST_GSEQ}\HOLCongruenceTheoremsWEAKXXEQUIVXXSUBSTXXGSEQ
\HOLThmTag{Congruence}{WG1}\HOLCongruenceTheoremsWGOne
\HOLThmTag{Congruence}{WG2}\HOLCongruenceTheoremsWGTwo
\HOLThmTag{Congruence}{WG3}\HOLCongruenceTheoremsWGThree
\HOLThmTag{Congruence}{WG4}\HOLCongruenceTheoremsWGFour
\HOLThmTag{Congruence}{WG5}\HOLCongruenceTheoremsWGFive
\HOLThmTag{Congruence}{WG6}\HOLCongruenceTheoremsWGSix
\HOLThmTag{Congruence}{WG7}\HOLCongruenceTheoremsWGSeven
\HOLThmTag{Congruence}{WG_cases}\HOLCongruenceTheoremsWGXXcases
\HOLThmTag{Congruence}{WG_ind}\HOLCongruenceTheoremsWGXXind
\HOLThmTag{Congruence}{WG_IS_CONTEXT}\HOLCongruenceTheoremsWGXXISXXCONTEXT
\HOLThmTag{Congruence}{WG_rules}\HOLCongruenceTheoremsWGXXrules
\HOLThmTag{Congruence}{WG_strongind}\HOLCongruenceTheoremsWGXXstrongind
\HOLThmTag{Congruence}{WGS1}\HOLCongruenceTheoremsWGSOne
\HOLThmTag{Congruence}{WGS2}\HOLCongruenceTheoremsWGSTwo
\HOLThmTag{Congruence}{WGS3}\HOLCongruenceTheoremsWGSThree
\HOLThmTag{Congruence}{WGS4}\HOLCongruenceTheoremsWGSFour
\HOLThmTag{Congruence}{WGS5}\HOLCongruenceTheoremsWGSFive
\HOLThmTag{Congruence}{WGS6}\HOLCongruenceTheoremsWGSSix
\HOLThmTag{Congruence}{WGS7}\HOLCongruenceTheoremsWGSSeven
\HOLThmTag{Congruence}{WGS_cases}\HOLCongruenceTheoremsWGSXXcases
\HOLThmTag{Congruence}{WGS_ind}\HOLCongruenceTheoremsWGSXXind
\HOLThmTag{Congruence}{WGS_IS_CONTEXT}\HOLCongruenceTheoremsWGSXXISXXCONTEXT
\HOLThmTag{Congruence}{WGS_IS_GCONTEXT}\HOLCongruenceTheoremsWGSXXISXXGCONTEXT
\HOLThmTag{Congruence}{WGS_rules}\HOLCongruenceTheoremsWGSXXrules
\HOLThmTag{Congruence}{WGS_strongind}\HOLCongruenceTheoremsWGSXXstrongind
}

\newcommand{\HOLTraceDate}{02 Dicembre 2017}
\newcommand{\HOLTraceTime}{13:31}
\begin{SaveVerbatim}{HOLTraceDefinitionsfiniteXXstateXXdef}
\HOLTokenTurnstile{} \HOLSymConst{\HOLTokenForall{}}\HOLBoundVar{p}. \HOLConst{finite_state} \HOLBoundVar{p} \HOLSymConst{\HOLTokenEquiv{}} \HOLConst{FINITE} (\HOLConst{NODES} \HOLBoundVar{p})
\end{SaveVerbatim}
\newcommand{\HOLTraceDefinitionsfiniteXXstateXXdef}{\UseVerbatim{HOLTraceDefinitionsfiniteXXstateXXdef}}
\begin{SaveVerbatim}{HOLTraceDefinitionsLRTCXXDEF}
\HOLTokenTurnstile{} \HOLSymConst{\HOLTokenForall{}}\HOLBoundVar{R} \HOLBoundVar{a} \HOLBoundVar{l} \HOLBoundVar{b}.
     \HOLConst{LRTC} \HOLBoundVar{R} \HOLBoundVar{a} \HOLBoundVar{l} \HOLBoundVar{b} \HOLSymConst{\HOLTokenEquiv{}}
     \HOLSymConst{\HOLTokenForall{}}\HOLBoundVar{P}.
       (\HOLSymConst{\HOLTokenForall{}}\HOLBoundVar{x}. \HOLBoundVar{P} \HOLBoundVar{x} [] \HOLBoundVar{x}) \HOLSymConst{\HOLTokenConj{}}
       (\HOLSymConst{\HOLTokenForall{}}\HOLBoundVar{x} \HOLBoundVar{h} \HOLBoundVar{y} \HOLBoundVar{t} \HOLBoundVar{z}. \HOLBoundVar{R} \HOLBoundVar{x} \HOLBoundVar{h} \HOLBoundVar{y} \HOLSymConst{\HOLTokenConj{}} \HOLBoundVar{P} \HOLBoundVar{y} \HOLBoundVar{t} \HOLBoundVar{z} \HOLSymConst{\HOLTokenImp{}} \HOLBoundVar{P} \HOLBoundVar{x} (\HOLBoundVar{h}\HOLSymConst{::}\HOLBoundVar{t}) \HOLBoundVar{z}) \HOLSymConst{\HOLTokenImp{}}
       \HOLBoundVar{P} \HOLBoundVar{a} \HOLBoundVar{l} \HOLBoundVar{b}
\end{SaveVerbatim}
\newcommand{\HOLTraceDefinitionsLRTCXXDEF}{\UseVerbatim{HOLTraceDefinitionsLRTCXXDEF}}
\begin{SaveVerbatim}{HOLTraceDefinitionsNOXXLABELXXdef}
\HOLTokenTurnstile{} \HOLSymConst{\HOLTokenForall{}}\HOLBoundVar{L}. \HOLConst{NO_LABEL} \HOLBoundVar{L} \HOLSymConst{\HOLTokenEquiv{}} \HOLSymConst{\HOLTokenNeg{}}\HOLSymConst{\HOLTokenExists{}}\HOLBoundVar{l}. \HOLConst{MEM} (\HOLConst{label} \HOLBoundVar{l}) \HOLBoundVar{L}
\end{SaveVerbatim}
\newcommand{\HOLTraceDefinitionsNOXXLABELXXdef}{\UseVerbatim{HOLTraceDefinitionsNOXXLABELXXdef}}
\begin{SaveVerbatim}{HOLTraceDefinitionsNODESXXdef}
\HOLTokenTurnstile{} \HOLSymConst{\HOLTokenForall{}}\HOLBoundVar{p}. \HOLConst{NODES} \HOLBoundVar{p} \HOLSymConst{=} \HOLTokenLeftbrace{}\HOLBoundVar{q} \HOLTokenBar{} \HOLConst{Reach} \HOLBoundVar{p} \HOLBoundVar{q}\HOLTokenRightbrace{}
\end{SaveVerbatim}
\newcommand{\HOLTraceDefinitionsNODESXXdef}{\UseVerbatim{HOLTraceDefinitionsNODESXXdef}}
\begin{SaveVerbatim}{HOLTraceDefinitionsReachXXdef}
\HOLTokenTurnstile{} \HOLConst{Reach} \HOLSymConst{=} (\HOLTokenLambda{}\HOLBoundVar{E} \HOLBoundVar{E\sp{\prime}}. \HOLSymConst{\HOLTokenExists{}}\HOLBoundVar{u}. \HOLBoundVar{E} \HOLTokenTransBegin\HOLBoundVar{u}\HOLTokenTransEnd \HOLBoundVar{E\sp{\prime}})\HOLSymConst{\HOLTokenSupStar{}}
\end{SaveVerbatim}
\newcommand{\HOLTraceDefinitionsReachXXdef}{\UseVerbatim{HOLTraceDefinitionsReachXXdef}}
\begin{SaveVerbatim}{HOLTraceDefinitionsSTEPXXdef}
\HOLTokenTurnstile{} \HOLSymConst{\HOLTokenForall{}}\HOLBoundVar{P} \HOLBoundVar{n} \HOLBoundVar{Q}. \HOLConst{STEP} \HOLBoundVar{P} \HOLBoundVar{n} \HOLBoundVar{Q} \HOLSymConst{\HOLTokenEquiv{}} \HOLConst{NRC} (\HOLTokenLambda{}\HOLBoundVar{E} \HOLBoundVar{E\sp{\prime}}. \HOLSymConst{\HOLTokenExists{}}\HOLBoundVar{u}. \HOLBoundVar{E} \HOLTokenTransBegin\HOLBoundVar{u}\HOLTokenTransEnd \HOLBoundVar{E\sp{\prime}}) \HOLBoundVar{n} \HOLBoundVar{P} \HOLBoundVar{Q}
\end{SaveVerbatim}
\newcommand{\HOLTraceDefinitionsSTEPXXdef}{\UseVerbatim{HOLTraceDefinitionsSTEPXXdef}}
\begin{SaveVerbatim}{HOLTraceDefinitionsTRACEXXdef}
\HOLTokenTurnstile{} \HOLConst{TRACE} \HOLSymConst{=} \HOLConst{LRTC} \HOLConst{TRANS}
\end{SaveVerbatim}
\newcommand{\HOLTraceDefinitionsTRACEXXdef}{\UseVerbatim{HOLTraceDefinitionsTRACEXXdef}}
\begin{SaveVerbatim}{HOLTraceDefinitionsUNIQUEXXLABELXXdef}
\HOLTokenTurnstile{} \HOLSymConst{\HOLTokenForall{}}\HOLBoundVar{u} \HOLBoundVar{L}.
     \HOLConst{UNIQUE_LABEL} \HOLBoundVar{u} \HOLBoundVar{L} \HOLSymConst{\HOLTokenEquiv{}}
     \HOLSymConst{\HOLTokenExists{}}\HOLBoundVar{L\sb{\mathrm{1}}} \HOLBoundVar{L\sb{\mathrm{2}}}.
       (\HOLBoundVar{L\sb{\mathrm{1}}} \HOLSymConst{++} [\HOLBoundVar{u}] \HOLSymConst{++} \HOLBoundVar{L\sb{\mathrm{2}}} \HOLSymConst{=} \HOLBoundVar{L}) \HOLSymConst{\HOLTokenConj{}}
       \HOLSymConst{\HOLTokenNeg{}}\HOLSymConst{\HOLTokenExists{}}\HOLBoundVar{l}. \HOLConst{MEM} (\HOLConst{label} \HOLBoundVar{l}) \HOLBoundVar{L\sb{\mathrm{1}}} \HOLSymConst{\HOLTokenDisj{}} \HOLConst{MEM} (\HOLConst{label} \HOLBoundVar{l}) \HOLBoundVar{L\sb{\mathrm{2}}}
\end{SaveVerbatim}
\newcommand{\HOLTraceDefinitionsUNIQUEXXLABELXXdef}{\UseVerbatim{HOLTraceDefinitionsUNIQUEXXLABELXXdef}}
\newcommand{\HOLTraceDefinitions}{
\HOLDfnTag{Trace}{finite_state_def}\HOLTraceDefinitionsfiniteXXstateXXdef
\HOLDfnTag{Trace}{LRTC_DEF}\HOLTraceDefinitionsLRTCXXDEF
\HOLDfnTag{Trace}{NO_LABEL_def}\HOLTraceDefinitionsNOXXLABELXXdef
\HOLDfnTag{Trace}{NODES_def}\HOLTraceDefinitionsNODESXXdef
\HOLDfnTag{Trace}{Reach_def}\HOLTraceDefinitionsReachXXdef
\HOLDfnTag{Trace}{STEP_def}\HOLTraceDefinitionsSTEPXXdef
\HOLDfnTag{Trace}{TRACE_def}\HOLTraceDefinitionsTRACEXXdef
\HOLDfnTag{Trace}{UNIQUE_LABEL_def}\HOLTraceDefinitionsUNIQUEXXLABELXXdef
}
\begin{SaveVerbatim}{HOLTraceTheoremsEPSXXANDXXSTEP}
\HOLTokenTurnstile{} \HOLSymConst{\HOLTokenForall{}}\HOLBoundVar{E} \HOLBoundVar{E\sp{\prime}}. \HOLConst{EPS} \HOLBoundVar{E} \HOLBoundVar{E\sp{\prime}} \HOLSymConst{\HOLTokenImp{}} \HOLSymConst{\HOLTokenExists{}}\HOLBoundVar{n}. \HOLConst{STEP} \HOLBoundVar{E} \HOLBoundVar{n} \HOLBoundVar{E\sp{\prime}}
\end{SaveVerbatim}
\newcommand{\HOLTraceTheoremsEPSXXANDXXSTEP}{\UseVerbatim{HOLTraceTheoremsEPSXXANDXXSTEP}}
\begin{SaveVerbatim}{HOLTraceTheoremsEPSXXANDXXTRACE}
\HOLTokenTurnstile{} \HOLSymConst{\HOLTokenForall{}}\HOLBoundVar{E} \HOLBoundVar{E\sp{\prime}}. \HOLConst{EPS} \HOLBoundVar{E} \HOLBoundVar{E\sp{\prime}} \HOLSymConst{\HOLTokenEquiv{}} \HOLSymConst{\HOLTokenExists{}}\HOLBoundVar{xs}. \HOLConst{TRACE} \HOLBoundVar{E} \HOLBoundVar{xs} \HOLBoundVar{E\sp{\prime}} \HOLSymConst{\HOLTokenConj{}} \HOLConst{NO_LABEL} \HOLBoundVar{xs}
\end{SaveVerbatim}
\newcommand{\HOLTraceTheoremsEPSXXANDXXTRACE}{\UseVerbatim{HOLTraceTheoremsEPSXXANDXXTRACE}}
\begin{SaveVerbatim}{HOLTraceTheoremsEPSXXINXXNODES}
\HOLTokenTurnstile{} \HOLSymConst{\HOLTokenForall{}}\HOLBoundVar{p} \HOLBoundVar{q}. \HOLConst{EPS} \HOLBoundVar{p} \HOLBoundVar{q} \HOLSymConst{\HOLTokenImp{}} \HOLBoundVar{q} \HOLConst{\HOLTokenIn{}} \HOLConst{NODES} \HOLBoundVar{p}
\end{SaveVerbatim}
\newcommand{\HOLTraceTheoremsEPSXXINXXNODES}{\UseVerbatim{HOLTraceTheoremsEPSXXINXXNODES}}
\begin{SaveVerbatim}{HOLTraceTheoremsEPSXXReach}
\HOLTokenTurnstile{} \HOLSymConst{\HOLTokenForall{}}\HOLBoundVar{p} \HOLBoundVar{q}. \HOLConst{EPS} \HOLBoundVar{p} \HOLBoundVar{q} \HOLSymConst{\HOLTokenImp{}} \HOLConst{Reach} \HOLBoundVar{p} \HOLBoundVar{q}
\end{SaveVerbatim}
\newcommand{\HOLTraceTheoremsEPSXXReach}{\UseVerbatim{HOLTraceTheoremsEPSXXReach}}
\begin{SaveVerbatim}{HOLTraceTheoremsEPSXXTRACE}
\HOLTokenTurnstile{} \HOLSymConst{\HOLTokenForall{}}\HOLBoundVar{E} \HOLBoundVar{E\sp{\prime}}. \HOLConst{EPS} \HOLBoundVar{E} \HOLBoundVar{E\sp{\prime}} \HOLSymConst{\HOLTokenImp{}} \HOLSymConst{\HOLTokenExists{}}\HOLBoundVar{xs}. \HOLConst{TRACE} \HOLBoundVar{E} \HOLBoundVar{xs} \HOLBoundVar{E\sp{\prime}}
\end{SaveVerbatim}
\newcommand{\HOLTraceTheoremsEPSXXTRACE}{\UseVerbatim{HOLTraceTheoremsEPSXXTRACE}}
\begin{SaveVerbatim}{HOLTraceTheoremsLRTCXXAPPENDXXCASES}
\HOLTokenTurnstile{} \HOLSymConst{\HOLTokenForall{}}\HOLBoundVar{R} \HOLBoundVar{l\sb{\mathrm{1}}} \HOLBoundVar{l\sb{\mathrm{2}}} \HOLBoundVar{x} \HOLBoundVar{y}.
     \HOLConst{LRTC} \HOLBoundVar{R} \HOLBoundVar{x} (\HOLBoundVar{l\sb{\mathrm{1}}} \HOLSymConst{++} \HOLBoundVar{l\sb{\mathrm{2}}}) \HOLBoundVar{y} \HOLSymConst{\HOLTokenEquiv{}} \HOLSymConst{\HOLTokenExists{}}\HOLBoundVar{u}. \HOLConst{LRTC} \HOLBoundVar{R} \HOLBoundVar{x} \HOLBoundVar{l\sb{\mathrm{1}}} \HOLBoundVar{u} \HOLSymConst{\HOLTokenConj{}} \HOLConst{LRTC} \HOLBoundVar{R} \HOLBoundVar{u} \HOLBoundVar{l\sb{\mathrm{2}}} \HOLBoundVar{y}
\end{SaveVerbatim}
\newcommand{\HOLTraceTheoremsLRTCXXAPPENDXXCASES}{\UseVerbatim{HOLTraceTheoremsLRTCXXAPPENDXXCASES}}
\begin{SaveVerbatim}{HOLTraceTheoremsLRTCXXCASESOne}
\HOLTokenTurnstile{} \HOLSymConst{\HOLTokenForall{}}\HOLBoundVar{R} \HOLBoundVar{x} \HOLBoundVar{l} \HOLBoundVar{y}.
     \HOLConst{LRTC} \HOLBoundVar{R} \HOLBoundVar{x} \HOLBoundVar{l} \HOLBoundVar{y} \HOLSymConst{\HOLTokenEquiv{}}
     \HOLKeyword{if} \HOLConst{NULL} \HOLBoundVar{l} \HOLKeyword{then} \HOLBoundVar{x} \HOLSymConst{=} \HOLBoundVar{y}
     \HOLKeyword{else} \HOLSymConst{\HOLTokenExists{}}\HOLBoundVar{u}. \HOLBoundVar{R} \HOLBoundVar{x} (\HOLConst{HD} \HOLBoundVar{l}) \HOLBoundVar{u} \HOLSymConst{\HOLTokenConj{}} \HOLConst{LRTC} \HOLBoundVar{R} \HOLBoundVar{u} (\HOLConst{TL} \HOLBoundVar{l}) \HOLBoundVar{y}
\end{SaveVerbatim}
\newcommand{\HOLTraceTheoremsLRTCXXCASESOne}{\UseVerbatim{HOLTraceTheoremsLRTCXXCASESOne}}
\begin{SaveVerbatim}{HOLTraceTheoremsLRTCXXCASESTwo}
\HOLTokenTurnstile{} \HOLSymConst{\HOLTokenForall{}}\HOLBoundVar{R} \HOLBoundVar{x} \HOLBoundVar{l} \HOLBoundVar{y}.
     \HOLConst{LRTC} \HOLBoundVar{R} \HOLBoundVar{x} \HOLBoundVar{l} \HOLBoundVar{y} \HOLSymConst{\HOLTokenEquiv{}}
     \HOLKeyword{if} \HOLConst{NULL} \HOLBoundVar{l} \HOLKeyword{then} \HOLBoundVar{x} \HOLSymConst{=} \HOLBoundVar{y}
     \HOLKeyword{else} \HOLSymConst{\HOLTokenExists{}}\HOLBoundVar{u}. \HOLConst{LRTC} \HOLBoundVar{R} \HOLBoundVar{x} (\HOLConst{FRONT} \HOLBoundVar{l}) \HOLBoundVar{u} \HOLSymConst{\HOLTokenConj{}} \HOLBoundVar{R} \HOLBoundVar{u} (\HOLConst{LAST} \HOLBoundVar{l}) \HOLBoundVar{y}
\end{SaveVerbatim}
\newcommand{\HOLTraceTheoremsLRTCXXCASESTwo}{\UseVerbatim{HOLTraceTheoremsLRTCXXCASESTwo}}
\begin{SaveVerbatim}{HOLTraceTheoremsLRTCXXCASESXXLRTCXXTWICE}
\HOLTokenTurnstile{} \HOLSymConst{\HOLTokenForall{}}\HOLBoundVar{R} \HOLBoundVar{x} \HOLBoundVar{l} \HOLBoundVar{y}.
     \HOLConst{LRTC} \HOLBoundVar{R} \HOLBoundVar{x} \HOLBoundVar{l} \HOLBoundVar{y} \HOLSymConst{\HOLTokenEquiv{}}
     \HOLSymConst{\HOLTokenExists{}}\HOLBoundVar{u} \HOLBoundVar{l\sb{\mathrm{1}}} \HOLBoundVar{l\sb{\mathrm{2}}}. \HOLConst{LRTC} \HOLBoundVar{R} \HOLBoundVar{x} \HOLBoundVar{l\sb{\mathrm{1}}} \HOLBoundVar{u} \HOLSymConst{\HOLTokenConj{}} \HOLConst{LRTC} \HOLBoundVar{R} \HOLBoundVar{u} \HOLBoundVar{l\sb{\mathrm{2}}} \HOLBoundVar{y} \HOLSymConst{\HOLTokenConj{}} (\HOLBoundVar{l} \HOLSymConst{=} \HOLBoundVar{l\sb{\mathrm{1}}} \HOLSymConst{++} \HOLBoundVar{l\sb{\mathrm{2}}})
\end{SaveVerbatim}
\newcommand{\HOLTraceTheoremsLRTCXXCASESXXLRTCXXTWICE}{\UseVerbatim{HOLTraceTheoremsLRTCXXCASESXXLRTCXXTWICE}}
\begin{SaveVerbatim}{HOLTraceTheoremsLRTCXXINDUCT}
\HOLTokenTurnstile{} \HOLSymConst{\HOLTokenForall{}}\HOLBoundVar{R} \HOLBoundVar{P}.
     (\HOLSymConst{\HOLTokenForall{}}\HOLBoundVar{x}. \HOLBoundVar{P} \HOLBoundVar{x} [] \HOLBoundVar{x}) \HOLSymConst{\HOLTokenConj{}}
     (\HOLSymConst{\HOLTokenForall{}}\HOLBoundVar{x} \HOLBoundVar{h} \HOLBoundVar{y} \HOLBoundVar{t} \HOLBoundVar{z}. \HOLBoundVar{R} \HOLBoundVar{x} \HOLBoundVar{h} \HOLBoundVar{y} \HOLSymConst{\HOLTokenConj{}} \HOLBoundVar{P} \HOLBoundVar{y} \HOLBoundVar{t} \HOLBoundVar{z} \HOLSymConst{\HOLTokenImp{}} \HOLBoundVar{P} \HOLBoundVar{x} (\HOLBoundVar{h}\HOLSymConst{::}\HOLBoundVar{t}) \HOLBoundVar{z}) \HOLSymConst{\HOLTokenImp{}}
     \HOLSymConst{\HOLTokenForall{}}\HOLBoundVar{x} \HOLBoundVar{l} \HOLBoundVar{y}. \HOLConst{LRTC} \HOLBoundVar{R} \HOLBoundVar{x} \HOLBoundVar{l} \HOLBoundVar{y} \HOLSymConst{\HOLTokenImp{}} \HOLBoundVar{P} \HOLBoundVar{x} \HOLBoundVar{l} \HOLBoundVar{y}
\end{SaveVerbatim}
\newcommand{\HOLTraceTheoremsLRTCXXINDUCT}{\UseVerbatim{HOLTraceTheoremsLRTCXXINDUCT}}
\begin{SaveVerbatim}{HOLTraceTheoremsLRTCXXLRTC}
\HOLTokenTurnstile{} \HOLSymConst{\HOLTokenForall{}}\HOLBoundVar{R} \HOLBoundVar{x} \HOLBoundVar{m} \HOLBoundVar{y}.
     \HOLConst{LRTC} \HOLBoundVar{R} \HOLBoundVar{x} \HOLBoundVar{m} \HOLBoundVar{y} \HOLSymConst{\HOLTokenImp{}} \HOLSymConst{\HOLTokenForall{}}\HOLBoundVar{n} \HOLBoundVar{z}. \HOLConst{LRTC} \HOLBoundVar{R} \HOLBoundVar{y} \HOLBoundVar{n} \HOLBoundVar{z} \HOLSymConst{\HOLTokenImp{}} \HOLConst{LRTC} \HOLBoundVar{R} \HOLBoundVar{x} (\HOLBoundVar{m} \HOLSymConst{++} \HOLBoundVar{n}) \HOLBoundVar{z}
\end{SaveVerbatim}
\newcommand{\HOLTraceTheoremsLRTCXXLRTC}{\UseVerbatim{HOLTraceTheoremsLRTCXXLRTC}}
\begin{SaveVerbatim}{HOLTraceTheoremsLRTCXXNIL}
\HOLTokenTurnstile{} \HOLSymConst{\HOLTokenForall{}}\HOLBoundVar{R} \HOLBoundVar{x} \HOLBoundVar{y}. \HOLConst{LRTC} \HOLBoundVar{R} \HOLBoundVar{x} [] \HOLBoundVar{y} \HOLSymConst{\HOLTokenEquiv{}} (\HOLBoundVar{x} \HOLSymConst{=} \HOLBoundVar{y})
\end{SaveVerbatim}
\newcommand{\HOLTraceTheoremsLRTCXXNIL}{\UseVerbatim{HOLTraceTheoremsLRTCXXNIL}}
\begin{SaveVerbatim}{HOLTraceTheoremsLRTCXXONE}
\HOLTokenTurnstile{} \HOLSymConst{\HOLTokenForall{}}\HOLBoundVar{R} \HOLBoundVar{x} \HOLBoundVar{t} \HOLBoundVar{y}. \HOLConst{LRTC} \HOLBoundVar{R} \HOLBoundVar{x} [\HOLBoundVar{t}] \HOLBoundVar{y} \HOLSymConst{\HOLTokenEquiv{}} \HOLBoundVar{R} \HOLBoundVar{x} \HOLBoundVar{t} \HOLBoundVar{y}
\end{SaveVerbatim}
\newcommand{\HOLTraceTheoremsLRTCXXONE}{\UseVerbatim{HOLTraceTheoremsLRTCXXONE}}
\begin{SaveVerbatim}{HOLTraceTheoremsLRTCXXREFL}
\HOLTokenTurnstile{} \HOLSymConst{\HOLTokenForall{}}\HOLBoundVar{R}. \HOLConst{LRTC} \HOLBoundVar{R} \HOLFreeVar{x} [] \HOLFreeVar{x}
\end{SaveVerbatim}
\newcommand{\HOLTraceTheoremsLRTCXXREFL}{\UseVerbatim{HOLTraceTheoremsLRTCXXREFL}}
\begin{SaveVerbatim}{HOLTraceTheoremsLRTCXXRULES}
\HOLTokenTurnstile{} \HOLSymConst{\HOLTokenForall{}}\HOLBoundVar{R}.
     (\HOLSymConst{\HOLTokenForall{}}\HOLBoundVar{x}. \HOLConst{LRTC} \HOLBoundVar{R} \HOLBoundVar{x} [] \HOLBoundVar{x}) \HOLSymConst{\HOLTokenConj{}}
     \HOLSymConst{\HOLTokenForall{}}\HOLBoundVar{x} \HOLBoundVar{h} \HOLBoundVar{y} \HOLBoundVar{t} \HOLBoundVar{z}. \HOLBoundVar{R} \HOLBoundVar{x} \HOLBoundVar{h} \HOLBoundVar{y} \HOLSymConst{\HOLTokenConj{}} \HOLConst{LRTC} \HOLBoundVar{R} \HOLBoundVar{y} \HOLBoundVar{t} \HOLBoundVar{z} \HOLSymConst{\HOLTokenImp{}} \HOLConst{LRTC} \HOLBoundVar{R} \HOLBoundVar{x} (\HOLBoundVar{h}\HOLSymConst{::}\HOLBoundVar{t}) \HOLBoundVar{z}
\end{SaveVerbatim}
\newcommand{\HOLTraceTheoremsLRTCXXRULES}{\UseVerbatim{HOLTraceTheoremsLRTCXXRULES}}
\begin{SaveVerbatim}{HOLTraceTheoremsLRTCXXSINGLE}
\HOLTokenTurnstile{} \HOLSymConst{\HOLTokenForall{}}\HOLBoundVar{R} \HOLBoundVar{x} \HOLBoundVar{t} \HOLBoundVar{y}. \HOLBoundVar{R} \HOLBoundVar{x} \HOLBoundVar{t} \HOLBoundVar{y} \HOLSymConst{\HOLTokenImp{}} \HOLConst{LRTC} \HOLBoundVar{R} \HOLBoundVar{x} [\HOLBoundVar{t}] \HOLBoundVar{y}
\end{SaveVerbatim}
\newcommand{\HOLTraceTheoremsLRTCXXSINGLE}{\UseVerbatim{HOLTraceTheoremsLRTCXXSINGLE}}
\begin{SaveVerbatim}{HOLTraceTheoremsLRTCXXSTRONGXXINDUCT}
\HOLTokenTurnstile{} \HOLSymConst{\HOLTokenForall{}}\HOLBoundVar{R} \HOLBoundVar{P}.
     (\HOLSymConst{\HOLTokenForall{}}\HOLBoundVar{x}. \HOLBoundVar{P} \HOLBoundVar{x} [] \HOLBoundVar{x}) \HOLSymConst{\HOLTokenConj{}}
     (\HOLSymConst{\HOLTokenForall{}}\HOLBoundVar{x} \HOLBoundVar{h} \HOLBoundVar{y} \HOLBoundVar{t} \HOLBoundVar{z}.
        \HOLBoundVar{R} \HOLBoundVar{x} \HOLBoundVar{h} \HOLBoundVar{y} \HOLSymConst{\HOLTokenConj{}} \HOLConst{LRTC} \HOLBoundVar{R} \HOLBoundVar{y} \HOLBoundVar{t} \HOLBoundVar{z} \HOLSymConst{\HOLTokenConj{}} \HOLBoundVar{P} \HOLBoundVar{y} \HOLBoundVar{t} \HOLBoundVar{z} \HOLSymConst{\HOLTokenImp{}} \HOLBoundVar{P} \HOLBoundVar{x} (\HOLBoundVar{h}\HOLSymConst{::}\HOLBoundVar{t}) \HOLBoundVar{z}) \HOLSymConst{\HOLTokenImp{}}
     \HOLSymConst{\HOLTokenForall{}}\HOLBoundVar{x} \HOLBoundVar{l} \HOLBoundVar{y}. \HOLConst{LRTC} \HOLBoundVar{R} \HOLBoundVar{x} \HOLBoundVar{l} \HOLBoundVar{y} \HOLSymConst{\HOLTokenImp{}} \HOLBoundVar{P} \HOLBoundVar{x} \HOLBoundVar{l} \HOLBoundVar{y}
\end{SaveVerbatim}
\newcommand{\HOLTraceTheoremsLRTCXXSTRONGXXINDUCT}{\UseVerbatim{HOLTraceTheoremsLRTCXXSTRONGXXINDUCT}}
\begin{SaveVerbatim}{HOLTraceTheoremsLRTCXXTRANS}
\HOLTokenTurnstile{} \HOLSymConst{\HOLTokenForall{}}\HOLBoundVar{R} \HOLBoundVar{x} \HOLBoundVar{m} \HOLBoundVar{y} \HOLBoundVar{n} \HOLBoundVar{z}.
     \HOLConst{LRTC} \HOLBoundVar{R} \HOLBoundVar{x} \HOLBoundVar{m} \HOLBoundVar{y} \HOLSymConst{\HOLTokenConj{}} \HOLConst{LRTC} \HOLBoundVar{R} \HOLBoundVar{y} \HOLBoundVar{n} \HOLBoundVar{z} \HOLSymConst{\HOLTokenImp{}} \HOLConst{LRTC} \HOLBoundVar{R} \HOLBoundVar{x} (\HOLBoundVar{m} \HOLSymConst{++} \HOLBoundVar{n}) \HOLBoundVar{z}
\end{SaveVerbatim}
\newcommand{\HOLTraceTheoremsLRTCXXTRANS}{\UseVerbatim{HOLTraceTheoremsLRTCXXTRANS}}
\begin{SaveVerbatim}{HOLTraceTheoremsMOREXXNODES}
\HOLTokenTurnstile{} \HOLSymConst{\HOLTokenForall{}}\HOLBoundVar{p} \HOLBoundVar{q} \HOLBoundVar{q\sp{\prime}}. \HOLBoundVar{q} \HOLConst{\HOLTokenIn{}} \HOLConst{NODES} \HOLBoundVar{p} \HOLSymConst{\HOLTokenConj{}} \HOLConst{Reach} \HOLBoundVar{q} \HOLBoundVar{q\sp{\prime}} \HOLSymConst{\HOLTokenImp{}} \HOLBoundVar{q\sp{\prime}} \HOLConst{\HOLTokenIn{}} \HOLConst{NODES} \HOLBoundVar{p}
\end{SaveVerbatim}
\newcommand{\HOLTraceTheoremsMOREXXNODES}{\UseVerbatim{HOLTraceTheoremsMOREXXNODES}}
\begin{SaveVerbatim}{HOLTraceTheoremsNOXXLABELXXcases}
\HOLTokenTurnstile{} \HOLSymConst{\HOLTokenForall{}}\HOLBoundVar{x} \HOLBoundVar{xs}. \HOLConst{NO_LABEL} (\HOLBoundVar{x}\HOLSymConst{::}\HOLBoundVar{xs}) \HOLSymConst{\HOLTokenEquiv{}} (\HOLBoundVar{x} \HOLSymConst{=} \HOLConst{\ensuremath{\tau}}) \HOLSymConst{\HOLTokenConj{}} \HOLConst{NO_LABEL} \HOLBoundVar{xs}
\end{SaveVerbatim}
\newcommand{\HOLTraceTheoremsNOXXLABELXXcases}{\UseVerbatim{HOLTraceTheoremsNOXXLABELXXcases}}
\begin{SaveVerbatim}{HOLTraceTheoremsReachXXcasesOne}
\HOLTokenTurnstile{} \HOLSymConst{\HOLTokenForall{}}\HOLBoundVar{x} \HOLBoundVar{y}. \HOLConst{Reach} \HOLBoundVar{x} \HOLBoundVar{y} \HOLSymConst{\HOLTokenEquiv{}} (\HOLBoundVar{x} \HOLSymConst{=} \HOLBoundVar{y}) \HOLSymConst{\HOLTokenDisj{}} \HOLSymConst{\HOLTokenExists{}}\HOLBoundVar{u}. (\HOLSymConst{\HOLTokenExists{}}\HOLBoundVar{u\sp{\prime}}. \HOLBoundVar{x} \HOLTokenTransBegin\HOLBoundVar{u\sp{\prime}}\HOLTokenTransEnd \HOLBoundVar{u}) \HOLSymConst{\HOLTokenConj{}} \HOLConst{Reach} \HOLBoundVar{u} \HOLBoundVar{y}
\end{SaveVerbatim}
\newcommand{\HOLTraceTheoremsReachXXcasesOne}{\UseVerbatim{HOLTraceTheoremsReachXXcasesOne}}
\begin{SaveVerbatim}{HOLTraceTheoremsReachXXcasesTwo}
\HOLTokenTurnstile{} \HOLSymConst{\HOLTokenForall{}}\HOLBoundVar{x} \HOLBoundVar{y}. \HOLConst{Reach} \HOLBoundVar{x} \HOLBoundVar{y} \HOLSymConst{\HOLTokenEquiv{}} (\HOLBoundVar{x} \HOLSymConst{=} \HOLBoundVar{y}) \HOLSymConst{\HOLTokenDisj{}} \HOLSymConst{\HOLTokenExists{}}\HOLBoundVar{u}. \HOLConst{Reach} \HOLBoundVar{x} \HOLBoundVar{u} \HOLSymConst{\HOLTokenConj{}} \HOLSymConst{\HOLTokenExists{}}\HOLBoundVar{u\sp{\prime}}. \HOLBoundVar{u} \HOLTokenTransBegin\HOLBoundVar{u\sp{\prime}}\HOLTokenTransEnd \HOLBoundVar{y}
\end{SaveVerbatim}
\newcommand{\HOLTraceTheoremsReachXXcasesTwo}{\UseVerbatim{HOLTraceTheoremsReachXXcasesTwo}}
\begin{SaveVerbatim}{HOLTraceTheoremsReachXXind}
\HOLTokenTurnstile{} \HOLSymConst{\HOLTokenForall{}}\HOLBoundVar{P}.
     (\HOLSymConst{\HOLTokenForall{}}\HOLBoundVar{x}. \HOLBoundVar{P} \HOLBoundVar{x} \HOLBoundVar{x}) \HOLSymConst{\HOLTokenConj{}} (\HOLSymConst{\HOLTokenForall{}}\HOLBoundVar{x} \HOLBoundVar{y} \HOLBoundVar{z}. (\HOLSymConst{\HOLTokenExists{}}\HOLBoundVar{u}. \HOLBoundVar{x} \HOLTokenTransBegin\HOLBoundVar{u}\HOLTokenTransEnd \HOLBoundVar{y}) \HOLSymConst{\HOLTokenConj{}} \HOLBoundVar{P} \HOLBoundVar{y} \HOLBoundVar{z} \HOLSymConst{\HOLTokenImp{}} \HOLBoundVar{P} \HOLBoundVar{x} \HOLBoundVar{z}) \HOLSymConst{\HOLTokenImp{}}
     \HOLSymConst{\HOLTokenForall{}}\HOLBoundVar{x} \HOLBoundVar{y}. \HOLConst{Reach} \HOLBoundVar{x} \HOLBoundVar{y} \HOLSymConst{\HOLTokenImp{}} \HOLBoundVar{P} \HOLBoundVar{x} \HOLBoundVar{y}
\end{SaveVerbatim}
\newcommand{\HOLTraceTheoremsReachXXind}{\UseVerbatim{HOLTraceTheoremsReachXXind}}
\begin{SaveVerbatim}{HOLTraceTheoremsReachXXindXXright}
\HOLTokenTurnstile{} \HOLSymConst{\HOLTokenForall{}}\HOLBoundVar{P}.
     (\HOLSymConst{\HOLTokenForall{}}\HOLBoundVar{x}. \HOLBoundVar{P} \HOLBoundVar{x} \HOLBoundVar{x}) \HOLSymConst{\HOLTokenConj{}} (\HOLSymConst{\HOLTokenForall{}}\HOLBoundVar{x} \HOLBoundVar{y} \HOLBoundVar{z}. \HOLBoundVar{P} \HOLBoundVar{x} \HOLBoundVar{y} \HOLSymConst{\HOLTokenConj{}} (\HOLSymConst{\HOLTokenExists{}}\HOLBoundVar{u}. \HOLBoundVar{y} \HOLTokenTransBegin\HOLBoundVar{u}\HOLTokenTransEnd \HOLBoundVar{z}) \HOLSymConst{\HOLTokenImp{}} \HOLBoundVar{P} \HOLBoundVar{x} \HOLBoundVar{z}) \HOLSymConst{\HOLTokenImp{}}
     \HOLSymConst{\HOLTokenForall{}}\HOLBoundVar{x} \HOLBoundVar{y}. \HOLConst{Reach} \HOLBoundVar{x} \HOLBoundVar{y} \HOLSymConst{\HOLTokenImp{}} \HOLBoundVar{P} \HOLBoundVar{x} \HOLBoundVar{y}
\end{SaveVerbatim}
\newcommand{\HOLTraceTheoremsReachXXindXXright}{\UseVerbatim{HOLTraceTheoremsReachXXindXXright}}
\begin{SaveVerbatim}{HOLTraceTheoremsReachXXNODES}
\HOLTokenTurnstile{} \HOLSymConst{\HOLTokenForall{}}\HOLBoundVar{p} \HOLBoundVar{q}. \HOLConst{Reach} \HOLBoundVar{p} \HOLBoundVar{q} \HOLSymConst{\HOLTokenImp{}} \HOLBoundVar{q} \HOLConst{\HOLTokenIn{}} \HOLConst{NODES} \HOLBoundVar{p}
\end{SaveVerbatim}
\newcommand{\HOLTraceTheoremsReachXXNODES}{\UseVerbatim{HOLTraceTheoremsReachXXNODES}}
\begin{SaveVerbatim}{HOLTraceTheoremsReachXXone}
\HOLTokenTurnstile{} \HOLSymConst{\HOLTokenForall{}}\HOLBoundVar{E} \HOLBoundVar{E\sp{\prime}}. (\HOLSymConst{\HOLTokenExists{}}\HOLBoundVar{u}. \HOLBoundVar{E} \HOLTokenTransBegin\HOLBoundVar{u}\HOLTokenTransEnd \HOLBoundVar{E\sp{\prime}}) \HOLSymConst{\HOLTokenImp{}} \HOLConst{Reach} \HOLBoundVar{E} \HOLBoundVar{E\sp{\prime}}
\end{SaveVerbatim}
\newcommand{\HOLTraceTheoremsReachXXone}{\UseVerbatim{HOLTraceTheoremsReachXXone}}
\begin{SaveVerbatim}{HOLTraceTheoremsReachXXself}
\HOLTokenTurnstile{} \HOLSymConst{\HOLTokenForall{}}\HOLBoundVar{E}. \HOLConst{Reach} \HOLBoundVar{E} \HOLBoundVar{E}
\end{SaveVerbatim}
\newcommand{\HOLTraceTheoremsReachXXself}{\UseVerbatim{HOLTraceTheoremsReachXXself}}
\begin{SaveVerbatim}{HOLTraceTheoremsReachXXstrongind}
\HOLTokenTurnstile{} \HOLSymConst{\HOLTokenForall{}}\HOLBoundVar{P}.
     (\HOLSymConst{\HOLTokenForall{}}\HOLBoundVar{x}. \HOLBoundVar{P} \HOLBoundVar{x} \HOLBoundVar{x}) \HOLSymConst{\HOLTokenConj{}}
     (\HOLSymConst{\HOLTokenForall{}}\HOLBoundVar{x} \HOLBoundVar{y} \HOLBoundVar{z}. (\HOLSymConst{\HOLTokenExists{}}\HOLBoundVar{u}. \HOLBoundVar{x} \HOLTokenTransBegin\HOLBoundVar{u}\HOLTokenTransEnd \HOLBoundVar{y}) \HOLSymConst{\HOLTokenConj{}} \HOLConst{Reach} \HOLBoundVar{y} \HOLBoundVar{z} \HOLSymConst{\HOLTokenConj{}} \HOLBoundVar{P} \HOLBoundVar{y} \HOLBoundVar{z} \HOLSymConst{\HOLTokenImp{}} \HOLBoundVar{P} \HOLBoundVar{x} \HOLBoundVar{z}) \HOLSymConst{\HOLTokenImp{}}
     \HOLSymConst{\HOLTokenForall{}}\HOLBoundVar{x} \HOLBoundVar{y}. \HOLConst{Reach} \HOLBoundVar{x} \HOLBoundVar{y} \HOLSymConst{\HOLTokenImp{}} \HOLBoundVar{P} \HOLBoundVar{x} \HOLBoundVar{y}
\end{SaveVerbatim}
\newcommand{\HOLTraceTheoremsReachXXstrongind}{\UseVerbatim{HOLTraceTheoremsReachXXstrongind}}
\begin{SaveVerbatim}{HOLTraceTheoremsReachXXstrongindXXright}
\HOLTokenTurnstile{} \HOLSymConst{\HOLTokenForall{}}\HOLBoundVar{P}.
     (\HOLSymConst{\HOLTokenForall{}}\HOLBoundVar{x}. \HOLBoundVar{P} \HOLBoundVar{x} \HOLBoundVar{x}) \HOLSymConst{\HOLTokenConj{}}
     (\HOLSymConst{\HOLTokenForall{}}\HOLBoundVar{x} \HOLBoundVar{y} \HOLBoundVar{z}. \HOLBoundVar{P} \HOLBoundVar{x} \HOLBoundVar{y} \HOLSymConst{\HOLTokenConj{}} \HOLConst{Reach} \HOLBoundVar{x} \HOLBoundVar{y} \HOLSymConst{\HOLTokenConj{}} (\HOLSymConst{\HOLTokenExists{}}\HOLBoundVar{u}. \HOLBoundVar{y} \HOLTokenTransBegin\HOLBoundVar{u}\HOLTokenTransEnd \HOLBoundVar{z}) \HOLSymConst{\HOLTokenImp{}} \HOLBoundVar{P} \HOLBoundVar{x} \HOLBoundVar{z}) \HOLSymConst{\HOLTokenImp{}}
     \HOLSymConst{\HOLTokenForall{}}\HOLBoundVar{x} \HOLBoundVar{y}. \HOLConst{Reach} \HOLBoundVar{x} \HOLBoundVar{y} \HOLSymConst{\HOLTokenImp{}} \HOLBoundVar{P} \HOLBoundVar{x} \HOLBoundVar{y}
\end{SaveVerbatim}
\newcommand{\HOLTraceTheoremsReachXXstrongindXXright}{\UseVerbatim{HOLTraceTheoremsReachXXstrongindXXright}}
\begin{SaveVerbatim}{HOLTraceTheoremsReachXXtrans}
\HOLTokenTurnstile{} \HOLSymConst{\HOLTokenForall{}}\HOLBoundVar{x} \HOLBoundVar{y} \HOLBoundVar{z}. \HOLConst{Reach} \HOLBoundVar{x} \HOLBoundVar{y} \HOLSymConst{\HOLTokenConj{}} \HOLConst{Reach} \HOLBoundVar{y} \HOLBoundVar{z} \HOLSymConst{\HOLTokenImp{}} \HOLConst{Reach} \HOLBoundVar{x} \HOLBoundVar{z}
\end{SaveVerbatim}
\newcommand{\HOLTraceTheoremsReachXXtrans}{\UseVerbatim{HOLTraceTheoremsReachXXtrans}}
\begin{SaveVerbatim}{HOLTraceTheoremsSELFXXNODES}
\HOLTokenTurnstile{} \HOLSymConst{\HOLTokenForall{}}\HOLBoundVar{p}. \HOLBoundVar{p} \HOLConst{\HOLTokenIn{}} \HOLConst{NODES} \HOLBoundVar{p}
\end{SaveVerbatim}
\newcommand{\HOLTraceTheoremsSELFXXNODES}{\UseVerbatim{HOLTraceTheoremsSELFXXNODES}}
\begin{SaveVerbatim}{HOLTraceTheoremsSTEPZero}
\HOLTokenTurnstile{} \HOLSymConst{\HOLTokenForall{}}\HOLBoundVar{x} \HOLBoundVar{y}. \HOLConst{STEP} \HOLBoundVar{x} \HOLNumLit{0} \HOLBoundVar{y} \HOLSymConst{\HOLTokenEquiv{}} (\HOLBoundVar{x} \HOLSymConst{=} \HOLBoundVar{y})
\end{SaveVerbatim}
\newcommand{\HOLTraceTheoremsSTEPZero}{\UseVerbatim{HOLTraceTheoremsSTEPZero}}
\begin{SaveVerbatim}{HOLTraceTheoremsSTEPOne}
\HOLTokenTurnstile{} \HOLConst{STEP} \HOLFreeVar{x} \HOLNumLit{1} \HOLFreeVar{y} \HOLSymConst{\HOLTokenEquiv{}} \HOLSymConst{\HOLTokenExists{}}\HOLBoundVar{u}. \HOLFreeVar{x} \HOLTokenTransBegin\HOLBoundVar{u}\HOLTokenTransEnd \HOLFreeVar{y}
\end{SaveVerbatim}
\newcommand{\HOLTraceTheoremsSTEPOne}{\UseVerbatim{HOLTraceTheoremsSTEPOne}}
\begin{SaveVerbatim}{HOLTraceTheoremsSTEPXXADDXXE}
\HOLTokenTurnstile{} \HOLSymConst{\HOLTokenForall{}}\HOLBoundVar{m} \HOLBoundVar{n} \HOLBoundVar{x} \HOLBoundVar{z}. \HOLConst{STEP} \HOLBoundVar{x} (\HOLBoundVar{m} \HOLSymConst{+} \HOLBoundVar{n}) \HOLBoundVar{z} \HOLSymConst{\HOLTokenImp{}} \HOLSymConst{\HOLTokenExists{}}\HOLBoundVar{y}. \HOLConst{STEP} \HOLBoundVar{x} \HOLBoundVar{m} \HOLBoundVar{y} \HOLSymConst{\HOLTokenConj{}} \HOLConst{STEP} \HOLBoundVar{y} \HOLBoundVar{n} \HOLBoundVar{z}
\end{SaveVerbatim}
\newcommand{\HOLTraceTheoremsSTEPXXADDXXE}{\UseVerbatim{HOLTraceTheoremsSTEPXXADDXXE}}
\begin{SaveVerbatim}{HOLTraceTheoremsSTEPXXADDXXEQN}
\HOLTokenTurnstile{} \HOLSymConst{\HOLTokenForall{}}\HOLBoundVar{m} \HOLBoundVar{n} \HOLBoundVar{x} \HOLBoundVar{z}. \HOLConst{STEP} \HOLBoundVar{x} (\HOLBoundVar{m} \HOLSymConst{+} \HOLBoundVar{n}) \HOLBoundVar{z} \HOLSymConst{\HOLTokenEquiv{}} \HOLSymConst{\HOLTokenExists{}}\HOLBoundVar{y}. \HOLConst{STEP} \HOLBoundVar{x} \HOLBoundVar{m} \HOLBoundVar{y} \HOLSymConst{\HOLTokenConj{}} \HOLConst{STEP} \HOLBoundVar{y} \HOLBoundVar{n} \HOLBoundVar{z}
\end{SaveVerbatim}
\newcommand{\HOLTraceTheoremsSTEPXXADDXXEQN}{\UseVerbatim{HOLTraceTheoremsSTEPXXADDXXEQN}}
\begin{SaveVerbatim}{HOLTraceTheoremsSTEPXXADDXXI}
\HOLTokenTurnstile{} \HOLSymConst{\HOLTokenForall{}}\HOLBoundVar{m} \HOLBoundVar{n} \HOLBoundVar{x} \HOLBoundVar{y} \HOLBoundVar{z}. \HOLConst{STEP} \HOLBoundVar{x} \HOLBoundVar{m} \HOLBoundVar{y} \HOLSymConst{\HOLTokenConj{}} \HOLConst{STEP} \HOLBoundVar{y} \HOLBoundVar{n} \HOLBoundVar{z} \HOLSymConst{\HOLTokenImp{}} \HOLConst{STEP} \HOLBoundVar{x} (\HOLBoundVar{m} \HOLSymConst{+} \HOLBoundVar{n}) \HOLBoundVar{z}
\end{SaveVerbatim}
\newcommand{\HOLTraceTheoremsSTEPXXADDXXI}{\UseVerbatim{HOLTraceTheoremsSTEPXXADDXXI}}
\begin{SaveVerbatim}{HOLTraceTheoremsSTEPXXSUC}
\HOLTokenTurnstile{} \HOLSymConst{\HOLTokenForall{}}\HOLBoundVar{n} \HOLBoundVar{x} \HOLBoundVar{y}. \HOLConst{STEP} \HOLBoundVar{x} (\HOLConst{SUC} \HOLBoundVar{n}) \HOLBoundVar{y} \HOLSymConst{\HOLTokenEquiv{}} \HOLSymConst{\HOLTokenExists{}}\HOLBoundVar{z}. (\HOLSymConst{\HOLTokenExists{}}\HOLBoundVar{u}. \HOLBoundVar{x} \HOLTokenTransBegin\HOLBoundVar{u}\HOLTokenTransEnd \HOLBoundVar{z}) \HOLSymConst{\HOLTokenConj{}} \HOLConst{STEP} \HOLBoundVar{z} \HOLBoundVar{n} \HOLBoundVar{y}
\end{SaveVerbatim}
\newcommand{\HOLTraceTheoremsSTEPXXSUC}{\UseVerbatim{HOLTraceTheoremsSTEPXXSUC}}
\begin{SaveVerbatim}{HOLTraceTheoremsSTEPXXSUCXXLEFT}
\HOLTokenTurnstile{} \HOLSymConst{\HOLTokenForall{}}\HOLBoundVar{n} \HOLBoundVar{x} \HOLBoundVar{y}. \HOLConst{STEP} \HOLBoundVar{x} (\HOLConst{SUC} \HOLBoundVar{n}) \HOLBoundVar{y} \HOLSymConst{\HOLTokenEquiv{}} \HOLSymConst{\HOLTokenExists{}}\HOLBoundVar{z}. \HOLConst{STEP} \HOLBoundVar{x} \HOLBoundVar{n} \HOLBoundVar{z} \HOLSymConst{\HOLTokenConj{}} \HOLSymConst{\HOLTokenExists{}}\HOLBoundVar{u}. \HOLBoundVar{z} \HOLTokenTransBegin\HOLBoundVar{u}\HOLTokenTransEnd \HOLBoundVar{y}
\end{SaveVerbatim}
\newcommand{\HOLTraceTheoremsSTEPXXSUCXXLEFT}{\UseVerbatim{HOLTraceTheoremsSTEPXXSUCXXLEFT}}
\begin{SaveVerbatim}{HOLTraceTheoremsTRACEOne}
\HOLTokenTurnstile{} \HOLSymConst{\HOLTokenForall{}}\HOLBoundVar{x} \HOLBoundVar{t} \HOLBoundVar{y}. \HOLBoundVar{x} \HOLTokenTransBegin\HOLBoundVar{t}\HOLTokenTransEnd \HOLBoundVar{y} \HOLSymConst{\HOLTokenImp{}} \HOLConst{TRACE} \HOLBoundVar{x} [\HOLBoundVar{t}] \HOLBoundVar{y}
\end{SaveVerbatim}
\newcommand{\HOLTraceTheoremsTRACEOne}{\UseVerbatim{HOLTraceTheoremsTRACEOne}}
\begin{SaveVerbatim}{HOLTraceTheoremsTRACETwo}
\HOLTokenTurnstile{} \HOLSymConst{\HOLTokenForall{}}\HOLBoundVar{E} \HOLBoundVar{x} \HOLBoundVar{E\sb{\mathrm{1}}} \HOLBoundVar{xs} \HOLBoundVar{E\sp{\prime}}. \HOLBoundVar{E} \HOLTokenTransBegin\HOLBoundVar{x}\HOLTokenTransEnd \HOLBoundVar{E\sb{\mathrm{1}}} \HOLSymConst{\HOLTokenConj{}} \HOLConst{TRACE} \HOLBoundVar{E\sb{\mathrm{1}}} \HOLBoundVar{xs} \HOLBoundVar{E\sp{\prime}} \HOLSymConst{\HOLTokenImp{}} \HOLConst{TRACE} \HOLBoundVar{E} (\HOLBoundVar{x}\HOLSymConst{::}\HOLBoundVar{xs}) \HOLBoundVar{E\sp{\prime}}
\end{SaveVerbatim}
\newcommand{\HOLTraceTheoremsTRACETwo}{\UseVerbatim{HOLTraceTheoremsTRACETwo}}
\begin{SaveVerbatim}{HOLTraceTheoremsTRACEXXAPPENDXXcases}
\HOLTokenTurnstile{} \HOLSymConst{\HOLTokenForall{}}\HOLBoundVar{l\sb{\mathrm{1}}} \HOLBoundVar{l\sb{\mathrm{2}}} \HOLBoundVar{x} \HOLBoundVar{y}.
     \HOLConst{TRACE} \HOLBoundVar{x} (\HOLBoundVar{l\sb{\mathrm{1}}} \HOLSymConst{++} \HOLBoundVar{l\sb{\mathrm{2}}}) \HOLBoundVar{y} \HOLSymConst{\HOLTokenEquiv{}} \HOLSymConst{\HOLTokenExists{}}\HOLBoundVar{u}. \HOLConst{TRACE} \HOLBoundVar{x} \HOLBoundVar{l\sb{\mathrm{1}}} \HOLBoundVar{u} \HOLSymConst{\HOLTokenConj{}} \HOLConst{TRACE} \HOLBoundVar{u} \HOLBoundVar{l\sb{\mathrm{2}}} \HOLBoundVar{y}
\end{SaveVerbatim}
\newcommand{\HOLTraceTheoremsTRACEXXAPPENDXXcases}{\UseVerbatim{HOLTraceTheoremsTRACEXXAPPENDXXcases}}
\begin{SaveVerbatim}{HOLTraceTheoremsTRACEXXcasesOne}
\HOLTokenTurnstile{} \HOLSymConst{\HOLTokenForall{}}\HOLBoundVar{x} \HOLBoundVar{l} \HOLBoundVar{y}.
     \HOLConst{TRACE} \HOLBoundVar{x} \HOLBoundVar{l} \HOLBoundVar{y} \HOLSymConst{\HOLTokenEquiv{}}
     \HOLKeyword{if} \HOLConst{NULL} \HOLBoundVar{l} \HOLKeyword{then} \HOLBoundVar{x} \HOLSymConst{=} \HOLBoundVar{y} \HOLKeyword{else} \HOLSymConst{\HOLTokenExists{}}\HOLBoundVar{u}. \HOLBoundVar{x} \HOLTokenTransBegin\HOLConst{HD} \HOLBoundVar{l}\HOLTokenTransEnd \HOLBoundVar{u} \HOLSymConst{\HOLTokenConj{}} \HOLConst{TRACE} \HOLBoundVar{u} (\HOLConst{TL} \HOLBoundVar{l}) \HOLBoundVar{y}
\end{SaveVerbatim}
\newcommand{\HOLTraceTheoremsTRACEXXcasesOne}{\UseVerbatim{HOLTraceTheoremsTRACEXXcasesOne}}
\begin{SaveVerbatim}{HOLTraceTheoremsTRACEXXcasesTwo}
\HOLTokenTurnstile{} \HOLSymConst{\HOLTokenForall{}}\HOLBoundVar{x} \HOLBoundVar{l} \HOLBoundVar{y}.
     \HOLConst{TRACE} \HOLBoundVar{x} \HOLBoundVar{l} \HOLBoundVar{y} \HOLSymConst{\HOLTokenEquiv{}}
     \HOLKeyword{if} \HOLConst{NULL} \HOLBoundVar{l} \HOLKeyword{then} \HOLBoundVar{x} \HOLSymConst{=} \HOLBoundVar{y}
     \HOLKeyword{else} \HOLSymConst{\HOLTokenExists{}}\HOLBoundVar{u}. \HOLConst{TRACE} \HOLBoundVar{x} (\HOLConst{FRONT} \HOLBoundVar{l}) \HOLBoundVar{u} \HOLSymConst{\HOLTokenConj{}} \HOLBoundVar{u} \HOLTokenTransBegin\HOLConst{LAST} \HOLBoundVar{l}\HOLTokenTransEnd \HOLBoundVar{y}
\end{SaveVerbatim}
\newcommand{\HOLTraceTheoremsTRACEXXcasesTwo}{\UseVerbatim{HOLTraceTheoremsTRACEXXcasesTwo}}
\begin{SaveVerbatim}{HOLTraceTheoremsTRACEXXcasesXXtwice}
\HOLTokenTurnstile{} \HOLSymConst{\HOLTokenForall{}}\HOLBoundVar{x} \HOLBoundVar{l} \HOLBoundVar{y}.
     \HOLConst{TRACE} \HOLBoundVar{x} \HOLBoundVar{l} \HOLBoundVar{y} \HOLSymConst{\HOLTokenEquiv{}}
     \HOLSymConst{\HOLTokenExists{}}\HOLBoundVar{u} \HOLBoundVar{l\sb{\mathrm{1}}} \HOLBoundVar{l\sb{\mathrm{2}}}. \HOLConst{TRACE} \HOLBoundVar{x} \HOLBoundVar{l\sb{\mathrm{1}}} \HOLBoundVar{u} \HOLSymConst{\HOLTokenConj{}} \HOLConst{TRACE} \HOLBoundVar{u} \HOLBoundVar{l\sb{\mathrm{2}}} \HOLBoundVar{y} \HOLSymConst{\HOLTokenConj{}} (\HOLBoundVar{l} \HOLSymConst{=} \HOLBoundVar{l\sb{\mathrm{1}}} \HOLSymConst{++} \HOLBoundVar{l\sb{\mathrm{2}}})
\end{SaveVerbatim}
\newcommand{\HOLTraceTheoremsTRACEXXcasesXXtwice}{\UseVerbatim{HOLTraceTheoremsTRACEXXcasesXXtwice}}
\begin{SaveVerbatim}{HOLTraceTheoremsTRACEXXind}
\HOLTokenTurnstile{} \HOLSymConst{\HOLTokenForall{}}\HOLBoundVar{P}.
     (\HOLSymConst{\HOLTokenForall{}}\HOLBoundVar{x}. \HOLBoundVar{P} \HOLBoundVar{x} \HOLConst{\ensuremath{\epsilon}} \HOLBoundVar{x}) \HOLSymConst{\HOLTokenConj{}}
     (\HOLSymConst{\HOLTokenForall{}}\HOLBoundVar{x} \HOLBoundVar{h} \HOLBoundVar{y} \HOLBoundVar{t} \HOLBoundVar{z}. \HOLBoundVar{x} \HOLTokenTransBegin\HOLBoundVar{h}\HOLTokenTransEnd \HOLBoundVar{y} \HOLSymConst{\HOLTokenConj{}} \HOLBoundVar{P} \HOLBoundVar{y} \HOLBoundVar{t} \HOLBoundVar{z} \HOLSymConst{\HOLTokenImp{}} \HOLBoundVar{P} \HOLBoundVar{x} (\HOLBoundVar{h}\HOLSymConst{::}\HOLBoundVar{t}) \HOLBoundVar{z}) \HOLSymConst{\HOLTokenImp{}}
     \HOLSymConst{\HOLTokenForall{}}\HOLBoundVar{x} \HOLBoundVar{l} \HOLBoundVar{y}. \HOLConst{TRACE} \HOLBoundVar{x} \HOLBoundVar{l} \HOLBoundVar{y} \HOLSymConst{\HOLTokenImp{}} \HOLBoundVar{P} \HOLBoundVar{x} \HOLBoundVar{l} \HOLBoundVar{y}
\end{SaveVerbatim}
\newcommand{\HOLTraceTheoremsTRACEXXind}{\UseVerbatim{HOLTraceTheoremsTRACEXXind}}
\begin{SaveVerbatim}{HOLTraceTheoremsTRACEXXNIL}
\HOLTokenTurnstile{} \HOLSymConst{\HOLTokenForall{}}\HOLBoundVar{x} \HOLBoundVar{y}. \HOLConst{TRACE} \HOLBoundVar{x} \HOLConst{\ensuremath{\epsilon}} \HOLBoundVar{y} \HOLSymConst{\HOLTokenEquiv{}} (\HOLBoundVar{x} \HOLSymConst{=} \HOLBoundVar{y})
\end{SaveVerbatim}
\newcommand{\HOLTraceTheoremsTRACEXXNIL}{\UseVerbatim{HOLTraceTheoremsTRACEXXNIL}}
\begin{SaveVerbatim}{HOLTraceTheoremsTRACEXXONE}
\HOLTokenTurnstile{} \HOLSymConst{\HOLTokenForall{}}\HOLBoundVar{x} \HOLBoundVar{t} \HOLBoundVar{y}. \HOLConst{TRACE} \HOLBoundVar{x} [\HOLBoundVar{t}] \HOLBoundVar{y} \HOLSymConst{\HOLTokenEquiv{}} \HOLBoundVar{x} \HOLTokenTransBegin\HOLBoundVar{t}\HOLTokenTransEnd \HOLBoundVar{y}
\end{SaveVerbatim}
\newcommand{\HOLTraceTheoremsTRACEXXONE}{\UseVerbatim{HOLTraceTheoremsTRACEXXONE}}
\begin{SaveVerbatim}{HOLTraceTheoremsTRACEXXREFL}
\HOLTokenTurnstile{} \HOLSymConst{\HOLTokenForall{}}\HOLBoundVar{E}. \HOLConst{TRACE} \HOLBoundVar{E} \HOLConst{\ensuremath{\epsilon}} \HOLBoundVar{E}
\end{SaveVerbatim}
\newcommand{\HOLTraceTheoremsTRACEXXREFL}{\UseVerbatim{HOLTraceTheoremsTRACEXXREFL}}
\begin{SaveVerbatim}{HOLTraceTheoremsTRACEXXruleZero}
\HOLTokenTurnstile{} \HOLSymConst{\HOLTokenForall{}}\HOLBoundVar{x}. \HOLConst{TRACE} \HOLBoundVar{x} \HOLConst{\ensuremath{\epsilon}} \HOLBoundVar{x}
\end{SaveVerbatim}
\newcommand{\HOLTraceTheoremsTRACEXXruleZero}{\UseVerbatim{HOLTraceTheoremsTRACEXXruleZero}}
\begin{SaveVerbatim}{HOLTraceTheoremsTRACEXXruleOne}
\HOLTokenTurnstile{} \HOLSymConst{\HOLTokenForall{}}\HOLBoundVar{x} \HOLBoundVar{h} \HOLBoundVar{y} \HOLBoundVar{t} \HOLBoundVar{z}. \HOLBoundVar{x} \HOLTokenTransBegin\HOLBoundVar{h}\HOLTokenTransEnd \HOLBoundVar{y} \HOLSymConst{\HOLTokenConj{}} \HOLConst{TRACE} \HOLBoundVar{y} \HOLBoundVar{t} \HOLBoundVar{z} \HOLSymConst{\HOLTokenImp{}} \HOLConst{TRACE} \HOLBoundVar{x} (\HOLBoundVar{h}\HOLSymConst{::}\HOLBoundVar{t}) \HOLBoundVar{z}
\end{SaveVerbatim}
\newcommand{\HOLTraceTheoremsTRACEXXruleOne}{\UseVerbatim{HOLTraceTheoremsTRACEXXruleOne}}
\begin{SaveVerbatim}{HOLTraceTheoremsTRACEXXrules}
\HOLTokenTurnstile{} (\HOLSymConst{\HOLTokenForall{}}\HOLBoundVar{x}. \HOLConst{TRACE} \HOLBoundVar{x} \HOLConst{\ensuremath{\epsilon}} \HOLBoundVar{x}) \HOLSymConst{\HOLTokenConj{}}
   \HOLSymConst{\HOLTokenForall{}}\HOLBoundVar{x} \HOLBoundVar{h} \HOLBoundVar{y} \HOLBoundVar{t} \HOLBoundVar{z}. \HOLBoundVar{x} \HOLTokenTransBegin\HOLBoundVar{h}\HOLTokenTransEnd \HOLBoundVar{y} \HOLSymConst{\HOLTokenConj{}} \HOLConst{TRACE} \HOLBoundVar{y} \HOLBoundVar{t} \HOLBoundVar{z} \HOLSymConst{\HOLTokenImp{}} \HOLConst{TRACE} \HOLBoundVar{x} (\HOLBoundVar{h}\HOLSymConst{::}\HOLBoundVar{t}) \HOLBoundVar{z}
\end{SaveVerbatim}
\newcommand{\HOLTraceTheoremsTRACEXXrules}{\UseVerbatim{HOLTraceTheoremsTRACEXXrules}}
\begin{SaveVerbatim}{HOLTraceTheoremsTRACEXXstrongind}
\HOLTokenTurnstile{} \HOLSymConst{\HOLTokenForall{}}\HOLBoundVar{P}.
     (\HOLSymConst{\HOLTokenForall{}}\HOLBoundVar{x}. \HOLBoundVar{P} \HOLBoundVar{x} \HOLConst{\ensuremath{\epsilon}} \HOLBoundVar{x}) \HOLSymConst{\HOLTokenConj{}}
     (\HOLSymConst{\HOLTokenForall{}}\HOLBoundVar{x} \HOLBoundVar{h} \HOLBoundVar{y} \HOLBoundVar{t} \HOLBoundVar{z}.
        \HOLBoundVar{x} \HOLTokenTransBegin\HOLBoundVar{h}\HOLTokenTransEnd \HOLBoundVar{y} \HOLSymConst{\HOLTokenConj{}} \HOLConst{TRACE} \HOLBoundVar{y} \HOLBoundVar{t} \HOLBoundVar{z} \HOLSymConst{\HOLTokenConj{}} \HOLBoundVar{P} \HOLBoundVar{y} \HOLBoundVar{t} \HOLBoundVar{z} \HOLSymConst{\HOLTokenImp{}} \HOLBoundVar{P} \HOLBoundVar{x} (\HOLBoundVar{h}\HOLSymConst{::}\HOLBoundVar{t}) \HOLBoundVar{z}) \HOLSymConst{\HOLTokenImp{}}
     \HOLSymConst{\HOLTokenForall{}}\HOLBoundVar{x} \HOLBoundVar{l} \HOLBoundVar{y}. \HOLConst{TRACE} \HOLBoundVar{x} \HOLBoundVar{l} \HOLBoundVar{y} \HOLSymConst{\HOLTokenImp{}} \HOLBoundVar{P} \HOLBoundVar{x} \HOLBoundVar{l} \HOLBoundVar{y}
\end{SaveVerbatim}
\newcommand{\HOLTraceTheoremsTRACEXXstrongind}{\UseVerbatim{HOLTraceTheoremsTRACEXXstrongind}}
\begin{SaveVerbatim}{HOLTraceTheoremsTRACEXXtrans}
\HOLTokenTurnstile{} \HOLSymConst{\HOLTokenForall{}}\HOLBoundVar{x} \HOLBoundVar{m} \HOLBoundVar{y}. \HOLConst{TRACE} \HOLBoundVar{x} \HOLBoundVar{m} \HOLBoundVar{y} \HOLSymConst{\HOLTokenImp{}} \HOLSymConst{\HOLTokenForall{}}\HOLBoundVar{n} \HOLBoundVar{z}. \HOLConst{TRACE} \HOLBoundVar{y} \HOLBoundVar{n} \HOLBoundVar{z} \HOLSymConst{\HOLTokenImp{}} \HOLConst{TRACE} \HOLBoundVar{x} (\HOLBoundVar{m} \HOLSymConst{++} \HOLBoundVar{n}) \HOLBoundVar{z}
\end{SaveVerbatim}
\newcommand{\HOLTraceTheoremsTRACEXXtrans}{\UseVerbatim{HOLTraceTheoremsTRACEXXtrans}}
\begin{SaveVerbatim}{HOLTraceTheoremsTRACEXXtransXXapplied}
\HOLTokenTurnstile{} \HOLSymConst{\HOLTokenForall{}}\HOLBoundVar{xs} \HOLBoundVar{ys} \HOLBoundVar{E} \HOLBoundVar{E\sb{\mathrm{1}}} \HOLBoundVar{E\sp{\prime}}.
     \HOLConst{TRACE} \HOLBoundVar{E} \HOLBoundVar{xs} \HOLBoundVar{E\sb{\mathrm{1}}} \HOLSymConst{\HOLTokenConj{}} \HOLConst{TRACE} \HOLBoundVar{E\sb{\mathrm{1}}} \HOLBoundVar{ys} \HOLBoundVar{E\sp{\prime}} \HOLSymConst{\HOLTokenImp{}} \HOLConst{TRACE} \HOLBoundVar{E} (\HOLBoundVar{xs} \HOLSymConst{++} \HOLBoundVar{ys}) \HOLBoundVar{E\sp{\prime}}
\end{SaveVerbatim}
\newcommand{\HOLTraceTheoremsTRACEXXtransXXapplied}{\UseVerbatim{HOLTraceTheoremsTRACEXXtransXXapplied}}
\begin{SaveVerbatim}{HOLTraceTheoremsTRANSXXINXXNODES}
\HOLTokenTurnstile{} \HOLSymConst{\HOLTokenForall{}}\HOLBoundVar{p} \HOLBoundVar{q} \HOLBoundVar{u}. \HOLBoundVar{p} \HOLTokenTransBegin\HOLBoundVar{u}\HOLTokenTransEnd \HOLBoundVar{q} \HOLSymConst{\HOLTokenImp{}} \HOLBoundVar{q} \HOLConst{\HOLTokenIn{}} \HOLConst{NODES} \HOLBoundVar{p}
\end{SaveVerbatim}
\newcommand{\HOLTraceTheoremsTRANSXXINXXNODES}{\UseVerbatim{HOLTraceTheoremsTRANSXXINXXNODES}}
\begin{SaveVerbatim}{HOLTraceTheoremsUNIQUEXXLABELXXcasesOne}
\HOLTokenTurnstile{} \HOLSymConst{\HOLTokenForall{}}\HOLBoundVar{l} \HOLBoundVar{xs}.
     \HOLConst{UNIQUE_LABEL} (\HOLConst{label} \HOLBoundVar{l}) (\HOLConst{\ensuremath{\tau}}\HOLSymConst{::}\HOLBoundVar{xs}) \HOLSymConst{\HOLTokenEquiv{}} \HOLConst{UNIQUE_LABEL} (\HOLConst{label} \HOLBoundVar{l}) \HOLBoundVar{xs}
\end{SaveVerbatim}
\newcommand{\HOLTraceTheoremsUNIQUEXXLABELXXcasesOne}{\UseVerbatim{HOLTraceTheoremsUNIQUEXXLABELXXcasesOne}}
\begin{SaveVerbatim}{HOLTraceTheoremsUNIQUEXXLABELXXcasesTwo}
\HOLTokenTurnstile{} \HOLSymConst{\HOLTokenForall{}}\HOLBoundVar{l} \HOLBoundVar{l\sp{\prime}} \HOLBoundVar{xs}.
     \HOLConst{UNIQUE_LABEL} (\HOLConst{label} \HOLBoundVar{l}) (\HOLConst{label} \HOLBoundVar{l\sp{\prime}}\HOLSymConst{::}\HOLBoundVar{xs}) \HOLSymConst{\HOLTokenEquiv{}}
     (\HOLBoundVar{l} \HOLSymConst{=} \HOLBoundVar{l\sp{\prime}}) \HOLSymConst{\HOLTokenConj{}} \HOLConst{NO_LABEL} \HOLBoundVar{xs}
\end{SaveVerbatim}
\newcommand{\HOLTraceTheoremsUNIQUEXXLABELXXcasesTwo}{\UseVerbatim{HOLTraceTheoremsUNIQUEXXLABELXXcasesTwo}}
\begin{SaveVerbatim}{HOLTraceTheoremsUNIQUEXXLABELXXIMPXXMEM}
\HOLTokenTurnstile{} \HOLSymConst{\HOLTokenForall{}}\HOLBoundVar{u} \HOLBoundVar{L}. \HOLConst{UNIQUE_LABEL} \HOLBoundVar{u} \HOLBoundVar{L} \HOLSymConst{\HOLTokenImp{}} \HOLConst{MEM} \HOLBoundVar{u} \HOLBoundVar{L}
\end{SaveVerbatim}
\newcommand{\HOLTraceTheoremsUNIQUEXXLABELXXIMPXXMEM}{\UseVerbatim{HOLTraceTheoremsUNIQUEXXLABELXXIMPXXMEM}}
\begin{SaveVerbatim}{HOLTraceTheoremsUNIQUEXXLABELXXNOTXXNULL}
\HOLTokenTurnstile{} \HOLSymConst{\HOLTokenForall{}}\HOLBoundVar{u} \HOLBoundVar{L}. \HOLConst{UNIQUE_LABEL} \HOLBoundVar{u} \HOLBoundVar{L} \HOLSymConst{\HOLTokenImp{}} \HOLSymConst{\HOLTokenNeg{}}\HOLConst{NULL} \HOLBoundVar{L}
\end{SaveVerbatim}
\newcommand{\HOLTraceTheoremsUNIQUEXXLABELXXNOTXXNULL}{\UseVerbatim{HOLTraceTheoremsUNIQUEXXLABELXXNOTXXNULL}}
\begin{SaveVerbatim}{HOLTraceTheoremsWEAKXXTRANSXXANDXXSTEP}
\HOLTokenTurnstile{} \HOLSymConst{\HOLTokenForall{}}\HOLBoundVar{E} \HOLBoundVar{u} \HOLBoundVar{E\sp{\prime}}. \HOLBoundVar{E} \HOLTokenWeakTransBegin\HOLBoundVar{u}\HOLTokenWeakTransEnd \HOLBoundVar{E\sp{\prime}} \HOLSymConst{\HOLTokenImp{}} \HOLSymConst{\HOLTokenExists{}}\HOLBoundVar{n}. \HOLConst{STEP} \HOLBoundVar{E} \HOLBoundVar{n} \HOLBoundVar{E\sp{\prime}}
\end{SaveVerbatim}
\newcommand{\HOLTraceTheoremsWEAKXXTRANSXXANDXXSTEP}{\UseVerbatim{HOLTraceTheoremsWEAKXXTRANSXXANDXXSTEP}}
\begin{SaveVerbatim}{HOLTraceTheoremsWEAKXXTRANSXXANDXXTRACE}
\HOLTokenTurnstile{} \HOLSymConst{\HOLTokenForall{}}\HOLBoundVar{E} \HOLBoundVar{u} \HOLBoundVar{E\sp{\prime}}.
     \HOLBoundVar{E} \HOLTokenWeakTransBegin\HOLBoundVar{u}\HOLTokenWeakTransEnd \HOLBoundVar{E\sp{\prime}} \HOLSymConst{\HOLTokenEquiv{}}
     \HOLSymConst{\HOLTokenExists{}}\HOLBoundVar{us}.
       \HOLConst{TRACE} \HOLBoundVar{E} \HOLBoundVar{us} \HOLBoundVar{E\sp{\prime}} \HOLSymConst{\HOLTokenConj{}} \HOLSymConst{\HOLTokenNeg{}}\HOLConst{NULL} \HOLBoundVar{us} \HOLSymConst{\HOLTokenConj{}}
       \HOLKeyword{if} \HOLBoundVar{u} \HOLSymConst{=} \HOLConst{\ensuremath{\tau}} \HOLKeyword{then} \HOLConst{NO_LABEL} \HOLBoundVar{us} \HOLKeyword{else} \HOLConst{UNIQUE_LABEL} \HOLBoundVar{u} \HOLBoundVar{us}
\end{SaveVerbatim}
\newcommand{\HOLTraceTheoremsWEAKXXTRANSXXANDXXTRACE}{\UseVerbatim{HOLTraceTheoremsWEAKXXTRANSXXANDXXTRACE}}
\begin{SaveVerbatim}{HOLTraceTheoremsWEAKXXTRANSXXINXXNODES}
\HOLTokenTurnstile{} \HOLSymConst{\HOLTokenForall{}}\HOLBoundVar{p} \HOLBoundVar{q} \HOLBoundVar{u}. \HOLBoundVar{p} \HOLTokenWeakTransBegin\HOLBoundVar{u}\HOLTokenWeakTransEnd \HOLBoundVar{q} \HOLSymConst{\HOLTokenImp{}} \HOLBoundVar{q} \HOLConst{\HOLTokenIn{}} \HOLConst{NODES} \HOLBoundVar{p}
\end{SaveVerbatim}
\newcommand{\HOLTraceTheoremsWEAKXXTRANSXXINXXNODES}{\UseVerbatim{HOLTraceTheoremsWEAKXXTRANSXXINXXNODES}}
\begin{SaveVerbatim}{HOLTraceTheoremsWEAKXXTRANSXXReach}
\HOLTokenTurnstile{} \HOLSymConst{\HOLTokenForall{}}\HOLBoundVar{p} \HOLBoundVar{q} \HOLBoundVar{u}. \HOLBoundVar{p} \HOLTokenWeakTransBegin\HOLBoundVar{u}\HOLTokenWeakTransEnd \HOLBoundVar{q} \HOLSymConst{\HOLTokenImp{}} \HOLConst{Reach} \HOLBoundVar{p} \HOLBoundVar{q}
\end{SaveVerbatim}
\newcommand{\HOLTraceTheoremsWEAKXXTRANSXXReach}{\UseVerbatim{HOLTraceTheoremsWEAKXXTRANSXXReach}}
\begin{SaveVerbatim}{HOLTraceTheoremsWEAKXXTRANSXXTRACE}
\HOLTokenTurnstile{} \HOLSymConst{\HOLTokenForall{}}\HOLBoundVar{E} \HOLBoundVar{u} \HOLBoundVar{E\sp{\prime}}. \HOLBoundVar{E} \HOLTokenWeakTransBegin\HOLBoundVar{u}\HOLTokenWeakTransEnd \HOLBoundVar{E\sp{\prime}} \HOLSymConst{\HOLTokenImp{}} \HOLSymConst{\HOLTokenExists{}}\HOLBoundVar{xs}. \HOLConst{TRACE} \HOLBoundVar{E} \HOLBoundVar{xs} \HOLBoundVar{E\sp{\prime}}
\end{SaveVerbatim}
\newcommand{\HOLTraceTheoremsWEAKXXTRANSXXTRACE}{\UseVerbatim{HOLTraceTheoremsWEAKXXTRANSXXTRACE}}
\newcommand{\HOLTraceTheorems}{
\HOLThmTag{Trace}{EPS_AND_STEP}\HOLTraceTheoremsEPSXXANDXXSTEP
\HOLThmTag{Trace}{EPS_AND_TRACE}\HOLTraceTheoremsEPSXXANDXXTRACE
\HOLThmTag{Trace}{EPS_IN_NODES}\HOLTraceTheoremsEPSXXINXXNODES
\HOLThmTag{Trace}{EPS_Reach}\HOLTraceTheoremsEPSXXReach
\HOLThmTag{Trace}{EPS_TRACE}\HOLTraceTheoremsEPSXXTRACE
\HOLThmTag{Trace}{LRTC_APPEND_CASES}\HOLTraceTheoremsLRTCXXAPPENDXXCASES
\HOLThmTag{Trace}{LRTC_CASES1}\HOLTraceTheoremsLRTCXXCASESOne
\HOLThmTag{Trace}{LRTC_CASES2}\HOLTraceTheoremsLRTCXXCASESTwo
\HOLThmTag{Trace}{LRTC_CASES_LRTC_TWICE}\HOLTraceTheoremsLRTCXXCASESXXLRTCXXTWICE
\HOLThmTag{Trace}{LRTC_INDUCT}\HOLTraceTheoremsLRTCXXINDUCT
\HOLThmTag{Trace}{LRTC_LRTC}\HOLTraceTheoremsLRTCXXLRTC
\HOLThmTag{Trace}{LRTC_NIL}\HOLTraceTheoremsLRTCXXNIL
\HOLThmTag{Trace}{LRTC_ONE}\HOLTraceTheoremsLRTCXXONE
\HOLThmTag{Trace}{LRTC_REFL}\HOLTraceTheoremsLRTCXXREFL
\HOLThmTag{Trace}{LRTC_RULES}\HOLTraceTheoremsLRTCXXRULES
\HOLThmTag{Trace}{LRTC_SINGLE}\HOLTraceTheoremsLRTCXXSINGLE
\HOLThmTag{Trace}{LRTC_STRONG_INDUCT}\HOLTraceTheoremsLRTCXXSTRONGXXINDUCT
\HOLThmTag{Trace}{LRTC_TRANS}\HOLTraceTheoremsLRTCXXTRANS
\HOLThmTag{Trace}{MORE_NODES}\HOLTraceTheoremsMOREXXNODES
\HOLThmTag{Trace}{NO_LABEL_cases}\HOLTraceTheoremsNOXXLABELXXcases
\HOLThmTag{Trace}{Reach_cases1}\HOLTraceTheoremsReachXXcasesOne
\HOLThmTag{Trace}{Reach_cases2}\HOLTraceTheoremsReachXXcasesTwo
\HOLThmTag{Trace}{Reach_ind}\HOLTraceTheoremsReachXXind
\HOLThmTag{Trace}{Reach_ind_right}\HOLTraceTheoremsReachXXindXXright
\HOLThmTag{Trace}{Reach_NODES}\HOLTraceTheoremsReachXXNODES
\HOLThmTag{Trace}{Reach_one}\HOLTraceTheoremsReachXXone
\HOLThmTag{Trace}{Reach_self}\HOLTraceTheoremsReachXXself
\HOLThmTag{Trace}{Reach_strongind}\HOLTraceTheoremsReachXXstrongind
\HOLThmTag{Trace}{Reach_strongind_right}\HOLTraceTheoremsReachXXstrongindXXright
\HOLThmTag{Trace}{Reach_trans}\HOLTraceTheoremsReachXXtrans
\HOLThmTag{Trace}{SELF_NODES}\HOLTraceTheoremsSELFXXNODES
\HOLThmTag{Trace}{STEP0}\HOLTraceTheoremsSTEPZero
\HOLThmTag{Trace}{STEP1}\HOLTraceTheoremsSTEPOne
\HOLThmTag{Trace}{STEP_ADD_E}\HOLTraceTheoremsSTEPXXADDXXE
\HOLThmTag{Trace}{STEP_ADD_EQN}\HOLTraceTheoremsSTEPXXADDXXEQN
\HOLThmTag{Trace}{STEP_ADD_I}\HOLTraceTheoremsSTEPXXADDXXI
\HOLThmTag{Trace}{STEP_SUC}\HOLTraceTheoremsSTEPXXSUC
\HOLThmTag{Trace}{STEP_SUC_LEFT}\HOLTraceTheoremsSTEPXXSUCXXLEFT
\HOLThmTag{Trace}{TRACE1}\HOLTraceTheoremsTRACEOne
\HOLThmTag{Trace}{TRACE2}\HOLTraceTheoremsTRACETwo
\HOLThmTag{Trace}{TRACE_APPEND_cases}\HOLTraceTheoremsTRACEXXAPPENDXXcases
\HOLThmTag{Trace}{TRACE_cases1}\HOLTraceTheoremsTRACEXXcasesOne
\HOLThmTag{Trace}{TRACE_cases2}\HOLTraceTheoremsTRACEXXcasesTwo
\HOLThmTag{Trace}{TRACE_cases_twice}\HOLTraceTheoremsTRACEXXcasesXXtwice
\HOLThmTag{Trace}{TRACE_ind}\HOLTraceTheoremsTRACEXXind
\HOLThmTag{Trace}{TRACE_NIL}\HOLTraceTheoremsTRACEXXNIL
\HOLThmTag{Trace}{TRACE_ONE}\HOLTraceTheoremsTRACEXXONE
\HOLThmTag{Trace}{TRACE_REFL}\HOLTraceTheoremsTRACEXXREFL
\HOLThmTag{Trace}{TRACE_rule0}\HOLTraceTheoremsTRACEXXruleZero
\HOLThmTag{Trace}{TRACE_rule1}\HOLTraceTheoremsTRACEXXruleOne
\HOLThmTag{Trace}{TRACE_rules}\HOLTraceTheoremsTRACEXXrules
\HOLThmTag{Trace}{TRACE_strongind}\HOLTraceTheoremsTRACEXXstrongind
\HOLThmTag{Trace}{TRACE_trans}\HOLTraceTheoremsTRACEXXtrans
\HOLThmTag{Trace}{TRACE_trans_applied}\HOLTraceTheoremsTRACEXXtransXXapplied
\HOLThmTag{Trace}{TRANS_IN_NODES}\HOLTraceTheoremsTRANSXXINXXNODES
\HOLThmTag{Trace}{UNIQUE_LABEL_cases1}\HOLTraceTheoremsUNIQUEXXLABELXXcasesOne
\HOLThmTag{Trace}{UNIQUE_LABEL_cases2}\HOLTraceTheoremsUNIQUEXXLABELXXcasesTwo
\HOLThmTag{Trace}{UNIQUE_LABEL_IMP_MEM}\HOLTraceTheoremsUNIQUEXXLABELXXIMPXXMEM
\HOLThmTag{Trace}{UNIQUE_LABEL_NOT_NULL}\HOLTraceTheoremsUNIQUEXXLABELXXNOTXXNULL
\HOLThmTag{Trace}{WEAK_TRANS_AND_STEP}\HOLTraceTheoremsWEAKXXTRANSXXANDXXSTEP
\HOLThmTag{Trace}{WEAK_TRANS_AND_TRACE}\HOLTraceTheoremsWEAKXXTRANSXXANDXXTRACE
\HOLThmTag{Trace}{WEAK_TRANS_IN_NODES}\HOLTraceTheoremsWEAKXXTRANSXXINXXNODES
\HOLThmTag{Trace}{WEAK_TRANS_Reach}\HOLTraceTheoremsWEAKXXTRANSXXReach
\HOLThmTag{Trace}{WEAK_TRANS_TRACE}\HOLTraceTheoremsWEAKXXTRANSXXTRACE
}

\newcommand{\HOLCoarsestCongrDate}{02 Dicembre 2017}
\newcommand{\HOLCoarsestCongrTime}{13:31}
\begin{SaveVerbatim}{HOLCoarsestCongrDefinitionsfreeXXactionXXdef}
\HOLTokenTurnstile{} \HOLSymConst{\HOLTokenForall{}}\HOLBoundVar{p}. \HOLConst{free_action} \HOLBoundVar{p} \HOLSymConst{\HOLTokenEquiv{}} \HOLSymConst{\HOLTokenExists{}}\HOLBoundVar{a}. \HOLSymConst{\HOLTokenForall{}}\HOLBoundVar{p\sp{\prime}}. \HOLSymConst{\HOLTokenNeg{}}(\HOLBoundVar{p} \HOLTokenWeakTransBegin\HOLConst{label} \HOLBoundVar{a}\HOLTokenWeakTransEnd \HOLBoundVar{p\sp{\prime}})
\end{SaveVerbatim}
\newcommand{\HOLCoarsestCongrDefinitionsfreeXXactionXXdef}{\UseVerbatim{HOLCoarsestCongrDefinitionsfreeXXactionXXdef}}
\begin{SaveVerbatim}{HOLCoarsestCongrDefinitionsKLOPXXdef}
\HOLTokenTurnstile{} (\HOLSymConst{\HOLTokenForall{}}\HOLBoundVar{a}. \HOLConst{KLOP} \HOLBoundVar{a} \HOLNumLit{0} \HOLSymConst{=} \HOLConst{nil}) \HOLSymConst{\HOLTokenConj{}}
   \HOLSymConst{\HOLTokenForall{}}\HOLBoundVar{a} \HOLBoundVar{n}. \HOLConst{KLOP} \HOLBoundVar{a} (\HOLConst{SUC} \HOLBoundVar{n}) \HOLSymConst{=} \HOLConst{KLOP} \HOLBoundVar{a} \HOLBoundVar{n} \HOLSymConst{+} \HOLConst{label} \HOLBoundVar{a}\HOLSymConst{..}\HOLConst{KLOP} \HOLBoundVar{a} \HOLBoundVar{n}
\end{SaveVerbatim}
\newcommand{\HOLCoarsestCongrDefinitionsKLOPXXdef}{\UseVerbatim{HOLCoarsestCongrDefinitionsKLOPXXdef}}
\begin{SaveVerbatim}{HOLCoarsestCongrDefinitionsSUMXXEQUIV}
\HOLTokenTurnstile{} \HOLConst{SUM_EQUIV} \HOLSymConst{=} (\HOLTokenLambda{}\HOLBoundVar{p} \HOLBoundVar{q}. \HOLSymConst{\HOLTokenForall{}}\HOLBoundVar{r}. \HOLConst{WEAK_EQUIV} (\HOLBoundVar{p} \HOLSymConst{+} \HOLBoundVar{r}) (\HOLBoundVar{q} \HOLSymConst{+} \HOLBoundVar{r}))
\end{SaveVerbatim}
\newcommand{\HOLCoarsestCongrDefinitionsSUMXXEQUIV}{\UseVerbatim{HOLCoarsestCongrDefinitionsSUMXXEQUIV}}
\begin{SaveVerbatim}{HOLCoarsestCongrDefinitionsWEAKXXCONGR}
\HOLTokenTurnstile{} \HOLConst{WEAK_CONGR} \HOLSymConst{=} \HOLConst{CC} \HOLConst{WEAK_EQUIV}
\end{SaveVerbatim}
\newcommand{\HOLCoarsestCongrDefinitionsWEAKXXCONGR}{\UseVerbatim{HOLCoarsestCongrDefinitionsWEAKXXCONGR}}
\newcommand{\HOLCoarsestCongrDefinitions}{
\HOLDfnTag{CoarsestCongr}{free_action_def}\HOLCoarsestCongrDefinitionsfreeXXactionXXdef
\HOLDfnTag{CoarsestCongr}{KLOP_def}\HOLCoarsestCongrDefinitionsKLOPXXdef
\HOLDfnTag{CoarsestCongr}{SUM_EQUIV}\HOLCoarsestCongrDefinitionsSUMXXEQUIV
\HOLDfnTag{CoarsestCongr}{WEAK_CONGR}\HOLCoarsestCongrDefinitionsWEAKXXCONGR
}
\begin{SaveVerbatim}{HOLCoarsestCongrTheoremsCOARSESTXXCONGRXXFINITE}
\HOLTokenTurnstile{} \HOLSymConst{\HOLTokenForall{}}\HOLBoundVar{p} \HOLBoundVar{q}.
     \HOLConst{finite_state} \HOLBoundVar{p} \HOLSymConst{\HOLTokenConj{}} \HOLConst{finite_state} \HOLBoundVar{q} \HOLSymConst{\HOLTokenImp{}}
     (\HOLConst{OBS_CONGR} \HOLBoundVar{p} \HOLBoundVar{q} \HOLSymConst{\HOLTokenEquiv{}} \HOLSymConst{\HOLTokenForall{}}\HOLBoundVar{r}. \HOLConst{WEAK_EQUIV} (\HOLBoundVar{p} \HOLSymConst{+} \HOLBoundVar{r}) (\HOLBoundVar{q} \HOLSymConst{+} \HOLBoundVar{r}))
\end{SaveVerbatim}
\newcommand{\HOLCoarsestCongrTheoremsCOARSESTXXCONGRXXFINITE}{\UseVerbatim{HOLCoarsestCongrTheoremsCOARSESTXXCONGRXXFINITE}}
\begin{SaveVerbatim}{HOLCoarsestCongrTheoremsCOARSESTXXCONGRXXLR}
\HOLTokenTurnstile{} \HOLSymConst{\HOLTokenForall{}}\HOLBoundVar{p} \HOLBoundVar{q}. \HOLConst{OBS_CONGR} \HOLBoundVar{p} \HOLBoundVar{q} \HOLSymConst{\HOLTokenImp{}} \HOLSymConst{\HOLTokenForall{}}\HOLBoundVar{r}. \HOLConst{WEAK_EQUIV} (\HOLBoundVar{p} \HOLSymConst{+} \HOLBoundVar{r}) (\HOLBoundVar{q} \HOLSymConst{+} \HOLBoundVar{r})
\end{SaveVerbatim}
\newcommand{\HOLCoarsestCongrTheoremsCOARSESTXXCONGRXXLR}{\UseVerbatim{HOLCoarsestCongrTheoremsCOARSESTXXCONGRXXLR}}
\begin{SaveVerbatim}{HOLCoarsestCongrTheoremsCOARSESTXXCONGRXXRL}
\HOLTokenTurnstile{} \HOLSymConst{\HOLTokenForall{}}\HOLBoundVar{p} \HOLBoundVar{q}.
     \HOLConst{free_action} \HOLBoundVar{p} \HOLSymConst{\HOLTokenConj{}} \HOLConst{free_action} \HOLBoundVar{q} \HOLSymConst{\HOLTokenImp{}}
     (\HOLSymConst{\HOLTokenForall{}}\HOLBoundVar{r}. \HOLConst{WEAK_EQUIV} (\HOLBoundVar{p} \HOLSymConst{+} \HOLBoundVar{r}) (\HOLBoundVar{q} \HOLSymConst{+} \HOLBoundVar{r})) \HOLSymConst{\HOLTokenImp{}}
     \HOLConst{OBS_CONGR} \HOLBoundVar{p} \HOLBoundVar{q}
\end{SaveVerbatim}
\newcommand{\HOLCoarsestCongrTheoremsCOARSESTXXCONGRXXRL}{\UseVerbatim{HOLCoarsestCongrTheoremsCOARSESTXXCONGRXXRL}}
\begin{SaveVerbatim}{HOLCoarsestCongrTheoremsCOARSESTXXCONGRXXTHM}
\HOLTokenTurnstile{} \HOLSymConst{\HOLTokenForall{}}\HOLBoundVar{p} \HOLBoundVar{q}.
     \HOLConst{free_action} \HOLBoundVar{p} \HOLSymConst{\HOLTokenConj{}} \HOLConst{free_action} \HOLBoundVar{q} \HOLSymConst{\HOLTokenImp{}}
     (\HOLConst{OBS_CONGR} \HOLBoundVar{p} \HOLBoundVar{q} \HOLSymConst{\HOLTokenEquiv{}} \HOLSymConst{\HOLTokenForall{}}\HOLBoundVar{r}. \HOLConst{WEAK_EQUIV} (\HOLBoundVar{p} \HOLSymConst{+} \HOLBoundVar{r}) (\HOLBoundVar{q} \HOLSymConst{+} \HOLBoundVar{r}))
\end{SaveVerbatim}
\newcommand{\HOLCoarsestCongrTheoremsCOARSESTXXCONGRXXTHM}{\UseVerbatim{HOLCoarsestCongrTheoremsCOARSESTXXCONGRXXTHM}}
\begin{SaveVerbatim}{HOLCoarsestCongrTheoremsDENGXXLEMMA}
\HOLTokenTurnstile{} \HOLSymConst{\HOLTokenForall{}}\HOLBoundVar{p} \HOLBoundVar{q}.
     \HOLConst{WEAK_EQUIV} \HOLBoundVar{p} \HOLBoundVar{q} \HOLSymConst{\HOLTokenImp{}}
     (\HOLSymConst{\HOLTokenExists{}}\HOLBoundVar{p\sp{\prime}}. \HOLBoundVar{p} \HOLTokenTransBegin\HOLConst{\ensuremath{\tau}}\HOLTokenTransEnd \HOLBoundVar{p\sp{\prime}} \HOLSymConst{\HOLTokenConj{}} \HOLConst{WEAK_EQUIV} \HOLBoundVar{p\sp{\prime}} \HOLBoundVar{q}) \HOLSymConst{\HOLTokenDisj{}}
     (\HOLSymConst{\HOLTokenExists{}}\HOLBoundVar{q\sp{\prime}}. \HOLBoundVar{q} \HOLTokenTransBegin\HOLConst{\ensuremath{\tau}}\HOLTokenTransEnd \HOLBoundVar{q\sp{\prime}} \HOLSymConst{\HOLTokenConj{}} \HOLConst{WEAK_EQUIV} \HOLBoundVar{p} \HOLBoundVar{q\sp{\prime}}) \HOLSymConst{\HOLTokenDisj{}} \HOLConst{OBS_CONGR} \HOLBoundVar{p} \HOLBoundVar{q}
\end{SaveVerbatim}
\newcommand{\HOLCoarsestCongrTheoremsDENGXXLEMMA}{\UseVerbatim{HOLCoarsestCongrTheoremsDENGXXLEMMA}}
\begin{SaveVerbatim}{HOLCoarsestCongrTheoremsHENNESSYXXLEMMA}
\HOLTokenTurnstile{} \HOLSymConst{\HOLTokenForall{}}\HOLBoundVar{p} \HOLBoundVar{q}.
     \HOLConst{WEAK_EQUIV} \HOLBoundVar{p} \HOLBoundVar{q} \HOLSymConst{\HOLTokenEquiv{}}
     \HOLConst{OBS_CONGR} \HOLBoundVar{p} \HOLBoundVar{q} \HOLSymConst{\HOLTokenDisj{}} \HOLConst{OBS_CONGR} \HOLBoundVar{p} (\HOLConst{\ensuremath{\tau}}\HOLSymConst{..}\HOLBoundVar{q}) \HOLSymConst{\HOLTokenDisj{}} \HOLConst{OBS_CONGR} (\HOLConst{\ensuremath{\tau}}\HOLSymConst{..}\HOLBoundVar{p}) \HOLBoundVar{q}
\end{SaveVerbatim}
\newcommand{\HOLCoarsestCongrTheoremsHENNESSYXXLEMMA}{\UseVerbatim{HOLCoarsestCongrTheoremsHENNESSYXXLEMMA}}
\begin{SaveVerbatim}{HOLCoarsestCongrTheoremsHENNESSYXXLEMMAXXLR}
\HOLTokenTurnstile{} \HOLSymConst{\HOLTokenForall{}}\HOLBoundVar{p} \HOLBoundVar{q}.
     \HOLConst{WEAK_EQUIV} \HOLBoundVar{p} \HOLBoundVar{q} \HOLSymConst{\HOLTokenImp{}}
     \HOLConst{OBS_CONGR} \HOLBoundVar{p} \HOLBoundVar{q} \HOLSymConst{\HOLTokenDisj{}} \HOLConst{OBS_CONGR} \HOLBoundVar{p} (\HOLConst{\ensuremath{\tau}}\HOLSymConst{..}\HOLBoundVar{q}) \HOLSymConst{\HOLTokenDisj{}} \HOLConst{OBS_CONGR} (\HOLConst{\ensuremath{\tau}}\HOLSymConst{..}\HOLBoundVar{p}) \HOLBoundVar{q}
\end{SaveVerbatim}
\newcommand{\HOLCoarsestCongrTheoremsHENNESSYXXLEMMAXXLR}{\UseVerbatim{HOLCoarsestCongrTheoremsHENNESSYXXLEMMAXXLR}}
\begin{SaveVerbatim}{HOLCoarsestCongrTheoremsHENNESSYXXLEMMAXXRL}
\HOLTokenTurnstile{} \HOLSymConst{\HOLTokenForall{}}\HOLBoundVar{p} \HOLBoundVar{q}.
     \HOLConst{OBS_CONGR} \HOLBoundVar{p} \HOLBoundVar{q} \HOLSymConst{\HOLTokenDisj{}} \HOLConst{OBS_CONGR} \HOLBoundVar{p} (\HOLConst{\ensuremath{\tau}}\HOLSymConst{..}\HOLBoundVar{q}) \HOLSymConst{\HOLTokenDisj{}} \HOLConst{OBS_CONGR} (\HOLConst{\ensuremath{\tau}}\HOLSymConst{..}\HOLBoundVar{p}) \HOLBoundVar{q} \HOLSymConst{\HOLTokenImp{}}
     \HOLConst{WEAK_EQUIV} \HOLBoundVar{p} \HOLBoundVar{q}
\end{SaveVerbatim}
\newcommand{\HOLCoarsestCongrTheoremsHENNESSYXXLEMMAXXRL}{\UseVerbatim{HOLCoarsestCongrTheoremsHENNESSYXXLEMMAXXRL}}
\begin{SaveVerbatim}{HOLCoarsestCongrTheoremsINFINITEXXEXISTSXXLEMMA}
\HOLTokenTurnstile{} \HOLSymConst{\HOLTokenForall{}}\HOLBoundVar{R} \HOLBoundVar{A} \HOLBoundVar{B}.
     \HOLConst{equivalence} \HOLBoundVar{R} \HOLSymConst{\HOLTokenImp{}}
     \HOLConst{FINITE} \HOLBoundVar{A} \HOLSymConst{\HOLTokenConj{}} \HOLConst{INFINITE} \HOLBoundVar{B} \HOLSymConst{\HOLTokenConj{}}
     (\HOLSymConst{\HOLTokenForall{}}\HOLBoundVar{x} \HOLBoundVar{y}. \HOLBoundVar{x} \HOLConst{\HOLTokenIn{}} \HOLBoundVar{B} \HOLSymConst{\HOLTokenConj{}} \HOLBoundVar{y} \HOLConst{\HOLTokenIn{}} \HOLBoundVar{B} \HOLSymConst{\HOLTokenConj{}} \HOLBoundVar{x} \HOLSymConst{\HOLTokenNotEqual{}} \HOLBoundVar{y} \HOLSymConst{\HOLTokenImp{}} \HOLSymConst{\HOLTokenNeg{}}\HOLBoundVar{R} \HOLBoundVar{x} \HOLBoundVar{y}) \HOLSymConst{\HOLTokenImp{}}
     \HOLSymConst{\HOLTokenExists{}}\HOLBoundVar{k}. \HOLBoundVar{k} \HOLConst{\HOLTokenIn{}} \HOLBoundVar{B} \HOLSymConst{\HOLTokenConj{}} \HOLSymConst{\HOLTokenForall{}}\HOLBoundVar{n}. \HOLBoundVar{n} \HOLConst{\HOLTokenIn{}} \HOLBoundVar{A} \HOLSymConst{\HOLTokenImp{}} \HOLSymConst{\HOLTokenNeg{}}\HOLBoundVar{R} \HOLBoundVar{n} \HOLBoundVar{k}
\end{SaveVerbatim}
\newcommand{\HOLCoarsestCongrTheoremsINFINITEXXEXISTSXXLEMMA}{\UseVerbatim{HOLCoarsestCongrTheoremsINFINITEXXEXISTSXXLEMMA}}
\begin{SaveVerbatim}{HOLCoarsestCongrTheoremsKZeroXXNOXXTRANS}
\HOLTokenTurnstile{} \HOLSymConst{\HOLTokenForall{}}\HOLBoundVar{a} \HOLBoundVar{u} \HOLBoundVar{E}. \HOLSymConst{\HOLTokenNeg{}}(\HOLConst{KLOP} \HOLBoundVar{a} \HOLNumLit{0} \HOLTokenTransBegin\HOLBoundVar{u}\HOLTokenTransEnd \HOLBoundVar{E})
\end{SaveVerbatim}
\newcommand{\HOLCoarsestCongrTheoremsKZeroXXNOXXTRANS}{\UseVerbatim{HOLCoarsestCongrTheoremsKZeroXXNOXXTRANS}}
\begin{SaveVerbatim}{HOLCoarsestCongrTheoremsKLOPXXdefXXcompute}
\HOLTokenTurnstile{} (\HOLSymConst{\HOLTokenForall{}}\HOLBoundVar{a}. \HOLConst{KLOP} \HOLBoundVar{a} \HOLNumLit{0} \HOLSymConst{=} \HOLConst{nil}) \HOLSymConst{\HOLTokenConj{}}
   (\HOLSymConst{\HOLTokenForall{}}\HOLBoundVar{a} \HOLBoundVar{n}.
      \HOLConst{KLOP} \HOLBoundVar{a} (\HOLConst{NUMERAL} (\HOLConst{BIT1} \HOLBoundVar{n})) \HOLSymConst{=}
      \HOLConst{KLOP} \HOLBoundVar{a} (\HOLConst{NUMERAL} (\HOLConst{BIT1} \HOLBoundVar{n}) \HOLSymConst{-} \HOLNumLit{1}) \HOLSymConst{+}
      \HOLConst{label} \HOLBoundVar{a}\HOLSymConst{..}\HOLConst{KLOP} \HOLBoundVar{a} (\HOLConst{NUMERAL} (\HOLConst{BIT1} \HOLBoundVar{n}) \HOLSymConst{-} \HOLNumLit{1})) \HOLSymConst{\HOLTokenConj{}}
   \HOLSymConst{\HOLTokenForall{}}\HOLBoundVar{a} \HOLBoundVar{n}.
     \HOLConst{KLOP} \HOLBoundVar{a} (\HOLConst{NUMERAL} (\HOLConst{BIT2} \HOLBoundVar{n})) \HOLSymConst{=}
     \HOLConst{KLOP} \HOLBoundVar{a} (\HOLConst{NUMERAL} (\HOLConst{BIT1} \HOLBoundVar{n})) \HOLSymConst{+}
     \HOLConst{label} \HOLBoundVar{a}\HOLSymConst{..}\HOLConst{KLOP} \HOLBoundVar{a} (\HOLConst{NUMERAL} (\HOLConst{BIT1} \HOLBoundVar{n}))
\end{SaveVerbatim}
\newcommand{\HOLCoarsestCongrTheoremsKLOPXXdefXXcompute}{\UseVerbatim{HOLCoarsestCongrTheoremsKLOPXXdefXXcompute}}
\begin{SaveVerbatim}{HOLCoarsestCongrTheoremsKLOPXXLEMMAXXFINITE}
\HOLTokenTurnstile{} \HOLSymConst{\HOLTokenForall{}}\HOLBoundVar{p} \HOLBoundVar{q}.
     \HOLConst{finite_state} \HOLBoundVar{p} \HOLSymConst{\HOLTokenConj{}} \HOLConst{finite_state} \HOLBoundVar{q} \HOLSymConst{\HOLTokenImp{}}
     \HOLSymConst{\HOLTokenExists{}}\HOLBoundVar{k}.
       \HOLConst{STABLE} \HOLBoundVar{k} \HOLSymConst{\HOLTokenConj{}} (\HOLSymConst{\HOLTokenForall{}}\HOLBoundVar{p\sp{\prime}} \HOLBoundVar{u}. \HOLBoundVar{p} \HOLTokenWeakTransBegin\HOLBoundVar{u}\HOLTokenWeakTransEnd \HOLBoundVar{p\sp{\prime}} \HOLSymConst{\HOLTokenImp{}} \HOLSymConst{\HOLTokenNeg{}}\HOLConst{WEAK_EQUIV} \HOLBoundVar{p\sp{\prime}} \HOLBoundVar{k}) \HOLSymConst{\HOLTokenConj{}}
       \HOLSymConst{\HOLTokenForall{}}\HOLBoundVar{q\sp{\prime}} \HOLBoundVar{u}. \HOLBoundVar{q} \HOLTokenWeakTransBegin\HOLBoundVar{u}\HOLTokenWeakTransEnd \HOLBoundVar{q\sp{\prime}} \HOLSymConst{\HOLTokenImp{}} \HOLSymConst{\HOLTokenNeg{}}\HOLConst{WEAK_EQUIV} \HOLBoundVar{q\sp{\prime}} \HOLBoundVar{k}
\end{SaveVerbatim}
\newcommand{\HOLCoarsestCongrTheoremsKLOPXXLEMMAXXFINITE}{\UseVerbatim{HOLCoarsestCongrTheoremsKLOPXXLEMMAXXFINITE}}
\begin{SaveVerbatim}{HOLCoarsestCongrTheoremsKLOPXXONEXXONE}
\HOLTokenTurnstile{} \HOLSymConst{\HOLTokenForall{}}\HOLBoundVar{a}. \HOLConst{ONE_ONE} (\HOLConst{KLOP} \HOLBoundVar{a})
\end{SaveVerbatim}
\newcommand{\HOLCoarsestCongrTheoremsKLOPXXONEXXONE}{\UseVerbatim{HOLCoarsestCongrTheoremsKLOPXXONEXXONE}}
\begin{SaveVerbatim}{HOLCoarsestCongrTheoremsKLOPXXPROPZero}
\HOLTokenTurnstile{} \HOLSymConst{\HOLTokenForall{}}\HOLBoundVar{a} \HOLBoundVar{n}. \HOLConst{STABLE} (\HOLConst{KLOP} \HOLBoundVar{a} \HOLBoundVar{n})
\end{SaveVerbatim}
\newcommand{\HOLCoarsestCongrTheoremsKLOPXXPROPZero}{\UseVerbatim{HOLCoarsestCongrTheoremsKLOPXXPROPZero}}
\begin{SaveVerbatim}{HOLCoarsestCongrTheoremsKLOPXXPROPOne}
\HOLTokenTurnstile{} \HOLSymConst{\HOLTokenForall{}}\HOLBoundVar{a} \HOLBoundVar{n} \HOLBoundVar{E}. \HOLConst{KLOP} \HOLBoundVar{a} \HOLBoundVar{n} \HOLTokenTransBegin\HOLConst{label} \HOLBoundVar{a}\HOLTokenTransEnd \HOLBoundVar{E} \HOLSymConst{\HOLTokenEquiv{}} \HOLSymConst{\HOLTokenExists{}}\HOLBoundVar{m}. \HOLBoundVar{m} \HOLSymConst{\HOLTokenLt{}} \HOLBoundVar{n} \HOLSymConst{\HOLTokenConj{}} (\HOLBoundVar{E} \HOLSymConst{=} \HOLConst{KLOP} \HOLBoundVar{a} \HOLBoundVar{m})
\end{SaveVerbatim}
\newcommand{\HOLCoarsestCongrTheoremsKLOPXXPROPOne}{\UseVerbatim{HOLCoarsestCongrTheoremsKLOPXXPROPOne}}
\begin{SaveVerbatim}{HOLCoarsestCongrTheoremsKLOPXXPROPOneYY}
\HOLTokenTurnstile{} \HOLSymConst{\HOLTokenForall{}}\HOLBoundVar{a} \HOLBoundVar{n} \HOLBoundVar{E}. \HOLConst{KLOP} \HOLBoundVar{a} \HOLBoundVar{n} \HOLTokenWeakTransBegin\HOLConst{label} \HOLBoundVar{a}\HOLTokenWeakTransEnd \HOLBoundVar{E} \HOLSymConst{\HOLTokenEquiv{}} \HOLSymConst{\HOLTokenExists{}}\HOLBoundVar{m}. \HOLBoundVar{m} \HOLSymConst{\HOLTokenLt{}} \HOLBoundVar{n} \HOLSymConst{\HOLTokenConj{}} (\HOLBoundVar{E} \HOLSymConst{=} \HOLConst{KLOP} \HOLBoundVar{a} \HOLBoundVar{m})
\end{SaveVerbatim}
\newcommand{\HOLCoarsestCongrTheoremsKLOPXXPROPOneYY}{\UseVerbatim{HOLCoarsestCongrTheoremsKLOPXXPROPOneYY}}
\begin{SaveVerbatim}{HOLCoarsestCongrTheoremsKLOPXXPROPOneXXLR}
\HOLTokenTurnstile{} \HOLSymConst{\HOLTokenForall{}}\HOLBoundVar{a} \HOLBoundVar{n} \HOLBoundVar{E}. \HOLConst{KLOP} \HOLBoundVar{a} \HOLBoundVar{n} \HOLTokenTransBegin\HOLConst{label} \HOLBoundVar{a}\HOLTokenTransEnd \HOLBoundVar{E} \HOLSymConst{\HOLTokenImp{}} \HOLSymConst{\HOLTokenExists{}}\HOLBoundVar{m}. \HOLBoundVar{m} \HOLSymConst{\HOLTokenLt{}} \HOLBoundVar{n} \HOLSymConst{\HOLTokenConj{}} (\HOLBoundVar{E} \HOLSymConst{=} \HOLConst{KLOP} \HOLBoundVar{a} \HOLBoundVar{m})
\end{SaveVerbatim}
\newcommand{\HOLCoarsestCongrTheoremsKLOPXXPROPOneXXLR}{\UseVerbatim{HOLCoarsestCongrTheoremsKLOPXXPROPOneXXLR}}
\begin{SaveVerbatim}{HOLCoarsestCongrTheoremsKLOPXXPROPOneXXRL}
\HOLTokenTurnstile{} \HOLSymConst{\HOLTokenForall{}}\HOLBoundVar{a} \HOLBoundVar{n} \HOLBoundVar{E}. (\HOLSymConst{\HOLTokenExists{}}\HOLBoundVar{m}. \HOLBoundVar{m} \HOLSymConst{\HOLTokenLt{}} \HOLBoundVar{n} \HOLSymConst{\HOLTokenConj{}} (\HOLBoundVar{E} \HOLSymConst{=} \HOLConst{KLOP} \HOLBoundVar{a} \HOLBoundVar{m})) \HOLSymConst{\HOLTokenImp{}} \HOLConst{KLOP} \HOLBoundVar{a} \HOLBoundVar{n} \HOLTokenTransBegin\HOLConst{label} \HOLBoundVar{a}\HOLTokenTransEnd \HOLBoundVar{E}
\end{SaveVerbatim}
\newcommand{\HOLCoarsestCongrTheoremsKLOPXXPROPOneXXRL}{\UseVerbatim{HOLCoarsestCongrTheoremsKLOPXXPROPOneXXRL}}
\begin{SaveVerbatim}{HOLCoarsestCongrTheoremsKLOPXXPROPTwo}
\HOLTokenTurnstile{} \HOLSymConst{\HOLTokenForall{}}\HOLBoundVar{a} \HOLBoundVar{n} \HOLBoundVar{m}. \HOLBoundVar{m} \HOLSymConst{\HOLTokenLt{}} \HOLBoundVar{n} \HOLSymConst{\HOLTokenImp{}} \HOLSymConst{\HOLTokenNeg{}}\HOLConst{STRONG_EQUIV} (\HOLConst{KLOP} \HOLBoundVar{a} \HOLBoundVar{m}) (\HOLConst{KLOP} \HOLBoundVar{a} \HOLBoundVar{n})
\end{SaveVerbatim}
\newcommand{\HOLCoarsestCongrTheoremsKLOPXXPROPTwo}{\UseVerbatim{HOLCoarsestCongrTheoremsKLOPXXPROPTwo}}
\begin{SaveVerbatim}{HOLCoarsestCongrTheoremsKLOPXXPROPTwoYY}
\HOLTokenTurnstile{} \HOLSymConst{\HOLTokenForall{}}\HOLBoundVar{a} \HOLBoundVar{n} \HOLBoundVar{m}. \HOLBoundVar{m} \HOLSymConst{\HOLTokenLt{}} \HOLBoundVar{n} \HOLSymConst{\HOLTokenImp{}} \HOLSymConst{\HOLTokenNeg{}}\HOLConst{WEAK_EQUIV} (\HOLConst{KLOP} \HOLBoundVar{a} \HOLBoundVar{m}) (\HOLConst{KLOP} \HOLBoundVar{a} \HOLBoundVar{n})
\end{SaveVerbatim}
\newcommand{\HOLCoarsestCongrTheoremsKLOPXXPROPTwoYY}{\UseVerbatim{HOLCoarsestCongrTheoremsKLOPXXPROPTwoYY}}
\begin{SaveVerbatim}{HOLCoarsestCongrTheoremsOBSXXCONGRXXIMPXXWEAKXXCONGR}
\HOLTokenTurnstile{} \HOLSymConst{\HOLTokenForall{}}\HOLBoundVar{p} \HOLBoundVar{q}. \HOLConst{OBS_CONGR} \HOLBoundVar{p} \HOLBoundVar{q} \HOLSymConst{\HOLTokenImp{}} \HOLConst{WEAK_CONGR} \HOLBoundVar{p} \HOLBoundVar{q}
\end{SaveVerbatim}
\newcommand{\HOLCoarsestCongrTheoremsOBSXXCONGRXXIMPXXWEAKXXCONGR}{\UseVerbatim{HOLCoarsestCongrTheoremsOBSXXCONGRXXIMPXXWEAKXXCONGR}}
\begin{SaveVerbatim}{HOLCoarsestCongrTheoremsPROPThreeXXCOMMON}
\HOLTokenTurnstile{} \HOLSymConst{\HOLTokenForall{}}\HOLBoundVar{p} \HOLBoundVar{q}.
     (\HOLSymConst{\HOLTokenExists{}}\HOLBoundVar{k}.
        \HOLConst{STABLE} \HOLBoundVar{k} \HOLSymConst{\HOLTokenConj{}} (\HOLSymConst{\HOLTokenForall{}}\HOLBoundVar{p\sp{\prime}} \HOLBoundVar{u}. \HOLBoundVar{p} \HOLTokenWeakTransBegin\HOLBoundVar{u}\HOLTokenWeakTransEnd \HOLBoundVar{p\sp{\prime}} \HOLSymConst{\HOLTokenImp{}} \HOLSymConst{\HOLTokenNeg{}}\HOLConst{WEAK_EQUIV} \HOLBoundVar{p\sp{\prime}} \HOLBoundVar{k}) \HOLSymConst{\HOLTokenConj{}}
        \HOLSymConst{\HOLTokenForall{}}\HOLBoundVar{q\sp{\prime}} \HOLBoundVar{u}. \HOLBoundVar{q} \HOLTokenWeakTransBegin\HOLBoundVar{u}\HOLTokenWeakTransEnd \HOLBoundVar{q\sp{\prime}} \HOLSymConst{\HOLTokenImp{}} \HOLSymConst{\HOLTokenNeg{}}\HOLConst{WEAK_EQUIV} \HOLBoundVar{q\sp{\prime}} \HOLBoundVar{k}) \HOLSymConst{\HOLTokenImp{}}
     (\HOLSymConst{\HOLTokenForall{}}\HOLBoundVar{r}. \HOLConst{WEAK_EQUIV} (\HOLBoundVar{p} \HOLSymConst{+} \HOLBoundVar{r}) (\HOLBoundVar{q} \HOLSymConst{+} \HOLBoundVar{r})) \HOLSymConst{\HOLTokenImp{}}
     \HOLConst{OBS_CONGR} \HOLBoundVar{p} \HOLBoundVar{q}
\end{SaveVerbatim}
\newcommand{\HOLCoarsestCongrTheoremsPROPThreeXXCOMMON}{\UseVerbatim{HOLCoarsestCongrTheoremsPROPThreeXXCOMMON}}
\begin{SaveVerbatim}{HOLCoarsestCongrTheoremsTAUXXSTRAT}
\HOLTokenTurnstile{} \HOLSymConst{\HOLTokenForall{}}\HOLBoundVar{E} \HOLBoundVar{E\sp{\prime}}. \HOLConst{OBS_CONGR} (\HOLBoundVar{E} \HOLSymConst{+} \HOLConst{\ensuremath{\tau}}\HOLSymConst{..}(\HOLBoundVar{E\sp{\prime}} \HOLSymConst{+} \HOLBoundVar{E})) (\HOLConst{\ensuremath{\tau}}\HOLSymConst{..}(\HOLBoundVar{E\sp{\prime}} \HOLSymConst{+} \HOLBoundVar{E}))
\end{SaveVerbatim}
\newcommand{\HOLCoarsestCongrTheoremsTAUXXSTRAT}{\UseVerbatim{HOLCoarsestCongrTheoremsTAUXXSTRAT}}
\begin{SaveVerbatim}{HOLCoarsestCongrTheoremsWEAKXXCONGRXXcongruence}
\HOLTokenTurnstile{} \HOLConst{congruence} \HOLConst{WEAK_CONGR}
\end{SaveVerbatim}
\newcommand{\HOLCoarsestCongrTheoremsWEAKXXCONGRXXcongruence}{\UseVerbatim{HOLCoarsestCongrTheoremsWEAKXXCONGRXXcongruence}}
\begin{SaveVerbatim}{HOLCoarsestCongrTheoremsWEAKXXCONGRXXIMPXXSUMXXEQUIV}
\HOLTokenTurnstile{} \HOLSymConst{\HOLTokenForall{}}\HOLBoundVar{p} \HOLBoundVar{q}. \HOLConst{WEAK_CONGR} \HOLBoundVar{p} \HOLBoundVar{q} \HOLSymConst{\HOLTokenImp{}} \HOLConst{SUM_EQUIV} \HOLBoundVar{p} \HOLBoundVar{q}
\end{SaveVerbatim}
\newcommand{\HOLCoarsestCongrTheoremsWEAKXXCONGRXXIMPXXSUMXXEQUIV}{\UseVerbatim{HOLCoarsestCongrTheoremsWEAKXXCONGRXXIMPXXSUMXXEQUIV}}
\begin{SaveVerbatim}{HOLCoarsestCongrTheoremsWEAKXXCONGRXXTHM}
\HOLTokenTurnstile{} \HOLConst{WEAK_CONGR} \HOLSymConst{=} (\HOLTokenLambda{}\HOLBoundVar{g} \HOLBoundVar{h}. \HOLSymConst{\HOLTokenForall{}}\HOLBoundVar{c}. \HOLConst{CONTEXT} \HOLBoundVar{c} \HOLSymConst{\HOLTokenImp{}} \HOLConst{WEAK_EQUIV} (\HOLBoundVar{c} \HOLBoundVar{g}) (\HOLBoundVar{c} \HOLBoundVar{h}))
\end{SaveVerbatim}
\newcommand{\HOLCoarsestCongrTheoremsWEAKXXCONGRXXTHM}{\UseVerbatim{HOLCoarsestCongrTheoremsWEAKXXCONGRXXTHM}}
\newcommand{\HOLCoarsestCongrTheorems}{
\HOLThmTag{CoarsestCongr}{COARSEST_CONGR_FINITE}\HOLCoarsestCongrTheoremsCOARSESTXXCONGRXXFINITE
\HOLThmTag{CoarsestCongr}{COARSEST_CONGR_LR}\HOLCoarsestCongrTheoremsCOARSESTXXCONGRXXLR
\HOLThmTag{CoarsestCongr}{COARSEST_CONGR_RL}\HOLCoarsestCongrTheoremsCOARSESTXXCONGRXXRL
\HOLThmTag{CoarsestCongr}{COARSEST_CONGR_THM}\HOLCoarsestCongrTheoremsCOARSESTXXCONGRXXTHM
\HOLThmTag{CoarsestCongr}{DENG_LEMMA}\HOLCoarsestCongrTheoremsDENGXXLEMMA
\HOLThmTag{CoarsestCongr}{HENNESSY_LEMMA}\HOLCoarsestCongrTheoremsHENNESSYXXLEMMA
\HOLThmTag{CoarsestCongr}{HENNESSY_LEMMA_LR}\HOLCoarsestCongrTheoremsHENNESSYXXLEMMAXXLR
\HOLThmTag{CoarsestCongr}{HENNESSY_LEMMA_RL}\HOLCoarsestCongrTheoremsHENNESSYXXLEMMAXXRL
\HOLThmTag{CoarsestCongr}{INFINITE_EXISTS_LEMMA}\HOLCoarsestCongrTheoremsINFINITEXXEXISTSXXLEMMA
\HOLThmTag{CoarsestCongr}{K0_NO_TRANS}\HOLCoarsestCongrTheoremsKZeroXXNOXXTRANS
\HOLThmTag{CoarsestCongr}{KLOP_def_compute}\HOLCoarsestCongrTheoremsKLOPXXdefXXcompute
\HOLThmTag{CoarsestCongr}{KLOP_LEMMA_FINITE}\HOLCoarsestCongrTheoremsKLOPXXLEMMAXXFINITE
\HOLThmTag{CoarsestCongr}{KLOP_ONE_ONE}\HOLCoarsestCongrTheoremsKLOPXXONEXXONE
\HOLThmTag{CoarsestCongr}{KLOP_PROP0}\HOLCoarsestCongrTheoremsKLOPXXPROPZero
\HOLThmTag{CoarsestCongr}{KLOP_PROP1}\HOLCoarsestCongrTheoremsKLOPXXPROPOne
\HOLThmTag{CoarsestCongr}{KLOP_PROP1'}\HOLCoarsestCongrTheoremsKLOPXXPROPOneYY
\HOLThmTag{CoarsestCongr}{KLOP_PROP1_LR}\HOLCoarsestCongrTheoremsKLOPXXPROPOneXXLR
\HOLThmTag{CoarsestCongr}{KLOP_PROP1_RL}\HOLCoarsestCongrTheoremsKLOPXXPROPOneXXRL
\HOLThmTag{CoarsestCongr}{KLOP_PROP2}\HOLCoarsestCongrTheoremsKLOPXXPROPTwo
\HOLThmTag{CoarsestCongr}{KLOP_PROP2'}\HOLCoarsestCongrTheoremsKLOPXXPROPTwoYY
\HOLThmTag{CoarsestCongr}{OBS_CONGR_IMP_WEAK_CONGR}\HOLCoarsestCongrTheoremsOBSXXCONGRXXIMPXXWEAKXXCONGR
\HOLThmTag{CoarsestCongr}{PROP3_COMMON}\HOLCoarsestCongrTheoremsPROPThreeXXCOMMON
\HOLThmTag{CoarsestCongr}{TAU_STRAT}\HOLCoarsestCongrTheoremsTAUXXSTRAT
\HOLThmTag{CoarsestCongr}{WEAK_CONGR_congruence}\HOLCoarsestCongrTheoremsWEAKXXCONGRXXcongruence
\HOLThmTag{CoarsestCongr}{WEAK_CONGR_IMP_SUM_EQUIV}\HOLCoarsestCongrTheoremsWEAKXXCONGRXXIMPXXSUMXXEQUIV
\HOLThmTag{CoarsestCongr}{WEAK_CONGR_THM}\HOLCoarsestCongrTheoremsWEAKXXCONGRXXTHM
}

\newcommand{\HOLBisimulationUptoDate}{02 Dicembre 2017}
\newcommand{\HOLBisimulationUptoTime}{13:31}
\begin{SaveVerbatim}{HOLBisimulationUptoDefinitionsOBSXXBISIMXXUPTO}
\HOLTokenTurnstile{} \HOLSymConst{\HOLTokenForall{}}\HOLBoundVar{Obsm}.
     \HOLConst{OBS_BISIM_UPTO} \HOLBoundVar{Obsm} \HOLSymConst{\HOLTokenEquiv{}}
     \HOLSymConst{\HOLTokenForall{}}\HOLBoundVar{E} \HOLBoundVar{E\sp{\prime}}.
       \HOLBoundVar{Obsm} \HOLBoundVar{E} \HOLBoundVar{E\sp{\prime}} \HOLSymConst{\HOLTokenImp{}}
       \HOLSymConst{\HOLTokenForall{}}\HOLBoundVar{u}.
         (\HOLSymConst{\HOLTokenForall{}}\HOLBoundVar{E\sb{\mathrm{1}}}.
            \HOLBoundVar{E} \HOLTokenTransBegin\HOLBoundVar{u}\HOLTokenTransEnd \HOLBoundVar{E\sb{\mathrm{1}}} \HOLSymConst{\HOLTokenImp{}}
            \HOLSymConst{\HOLTokenExists{}}\HOLBoundVar{E\sb{\mathrm{2}}}.
              \HOLBoundVar{E\sp{\prime}} \HOLTokenWeakTransBegin\HOLBoundVar{u}\HOLTokenWeakTransEnd \HOLBoundVar{E\sb{\mathrm{2}}} \HOLSymConst{\HOLTokenConj{}}
              (\HOLConst{WEAK_EQUIV} \HOLConst{O} \HOLBoundVar{Obsm} \HOLConst{O} \HOLConst{STRONG_EQUIV}) \HOLBoundVar{E\sb{\mathrm{1}}} \HOLBoundVar{E\sb{\mathrm{2}}}) \HOLSymConst{\HOLTokenConj{}}
         \HOLSymConst{\HOLTokenForall{}}\HOLBoundVar{E\sb{\mathrm{2}}}.
           \HOLBoundVar{E\sp{\prime}} \HOLTokenTransBegin\HOLBoundVar{u}\HOLTokenTransEnd \HOLBoundVar{E\sb{\mathrm{2}}} \HOLSymConst{\HOLTokenImp{}}
           \HOLSymConst{\HOLTokenExists{}}\HOLBoundVar{E\sb{\mathrm{1}}}.
             \HOLBoundVar{E} \HOLTokenWeakTransBegin\HOLBoundVar{u}\HOLTokenWeakTransEnd \HOLBoundVar{E\sb{\mathrm{1}}} \HOLSymConst{\HOLTokenConj{}} (\HOLConst{STRONG_EQUIV} \HOLConst{O} \HOLBoundVar{Obsm} \HOLConst{O} \HOLConst{WEAK_EQUIV}) \HOLBoundVar{E\sb{\mathrm{1}}} \HOLBoundVar{E\sb{\mathrm{2}}}
\end{SaveVerbatim}
\newcommand{\HOLBisimulationUptoDefinitionsOBSXXBISIMXXUPTO}{\UseVerbatim{HOLBisimulationUptoDefinitionsOBSXXBISIMXXUPTO}}
\begin{SaveVerbatim}{HOLBisimulationUptoDefinitionsSTRONGXXBISIMXXUPTO}
\HOLTokenTurnstile{} \HOLSymConst{\HOLTokenForall{}}\HOLBoundVar{Bsm}.
     \HOLConst{STRONG_BISIM_UPTO} \HOLBoundVar{Bsm} \HOLSymConst{\HOLTokenEquiv{}}
     \HOLSymConst{\HOLTokenForall{}}\HOLBoundVar{E} \HOLBoundVar{E\sp{\prime}}.
       \HOLBoundVar{Bsm} \HOLBoundVar{E} \HOLBoundVar{E\sp{\prime}} \HOLSymConst{\HOLTokenImp{}}
       \HOLSymConst{\HOLTokenForall{}}\HOLBoundVar{u}.
         (\HOLSymConst{\HOLTokenForall{}}\HOLBoundVar{E\sb{\mathrm{1}}}.
            \HOLBoundVar{E} \HOLTokenTransBegin\HOLBoundVar{u}\HOLTokenTransEnd \HOLBoundVar{E\sb{\mathrm{1}}} \HOLSymConst{\HOLTokenImp{}}
            \HOLSymConst{\HOLTokenExists{}}\HOLBoundVar{E\sb{\mathrm{2}}}.
              \HOLBoundVar{E\sp{\prime}} \HOLTokenTransBegin\HOLBoundVar{u}\HOLTokenTransEnd \HOLBoundVar{E\sb{\mathrm{2}}} \HOLSymConst{\HOLTokenConj{}}
              (\HOLConst{STRONG_EQUIV} \HOLConst{O} \HOLBoundVar{Bsm} \HOLConst{O} \HOLConst{STRONG_EQUIV}) \HOLBoundVar{E\sb{\mathrm{1}}} \HOLBoundVar{E\sb{\mathrm{2}}}) \HOLSymConst{\HOLTokenConj{}}
         \HOLSymConst{\HOLTokenForall{}}\HOLBoundVar{E\sb{\mathrm{2}}}.
           \HOLBoundVar{E\sp{\prime}} \HOLTokenTransBegin\HOLBoundVar{u}\HOLTokenTransEnd \HOLBoundVar{E\sb{\mathrm{2}}} \HOLSymConst{\HOLTokenImp{}}
           \HOLSymConst{\HOLTokenExists{}}\HOLBoundVar{E\sb{\mathrm{1}}}.
             \HOLBoundVar{E} \HOLTokenTransBegin\HOLBoundVar{u}\HOLTokenTransEnd \HOLBoundVar{E\sb{\mathrm{1}}} \HOLSymConst{\HOLTokenConj{}}
             (\HOLConst{STRONG_EQUIV} \HOLConst{O} \HOLBoundVar{Bsm} \HOLConst{O} \HOLConst{STRONG_EQUIV}) \HOLBoundVar{E\sb{\mathrm{1}}} \HOLBoundVar{E\sb{\mathrm{2}}}
\end{SaveVerbatim}
\newcommand{\HOLBisimulationUptoDefinitionsSTRONGXXBISIMXXUPTO}{\UseVerbatim{HOLBisimulationUptoDefinitionsSTRONGXXBISIMXXUPTO}}
\begin{SaveVerbatim}{HOLBisimulationUptoDefinitionsWEAKXXBISIMXXUPTO}
\HOLTokenTurnstile{} \HOLSymConst{\HOLTokenForall{}}\HOLBoundVar{Wbsm}.
     \HOLConst{WEAK_BISIM_UPTO} \HOLBoundVar{Wbsm} \HOLSymConst{\HOLTokenEquiv{}}
     \HOLSymConst{\HOLTokenForall{}}\HOLBoundVar{E} \HOLBoundVar{E\sp{\prime}}.
       \HOLBoundVar{Wbsm} \HOLBoundVar{E} \HOLBoundVar{E\sp{\prime}} \HOLSymConst{\HOLTokenImp{}}
       (\HOLSymConst{\HOLTokenForall{}}\HOLBoundVar{l}.
          (\HOLSymConst{\HOLTokenForall{}}\HOLBoundVar{E\sb{\mathrm{1}}}.
             \HOLBoundVar{E} \HOLTokenTransBegin\HOLConst{label} \HOLBoundVar{l}\HOLTokenTransEnd \HOLBoundVar{E\sb{\mathrm{1}}} \HOLSymConst{\HOLTokenImp{}}
             \HOLSymConst{\HOLTokenExists{}}\HOLBoundVar{E\sb{\mathrm{2}}}.
               \HOLBoundVar{E\sp{\prime}} \HOLTokenWeakTransBegin\HOLConst{label} \HOLBoundVar{l}\HOLTokenWeakTransEnd \HOLBoundVar{E\sb{\mathrm{2}}} \HOLSymConst{\HOLTokenConj{}}
               (\HOLConst{WEAK_EQUIV} \HOLConst{O} \HOLBoundVar{Wbsm} \HOLConst{O} \HOLConst{STRONG_EQUIV}) \HOLBoundVar{E\sb{\mathrm{1}}} \HOLBoundVar{E\sb{\mathrm{2}}}) \HOLSymConst{\HOLTokenConj{}}
          \HOLSymConst{\HOLTokenForall{}}\HOLBoundVar{E\sb{\mathrm{2}}}.
            \HOLBoundVar{E\sp{\prime}} \HOLTokenTransBegin\HOLConst{label} \HOLBoundVar{l}\HOLTokenTransEnd \HOLBoundVar{E\sb{\mathrm{2}}} \HOLSymConst{\HOLTokenImp{}}
            \HOLSymConst{\HOLTokenExists{}}\HOLBoundVar{E\sb{\mathrm{1}}}.
              \HOLBoundVar{E} \HOLTokenWeakTransBegin\HOLConst{label} \HOLBoundVar{l}\HOLTokenWeakTransEnd \HOLBoundVar{E\sb{\mathrm{1}}} \HOLSymConst{\HOLTokenConj{}}
              (\HOLConst{STRONG_EQUIV} \HOLConst{O} \HOLBoundVar{Wbsm} \HOLConst{O} \HOLConst{WEAK_EQUIV}) \HOLBoundVar{E\sb{\mathrm{1}}} \HOLBoundVar{E\sb{\mathrm{2}}}) \HOLSymConst{\HOLTokenConj{}}
       (\HOLSymConst{\HOLTokenForall{}}\HOLBoundVar{E\sb{\mathrm{1}}}.
          \HOLBoundVar{E} \HOLTokenTransBegin\HOLConst{\ensuremath{\tau}}\HOLTokenTransEnd \HOLBoundVar{E\sb{\mathrm{1}}} \HOLSymConst{\HOLTokenImp{}}
          \HOLSymConst{\HOLTokenExists{}}\HOLBoundVar{E\sb{\mathrm{2}}}.
            \HOLConst{EPS} \HOLBoundVar{E\sp{\prime}} \HOLBoundVar{E\sb{\mathrm{2}}} \HOLSymConst{\HOLTokenConj{}}
            (\HOLConst{WEAK_EQUIV} \HOLConst{O} \HOLBoundVar{Wbsm} \HOLConst{O} \HOLConst{STRONG_EQUIV}) \HOLBoundVar{E\sb{\mathrm{1}}} \HOLBoundVar{E\sb{\mathrm{2}}}) \HOLSymConst{\HOLTokenConj{}}
       \HOLSymConst{\HOLTokenForall{}}\HOLBoundVar{E\sb{\mathrm{2}}}.
         \HOLBoundVar{E\sp{\prime}} \HOLTokenTransBegin\HOLConst{\ensuremath{\tau}}\HOLTokenTransEnd \HOLBoundVar{E\sb{\mathrm{2}}} \HOLSymConst{\HOLTokenImp{}}
         \HOLSymConst{\HOLTokenExists{}}\HOLBoundVar{E\sb{\mathrm{1}}}.
           \HOLConst{EPS} \HOLBoundVar{E} \HOLBoundVar{E\sb{\mathrm{1}}} \HOLSymConst{\HOLTokenConj{}} (\HOLConst{STRONG_EQUIV} \HOLConst{O} \HOLBoundVar{Wbsm} \HOLConst{O} \HOLConst{WEAK_EQUIV}) \HOLBoundVar{E\sb{\mathrm{1}}} \HOLBoundVar{E\sb{\mathrm{2}}}
\end{SaveVerbatim}
\newcommand{\HOLBisimulationUptoDefinitionsWEAKXXBISIMXXUPTO}{\UseVerbatim{HOLBisimulationUptoDefinitionsWEAKXXBISIMXXUPTO}}
\begin{SaveVerbatim}{HOLBisimulationUptoDefinitionsWEAKXXBISIMXXUPTOXXALT}
\HOLTokenTurnstile{} \HOLSymConst{\HOLTokenForall{}}\HOLBoundVar{Wbsm}.
     \HOLConst{WEAK_BISIM_UPTO_ALT} \HOLBoundVar{Wbsm} \HOLSymConst{\HOLTokenEquiv{}}
     \HOLSymConst{\HOLTokenForall{}}\HOLBoundVar{E} \HOLBoundVar{E\sp{\prime}}.
       \HOLBoundVar{Wbsm} \HOLBoundVar{E} \HOLBoundVar{E\sp{\prime}} \HOLSymConst{\HOLTokenImp{}}
       (\HOLSymConst{\HOLTokenForall{}}\HOLBoundVar{l}.
          (\HOLSymConst{\HOLTokenForall{}}\HOLBoundVar{E\sb{\mathrm{1}}}.
             \HOLBoundVar{E} \HOLTokenWeakTransBegin\HOLConst{label} \HOLBoundVar{l}\HOLTokenWeakTransEnd \HOLBoundVar{E\sb{\mathrm{1}}} \HOLSymConst{\HOLTokenImp{}}
             \HOLSymConst{\HOLTokenExists{}}\HOLBoundVar{E\sb{\mathrm{2}}}.
               \HOLBoundVar{E\sp{\prime}} \HOLTokenWeakTransBegin\HOLConst{label} \HOLBoundVar{l}\HOLTokenWeakTransEnd \HOLBoundVar{E\sb{\mathrm{2}}} \HOLSymConst{\HOLTokenConj{}}
               (\HOLConst{WEAK_EQUIV} \HOLConst{O} \HOLBoundVar{Wbsm} \HOLConst{O} \HOLConst{WEAK_EQUIV}) \HOLBoundVar{E\sb{\mathrm{1}}} \HOLBoundVar{E\sb{\mathrm{2}}}) \HOLSymConst{\HOLTokenConj{}}
          \HOLSymConst{\HOLTokenForall{}}\HOLBoundVar{E\sb{\mathrm{2}}}.
            \HOLBoundVar{E\sp{\prime}} \HOLTokenWeakTransBegin\HOLConst{label} \HOLBoundVar{l}\HOLTokenWeakTransEnd \HOLBoundVar{E\sb{\mathrm{2}}} \HOLSymConst{\HOLTokenImp{}}
            \HOLSymConst{\HOLTokenExists{}}\HOLBoundVar{E\sb{\mathrm{1}}}.
              \HOLBoundVar{E} \HOLTokenWeakTransBegin\HOLConst{label} \HOLBoundVar{l}\HOLTokenWeakTransEnd \HOLBoundVar{E\sb{\mathrm{1}}} \HOLSymConst{\HOLTokenConj{}}
              (\HOLConst{WEAK_EQUIV} \HOLConst{O} \HOLBoundVar{Wbsm} \HOLConst{O} \HOLConst{WEAK_EQUIV}) \HOLBoundVar{E\sb{\mathrm{1}}} \HOLBoundVar{E\sb{\mathrm{2}}}) \HOLSymConst{\HOLTokenConj{}}
       (\HOLSymConst{\HOLTokenForall{}}\HOLBoundVar{E\sb{\mathrm{1}}}.
          \HOLBoundVar{E} \HOLTokenWeakTransBegin\HOLConst{\ensuremath{\tau}}\HOLTokenWeakTransEnd \HOLBoundVar{E\sb{\mathrm{1}}} \HOLSymConst{\HOLTokenImp{}}
          \HOLSymConst{\HOLTokenExists{}}\HOLBoundVar{E\sb{\mathrm{2}}}.
            \HOLConst{EPS} \HOLBoundVar{E\sp{\prime}} \HOLBoundVar{E\sb{\mathrm{2}}} \HOLSymConst{\HOLTokenConj{}}
            (\HOLConst{WEAK_EQUIV} \HOLConst{O} \HOLBoundVar{Wbsm} \HOLConst{O} \HOLConst{WEAK_EQUIV}) \HOLBoundVar{E\sb{\mathrm{1}}} \HOLBoundVar{E\sb{\mathrm{2}}}) \HOLSymConst{\HOLTokenConj{}}
       \HOLSymConst{\HOLTokenForall{}}\HOLBoundVar{E\sb{\mathrm{2}}}.
         \HOLBoundVar{E\sp{\prime}} \HOLTokenWeakTransBegin\HOLConst{\ensuremath{\tau}}\HOLTokenWeakTransEnd \HOLBoundVar{E\sb{\mathrm{2}}} \HOLSymConst{\HOLTokenImp{}}
         \HOLSymConst{\HOLTokenExists{}}\HOLBoundVar{E\sb{\mathrm{1}}}. \HOLConst{EPS} \HOLBoundVar{E} \HOLBoundVar{E\sb{\mathrm{1}}} \HOLSymConst{\HOLTokenConj{}} (\HOLConst{WEAK_EQUIV} \HOLConst{O} \HOLBoundVar{Wbsm} \HOLConst{O} \HOLConst{WEAK_EQUIV}) \HOLBoundVar{E\sb{\mathrm{1}}} \HOLBoundVar{E\sb{\mathrm{2}}}
\end{SaveVerbatim}
\newcommand{\HOLBisimulationUptoDefinitionsWEAKXXBISIMXXUPTOXXALT}{\UseVerbatim{HOLBisimulationUptoDefinitionsWEAKXXBISIMXXUPTOXXALT}}
\newcommand{\HOLBisimulationUptoDefinitions}{
\HOLDfnTag{BisimulationUpto}{OBS_BISIM_UPTO}\HOLBisimulationUptoDefinitionsOBSXXBISIMXXUPTO
\HOLDfnTag{BisimulationUpto}{STRONG_BISIM_UPTO}\HOLBisimulationUptoDefinitionsSTRONGXXBISIMXXUPTO
\HOLDfnTag{BisimulationUpto}{WEAK_BISIM_UPTO}\HOLBisimulationUptoDefinitionsWEAKXXBISIMXXUPTO
\HOLDfnTag{BisimulationUpto}{WEAK_BISIM_UPTO_ALT}\HOLBisimulationUptoDefinitionsWEAKXXBISIMXXUPTOXXALT
}
\begin{SaveVerbatim}{HOLBisimulationUptoTheoremsCONVERSEXXOBSXXBISIMXXUPTO}
\HOLTokenTurnstile{} \HOLSymConst{\HOLTokenForall{}}\HOLBoundVar{Obsm}. \HOLConst{OBS_BISIM_UPTO} \HOLBoundVar{Obsm} \HOLSymConst{\HOLTokenImp{}} \HOLConst{OBS_BISIM_UPTO} (\HOLConst{relinv} \HOLBoundVar{Obsm})
\end{SaveVerbatim}
\newcommand{\HOLBisimulationUptoTheoremsCONVERSEXXOBSXXBISIMXXUPTO}{\UseVerbatim{HOLBisimulationUptoTheoremsCONVERSEXXOBSXXBISIMXXUPTO}}
\begin{SaveVerbatim}{HOLBisimulationUptoTheoremsCONVERSEXXSTRONGXXBISIMXXUPTO}
\HOLTokenTurnstile{} \HOLSymConst{\HOLTokenForall{}}\HOLBoundVar{Wbsm}.
     \HOLConst{STRONG_BISIM_UPTO} \HOLBoundVar{Wbsm} \HOLSymConst{\HOLTokenImp{}} \HOLConst{STRONG_BISIM_UPTO} (\HOLConst{relinv} \HOLBoundVar{Wbsm})
\end{SaveVerbatim}
\newcommand{\HOLBisimulationUptoTheoremsCONVERSEXXSTRONGXXBISIMXXUPTO}{\UseVerbatim{HOLBisimulationUptoTheoremsCONVERSEXXSTRONGXXBISIMXXUPTO}}
\begin{SaveVerbatim}{HOLBisimulationUptoTheoremsCONVERSEXXWEAKXXBISIMXXUPTO}
\HOLTokenTurnstile{} \HOLSymConst{\HOLTokenForall{}}\HOLBoundVar{Wbsm}. \HOLConst{WEAK_BISIM_UPTO} \HOLBoundVar{Wbsm} \HOLSymConst{\HOLTokenImp{}} \HOLConst{WEAK_BISIM_UPTO} (\HOLConst{relinv} \HOLBoundVar{Wbsm})
\end{SaveVerbatim}
\newcommand{\HOLBisimulationUptoTheoremsCONVERSEXXWEAKXXBISIMXXUPTO}{\UseVerbatim{HOLBisimulationUptoTheoremsCONVERSEXXWEAKXXBISIMXXUPTO}}
\begin{SaveVerbatim}{HOLBisimulationUptoTheoremsCONVERSEXXWEAKXXBISIMXXUPTOXXlemma}
\HOLTokenTurnstile{} \HOLSymConst{\HOLTokenForall{}}\HOLBoundVar{Wbsm} \HOLBoundVar{E} \HOLBoundVar{E\sp{\prime}}.
     (\HOLConst{WEAK_EQUIV} \HOLConst{O} \HOLConst{relinv} \HOLBoundVar{Wbsm} \HOLConst{O} \HOLConst{STRONG_EQUIV}) \HOLBoundVar{E} \HOLBoundVar{E\sp{\prime}} \HOLSymConst{\HOLTokenEquiv{}}
     (\HOLConst{STRONG_EQUIV} \HOLConst{O} \HOLBoundVar{Wbsm} \HOLConst{O} \HOLConst{WEAK_EQUIV}) \HOLBoundVar{E\sp{\prime}} \HOLBoundVar{E}
\end{SaveVerbatim}
\newcommand{\HOLBisimulationUptoTheoremsCONVERSEXXWEAKXXBISIMXXUPTOXXlemma}{\UseVerbatim{HOLBisimulationUptoTheoremsCONVERSEXXWEAKXXBISIMXXUPTOXXlemma}}
\begin{SaveVerbatim}{HOLBisimulationUptoTheoremsCONVERSEXXWEAKXXBISIMXXUPTOXXlemmaYY}
\HOLTokenTurnstile{} \HOLSymConst{\HOLTokenForall{}}\HOLBoundVar{Wbsm} \HOLBoundVar{E} \HOLBoundVar{E\sp{\prime}}.
     (\HOLConst{STRONG_EQUIV} \HOLConst{O} \HOLConst{relinv} \HOLBoundVar{Wbsm} \HOLConst{O} \HOLConst{WEAK_EQUIV}) \HOLBoundVar{E} \HOLBoundVar{E\sp{\prime}} \HOLSymConst{\HOLTokenEquiv{}}
     (\HOLConst{WEAK_EQUIV} \HOLConst{O} \HOLBoundVar{Wbsm} \HOLConst{O} \HOLConst{STRONG_EQUIV}) \HOLBoundVar{E\sp{\prime}} \HOLBoundVar{E}
\end{SaveVerbatim}
\newcommand{\HOLBisimulationUptoTheoremsCONVERSEXXWEAKXXBISIMXXUPTOXXlemmaYY}{\UseVerbatim{HOLBisimulationUptoTheoremsCONVERSEXXWEAKXXBISIMXXUPTOXXlemmaYY}}
\begin{SaveVerbatim}{HOLBisimulationUptoTheoremsIDENTITYXXOBSXXBISIMXXUPTO}
\HOLTokenTurnstile{} \HOLConst{OBS_BISIM_UPTO} (\HOLSymConst{=})
\end{SaveVerbatim}
\newcommand{\HOLBisimulationUptoTheoremsIDENTITYXXOBSXXBISIMXXUPTO}{\UseVerbatim{HOLBisimulationUptoTheoremsIDENTITYXXOBSXXBISIMXXUPTO}}
\begin{SaveVerbatim}{HOLBisimulationUptoTheoremsIDENTITYXXSTRONGXXBISIMXXUPTO}
\HOLTokenTurnstile{} \HOLConst{STRONG_BISIM_UPTO} (\HOLSymConst{=})
\end{SaveVerbatim}
\newcommand{\HOLBisimulationUptoTheoremsIDENTITYXXSTRONGXXBISIMXXUPTO}{\UseVerbatim{HOLBisimulationUptoTheoremsIDENTITYXXSTRONGXXBISIMXXUPTO}}
\begin{SaveVerbatim}{HOLBisimulationUptoTheoremsIDENTITYXXSTRONGXXBISIMXXUPTOXXlemma}
\HOLTokenTurnstile{} \HOLSymConst{\HOLTokenForall{}}\HOLBoundVar{E}. (\HOLConst{STRONG_EQUIV} \HOLConst{O} (\HOLSymConst{=}) \HOLConst{O} \HOLConst{STRONG_EQUIV}) \HOLBoundVar{E} \HOLBoundVar{E}
\end{SaveVerbatim}
\newcommand{\HOLBisimulationUptoTheoremsIDENTITYXXSTRONGXXBISIMXXUPTOXXlemma}{\UseVerbatim{HOLBisimulationUptoTheoremsIDENTITYXXSTRONGXXBISIMXXUPTOXXlemma}}
\begin{SaveVerbatim}{HOLBisimulationUptoTheoremsIDENTITYXXWEAKXXBISIMXXUPTO}
\HOLTokenTurnstile{} \HOLConst{WEAK_BISIM_UPTO} (\HOLSymConst{=})
\end{SaveVerbatim}
\newcommand{\HOLBisimulationUptoTheoremsIDENTITYXXWEAKXXBISIMXXUPTO}{\UseVerbatim{HOLBisimulationUptoTheoremsIDENTITYXXWEAKXXBISIMXXUPTO}}
\begin{SaveVerbatim}{HOLBisimulationUptoTheoremsIDENTITYXXWEAKXXBISIMXXUPTOXXlemma}
\HOLTokenTurnstile{} \HOLSymConst{\HOLTokenForall{}}\HOLBoundVar{E}. (\HOLConst{WEAK_EQUIV} \HOLConst{O} (\HOLSymConst{=}) \HOLConst{O} \HOLConst{STRONG_EQUIV}) \HOLBoundVar{E} \HOLBoundVar{E}
\end{SaveVerbatim}
\newcommand{\HOLBisimulationUptoTheoremsIDENTITYXXWEAKXXBISIMXXUPTOXXlemma}{\UseVerbatim{HOLBisimulationUptoTheoremsIDENTITYXXWEAKXXBISIMXXUPTOXXlemma}}
\begin{SaveVerbatim}{HOLBisimulationUptoTheoremsIDENTITYXXWEAKXXBISIMXXUPTOXXlemmaYY}
\HOLTokenTurnstile{} \HOLSymConst{\HOLTokenForall{}}\HOLBoundVar{E}. (\HOLConst{STRONG_EQUIV} \HOLConst{O} (\HOLSymConst{=}) \HOLConst{O} \HOLConst{WEAK_EQUIV}) \HOLBoundVar{E} \HOLBoundVar{E}
\end{SaveVerbatim}
\newcommand{\HOLBisimulationUptoTheoremsIDENTITYXXWEAKXXBISIMXXUPTOXXlemmaYY}{\UseVerbatim{HOLBisimulationUptoTheoremsIDENTITYXXWEAKXXBISIMXXUPTOXXlemmaYY}}
\begin{SaveVerbatim}{HOLBisimulationUptoTheoremsOBSXXBISIMXXUPTOXXEPS}
\HOLTokenTurnstile{} \HOLSymConst{\HOLTokenForall{}}\HOLBoundVar{Obsm}.
     \HOLConst{OBS_BISIM_UPTO} \HOLBoundVar{Obsm} \HOLSymConst{\HOLTokenImp{}}
     \HOLSymConst{\HOLTokenForall{}}\HOLBoundVar{E} \HOLBoundVar{E\sp{\prime}}.
       \HOLBoundVar{Obsm} \HOLBoundVar{E} \HOLBoundVar{E\sp{\prime}} \HOLSymConst{\HOLTokenImp{}}
       \HOLSymConst{\HOLTokenForall{}}\HOLBoundVar{E\sb{\mathrm{1}}}.
         \HOLConst{EPS} \HOLBoundVar{E} \HOLBoundVar{E\sb{\mathrm{1}}} \HOLSymConst{\HOLTokenImp{}}
         \HOLSymConst{\HOLTokenExists{}}\HOLBoundVar{E\sb{\mathrm{2}}}.
           \HOLConst{EPS} \HOLBoundVar{E\sp{\prime}} \HOLBoundVar{E\sb{\mathrm{2}}} \HOLSymConst{\HOLTokenConj{}} (\HOLConst{WEAK_EQUIV} \HOLConst{O} \HOLBoundVar{Obsm} \HOLConst{O} \HOLConst{STRONG_EQUIV}) \HOLBoundVar{E\sb{\mathrm{1}}} \HOLBoundVar{E\sb{\mathrm{2}}}
\end{SaveVerbatim}
\newcommand{\HOLBisimulationUptoTheoremsOBSXXBISIMXXUPTOXXEPS}{\UseVerbatim{HOLBisimulationUptoTheoremsOBSXXBISIMXXUPTOXXEPS}}
\begin{SaveVerbatim}{HOLBisimulationUptoTheoremsOBSXXBISIMXXUPTOXXEPSYY}
\HOLTokenTurnstile{} \HOLSymConst{\HOLTokenForall{}}\HOLBoundVar{Obsm}.
     \HOLConst{OBS_BISIM_UPTO} \HOLBoundVar{Obsm} \HOLSymConst{\HOLTokenImp{}}
     \HOLSymConst{\HOLTokenForall{}}\HOLBoundVar{E} \HOLBoundVar{E\sp{\prime}}.
       \HOLBoundVar{Obsm} \HOLBoundVar{E} \HOLBoundVar{E\sp{\prime}} \HOLSymConst{\HOLTokenImp{}}
       \HOLSymConst{\HOLTokenForall{}}\HOLBoundVar{E\sb{\mathrm{2}}}.
         \HOLConst{EPS} \HOLBoundVar{E\sp{\prime}} \HOLBoundVar{E\sb{\mathrm{2}}} \HOLSymConst{\HOLTokenImp{}}
         \HOLSymConst{\HOLTokenExists{}}\HOLBoundVar{E\sb{\mathrm{1}}}.
           \HOLConst{EPS} \HOLBoundVar{E} \HOLBoundVar{E\sb{\mathrm{1}}} \HOLSymConst{\HOLTokenConj{}} (\HOLConst{STRONG_EQUIV} \HOLConst{O} \HOLBoundVar{Obsm} \HOLConst{O} \HOLConst{WEAK_EQUIV}) \HOLBoundVar{E\sb{\mathrm{1}}} \HOLBoundVar{E\sb{\mathrm{2}}}
\end{SaveVerbatim}
\newcommand{\HOLBisimulationUptoTheoremsOBSXXBISIMXXUPTOXXEPSYY}{\UseVerbatim{HOLBisimulationUptoTheoremsOBSXXBISIMXXUPTOXXEPSYY}}
\begin{SaveVerbatim}{HOLBisimulationUptoTheoremsOBSXXBISIMXXUPTOXXTHM}
\HOLTokenTurnstile{} \HOLSymConst{\HOLTokenForall{}}\HOLBoundVar{Obsm}. \HOLConst{OBS_BISIM_UPTO} \HOLBoundVar{Obsm} \HOLSymConst{\HOLTokenImp{}} \HOLBoundVar{Obsm} \HOLConst{RSUBSET} \HOLConst{OBS_CONGR}
\end{SaveVerbatim}
\newcommand{\HOLBisimulationUptoTheoremsOBSXXBISIMXXUPTOXXTHM}{\UseVerbatim{HOLBisimulationUptoTheoremsOBSXXBISIMXXUPTOXXTHM}}
\begin{SaveVerbatim}{HOLBisimulationUptoTheoremsOBSXXBISIMXXUPTOXXTRANS}
\HOLTokenTurnstile{} \HOLSymConst{\HOLTokenForall{}}\HOLBoundVar{Obsm}.
     \HOLConst{OBS_BISIM_UPTO} \HOLBoundVar{Obsm} \HOLSymConst{\HOLTokenImp{}}
     \HOLSymConst{\HOLTokenForall{}}\HOLBoundVar{E} \HOLBoundVar{E\sp{\prime}}.
       \HOLBoundVar{Obsm} \HOLBoundVar{E} \HOLBoundVar{E\sp{\prime}} \HOLSymConst{\HOLTokenImp{}}
       \HOLSymConst{\HOLTokenForall{}}\HOLBoundVar{u} \HOLBoundVar{E\sb{\mathrm{1}}}.
         \HOLBoundVar{E} \HOLTokenTransBegin\HOLBoundVar{u}\HOLTokenTransEnd \HOLBoundVar{E\sb{\mathrm{1}}} \HOLSymConst{\HOLTokenImp{}}
         \HOLSymConst{\HOLTokenExists{}}\HOLBoundVar{E\sb{\mathrm{2}}}.
           \HOLBoundVar{E\sp{\prime}} \HOLTokenWeakTransBegin\HOLBoundVar{u}\HOLTokenWeakTransEnd \HOLBoundVar{E\sb{\mathrm{2}}} \HOLSymConst{\HOLTokenConj{}} (\HOLConst{WEAK_EQUIV} \HOLConst{O} \HOLBoundVar{Obsm} \HOLConst{O} \HOLConst{STRONG_EQUIV}) \HOLBoundVar{E\sb{\mathrm{1}}} \HOLBoundVar{E\sb{\mathrm{2}}}
\end{SaveVerbatim}
\newcommand{\HOLBisimulationUptoTheoremsOBSXXBISIMXXUPTOXXTRANS}{\UseVerbatim{HOLBisimulationUptoTheoremsOBSXXBISIMXXUPTOXXTRANS}}
\begin{SaveVerbatim}{HOLBisimulationUptoTheoremsOBSXXBISIMXXUPTOXXTRANSYY}
\HOLTokenTurnstile{} \HOLSymConst{\HOLTokenForall{}}\HOLBoundVar{Obsm}.
     \HOLConst{OBS_BISIM_UPTO} \HOLBoundVar{Obsm} \HOLSymConst{\HOLTokenImp{}}
     \HOLSymConst{\HOLTokenForall{}}\HOLBoundVar{E} \HOLBoundVar{E\sp{\prime}}.
       \HOLBoundVar{Obsm} \HOLBoundVar{E} \HOLBoundVar{E\sp{\prime}} \HOLSymConst{\HOLTokenImp{}}
       \HOLSymConst{\HOLTokenForall{}}\HOLBoundVar{u} \HOLBoundVar{E\sb{\mathrm{2}}}.
         \HOLBoundVar{E\sp{\prime}} \HOLTokenTransBegin\HOLBoundVar{u}\HOLTokenTransEnd \HOLBoundVar{E\sb{\mathrm{2}}} \HOLSymConst{\HOLTokenImp{}}
         \HOLSymConst{\HOLTokenExists{}}\HOLBoundVar{E\sb{\mathrm{1}}}.
           \HOLBoundVar{E} \HOLTokenWeakTransBegin\HOLBoundVar{u}\HOLTokenWeakTransEnd \HOLBoundVar{E\sb{\mathrm{1}}} \HOLSymConst{\HOLTokenConj{}} (\HOLConst{STRONG_EQUIV} \HOLConst{O} \HOLBoundVar{Obsm} \HOLConst{O} \HOLConst{WEAK_EQUIV}) \HOLBoundVar{E\sb{\mathrm{1}}} \HOLBoundVar{E\sb{\mathrm{2}}}
\end{SaveVerbatim}
\newcommand{\HOLBisimulationUptoTheoremsOBSXXBISIMXXUPTOXXTRANSYY}{\UseVerbatim{HOLBisimulationUptoTheoremsOBSXXBISIMXXUPTOXXTRANSYY}}
\begin{SaveVerbatim}{HOLBisimulationUptoTheoremsOBSXXBISIMXXUPTOXXWEAKXXTRANSXXlabel}
\HOLTokenTurnstile{} \HOLSymConst{\HOLTokenForall{}}\HOLBoundVar{Obsm}.
     \HOLConst{OBS_BISIM_UPTO} \HOLBoundVar{Obsm} \HOLSymConst{\HOLTokenImp{}}
     \HOLSymConst{\HOLTokenForall{}}\HOLBoundVar{E} \HOLBoundVar{E\sp{\prime}}.
       \HOLBoundVar{Obsm} \HOLBoundVar{E} \HOLBoundVar{E\sp{\prime}} \HOLSymConst{\HOLTokenImp{}}
       \HOLSymConst{\HOLTokenForall{}}\HOLBoundVar{l} \HOLBoundVar{E\sb{\mathrm{1}}}.
         \HOLBoundVar{E} \HOLTokenWeakTransBegin\HOLConst{label} \HOLBoundVar{l}\HOLTokenWeakTransEnd \HOLBoundVar{E\sb{\mathrm{1}}} \HOLSymConst{\HOLTokenImp{}}
         \HOLSymConst{\HOLTokenExists{}}\HOLBoundVar{E\sb{\mathrm{2}}}.
           \HOLBoundVar{E\sp{\prime}} \HOLTokenWeakTransBegin\HOLConst{label} \HOLBoundVar{l}\HOLTokenWeakTransEnd \HOLBoundVar{E\sb{\mathrm{2}}} \HOLSymConst{\HOLTokenConj{}}
           (\HOLConst{WEAK_EQUIV} \HOLConst{O} \HOLBoundVar{Obsm} \HOLConst{O} \HOLConst{STRONG_EQUIV}) \HOLBoundVar{E\sb{\mathrm{1}}} \HOLBoundVar{E\sb{\mathrm{2}}}
\end{SaveVerbatim}
\newcommand{\HOLBisimulationUptoTheoremsOBSXXBISIMXXUPTOXXWEAKXXTRANSXXlabel}{\UseVerbatim{HOLBisimulationUptoTheoremsOBSXXBISIMXXUPTOXXWEAKXXTRANSXXlabel}}
\begin{SaveVerbatim}{HOLBisimulationUptoTheoremsOBSXXBISIMXXUPTOXXWEAKXXTRANSXXlabelYY}
\HOLTokenTurnstile{} \HOLSymConst{\HOLTokenForall{}}\HOLBoundVar{Obsm}.
     \HOLConst{OBS_BISIM_UPTO} \HOLBoundVar{Obsm} \HOLSymConst{\HOLTokenImp{}}
     \HOLSymConst{\HOLTokenForall{}}\HOLBoundVar{E} \HOLBoundVar{E\sp{\prime}}.
       \HOLBoundVar{Obsm} \HOLBoundVar{E} \HOLBoundVar{E\sp{\prime}} \HOLSymConst{\HOLTokenImp{}}
       \HOLSymConst{\HOLTokenForall{}}\HOLBoundVar{l} \HOLBoundVar{E\sb{\mathrm{2}}}.
         \HOLBoundVar{E\sp{\prime}} \HOLTokenWeakTransBegin\HOLConst{label} \HOLBoundVar{l}\HOLTokenWeakTransEnd \HOLBoundVar{E\sb{\mathrm{2}}} \HOLSymConst{\HOLTokenImp{}}
         \HOLSymConst{\HOLTokenExists{}}\HOLBoundVar{E\sb{\mathrm{1}}}.
           \HOLBoundVar{E} \HOLTokenWeakTransBegin\HOLConst{label} \HOLBoundVar{l}\HOLTokenWeakTransEnd \HOLBoundVar{E\sb{\mathrm{1}}} \HOLSymConst{\HOLTokenConj{}}
           (\HOLConst{STRONG_EQUIV} \HOLConst{O} \HOLBoundVar{Obsm} \HOLConst{O} \HOLConst{WEAK_EQUIV}) \HOLBoundVar{E\sb{\mathrm{1}}} \HOLBoundVar{E\sb{\mathrm{2}}}
\end{SaveVerbatim}
\newcommand{\HOLBisimulationUptoTheoremsOBSXXBISIMXXUPTOXXWEAKXXTRANSXXlabelYY}{\UseVerbatim{HOLBisimulationUptoTheoremsOBSXXBISIMXXUPTOXXWEAKXXTRANSXXlabelYY}}
\begin{SaveVerbatim}{HOLBisimulationUptoTheoremsSTRONGXXBISIMXXUPTOXXLEMMA}
\HOLTokenTurnstile{} \HOLSymConst{\HOLTokenForall{}}\HOLBoundVar{Bsm}.
     \HOLConst{STRONG_BISIM_UPTO} \HOLBoundVar{Bsm} \HOLSymConst{\HOLTokenImp{}}
     \HOLConst{STRONG_BISIM} (\HOLConst{STRONG_EQUIV} \HOLConst{O} \HOLBoundVar{Bsm} \HOLConst{O} \HOLConst{STRONG_EQUIV})
\end{SaveVerbatim}
\newcommand{\HOLBisimulationUptoTheoremsSTRONGXXBISIMXXUPTOXXLEMMA}{\UseVerbatim{HOLBisimulationUptoTheoremsSTRONGXXBISIMXXUPTOXXLEMMA}}
\begin{SaveVerbatim}{HOLBisimulationUptoTheoremsSTRONGXXBISIMXXUPTOXXTHM}
\HOLTokenTurnstile{} \HOLSymConst{\HOLTokenForall{}}\HOLBoundVar{Bsm}. \HOLConst{STRONG_BISIM_UPTO} \HOLBoundVar{Bsm} \HOLSymConst{\HOLTokenImp{}} \HOLBoundVar{Bsm} \HOLConst{RSUBSET} \HOLConst{STRONG_EQUIV}
\end{SaveVerbatim}
\newcommand{\HOLBisimulationUptoTheoremsSTRONGXXBISIMXXUPTOXXTHM}{\UseVerbatim{HOLBisimulationUptoTheoremsSTRONGXXBISIMXXUPTOXXTHM}}
\begin{SaveVerbatim}{HOLBisimulationUptoTheoremsWEAKXXBISIMXXUPTOXXALTXXEPS}
\HOLTokenTurnstile{} \HOLSymConst{\HOLTokenForall{}}\HOLBoundVar{Wbsm}.
     \HOLConst{WEAK_BISIM_UPTO_ALT} \HOLBoundVar{Wbsm} \HOLSymConst{\HOLTokenImp{}}
     \HOLSymConst{\HOLTokenForall{}}\HOLBoundVar{E} \HOLBoundVar{E\sp{\prime}}.
       \HOLBoundVar{Wbsm} \HOLBoundVar{E} \HOLBoundVar{E\sp{\prime}} \HOLSymConst{\HOLTokenImp{}}
       \HOLSymConst{\HOLTokenForall{}}\HOLBoundVar{E\sb{\mathrm{1}}}.
         \HOLConst{EPS} \HOLBoundVar{E} \HOLBoundVar{E\sb{\mathrm{1}}} \HOLSymConst{\HOLTokenImp{}}
         \HOLSymConst{\HOLTokenExists{}}\HOLBoundVar{E\sb{\mathrm{2}}}. \HOLConst{EPS} \HOLBoundVar{E\sp{\prime}} \HOLBoundVar{E\sb{\mathrm{2}}} \HOLSymConst{\HOLTokenConj{}} (\HOLConst{WEAK_EQUIV} \HOLConst{O} \HOLBoundVar{Wbsm} \HOLConst{O} \HOLConst{WEAK_EQUIV}) \HOLBoundVar{E\sb{\mathrm{1}}} \HOLBoundVar{E\sb{\mathrm{2}}}
\end{SaveVerbatim}
\newcommand{\HOLBisimulationUptoTheoremsWEAKXXBISIMXXUPTOXXALTXXEPS}{\UseVerbatim{HOLBisimulationUptoTheoremsWEAKXXBISIMXXUPTOXXALTXXEPS}}
\begin{SaveVerbatim}{HOLBisimulationUptoTheoremsWEAKXXBISIMXXUPTOXXALTXXEPSYY}
\HOLTokenTurnstile{} \HOLSymConst{\HOLTokenForall{}}\HOLBoundVar{Wbsm}.
     \HOLConst{WEAK_BISIM_UPTO_ALT} \HOLBoundVar{Wbsm} \HOLSymConst{\HOLTokenImp{}}
     \HOLSymConst{\HOLTokenForall{}}\HOLBoundVar{E} \HOLBoundVar{E\sp{\prime}}.
       \HOLBoundVar{Wbsm} \HOLBoundVar{E} \HOLBoundVar{E\sp{\prime}} \HOLSymConst{\HOLTokenImp{}}
       \HOLSymConst{\HOLTokenForall{}}\HOLBoundVar{E\sb{\mathrm{2}}}.
         \HOLConst{EPS} \HOLBoundVar{E\sp{\prime}} \HOLBoundVar{E\sb{\mathrm{2}}} \HOLSymConst{\HOLTokenImp{}}
         \HOLSymConst{\HOLTokenExists{}}\HOLBoundVar{E\sb{\mathrm{1}}}. \HOLConst{EPS} \HOLBoundVar{E} \HOLBoundVar{E\sb{\mathrm{1}}} \HOLSymConst{\HOLTokenConj{}} (\HOLConst{WEAK_EQUIV} \HOLConst{O} \HOLBoundVar{Wbsm} \HOLConst{O} \HOLConst{WEAK_EQUIV}) \HOLBoundVar{E\sb{\mathrm{1}}} \HOLBoundVar{E\sb{\mathrm{2}}}
\end{SaveVerbatim}
\newcommand{\HOLBisimulationUptoTheoremsWEAKXXBISIMXXUPTOXXALTXXEPSYY}{\UseVerbatim{HOLBisimulationUptoTheoremsWEAKXXBISIMXXUPTOXXALTXXEPSYY}}
\begin{SaveVerbatim}{HOLBisimulationUptoTheoremsWEAKXXBISIMXXUPTOXXALTXXLEMMA}
\HOLTokenTurnstile{} \HOLSymConst{\HOLTokenForall{}}\HOLBoundVar{Wbsm}.
     \HOLConst{WEAK_BISIM_UPTO_ALT} \HOLBoundVar{Wbsm} \HOLSymConst{\HOLTokenImp{}}
     \HOLConst{WEAK_BISIM} (\HOLConst{WEAK_EQUIV} \HOLConst{O} \HOLBoundVar{Wbsm} \HOLConst{O} \HOLConst{WEAK_EQUIV})
\end{SaveVerbatim}
\newcommand{\HOLBisimulationUptoTheoremsWEAKXXBISIMXXUPTOXXALTXXLEMMA}{\UseVerbatim{HOLBisimulationUptoTheoremsWEAKXXBISIMXXUPTOXXALTXXLEMMA}}
\begin{SaveVerbatim}{HOLBisimulationUptoTheoremsWEAKXXBISIMXXUPTOXXALTXXTHM}
\HOLTokenTurnstile{} \HOLSymConst{\HOLTokenForall{}}\HOLBoundVar{Wbsm}. \HOLConst{WEAK_BISIM_UPTO_ALT} \HOLBoundVar{Wbsm} \HOLSymConst{\HOLTokenImp{}} \HOLBoundVar{Wbsm} \HOLConst{RSUBSET} \HOLConst{WEAK_EQUIV}
\end{SaveVerbatim}
\newcommand{\HOLBisimulationUptoTheoremsWEAKXXBISIMXXUPTOXXALTXXTHM}{\UseVerbatim{HOLBisimulationUptoTheoremsWEAKXXBISIMXXUPTOXXALTXXTHM}}
\begin{SaveVerbatim}{HOLBisimulationUptoTheoremsWEAKXXBISIMXXUPTOXXALTXXWEAKXXTRANSXXlabel}
\HOLTokenTurnstile{} \HOLSymConst{\HOLTokenForall{}}\HOLBoundVar{Wbsm}.
     \HOLConst{WEAK_BISIM_UPTO_ALT} \HOLBoundVar{Wbsm} \HOLSymConst{\HOLTokenImp{}}
     \HOLSymConst{\HOLTokenForall{}}\HOLBoundVar{E} \HOLBoundVar{E\sp{\prime}}.
       \HOLBoundVar{Wbsm} \HOLBoundVar{E} \HOLBoundVar{E\sp{\prime}} \HOLSymConst{\HOLTokenImp{}}
       \HOLSymConst{\HOLTokenForall{}}\HOLBoundVar{l} \HOLBoundVar{E\sb{\mathrm{1}}}.
         \HOLBoundVar{E} \HOLTokenWeakTransBegin\HOLConst{label} \HOLBoundVar{l}\HOLTokenWeakTransEnd \HOLBoundVar{E\sb{\mathrm{1}}} \HOLSymConst{\HOLTokenImp{}}
         \HOLSymConst{\HOLTokenExists{}}\HOLBoundVar{E\sb{\mathrm{2}}}.
           \HOLBoundVar{E\sp{\prime}} \HOLTokenWeakTransBegin\HOLConst{label} \HOLBoundVar{l}\HOLTokenWeakTransEnd \HOLBoundVar{E\sb{\mathrm{2}}} \HOLSymConst{\HOLTokenConj{}}
           (\HOLConst{WEAK_EQUIV} \HOLConst{O} \HOLBoundVar{Wbsm} \HOLConst{O} \HOLConst{WEAK_EQUIV}) \HOLBoundVar{E\sb{\mathrm{1}}} \HOLBoundVar{E\sb{\mathrm{2}}}
\end{SaveVerbatim}
\newcommand{\HOLBisimulationUptoTheoremsWEAKXXBISIMXXUPTOXXALTXXWEAKXXTRANSXXlabel}{\UseVerbatim{HOLBisimulationUptoTheoremsWEAKXXBISIMXXUPTOXXALTXXWEAKXXTRANSXXlabel}}
\begin{SaveVerbatim}{HOLBisimulationUptoTheoremsWEAKXXBISIMXXUPTOXXALTXXWEAKXXTRANSXXlabelYY}
\HOLTokenTurnstile{} \HOLSymConst{\HOLTokenForall{}}\HOLBoundVar{Wbsm}.
     \HOLConst{WEAK_BISIM_UPTO_ALT} \HOLBoundVar{Wbsm} \HOLSymConst{\HOLTokenImp{}}
     \HOLSymConst{\HOLTokenForall{}}\HOLBoundVar{E} \HOLBoundVar{E\sp{\prime}}.
       \HOLBoundVar{Wbsm} \HOLBoundVar{E} \HOLBoundVar{E\sp{\prime}} \HOLSymConst{\HOLTokenImp{}}
       \HOLSymConst{\HOLTokenForall{}}\HOLBoundVar{l} \HOLBoundVar{E\sb{\mathrm{2}}}.
         \HOLBoundVar{E\sp{\prime}} \HOLTokenWeakTransBegin\HOLConst{label} \HOLBoundVar{l}\HOLTokenWeakTransEnd \HOLBoundVar{E\sb{\mathrm{2}}} \HOLSymConst{\HOLTokenImp{}}
         \HOLSymConst{\HOLTokenExists{}}\HOLBoundVar{E\sb{\mathrm{1}}}.
           \HOLBoundVar{E} \HOLTokenWeakTransBegin\HOLConst{label} \HOLBoundVar{l}\HOLTokenWeakTransEnd \HOLBoundVar{E\sb{\mathrm{1}}} \HOLSymConst{\HOLTokenConj{}}
           (\HOLConst{WEAK_EQUIV} \HOLConst{O} \HOLBoundVar{Wbsm} \HOLConst{O} \HOLConst{WEAK_EQUIV}) \HOLBoundVar{E\sb{\mathrm{1}}} \HOLBoundVar{E\sb{\mathrm{2}}}
\end{SaveVerbatim}
\newcommand{\HOLBisimulationUptoTheoremsWEAKXXBISIMXXUPTOXXALTXXWEAKXXTRANSXXlabelYY}{\UseVerbatim{HOLBisimulationUptoTheoremsWEAKXXBISIMXXUPTOXXALTXXWEAKXXTRANSXXlabelYY}}
\begin{SaveVerbatim}{HOLBisimulationUptoTheoremsWEAKXXBISIMXXUPTOXXALTXXWEAKXXTRANSXXtau}
\HOLTokenTurnstile{} \HOLSymConst{\HOLTokenForall{}}\HOLBoundVar{Wbsm}.
     \HOLConst{WEAK_BISIM_UPTO_ALT} \HOLBoundVar{Wbsm} \HOLSymConst{\HOLTokenImp{}}
     \HOLSymConst{\HOLTokenForall{}}\HOLBoundVar{E} \HOLBoundVar{E\sp{\prime}}.
       \HOLBoundVar{Wbsm} \HOLBoundVar{E} \HOLBoundVar{E\sp{\prime}} \HOLSymConst{\HOLTokenImp{}}
       \HOLSymConst{\HOLTokenForall{}}\HOLBoundVar{E\sb{\mathrm{1}}}.
         \HOLBoundVar{E} \HOLTokenWeakTransBegin\HOLConst{\ensuremath{\tau}}\HOLTokenWeakTransEnd \HOLBoundVar{E\sb{\mathrm{1}}} \HOLSymConst{\HOLTokenImp{}}
         \HOLSymConst{\HOLTokenExists{}}\HOLBoundVar{E\sb{\mathrm{2}}}. \HOLConst{EPS} \HOLBoundVar{E\sp{\prime}} \HOLBoundVar{E\sb{\mathrm{2}}} \HOLSymConst{\HOLTokenConj{}} (\HOLConst{WEAK_EQUIV} \HOLConst{O} \HOLBoundVar{Wbsm} \HOLConst{O} \HOLConst{WEAK_EQUIV}) \HOLBoundVar{E\sb{\mathrm{1}}} \HOLBoundVar{E\sb{\mathrm{2}}}
\end{SaveVerbatim}
\newcommand{\HOLBisimulationUptoTheoremsWEAKXXBISIMXXUPTOXXALTXXWEAKXXTRANSXXtau}{\UseVerbatim{HOLBisimulationUptoTheoremsWEAKXXBISIMXXUPTOXXALTXXWEAKXXTRANSXXtau}}
\begin{SaveVerbatim}{HOLBisimulationUptoTheoremsWEAKXXBISIMXXUPTOXXALTXXWEAKXXTRANSXXtauYY}
\HOLTokenTurnstile{} \HOLSymConst{\HOLTokenForall{}}\HOLBoundVar{Wbsm}.
     \HOLConst{WEAK_BISIM_UPTO_ALT} \HOLBoundVar{Wbsm} \HOLSymConst{\HOLTokenImp{}}
     \HOLSymConst{\HOLTokenForall{}}\HOLBoundVar{E} \HOLBoundVar{E\sp{\prime}}.
       \HOLBoundVar{Wbsm} \HOLBoundVar{E} \HOLBoundVar{E\sp{\prime}} \HOLSymConst{\HOLTokenImp{}}
       \HOLSymConst{\HOLTokenForall{}}\HOLBoundVar{E\sb{\mathrm{2}}}.
         \HOLBoundVar{E\sp{\prime}} \HOLTokenWeakTransBegin\HOLConst{\ensuremath{\tau}}\HOLTokenWeakTransEnd \HOLBoundVar{E\sb{\mathrm{2}}} \HOLSymConst{\HOLTokenImp{}}
         \HOLSymConst{\HOLTokenExists{}}\HOLBoundVar{E\sb{\mathrm{1}}}. \HOLConst{EPS} \HOLBoundVar{E} \HOLBoundVar{E\sb{\mathrm{1}}} \HOLSymConst{\HOLTokenConj{}} (\HOLConst{WEAK_EQUIV} \HOLConst{O} \HOLBoundVar{Wbsm} \HOLConst{O} \HOLConst{WEAK_EQUIV}) \HOLBoundVar{E\sb{\mathrm{1}}} \HOLBoundVar{E\sb{\mathrm{2}}}
\end{SaveVerbatim}
\newcommand{\HOLBisimulationUptoTheoremsWEAKXXBISIMXXUPTOXXALTXXWEAKXXTRANSXXtauYY}{\UseVerbatim{HOLBisimulationUptoTheoremsWEAKXXBISIMXXUPTOXXALTXXWEAKXXTRANSXXtauYY}}
\begin{SaveVerbatim}{HOLBisimulationUptoTheoremsWEAKXXBISIMXXUPTOXXEPS}
\HOLTokenTurnstile{} \HOLSymConst{\HOLTokenForall{}}\HOLBoundVar{Wbsm}.
     \HOLConst{WEAK_BISIM_UPTO} \HOLBoundVar{Wbsm} \HOLSymConst{\HOLTokenImp{}}
     \HOLSymConst{\HOLTokenForall{}}\HOLBoundVar{E} \HOLBoundVar{E\sp{\prime}}.
       \HOLBoundVar{Wbsm} \HOLBoundVar{E} \HOLBoundVar{E\sp{\prime}} \HOLSymConst{\HOLTokenImp{}}
       \HOLSymConst{\HOLTokenForall{}}\HOLBoundVar{E\sb{\mathrm{1}}}.
         \HOLConst{EPS} \HOLBoundVar{E} \HOLBoundVar{E\sb{\mathrm{1}}} \HOLSymConst{\HOLTokenImp{}}
         \HOLSymConst{\HOLTokenExists{}}\HOLBoundVar{E\sb{\mathrm{2}}}.
           \HOLConst{EPS} \HOLBoundVar{E\sp{\prime}} \HOLBoundVar{E\sb{\mathrm{2}}} \HOLSymConst{\HOLTokenConj{}} (\HOLConst{WEAK_EQUIV} \HOLConst{O} \HOLBoundVar{Wbsm} \HOLConst{O} \HOLConst{STRONG_EQUIV}) \HOLBoundVar{E\sb{\mathrm{1}}} \HOLBoundVar{E\sb{\mathrm{2}}}
\end{SaveVerbatim}
\newcommand{\HOLBisimulationUptoTheoremsWEAKXXBISIMXXUPTOXXEPS}{\UseVerbatim{HOLBisimulationUptoTheoremsWEAKXXBISIMXXUPTOXXEPS}}
\begin{SaveVerbatim}{HOLBisimulationUptoTheoremsWEAKXXBISIMXXUPTOXXEPSYY}
\HOLTokenTurnstile{} \HOLSymConst{\HOLTokenForall{}}\HOLBoundVar{Wbsm}.
     \HOLConst{WEAK_BISIM_UPTO} \HOLBoundVar{Wbsm} \HOLSymConst{\HOLTokenImp{}}
     \HOLSymConst{\HOLTokenForall{}}\HOLBoundVar{E} \HOLBoundVar{E\sp{\prime}}.
       \HOLBoundVar{Wbsm} \HOLBoundVar{E} \HOLBoundVar{E\sp{\prime}} \HOLSymConst{\HOLTokenImp{}}
       \HOLSymConst{\HOLTokenForall{}}\HOLBoundVar{E\sb{\mathrm{2}}}.
         \HOLConst{EPS} \HOLBoundVar{E\sp{\prime}} \HOLBoundVar{E\sb{\mathrm{2}}} \HOLSymConst{\HOLTokenImp{}}
         \HOLSymConst{\HOLTokenExists{}}\HOLBoundVar{E\sb{\mathrm{1}}}.
           \HOLConst{EPS} \HOLBoundVar{E} \HOLBoundVar{E\sb{\mathrm{1}}} \HOLSymConst{\HOLTokenConj{}} (\HOLConst{STRONG_EQUIV} \HOLConst{O} \HOLBoundVar{Wbsm} \HOLConst{O} \HOLConst{WEAK_EQUIV}) \HOLBoundVar{E\sb{\mathrm{1}}} \HOLBoundVar{E\sb{\mathrm{2}}}
\end{SaveVerbatim}
\newcommand{\HOLBisimulationUptoTheoremsWEAKXXBISIMXXUPTOXXEPSYY}{\UseVerbatim{HOLBisimulationUptoTheoremsWEAKXXBISIMXXUPTOXXEPSYY}}
\begin{SaveVerbatim}{HOLBisimulationUptoTheoremsWEAKXXBISIMXXUPTOXXLEMMA}
\HOLTokenTurnstile{} \HOLSymConst{\HOLTokenForall{}}\HOLBoundVar{Wbsm}.
     \HOLConst{WEAK_BISIM_UPTO} \HOLBoundVar{Wbsm} \HOLSymConst{\HOLTokenImp{}}
     \HOLConst{WEAK_BISIM} (\HOLConst{WEAK_EQUIV} \HOLConst{O} \HOLBoundVar{Wbsm} \HOLConst{O} \HOLConst{WEAK_EQUIV})
\end{SaveVerbatim}
\newcommand{\HOLBisimulationUptoTheoremsWEAKXXBISIMXXUPTOXXLEMMA}{\UseVerbatim{HOLBisimulationUptoTheoremsWEAKXXBISIMXXUPTOXXLEMMA}}
\begin{SaveVerbatim}{HOLBisimulationUptoTheoremsWEAKXXBISIMXXUPTOXXTHM}
\HOLTokenTurnstile{} \HOLSymConst{\HOLTokenForall{}}\HOLBoundVar{Wbsm}. \HOLConst{WEAK_BISIM_UPTO} \HOLBoundVar{Wbsm} \HOLSymConst{\HOLTokenImp{}} \HOLBoundVar{Wbsm} \HOLConst{RSUBSET} \HOLConst{WEAK_EQUIV}
\end{SaveVerbatim}
\newcommand{\HOLBisimulationUptoTheoremsWEAKXXBISIMXXUPTOXXTHM}{\UseVerbatim{HOLBisimulationUptoTheoremsWEAKXXBISIMXXUPTOXXTHM}}
\begin{SaveVerbatim}{HOLBisimulationUptoTheoremsWEAKXXBISIMXXUPTOXXTRANSXXlabel}
\HOLTokenTurnstile{} \HOLSymConst{\HOLTokenForall{}}\HOLBoundVar{Wbsm}.
     \HOLConst{WEAK_BISIM_UPTO} \HOLBoundVar{Wbsm} \HOLSymConst{\HOLTokenImp{}}
     \HOLSymConst{\HOLTokenForall{}}\HOLBoundVar{E} \HOLBoundVar{E\sp{\prime}}.
       \HOLBoundVar{Wbsm} \HOLBoundVar{E} \HOLBoundVar{E\sp{\prime}} \HOLSymConst{\HOLTokenImp{}}
       \HOLSymConst{\HOLTokenForall{}}\HOLBoundVar{l} \HOLBoundVar{E\sb{\mathrm{1}}}.
         \HOLBoundVar{E} \HOLTokenTransBegin\HOLConst{label} \HOLBoundVar{l}\HOLTokenTransEnd \HOLBoundVar{E\sb{\mathrm{1}}} \HOLSymConst{\HOLTokenImp{}}
         \HOLSymConst{\HOLTokenExists{}}\HOLBoundVar{E\sb{\mathrm{2}}}.
           \HOLBoundVar{E\sp{\prime}} \HOLTokenWeakTransBegin\HOLConst{label} \HOLBoundVar{l}\HOLTokenWeakTransEnd \HOLBoundVar{E\sb{\mathrm{2}}} \HOLSymConst{\HOLTokenConj{}}
           (\HOLConst{WEAK_EQUIV} \HOLConst{O} \HOLBoundVar{Wbsm} \HOLConst{O} \HOLConst{STRONG_EQUIV}) \HOLBoundVar{E\sb{\mathrm{1}}} \HOLBoundVar{E\sb{\mathrm{2}}}
\end{SaveVerbatim}
\newcommand{\HOLBisimulationUptoTheoremsWEAKXXBISIMXXUPTOXXTRANSXXlabel}{\UseVerbatim{HOLBisimulationUptoTheoremsWEAKXXBISIMXXUPTOXXTRANSXXlabel}}
\begin{SaveVerbatim}{HOLBisimulationUptoTheoremsWEAKXXBISIMXXUPTOXXTRANSXXlabelYY}
\HOLTokenTurnstile{} \HOLSymConst{\HOLTokenForall{}}\HOLBoundVar{Wbsm}.
     \HOLConst{WEAK_BISIM_UPTO} \HOLBoundVar{Wbsm} \HOLSymConst{\HOLTokenImp{}}
     \HOLSymConst{\HOLTokenForall{}}\HOLBoundVar{E} \HOLBoundVar{E\sp{\prime}}.
       \HOLBoundVar{Wbsm} \HOLBoundVar{E} \HOLBoundVar{E\sp{\prime}} \HOLSymConst{\HOLTokenImp{}}
       \HOLSymConst{\HOLTokenForall{}}\HOLBoundVar{l} \HOLBoundVar{E\sb{\mathrm{2}}}.
         \HOLBoundVar{E\sp{\prime}} \HOLTokenTransBegin\HOLConst{label} \HOLBoundVar{l}\HOLTokenTransEnd \HOLBoundVar{E\sb{\mathrm{2}}} \HOLSymConst{\HOLTokenImp{}}
         \HOLSymConst{\HOLTokenExists{}}\HOLBoundVar{E\sb{\mathrm{1}}}.
           \HOLBoundVar{E} \HOLTokenWeakTransBegin\HOLConst{label} \HOLBoundVar{l}\HOLTokenWeakTransEnd \HOLBoundVar{E\sb{\mathrm{1}}} \HOLSymConst{\HOLTokenConj{}}
           (\HOLConst{STRONG_EQUIV} \HOLConst{O} \HOLBoundVar{Wbsm} \HOLConst{O} \HOLConst{WEAK_EQUIV}) \HOLBoundVar{E\sb{\mathrm{1}}} \HOLBoundVar{E\sb{\mathrm{2}}}
\end{SaveVerbatim}
\newcommand{\HOLBisimulationUptoTheoremsWEAKXXBISIMXXUPTOXXTRANSXXlabelYY}{\UseVerbatim{HOLBisimulationUptoTheoremsWEAKXXBISIMXXUPTOXXTRANSXXlabelYY}}
\begin{SaveVerbatim}{HOLBisimulationUptoTheoremsWEAKXXBISIMXXUPTOXXTRANSXXtau}
\HOLTokenTurnstile{} \HOLSymConst{\HOLTokenForall{}}\HOLBoundVar{Wbsm}.
     \HOLConst{WEAK_BISIM_UPTO} \HOLBoundVar{Wbsm} \HOLSymConst{\HOLTokenImp{}}
     \HOLSymConst{\HOLTokenForall{}}\HOLBoundVar{E} \HOLBoundVar{E\sp{\prime}}.
       \HOLBoundVar{Wbsm} \HOLBoundVar{E} \HOLBoundVar{E\sp{\prime}} \HOLSymConst{\HOLTokenImp{}}
       \HOLSymConst{\HOLTokenForall{}}\HOLBoundVar{E\sb{\mathrm{1}}}.
         \HOLBoundVar{E} \HOLTokenTransBegin\HOLConst{\ensuremath{\tau}}\HOLTokenTransEnd \HOLBoundVar{E\sb{\mathrm{1}}} \HOLSymConst{\HOLTokenImp{}}
         \HOLSymConst{\HOLTokenExists{}}\HOLBoundVar{E\sb{\mathrm{2}}}.
           \HOLConst{EPS} \HOLBoundVar{E\sp{\prime}} \HOLBoundVar{E\sb{\mathrm{2}}} \HOLSymConst{\HOLTokenConj{}} (\HOLConst{WEAK_EQUIV} \HOLConst{O} \HOLBoundVar{Wbsm} \HOLConst{O} \HOLConst{STRONG_EQUIV}) \HOLBoundVar{E\sb{\mathrm{1}}} \HOLBoundVar{E\sb{\mathrm{2}}}
\end{SaveVerbatim}
\newcommand{\HOLBisimulationUptoTheoremsWEAKXXBISIMXXUPTOXXTRANSXXtau}{\UseVerbatim{HOLBisimulationUptoTheoremsWEAKXXBISIMXXUPTOXXTRANSXXtau}}
\begin{SaveVerbatim}{HOLBisimulationUptoTheoremsWEAKXXBISIMXXUPTOXXTRANSXXtauYY}
\HOLTokenTurnstile{} \HOLSymConst{\HOLTokenForall{}}\HOLBoundVar{Wbsm}.
     \HOLConst{WEAK_BISIM_UPTO} \HOLBoundVar{Wbsm} \HOLSymConst{\HOLTokenImp{}}
     \HOLSymConst{\HOLTokenForall{}}\HOLBoundVar{E} \HOLBoundVar{E\sp{\prime}}.
       \HOLBoundVar{Wbsm} \HOLBoundVar{E} \HOLBoundVar{E\sp{\prime}} \HOLSymConst{\HOLTokenImp{}}
       \HOLSymConst{\HOLTokenForall{}}\HOLBoundVar{E\sb{\mathrm{2}}}.
         \HOLBoundVar{E\sp{\prime}} \HOLTokenTransBegin\HOLConst{\ensuremath{\tau}}\HOLTokenTransEnd \HOLBoundVar{E\sb{\mathrm{2}}} \HOLSymConst{\HOLTokenImp{}}
         \HOLSymConst{\HOLTokenExists{}}\HOLBoundVar{E\sb{\mathrm{1}}}.
           \HOLConst{EPS} \HOLBoundVar{E} \HOLBoundVar{E\sb{\mathrm{1}}} \HOLSymConst{\HOLTokenConj{}} (\HOLConst{STRONG_EQUIV} \HOLConst{O} \HOLBoundVar{Wbsm} \HOLConst{O} \HOLConst{WEAK_EQUIV}) \HOLBoundVar{E\sb{\mathrm{1}}} \HOLBoundVar{E\sb{\mathrm{2}}}
\end{SaveVerbatim}
\newcommand{\HOLBisimulationUptoTheoremsWEAKXXBISIMXXUPTOXXTRANSXXtauYY}{\UseVerbatim{HOLBisimulationUptoTheoremsWEAKXXBISIMXXUPTOXXTRANSXXtauYY}}
\begin{SaveVerbatim}{HOLBisimulationUptoTheoremsWEAKXXBISIMXXUPTOXXWEAKXXTRANSXXlabel}
\HOLTokenTurnstile{} \HOLSymConst{\HOLTokenForall{}}\HOLBoundVar{Wbsm}.
     \HOLConst{WEAK_BISIM_UPTO} \HOLBoundVar{Wbsm} \HOLSymConst{\HOLTokenImp{}}
     \HOLSymConst{\HOLTokenForall{}}\HOLBoundVar{E} \HOLBoundVar{E\sp{\prime}}.
       \HOLBoundVar{Wbsm} \HOLBoundVar{E} \HOLBoundVar{E\sp{\prime}} \HOLSymConst{\HOLTokenImp{}}
       \HOLSymConst{\HOLTokenForall{}}\HOLBoundVar{l} \HOLBoundVar{E\sb{\mathrm{1}}}.
         \HOLBoundVar{E} \HOLTokenWeakTransBegin\HOLConst{label} \HOLBoundVar{l}\HOLTokenWeakTransEnd \HOLBoundVar{E\sb{\mathrm{1}}} \HOLSymConst{\HOLTokenImp{}}
         \HOLSymConst{\HOLTokenExists{}}\HOLBoundVar{E\sb{\mathrm{2}}}.
           \HOLBoundVar{E\sp{\prime}} \HOLTokenWeakTransBegin\HOLConst{label} \HOLBoundVar{l}\HOLTokenWeakTransEnd \HOLBoundVar{E\sb{\mathrm{2}}} \HOLSymConst{\HOLTokenConj{}}
           (\HOLConst{WEAK_EQUIV} \HOLConst{O} \HOLBoundVar{Wbsm} \HOLConst{O} \HOLConst{STRONG_EQUIV}) \HOLBoundVar{E\sb{\mathrm{1}}} \HOLBoundVar{E\sb{\mathrm{2}}}
\end{SaveVerbatim}
\newcommand{\HOLBisimulationUptoTheoremsWEAKXXBISIMXXUPTOXXWEAKXXTRANSXXlabel}{\UseVerbatim{HOLBisimulationUptoTheoremsWEAKXXBISIMXXUPTOXXWEAKXXTRANSXXlabel}}
\begin{SaveVerbatim}{HOLBisimulationUptoTheoremsWEAKXXBISIMXXUPTOXXWEAKXXTRANSXXlabelYY}
\HOLTokenTurnstile{} \HOLSymConst{\HOLTokenForall{}}\HOLBoundVar{Wbsm}.
     \HOLConst{WEAK_BISIM_UPTO} \HOLBoundVar{Wbsm} \HOLSymConst{\HOLTokenImp{}}
     \HOLSymConst{\HOLTokenForall{}}\HOLBoundVar{E} \HOLBoundVar{E\sp{\prime}}.
       \HOLBoundVar{Wbsm} \HOLBoundVar{E} \HOLBoundVar{E\sp{\prime}} \HOLSymConst{\HOLTokenImp{}}
       \HOLSymConst{\HOLTokenForall{}}\HOLBoundVar{l} \HOLBoundVar{E\sb{\mathrm{2}}}.
         \HOLBoundVar{E\sp{\prime}} \HOLTokenWeakTransBegin\HOLConst{label} \HOLBoundVar{l}\HOLTokenWeakTransEnd \HOLBoundVar{E\sb{\mathrm{2}}} \HOLSymConst{\HOLTokenImp{}}
         \HOLSymConst{\HOLTokenExists{}}\HOLBoundVar{E\sb{\mathrm{1}}}.
           \HOLBoundVar{E} \HOLTokenWeakTransBegin\HOLConst{label} \HOLBoundVar{l}\HOLTokenWeakTransEnd \HOLBoundVar{E\sb{\mathrm{1}}} \HOLSymConst{\HOLTokenConj{}}
           (\HOLConst{STRONG_EQUIV} \HOLConst{O} \HOLBoundVar{Wbsm} \HOLConst{O} \HOLConst{WEAK_EQUIV}) \HOLBoundVar{E\sb{\mathrm{1}}} \HOLBoundVar{E\sb{\mathrm{2}}}
\end{SaveVerbatim}
\newcommand{\HOLBisimulationUptoTheoremsWEAKXXBISIMXXUPTOXXWEAKXXTRANSXXlabelYY}{\UseVerbatim{HOLBisimulationUptoTheoremsWEAKXXBISIMXXUPTOXXWEAKXXTRANSXXlabelYY}}
\newcommand{\HOLBisimulationUptoTheorems}{
\HOLThmTag{BisimulationUpto}{CONVERSE_OBS_BISIM_UPTO}\HOLBisimulationUptoTheoremsCONVERSEXXOBSXXBISIMXXUPTO
\HOLThmTag{BisimulationUpto}{CONVERSE_STRONG_BISIM_UPTO}\HOLBisimulationUptoTheoremsCONVERSEXXSTRONGXXBISIMXXUPTO
\HOLThmTag{BisimulationUpto}{CONVERSE_WEAK_BISIM_UPTO}\HOLBisimulationUptoTheoremsCONVERSEXXWEAKXXBISIMXXUPTO
\HOLThmTag{BisimulationUpto}{CONVERSE_WEAK_BISIM_UPTO_lemma}\HOLBisimulationUptoTheoremsCONVERSEXXWEAKXXBISIMXXUPTOXXlemma
\HOLThmTag{BisimulationUpto}{CONVERSE_WEAK_BISIM_UPTO_lemma'}\HOLBisimulationUptoTheoremsCONVERSEXXWEAKXXBISIMXXUPTOXXlemmaYY
\HOLThmTag{BisimulationUpto}{IDENTITY_OBS_BISIM_UPTO}\HOLBisimulationUptoTheoremsIDENTITYXXOBSXXBISIMXXUPTO
\HOLThmTag{BisimulationUpto}{IDENTITY_STRONG_BISIM_UPTO}\HOLBisimulationUptoTheoremsIDENTITYXXSTRONGXXBISIMXXUPTO
\HOLThmTag{BisimulationUpto}{IDENTITY_STRONG_BISIM_UPTO_lemma}\HOLBisimulationUptoTheoremsIDENTITYXXSTRONGXXBISIMXXUPTOXXlemma
\HOLThmTag{BisimulationUpto}{IDENTITY_WEAK_BISIM_UPTO}\HOLBisimulationUptoTheoremsIDENTITYXXWEAKXXBISIMXXUPTO
\HOLThmTag{BisimulationUpto}{IDENTITY_WEAK_BISIM_UPTO_lemma}\HOLBisimulationUptoTheoremsIDENTITYXXWEAKXXBISIMXXUPTOXXlemma
\HOLThmTag{BisimulationUpto}{IDENTITY_WEAK_BISIM_UPTO_lemma'}\HOLBisimulationUptoTheoremsIDENTITYXXWEAKXXBISIMXXUPTOXXlemmaYY
\HOLThmTag{BisimulationUpto}{OBS_BISIM_UPTO_EPS}\HOLBisimulationUptoTheoremsOBSXXBISIMXXUPTOXXEPS
\HOLThmTag{BisimulationUpto}{OBS_BISIM_UPTO_EPS'}\HOLBisimulationUptoTheoremsOBSXXBISIMXXUPTOXXEPSYY
\HOLThmTag{BisimulationUpto}{OBS_BISIM_UPTO_THM}\HOLBisimulationUptoTheoremsOBSXXBISIMXXUPTOXXTHM
\HOLThmTag{BisimulationUpto}{OBS_BISIM_UPTO_TRANS}\HOLBisimulationUptoTheoremsOBSXXBISIMXXUPTOXXTRANS
\HOLThmTag{BisimulationUpto}{OBS_BISIM_UPTO_TRANS'}\HOLBisimulationUptoTheoremsOBSXXBISIMXXUPTOXXTRANSYY
\HOLThmTag{BisimulationUpto}{OBS_BISIM_UPTO_WEAK_TRANS_label}\HOLBisimulationUptoTheoremsOBSXXBISIMXXUPTOXXWEAKXXTRANSXXlabel
\HOLThmTag{BisimulationUpto}{OBS_BISIM_UPTO_WEAK_TRANS_label'}\HOLBisimulationUptoTheoremsOBSXXBISIMXXUPTOXXWEAKXXTRANSXXlabelYY
\HOLThmTag{BisimulationUpto}{STRONG_BISIM_UPTO_LEMMA}\HOLBisimulationUptoTheoremsSTRONGXXBISIMXXUPTOXXLEMMA
\HOLThmTag{BisimulationUpto}{STRONG_BISIM_UPTO_THM}\HOLBisimulationUptoTheoremsSTRONGXXBISIMXXUPTOXXTHM
\HOLThmTag{BisimulationUpto}{WEAK_BISIM_UPTO_ALT_EPS}\HOLBisimulationUptoTheoremsWEAKXXBISIMXXUPTOXXALTXXEPS
\HOLThmTag{BisimulationUpto}{WEAK_BISIM_UPTO_ALT_EPS'}\HOLBisimulationUptoTheoremsWEAKXXBISIMXXUPTOXXALTXXEPSYY
\HOLThmTag{BisimulationUpto}{WEAK_BISIM_UPTO_ALT_LEMMA}\HOLBisimulationUptoTheoremsWEAKXXBISIMXXUPTOXXALTXXLEMMA
\HOLThmTag{BisimulationUpto}{WEAK_BISIM_UPTO_ALT_THM}\HOLBisimulationUptoTheoremsWEAKXXBISIMXXUPTOXXALTXXTHM
\HOLThmTag{BisimulationUpto}{WEAK_BISIM_UPTO_ALT_WEAK_TRANS_label}\HOLBisimulationUptoTheoremsWEAKXXBISIMXXUPTOXXALTXXWEAKXXTRANSXXlabel
\HOLThmTag{BisimulationUpto}{WEAK_BISIM_UPTO_ALT_WEAK_TRANS_label'}\HOLBisimulationUptoTheoremsWEAKXXBISIMXXUPTOXXALTXXWEAKXXTRANSXXlabelYY
\HOLThmTag{BisimulationUpto}{WEAK_BISIM_UPTO_ALT_WEAK_TRANS_tau}\HOLBisimulationUptoTheoremsWEAKXXBISIMXXUPTOXXALTXXWEAKXXTRANSXXtau
\HOLThmTag{BisimulationUpto}{WEAK_BISIM_UPTO_ALT_WEAK_TRANS_tau'}\HOLBisimulationUptoTheoremsWEAKXXBISIMXXUPTOXXALTXXWEAKXXTRANSXXtauYY
\HOLThmTag{BisimulationUpto}{WEAK_BISIM_UPTO_EPS}\HOLBisimulationUptoTheoremsWEAKXXBISIMXXUPTOXXEPS
\HOLThmTag{BisimulationUpto}{WEAK_BISIM_UPTO_EPS'}\HOLBisimulationUptoTheoremsWEAKXXBISIMXXUPTOXXEPSYY
\HOLThmTag{BisimulationUpto}{WEAK_BISIM_UPTO_LEMMA}\HOLBisimulationUptoTheoremsWEAKXXBISIMXXUPTOXXLEMMA
\HOLThmTag{BisimulationUpto}{WEAK_BISIM_UPTO_THM}\HOLBisimulationUptoTheoremsWEAKXXBISIMXXUPTOXXTHM
\HOLThmTag{BisimulationUpto}{WEAK_BISIM_UPTO_TRANS_label}\HOLBisimulationUptoTheoremsWEAKXXBISIMXXUPTOXXTRANSXXlabel
\HOLThmTag{BisimulationUpto}{WEAK_BISIM_UPTO_TRANS_label'}\HOLBisimulationUptoTheoremsWEAKXXBISIMXXUPTOXXTRANSXXlabelYY
\HOLThmTag{BisimulationUpto}{WEAK_BISIM_UPTO_TRANS_tau}\HOLBisimulationUptoTheoremsWEAKXXBISIMXXUPTOXXTRANSXXtau
\HOLThmTag{BisimulationUpto}{WEAK_BISIM_UPTO_TRANS_tau'}\HOLBisimulationUptoTheoremsWEAKXXBISIMXXUPTOXXTRANSXXtauYY
\HOLThmTag{BisimulationUpto}{WEAK_BISIM_UPTO_WEAK_TRANS_label}\HOLBisimulationUptoTheoremsWEAKXXBISIMXXUPTOXXWEAKXXTRANSXXlabel
\HOLThmTag{BisimulationUpto}{WEAK_BISIM_UPTO_WEAK_TRANS_label'}\HOLBisimulationUptoTheoremsWEAKXXBISIMXXUPTOXXWEAKXXTRANSXXlabelYY
}

\newcommand{\HOLExpansionDate}{02 Dicembre 2017}
\newcommand{\HOLExpansionTime}{13:31}
\begin{SaveVerbatim}{HOLExpansionDefinitionsexpandsXXdef}
\HOLTokenTurnstile{} (\HOLConst{expands}) \HOLSymConst{=}
   (\HOLTokenLambda{}\HOLBoundVar{a\sb{\mathrm{0}}} \HOLBoundVar{a\sb{\mathrm{1}}}.
      \HOLSymConst{\HOLTokenExists{}}\HOLBoundVar{expands\sp{\prime}}.
        \HOLBoundVar{expands\sp{\prime}} \HOLBoundVar{a\sb{\mathrm{0}}} \HOLBoundVar{a\sb{\mathrm{1}}} \HOLSymConst{\HOLTokenConj{}}
        \HOLSymConst{\HOLTokenForall{}}\HOLBoundVar{a\sb{\mathrm{0}}} \HOLBoundVar{a\sb{\mathrm{1}}}.
          \HOLBoundVar{expands\sp{\prime}} \HOLBoundVar{a\sb{\mathrm{0}}} \HOLBoundVar{a\sb{\mathrm{1}}} \HOLSymConst{\HOLTokenImp{}}
          (\HOLSymConst{\HOLTokenForall{}}\HOLBoundVar{l}.
             (\HOLSymConst{\HOLTokenForall{}}\HOLBoundVar{E\sb{\mathrm{1}}}.
                \HOLBoundVar{a\sb{\mathrm{0}}} \HOLTokenTransBegin\HOLConst{label} \HOLBoundVar{l}\HOLTokenTransEnd \HOLBoundVar{E\sb{\mathrm{1}}} \HOLSymConst{\HOLTokenImp{}}
                \HOLSymConst{\HOLTokenExists{}}\HOLBoundVar{E\sb{\mathrm{2}}}. \HOLBoundVar{a\sb{\mathrm{1}}} \HOLTokenTransBegin\HOLConst{label} \HOLBoundVar{l}\HOLTokenTransEnd \HOLBoundVar{E\sb{\mathrm{2}}} \HOLSymConst{\HOLTokenConj{}} \HOLBoundVar{expands\sp{\prime}} \HOLBoundVar{E\sb{\mathrm{1}}} \HOLBoundVar{E\sb{\mathrm{2}}}) \HOLSymConst{\HOLTokenConj{}}
             \HOLSymConst{\HOLTokenForall{}}\HOLBoundVar{E\sb{\mathrm{2}}}.
               \HOLBoundVar{a\sb{\mathrm{1}}} \HOLTokenTransBegin\HOLConst{label} \HOLBoundVar{l}\HOLTokenTransEnd \HOLBoundVar{E\sb{\mathrm{2}}} \HOLSymConst{\HOLTokenImp{}}
               \HOLSymConst{\HOLTokenExists{}}\HOLBoundVar{E\sb{\mathrm{1}}}. \HOLBoundVar{a\sb{\mathrm{0}}} \HOLTokenWeakTransBegin\HOLConst{label} \HOLBoundVar{l}\HOLTokenWeakTransEnd \HOLBoundVar{E\sb{\mathrm{1}}} \HOLSymConst{\HOLTokenConj{}} \HOLBoundVar{expands\sp{\prime}} \HOLBoundVar{E\sb{\mathrm{1}}} \HOLBoundVar{E\sb{\mathrm{2}}}) \HOLSymConst{\HOLTokenConj{}}
          (\HOLSymConst{\HOLTokenForall{}}\HOLBoundVar{E\sb{\mathrm{1}}}.
             \HOLBoundVar{a\sb{\mathrm{0}}} \HOLTokenTransBegin\HOLConst{\ensuremath{\tau}}\HOLTokenTransEnd \HOLBoundVar{E\sb{\mathrm{1}}} \HOLSymConst{\HOLTokenImp{}}
             \HOLBoundVar{expands\sp{\prime}} \HOLBoundVar{E\sb{\mathrm{1}}} \HOLBoundVar{a\sb{\mathrm{1}}} \HOLSymConst{\HOLTokenDisj{}} \HOLSymConst{\HOLTokenExists{}}\HOLBoundVar{E\sb{\mathrm{2}}}. \HOLBoundVar{a\sb{\mathrm{1}}} \HOLTokenTransBegin\HOLConst{\ensuremath{\tau}}\HOLTokenTransEnd \HOLBoundVar{E\sb{\mathrm{2}}} \HOLSymConst{\HOLTokenConj{}} \HOLBoundVar{expands\sp{\prime}} \HOLBoundVar{E\sb{\mathrm{1}}} \HOLBoundVar{E\sb{\mathrm{2}}}) \HOLSymConst{\HOLTokenConj{}}
          \HOLSymConst{\HOLTokenForall{}}\HOLBoundVar{E\sb{\mathrm{2}}}. \HOLBoundVar{a\sb{\mathrm{1}}} \HOLTokenTransBegin\HOLConst{\ensuremath{\tau}}\HOLTokenTransEnd \HOLBoundVar{E\sb{\mathrm{2}}} \HOLSymConst{\HOLTokenImp{}} \HOLSymConst{\HOLTokenExists{}}\HOLBoundVar{E\sb{\mathrm{1}}}. \HOLBoundVar{a\sb{\mathrm{0}}} \HOLTokenWeakTransBegin\HOLConst{\ensuremath{\tau}}\HOLTokenWeakTransEnd \HOLBoundVar{E\sb{\mathrm{1}}} \HOLSymConst{\HOLTokenConj{}} \HOLBoundVar{expands\sp{\prime}} \HOLBoundVar{E\sb{\mathrm{1}}} \HOLBoundVar{E\sb{\mathrm{2}}})
\end{SaveVerbatim}
\newcommand{\HOLExpansionDefinitionsexpandsXXdef}{\UseVerbatim{HOLExpansionDefinitionsexpandsXXdef}}
\begin{SaveVerbatim}{HOLExpansionDefinitionsEXPANSION}
\HOLTokenTurnstile{} \HOLSymConst{\HOLTokenForall{}}\HOLBoundVar{Exp}.
     \HOLConst{EXPANSION} \HOLBoundVar{Exp} \HOLSymConst{\HOLTokenEquiv{}}
     \HOLSymConst{\HOLTokenForall{}}\HOLBoundVar{E} \HOLBoundVar{E\sp{\prime}}.
       \HOLBoundVar{Exp} \HOLBoundVar{E} \HOLBoundVar{E\sp{\prime}} \HOLSymConst{\HOLTokenImp{}}
       (\HOLSymConst{\HOLTokenForall{}}\HOLBoundVar{l}.
          (\HOLSymConst{\HOLTokenForall{}}\HOLBoundVar{E\sb{\mathrm{1}}}.
             \HOLBoundVar{E} \HOLTokenTransBegin\HOLConst{label} \HOLBoundVar{l}\HOLTokenTransEnd \HOLBoundVar{E\sb{\mathrm{1}}} \HOLSymConst{\HOLTokenImp{}}
             \HOLSymConst{\HOLTokenExists{}}\HOLBoundVar{E\sb{\mathrm{2}}}. \HOLBoundVar{E\sp{\prime}} \HOLTokenTransBegin\HOLConst{label} \HOLBoundVar{l}\HOLTokenTransEnd \HOLBoundVar{E\sb{\mathrm{2}}} \HOLSymConst{\HOLTokenConj{}} \HOLBoundVar{Exp} \HOLBoundVar{E\sb{\mathrm{1}}} \HOLBoundVar{E\sb{\mathrm{2}}}) \HOLSymConst{\HOLTokenConj{}}
          \HOLSymConst{\HOLTokenForall{}}\HOLBoundVar{E\sb{\mathrm{2}}}.
            \HOLBoundVar{E\sp{\prime}} \HOLTokenTransBegin\HOLConst{label} \HOLBoundVar{l}\HOLTokenTransEnd \HOLBoundVar{E\sb{\mathrm{2}}} \HOLSymConst{\HOLTokenImp{}} \HOLSymConst{\HOLTokenExists{}}\HOLBoundVar{E\sb{\mathrm{1}}}. \HOLBoundVar{E} \HOLTokenWeakTransBegin\HOLConst{label} \HOLBoundVar{l}\HOLTokenWeakTransEnd \HOLBoundVar{E\sb{\mathrm{1}}} \HOLSymConst{\HOLTokenConj{}} \HOLBoundVar{Exp} \HOLBoundVar{E\sb{\mathrm{1}}} \HOLBoundVar{E\sb{\mathrm{2}}}) \HOLSymConst{\HOLTokenConj{}}
       (\HOLSymConst{\HOLTokenForall{}}\HOLBoundVar{E\sb{\mathrm{1}}}.
          \HOLBoundVar{E} \HOLTokenTransBegin\HOLConst{\ensuremath{\tau}}\HOLTokenTransEnd \HOLBoundVar{E\sb{\mathrm{1}}} \HOLSymConst{\HOLTokenImp{}} \HOLBoundVar{Exp} \HOLBoundVar{E\sb{\mathrm{1}}} \HOLBoundVar{E\sp{\prime}} \HOLSymConst{\HOLTokenDisj{}} \HOLSymConst{\HOLTokenExists{}}\HOLBoundVar{E\sb{\mathrm{2}}}. \HOLBoundVar{E\sp{\prime}} \HOLTokenTransBegin\HOLConst{\ensuremath{\tau}}\HOLTokenTransEnd \HOLBoundVar{E\sb{\mathrm{2}}} \HOLSymConst{\HOLTokenConj{}} \HOLBoundVar{Exp} \HOLBoundVar{E\sb{\mathrm{1}}} \HOLBoundVar{E\sb{\mathrm{2}}}) \HOLSymConst{\HOLTokenConj{}}
       \HOLSymConst{\HOLTokenForall{}}\HOLBoundVar{E\sb{\mathrm{2}}}. \HOLBoundVar{E\sp{\prime}} \HOLTokenTransBegin\HOLConst{\ensuremath{\tau}}\HOLTokenTransEnd \HOLBoundVar{E\sb{\mathrm{2}}} \HOLSymConst{\HOLTokenImp{}} \HOLSymConst{\HOLTokenExists{}}\HOLBoundVar{E\sb{\mathrm{1}}}. \HOLBoundVar{E} \HOLTokenWeakTransBegin\HOLConst{\ensuremath{\tau}}\HOLTokenWeakTransEnd \HOLBoundVar{E\sb{\mathrm{1}}} \HOLSymConst{\HOLTokenConj{}} \HOLBoundVar{Exp} \HOLBoundVar{E\sb{\mathrm{1}}} \HOLBoundVar{E\sb{\mathrm{2}}}
\end{SaveVerbatim}
\newcommand{\HOLExpansionDefinitionsEXPANSION}{\UseVerbatim{HOLExpansionDefinitionsEXPANSION}}
\newcommand{\HOLExpansionDefinitions}{
\HOLDfnTag{Expansion}{expands_def}\HOLExpansionDefinitionsexpandsXXdef
\HOLDfnTag{Expansion}{EXPANSION}\HOLExpansionDefinitionsEXPANSION
}
\begin{SaveVerbatim}{HOLExpansionTheoremsCOMPXXEXPANSION}
\HOLTokenTurnstile{} \HOLSymConst{\HOLTokenForall{}}\HOLBoundVar{Exp\sb{\mathrm{1}}} \HOLBoundVar{Exp\sb{\mathrm{2}}}.
     \HOLConst{EXPANSION} \HOLBoundVar{Exp\sb{\mathrm{1}}} \HOLSymConst{\HOLTokenConj{}} \HOLConst{EXPANSION} \HOLBoundVar{Exp\sb{\mathrm{2}}} \HOLSymConst{\HOLTokenImp{}} \HOLConst{EXPANSION} (\HOLBoundVar{Exp\sb{\mathrm{2}}} \HOLConst{O} \HOLBoundVar{Exp\sb{\mathrm{1}}})
\end{SaveVerbatim}
\newcommand{\HOLExpansionTheoremsCOMPXXEXPANSION}{\UseVerbatim{HOLExpansionTheoremsCOMPXXEXPANSION}}
\begin{SaveVerbatim}{HOLExpansionTheoremsEQXXIMPXXexpands}
\HOLTokenTurnstile{} \HOLSymConst{\HOLTokenForall{}}\HOLBoundVar{E} \HOLBoundVar{E\sp{\prime}}. (\HOLBoundVar{E} \HOLSymConst{=} \HOLBoundVar{E\sp{\prime}}) \HOLSymConst{\HOLTokenImp{}} \HOLBoundVar{E} \HOLConst{expands} \HOLBoundVar{E\sp{\prime}}
\end{SaveVerbatim}
\newcommand{\HOLExpansionTheoremsEQXXIMPXXexpands}{\UseVerbatim{HOLExpansionTheoremsEQXXIMPXXexpands}}
\begin{SaveVerbatim}{HOLExpansionTheoremsexpandsXXANDXXTRACEXXlabel}
\HOLTokenTurnstile{} \HOLSymConst{\HOLTokenForall{}}\HOLBoundVar{E} \HOLBoundVar{E\sp{\prime}}.
     \HOLBoundVar{E} \HOLConst{expands} \HOLBoundVar{E\sp{\prime}} \HOLSymConst{\HOLTokenImp{}}
     \HOLSymConst{\HOLTokenForall{}}\HOLBoundVar{xs} \HOLBoundVar{l} \HOLBoundVar{E\sb{\mathrm{1}}}.
       \HOLConst{TRACE} \HOLBoundVar{E} \HOLBoundVar{xs} \HOLBoundVar{E\sb{\mathrm{1}}} \HOLSymConst{\HOLTokenConj{}} \HOLConst{UNIQUE_LABEL} (\HOLConst{label} \HOLBoundVar{l}) \HOLBoundVar{xs} \HOLSymConst{\HOLTokenImp{}}
       \HOLSymConst{\HOLTokenExists{}}\HOLBoundVar{xs\sp{\prime}} \HOLBoundVar{E\sb{\mathrm{2}}}.
         \HOLConst{TRACE} \HOLBoundVar{E\sp{\prime}} \HOLBoundVar{xs\sp{\prime}} \HOLBoundVar{E\sb{\mathrm{2}}} \HOLSymConst{\HOLTokenConj{}} \HOLBoundVar{E\sb{\mathrm{1}}} \HOLConst{expands} \HOLBoundVar{E\sb{\mathrm{2}}} \HOLSymConst{\HOLTokenConj{}}
         \HOLConst{LENGTH} \HOLBoundVar{xs\sp{\prime}} \HOLSymConst{\HOLTokenLeq{}} \HOLConst{LENGTH} \HOLBoundVar{xs} \HOLSymConst{\HOLTokenConj{}} \HOLConst{UNIQUE_LABEL} (\HOLConst{label} \HOLBoundVar{l}) \HOLBoundVar{xs\sp{\prime}}
\end{SaveVerbatim}
\newcommand{\HOLExpansionTheoremsexpandsXXANDXXTRACEXXlabel}{\UseVerbatim{HOLExpansionTheoremsexpandsXXANDXXTRACEXXlabel}}
\begin{SaveVerbatim}{HOLExpansionTheoremsexpandsXXANDXXTRACEXXtau}
\HOLTokenTurnstile{} \HOLSymConst{\HOLTokenForall{}}\HOLBoundVar{E} \HOLBoundVar{E\sp{\prime}}.
     \HOLBoundVar{E} \HOLConst{expands} \HOLBoundVar{E\sp{\prime}} \HOLSymConst{\HOLTokenImp{}}
     \HOLSymConst{\HOLTokenForall{}}\HOLBoundVar{xs} \HOLBoundVar{l} \HOLBoundVar{E\sb{\mathrm{1}}}.
       \HOLConst{TRACE} \HOLBoundVar{E} \HOLBoundVar{xs} \HOLBoundVar{E\sb{\mathrm{1}}} \HOLSymConst{\HOLTokenConj{}} \HOLConst{NO_LABEL} \HOLBoundVar{xs} \HOLSymConst{\HOLTokenImp{}}
       \HOLSymConst{\HOLTokenExists{}}\HOLBoundVar{xs\sp{\prime}} \HOLBoundVar{E\sb{\mathrm{2}}}.
         \HOLConst{TRACE} \HOLBoundVar{E\sp{\prime}} \HOLBoundVar{xs\sp{\prime}} \HOLBoundVar{E\sb{\mathrm{2}}} \HOLSymConst{\HOLTokenConj{}} \HOLBoundVar{E\sb{\mathrm{1}}} \HOLConst{expands} \HOLBoundVar{E\sb{\mathrm{2}}} \HOLSymConst{\HOLTokenConj{}}
         \HOLConst{LENGTH} \HOLBoundVar{xs\sp{\prime}} \HOLSymConst{\HOLTokenLeq{}} \HOLConst{LENGTH} \HOLBoundVar{xs} \HOLSymConst{\HOLTokenConj{}} \HOLConst{NO_LABEL} \HOLBoundVar{xs\sp{\prime}}
\end{SaveVerbatim}
\newcommand{\HOLExpansionTheoremsexpandsXXANDXXTRACEXXtau}{\UseVerbatim{HOLExpansionTheoremsexpandsXXANDXXTRACEXXtau}}
\begin{SaveVerbatim}{HOLExpansionTheoremsexpandsXXcases}
\HOLTokenTurnstile{} \HOLSymConst{\HOLTokenForall{}}\HOLBoundVar{a\sb{\mathrm{0}}} \HOLBoundVar{a\sb{\mathrm{1}}}.
     \HOLBoundVar{a\sb{\mathrm{0}}} \HOLConst{expands} \HOLBoundVar{a\sb{\mathrm{1}}} \HOLSymConst{\HOLTokenEquiv{}}
     (\HOLSymConst{\HOLTokenForall{}}\HOLBoundVar{l}.
        (\HOLSymConst{\HOLTokenForall{}}\HOLBoundVar{E\sb{\mathrm{1}}}.
           \HOLBoundVar{a\sb{\mathrm{0}}} \HOLTokenTransBegin\HOLConst{label} \HOLBoundVar{l}\HOLTokenTransEnd \HOLBoundVar{E\sb{\mathrm{1}}} \HOLSymConst{\HOLTokenImp{}}
           \HOLSymConst{\HOLTokenExists{}}\HOLBoundVar{E\sb{\mathrm{2}}}. \HOLBoundVar{a\sb{\mathrm{1}}} \HOLTokenTransBegin\HOLConst{label} \HOLBoundVar{l}\HOLTokenTransEnd \HOLBoundVar{E\sb{\mathrm{2}}} \HOLSymConst{\HOLTokenConj{}} \HOLBoundVar{E\sb{\mathrm{1}}} \HOLConst{expands} \HOLBoundVar{E\sb{\mathrm{2}}}) \HOLSymConst{\HOLTokenConj{}}
        \HOLSymConst{\HOLTokenForall{}}\HOLBoundVar{E\sb{\mathrm{2}}}.
          \HOLBoundVar{a\sb{\mathrm{1}}} \HOLTokenTransBegin\HOLConst{label} \HOLBoundVar{l}\HOLTokenTransEnd \HOLBoundVar{E\sb{\mathrm{2}}} \HOLSymConst{\HOLTokenImp{}}
          \HOLSymConst{\HOLTokenExists{}}\HOLBoundVar{E\sb{\mathrm{1}}}. \HOLBoundVar{a\sb{\mathrm{0}}} \HOLTokenWeakTransBegin\HOLConst{label} \HOLBoundVar{l}\HOLTokenWeakTransEnd \HOLBoundVar{E\sb{\mathrm{1}}} \HOLSymConst{\HOLTokenConj{}} \HOLBoundVar{E\sb{\mathrm{1}}} \HOLConst{expands} \HOLBoundVar{E\sb{\mathrm{2}}}) \HOLSymConst{\HOLTokenConj{}}
     (\HOLSymConst{\HOLTokenForall{}}\HOLBoundVar{E\sb{\mathrm{1}}}.
        \HOLBoundVar{a\sb{\mathrm{0}}} \HOLTokenTransBegin\HOLConst{\ensuremath{\tau}}\HOLTokenTransEnd \HOLBoundVar{E\sb{\mathrm{1}}} \HOLSymConst{\HOLTokenImp{}}
        \HOLBoundVar{E\sb{\mathrm{1}}} \HOLConst{expands} \HOLBoundVar{a\sb{\mathrm{1}}} \HOLSymConst{\HOLTokenDisj{}} \HOLSymConst{\HOLTokenExists{}}\HOLBoundVar{E\sb{\mathrm{2}}}. \HOLBoundVar{a\sb{\mathrm{1}}} \HOLTokenTransBegin\HOLConst{\ensuremath{\tau}}\HOLTokenTransEnd \HOLBoundVar{E\sb{\mathrm{2}}} \HOLSymConst{\HOLTokenConj{}} \HOLBoundVar{E\sb{\mathrm{1}}} \HOLConst{expands} \HOLBoundVar{E\sb{\mathrm{2}}}) \HOLSymConst{\HOLTokenConj{}}
     \HOLSymConst{\HOLTokenForall{}}\HOLBoundVar{E\sb{\mathrm{2}}}. \HOLBoundVar{a\sb{\mathrm{1}}} \HOLTokenTransBegin\HOLConst{\ensuremath{\tau}}\HOLTokenTransEnd \HOLBoundVar{E\sb{\mathrm{2}}} \HOLSymConst{\HOLTokenImp{}} \HOLSymConst{\HOLTokenExists{}}\HOLBoundVar{E\sb{\mathrm{1}}}. \HOLBoundVar{a\sb{\mathrm{0}}} \HOLTokenWeakTransBegin\HOLConst{\ensuremath{\tau}}\HOLTokenWeakTransEnd \HOLBoundVar{E\sb{\mathrm{1}}} \HOLSymConst{\HOLTokenConj{}} \HOLBoundVar{E\sb{\mathrm{1}}} \HOLConst{expands} \HOLBoundVar{E\sb{\mathrm{2}}}
\end{SaveVerbatim}
\newcommand{\HOLExpansionTheoremsexpandsXXcases}{\UseVerbatim{HOLExpansionTheoremsexpandsXXcases}}
\begin{SaveVerbatim}{HOLExpansionTheoremsexpandsXXcasesYY}
\HOLTokenTurnstile{} \HOLSymConst{\HOLTokenForall{}}\HOLBoundVar{E} \HOLBoundVar{E\sp{\prime}}.
     \HOLBoundVar{E} \HOLConst{expands} \HOLBoundVar{E\sp{\prime}} \HOLSymConst{\HOLTokenEquiv{}}
     (\HOLSymConst{\HOLTokenForall{}}\HOLBoundVar{l} \HOLBoundVar{E\sb{\mathrm{1}}}.
        \HOLBoundVar{E} \HOLTokenTransBegin\HOLConst{label} \HOLBoundVar{l}\HOLTokenTransEnd \HOLBoundVar{E\sb{\mathrm{1}}} \HOLSymConst{\HOLTokenImp{}} \HOLSymConst{\HOLTokenExists{}}\HOLBoundVar{E\sb{\mathrm{2}}}. \HOLBoundVar{E\sp{\prime}} \HOLTokenTransBegin\HOLConst{label} \HOLBoundVar{l}\HOLTokenTransEnd \HOLBoundVar{E\sb{\mathrm{2}}} \HOLSymConst{\HOLTokenConj{}} \HOLBoundVar{E\sb{\mathrm{1}}} \HOLConst{expands} \HOLBoundVar{E\sb{\mathrm{2}}}) \HOLSymConst{\HOLTokenConj{}}
     (\HOLSymConst{\HOLTokenForall{}}\HOLBoundVar{E\sb{\mathrm{1}}}.
        \HOLBoundVar{E} \HOLTokenTransBegin\HOLConst{\ensuremath{\tau}}\HOLTokenTransEnd \HOLBoundVar{E\sb{\mathrm{1}}} \HOLSymConst{\HOLTokenImp{}}
        \HOLBoundVar{E\sb{\mathrm{1}}} \HOLConst{expands} \HOLBoundVar{E\sp{\prime}} \HOLSymConst{\HOLTokenDisj{}} \HOLSymConst{\HOLTokenExists{}}\HOLBoundVar{E\sb{\mathrm{2}}}. \HOLBoundVar{E\sp{\prime}} \HOLTokenTransBegin\HOLConst{\ensuremath{\tau}}\HOLTokenTransEnd \HOLBoundVar{E\sb{\mathrm{2}}} \HOLSymConst{\HOLTokenConj{}} \HOLBoundVar{E\sb{\mathrm{1}}} \HOLConst{expands} \HOLBoundVar{E\sb{\mathrm{2}}}) \HOLSymConst{\HOLTokenConj{}}
     \HOLSymConst{\HOLTokenForall{}}\HOLBoundVar{u} \HOLBoundVar{E\sb{\mathrm{2}}}. \HOLBoundVar{E\sp{\prime}} \HOLTokenTransBegin\HOLBoundVar{u}\HOLTokenTransEnd \HOLBoundVar{E\sb{\mathrm{2}}} \HOLSymConst{\HOLTokenImp{}} \HOLSymConst{\HOLTokenExists{}}\HOLBoundVar{E\sb{\mathrm{1}}}. \HOLBoundVar{E} \HOLTokenWeakTransBegin\HOLBoundVar{u}\HOLTokenWeakTransEnd \HOLBoundVar{E\sb{\mathrm{1}}} \HOLSymConst{\HOLTokenConj{}} \HOLBoundVar{E\sb{\mathrm{1}}} \HOLConst{expands} \HOLBoundVar{E\sb{\mathrm{2}}}
\end{SaveVerbatim}
\newcommand{\HOLExpansionTheoremsexpandsXXcasesYY}{\UseVerbatim{HOLExpansionTheoremsexpandsXXcasesYY}}
\begin{SaveVerbatim}{HOLExpansionTheoremsexpandsXXcoind}
\HOLTokenTurnstile{} \HOLSymConst{\HOLTokenForall{}}\HOLBoundVar{expands\sp{\prime}}.
     (\HOLSymConst{\HOLTokenForall{}}\HOLBoundVar{a\sb{\mathrm{0}}} \HOLBoundVar{a\sb{\mathrm{1}}}.
        \HOLBoundVar{expands\sp{\prime}} \HOLBoundVar{a\sb{\mathrm{0}}} \HOLBoundVar{a\sb{\mathrm{1}}} \HOLSymConst{\HOLTokenImp{}}
        (\HOLSymConst{\HOLTokenForall{}}\HOLBoundVar{l}.
           (\HOLSymConst{\HOLTokenForall{}}\HOLBoundVar{E\sb{\mathrm{1}}}.
              \HOLBoundVar{a\sb{\mathrm{0}}} \HOLTokenTransBegin\HOLConst{label} \HOLBoundVar{l}\HOLTokenTransEnd \HOLBoundVar{E\sb{\mathrm{1}}} \HOLSymConst{\HOLTokenImp{}}
              \HOLSymConst{\HOLTokenExists{}}\HOLBoundVar{E\sb{\mathrm{2}}}. \HOLBoundVar{a\sb{\mathrm{1}}} \HOLTokenTransBegin\HOLConst{label} \HOLBoundVar{l}\HOLTokenTransEnd \HOLBoundVar{E\sb{\mathrm{2}}} \HOLSymConst{\HOLTokenConj{}} \HOLBoundVar{expands\sp{\prime}} \HOLBoundVar{E\sb{\mathrm{1}}} \HOLBoundVar{E\sb{\mathrm{2}}}) \HOLSymConst{\HOLTokenConj{}}
           \HOLSymConst{\HOLTokenForall{}}\HOLBoundVar{E\sb{\mathrm{2}}}.
             \HOLBoundVar{a\sb{\mathrm{1}}} \HOLTokenTransBegin\HOLConst{label} \HOLBoundVar{l}\HOLTokenTransEnd \HOLBoundVar{E\sb{\mathrm{2}}} \HOLSymConst{\HOLTokenImp{}}
             \HOLSymConst{\HOLTokenExists{}}\HOLBoundVar{E\sb{\mathrm{1}}}. \HOLBoundVar{a\sb{\mathrm{0}}} \HOLTokenWeakTransBegin\HOLConst{label} \HOLBoundVar{l}\HOLTokenWeakTransEnd \HOLBoundVar{E\sb{\mathrm{1}}} \HOLSymConst{\HOLTokenConj{}} \HOLBoundVar{expands\sp{\prime}} \HOLBoundVar{E\sb{\mathrm{1}}} \HOLBoundVar{E\sb{\mathrm{2}}}) \HOLSymConst{\HOLTokenConj{}}
        (\HOLSymConst{\HOLTokenForall{}}\HOLBoundVar{E\sb{\mathrm{1}}}.
           \HOLBoundVar{a\sb{\mathrm{0}}} \HOLTokenTransBegin\HOLConst{\ensuremath{\tau}}\HOLTokenTransEnd \HOLBoundVar{E\sb{\mathrm{1}}} \HOLSymConst{\HOLTokenImp{}}
           \HOLBoundVar{expands\sp{\prime}} \HOLBoundVar{E\sb{\mathrm{1}}} \HOLBoundVar{a\sb{\mathrm{1}}} \HOLSymConst{\HOLTokenDisj{}} \HOLSymConst{\HOLTokenExists{}}\HOLBoundVar{E\sb{\mathrm{2}}}. \HOLBoundVar{a\sb{\mathrm{1}}} \HOLTokenTransBegin\HOLConst{\ensuremath{\tau}}\HOLTokenTransEnd \HOLBoundVar{E\sb{\mathrm{2}}} \HOLSymConst{\HOLTokenConj{}} \HOLBoundVar{expands\sp{\prime}} \HOLBoundVar{E\sb{\mathrm{1}}} \HOLBoundVar{E\sb{\mathrm{2}}}) \HOLSymConst{\HOLTokenConj{}}
        \HOLSymConst{\HOLTokenForall{}}\HOLBoundVar{E\sb{\mathrm{2}}}. \HOLBoundVar{a\sb{\mathrm{1}}} \HOLTokenTransBegin\HOLConst{\ensuremath{\tau}}\HOLTokenTransEnd \HOLBoundVar{E\sb{\mathrm{2}}} \HOLSymConst{\HOLTokenImp{}} \HOLSymConst{\HOLTokenExists{}}\HOLBoundVar{E\sb{\mathrm{1}}}. \HOLBoundVar{a\sb{\mathrm{0}}} \HOLTokenWeakTransBegin\HOLConst{\ensuremath{\tau}}\HOLTokenWeakTransEnd \HOLBoundVar{E\sb{\mathrm{1}}} \HOLSymConst{\HOLTokenConj{}} \HOLBoundVar{expands\sp{\prime}} \HOLBoundVar{E\sb{\mathrm{1}}} \HOLBoundVar{E\sb{\mathrm{2}}}) \HOLSymConst{\HOLTokenImp{}}
     \HOLSymConst{\HOLTokenForall{}}\HOLBoundVar{a\sb{\mathrm{0}}} \HOLBoundVar{a\sb{\mathrm{1}}}. \HOLBoundVar{expands\sp{\prime}} \HOLBoundVar{a\sb{\mathrm{0}}} \HOLBoundVar{a\sb{\mathrm{1}}} \HOLSymConst{\HOLTokenImp{}} \HOLBoundVar{a\sb{\mathrm{0}}} \HOLConst{expands} \HOLBoundVar{a\sb{\mathrm{1}}}
\end{SaveVerbatim}
\newcommand{\HOLExpansionTheoremsexpandsXXcoind}{\UseVerbatim{HOLExpansionTheoremsexpandsXXcoind}}
\begin{SaveVerbatim}{HOLExpansionTheoremsexpandsXXEPS}
\HOLTokenTurnstile{} \HOLSymConst{\HOLTokenForall{}}\HOLBoundVar{E} \HOLBoundVar{E\sp{\prime}}.
     \HOLBoundVar{E} \HOLConst{expands} \HOLBoundVar{E\sp{\prime}} \HOLSymConst{\HOLTokenImp{}}
     \HOLSymConst{\HOLTokenForall{}}\HOLBoundVar{E\sb{\mathrm{1}}}. \HOLConst{EPS} \HOLBoundVar{E} \HOLBoundVar{E\sb{\mathrm{1}}} \HOLSymConst{\HOLTokenImp{}} \HOLSymConst{\HOLTokenExists{}}\HOLBoundVar{E\sb{\mathrm{2}}}. \HOLConst{EPS} \HOLBoundVar{E\sp{\prime}} \HOLBoundVar{E\sb{\mathrm{2}}} \HOLSymConst{\HOLTokenConj{}} \HOLBoundVar{E\sb{\mathrm{1}}} \HOLConst{expands} \HOLBoundVar{E\sb{\mathrm{2}}}
\end{SaveVerbatim}
\newcommand{\HOLExpansionTheoremsexpandsXXEPS}{\UseVerbatim{HOLExpansionTheoremsexpandsXXEPS}}
\begin{SaveVerbatim}{HOLExpansionTheoremsexpandsXXEPSYY}
\HOLTokenTurnstile{} \HOLSymConst{\HOLTokenForall{}}\HOLBoundVar{E} \HOLBoundVar{E\sp{\prime}}.
     \HOLBoundVar{E} \HOLConst{expands} \HOLBoundVar{E\sp{\prime}} \HOLSymConst{\HOLTokenImp{}}
     \HOLSymConst{\HOLTokenForall{}}\HOLBoundVar{E\sb{\mathrm{2}}}. \HOLConst{EPS} \HOLBoundVar{E\sp{\prime}} \HOLBoundVar{E\sb{\mathrm{2}}} \HOLSymConst{\HOLTokenImp{}} \HOLSymConst{\HOLTokenExists{}}\HOLBoundVar{E\sb{\mathrm{1}}}. \HOLConst{EPS} \HOLBoundVar{E} \HOLBoundVar{E\sb{\mathrm{1}}} \HOLSymConst{\HOLTokenConj{}} \HOLBoundVar{E\sb{\mathrm{1}}} \HOLConst{expands} \HOLBoundVar{E\sb{\mathrm{2}}}
\end{SaveVerbatim}
\newcommand{\HOLExpansionTheoremsexpandsXXEPSYY}{\UseVerbatim{HOLExpansionTheoremsexpandsXXEPSYY}}
\begin{SaveVerbatim}{HOLExpansionTheoremsexpandsXXIMPXXWEAKXXEQUIV}
\HOLTokenTurnstile{} \HOLSymConst{\HOLTokenForall{}}\HOLBoundVar{P} \HOLBoundVar{Q}. \HOLBoundVar{P} \HOLConst{expands} \HOLBoundVar{Q} \HOLSymConst{\HOLTokenImp{}} \HOLConst{WEAK_EQUIV} \HOLBoundVar{P} \HOLBoundVar{Q}
\end{SaveVerbatim}
\newcommand{\HOLExpansionTheoremsexpandsXXIMPXXWEAKXXEQUIV}{\UseVerbatim{HOLExpansionTheoremsexpandsXXIMPXXWEAKXXEQUIV}}
\begin{SaveVerbatim}{HOLExpansionTheoremsexpandsXXisXXEXPANSION}
\HOLTokenTurnstile{} \HOLConst{EXPANSION} (\HOLConst{expands})
\end{SaveVerbatim}
\newcommand{\HOLExpansionTheoremsexpandsXXisXXEXPANSION}{\UseVerbatim{HOLExpansionTheoremsexpandsXXisXXEXPANSION}}
\begin{SaveVerbatim}{HOLExpansionTheoremsexpandsXXprecongruence}
\HOLTokenTurnstile{} \HOLConst{precongruence1} (\HOLConst{expands})
\end{SaveVerbatim}
\newcommand{\HOLExpansionTheoremsexpandsXXprecongruence}{\UseVerbatim{HOLExpansionTheoremsexpandsXXprecongruence}}
\begin{SaveVerbatim}{HOLExpansionTheoremsexpandsXXPreOrder}
\HOLTokenTurnstile{} \HOLConst{PreOrder} (\HOLConst{expands})
\end{SaveVerbatim}
\newcommand{\HOLExpansionTheoremsexpandsXXPreOrder}{\UseVerbatim{HOLExpansionTheoremsexpandsXXPreOrder}}
\begin{SaveVerbatim}{HOLExpansionTheoremsexpandsXXPRESDXXBYXXGUARDEDXXSUM}
\HOLTokenTurnstile{} \HOLSymConst{\HOLTokenForall{}}\HOLBoundVar{E\sb{\mathrm{1}}} \HOLBoundVar{E\sb{\mathrm{1}}\sp{\prime}} \HOLBoundVar{E\sb{\mathrm{2}}} \HOLBoundVar{E\sb{\mathrm{2}}\sp{\prime}} \HOLBoundVar{a\sb{\mathrm{1}}} \HOLBoundVar{a\sb{\mathrm{2}}}.
     \HOLBoundVar{E\sb{\mathrm{1}}} \HOLConst{expands} \HOLBoundVar{E\sb{\mathrm{1}}\sp{\prime}} \HOLSymConst{\HOLTokenConj{}} \HOLBoundVar{E\sb{\mathrm{2}}} \HOLConst{expands} \HOLBoundVar{E\sb{\mathrm{2}}\sp{\prime}} \HOLSymConst{\HOLTokenImp{}}
     \HOLBoundVar{a\sb{\mathrm{1}}}\HOLSymConst{..}\HOLBoundVar{E\sb{\mathrm{1}}} \HOLSymConst{+} \HOLBoundVar{a\sb{\mathrm{2}}}\HOLSymConst{..}\HOLBoundVar{E\sb{\mathrm{2}}} \HOLConst{expands} (\HOLBoundVar{a\sb{\mathrm{1}}}\HOLSymConst{..}\HOLBoundVar{E\sb{\mathrm{1}}\sp{\prime}} \HOLSymConst{+} \HOLBoundVar{a\sb{\mathrm{2}}}\HOLSymConst{..}\HOLBoundVar{E\sb{\mathrm{2}}\sp{\prime}})
\end{SaveVerbatim}
\newcommand{\HOLExpansionTheoremsexpandsXXPRESDXXBYXXGUARDEDXXSUM}{\UseVerbatim{HOLExpansionTheoremsexpandsXXPRESDXXBYXXGUARDEDXXSUM}}
\begin{SaveVerbatim}{HOLExpansionTheoremsexpandsXXPRESDXXBYXXPAR}
\HOLTokenTurnstile{} \HOLSymConst{\HOLTokenForall{}}\HOLBoundVar{E\sb{\mathrm{1}}} \HOLBoundVar{E\sb{\mathrm{1}}\sp{\prime}} \HOLBoundVar{E\sb{\mathrm{2}}} \HOLBoundVar{E\sb{\mathrm{2}}\sp{\prime}}.
     \HOLBoundVar{E\sb{\mathrm{1}}} \HOLConst{expands} \HOLBoundVar{E\sb{\mathrm{1}}\sp{\prime}} \HOLSymConst{\HOLTokenConj{}} \HOLBoundVar{E\sb{\mathrm{2}}} \HOLConst{expands} \HOLBoundVar{E\sb{\mathrm{2}}\sp{\prime}} \HOLSymConst{\HOLTokenImp{}} \HOLBoundVar{E\sb{\mathrm{1}}} \HOLSymConst{\ensuremath{\parallel}} \HOLBoundVar{E\sb{\mathrm{2}}} \HOLConst{expands} \HOLBoundVar{E\sb{\mathrm{1}}\sp{\prime}} \HOLSymConst{\ensuremath{\parallel}} \HOLBoundVar{E\sb{\mathrm{2}}\sp{\prime}}
\end{SaveVerbatim}
\newcommand{\HOLExpansionTheoremsexpandsXXPRESDXXBYXXPAR}{\UseVerbatim{HOLExpansionTheoremsexpandsXXPRESDXXBYXXPAR}}
\begin{SaveVerbatim}{HOLExpansionTheoremsexpandsXXREFL}
\HOLTokenTurnstile{} \HOLSymConst{\HOLTokenForall{}}\HOLBoundVar{x}. \HOLBoundVar{x} \HOLConst{expands} \HOLBoundVar{x}
\end{SaveVerbatim}
\newcommand{\HOLExpansionTheoremsexpandsXXREFL}{\UseVerbatim{HOLExpansionTheoremsexpandsXXREFL}}
\begin{SaveVerbatim}{HOLExpansionTheoremsexpandsXXreflexive}
\HOLTokenTurnstile{} \HOLConst{reflexive} (\HOLConst{expands})
\end{SaveVerbatim}
\newcommand{\HOLExpansionTheoremsexpandsXXreflexive}{\UseVerbatim{HOLExpansionTheoremsexpandsXXreflexive}}
\begin{SaveVerbatim}{HOLExpansionTheoremsexpandsXXrules}
\HOLTokenTurnstile{} \HOLSymConst{\HOLTokenForall{}}\HOLBoundVar{E} \HOLBoundVar{E\sp{\prime}}.
     (\HOLSymConst{\HOLTokenForall{}}\HOLBoundVar{l}.
        (\HOLSymConst{\HOLTokenForall{}}\HOLBoundVar{E\sb{\mathrm{1}}}.
           \HOLBoundVar{E} \HOLTokenTransBegin\HOLConst{label} \HOLBoundVar{l}\HOLTokenTransEnd \HOLBoundVar{E\sb{\mathrm{1}}} \HOLSymConst{\HOLTokenImp{}}
           \HOLSymConst{\HOLTokenExists{}}\HOLBoundVar{E\sb{\mathrm{2}}}. \HOLBoundVar{E\sp{\prime}} \HOLTokenTransBegin\HOLConst{label} \HOLBoundVar{l}\HOLTokenTransEnd \HOLBoundVar{E\sb{\mathrm{2}}} \HOLSymConst{\HOLTokenConj{}} \HOLBoundVar{E\sb{\mathrm{1}}} \HOLConst{expands} \HOLBoundVar{E\sb{\mathrm{2}}}) \HOLSymConst{\HOLTokenConj{}}
        \HOLSymConst{\HOLTokenForall{}}\HOLBoundVar{E\sb{\mathrm{2}}}.
          \HOLBoundVar{E\sp{\prime}} \HOLTokenTransBegin\HOLConst{label} \HOLBoundVar{l}\HOLTokenTransEnd \HOLBoundVar{E\sb{\mathrm{2}}} \HOLSymConst{\HOLTokenImp{}}
          \HOLSymConst{\HOLTokenExists{}}\HOLBoundVar{E\sb{\mathrm{1}}}. \HOLBoundVar{E} \HOLTokenWeakTransBegin\HOLConst{label} \HOLBoundVar{l}\HOLTokenWeakTransEnd \HOLBoundVar{E\sb{\mathrm{1}}} \HOLSymConst{\HOLTokenConj{}} \HOLBoundVar{E\sb{\mathrm{1}}} \HOLConst{expands} \HOLBoundVar{E\sb{\mathrm{2}}}) \HOLSymConst{\HOLTokenConj{}}
     (\HOLSymConst{\HOLTokenForall{}}\HOLBoundVar{E\sb{\mathrm{1}}}.
        \HOLBoundVar{E} \HOLTokenTransBegin\HOLConst{\ensuremath{\tau}}\HOLTokenTransEnd \HOLBoundVar{E\sb{\mathrm{1}}} \HOLSymConst{\HOLTokenImp{}}
        \HOLBoundVar{E\sb{\mathrm{1}}} \HOLConst{expands} \HOLBoundVar{E\sp{\prime}} \HOLSymConst{\HOLTokenDisj{}} \HOLSymConst{\HOLTokenExists{}}\HOLBoundVar{E\sb{\mathrm{2}}}. \HOLBoundVar{E\sp{\prime}} \HOLTokenTransBegin\HOLConst{\ensuremath{\tau}}\HOLTokenTransEnd \HOLBoundVar{E\sb{\mathrm{2}}} \HOLSymConst{\HOLTokenConj{}} \HOLBoundVar{E\sb{\mathrm{1}}} \HOLConst{expands} \HOLBoundVar{E\sb{\mathrm{2}}}) \HOLSymConst{\HOLTokenConj{}}
     (\HOLSymConst{\HOLTokenForall{}}\HOLBoundVar{E\sb{\mathrm{2}}}. \HOLBoundVar{E\sp{\prime}} \HOLTokenTransBegin\HOLConst{\ensuremath{\tau}}\HOLTokenTransEnd \HOLBoundVar{E\sb{\mathrm{2}}} \HOLSymConst{\HOLTokenImp{}} \HOLSymConst{\HOLTokenExists{}}\HOLBoundVar{E\sb{\mathrm{1}}}. \HOLBoundVar{E} \HOLTokenWeakTransBegin\HOLConst{\ensuremath{\tau}}\HOLTokenWeakTransEnd \HOLBoundVar{E\sb{\mathrm{1}}} \HOLSymConst{\HOLTokenConj{}} \HOLBoundVar{E\sb{\mathrm{1}}} \HOLConst{expands} \HOLBoundVar{E\sb{\mathrm{2}}}) \HOLSymConst{\HOLTokenImp{}}
     \HOLBoundVar{E} \HOLConst{expands} \HOLBoundVar{E\sp{\prime}}
\end{SaveVerbatim}
\newcommand{\HOLExpansionTheoremsexpandsXXrules}{\UseVerbatim{HOLExpansionTheoremsexpandsXXrules}}
\begin{SaveVerbatim}{HOLExpansionTheoremsexpandsXXSUBSTXXGCONTEXT}
\HOLTokenTurnstile{} \HOLSymConst{\HOLTokenForall{}}\HOLBoundVar{P} \HOLBoundVar{Q}. \HOLBoundVar{P} \HOLConst{expands} \HOLBoundVar{Q} \HOLSymConst{\HOLTokenImp{}} \HOLSymConst{\HOLTokenForall{}}\HOLBoundVar{E}. \HOLConst{GCONTEXT} \HOLBoundVar{E} \HOLSymConst{\HOLTokenImp{}} \HOLBoundVar{E} \HOLBoundVar{P} \HOLConst{expands} \HOLBoundVar{E} \HOLBoundVar{Q}
\end{SaveVerbatim}
\newcommand{\HOLExpansionTheoremsexpandsXXSUBSTXXGCONTEXT}{\UseVerbatim{HOLExpansionTheoremsexpandsXXSUBSTXXGCONTEXT}}
\begin{SaveVerbatim}{HOLExpansionTheoremsexpandsXXSUBSTXXPREFIX}
\HOLTokenTurnstile{} \HOLSymConst{\HOLTokenForall{}}\HOLBoundVar{E} \HOLBoundVar{E\sp{\prime}}. \HOLBoundVar{E} \HOLConst{expands} \HOLBoundVar{E\sp{\prime}} \HOLSymConst{\HOLTokenImp{}} \HOLSymConst{\HOLTokenForall{}}\HOLBoundVar{u}. \HOLBoundVar{u}\HOLSymConst{..}\HOLBoundVar{E} \HOLConst{expands} \HOLBoundVar{u}\HOLSymConst{..}\HOLBoundVar{E\sp{\prime}}
\end{SaveVerbatim}
\newcommand{\HOLExpansionTheoremsexpandsXXSUBSTXXPREFIX}{\UseVerbatim{HOLExpansionTheoremsexpandsXXSUBSTXXPREFIX}}
\begin{SaveVerbatim}{HOLExpansionTheoremsexpandsXXSUBSTXXRELAB}
\HOLTokenTurnstile{} \HOLSymConst{\HOLTokenForall{}}\HOLBoundVar{E} \HOLBoundVar{E\sp{\prime}}. \HOLBoundVar{E} \HOLConst{expands} \HOLBoundVar{E\sp{\prime}} \HOLSymConst{\HOLTokenImp{}} \HOLSymConst{\HOLTokenForall{}}\HOLBoundVar{rf}. \HOLConst{relab} \HOLBoundVar{E} \HOLBoundVar{rf} \HOLConst{expands} \HOLConst{relab} \HOLBoundVar{E\sp{\prime}} \HOLBoundVar{rf}
\end{SaveVerbatim}
\newcommand{\HOLExpansionTheoremsexpandsXXSUBSTXXRELAB}{\UseVerbatim{HOLExpansionTheoremsexpandsXXSUBSTXXRELAB}}
\begin{SaveVerbatim}{HOLExpansionTheoremsexpandsXXSUBSTXXRESTR}
\HOLTokenTurnstile{} \HOLSymConst{\HOLTokenForall{}}\HOLBoundVar{E} \HOLBoundVar{E\sp{\prime}}. \HOLBoundVar{E} \HOLConst{expands} \HOLBoundVar{E\sp{\prime}} \HOLSymConst{\HOLTokenImp{}} \HOLSymConst{\HOLTokenForall{}}\HOLBoundVar{L}. \HOLConst{\ensuremath{\nu}} \HOLBoundVar{L} \HOLBoundVar{E} \HOLConst{expands} \HOLConst{\ensuremath{\nu}} \HOLBoundVar{L} \HOLBoundVar{E\sp{\prime}}
\end{SaveVerbatim}
\newcommand{\HOLExpansionTheoremsexpandsXXSUBSTXXRESTR}{\UseVerbatim{HOLExpansionTheoremsexpandsXXSUBSTXXRESTR}}
\begin{SaveVerbatim}{HOLExpansionTheoremsexpandsXXthm}
\HOLTokenTurnstile{} \HOLSymConst{\HOLTokenForall{}}\HOLBoundVar{P} \HOLBoundVar{Q}. \HOLBoundVar{P} \HOLConst{expands} \HOLBoundVar{Q} \HOLSymConst{\HOLTokenEquiv{}} \HOLSymConst{\HOLTokenExists{}}\HOLBoundVar{Exp}. \HOLBoundVar{Exp} \HOLBoundVar{P} \HOLBoundVar{Q} \HOLSymConst{\HOLTokenConj{}} \HOLConst{EXPANSION} \HOLBoundVar{Exp}
\end{SaveVerbatim}
\newcommand{\HOLExpansionTheoremsexpandsXXthm}{\UseVerbatim{HOLExpansionTheoremsexpandsXXthm}}
\begin{SaveVerbatim}{HOLExpansionTheoremsexpandsXXTRANS}
\HOLTokenTurnstile{} \HOLSymConst{\HOLTokenForall{}}\HOLBoundVar{x} \HOLBoundVar{y} \HOLBoundVar{z}. \HOLBoundVar{x} \HOLConst{expands} \HOLBoundVar{y} \HOLSymConst{\HOLTokenConj{}} \HOLBoundVar{y} \HOLConst{expands} \HOLBoundVar{z} \HOLSymConst{\HOLTokenImp{}} \HOLBoundVar{x} \HOLConst{expands} \HOLBoundVar{z}
\end{SaveVerbatim}
\newcommand{\HOLExpansionTheoremsexpandsXXTRANS}{\UseVerbatim{HOLExpansionTheoremsexpandsXXTRANS}}
\begin{SaveVerbatim}{HOLExpansionTheoremsexpandsXXTRANSXXactionYY}
\HOLTokenTurnstile{} \HOLSymConst{\HOLTokenForall{}}\HOLBoundVar{E} \HOLBoundVar{E\sp{\prime}}.
     \HOLBoundVar{E} \HOLConst{expands} \HOLBoundVar{E\sp{\prime}} \HOLSymConst{\HOLTokenImp{}}
     \HOLSymConst{\HOLTokenForall{}}\HOLBoundVar{u} \HOLBoundVar{E\sb{\mathrm{2}}}. \HOLBoundVar{E\sp{\prime}} \HOLTokenTransBegin\HOLBoundVar{u}\HOLTokenTransEnd \HOLBoundVar{E\sb{\mathrm{2}}} \HOLSymConst{\HOLTokenImp{}} \HOLSymConst{\HOLTokenExists{}}\HOLBoundVar{E\sb{\mathrm{1}}}. \HOLBoundVar{E} \HOLTokenWeakTransBegin\HOLBoundVar{u}\HOLTokenWeakTransEnd \HOLBoundVar{E\sb{\mathrm{1}}} \HOLSymConst{\HOLTokenConj{}} \HOLBoundVar{E\sb{\mathrm{1}}} \HOLConst{expands} \HOLBoundVar{E\sb{\mathrm{2}}}
\end{SaveVerbatim}
\newcommand{\HOLExpansionTheoremsexpandsXXTRANSXXactionYY}{\UseVerbatim{HOLExpansionTheoremsexpandsXXTRANSXXactionYY}}
\begin{SaveVerbatim}{HOLExpansionTheoremsexpandsXXTRANSXXlabel}
\HOLTokenTurnstile{} \HOLSymConst{\HOLTokenForall{}}\HOLBoundVar{E} \HOLBoundVar{E\sp{\prime}}.
     \HOLBoundVar{E} \HOLConst{expands} \HOLBoundVar{E\sp{\prime}} \HOLSymConst{\HOLTokenImp{}}
     \HOLSymConst{\HOLTokenForall{}}\HOLBoundVar{l} \HOLBoundVar{E\sb{\mathrm{1}}}.
       \HOLBoundVar{E} \HOLTokenTransBegin\HOLConst{label} \HOLBoundVar{l}\HOLTokenTransEnd \HOLBoundVar{E\sb{\mathrm{1}}} \HOLSymConst{\HOLTokenImp{}} \HOLSymConst{\HOLTokenExists{}}\HOLBoundVar{E\sb{\mathrm{2}}}. \HOLBoundVar{E\sp{\prime}} \HOLTokenTransBegin\HOLConst{label} \HOLBoundVar{l}\HOLTokenTransEnd \HOLBoundVar{E\sb{\mathrm{2}}} \HOLSymConst{\HOLTokenConj{}} \HOLBoundVar{E\sb{\mathrm{1}}} \HOLConst{expands} \HOLBoundVar{E\sb{\mathrm{2}}}
\end{SaveVerbatim}
\newcommand{\HOLExpansionTheoremsexpandsXXTRANSXXlabel}{\UseVerbatim{HOLExpansionTheoremsexpandsXXTRANSXXlabel}}
\begin{SaveVerbatim}{HOLExpansionTheoremsexpandsXXTRANSXXlabelYY}
\HOLTokenTurnstile{} \HOLSymConst{\HOLTokenForall{}}\HOLBoundVar{E} \HOLBoundVar{E\sp{\prime}}.
     \HOLBoundVar{E} \HOLConst{expands} \HOLBoundVar{E\sp{\prime}} \HOLSymConst{\HOLTokenImp{}}
     \HOLSymConst{\HOLTokenForall{}}\HOLBoundVar{l} \HOLBoundVar{E\sb{\mathrm{2}}}.
       \HOLBoundVar{E\sp{\prime}} \HOLTokenTransBegin\HOLConst{label} \HOLBoundVar{l}\HOLTokenTransEnd \HOLBoundVar{E\sb{\mathrm{2}}} \HOLSymConst{\HOLTokenImp{}} \HOLSymConst{\HOLTokenExists{}}\HOLBoundVar{E\sb{\mathrm{1}}}. \HOLBoundVar{E} \HOLTokenWeakTransBegin\HOLConst{label} \HOLBoundVar{l}\HOLTokenWeakTransEnd \HOLBoundVar{E\sb{\mathrm{1}}} \HOLSymConst{\HOLTokenConj{}} \HOLBoundVar{E\sb{\mathrm{1}}} \HOLConst{expands} \HOLBoundVar{E\sb{\mathrm{2}}}
\end{SaveVerbatim}
\newcommand{\HOLExpansionTheoremsexpandsXXTRANSXXlabelYY}{\UseVerbatim{HOLExpansionTheoremsexpandsXXTRANSXXlabelYY}}
\begin{SaveVerbatim}{HOLExpansionTheoremsexpandsXXTRANSXXtau}
\HOLTokenTurnstile{} \HOLSymConst{\HOLTokenForall{}}\HOLBoundVar{E} \HOLBoundVar{E\sp{\prime}}.
     \HOLBoundVar{E} \HOLConst{expands} \HOLBoundVar{E\sp{\prime}} \HOLSymConst{\HOLTokenImp{}}
     \HOLSymConst{\HOLTokenForall{}}\HOLBoundVar{E\sb{\mathrm{1}}}.
       \HOLBoundVar{E} \HOLTokenTransBegin\HOLConst{\ensuremath{\tau}}\HOLTokenTransEnd \HOLBoundVar{E\sb{\mathrm{1}}} \HOLSymConst{\HOLTokenImp{}} \HOLBoundVar{E\sb{\mathrm{1}}} \HOLConst{expands} \HOLBoundVar{E\sp{\prime}} \HOLSymConst{\HOLTokenDisj{}} \HOLSymConst{\HOLTokenExists{}}\HOLBoundVar{E\sb{\mathrm{2}}}. \HOLBoundVar{E\sp{\prime}} \HOLTokenTransBegin\HOLConst{\ensuremath{\tau}}\HOLTokenTransEnd \HOLBoundVar{E\sb{\mathrm{2}}} \HOLSymConst{\HOLTokenConj{}} \HOLBoundVar{E\sb{\mathrm{1}}} \HOLConst{expands} \HOLBoundVar{E\sb{\mathrm{2}}}
\end{SaveVerbatim}
\newcommand{\HOLExpansionTheoremsexpandsXXTRANSXXtau}{\UseVerbatim{HOLExpansionTheoremsexpandsXXTRANSXXtau}}
\begin{SaveVerbatim}{HOLExpansionTheoremsexpandsXXTRANSXXtauYY}
\HOLTokenTurnstile{} \HOLSymConst{\HOLTokenForall{}}\HOLBoundVar{E} \HOLBoundVar{E\sp{\prime}}.
     \HOLBoundVar{E} \HOLConst{expands} \HOLBoundVar{E\sp{\prime}} \HOLSymConst{\HOLTokenImp{}}
     \HOLSymConst{\HOLTokenForall{}}\HOLBoundVar{E\sb{\mathrm{2}}}. \HOLBoundVar{E\sp{\prime}} \HOLTokenTransBegin\HOLConst{\ensuremath{\tau}}\HOLTokenTransEnd \HOLBoundVar{E\sb{\mathrm{2}}} \HOLSymConst{\HOLTokenImp{}} \HOLSymConst{\HOLTokenExists{}}\HOLBoundVar{E\sb{\mathrm{1}}}. \HOLBoundVar{E} \HOLTokenWeakTransBegin\HOLConst{\ensuremath{\tau}}\HOLTokenWeakTransEnd \HOLBoundVar{E\sb{\mathrm{1}}} \HOLSymConst{\HOLTokenConj{}} \HOLBoundVar{E\sb{\mathrm{1}}} \HOLConst{expands} \HOLBoundVar{E\sb{\mathrm{2}}}
\end{SaveVerbatim}
\newcommand{\HOLExpansionTheoremsexpandsXXTRANSXXtauYY}{\UseVerbatim{HOLExpansionTheoremsexpandsXXTRANSXXtauYY}}
\begin{SaveVerbatim}{HOLExpansionTheoremsexpandsXXtransitive}
\HOLTokenTurnstile{} \HOLConst{transitive} (\HOLConst{expands})
\end{SaveVerbatim}
\newcommand{\HOLExpansionTheoremsexpandsXXtransitive}{\UseVerbatim{HOLExpansionTheoremsexpandsXXtransitive}}
\begin{SaveVerbatim}{HOLExpansionTheoremsexpandsXXWEAKXXTRANSYY}
\HOLTokenTurnstile{} \HOLSymConst{\HOLTokenForall{}}\HOLBoundVar{E} \HOLBoundVar{E\sp{\prime}}.
     \HOLBoundVar{E} \HOLConst{expands} \HOLBoundVar{E\sp{\prime}} \HOLSymConst{\HOLTokenImp{}}
     \HOLSymConst{\HOLTokenForall{}}\HOLBoundVar{u} \HOLBoundVar{E\sb{\mathrm{2}}}. \HOLBoundVar{E\sp{\prime}} \HOLTokenWeakTransBegin\HOLBoundVar{u}\HOLTokenWeakTransEnd \HOLBoundVar{E\sb{\mathrm{2}}} \HOLSymConst{\HOLTokenImp{}} \HOLSymConst{\HOLTokenExists{}}\HOLBoundVar{E\sb{\mathrm{1}}}. \HOLBoundVar{E} \HOLTokenWeakTransBegin\HOLBoundVar{u}\HOLTokenWeakTransEnd \HOLBoundVar{E\sb{\mathrm{1}}} \HOLSymConst{\HOLTokenConj{}} \HOLBoundVar{E\sb{\mathrm{1}}} \HOLConst{expands} \HOLBoundVar{E\sb{\mathrm{2}}}
\end{SaveVerbatim}
\newcommand{\HOLExpansionTheoremsexpandsXXWEAKXXTRANSYY}{\UseVerbatim{HOLExpansionTheoremsexpandsXXWEAKXXTRANSYY}}
\begin{SaveVerbatim}{HOLExpansionTheoremsexpandsXXWEAKXXTRANSXXlabel}
\HOLTokenTurnstile{} \HOLSymConst{\HOLTokenForall{}}\HOLBoundVar{E} \HOLBoundVar{E\sp{\prime}}.
     \HOLBoundVar{E} \HOLConst{expands} \HOLBoundVar{E\sp{\prime}} \HOLSymConst{\HOLTokenImp{}}
     \HOLSymConst{\HOLTokenForall{}}\HOLBoundVar{l} \HOLBoundVar{E\sb{\mathrm{1}}}.
       \HOLBoundVar{E} \HOLTokenWeakTransBegin\HOLConst{label} \HOLBoundVar{l}\HOLTokenWeakTransEnd \HOLBoundVar{E\sb{\mathrm{1}}} \HOLSymConst{\HOLTokenImp{}} \HOLSymConst{\HOLTokenExists{}}\HOLBoundVar{E\sb{\mathrm{2}}}. \HOLBoundVar{E\sp{\prime}} \HOLTokenWeakTransBegin\HOLConst{label} \HOLBoundVar{l}\HOLTokenWeakTransEnd \HOLBoundVar{E\sb{\mathrm{2}}} \HOLSymConst{\HOLTokenConj{}} \HOLBoundVar{E\sb{\mathrm{1}}} \HOLConst{expands} \HOLBoundVar{E\sb{\mathrm{2}}}
\end{SaveVerbatim}
\newcommand{\HOLExpansionTheoremsexpandsXXWEAKXXTRANSXXlabel}{\UseVerbatim{HOLExpansionTheoremsexpandsXXWEAKXXTRANSXXlabel}}
\begin{SaveVerbatim}{HOLExpansionTheoremsexpandsXXWEAKXXTRANSXXtau}
\HOLTokenTurnstile{} \HOLSymConst{\HOLTokenForall{}}\HOLBoundVar{E} \HOLBoundVar{E\sp{\prime}}.
     \HOLBoundVar{E} \HOLConst{expands} \HOLBoundVar{E\sp{\prime}} \HOLSymConst{\HOLTokenImp{}}
     \HOLSymConst{\HOLTokenForall{}}\HOLBoundVar{E\sb{\mathrm{1}}}. \HOLBoundVar{E} \HOLTokenWeakTransBegin\HOLConst{\ensuremath{\tau}}\HOLTokenWeakTransEnd \HOLBoundVar{E\sb{\mathrm{1}}} \HOLSymConst{\HOLTokenImp{}} \HOLSymConst{\HOLTokenExists{}}\HOLBoundVar{E\sb{\mathrm{2}}}. \HOLConst{EPS} \HOLBoundVar{E\sp{\prime}} \HOLBoundVar{E\sb{\mathrm{2}}} \HOLSymConst{\HOLTokenConj{}} \HOLBoundVar{E\sb{\mathrm{1}}} \HOLConst{expands} \HOLBoundVar{E\sb{\mathrm{2}}}
\end{SaveVerbatim}
\newcommand{\HOLExpansionTheoremsexpandsXXWEAKXXTRANSXXtau}{\UseVerbatim{HOLExpansionTheoremsexpandsXXWEAKXXTRANSXXtau}}
\begin{SaveVerbatim}{HOLExpansionTheoremsEXPANSIONXXALT}
\HOLTokenTurnstile{} \HOLConst{EXPANSION} \HOLFreeVar{Exp} \HOLSymConst{\HOLTokenEquiv{}}
   \HOLSymConst{\HOLTokenForall{}}\HOLBoundVar{E} \HOLBoundVar{E\sp{\prime}}.
     \HOLFreeVar{Exp} \HOLBoundVar{E} \HOLBoundVar{E\sp{\prime}} \HOLSymConst{\HOLTokenImp{}}
     (\HOLSymConst{\HOLTokenForall{}}\HOLBoundVar{l} \HOLBoundVar{E\sb{\mathrm{1}}}.
        \HOLBoundVar{E} \HOLTokenTransBegin\HOLConst{label} \HOLBoundVar{l}\HOLTokenTransEnd \HOLBoundVar{E\sb{\mathrm{1}}} \HOLSymConst{\HOLTokenImp{}} \HOLSymConst{\HOLTokenExists{}}\HOLBoundVar{E\sb{\mathrm{2}}}. \HOLBoundVar{E\sp{\prime}} \HOLTokenTransBegin\HOLConst{label} \HOLBoundVar{l}\HOLTokenTransEnd \HOLBoundVar{E\sb{\mathrm{2}}} \HOLSymConst{\HOLTokenConj{}} \HOLFreeVar{Exp} \HOLBoundVar{E\sb{\mathrm{1}}} \HOLBoundVar{E\sb{\mathrm{2}}}) \HOLSymConst{\HOLTokenConj{}}
     (\HOLSymConst{\HOLTokenForall{}}\HOLBoundVar{E\sb{\mathrm{1}}}. \HOLBoundVar{E} \HOLTokenTransBegin\HOLConst{\ensuremath{\tau}}\HOLTokenTransEnd \HOLBoundVar{E\sb{\mathrm{1}}} \HOLSymConst{\HOLTokenImp{}} \HOLFreeVar{Exp} \HOLBoundVar{E\sb{\mathrm{1}}} \HOLBoundVar{E\sp{\prime}} \HOLSymConst{\HOLTokenDisj{}} \HOLSymConst{\HOLTokenExists{}}\HOLBoundVar{E\sb{\mathrm{2}}}. \HOLBoundVar{E\sp{\prime}} \HOLTokenTransBegin\HOLConst{\ensuremath{\tau}}\HOLTokenTransEnd \HOLBoundVar{E\sb{\mathrm{2}}} \HOLSymConst{\HOLTokenConj{}} \HOLFreeVar{Exp} \HOLBoundVar{E\sb{\mathrm{1}}} \HOLBoundVar{E\sb{\mathrm{2}}}) \HOLSymConst{\HOLTokenConj{}}
     \HOLSymConst{\HOLTokenForall{}}\HOLBoundVar{u} \HOLBoundVar{E\sb{\mathrm{2}}}. \HOLBoundVar{E\sp{\prime}} \HOLTokenTransBegin\HOLBoundVar{u}\HOLTokenTransEnd \HOLBoundVar{E\sb{\mathrm{2}}} \HOLSymConst{\HOLTokenImp{}} \HOLSymConst{\HOLTokenExists{}}\HOLBoundVar{E\sb{\mathrm{1}}}. \HOLBoundVar{E} \HOLTokenWeakTransBegin\HOLBoundVar{u}\HOLTokenWeakTransEnd \HOLBoundVar{E\sb{\mathrm{1}}} \HOLSymConst{\HOLTokenConj{}} \HOLFreeVar{Exp} \HOLBoundVar{E\sb{\mathrm{1}}} \HOLBoundVar{E\sb{\mathrm{2}}}
\end{SaveVerbatim}
\newcommand{\HOLExpansionTheoremsEXPANSIONXXALT}{\UseVerbatim{HOLExpansionTheoremsEXPANSIONXXALT}}
\begin{SaveVerbatim}{HOLExpansionTheoremsEXPANSIONXXEPS}
\HOLTokenTurnstile{} \HOLSymConst{\HOLTokenForall{}}\HOLBoundVar{Exp}.
     \HOLConst{EXPANSION} \HOLBoundVar{Exp} \HOLSymConst{\HOLTokenImp{}}
     \HOLSymConst{\HOLTokenForall{}}\HOLBoundVar{E} \HOLBoundVar{E\sp{\prime}}.
       \HOLBoundVar{Exp} \HOLBoundVar{E} \HOLBoundVar{E\sp{\prime}} \HOLSymConst{\HOLTokenImp{}} \HOLSymConst{\HOLTokenForall{}}\HOLBoundVar{E\sb{\mathrm{1}}}. \HOLConst{EPS} \HOLBoundVar{E} \HOLBoundVar{E\sb{\mathrm{1}}} \HOLSymConst{\HOLTokenImp{}} \HOLSymConst{\HOLTokenExists{}}\HOLBoundVar{E\sb{\mathrm{2}}}. \HOLConst{EPS} \HOLBoundVar{E\sp{\prime}} \HOLBoundVar{E\sb{\mathrm{2}}} \HOLSymConst{\HOLTokenConj{}} \HOLBoundVar{Exp} \HOLBoundVar{E\sb{\mathrm{1}}} \HOLBoundVar{E\sb{\mathrm{2}}}
\end{SaveVerbatim}
\newcommand{\HOLExpansionTheoremsEXPANSIONXXEPS}{\UseVerbatim{HOLExpansionTheoremsEXPANSIONXXEPS}}
\begin{SaveVerbatim}{HOLExpansionTheoremsEXPANSIONXXEPSYY}
\HOLTokenTurnstile{} \HOLSymConst{\HOLTokenForall{}}\HOLBoundVar{Exp}.
     \HOLConst{EXPANSION} \HOLBoundVar{Exp} \HOLSymConst{\HOLTokenImp{}}
     \HOLSymConst{\HOLTokenForall{}}\HOLBoundVar{E} \HOLBoundVar{E\sp{\prime}}.
       \HOLBoundVar{Exp} \HOLBoundVar{E} \HOLBoundVar{E\sp{\prime}} \HOLSymConst{\HOLTokenImp{}} \HOLSymConst{\HOLTokenForall{}}\HOLBoundVar{E\sb{\mathrm{2}}}. \HOLConst{EPS} \HOLBoundVar{E\sp{\prime}} \HOLBoundVar{E\sb{\mathrm{2}}} \HOLSymConst{\HOLTokenImp{}} \HOLSymConst{\HOLTokenExists{}}\HOLBoundVar{E\sb{\mathrm{1}}}. \HOLConst{EPS} \HOLBoundVar{E} \HOLBoundVar{E\sb{\mathrm{1}}} \HOLSymConst{\HOLTokenConj{}} \HOLBoundVar{Exp} \HOLBoundVar{E\sb{\mathrm{1}}} \HOLBoundVar{E\sb{\mathrm{2}}}
\end{SaveVerbatim}
\newcommand{\HOLExpansionTheoremsEXPANSIONXXEPSYY}{\UseVerbatim{HOLExpansionTheoremsEXPANSIONXXEPSYY}}
\begin{SaveVerbatim}{HOLExpansionTheoremsEXPANSIONXXIMPXXWEAKXXBISIM}
\HOLTokenTurnstile{} \HOLSymConst{\HOLTokenForall{}}\HOLBoundVar{Exp}. \HOLConst{EXPANSION} \HOLBoundVar{Exp} \HOLSymConst{\HOLTokenImp{}} \HOLConst{WEAK_BISIM} \HOLBoundVar{Exp}
\end{SaveVerbatim}
\newcommand{\HOLExpansionTheoremsEXPANSIONXXIMPXXWEAKXXBISIM}{\UseVerbatim{HOLExpansionTheoremsEXPANSIONXXIMPXXWEAKXXBISIM}}
\begin{SaveVerbatim}{HOLExpansionTheoremsEXPANSIONXXSUBSETXXexpands}
\HOLTokenTurnstile{} \HOLSymConst{\HOLTokenForall{}}\HOLBoundVar{Exp}. \HOLConst{EXPANSION} \HOLBoundVar{Exp} \HOLSymConst{\HOLTokenImp{}} \HOLBoundVar{Exp} \HOLConst{RSUBSET} (\HOLConst{expands})
\end{SaveVerbatim}
\newcommand{\HOLExpansionTheoremsEXPANSIONXXSUBSETXXexpands}{\UseVerbatim{HOLExpansionTheoremsEXPANSIONXXSUBSETXXexpands}}
\begin{SaveVerbatim}{HOLExpansionTheoremsEXPANSIONXXWEAKXXTRANSYY}
\HOLTokenTurnstile{} \HOLSymConst{\HOLTokenForall{}}\HOLBoundVar{Exp}.
     \HOLConst{EXPANSION} \HOLBoundVar{Exp} \HOLSymConst{\HOLTokenImp{}}
     \HOLSymConst{\HOLTokenForall{}}\HOLBoundVar{E} \HOLBoundVar{E\sp{\prime}}.
       \HOLBoundVar{Exp} \HOLBoundVar{E} \HOLBoundVar{E\sp{\prime}} \HOLSymConst{\HOLTokenImp{}} \HOLSymConst{\HOLTokenForall{}}\HOLBoundVar{u} \HOLBoundVar{E\sb{\mathrm{2}}}. \HOLBoundVar{E\sp{\prime}} \HOLTokenWeakTransBegin\HOLBoundVar{u}\HOLTokenWeakTransEnd \HOLBoundVar{E\sb{\mathrm{2}}} \HOLSymConst{\HOLTokenImp{}} \HOLSymConst{\HOLTokenExists{}}\HOLBoundVar{E\sb{\mathrm{1}}}. \HOLBoundVar{E} \HOLTokenWeakTransBegin\HOLBoundVar{u}\HOLTokenWeakTransEnd \HOLBoundVar{E\sb{\mathrm{1}}} \HOLSymConst{\HOLTokenConj{}} \HOLBoundVar{Exp} \HOLBoundVar{E\sb{\mathrm{1}}} \HOLBoundVar{E\sb{\mathrm{2}}}
\end{SaveVerbatim}
\newcommand{\HOLExpansionTheoremsEXPANSIONXXWEAKXXTRANSYY}{\UseVerbatim{HOLExpansionTheoremsEXPANSIONXXWEAKXXTRANSYY}}
\begin{SaveVerbatim}{HOLExpansionTheoremsIDENTITYXXEXPANSION}
\HOLTokenTurnstile{} \HOLConst{EXPANSION} (\HOLSymConst{=})
\end{SaveVerbatim}
\newcommand{\HOLExpansionTheoremsIDENTITYXXEXPANSION}{\UseVerbatim{HOLExpansionTheoremsIDENTITYXXEXPANSION}}
\begin{SaveVerbatim}{HOLExpansionTheoremsSTRONGXXBISIMXXIMPXXEXPANSION}
\HOLTokenTurnstile{} \HOLSymConst{\HOLTokenForall{}}\HOLBoundVar{Exp}. \HOLConst{STRONG_BISIM} \HOLBoundVar{Exp} \HOLSymConst{\HOLTokenImp{}} \HOLConst{EXPANSION} \HOLBoundVar{Exp}
\end{SaveVerbatim}
\newcommand{\HOLExpansionTheoremsSTRONGXXBISIMXXIMPXXEXPANSION}{\UseVerbatim{HOLExpansionTheoremsSTRONGXXBISIMXXIMPXXEXPANSION}}
\begin{SaveVerbatim}{HOLExpansionTheoremsSTRONGXXEQUIVXXIMPXXexpands}
\HOLTokenTurnstile{} \HOLSymConst{\HOLTokenForall{}}\HOLBoundVar{P} \HOLBoundVar{Q}. \HOLConst{STRONG_EQUIV} \HOLBoundVar{P} \HOLBoundVar{Q} \HOLSymConst{\HOLTokenImp{}} \HOLBoundVar{P} \HOLConst{expands} \HOLBoundVar{Q}
\end{SaveVerbatim}
\newcommand{\HOLExpansionTheoremsSTRONGXXEQUIVXXIMPXXexpands}{\UseVerbatim{HOLExpansionTheoremsSTRONGXXEQUIVXXIMPXXexpands}}
\newcommand{\HOLExpansionTheorems}{
\HOLThmTag{Expansion}{COMP_EXPANSION}\HOLExpansionTheoremsCOMPXXEXPANSION
\HOLThmTag{Expansion}{EQ_IMP_expands}\HOLExpansionTheoremsEQXXIMPXXexpands
\HOLThmTag{Expansion}{expands_AND_TRACE_label}\HOLExpansionTheoremsexpandsXXANDXXTRACEXXlabel
\HOLThmTag{Expansion}{expands_AND_TRACE_tau}\HOLExpansionTheoremsexpandsXXANDXXTRACEXXtau
\HOLThmTag{Expansion}{expands_cases}\HOLExpansionTheoremsexpandsXXcases
\HOLThmTag{Expansion}{expands_cases'}\HOLExpansionTheoremsexpandsXXcasesYY
\HOLThmTag{Expansion}{expands_coind}\HOLExpansionTheoremsexpandsXXcoind
\HOLThmTag{Expansion}{expands_EPS}\HOLExpansionTheoremsexpandsXXEPS
\HOLThmTag{Expansion}{expands_EPS'}\HOLExpansionTheoremsexpandsXXEPSYY
\HOLThmTag{Expansion}{expands_IMP_WEAK_EQUIV}\HOLExpansionTheoremsexpandsXXIMPXXWEAKXXEQUIV
\HOLThmTag{Expansion}{expands_is_EXPANSION}\HOLExpansionTheoremsexpandsXXisXXEXPANSION
\HOLThmTag{Expansion}{expands_precongruence}\HOLExpansionTheoremsexpandsXXprecongruence
\HOLThmTag{Expansion}{expands_PreOrder}\HOLExpansionTheoremsexpandsXXPreOrder
\HOLThmTag{Expansion}{expands_PRESD_BY_GUARDED_SUM}\HOLExpansionTheoremsexpandsXXPRESDXXBYXXGUARDEDXXSUM
\HOLThmTag{Expansion}{expands_PRESD_BY_PAR}\HOLExpansionTheoremsexpandsXXPRESDXXBYXXPAR
\HOLThmTag{Expansion}{expands_REFL}\HOLExpansionTheoremsexpandsXXREFL
\HOLThmTag{Expansion}{expands_reflexive}\HOLExpansionTheoremsexpandsXXreflexive
\HOLThmTag{Expansion}{expands_rules}\HOLExpansionTheoremsexpandsXXrules
\HOLThmTag{Expansion}{expands_SUBST_GCONTEXT}\HOLExpansionTheoremsexpandsXXSUBSTXXGCONTEXT
\HOLThmTag{Expansion}{expands_SUBST_PREFIX}\HOLExpansionTheoremsexpandsXXSUBSTXXPREFIX
\HOLThmTag{Expansion}{expands_SUBST_RELAB}\HOLExpansionTheoremsexpandsXXSUBSTXXRELAB
\HOLThmTag{Expansion}{expands_SUBST_RESTR}\HOLExpansionTheoremsexpandsXXSUBSTXXRESTR
\HOLThmTag{Expansion}{expands_thm}\HOLExpansionTheoremsexpandsXXthm
\HOLThmTag{Expansion}{expands_TRANS}\HOLExpansionTheoremsexpandsXXTRANS
\HOLThmTag{Expansion}{expands_TRANS_action'}\HOLExpansionTheoremsexpandsXXTRANSXXactionYY
\HOLThmTag{Expansion}{expands_TRANS_label}\HOLExpansionTheoremsexpandsXXTRANSXXlabel
\HOLThmTag{Expansion}{expands_TRANS_label'}\HOLExpansionTheoremsexpandsXXTRANSXXlabelYY
\HOLThmTag{Expansion}{expands_TRANS_tau}\HOLExpansionTheoremsexpandsXXTRANSXXtau
\HOLThmTag{Expansion}{expands_TRANS_tau'}\HOLExpansionTheoremsexpandsXXTRANSXXtauYY
\HOLThmTag{Expansion}{expands_transitive}\HOLExpansionTheoremsexpandsXXtransitive
\HOLThmTag{Expansion}{expands_WEAK_TRANS'}\HOLExpansionTheoremsexpandsXXWEAKXXTRANSYY
\HOLThmTag{Expansion}{expands_WEAK_TRANS_label}\HOLExpansionTheoremsexpandsXXWEAKXXTRANSXXlabel
\HOLThmTag{Expansion}{expands_WEAK_TRANS_tau}\HOLExpansionTheoremsexpandsXXWEAKXXTRANSXXtau
\HOLThmTag{Expansion}{EXPANSION_ALT}\HOLExpansionTheoremsEXPANSIONXXALT
\HOLThmTag{Expansion}{EXPANSION_EPS}\HOLExpansionTheoremsEXPANSIONXXEPS
\HOLThmTag{Expansion}{EXPANSION_EPS'}\HOLExpansionTheoremsEXPANSIONXXEPSYY
\HOLThmTag{Expansion}{EXPANSION_IMP_WEAK_BISIM}\HOLExpansionTheoremsEXPANSIONXXIMPXXWEAKXXBISIM
\HOLThmTag{Expansion}{EXPANSION_SUBSET_expands}\HOLExpansionTheoremsEXPANSIONXXSUBSETXXexpands
\HOLThmTag{Expansion}{EXPANSION_WEAK_TRANS'}\HOLExpansionTheoremsEXPANSIONXXWEAKXXTRANSYY
\HOLThmTag{Expansion}{IDENTITY_EXPANSION}\HOLExpansionTheoremsIDENTITYXXEXPANSION
\HOLThmTag{Expansion}{STRONG_BISIM_IMP_EXPANSION}\HOLExpansionTheoremsSTRONGXXBISIMXXIMPXXEXPANSION
\HOLThmTag{Expansion}{STRONG_EQUIV_IMP_expands}\HOLExpansionTheoremsSTRONGXXEQUIVXXIMPXXexpands
}

\newcommand{\HOLContractionDate}{02 Dicembre 2017}
\newcommand{\HOLContractionTime}{13:31}
\begin{SaveVerbatim}{HOLContractionDefinitionsCXXcontracts}
\HOLTokenTurnstile{} \HOLConst{C_contracts} \HOLSymConst{=} \HOLConst{CC} (\HOLConst{contracts})
\end{SaveVerbatim}
\newcommand{\HOLContractionDefinitionsCXXcontracts}{\UseVerbatim{HOLContractionDefinitionsCXXcontracts}}
\begin{SaveVerbatim}{HOLContractionDefinitionsCONTRACTION}
\HOLTokenTurnstile{} \HOLSymConst{\HOLTokenForall{}}\HOLBoundVar{Con}.
     \HOLConst{CONTRACTION} \HOLBoundVar{Con} \HOLSymConst{\HOLTokenEquiv{}}
     \HOLSymConst{\HOLTokenForall{}}\HOLBoundVar{E} \HOLBoundVar{E\sp{\prime}}.
       \HOLBoundVar{Con} \HOLBoundVar{E} \HOLBoundVar{E\sp{\prime}} \HOLSymConst{\HOLTokenImp{}}
       (\HOLSymConst{\HOLTokenForall{}}\HOLBoundVar{l}.
          (\HOLSymConst{\HOLTokenForall{}}\HOLBoundVar{E\sb{\mathrm{1}}}.
             \HOLBoundVar{E} \HOLTokenTransBegin\HOLConst{label} \HOLBoundVar{l}\HOLTokenTransEnd \HOLBoundVar{E\sb{\mathrm{1}}} \HOLSymConst{\HOLTokenImp{}}
             \HOLSymConst{\HOLTokenExists{}}\HOLBoundVar{E\sb{\mathrm{2}}}. \HOLBoundVar{E\sp{\prime}} \HOLTokenTransBegin\HOLConst{label} \HOLBoundVar{l}\HOLTokenTransEnd \HOLBoundVar{E\sb{\mathrm{2}}} \HOLSymConst{\HOLTokenConj{}} \HOLBoundVar{Con} \HOLBoundVar{E\sb{\mathrm{1}}} \HOLBoundVar{E\sb{\mathrm{2}}}) \HOLSymConst{\HOLTokenConj{}}
          \HOLSymConst{\HOLTokenForall{}}\HOLBoundVar{E\sb{\mathrm{2}}}.
            \HOLBoundVar{E\sp{\prime}} \HOLTokenTransBegin\HOLConst{label} \HOLBoundVar{l}\HOLTokenTransEnd \HOLBoundVar{E\sb{\mathrm{2}}} \HOLSymConst{\HOLTokenImp{}}
            \HOLSymConst{\HOLTokenExists{}}\HOLBoundVar{E\sb{\mathrm{1}}}. \HOLBoundVar{E} \HOLTokenWeakTransBegin\HOLConst{label} \HOLBoundVar{l}\HOLTokenWeakTransEnd \HOLBoundVar{E\sb{\mathrm{1}}} \HOLSymConst{\HOLTokenConj{}} \HOLConst{WEAK_EQUIV} \HOLBoundVar{E\sb{\mathrm{1}}} \HOLBoundVar{E\sb{\mathrm{2}}}) \HOLSymConst{\HOLTokenConj{}}
       (\HOLSymConst{\HOLTokenForall{}}\HOLBoundVar{E\sb{\mathrm{1}}}.
          \HOLBoundVar{E} \HOLTokenTransBegin\HOLConst{\ensuremath{\tau}}\HOLTokenTransEnd \HOLBoundVar{E\sb{\mathrm{1}}} \HOLSymConst{\HOLTokenImp{}} \HOLBoundVar{Con} \HOLBoundVar{E\sb{\mathrm{1}}} \HOLBoundVar{E\sp{\prime}} \HOLSymConst{\HOLTokenDisj{}} \HOLSymConst{\HOLTokenExists{}}\HOLBoundVar{E\sb{\mathrm{2}}}. \HOLBoundVar{E\sp{\prime}} \HOLTokenTransBegin\HOLConst{\ensuremath{\tau}}\HOLTokenTransEnd \HOLBoundVar{E\sb{\mathrm{2}}} \HOLSymConst{\HOLTokenConj{}} \HOLBoundVar{Con} \HOLBoundVar{E\sb{\mathrm{1}}} \HOLBoundVar{E\sb{\mathrm{2}}}) \HOLSymConst{\HOLTokenConj{}}
       \HOLSymConst{\HOLTokenForall{}}\HOLBoundVar{E\sb{\mathrm{2}}}. \HOLBoundVar{E\sp{\prime}} \HOLTokenTransBegin\HOLConst{\ensuremath{\tau}}\HOLTokenTransEnd \HOLBoundVar{E\sb{\mathrm{2}}} \HOLSymConst{\HOLTokenImp{}} \HOLSymConst{\HOLTokenExists{}}\HOLBoundVar{E\sb{\mathrm{1}}}. \HOLConst{EPS} \HOLBoundVar{E} \HOLBoundVar{E\sb{\mathrm{1}}} \HOLSymConst{\HOLTokenConj{}} \HOLConst{WEAK_EQUIV} \HOLBoundVar{E\sb{\mathrm{1}}} \HOLBoundVar{E\sb{\mathrm{2}}}
\end{SaveVerbatim}
\newcommand{\HOLContractionDefinitionsCONTRACTION}{\UseVerbatim{HOLContractionDefinitionsCONTRACTION}}
\begin{SaveVerbatim}{HOLContractionDefinitionscontractsXXdef}
\HOLTokenTurnstile{} (\HOLConst{contracts}) \HOLSymConst{=}
   (\HOLTokenLambda{}\HOLBoundVar{a\sb{\mathrm{0}}} \HOLBoundVar{a\sb{\mathrm{1}}}.
      \HOLSymConst{\HOLTokenExists{}}\HOLBoundVar{contracts\sp{\prime}}.
        \HOLBoundVar{contracts\sp{\prime}} \HOLBoundVar{a\sb{\mathrm{0}}} \HOLBoundVar{a\sb{\mathrm{1}}} \HOLSymConst{\HOLTokenConj{}}
        \HOLSymConst{\HOLTokenForall{}}\HOLBoundVar{a\sb{\mathrm{0}}} \HOLBoundVar{a\sb{\mathrm{1}}}.
          \HOLBoundVar{contracts\sp{\prime}} \HOLBoundVar{a\sb{\mathrm{0}}} \HOLBoundVar{a\sb{\mathrm{1}}} \HOLSymConst{\HOLTokenImp{}}
          (\HOLSymConst{\HOLTokenForall{}}\HOLBoundVar{l}.
             (\HOLSymConst{\HOLTokenForall{}}\HOLBoundVar{E\sb{\mathrm{1}}}.
                \HOLBoundVar{a\sb{\mathrm{0}}} \HOLTokenTransBegin\HOLConst{label} \HOLBoundVar{l}\HOLTokenTransEnd \HOLBoundVar{E\sb{\mathrm{1}}} \HOLSymConst{\HOLTokenImp{}}
                \HOLSymConst{\HOLTokenExists{}}\HOLBoundVar{E\sb{\mathrm{2}}}. \HOLBoundVar{a\sb{\mathrm{1}}} \HOLTokenTransBegin\HOLConst{label} \HOLBoundVar{l}\HOLTokenTransEnd \HOLBoundVar{E\sb{\mathrm{2}}} \HOLSymConst{\HOLTokenConj{}} \HOLBoundVar{contracts\sp{\prime}} \HOLBoundVar{E\sb{\mathrm{1}}} \HOLBoundVar{E\sb{\mathrm{2}}}) \HOLSymConst{\HOLTokenConj{}}
             \HOLSymConst{\HOLTokenForall{}}\HOLBoundVar{E\sb{\mathrm{2}}}.
               \HOLBoundVar{a\sb{\mathrm{1}}} \HOLTokenTransBegin\HOLConst{label} \HOLBoundVar{l}\HOLTokenTransEnd \HOLBoundVar{E\sb{\mathrm{2}}} \HOLSymConst{\HOLTokenImp{}}
               \HOLSymConst{\HOLTokenExists{}}\HOLBoundVar{E\sb{\mathrm{1}}}. \HOLBoundVar{a\sb{\mathrm{0}}} \HOLTokenWeakTransBegin\HOLConst{label} \HOLBoundVar{l}\HOLTokenWeakTransEnd \HOLBoundVar{E\sb{\mathrm{1}}} \HOLSymConst{\HOLTokenConj{}} \HOLConst{WEAK_EQUIV} \HOLBoundVar{E\sb{\mathrm{1}}} \HOLBoundVar{E\sb{\mathrm{2}}}) \HOLSymConst{\HOLTokenConj{}}
          (\HOLSymConst{\HOLTokenForall{}}\HOLBoundVar{E\sb{\mathrm{1}}}.
             \HOLBoundVar{a\sb{\mathrm{0}}} \HOLTokenTransBegin\HOLConst{\ensuremath{\tau}}\HOLTokenTransEnd \HOLBoundVar{E\sb{\mathrm{1}}} \HOLSymConst{\HOLTokenImp{}}
             \HOLBoundVar{contracts\sp{\prime}} \HOLBoundVar{E\sb{\mathrm{1}}} \HOLBoundVar{a\sb{\mathrm{1}}} \HOLSymConst{\HOLTokenDisj{}}
             \HOLSymConst{\HOLTokenExists{}}\HOLBoundVar{E\sb{\mathrm{2}}}. \HOLBoundVar{a\sb{\mathrm{1}}} \HOLTokenTransBegin\HOLConst{\ensuremath{\tau}}\HOLTokenTransEnd \HOLBoundVar{E\sb{\mathrm{2}}} \HOLSymConst{\HOLTokenConj{}} \HOLBoundVar{contracts\sp{\prime}} \HOLBoundVar{E\sb{\mathrm{1}}} \HOLBoundVar{E\sb{\mathrm{2}}}) \HOLSymConst{\HOLTokenConj{}}
          \HOLSymConst{\HOLTokenForall{}}\HOLBoundVar{E\sb{\mathrm{2}}}. \HOLBoundVar{a\sb{\mathrm{1}}} \HOLTokenTransBegin\HOLConst{\ensuremath{\tau}}\HOLTokenTransEnd \HOLBoundVar{E\sb{\mathrm{2}}} \HOLSymConst{\HOLTokenImp{}} \HOLSymConst{\HOLTokenExists{}}\HOLBoundVar{E\sb{\mathrm{1}}}. \HOLConst{EPS} \HOLBoundVar{a\sb{\mathrm{0}}} \HOLBoundVar{E\sb{\mathrm{1}}} \HOLSymConst{\HOLTokenConj{}} \HOLConst{WEAK_EQUIV} \HOLBoundVar{E\sb{\mathrm{1}}} \HOLBoundVar{E\sb{\mathrm{2}}})
\end{SaveVerbatim}
\newcommand{\HOLContractionDefinitionscontractsXXdef}{\UseVerbatim{HOLContractionDefinitionscontractsXXdef}}
\begin{SaveVerbatim}{HOLContractionDefinitionsOBSXXcontracts}
\HOLTokenTurnstile{} \HOLSymConst{\HOLTokenForall{}}\HOLBoundVar{E} \HOLBoundVar{E\sp{\prime}}.
     \HOLConst{OBS_contracts} \HOLBoundVar{E} \HOLBoundVar{E\sp{\prime}} \HOLSymConst{\HOLTokenEquiv{}}
     \HOLSymConst{\HOLTokenForall{}}\HOLBoundVar{u}.
       (\HOLSymConst{\HOLTokenForall{}}\HOLBoundVar{E\sb{\mathrm{1}}}. \HOLBoundVar{E} \HOLTokenTransBegin\HOLBoundVar{u}\HOLTokenTransEnd \HOLBoundVar{E\sb{\mathrm{1}}} \HOLSymConst{\HOLTokenImp{}} \HOLSymConst{\HOLTokenExists{}}\HOLBoundVar{E\sb{\mathrm{2}}}. \HOLBoundVar{E\sp{\prime}} \HOLTokenTransBegin\HOLBoundVar{u}\HOLTokenTransEnd \HOLBoundVar{E\sb{\mathrm{2}}} \HOLSymConst{\HOLTokenConj{}} \HOLBoundVar{E\sb{\mathrm{1}}} \HOLConst{contracts} \HOLBoundVar{E\sb{\mathrm{2}}}) \HOLSymConst{\HOLTokenConj{}}
       \HOLSymConst{\HOLTokenForall{}}\HOLBoundVar{E\sb{\mathrm{2}}}. \HOLBoundVar{E\sp{\prime}} \HOLTokenTransBegin\HOLBoundVar{u}\HOLTokenTransEnd \HOLBoundVar{E\sb{\mathrm{2}}} \HOLSymConst{\HOLTokenImp{}} \HOLSymConst{\HOLTokenExists{}}\HOLBoundVar{E\sb{\mathrm{1}}}. \HOLBoundVar{E} \HOLTokenWeakTransBegin\HOLBoundVar{u}\HOLTokenWeakTransEnd \HOLBoundVar{E\sb{\mathrm{1}}} \HOLSymConst{\HOLTokenConj{}} \HOLConst{WEAK_EQUIV} \HOLBoundVar{E\sb{\mathrm{1}}} \HOLBoundVar{E\sb{\mathrm{2}}}
\end{SaveVerbatim}
\newcommand{\HOLContractionDefinitionsOBSXXcontracts}{\UseVerbatim{HOLContractionDefinitionsOBSXXcontracts}}
\begin{SaveVerbatim}{HOLContractionDefinitionsSUMXXcontracts}
\HOLTokenTurnstile{} \HOLConst{SUM_contracts} \HOLSymConst{=} (\HOLTokenLambda{}\HOLBoundVar{p} \HOLBoundVar{q}. \HOLSymConst{\HOLTokenForall{}}\HOLBoundVar{r}. \HOLBoundVar{p} \HOLSymConst{+} \HOLBoundVar{r} \HOLConst{contracts} \HOLBoundVar{q} \HOLSymConst{+} \HOLBoundVar{r})
\end{SaveVerbatim}
\newcommand{\HOLContractionDefinitionsSUMXXcontracts}{\UseVerbatim{HOLContractionDefinitionsSUMXXcontracts}}
\newcommand{\HOLContractionDefinitions}{
\HOLDfnTag{Contraction}{C_contracts}\HOLContractionDefinitionsCXXcontracts
\HOLDfnTag{Contraction}{CONTRACTION}\HOLContractionDefinitionsCONTRACTION
\HOLDfnTag{Contraction}{contracts_def}\HOLContractionDefinitionscontractsXXdef
\HOLDfnTag{Contraction}{OBS_contracts}\HOLContractionDefinitionsOBSXXcontracts
\HOLDfnTag{Contraction}{SUM_contracts}\HOLContractionDefinitionsSUMXXcontracts
}
\begin{SaveVerbatim}{HOLContractionTheoremsCXXcontractsXXIMPXXSUMXXcontracts}
\HOLTokenTurnstile{} \HOLSymConst{\HOLTokenForall{}}\HOLBoundVar{p} \HOLBoundVar{q}. \HOLConst{C_contracts} \HOLBoundVar{p} \HOLBoundVar{q} \HOLSymConst{\HOLTokenImp{}} \HOLConst{SUM_contracts} \HOLBoundVar{p} \HOLBoundVar{q}
\end{SaveVerbatim}
\newcommand{\HOLContractionTheoremsCXXcontractsXXIMPXXSUMXXcontracts}{\UseVerbatim{HOLContractionTheoremsCXXcontractsXXIMPXXSUMXXcontracts}}
\begin{SaveVerbatim}{HOLContractionTheoremsCXXcontractsXXprecongruence}
\HOLTokenTurnstile{} \HOLConst{precongruence} \HOLConst{C_contracts}
\end{SaveVerbatim}
\newcommand{\HOLContractionTheoremsCXXcontractsXXprecongruence}{\UseVerbatim{HOLContractionTheoremsCXXcontractsXXprecongruence}}
\begin{SaveVerbatim}{HOLContractionTheoremsCXXcontractsXXthm}
\HOLTokenTurnstile{} \HOLConst{C_contracts} \HOLSymConst{=} (\HOLTokenLambda{}\HOLBoundVar{g} \HOLBoundVar{h}. \HOLSymConst{\HOLTokenForall{}}\HOLBoundVar{c}. \HOLConst{CONTEXT} \HOLBoundVar{c} \HOLSymConst{\HOLTokenImp{}} \HOLBoundVar{c} \HOLBoundVar{g} \HOLConst{contracts} \HOLBoundVar{c} \HOLBoundVar{h})
\end{SaveVerbatim}
\newcommand{\HOLContractionTheoremsCXXcontractsXXthm}{\UseVerbatim{HOLContractionTheoremsCXXcontractsXXthm}}
\begin{SaveVerbatim}{HOLContractionTheoremsCOARSESTXXPRECONGRXXTHM}
\HOLTokenTurnstile{} \HOLSymConst{\HOLTokenForall{}}\HOLBoundVar{p} \HOLBoundVar{q}.
     \HOLConst{free_action} \HOLBoundVar{p} \HOLSymConst{\HOLTokenConj{}} \HOLConst{free_action} \HOLBoundVar{q} \HOLSymConst{\HOLTokenImp{}}
     (\HOLConst{OBS_contracts} \HOLBoundVar{p} \HOLBoundVar{q} \HOLSymConst{\HOLTokenEquiv{}} \HOLConst{SUM_contracts} \HOLBoundVar{p} \HOLBoundVar{q})
\end{SaveVerbatim}
\newcommand{\HOLContractionTheoremsCOARSESTXXPRECONGRXXTHM}{\UseVerbatim{HOLContractionTheoremsCOARSESTXXPRECONGRXXTHM}}
\begin{SaveVerbatim}{HOLContractionTheoremsCOARSESTXXPRECONGRXXTHMYY}
\HOLTokenTurnstile{} \HOLSymConst{\HOLTokenForall{}}\HOLBoundVar{p} \HOLBoundVar{q}.
     \HOLConst{free_action} \HOLBoundVar{p} \HOLSymConst{\HOLTokenConj{}} \HOLConst{free_action} \HOLBoundVar{q} \HOLSymConst{\HOLTokenImp{}}
     (\HOLConst{OBS_contracts} \HOLBoundVar{p} \HOLBoundVar{q} \HOLSymConst{\HOLTokenEquiv{}} \HOLSymConst{\HOLTokenForall{}}\HOLBoundVar{r}. \HOLBoundVar{p} \HOLSymConst{+} \HOLBoundVar{r} \HOLConst{contracts} \HOLBoundVar{q} \HOLSymConst{+} \HOLBoundVar{r})
\end{SaveVerbatim}
\newcommand{\HOLContractionTheoremsCOARSESTXXPRECONGRXXTHMYY}{\UseVerbatim{HOLContractionTheoremsCOARSESTXXPRECONGRXXTHMYY}}
\begin{SaveVerbatim}{HOLContractionTheoremsCOMPXXCONTRACTION}
\HOLTokenTurnstile{} \HOLSymConst{\HOLTokenForall{}}\HOLBoundVar{Con\sb{\mathrm{1}}} \HOLBoundVar{Con\sb{\mathrm{2}}}.
     \HOLConst{CONTRACTION} \HOLBoundVar{Con\sb{\mathrm{1}}} \HOLSymConst{\HOLTokenConj{}} \HOLConst{CONTRACTION} \HOLBoundVar{Con\sb{\mathrm{2}}} \HOLSymConst{\HOLTokenImp{}}
     \HOLConst{CONTRACTION} (\HOLBoundVar{Con\sb{\mathrm{2}}} \HOLConst{O} \HOLBoundVar{Con\sb{\mathrm{1}}})
\end{SaveVerbatim}
\newcommand{\HOLContractionTheoremsCOMPXXCONTRACTION}{\UseVerbatim{HOLContractionTheoremsCOMPXXCONTRACTION}}
\begin{SaveVerbatim}{HOLContractionTheoremsCONTRACTIONXXEPS}
\HOLTokenTurnstile{} \HOLSymConst{\HOLTokenForall{}}\HOLBoundVar{Con}.
     \HOLConst{CONTRACTION} \HOLBoundVar{Con} \HOLSymConst{\HOLTokenImp{}}
     \HOLSymConst{\HOLTokenForall{}}\HOLBoundVar{E} \HOLBoundVar{E\sp{\prime}}.
       \HOLBoundVar{Con} \HOLBoundVar{E} \HOLBoundVar{E\sp{\prime}} \HOLSymConst{\HOLTokenImp{}} \HOLSymConst{\HOLTokenForall{}}\HOLBoundVar{E\sb{\mathrm{1}}}. \HOLConst{EPS} \HOLBoundVar{E} \HOLBoundVar{E\sb{\mathrm{1}}} \HOLSymConst{\HOLTokenImp{}} \HOLSymConst{\HOLTokenExists{}}\HOLBoundVar{E\sb{\mathrm{2}}}. \HOLConst{EPS} \HOLBoundVar{E\sp{\prime}} \HOLBoundVar{E\sb{\mathrm{2}}} \HOLSymConst{\HOLTokenConj{}} \HOLBoundVar{Con} \HOLBoundVar{E\sb{\mathrm{1}}} \HOLBoundVar{E\sb{\mathrm{2}}}
\end{SaveVerbatim}
\newcommand{\HOLContractionTheoremsCONTRACTIONXXEPS}{\UseVerbatim{HOLContractionTheoremsCONTRACTIONXXEPS}}
\begin{SaveVerbatim}{HOLContractionTheoremsCONTRACTIONXXEPSYY}
\HOLTokenTurnstile{} \HOLSymConst{\HOLTokenForall{}}\HOLBoundVar{Con}.
     \HOLConst{CONTRACTION} \HOLBoundVar{Con} \HOLSymConst{\HOLTokenImp{}}
     \HOLSymConst{\HOLTokenForall{}}\HOLBoundVar{E} \HOLBoundVar{E\sp{\prime}}.
       \HOLBoundVar{Con} \HOLBoundVar{E} \HOLBoundVar{E\sp{\prime}} \HOLSymConst{\HOLTokenImp{}}
       \HOLSymConst{\HOLTokenForall{}}\HOLBoundVar{u} \HOLBoundVar{E\sb{\mathrm{2}}}. \HOLConst{EPS} \HOLBoundVar{E\sp{\prime}} \HOLBoundVar{E\sb{\mathrm{2}}} \HOLSymConst{\HOLTokenImp{}} \HOLSymConst{\HOLTokenExists{}}\HOLBoundVar{E\sb{\mathrm{1}}}. \HOLConst{EPS} \HOLBoundVar{E} \HOLBoundVar{E\sb{\mathrm{1}}} \HOLSymConst{\HOLTokenConj{}} \HOLConst{WEAK_EQUIV} \HOLBoundVar{E\sb{\mathrm{1}}} \HOLBoundVar{E\sb{\mathrm{2}}}
\end{SaveVerbatim}
\newcommand{\HOLContractionTheoremsCONTRACTIONXXEPSYY}{\UseVerbatim{HOLContractionTheoremsCONTRACTIONXXEPSYY}}
\begin{SaveVerbatim}{HOLContractionTheoremsCONTRACTIONXXSUBSETXXcontracts}
\HOLTokenTurnstile{} \HOLSymConst{\HOLTokenForall{}}\HOLBoundVar{Con}. \HOLConst{CONTRACTION} \HOLBoundVar{Con} \HOLSymConst{\HOLTokenImp{}} \HOLBoundVar{Con} \HOLConst{RSUBSET} (\HOLConst{contracts})
\end{SaveVerbatim}
\newcommand{\HOLContractionTheoremsCONTRACTIONXXSUBSETXXcontracts}{\UseVerbatim{HOLContractionTheoremsCONTRACTIONXXSUBSETXXcontracts}}
\begin{SaveVerbatim}{HOLContractionTheoremsCONTRACTIONXXWEAKXXTRANSXXlabelYY}
\HOLTokenTurnstile{} \HOLSymConst{\HOLTokenForall{}}\HOLBoundVar{Con}.
     \HOLConst{CONTRACTION} \HOLBoundVar{Con} \HOLSymConst{\HOLTokenImp{}}
     \HOLSymConst{\HOLTokenForall{}}\HOLBoundVar{E} \HOLBoundVar{E\sp{\prime}}.
       \HOLBoundVar{Con} \HOLBoundVar{E} \HOLBoundVar{E\sp{\prime}} \HOLSymConst{\HOLTokenImp{}}
       \HOLSymConst{\HOLTokenForall{}}\HOLBoundVar{l} \HOLBoundVar{E\sb{\mathrm{2}}}.
         \HOLBoundVar{E\sp{\prime}} \HOLTokenWeakTransBegin\HOLConst{label} \HOLBoundVar{l}\HOLTokenWeakTransEnd \HOLBoundVar{E\sb{\mathrm{2}}} \HOLSymConst{\HOLTokenImp{}}
         \HOLSymConst{\HOLTokenExists{}}\HOLBoundVar{E\sb{\mathrm{1}}}. \HOLBoundVar{E} \HOLTokenWeakTransBegin\HOLConst{label} \HOLBoundVar{l}\HOLTokenWeakTransEnd \HOLBoundVar{E\sb{\mathrm{1}}} \HOLSymConst{\HOLTokenConj{}} \HOLConst{WEAK_EQUIV} \HOLBoundVar{E\sb{\mathrm{1}}} \HOLBoundVar{E\sb{\mathrm{2}}}
\end{SaveVerbatim}
\newcommand{\HOLContractionTheoremsCONTRACTIONXXWEAKXXTRANSXXlabelYY}{\UseVerbatim{HOLContractionTheoremsCONTRACTIONXXWEAKXXTRANSXXlabelYY}}
\begin{SaveVerbatim}{HOLContractionTheoremscontractsXXANDXXTRACEOne}
\HOLTokenTurnstile{} \HOLSymConst{\HOLTokenForall{}}\HOLBoundVar{E} \HOLBoundVar{E\sp{\prime}}.
     \HOLBoundVar{E} \HOLConst{contracts} \HOLBoundVar{E\sp{\prime}} \HOLSymConst{\HOLTokenImp{}}
     \HOLSymConst{\HOLTokenForall{}}\HOLBoundVar{xs} \HOLBoundVar{E\sb{\mathrm{1}}}.
       \HOLConst{TRACE} \HOLBoundVar{E} \HOLBoundVar{xs} \HOLBoundVar{E\sb{\mathrm{1}}} \HOLSymConst{\HOLTokenImp{}}
       \HOLSymConst{\HOLTokenExists{}}\HOLBoundVar{xs\sp{\prime}} \HOLBoundVar{E\sb{\mathrm{2}}}. \HOLConst{TRACE} \HOLBoundVar{E\sp{\prime}} \HOLBoundVar{xs\sp{\prime}} \HOLBoundVar{E\sb{\mathrm{2}}} \HOLSymConst{\HOLTokenConj{}} \HOLBoundVar{E\sb{\mathrm{1}}} \HOLConst{contracts} \HOLBoundVar{E\sb{\mathrm{2}}}
\end{SaveVerbatim}
\newcommand{\HOLContractionTheoremscontractsXXANDXXTRACEOne}{\UseVerbatim{HOLContractionTheoremscontractsXXANDXXTRACEOne}}
\begin{SaveVerbatim}{HOLContractionTheoremscontractsXXANDXXTRACETwo}
\HOLTokenTurnstile{} \HOLSymConst{\HOLTokenForall{}}\HOLBoundVar{E} \HOLBoundVar{E\sp{\prime}}.
     \HOLBoundVar{E} \HOLConst{contracts} \HOLBoundVar{E\sp{\prime}} \HOLSymConst{\HOLTokenImp{}}
     \HOLSymConst{\HOLTokenForall{}}\HOLBoundVar{xs} \HOLBoundVar{E\sb{\mathrm{1}}}.
       \HOLConst{TRACE} \HOLBoundVar{E} \HOLBoundVar{xs} \HOLBoundVar{E\sb{\mathrm{1}}} \HOLSymConst{\HOLTokenImp{}}
       \HOLSymConst{\HOLTokenExists{}}\HOLBoundVar{xs\sp{\prime}} \HOLBoundVar{E\sb{\mathrm{2}}}.
         \HOLConst{TRACE} \HOLBoundVar{E\sp{\prime}} \HOLBoundVar{xs\sp{\prime}} \HOLBoundVar{E\sb{\mathrm{2}}} \HOLSymConst{\HOLTokenConj{}} \HOLBoundVar{E\sb{\mathrm{1}}} \HOLConst{contracts} \HOLBoundVar{E\sb{\mathrm{2}}} \HOLSymConst{\HOLTokenConj{}}
         \HOLConst{LENGTH} \HOLBoundVar{xs\sp{\prime}} \HOLSymConst{\HOLTokenLeq{}} \HOLConst{LENGTH} \HOLBoundVar{xs}
\end{SaveVerbatim}
\newcommand{\HOLContractionTheoremscontractsXXANDXXTRACETwo}{\UseVerbatim{HOLContractionTheoremscontractsXXANDXXTRACETwo}}
\begin{SaveVerbatim}{HOLContractionTheoremscontractsXXANDXXTRACEXXlabel}
\HOLTokenTurnstile{} \HOLSymConst{\HOLTokenForall{}}\HOLBoundVar{E} \HOLBoundVar{E\sp{\prime}}.
     \HOLBoundVar{E} \HOLConst{contracts} \HOLBoundVar{E\sp{\prime}} \HOLSymConst{\HOLTokenImp{}}
     \HOLSymConst{\HOLTokenForall{}}\HOLBoundVar{xs} \HOLBoundVar{l} \HOLBoundVar{E\sb{\mathrm{1}}}.
       \HOLConst{TRACE} \HOLBoundVar{E} \HOLBoundVar{xs} \HOLBoundVar{E\sb{\mathrm{1}}} \HOLSymConst{\HOLTokenConj{}} \HOLConst{UNIQUE_LABEL} (\HOLConst{label} \HOLBoundVar{l}) \HOLBoundVar{xs} \HOLSymConst{\HOLTokenImp{}}
       \HOLSymConst{\HOLTokenExists{}}\HOLBoundVar{xs\sp{\prime}} \HOLBoundVar{E\sb{\mathrm{2}}}.
         \HOLConst{TRACE} \HOLBoundVar{E\sp{\prime}} \HOLBoundVar{xs\sp{\prime}} \HOLBoundVar{E\sb{\mathrm{2}}} \HOLSymConst{\HOLTokenConj{}} \HOLBoundVar{E\sb{\mathrm{1}}} \HOLConst{contracts} \HOLBoundVar{E\sb{\mathrm{2}}} \HOLSymConst{\HOLTokenConj{}}
         \HOLConst{LENGTH} \HOLBoundVar{xs\sp{\prime}} \HOLSymConst{\HOLTokenLeq{}} \HOLConst{LENGTH} \HOLBoundVar{xs} \HOLSymConst{\HOLTokenConj{}} \HOLConst{UNIQUE_LABEL} (\HOLConst{label} \HOLBoundVar{l}) \HOLBoundVar{xs\sp{\prime}}
\end{SaveVerbatim}
\newcommand{\HOLContractionTheoremscontractsXXANDXXTRACEXXlabel}{\UseVerbatim{HOLContractionTheoremscontractsXXANDXXTRACEXXlabel}}
\begin{SaveVerbatim}{HOLContractionTheoremscontractsXXANDXXTRACEXXtau}
\HOLTokenTurnstile{} \HOLSymConst{\HOLTokenForall{}}\HOLBoundVar{E} \HOLBoundVar{E\sp{\prime}}.
     \HOLBoundVar{E} \HOLConst{contracts} \HOLBoundVar{E\sp{\prime}} \HOLSymConst{\HOLTokenImp{}}
     \HOLSymConst{\HOLTokenForall{}}\HOLBoundVar{xs} \HOLBoundVar{E\sb{\mathrm{1}}}.
       \HOLConst{TRACE} \HOLBoundVar{E} \HOLBoundVar{xs} \HOLBoundVar{E\sb{\mathrm{1}}} \HOLSymConst{\HOLTokenConj{}} \HOLConst{NO_LABEL} \HOLBoundVar{xs} \HOLSymConst{\HOLTokenImp{}}
       \HOLSymConst{\HOLTokenExists{}}\HOLBoundVar{xs\sp{\prime}} \HOLBoundVar{E\sb{\mathrm{2}}}.
         \HOLConst{TRACE} \HOLBoundVar{E\sp{\prime}} \HOLBoundVar{xs\sp{\prime}} \HOLBoundVar{E\sb{\mathrm{2}}} \HOLSymConst{\HOLTokenConj{}} \HOLBoundVar{E\sb{\mathrm{1}}} \HOLConst{contracts} \HOLBoundVar{E\sb{\mathrm{2}}} \HOLSymConst{\HOLTokenConj{}}
         \HOLConst{LENGTH} \HOLBoundVar{xs\sp{\prime}} \HOLSymConst{\HOLTokenLeq{}} \HOLConst{LENGTH} \HOLBoundVar{xs} \HOLSymConst{\HOLTokenConj{}} \HOLConst{NO_LABEL} \HOLBoundVar{xs\sp{\prime}}
\end{SaveVerbatim}
\newcommand{\HOLContractionTheoremscontractsXXANDXXTRACEXXtau}{\UseVerbatim{HOLContractionTheoremscontractsXXANDXXTRACEXXtau}}
\begin{SaveVerbatim}{HOLContractionTheoremscontractsXXcases}
\HOLTokenTurnstile{} \HOLSymConst{\HOLTokenForall{}}\HOLBoundVar{a\sb{\mathrm{0}}} \HOLBoundVar{a\sb{\mathrm{1}}}.
     \HOLBoundVar{a\sb{\mathrm{0}}} \HOLConst{contracts} \HOLBoundVar{a\sb{\mathrm{1}}} \HOLSymConst{\HOLTokenEquiv{}}
     (\HOLSymConst{\HOLTokenForall{}}\HOLBoundVar{l}.
        (\HOLSymConst{\HOLTokenForall{}}\HOLBoundVar{E\sb{\mathrm{1}}}.
           \HOLBoundVar{a\sb{\mathrm{0}}} \HOLTokenTransBegin\HOLConst{label} \HOLBoundVar{l}\HOLTokenTransEnd \HOLBoundVar{E\sb{\mathrm{1}}} \HOLSymConst{\HOLTokenImp{}}
           \HOLSymConst{\HOLTokenExists{}}\HOLBoundVar{E\sb{\mathrm{2}}}. \HOLBoundVar{a\sb{\mathrm{1}}} \HOLTokenTransBegin\HOLConst{label} \HOLBoundVar{l}\HOLTokenTransEnd \HOLBoundVar{E\sb{\mathrm{2}}} \HOLSymConst{\HOLTokenConj{}} \HOLBoundVar{E\sb{\mathrm{1}}} \HOLConst{contracts} \HOLBoundVar{E\sb{\mathrm{2}}}) \HOLSymConst{\HOLTokenConj{}}
        \HOLSymConst{\HOLTokenForall{}}\HOLBoundVar{E\sb{\mathrm{2}}}.
          \HOLBoundVar{a\sb{\mathrm{1}}} \HOLTokenTransBegin\HOLConst{label} \HOLBoundVar{l}\HOLTokenTransEnd \HOLBoundVar{E\sb{\mathrm{2}}} \HOLSymConst{\HOLTokenImp{}}
          \HOLSymConst{\HOLTokenExists{}}\HOLBoundVar{E\sb{\mathrm{1}}}. \HOLBoundVar{a\sb{\mathrm{0}}} \HOLTokenWeakTransBegin\HOLConst{label} \HOLBoundVar{l}\HOLTokenWeakTransEnd \HOLBoundVar{E\sb{\mathrm{1}}} \HOLSymConst{\HOLTokenConj{}} \HOLConst{WEAK_EQUIV} \HOLBoundVar{E\sb{\mathrm{1}}} \HOLBoundVar{E\sb{\mathrm{2}}}) \HOLSymConst{\HOLTokenConj{}}
     (\HOLSymConst{\HOLTokenForall{}}\HOLBoundVar{E\sb{\mathrm{1}}}.
        \HOLBoundVar{a\sb{\mathrm{0}}} \HOLTokenTransBegin\HOLConst{\ensuremath{\tau}}\HOLTokenTransEnd \HOLBoundVar{E\sb{\mathrm{1}}} \HOLSymConst{\HOLTokenImp{}}
        \HOLBoundVar{E\sb{\mathrm{1}}} \HOLConst{contracts} \HOLBoundVar{a\sb{\mathrm{1}}} \HOLSymConst{\HOLTokenDisj{}} \HOLSymConst{\HOLTokenExists{}}\HOLBoundVar{E\sb{\mathrm{2}}}. \HOLBoundVar{a\sb{\mathrm{1}}} \HOLTokenTransBegin\HOLConst{\ensuremath{\tau}}\HOLTokenTransEnd \HOLBoundVar{E\sb{\mathrm{2}}} \HOLSymConst{\HOLTokenConj{}} \HOLBoundVar{E\sb{\mathrm{1}}} \HOLConst{contracts} \HOLBoundVar{E\sb{\mathrm{2}}}) \HOLSymConst{\HOLTokenConj{}}
     \HOLSymConst{\HOLTokenForall{}}\HOLBoundVar{E\sb{\mathrm{2}}}. \HOLBoundVar{a\sb{\mathrm{1}}} \HOLTokenTransBegin\HOLConst{\ensuremath{\tau}}\HOLTokenTransEnd \HOLBoundVar{E\sb{\mathrm{2}}} \HOLSymConst{\HOLTokenImp{}} \HOLSymConst{\HOLTokenExists{}}\HOLBoundVar{E\sb{\mathrm{1}}}. \HOLConst{EPS} \HOLBoundVar{a\sb{\mathrm{0}}} \HOLBoundVar{E\sb{\mathrm{1}}} \HOLSymConst{\HOLTokenConj{}} \HOLConst{WEAK_EQUIV} \HOLBoundVar{E\sb{\mathrm{1}}} \HOLBoundVar{E\sb{\mathrm{2}}}
\end{SaveVerbatim}
\newcommand{\HOLContractionTheoremscontractsXXcases}{\UseVerbatim{HOLContractionTheoremscontractsXXcases}}
\begin{SaveVerbatim}{HOLContractionTheoremscontractsXXcoind}
\HOLTokenTurnstile{} \HOLSymConst{\HOLTokenForall{}}\HOLBoundVar{contracts\sp{\prime}}.
     (\HOLSymConst{\HOLTokenForall{}}\HOLBoundVar{a\sb{\mathrm{0}}} \HOLBoundVar{a\sb{\mathrm{1}}}.
        \HOLBoundVar{contracts\sp{\prime}} \HOLBoundVar{a\sb{\mathrm{0}}} \HOLBoundVar{a\sb{\mathrm{1}}} \HOLSymConst{\HOLTokenImp{}}
        (\HOLSymConst{\HOLTokenForall{}}\HOLBoundVar{l}.
           (\HOLSymConst{\HOLTokenForall{}}\HOLBoundVar{E\sb{\mathrm{1}}}.
              \HOLBoundVar{a\sb{\mathrm{0}}} \HOLTokenTransBegin\HOLConst{label} \HOLBoundVar{l}\HOLTokenTransEnd \HOLBoundVar{E\sb{\mathrm{1}}} \HOLSymConst{\HOLTokenImp{}}
              \HOLSymConst{\HOLTokenExists{}}\HOLBoundVar{E\sb{\mathrm{2}}}. \HOLBoundVar{a\sb{\mathrm{1}}} \HOLTokenTransBegin\HOLConst{label} \HOLBoundVar{l}\HOLTokenTransEnd \HOLBoundVar{E\sb{\mathrm{2}}} \HOLSymConst{\HOLTokenConj{}} \HOLBoundVar{contracts\sp{\prime}} \HOLBoundVar{E\sb{\mathrm{1}}} \HOLBoundVar{E\sb{\mathrm{2}}}) \HOLSymConst{\HOLTokenConj{}}
           \HOLSymConst{\HOLTokenForall{}}\HOLBoundVar{E\sb{\mathrm{2}}}.
             \HOLBoundVar{a\sb{\mathrm{1}}} \HOLTokenTransBegin\HOLConst{label} \HOLBoundVar{l}\HOLTokenTransEnd \HOLBoundVar{E\sb{\mathrm{2}}} \HOLSymConst{\HOLTokenImp{}}
             \HOLSymConst{\HOLTokenExists{}}\HOLBoundVar{E\sb{\mathrm{1}}}. \HOLBoundVar{a\sb{\mathrm{0}}} \HOLTokenWeakTransBegin\HOLConst{label} \HOLBoundVar{l}\HOLTokenWeakTransEnd \HOLBoundVar{E\sb{\mathrm{1}}} \HOLSymConst{\HOLTokenConj{}} \HOLConst{WEAK_EQUIV} \HOLBoundVar{E\sb{\mathrm{1}}} \HOLBoundVar{E\sb{\mathrm{2}}}) \HOLSymConst{\HOLTokenConj{}}
        (\HOLSymConst{\HOLTokenForall{}}\HOLBoundVar{E\sb{\mathrm{1}}}.
           \HOLBoundVar{a\sb{\mathrm{0}}} \HOLTokenTransBegin\HOLConst{\ensuremath{\tau}}\HOLTokenTransEnd \HOLBoundVar{E\sb{\mathrm{1}}} \HOLSymConst{\HOLTokenImp{}}
           \HOLBoundVar{contracts\sp{\prime}} \HOLBoundVar{E\sb{\mathrm{1}}} \HOLBoundVar{a\sb{\mathrm{1}}} \HOLSymConst{\HOLTokenDisj{}}
           \HOLSymConst{\HOLTokenExists{}}\HOLBoundVar{E\sb{\mathrm{2}}}. \HOLBoundVar{a\sb{\mathrm{1}}} \HOLTokenTransBegin\HOLConst{\ensuremath{\tau}}\HOLTokenTransEnd \HOLBoundVar{E\sb{\mathrm{2}}} \HOLSymConst{\HOLTokenConj{}} \HOLBoundVar{contracts\sp{\prime}} \HOLBoundVar{E\sb{\mathrm{1}}} \HOLBoundVar{E\sb{\mathrm{2}}}) \HOLSymConst{\HOLTokenConj{}}
        \HOLSymConst{\HOLTokenForall{}}\HOLBoundVar{E\sb{\mathrm{2}}}. \HOLBoundVar{a\sb{\mathrm{1}}} \HOLTokenTransBegin\HOLConst{\ensuremath{\tau}}\HOLTokenTransEnd \HOLBoundVar{E\sb{\mathrm{2}}} \HOLSymConst{\HOLTokenImp{}} \HOLSymConst{\HOLTokenExists{}}\HOLBoundVar{E\sb{\mathrm{1}}}. \HOLConst{EPS} \HOLBoundVar{a\sb{\mathrm{0}}} \HOLBoundVar{E\sb{\mathrm{1}}} \HOLSymConst{\HOLTokenConj{}} \HOLConst{WEAK_EQUIV} \HOLBoundVar{E\sb{\mathrm{1}}} \HOLBoundVar{E\sb{\mathrm{2}}}) \HOLSymConst{\HOLTokenImp{}}
     \HOLSymConst{\HOLTokenForall{}}\HOLBoundVar{a\sb{\mathrm{0}}} \HOLBoundVar{a\sb{\mathrm{1}}}. \HOLBoundVar{contracts\sp{\prime}} \HOLBoundVar{a\sb{\mathrm{0}}} \HOLBoundVar{a\sb{\mathrm{1}}} \HOLSymConst{\HOLTokenImp{}} \HOLBoundVar{a\sb{\mathrm{0}}} \HOLConst{contracts} \HOLBoundVar{a\sb{\mathrm{1}}}
\end{SaveVerbatim}
\newcommand{\HOLContractionTheoremscontractsXXcoind}{\UseVerbatim{HOLContractionTheoremscontractsXXcoind}}
\begin{SaveVerbatim}{HOLContractionTheoremscontractsXXEPS}
\HOLTokenTurnstile{} \HOLSymConst{\HOLTokenForall{}}\HOLBoundVar{E} \HOLBoundVar{E\sp{\prime}}.
     \HOLBoundVar{E} \HOLConst{contracts} \HOLBoundVar{E\sp{\prime}} \HOLSymConst{\HOLTokenImp{}}
     \HOLSymConst{\HOLTokenForall{}}\HOLBoundVar{E\sb{\mathrm{1}}}. \HOLConst{EPS} \HOLBoundVar{E} \HOLBoundVar{E\sb{\mathrm{1}}} \HOLSymConst{\HOLTokenImp{}} \HOLSymConst{\HOLTokenExists{}}\HOLBoundVar{E\sb{\mathrm{2}}}. \HOLConst{EPS} \HOLBoundVar{E\sp{\prime}} \HOLBoundVar{E\sb{\mathrm{2}}} \HOLSymConst{\HOLTokenConj{}} \HOLBoundVar{E\sb{\mathrm{1}}} \HOLConst{contracts} \HOLBoundVar{E\sb{\mathrm{2}}}
\end{SaveVerbatim}
\newcommand{\HOLContractionTheoremscontractsXXEPS}{\UseVerbatim{HOLContractionTheoremscontractsXXEPS}}
\begin{SaveVerbatim}{HOLContractionTheoremscontractsXXEPSYY}
\HOLTokenTurnstile{} \HOLSymConst{\HOLTokenForall{}}\HOLBoundVar{E} \HOLBoundVar{E\sp{\prime}}.
     \HOLBoundVar{E} \HOLConst{contracts} \HOLBoundVar{E\sp{\prime}} \HOLSymConst{\HOLTokenImp{}}
     \HOLSymConst{\HOLTokenForall{}}\HOLBoundVar{E\sb{\mathrm{2}}}. \HOLConst{EPS} \HOLBoundVar{E\sp{\prime}} \HOLBoundVar{E\sb{\mathrm{2}}} \HOLSymConst{\HOLTokenImp{}} \HOLSymConst{\HOLTokenExists{}}\HOLBoundVar{E\sb{\mathrm{1}}}. \HOLConst{EPS} \HOLBoundVar{E} \HOLBoundVar{E\sb{\mathrm{1}}} \HOLSymConst{\HOLTokenConj{}} \HOLConst{WEAK_EQUIV} \HOLBoundVar{E\sb{\mathrm{1}}} \HOLBoundVar{E\sb{\mathrm{2}}}
\end{SaveVerbatim}
\newcommand{\HOLContractionTheoremscontractsXXEPSYY}{\UseVerbatim{HOLContractionTheoremscontractsXXEPSYY}}
\begin{SaveVerbatim}{HOLContractionTheoremscontractsXXIMPXXWEAKXXEQUIV}
\HOLTokenTurnstile{} \HOLSymConst{\HOLTokenForall{}}\HOLBoundVar{P} \HOLBoundVar{Q}. \HOLBoundVar{P} \HOLConst{contracts} \HOLBoundVar{Q} \HOLSymConst{\HOLTokenImp{}} \HOLConst{WEAK_EQUIV} \HOLBoundVar{P} \HOLBoundVar{Q}
\end{SaveVerbatim}
\newcommand{\HOLContractionTheoremscontractsXXIMPXXWEAKXXEQUIV}{\UseVerbatim{HOLContractionTheoremscontractsXXIMPXXWEAKXXEQUIV}}
\begin{SaveVerbatim}{HOLContractionTheoremscontractsXXisXXCONTRACTION}
\HOLTokenTurnstile{} \HOLConst{CONTRACTION} (\HOLConst{contracts})
\end{SaveVerbatim}
\newcommand{\HOLContractionTheoremscontractsXXisXXCONTRACTION}{\UseVerbatim{HOLContractionTheoremscontractsXXisXXCONTRACTION}}
\begin{SaveVerbatim}{HOLContractionTheoremscontractsXXprecongruence}
\HOLTokenTurnstile{} \HOLConst{precongruence1} (\HOLConst{contracts})
\end{SaveVerbatim}
\newcommand{\HOLContractionTheoremscontractsXXprecongruence}{\UseVerbatim{HOLContractionTheoremscontractsXXprecongruence}}
\begin{SaveVerbatim}{HOLContractionTheoremscontractsXXPreOrder}
\HOLTokenTurnstile{} \HOLConst{PreOrder} (\HOLConst{contracts})
\end{SaveVerbatim}
\newcommand{\HOLContractionTheoremscontractsXXPreOrder}{\UseVerbatim{HOLContractionTheoremscontractsXXPreOrder}}
\begin{SaveVerbatim}{HOLContractionTheoremscontractsXXPRESDXXBYXXGUARDEDXXSUM}
\HOLTokenTurnstile{} \HOLSymConst{\HOLTokenForall{}}\HOLBoundVar{E\sb{\mathrm{1}}} \HOLBoundVar{E\sb{\mathrm{1}}\sp{\prime}} \HOLBoundVar{E\sb{\mathrm{2}}} \HOLBoundVar{E\sb{\mathrm{2}}\sp{\prime}} \HOLBoundVar{a\sb{\mathrm{1}}} \HOLBoundVar{a\sb{\mathrm{2}}}.
     \HOLBoundVar{E\sb{\mathrm{1}}} \HOLConst{contracts} \HOLBoundVar{E\sb{\mathrm{1}}\sp{\prime}} \HOLSymConst{\HOLTokenConj{}} \HOLBoundVar{E\sb{\mathrm{2}}} \HOLConst{contracts} \HOLBoundVar{E\sb{\mathrm{2}}\sp{\prime}} \HOLSymConst{\HOLTokenImp{}}
     \HOLBoundVar{a\sb{\mathrm{1}}}\HOLSymConst{..}\HOLBoundVar{E\sb{\mathrm{1}}} \HOLSymConst{+} \HOLBoundVar{a\sb{\mathrm{2}}}\HOLSymConst{..}\HOLBoundVar{E\sb{\mathrm{2}}} \HOLConst{contracts} \HOLBoundVar{a\sb{\mathrm{1}}}\HOLSymConst{..}\HOLBoundVar{E\sb{\mathrm{1}}\sp{\prime}} \HOLSymConst{+} \HOLBoundVar{a\sb{\mathrm{2}}}\HOLSymConst{..}\HOLBoundVar{E\sb{\mathrm{2}}\sp{\prime}}
\end{SaveVerbatim}
\newcommand{\HOLContractionTheoremscontractsXXPRESDXXBYXXGUARDEDXXSUM}{\UseVerbatim{HOLContractionTheoremscontractsXXPRESDXXBYXXGUARDEDXXSUM}}
\begin{SaveVerbatim}{HOLContractionTheoremscontractsXXPRESDXXBYXXPAR}
\HOLTokenTurnstile{} \HOLSymConst{\HOLTokenForall{}}\HOLBoundVar{E\sb{\mathrm{1}}} \HOLBoundVar{E\sb{\mathrm{1}}\sp{\prime}} \HOLBoundVar{E\sb{\mathrm{2}}} \HOLBoundVar{E\sb{\mathrm{2}}\sp{\prime}}.
     \HOLBoundVar{E\sb{\mathrm{1}}} \HOLConst{contracts} \HOLBoundVar{E\sb{\mathrm{1}}\sp{\prime}} \HOLSymConst{\HOLTokenConj{}} \HOLBoundVar{E\sb{\mathrm{2}}} \HOLConst{contracts} \HOLBoundVar{E\sb{\mathrm{2}}\sp{\prime}} \HOLSymConst{\HOLTokenImp{}}
     \HOLBoundVar{E\sb{\mathrm{1}}} \HOLSymConst{\ensuremath{\parallel}} \HOLBoundVar{E\sb{\mathrm{2}}} \HOLConst{contracts} \HOLBoundVar{E\sb{\mathrm{1}}\sp{\prime}} \HOLSymConst{\ensuremath{\parallel}} \HOLBoundVar{E\sb{\mathrm{2}}\sp{\prime}}
\end{SaveVerbatim}
\newcommand{\HOLContractionTheoremscontractsXXPRESDXXBYXXPAR}{\UseVerbatim{HOLContractionTheoremscontractsXXPRESDXXBYXXPAR}}
\begin{SaveVerbatim}{HOLContractionTheoremscontractsXXPROPSix}
\HOLTokenTurnstile{} \HOLSymConst{\HOLTokenForall{}}\HOLBoundVar{E} \HOLBoundVar{E\sp{\prime}}. \HOLBoundVar{E} \HOLConst{contracts} \HOLBoundVar{E\sp{\prime}} \HOLSymConst{\HOLTokenImp{}} \HOLSymConst{\HOLTokenForall{}}\HOLBoundVar{u}. \HOLConst{OBS_contracts} (\HOLBoundVar{u}\HOLSymConst{..}\HOLBoundVar{E}) (\HOLBoundVar{u}\HOLSymConst{..}\HOLBoundVar{E\sp{\prime}})
\end{SaveVerbatim}
\newcommand{\HOLContractionTheoremscontractsXXPROPSix}{\UseVerbatim{HOLContractionTheoremscontractsXXPROPSix}}
\begin{SaveVerbatim}{HOLContractionTheoremscontractsXXREFL}
\HOLTokenTurnstile{} \HOLSymConst{\HOLTokenForall{}}\HOLBoundVar{x}. \HOLBoundVar{x} \HOLConst{contracts} \HOLBoundVar{x}
\end{SaveVerbatim}
\newcommand{\HOLContractionTheoremscontractsXXREFL}{\UseVerbatim{HOLContractionTheoremscontractsXXREFL}}
\begin{SaveVerbatim}{HOLContractionTheoremscontractsXXreflexive}
\HOLTokenTurnstile{} \HOLConst{reflexive} (\HOLConst{contracts})
\end{SaveVerbatim}
\newcommand{\HOLContractionTheoremscontractsXXreflexive}{\UseVerbatim{HOLContractionTheoremscontractsXXreflexive}}
\begin{SaveVerbatim}{HOLContractionTheoremscontractsXXrules}
\HOLTokenTurnstile{} \HOLSymConst{\HOLTokenForall{}}\HOLBoundVar{E} \HOLBoundVar{E\sp{\prime}}.
     (\HOLSymConst{\HOLTokenForall{}}\HOLBoundVar{l}.
        (\HOLSymConst{\HOLTokenForall{}}\HOLBoundVar{E\sb{\mathrm{1}}}.
           \HOLBoundVar{E} \HOLTokenTransBegin\HOLConst{label} \HOLBoundVar{l}\HOLTokenTransEnd \HOLBoundVar{E\sb{\mathrm{1}}} \HOLSymConst{\HOLTokenImp{}}
           \HOLSymConst{\HOLTokenExists{}}\HOLBoundVar{E\sb{\mathrm{2}}}. \HOLBoundVar{E\sp{\prime}} \HOLTokenTransBegin\HOLConst{label} \HOLBoundVar{l}\HOLTokenTransEnd \HOLBoundVar{E\sb{\mathrm{2}}} \HOLSymConst{\HOLTokenConj{}} \HOLBoundVar{E\sb{\mathrm{1}}} \HOLConst{contracts} \HOLBoundVar{E\sb{\mathrm{2}}}) \HOLSymConst{\HOLTokenConj{}}
        \HOLSymConst{\HOLTokenForall{}}\HOLBoundVar{E\sb{\mathrm{2}}}.
          \HOLBoundVar{E\sp{\prime}} \HOLTokenTransBegin\HOLConst{label} \HOLBoundVar{l}\HOLTokenTransEnd \HOLBoundVar{E\sb{\mathrm{2}}} \HOLSymConst{\HOLTokenImp{}}
          \HOLSymConst{\HOLTokenExists{}}\HOLBoundVar{E\sb{\mathrm{1}}}. \HOLBoundVar{E} \HOLTokenWeakTransBegin\HOLConst{label} \HOLBoundVar{l}\HOLTokenWeakTransEnd \HOLBoundVar{E\sb{\mathrm{1}}} \HOLSymConst{\HOLTokenConj{}} \HOLConst{WEAK_EQUIV} \HOLBoundVar{E\sb{\mathrm{1}}} \HOLBoundVar{E\sb{\mathrm{2}}}) \HOLSymConst{\HOLTokenConj{}}
     (\HOLSymConst{\HOLTokenForall{}}\HOLBoundVar{E\sb{\mathrm{1}}}.
        \HOLBoundVar{E} \HOLTokenTransBegin\HOLConst{\ensuremath{\tau}}\HOLTokenTransEnd \HOLBoundVar{E\sb{\mathrm{1}}} \HOLSymConst{\HOLTokenImp{}}
        \HOLBoundVar{E\sb{\mathrm{1}}} \HOLConst{contracts} \HOLBoundVar{E\sp{\prime}} \HOLSymConst{\HOLTokenDisj{}} \HOLSymConst{\HOLTokenExists{}}\HOLBoundVar{E\sb{\mathrm{2}}}. \HOLBoundVar{E\sp{\prime}} \HOLTokenTransBegin\HOLConst{\ensuremath{\tau}}\HOLTokenTransEnd \HOLBoundVar{E\sb{\mathrm{2}}} \HOLSymConst{\HOLTokenConj{}} \HOLBoundVar{E\sb{\mathrm{1}}} \HOLConst{contracts} \HOLBoundVar{E\sb{\mathrm{2}}}) \HOLSymConst{\HOLTokenConj{}}
     (\HOLSymConst{\HOLTokenForall{}}\HOLBoundVar{E\sb{\mathrm{2}}}. \HOLBoundVar{E\sp{\prime}} \HOLTokenTransBegin\HOLConst{\ensuremath{\tau}}\HOLTokenTransEnd \HOLBoundVar{E\sb{\mathrm{2}}} \HOLSymConst{\HOLTokenImp{}} \HOLSymConst{\HOLTokenExists{}}\HOLBoundVar{E\sb{\mathrm{1}}}. \HOLConst{EPS} \HOLBoundVar{E} \HOLBoundVar{E\sb{\mathrm{1}}} \HOLSymConst{\HOLTokenConj{}} \HOLConst{WEAK_EQUIV} \HOLBoundVar{E\sb{\mathrm{1}}} \HOLBoundVar{E\sb{\mathrm{2}}}) \HOLSymConst{\HOLTokenImp{}}
     \HOLBoundVar{E} \HOLConst{contracts} \HOLBoundVar{E\sp{\prime}}
\end{SaveVerbatim}
\newcommand{\HOLContractionTheoremscontractsXXrules}{\UseVerbatim{HOLContractionTheoremscontractsXXrules}}
\begin{SaveVerbatim}{HOLContractionTheoremscontractsXXSUBSTXXGCONTEXT}
\HOLTokenTurnstile{} \HOLSymConst{\HOLTokenForall{}}\HOLBoundVar{P} \HOLBoundVar{Q}. \HOLBoundVar{P} \HOLConst{contracts} \HOLBoundVar{Q} \HOLSymConst{\HOLTokenImp{}} \HOLSymConst{\HOLTokenForall{}}\HOLBoundVar{E}. \HOLConst{GCONTEXT} \HOLBoundVar{E} \HOLSymConst{\HOLTokenImp{}} \HOLBoundVar{E} \HOLBoundVar{P} \HOLConst{contracts} \HOLBoundVar{E} \HOLBoundVar{Q}
\end{SaveVerbatim}
\newcommand{\HOLContractionTheoremscontractsXXSUBSTXXGCONTEXT}{\UseVerbatim{HOLContractionTheoremscontractsXXSUBSTXXGCONTEXT}}
\begin{SaveVerbatim}{HOLContractionTheoremscontractsXXSUBSTXXPARXXL}
\HOLTokenTurnstile{} \HOLSymConst{\HOLTokenForall{}}\HOLBoundVar{E} \HOLBoundVar{E\sp{\prime}}. \HOLBoundVar{E} \HOLConst{contracts} \HOLBoundVar{E\sp{\prime}} \HOLSymConst{\HOLTokenImp{}} \HOLSymConst{\HOLTokenForall{}}\HOLBoundVar{E\sp{\prime\prime}}. \HOLBoundVar{E\sp{\prime\prime}} \HOLSymConst{\ensuremath{\parallel}} \HOLBoundVar{E} \HOLConst{contracts} \HOLBoundVar{E\sp{\prime\prime}} \HOLSymConst{\ensuremath{\parallel}} \HOLBoundVar{E\sp{\prime}}
\end{SaveVerbatim}
\newcommand{\HOLContractionTheoremscontractsXXSUBSTXXPARXXL}{\UseVerbatim{HOLContractionTheoremscontractsXXSUBSTXXPARXXL}}
\begin{SaveVerbatim}{HOLContractionTheoremscontractsXXSUBSTXXPARXXR}
\HOLTokenTurnstile{} \HOLSymConst{\HOLTokenForall{}}\HOLBoundVar{E} \HOLBoundVar{E\sp{\prime}}. \HOLBoundVar{E} \HOLConst{contracts} \HOLBoundVar{E\sp{\prime}} \HOLSymConst{\HOLTokenImp{}} \HOLSymConst{\HOLTokenForall{}}\HOLBoundVar{E\sp{\prime\prime}}. \HOLBoundVar{E} \HOLSymConst{\ensuremath{\parallel}} \HOLBoundVar{E\sp{\prime\prime}} \HOLConst{contracts} \HOLBoundVar{E\sp{\prime}} \HOLSymConst{\ensuremath{\parallel}} \HOLBoundVar{E\sp{\prime\prime}}
\end{SaveVerbatim}
\newcommand{\HOLContractionTheoremscontractsXXSUBSTXXPARXXR}{\UseVerbatim{HOLContractionTheoremscontractsXXSUBSTXXPARXXR}}
\begin{SaveVerbatim}{HOLContractionTheoremscontractsXXSUBSTXXPREFIX}
\HOLTokenTurnstile{} \HOLSymConst{\HOLTokenForall{}}\HOLBoundVar{E} \HOLBoundVar{E\sp{\prime}}. \HOLBoundVar{E} \HOLConst{contracts} \HOLBoundVar{E\sp{\prime}} \HOLSymConst{\HOLTokenImp{}} \HOLSymConst{\HOLTokenForall{}}\HOLBoundVar{u}. \HOLBoundVar{u}\HOLSymConst{..}\HOLBoundVar{E} \HOLConst{contracts} \HOLBoundVar{u}\HOLSymConst{..}\HOLBoundVar{E\sp{\prime}}
\end{SaveVerbatim}
\newcommand{\HOLContractionTheoremscontractsXXSUBSTXXPREFIX}{\UseVerbatim{HOLContractionTheoremscontractsXXSUBSTXXPREFIX}}
\begin{SaveVerbatim}{HOLContractionTheoremscontractsXXSUBSTXXRELAB}
\HOLTokenTurnstile{} \HOLSymConst{\HOLTokenForall{}}\HOLBoundVar{E} \HOLBoundVar{E\sp{\prime}}. \HOLBoundVar{E} \HOLConst{contracts} \HOLBoundVar{E\sp{\prime}} \HOLSymConst{\HOLTokenImp{}} \HOLSymConst{\HOLTokenForall{}}\HOLBoundVar{rf}. \HOLConst{relab} \HOLBoundVar{E} \HOLBoundVar{rf} \HOLConst{contracts} \HOLConst{relab} \HOLBoundVar{E\sp{\prime}} \HOLBoundVar{rf}
\end{SaveVerbatim}
\newcommand{\HOLContractionTheoremscontractsXXSUBSTXXRELAB}{\UseVerbatim{HOLContractionTheoremscontractsXXSUBSTXXRELAB}}
\begin{SaveVerbatim}{HOLContractionTheoremscontractsXXSUBSTXXRESTR}
\HOLTokenTurnstile{} \HOLSymConst{\HOLTokenForall{}}\HOLBoundVar{E} \HOLBoundVar{E\sp{\prime}}. \HOLBoundVar{E} \HOLConst{contracts} \HOLBoundVar{E\sp{\prime}} \HOLSymConst{\HOLTokenImp{}} \HOLSymConst{\HOLTokenForall{}}\HOLBoundVar{L}. \HOLConst{\ensuremath{\nu}} \HOLBoundVar{L} \HOLBoundVar{E} \HOLConst{contracts} \HOLConst{\ensuremath{\nu}} \HOLBoundVar{L} \HOLBoundVar{E\sp{\prime}}
\end{SaveVerbatim}
\newcommand{\HOLContractionTheoremscontractsXXSUBSTXXRESTR}{\UseVerbatim{HOLContractionTheoremscontractsXXSUBSTXXRESTR}}
\begin{SaveVerbatim}{HOLContractionTheoremscontractsXXthm}
\HOLTokenTurnstile{} \HOLSymConst{\HOLTokenForall{}}\HOLBoundVar{P} \HOLBoundVar{Q}. \HOLBoundVar{P} \HOLConst{contracts} \HOLBoundVar{Q} \HOLSymConst{\HOLTokenEquiv{}} \HOLSymConst{\HOLTokenExists{}}\HOLBoundVar{Con}. \HOLBoundVar{Con} \HOLBoundVar{P} \HOLBoundVar{Q} \HOLSymConst{\HOLTokenConj{}} \HOLConst{CONTRACTION} \HOLBoundVar{Con}
\end{SaveVerbatim}
\newcommand{\HOLContractionTheoremscontractsXXthm}{\UseVerbatim{HOLContractionTheoremscontractsXXthm}}
\begin{SaveVerbatim}{HOLContractionTheoremscontractsXXTRANS}
\HOLTokenTurnstile{} \HOLSymConst{\HOLTokenForall{}}\HOLBoundVar{x} \HOLBoundVar{y} \HOLBoundVar{z}. \HOLBoundVar{x} \HOLConst{contracts} \HOLBoundVar{y} \HOLSymConst{\HOLTokenConj{}} \HOLBoundVar{y} \HOLConst{contracts} \HOLBoundVar{z} \HOLSymConst{\HOLTokenImp{}} \HOLBoundVar{x} \HOLConst{contracts} \HOLBoundVar{z}
\end{SaveVerbatim}
\newcommand{\HOLContractionTheoremscontractsXXTRANS}{\UseVerbatim{HOLContractionTheoremscontractsXXTRANS}}
\begin{SaveVerbatim}{HOLContractionTheoremscontractsXXTRANSXXlabel}
\HOLTokenTurnstile{} \HOLSymConst{\HOLTokenForall{}}\HOLBoundVar{E} \HOLBoundVar{E\sp{\prime}}.
     \HOLBoundVar{E} \HOLConst{contracts} \HOLBoundVar{E\sp{\prime}} \HOLSymConst{\HOLTokenImp{}}
     \HOLSymConst{\HOLTokenForall{}}\HOLBoundVar{l} \HOLBoundVar{E\sb{\mathrm{1}}}.
       \HOLBoundVar{E} \HOLTokenTransBegin\HOLConst{label} \HOLBoundVar{l}\HOLTokenTransEnd \HOLBoundVar{E\sb{\mathrm{1}}} \HOLSymConst{\HOLTokenImp{}} \HOLSymConst{\HOLTokenExists{}}\HOLBoundVar{E\sb{\mathrm{2}}}. \HOLBoundVar{E\sp{\prime}} \HOLTokenTransBegin\HOLConst{label} \HOLBoundVar{l}\HOLTokenTransEnd \HOLBoundVar{E\sb{\mathrm{2}}} \HOLSymConst{\HOLTokenConj{}} \HOLBoundVar{E\sb{\mathrm{1}}} \HOLConst{contracts} \HOLBoundVar{E\sb{\mathrm{2}}}
\end{SaveVerbatim}
\newcommand{\HOLContractionTheoremscontractsXXTRANSXXlabel}{\UseVerbatim{HOLContractionTheoremscontractsXXTRANSXXlabel}}
\begin{SaveVerbatim}{HOLContractionTheoremscontractsXXTRANSXXlabelYY}
\HOLTokenTurnstile{} \HOLSymConst{\HOLTokenForall{}}\HOLBoundVar{E} \HOLBoundVar{E\sp{\prime}}.
     \HOLBoundVar{E} \HOLConst{contracts} \HOLBoundVar{E\sp{\prime}} \HOLSymConst{\HOLTokenImp{}}
     \HOLSymConst{\HOLTokenForall{}}\HOLBoundVar{l} \HOLBoundVar{E\sb{\mathrm{2}}}.
       \HOLBoundVar{E\sp{\prime}} \HOLTokenTransBegin\HOLConst{label} \HOLBoundVar{l}\HOLTokenTransEnd \HOLBoundVar{E\sb{\mathrm{2}}} \HOLSymConst{\HOLTokenImp{}} \HOLSymConst{\HOLTokenExists{}}\HOLBoundVar{E\sb{\mathrm{1}}}. \HOLBoundVar{E} \HOLTokenWeakTransBegin\HOLConst{label} \HOLBoundVar{l}\HOLTokenWeakTransEnd \HOLBoundVar{E\sb{\mathrm{1}}} \HOLSymConst{\HOLTokenConj{}} \HOLConst{WEAK_EQUIV} \HOLBoundVar{E\sb{\mathrm{1}}} \HOLBoundVar{E\sb{\mathrm{2}}}
\end{SaveVerbatim}
\newcommand{\HOLContractionTheoremscontractsXXTRANSXXlabelYY}{\UseVerbatim{HOLContractionTheoremscontractsXXTRANSXXlabelYY}}
\begin{SaveVerbatim}{HOLContractionTheoremscontractsXXTRANSXXtau}
\HOLTokenTurnstile{} \HOLSymConst{\HOLTokenForall{}}\HOLBoundVar{E} \HOLBoundVar{E\sp{\prime}}.
     \HOLBoundVar{E} \HOLConst{contracts} \HOLBoundVar{E\sp{\prime}} \HOLSymConst{\HOLTokenImp{}}
     \HOLSymConst{\HOLTokenForall{}}\HOLBoundVar{E\sb{\mathrm{1}}}.
       \HOLBoundVar{E} \HOLTokenTransBegin\HOLConst{\ensuremath{\tau}}\HOLTokenTransEnd \HOLBoundVar{E\sb{\mathrm{1}}} \HOLSymConst{\HOLTokenImp{}}
       \HOLBoundVar{E\sb{\mathrm{1}}} \HOLConst{contracts} \HOLBoundVar{E\sp{\prime}} \HOLSymConst{\HOLTokenDisj{}} \HOLSymConst{\HOLTokenExists{}}\HOLBoundVar{E\sb{\mathrm{2}}}. \HOLBoundVar{E\sp{\prime}} \HOLTokenTransBegin\HOLConst{\ensuremath{\tau}}\HOLTokenTransEnd \HOLBoundVar{E\sb{\mathrm{2}}} \HOLSymConst{\HOLTokenConj{}} \HOLBoundVar{E\sb{\mathrm{1}}} \HOLConst{contracts} \HOLBoundVar{E\sb{\mathrm{2}}}
\end{SaveVerbatim}
\newcommand{\HOLContractionTheoremscontractsXXTRANSXXtau}{\UseVerbatim{HOLContractionTheoremscontractsXXTRANSXXtau}}
\begin{SaveVerbatim}{HOLContractionTheoremscontractsXXTRANSXXtauYY}
\HOLTokenTurnstile{} \HOLSymConst{\HOLTokenForall{}}\HOLBoundVar{E} \HOLBoundVar{E\sp{\prime}}.
     \HOLBoundVar{E} \HOLConst{contracts} \HOLBoundVar{E\sp{\prime}} \HOLSymConst{\HOLTokenImp{}}
     \HOLSymConst{\HOLTokenForall{}}\HOLBoundVar{E\sb{\mathrm{2}}}. \HOLBoundVar{E\sp{\prime}} \HOLTokenTransBegin\HOLConst{\ensuremath{\tau}}\HOLTokenTransEnd \HOLBoundVar{E\sb{\mathrm{2}}} \HOLSymConst{\HOLTokenImp{}} \HOLSymConst{\HOLTokenExists{}}\HOLBoundVar{E\sb{\mathrm{1}}}. \HOLConst{EPS} \HOLBoundVar{E} \HOLBoundVar{E\sb{\mathrm{1}}} \HOLSymConst{\HOLTokenConj{}} \HOLConst{WEAK_EQUIV} \HOLBoundVar{E\sb{\mathrm{1}}} \HOLBoundVar{E\sb{\mathrm{2}}}
\end{SaveVerbatim}
\newcommand{\HOLContractionTheoremscontractsXXTRANSXXtauYY}{\UseVerbatim{HOLContractionTheoremscontractsXXTRANSXXtauYY}}
\begin{SaveVerbatim}{HOLContractionTheoremscontractsXXtransitive}
\HOLTokenTurnstile{} \HOLConst{transitive} (\HOLConst{contracts})
\end{SaveVerbatim}
\newcommand{\HOLContractionTheoremscontractsXXtransitive}{\UseVerbatim{HOLContractionTheoremscontractsXXtransitive}}
\begin{SaveVerbatim}{HOLContractionTheoremscontractsXXWEAKXXTRANSXXlabel}
\HOLTokenTurnstile{} \HOLSymConst{\HOLTokenForall{}}\HOLBoundVar{E} \HOLBoundVar{E\sp{\prime}}.
     \HOLBoundVar{E} \HOLConst{contracts} \HOLBoundVar{E\sp{\prime}} \HOLSymConst{\HOLTokenImp{}}
     \HOLSymConst{\HOLTokenForall{}}\HOLBoundVar{l} \HOLBoundVar{E\sb{\mathrm{1}}}.
       \HOLBoundVar{E} \HOLTokenWeakTransBegin\HOLConst{label} \HOLBoundVar{l}\HOLTokenWeakTransEnd \HOLBoundVar{E\sb{\mathrm{1}}} \HOLSymConst{\HOLTokenImp{}} \HOLSymConst{\HOLTokenExists{}}\HOLBoundVar{E\sb{\mathrm{2}}}. \HOLBoundVar{E\sp{\prime}} \HOLTokenWeakTransBegin\HOLConst{label} \HOLBoundVar{l}\HOLTokenWeakTransEnd \HOLBoundVar{E\sb{\mathrm{2}}} \HOLSymConst{\HOLTokenConj{}} \HOLBoundVar{E\sb{\mathrm{1}}} \HOLConst{contracts} \HOLBoundVar{E\sb{\mathrm{2}}}
\end{SaveVerbatim}
\newcommand{\HOLContractionTheoremscontractsXXWEAKXXTRANSXXlabel}{\UseVerbatim{HOLContractionTheoremscontractsXXWEAKXXTRANSXXlabel}}
\begin{SaveVerbatim}{HOLContractionTheoremscontractsXXWEAKXXTRANSXXlabelYY}
\HOLTokenTurnstile{} \HOLSymConst{\HOLTokenForall{}}\HOLBoundVar{E} \HOLBoundVar{E\sp{\prime}}.
     \HOLBoundVar{E} \HOLConst{contracts} \HOLBoundVar{E\sp{\prime}} \HOLSymConst{\HOLTokenImp{}}
     \HOLSymConst{\HOLTokenForall{}}\HOLBoundVar{l} \HOLBoundVar{E\sb{\mathrm{2}}}.
       \HOLBoundVar{E\sp{\prime}} \HOLTokenWeakTransBegin\HOLConst{label} \HOLBoundVar{l}\HOLTokenWeakTransEnd \HOLBoundVar{E\sb{\mathrm{2}}} \HOLSymConst{\HOLTokenImp{}} \HOLSymConst{\HOLTokenExists{}}\HOLBoundVar{E\sb{\mathrm{1}}}. \HOLBoundVar{E} \HOLTokenWeakTransBegin\HOLConst{label} \HOLBoundVar{l}\HOLTokenWeakTransEnd \HOLBoundVar{E\sb{\mathrm{1}}} \HOLSymConst{\HOLTokenConj{}} \HOLConst{WEAK_EQUIV} \HOLBoundVar{E\sb{\mathrm{1}}} \HOLBoundVar{E\sb{\mathrm{2}}}
\end{SaveVerbatim}
\newcommand{\HOLContractionTheoremscontractsXXWEAKXXTRANSXXlabelYY}{\UseVerbatim{HOLContractionTheoremscontractsXXWEAKXXTRANSXXlabelYY}}
\begin{SaveVerbatim}{HOLContractionTheoremscontractsXXWEAKXXTRANSXXtau}
\HOLTokenTurnstile{} \HOLSymConst{\HOLTokenForall{}}\HOLBoundVar{E} \HOLBoundVar{E\sp{\prime}}.
     \HOLBoundVar{E} \HOLConst{contracts} \HOLBoundVar{E\sp{\prime}} \HOLSymConst{\HOLTokenImp{}}
     \HOLSymConst{\HOLTokenForall{}}\HOLBoundVar{E\sb{\mathrm{1}}}. \HOLBoundVar{E} \HOLTokenWeakTransBegin\HOLConst{\ensuremath{\tau}}\HOLTokenWeakTransEnd \HOLBoundVar{E\sb{\mathrm{1}}} \HOLSymConst{\HOLTokenImp{}} \HOLSymConst{\HOLTokenExists{}}\HOLBoundVar{E\sb{\mathrm{2}}}. \HOLConst{EPS} \HOLBoundVar{E\sp{\prime}} \HOLBoundVar{E\sb{\mathrm{2}}} \HOLSymConst{\HOLTokenConj{}} \HOLBoundVar{E\sb{\mathrm{1}}} \HOLConst{contracts} \HOLBoundVar{E\sb{\mathrm{2}}}
\end{SaveVerbatim}
\newcommand{\HOLContractionTheoremscontractsXXWEAKXXTRANSXXtau}{\UseVerbatim{HOLContractionTheoremscontractsXXWEAKXXTRANSXXtau}}
\begin{SaveVerbatim}{HOLContractionTheoremscontractsXXWEAKXXTRANSXXtauYY}
\HOLTokenTurnstile{} \HOLSymConst{\HOLTokenForall{}}\HOLBoundVar{E} \HOLBoundVar{E\sp{\prime}}.
     \HOLBoundVar{E} \HOLConst{contracts} \HOLBoundVar{E\sp{\prime}} \HOLSymConst{\HOLTokenImp{}}
     \HOLSymConst{\HOLTokenForall{}}\HOLBoundVar{l} \HOLBoundVar{E\sb{\mathrm{2}}}. \HOLBoundVar{E\sp{\prime}} \HOLTokenWeakTransBegin\HOLConst{\ensuremath{\tau}}\HOLTokenWeakTransEnd \HOLBoundVar{E\sb{\mathrm{2}}} \HOLSymConst{\HOLTokenImp{}} \HOLSymConst{\HOLTokenExists{}}\HOLBoundVar{E\sb{\mathrm{1}}}. \HOLConst{EPS} \HOLBoundVar{E} \HOLBoundVar{E\sb{\mathrm{1}}} \HOLSymConst{\HOLTokenConj{}} \HOLConst{WEAK_EQUIV} \HOLBoundVar{E\sb{\mathrm{1}}} \HOLBoundVar{E\sb{\mathrm{2}}}
\end{SaveVerbatim}
\newcommand{\HOLContractionTheoremscontractsXXWEAKXXTRANSXXtauYY}{\UseVerbatim{HOLContractionTheoremscontractsXXWEAKXXTRANSXXtauYY}}
\begin{SaveVerbatim}{HOLContractionTheoremsexpandsXXIMPXXcontracts}
\HOLTokenTurnstile{} \HOLSymConst{\HOLTokenForall{}}\HOLBoundVar{P} \HOLBoundVar{Q}. \HOLBoundVar{P} \HOLConst{expands} \HOLBoundVar{Q} \HOLSymConst{\HOLTokenImp{}} \HOLBoundVar{P} \HOLConst{contracts} \HOLBoundVar{Q}
\end{SaveVerbatim}
\newcommand{\HOLContractionTheoremsexpandsXXIMPXXcontracts}{\UseVerbatim{HOLContractionTheoremsexpandsXXIMPXXcontracts}}
\begin{SaveVerbatim}{HOLContractionTheoremsEXPANSIONXXIMPXXCONTRACTION}
\HOLTokenTurnstile{} \HOLSymConst{\HOLTokenForall{}}\HOLBoundVar{Con}. \HOLConst{EXPANSION} \HOLBoundVar{Con} \HOLSymConst{\HOLTokenImp{}} \HOLConst{CONTRACTION} \HOLBoundVar{Con}
\end{SaveVerbatim}
\newcommand{\HOLContractionTheoremsEXPANSIONXXIMPXXCONTRACTION}{\UseVerbatim{HOLContractionTheoremsEXPANSIONXXIMPXXCONTRACTION}}
\begin{SaveVerbatim}{HOLContractionTheoremsIDENTITYXXCONTRACTION}
\HOLTokenTurnstile{} \HOLConst{CONTRACTION} (\HOLSymConst{=})
\end{SaveVerbatim}
\newcommand{\HOLContractionTheoremsIDENTITYXXCONTRACTION}{\UseVerbatim{HOLContractionTheoremsIDENTITYXXCONTRACTION}}
\begin{SaveVerbatim}{HOLContractionTheoremsOBSXXcontractsXXANDXXTRACEXXlabel}
\HOLTokenTurnstile{} \HOLSymConst{\HOLTokenForall{}}\HOLBoundVar{E} \HOLBoundVar{E\sp{\prime}}.
     \HOLConst{OBS_contracts} \HOLBoundVar{E} \HOLBoundVar{E\sp{\prime}} \HOLSymConst{\HOLTokenImp{}}
     \HOLSymConst{\HOLTokenForall{}}\HOLBoundVar{xs} \HOLBoundVar{l} \HOLBoundVar{E\sb{\mathrm{1}}}.
       \HOLConst{TRACE} \HOLBoundVar{E} \HOLBoundVar{xs} \HOLBoundVar{E\sb{\mathrm{1}}} \HOLSymConst{\HOLTokenConj{}} \HOLConst{UNIQUE_LABEL} (\HOLConst{label} \HOLBoundVar{l}) \HOLBoundVar{xs} \HOLSymConst{\HOLTokenImp{}}
       \HOLSymConst{\HOLTokenExists{}}\HOLBoundVar{xs\sp{\prime}} \HOLBoundVar{E\sb{\mathrm{2}}}.
         \HOLConst{TRACE} \HOLBoundVar{E\sp{\prime}} \HOLBoundVar{xs\sp{\prime}} \HOLBoundVar{E\sb{\mathrm{2}}} \HOLSymConst{\HOLTokenConj{}} \HOLBoundVar{E\sb{\mathrm{1}}} \HOLConst{contracts} \HOLBoundVar{E\sb{\mathrm{2}}} \HOLSymConst{\HOLTokenConj{}}
         \HOLConst{LENGTH} \HOLBoundVar{xs\sp{\prime}} \HOLSymConst{\HOLTokenLeq{}} \HOLConst{LENGTH} \HOLBoundVar{xs} \HOLSymConst{\HOLTokenConj{}} \HOLConst{UNIQUE_LABEL} (\HOLConst{label} \HOLBoundVar{l}) \HOLBoundVar{xs\sp{\prime}}
\end{SaveVerbatim}
\newcommand{\HOLContractionTheoremsOBSXXcontractsXXANDXXTRACEXXlabel}{\UseVerbatim{HOLContractionTheoremsOBSXXcontractsXXANDXXTRACEXXlabel}}
\begin{SaveVerbatim}{HOLContractionTheoremsOBSXXcontractsXXANDXXTRACEXXtau}
\HOLTokenTurnstile{} \HOLSymConst{\HOLTokenForall{}}\HOLBoundVar{E} \HOLBoundVar{E\sp{\prime}}.
     \HOLConst{OBS_contracts} \HOLBoundVar{E} \HOLBoundVar{E\sp{\prime}} \HOLSymConst{\HOLTokenImp{}}
     \HOLSymConst{\HOLTokenForall{}}\HOLBoundVar{xs} \HOLBoundVar{l} \HOLBoundVar{E\sb{\mathrm{1}}}.
       \HOLConst{TRACE} \HOLBoundVar{E} \HOLBoundVar{xs} \HOLBoundVar{E\sb{\mathrm{1}}} \HOLSymConst{\HOLTokenConj{}} \HOLConst{NO_LABEL} \HOLBoundVar{xs} \HOLSymConst{\HOLTokenImp{}}
       \HOLSymConst{\HOLTokenExists{}}\HOLBoundVar{xs\sp{\prime}} \HOLBoundVar{E\sb{\mathrm{2}}}.
         \HOLConst{TRACE} \HOLBoundVar{E\sp{\prime}} \HOLBoundVar{xs\sp{\prime}} \HOLBoundVar{E\sb{\mathrm{2}}} \HOLSymConst{\HOLTokenConj{}} \HOLBoundVar{E\sb{\mathrm{1}}} \HOLConst{contracts} \HOLBoundVar{E\sb{\mathrm{2}}} \HOLSymConst{\HOLTokenConj{}}
         \HOLConst{LENGTH} \HOLBoundVar{xs\sp{\prime}} \HOLSymConst{\HOLTokenLeq{}} \HOLConst{LENGTH} \HOLBoundVar{xs} \HOLSymConst{\HOLTokenConj{}} \HOLConst{NO_LABEL} \HOLBoundVar{xs\sp{\prime}}
\end{SaveVerbatim}
\newcommand{\HOLContractionTheoremsOBSXXcontractsXXANDXXTRACEXXtau}{\UseVerbatim{HOLContractionTheoremsOBSXXcontractsXXANDXXTRACEXXtau}}
\begin{SaveVerbatim}{HOLContractionTheoremsOBSXXcontractsXXBYXXCONTRACTION}
\HOLTokenTurnstile{} \HOLSymConst{\HOLTokenForall{}}\HOLBoundVar{Con}.
     \HOLConst{CONTRACTION} \HOLBoundVar{Con} \HOLSymConst{\HOLTokenImp{}}
     \HOLSymConst{\HOLTokenForall{}}\HOLBoundVar{E} \HOLBoundVar{E\sp{\prime}}.
       (\HOLSymConst{\HOLTokenForall{}}\HOLBoundVar{u}.
          (\HOLSymConst{\HOLTokenForall{}}\HOLBoundVar{E\sb{\mathrm{1}}}. \HOLBoundVar{E} \HOLTokenTransBegin\HOLBoundVar{u}\HOLTokenTransEnd \HOLBoundVar{E\sb{\mathrm{1}}} \HOLSymConst{\HOLTokenImp{}} \HOLSymConst{\HOLTokenExists{}}\HOLBoundVar{E\sb{\mathrm{2}}}. \HOLBoundVar{E\sp{\prime}} \HOLTokenTransBegin\HOLBoundVar{u}\HOLTokenTransEnd \HOLBoundVar{E\sb{\mathrm{2}}} \HOLSymConst{\HOLTokenConj{}} \HOLBoundVar{Con} \HOLBoundVar{E\sb{\mathrm{1}}} \HOLBoundVar{E\sb{\mathrm{2}}}) \HOLSymConst{\HOLTokenConj{}}
          \HOLSymConst{\HOLTokenForall{}}\HOLBoundVar{E\sb{\mathrm{2}}}. \HOLBoundVar{E\sp{\prime}} \HOLTokenTransBegin\HOLBoundVar{u}\HOLTokenTransEnd \HOLBoundVar{E\sb{\mathrm{2}}} \HOLSymConst{\HOLTokenImp{}} \HOLSymConst{\HOLTokenExists{}}\HOLBoundVar{E\sb{\mathrm{1}}}. \HOLBoundVar{E} \HOLTokenWeakTransBegin\HOLBoundVar{u}\HOLTokenWeakTransEnd \HOLBoundVar{E\sb{\mathrm{1}}} \HOLSymConst{\HOLTokenConj{}} \HOLBoundVar{Con} \HOLBoundVar{E\sb{\mathrm{1}}} \HOLBoundVar{E\sb{\mathrm{2}}}) \HOLSymConst{\HOLTokenImp{}}
       \HOLConst{OBS_contracts} \HOLBoundVar{E} \HOLBoundVar{E\sp{\prime}}
\end{SaveVerbatim}
\newcommand{\HOLContractionTheoremsOBSXXcontractsXXBYXXCONTRACTION}{\UseVerbatim{HOLContractionTheoremsOBSXXcontractsXXBYXXCONTRACTION}}
\begin{SaveVerbatim}{HOLContractionTheoremsOBSXXcontractsXXEPSYY}
\HOLTokenTurnstile{} \HOLSymConst{\HOLTokenForall{}}\HOLBoundVar{E} \HOLBoundVar{E\sp{\prime}}.
     \HOLConst{OBS_contracts} \HOLBoundVar{E} \HOLBoundVar{E\sp{\prime}} \HOLSymConst{\HOLTokenImp{}}
     \HOLSymConst{\HOLTokenForall{}}\HOLBoundVar{E\sb{\mathrm{2}}}. \HOLConst{EPS} \HOLBoundVar{E\sp{\prime}} \HOLBoundVar{E\sb{\mathrm{2}}} \HOLSymConst{\HOLTokenImp{}} \HOLSymConst{\HOLTokenExists{}}\HOLBoundVar{E\sb{\mathrm{1}}}. \HOLConst{EPS} \HOLBoundVar{E} \HOLBoundVar{E\sb{\mathrm{1}}} \HOLSymConst{\HOLTokenConj{}} \HOLConst{WEAK_EQUIV} \HOLBoundVar{E\sb{\mathrm{1}}} \HOLBoundVar{E\sb{\mathrm{2}}}
\end{SaveVerbatim}
\newcommand{\HOLContractionTheoremsOBSXXcontractsXXEPSYY}{\UseVerbatim{HOLContractionTheoremsOBSXXcontractsXXEPSYY}}
\begin{SaveVerbatim}{HOLContractionTheoremsOBSXXcontractsXXIMPXXCXXcontracts}
\HOLTokenTurnstile{} \HOLSymConst{\HOLTokenForall{}}\HOLBoundVar{p} \HOLBoundVar{q}. \HOLConst{OBS_contracts} \HOLBoundVar{p} \HOLBoundVar{q} \HOLSymConst{\HOLTokenImp{}} \HOLConst{C_contracts} \HOLBoundVar{p} \HOLBoundVar{q}
\end{SaveVerbatim}
\newcommand{\HOLContractionTheoremsOBSXXcontractsXXIMPXXCXXcontracts}{\UseVerbatim{HOLContractionTheoremsOBSXXcontractsXXIMPXXCXXcontracts}}
\begin{SaveVerbatim}{HOLContractionTheoremsOBSXXcontractsXXIMPXXcontracts}
\HOLTokenTurnstile{} \HOLSymConst{\HOLTokenForall{}}\HOLBoundVar{E} \HOLBoundVar{E\sp{\prime}}. \HOLConst{OBS_contracts} \HOLBoundVar{E} \HOLBoundVar{E\sp{\prime}} \HOLSymConst{\HOLTokenImp{}} \HOLBoundVar{E} \HOLConst{contracts} \HOLBoundVar{E\sp{\prime}}
\end{SaveVerbatim}
\newcommand{\HOLContractionTheoremsOBSXXcontractsXXIMPXXcontracts}{\UseVerbatim{HOLContractionTheoremsOBSXXcontractsXXIMPXXcontracts}}
\begin{SaveVerbatim}{HOLContractionTheoremsOBSXXcontractsXXIMPXXOBSXXCONGR}
\HOLTokenTurnstile{} \HOLSymConst{\HOLTokenForall{}}\HOLBoundVar{E} \HOLBoundVar{E\sp{\prime}}. \HOLConst{OBS_contracts} \HOLBoundVar{E} \HOLBoundVar{E\sp{\prime}} \HOLSymConst{\HOLTokenImp{}} \HOLConst{OBS_CONGR} \HOLBoundVar{E} \HOLBoundVar{E\sp{\prime}}
\end{SaveVerbatim}
\newcommand{\HOLContractionTheoremsOBSXXcontractsXXIMPXXOBSXXCONGR}{\UseVerbatim{HOLContractionTheoremsOBSXXcontractsXXIMPXXOBSXXCONGR}}
\begin{SaveVerbatim}{HOLContractionTheoremsOBSXXcontractsXXIMPXXSUMXXcontracts}
\HOLTokenTurnstile{} \HOLSymConst{\HOLTokenForall{}}\HOLBoundVar{p} \HOLBoundVar{q}. \HOLConst{OBS_contracts} \HOLBoundVar{p} \HOLBoundVar{q} \HOLSymConst{\HOLTokenImp{}} \HOLConst{SUM_contracts} \HOLBoundVar{p} \HOLBoundVar{q}
\end{SaveVerbatim}
\newcommand{\HOLContractionTheoremsOBSXXcontractsXXIMPXXSUMXXcontracts}{\UseVerbatim{HOLContractionTheoremsOBSXXcontractsXXIMPXXSUMXXcontracts}}
\begin{SaveVerbatim}{HOLContractionTheoremsOBSXXcontractsXXIMPXXWEAKXXEQUIV}
\HOLTokenTurnstile{} \HOLSymConst{\HOLTokenForall{}}\HOLBoundVar{E} \HOLBoundVar{E\sp{\prime}}. \HOLConst{OBS_contracts} \HOLBoundVar{E} \HOLBoundVar{E\sp{\prime}} \HOLSymConst{\HOLTokenImp{}} \HOLConst{WEAK_EQUIV} \HOLBoundVar{E} \HOLBoundVar{E\sp{\prime}}
\end{SaveVerbatim}
\newcommand{\HOLContractionTheoremsOBSXXcontractsXXIMPXXWEAKXXEQUIV}{\UseVerbatim{HOLContractionTheoremsOBSXXcontractsXXIMPXXWEAKXXEQUIV}}
\begin{SaveVerbatim}{HOLContractionTheoremsOBSXXcontractsXXIMPXXWEAKXXEQUIVYY}
\HOLTokenTurnstile{} \HOLSymConst{\HOLTokenForall{}}\HOLBoundVar{E} \HOLBoundVar{E\sp{\prime}}. \HOLConst{OBS_contracts} \HOLBoundVar{E} \HOLBoundVar{E\sp{\prime}} \HOLSymConst{\HOLTokenImp{}} \HOLConst{WEAK_EQUIV} \HOLBoundVar{E} \HOLBoundVar{E\sp{\prime}}
\end{SaveVerbatim}
\newcommand{\HOLContractionTheoremsOBSXXcontractsXXIMPXXWEAKXXEQUIVYY}{\UseVerbatim{HOLContractionTheoremsOBSXXcontractsXXIMPXXWEAKXXEQUIVYY}}
\begin{SaveVerbatim}{HOLContractionTheoremsOBSXXcontractsXXprecongruence}
\HOLTokenTurnstile{} \HOLConst{precongruence} \HOLConst{OBS_contracts}
\end{SaveVerbatim}
\newcommand{\HOLContractionTheoremsOBSXXcontractsXXprecongruence}{\UseVerbatim{HOLContractionTheoremsOBSXXcontractsXXprecongruence}}
\begin{SaveVerbatim}{HOLContractionTheoremsOBSXXcontractsXXPreOrder}
\HOLTokenTurnstile{} \HOLConst{PreOrder} \HOLConst{OBS_contracts}
\end{SaveVerbatim}
\newcommand{\HOLContractionTheoremsOBSXXcontractsXXPreOrder}{\UseVerbatim{HOLContractionTheoremsOBSXXcontractsXXPreOrder}}
\begin{SaveVerbatim}{HOLContractionTheoremsOBSXXcontractsXXPRESDXXBYXXPAR}
\HOLTokenTurnstile{} \HOLSymConst{\HOLTokenForall{}}\HOLBoundVar{E\sb{\mathrm{1}}} \HOLBoundVar{E\sb{\mathrm{1}}\sp{\prime}} \HOLBoundVar{E\sb{\mathrm{2}}} \HOLBoundVar{E\sb{\mathrm{2}}\sp{\prime}}.
     \HOLConst{OBS_contracts} \HOLBoundVar{E\sb{\mathrm{1}}} \HOLBoundVar{E\sb{\mathrm{1}}\sp{\prime}} \HOLSymConst{\HOLTokenConj{}} \HOLConst{OBS_contracts} \HOLBoundVar{E\sb{\mathrm{2}}} \HOLBoundVar{E\sb{\mathrm{2}}\sp{\prime}} \HOLSymConst{\HOLTokenImp{}}
     \HOLConst{OBS_contracts} (\HOLBoundVar{E\sb{\mathrm{1}}} \HOLSymConst{\ensuremath{\parallel}} \HOLBoundVar{E\sb{\mathrm{2}}}) (\HOLBoundVar{E\sb{\mathrm{1}}\sp{\prime}} \HOLSymConst{\ensuremath{\parallel}} \HOLBoundVar{E\sb{\mathrm{2}}\sp{\prime}})
\end{SaveVerbatim}
\newcommand{\HOLContractionTheoremsOBSXXcontractsXXPRESDXXBYXXPAR}{\UseVerbatim{HOLContractionTheoremsOBSXXcontractsXXPRESDXXBYXXPAR}}
\begin{SaveVerbatim}{HOLContractionTheoremsOBSXXcontractsXXPRESDXXBYXXSUM}
\HOLTokenTurnstile{} \HOLSymConst{\HOLTokenForall{}}\HOLBoundVar{E\sb{\mathrm{1}}} \HOLBoundVar{E\sb{\mathrm{1}}\sp{\prime}} \HOLBoundVar{E\sb{\mathrm{2}}} \HOLBoundVar{E\sb{\mathrm{2}}\sp{\prime}}.
     \HOLConst{OBS_contracts} \HOLBoundVar{E\sb{\mathrm{1}}} \HOLBoundVar{E\sb{\mathrm{1}}\sp{\prime}} \HOLSymConst{\HOLTokenConj{}} \HOLConst{OBS_contracts} \HOLBoundVar{E\sb{\mathrm{2}}} \HOLBoundVar{E\sb{\mathrm{2}}\sp{\prime}} \HOLSymConst{\HOLTokenImp{}}
     \HOLConst{OBS_contracts} (\HOLBoundVar{E\sb{\mathrm{1}}} \HOLSymConst{+} \HOLBoundVar{E\sb{\mathrm{2}}}) (\HOLBoundVar{E\sb{\mathrm{1}}\sp{\prime}} \HOLSymConst{+} \HOLBoundVar{E\sb{\mathrm{2}}\sp{\prime}})
\end{SaveVerbatim}
\newcommand{\HOLContractionTheoremsOBSXXcontractsXXPRESDXXBYXXSUM}{\UseVerbatim{HOLContractionTheoremsOBSXXcontractsXXPRESDXXBYXXSUM}}
\begin{SaveVerbatim}{HOLContractionTheoremsOBSXXcontractsXXREFL}
\HOLTokenTurnstile{} \HOLSymConst{\HOLTokenForall{}}\HOLBoundVar{E}. \HOLConst{OBS_contracts} \HOLBoundVar{E} \HOLBoundVar{E}
\end{SaveVerbatim}
\newcommand{\HOLContractionTheoremsOBSXXcontractsXXREFL}{\UseVerbatim{HOLContractionTheoremsOBSXXcontractsXXREFL}}
\begin{SaveVerbatim}{HOLContractionTheoremsOBSXXcontractsXXSUBSTXXCONTEXT}
\HOLTokenTurnstile{} \HOLSymConst{\HOLTokenForall{}}\HOLBoundVar{P} \HOLBoundVar{Q}.
     \HOLConst{OBS_contracts} \HOLBoundVar{P} \HOLBoundVar{Q} \HOLSymConst{\HOLTokenImp{}}
     \HOLSymConst{\HOLTokenForall{}}\HOLBoundVar{E}. \HOLConst{CONTEXT} \HOLBoundVar{E} \HOLSymConst{\HOLTokenImp{}} \HOLConst{OBS_contracts} (\HOLBoundVar{E} \HOLBoundVar{P}) (\HOLBoundVar{E} \HOLBoundVar{Q})
\end{SaveVerbatim}
\newcommand{\HOLContractionTheoremsOBSXXcontractsXXSUBSTXXCONTEXT}{\UseVerbatim{HOLContractionTheoremsOBSXXcontractsXXSUBSTXXCONTEXT}}
\begin{SaveVerbatim}{HOLContractionTheoremsOBSXXcontractsXXSUBSTXXPARXXL}
\HOLTokenTurnstile{} \HOLSymConst{\HOLTokenForall{}}\HOLBoundVar{E} \HOLBoundVar{E\sp{\prime}}.
     \HOLConst{OBS_contracts} \HOLBoundVar{E} \HOLBoundVar{E\sp{\prime}} \HOLSymConst{\HOLTokenImp{}}
     \HOLSymConst{\HOLTokenForall{}}\HOLBoundVar{E\sp{\prime\prime}}. \HOLConst{OBS_contracts} (\HOLBoundVar{E\sp{\prime\prime}} \HOLSymConst{\ensuremath{\parallel}} \HOLBoundVar{E}) (\HOLBoundVar{E\sp{\prime\prime}} \HOLSymConst{\ensuremath{\parallel}} \HOLBoundVar{E\sp{\prime}})
\end{SaveVerbatim}
\newcommand{\HOLContractionTheoremsOBSXXcontractsXXSUBSTXXPARXXL}{\UseVerbatim{HOLContractionTheoremsOBSXXcontractsXXSUBSTXXPARXXL}}
\begin{SaveVerbatim}{HOLContractionTheoremsOBSXXcontractsXXSUBSTXXPARXXR}
\HOLTokenTurnstile{} \HOLSymConst{\HOLTokenForall{}}\HOLBoundVar{E} \HOLBoundVar{E\sp{\prime}}.
     \HOLConst{OBS_contracts} \HOLBoundVar{E} \HOLBoundVar{E\sp{\prime}} \HOLSymConst{\HOLTokenImp{}}
     \HOLSymConst{\HOLTokenForall{}}\HOLBoundVar{E\sp{\prime\prime}}. \HOLConst{OBS_contracts} (\HOLBoundVar{E} \HOLSymConst{\ensuremath{\parallel}} \HOLBoundVar{E\sp{\prime\prime}}) (\HOLBoundVar{E\sp{\prime}} \HOLSymConst{\ensuremath{\parallel}} \HOLBoundVar{E\sp{\prime\prime}})
\end{SaveVerbatim}
\newcommand{\HOLContractionTheoremsOBSXXcontractsXXSUBSTXXPARXXR}{\UseVerbatim{HOLContractionTheoremsOBSXXcontractsXXSUBSTXXPARXXR}}
\begin{SaveVerbatim}{HOLContractionTheoremsOBSXXcontractsXXSUBSTXXPREFIX}
\HOLTokenTurnstile{} \HOLSymConst{\HOLTokenForall{}}\HOLBoundVar{E} \HOLBoundVar{E\sp{\prime}}. \HOLConst{OBS_contracts} \HOLBoundVar{E} \HOLBoundVar{E\sp{\prime}} \HOLSymConst{\HOLTokenImp{}} \HOLSymConst{\HOLTokenForall{}}\HOLBoundVar{u}. \HOLConst{OBS_contracts} (\HOLBoundVar{u}\HOLSymConst{..}\HOLBoundVar{E}) (\HOLBoundVar{u}\HOLSymConst{..}\HOLBoundVar{E\sp{\prime}})
\end{SaveVerbatim}
\newcommand{\HOLContractionTheoremsOBSXXcontractsXXSUBSTXXPREFIX}{\UseVerbatim{HOLContractionTheoremsOBSXXcontractsXXSUBSTXXPREFIX}}
\begin{SaveVerbatim}{HOLContractionTheoremsOBSXXcontractsXXSUBSTXXRELAB}
\HOLTokenTurnstile{} \HOLSymConst{\HOLTokenForall{}}\HOLBoundVar{E} \HOLBoundVar{E\sp{\prime}}.
     \HOLConst{OBS_contracts} \HOLBoundVar{E} \HOLBoundVar{E\sp{\prime}} \HOLSymConst{\HOLTokenImp{}}
     \HOLSymConst{\HOLTokenForall{}}\HOLBoundVar{rf}. \HOLConst{OBS_contracts} (\HOLConst{relab} \HOLBoundVar{E} \HOLBoundVar{rf}) (\HOLConst{relab} \HOLBoundVar{E\sp{\prime}} \HOLBoundVar{rf})
\end{SaveVerbatim}
\newcommand{\HOLContractionTheoremsOBSXXcontractsXXSUBSTXXRELAB}{\UseVerbatim{HOLContractionTheoremsOBSXXcontractsXXSUBSTXXRELAB}}
\begin{SaveVerbatim}{HOLContractionTheoremsOBSXXcontractsXXSUBSTXXRESTR}
\HOLTokenTurnstile{} \HOLSymConst{\HOLTokenForall{}}\HOLBoundVar{E} \HOLBoundVar{E\sp{\prime}}.
     \HOLConst{OBS_contracts} \HOLBoundVar{E} \HOLBoundVar{E\sp{\prime}} \HOLSymConst{\HOLTokenImp{}} \HOLSymConst{\HOLTokenForall{}}\HOLBoundVar{L}. \HOLConst{OBS_contracts} (\HOLConst{\ensuremath{\nu}} \HOLBoundVar{L} \HOLBoundVar{E}) (\HOLConst{\ensuremath{\nu}} \HOLBoundVar{L} \HOLBoundVar{E\sp{\prime}})
\end{SaveVerbatim}
\newcommand{\HOLContractionTheoremsOBSXXcontractsXXSUBSTXXRESTR}{\UseVerbatim{HOLContractionTheoremsOBSXXcontractsXXSUBSTXXRESTR}}
\begin{SaveVerbatim}{HOLContractionTheoremsOBSXXcontractsXXSUBSTXXSUMXXL}
\HOLTokenTurnstile{} \HOLSymConst{\HOLTokenForall{}}\HOLBoundVar{E} \HOLBoundVar{E\sp{\prime}}.
     \HOLConst{OBS_contracts} \HOLBoundVar{E} \HOLBoundVar{E\sp{\prime}} \HOLSymConst{\HOLTokenImp{}}
     \HOLSymConst{\HOLTokenForall{}}\HOLBoundVar{E\sp{\prime\prime}}. \HOLConst{OBS_contracts} (\HOLBoundVar{E\sp{\prime\prime}} \HOLSymConst{+} \HOLBoundVar{E}) (\HOLBoundVar{E\sp{\prime\prime}} \HOLSymConst{+} \HOLBoundVar{E\sp{\prime}})
\end{SaveVerbatim}
\newcommand{\HOLContractionTheoremsOBSXXcontractsXXSUBSTXXSUMXXL}{\UseVerbatim{HOLContractionTheoremsOBSXXcontractsXXSUBSTXXSUMXXL}}
\begin{SaveVerbatim}{HOLContractionTheoremsOBSXXcontractsXXSUBSTXXSUMXXR}
\HOLTokenTurnstile{} \HOLSymConst{\HOLTokenForall{}}\HOLBoundVar{E} \HOLBoundVar{E\sp{\prime}}.
     \HOLConst{OBS_contracts} \HOLBoundVar{E} \HOLBoundVar{E\sp{\prime}} \HOLSymConst{\HOLTokenImp{}}
     \HOLSymConst{\HOLTokenForall{}}\HOLBoundVar{E\sp{\prime\prime}}. \HOLConst{OBS_contracts} (\HOLBoundVar{E} \HOLSymConst{+} \HOLBoundVar{E\sp{\prime\prime}}) (\HOLBoundVar{E\sp{\prime}} \HOLSymConst{+} \HOLBoundVar{E\sp{\prime\prime}})
\end{SaveVerbatim}
\newcommand{\HOLContractionTheoremsOBSXXcontractsXXSUBSTXXSUMXXR}{\UseVerbatim{HOLContractionTheoremsOBSXXcontractsXXSUBSTXXSUMXXR}}
\begin{SaveVerbatim}{HOLContractionTheoremsOBSXXcontractsXXTRANS}
\HOLTokenTurnstile{} \HOLSymConst{\HOLTokenForall{}}\HOLBoundVar{E} \HOLBoundVar{E\sp{\prime}} \HOLBoundVar{E\sp{\prime\prime}}.
     \HOLConst{OBS_contracts} \HOLBoundVar{E} \HOLBoundVar{E\sp{\prime}} \HOLSymConst{\HOLTokenConj{}} \HOLConst{OBS_contracts} \HOLBoundVar{E\sp{\prime}} \HOLBoundVar{E\sp{\prime\prime}} \HOLSymConst{\HOLTokenImp{}}
     \HOLConst{OBS_contracts} \HOLBoundVar{E} \HOLBoundVar{E\sp{\prime\prime}}
\end{SaveVerbatim}
\newcommand{\HOLContractionTheoremsOBSXXcontractsXXTRANS}{\UseVerbatim{HOLContractionTheoremsOBSXXcontractsXXTRANS}}
\begin{SaveVerbatim}{HOLContractionTheoremsOBSXXcontractsXXTRANSXXLEFT}
\HOLTokenTurnstile{} \HOLSymConst{\HOLTokenForall{}}\HOLBoundVar{E} \HOLBoundVar{E\sp{\prime}}.
     \HOLConst{OBS_contracts} \HOLBoundVar{E} \HOLBoundVar{E\sp{\prime}} \HOLSymConst{\HOLTokenImp{}}
     \HOLSymConst{\HOLTokenForall{}}\HOLBoundVar{u} \HOLBoundVar{E\sb{\mathrm{1}}}. \HOLBoundVar{E} \HOLTokenTransBegin\HOLBoundVar{u}\HOLTokenTransEnd \HOLBoundVar{E\sb{\mathrm{1}}} \HOLSymConst{\HOLTokenImp{}} \HOLSymConst{\HOLTokenExists{}}\HOLBoundVar{E\sb{\mathrm{2}}}. \HOLBoundVar{E\sp{\prime}} \HOLTokenTransBegin\HOLBoundVar{u}\HOLTokenTransEnd \HOLBoundVar{E\sb{\mathrm{2}}} \HOLSymConst{\HOLTokenConj{}} \HOLBoundVar{E\sb{\mathrm{1}}} \HOLConst{contracts} \HOLBoundVar{E\sb{\mathrm{2}}}
\end{SaveVerbatim}
\newcommand{\HOLContractionTheoremsOBSXXcontractsXXTRANSXXLEFT}{\UseVerbatim{HOLContractionTheoremsOBSXXcontractsXXTRANSXXLEFT}}
\begin{SaveVerbatim}{HOLContractionTheoremsOBSXXcontractsXXTRANSXXRIGHT}
\HOLTokenTurnstile{} \HOLSymConst{\HOLTokenForall{}}\HOLBoundVar{E} \HOLBoundVar{E\sp{\prime}}.
     \HOLConst{OBS_contracts} \HOLBoundVar{E} \HOLBoundVar{E\sp{\prime}} \HOLSymConst{\HOLTokenImp{}}
     \HOLSymConst{\HOLTokenForall{}}\HOLBoundVar{u} \HOLBoundVar{E\sb{\mathrm{2}}}. \HOLBoundVar{E\sp{\prime}} \HOLTokenTransBegin\HOLBoundVar{u}\HOLTokenTransEnd \HOLBoundVar{E\sb{\mathrm{2}}} \HOLSymConst{\HOLTokenImp{}} \HOLSymConst{\HOLTokenExists{}}\HOLBoundVar{E\sb{\mathrm{1}}}. \HOLBoundVar{E} \HOLTokenWeakTransBegin\HOLBoundVar{u}\HOLTokenWeakTransEnd \HOLBoundVar{E\sb{\mathrm{1}}} \HOLSymConst{\HOLTokenConj{}} \HOLConst{WEAK_EQUIV} \HOLBoundVar{E\sb{\mathrm{1}}} \HOLBoundVar{E\sb{\mathrm{2}}}
\end{SaveVerbatim}
\newcommand{\HOLContractionTheoremsOBSXXcontractsXXTRANSXXRIGHT}{\UseVerbatim{HOLContractionTheoremsOBSXXcontractsXXTRANSXXRIGHT}}
\begin{SaveVerbatim}{HOLContractionTheoremsOBSXXcontractsXXWEAKXXTRANSYY}
\HOLTokenTurnstile{} \HOLSymConst{\HOLTokenForall{}}\HOLBoundVar{E} \HOLBoundVar{E\sp{\prime}}.
     \HOLConst{OBS_contracts} \HOLBoundVar{E} \HOLBoundVar{E\sp{\prime}} \HOLSymConst{\HOLTokenImp{}}
     \HOLSymConst{\HOLTokenForall{}}\HOLBoundVar{u} \HOLBoundVar{E\sb{\mathrm{2}}}. \HOLBoundVar{E\sp{\prime}} \HOLTokenWeakTransBegin\HOLBoundVar{u}\HOLTokenWeakTransEnd \HOLBoundVar{E\sb{\mathrm{2}}} \HOLSymConst{\HOLTokenImp{}} \HOLSymConst{\HOLTokenExists{}}\HOLBoundVar{E\sb{\mathrm{1}}}. \HOLBoundVar{E} \HOLTokenWeakTransBegin\HOLBoundVar{u}\HOLTokenWeakTransEnd \HOLBoundVar{E\sb{\mathrm{1}}} \HOLSymConst{\HOLTokenConj{}} \HOLConst{WEAK_EQUIV} \HOLBoundVar{E\sb{\mathrm{1}}} \HOLBoundVar{E\sb{\mathrm{2}}}
\end{SaveVerbatim}
\newcommand{\HOLContractionTheoremsOBSXXcontractsXXWEAKXXTRANSYY}{\UseVerbatim{HOLContractionTheoremsOBSXXcontractsXXWEAKXXTRANSYY}}
\begin{SaveVerbatim}{HOLContractionTheoremsOBSXXcontractsXXWEAKXXTRANSXXlabelYY}
\HOLTokenTurnstile{} \HOLSymConst{\HOLTokenForall{}}\HOLBoundVar{E} \HOLBoundVar{E\sp{\prime}}.
     \HOLConst{OBS_contracts} \HOLBoundVar{E} \HOLBoundVar{E\sp{\prime}} \HOLSymConst{\HOLTokenImp{}}
     \HOLSymConst{\HOLTokenForall{}}\HOLBoundVar{l} \HOLBoundVar{E\sb{\mathrm{2}}}.
       \HOLBoundVar{E\sp{\prime}} \HOLTokenWeakTransBegin\HOLConst{label} \HOLBoundVar{l}\HOLTokenWeakTransEnd \HOLBoundVar{E\sb{\mathrm{2}}} \HOLSymConst{\HOLTokenImp{}} \HOLSymConst{\HOLTokenExists{}}\HOLBoundVar{E\sb{\mathrm{1}}}. \HOLBoundVar{E} \HOLTokenWeakTransBegin\HOLConst{label} \HOLBoundVar{l}\HOLTokenWeakTransEnd \HOLBoundVar{E\sb{\mathrm{1}}} \HOLSymConst{\HOLTokenConj{}} \HOLConst{WEAK_EQUIV} \HOLBoundVar{E\sb{\mathrm{1}}} \HOLBoundVar{E\sb{\mathrm{2}}}
\end{SaveVerbatim}
\newcommand{\HOLContractionTheoremsOBSXXcontractsXXWEAKXXTRANSXXlabelYY}{\UseVerbatim{HOLContractionTheoremsOBSXXcontractsXXWEAKXXTRANSXXlabelYY}}
\begin{SaveVerbatim}{HOLContractionTheoremsSUMXXcontractsXXIMPXXOBSXXcontracts}
\HOLTokenTurnstile{} \HOLSymConst{\HOLTokenForall{}}\HOLBoundVar{p} \HOLBoundVar{q}.
     \HOLConst{free_action} \HOLBoundVar{p} \HOLSymConst{\HOLTokenConj{}} \HOLConst{free_action} \HOLBoundVar{q} \HOLSymConst{\HOLTokenImp{}}
     \HOLConst{SUM_contracts} \HOLBoundVar{p} \HOLBoundVar{q} \HOLSymConst{\HOLTokenImp{}}
     \HOLConst{OBS_contracts} \HOLBoundVar{p} \HOLBoundVar{q}
\end{SaveVerbatim}
\newcommand{\HOLContractionTheoremsSUMXXcontractsXXIMPXXOBSXXcontracts}{\UseVerbatim{HOLContractionTheoremsSUMXXcontractsXXIMPXXOBSXXcontracts}}
\newcommand{\HOLContractionTheorems}{
\HOLThmTag{Contraction}{C_contracts_IMP_SUM_contracts}\HOLContractionTheoremsCXXcontractsXXIMPXXSUMXXcontracts
\HOLThmTag{Contraction}{C_contracts_precongruence}\HOLContractionTheoremsCXXcontractsXXprecongruence
\HOLThmTag{Contraction}{C_contracts_thm}\HOLContractionTheoremsCXXcontractsXXthm
\HOLThmTag{Contraction}{COARSEST_PRECONGR_THM}\HOLContractionTheoremsCOARSESTXXPRECONGRXXTHM
\HOLThmTag{Contraction}{COARSEST_PRECONGR_THM'}\HOLContractionTheoremsCOARSESTXXPRECONGRXXTHMYY
\HOLThmTag{Contraction}{COMP_CONTRACTION}\HOLContractionTheoremsCOMPXXCONTRACTION
\HOLThmTag{Contraction}{CONTRACTION_EPS}\HOLContractionTheoremsCONTRACTIONXXEPS
\HOLThmTag{Contraction}{CONTRACTION_EPS'}\HOLContractionTheoremsCONTRACTIONXXEPSYY
\HOLThmTag{Contraction}{CONTRACTION_SUBSET_contracts}\HOLContractionTheoremsCONTRACTIONXXSUBSETXXcontracts
\HOLThmTag{Contraction}{CONTRACTION_WEAK_TRANS_label'}\HOLContractionTheoremsCONTRACTIONXXWEAKXXTRANSXXlabelYY
\HOLThmTag{Contraction}{contracts_AND_TRACE1}\HOLContractionTheoremscontractsXXANDXXTRACEOne
\HOLThmTag{Contraction}{contracts_AND_TRACE2}\HOLContractionTheoremscontractsXXANDXXTRACETwo
\HOLThmTag{Contraction}{contracts_AND_TRACE_label}\HOLContractionTheoremscontractsXXANDXXTRACEXXlabel
\HOLThmTag{Contraction}{contracts_AND_TRACE_tau}\HOLContractionTheoremscontractsXXANDXXTRACEXXtau
\HOLThmTag{Contraction}{contracts_cases}\HOLContractionTheoremscontractsXXcases
\HOLThmTag{Contraction}{contracts_coind}\HOLContractionTheoremscontractsXXcoind
\HOLThmTag{Contraction}{contracts_EPS}\HOLContractionTheoremscontractsXXEPS
\HOLThmTag{Contraction}{contracts_EPS'}\HOLContractionTheoremscontractsXXEPSYY
\HOLThmTag{Contraction}{contracts_IMP_WEAK_EQUIV}\HOLContractionTheoremscontractsXXIMPXXWEAKXXEQUIV
\HOLThmTag{Contraction}{contracts_is_CONTRACTION}\HOLContractionTheoremscontractsXXisXXCONTRACTION
\HOLThmTag{Contraction}{contracts_precongruence}\HOLContractionTheoremscontractsXXprecongruence
\HOLThmTag{Contraction}{contracts_PreOrder}\HOLContractionTheoremscontractsXXPreOrder
\HOLThmTag{Contraction}{contracts_PRESD_BY_GUARDED_SUM}\HOLContractionTheoremscontractsXXPRESDXXBYXXGUARDEDXXSUM
\HOLThmTag{Contraction}{contracts_PRESD_BY_PAR}\HOLContractionTheoremscontractsXXPRESDXXBYXXPAR
\HOLThmTag{Contraction}{contracts_PROP6}\HOLContractionTheoremscontractsXXPROPSix
\HOLThmTag{Contraction}{contracts_REFL}\HOLContractionTheoremscontractsXXREFL
\HOLThmTag{Contraction}{contracts_reflexive}\HOLContractionTheoremscontractsXXreflexive
\HOLThmTag{Contraction}{contracts_rules}\HOLContractionTheoremscontractsXXrules
\HOLThmTag{Contraction}{contracts_SUBST_GCONTEXT}\HOLContractionTheoremscontractsXXSUBSTXXGCONTEXT
\HOLThmTag{Contraction}{contracts_SUBST_PAR_L}\HOLContractionTheoremscontractsXXSUBSTXXPARXXL
\HOLThmTag{Contraction}{contracts_SUBST_PAR_R}\HOLContractionTheoremscontractsXXSUBSTXXPARXXR
\HOLThmTag{Contraction}{contracts_SUBST_PREFIX}\HOLContractionTheoremscontractsXXSUBSTXXPREFIX
\HOLThmTag{Contraction}{contracts_SUBST_RELAB}\HOLContractionTheoremscontractsXXSUBSTXXRELAB
\HOLThmTag{Contraction}{contracts_SUBST_RESTR}\HOLContractionTheoremscontractsXXSUBSTXXRESTR
\HOLThmTag{Contraction}{contracts_thm}\HOLContractionTheoremscontractsXXthm
\HOLThmTag{Contraction}{contracts_TRANS}\HOLContractionTheoremscontractsXXTRANS
\HOLThmTag{Contraction}{contracts_TRANS_label}\HOLContractionTheoremscontractsXXTRANSXXlabel
\HOLThmTag{Contraction}{contracts_TRANS_label'}\HOLContractionTheoremscontractsXXTRANSXXlabelYY
\HOLThmTag{Contraction}{contracts_TRANS_tau}\HOLContractionTheoremscontractsXXTRANSXXtau
\HOLThmTag{Contraction}{contracts_TRANS_tau'}\HOLContractionTheoremscontractsXXTRANSXXtauYY
\HOLThmTag{Contraction}{contracts_transitive}\HOLContractionTheoremscontractsXXtransitive
\HOLThmTag{Contraction}{contracts_WEAK_TRANS_label}\HOLContractionTheoremscontractsXXWEAKXXTRANSXXlabel
\HOLThmTag{Contraction}{contracts_WEAK_TRANS_label'}\HOLContractionTheoremscontractsXXWEAKXXTRANSXXlabelYY
\HOLThmTag{Contraction}{contracts_WEAK_TRANS_tau}\HOLContractionTheoremscontractsXXWEAKXXTRANSXXtau
\HOLThmTag{Contraction}{contracts_WEAK_TRANS_tau'}\HOLContractionTheoremscontractsXXWEAKXXTRANSXXtauYY
\HOLThmTag{Contraction}{expands_IMP_contracts}\HOLContractionTheoremsexpandsXXIMPXXcontracts
\HOLThmTag{Contraction}{EXPANSION_IMP_CONTRACTION}\HOLContractionTheoremsEXPANSIONXXIMPXXCONTRACTION
\HOLThmTag{Contraction}{IDENTITY_CONTRACTION}\HOLContractionTheoremsIDENTITYXXCONTRACTION
\HOLThmTag{Contraction}{OBS_contracts_AND_TRACE_label}\HOLContractionTheoremsOBSXXcontractsXXANDXXTRACEXXlabel
\HOLThmTag{Contraction}{OBS_contracts_AND_TRACE_tau}\HOLContractionTheoremsOBSXXcontractsXXANDXXTRACEXXtau
\HOLThmTag{Contraction}{OBS_contracts_BY_CONTRACTION}\HOLContractionTheoremsOBSXXcontractsXXBYXXCONTRACTION
\HOLThmTag{Contraction}{OBS_contracts_EPS'}\HOLContractionTheoremsOBSXXcontractsXXEPSYY
\HOLThmTag{Contraction}{OBS_contracts_IMP_C_contracts}\HOLContractionTheoremsOBSXXcontractsXXIMPXXCXXcontracts
\HOLThmTag{Contraction}{OBS_contracts_IMP_contracts}\HOLContractionTheoremsOBSXXcontractsXXIMPXXcontracts
\HOLThmTag{Contraction}{OBS_contracts_IMP_OBS_CONGR}\HOLContractionTheoremsOBSXXcontractsXXIMPXXOBSXXCONGR
\HOLThmTag{Contraction}{OBS_contracts_IMP_SUM_contracts}\HOLContractionTheoremsOBSXXcontractsXXIMPXXSUMXXcontracts
\HOLThmTag{Contraction}{OBS_contracts_IMP_WEAK_EQUIV}\HOLContractionTheoremsOBSXXcontractsXXIMPXXWEAKXXEQUIV
\HOLThmTag{Contraction}{OBS_contracts_IMP_WEAK_EQUIV'}\HOLContractionTheoremsOBSXXcontractsXXIMPXXWEAKXXEQUIVYY
\HOLThmTag{Contraction}{OBS_contracts_precongruence}\HOLContractionTheoremsOBSXXcontractsXXprecongruence
\HOLThmTag{Contraction}{OBS_contracts_PreOrder}\HOLContractionTheoremsOBSXXcontractsXXPreOrder
\HOLThmTag{Contraction}{OBS_contracts_PRESD_BY_PAR}\HOLContractionTheoremsOBSXXcontractsXXPRESDXXBYXXPAR
\HOLThmTag{Contraction}{OBS_contracts_PRESD_BY_SUM}\HOLContractionTheoremsOBSXXcontractsXXPRESDXXBYXXSUM
\HOLThmTag{Contraction}{OBS_contracts_REFL}\HOLContractionTheoremsOBSXXcontractsXXREFL
\HOLThmTag{Contraction}{OBS_contracts_SUBST_CONTEXT}\HOLContractionTheoremsOBSXXcontractsXXSUBSTXXCONTEXT
\HOLThmTag{Contraction}{OBS_contracts_SUBST_PAR_L}\HOLContractionTheoremsOBSXXcontractsXXSUBSTXXPARXXL
\HOLThmTag{Contraction}{OBS_contracts_SUBST_PAR_R}\HOLContractionTheoremsOBSXXcontractsXXSUBSTXXPARXXR
\HOLThmTag{Contraction}{OBS_contracts_SUBST_PREFIX}\HOLContractionTheoremsOBSXXcontractsXXSUBSTXXPREFIX
\HOLThmTag{Contraction}{OBS_contracts_SUBST_RELAB}\HOLContractionTheoremsOBSXXcontractsXXSUBSTXXRELAB
\HOLThmTag{Contraction}{OBS_contracts_SUBST_RESTR}\HOLContractionTheoremsOBSXXcontractsXXSUBSTXXRESTR
\HOLThmTag{Contraction}{OBS_contracts_SUBST_SUM_L}\HOLContractionTheoremsOBSXXcontractsXXSUBSTXXSUMXXL
\HOLThmTag{Contraction}{OBS_contracts_SUBST_SUM_R}\HOLContractionTheoremsOBSXXcontractsXXSUBSTXXSUMXXR
\HOLThmTag{Contraction}{OBS_contracts_TRANS}\HOLContractionTheoremsOBSXXcontractsXXTRANS
\HOLThmTag{Contraction}{OBS_contracts_TRANS_LEFT}\HOLContractionTheoremsOBSXXcontractsXXTRANSXXLEFT
\HOLThmTag{Contraction}{OBS_contracts_TRANS_RIGHT}\HOLContractionTheoremsOBSXXcontractsXXTRANSXXRIGHT
\HOLThmTag{Contraction}{OBS_contracts_WEAK_TRANS'}\HOLContractionTheoremsOBSXXcontractsXXWEAKXXTRANSYY
\HOLThmTag{Contraction}{OBS_contracts_WEAK_TRANS_label'}\HOLContractionTheoremsOBSXXcontractsXXWEAKXXTRANSXXlabelYY
\HOLThmTag{Contraction}{SUM_contracts_IMP_OBS_contracts}\HOLContractionTheoremsSUMXXcontractsXXIMPXXOBSXXcontracts
}

\newcommand{\HOLUniqueSolutionsDate}{02 Dicembre 2017}
\newcommand{\HOLUniqueSolutionsTime}{13:31}
\begin{SaveVerbatim}{HOLUniqueSolutionsTheoremsGSEQXXEPSXXlemma}
\HOLTokenTurnstile{} \HOLSymConst{\HOLTokenForall{}}\HOLBoundVar{E} \HOLBoundVar{P} \HOLBoundVar{Q} \HOLBoundVar{R} \HOLBoundVar{H}.
     \HOLConst{SG} \HOLBoundVar{E} \HOLSymConst{\HOLTokenConj{}} \HOLConst{GSEQ} \HOLBoundVar{E} \HOLSymConst{\HOLTokenConj{}} \HOLConst{WEAK_EQUIV} \HOLBoundVar{P} (\HOLBoundVar{E} \HOLBoundVar{P}) \HOLSymConst{\HOLTokenConj{}} \HOLConst{WEAK_EQUIV} \HOLBoundVar{Q} (\HOLBoundVar{E} \HOLBoundVar{Q}) \HOLSymConst{\HOLTokenConj{}}
     \HOLConst{GSEQ} \HOLBoundVar{H} \HOLSymConst{\HOLTokenConj{}} (\HOLBoundVar{R} \HOLSymConst{=} (\HOLTokenLambda{}\HOLBoundVar{x} \HOLBoundVar{y}. \HOLSymConst{\HOLTokenExists{}}\HOLBoundVar{H}. \HOLConst{GSEQ} \HOLBoundVar{H} \HOLSymConst{\HOLTokenConj{}} (\HOLBoundVar{x} \HOLSymConst{=} \HOLBoundVar{H} \HOLBoundVar{P}) \HOLSymConst{\HOLTokenConj{}} (\HOLBoundVar{y} \HOLSymConst{=} \HOLBoundVar{H} \HOLBoundVar{Q}))) \HOLSymConst{\HOLTokenImp{}}
     (\HOLSymConst{\HOLTokenForall{}}\HOLBoundVar{P\sp{\prime}}.
        \HOLConst{EPS} (\HOLBoundVar{H} \HOLBoundVar{P}) \HOLBoundVar{P\sp{\prime}} \HOLSymConst{\HOLTokenImp{}}
        \HOLSymConst{\HOLTokenExists{}}\HOLBoundVar{Q\sp{\prime}}.
          \HOLConst{EPS} (\HOLBoundVar{H} \HOLBoundVar{Q}) \HOLBoundVar{Q\sp{\prime}} \HOLSymConst{\HOLTokenConj{}} (\HOLConst{WEAK_EQUIV} \HOLConst{O} \HOLBoundVar{R} \HOLConst{O} \HOLConst{WEAK_EQUIV}) \HOLBoundVar{P\sp{\prime}} \HOLBoundVar{Q\sp{\prime}}) \HOLSymConst{\HOLTokenConj{}}
     \HOLSymConst{\HOLTokenForall{}}\HOLBoundVar{Q\sp{\prime}}.
       \HOLConst{EPS} (\HOLBoundVar{H} \HOLBoundVar{Q}) \HOLBoundVar{Q\sp{\prime}} \HOLSymConst{\HOLTokenImp{}}
       \HOLSymConst{\HOLTokenExists{}}\HOLBoundVar{P\sp{\prime}}. \HOLConst{EPS} (\HOLBoundVar{H} \HOLBoundVar{P}) \HOLBoundVar{P\sp{\prime}} \HOLSymConst{\HOLTokenConj{}} (\HOLConst{WEAK_EQUIV} \HOLConst{O} \HOLBoundVar{R} \HOLConst{O} \HOLConst{WEAK_EQUIV}) \HOLBoundVar{P\sp{\prime}} \HOLBoundVar{Q\sp{\prime}}
\end{SaveVerbatim}
\newcommand{\HOLUniqueSolutionsTheoremsGSEQXXEPSXXlemma}{\UseVerbatim{HOLUniqueSolutionsTheoremsGSEQXXEPSXXlemma}}
\begin{SaveVerbatim}{HOLUniqueSolutionsTheoremsOBSXXunfoldingXXlemmaOne}
\HOLTokenTurnstile{} \HOLSymConst{\HOLTokenForall{}}\HOLBoundVar{E} \HOLBoundVar{C} \HOLBoundVar{P}.
     \HOLConst{CONTEXT} \HOLBoundVar{E} \HOLSymConst{\HOLTokenConj{}} \HOLConst{CONTEXT} \HOLBoundVar{C} \HOLSymConst{\HOLTokenConj{}} \HOLConst{OBS_contracts} \HOLBoundVar{P} (\HOLBoundVar{E} \HOLBoundVar{P}) \HOLSymConst{\HOLTokenImp{}}
     \HOLSymConst{\HOLTokenForall{}}\HOLBoundVar{n}. \HOLConst{OBS_contracts} (\HOLBoundVar{C} \HOLBoundVar{P}) ((\HOLBoundVar{C} \HOLConst{\HOLTokenCompose} \HOLConst{FUNPOW} \HOLBoundVar{E} \HOLBoundVar{n}) \HOLBoundVar{P})
\end{SaveVerbatim}
\newcommand{\HOLUniqueSolutionsTheoremsOBSXXunfoldingXXlemmaOne}{\UseVerbatim{HOLUniqueSolutionsTheoremsOBSXXunfoldingXXlemmaOne}}
\begin{SaveVerbatim}{HOLUniqueSolutionsTheoremsOBSXXunfoldingXXlemmaTwo}
\HOLTokenTurnstile{} \HOLSymConst{\HOLTokenForall{}}\HOLBoundVar{E}.
     \HOLConst{WG} \HOLBoundVar{E} \HOLSymConst{\HOLTokenImp{}}
     \HOLSymConst{\HOLTokenForall{}}\HOLBoundVar{P} \HOLBoundVar{u} \HOLBoundVar{P\sp{\prime}}.
       \HOLBoundVar{E} \HOLBoundVar{P} \HOLTokenTransBegin\HOLBoundVar{u}\HOLTokenTransEnd \HOLBoundVar{P\sp{\prime}} \HOLSymConst{\HOLTokenImp{}}
       \HOLSymConst{\HOLTokenExists{}}\HOLBoundVar{C\sp{\prime}}. \HOLConst{CONTEXT} \HOLBoundVar{C\sp{\prime}} \HOLSymConst{\HOLTokenConj{}} (\HOLBoundVar{P\sp{\prime}} \HOLSymConst{=} \HOLBoundVar{C\sp{\prime}} \HOLBoundVar{P}) \HOLSymConst{\HOLTokenConj{}} \HOLSymConst{\HOLTokenForall{}}\HOLBoundVar{Q}. \HOLBoundVar{E} \HOLBoundVar{Q} \HOLTokenTransBegin\HOLBoundVar{u}\HOLTokenTransEnd \HOLBoundVar{C\sp{\prime}} \HOLBoundVar{Q}
\end{SaveVerbatim}
\newcommand{\HOLUniqueSolutionsTheoremsOBSXXunfoldingXXlemmaTwo}{\UseVerbatim{HOLUniqueSolutionsTheoremsOBSXXunfoldingXXlemmaTwo}}
\begin{SaveVerbatim}{HOLUniqueSolutionsTheoremsOBSXXunfoldingXXlemmaThree}
\HOLTokenTurnstile{} \HOLSymConst{\HOLTokenForall{}}\HOLBoundVar{C} \HOLBoundVar{E}.
     \HOLConst{CONTEXT} \HOLBoundVar{C} \HOLSymConst{\HOLTokenConj{}} \HOLConst{WG} \HOLBoundVar{E} \HOLSymConst{\HOLTokenImp{}}
     \HOLSymConst{\HOLTokenForall{}}\HOLBoundVar{P} \HOLBoundVar{x} \HOLBoundVar{P\sp{\prime}}.
       \HOLBoundVar{C} (\HOLBoundVar{E} \HOLBoundVar{P}) \HOLTokenTransBegin\HOLBoundVar{x}\HOLTokenTransEnd \HOLBoundVar{P\sp{\prime}} \HOLSymConst{\HOLTokenImp{}}
       \HOLSymConst{\HOLTokenExists{}}\HOLBoundVar{C\sp{\prime}}. \HOLConst{CONTEXT} \HOLBoundVar{C\sp{\prime}} \HOLSymConst{\HOLTokenConj{}} (\HOLBoundVar{P\sp{\prime}} \HOLSymConst{=} \HOLBoundVar{C\sp{\prime}} \HOLBoundVar{P}) \HOLSymConst{\HOLTokenConj{}} \HOLSymConst{\HOLTokenForall{}}\HOLBoundVar{Q}. \HOLBoundVar{C} (\HOLBoundVar{E} \HOLBoundVar{Q}) \HOLTokenTransBegin\HOLBoundVar{x}\HOLTokenTransEnd \HOLBoundVar{C\sp{\prime}} \HOLBoundVar{Q}
\end{SaveVerbatim}
\newcommand{\HOLUniqueSolutionsTheoremsOBSXXunfoldingXXlemmaThree}{\UseVerbatim{HOLUniqueSolutionsTheoremsOBSXXunfoldingXXlemmaThree}}
\begin{SaveVerbatim}{HOLUniqueSolutionsTheoremsOBSXXunfoldingXXlemmaFour}
\HOLTokenTurnstile{} \HOLSymConst{\HOLTokenForall{}}\HOLBoundVar{C} \HOLBoundVar{E} \HOLBoundVar{n} \HOLBoundVar{xs} \HOLBoundVar{P\sp{\prime}} \HOLBoundVar{P}.
     \HOLConst{CONTEXT} \HOLBoundVar{C} \HOLSymConst{\HOLTokenConj{}} \HOLConst{WG} \HOLBoundVar{E} \HOLSymConst{\HOLTokenConj{}} \HOLConst{TRACE} ((\HOLBoundVar{C} \HOLConst{\HOLTokenCompose} \HOLConst{FUNPOW} \HOLBoundVar{E} \HOLBoundVar{n}) \HOLBoundVar{P}) \HOLBoundVar{xs} \HOLBoundVar{P\sp{\prime}} \HOLSymConst{\HOLTokenConj{}}
     \HOLConst{LENGTH} \HOLBoundVar{xs} \HOLSymConst{\HOLTokenLeq{}} \HOLBoundVar{n} \HOLSymConst{\HOLTokenImp{}}
     \HOLSymConst{\HOLTokenExists{}}\HOLBoundVar{C\sp{\prime}}.
       \HOLConst{CONTEXT} \HOLBoundVar{C\sp{\prime}} \HOLSymConst{\HOLTokenConj{}} (\HOLBoundVar{P\sp{\prime}} \HOLSymConst{=} \HOLBoundVar{C\sp{\prime}} \HOLBoundVar{P}) \HOLSymConst{\HOLTokenConj{}}
       \HOLSymConst{\HOLTokenForall{}}\HOLBoundVar{Q}. \HOLConst{TRACE} ((\HOLBoundVar{C} \HOLConst{\HOLTokenCompose} \HOLConst{FUNPOW} \HOLBoundVar{E} \HOLBoundVar{n}) \HOLBoundVar{Q}) \HOLBoundVar{xs} (\HOLBoundVar{C\sp{\prime}} \HOLBoundVar{Q})
\end{SaveVerbatim}
\newcommand{\HOLUniqueSolutionsTheoremsOBSXXunfoldingXXlemmaFour}{\UseVerbatim{HOLUniqueSolutionsTheoremsOBSXXunfoldingXXlemmaFour}}
\begin{SaveVerbatim}{HOLUniqueSolutionsTheoremsOBSXXUNIQUEXXSOLUTIONS}
\HOLTokenTurnstile{} \HOLSymConst{\HOLTokenForall{}}\HOLBoundVar{E}.
     \HOLConst{SG} \HOLBoundVar{E} \HOLSymConst{\HOLTokenConj{}} \HOLConst{SEQ} \HOLBoundVar{E} \HOLSymConst{\HOLTokenImp{}}
     \HOLSymConst{\HOLTokenForall{}}\HOLBoundVar{P} \HOLBoundVar{Q}. \HOLConst{OBS_CONGR} \HOLBoundVar{P} (\HOLBoundVar{E} \HOLBoundVar{P}) \HOLSymConst{\HOLTokenConj{}} \HOLConst{OBS_CONGR} \HOLBoundVar{Q} (\HOLBoundVar{E} \HOLBoundVar{Q}) \HOLSymConst{\HOLTokenImp{}} \HOLConst{OBS_CONGR} \HOLBoundVar{P} \HOLBoundVar{Q}
\end{SaveVerbatim}
\newcommand{\HOLUniqueSolutionsTheoremsOBSXXUNIQUEXXSOLUTIONS}{\UseVerbatim{HOLUniqueSolutionsTheoremsOBSXXUNIQUEXXSOLUTIONS}}
\begin{SaveVerbatim}{HOLUniqueSolutionsTheoremsOBSXXUNIQUEXXSOLUTIONSXXLEMMA}
\HOLTokenTurnstile{} \HOLSymConst{\HOLTokenForall{}}\HOLBoundVar{G}.
     \HOLConst{SG} \HOLBoundVar{G} \HOLSymConst{\HOLTokenConj{}} \HOLConst{SEQ} \HOLBoundVar{G} \HOLSymConst{\HOLTokenImp{}}
     \HOLSymConst{\HOLTokenForall{}}\HOLBoundVar{P} \HOLBoundVar{a} \HOLBoundVar{P\sp{\prime}}.
       \HOLBoundVar{G} \HOLBoundVar{P} \HOLTokenTransBegin\HOLBoundVar{a}\HOLTokenTransEnd \HOLBoundVar{P\sp{\prime}} \HOLSymConst{\HOLTokenImp{}}
       \HOLSymConst{\HOLTokenExists{}}\HOLBoundVar{H}.
         \HOLConst{SEQ} \HOLBoundVar{H} \HOLSymConst{\HOLTokenConj{}} ((\HOLBoundVar{a} \HOLSymConst{=} \HOLConst{\ensuremath{\tau}}) \HOLSymConst{\HOLTokenImp{}} \HOLConst{SG} \HOLBoundVar{H}) \HOLSymConst{\HOLTokenConj{}} (\HOLBoundVar{P\sp{\prime}} \HOLSymConst{=} \HOLBoundVar{H} \HOLBoundVar{P}) \HOLSymConst{\HOLTokenConj{}} \HOLSymConst{\HOLTokenForall{}}\HOLBoundVar{Q}. \HOLBoundVar{G} \HOLBoundVar{Q} \HOLTokenTransBegin\HOLBoundVar{a}\HOLTokenTransEnd \HOLBoundVar{H} \HOLBoundVar{Q}
\end{SaveVerbatim}
\newcommand{\HOLUniqueSolutionsTheoremsOBSXXUNIQUEXXSOLUTIONSXXLEMMA}{\UseVerbatim{HOLUniqueSolutionsTheoremsOBSXXUNIQUEXXSOLUTIONSXXLEMMA}}
\begin{SaveVerbatim}{HOLUniqueSolutionsTheoremsOBSXXUNIQUEXXSOLUTIONSXXLEMMAXXEPS}
\HOLTokenTurnstile{} \HOLSymConst{\HOLTokenForall{}}\HOLBoundVar{G}.
     \HOLConst{SG} \HOLBoundVar{G} \HOLSymConst{\HOLTokenConj{}} \HOLConst{SEQ} \HOLBoundVar{G} \HOLSymConst{\HOLTokenImp{}}
     \HOLSymConst{\HOLTokenForall{}}\HOLBoundVar{P} \HOLBoundVar{P\sp{\prime}}.
       \HOLConst{EPS} (\HOLBoundVar{G} \HOLBoundVar{P}) \HOLBoundVar{P\sp{\prime}} \HOLSymConst{\HOLTokenImp{}}
       \HOLSymConst{\HOLTokenExists{}}\HOLBoundVar{H}. \HOLConst{SG} \HOLBoundVar{H} \HOLSymConst{\HOLTokenConj{}} \HOLConst{SEQ} \HOLBoundVar{H} \HOLSymConst{\HOLTokenConj{}} (\HOLBoundVar{P\sp{\prime}} \HOLSymConst{=} \HOLBoundVar{H} \HOLBoundVar{P}) \HOLSymConst{\HOLTokenConj{}} \HOLSymConst{\HOLTokenForall{}}\HOLBoundVar{Q}. \HOLConst{EPS} (\HOLBoundVar{G} \HOLBoundVar{Q}) (\HOLBoundVar{H} \HOLBoundVar{Q})
\end{SaveVerbatim}
\newcommand{\HOLUniqueSolutionsTheoremsOBSXXUNIQUEXXSOLUTIONSXXLEMMAXXEPS}{\UseVerbatim{HOLUniqueSolutionsTheoremsOBSXXUNIQUEXXSOLUTIONSXXLEMMAXXEPS}}
\begin{SaveVerbatim}{HOLUniqueSolutionsTheoremsSTRONGXXUNIQUEXXSOLUTIONS}
\HOLTokenTurnstile{} \HOLSymConst{\HOLTokenForall{}}\HOLBoundVar{E}.
     \HOLConst{WG} \HOLBoundVar{E} \HOLSymConst{\HOLTokenImp{}}
     \HOLSymConst{\HOLTokenForall{}}\HOLBoundVar{P} \HOLBoundVar{Q}.
       \HOLConst{STRONG_EQUIV} \HOLBoundVar{P} (\HOLBoundVar{E} \HOLBoundVar{P}) \HOLSymConst{\HOLTokenConj{}} \HOLConst{STRONG_EQUIV} \HOLBoundVar{Q} (\HOLBoundVar{E} \HOLBoundVar{Q}) \HOLSymConst{\HOLTokenImp{}}
       \HOLConst{STRONG_EQUIV} \HOLBoundVar{P} \HOLBoundVar{Q}
\end{SaveVerbatim}
\newcommand{\HOLUniqueSolutionsTheoremsSTRONGXXUNIQUEXXSOLUTIONS}{\UseVerbatim{HOLUniqueSolutionsTheoremsSTRONGXXUNIQUEXXSOLUTIONS}}
\begin{SaveVerbatim}{HOLUniqueSolutionsTheoremsSTRONGXXUNIQUEXXSOLUTIONSXXLEMMA}
\HOLTokenTurnstile{} \HOLSymConst{\HOLTokenForall{}}\HOLBoundVar{E}.
     \HOLConst{WG} \HOLBoundVar{E} \HOLSymConst{\HOLTokenImp{}}
     \HOLSymConst{\HOLTokenForall{}}\HOLBoundVar{P} \HOLBoundVar{a} \HOLBoundVar{P\sp{\prime}}.
       \HOLBoundVar{E} \HOLBoundVar{P} \HOLTokenTransBegin\HOLBoundVar{a}\HOLTokenTransEnd \HOLBoundVar{P\sp{\prime}} \HOLSymConst{\HOLTokenImp{}}
       \HOLSymConst{\HOLTokenExists{}}\HOLBoundVar{E\sp{\prime}}. \HOLConst{CONTEXT} \HOLBoundVar{E\sp{\prime}} \HOLSymConst{\HOLTokenConj{}} (\HOLBoundVar{P\sp{\prime}} \HOLSymConst{=} \HOLBoundVar{E\sp{\prime}} \HOLBoundVar{P}) \HOLSymConst{\HOLTokenConj{}} \HOLSymConst{\HOLTokenForall{}}\HOLBoundVar{Q}. \HOLBoundVar{E} \HOLBoundVar{Q} \HOLTokenTransBegin\HOLBoundVar{a}\HOLTokenTransEnd \HOLBoundVar{E\sp{\prime}} \HOLBoundVar{Q}
\end{SaveVerbatim}
\newcommand{\HOLUniqueSolutionsTheoremsSTRONGXXUNIQUEXXSOLUTIONSXXLEMMA}{\UseVerbatim{HOLUniqueSolutionsTheoremsSTRONGXXUNIQUEXXSOLUTIONSXXLEMMA}}
\begin{SaveVerbatim}{HOLUniqueSolutionsTheoremsunfoldingXXlemmaOne}
\HOLTokenTurnstile{} \HOLSymConst{\HOLTokenForall{}}\HOLBoundVar{E} \HOLBoundVar{C} \HOLBoundVar{P}.
     \HOLConst{GCONTEXT} \HOLBoundVar{E} \HOLSymConst{\HOLTokenConj{}} \HOLConst{GCONTEXT} \HOLBoundVar{C} \HOLSymConst{\HOLTokenConj{}} \HOLBoundVar{P} \HOLConst{contracts} \HOLBoundVar{E} \HOLBoundVar{P} \HOLSymConst{\HOLTokenImp{}}
     \HOLSymConst{\HOLTokenForall{}}\HOLBoundVar{n}. \HOLBoundVar{C} \HOLBoundVar{P} \HOLConst{contracts} (\HOLBoundVar{C} \HOLConst{\HOLTokenCompose} \HOLConst{FUNPOW} \HOLBoundVar{E} \HOLBoundVar{n}) \HOLBoundVar{P}
\end{SaveVerbatim}
\newcommand{\HOLUniqueSolutionsTheoremsunfoldingXXlemmaOne}{\UseVerbatim{HOLUniqueSolutionsTheoremsunfoldingXXlemmaOne}}
\begin{SaveVerbatim}{HOLUniqueSolutionsTheoremsunfoldingXXlemmaOneYY}
\HOLTokenTurnstile{} \HOLSymConst{\HOLTokenForall{}}\HOLBoundVar{E} \HOLBoundVar{C} \HOLBoundVar{P}.
     \HOLConst{GCONTEXT} \HOLBoundVar{E} \HOLSymConst{\HOLTokenConj{}} \HOLConst{GCONTEXT} \HOLBoundVar{C} \HOLSymConst{\HOLTokenConj{}} \HOLBoundVar{P} \HOLConst{expands} \HOLBoundVar{E} \HOLBoundVar{P} \HOLSymConst{\HOLTokenImp{}}
     \HOLSymConst{\HOLTokenForall{}}\HOLBoundVar{n}. \HOLBoundVar{C} \HOLBoundVar{P} \HOLConst{expands} (\HOLBoundVar{C} \HOLConst{\HOLTokenCompose} \HOLConst{FUNPOW} \HOLBoundVar{E} \HOLBoundVar{n}) \HOLBoundVar{P}
\end{SaveVerbatim}
\newcommand{\HOLUniqueSolutionsTheoremsunfoldingXXlemmaOneYY}{\UseVerbatim{HOLUniqueSolutionsTheoremsunfoldingXXlemmaOneYY}}
\begin{SaveVerbatim}{HOLUniqueSolutionsTheoremsunfoldingXXlemmaTwo}
\HOLTokenTurnstile{} \HOLSymConst{\HOLTokenForall{}}\HOLBoundVar{E}.
     \HOLConst{WGS} \HOLBoundVar{E} \HOLSymConst{\HOLTokenImp{}}
     \HOLSymConst{\HOLTokenForall{}}\HOLBoundVar{P} \HOLBoundVar{u} \HOLBoundVar{P\sp{\prime}}.
       \HOLBoundVar{E} \HOLBoundVar{P} \HOLTokenTransBegin\HOLBoundVar{u}\HOLTokenTransEnd \HOLBoundVar{P\sp{\prime}} \HOLSymConst{\HOLTokenImp{}}
       \HOLSymConst{\HOLTokenExists{}}\HOLBoundVar{C\sp{\prime}}. \HOLConst{GCONTEXT} \HOLBoundVar{C\sp{\prime}} \HOLSymConst{\HOLTokenConj{}} (\HOLBoundVar{P\sp{\prime}} \HOLSymConst{=} \HOLBoundVar{C\sp{\prime}} \HOLBoundVar{P}) \HOLSymConst{\HOLTokenConj{}} \HOLSymConst{\HOLTokenForall{}}\HOLBoundVar{Q}. \HOLBoundVar{E} \HOLBoundVar{Q} \HOLTokenTransBegin\HOLBoundVar{u}\HOLTokenTransEnd \HOLBoundVar{C\sp{\prime}} \HOLBoundVar{Q}
\end{SaveVerbatim}
\newcommand{\HOLUniqueSolutionsTheoremsunfoldingXXlemmaTwo}{\UseVerbatim{HOLUniqueSolutionsTheoremsunfoldingXXlemmaTwo}}
\begin{SaveVerbatim}{HOLUniqueSolutionsTheoremsunfoldingXXlemmaThree}
\HOLTokenTurnstile{} \HOLSymConst{\HOLTokenForall{}}\HOLBoundVar{C} \HOLBoundVar{E}.
     \HOLConst{GCONTEXT} \HOLBoundVar{C} \HOLSymConst{\HOLTokenConj{}} \HOLConst{WGS} \HOLBoundVar{E} \HOLSymConst{\HOLTokenImp{}}
     \HOLSymConst{\HOLTokenForall{}}\HOLBoundVar{P} \HOLBoundVar{x} \HOLBoundVar{P\sp{\prime}}.
       \HOLBoundVar{C} (\HOLBoundVar{E} \HOLBoundVar{P}) \HOLTokenTransBegin\HOLBoundVar{x}\HOLTokenTransEnd \HOLBoundVar{P\sp{\prime}} \HOLSymConst{\HOLTokenImp{}}
       \HOLSymConst{\HOLTokenExists{}}\HOLBoundVar{C\sp{\prime}}. \HOLConst{GCONTEXT} \HOLBoundVar{C\sp{\prime}} \HOLSymConst{\HOLTokenConj{}} (\HOLBoundVar{P\sp{\prime}} \HOLSymConst{=} \HOLBoundVar{C\sp{\prime}} \HOLBoundVar{P}) \HOLSymConst{\HOLTokenConj{}} \HOLSymConst{\HOLTokenForall{}}\HOLBoundVar{Q}. \HOLBoundVar{C} (\HOLBoundVar{E} \HOLBoundVar{Q}) \HOLTokenTransBegin\HOLBoundVar{x}\HOLTokenTransEnd \HOLBoundVar{C\sp{\prime}} \HOLBoundVar{Q}
\end{SaveVerbatim}
\newcommand{\HOLUniqueSolutionsTheoremsunfoldingXXlemmaThree}{\UseVerbatim{HOLUniqueSolutionsTheoremsunfoldingXXlemmaThree}}
\begin{SaveVerbatim}{HOLUniqueSolutionsTheoremsunfoldingXXlemmaFour}
\HOLTokenTurnstile{} \HOLSymConst{\HOLTokenForall{}}\HOLBoundVar{C} \HOLBoundVar{E} \HOLBoundVar{n} \HOLBoundVar{xs} \HOLBoundVar{P\sp{\prime}} \HOLBoundVar{P}.
     \HOLConst{GCONTEXT} \HOLBoundVar{C} \HOLSymConst{\HOLTokenConj{}} \HOLConst{WGS} \HOLBoundVar{E} \HOLSymConst{\HOLTokenConj{}} \HOLConst{TRACE} ((\HOLBoundVar{C} \HOLConst{\HOLTokenCompose} \HOLConst{FUNPOW} \HOLBoundVar{E} \HOLBoundVar{n}) \HOLBoundVar{P}) \HOLBoundVar{xs} \HOLBoundVar{P\sp{\prime}} \HOLSymConst{\HOLTokenConj{}}
     \HOLConst{LENGTH} \HOLBoundVar{xs} \HOLSymConst{\HOLTokenLeq{}} \HOLBoundVar{n} \HOLSymConst{\HOLTokenImp{}}
     \HOLSymConst{\HOLTokenExists{}}\HOLBoundVar{C\sp{\prime}}.
       \HOLConst{GCONTEXT} \HOLBoundVar{C\sp{\prime}} \HOLSymConst{\HOLTokenConj{}} (\HOLBoundVar{P\sp{\prime}} \HOLSymConst{=} \HOLBoundVar{C\sp{\prime}} \HOLBoundVar{P}) \HOLSymConst{\HOLTokenConj{}}
       \HOLSymConst{\HOLTokenForall{}}\HOLBoundVar{Q}. \HOLConst{TRACE} ((\HOLBoundVar{C} \HOLConst{\HOLTokenCompose} \HOLConst{FUNPOW} \HOLBoundVar{E} \HOLBoundVar{n}) \HOLBoundVar{Q}) \HOLBoundVar{xs} (\HOLBoundVar{C\sp{\prime}} \HOLBoundVar{Q})
\end{SaveVerbatim}
\newcommand{\HOLUniqueSolutionsTheoremsunfoldingXXlemmaFour}{\UseVerbatim{HOLUniqueSolutionsTheoremsunfoldingXXlemmaFour}}
\begin{SaveVerbatim}{HOLUniqueSolutionsTheoremsUNIQUEXXSOLUTIONSXXOFXXCONTRACTIONS}
\HOLTokenTurnstile{} \HOLSymConst{\HOLTokenForall{}}\HOLBoundVar{E}.
     \HOLConst{WGS} \HOLBoundVar{E} \HOLSymConst{\HOLTokenImp{}}
     \HOLSymConst{\HOLTokenForall{}}\HOLBoundVar{P} \HOLBoundVar{Q}. \HOLBoundVar{P} \HOLConst{contracts} \HOLBoundVar{E} \HOLBoundVar{P} \HOLSymConst{\HOLTokenConj{}} \HOLBoundVar{Q} \HOLConst{contracts} \HOLBoundVar{E} \HOLBoundVar{Q} \HOLSymConst{\HOLTokenImp{}} \HOLConst{WEAK_EQUIV} \HOLBoundVar{P} \HOLBoundVar{Q}
\end{SaveVerbatim}
\newcommand{\HOLUniqueSolutionsTheoremsUNIQUEXXSOLUTIONSXXOFXXCONTRACTIONS}{\UseVerbatim{HOLUniqueSolutionsTheoremsUNIQUEXXSOLUTIONSXXOFXXCONTRACTIONS}}
\begin{SaveVerbatim}{HOLUniqueSolutionsTheoremsUNIQUEXXSOLUTIONSXXOFXXCONTRACTIONSXXLEMMA}
\HOLTokenTurnstile{} \HOLSymConst{\HOLTokenForall{}}\HOLBoundVar{P} \HOLBoundVar{Q}.
     (\HOLSymConst{\HOLTokenExists{}}\HOLBoundVar{E}. \HOLConst{WGS} \HOLBoundVar{E} \HOLSymConst{\HOLTokenConj{}} \HOLBoundVar{P} \HOLConst{contracts} \HOLBoundVar{E} \HOLBoundVar{P} \HOLSymConst{\HOLTokenConj{}} \HOLBoundVar{Q} \HOLConst{contracts} \HOLBoundVar{E} \HOLBoundVar{Q}) \HOLSymConst{\HOLTokenImp{}}
     \HOLSymConst{\HOLTokenForall{}}\HOLBoundVar{C}.
       \HOLConst{GCONTEXT} \HOLBoundVar{C} \HOLSymConst{\HOLTokenImp{}}
       (\HOLSymConst{\HOLTokenForall{}}\HOLBoundVar{l} \HOLBoundVar{R}.
          \HOLBoundVar{C} \HOLBoundVar{P} \HOLTokenWeakTransBegin\HOLConst{label} \HOLBoundVar{l}\HOLTokenWeakTransEnd \HOLBoundVar{R} \HOLSymConst{\HOLTokenImp{}}
          \HOLSymConst{\HOLTokenExists{}}\HOLBoundVar{C\sp{\prime}}.
            \HOLConst{GCONTEXT} \HOLBoundVar{C\sp{\prime}} \HOLSymConst{\HOLTokenConj{}} \HOLBoundVar{R} \HOLConst{contracts} \HOLBoundVar{C\sp{\prime}} \HOLBoundVar{P} \HOLSymConst{\HOLTokenConj{}}
            (\HOLConst{WEAK_EQUIV} \HOLConst{O} (\HOLTokenLambda{}\HOLBoundVar{x} \HOLBoundVar{y}. \HOLBoundVar{x} \HOLTokenWeakTransBegin\HOLConst{label} \HOLBoundVar{l}\HOLTokenWeakTransEnd \HOLBoundVar{y})) (\HOLBoundVar{C} \HOLBoundVar{Q}) (\HOLBoundVar{C\sp{\prime}} \HOLBoundVar{Q})) \HOLSymConst{\HOLTokenConj{}}
       \HOLSymConst{\HOLTokenForall{}}\HOLBoundVar{R}.
         \HOLBoundVar{C} \HOLBoundVar{P} \HOLTokenWeakTransBegin\HOLConst{\ensuremath{\tau}}\HOLTokenWeakTransEnd \HOLBoundVar{R} \HOLSymConst{\HOLTokenImp{}}
         \HOLSymConst{\HOLTokenExists{}}\HOLBoundVar{C\sp{\prime}}.
           \HOLConst{GCONTEXT} \HOLBoundVar{C\sp{\prime}} \HOLSymConst{\HOLTokenConj{}} \HOLBoundVar{R} \HOLConst{contracts} \HOLBoundVar{C\sp{\prime}} \HOLBoundVar{P} \HOLSymConst{\HOLTokenConj{}}
           (\HOLConst{WEAK_EQUIV} \HOLConst{O} \HOLConst{EPS}) (\HOLBoundVar{C} \HOLBoundVar{Q}) (\HOLBoundVar{C\sp{\prime}} \HOLBoundVar{Q})
\end{SaveVerbatim}
\newcommand{\HOLUniqueSolutionsTheoremsUNIQUEXXSOLUTIONSXXOFXXCONTRACTIONSXXLEMMA}{\UseVerbatim{HOLUniqueSolutionsTheoremsUNIQUEXXSOLUTIONSXXOFXXCONTRACTIONSXXLEMMA}}
\begin{SaveVerbatim}{HOLUniqueSolutionsTheoremsUNIQUEXXSOLUTIONSXXOFXXEXPANSIONS}
\HOLTokenTurnstile{} \HOLSymConst{\HOLTokenForall{}}\HOLBoundVar{E}.
     \HOLConst{WGS} \HOLBoundVar{E} \HOLSymConst{\HOLTokenImp{}}
     \HOLSymConst{\HOLTokenForall{}}\HOLBoundVar{P} \HOLBoundVar{Q}. \HOLBoundVar{P} \HOLConst{expands} \HOLBoundVar{E} \HOLBoundVar{P} \HOLSymConst{\HOLTokenConj{}} \HOLBoundVar{Q} \HOLConst{expands} \HOLBoundVar{E} \HOLBoundVar{Q} \HOLSymConst{\HOLTokenImp{}} \HOLConst{WEAK_EQUIV} \HOLBoundVar{P} \HOLBoundVar{Q}
\end{SaveVerbatim}
\newcommand{\HOLUniqueSolutionsTheoremsUNIQUEXXSOLUTIONSXXOFXXEXPANSIONS}{\UseVerbatim{HOLUniqueSolutionsTheoremsUNIQUEXXSOLUTIONSXXOFXXEXPANSIONS}}
\begin{SaveVerbatim}{HOLUniqueSolutionsTheoremsUNIQUEXXSOLUTIONSXXOFXXEXPANSIONSYY}
\HOLTokenTurnstile{} \HOLSymConst{\HOLTokenForall{}}\HOLBoundVar{E}.
     \HOLConst{WGS} \HOLBoundVar{E} \HOLSymConst{\HOLTokenImp{}}
     \HOLSymConst{\HOLTokenForall{}}\HOLBoundVar{P} \HOLBoundVar{Q}. \HOLBoundVar{P} \HOLConst{expands} \HOLBoundVar{E} \HOLBoundVar{P} \HOLSymConst{\HOLTokenConj{}} \HOLBoundVar{Q} \HOLConst{expands} \HOLBoundVar{E} \HOLBoundVar{Q} \HOLSymConst{\HOLTokenImp{}} \HOLConst{WEAK_EQUIV} \HOLBoundVar{P} \HOLBoundVar{Q}
\end{SaveVerbatim}
\newcommand{\HOLUniqueSolutionsTheoremsUNIQUEXXSOLUTIONSXXOFXXEXPANSIONSYY}{\UseVerbatim{HOLUniqueSolutionsTheoremsUNIQUEXXSOLUTIONSXXOFXXEXPANSIONSYY}}
\begin{SaveVerbatim}{HOLUniqueSolutionsTheoremsUNIQUEXXSOLUTIONSXXOFXXEXPANSIONSXXLEMMA}
\HOLTokenTurnstile{} \HOLSymConst{\HOLTokenForall{}}\HOLBoundVar{P} \HOLBoundVar{Q}.
     (\HOLSymConst{\HOLTokenExists{}}\HOLBoundVar{E}. \HOLConst{WGS} \HOLBoundVar{E} \HOLSymConst{\HOLTokenConj{}} \HOLBoundVar{P} \HOLConst{expands} \HOLBoundVar{E} \HOLBoundVar{P} \HOLSymConst{\HOLTokenConj{}} \HOLBoundVar{Q} \HOLConst{expands} \HOLBoundVar{E} \HOLBoundVar{Q}) \HOLSymConst{\HOLTokenImp{}}
     \HOLSymConst{\HOLTokenForall{}}\HOLBoundVar{C}.
       \HOLConst{GCONTEXT} \HOLBoundVar{C} \HOLSymConst{\HOLTokenImp{}}
       (\HOLSymConst{\HOLTokenForall{}}\HOLBoundVar{l} \HOLBoundVar{R}.
          \HOLBoundVar{C} \HOLBoundVar{P} \HOLTokenWeakTransBegin\HOLConst{label} \HOLBoundVar{l}\HOLTokenWeakTransEnd \HOLBoundVar{R} \HOLSymConst{\HOLTokenImp{}}
          \HOLSymConst{\HOLTokenExists{}}\HOLBoundVar{C\sp{\prime}}.
            \HOLConst{GCONTEXT} \HOLBoundVar{C\sp{\prime}} \HOLSymConst{\HOLTokenConj{}} \HOLBoundVar{R} \HOLConst{expands} \HOLBoundVar{C\sp{\prime}} \HOLBoundVar{P} \HOLSymConst{\HOLTokenConj{}}
            (\HOLConst{WEAK_EQUIV} \HOLConst{O} (\HOLTokenLambda{}\HOLBoundVar{x} \HOLBoundVar{y}. \HOLBoundVar{x} \HOLTokenWeakTransBegin\HOLConst{label} \HOLBoundVar{l}\HOLTokenWeakTransEnd \HOLBoundVar{y})) (\HOLBoundVar{C} \HOLBoundVar{Q}) (\HOLBoundVar{C\sp{\prime}} \HOLBoundVar{Q})) \HOLSymConst{\HOLTokenConj{}}
       \HOLSymConst{\HOLTokenForall{}}\HOLBoundVar{R}.
         \HOLBoundVar{C} \HOLBoundVar{P} \HOLTokenWeakTransBegin\HOLConst{\ensuremath{\tau}}\HOLTokenWeakTransEnd \HOLBoundVar{R} \HOLSymConst{\HOLTokenImp{}}
         \HOLSymConst{\HOLTokenExists{}}\HOLBoundVar{C\sp{\prime}}.
           \HOLConst{GCONTEXT} \HOLBoundVar{C\sp{\prime}} \HOLSymConst{\HOLTokenConj{}} \HOLBoundVar{R} \HOLConst{expands} \HOLBoundVar{C\sp{\prime}} \HOLBoundVar{P} \HOLSymConst{\HOLTokenConj{}}
           (\HOLConst{WEAK_EQUIV} \HOLConst{O} \HOLConst{EPS}) (\HOLBoundVar{C} \HOLBoundVar{Q}) (\HOLBoundVar{C\sp{\prime}} \HOLBoundVar{Q})
\end{SaveVerbatim}
\newcommand{\HOLUniqueSolutionsTheoremsUNIQUEXXSOLUTIONSXXOFXXEXPANSIONSXXLEMMA}{\UseVerbatim{HOLUniqueSolutionsTheoremsUNIQUEXXSOLUTIONSXXOFXXEXPANSIONSXXLEMMA}}
\begin{SaveVerbatim}{HOLUniqueSolutionsTheoremsUNIQUEXXSOLUTIONSXXOFXXOBSXXCONTRACTIONS}
\HOLTokenTurnstile{} \HOLSymConst{\HOLTokenForall{}}\HOLBoundVar{E}.
     \HOLConst{WG} \HOLBoundVar{E} \HOLSymConst{\HOLTokenImp{}}
     \HOLSymConst{\HOLTokenForall{}}\HOLBoundVar{P} \HOLBoundVar{Q}.
       \HOLConst{OBS_contracts} \HOLBoundVar{P} (\HOLBoundVar{E} \HOLBoundVar{P}) \HOLSymConst{\HOLTokenConj{}} \HOLConst{OBS_contracts} \HOLBoundVar{Q} (\HOLBoundVar{E} \HOLBoundVar{Q}) \HOLSymConst{\HOLTokenImp{}}
       \HOLConst{WEAK_EQUIV} \HOLBoundVar{P} \HOLBoundVar{Q}
\end{SaveVerbatim}
\newcommand{\HOLUniqueSolutionsTheoremsUNIQUEXXSOLUTIONSXXOFXXOBSXXCONTRACTIONS}{\UseVerbatim{HOLUniqueSolutionsTheoremsUNIQUEXXSOLUTIONSXXOFXXOBSXXCONTRACTIONS}}
\begin{SaveVerbatim}{HOLUniqueSolutionsTheoremsUNIQUEXXSOLUTIONSXXOFXXOBSXXCONTRACTIONSYY}
\HOLTokenTurnstile{} \HOLSymConst{\HOLTokenForall{}}\HOLBoundVar{E}.
     \HOLConst{WGS} \HOLBoundVar{E} \HOLSymConst{\HOLTokenImp{}}
     \HOLSymConst{\HOLTokenForall{}}\HOLBoundVar{P} \HOLBoundVar{Q}.
       \HOLConst{OBS_contracts} \HOLBoundVar{P} (\HOLBoundVar{E} \HOLBoundVar{P}) \HOLSymConst{\HOLTokenConj{}} \HOLConst{OBS_contracts} \HOLBoundVar{Q} (\HOLBoundVar{E} \HOLBoundVar{Q}) \HOLSymConst{\HOLTokenImp{}}
       \HOLConst{WEAK_EQUIV} \HOLBoundVar{P} \HOLBoundVar{Q}
\end{SaveVerbatim}
\newcommand{\HOLUniqueSolutionsTheoremsUNIQUEXXSOLUTIONSXXOFXXOBSXXCONTRACTIONSYY}{\UseVerbatim{HOLUniqueSolutionsTheoremsUNIQUEXXSOLUTIONSXXOFXXOBSXXCONTRACTIONSYY}}
\begin{SaveVerbatim}{HOLUniqueSolutionsTheoremsUNIQUEXXSOLUTIONSXXOFXXOBSXXCONTRACTIONSXXLEMMA}
\HOLTokenTurnstile{} \HOLSymConst{\HOLTokenForall{}}\HOLBoundVar{P} \HOLBoundVar{Q}.
     (\HOLSymConst{\HOLTokenExists{}}\HOLBoundVar{E}.
        \HOLConst{WG} \HOLBoundVar{E} \HOLSymConst{\HOLTokenConj{}} \HOLConst{OBS_contracts} \HOLBoundVar{P} (\HOLBoundVar{E} \HOLBoundVar{P}) \HOLSymConst{\HOLTokenConj{}} \HOLConst{OBS_contracts} \HOLBoundVar{Q} (\HOLBoundVar{E} \HOLBoundVar{Q})) \HOLSymConst{\HOLTokenImp{}}
     \HOLSymConst{\HOLTokenForall{}}\HOLBoundVar{C}.
       \HOLConst{CONTEXT} \HOLBoundVar{C} \HOLSymConst{\HOLTokenImp{}}
       (\HOLSymConst{\HOLTokenForall{}}\HOLBoundVar{l} \HOLBoundVar{R}.
          \HOLBoundVar{C} \HOLBoundVar{P} \HOLTokenWeakTransBegin\HOLConst{label} \HOLBoundVar{l}\HOLTokenWeakTransEnd \HOLBoundVar{R} \HOLSymConst{\HOLTokenImp{}}
          \HOLSymConst{\HOLTokenExists{}}\HOLBoundVar{C\sp{\prime}}.
            \HOLConst{CONTEXT} \HOLBoundVar{C\sp{\prime}} \HOLSymConst{\HOLTokenConj{}} \HOLBoundVar{R} \HOLConst{contracts} \HOLBoundVar{C\sp{\prime}} \HOLBoundVar{P} \HOLSymConst{\HOLTokenConj{}}
            (\HOLConst{WEAK_EQUIV} \HOLConst{O} (\HOLTokenLambda{}\HOLBoundVar{x} \HOLBoundVar{y}. \HOLBoundVar{x} \HOLTokenWeakTransBegin\HOLConst{label} \HOLBoundVar{l}\HOLTokenWeakTransEnd \HOLBoundVar{y})) (\HOLBoundVar{C} \HOLBoundVar{Q}) (\HOLBoundVar{C\sp{\prime}} \HOLBoundVar{Q})) \HOLSymConst{\HOLTokenConj{}}
       \HOLSymConst{\HOLTokenForall{}}\HOLBoundVar{R}.
         \HOLBoundVar{C} \HOLBoundVar{P} \HOLTokenWeakTransBegin\HOLConst{\ensuremath{\tau}}\HOLTokenWeakTransEnd \HOLBoundVar{R} \HOLSymConst{\HOLTokenImp{}}
         \HOLSymConst{\HOLTokenExists{}}\HOLBoundVar{C\sp{\prime}}.
           \HOLConst{CONTEXT} \HOLBoundVar{C\sp{\prime}} \HOLSymConst{\HOLTokenConj{}} \HOLBoundVar{R} \HOLConst{contracts} \HOLBoundVar{C\sp{\prime}} \HOLBoundVar{P} \HOLSymConst{\HOLTokenConj{}}
           (\HOLConst{WEAK_EQUIV} \HOLConst{O} \HOLConst{EPS}) (\HOLBoundVar{C} \HOLBoundVar{Q}) (\HOLBoundVar{C\sp{\prime}} \HOLBoundVar{Q})
\end{SaveVerbatim}
\newcommand{\HOLUniqueSolutionsTheoremsUNIQUEXXSOLUTIONSXXOFXXOBSXXCONTRACTIONSXXLEMMA}{\UseVerbatim{HOLUniqueSolutionsTheoremsUNIQUEXXSOLUTIONSXXOFXXOBSXXCONTRACTIONSXXLEMMA}}
\begin{SaveVerbatim}{HOLUniqueSolutionsTheoremsWEAKXXUNIQUEXXSOLUTIONS}
\HOLTokenTurnstile{} \HOLSymConst{\HOLTokenForall{}}\HOLBoundVar{E}.
     \HOLConst{SG} \HOLBoundVar{E} \HOLSymConst{\HOLTokenConj{}} \HOLConst{GSEQ} \HOLBoundVar{E} \HOLSymConst{\HOLTokenImp{}}
     \HOLSymConst{\HOLTokenForall{}}\HOLBoundVar{P} \HOLBoundVar{Q}.
       \HOLConst{WEAK_EQUIV} \HOLBoundVar{P} (\HOLBoundVar{E} \HOLBoundVar{P}) \HOLSymConst{\HOLTokenConj{}} \HOLConst{WEAK_EQUIV} \HOLBoundVar{Q} (\HOLBoundVar{E} \HOLBoundVar{Q}) \HOLSymConst{\HOLTokenImp{}} \HOLConst{WEAK_EQUIV} \HOLBoundVar{P} \HOLBoundVar{Q}
\end{SaveVerbatim}
\newcommand{\HOLUniqueSolutionsTheoremsWEAKXXUNIQUEXXSOLUTIONS}{\UseVerbatim{HOLUniqueSolutionsTheoremsWEAKXXUNIQUEXXSOLUTIONS}}
\begin{SaveVerbatim}{HOLUniqueSolutionsTheoremsWEAKXXUNIQUEXXSOLUTIONSXXLEMMA}
\HOLTokenTurnstile{} \HOLSymConst{\HOLTokenForall{}}\HOLBoundVar{G}.
     \HOLConst{SG} \HOLBoundVar{G} \HOLSymConst{\HOLTokenConj{}} \HOLConst{GSEQ} \HOLBoundVar{G} \HOLSymConst{\HOLTokenImp{}}
     \HOLSymConst{\HOLTokenForall{}}\HOLBoundVar{P} \HOLBoundVar{a} \HOLBoundVar{P\sp{\prime}}.
       \HOLBoundVar{G} \HOLBoundVar{P} \HOLTokenTransBegin\HOLBoundVar{a}\HOLTokenTransEnd \HOLBoundVar{P\sp{\prime}} \HOLSymConst{\HOLTokenImp{}}
       \HOLSymConst{\HOLTokenExists{}}\HOLBoundVar{H}.
         \HOLConst{GSEQ} \HOLBoundVar{H} \HOLSymConst{\HOLTokenConj{}} ((\HOLBoundVar{a} \HOLSymConst{=} \HOLConst{\ensuremath{\tau}}) \HOLSymConst{\HOLTokenImp{}} \HOLConst{SG} \HOLBoundVar{H}) \HOLSymConst{\HOLTokenConj{}} (\HOLBoundVar{P\sp{\prime}} \HOLSymConst{=} \HOLBoundVar{H} \HOLBoundVar{P}) \HOLSymConst{\HOLTokenConj{}}
         \HOLSymConst{\HOLTokenForall{}}\HOLBoundVar{Q}. \HOLBoundVar{G} \HOLBoundVar{Q} \HOLTokenTransBegin\HOLBoundVar{a}\HOLTokenTransEnd \HOLBoundVar{H} \HOLBoundVar{Q}
\end{SaveVerbatim}
\newcommand{\HOLUniqueSolutionsTheoremsWEAKXXUNIQUEXXSOLUTIONSXXLEMMA}{\UseVerbatim{HOLUniqueSolutionsTheoremsWEAKXXUNIQUEXXSOLUTIONSXXLEMMA}}
\begin{SaveVerbatim}{HOLUniqueSolutionsTheoremsWEAKXXUNIQUEXXSOLUTIONSXXLEMMAXXEPS}
\HOLTokenTurnstile{} \HOLSymConst{\HOLTokenForall{}}\HOLBoundVar{G}.
     \HOLConst{SG} \HOLBoundVar{G} \HOLSymConst{\HOLTokenConj{}} \HOLConst{GSEQ} \HOLBoundVar{G} \HOLSymConst{\HOLTokenImp{}}
     \HOLSymConst{\HOLTokenForall{}}\HOLBoundVar{P} \HOLBoundVar{P\sp{\prime}}.
       \HOLConst{EPS} (\HOLBoundVar{G} \HOLBoundVar{P}) \HOLBoundVar{P\sp{\prime}} \HOLSymConst{\HOLTokenImp{}}
       \HOLSymConst{\HOLTokenExists{}}\HOLBoundVar{H}. \HOLConst{SG} \HOLBoundVar{H} \HOLSymConst{\HOLTokenConj{}} \HOLConst{GSEQ} \HOLBoundVar{H} \HOLSymConst{\HOLTokenConj{}} (\HOLBoundVar{P\sp{\prime}} \HOLSymConst{=} \HOLBoundVar{H} \HOLBoundVar{P}) \HOLSymConst{\HOLTokenConj{}} \HOLSymConst{\HOLTokenForall{}}\HOLBoundVar{Q}. \HOLConst{EPS} (\HOLBoundVar{G} \HOLBoundVar{Q}) (\HOLBoundVar{H} \HOLBoundVar{Q})
\end{SaveVerbatim}
\newcommand{\HOLUniqueSolutionsTheoremsWEAKXXUNIQUEXXSOLUTIONSXXLEMMAXXEPS}{\UseVerbatim{HOLUniqueSolutionsTheoremsWEAKXXUNIQUEXXSOLUTIONSXXLEMMAXXEPS}}
\newcommand{\HOLUniqueSolutionsTheorems}{
\HOLThmTag{UniqueSolutions}{GSEQ_EPS_lemma}\HOLUniqueSolutionsTheoremsGSEQXXEPSXXlemma
\HOLThmTag{UniqueSolutions}{OBS_unfolding_lemma1}\HOLUniqueSolutionsTheoremsOBSXXunfoldingXXlemmaOne
\HOLThmTag{UniqueSolutions}{OBS_unfolding_lemma2}\HOLUniqueSolutionsTheoremsOBSXXunfoldingXXlemmaTwo
\HOLThmTag{UniqueSolutions}{OBS_unfolding_lemma3}\HOLUniqueSolutionsTheoremsOBSXXunfoldingXXlemmaThree
\HOLThmTag{UniqueSolutions}{OBS_unfolding_lemma4}\HOLUniqueSolutionsTheoremsOBSXXunfoldingXXlemmaFour
\HOLThmTag{UniqueSolutions}{OBS_UNIQUE_SOLUTIONS}\HOLUniqueSolutionsTheoremsOBSXXUNIQUEXXSOLUTIONS
\HOLThmTag{UniqueSolutions}{OBS_UNIQUE_SOLUTIONS_LEMMA}\HOLUniqueSolutionsTheoremsOBSXXUNIQUEXXSOLUTIONSXXLEMMA
\HOLThmTag{UniqueSolutions}{OBS_UNIQUE_SOLUTIONS_LEMMA_EPS}\HOLUniqueSolutionsTheoremsOBSXXUNIQUEXXSOLUTIONSXXLEMMAXXEPS
\HOLThmTag{UniqueSolutions}{STRONG_UNIQUE_SOLUTIONS}\HOLUniqueSolutionsTheoremsSTRONGXXUNIQUEXXSOLUTIONS
\HOLThmTag{UniqueSolutions}{STRONG_UNIQUE_SOLUTIONS_LEMMA}\HOLUniqueSolutionsTheoremsSTRONGXXUNIQUEXXSOLUTIONSXXLEMMA
\HOLThmTag{UniqueSolutions}{unfolding_lemma1}\HOLUniqueSolutionsTheoremsunfoldingXXlemmaOne
\HOLThmTag{UniqueSolutions}{unfolding_lemma1'}\HOLUniqueSolutionsTheoremsunfoldingXXlemmaOneYY
\HOLThmTag{UniqueSolutions}{unfolding_lemma2}\HOLUniqueSolutionsTheoremsunfoldingXXlemmaTwo
\HOLThmTag{UniqueSolutions}{unfolding_lemma3}\HOLUniqueSolutionsTheoremsunfoldingXXlemmaThree
\HOLThmTag{UniqueSolutions}{unfolding_lemma4}\HOLUniqueSolutionsTheoremsunfoldingXXlemmaFour
\HOLThmTag{UniqueSolutions}{UNIQUE_SOLUTIONS_OF_CONTRACTIONS}\HOLUniqueSolutionsTheoremsUNIQUEXXSOLUTIONSXXOFXXCONTRACTIONS
\HOLThmTag{UniqueSolutions}{UNIQUE_SOLUTIONS_OF_CONTRACTIONS_LEMMA}\HOLUniqueSolutionsTheoremsUNIQUEXXSOLUTIONSXXOFXXCONTRACTIONSXXLEMMA
\HOLThmTag{UniqueSolutions}{UNIQUE_SOLUTIONS_OF_EXPANSIONS}\HOLUniqueSolutionsTheoremsUNIQUEXXSOLUTIONSXXOFXXEXPANSIONS
\HOLThmTag{UniqueSolutions}{UNIQUE_SOLUTIONS_OF_EXPANSIONS'}\HOLUniqueSolutionsTheoremsUNIQUEXXSOLUTIONSXXOFXXEXPANSIONSYY
\HOLThmTag{UniqueSolutions}{UNIQUE_SOLUTIONS_OF_EXPANSIONS_LEMMA}\HOLUniqueSolutionsTheoremsUNIQUEXXSOLUTIONSXXOFXXEXPANSIONSXXLEMMA
\HOLThmTag{UniqueSolutions}{UNIQUE_SOLUTIONS_OF_OBS_CONTRACTIONS}\HOLUniqueSolutionsTheoremsUNIQUEXXSOLUTIONSXXOFXXOBSXXCONTRACTIONS
\HOLThmTag{UniqueSolutions}{UNIQUE_SOLUTIONS_OF_OBS_CONTRACTIONS'}\HOLUniqueSolutionsTheoremsUNIQUEXXSOLUTIONSXXOFXXOBSXXCONTRACTIONSYY
\HOLThmTag{UniqueSolutions}{UNIQUE_SOLUTIONS_OF_OBS_CONTRACTIONS_LEMMA}\HOLUniqueSolutionsTheoremsUNIQUEXXSOLUTIONSXXOFXXOBSXXCONTRACTIONSXXLEMMA
\HOLThmTag{UniqueSolutions}{WEAK_UNIQUE_SOLUTIONS}\HOLUniqueSolutionsTheoremsWEAKXXUNIQUEXXSOLUTIONS
\HOLThmTag{UniqueSolutions}{WEAK_UNIQUE_SOLUTIONS_LEMMA}\HOLUniqueSolutionsTheoremsWEAKXXUNIQUEXXSOLUTIONSXXLEMMA
\HOLThmTag{UniqueSolutions}{WEAK_UNIQUE_SOLUTIONS_LEMMA_EPS}\HOLUniqueSolutionsTheoremsWEAKXXUNIQUEXXSOLUTIONSXXLEMMAXXEPS
}

%%%% the following contents should be manually updated from references.tex
% ::::::::::::::::::::::::::::::::::::::::::::::::::::::::::::::::::::::::::
\section{CCS Theory}
\index{CCS Theory@\textbf  {CCS Theory}}
\begin{flushleft}
\textbf{Built:} \HOLCCSDate \\[2pt]
\textbf{Parent Theories:} real
\end{flushleft}
% ::::::::::::::::::::::::::::::::::::::::::::::::::::::::::::::::::::::::::

\subsection{Datatypes}
\index{CCS Theory@\textbf  {CCS Theory}!Datatypes}
% .....................................

\HOLCCSDatatypes

\subsection{Definitions}
\index{CCS Theory@\textbf  {CCS Theory}!Definitions}
% .....................................

\HOLCCSDefinitions

\subsection{Theorems}
\index{CCS Theory@\textbf  {CCS Theory}!Theorems}
% .....................................

\HOLCCSTheorems

% ::::::::::::::::::::::::::::::::::::::::::::::::::::::::::::::::::::::::::
\section{StrongEQ Theory}
\index{StrongEQ Theory@\textbf  {StrongEQ Theory}}
\begin{flushleft}
\textbf{Built:} \HOLStrongEQDate \\[2pt]
\textbf{Parent Theories:} CCS
\end{flushleft}
% ::::::::::::::::::::::::::::::::::::::::::::::::::::::::::::::::::::::::::

% No datatypes

\subsection{Definitions}
\index{StrongEQ Theory@\textbf  {StrongEQ Theory}!Definitions}
% .....................................

\HOLStrongEQDefinitions

\subsection{Theorems}
\index{StrongEQ Theory@\textbf  {StrongEQ Theory}!Theorems}
% .....................................

\HOLStrongEQTheorems

% ::::::::::::::::::::::::::::::::::::::::::::::::::::::::::::::::::::::::::
\section{StrongLaws Theory}
\index{StrongLaws Theory@\textbf  {StrongLaws Theory}}
\begin{flushleft}
\textbf{Built:} \HOLStrongLawsDate \\[2pt]
\textbf{Parent Theories:} StrongEQ
\end{flushleft}
% ::::::::::::::::::::::::::::::::::::::::::::::::::::::::::::::::::::::::::

% No datatypes

\subsection{Definitions}
\index{StrongLaws Theory@\textbf  {StrongLaws Theory}!Definitions}
% .....................................

\HOLStrongLawsDefinitions

\subsection{Theorems}
\index{StrongLaws Theory@\textbf  {StrongLaws Theory}!Theorems}
% .....................................

\HOLStrongLawsTheorems

% ::::::::::::::::::::::::::::::::::::::::::::::::::::::::::::::::::::::::::
\section{WeakEQ Theory}
\index{WeakEQ Theory@\textbf  {WeakEQ Theory}}
\begin{flushleft}
\textbf{Built:} \HOLWeakEQDate \\[2pt]
\textbf{Parent Theories:} StrongEQ
\end{flushleft}
% ::::::::::::::::::::::::::::::::::::::::::::::::::::::::::::::::::::::::::

% No datatypes

\subsection{Definitions}
\index{WeakEQ Theory@\textbf  {WeakEQ Theory}!Definitions}
% .....................................

\HOLWeakEQDefinitions

\subsection{Theorems}
\index{WeakEQ Theory@\textbf  {WeakEQ Theory}!Theorems}
% .....................................

\HOLWeakEQTheorems

% ::::::::::::::::::::::::::::::::::::::::::::::::::::::::::::::::::::::::::
\section{WeakLaws Theory}
\index{WeakLaws Theory@\textbf  {WeakLaws Theory}}
\begin{flushleft}
\textbf{Built:} \HOLWeakLawsDate \\[2pt]
\textbf{Parent Theories:} WeakEQ, StrongLaws
\end{flushleft}
% ::::::::::::::::::::::::::::::::::::::::::::::::::::::::::::::::::::::::::

% No datatypes

% No definitions

\subsection{Theorems}
\index{WeakLaws Theory@\textbf  {WeakLaws Theory}!Theorems}
% .....................................

\HOLWeakLawsTheorems

% ::::::::::::::::::::::::::::::::::::::::::::::::::::::::::::::::::::::::::
\section{ObsCongr Theory}
\index{ObsCongr Theory@\textbf  {ObsCongr Theory}}
\begin{flushleft}
\textbf{Built:} \HOLObsCongrDate \\[2pt]
\textbf{Parent Theories:} WeakLaws
\end{flushleft}
% ::::::::::::::::::::::::::::::::::::::::::::::::::::::::::::::::::::::::::

% No datatypes

\subsection{Definitions}
\index{ObsCongr Theory@\textbf  {ObsCongr Theory}!Definitions}
% .....................................

\HOLObsCongrDefinitions

\subsection{Theorems}
\index{ObsCongr Theory@\textbf  {ObsCongr Theory}!Theorems}
% .....................................

\HOLObsCongrTheorems

% ::::::::::::::::::::::::::::::::::::::::::::::::::::::::::::::::::::::::::
\section{ObsCongrLaws Theory}
\index{ObsCongrLaws Theory@\textbf  {ObsCongrLaws Theory}}
\begin{flushleft}
\textbf{Built:} \HOLObsCongrLawsDate \\[2pt]
\textbf{Parent Theories:} ObsCongr
\end{flushleft}
% ::::::::::::::::::::::::::::::::::::::::::::::::::::::::::::::::::::::::::

% No datatypes

% No definitions

\subsection{Theorems}
\index{ObsCongrLaws Theory@\textbf  {ObsCongrLaws Theory}!Theorems}
% .....................................

\HOLObsCongrLawsTheorems

% ::::::::::::::::::::::::::::::::::::::::::::::::::::::::::::::::::::::::::
\section{Congruence Theory}
\index{Congruence Theory@\textbf  {Congruence Theory}}
\begin{flushleft}
\textbf{Built:} \HOLCongruenceDate \\[2pt]
\textbf{Parent Theories:} ObsCongrLaws, string
\end{flushleft}
% ::::::::::::::::::::::::::::::::::::::::::::::::::::::::::::::::::::::::::

% No datatypes

\subsection{Definitions}
\index{Congruence Theory@\textbf  {Congruence Theory}!Definitions}
% .....................................

\HOLCongruenceDefinitions

\subsection{Theorems}
\index{Congruence Theory@\textbf  {Congruence Theory}!Theorems}
% .....................................

\HOLCongruenceTheorems

% ::::::::::::::::::::::::::::::::::::::::::::::::::::::::::::::::::::::::::
\section{CoarsestCongr Theory}
\index{CoarsestCongr Theory@\textbf  {CoarsestCongr Theory}}
\begin{flushleft}
\textbf{Built:} \HOLCoarsestCongrDate \\[2pt]
\textbf{Parent Theories:} Trace, Congruence
\end{flushleft}
% ::::::::::::::::::::::::::::::::::::::::::::::::::::::::::::::::::::::::::

% No datatypes

\subsection{Definitions}
\index{CoarsestCongr Theory@\textbf  {CoarsestCongr Theory}!Definitions}
% .....................................

\HOLCoarsestCongrDefinitions

\subsection{Theorems}
\index{CoarsestCongr Theory@\textbf  {CoarsestCongr Theory}!Theorems}
% .....................................

\HOLCoarsestCongrTheorems

% ::::::::::::::::::::::::::::::::::::::::::::::::::::::::::::::::::::::::::
\section{BisimulationUpto Theory}
\index{BisimulationUpto Theory@\textbf  {BisimulationUpto Theory}}
\begin{flushleft}
\textbf{Built:} \HOLBisimulationUptoDate \\[2pt]
\textbf{Parent Theories:} Congruence
\end{flushleft}
% ::::::::::::::::::::::::::::::::::::::::::::::::::::::::::::::::::::::::::

% No datatypes

\subsection{Definitions}
\index{BisimulationUpto Theory@\textbf  {BisimulationUpto Theory}!Definitions}
% .....................................

\HOLBisimulationUptoDefinitions

\subsection{Theorems}
\index{BisimulationUpto Theory@\textbf  {BisimulationUpto Theory}!Theorems}
% .....................................

\HOLBisimulationUptoTheorems

% ::::::::::::::::::::::::::::::::::::::::::::::::::::::::::::::::::::::::::
\section{Trace Theory}
\index{Trace Theory@\textbf  {Trace Theory}}
\begin{flushleft}
\textbf{Built:} \HOLTraceDate \\[2pt]
\textbf{Parent Theories:} WeakEQ
\end{flushleft}
% ::::::::::::::::::::::::::::::::::::::::::::::::::::::::::::::::::::::::::

% No datatypes

\subsection{Definitions}
\index{Trace Theory@\textbf  {Trace Theory}!Definitions}
% .....................................

\HOLTraceDefinitions

\subsection{Theorems}
\index{Trace Theory@\textbf  {Trace Theory}!Theorems}
% .....................................

\HOLTraceTheorems

% ::::::::::::::::::::::::::::::::::::::::::::::::::::::::::::::::::::::::::
\section{Expansion Theory}
\index{Expansion Theory@\textbf  {Expansion Theory}}
\begin{flushleft}
\textbf{Built:} \HOLExpansionDate \\[2pt]
\textbf{Parent Theories:} Trace, Congruence
\end{flushleft}
% ::::::::::::::::::::::::::::::::::::::::::::::::::::::::::::::::::::::::::

% No datatypes

\subsection{Definitions}
\index{Expansion Theory@\textbf  {Expansion Theory}!Definitions}
% .....................................

\HOLExpansionDefinitions

\subsection{Theorems}
\index{Expansion Theory@\textbf  {Expansion Theory}!Theorems}
% .....................................

\HOLExpansionTheorems

% ::::::::::::::::::::::::::::::::::::::::::::::::::::::::::::::::::::::::::
\section{Contraction Theory}
\index{Contraction Theory@\textbf  {Contraction Theory}}
\begin{flushleft}
\textbf{Built:} \HOLContractionDate \\[2pt]
\textbf{Parent Theories:} Expansion
\end{flushleft}
% ::::::::::::::::::::::::::::::::::::::::::::::::::::::::::::::::::::::::::

% No datatypes

\subsection{Definitions}
\index{Contraction Theory@\textbf  {Contraction Theory}!Definitions}
% .....................................

\HOLContractionDefinitions

\subsection{Theorems}
\index{Contraction Theory@\textbf  {Contraction Theory}!Theorems}
% .....................................

\HOLContractionTheorems

% ::::::::::::::::::::::::::::::::::::::::::::::::::::::::::::::::::::::::::
\section{UniqueSolutions Theory}
\index{UniqueSolutions Theory@\textbf  {UniqueSolutions Theory}}
\begin{flushleft}
\textbf{Built:} \HOLUniqueSolutionsDate \\[2pt]
\textbf{Parent Theories:} Contraction, BisimulationUpto
\end{flushleft}
% ::::::::::::::::::::::::::::::::::::::::::::::::::::::::::::::::::::::::::

% No datatypes

% No definitions

\subsection{Theorems}
\index{UniqueSolutions Theory@\textbf  {UniqueSolutions Theory}!Theorems}
% .....................................

\HOLUniqueSolutionsTheorems

\bibliography{thesis}{}
\bibliographystyle{splncs}

\HOLindex
\end{document}